\documentclass[]{aa}                 % (traditional abstract)
\usepackage{amsmath,amsfonts,amsbsy,mathrsfs,amssymb}
\usepackage{graphicx}
\usepackage[T1]{fontenc}
\usepackage{lmodern}
\usepackage[utf8]{inputenc}
\usepackage[english]{babel}
\usepackage{lscape}
\usepackage{supertabular,booktabs}
\usepackage{subcaption}
\usepackage{float}
\usepackage[flushleft]{threeparttable}
\usepackage{array}
\usepackage{longtable}
\usepackage[table]{xcolor}
\usepackage{rotating}
\usepackage{tablefootnote} % for table footnotes
\usepackage{afterpage}
\usepackage{chngcntr}
\usepackage{needspace}
\usepackage{placeins}
\usepackage[pdftex=true, colorlinks=true, linkcolor=blue, citecolor=blue, pdfborder={0 0 0}, bookmarks=false, urlcolor=blue, breaklinks=true, pdftitle={The physical and chemical structure of Sagittarius B2, VIII. Full molecular line survey of hot cores}, pdfsubject={}, pdfauthor={}, pdfkeywords={}]{hyperref}

\usepackage{array}
\newsavebox\dummy
\newcolumntype{H}{>{\begin{lrbox}{\dummy}}c<{\end{lrbox}}@{}}

\usepackage{natbib,twoopt}
\bibpunct{(}{)}{;}{a}{}{,}          % to follow the A&A style

\makeatletter
\newcommandtwoopt{\citeads}[3][][]{\href{http://adsabs.harvard.edu/abs/#3}%
{\def\hyper@linkstart##1##2{}%
\let\hyper@linkend\@empty\citealp[#1][#2]{#3}}}
\newcommandtwoopt{\citepads}[3][][]{\href{http://adsabs.harvard.edu/abs/#3}%
{\def\hyper@linkstart##1##2{}%
\let\hyper@linkend\@empty\citep[#1][#2]{#3}}}
\newcommandtwoopt{\citetads}[3][][]{\href{http://adsabs.harvard.edu/abs/#3}%
{\def\hyper@linkstart##1##2{}%
\let\hyper@linkend\@empty\citet[#1][#2]{#3}}}
\newcommandtwoopt{\citeyearads}[3][][]%
{\href{http://adsabs.harvard.edu/abs/#3}
{\def\hyper@linkstart##1##2{}%
\let\hyper@linkend\@empty\citeyear[#1][#2]{#3}}}
\makeatother

\newcolumntype{C}[1]{>{\centering\arraybackslash}p{#1}}
\newcolumntype{R}[1]{>{\flushright\arraybackslash}p{#1}}

\newcommand{\kms}     {km~s$^{-1}$}
\newcommand{\gcm}     {g~cm$^{-2}$}
\newcommand{\hii}     {\ion{H}{ii}}
\newcommand{\mo}      {$M_\odot$}
\newcommand{\lo}      {$L_\odot$}
\newcommand{\farcd}   {.\!\!^{\circ}}
\definecolor{lightgray}{gray}{0.9}
\definecolor{darkgray}{gray}{0.7}

\begin{document}
    \title{The physical and chemical structure of Sagittarius B2\\
           VIII. Full molecular line survey of hot cores}

% \subtitle{}

    \author{T.~M\"{o}ller\inst{1}
            \and
            P.~Schilke\inst{1}
            \and
            \'{A}.~S\'{a}nchez-Monge\inst{2,3}
            \and
            A.~Schmiedeke\inst{4}
            }

    \institute{I. Physikalisches Institut, Universit\"{a}t zu K\"{o}ln,
               Z\"{u}lpicher Str. 77, D-50937 K\"{o}ln, Germany\\
               \email{moeller@ph1.uni-koeln.de}
               \and
               Institut de Ci\`encies de l'Espai (ICE, CSIC), Campus UAB, Carrer de Can Magrans s/n, 08193, Bellaterra (Barcelona), Spain
               \and
               Institut d'Estudis Espacials de Catalunya (IEEC), 08860, Castelldefels (Barcelona), Spain
               \and
               Green Bank Observatory, 155 Observatory Rd,
               Green Bank, WV 24944 (USA)
              }

    \date{Received 26 June 2024 / Accepted 20 November 2024}

    %___________________________________________________________________________
    % \abstract {context} {aims} {methods} {results} {conclusion}
    \abstract
    %
    % context
    {The giant molecular cloud complex Sagittarius B2 (Sgr~B2) in the central molecular zone of our Galaxy hosts several high-mass star formation sites, with Sgr~B2(M) and Sgr~B2(N) being the main centers of activity. This analysis aims to comprehensively model each core spectrum, considering molecular lines, dust attenuation, and free-free emission interactions. We describe the molecular content analysis of each hot core and identify the chemical composition of detected sources.}
    % aims
    {Using ALMA's high sensitivity, we aim to characterize the hot core population in Sgr~B2(M) and N, gaining a better understanding of the different evolutionary phases of star formation processes in this complex.}
    % methods
    {We conducted an unbiased ALMA spectral line survey of 47 sources in band 6 (211-275 GHz). Chemical composition and column densities were derived using XCLASS, assuming local thermodynamic equilibrium. Quantitative descriptions for each molecule were determined, considering all emission and absorption features across the spectral range. Temperature and velocity distributions were analyzed, and derived abundances were compared with other spectral line surveys.}
    % results
    {We identified 65 isotopologs from 41 different molecules, ranging from light molecules to complex organic compounds, originating from various environments. Most sources in the Sgr~B2 complex were assigned different evolutionary phases of high-mass star formation.}
    % conclusion
    {Sgr~B2(N) hot cores show more complex molecules such as CH$_3$OH, CH$_3$OCHO, and CH$_3$OCH$_3$, while M cores contain lighter molecules such as SO$_2$, SO, and NO. Some sulfur-bearing molecules are more abundant in N than in M. The derived molecular abundances can be used for comparison and to constrain astrochemical models. Inner sources in both regions were generally more developed than outer sources, with some exceptions.}

    % highest summed abundances:
    % M:  SO$_2$, SO, CH$_3$CN, NO
    % N:  CH$_3$OH, CH$_3$CN, H$_2$CCO, CH$_3$OCHO
    % evolutionary phases:
    % M: 2 x V, 7 x IV, 3 x III, 3 x II, 11 x I, 1 x ?
    % N: 2 x V, 2 x IV, 4 x III, 1 x II,  9 x I, 2 x ?

    \keywords{ISM: clouds – dust, extinction – evolution – ISM: molecules – ISM: individual objects: Sagittarius B2(M) – ISM: individual objects: Sagittarius B2(N)}

    \titlerunning{Full molecular line survey of hot cores}
    \authorrunning{T.~M\"{o}ller \textit{et al.}}

    \maketitle

%===============================================================================
% Introduction
\section{Introduction}\label{sec:Introduction}

% Sgr~B2
With a mass of 10$^7$~\mo~and H$_2$ densities of $10^3$ -- $10^5$~cm$^{-3}$ (\citetads{2016A&A...588A.143S}, \citetads{1995A&A...294..667H}, \citetads{1989ApJ...337..704L}), Sagittarius B2 (Sgr~B2) is one of the most massive molecular clouds in our Galaxy. Sgr~B2 is part of the central molecular zone \citepads[CMZ,][]{2023ASPC..534...83H}). It is located at a distance\footnote{In order to be consistent with previous models we use a distance of 8.5~kpc.} of $8.178 \pm 0.013_{\rm stat.} \pm 0.022_{\rm sys.}$~kpc \citepads{2019A&A...625L..10G} and is situated at a projected distance of 107~pc from Sgr~A$^{*}$, the compact radio source associated with the supermassive black hole in the Galactic center.

% Sgr~B2(M) and Sgr~B2(N)
The Sgr~B2 complex contains two main sites of active high-mass star formation, Sgr~B2 Main (M) and North (N), which are separated by $\sim$48$\arcsec$ ($\sim$1.9~pc in projection). With comparable luminosities of $2 - 10 \times 10^6$~\lo, masses of $5 \times 10^4$~\mo,~and sizes of $\sim$0.5~pc \citepads[see][]{2016A&A...588A.143S} these two sites are located at the center of an envelope with a radius of 2~pc, which contains at least $\sim$70 high-mass stars with spectral types in the range from O5 to B0 (see e.g., \citetads{1995ApJ...449..663G}, \citetads{2014ApJ...781L..36D}). All of this together is embedded in another envelope with a radius of 20~pc, which contains more than 99~\% of the total mass of Sgr~B2. Compared to Sgr~B2(N), Sgr~B2(M) shows a higher degree of fragmentation, has a higher luminosity, and contains more ultracompact \hbox{H\,{\sc ii}} regions (see e.g., \citetads{1992ApJ...389..338G}, \citetads{2011A&A...530L...9Q}, \citetads{2017A&A...604A...6S}, \citetads{2019A&A...628A...6S}, \citetads{2019A&A...630A..73M}), which suggests a more evolved stage and larger amount of feedback. Moreover, previous studies show that M is very rich in sulfur-bearing molecules, while N is dominated by organics (\citetads{1991ApJS...77..255S}, \citetads{1998ApJS..117..427N}, \citetads{2004ApJ...600..234F}, \citetads{2013A&A...559A..47B}, \citetads{2014ApJ...789....8N}). The complex and multilayered structures of high-mass proto-clusters such as Sgr~B2 are difficult to study. The high density of molecular lines and the continuum emission detected toward the two main sites indicate the presence of a large amount of material to form new stars. Spectral line studies provide the opportunity to gain a better insight into their thermal excitation conditions and dynamics by examining line intensities and profiles, allowing for the separation of different physical components and identification of chemical patterns. Additionally, line surveys offer the possibility to study the distribution of exciting sources by looking at vibrationally excited molecules. Since IR radiation is needed to excite them \citepads[e.g.,][]{2010A&A...515A..71C}, they should delineate the exact position of the exciting sources, which are otherwise difficult to see due to extinction even at millimeter wavelengths.

% Why do we analyze another spectral line survey?
Although Sgr~B2 has been the target of numerous spectral line surveys before (see e.g., \citetads{1986ApJS...60..819C}, \citetads{1989ApJS...70..539T}, \citetads{1991ApJS...77..255S}, \citetads{1998ApJS..117..427N}, \citetads{2004ApJ...600..234F}, \citetads{2013A&A...559A..47B}, \citetads{2014ApJ...789....8N},  \citetads{2016A&A...587A..91B},  \citetads{2019A&A...628A..10B}, \citetads{2021A&A...651A...9M}), the unique capabilities of the Atacama Large Millimeter/submillimeter Array (ALMA) now offer further insight into the process of star formation.
%
% something about hot cores
Hot cores have been identified as birthplaces of high-mass stars where the central protostar(s) heats up the surrounding dense envelope. These regions are compact (diameters $\leq 0.1$~pc), dense ($n \geq 10^7$~cm$^{-3}$), hot ($T \geq 100$~K), and dark ($A_v\geq 100$~mag) molecular cloud cores \citepads[see e.g., ]{2004IAUS..221...59V}. The high temperatures of the hot cores lead to the sublimation of icy dust mantles, which are the main sites for the formation of complex organic molecules. Additionally, the molecular line emission in hot cores provides powerful constraints on both the physical and chemical conditions in these regions (\citetads{2000prpl.conf..299K}, \citetads{2007A&A...465..913B}). Although many hot cores show similarities in their molecular content, they possess a great diversity in terms of relative molecular abundances (see e.g., \citetads{1993dca..book...37W}, \citetads{2007A&A...465..913B}, \citetads{2013A&A...559A..47B}, \citetads{2018ApJ...864..102M}). Furthermore, hot cores are often associated with outflows, infall, and rotation. In later stages of the hot cores, ultracompact \hii~regions (UC\hii) are formed, where the protostars ionize their surrounding envelope.

In this paper, which continues our series of studies on Sgr~B2 (\citetads{2016A&A...588A.143S}, \citetads{2017A&A...604A...6S}, \citetads{2018A&A...614A.123P}, \citetads{2019A&A...628A...6S}, \citetads{2019A&A...630A..73M}, \citetads{2022A&A...666A..31M}, \citetads{2023A&A...676A.121M}), we describe the full analysis of broadband spectral line surveys of hot cores, identified by \citetads{2017A&A...604A...6S} in Sgr~B2(M) and N, to characterize the hot core population in the Sgr~B2 complex, where we take the complex interactions between molecular lines, dust attenuation, and free-free emission arising from \hbox{H\,{\sc ii}} regions into account. In the first half of our analysis, described in the seventh paper of our series on the analysis of the Sgr~B2 complex, Paper~VII \citepads{2023A&A...676A.121M}, we have quantified the dust and, if contained, the free-free contributions to the continuum levels of each core, and we have derived the corresponding parameters not only for each core but also for their local surrounding envelope, and determined their physical properties. In this paper, we describe the analysis of the molecular content of each hot core and identify the chemical composition of the detected sources.

% structure of paper
This paper is laid out as follows. We start with Sect.~\ref{sec:DataRed}, where we describe the observations and outline the data reduction procedure, followed by Sect.~\ref{sec:DataAnalysis} presenting the modeling methodology used to analyze the dataset. In Sect.~\ref{sec:Results}, in which we present the results of our analysis and then discuss them in Sect.~\ref{sec:Discussion}. We end with a summary and conclusions in Sect.~\ref{sec:Conclusions}.

%*******************************************************************************
% Figure: Hot cores in Sgr B2(M)
\begin{figure*}[ht!]
   \centering
   \includegraphics[width=0.86\textwidth]{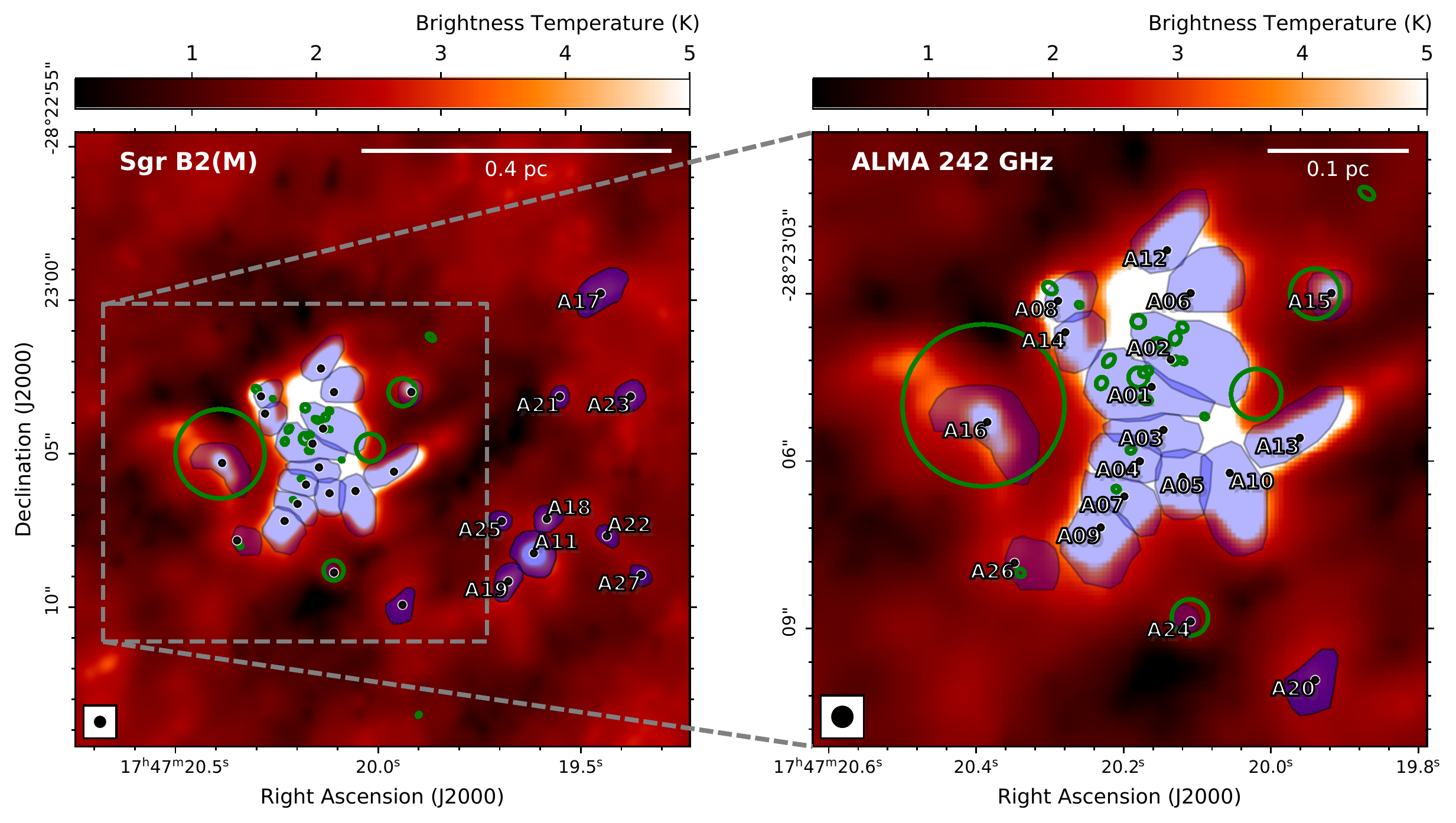}\\
   \caption{Partly taken from \citetads{2023A&A...676A.121M}. Continuum emission toward Sgr~B2(M) at 242~GHz. A close-up of the central part is presented in the right panel. The identified sources are marked with shaded blue polygons and indicated with the corresponding source ID. The shaded light blue polygons describe the inner cores. The black points indicate the position of each core, described by \citetads{2017A&A...604A...6S}. The intensity color scale is shown in units of brightness temperature, and the synthesized beam of $0 \farcs 4$ is described in the lower left corner. The green ellipses describe the \hii~regions identified by \citetads{2015ApJ...815..123D}, where the size of each ellipse indicates to the size of the corresponding \hii~region.}
   \label{fig:CorePosSgrB2M}
\end{figure*}

%*******************************************************************************
% Figure: Hot cores in Sgr B2(N)
\begin{figure*}[ht!]
   \centering
   \includegraphics[width=0.86\textwidth]{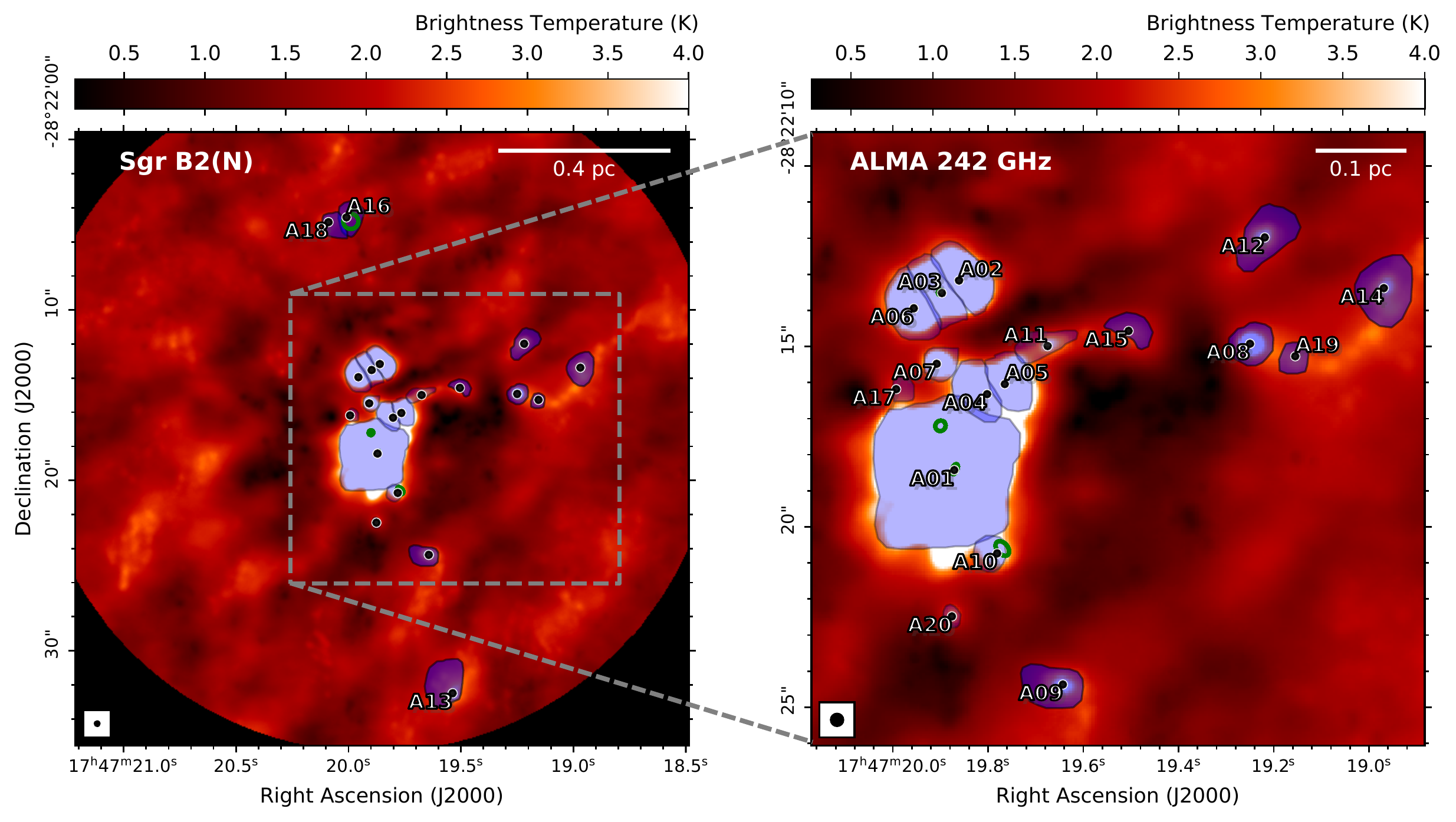}\\
   \caption{Partly taken from \citetads{2023A&A...676A.121M}. Continuum emission toward Sgr~B2(N) at 242~GHz. The right panel describes a close-up of the central part. The shaded blue polygons together with the corresponding source ID indicate the identified hot cores. The shaded light blue polygons describe the inner cores. The black points indicate the position of each core described by \citetads{2017A&A...604A...6S}. The intensity color scale is shown in units of brightness temperature, while the synthesized beam of $0 \farcs 4$ is indicated in the lower left corner. The green ellipses describe the \hii~regions identified by \citetads{2015ApJ...815..123D}, where the size of each ellipse indicates to the size of the corresponding \hii~region.}
   \label{fig:CorePosSgrB2N}
\end{figure*}

%===============================================================================
% Observations and data reduction
\section{Observations and data reduction}\label{sec:DataRed}

As is described in \citetads{2017A&A...604A...6S}, Sgr~B2 was observed with ALMA \citepads[Atacama Large Millimeter/submillimeter Array;][]{2015ApJ...808L...1A} during Cycle~2 in June~2014 and June~2015, using 34 -- 36 antennas in an extended configuration with baselines in the range from 30~m to 650~m, which results in an angular resolution of $0 \farcs 3 - 0 \farcs 7$ (corresponding to $\sim$3300~au). The observations were carried out in the spectral scan mode covering the whole ALMA band 6 (211 to 275~GHz) with 10 different spectral tunings, providing a resolution of 0.5 -- 0.7~km~s$^{-1}$ across the full frequency band. The two sources Sgr~B2(M) and Sgr~B2(N) were observed in track-sharing mode, with phase centers at $\alpha_{\rm J2000} = 17^{\rm h} 47^{\rm m} 20 \fs 157$, $\delta_{\rm J2000} = -28^\circ 23' 04 \farcs 53$ for Sgr~B2(M), and at $\alpha_{\rm J2000} = 17^{\rm h} 47^{\rm m} 19 \fs 887$, $\delta_{\rm J2000} = -28^\circ 22' 15 \farcs 76$ for Sgr~B2(N). Calibration and imaging were carried out with CASA\footnote{The Common Astronomy Software Applications \citepads[CASA,][]{2007ASPC..376..127M} is available at \url{https://casa.nrao.edu}.} version 4.4.0. Finally, all images were restored with a common Gaussian beam of $0 \farcs 4$. The continuum emission was determine using STATCONT \citepads[][]{2018A&A...609A.101S}. Details of the observations, calibration and imaging procedures are described in \citetads{2017A&A...604A...6S} and \citetads{2019A&A...628A...6S}.

%===============================================================================
% Data analysis
\section{Data analysis}\label{sec:DataAnalysis}

% definition of core
\citetads{2017A&A...604A...6S} identified a core within continuum emission maps of Sgr~B2(M) and N, if at least one closed contour (polygon) above the 3$\sigma$ level (with $\sigma$ the root mean square (rms) noise level of the map: 8~mJy~beam$^{-1}$) was found, see Figs.~\ref{fig:CorePosSgrB2M}~-~\ref{fig:CorePosSgrB2N}. The spectra for each core, described in Figs.~\ref{fig:CoreSpecSgrB2M}~-~\ref{fig:CoreSpecSgrB2N}, are obtained by averaging over all pixels contained in a polygon to improve the signal-to-noise and detection of weak lines. The rms noise level\footnote{rms noise level was determined using an iterative sigma clipping approach. First, the sigma value was optimized so that the clipped spectrum contained approximately 68~\% of the data points. This optimal sigma was then used in a final sigma clipping procedure to isolate the noise and determine the rms noise level of the corresponding source. Although this method provides a robust estimate of the noise level by excluding potential spectral features or artifacts, we cannot exclude the possibility that the determined noise level is still dominated by weak lines instead of thermal noise.} of sources in Sgr~B2(M) ranges from 1.07~mJy~beam$^{-1}$ to 106.17~mJy~beam$^{-1}$, while for the cores in N, it varies between 2.24~mJy~beam$^{-1}$ and 307.11~mJy~beam$^{-1}$.

% XCLASS general
The spectra of each hot core were modeled using the eXtended CASA Line Analysis Software Suite \citepads[XCLASS\footnote{\url{https://xclass.astro.uni-koeln.de/}},][]{2017A&A...598A...7M} with additional extensions (M\"{o}ller in prep.). XCLASS allows modeling and fitting of molecular and recombination lines by solving the 1D radiative transfer equation assuming local thermal equilibrium (LTE) conditions and an isothermal source. Because of the high hydrogen column densities (n$_{{\rm H}_2} \ge 4.0 \cdot 10^6$~cm$^{-3}$), which we found in the first paper even in the outer sources, LTE is a valid approximation and the kinetic temperature of the gas can be estimated from the rotation temperature: $T_{\rm rot} \approx T_{\rm kin}$. Additionally, finite source size, dust attenuation, and optical depth effects are taken into account as well. Details of the calculation procedure are described in Paper~VII.

% model parameters
The contribution of each molecule is described by a certain number of components, where each component is defined by the source size $\theta_{\rm source}$, the rotation temperature $T_{\rm rot}$, the column density $N_{\rm tot}$, the Gaussian line width $\Delta v_G$, and the velocity offset from the systemic velocity $v_{\rm off}$. Additionally, each component is located at a certain distance along the line of sight to recreate a layered structure, with some components situated in front of or behind others, mimicking the relative depths of the original sources. In addition to molecules, \textsc{XCLASS} offers the possibility to model the contribution of radio recombination lines (RRLs) up to $\Delta n = 6$ ($\zeta$-transitions) as well, using Voigt line profiles instead of Gaussian line profiles\footnote{Here, the Gaussian line profile is caused by the thermal motion of the gas particles and micro-turbulence, see \citetads{2017A&A...598A...7M} for more details.} as for molecules. Similar to molecules, the contribution of each RRL is described by multiple components, defining the source size, the electron temperature $T_e$, the emission measure EM, the Gaussian $\Delta v_G$ and Lorentzian $\Delta v_L$ line width, the velocity offset $v_{\rm off}$, and the distance to the observer along the line of sight.

% MAGIX and isotopologs
\textsc{XCLASS} includes the MAGIX \citepads{2013A&A...549A..21M} optimization package, which is used to fit these model parameters to observational data. In order to reduce the number of fit parameters, the modeling can be done simultaneously with corresponding isotopologs, where both molecules (isotopolog and main species) are described by the same parameters, except that the column density of each component is scaled by a user defined ratio. For all sources in Sgr~B2(M) and N the $^{12}$C/$^{13}$C ratio was taken to be 20 (\citetads{1994ARA&A..32..191W}, \citetads{2020A&A...642A.222H}), the $^{16}$O/$^{18}$O to be 250 and $^{32}$S/$^{34}$S to be 13. Additionally, we take the ratios described by \citetads{2014ApJ...789....8N} for $^{16}$O/$^{17}$O to be 800, for $^{32}$S/$^{33}$S to be 75, for $^{14}$N/$^{15}$N to be 182, and for $^{35}$Cl/$^{37}$Cl to be 3. If deuterated isotopologs are identified, we leave the ratio as an additional fit parameter.

% CDMS + JPL
All molecular parameters (e.g., transition frequencies, Einstein~A coefficients, etc.) are taken from an SQLite database embedded in XCLASS containing entries from the Cologne Database for Molecular Spectroscopy (CDMS, \citetads{2001A&A...370L..49M}, \citetads{2005JMoSt.742..215M}) and Jet Propulsion Laboratory database (JPL, \citetads{1998JQSRT..60..883P}) using the Virtual Atomic and Molecular Data Center (VAMDC, \citetads{2016JMoSp.327...95E}).

% local overlap
In line-crowded sources such as Sgr~B2(M) and N, line intensities from two partly overlapping lines do not simply add up if at least one line is optically thick, because photons emitted from one line are absorbed by the other line. XCLASS takes the local line overlap \citepads[described by][]{1991A&A...241..537C} into account, by computing an average source function $S_l (\nu)$ for each frequency $\nu$ and distance $l$. Details of this procedure are described in \citetads{2021A&A...651A...9M}.

% general model setup
For all cores in Sgr~B2(M) and N, we assume a two-layer model, where all components belonging to a layer have the same distance to the observer. The first layer (hereafter called core-layer) contains all components describing emission features in the corresponding core spectrum, because these lines require in general excitation temperatures larger than the continuum brightness temperature, which occur only in the hot cores. The analysis of the continuum levels described in the first paper has shown that some local surrounding envelopes contain warm molecules as well, whose emission features could also be included in the corresponding core spectra leading to falsified results. However, since the dust temperatures of the various sources range from 190 to 354~K, the contributions of these molecules are small and have no further influence on our analysis. The source size $\theta_{\rm core}$ for each component in the core layer is given by the diameter of a circle that has the same area, $A_{\rm core}$, as the corresponding polygon of the source; that is, we determined the source size, $\theta_{\rm core}$, using
\begin{align}\label{myxclass:dustOpacity}
  A_{\rm core} &= \pi \, \left(\frac{\theta_{\rm core}}{2} \right)^2 \nonumber\\
  \Rightarrow \theta_{\rm core} &= 2 \, \sqrt{\frac{A_{\rm core}}{\pi}}.
\end{align}
The second layer (envelope-layer) contains components describing absorption features, since low excitation temperatures below the continuum temperature are needed. Additionally, we assume beam filling for all components located in the envelope layer; in other words, all components located in this layer cover the full beam.

% fitting procedure
In the first step of the analysis of the line survey, we used parameters from previous surveys of Sgr~B2(M) and N (\citetads{2013A&A...559A..47B}, \citetads{2014ApJ...789....8N}, \citetads{2017A&A...604A..60B}), to estimate initial parameter sets for each molecule showing at least one transition within the frequency ranges covered by the ALMA observations. Subsequent to adjusting of the parameters by eye, we made use of the Levenberg-Marquardt algorithm \citepads{doi:10.1137/0111030} to further improve the description. A molecule is claimed as identified if a sufficient number of lines are stronger than the 5$\sigma$ level of the random noise, are not blended, and if the model does not predict a strong line that is not detected. In general, the spectra of many sources located further out in Sgr~B2(M) and N show a significantly lower signal-to-noise ratio than the inner sources, which makes the identification of molecules and the determination of physical parameters more difficult.

% unidentified lines and ghost lines
For sources in Sgr~B2(M), the percentage of unidentified lines\footnote{In this analysis, we declare a line as unidentified if the model describes less than 20~\% of the height of the line.} varies between 7~\% (A10) and 31~\% (A03), with some weak, outlying sources (A18, A19, A21, and A27) having significantly higher proportions of unidentified lines (42~-~58~\%). We find a similar behavior for sources in Sgr B2(N), where the proportion of unidentified lines ranges from 14~\% (A06) and 37~\% (A03), with some weak, far-out sources (A12, A13, A15, A16, and A19) having significantly higher fraction (41~-~67~\%). Many of the unidentified spectral features toward the weaker cores are actually due to imaging artifacts in this high dynamic range cube. These artifacts can also affect intensities and shapes of real lines, adding another source of error.

% species with only one transition
Since line overlap plays a major role in many sources, it is necessary to also model those molecules, such as CO and HCN, which show only one transition within the survey. However, for these molecules a quantitative analysis is not possible, which is why we have fixed the excitation temperatures for components describing emission features of these species to a value of 200~K. For components describing absorption features, we assume a temperature of 2.7~K. (The exact values of the temperatures have no further meaning for our analysis, since we only need a phenomenological descriptions of the line shapes of the corresponding molecules.) Afterward, we fit the column densities, line widths and velocity offsets of all components to obtain a good description of their contribution.
%
%
% final step
Finally, we fit again all parameters describing the contributions of all identified molecules and RRLs simultaneously, where we take the continuum parameters derived in the first paper into account (see Figs.~\ref{fig:SgrB2MExcerpt}~-~\ref{fig:SgrB2NExcerpt}).

% error analysis
Due to the large number of model parameters, a reliable estimation of the errors is not feasible within an acceptable time. We have therefore determined the errors of the model parameters for OCS and CH$_3$CN of core A03 in Sgr~B2(M) as a proxy for the other parameters to give the reader some guidance as to the reliability of the model parameters. Here, the contributions of all other species are taken into account as well. Details of the fitting process are described in \citetads{2021A&A...651A...9M}. The errors of the model parameters were estimated using the \texttt{emcee}\footnote{\url{https://emcee.readthedocs.io/en/stable/}} package \citepads{2013PASP..125..306F}, which implements the affine-invariant ensemble sampler of \citetads{2010CAMCS...5...65G}, to perform a Markov chain Monte Carlo (MCMC) algorithm that approximates the posterior distribution of model parameters by random sampling in a probabilistic space. Here, the MCMC algorithm starts at the calculated maximum of the likelihood function -- that is, the parameters of the best-fit LTE model of OCS (CH$_3$CN) -- and draws 40 samples (walkers) of model parameters from the likelihood function in a ball around this position. For each parameter we used 1500 steps to sample the posterior. The probability distribution and the corresponding highest posterior density (HPD) interval of each model parameter are calculated afterward. In order to get a more reliable error estimation, the errors for the column densities were calculated on a log scale; that is, these parameters were converted to the log10 value before applying the MCMC algorithm and converted back to linear scale after finishing the error estimation procedure. Details of the calculation procedure as well as the HPD interval are described in \citetads{2021A&A...651A...9M}. The posterior distributions of the individual parameters of OCS and CH$_3$CN are shown in Figs.~\ref{fig:ErrorEstimOCS} and \ref{fig:ErrorEstimCH3CN}. For all parameters we find a unimodal distribution, which means that there is no other fit within the given parameter ranges that describes the data so well.

%===============================================================================
% Results
\section{Results}\label{sec:Results}

%-------------------------------------------------------------------------------
% Detected molecules
% \subsection{Detected molecules}\label{subsec:DetectedMolecules}

% introduction for section
In this section we describe the molecules and their isotopologs and vibrational excited states identified in the line surveys of each source in Sgr~B2(M) and N. Due to their chemical relationships, for example having the same heavy-atom backbone or the same functional group, the detected molecules are divided into eight families: Simple O-bearing molecules (Sect.~\ref{subsec:SimpleOBearingMol}), complex O-bearing molecules (Sect.~\ref{subsec:ComplexOBearingMol}), NH-bearing molecules (Sect.~\ref{subsec:NHBearingMol}), N- and O-bearing molecules (Sect.~\ref{subsec:NOBearingMol}), cyanide molecules (Sect.~\ref{subsec:CyanideMol}), S-bearing molecules (Sect.~\ref{subsec:SBearingMol}), carbon and hydrocarbons (Sect.~\ref{subsec:CHMol}), and other molecules (Sect.~\ref{subsec:OtherMol}).

% something about chemistry in hot cores
Cold gas in the interstellar medium is often made up of simple molecules (e.g., CO, HCN, N$_2$, O$_2$ etc), which are frozen onto dust grains. Both hydrogenation and reactions with CO produce more complex molecules on the dust surface; for example, CO$_2$, CH$_3$OH, and H$_2$O. UV radiation can cause some of these molecules to dissociate into radicals. With increasing temperature, these radicals become mobile and form new species. When the temperature is high enough (T~$\sim$~100-150~K), the ice mantles sublimate and the resulting molecules enter the gas phase and form, among other things, precursors of complex organic molecules (COMs) \citepads{2016ApJ...821...46T}. In addition, shocks generated by the interaction of jets and outflows with the surrounding envelope can also sublimate or sputter icy grain mantles, releasing molecules such as CH$_3$OH and other COMs into the gas-phase. Furthermore, shocks can also sputter the grain cores themselves, releasing the embedded Si and S atoms and increasing the production of Si- and S-bearing species such as SiO, SO$_2$ and SO (see e.g., \citetads{2008A&A...482..809G}, \citetads{2018IAUS..332....3V}). It is important to note that while grain-surface chemistry plays a significant role in the formation of COMs, these molecules can also form in the gas phase. Various models, such as those by \citetads{2009ApJ...691.1459V} and \citetads{2021A&A...648A..72W}, suggest that gas-phase reactions contribute significantly to the synthesis of COMs.

% description of temperatures and abundances
The derived excitation temperatures and abundances for each detected molecule and source in Sgr~B2(M) and N, respectively, are shown in Appendix~\ref{app:temp:SgrB2} and \ref{app:abund:SgrB2}. For molecules for which more than one component was required, we give the column density weighted mean temperature and the summed abundances. We calculated the abundances relative to H$_2$ using the hydrogen column densities derived in our first paper for each source. A comparison between the obtained temperature and abundance ranges for each source in Sgr~B2(M) and N and literature values are given in Table~\ref{Tab:ParamRangesSgrB2M}~-~\ref{Tab:ParamRangesSgrB2N}, respectively. In general, a direct comparison of the obtained temperatures and abundances with previous analysis is of limited value. First, compared to \citetads{1991ApJS...77..255S}, \citetads{2013A&A...559A..47B}, \citetads{2014ApJ...789....8N}, and \citetads{2021A&A...651A...9M}, our observations have a much higher spatial resolution, which is why we can detect structures that could not be included in these analyses. Only the observations of \citetads{2017A&A...604A..60B} have a similar resolution. Nevertheless, our analysis differs significantly from previous analyses in some aspects, because we determined not only the dust and free-free contributions to each source but in addition considered effects such as local overlap effects, which are important for line-rich sources such as Sgr~B2(M) and N.

% figures of selected transitions for each molecule
A detailed description of all identified species together with figures showing the spectra of each detected molecule is provided in Appendix~\ref{app:identifiedspecies}~-~\ref{app:transplots}. For molecules with a large number of transitions, a representative selection of the detected transitions is shown\footnote{The fitted spectra are published as online material}.

%===============================================================================
% Discussion
\section{Discussion}\label{sec:Discussion}

%-------------------------------------------------------------------------------
% Physical parameters
\subsection{Physical parameters}\label{subsec:PhysParam}

%*******************************************************************************
% Table: Velocity ranges corresponding to different kinematic components
\begin{table}[!tb]
    \centering
    \caption{Velocity ranges}
    \begin{tabular}{lcc}
        \hline
        \hline
        Component            & $v_{\rm LSR}$~(km~s$^{-1}$) & $v_{\rm off}$~(km~s$^{-1}$)\\
        \hline
        Galactic Center (GC) &              $-$92 to $-$69 &            $-$156 to $-$133\\
        Norma Arm            &              $-$47 to $-$13 &            $-$111 to  $-$77\\
        Galactic Center (GC) &                   $-$9 to 8 &             $-$73 to  $-$56\\
        Scutum Arm           &                    12 to 22 &             $-$52 to  $-$42\\
        Sagittarius B2       &                       $>$22 &                      $>-$42\\
        \hline
    \end{tabular}
    \tablefoot{Velocity ranges corresponding to different kinematic components, taken from \citetads{2014ApJ...785..135L}. Velocity offsets $v_{\rm off}$ are computed relative to the source velocity of 64~(km~s$^{-1}$).}
    \label{Tab:VelStruc}
\end{table}

% general overview, motivation to use KDEs
To better understand the distribution of the derived physical parameters, we computed in addition to the mean temperatures of each source and layer shown in Figs.~\ref{fig:TempEnvSgrB2M}~-~\ref{fig:TempCoreSgrB2N}, kernel density estimations (KDE, \citetads{rosenblatt1956}, \citetads{parzen1962}) with a Gaussian kernel for each source, layer and molecular family, see Figs.~\ref{fig:KDETrotEnvM}~-~\ref{fig:KDEvoffCoreSgrB2N}. Here, we consider only those components that clearly belong to Sgr~B2; that is, whose source velocity is greater than $22$~km~s$^{-1}$ (see Table~\ref{Tab:VelStruc}). In addition, we take only those molecules into account, which show more than one transition within the frequency ranges covered by our observations. Furthermore, Silverman's rule \citepads{1986desd.book.....S} as scipy implementation \citepads{2020SciPy-NMeth} is used to compute the bandwidth of each KDE. For a better visibility, we have normalized the maximum of each KDE to one. In general, the KDE is a nonparametric way to estimate the probability density function of a given parameter and offers a visual representation of the individual parameters with the aim to see their spread and to search for general trends. The shape of the KDE plot can provide insights into the underlying distribution of the data. For example, a unimodal KDE plot with a single peak suggests that the data is distributed around a central value, while a bimodal KDE plot with two peaks suggests that the data may be drawn from two different distributions.

% T_rot, envelope layer, Sgr~B2(M) and N
The KDEs of excitation temperatures of components belonging to the envelope layer in both regions mostly show an unimodal distribution with maxima located at around 20~K for most sources and molecular families, see Fig.~\ref{fig:KDETrotEnvM} and Fig.~\ref{fig:KDETrotEnvN}. This is remarkable because in our analysis of continuum contributions (see \citetads{2023A&A...676A.121M}) we found excitation temperatures in the envelopes of around 73~K (for sources in Sgr~B2(M)) and 59~K (for sources in Sgr~B2(N)). A possible explanation for this may lie in the way the excitation temperatures were determined in \citetads{2023A&A...676A.121M}, where we selected pixels around each source and used the spectra at these positions to compute an averaged envelope spectrum. Afterward, we analyzed the contributions of CH$_3$CN, H$_2$CCO, H$_2$CO, H$_2$CS, HNCO, and SO to these envelope spectra to determine the averaged gas temperature in each envelope. Details of the calculation procedure are described in \citetads{2023A&A...676A.121M}. In contrast to the molecules in the current analysis located in the envelope layer, these molecules are seen more or less in emission only. But emission features from the envelope toward the sources cannot be separated from the core. Molecules in emission have mostly higher temperatures than molecules in absorption. Therefore, we underestimate in this analysis the real excitation temperatures in the envelopes around each source.

For the inner sources in Sgr~B2(M) and N -- those located less than 50~au from the center of the central core A01 for both regions -- we find broader distributions of simple O-bearing and cyanide molecules, while the NH-bearing molecules are usually at lower temperatures. A possible explanation for this behavior could be the high contributions of H$_2$CO, CH$_3$OH, HCN, and HNC in the inner sources, which are hot and very close to each other heating up the surrounding envelope. For sources A08, A09, A20, and A22 in Sgr~B2(M), the maximum of the KDE of sulfur-bearing molecules is shifted toward slightly higher temperatures. The formation and abundance of certain sulphur-bearing molecules, such as SO and SO$_2$, are positively correlated with the kinetic temperature \citepads{2023A&A...680A..58F}, suggesting that these molecules become more abundant as the temperature increases through the evolutionary stages. Therefore these molecules are usually located in the warmer regions of the envelopes. For some inner sources in Sgr~B2(N), we find slightly broader distributions of simple O-bearing and NH-bearing molecules. Additionally, for source A07 the different molecular families show clear separated maxima, where the sulfur-bearing molecules, as in source A14, are the hottest.

% T_rot, core layer, Sgr~B2(M) and N
The KDEs of temperatures whose components are located in the core layer of Sgr~B2(M) show a very different behavior, see Fig.~\ref{fig:KDETrotCoreM}. While the inner sources show broad distributions for many molecular families, we see narrower and sometimes spike-like\footnote{A spike-like KDE has only a single data point and thus shows no variation.} KDEs in the outer sources. This is due to the fact that many outer sources are mainly seen in absorption and have fewer emission features. In many sources, the sulfur-bearing molecules exhibit the highest temperatures, which, in addition to the fact that these molecules require high temperatures for formation, can also be explained by the fact that molecules of this family often exhibit vibrationally excited states with high excitation temperatures. For sources in Sgr~B2(N) the temperatures in the core layer show, in contrast to sources in M, much narrower distributions, see Fig.~\ref{fig:KDETrotCoreN}. The sources in N thus seem to have a more homogeneous structure.

% v_width, envelope and core layer, Sgr~B2(M) and N
For the line widths, we obtain KDEs with mostly unimodal distributions in all sources, layers, and molecular families, see Figs.~\ref{fig:KDEwidthEnvSgrB2M}~-~\ref{fig:KDEwidthCoreSgrB2N}. However, the distribution between the different families of molecules and sources is sometimes very different. As for the excitation temperatures, in both regions the outer sources show narrower distributions than the inner sources. In general, the line width contains among other things a thermal and a turbulent proportion, where the thermal one depends on the gas temperature and the respective molecular weight. For all molecular families and sources the thermal line widths range from 0.1 to 0.6~km~s$^{-1}$. Therefore, the distribution of line widths mainly shows the turbulent contribution. Line widths above 10~km~s$^{-1}$ indicate outflows or large-scale motions. Some sources show significant differences between molecular families. For example, in source A20 in Sgr~B2(M), the N- and O-bearing molecules in the core layer show a significantly larger line width, see Fig.~\ref{fig:KDEwidthCoreSgrB2M}, and a larger velocity offset, see Fig.~\ref{fig:KDEvoffCoreSgrB2M}, than the other molecules, which could indicate that it cannot be directly assigned to the hot core A20 but to a filament. In addition, some molecules show high asymmetric line shapes, which can be associated with outflows. The description of these line shapes requires many components, which falsifies the true distribution of line widths.

% v_off, envelope layer, Sgr~B2(M) and N
By using the distributions of velocity offsets, it is possible not only to determine the velocity of each source, but also whether certain families of molecules show deviations from it, which in turn can indicate a different affiliation. A deviating source velocity could, for example, show that the associated molecular family does not belong to the source in question, but to a filament located in front of the source. For example, the distributions of velocity offsets of the different molecular families in the core layer of source A14 in Sgr~B2(M) show clear differences. While the N- and O-bearing and sulfur-bearing molecules show more or less the same velocity offsets, we see clear shifts toward higher velocities for simple O-bearing, cyanide, and NH-bearing molecules. A similar but less pronounced behavior can be found among others for source A08, A15, and A22. In general, we find, in both regions, broader distributions in the envelope than in the core layer for most sources.

%-------------------------------------------------------------------------------
% Correlations between molecules
\subsection{Correlations between sources}

To find correlations between different sources in Sgr~B2(M) and N that might indicate similar evolutionary phases, we used the principal component analysis (PCA). PCA is a widely used multivariate analysis method that uses an orthogonal transformation to convert observed variables into a set of linearly uncorrelated new variables called principal components. The first principal component contains the largest possible variance, and the subsequent principal component has the next highest variance, provided that it is orthogonal to the previous component. In contrast to other PCA analyses (e.g., \citetads{2017A&A...599A.100G}; \citetads{2023A&A...670A.111K}), we do not apply PCA directly to the observed spectra, but use the calculated abundances relative to the hydrogen column densities derived in our first paper for each source of the core components\footnote{Here we consider only those molecules that have more than one transition in the frequency ranges covered by our observations.} for the different sources (see Figs.~\ref{fig:AbundCoreSgrB2M} and \ref{fig:AbundCoreSgrB2N}) whose source velocity is greater than $22$~km~s$^{-1}$. By using abundances, the results of the PCA are not distorted by line overlap or absorption features. Furthermore, we do not need to normalize the spectra, nor do the spectra of the different sources need to be corrected due to the different source velocities. Here, we assume that two sources of the same age also have great similarities in the sequence of their abundances. However, the abundances can be distorted by contributions from filaments, foreground objects and imaging artifacts. Furthermore, many spectra from sources located further out in Sgr~B2(M) and N have a significantly lower signal-to-noise ratio than the inner sources, which makes the identification of molecules and the determination of the physical parameters more difficult. This in turn can lead to incorrect PCA results, whereby correlations that are not real are identified. For our analysis, we have used the PCA implementation available in the Python package \texttt{scikit-learn} \citepads{scikit-learn}.

\begin{figure}[!tb]
   \begin{subfigure}[t]{0.49\textwidth}
   \centering
   \includegraphics[width=0.99\textwidth]{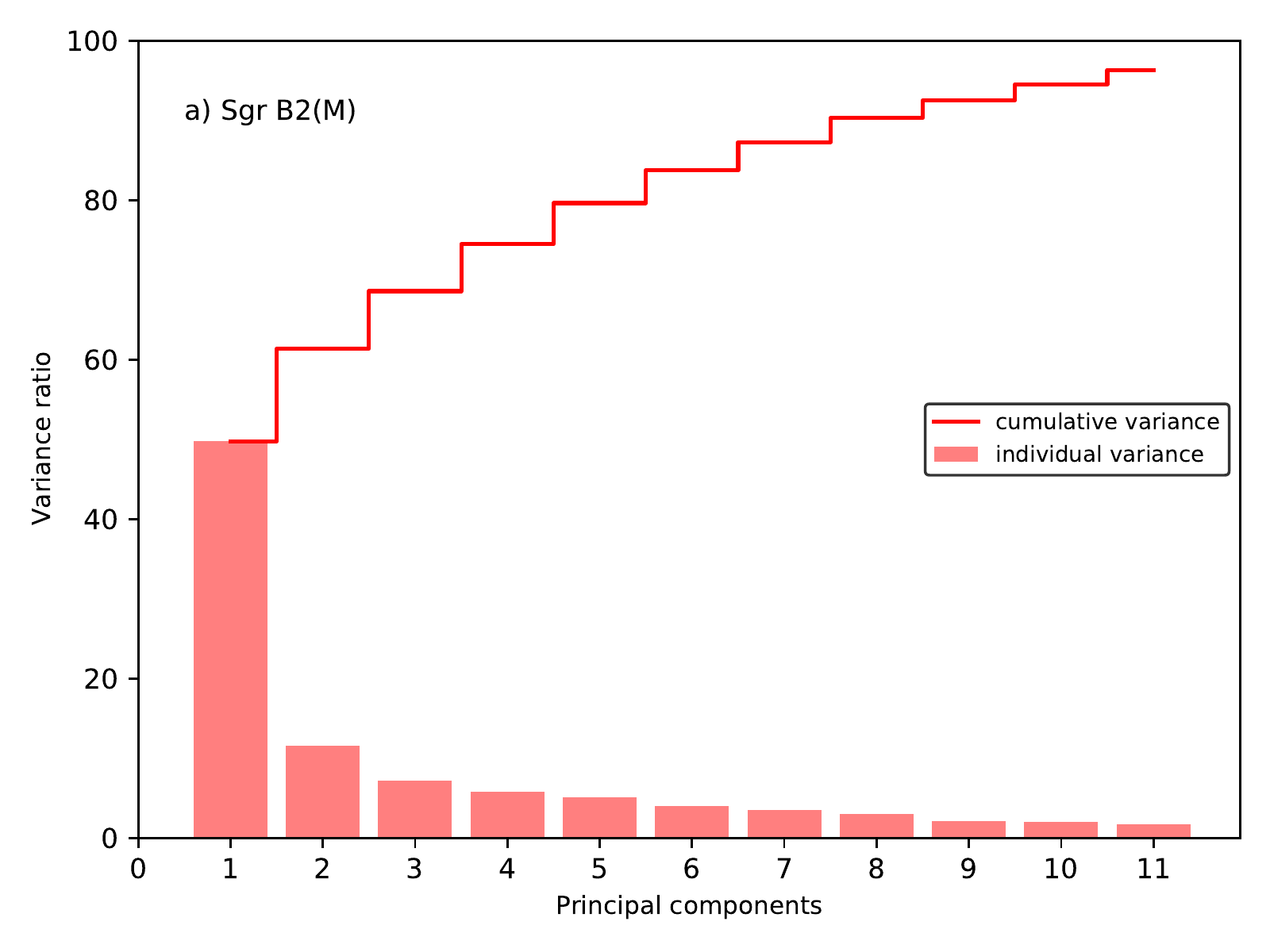}\\
   \end{subfigure}
\quad
   \begin{subfigure}[b]{0.49\textwidth}
   \centering
   \includegraphics[width=0.99\textwidth]{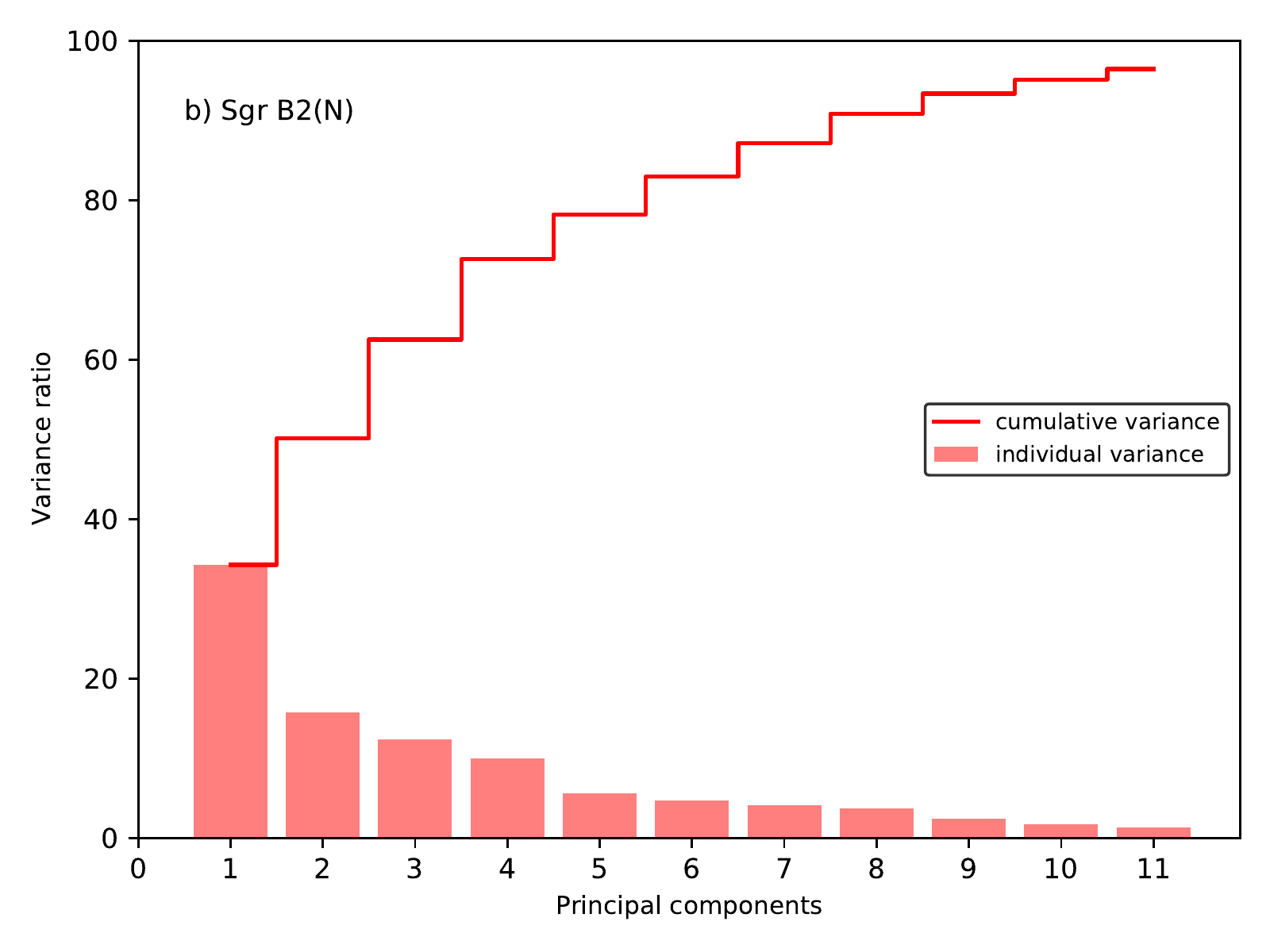}\\
   \end{subfigure}
   \caption{The percentages of the first eleven eigenvalues in descending order for sources in a) Sgr~B2(M) and b) Sgr~B2(N).}
   \label{fig:PCAStrength}
\end{figure}

% percentage of eigenvalues
The percentages of eigenvalues are shown in Fig.~\ref{fig:PCAStrength}. For both regions, the first eigenvalue has the highest proportion (49.7\% for Sgr~B2(M) and 34.2\% for Sgr~B2(N)), while the first six principal components together account for more than 80\% of the total variance for both regions.
%
% Eigenvectors
The first six eigenvectors for each region are described in Figs.~\ref{fig:PCAEVSgrB2M}~-~\ref{fig:PCAEVSgrB2N}. For sources in Sgr~B2(M), the first eigenvector is mainly dominated by contributions from outer sources, in particular by A20, while the remaining eigenvectors exhibit relatively uniform distributions. Similarly, for Sgr~B2(N), the first eigenvector again shows the most significant contributions, with the outer sources A09, A12, and A13 being dominant, while the other eigenvectors display a more consistent structure.

\begin{figure}[!htb]
   \begin{subfigure}[t]{0.49\textwidth}
   \centering
   \includegraphics[width=0.99\textwidth]{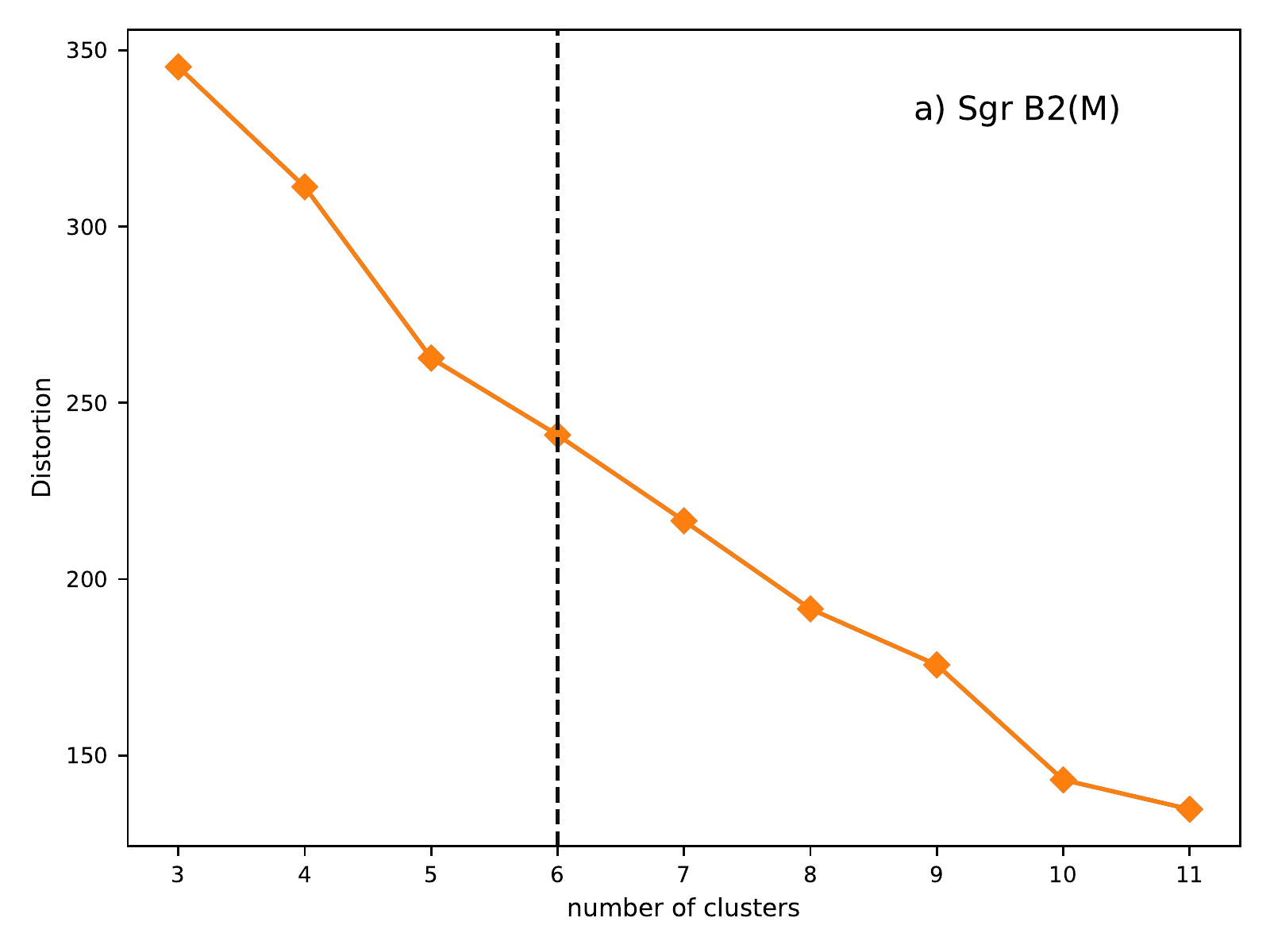}\\
   \end{subfigure}
\quad
   \begin{subfigure}[b]{0.49\textwidth}
   \centering
   \includegraphics[width=0.99\textwidth]{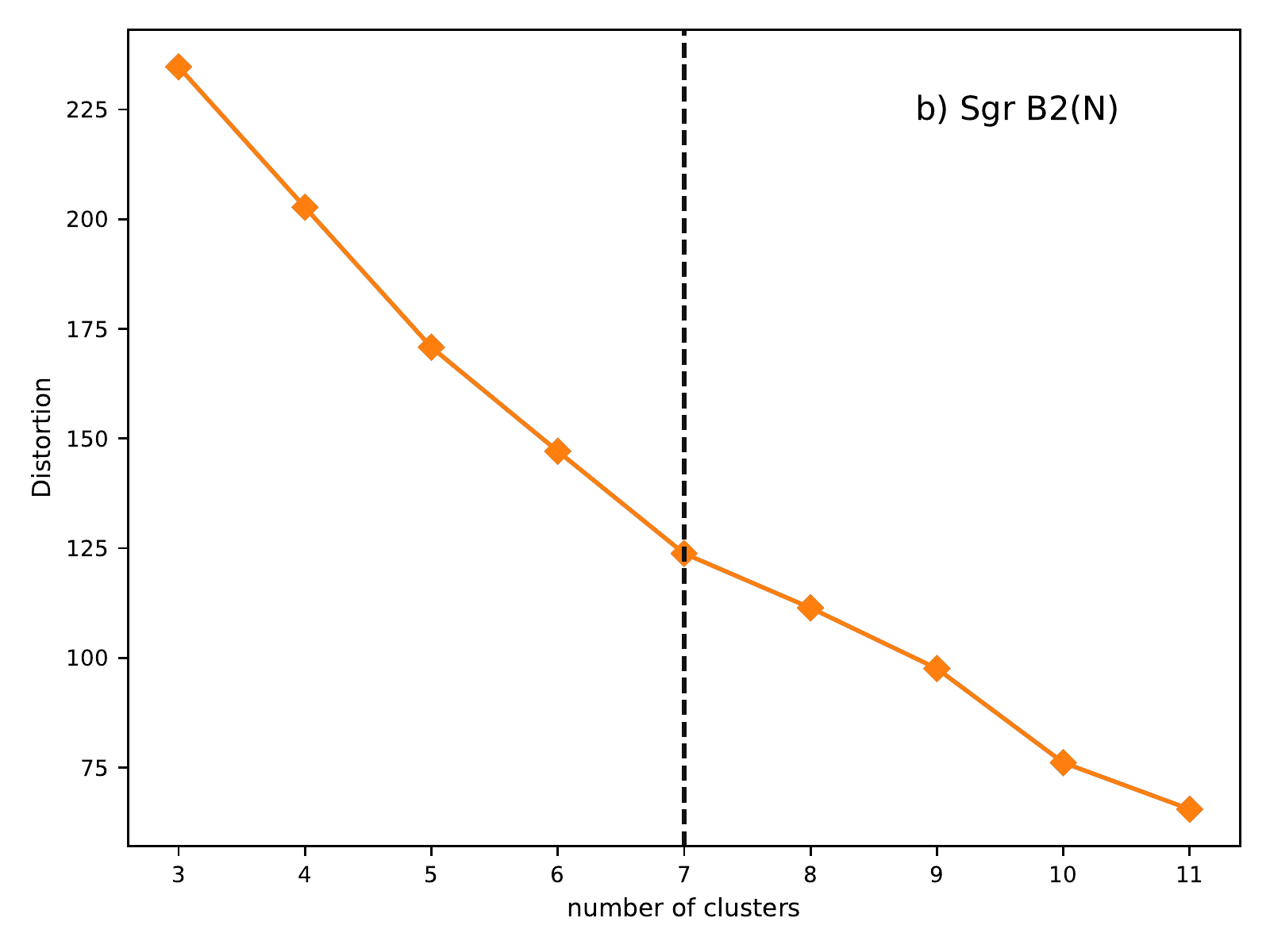}\\
   \end{subfigure}
   \caption{Elbow plot for sources in a) Sgr~B2(M) and b) Sgr~B2(N).}
   \label{fig:PCAElbow}
\end{figure}

% projection onto a pair of principal components
The abundances of each source in Sgr~B2(M) and N can now be projected onto a pair of principal components (see Figs.~\ref{fig:PCAProjectionSgrB2M}~-~\ref{fig:PCAProjectionSgrB2N}). After transforming the data into the principal component space, the k-means \citepads{hartigan1979} function from the scikit-learn package is used to categorize the cores into a predefined number of distinct, non-overlapping clusters. In order to obtain the best number of clusters, we apply the elbow method \citepads{Thorndike1953} using the \texttt{yellowbrick} Python package \citepads{bengfort_yellowbrick_2018}, which involves plotting the within-cluster sum of squares (WCSS) against the number of clusters (see Fig.~\ref{fig:PCAElbow}). The point of maximum curvature in the plot called elbow or knee represents the point where the WCSS starts to decrease at a slower rate. This point indicates the appropriate number of clusters. For sources in Sgr~B2(M), the elbow is at six and for N at seven clusters. The identified cluster together with the contained sources for both regions are shown in Table~\ref{Tab:PCACluster}.

%*******************************************************************************
% Table: Identified cluster and corresponding sources
\begin{table}[!tb]
    \centering
    \caption{Clusters identified by the k-means algorithm.}
    \begin{tabular}{ll}
        \hline
        \hline
        Cluster:      & Sources: \\
        \hline
        \hline
        \multicolumn{2}{l}{Sgr~B2(M)} \\
        \hline
             CM1      & A01, A03, A04, A05, A06, A07, A10 \\
             CM2      & A11, A23 \\
             CM3      & A17, A22, A25 \\
             CM4      & A02, A08, A09, A12, A13, A14, A15 \\
             CM5      & A20 \\
             CM6      & A16, A18, A19, A21, A24, A26, A27 \\
        \hline
        \multicolumn{2}{l}{Sgr~B2(N)} \\
        \hline
             CN1      & A16, A19, A20 \\
             CN2      & A05, A06, A07, A08, A10 \\
             CN3      & A01, A02, A17 \\
             CN4      & A03, A04 \\
             CN5      & A13 \\
             CN6      & A09, A14, A15, A18 \\
             CN7      & A11, A12 \\
        \hline
    \end{tabular}
    \tablefoot{Clusters identified by the k-means algorithm with the contained sources for both regions in the projection of the first two eigenvectors PC1 and PC2.}
    \label{Tab:PCACluster}
\end{table}

The projection of the first two eigenvectors PC1 and PC2, which together have a variance of more than 57\% for both regions, shows a clear separation between different sources in Sgr~B2(M). For each identified cluster and region in this projection of PC1 and PC2, we calculated the mean abundances (see Fig.~\ref{fig:PCAMeanXSgrB2M} (for Sgr~B2(M)) and Fig.~\ref{fig:PCAMeanXSgrB2N} (for Sgr~B2(N)), and the mean temperatures, see Fig.~\ref{fig:PCAMeanTSgrB2M} (for Sgr~B2(M)) and Fig.~\ref{fig:PCAMeanTSgrB2N} (for Sgr~B2(N))) for each identified molecule. In addition to the level of evolution of the source, other contributions such as filaments can also play a role in the classification of the sources into the various clusters.

%*******************************************************************************
% Figure: Strength of Eigenvalues in Sgr~B2

%*******************************************************************************
% Figure: Elbow method

%*******************************************************************************
% Figure: violin plots
\begin{figure}[!htb]
   \begin{subfigure}[t]{0.49\textwidth}
   \centering
   \includegraphics[width=0.99\textwidth]{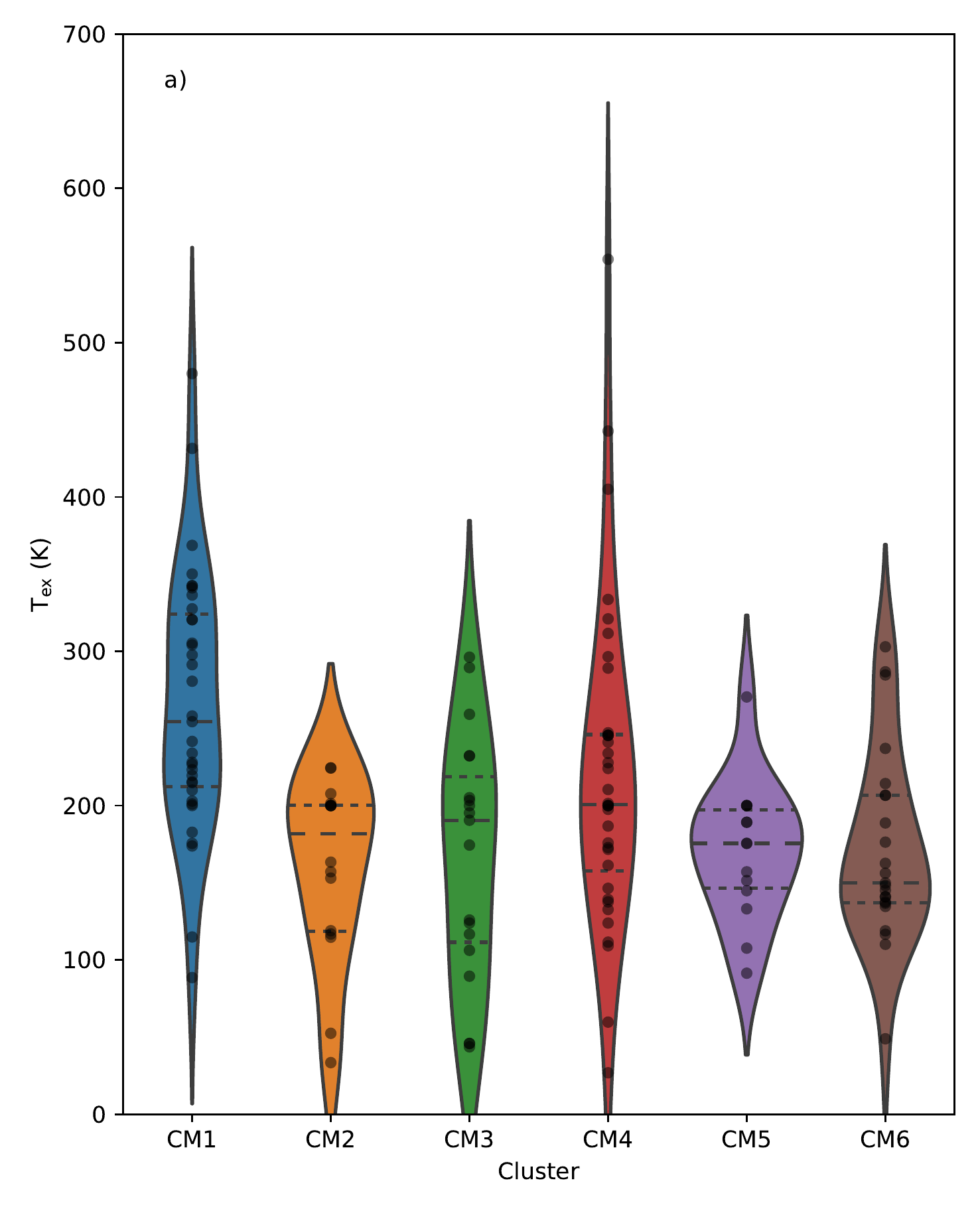}\\
   \end{subfigure}
\qquad
   \begin{subfigure}[b]{0.47\textwidth}
   \centering
   \includegraphics[width=0.99\textwidth]{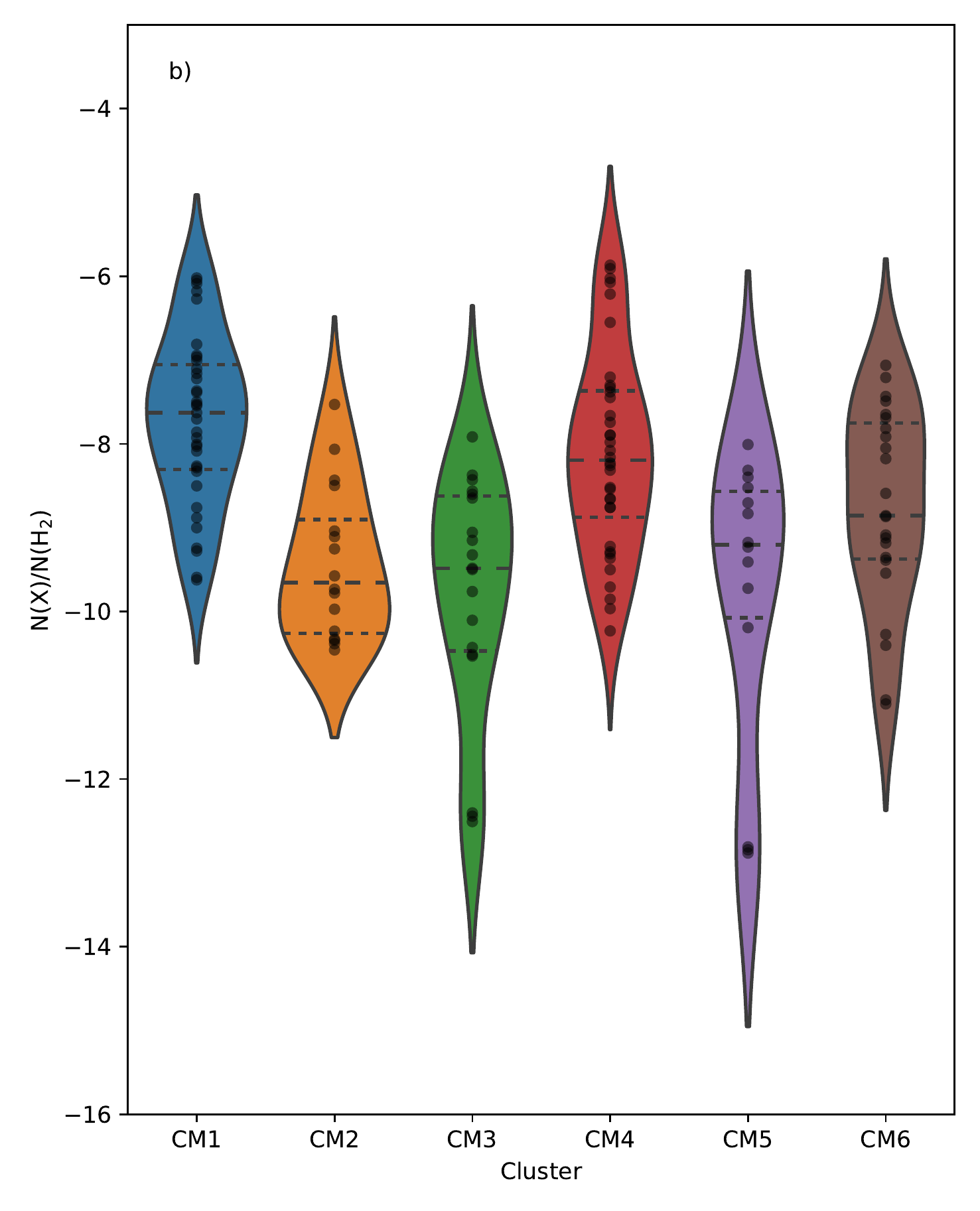}\\
   \end{subfigure}
   \caption{Violin plots describing a) temperature and b) abundance distributions for each cluster in Sgr~B2(M). Here, the long dashed lines describe the respective medians, while the short dashed lines represent the quartiles of the data, which are also described as black dots.}
   \label{fig:PCAVioloinSgrB2M}
\end{figure}

\begin{figure}[!htb]
   \begin{subfigure}[t]{0.49\textwidth}
   \centering
   \includegraphics[width=0.99\textwidth]{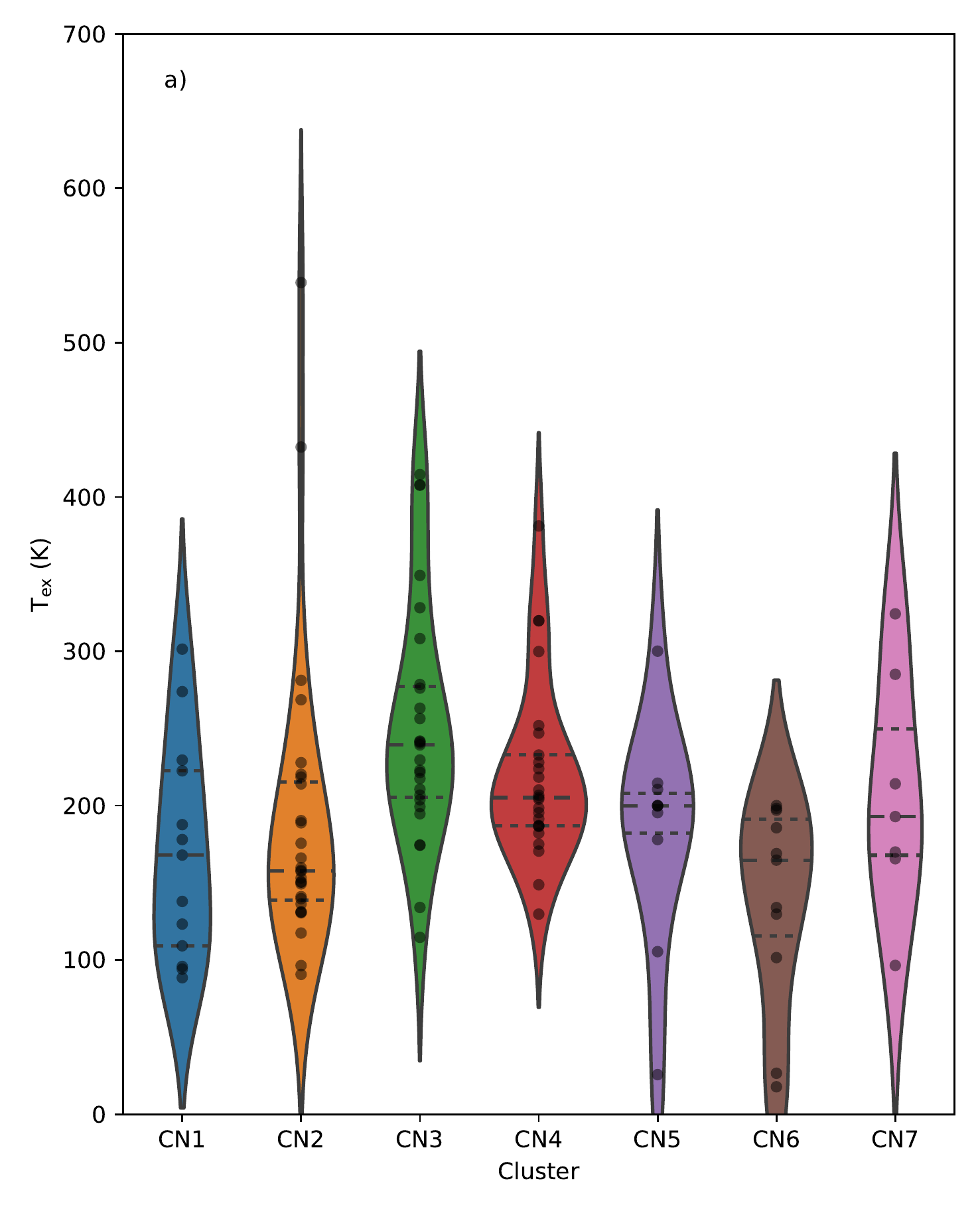}\\
   \end{subfigure}
\qquad
   \begin{subfigure}[b]{0.47\textwidth}
   \centering
   \includegraphics[width=0.99\textwidth]{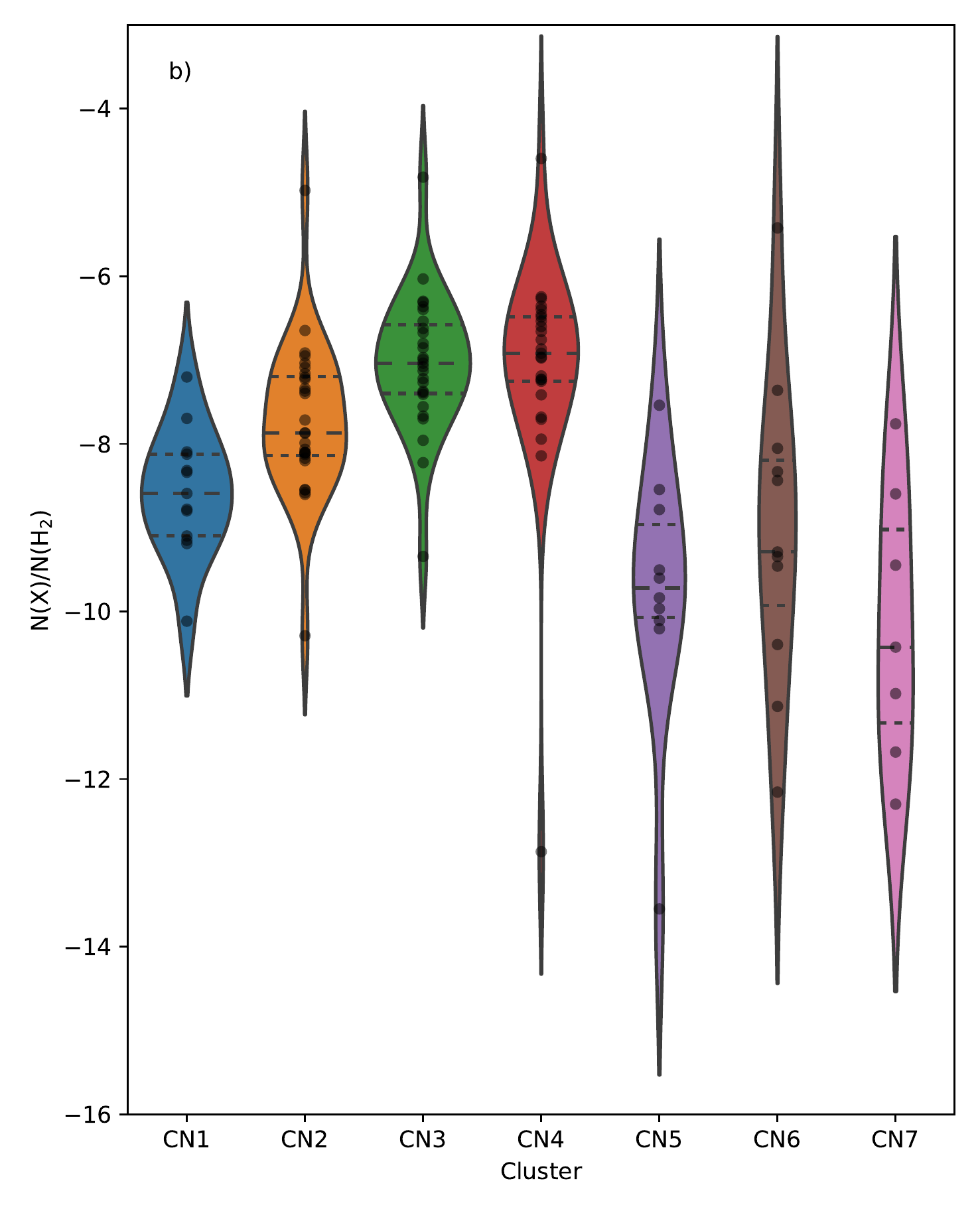}\\
   \end{subfigure}
   \caption{Violin plots show a) temperature and b) abundance distributions for clusters in  Sgr~B2(N). Long dashes represent medians, short dashes mark quartiles, and black dots depict data points.\\}
   \label{fig:PCAVioloinSgrB2N}
\end{figure}

% cluster in Sgr B2(M)
The temperature and abundances dispersions for each cluster in Sgr~B2(M) are described in Fig.~\ref{fig:PCAVioloinSgrB2M}, which show more a less a unimodal behavior.
% CM1, CM4
Clusters~CM1 and CM4 in Sgr~B2(M) have similar abundances for most molecules, with t-HCOOH and NH$_2$CN found exclusively in cluster~CM1, and c-C$_3$H$_2$ only in cluster~CM4. Furthermore, cluster~CM1, which contains most of the inner sources, has the highest mean temperatures of all species and clusters. In addition, the cluster has the highest mean temperature for the sulfur-bearing molecules except NS. This could indicate that the sources contained in these two clusters are in later stages of evolution, where more complex molecules such as C$_2$H$_5$OH can also form. Furthermore, the presence of vibrationally excited molecules, such as C$_2$H$_3$CN,v$_{10}$=1 (E$_{\rm low}$ = 1662.9~K) and C$_2$H$_5$CN,v$_{20}$=1 (E$_{\rm low}$ = 694.0~K), indicates that the protostars have already reached high temperatures, which in turn suggests an advanced state of evolution. As cluster CM1 shows slightly higher mean temperatures than cluster CM4, the majority of sources in this cluster appear to have progressed furthest in their development.
% CM2
In contrast to that, sources in cluster~CM2 exhibits the lowest mean abundances of CH$_3$OH, H$_2$CCO, CH$_3$CN, C$_2$H$_5$CN, HCCCN, OCS, and its vibrational state OCS,v$_2$=1 and the highest abundance of CCH. Furthermore, this cluster also shows the lowest mean temperature of methanol. Both together suggest that the sources in this cluster are in earlier stages of evolution.
% CM3
In addition, cluster~CM3 has the lowest mean abundances for SO, H$_2$CCN, H$_2$CS, HNCO, NH$_2$D, CH$_3$OCH$_3$, and H$_2$CO, while OCS, OCS,v$_2$=1, H$_2$CS, and HCCCN show the lowest mean temperatures, again indicating a fairly early stage of development.
% CM5
Cluster~CM5 containing only source A20 has the lowest abundances of SO$_2$ and its vibrational state SO$_2$,v$_2$=1, NO, and CH$_3$OCHO. At the same time, this source shows the lowest mean temperatures for SO, NO, and CH$_3$OCHO, which again suggests an early phase of evolution.
% CM6
Finally, cluster~CM6 has the lowest mean abundances for SiO, CH$_2$NH, CN, and NS of all clusters. In addition, this cluster exhibits the highest mean temperatures for NO on the one hand and the lowest mean temperatures for NS, H$_2 \! ^{34}$S, HCS$^+$, SO$_2$, SO$_2$,v$_2$=1, HC(O)NH$_2$, HNCO and H$_2$CO on the other, which makes it difficult to classify the cluster into an evolutionary stage.

% cluster in Sgr B2(N)
For Sgr~B2(N), we find a somewhat larger number of clusters, two of which consist of only two elements and one of which consists of only one element. The temperature and abundances dispersions for each cluster in Sgr~B2(N) are described in Fig.~\ref{fig:PCAVioloinSgrB2N}.
% CN1
Cluster~CN1 exhibits the lowest mean abundances of CH$_3$OH, H$_2$CCO, CH$_3$OCHO, and CN, as well as the lowest mean temperatures for CH$_3$OH, H$_2$CCO, and SiO. This could indicate that the sources in this cluster are in early stages of evolution.
% CN2
The cluster~CN2 shows the lowest mean abundances for C$_2$H$_5$OH, CH$_3$CHO, CH$_3$NH$_2$, HC(O)NH$_2$, C$_2$H$_3$CN, C$_2$H$_3$CN,v$_{10}$=1, C$_2$H$_5$CN, C$_2$H$_5$CN,v$_{20}$=1, HCCCN,v$_7$=2, SO$_2$, and SO$_2$,v$_2$=1. Furthermore, the lowest temperatures are also found in this cluster for many of these molecules; that is, CH$_3$OCH$_3$, C$_2$H$_5$OH, CH$_3$OCHO, CH$_3$CHO, CH$_2$NH, HC(O)NH$_2$, C$_2$H$_3$CN, C$_2$H$_5$CN, H$_2 \! ^{34}$S, SO$_2$, SO$_2$,v$_2$=1, and OCS,v$_2$=1. At the same time, this cluster contains the highest abundances for NO and SO and the highest mean temperatures for NO, C$_2$H$_3$CN,v$_{10}$=1, and HCCCN,v$_7$=2. In addition, cluster~CN2 is the only cluster that contains NH$_2$D. This behavior indicates a very heterogeneous source structure, so that no statement can be made about the state of development.
% CN3
For cluster~CN3, we found the highest mean abundances for several molecules -- H$_2$CO, SiO, CH$_3$OCH$_3$, CH$_2$NH, C$_2$H$_3$CN, C$_2$H$_3$CN,v$_{10}$=1, C$_2$H$_5$CN, C$_2$H$_5$CN,v$_{20}$=1, HCCCN,v$_7$=2, H$_2 \! ^{34}$S, SO$_2$, and NS -- and the highest mean temperatures for CH$_3$OH, H$_2$CCO, SiO, C$_2$H$_5$OH, CH$_3$CHO, CH$_2$NH, HNCO, C$_2$H$_3$CN, C$_2$H$_5$CN, C$_2$H$_5$CN,v$_{20}$=1, SO, SO$_2$,v$_2$=1, OCS, and OCS,v$_2$=1. The lowest temperature of H$_2$CO, CH$_3$NH$_2$, and CH$_3$NH$_2$ is also found in this cluster. In summary, the high temperatures and the high abundances of some vibrationally excited states indicate that the majority of the sources in this cluster are at higher evolutionary stages.
% CN4
Cluster~CN4 exhibits the highest mean abundances of CH$_3$OH, H$_2$CCO, C$_2$H$_5$OH, CH$_3$OCHO, CH$_3$CHO, CH$_3$NH$_2$, HNCO, HC(O)NH$_2$, HCCCN, H$_2$CS, and OCS,v$_2$=1, as well as the highest mean temperatures for H$_2$CO, CH$_3$OCHO, CH$_3$NH$_2$, HC(O)NH$_2$, HCCCN, H$_2 \! ^{34}$S, SO$_2$, and H$_2$CS. In contrast, NS has the lowest concentration, and C$_2$H$_3$CN,v$_{10}$=1, C$_2$H$_5$CN,v$_{20}$=1, HCCCN,v$_7$=2, and NS have the lowest mean temperatures. This could indicate that the sources in this cluster are in an intermediate stage of development, at which the temperature is still too low to excite vibrationally excited states sufficiently.
% CN5
The cluster~CN5, which contains only source A13, shows the lowest mean abundances for CH$_3$OCH$_3$, CH$_2$NH, H$_2 \! ^{34}$S, and OCS on the one hand and the highest mean temperatures for CH$_3$OCH$_3$ and NS on the other. In addition, we find that HCCCN has by far the lowest excitation temperature of all clusters. This unusual behavior could indicate a more complex source structure.
% CN6
In the case of cluster~CN6, the mean abundance is lowest for H$_2$CO, HNCO, SO, and OCS,v$_2$=1, and highest for CH$_3$CN, CN, and CCH. Additionally, the lowest mean temperatures are found for CH$_3$CN, H$_2$CS, and CCH, while the highest temperature is found for CN within this cluster. As with the previous cluster~CN5, this behavior indicates again a more complex source structure.
% CN7
Finally, cluster~CN7, shows the lowest mean abundances for SiO, NO, CH$_3$CN, HCCCN, and H$_2$CS, and the lowest mean temperatures for NO and OCS. However, this cluster also shows the highest temperature for CH$_3$CN, which may be due to inaccuracies in the fit. The sources contained in this cluster have been found to contain only a small number of molecules, which also show relatively low mean temperatures. This indicates that the sources in this cluster are still in a relatively young phase of evolution.

% Kendall coefficient
The Kendall correlation coefficient \citepads{kendall1938measure} analysis offers another possibility to find correlations between different sources in a more rigorous and unbiased way. The coefficient, commonly referred to as Kendall's $\tau$ coefficient\footnote{Here, we use the Kendall instead of the Pearson \citepads{pearson1901} or Spearman \citepads{Spearman1904} correlation coefficient, because the Kendall correlation is more robust and efficient, making it preferred when dealing with small samples or outliers. In addition, the Kendall coefficient does not require continuous data and can be used with ordinal data, making it suitable for variables with inconsistent intervals between ranks.}, is applied to the calculated abundances of the core components for the different sources for both regions, whose source velocities are greater than 22~km~s$^{-1}$. The coefficient is calculated for each pair of sources and is given by
\begin{align}\label{pca:Kendall}
    \tau = \frac{2}{n(n-1)} \sum_{i=1}^{n-1} \sum_{j=i+1}^{n} \text{sign}(x_i - x_j) \, \text{sign}(y_i - y_j),
\end{align}
where $n$ represents the number of paired observations and $x_i$, $x_j$ are the observations of the first and $y_i$, $y_j$ are the observations of the second variable, respectively. The coefficient is used to measure the ordinal association between two measured quantities. It is particularly suitable for ordinal variables and is a non-parametric measure of the strength and direction of association that exists between two variables. The coefficient ranges from -1 to 1, where 1 indicates perfect agreement, -1 indicates perfect disagreement, and 0 indicates no association. Setting an (arbitrary) threshold at $\ge|0.9|$ to indicate a good correlation (see Figs.~\ref{fig:KendallSgrB2M}~-~\ref{fig:KendallSgrB2N}).
% Kendall for Sgr B2(M)
% A08 - A14 (both cluster CM4)
% A18 - A19 (both cluster CM6)
% A18 - A21 (both cluster CM6)
% A19 - A21 (both cluster CM6)
% A21 - A24 (both cluster CM6)
For sources in Sgr~B2(M) we find strong correlations between A08 and A14 (both are located in cluster~CM4 and are next to each other), A18 and A19, A18 and A21, A19 and A21, and between A21 and A24 (all contained in cluster~CM6). Source A20, on the other hand, shows a remarkably low correlation ($\le|0.5|$) with most of the other sources. There appears to be a somewhat greater correlation only with sources A16, A17, A22 and A25, although these sources are not located in the vicinity of A20.
% Kendall for Sgr B2(N)
% A09 - A11 (A09: CN6, A11: CN7)
% A09 - A14 (both cluster CN6)
% A09 - A15 (both cluster CN6)
% A11 - A12 (both cluster CN7)
% A11 - A16 (A11: CN7, A16: CN1)
The correlations for sources in Sgr B2(N) are generally smaller than in Sgr B2(M). Strong correlations are found between sources A09 and A14, and between A09 and A15 (all sources are located in cluster~CN6) and between A11 and A12 (both part of cluster~CN7). In addition, there are also strong correlations between sources A09 and A11 and between A11 and A16 although the respective sources are not in the same cluster. For example, source A09 belongs to cluster~CN6 and source A11 to cluster~CN7. The correlation between sources A11 and A16 is remarkable, as neither source is neighboring and A16, in contrast to source A11, contains an \hii~region, which is why both sources are in different evolutionary stages. There is also a particularly low correlation between sources A13 and A17.

%*******************************************************************************
% Figure: evolutionary sequence
\begin{figure*}[!tb]
   \centering
   \includegraphics[scale=0.90]{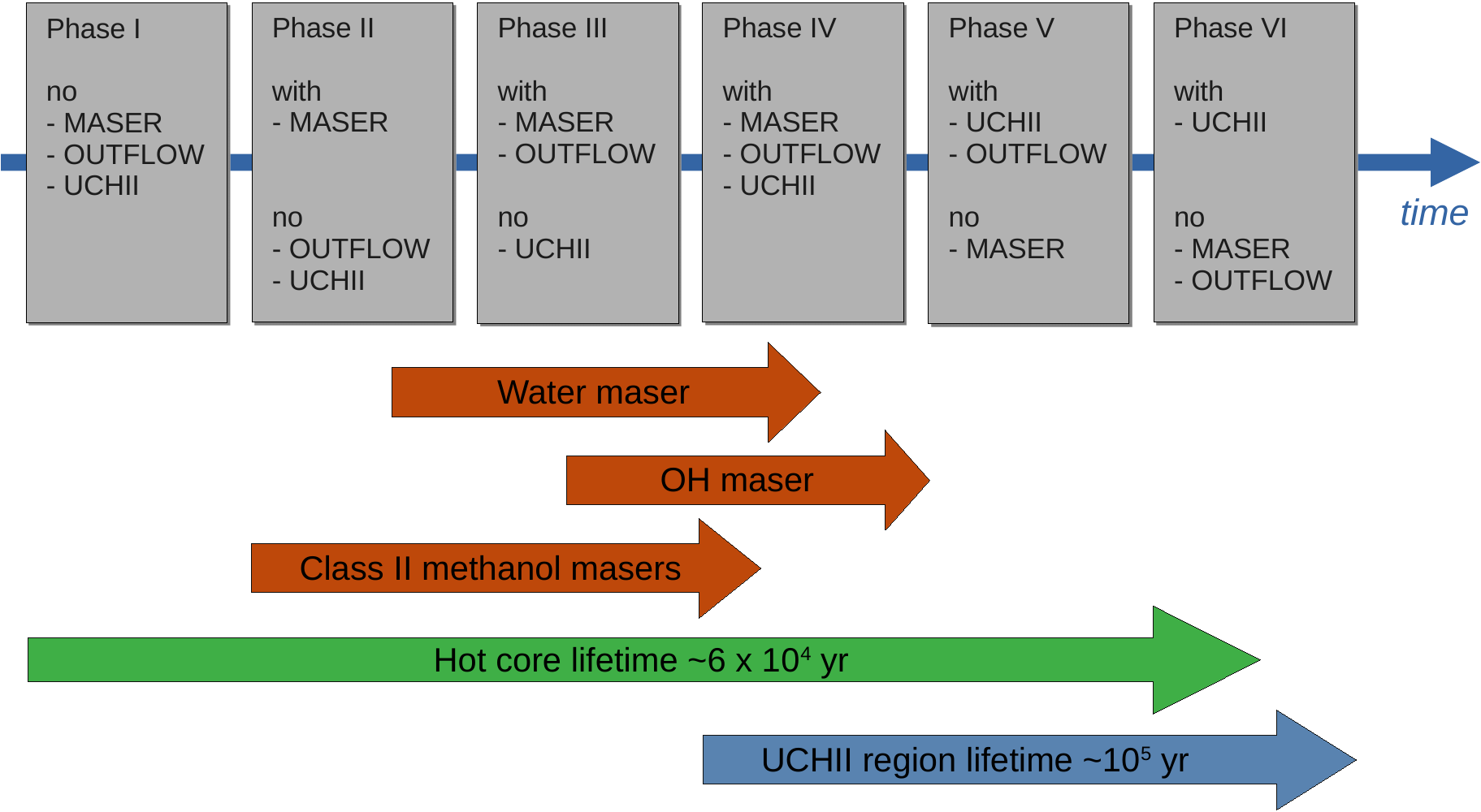}\\
   \caption{Evolutionary sequence as proposed by \citetads{2017A&A...604A..60B} for the hot cores embedded in the Sgr~B2 complex based on Fig. 5 of \citetads{2004A&A...417..615C}, along with the “straw man” model \citepads{2007IAUS..242..213E} that shows the lifetimes of various maser species.}
   \label{fig:EvolutionarySequence}
\end{figure*}

%-------------------------------------------------------------------------------
% Evolutionary sequence in Sgr~B2(M) and N
\subsection{Evolutionary sequence in Sgr~B2(M) and N}\label{subsec:EvolSeq}

Following \citetads{2004A&A...417..615C} and \citetads{2017A&A...604A..60B}, we used the distributions of masers, outflows and \hii~regions to determine the evolutionary sequence and age of the different hot cores in the Sgr~B2 complex (see Fig.~\ref{fig:EvolutionarySequence}).

\citetads{2004A&A...417..615C} proposed an evolutionary scheme for high-mass star forming regions, where class~II methanol masers are formed before \hii~regions can be observed, because the masers are thought to be associated with deeply embedded, high-mass protostars that are not evolved enough to ionize the surrounding gas and produce a detectable \hii~region. In the next phase, maser and \hii~region coexist, where the methanol molecules are shielded from the central UV radiation by the warm dust in the slowly expanding molecular envelope of the UC\hii~region, whose emission also provides its pumping photons. After $2.5 \times 10^4$ to $4.5 \times 10^4$~yr \citepads[depending on the assumed initial mass function (IMF), ][]{2005MNRAS.360..153V}, the maser emission stops as the UC\hii~region continues to expand. However, this analysis is complicated by the fact that we have averaged the spectra of each source over several pixels in order to obtain a better signal-to-noise ratio. This means that the averaged spectra may contain contributions from different protostars. For example, the polygon of core A01 in Sgr~B2(M) contains six \hii~regions and more than 33 masers (see Table~\ref{Tab:EvolPhases}), indicating that core A01 includes multiple protostars.
% Sources that also contain multiple \hii~regions and/or multiple masers are A02, A03, A05, A06, A07, A08, A10, A11 and A16 in Sgr~B2(M) and A01, A02, A03, A04 and A08 in Sgr~B2(N).

%-------------------------------------------------------------------------------
% maser
\subsubsection{Maser}\label{subsec:maser}

%*******************************************************************************
% Figure: Maser in Sgr~B2
\begin{figure*}[p]%\ContinuedFloat
   \begin{subfigure}[t]{0.99\textwidth}
   \centering
   \includegraphics[width=0.90\textwidth]{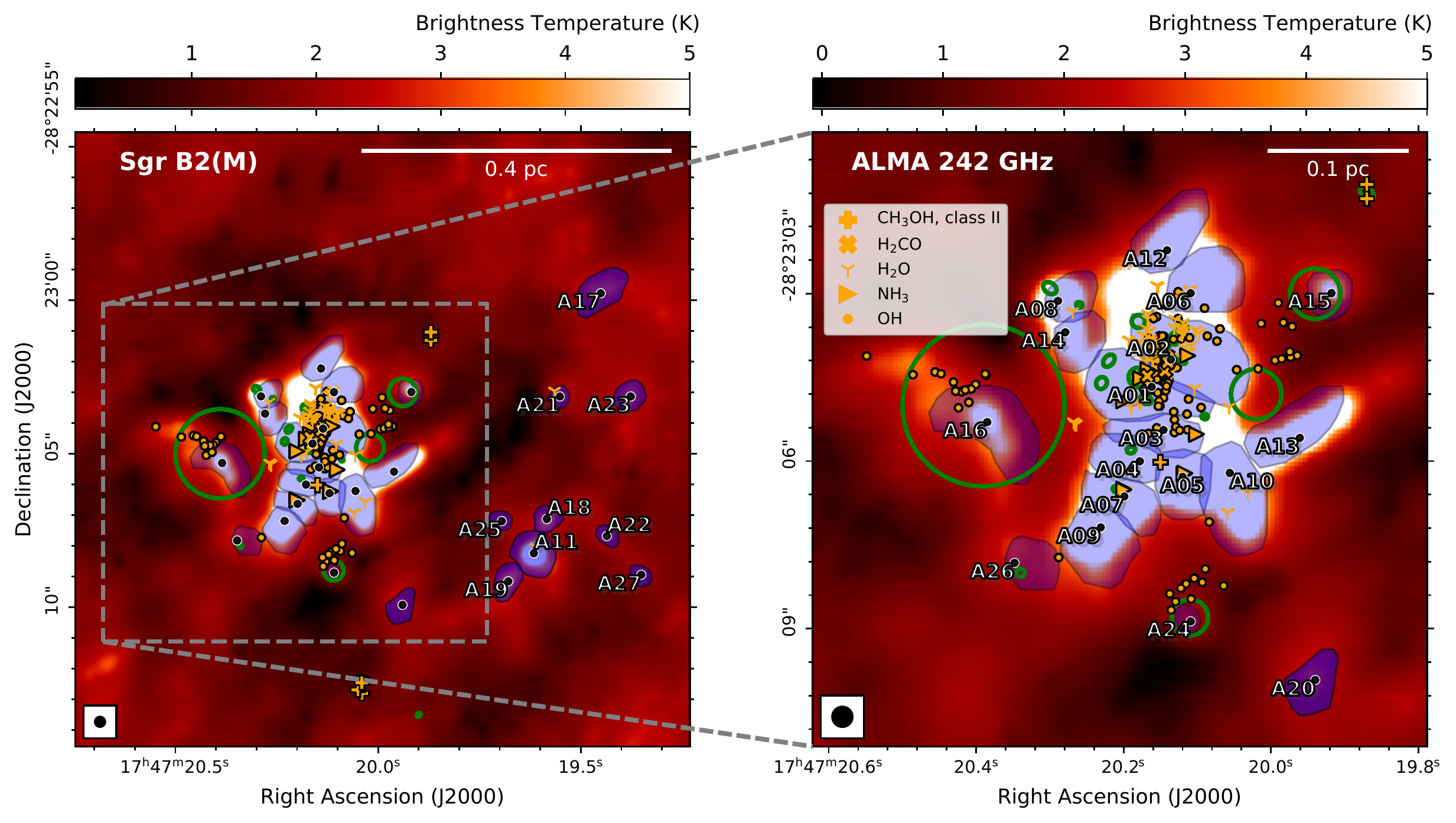}\\
   \caption{Masers in Sgr~B2(M).}
   \label{fig:MaserPosSgrB2M}
   \end{subfigure}
\quad
   \begin{subfigure}[b]{0.99\textwidth}
   \centering
   \includegraphics[width=0.90\textwidth]{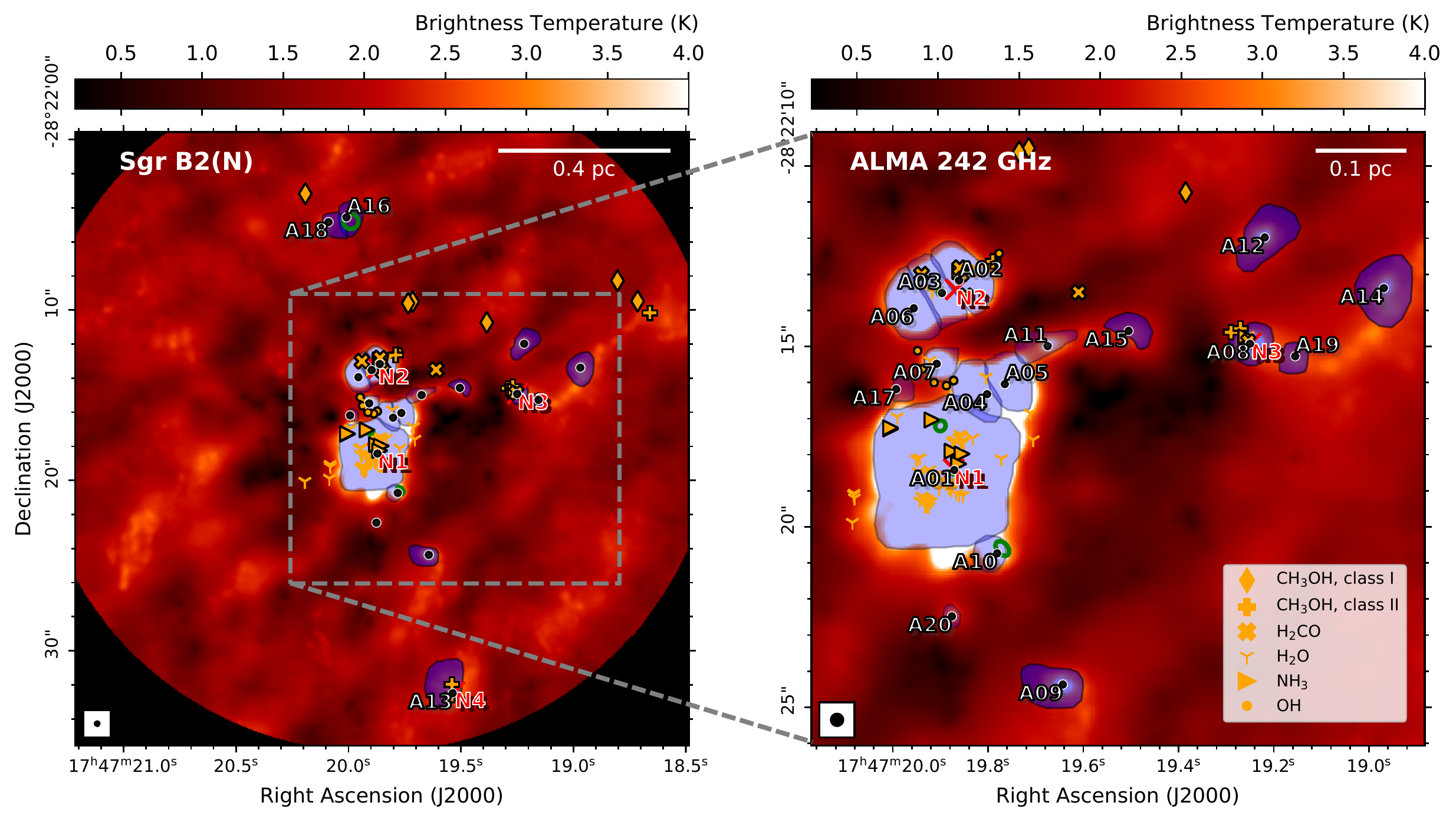}\\
   \caption{Masers in Sgr~B2(N).}
   \label{fig:MaserPosSgrB2N}
   \end{subfigure}
   \caption{Sources in Sgr~B2(M) and N together with the positions of the different masers. A close-up of the central part is presented in the right panel. The identified sources are marked with shaded blue polygons and indicated with the corresponding source ID. Similar to Figs.~\ref{fig:CorePosSgrB2M}~-~\ref{fig:CorePosSgrB2N}, the shaded light blue polygons describe the inner cores of the corresponding region. The black points indicate the positions of each source described by \citetads{2017A&A...604A...6S}. The intensity color scale is shown in units of brightness temperature and the synthesized beam of $0 \farcs 4$ is described in the lower left corner. The green ellipses describe the \hii~regions identified by \citetads{2015ApJ...815..123D}. The markers indicate the positions of published maser detections from the 44~GHz class~I \citepads{1997ApJ...474..346M} and from the 6.7~GHz class~II methanol masers (\citetads{1996MNRAS.283..606C}, \citetads{2016ApJ...833...18H}, \citetads{2019ApJS..244...35L}). Additionally, the positions from H$_2$CO (\citetads{1987IAUS..115..161W}, \citetads{1994ApJ...434..237M}, \citetads{2017IAUS..322...99L}), H$_2$O (\citetads{2004ApJS..155..577M}), NH$_3$ (\citetads{1999ApJ...519..667M}, \citetads{2018ApJ...869L..14M}, \citetads{2022A&A...666L..15Y}), OH (\citetads{1990ApJ...351..538G}, \citetads{2016ApJ...819L..35Y}), and SiO (\citetads{1992PASJ...44..373M}, \citetads{1997ApJ...478..206S}, \citetads{2009ApJ...691..332Z}) masers are described as well. The red crosses and names indicate the hot cores in Sgr~B2(N) identified by \citetads{2017A&A...604A..60B}.}
   \label{fig:MaserPosSgrB2}
\end{figure*}

%*******************************************************************************
% Figure: Outflows in Sgr~B2
\begin{figure*}[!htb]
   \centering
   \includegraphics[width=0.99\textwidth]{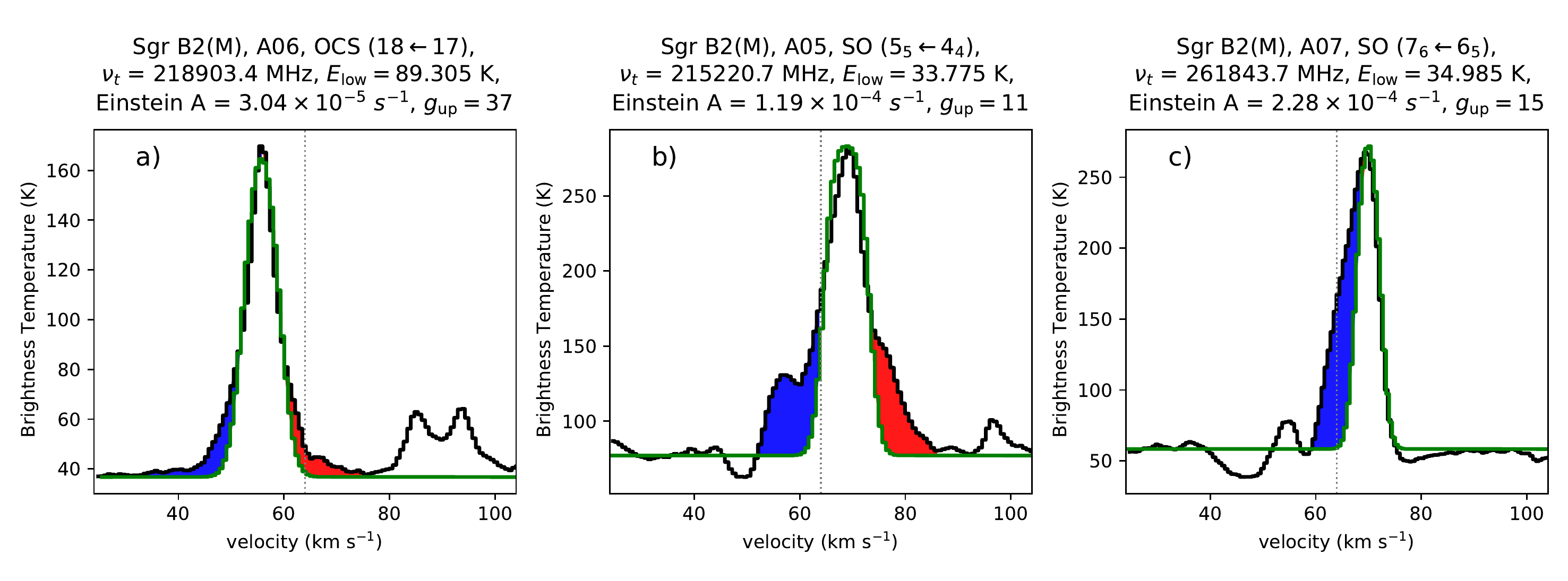}\\
   \caption{Example of transitions used to identify outflows toward Sgr~B2(M) and N. The vertical dashed line marks the systemic velocity of the Sgr~B2 complex and the high velocity wings are indicated in blue and red. The black line describes the observational data, while the green line represents a fit with a single Gaussian to emphasize the non-Gaussian wings of each peak. a) Some transitions show more or less symmetric, non-Gaussian line shapes. b) In addition, some transitions contain contributions from an isotopolog or another molecule. c) There are also some transitions that show a strongly asymmetric line shape.}
   \label{fig:Outflow}
\end{figure*}

As is described by Fig.~\ref{fig:MaserPosSgrB2} many sources in Sgr~B2(M) and N contain a variety of masers.
%
% methanol masers
Methanol masers are commonly found in the early stages of high-mass star formation and are associated with the presence of hot molecular cores (\citetads{2010A&A...517A..56F}, \citetads{2023A&A...675A.112Y}). They are known to trace the processes of mass accretion and outflow in these region. They are classified into two series of transitions called class~I and class~II.
%
% general overview of class~II masers
Class~I methanol masers are excited by collisional processes, such that they are often found in outflows and interacting regions between outflows and dense ambient gases, usually offset from the associated protostellar object \citepads{2014ApJ...789L...1M}. The $7_0 - 6_1 A^+$ methanol maser at 44~GHz is representative of the class~I methanol maser and was found only in the envelope of Sgr~B2(N), see Fig.~\ref{fig:MaserPosSgrB2}. Class~II methanol masers are radiatively excited and closely associated with high-mass protostars, because the pumping requires dust temperatures \mbox{$>150$}~K, high methanol column densities ($>2 \times 10^{15}$~cm$^{-2}$), and moderate hydrogen densities ($n_{\rm H} < 10^8$~cm$^{-3}$). The $5_1 - 6_0 A^+$ class~II methanol transition at 6.7~GHz in particular is the strongest and most widespread of the methanol masers \citepads{1997A&A...324..211S}. The sources\footnote{The class~II methanol maser is located at the edge of sources A04 and A05 and cannot be assigned to one of the two sources alone.} A04 / A05 in Sgr~B2(M) as well as A08 and A13 in Sgr~B2(N) contain a single class~II methanol maser, respectively.
%
% H2O masers
% Sgr~B2(M): A01, A02, A06, A07, A08, A10, A11, and A21
% Sgr~B2(N): A01, A02, A03, A04, A07, and A08
Moreover, H$_2$O masers, are primarily located in the inner sources of Sgr~B2(M) (A01, A02, A06, A07, A08, A10, A11, and A21) and N (A01, A02, A03, A04, A07, and A08). Similar to methanol masers, water masers tend to disappear as an UC\hii~region develops. These masers are also directly connected to molecular outflows \citepads{2004A&A...417..615C}. Furthermore, they are often associated with shocks resulting from the interaction of dense molecular outflows with the surrounding medium \citepads{1989ApJ...346..983E}. The characteristics of H$_2$O masers can help to understand the distribution of water vapor and the impact of shocks and radiation on its abundance \citepads{2021A&A...648A..24V}.
%
% NH3 masers
% Sgr~B2(M): A01, A02, A03, A05, and A07
% Sgr~B2(N): A01, A02, and A08
Moreover, ammonia (NH$_3$) masers are also found in the early stages of high-mass star formation and are associated with the presence of dense, hot molecular cores (\citetads{2020ApJ...898..157M}, \citetads{2022A&A...659A...5Y}). They are known to trace the presence of high-density gas and strong shocks and are often used to study the processes of mass accretion and outflow. Similar to H$_2$O masers, NH$_3$ masers are predominantly found in the inner sources of Sgr~B2(M) (A01, A02, A03, A05, and A07) and N (A01, A02, and A08).
%
% H2CO masers
% Sgr~B2(M): A01, and A02
% Sgr~B2(N): A02, A03, and A08
Additionally, formaldehyde (H$_2$CO) masers are relatively rare and have been observed in a few high-mass star-forming regions. They are known to indicate the presence of high-velocity shocks and are found in many sources, especially in the inner parts of Sgr~B2(M) (A01, and A02) and N (A02, A03, and A08), respectively.
%
% OH maser
% Sgr~B2(M): A01, A02, A03, A06, A10, A16, and A26
% Sgr~B2(N): A04
% We therefore use these masers in our analysis as a further characteristic for evolutionary phases II, III and IV.
Finally, OH masers are typically associated with the late stages of high-mass star formation and are often found in regions with evolved stars or supernova remnants \citepads{2006A&A...450..607W}. Additionally, OH masers are known to detect the presence of high-speed shocks and are often used to study the magnetic fields in these regions, among other things. The sources in Sgr~B2(M) in particular contain a large number of OH masers (A01, A02, A03, A06, A10, A16, and A26), whereas only source A04 in Sgr~B2(N) contains such a maser.

%-------------------------------------------------------------------------------
% outflows
\subsubsection{Outflows}\label{subsec:outflows}

The evolutionary sequence, as is shown in Fig.~\ref{fig:EvolutionarySequence}, requires the identification of outflows. In general, the study of outflows in high-mass star-forming regions can provide crucial information about the formation and evolution of massive stars. The momentum injected by outflows may limit accretion onto the massive star, potentially setting an upper mass limit to the stellar initial mass function. Additionally, there is a decrease in jet activity with evolutionary time in high-mass star-forming regions \citepads{2007prpl.conf..245A}.

% outflow contributions (has to be re-analyzed)
The line profiles of the outflow tracers (e.g., SiO, SO, and OCS) show high velocity emissions and non-Gaussian line-wings that are likely associated with out flowing gas at blue- and red-shifted velocities compared to the systemic velocity of the source (see Fig.~\ref{fig:Outflow}). In addition to outflows, rotation and infall motions can also lead to red and blue asymmetric line profiles, where infall usually produces stronger blue- than red-shifted emission \citepads{2005A&A...442..949F}. Here, we assume that all non-Gaussian line-wings indicate outflows. (Due to its large optical depth, SiO shows self-absorbing line profiles, which is why we do not consider it further in the analysis.)

% What are the advantages of using SO as an outflow tracer
Sulfur monoxide (SO) and carbonyl sulfide (OCS) are excellent outflow tracers due to their strong, high-velocity line wings, and spatial distributions. In high-mass star-forming regions, SO and OCS have several advantages as outflow tracers. Unlike other molecules, SO and OCS are less confused by emission and absorption by unrelated molecular clouds, making these valuable tracers of outflow activity in these regions \citepads{2001A&A...372..281H}.

% why are blue- and red wings sometimes not visible, explain sources of errors and problems
When analyzing the outflow components, it is important to consider that the line shape may contain other contributions. For instance, a filament in front of the source, which also contains an outflow tracer such as SO, can impact the line shape. Depending on the temperature of this gas, it can result in a lateral peak or an additional absorption structure in the line shape, as illustrated in panel~b) and panel~c) in Fig.~\ref{fig:Outflow}, respectively. The detection of outflows in the inner sources in Sgr~B2(N) is complicated by strong overlap of other molecular lines. In order to be able to estimate these contributions, the gas would have to be examined in detail not only in the hot cores but also between the cores in order to precisely determine the expansions of the filaments and the outflows.

%-------------------------------------------------------------------------------
% Classification according to evolutionary phase
\subsubsection{Classification according to evolutionary phase}\label{subsec:class:phase}

\begin{figure}[!htb]
   \centering
   \includegraphics[width=0.49\textwidth]{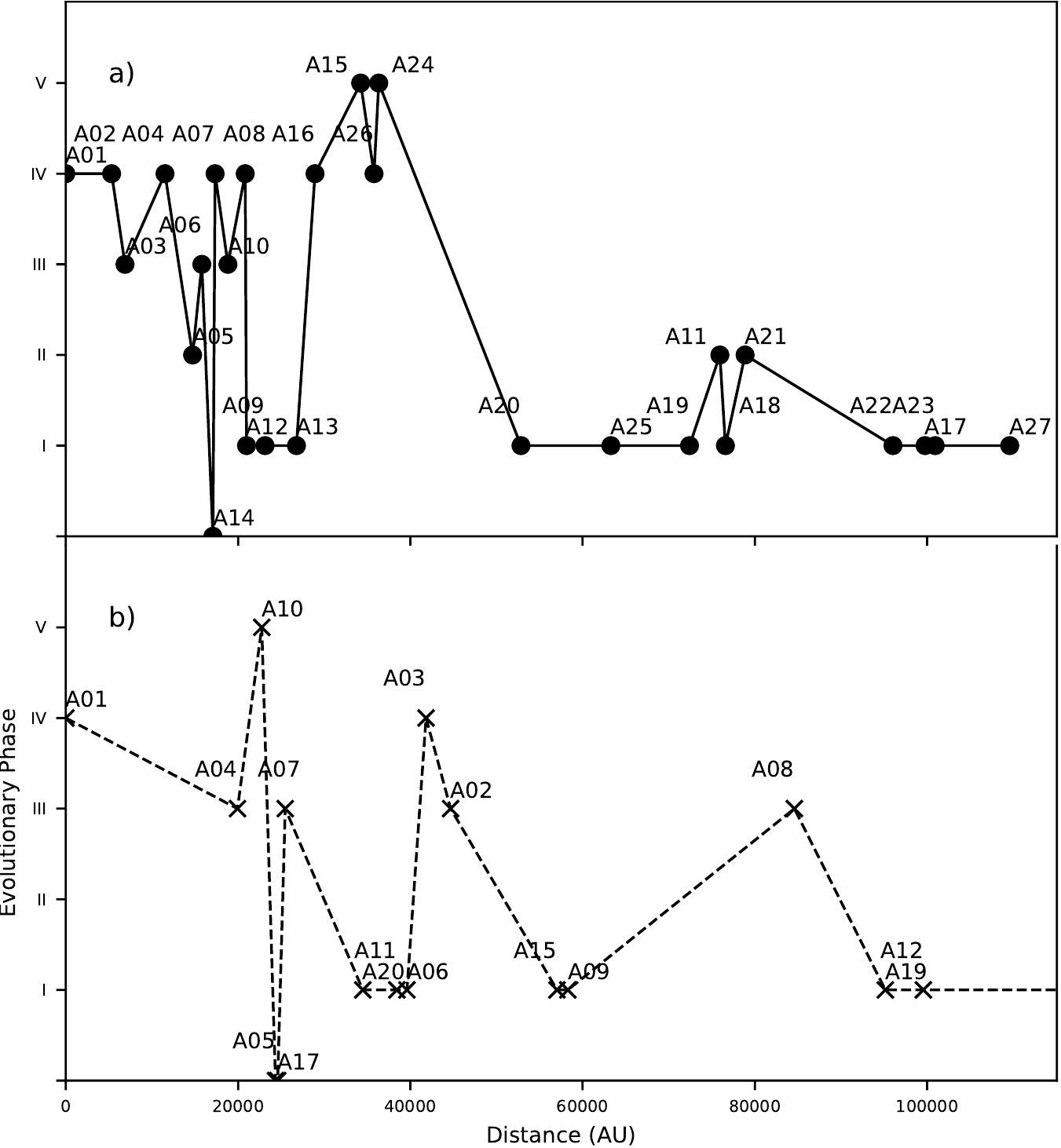}\\
   \caption{Evolutionary phases as function of the distance from the center from the respective region for a) Sgr~B2(M) and b) Sgr~B2(N).}
   \label{fig:EvolPhaseDistance}
\end{figure}

The classification of each sources in Sgr~B2(M) and N into the individual evolutionary phases described in Sect.~\ref{subsec:EvolSeq}, is shown in Table~\ref{Tab:EvolPhases}. For sources A14 in Sgr~B2(M) and A05 and A17 in Sgr~B2(N) we found evidences of outflows, but neither masers nor \hii~regions. Perhaps the associated \hii~regions are too small, which is why they have not yet been discovered. However, the asymmetrical line profiles could also have other causes, such as infall motions. Sources A15 and A24 in Sgr~B2(M), which roughly correspond in position and size to a single \hii~region, and sources A10 and A16 in Sgr~B2(N) show the highest evolutionary phase. In general, the inner sources in the two regions Sgr~B2(M) and N, which are located within a distance of less than 50~au around the central core A01, are usually in a more advanced phase than the outer sources (see Fig.~\ref{fig:EvolPhaseDistance}).

Exceptions are the sources A24 and A26 in Sgr~B2(M) and A16 in Sgr~B2(N). We have identified the same evolutionary stages for sources A08 (associated with N3) and A13 (associated with N4) in Sgr~B2(N) as are reported by \citetads{2017A&A...604A..60B}. However, we assign the inner source A01 (N1) to an earlier evolutionary stage due to the presence of H$_2$O masers. According to our analysis, source A02 (N2) is at a significantly later evolutionary stage as classified by \citetads{2017A&A...604A..60B}, as we not only consider the occurrence of H$_2$O maser, but also identify outflows in this source. This difference may be attributed to the fact that Bonfand's source N2 in our analysis comprises two sources (A02 and A03).

%*******************************************************************************
% Table: Phase of different cores
\begin{table*}[!htb]
    \centering
    \caption{Evolutionary phases of the individual sources in Sgr~B2(M) and N.}
    \rowcolors{2}{gray!25}{white} % Alternate row colors
    \begin{tabular}{lrrrrr}
        \hline
        \hline
        Source: & Number of     & Number and type of maser:                                                                       & Outflow:  & Evolution-Phase: & Cluster:\\
                & \hii~regions: &                                                                                                 &           &           &                  \\
        \hline
        \rowcolor{white} % Manually set the background color of the first row of Sgr~B2(N) to white
        \multicolumn{6}{l}{Sgr~B2(M)} \\
        \hline
            A01 &           6   & 1 $\times$ H$_2$CO; 5 $\times$ H$_2$O; 3 $\times$ NH$_3$; $>$24 $\times$ OH                     &  OCS; SO  & IV & CM1 \\
            A02 &         6.5   & 2 $\times$ H$_2$CO; $>$33 $\times$ H$_2$O; 1 $\times$ NH$_3$; $>$36 $\times$ OH                 &  OCS; SO  & IV & CM4 \\
            A03 &           -   & 1 $\times$ NH$_3$; 7 $\times$ OH                                                                &  OCS; SO  & III  & CM1 \\
            A04 &           1   & 1 $\times$ CH3OH, class II$^*$                                                                  &  SO       & IV  & CM1 \\
            A05 &           -   & 2 $\times$ NH$_3$; 1 $\times$ CH$_3$OH, class II$^*$                                            &  -        & II  & CM1 \\
            A06 &           -   & 3 $\times$ H$_2$O; 2 $\times$ OH                                                                &  OCS; SO  & III  & CM1 \\
            A07 &           1   & 1 $\times$ H$_2$O; 2 $\times$ NH$_3$                                                            &  OCS; SO  & IV  & CM1 \\
            A08 &         1.5   & 1 $\times$ H$_2$O                                                                               &  OCS; SO  & IV  & CM4 \\
            A09 &           -   & -                                                                                               &  -        & I  & CM4 \\
            A10 &           -   & 2 $\times$ H$_2$O; 1 $\times$ OH                                                                &  OCS      & III  & CM1 \\
            A11 &           -   & 2 $\times$ H$_2$O                                                                               &  -        & II  & CM2 \\
            A12 &           -   & -                                                                                               &  -        & I  & CM4 \\
            A13 &           -   & -                                                                                               &  -        & I  & CM4 \\
            A14 &           -   & -                                                                                               &  OCS; SO  & ?  & CM4 \\
            A15 &           1   & -                                                                                               &  SO       & V  & CM4 \\
            A16 &           1   & 7 $\times$ OH                                                                                   &  OCS      & IV  & CM6 \\
            A17 &           -   & -                                                                                               &  -        & I  & CM3 \\
            A18 &           -   & -                                                                                               &  -        & I  & CM6 \\
            A19 &           -   & -                                                                                               &  -        & I  & CM6 \\
            A20 &           -   & -                                                                                               &  -        & I  & CM5 \\
            A21 &           -   & 1 $\times$ H$_2$O                                                                               &  -        & II  & CM6 \\
            A22 &           -   & -                                                                                               &  -        & I  & CM3 \\
            A23 &           -   & -                                                                                               &  -        & I  & CM2 \\
            A24 &           1   & -                                                                                               &  SO       & V  & CM6 \\
            A25 &           -   & -                                                                                               &  -        & I  & CM3 \\
            A26 &           1   & 1 $\times$ OH                                                                                   &  OCS      & IV  & CM6 \\
            A27 &           -   & -                                                                                               &  -        & I  & CM6 \\
        \hline
        \rowcolor{white} % Manually set the background color of the first row of Sgr~B2(N) to white
        \multicolumn{6}{l}{Sgr~B2(N)} \\
        \hline
            A01 &           2   & $>$47 $\times$ H$_2$O; 5 $\times$ NH$_3$                                                        & OCS; SO   & IV  & CN3 \\
            A02 &           -   & 3 $\times$ H$_2$CO; 3 $\times$ H$_2$O; 1 $\times$ NH$_3$                                        & OCS; SO   & III  & CN3 \\
            A03 &           1   & 1 $\times$ H$_2$CO; 1 $\times$ H$_2$O                                                           & OCS       & IV  & CN4 \\
            A04 &           -   & 1 $\times$ H$_2$O; 1 $\times$ OH                                                                & OCS       & III  & CN4 \\
            A05 &           -   & -                                                                                               & OCS       & ?  & CN2 \\
            A06 &           -   & -                                                                                               & -         & I  & CN2 \\
            A07 &           -   & 1 $\times$ H$_2$O                                                                               & OCS       & III  & CN2 \\
            A08 &           -   & 1 $\times$ CH$_3$OH, class II; 1 $\times$ H$_2$CO; 1 $\times$ H$_2$O; 1 $\times$ NH$_3$         & OCS       & III  & CN2 \\
            A09 &           -   & -                                                                                               & -         & I  & CN6 \\
            A10 &           1   & -                                                                                               & OCS; SO   & V  & CN2 \\
            A11 &           -   & -                                                                                               & -         & I  & CN7 \\
            A12 &           -   & -                                                                                               & -         & I  & CN7 \\
            A13 &           -   & 1 $\times$ CH$_3$OH, class II                                                                   & -         & II  & CN5 \\
            A14 &           -   & -                                                                                               & -         & I  & CN6 \\
            A15 &           -   & -                                                                                               & -         & I  & CN6 \\
            A16 &           1   & -                                                                                               & OCS       & V  & CN1 \\
            A17 &           -   & -                                                                                               & OCS; SO   & ?  & CN3 \\
            A18 &           -   & -                                                                                               & -         & I  & CN6 \\
            A19 &           -   & -                                                                                               & -         & I  & CN1 \\
            A20 &           -   & -                                                                                               & -         & I  & CN1 \\
        \hline
        \rowcolors{0}{} \\
    \end{tabular}
    \tablefoot{When an \hii~region is not entirely contained within a source, it results in a non-integer quantity of \hii~regions. The CH$_3$OH class~II maser, which is marked with an asterisk ($^*$), is located at the edge of sources A04 and A05 and cannot be assigned to one of the two sources alone.}
    \label{Tab:EvolPhases}
\end{table*}

%-------------------------------------------------------------------------------
% chemical clocks
\subsection{Chemical clocks}\label{subsec:ChemicalClocks}

%*******************************************************************************
% Table: Evolutionary sequences according to chemical clocks
\begin{table*}[!htb]
    \centering
    \caption{Evolutionary sequences of the individual sources in Sgr~B2(M) and N.}
    \tiny
    \begin{tabular}
{{l@{{\hspace{{0.2cm}}}}l@{{\hspace{{0.2cm}}}}l@{{\hspace{{0.2cm}}}}l@{{\hspace{{0.2cm}}}}l@{{\hspace{{0.2cm}}}}
  l@{{\hspace{{0.2cm}}}}l@{{\hspace{{0.2cm}}}}l@{{\hspace{{0.2cm}}}}l@{{\hspace{{0.2cm}}}}l@{{\hspace{{0.2cm}}}}
  l@{{\hspace{{0.2cm}}}}l@{{\hspace{{0.2cm}}}}l@{{\hspace{{0.2cm}}}}l@{{\hspace{{0.2cm}}}}l@{{\hspace{{0.2cm}}}}
  l@{{\hspace{{0.2cm}}}}l@{{\hspace{{0.2cm}}}}l@{{\hspace{{0.2cm}}}}l@{{\hspace{{0.2cm}}}}l@{{\hspace{{0.2cm}}}}}}
        \hline
        \hline
        Abundance Ratio: & \multicolumn{18}{l@{\hspace{0.2cm}}}{Sources (young $\longrightarrow$ old):} \\
        \hline
        \hline
        \multicolumn{19}{l}{Sgr~B2(M)} \\
        \hline
        SO / OCS & A24 & A09 & A02 & A20 & A26 & A05 & A07 & A10 & A13 & A03 & A01 & A12 & A06 & A04 & A14 & A08 \\
        SO$_2$ / SO & A20 & A17 & A26 & A04 & A10 & A08 & A05 & A14 & A09 & A13 & A12 & A16 & A06 & A07 & A03 & A15 & A01 & A24 & A02 \\
        SO$_2$ / OCS & A22 & A20 & A26 & A09 & A24 & A10 & A05 & A07 & A13 & A03 & A04 & A01 & A12 & A02 & A06 & A08 & A14 \\
        SO$_2$, v$_2$=1 / OCS, v$_2$=1 & A22 & A20 & A26 & A07 & A24 & A09 & A05 & A03 & A13 & A04 & A02 & A12 & A06 & A08 & A14 \\
        CH$_3$CN / CH$_3$OH & A11 & A24 & A09 & A26 & A13 & A10 & A22 & A20 & A05 & A27 & A07 & A03 & A04 & A23 & A06 \\
        C$_2$H$_3$CN / HCCCN & A02 & A09 & A01 & A05 & A07 & A03 & A10 & A06 & A04 \\
        C$_2$H$_3$CN, v$_{10}$=1 / HCCCN, v$_7$=2 & A02 & A03 & A01 & A06 \\
        C$_2$H$_5$CN, v$_{20}$=1 / HCCCN, v$_7$=2 & A02 & A03 & A07 & A01 & A04 & A05 \\
        C$_2$H$_5$CN / HNCO & A06 & A04 & A13 & A01 & A02 & A03 & A07 & A05 & A09 & A10 \\
        CH$_3$CN / C$_2$H$_5$CN & A11 & A10 & A05 & A13 & A09 & A07 & A02 & A03 & A04 & A01 & A06 \\
        C$_2$H$_3$CN / C$_2$H$_5$CN & A07 & A05 & A10 & A09 & A04 & A02 & A03 & A06 & A01 \\
        C$_2$H$_3$CN, v$_{10}$=1 / C$_2$H$_5$CN, v$_{20}$=1 & A03 & A01 & A02 \\
        \hline
        \multicolumn{19}{l}{Sgr~B2(N)} \\
        \hline
        SO / OCS & A02 & A03 & A20 & A04 & A01 & A06 & A10 & A08 & A16 & A05 & A07 & A17 \\
        SO$_2$ / SO & A08 & A05 & A06 & A17 & A07 & A10 & A04 & A02 & A03 & A01 \\
        SO$_2$ / OCS & A08 & A02 & A06 & A05 & A03 & A04 & A10 & A17 & A07 & A01 \\
        SO$_2$, v$_2$=1 / OCS, v$_2$=1 & A08 & A06 & A05 & A02 & A03 & A07 & A04 & A01 & A10 \\
        CH$_3$CN / CH$_3$OH & A13 & A08 & A10 & A05 & A07 & A04 & A17 & A03 & A06 & A02 & A20 & A01 \\
        C$_2$H$_3$CN / HCCCN & A03 & A06 & A04 & A02 & A01 & A08 \\
        C$_2$H$_3$CN, v$_{10}$=1 / HCCCN, v$_7$=2 & A06 & A04 & A02 & A03 & A01 & A08 \\
        C$_2$H$_5$CN, v$_{20}$=1 / HCCCN, v$_7$=2 & A01 & A03 & A04 & A02 & A05 \\
        C$_2$H$_5$CN / HNCO & A10 & A01 & A03 & A08 & A04 & A06 & A17 & A02 \\
        CH$_3$CN / C$_2$H$_5$CN & A02 & A08 & A03 & A01 & A04 & A10 & A06 & A17 & A07 & A05 \\
        C$_2$H$_3$CN / C$_2$H$_5$CN & A02 & A06 & A04 & A03 & A08 & A01 \\
        C$_2$H$_3$CN, v$_{10}$=1 / C$_2$H$_5$CN, v$_{20}$=1 & A04 & A02 & A03 & A01 \\
        \hline
    \end{tabular}
    \tablefoot{Evolutionary sequences of the individual sources in Sgr~B2(M) and N based on abundance ratios of some chemical clocks. The sources in each row are ordered from young to old (evolved), whereby the absolute age is different for each row.}
    \label{Tab:ChemClocks}
\end{table*}

In addition to the methods described above, the abundances of certain molecules can be used to determine the age of a source. Chemical tracers that show empirical relations between their abundances and the different phases of the star formation process are often referred to as chemical clocks (see e.g., \citetads{1997ApJ...481..396C}; \citetads{1998A&A...338..713H}). Some molecules become abundant (or disappear) after a certain evolutionary phase, while other tracers may show a constant increase (or decrease) throughout the evolutionary sequence. Chemical clocks are based on the relative abundances ratios of different molecular species. In the following we use the abundances derived in Sect.~\ref{sec:Results} for the core components\footnote{Here we focus on the abundances of the core components, as we can identify them quite reliably with the hot cores and not with filaments or with the envelope.} (see Fig.~\ref{fig:AbundCoreSgrB2M} and Fig.~\ref{fig:AbundCoreSgrB2N}), to estimate the evolutionary sequence of each source in Sgr~B2(M) and N.

% S-bearing molecules
\citetads{1997ApJ...481..396C}, \citetads{1998A&A...338..713H} and \citetads{2009A&A...504..853H} propose the usage of the relative abundance ratios of OCS, SO, SO$_2$ and H$_2$S as chemical clocks, because the main reservoir of sulfur on grain mantles is H$_2$S. (Recent observations by \citetads{2023NatAs...7..431M} using the James Webb Space Telescope (JWST) have provided new insights into the composition of ices in dense molecular clouds. Contrary to previous assumptions, H$_2$S was not detected as a major component in these ices. \citetads{2020ApJ...888...52S} identified cyclic octatomic sulfur (S$_8$) as a potential alternative reservoir of sulfur on grain surfaces. However, for the purposes of the further analysis, we shall maintain the assumption that H$_2$S remains the primary reservoir for sulfur on dust mantles. This approach allows coherence with existing models while taking into account the need for future refinements based on these new observations.) When the dust temperature in the hot core exceeds the mantle evaporation temperature, all components of the grain mantles, including hydrogen sulfide, are suddenly injected into the gas phase \citepads{2004A&A...422..159W}. Subsequently, endothermic reactions in the hot gas transform H$_2$S into atomic sulfur and SO, leading to the formation of more SO and afterward to SO$_2$ on timescales of $\sim10^3$~yr. Therefore, the SO$_2$/SO and SO/H$_2$S ratios are functions of time and can thus indicate the age of the hot cores \citepads{2004A&A...422..159W}. In addition, \citetads{2011A&A...529A.112W} predict an increase in the SO$_2$/OCS and SO$_2$/H$_2$S ratios for sources with more evolved chemistry. Since H$_2$S has no transition in the frequency ranges covered by our observations, we only use the ratios of SO$_2$/OCS and SO$_2$/SO to estimate the age.

% cyanides as chemical clocks
In addition to sulfur-bearing molecules, complex cyanides, especially those containing the CN group, have also been identified as possible chemical clocks in high-mass star-forming regions. \citetads{2009A&A...499..215B} and \citetads{2017A&A...601A..48G} suggested that the formation of C$_2$H$_5$CN on grains occurs through the hydrogenation of C$_2$H$_3$CN, which is formed from HCCCN during the collapse and warm-up phase. Subsequently, C$_2$H$_5$CN sublimates to the gas phase at a temperature of 100~K \citepads{2017A&A...601A..48G}. The majority of the C$_2$H$_5$CN formed on the grains is subsequently eliminated by gas-phase reactions. In their chemical simulations, \citetads{2014Sci...345.1584B} observed a reduction in the gas phase abundance of C$_2$H$_5$CN and an increase in the abundance of C$_2$H$_3$CN after approximately $\sim3.5 \times 10^5$~yr. Here, we use the ratio of ethyl cyanide to vinyl cyanide as a chemical clock, similar to \citetads{2017A&A...604A..60B}. According to \citetads{2016ApJ...824...99M}, the formation of CH$_3$CN is anticipated to occur at a later stage than CH$_3$OH. Therefore, comparing the abundances of these two species can serve as a reliable indicator of the evolutionary timescale during the initial phases of massive star formation. The CH$_3$CN/CH$_3$OH abundance ratios are expected to be low during the very early phase of grain mantle evaporation and increases over time by the production of CH$_3$CN and the destruction of CH$_3$OH through its transformation into other complex species. As was discussed by \citetads{2001ApJ...546..324R}, CH$_3$OH maintains a high abundance until the age of around $10^4$~yr, after which it begins to be depleted. Over the next $10^5$~yr, the abundance of CH$_3$CN is expected to increase, reaching a level only a few times less than that of CH$_3$OH. However, \citetads{2022ApJS..259....1G} have revealed a more nuanced understanding of cyanide formation and abundance in these environments. While the formation of C$_2$H$_5$CN on grains through hydrogenation of C$_2$H$_3$CN remains a viable pathway, the gas-phase chemistry of cyanides at high temperatures has been shown to play a significant role. The revised models indicate that certain N-bearing molecules, especially cyanides, experience substantial enhancement through gas-phase chemistry at temperatures above 100~K. HCN and CH$_3$CN, in particular, see their greatest gas-phase production at very high temperatures exceeding 300~K, with considerable production also occurring between 120~-~160~K as ice-mantle material sublimates. This new understanding complicates the use of cyanide abundance ratios as simple chemical clocks. The CH$_3$CN/CH$_3$OH abundance ratio, previously thought to increase steadily over time due to CH$_3$CN production and CH$_3$OH destruction, may now be influenced by high-temperature gas-phase chemistry. This additional production pathway for CH$_3$CN could potentially accelerate its abundance increase relative to CH$_3$OH, altering the expected temporal evolution of their ratio.

% limitations
However, it is important to note that the ratios are also influenced by physical conditions and the composition of grain mantles, making their interpretation as standalone indicators of age challenging \citepads{2018MNRAS.474.5575V}. Determining the exact age of each source would require complex chemical modeling of each individual source, which would go beyond the scope of this paper. Furthermore, we cannot completely exclude the possibility that one or the other emission belongs to a filament rather than a hot core, which would falsify the determination of the abundances as well. It should also be noted that the sources identified by \citetads{2017A&A...604A...6S} sometimes consist of several individual hot cores.

% Description of results
The evolutionary sequences obtained from the different chemical clocks are described in Table~\ref{Tab:ChemClocks}, where the sources in each row are ordered from young to old (evolved). In addition to the abundance ratios described above, we also consider the abundance ratios of vibrationally excited molecules, as these species describe the innermost, hottest part of each source. The orders of the sources based on the ratios of vibrationally excited molecules are broadly consistent with the orders determined using the abundance ratios of non-vibrationally excited molecules. In general, we do not obtain a uniform order of the various sources in Sgr~B2(M) and N. In the following discussion, the evolution phase of a source, as is described in Table~\ref{Tab:EvolPhases}, is indicated in round brackets.

%*******************************************************************************
% Figure: Outflows in Sgr~B2

% Sgr B2(M)
% For sources in Sgr~B2(M), the abundance ratios containing sulfur-bearing molecules indicate A20 (I) and A26 (IV) mostly as young, A03 (III), A13 (I) and A12 (I) as intermediate, and A14 (no classification) and A08 (IV) as old sources. While we find a good agreement with the classification from the previous analysis for sources A20, A03 and A08, we obtain a clear deviation for source A26. A possible explanation for this deviation could lie in the fact that the \hii~region, which is decisive for the classification of the source into phase IV, only covers a small part of the source, so that it can be assumed that the source is not monolithic, but has a complex substructure consisting of several sources. Several chemical clocks based on cyanides classify A02 (IV), A01 (IV), and A10 (III) as predominantly young sources, A05 (II) and A07 (IV) as intermediate, and A03 (III), A04 (IV), and A06 (III) as evolved sources. However, there are notable discrepancies to the classifications of the previous analysis, which may be attributed to a more complex substructure. These sources sometimes exhibit a significant number of \hii~regions and/or multiple maser emissions.
For sources in Sgr~B2(M), sulfur-bearing molecule ratios suggest A20 (evolution phase I) and A26 (IV) as young sources, A03 (III), A13 (I) and A12 (I) as intermediate, and A14 (no classification) and A08 (IV) as evolved. These results largely align with previous analyses, with A26 being a notable exception, possibly due to its complex substructure. The \hii~region, which is decisive for the classification of the source A26 into phase IV, only covers a small part of the source, so that it can be assumed that the source is not monolithic, but has a complex substructure consisting of several sources. Cyanide-based ratios, however, present a different picture, classifying A0 (IV), A01 (IV), and A10 (III) as young, A05 (II) and A07 (IV) as intermediate, and A03 (III), A04 (IV), and A06 (III) as evolved. These discrepancies may be attributed to the potentially complex substructures within these sources, many of which contain multiple \hii~regions and/or maser emissions.

% Sgr B2(N)
% In Sgr~B2(N), the sulfur-bearing molecules indicate A08 (III) and A06 (I) as young, A04 (III) as intermediate, and A10 (V) as old sources, which agrees well with the classification of the previous analysis, except for source A08. In contrast, cyanide-based ratios classify A01 (IV) as a young, A06 (I), A04 (III), and A02 (III) as intermediate and A08 (III) as an evolved source. Here, the more complex substructure of sources A01 and A08 could explain the deviations from the previous analysis as well.
In Sgr~B2(N), our analysis using sulfur-bearing molecules indicates A08 (III) and A06 (I) as young sources, A04 (III) as intermediate, and A10 (V) as evolved, generally corroborating previous analyses except for A08. Conversely, cyanide-based ratios suggest A01 (IV) as young, A06 (I), A04 (III), and A02 (III) as intermediate, and A08 (III) as evolved. The disparities in classifications, particularly for A01 (IV) and A08 (III), may be attributed to their complex substructures.

% summary
Our study highlights the challenges in determining the evolutionary stages of sources within the Sgr~B2 complex. The results of our analyses for each source in Sgr~B2(M) and N are given in the Appendix~\ref{app:sec:Results}. The use of different chemical tracers can lead to varying age classifications, while the complex substructures of some sources contribute to discrepancies between different classification methods and previous analyses. These findings underscore the need for a multifaceted approach when studying the chemical evolution of star-forming regions, taking into account both molecular abundances and structural complexities.

%===============================================================================
% Conclusions
\section{Conclusions}\label{sec:Conclusions}

In this paper, we have described a comprehensive analysis of the molecular content of each hot core identified by \citetads{2017A&A...604A...6S}, where we took the complex interaction between molecular lines, dust attenuation, and free-free emission into account. We identified the chemical composition of the detected sources and derived physical parameters including column densities, temperatures, line widths, and velocities and determined their distributions, which can be used to draw conclusions about the source structure and possible connections to larger structures such as filaments. Finally, we estimated the age of the each hot core and identified correlations between sources using the PCA and the Kendall correlation coefficient analysis.

% final remarks
The hot cores in Sgr~B2(N) exhibit a greater variety of complex molecules such as CH$_3$OCH$_3$, CH$_3$NH$_2$, or NH$_2$CHO, while the cores in M contain more lighter molecules like NO, SO, and SO$_2$. However, some sulfur-bearing molecules such as H$_2$CS, CS, NS, and OCS are more abundant in N than in M. In both regions, the inner sources were found to be more evolved than the outer sources, with a few exceptions. Furthermore, some sources in Sgr~B2(N) seem to be embedded in filaments connecting different sources. These filaments seem to be quite warm, as we can find their possible contributions in the emission spectra of the individual sources. So far, however, such a filamentary structure has not yet been identified in Sgr~B2(M), but the distributions of the fit parameters suggest that filaments can also be found in Sgr~B2(M). The contribution of the filaments also influences the age determinations of each core. Assuming that sources that are at the same evolutionary phase also have a similar molecular composition, the contribution of filaments distorts this correlation and the results of the chemical clock, PCA and Kendall coefficient analysis may give incorrect results. In addition, the emission spectra of many sources located further out in Sgr~B2(M) and N show a significantly lower signal-to-noise ratio than the inner sources, which makes the identification of molecules and the determination of physical parameters more difficult. This in turn can also lead to incorrect conclusions about the age and possible correlations between individual sources. The classification of sources into different evolutionary stages based on characteristic features such as masers, outflows, or \hii~regions can also be falsified by various circumstances. For example, non-Gaussian line shapes can be caused not only by outflows but also by infall motions. Furthermore, the emission spectra of each source were averaged over several pixels, so that not every source is caused by a single protostar. This complex source structure can also be seen in the fact that some sources contain multiple masers and \hii~regions.

%___________________________________________________________________________
% acknowledgements
\begin{acknowledgements}
    This work was supported by the Deutsche Forschungsgemeinschaft (DFG) through grant Collaborative Research Centre 956 (subproject A6 and C3, project ID 184018867) and from BMBF/Verbundforschung through the projects ALMA-ARC 05A14PK1 and ALMA-ARC 05A20PK1. A.S.-M.\ acknowledges support from the RyC2021-032892-I grant funded by MCIN/AEI/10.13039/501100011033 and by the European Union `Next GenerationEU'/PRTR, as well as the program Unidad de Excelencia Mar\'ia de Maeztu CEX2020-001058-M, and support from the PID2020-117710GB-I00 (MCI-AEI-FEDER, UE). This paper makes use of the following ALMA data: ADS/JAO.ALMA\#2013.1.00332.S. ALMA is a partnership of ESO (representing its member states), NSF (USA) and NINS (Japan), together with NRC (Canada), MOST and ASIAA (Taiwan), and KASI (Republic of Korea), in cooperation with the Republic of Chile. The Joint ALMA Observatory is operated by ESO, AUI/NRAO and NAOJ. In order to do the analysis and plots, we used the Python packages \texttt{scipy}, \texttt{numpy}, \texttt{pandas}, \texttt{matplotlib}, \texttt{astropy}, \texttt{spectral\_cube}, \texttt{regions}, \texttt{aplpy}, \texttt{seaborn}, and \texttt{scikit-learn}.
\end{acknowledgements}

    %___________________________________________________________________________
    % bibliography

\bibliographystyle{aa}
\bibliography{SgrB2ADS.bib}

\begin{thebibliography}{201}
\expandafter\ifx\csname natexlab\endcsname\relax\def\natexlab#1{#1}\fi

\bibitem[{{Acharyya} \& {Herbst}(2018)}]{2018ApJ...859...51A}
{Acharyya}, K. \& {Herbst}, E. 2018, \apj, 859, 51

\bibitem[{{Allegrinia} {et~al.}(1979){Allegrinia}, {Johns}, \&
  {McKellar}}]{1979JChPh..70.2829A}
{Allegrinia}, M., {Johns}, J.~W.~C., \& {McKellar}, A.~R.~W. 1979, \jcp, 70,
  2829

\bibitem[{{ALMA Partnership} {et~al.}(2015){ALMA Partnership}, {Fomalont},
  {Vlahakis}, {Corder}, {Remijan}, {Barkats}, {Lucas}, {Hunter}, {Brogan},
  {Asaki}, {Matsushita}, {Dent}, {Hills}, {Phillips}, {Richards}, {Cox},
  {Amestica}, {Broguiere}, {Cotton}, {Hales}, {Hiriart}, {Hirota}, {Hodge},
  {Impellizzeri}, {Kern}, {Kneissl}, {Liuzzo}, {Marcelino}, {Marson},
  {Mignano}, {Nakanishi}, {Nikolic}, {Perez}, {P{\'e}rez}, {Toledo}, {Aladro},
  {Butler}, {Cortes}, {Cortes}, {Dhawan}, {Di Francesco}, {Espada}, {Galarza},
  {Garcia-Appadoo}, {Guzman-Ramirez}, {Humphreys}, {Jung}, {Kameno}, {Laing},
  {Leon}, {Mangum}, {Marconi}, {Nagai}, {Nyman}, {Radiszcz}, {Rod{\'o}n},
  {Sawada}, {Takahashi}, {Tilanus}, {van Kempen}, {Vila Vilaro}, {Watson},
  {Wiklind}, {Gueth}, {Tatematsu}, {Wootten}, {Castro-Carrizo}, {Chapillon},
  {Dumas}, {de Gregorio-Monsalvo}, {Francke}, {Gallardo}, {Garcia}, {Gonzalez},
  {Hibbard}, {Hill}, {Kaminski}, {Karim}, {Krips}, {Kurono}, {Lopez}, {Martin},
  {Maud}, {Morales}, {Pietu}, {Plarre}, {Schieven}, {Testi}, {Videla},
  {Villard}, {Whyborn}, {Zwaan}, {Alves}, {Andreani}, {Avison}, {Barta},
  {Bedosti}, {Bendo}, {Bertoldi}, {Bethermin}, {Biggs}, {Boissier}, {Brand},
  {Burkutean}, {Casasola}, {Conway}, {Cortese}, {Dabrowski}, {Davis}, {Diaz
  Trigo}, {Fontani}, {Franco-Hernandez}, {Fuller}, {Galvan Madrid},
  {Giannetti}, {Ginsburg}, {Graves}, {Hatziminaoglou}, {Hogerheijde}, {Jachym},
  {Jimenez Serra}, {Karlicky}, {Klaasen}, {Kraus}, {Kunneriath}, {Lagos},
  {Longmore}, {Leurini}, {Maercker}, {Magnelli}, {Marti Vidal}, {Massardi},
  {Maury}, {Muehle}, {Muller}, {Muxlow}, {O'Gorman}, {Paladino}, {Petry},
  {Pineda}, {Randall}, {Richer}, {Rossetti}, {Rushton}, {Rygl}, {Sanchez
  Monge}, {Schaaf}, {Schilke}, {Stanke}, {Schmalzl}, {Stoehr}, {Urban}, {van
  Kampen}, {Vlemmings}, {Wang}, {Wild}, {Yang}, {Iguchi}, {Hasegawa}, {Saito},
  {Inatani}, {Mizuno}, {Asayama}, {Kosugi}, {Morita}, {Chiba}, {Kawashima},
  {Okumura}, {Ohashi}, {Ogasawara}, {Sakamoto}, {Noguchi}, {Huang}, {Liu},
  {Kemper}, {Koch}, {Chen}, {Chikada}, {Hiramatsu}, {Iono}, {Shimojo},
  {Komugi}, {Kim}, {Lyo}, {Muller}, {Herrera}, {Miura}, {Ueda}, {Chibueze},
  {Su}, {Trejo-Cruz}, {Wang}, {Kiuchi}, {Ukita}, {Sugimoto}, {Kawabe},
  {Hayashi}, {Miyama}, {Ho}, {Kaifu}, {Ishiguro}, {Beasley}, {Bhatnagar},
  {Braatz}, {Brisbin}, {Brunetti}, {Carilli}, {Crossley}, {D'Addario}, {Donovan
  Meyer}, {Emerson}, {Evans}, {Fisher}, {Golap}, {Griffith}, {Hale},
  {Halstead}, {Hardy}, {Hatz}, {Holdaway}, {Indebetouw}, {Jewell}, {Kepley},
  {Kim}, {Lacy}, {Leroy}, {Liszt}, {Lonsdale}, {Matthews}, {McKinnon}, {Mason},
  {Moellenbrock}, {Moullet}, {Myers}, {Ott}, {Peck}, {Pisano}, {Radford},
  {Randolph}, {Rao Venkata}, {Rawlings}, {Rosen}, {Schnee}, {Scott}, {Sharp},
  {Sheth}, {Simon}, {Tsutsumi}, \& {Wood}}]{2015ApJ...808L...1A}
{ALMA Partnership}, {Fomalont}, E.~B., {Vlahakis}, C., {et~al.} 2015, \apjl,
  808, L1

\bibitem[{{Anderl} {et~al.}(2013){Anderl}, {Guillet}, {Pineau des For{\^e}ts},
  \& {Flower}}]{2013A&A...556A..69A}
{Anderl}, S., {Guillet}, V., {Pineau des For{\^e}ts}, G., \& {Flower}, D.~R.
  2013, \aap, 556, A69

\bibitem[{{Apponi} \& {Ziurys}(1997)}]{1997ApJ...481..800A}
{Apponi}, A.~J. \& {Ziurys}, L.~M. 1997, \apj, 481, 800

\bibitem[{{Arce} {et~al.}(2007){Arce}, {Shepherd}, {Gueth}, {Lee}, {Bachiller},
  {Rosen}, \& {Beuther}}]{2007prpl.conf..245A}
{Arce}, H.~G., {Shepherd}, D., {Gueth}, F., {et~al.} 2007, in Protostars and
  Planets V, ed. B.~{Reipurth}, D.~{Jewitt}, \& K.~{Keil}, 245

\bibitem[{{Bakker} {et~al.}(1996){Bakker}, {Waters}, {Lamers}, {Trams}, \& {van
  der Wolf}}]{1996A&A...310..893B}
{Bakker}, E.~J., {Waters}, L.~B.~F.~M., {Lamers}, H.~J.~G.~L.~M., {Trams},
  N.~R., \& {van der Wolf}, F.~L.~A. 1996, \aap, 310, 893

\bibitem[{{Balucani} {et~al.}(2015){Balucani}, {Ceccarelli}, \&
  {Taquet}}]{2015MNRAS.449L..16B}
{Balucani}, N., {Ceccarelli}, C., \& {Taquet}, V. 2015, \mnras, 449, L16

\bibitem[{{Barone} {et~al.}(2015){Barone}, {Latouche}, {Skouteris}, {Vazart},
  {Balucani}, {Ceccarelli}, \& {Lefloch}}]{2015MNRAS.453L..31B}
{Barone}, V., {Latouche}, C., {Skouteris}, D., {et~al.} 2015, \mnras, 453, L31

\bibitem[{{Barr} {et~al.}(2018){Barr}, {Boogert}, {DeWitt}, {Montiel},
  {Richter}, {Indriolo}, {Neufeld}, {Pendleton}, {Chiar}, {Dungee}, \&
  {Tielens}}]{2018ApJ...868L...2B}
{Barr}, A.~G., {Boogert}, A., {DeWitt}, C.~N., {et~al.} 2018, \apjl, 868, L2

\bibitem[{{Belloche} {et~al.}(2014){Belloche}, {Garrod}, {M{\"u}ller}, \&
  {Menten}}]{2014Sci...345.1584B}
{Belloche}, A., {Garrod}, R.~T., {M{\"u}ller}, H. S.~P., \& {Menten}, K.~M.
  2014, Science, 345, 1584

\bibitem[{{Belloche} {et~al.}(2009){Belloche}, {Garrod}, {M\"{u}ller},
  {Menten}, {Comito}, \& {Schilke}}]{2009A&A...499..215B}
{Belloche}, A., {Garrod}, R.~T., {M\"{u}ller}, H.~S.~P., {et~al.} 2009, \aap,
  499, 215

\bibitem[{{Belloche} {et~al.}(2019){Belloche}, {Garrod}, {M\"{u}ller},
  {Menten}, {Medvedev}, {Thomas}, \& {Kisiel}}]{2019A&A...628A..10B}
{Belloche}, A., {Garrod}, R.~T., {M\"{u}ller}, H.~S.~P., {et~al.} 2019, \aap,
  628, A10

\bibitem[{{Belloche} {et~al.}(2016){Belloche}, {M\"{u}ller}, {Garrod}, \&
  {Menten}}]{2016A&A...587A..91B}
{Belloche}, A., {M\"{u}ller}, H.~S.~P., {Garrod}, R.~T., \& {Menten}, K.~M.
  2016, \aap, 587, A91

\bibitem[{{Belloche} {et~al.}(2013){Belloche}, {M\"{u}ller}, {Menten},
  {Schilke}, \& {Comito}}]{2013A&A...559A..47B}
{Belloche}, A., {M\"{u}ller}, H.~S.~P., {Menten}, K.~M., {Schilke}, P., \&
  {Comito}, C. 2013, \aap, 559, A47

\bibitem[{Bengfort \& Bilbro(2019)}]{bengfort_yellowbrick_2018}
Bengfort, B. \& Bilbro, R. 2019, Journal of Open Source Software, 4, 1075

\bibitem[{{Beuther} {et~al.}(2008){Beuther}, {Semenov}, {Henning}, \&
  {Linz}}]{2008ApJ...675L..33B}
{Beuther}, H., {Semenov}, D., {Henning}, T., \& {Linz}, H. 2008, \apjl, 675,
  L33

\bibitem[{{Bisschop} {et~al.}(2007){Bisschop}, {J{\o}rgensen}, {van Dishoeck},
  \& {de Wachter}}]{2007A&A...465..913B}
{Bisschop}, S.~E., {J{\o}rgensen}, J.~K., {van Dishoeck}, E.~F., \& {de
  Wachter}, E.~B.~M. 2007, \aap, 465, 913

\bibitem[{{Blair} {et~al.}(2008){Blair}, {Magnani}, {Brand}, \&
  {Wouterloot}}]{2008AsBio...8...59B}
{Blair}, S.~K., {Magnani}, L., {Brand}, J., \& {Wouterloot}, J. G.~A. 2008,
  Astrobiology, 8, 59

\bibitem[{{Blake} {et~al.}(1996){Blake}, {Mundy}, {Carlstrom}, {Padin},
  {Scott}, {Scoville}, \& {Woody}}]{1996ApJ...472L..49B}
{Blake}, G.~A., {Mundy}, L.~G., {Carlstrom}, J.~E., {et~al.} 1996, \apjl, 472,
  L49

\bibitem[{{Boger} \& {Sternberg}(2005)}]{2005ApJ...632..302B}
{Boger}, G.~I. \& {Sternberg}, A. 2005, \apj, 632, 302

\bibitem[{{Bonfand} {et~al.}(2019){Bonfand}, {Belloche}, {Garrod}, {Menten},
  {Willis}, {St{\'e}phan}, \& {M\"{u}ller}}]{2019A&A...628A..27B}
{Bonfand}, M., {Belloche}, A., {Garrod}, R.~T., {et~al.} 2019, \aap, 628, A27

\bibitem[{{Bonfand} {et~al.}(2017){Bonfand}, {Belloche}, {Menten}, {Garrod}, \&
  {M\"{u}ller}}]{2017A&A...604A..60B}
{Bonfand}, M., {Belloche}, A., {Menten}, K.~M., {Garrod}, R.~T., \&
  {M\"{u}ller}, H.~S.~P. 2017, \aap, 604, A60

\bibitem[{{Cabezas} {et~al.}(2021){Cabezas}, {Endo}, \&
  {Cernicharo}}]{2021JMoSp.37711448C}
{Cabezas}, C., {Endo}, Y., \& {Cernicharo}, J. 2021, Journal of Molecular
  Spectroscopy, 377, 111448

\bibitem[{{Canelo} {et~al.}(2021){Canelo}, {Bronfman}, {Mendoza}, {Duronea},
  {Merello}, {Carvajal}, {Fria{\c{c}}a}, \& {Lepine}}]{2021MNRAS.504.4428C}
{Canelo}, C.~M., {Bronfman}, L., {Mendoza}, E., {et~al.} 2021, \mnras, 504,
  4428

\bibitem[{{Caswell}(1996)}]{1996MNRAS.283..606C}
{Caswell}, J.~L. 1996, \mnras, 283, 606

\bibitem[{{Ceccarelli} {et~al.}(2001){Ceccarelli}, {Loinard}, {Castets},
  {Tielens}, {Caux}, {Lefloch}, \& {Vastel}}]{2001A&A...372..998C}
{Ceccarelli}, C., {Loinard}, L., {Castets}, A., {et~al.} 2001, \aap, 372, 998

\bibitem[{{Cesaroni} \& {Walmsley}(1991)}]{1991A&A...241..537C}
{Cesaroni}, R. \& {Walmsley}, C.~M. 1991, \aap, 241, 537

\bibitem[{{Chantzos} {et~al.}(2020){Chantzos}, {Rivilla}, {Vasyunin},
  {Redaelli}, {Bizzocchi}, {Fontani}, \& {Caselli}}]{2020A&A...633A..54C}
{Chantzos}, J., {Rivilla}, V.~M., {Vasyunin}, A., {et~al.} 2020, \aap, 633, A54

\bibitem[{{Chapman} {et~al.}(2009){Chapman}, {Millar}, {Wardle}, {Burton}, \&
  {Walsh}}]{2009MNRAS.394..221C}
{Chapman}, J.~F., {Millar}, T.~J., {Wardle}, M., {Burton}, M.~G., \& {Walsh},
  A.~J. 2009, \mnras, 394, 221

\bibitem[{Charnley(2004)}]{CHARNLEY200423}
Charnley, S. 2004, Advances in Space Research, 33, 23, space Life Sciences:
  Steps Toward Origin(s) of Life

\bibitem[{{Charnley}(1997)}]{1997ApJ...481..396C}
{Charnley}, S.~B. 1997, \apj, 481, 396

\bibitem[{{Charnley} \& {Rodgers}(2008)}]{2008SSRv..138...59C}
{Charnley}, S.~B. \& {Rodgers}, S.~D. 2008, \ssr, 138, 59

\bibitem[{{Codella} {et~al.}(2004){Codella}, {Lorenzani}, {Gallego},
  {Cesaroni}, \& {Moscadelli}}]{2004A&A...417..615C}
{Codella}, C., {Lorenzani}, A., {Gallego}, A.~T., {Cesaroni}, R., \&
  {Moscadelli}, L. 2004, \aap, 417, 615

\bibitem[{{Corby} {et~al.}(2015){Corby}, {Jones}, {Cunningham}, {Menten},
  {Belloche}, {Schwab}, {Walsh}, {Balnozan}, {Bronfman}, {Lo}, \&
  {Remijan}}]{2015MNRAS.452.3969C}
{Corby}, J.~F., {Jones}, P.~A., {Cunningham}, M.~R., {et~al.} 2015, \mnras,
  452, 3969

\bibitem[{{Costagliola} \& {Aalto}(2010)}]{2010A&A...515A..71C}
{Costagliola}, F. \& {Aalto}, S. 2010, \aap, 515, A71

\bibitem[{{Coutens} {et~al.}(2018){Coutens}, {Willis}, {Garrod}, {M{\"u}ller},
  {Bourke}, {Calcutt}, {Drozdovskaya}, {J{\o}rgensen}, {Ligterink}, {Persson},
  {St{\'e}phan}, {van der Wiel}, {van Dishoeck}, \&
  {Wampfler}}]{2018A&A...612A.107C}
{Coutens}, A., {Willis}, E.~R., {Garrod}, R.~T., {et~al.} 2018, \aap, 612, A107

\bibitem[{{Cummins} {et~al.}(1986){Cummins}, {Linke}, \&
  {Thaddeus}}]{1986ApJS...60..819C}
{Cummins}, S.~E., {Linke}, R.~A., \& {Thaddeus}, P. 1986, \apjs, 60, 819

\bibitem[{{De Pree} {et~al.}(2014){De Pree}, {Peters}, {Mac Low}, {Wilner},
  {Goss}, {Galv{\'a}n-Madrid}, {Keto}, {Klessen}, \&
  {Monsrud}}]{2014ApJ...781L..36D}
{De Pree}, C.~G., {Peters}, T., {Mac Low}, M.~M., {et~al.} 2014, \apjl, 781,
  L36

\bibitem[{{De Pree} {et~al.}(2015){De Pree}, {Peters}, {Mac Low}, {Wilner},
  {Goss}, {Galv{\'a}n-Madrid}, {Keto}, {Klessen}, \&
  {Monsrud}}]{2015ApJ...815..123D}
{De Pree}, C.~G., {Peters}, T., {Mac Low}, M.~M., {et~al.} 2015, \apj, 815, 123

\bibitem[{{Dickinson} \& {Kuiper}(1981)}]{1981ApJ...247..112D}
{Dickinson}, D.~F. \& {Kuiper}, E.~N.~R. 1981, \apj, 247, 112

\bibitem[{{Doddipatla} {et~al.}(2021){Doddipatla}, {He}, {Goettl}, {Kaiser},
  {Galv{\~a}o}, \& {Millar}}]{2021SciA....7.7003D}
{Doddipatla}, S., {He}, C., {Goettl}, S.~J., {et~al.} 2021, Science Advances,
  7, eabg7003

\bibitem[{Duvernay {et~al.}(2005)Duvernay, Chiavassa, Borget, \&
  Aycard}]{doi:10.1021/jp0459256}
Duvernay, F., Chiavassa, T., Borget, F., \& Aycard, J.-P. 2005, The Journal of
  Physical Chemistry A, 109, 603, pMID: 16833385

\bibitem[{{Elitzur} {et~al.}(1989){Elitzur}, {Hollenbach}, \&
  {McKee}}]{1989ApJ...346..983E}
{Elitzur}, M., {Hollenbach}, D.~J., \& {McKee}, C.~F. 1989, \apj, 346, 983

\bibitem[{{Ellingsen} {et~al.}(2007){Ellingsen}, {Voronkov}, {Cragg},
  {Sobolev}, {Breen}, \& {Godfrey}}]{2007IAUS..242..213E}
{Ellingsen}, S.~P., {Voronkov}, M.~A., {Cragg}, D.~M., {et~al.} 2007, in
  Astrophysical Masers and their Environments, ed. J.~M. {Chapman} \& W.~A.
  {Baan}, Vol. 242, 213--217

\bibitem[{{Endres} {et~al.}(2016){Endres}, {Schlemmer}, {Schilke}, {Stutzki},
  \& {M\"{u}ller}}]{2016JMoSp.327...95E}
{Endres}, C.~P., {Schlemmer}, S., {Schilke}, P., {Stutzki}, J., \&
  {M\"{u}ller}, H. S.~P. 2016, Journal of Molecular Spectroscopy, 327, 95

\bibitem[{{Enrique-Romero} {et~al.}(2020){Enrique-Romero},
  {{\'A}lvarez-Barcia}, {Kolb}, {Rimola}, {Ceccarelli}, {Balucani}, {Meisner},
  {Ugliengo}, {Lamberts}, \& {K{\"a}stner}}]{2020MNRAS.493.2523E}
{Enrique-Romero}, J., {{\'A}lvarez-Barcia}, S., {Kolb}, F.~J., {et~al.} 2020,
  \mnras, 493, 2523

\bibitem[{{Enrique-Romero} {et~al.}(2016){Enrique-Romero}, {Rimola},
  {Ceccarelli}, \& {Balucani}}]{2016MNRAS.459L...6E}
{Enrique-Romero}, J., {Rimola}, A., {Ceccarelli}, C., \& {Balucani}, N. 2016,
  \mnras, 459, L6

\bibitem[{{Fabricant} {et~al.}(1977){Fabricant}, {Krieger}, \&
  {Muenter}}]{1977JChPh..67.1576F}
{Fabricant}, B., {Krieger}, D., \& {Muenter}, J.~S. 1977, \jcp, 67, 1576

\bibitem[{{Feketeov{\'a}} {et~al.}(2018){Feketeov{\'a}}, {Pelc}, {Ribar},
  {Huber}, \& {Denifl}}]{2018A&A...617A.102F}
{Feketeov{\'a}}, L., {Pelc}, A., {Ribar}, A., {Huber}, S.~E., \& {Denifl}, S.
  2018, \aap, 617, A102

\bibitem[{{Fontani} {et~al.}(2010){Fontani}, {Cesaroni}, \&
  {Furuya}}]{2010A&A...517A..56F}
{Fontani}, F., {Cesaroni}, R., \& {Furuya}, R.~S. 2010, \aap, 517, A56

\bibitem[{{Fontani} {et~al.}(2023){Fontani}, {Roueff}, {Colzi}, \&
  {Caselli}}]{2023A&A...680A..58F}
{Fontani}, F., {Roueff}, E., {Colzi}, L., \& {Caselli}, P. 2023, \aap, 680, A58

\bibitem[{Foreman-Mackey(2016)}]{corner}
Foreman-Mackey, D. 2016, The Journal of Open Source Software, 1, 24

\bibitem[{{Foreman-Mackey} {et~al.}(2013){Foreman-Mackey}, {Hogg}, {Lang}, \&
  {Goodman}}]{2013PASP..125..306F}
{Foreman-Mackey}, D., {Hogg}, D.~W., {Lang}, D., \& {Goodman}, J. 2013, \pasp,
  125, 306

\bibitem[{{Friedel} {et~al.}(2004){Friedel}, {Snyder}, {Turner}, \&
  {Remijan}}]{2004ApJ...600..234F}
{Friedel}, D.~N., {Snyder}, L.~E., {Turner}, B.~E., \& {Remijan}, A. 2004,
  \apj, 600, 234

\bibitem[{{Fuller} {et~al.}(2005){Fuller}, {Williams}, \&
  {Sridharan}}]{2005A&A...442..949F}
{Fuller}, G.~A., {Williams}, S.~J., \& {Sridharan}, T.~K. 2005, \aap, 442, 949

\bibitem[{Gans {et~al.}(2017)Gans, Boyé-Péronne, Garcia, Röder, Schleier,
  Halvick, \& Loison}]{doi:10.1021/acs.jpclett.7b01853}
Gans, B., Boyé-Péronne, S., Garcia, G.~A., {et~al.} 2017, The Journal of
  Physical Chemistry Letters, 8, 4038, pMID: 28796511

\bibitem[{{Garrod} {et~al.}(2017){Garrod}, {Belloche}, {M\"{u}ller}, \&
  {Menten}}]{2017A&A...601A..48G}
{Garrod}, R.~T., {Belloche}, A., {M\"{u}ller}, H.~S.~P., \& {Menten}, K.~M.
  2017, \aap, 601, A48

\bibitem[{{Garrod} \& {Herbst}(2006)}]{2006A&A...457..927G}
{Garrod}, R.~T. \& {Herbst}, E. 2006, \aap, 457, 927

\bibitem[{{Garrod} {et~al.}(2022){Garrod}, {Jin}, {Matis}, {Jones}, {Willis},
  \& {Herbst}}]{2022ApJS..259....1G}
{Garrod}, R.~T., {Jin}, M., {Matis}, K.~A., {et~al.} 2022, \apjs, 259, 1

\bibitem[{{Gaume} \& {Claussen}(1990)}]{1990ApJ...351..538G}
{Gaume}, R.~A. \& {Claussen}, M.~J. 1990, \apj, 351, 538

\bibitem[{{Gaume} {et~al.}(1995){Gaume}, {Claussen}, {de Pree}, {Goss}, \&
  {Mehringer}}]{1995ApJ...449..663G}
{Gaume}, R.~A., {Claussen}, M.~J., {de Pree}, C.~G., {Goss}, W.~M., \&
  {Mehringer}, D.~M. 1995, \apj, 449, 663

\bibitem[{{Gieser} {et~al.}(2021){Gieser}, {Beuther}, {Semenov}, {Ahmadi},
  {Suri}, {M{\"o}ller}, {Beltr{\'a}n}, {Klaassen}, {Zhang}, {Urquhart},
  {Henning}, {Feng}, {Galv{\'a}n-Madrid}, {de Souza Magalh{\~a}es},
  {Moscadelli}, {Longmore}, {Leurini}, {Kuiper}, {Peters}, {Menten},
  {Csengeri}, {Fuller}, {Wyrowski}, {Lumsden}, {S{\'a}nchez-Monge}, {Maud},
  {Linz}, {Palau}, {Schilke}, {Pety}, {Pudritz}, {Winters}, \&
  {Pi{\'e}tu}}]{2021A&A...648A..66G}
{Gieser}, C., {Beuther}, H., {Semenov}, D., {et~al.} 2021, \aap, 648, A66

\bibitem[{{Goldsmith} {et~al.}(1992){Goldsmith}, {Lis}, {Lester}, \&
  {Harvey}}]{1992ApJ...389..338G}
{Goldsmith}, P.~F., {Lis}, D.~C., {Lester}, D.~F., \& {Harvey}, P.~M. 1992,
  \apj, 389, 338

\bibitem[{{Goodman} \& {Weare}(2010)}]{2010CAMCS...5...65G}
{Goodman}, J. \& {Weare}, J. 2010, Communications in Applied Mathematics and
  Computational Science, 5, 65

\bibitem[{{Gratier} {et~al.}(2017){Gratier}, {Bron}, {Gerin}, {Pety}, {Guzman},
  {Orkisz}, {Bardeau}, {Goicoechea}, {Le Petit}, {Liszt}, {{\"O}berg},
  {Peretto}, {Roueff}, {Sievers}, \& {Tremblin}}]{2017A&A...599A.100G}
{Gratier}, P., {Bron}, E., {Gerin}, M., {et~al.} 2017, \aap, 599, A100

\bibitem[{{Gravity Collaboration} {et~al.}(2019){Gravity Collaboration},
  {Abuter}, {Amorim}, {Baub{\"o}ck}, {Berger}, {Bonnet}, {Brandner},
  {Cl{\'e}net}, {Coud{\'e} Du Foresto}, {de Zeeuw}, {Dexter}, {Duvert},
  {Eckart}, {Eisenhauer}, {F{\"o}rster Schreiber}, {Garcia}, {Gao}, {Gendron},
  {Genzel}, {Gerhard}, {Gillessen}, {Habibi}, {Haubois}, {Henning}, {Hippler},
  {Horrobin}, {Jim{\'e}nez-Rosales}, {Jocou}, {Kervella}, {Lacour},
  {Lapeyr{\`e}re}, {Le Bouquin}, {L{\'e}na}, {Ott}, {Paumard}, {Perraut},
  {Perrin}, {Pfuhl}, {Rabien}, {Rodriguez Coira}, {Rousset}, {Scheithauer},
  {Sternberg}, {Straub}, {Straubmeier}, {Sturm}, {Tacconi}, {Vincent}, {von
  Fellenberg}, {Waisberg}, {Widmann}, {Wieprecht}, {Wiezorrek}, {Woillez}, \&
  {Yazici}}]{2019A&A...625L..10G}
{Gravity Collaboration}, {Abuter}, R., {Amorim}, A., {et~al.} 2019, \aap, 625,
  L10

\bibitem[{{Gusdorf} {et~al.}(2008){Gusdorf}, {Cabrit}, {Flower}, \& {Pineau Des
  For{\^e}ts}}]{2008A&A...482..809G}
{Gusdorf}, A., {Cabrit}, S., {Flower}, D.~R., \& {Pineau Des For{\^e}ts}, G.
  2008, \aap, 482, 809

\bibitem[{Hartigan \& Wong(1979)}]{hartigan1979}
Hartigan, J. \& Wong, M. 1979, Applied Statistics, 28, 100

\bibitem[{{Hatchell} {et~al.}(2001){Hatchell}, {Fuller}, \&
  {Millar}}]{2001A&A...372..281H}
{Hatchell}, J., {Fuller}, G.~A., \& {Millar}, T.~J. 2001, \aap, 372, 281

\bibitem[{{Hatchell} {et~al.}(1998){Hatchell}, {Thompson}, {Millar}, \&
  {MacDonald}}]{1998A&A...338..713H}
{Hatchell}, J., {Thompson}, M.~A., {Millar}, T.~J., \& {MacDonald}, G.~H. 1998,
  \aap, 338, 713

\bibitem[{{Henshaw} {et~al.}(2023){Henshaw}, {Barnes}, {Battersby}, {Ginsburg},
  {Sormani}, \& {Walker}}]{2023ASPC..534...83H}
{Henshaw}, J.~D., {Barnes}, A.~T., {Battersby}, C., {et~al.} 2023, in
  Astronomical Society of the Pacific Conference Series, Vol. 534, Protostars
  and Planets VII, ed. S.~{Inutsuka}, Y.~{Aikawa}, T.~{Muto}, K.~{Tomida}, \&
  M.~{Tamura}, 83

\bibitem[{{Herbst} \& {Klemperer}(1973)}]{1973ApJ...185..505H}
{Herbst}, E. \& {Klemperer}, W. 1973, \apj, 185, 505

\bibitem[{{Herbst} \& {Leung}(1989)}]{1989ApJS...69..271H}
{Herbst}, E. \& {Leung}, C.~M. 1989, \apjs, 69, 271

\bibitem[{{Herbst} \& {van Dishoeck}(2009)}]{2009ARA&A..47..427H}
{Herbst}, E. \& {van Dishoeck}, E.~F. 2009, \araa, 47, 427

\bibitem[{{Herpin} {et~al.}(2009){Herpin}, {Marseille}, {Wakelam}, {Bontemps},
  \& {Lis}}]{2009A&A...504..853H}
{Herpin}, F., {Marseille}, M., {Wakelam}, V., {Bontemps}, S., \& {Lis}, D.~C.
  2009, \aap, 504, 853

\bibitem[{{Higuchi} {et~al.}(2014){Higuchi}, {Chibueze}, {Habe}, {Takahira}, \&
  {Takano}}]{2014AJ....147..141H}
{Higuchi}, A.~E., {Chibueze}, J.~O., {Habe}, A., {Takahira}, K., \& {Takano},
  S. 2014, \aj, 147, 141

\bibitem[{{Hoffman} {et~al.}(2007){Hoffman}, {Goss}, \&
  {Palmer}}]{2007ApJ...654..971H}
{Hoffman}, I.~M., {Goss}, W.~M., \& {Palmer}, P. 2007, \apj, 654, 971

\bibitem[{{Hollis} {et~al.}(2004){Hollis}, {Jewell}, {Lovas}, {Remijan}, \&
  {M{\o}llendal}}]{2004ApJ...610L..21H}
{Hollis}, J.~M., {Jewell}, P.~R., {Lovas}, F.~J., {Remijan}, A., \&
  {M{\o}llendal}, H. 2004, \apjl, 610, L21

\bibitem[{{Hu} {et~al.}(2016){Hu}, {Menten}, {Wu}, {Bartkiewicz}, {Rygl},
  {Reid}, {Urquhart}, \& {Zheng}}]{2016ApJ...833...18H}
{Hu}, B., {Menten}, K.~M., {Wu}, Y., {et~al.} 2016, \apj, 833, 18

\bibitem[{{Humire} {et~al.}(2020){Humire}, {Thiel}, {Henkel}, {Belloche},
  {Loison}, {Pillai}, {Riquelme}, {Wakelam}, {Langer},
  {Hern{\'a}ndez-G{\'o}mez}, {Mauersberger}, \& {Menten}}]{2020A&A...642A.222H}
{Humire}, P.~K., {Thiel}, V., {Henkel}, C., {et~al.} 2020, \aap, 642, A222

\bibitem[{{H{\"u}ttemeister} {et~al.}(1995){H{\"u}ttemeister}, {Henkel},
  {Mauersberger}, {Brouillet}, {Wiklind}, \& {Millar}}]{1995A&A...295..571H}
{H{\"u}ttemeister}, S., {Henkel}, C., {Mauersberger}, R., {et~al.} 1995, \aap,
  295, 571

\bibitem[{{H\"{u}ttemeister} {et~al.}(1995){H\"{u}ttemeister}, {Wilson},
  {Mauersberger}, {Lemme}, {Dahmen}, \& {Henkel}}]{1995A&A...294..667H}
{H\"{u}ttemeister}, S., {Wilson}, T.~L., {Mauersberger}, R., {et~al.} 1995,
  \aap, 294, 667

\bibitem[{{Ilyushin} \& {Lovas}(2007)}]{2007JPCRD..36.1141I}
{Ilyushin}, V. \& {Lovas}, F.~J. 2007, Journal of Physical and Chemical
  Reference Data, 36, 1141

\bibitem[{{Juvela} {et~al.}(2012){Juvela}, {Harju}, {Ysard}, \&
  {Lunttila}}]{2012A&A...538A.133J}
{Juvela}, M., {Harju}, J., {Ysard}, N., \& {Lunttila}, T. 2012, \aap, 538, A133

\bibitem[{Kendall(1938)}]{kendall1938measure}
Kendall, M.~G. 1938, Biometrika, 30, 81

\bibitem[{{Kim} {et~al.}(2023){Kim}, {Schilke}, {Neufeld}, {Jacob},
  {S{\'a}nchez-Monge}, {Seifried}, {Godard}, {Menten}, {Walch}, {Falgarone},
  {Veena}, {Bialy}, {M{\"o}ller}, \& {Wyrowski}}]{2023A&A...670A.111K}
{Kim}, W.~J., {Schilke}, P., {Neufeld}, D.~A., {et~al.} 2023, \aap, 670, A111

\bibitem[{{Kra{\'s}nicki} \& {Kisiel}(2011)}]{2011JMoSp.270...83K}
{Kra{\'s}nicki}, A. \& {Kisiel}, Z. 2011, Journal of Molecular Spectroscopy,
  270, 83

\bibitem[{{Kurtz} {et~al.}(2000){Kurtz}, {Cesaroni}, {Churchwell}, {Hofner}, \&
  {Walmsley}}]{2000prpl.conf..299K}
{Kurtz}, S., {Cesaroni}, R., {Churchwell}, E., {Hofner}, P., \& {Walmsley},
  C.~M. 2000, in Protostars and Planets IV, ed. V.~{Mannings}, A.~P. {Boss}, \&
  S.~S. {Russell}, 299--326

\bibitem[{{Ligterink} {et~al.}(2018){Ligterink}, {Calcutt}, {Coutens},
  {Kristensen}, {Bourke}, {Drozdovskaya}, {M\"{u}ller}, {Wampfler}, {van der
  Wiel}, {van Dishoeck}, \& {J{\o}rgensen}}]{2018A&A...619A..28L}
{Ligterink}, N.~F.~W., {Calcutt}, H., {Coutens}, A., {et~al.} 2018, \aap, 619,
  A28

\bibitem[{{Ligterink} {et~al.}(2015){Ligterink}, {Tenenbaum}, \& {van
  Dishoeck}}]{2015A&A...576A..35L}
{Ligterink}, N.~F.~W., {Tenenbaum}, E.~D., \& {van Dishoeck}, E.~F. 2015, \aap,
  576, A35

\bibitem[{{Lis} \& {Goldsmith}(1989)}]{1989ApJ...337..704L}
{Lis}, D.~C. \& {Goldsmith}, P.~F. 1989, \apj, 337, 704

\bibitem[{{Lis} {et~al.}(2014){Lis}, {Schilke}, {Bergin}, {Gerin}, {Black},
  {Comito}, {De Luca}, {Godard}, {Higgins}, {Le Petit}, {Pearson},
  {Pellegrini}, {Phillips}, \& {Yu}}]{2014ApJ...785..135L}
{Lis}, D.~C., {Schilke}, P., {Bergin}, E.~A., {et~al.} 2014, \apj, 785, 135

\bibitem[{{Liszt} \& {Lucas}(2001)}]{2001A&A...370..576L}
{Liszt}, H. \& {Lucas}, R. 2001, \aap, 370, 576

\bibitem[{{Liu} {et~al.}(2001){Liu}, {Mehringer}, \&
  {Snyder}}]{2001ApJ...552..654L}
{Liu}, S.-Y., {Mehringer}, D.~M., \& {Snyder}, L.~E. 2001, \apj, 552, 654

\bibitem[{{L{\'o}pez-Sepulcre} {et~al.}(2015){L{\'o}pez-Sepulcre}, {Jaber},
  {Mendoza}, {Lefloch}, {Ceccarelli}, {Vastel}, {Bachiller}, {Cernicharo},
  {Codella}, {Kahane}, {Kama}, \& {Tafalla}}]{2015MNRAS.449.2438L}
{L{\'o}pez-Sepulcre}, A., {Jaber}, A.~A., {Mendoza}, E., {et~al.} 2015, \mnras,
  449, 2438

\bibitem[{{Lu} {et~al.}(2019){Lu}, {Mills}, {Ginsburg}, {Walker}, {Barnes},
  {Butterfield}, {Henshaw}, {Battersby}, {Kruijssen}, {Longmore}, {Zhang},
  {Bally}, {Kauffmann}, {Ott}, {Rickert}, \& {Wang}}]{2019ApJS..244...35L}
{Lu}, X., {Mills}, E. A.~C., {Ginsburg}, A., {et~al.} 2019, \apjs, 244, 35

\bibitem[{{Lu} {et~al.}(2017){Lu}, {Zhang}, {Kauffmann}, {Pillai}, {Longmore},
  {Kruijssen}, \& {Battersby}}]{2017IAUS..322...99L}
{Lu}, X., {Zhang}, Q., {Kauffmann}, J., {et~al.} 2017, in IAU Symposium, Vol.
  322, The Multi-Messenger Astrophysics of the Galactic Centre, ed. R.~M.
  {Crocker}, S.~N. {Longmore}, \& G.~V. {Bicknell}, 99--102

\bibitem[{Maksyutenko {et~al.}(2011)Maksyutenko, Zhang, Gu, \&
  Kaiser}]{C0CP01529F}
Maksyutenko, P., Zhang, F., Gu, X., \& Kaiser, R.~I. 2011, Phys. Chem. Chem.
  Phys., 13, 240

\bibitem[{{Maluendes} {et~al.}(1993){Maluendes}, {McLean}, \&
  {Herbst}}]{1993ApJ...417..181M}
{Maluendes}, S.~A., {McLean}, A.~D., \& {Herbst}, E. 1993, \apj, 417, 181

\bibitem[{Marquardt(1963)}]{doi:10.1137/0111030}
Marquardt, D.~W. 1963, Journal of the Society for Industrial and Applied
  Mathematics, 11, 431

\bibitem[{{Mart{\'\i}n-Pintado} {et~al.}(1999){Mart{\'\i}n-Pintado}, {Gaume},
  {Rodr{\'\i}guez-Fern{\'a}ndez}, {de Vicente}, \&
  {Wilson}}]{1999ApJ...519..667M}
{Mart{\'\i}n-Pintado}, J., {Gaume}, R.~A., {Rodr{\'\i}guez-Fern{\'a}ndez}, N.,
  {de Vicente}, P., \& {Wilson}, T.~L. 1999, \apj, 519, 667

\bibitem[{{Matsumoto} {et~al.}(2014){Matsumoto}, {Hirota}, {Sugiyama}, {Kim},
  {Kim}, {Byun}, {Jung}, {Chibueze}, {Honma}, {Kameya}, {Kim}, {Lyo}, {Motogi},
  {Oh}, {Shino}, {Sunada}, {Bae}, {Chung}, {Chung}, {Cho}, {Han}, {Han},
  {Hwang}, {Je}, {Jike}, {Jung}, {Jung}, {Kang}, {Kang}, {Kang}, {Kan-ya},
  {Kawaguchi}, {Kim}, {Kim}, {Ryoung Kim}, {Kim}, {Kobayashi}, {Kono},
  {Kurayama}, {Lee}, {Lee}, {Lee}, {Lee}, {Lee}, {Lee}, {Minh}, {Miyazaki},
  {Oh}, {Oyama}, {Park}, {Roh}, {Sasao}, {Sawada-Satoh}, {Shibata}, {Sohn},
  {Song}, {Tamura}, {Wajima}, {Wi}, {Yeom}, \& {Yun}}]{2014ApJ...789L...1M}
{Matsumoto}, N., {Hirota}, T., {Sugiyama}, K., {et~al.} 2014, \apjl, 789, L1

\bibitem[{{McAllister}(1978)}]{1978ApJ...225..857M}
{McAllister}, T. 1978, \apj, 225, 857

\bibitem[{{McClure} {et~al.}(2023){McClure}, {Rocha}, {Pontoppidan}, {Crouzet},
  {Chu}, {Dartois}, {Lamberts}, {Noble}, {Pendleton}, {Perotti}, {Qasim},
  {Rachid}, {Smith}, {Sun}, {Beck}, {Boogert}, {Brown}, {Caselli}, {Charnley},
  {Cuppen}, {Dickinson}, {Drozdovskaya}, {Egami}, {Erkal}, {Fraser}, {Garrod},
  {Harsono}, {Ioppolo}, {Jim{\'e}nez-Serra}, {Jin}, {J{\o}rgensen},
  {Kristensen}, {Lis}, {McCoustra}, {McGuire}, {Melnick}, {{\~A}-berg},
  {Palumbo}, {Shimonishi}, {Sturm}, {van Dishoeck}, \&
  {Linnartz}}]{2023NatAs...7..431M}
{McClure}, M.~K., {Rocha}, W.~R.~M., {Pontoppidan}, K.~M., {et~al.} 2023,
  Nature Astronomy, 7, 431

\bibitem[{{McGonagle} {et~al.}(1994){McGonagle}, {Irvine}, \&
  {Ohishi}}]{1994ApJ...422..621M}
{McGonagle}, D., {Irvine}, W.~M., \& {Ohishi}, M. 1994, \apj, 422, 621

\bibitem[{{McGrath} {et~al.}(2004){McGrath}, {Goss}, \& {De
  Pree}}]{2004ApJS..155..577M}
{McGrath}, E.~J., {Goss}, W.~M., \& {De Pree}, C.~G. 2004, \apjs, 155, 577

\bibitem[{{McMullin} {et~al.}(2007){McMullin}, {Waters}, {Schiebel}, {Young},
  \& {Golap}}]{2007ASPC..376..127M}
{McMullin}, J.~P., {Waters}, B., {Schiebel}, D., {Young}, W., \& {Golap}, K.
  2007, in Astronomical Society of the Pacific Conference Series, Vol. 376,
  Astronomical Data Analysis Software and Systems XVI, ed. R.~A. {Shaw},
  F.~{Hill}, \& D.~J. {Bell}, 127

\bibitem[{{Mehringer} {et~al.}(1994){Mehringer}, {Goss}, \&
  {Palmer}}]{1994ApJ...434..237M}
{Mehringer}, D.~M., {Goss}, W.~M., \& {Palmer}, P. 1994, \apj, 434, 237

\bibitem[{{Mehringer} \& {Menten}(1997)}]{1997ApJ...474..346M}
{Mehringer}, D.~M. \& {Menten}, K.~M. 1997, \apj, 474, 346

\bibitem[{{Mehringer} {et~al.}(1997){Mehringer}, {Snyder}, {Miao}, \&
  {Lovas}}]{1997ApJ...480L..71M}
{Mehringer}, D.~M., {Snyder}, L.~E., {Miao}, Y., \& {Lovas}, F.~J. 1997, \apjl,
  480, L71

\bibitem[{{Mei} {et~al.}(2020){Mei}, {Chen}, {Shen}, \&
  {Li}}]{2020ApJ...898..157M}
{Mei}, Y., {Chen}, X., {Shen}, Z.-Q., \& {Li}, B. 2020, \apj, 898, 157

\bibitem[{{Meier} \& {Turner}(2005)}]{2005ApJ...618..259M}
{Meier}, D.~S. \& {Turner}, J.~L. 2005, \apj, 618, 259

\bibitem[{{Meng} {et~al.}(2022){Meng}, {S{\'a}nchez-Monge}, {Schilke},
  {Ginsburg}, {DePree}, {Budaiev}, {Jeff}, {Schmiedeke}, {Schw{\"o}rer},
  {Veena}, \& {M{\"o}ller}}]{2022A&A...666A..31M}
{Meng}, F., {S{\'a}nchez-Monge}, {\'A}., {Schilke}, P., {et~al.} 2022, \aap,
  666, A31

\bibitem[{{Meng} {et~al.}(2019){Meng}, {S{\'a}nchez-Monge}, {Schilke},
  {Padovani}, {Marcowith}, {Ginsburg}, {Schmiedeke}, {Schw{\"o}rer}, {DePree},
  {Veena}, \& {M{\"o}ller}}]{2019A&A...630A..73M}
{Meng}, F., {S{\'a}nchez-Monge}, {\'A}., {Schilke}, P., {et~al.} 2019, \aap,
  630, A73

\bibitem[{{Millar} {et~al.}(1985){Millar}, {Adams}, {Smith}, \&
  {Clary}}]{1985MNRAS.216.1025M}
{Millar}, T.~J., {Adams}, N.~G., {Smith}, D., \& {Clary}, D.~C. 1985, \mnras,
  216, 1025

\bibitem[{{Millar} {et~al.}(1986){Millar}, {Adams}, {Smith}, {Lindinger}, \&
  {Villinger}}]{1986MNRAS.221..673M}
{Millar}, T.~J., {Adams}, N.~G., {Smith}, D., {Lindinger}, W., \& {Villinger},
  H. 1986, \mnras, 221, 673

\bibitem[{{Millar} \& {Herbst}(1990)}]{1990A&A...231..466M}
{Millar}, T.~J. \& {Herbst}, E. 1990, \aap, 231, 466

\bibitem[{{Mills} {et~al.}(2018){Mills}, {Ginsburg}, {Clements}, {Schilke},
  {S{\'a}nchez-Monge}, {Menten}, {Butterfield}, {Goddi}, {Schmiedeke}, \& {De
  Pree}}]{2018ApJ...869L..14M}
{Mills}, E.~A.~C., {Ginsburg}, A., {Clements}, A.~R., {et~al.} 2018, \apjl,
  869, L14

\bibitem[{{Minh} {et~al.}(2016){Minh}, {Liu}, \&
  {Galva{\'n}-Madrid}}]{2016ApJ...824...99M}
{Minh}, Y.~C., {Liu}, H.~B., \& {Galva{\'n}-Madrid}, R. 2016, \apj, 824, 99

\bibitem[{{Minh} {et~al.}(2018){Minh}, {Liu}, {Galva{\'n}-Madrid}, {Sahu},
  {He}, \& {Hasegawa}}]{2018ApJ...864..102M}
{Minh}, Y.~C., {Liu}, H.~B., {Galva{\'n}-Madrid}, R., {et~al.} 2018, \apj, 864,
  102

\bibitem[{{Mininni} {et~al.}(2018){Mininni}, {Fontani}, {Rivilla},
  {Beltr{\'a}n}, {Caselli}, \& {Vasyunin}}]{2018MNRAS.476L..39M}
{Mininni}, C., {Fontani}, F., {Rivilla}, V.~M., {et~al.} 2018, \mnras, 476, L39

\bibitem[{{M{\"o}ller} {et~al.}(2013){M{\"o}ller}, {Bernst}, {Panoglou},
  {Muders}, {Ossenkopf}, {R{\"o}llig}, \& {Schilke}}]{2013A&A...549A..21M}
{M{\"o}ller}, T., {Bernst}, I., {Panoglou}, D., {et~al.} 2013, \aap, 549, A21

\bibitem[{{M\"{o}ller} {et~al.}(2017){M\"{o}ller}, {Endres}, \&
  {Schilke}}]{2017A&A...598A...7M}
{M\"{o}ller}, T., {Endres}, C., \& {Schilke}, P. 2017, \aap, 598, A7

\bibitem[{{M{\"o}ller} {et~al.}(2023){M{\"o}ller}, {Schilke},
  {S{\'a}nchez-Monge}, {Schmiedeke}, \& {Meng}}]{2023A&A...676A.121M}
{M{\"o}ller}, T., {Schilke}, P., {S{\'a}nchez-Monge}, {\'A}., {Schmiedeke}, A.,
  \& {Meng}, F. 2023, \aap, 676, A121

\bibitem[{{M{\"o}ller} {et~al.}(2021){M{\"o}ller}, {Schilke}, {Schmiedeke},
  {Bergin}, {Lis}, {S{\'a}nchez-Monge}, {Schw{\"o}rer}, \&
  {Comito}}]{2021A&A...651A...9M}
{M{\"o}ller}, T., {Schilke}, P., {Schmiedeke}, A., {et~al.} 2021, \aap, 651, A9

\bibitem[{{Morita} {et~al.}(1992){Morita}, {Hasegawa}, {Ukita}, {Okumura}, \&
  {Ishiguro}}]{1992PASJ...44..373M}
{Morita}, K.-I., {Hasegawa}, T., {Ukita}, N., {Okumura}, S.~K., \& {Ishiguro},
  M. 1992, \pasj, 44, 373

\bibitem[{{Mota} {et~al.}(2021){Mota}, {Varandas}, {Mendoza}, {Wakelam}, \&
  {Galv{\~a}o}}]{2021ApJ...920...37M}
{Mota}, V.~C., {Varandas}, A.~J.~C., {Mendoza}, E., {Wakelam}, V., \&
  {Galv{\~a}o}, B.~R.~L. 2021, \apj, 920, 37

\bibitem[{{M\"{u}ller} {et~al.}(2005){M\"{u}ller}, {Schl{\"o}der}, {Stutzki},
  \& {Winnewisser}}]{2005JMoSt.742..215M}
{M\"{u}ller}, H. S.~P., {Schl{\"o}der}, F., {Stutzki}, J., \& {Winnewisser}, G.
  2005, Journal of Molecular Structure, 742, 215

\bibitem[{{M\"{u}ller} {et~al.}(2001){M\"{u}ller}, {Thorwirth}, {Roth}, \&
  {Winnewisser}}]{2001A&A...370L..49M}
{M\"{u}ller}, H.~S.~P., {Thorwirth}, S., {Roth}, D.~A., \& {Winnewisser}, G.
  2001, \aap, 370, L49

\bibitem[{{Neill} {et~al.}(2014){Neill}, {Bergin}, {Lis}, {Schilke},
  {Crockett}, {Favre}, {Emprechtinger}, {Comito}, {Qin}, {Anderson},
  {Burkhardt}, {Chen}, {Harris}, {Lord}, {McGuire}, {McNeill}, {Monje},
  {Phillips}, {Steber}, {Vasyunina}, \& {Yu}}]{2014ApJ...789....8N}
{Neill}, J.~L., {Bergin}, E.~A., {Lis}, D.~C., {et~al.} 2014, \apj, 789, 8

\bibitem[{Ngo(2022)}]{kups63059}
Ngo, T.~L. 2022, PhD thesis, Universit{\"a}t zu K{\"o}ln

\bibitem[{{Niedenhoff} {et~al.}(1995){Niedenhoff}, {Yamada}, {Belov}, \&
  {Winnewisser}}]{1995JMoSp.174..151N}
{Niedenhoff}, M., {Yamada}, K.~M.~T., {Belov}, S.~P., \& {Winnewisser}, G.
  1995, Journal of Molecular Spectroscopy, 174, 151

\bibitem[{{Noble} {et~al.}(2015){Noble}, {Theule}, {Congiu}, {Dulieu},
  {Bonnin}, {Bassas}, {Duvernay}, {Danger}, \&
  {Chiavassa}}]{2015A&A...576A..91N}
{Noble}, J.~A., {Theule}, P., {Congiu}, E., {et~al.} 2015, \aap, 576, A91

\bibitem[{{Nummelin} {et~al.}(1998){Nummelin}, {Bergman}, {Hjalmarson},
  {Friberg}, {Irvine}, {Millar}, {Ohishi}, \& {Saito}}]{1998ApJS..117..427N}
{Nummelin}, A., {Bergman}, P., {Hjalmarson}, {\r{A}}., {et~al.} 1998, \apjs,
  117, 427

\bibitem[{{Ohashi} \& {Hougen}(1987)}]{1987JMoSp.121..474O}
{Ohashi}, N. \& {Hougen}, J.~T. 1987, Journal of Molecular Spectroscopy, 121,
  474

\bibitem[{Parzen(1962)}]{parzen1962}
Parzen, E. 1962, Ann. Math. Statist., 33, 1065

\bibitem[{Pearson(1901)}]{pearson1901}
Pearson, K. 1901, Philosophical Magazine, 2, 559

\bibitem[{Pedregosa {et~al.}(2011)Pedregosa, Varoquaux, Gramfort, Michel,
  Thirion, Grisel, Blondel, Prettenhofer, Weiss, Dubourg, Vanderplas, Passos,
  Cournapeau, Brucher, Perrot, \& Duchesnay}]{scikit-learn}
Pedregosa, F., Varoquaux, G., Gramfort, A., {et~al.} 2011, Journal of Machine
  Learning Research, 12, 2825

\bibitem[{{Pickett} {et~al.}(1998){Pickett}, {Poynter}, {Cohen}, {Delitsky},
  {Pearson}, \& {M\"{u}ller}}]{1998JQSRT..60..883P}
{Pickett}, H.~M., {Poynter}, R.~L., {Cohen}, E.~A., {et~al.} 1998, \jqsrt, 60,
  883

\bibitem[{{Pickles} \& {Williams}(1977)}]{1977Ap&SS..52..453P}
{Pickles}, J.~B. \& {Williams}, D.~A. 1977, \apss, 52, 453

\bibitem[{{Pilling} {et~al.}(2011){Pilling}, {Baptista}, {Boechat-Roberty}, \&
  {Andrade}}]{2011AsBio..11..883P}
{Pilling}, S., {Baptista}, L., {Boechat-Roberty}, H.~M., \& {Andrade}, D. P.~P.
  2011, Astrobiology, 11, 883

\bibitem[{{Podio} {et~al.}(2014){Podio}, {Lefloch}, {Ceccarelli}, {Codella}, \&
  {Bachiller}}]{2014A&A...565A..64P}
{Podio}, L., {Lefloch}, B., {Ceccarelli}, C., {Codella}, C., \& {Bachiller}, R.
  2014, \aap, 565, A64

\bibitem[{{Pols} {et~al.}(2018){Pols}, {Schw{\"o}rer}, {Schilke}, {Schmiedeke},
  {S{\'a}nchez-Monge}, \& {M\"{o}ller}}]{2018A&A...614A.123P}
{Pols}, S., {Schw{\"o}rer}, A., {Schilke}, P., {et~al.} 2018, \aap, 614, A123

\bibitem[{{Qin} {et~al.}(2011){Qin}, {Schilke}, {Rolffs}, {Comito}, {Lis}, \&
  {Zhang}}]{2011A&A...530L...9Q}
{Qin}, S.~L., {Schilke}, P., {Rolffs}, R., {et~al.} 2011, \aap, 530, L9

\bibitem[{{Ramal-Olmedo} {et~al.}(2021){Ramal-Olmedo}, {Menor-Salv{\'a}n}, \&
  {Fortenberry}}]{2021A&A...656A.148R}
{Ramal-Olmedo}, J.~C., {Menor-Salv{\'a}n}, C.~A., \& {Fortenberry}, R.~C. 2021,
  \aap, 656, A148

\bibitem[{{Rivilla} {et~al.}(2020){Rivilla}, {Drozdovskaya}, {Altwegg},
  {Caselli}, {Beltr{\'a}n}, {Fontani}, {van der Tak}, {Cesaroni}, {Vasyunin},
  {Rubin}, {Lique}, {Marinakis}, {Testi}, {Rosina Team}, {Balsiger},
  {Berthelier}, {de Keyser}, {Fiethe}, {Fuselier}, {Gasc}, {Gombosi},
  {S{\'e}mon}, \& {Tzou}}]{2020MNRAS.492.1180R}
{Rivilla}, V.~M., {Drozdovskaya}, M.~N., {Altwegg}, K., {et~al.} 2020, \mnras,
  492, 1180

\bibitem[{{Rodgers} \& {Charnley}(2001)}]{2001ApJ...546..324R}
{Rodgers}, S.~D. \& {Charnley}, S.~B. 2001, \apj, 546, 324

\bibitem[{Rosenblatt(1956)}]{rosenblatt1956}
Rosenblatt, M. 1956, Ann. Math. Statist., 27, 832

\bibitem[{{S{\'a}nchez-Monge} {et~al.}(2018){S{\'a}nchez-Monge}, {Schilke},
  {Ginsburg}, {Cesaroni}, \& {Schmiedeke}}]{2018A&A...609A.101S}
{S{\'a}nchez-Monge}, {\'A}., {Schilke}, P., {Ginsburg}, A., {Cesaroni}, R., \&
  {Schmiedeke}, A. 2018, \aap, 609, A101

\bibitem[{{S{\'a}nchez-Monge} {et~al.}(2017){S{\'a}nchez-Monge}, {Schilke},
  {Schmiedeke}, {Ginsburg}, {Cesaroni}, {Lis}, {Qin}, {M\"{u}ller}, {Bergin},
  {Comito}, \& {M\"{o}ller}}]{2017A&A...604A...6S}
{S{\'a}nchez-Monge}, {\'A}., {Schilke}, P., {Schmiedeke}, A., {et~al.} 2017,
  \aap, 604, A6

\bibitem[{{Schilke} {et~al.}(1997){Schilke}, {Groesbeck}, {Blake}, {Phillips},
  \& {T.~G.}}]{1997ApJS..108..301S}
{Schilke}, P., {Groesbeck}, T.~D., {Blake}, G.~A., {Phillips}, \& {T.~G.} 1997,
  \apjs, 108, 301

\bibitem[{{Schmiedeke} {et~al.}(2016){Schmiedeke}, {Schilke}, {M\"{o}ller},
  {S{\'a}nchez-Monge}, {Bergin}, {Comito}, {Csengeri}, {Lis}, {Molinari},
  {Qin}, \& {Rolffs}}]{2016A&A...588A.143S}
{Schmiedeke}, A., {Schilke}, P., {M\"{o}ller}, T., {et~al.} 2016, \aap, 588,
  A143

\bibitem[{{Schw{\"o}rer} {et~al.}(2019){Schw{\"o}rer}, {S{\'a}nchez-Monge},
  {Schilke}, {M\"{o}ller}, {Ginsburg}, {Meng}, {Schmiedeke}, {M\"{u}ller},
  {Lis}, \& {Qin}}]{2019A&A...628A...6S}
{Schw{\"o}rer}, A., {S{\'a}nchez-Monge}, {\'A}., {Schilke}, P., {et~al.} 2019,
  \aap, 628, A6

\bibitem[{{Shiki} \& {Deguchi}(1997)}]{1997ApJ...478..206S}
{Shiki}, S. \& {Deguchi}, S. 1997, \apj, 478, 206

\bibitem[{{Shingledecker} {et~al.}(2020){Shingledecker}, {Lamberts}, {Laas},
  {Vasyunin}, {Herbst}, {K{\"a}stner}, \& {Caselli}}]{2020ApJ...888...52S}
{Shingledecker}, C.~N., {Lamberts}, T., {Laas}, J.~C., {et~al.} 2020, \apj,
  888, 52

\bibitem[{{Sil} {et~al.}(2018){Sil}, {Gorai}, {Das}, {Bhat}, {Etim}, \&
  {Chakrabarti}}]{2018ApJ...853..139S}
{Sil}, M., {Gorai}, P., {Das}, A., {et~al.} 2018, \apj, 853, 139

\bibitem[{{Silverman}(1986)}]{1986desd.book.....S}
{Silverman}, B.~W. 1986, {Density estimation for statistics and data analysis}

\bibitem[{{Sinclair} {et~al.}(1992){Sinclair}, {Carrad}, {Caswell}, {Norris},
  \& {Whiteoak}}]{1992MNRAS.256P..33S}
{Sinclair}, M.~W., {Carrad}, G.~J., {Caswell}, J.~L., {Norris}, R.~P., \&
  {Whiteoak}, J.~B. 1992, \mnras, 256, 33P

\bibitem[{{Singh} \& {Maciel}(1980)}]{1980Ap&SS..68...87S}
{Singh}, P.~S. \& {Maciel}, W.~J. 1980, \apss, 68, 87

\bibitem[{{Smith} {et~al.}(2004){Smith}, {Herbst}, \&
  {Chang}}]{2004MNRAS.350..323S}
{Smith}, I. W.~M., {Herbst}, E., \& {Chang}, Q. 2004, \mnras, 350, 323

\bibitem[{{Sobolev} {et~al.}(1997){Sobolev}, {Cragg}, \&
  {Godfrey}}]{1997A&A...324..211S}
{Sobolev}, A.~M., {Cragg}, D.~M., \& {Godfrey}, P.~D. 1997, \aap, 324, 211

\bibitem[{Spearman(1904)}]{Spearman1904}
Spearman, C. 1904, The American Journal of Psychology, 15, 72

\bibitem[{{Sutton} {et~al.}(1991){Sutton}, {Jaminet}, {Danchi}, \&
  {Blake}}]{1991ApJS...77..255S}
{Sutton}, E.~C., {Jaminet}, P.~A., {Danchi}, W.~C., \& {Blake}, G.~A. 1991,
  \apjs, 77, 255

\bibitem[{{Taniguchi} {et~al.}(2020){Taniguchi}, {Guzm{\'a}n}, {Majumdar},
  {Saito}, \& {Tokuda}}]{2020ApJ...898...54T}
{Taniguchi}, K., {Guzm{\'a}n}, A.~E., {Majumdar}, L., {Saito}, M., \& {Tokuda},
  K. 2020, \apj, 898, 54

\bibitem[{{Taniguchi} {et~al.}(2018){Taniguchi}, {Miyamoto}, {Saito},
  {Sanhueza}, {Shimoikura}, {Dobashi}, {Nakamura}, \&
  {Ozeki}}]{2018ApJ...866...32T}
{Taniguchi}, K., {Miyamoto}, Y., {Saito}, M., {et~al.} 2018, \apj, 866, 32

\bibitem[{{Taniguchi} {et~al.}(2016){Taniguchi}, {Saito}, \&
  {Ozeki}}]{2016ApJ...830..106T}
{Taniguchi}, K., {Saito}, M., \& {Ozeki}, H. 2016, \apj, 830, 106

\bibitem[{{Taquet} {et~al.}(2016){Taquet}, {Wirstr{\"o}m}, \&
  {Charnley}}]{2016ApJ...821...46T}
{Taquet}, V., {Wirstr{\"o}m}, E.~S., \& {Charnley}, S.~B. 2016, \apj, 821, 46

\bibitem[{Thorndike(1953)}]{Thorndike1953}
Thorndike, R.~L. 1953, Psychometrika, 18, 267

\bibitem[{{Turner}(1989)}]{1989ApJS...70..539T}
{Turner}, B.~E. 1989, \apjs, 70, 539

\bibitem[{{Turner}(1992)}]{1992ApJ...396L.107T}
{Turner}, B.~E. 1992, \apjl, 396, L107

\bibitem[{{Turner}(1998)}]{1998ApJ...501..731T}
{Turner}, B.~E. 1998, \apj, 501, 731

\bibitem[{{Turner} {et~al.}(1990){Turner}, {Tsuji}, {Bally}, {Guelin}, \&
  {Cernicharo}}]{1990ApJ...365..569T}
{Turner}, B.~E., {Tsuji}, T., {Bally}, J., {Guelin}, M., \& {Cernicharo}, J.
  1990, \apj, 365, 569

\bibitem[{{van der Tak}(2004)}]{2004IAUS..221...59V}
{van der Tak}, F.~F.~S. 2004, in Star Formation at High Angular Resolution, ed.
  M.~G. {Burton}, R.~{Jayawardhana}, \& T.~L. {Bourke}, Vol. 221, 59

\bibitem[{{van der Walt}(2005)}]{2005MNRAS.360..153V}
{van der Walt}, J. 2005, \mnras, 360, 153

\bibitem[{{van Dishoeck}(2018)}]{2018IAUS..332....3V}
{van Dishoeck}, E.~F. 2018, IAU Symposium, 332, 3

\bibitem[{{van Dishoeck} {et~al.}(2021){van Dishoeck}, {Kristensen}, {Mottram},
  {Benz}, {Bergin}, {Caselli}, {Herpin}, {Hogerheijde}, {Johnstone}, {Liseau},
  {Nisini}, {Tafalla}, {van der Tak}, {Wyrowski}, {Baudry}, {Benedettini},
  {Bjerkeli}, {Blake}, {Braine}, {Bruderer}, {Cabrit}, {Cernicharo}, {Choi},
  {Coutens}, {de Graauw}, {Dominik}, {Fedele}, {Fich}, {Fuente}, {Furuya},
  {Goicoechea}, {Harsono}, {Helmich}, {Herczeg}, {Jacq}, {Karska}, {Kaufman},
  {Keto}, {Lamberts}, {Larsson}, {Leurini}, {Lis}, {Melnick}, {Neufeld},
  {Pagani}, {Persson}, {Shipman}, {Taquet}, {van Kempen}, {Walsh}, {Wampfler},
  {Y{\i}ld{\i}z}, \& {WISH Team}}]{2021A&A...648A..24V}
{van Dishoeck}, E.~F., {Kristensen}, L.~E., {Mottram}, J.~C., {et~al.} 2021,
  \aap, 648, A24

\bibitem[{{Vasyunin} {et~al.}(2009){Vasyunin}, {Semenov}, {Wiebe}, \&
  {Henning}}]{2009ApJ...691.1459V}
{Vasyunin}, A.~I., {Semenov}, D.~A., {Wiebe}, D.~S., \& {Henning}, T. 2009,
  \apj, 691, 1459

\bibitem[{{Vasyunina} {et~al.}(2011){Vasyunina}, {Linz}, {Henning},
  {Zinchenko}, {Beuther}, \& {Voronkov}}]{2011A&A...527A..88V}
{Vasyunina}, T., {Linz}, H., {Henning}, T., {et~al.} 2011, \aap, 527, A88

\bibitem[{{Vidal} \& {Wakelam}(2018)}]{2018MNRAS.474.5575V}
{Vidal}, T. H.~G. \& {Wakelam}, V. 2018, \mnras, 474, 5575

\bibitem[{Virtanen {et~al.}(2020)Virtanen, Gommers, Oliphant, Haberland, Reddy,
  Cournapeau, Burovski, Peterson, Weckesser, Bright, {van der Walt}, Brett,
  Wilson, Millman, Mayorov, Nelson, Jones, Kern, Larson, Carey, Polat, Feng,
  Moore, {VanderPlas}, Laxalde, Perktold, Cimrman, Henriksen, Quintero, Harris,
  Archibald, Ribeiro, Pedregosa, {van Mulbregt}, \& {SciPy 1.0
  Contributors}}]{2020SciPy-NMeth}
Virtanen, P., Gommers, R., Oliphant, T.~E., {et~al.} 2020, Nature Methods, 17,
  261

\bibitem[{{Wakelam} {et~al.}(2004{\natexlab{a}}){Wakelam}, {Caselli},
  {Ceccarelli}, {Herbst}, \& {Castets}}]{2004A&A...422..159W}
{Wakelam}, V., {Caselli}, P., {Ceccarelli}, C., {Herbst}, E., \& {Castets}, A.
  2004{\natexlab{a}}, \aap, 422, 159

\bibitem[{{Wakelam} {et~al.}(2004{\natexlab{b}}){Wakelam}, {Castets},
  {Ceccarelli}, {Lefloch}, {Caux}, \& {Pagani}}]{2004A&A...413..609W}
{Wakelam}, V., {Castets}, A., {Ceccarelli}, C., {et~al.} 2004{\natexlab{b}},
  \aap, 413, 609

\bibitem[{{Wakelam} {et~al.}(2011){Wakelam}, {Hersant}, \&
  {Herpin}}]{2011A&A...529A.112W}
{Wakelam}, V., {Hersant}, F., \& {Herpin}, F. 2011, \aap, 529, A112

\bibitem[{{Walmsley} \& {Schilke}(1993)}]{1993dca..book...37W}
{Walmsley}, C.~M. \& {Schilke}, P. 1993, in Dust and Chemistry in Astronomy,
  ed. T.~J. {Millar} \& D.~A. {Williams}, 37

\bibitem[{{Wang} {et~al.}(2021){Wang}, {Du}, {Semenov}, {Wang}, \&
  {Li}}]{2021A&A...648A..72W}
{Wang}, Y., {Du}, F., {Semenov}, D., {Wang}, H., \& {Li}, J. 2021, \aap, 648,
  A72

\bibitem[{{Watanabe} {et~al.}(2003){Watanabe}, {Shiraki}, \&
  {Kouchi}}]{2003ApJ...588L.121W}
{Watanabe}, N., {Shiraki}, T., \& {Kouchi}, A. 2003, \apjl, 588, L121

\bibitem[{{Whiteoak} {et~al.}(1987){Whiteoak}, {Gardner}, {Forster}, {Palmer},
  \& {Pankonin}}]{1987IAUS..115..161W}
{Whiteoak}, J.~B., {Gardner}, F.~F., {Forster}, J.~R., {Palmer}, P., \&
  {Pankonin}, V. 1987, in IAU Symposium, Vol. 115, Star Forming Regions, ed.
  M.~{Peimbert} \& J.~{Jugaku}, 161

\bibitem[{{Wiebe} {et~al.}(2008){Wiebe}, {Kirsanova}, {Shustov}, \&
  {Pavlyuchenkov}}]{2008ARep...52..976W}
{Wiebe}, D.~Z., {Kirsanova}, M.~S., {Shustov}, B.~M., \& {Pavlyuchenkov}, Y.~N.
  2008, Astronomy Reports, 52, 976

\bibitem[{Williams \& Ibrahim(1981)}]{doi:10.1021/cr00046a004}
Williams, A. \& Ibrahim, I.~T. 1981, Chemical Reviews, 81, 589

\bibitem[{{Wilson} {et~al.}(2012){Wilson}, {Parker}, {Zhang}, \&
  {Kaiser}}]{2012PCCP...14..477W}
{Wilson}, A.~V., {Parker}, D. S.~N., {Zhang}, F., \& {Kaiser}, R.~I. 2012,
  Physical Chemistry Chemical Physics (Incorporating Faraday Transactions), 14,
  477

\bibitem[{{Wilson} \& {Rood}(1994)}]{1994ARA&A..32..191W}
{Wilson}, T.~L. \& {Rood}, R. 1994, \araa, 32, 191

\bibitem[{{Wright} {et~al.}(1996){Wright}, {Plambeck}, \&
  {Wilner}}]{1996ApJ...469..216W}
{Wright}, M.~C.~H., {Plambeck}, R.~L., \& {Wilner}, D.~J. 1996, \apj, 469, 216

\bibitem[{{Wu} {et~al.}(2006){Wu}, {Zhang}, {Yu}, {Miller}, {Mao}, {Sun}, \&
  {Wang}}]{2006A&A...450..607W}
{Wu}, Y., {Zhang}, Q., {Yu}, W., {et~al.} 2006, \aap, 450, 607

\bibitem[{{Yamada} {et~al.}(2002){Yamada}, {Osamura}, \&
  {Kaiser}}]{2002A&A...395.1031Y}
{Yamada}, M., {Osamura}, Y., \& {Kaiser}, R.~I. 2002, \aap, 395, 1031

\bibitem[{{Yan} {et~al.}(2022{\natexlab{a}}){Yan}, {Henkel}, {Menten}, {Gong},
  {Nguyen}, {Ott}, {Ginsburg}, {Wilson}, {Brunthaler}, {Belloche}, {Zhang},
  {Budaiev}, \& {Jeff}}]{2022A&A...666L..15Y}
{Yan}, Y.~T., {Henkel}, C., {Menten}, K.~M., {et~al.} 2022{\natexlab{a}}, \aap,
  666, L15

\bibitem[{{Yan} {et~al.}(2022{\natexlab{b}}){Yan}, {Henkel}, {Menten}, {Gong},
  {Ott}, {Wilson}, {Wootten}, {Brunthaler}, {Zhang}, {Chen}, \&
  {Yang}}]{2022A&A...659A...5Y}
{Yan}, Y.~T., {Henkel}, C., {Menten}, K.~M., {et~al.} 2022{\natexlab{b}}, \aap,
  659, A5

\bibitem[{{Yang} {et~al.}(2023){Yang}, {Gong}, {Menten}, {Urquhart}, {Henkel},
  {Wyrowski}, {Csengeri}, {Ellingsen}, {Bemis}, \&
  {Jang}}]{2023A&A...675A.112Y}
{Yang}, W., {Gong}, Y., {Menten}, K.~M., {et~al.} 2023, \aap, 675, A112

\bibitem[{{Yusef-Zadeh} {et~al.}(2016){Yusef-Zadeh}, {Cotton}, {Wardle}, \&
  {Intema}}]{2016ApJ...819L..35Y}
{Yusef-Zadeh}, F., {Cotton}, W., {Wardle}, M., \& {Intema}, H. 2016, \apjl,
  819, L35

\bibitem[{{Zapata} {et~al.}(2009){Zapata}, {Menten}, {Reid}, \&
  {Beuther}}]{2009ApJ...691..332Z}
{Zapata}, L.~A., {Menten}, K., {Reid}, M., \& {Beuther}, H. 2009, \apj, 691,
  332

\bibitem[{{Zeng} {et~al.}(2018){Zeng}, {Jim{\'e}nez-Serra}, {Rivilla},
  {Mart{\'\i}n}, {Mart{\'\i}n-Pintado}, {Requena-Torres},
  {Armijos-Abenda{\~n}o}, {Riquelme}, \& {Aladro}}]{2018MNRAS.478.2962Z}
{Zeng}, S., {Jim{\'e}nez-Serra}, I., {Rivilla}, V.~M., {et~al.} 2018, \mnras,
  478, 2962

\end{thebibliography}

    %___________________________________________________________________________
    % appendix
    \newpage
    \Online
    \begin{appendix}

%===============================================================================
% plots describing spectra of each hot core
\onecolumn
\section{Spectra of hot cores}\label{app:hotcorespectra}

%*******************************************************************************
% Figure: observational spectrum for each core in Sgr~B2(M)
\begin{figure*}[!b]
   \centering
   \includegraphics[scale=0.85]{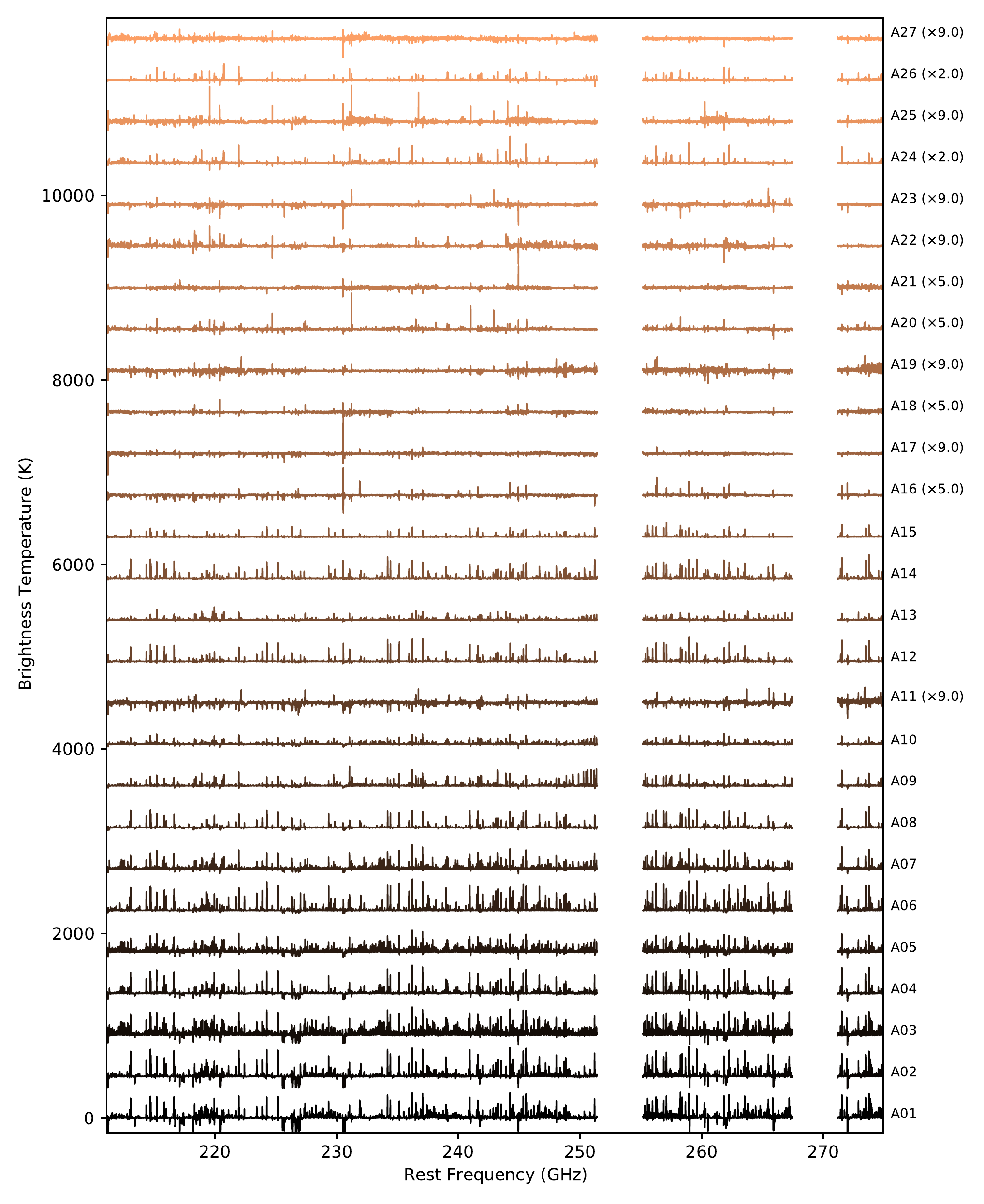}\\
   \caption{Continuum subtracted observational spectrum of each hot core A01 -- A27 in Sgr~B2(M). The scaling factor for each spectrum is given on the right side of the corresponding core name, if the factor is different from one. The offset between the different spectra is set to 450.}
   \label{fig:CoreSpecSgrB2M}
\end{figure*}
\clearpage
\hfill

%*******************************************************************************
% Figure: observational spectrum for each core in Sgr~B2(N)
\begin{figure*}[!b]
   \centering
   \includegraphics[scale=0.85]{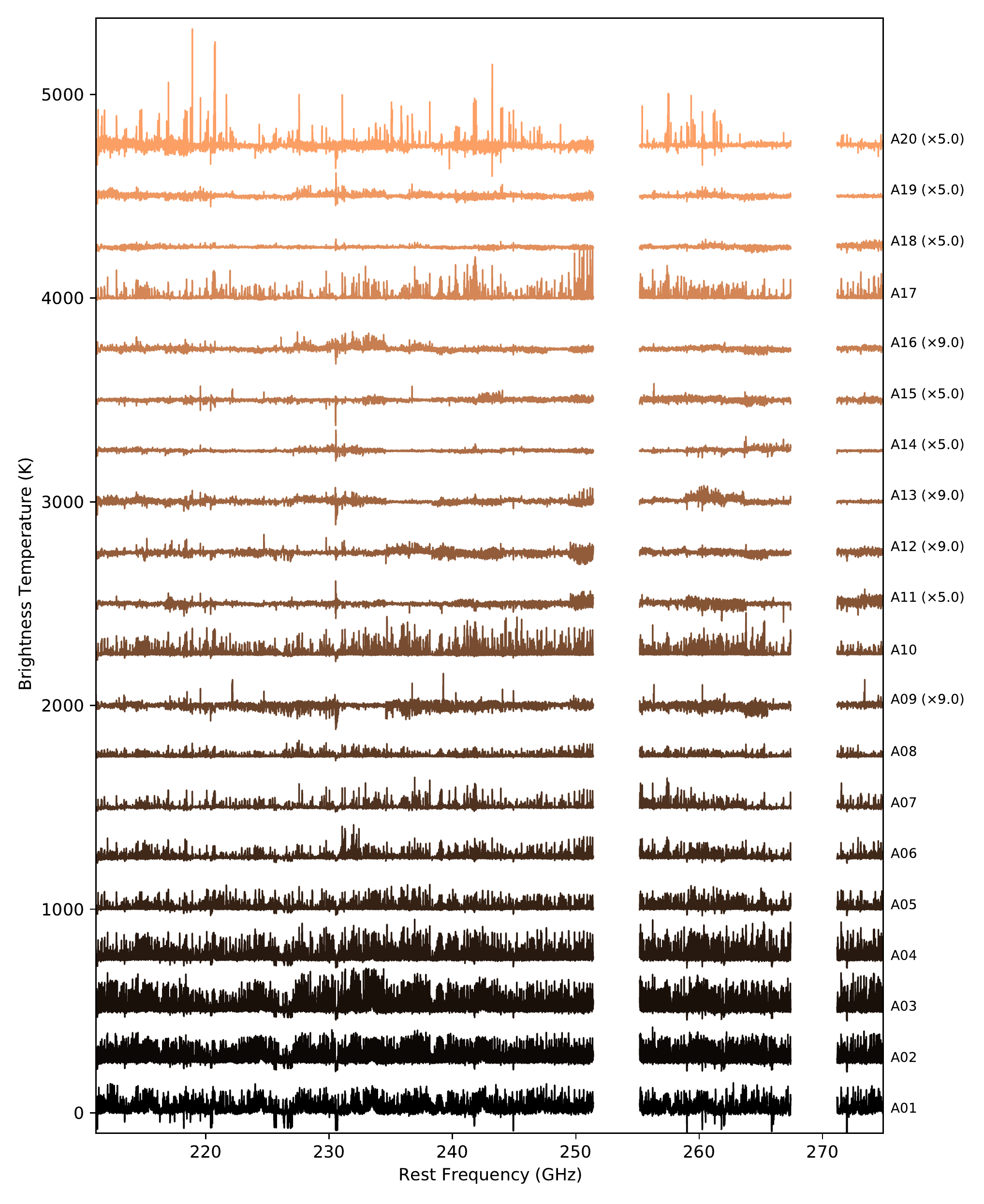}\\
   \caption{Continuum subtracted observational spectrum of each hot core A01 -- A20 in Sgr~B2(N). The scaling factor for each spectrum is given on the right side of the corresponding core name, if the factor is different from one. The offset between the different spectra is set to 250.}
   \label{fig:CoreSpecSgrB2N}
\end{figure*}
\clearpage
\newpage
\twocolumn

%===============================================================================
% plots describing spectra and fits
\onecolumn
\section{Fitted spectra of hot cores}\label{app:excerptsfit}

%*******************************************************************************
% Figure: excerpt of overall fit for Sgr~B2(M)
\begin{figure*}[!b]
   \centering
   \includegraphics[width=0.80\textwidth]{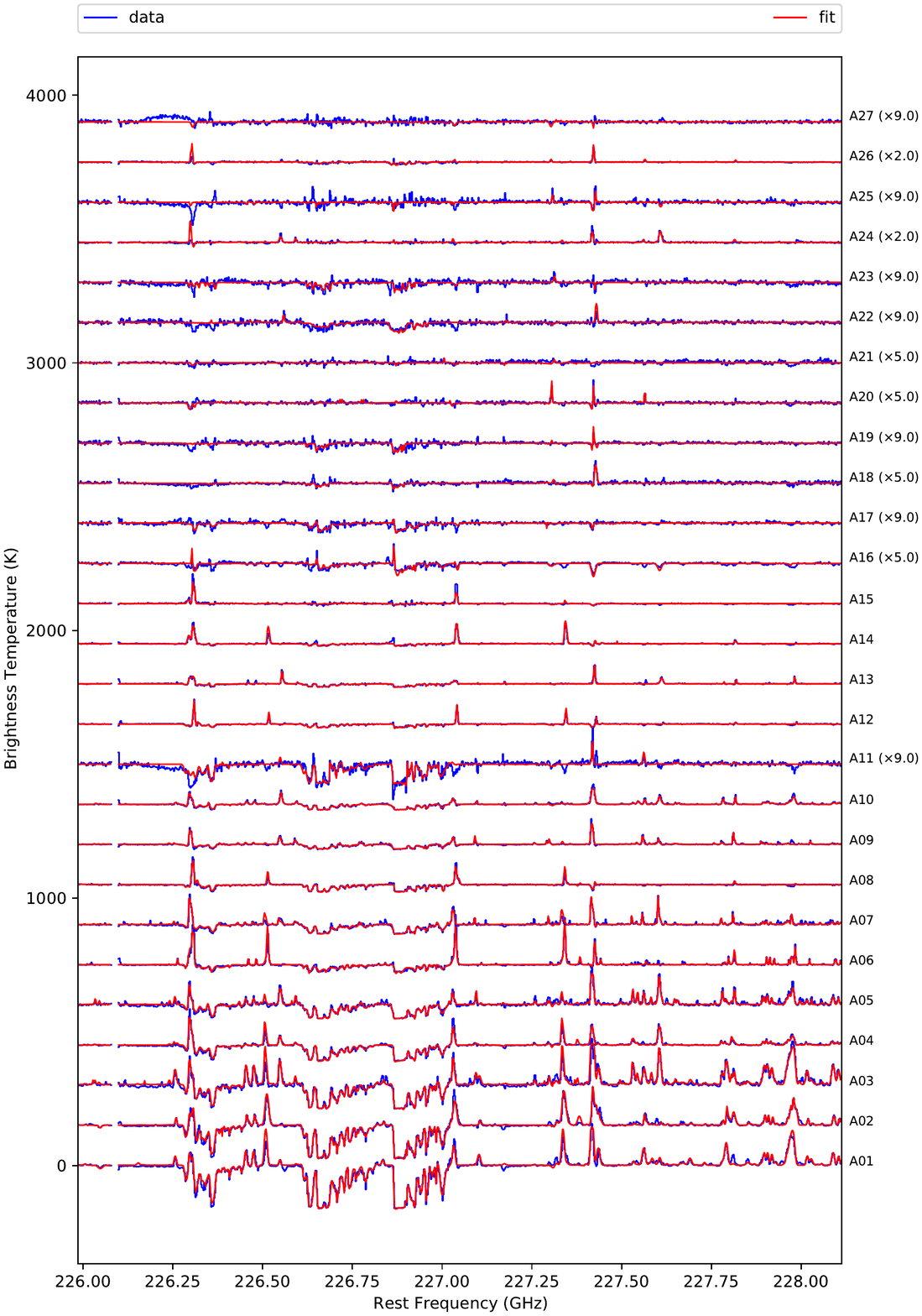}\\
   \caption{Excerpts from the ALMA survey of Sgr~B2(M). Here, the observed spectrum of each core is shown in blue together with the full model in red. On the right sight of each spectrum the name of the corresponding core is described together with the applied scaling factor (in round brackets).}
   \label{fig:SgrB2MExcerpt}
\end{figure*}
\hfill

%*******************************************************************************
% Figure: excerpt of overall fit for Sgr~B2(M)
\begin{figure*}[!b]
   \centering
   \includegraphics[scale=0.745]{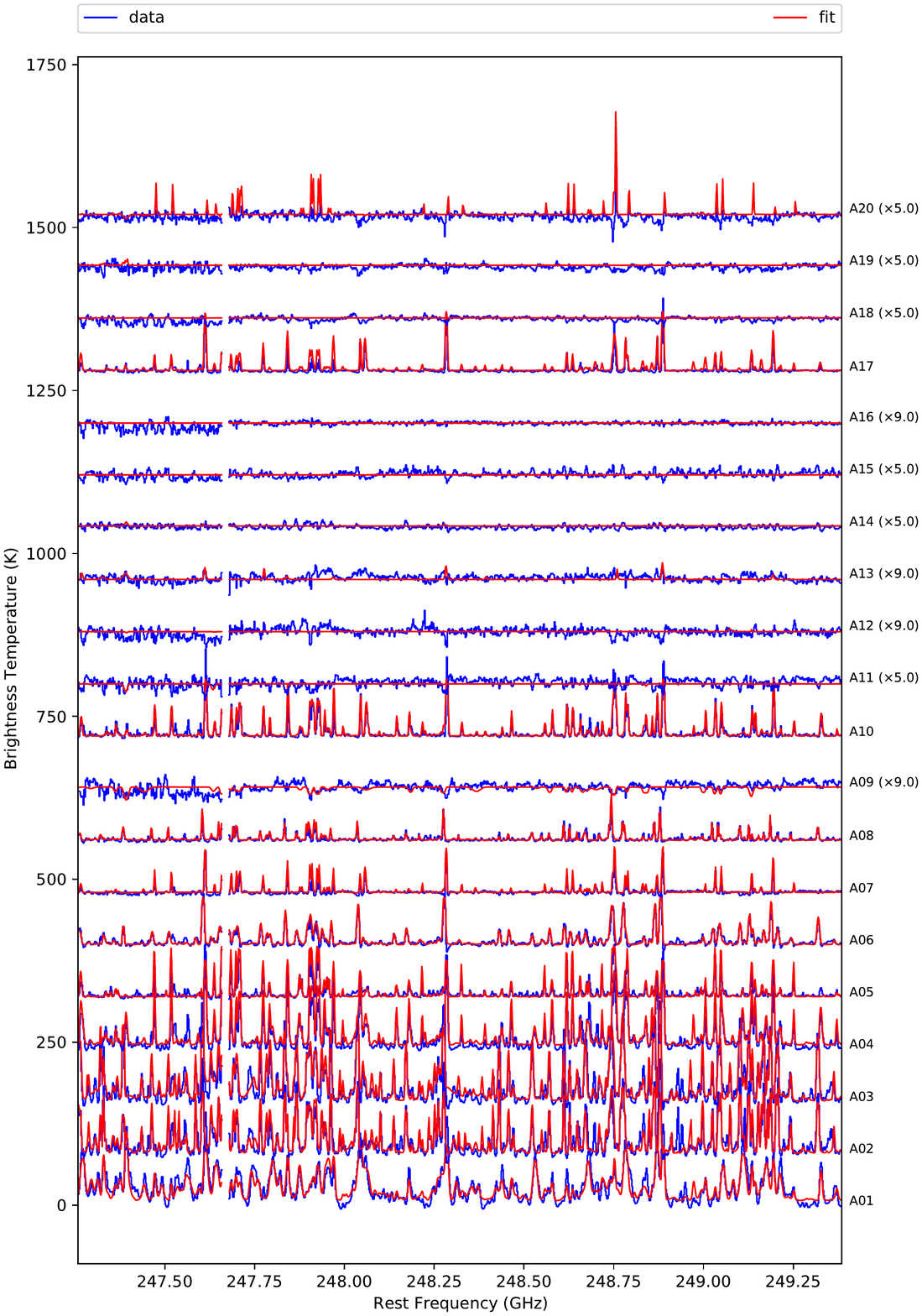}\\
   \caption{Excerpts from the ALMA survey of Sgr~B2(N). Here, the observed spectrum of each core is shown in blue together with the full model in red. On the right sight of each spectrum the name of the corresponding core is described together with the applied scaling factor (in round brackets).}
   \label{fig:SgrB2NExcerpt}
\end{figure*}
\clearpage
\newpage
\twocolumn

%===============================================================================
% plots describing error estimation
\onecolumn
\section{Error estimation}\label{app:errorestim}

%*******************************************************************************
% Figure: corner plot of MCMC error analysis
\begin{figure*}[!b]
   \centering
   \includegraphics[width=0.97\textwidth]{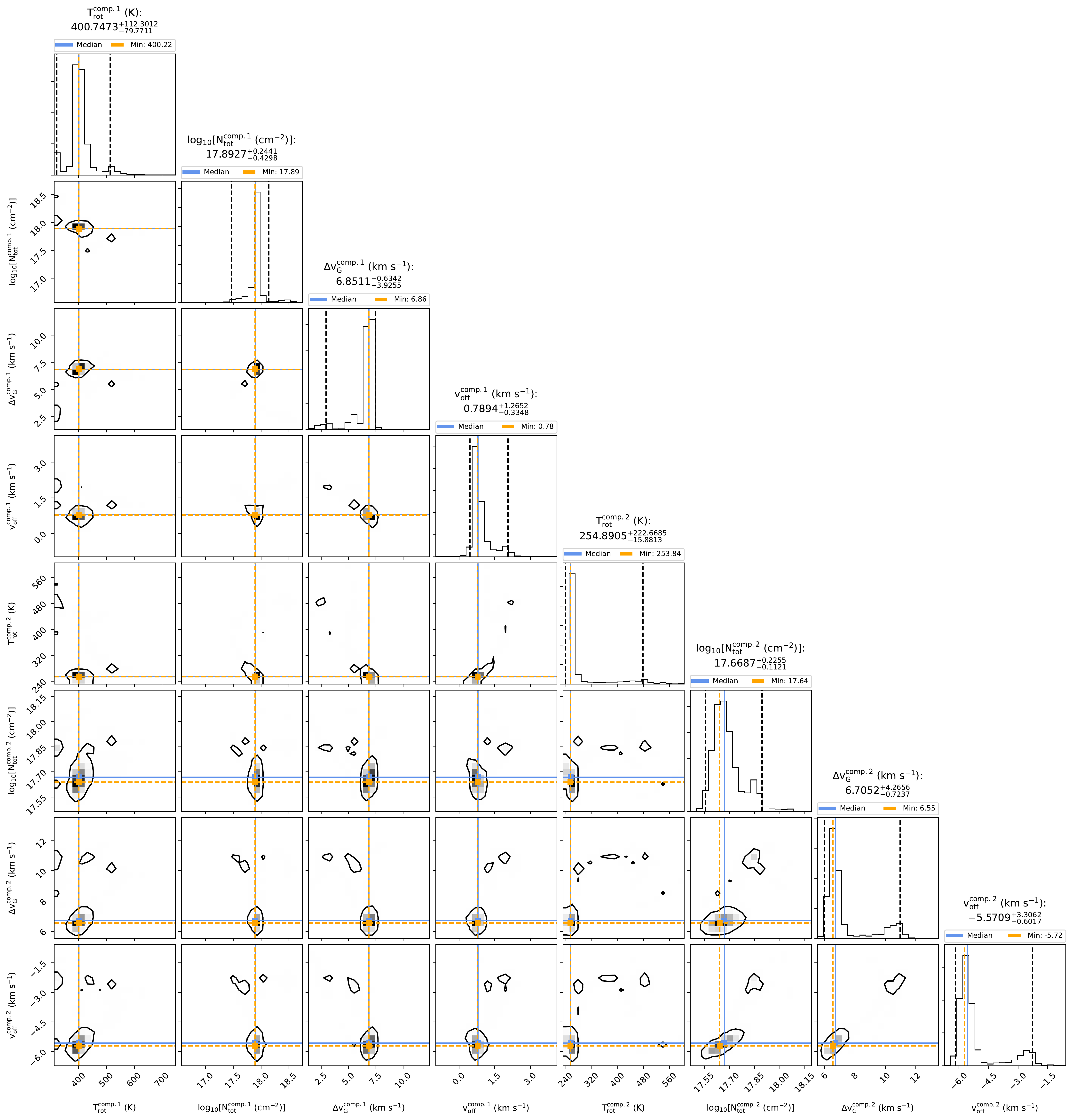}\\
   \caption{Corner plot \citepads{corner} showing the one and two dimensional projections of the posterior probability distributions of the model parameters for OCS of core A03 in Sgr~B2(M). On top of each column the probability distribution for each free parameter is shown together with the value of the 50~\% quantile (median) and the corresponding left and right errors. The black dashed lines indicate the lower and upper limits of the corresponding highest posterior density (HPD) interval, respectively. The median is shown in blue, while the dashed orange lines indicate the parameter values of the best fit. For very asymmetric distributions, the parameter values of the best fit are at quite some distance from the median, at or even beyond the limits of the HPD interval. The plots in the lower left corner describe the projected 2D histograms of two parameters and the contours the HPD regions, respectively. In order to get a better estimation of the errors, we determine the error of the column density on log scale and use the velocity offset (v$_{\rm off}$) related to the source velocity of v$_{\rm LSR}$ = 64~km~s$^{-1}$.\\}
   \label{fig:ErrorEstimOCS}
\end{figure*}

%*******************************************************************************
% Figure: corner plot of MCMC error analysis
\begin{figure*}[!b]
   \centering
   \includegraphics[width=0.99\textwidth]{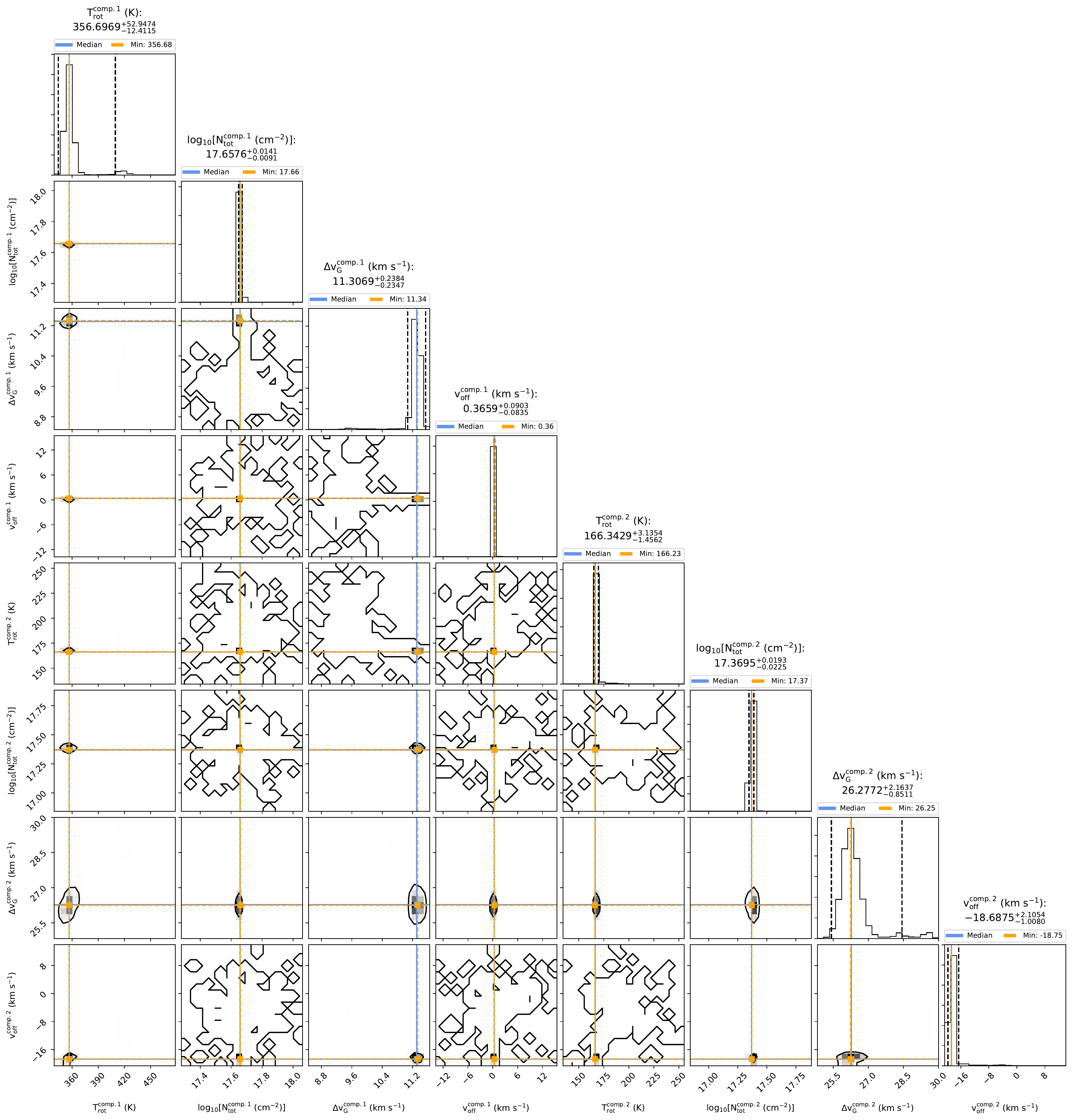}\\
   \caption{Corner plot \citepads{corner} showing the one and two dimensional projections of the posterior probability distributions of the model parameters for CH$_3$CN of core A03 in Sgr~B2(M). On top of each column the probability distribution for each free parameter is shown together with the value of the 50~\% quantile (median) and the corresponding left and right errors. The black dashed lines indicate the lower and upper limits of the corresponding highest posterior density (HPD) interval, respectively. The median is shown in blue, while the dashed orange lines indicate the parameter values of the best fit. For very asymmetric distributions, the parameter values of the best fit are at quite some distance from the median, at or even beyond the limits of the HPD interval. The plots in the lower left corner describe the projected 2D histograms of two parameters and the contours the HPD regions, respectively. In order to get a better estimation of the errors, we determine the error of the column density on log scale and use the velocity offset (v$_{\rm off}$) related to the source velocity of v$_{\rm LSR}$ = 64~km~s$^{-1}$.\\}
   \label{fig:ErrorEstimCH3CN}
\end{figure*}
\clearpage
\newpage
\twocolumn

%===============================================================================
% Principle component analysis (PCA) basics
\onecolumn
\section{Principle component analysis (PCA)}\label{app:sec:PCA}

%*******************************************************************************
% Figure: Projection onto a pair of principal components: Sgr B2(M)
\begin{figure*}[!b]
   \centering
   \includegraphics[width=0.85\textwidth]{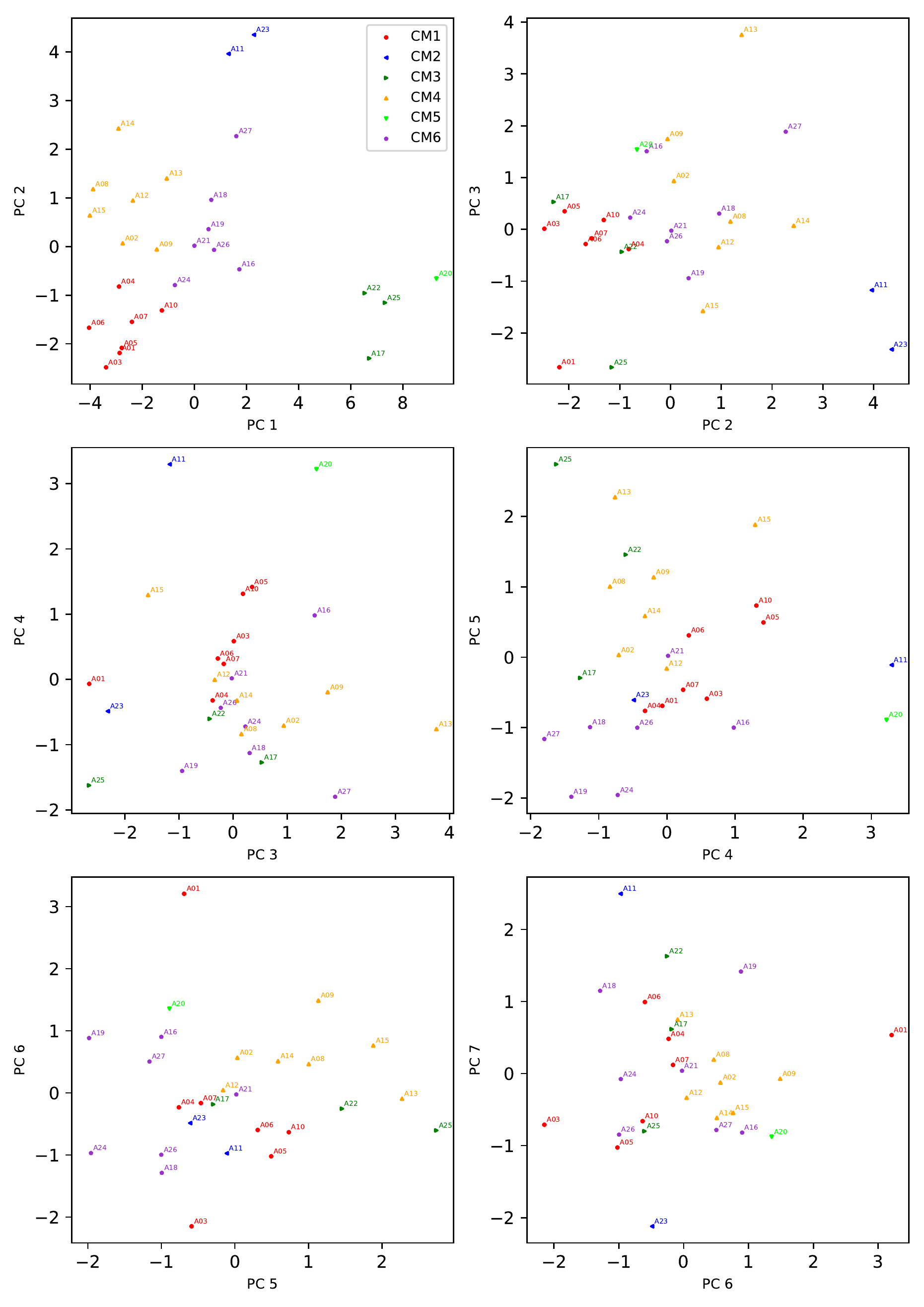}\\
   \caption{The abundances of each source in Sgr~B2(M) projected onto a pair of principal components. The colors indicate the different classes identified by the k-means clustering algorithm.}
   \label{fig:PCAProjectionSgrB2M}
\end{figure*}
\clearpage
\hfill

%*******************************************************************************
% Figure: Projection onto a pair of principal components: Sgr B2(N)
\begin{figure*}[!b]
   \centering
   \includegraphics[width=0.85\textwidth]{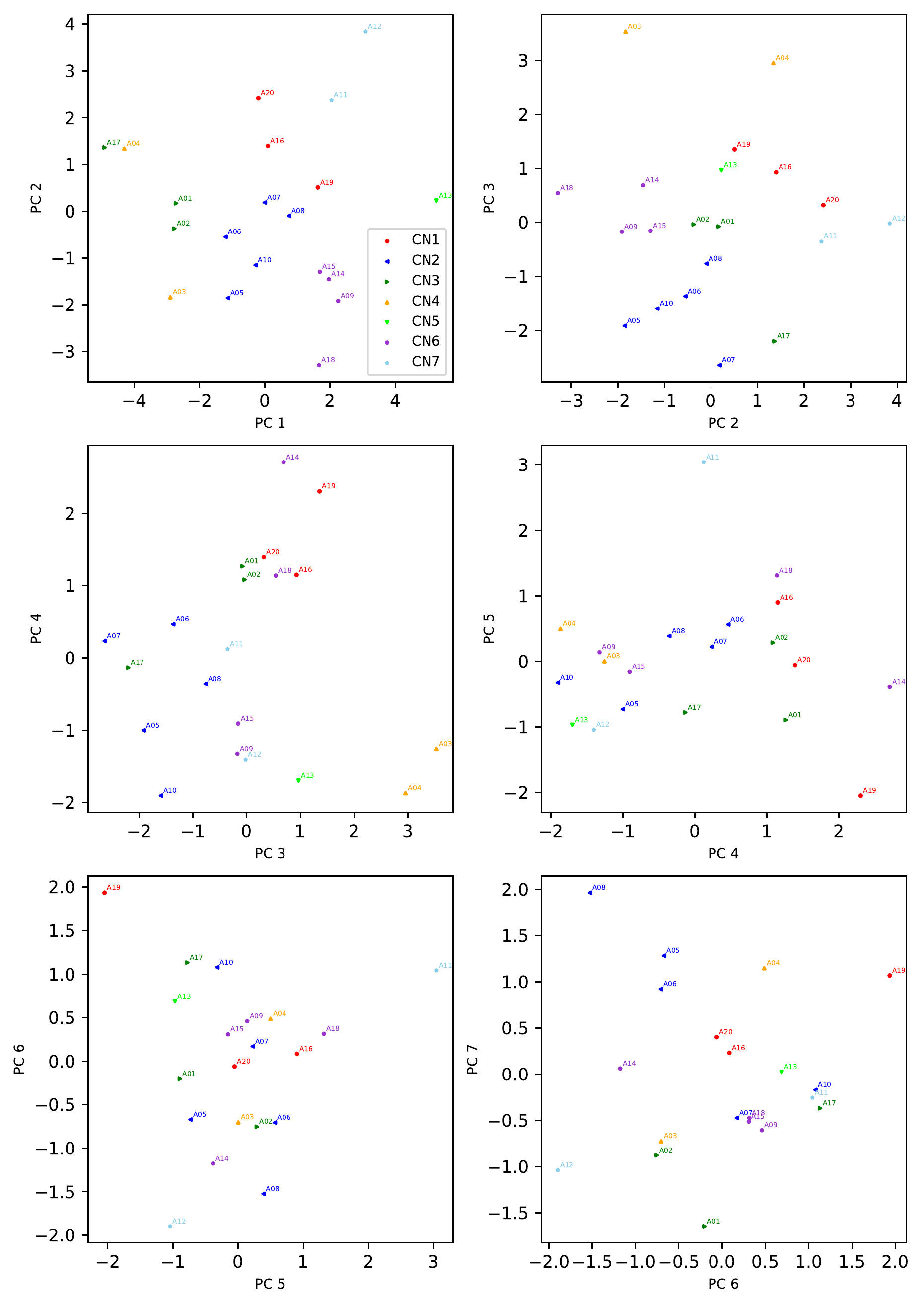}\\
   \caption{The abundances of each source in Sgr~B2(N) projected onto a pair of principal components. The colors indicate the different classes identified by the k-means clustering algorithm.}
   \label{fig:PCAProjectionSgrB2N}
\end{figure*}
\clearpage
\hfill

%:::::::::::::::::::::::::::::::::::::::::::::::::::::::::::::::::::::::::::::::
% Projection onto a pair of principal components
%:::::::::::::::::::::::::::::::::::::::::::::::::::::::::::::::::::::::::::::::

%*******************************************************************************
% Figure: first eight eigenvectors of abundances in Sgr B2(M)
\begin{figure*}[!b]
   \centering
   \includegraphics[width=0.85\textwidth]{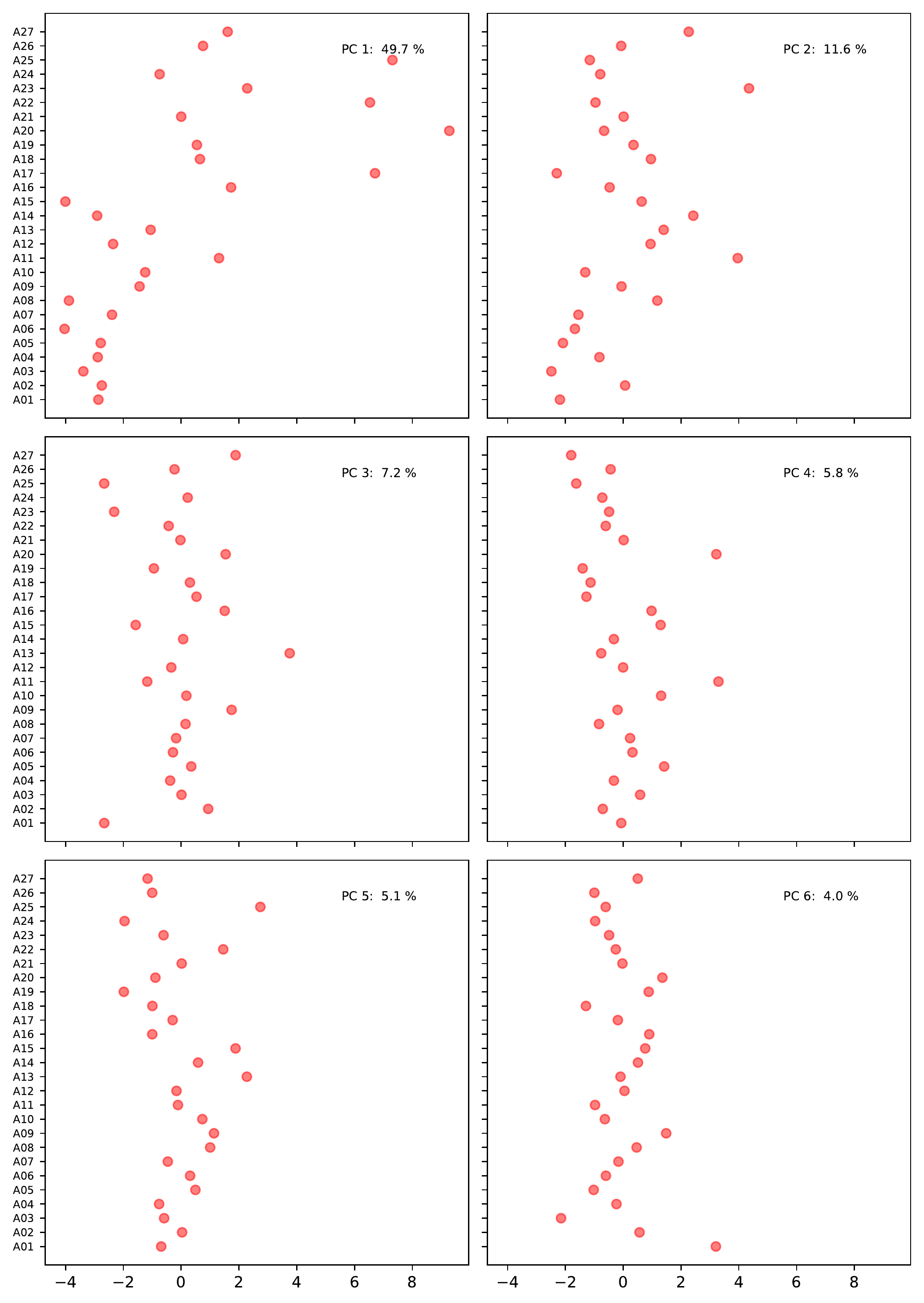}\\
   \caption{First six eigenvectors of abundances in Sgr B2(M).}
   \label{fig:PCAEVSgrB2M}
\end{figure*}
\clearpage
\hfill

%*******************************************************************************
% Figure: velocity offsets of sources in Sgr~B2(N), envelope layer
\begin{figure*}[!b]
   \centering
   \includegraphics[width=0.85\textwidth]{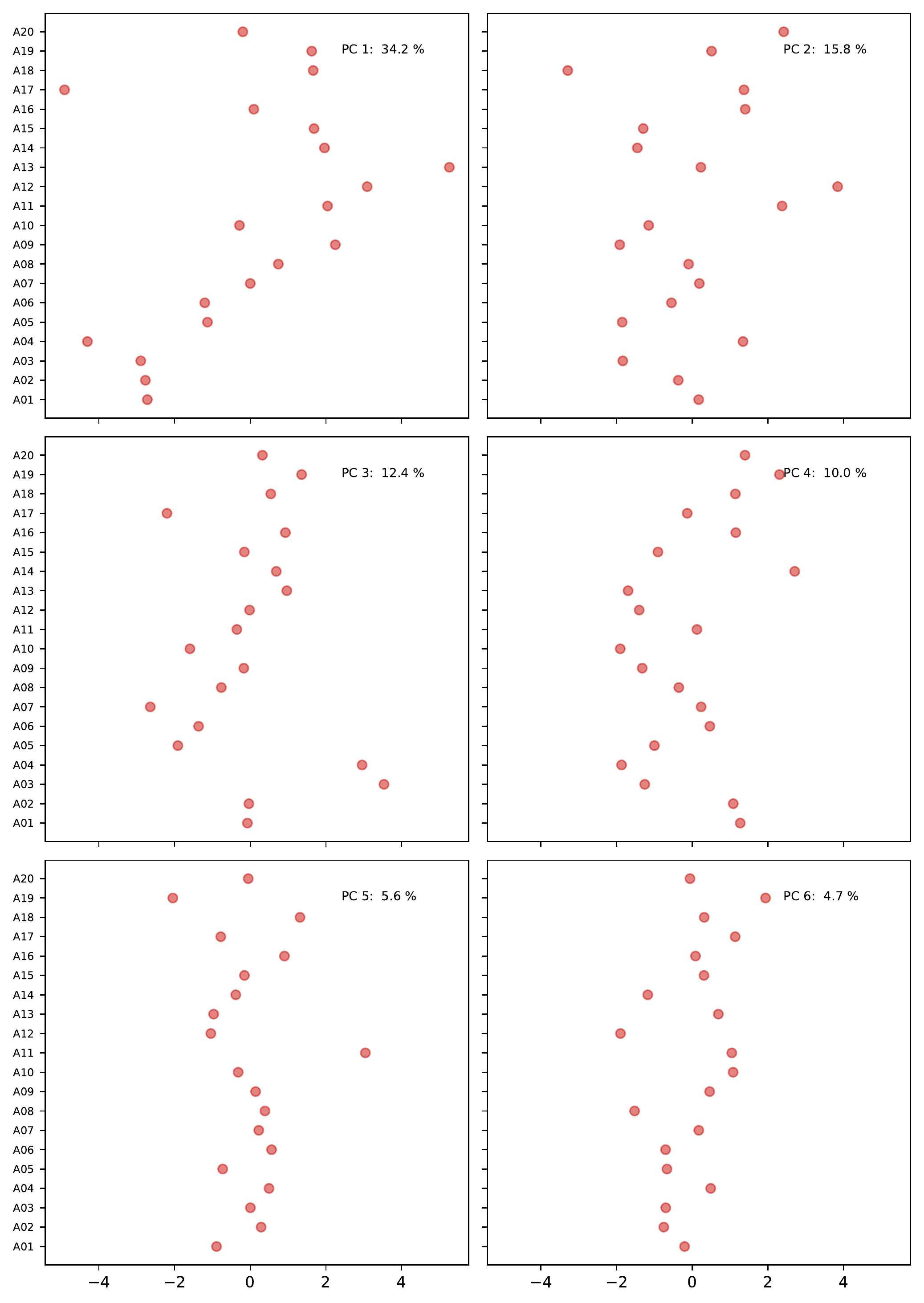}\\
   \caption{First six eigenvectors of abundances in Sgr B2(N).}
   \label{fig:PCAEVSgrB2N}
\end{figure*}
\clearpage
\hfill

%:::::::::::::::::::::::::::::::::::::::::::::::::::::::::::::::::::::::::::::::
% Projection onto a pair of principal components
%:::::::::::::::::::::::::::::::::::::::::::::::::::::::::::::::::::::::::::::::

%*******************************************************************************
% Figure: Averaged abundances for each identified cluster in Sgr B2(M)
\begin{figure*}[!htb]
   \centering
   \includegraphics[width=0.85\textwidth]{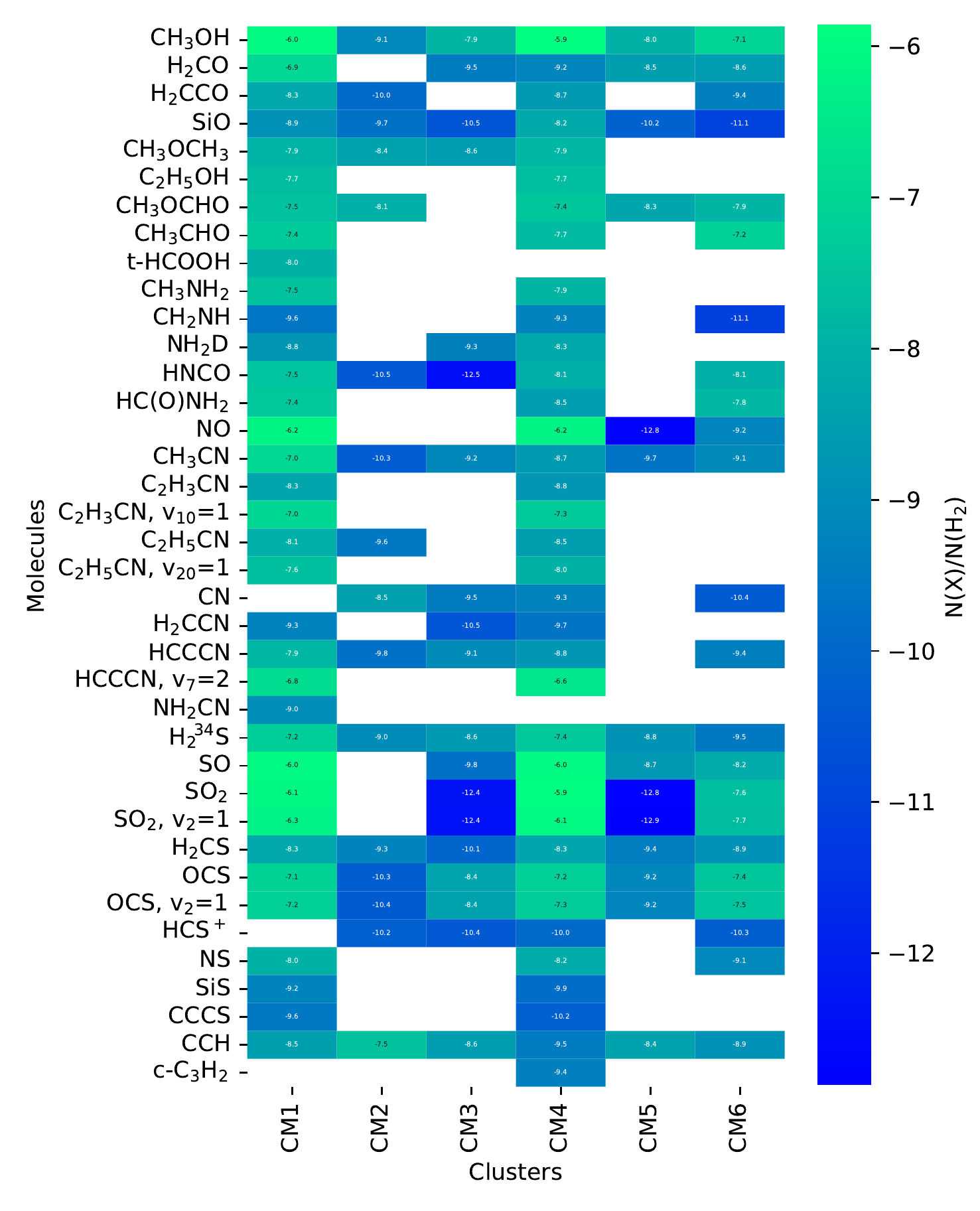}\\
   \caption{Mean abundances for each identified cluster in Sgr B2(M).}
   \label{fig:PCAMeanXSgrB2M}
\end{figure*}
\newpage
\clearpage

%*******************************************************************************
% Figure: Averaged abundances for each identified cluster in Sgr~B2(N), envelope layer
\begin{figure*}[!htb]
   \centering
   \includegraphics[width=0.85\textwidth]{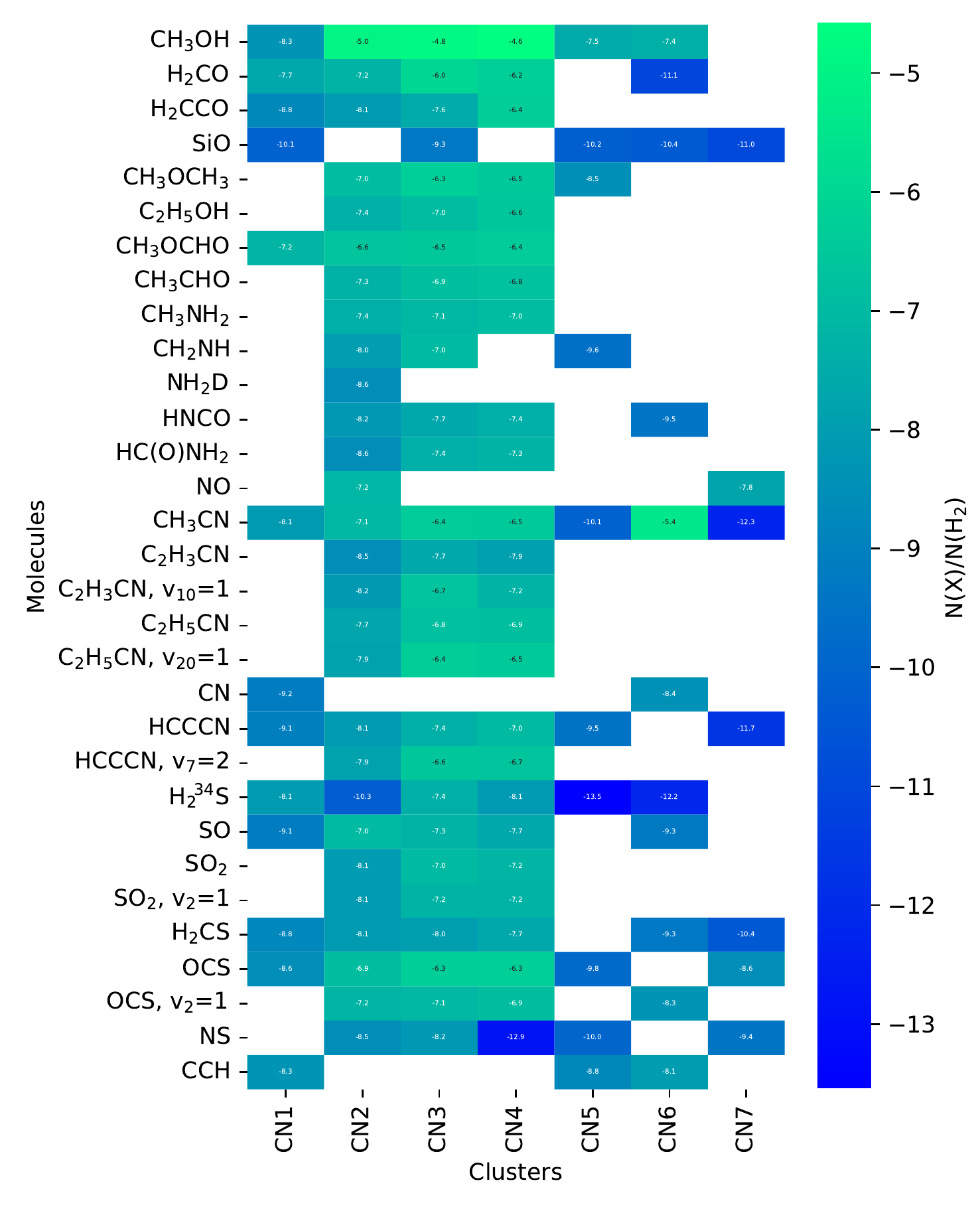}\\
   \caption{Mean abundances for each identified cluster in Sgr B2(N).}
   \label{fig:PCAMeanXSgrB2N}
\end{figure*}
\newpage
\clearpage

%*******************************************************************************
% Figure: Averaged temperature for each identified cluster in Sgr B2(M)
\begin{figure*}[!htb]
   \centering
   \includegraphics[width=0.85\textwidth]{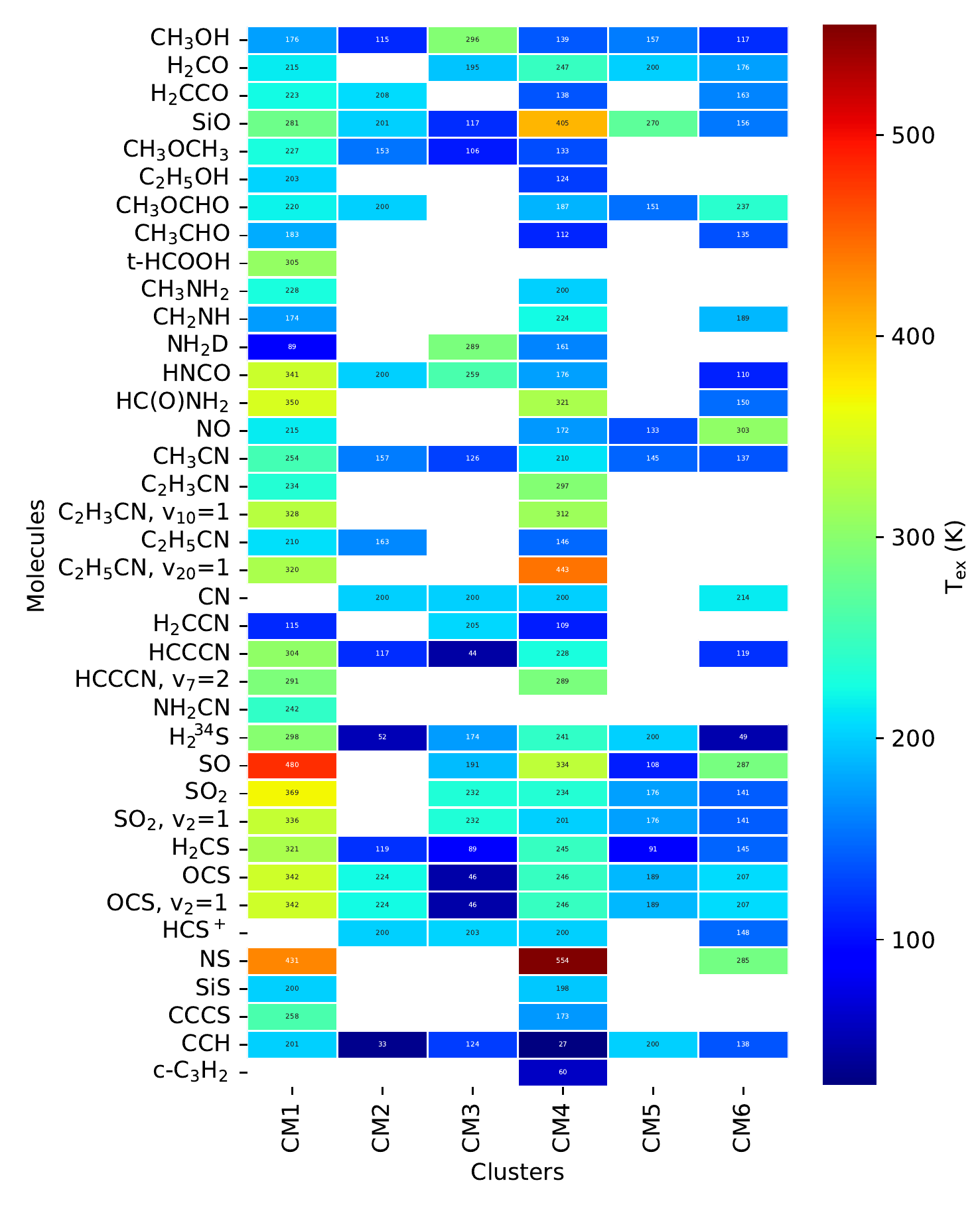}\\
   \caption{Mean temperature for each identified cluster in Sgr B2(M).}
   \label{fig:PCAMeanTSgrB2M}
\end{figure*}
\newpage
\clearpage

%*******************************************************************************
% Figure: Averaged temperature for each identified cluster in Sgr~B2(N), envelope layer
\begin{figure*}[!htb]
   \centering
   \includegraphics[width=0.85\textwidth]{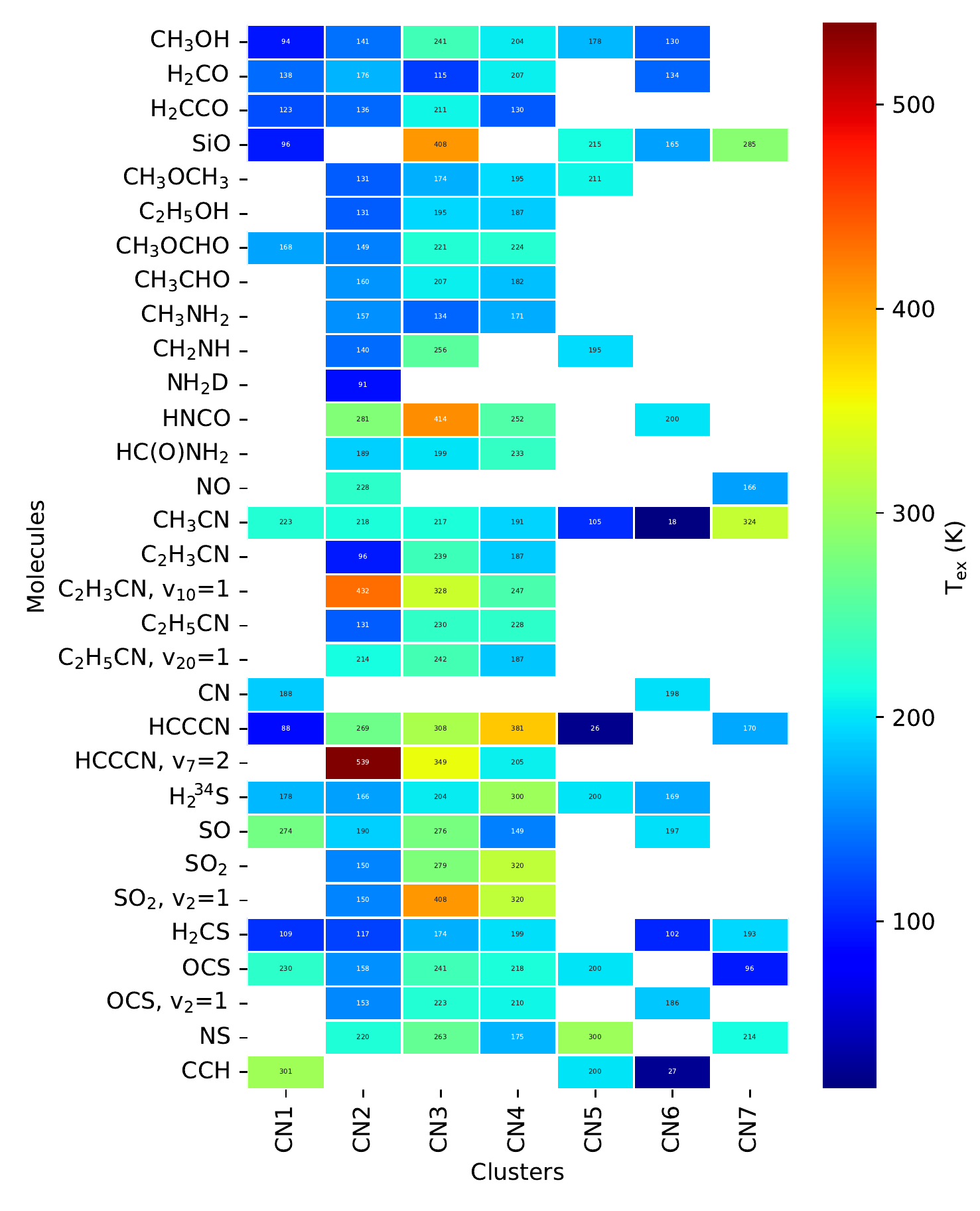}\\
   \caption{Mean temperature for each identified cluster in Sgr B2(N).}
   \label{fig:PCAMeanTSgrB2N}
\end{figure*}
\clearpage
\newpage
\twocolumn

%===============================================================================
% plots of selected transitions of identified molecules
\section{Identified species}\label{app:identifiedspecies}

% see one of the following paper for a similar analysis
% (path: /home/moeller/Projects/SgrB2/paper/others/hot-cores/)
% 2202.08621.pdf, 2201.09945.pdf

% Belloche et al. 2013:     \citepads{2013A&A...559A..47B}
% Bisschop et al. 2007:     \citepads{2007A&A...465..913B}
% Boegelund et al.2018:     \citepads{2019A&A...628A...2B}
% Bonfand et al. 2017:      \citepads{2017A&A...604A..60B}
% Moeller et al. 2021:      \citepads{2021A&A...651A...9M}
% Neill et al. 2014:        \citepads{2014ApJ...789....8N}
% Sutton et al. 1991:       \citepads{1991ApJS...77..255S}

%*******************************************************************************
% Table: Comparison of temperature and abundance for Sgr~B2(M)
\begin{table*}[!htb]
\caption{Comparison of temperature T$_{\rm rot}$ and abundance $X = N(x)/N(H_2)$ ranges for species identified in sources in Sgr~B2(M).}
\label{Tab:ParamRangesSgrB2M}
\centering
\tiny
\begin{tabular}{lrrrr}
    \hline
    \hline
molecule:                                    & T$_{\rm rot}^{\rm min}$~(K) -- T$_{\rm rot}^{\rm max}$~(K): & T$_{\rm rot, ref}^{\rm min}$~(K) -- T$_{\rm rot, ref}^{\rm max}$~(K): & X$^{\rm min}$ -- X$^{\rm max}$:                             & X$_{\rm ref}^{\rm min}$ -- X$_{\rm ref}^{\rm max}$:  \\
    \hline
CO                                           &                                                           - &                                                                     - &         $1 \cdot 10^{-11}$~(A01) -- $3 \cdot 10^{-5}$~(A15) &     $5 \cdot 10^{-10}$ -- $4 \cdot 10^{-3}$$^{(a)}$ \\
CH$_3$OH                                     &                                        8 (A10) -- 337 (A22) &                                                      3 -- 308$^{(a)}$ &         $7 \cdot 10^{-12}$~(A23) -- $2 \cdot 10^{-6}$~(A05) &     $7 \cdot 10^{-10}$ -- $2 \cdot 10^{-8}$$^{(a)}$ \\
H$_2$CO                                      &                                        5 (A22) -- 369 (A13) &                                                      43 -- 74$^{(a)}$ &         $2 \cdot 10^{-13}$~(A13) -- $3 \cdot 10^{-7}$~(A03) &    $2 \cdot 10^{-10}$ -- $5 \cdot 10^{-10}$$^{(a)}$ \\
H$_2$CCO                                     &                                       11 (A06) -- 342 (A03) &                                                            71$^{(b)}$ &         $1 \cdot 10^{-11}$~(A18) -- $2 \cdot 10^{-8}$~(A01) &                          $2 \cdot 10^{-10}$$^{(b)}$ \\
H$^{13}$CO$^+$                               &                                                           - &                                                                     - &         $4 \cdot 10^{-13}$~(A18) -- $2 \cdot 10^{-9}$~(A21) &    $2 \cdot 10^{-12}$ -- $7 \cdot 10^{-11}$$^{(a)}$ \\
HO$^{13}$C$^+$                               &                                                           - &                                                                     - &         $3 \cdot 10^{-13}$~(A05) -- $4 \cdot 10^{-9}$~(A15) &                                                   - \\
SiO                                          &                                       3 (A16) -- 1097 (A15) &                                                     25 -- 103$^{(a)}$ &         $2 \cdot 10^{-12}$~(A14) -- $4 \cdot 10^{-7}$~(A15) &    $2 \cdot 10^{-11}$ -- $9 \cdot 10^{-11}$$^{(a)}$ \\
CH$_3$OCH$_3$                                &                                        8 (A11) -- 302 (A04) &                                                           100$^{(a)}$ &         $1 \cdot 10^{-11}$~(A11) -- $4 \cdot 10^{-8}$~(A05) &                          $4 \cdot 10^{-10}$$^{(a)}$ \\
C$_2$H$_5$OH                                 &                                      124 (A09) -- 206 (A05) &                                                     20 -- 100$^{(c)}$ &          $1 \cdot 10^{-8}$~(A10) -- $3 \cdot 10^{-8}$~(A05) &     $5 \cdot 10^{-10}$ -- $1 \cdot 10^{-8}$$^{(c)}$ \\
CH$_3$OCHO                                   &                                      151 (A20) -- 266 (A26) &                                                            40$^{(c)}$ &          $4 \cdot 10^{-9}$~(A20) -- $3 \cdot 10^{-8}$~(A05) &                           $2 \cdot 10^{-8}$$^{(c)}$ \\
CH$_3$CHO                                    &                                       80 (A13) -- 273 (A03) &                                                            40$^{(c)}$ &          $7 \cdot 10^{-9}$~(A13) -- $9 \cdot 10^{-8}$~(A03) &                          $5 \cdot 10^{-10}$$^{(c)}$ \\
t-HCOOH                                      &                                      200 (A04) -- 392 (A03) &                                                                     - &          $5 \cdot 10^{-9}$~(A04) -- $1 \cdot 10^{-8}$~(A03) &                                                   - \\
CH$_3$NH$_2$                                 &                                        9 (A16) -- 311 (A03) &                                                            14$^{(a)}$ &         $1 \cdot 10^{-10}$~(A16) -- $7 \cdot 10^{-8}$~(A03) &                           $9 \cdot 10^{-9}$$^{(a)}$ \\
CH$_2$NH                                     &                                        3 (A27) -- 314 (A03) &                                                            11$^{(a)}$ &         $5 \cdot 10^{-14}$~(A01) -- $1 \cdot 10^{-8}$~(A03) &                          $1 \cdot 10^{-10}$$^{(a)}$ \\
NH$_2$D                                      &                                        9 (A18) -- 289 (A22) &                                                                     - &         $4 \cdot 10^{-10}$~(A22) -- $2 \cdot 10^{-7}$~(A18) &                                                   - \\
HNCO                                         &                                        3 (A17) -- 542 (A01) &                                                     12 -- 300$^{(a)}$ &         $3 \cdot 10^{-13}$~(A25) -- $4 \cdot 10^{-6}$~(A15) &    $7 \cdot 10^{-11}$ -- $4 \cdot 10^{-10}$$^{(a)}$ \\
HC(O)NH$_2$                                  &                                       19 (A25) -- 697 (A02) &                                                           300$^{(a)}$ &         $5 \cdot 10^{-10}$~(A16) -- $2 \cdot 10^{-7}$~(A03) &                          $1 \cdot 10^{-10}$$^{(a)}$ \\
NO                                           &                                        3 (A08) -- 314 (A03) &                                                     24 -- 325$^{(a)}$ &         $1 \cdot 10^{-13}$~(A20) -- $5 \cdot 10^{-6}$~(A15) &      $4 \cdot 10^{-9}$ -- $8 \cdot 10^{-8}$$^{(a)}$ \\
NO$^+$                                       &                                                           - &                                                                     - &         $1 \cdot 10^{-11}$~(A18) -- $4 \cdot 10^{-8}$~(A01) &                                                   - \\
HCN                                          &                                                           - &                                                                     - &         $1 \cdot 10^{-13}$~(A26) -- $6 \cdot 10^{-7}$~(A02) &     $1 \cdot 10^{-12}$ -- $3 \cdot 10^{-5}$$^{(a)}$ \\
HNC                                          &                                                           - &                                                                     - &         $1 \cdot 10^{-14}$~(A21) -- $1 \cdot 10^{-8}$~(A15) &     $1 \cdot 10^{-10}$ -- $3 \cdot 10^{-9}$$^{(a)}$ \\
CH$_3$CN                                     &                                        4 (A19) -- 432 (A01) &                                                     60 -- 200$^{(c)}$ &         $2 \cdot 10^{-13}$~(A05) -- $2 \cdot 10^{-5}$~(A06) &     $2 \cdot 10^{-10}$ -- $1 \cdot 10^{-7}$$^{(c)}$ \\
C$_2$H$_3$CN                                 &                                      162 (A10) -- 405 (A01) &                                                            50$^{(c)}$ &         $5 \cdot 10^{-10}$~(A09) -- $2 \cdot 10^{-8}$~(A01) &                           $3 \cdot 10^{-9}$$^{(c)}$ \\
C$_2$H$_3$CN, v$_{10}$=1                     &                                      307 (A03) -- 343 (A06) &                                                                     - &          $4 \cdot 10^{-8}$~(A02) -- $2 \cdot 10^{-7}$~(A01) &                                                   - \\
C$_2$H$_5$CN                                 &                                       10 (A06) -- 381 (A01) &                                                            70$^{(c)}$ &         $5 \cdot 10^{-12}$~(A06) -- $2 \cdot 10^{-8}$~(A03) &                           $9 \cdot 10^{-9}$$^{(c)}$ \\
C$_2$H$_5$CN, v$_{20}$=1                     &                                      188 (A07) -- 443 (A02) &                                                                     - &          $1 \cdot 10^{-8}$~(A04) -- $7 \cdot 10^{-8}$~(A01) &                                                   - \\
CN                                           &                                        3 (A25) -- 273 (A24) &                                                            33$^{(a)}$ &         $1 \cdot 10^{-13}$~(A26) -- $5 \cdot 10^{-8}$~(A14) &                           $1 \cdot 10^{-9}$$^{(a)}$ \\
H$_2$CCN                                     &                                       10 (A25) -- 205 (A25) &                                                                     - &        $1 \cdot 10^{-11}$~(A16) -- $5 \cdot 10^{-10}$~(A09) &                                                   - \\
HCCCN                                        &                                        3 (A12) -- 536 (A01) &                                                      20 -- 60$^{(c)}$ &         $8 \cdot 10^{-11}$~(A19) -- $6 \cdot 10^{-6}$~(A15) &    $2 \cdot 10^{-10}$ -- $4 \cdot 10^{-10}$$^{(c)}$ \\
HCCCN, v$_7$=2                               &                                      120 (A07) -- 568 (A01) &                                                           200$^{(c)}$ &          $7 \cdot 10^{-8}$~(A06) -- $5 \cdot 10^{-7}$~(A01) &                           $4 \cdot 10^{-7}$$^{(c)}$ \\
NH$_2$CN                                     &                                      175 (A05) -- 308 (A03) &                                                            20$^{(c)}$ &         $3 \cdot 10^{-10}$~(A05) -- $2 \cdot 10^{-9}$~(A03) &                          $3 \cdot 10^{-11}$$^{(c)}$ \\
H$_2 \! ^{34}$S                              &                                        5 (A27) -- 496 (A01) &                                                       2 -- 97$^{(a)}$ &         $2 \cdot 10^{-10}$~(A26) -- $3 \cdot 10^{-5}$~(A19) &     $9 \cdot 10^{-14}$ -- $5 \cdot 10^{-9}$$^{(a)}$ \\
SO                                           &                                        6 (A18) -- 866 (A01) &                                                     16 -- 174$^{(a)}$ &         $4 \cdot 10^{-11}$~(A27) -- $1 \cdot 10^{-5}$~(A15) &     $9 \cdot 10^{-11}$ -- $6 \cdot 10^{-8}$$^{(a)}$ \\
SO$^+$                                       &                                                           - &                                                                     - &         $8 \cdot 10^{-10}$~(A22) -- $5 \cdot 10^{-8}$~(A22) &    $3 \cdot 10^{-11}$ -- $6 \cdot 10^{-10}$$^{(a)}$ \\
SO$_2$                                       &                                       16 (A19) -- 648 (A01) &                                                     50 -- 200$^{(c)}$ &         $1 \cdot 10^{-13}$~(A20) -- $1 \cdot 10^{-5}$~(A15) &      $5 \cdot 10^{-9}$ -- $1 \cdot 10^{-5}$$^{(c)}$ \\
SO$_2$, v$_2$=1                              &                                       16 (A19) -- 434 (A02) &                                                           200$^{(c)}$ &         $1 \cdot 10^{-13}$~(A20) -- $5 \cdot 10^{-6}$~(A08) &                           $7 \cdot 10^{-5}$$^{(c)}$ \\
CS                                           &                                                           - &                                                                     - &         $3 \cdot 10^{-11}$~(A04) -- $2 \cdot 10^{-6}$~(A14) &     $7 \cdot 10^{-10}$ -- $4 \cdot 10^{-9}$$^{(a)}$ \\
H$_2$CS                                      &                                        3 (A15) -- 774 (A01) &                                                           120$^{(a)}$ &         $7 \cdot 10^{-13}$~(A17) -- $1 \cdot 10^{-5}$~(A15) &                          $2 \cdot 10^{-10}$$^{(a)}$ \\
OCS                                          &                                        6 (A22) -- 612 (A01) &                                                    134 -- 172$^{(a)}$ &         $3 \cdot 10^{-12}$~(A11) -- $1 \cdot 10^{-5}$~(A15) &      $1 \cdot 10^{-9}$ -- $5 \cdot 10^{-9}$$^{(a)}$ \\
OCS, v$_2$=1                                 &                                        6 (A22) -- 612 (A01) &                                                                     - &         $3 \cdot 10^{-12}$~(A11) -- $1 \cdot 10^{-5}$~(A15) &                                                   - \\
HCS$^+$                                      &                                        3 (A12) -- 203 (A22) &                                                            51$^{(a)}$ &        $1 \cdot 10^{-12}$~(A26) -- $9 \cdot 10^{-10}$~(A12) &                          $1 \cdot 10^{-10}$$^{(a)}$ \\
NS                                           &                                      285 (A18) -- 554 (A02) &                                                            70$^{(a)}$ &         $7 \cdot 10^{-10}$~(A18) -- $1 \cdot 10^{-8}$~(A01) &                          $2 \cdot 10^{-10}$$^{(a)}$ \\
SiS                                          &                                      198 (A09) -- 200 (A03) &                                                                     - &        $1 \cdot 10^{-10}$~(A09) -- $8 \cdot 10^{-10}$~(A04) &                                                   - \\
CCCS                                         &                                      173 (A02) -- 258 (A01) &                                                                     - &        $5 \cdot 10^{-11}$~(A02) -- $2 \cdot 10^{-10}$~(A01) &                                                   - \\
CCH                                          &                                        3 (A03) -- 201 (A07) &                                                            26$^{(a)}$ &         $6 \cdot 10^{-12}$~(A18) -- $1 \cdot 10^{-7}$~(A15) &                           $5 \cdot 10^{-9}$$^{(a)}$ \\
c-C$_3$H$_2$                                 &                                         5 (A04) -- 60 (A13) &                                                       2 -- 15$^{(c)}$ &         $1 \cdot 10^{-11}$~(A04) -- $1 \cdot 10^{-9}$~(A15) &    $1 \cdot 10^{-13}$ -- $2 \cdot 10^{-10}$$^{(c)}$ \\
PH$_3$                                       &                                                           - &                                                                     - &         $3 \cdot 10^{-12}$~(A09) -- $5 \cdot 10^{-6}$~(A13) &                                                   - \\
    \hline
    \hline
\end{tabular}
\begin{tablenotes}
   \item {The source ID in round brackets indicates the source where the minimum and maximum value of the corresponding parameter was determined. The abundances are computed relative to the hydrogen column densities derived in our first paper for each source. The abundance ranges for references $^{(a)}$\citetads{2021A&A...651A...9M} and $^{(c)}$\citetads{2013A&A...559A..47B} are given for an H$_2$ column density of $1.1 \times 10^{24}$~cm$^{-2}$ for core and $1.8 \times 10^{24}$~cm$^{-2}$ for foreground components. For reference $^{(b)}$\citetads{1991ApJS...77..255S} an H$_2$ column density of $2.6 \times 10^{24}$~cm$^{-2}$ was used. For species with only a single transition within the frequency ranges covered by our observations, a reliable temperature estimation is not possible.}
\end{tablenotes}
\end{table*}

%*******************************************************************************
% Table: Comparison of temperature and abundance for Sgr~B2(N)
\begin{table*}[!htb]
\caption{Comparison of temperature T$_{\rm rot}$ and abundance $X = N(x)/N(H_2)$ ranges for species identified in sources in Sgr~B2(N).}
\label{Tab:ParamRangesSgrB2N}
\centering
\tiny
\begin{tabular}{lrrrr}
    \hline
    \hline
molecule:                                    & T$_{\rm rot}^{\rm min}$~(K) -- T$_{\rm rot}^{\rm max}$~(K): & T$_{\rm rot, ref}^{\rm min}$~(K) -- T$_{\rm rot, ref}^{\rm max}$~(K): & X$^{\rm min}$ -- X$^{\rm max}$:                             & X$_{\rm ref}^{\rm min}$ -- X$_{\rm ref}^{\rm max}$:  \\
    \hline
CO                                           &                                                           - &                                                                     - &          $6 \cdot 10^{-9}$~(A12) -- $1 \cdot 10^{-5}$~(A03) &      $1 \cdot 10^{-9}$ -- $3 \cdot 10^{-9}$$^{(d)}$ \\
CH$_3$OH                                     &                                       11 (A19) -- 294 (A01) &                                                    160 -- 190$^{(e)}$ &         $9 \cdot 10^{-11}$~(A19) -- $5 \cdot 10^{-5}$~(A18) &      $8 \cdot 10^{-8}$ -- $2 \cdot 10^{-5}$$^{(e)}$ \\
H$_2$CO                                      &                                        4 (A14) -- 236 (A14) &                                                      2 -- 100$^{(d)}$ &         $8 \cdot 10^{-13}$~(A14) -- $9 \cdot 10^{-7}$~(A17) &      $2 \cdot 10^{-9}$ -- $6 \cdot 10^{-8}$$^{(d)}$ \\
H$_2$CCO                                     &                                       14 (A11) -- 309 (A01) &                                                           180$^{(f)}$ &          $1 \cdot 10^{-9}$~(A20) -- $6 \cdot 10^{-6}$~(A04) &                           $8 \cdot 10^{-9}$$^{(f)}$ \\
H$^{13}$CO$^+$                               &                                                           - &                                                                     - &         $2 \cdot 10^{-13}$~(A06) -- $1 \cdot 10^{-9}$~(A08) &    $3 \cdot 10^{-15}$ -- $3 \cdot 10^{-11}$$^{(d)}$ \\
HO$^{13}$C$^+$                               &                                                           - &                                                                     - &         $6 \cdot 10^{-13}$~(A16) -- $3 \cdot 10^{-7}$~(A04) &                                                   - \\
SiO                                          &                                        3 (A20) -- 408 (A17) &                                                       2 -- 16$^{(d)}$ &         $2 \cdot 10^{-12}$~(A11) -- $6 \cdot 10^{-9}$~(A17) &    $6 \cdot 10^{-14}$ -- $4 \cdot 10^{-11}$$^{(d)}$ \\
CH$_3$OCH$_3$                                &                                      120 (A08) -- 210 (A13) &                                                           130$^{(f)}$ &          $2 \cdot 10^{-9}$~(A13) -- $8 \cdot 10^{-7}$~(A17) &                           $3 \cdot 10^{-8}$$^{(f)}$ \\
C$_2$H$_5$OH                                 &                                       93 (A17) -- 318 (A01) &                                                    145 -- 150$^{(e)}$ &          $1 \cdot 10^{-8}$~(A07) -- $2 \cdot 10^{-7}$~(A03) &      $6 \cdot 10^{-9}$ -- $1 \cdot 10^{-6}$$^{(e)}$ \\
CH$_3$OCHO                                   &                                      109 (A06) -- 298 (A01) &                                                    145 -- 150$^{(e)}$ &          $6 \cdot 10^{-8}$~(A20) -- $8 \cdot 10^{-7}$~(A17) &      $2 \cdot 10^{-8}$ -- $2 \cdot 10^{-6}$$^{(e)}$ \\
CH$_3$CHO                                    &                                      144 (A10) -- 265 (A01) &                                                           100$^{(d)}$ &          $2 \cdot 10^{-8}$~(A08) -- $3 \cdot 10^{-7}$~(A03) &      $1 \cdot 10^{-8}$ -- $1 \cdot 10^{-8}$$^{(d)}$ \\
CH$_3$NH$_2$                                 &                                      103 (A07) -- 181 (A03) &                                                      2 -- 100$^{(d)}$ &          $1 \cdot 10^{-8}$~(A08) -- $1 \cdot 10^{-7}$~(A10) &     $1 \cdot 10^{-11}$ -- $7 \cdot 10^{-8}$$^{(d)}$ \\
CH$_2$NH                                     &                                        3 (A05) -- 256 (A17) &                                                           200$^{(d)}$ &         $1 \cdot 10^{-12}$~(A06) -- $1 \cdot 10^{-7}$~(A17) &      $3 \cdot 10^{-8}$ -- $1 \cdot 10^{-7}$$^{(d)}$ \\
NH$_2$D                                      &                                        91 (A05) -- 91 (A08) &                                                           200$^{(d)}$ &         $9 \cdot 10^{-10}$~(A08) -- $6 \cdot 10^{-9}$~(A05) &      $3 \cdot 10^{-9}$ -- $4 \cdot 10^{-9}$$^{(d)}$ \\
HNCO                                         &                                        3 (A16) -- 586 (A02) &                                                    145 -- 240$^{(e)}$ &         $1 \cdot 10^{-10}$~(A20) -- $5 \cdot 10^{-6}$~(A11) &     $8 \cdot 10^{-10}$ -- $1 \cdot 10^{-6}$$^{(e)}$ \\
HC(O)NH$_2$                                  &                                       13 (A11) -- 341 (A01) &                                                      2 -- 180$^{(d)}$ &         $1 \cdot 10^{-10}$~(A05) -- $2 \cdot 10^{-7}$~(A03) &     $3 \cdot 10^{-12}$ -- $1 \cdot 10^{-7}$$^{(d)}$ \\
NO                                           &                                        3 (A11) -- 268 (A07) &                                                     50 -- 180$^{(f)}$ &          $3 \cdot 10^{-9}$~(A11) -- $7 \cdot 10^{-8}$~(A07) &      $1 \cdot 10^{-8}$ -- $1 \cdot 10^{-7}$$^{(f)}$ \\
NO$^+$                                       &                                                           - &                                                                     - &        $5 \cdot 10^{-11}$~(A16) -- $7 \cdot 10^{-10}$~(A17) &                                                   - \\
HCN                                          &                                                           - &                                                                     - &         $5 \cdot 10^{-13}$~(A16) -- $7 \cdot 10^{-6}$~(A07) &    $2 \cdot 10^{-14}$ -- $6 \cdot 10^{-10}$$^{(d)}$ \\
HNC                                          &                                                           - &                                                                     - &         $1 \cdot 10^{-11}$~(A20) -- $4 \cdot 10^{-8}$~(A03) &    $9 \cdot 10^{-16}$ -- $3 \cdot 10^{-11}$$^{(d)}$ \\
CH$_3$CN                                     &                                        4 (A15) -- 518 (A08) &                                                     90 -- 300$^{(f)}$ &         $5 \cdot 10^{-13}$~(A12) -- $3 \cdot 10^{-6}$~(A18) &     $1 \cdot 10^{-11}$ -- $1 \cdot 10^{-9}$$^{(f)}$ \\
C$_2$H$_3$CN                                 &                                       86 (A08) -- 289 (A01) &                                                    145 -- 200$^{(e)}$ &          $2 \cdot 10^{-9}$~(A06) -- $4 \cdot 10^{-8}$~(A01) &     $5 \cdot 10^{-10}$ -- $2 \cdot 10^{-7}$$^{(e)}$ \\
C$_2$H$_3$CN, v$_{10}$=1                     &                                      240 (A03) -- 606 (A08) &                                                                     - &          $1 \cdot 10^{-9}$~(A06) -- $5 \cdot 10^{-7}$~(A01) &                                                   - \\
C$_2$H$_5$CN                                 &                                        8 (A13) -- 292 (A01) &                                                    150 -- 170$^{(e)}$ &         $1 \cdot 10^{-10}$~(A13) -- $3 \cdot 10^{-7}$~(A02) &      $4 \cdot 10^{-9}$ -- $3 \cdot 10^{-6}$$^{(e)}$ \\
C$_2$H$_5$CN, v$_{20}$=1                     &                                      156 (A05) -- 276 (A01) &                                                           170$^{(d)}$ &          $4 \cdot 10^{-9}$~(A07) -- $5 \cdot 10^{-7}$~(A02) &      $1 \cdot 10^{-7}$ -- $5 \cdot 10^{-7}$$^{(d)}$ \\
CN                                           &                                        3 (A11) -- 198 (A18) &                                                           170$^{(d)}$ &         $1 \cdot 10^{-11}$~(A11) -- $1 \cdot 10^{-6}$~(A17) &      $7 \cdot 10^{-9}$ -- $9 \cdot 10^{-9}$$^{(d)}$ \\
HCCCN                                        &                                       14 (A11) -- 460 (A03) &                                                      60 -- 80$^{(d)}$ &         $2 \cdot 10^{-13}$~(A11) -- $1 \cdot 10^{-5}$~(A18) &     $9 \cdot 10^{-10}$ -- $8 \cdot 10^{-9}$$^{(d)}$ \\
HCCCN, v$_7$=2                               &                                     177 (A04) -- 1100 (A05) &                                                           200$^{(d)}$ &          $4 \cdot 10^{-9}$~(A05) -- $3 \cdot 10^{-7}$~(A03) &      $1 \cdot 10^{-8}$ -- $3 \cdot 10^{-7}$$^{(d)}$ \\
H$_2 \! ^{34}$S                              &                                        3 (A12) -- 400 (A04) &                                                    160 -- 190$^{(f)}$ &         $2 \cdot 10^{-14}$~(A13) -- $3 \cdot 10^{-7}$~(A17) &     $5 \cdot 10^{-10}$ -- $1 \cdot 10^{-9}$$^{(f)}$ \\
SO                                           &                                        9 (A19) -- 500 (A01) &                                                      2 -- 150$^{(d)}$ &         $6 \cdot 10^{-12}$~(A18) -- $1 \cdot 10^{-6}$~(A17) &     $1 \cdot 10^{-11}$ -- $1 \cdot 10^{-7}$$^{(d)}$ \\
SO$_2$                                       &                                       13 (A18) -- 497 (A03) &                                                      2 -- 150$^{(d)}$ &         $1 \cdot 10^{-10}$~(A16) -- $3 \cdot 10^{-7}$~(A17) &     $1 \cdot 10^{-11}$ -- $1 \cdot 10^{-7}$$^{(d)}$ \\
SO$_2$, v$_2$=1                              &                                       13 (A18) -- 654 (A01) &                                                                     - &         $1 \cdot 10^{-10}$~(A16) -- $3 \cdot 10^{-7}$~(A17) &                                                   - \\
CS                                           &                                                           - &                                                                     - &         $3 \cdot 10^{-11}$~(A16) -- $5 \cdot 10^{-6}$~(A02) &     $1 \cdot 10^{-13}$ -- $3 \cdot 10^{-8}$$^{(d)}$ \\
H$_2$CS                                      &                                       10 (A06) -- 202 (A03) &                                                     15 -- 150$^{(d)}$ &         $2 \cdot 10^{-13}$~(A06) -- $3 \cdot 10^{-7}$~(A07) &     $1 \cdot 10^{-11}$ -- $5 \cdot 10^{-8}$$^{(d)}$ \\
OCS                                          &                                       15 (A19) -- 272 (A20) &                                                      15 -- 60$^{(d)}$ &         $1 \cdot 10^{-10}$~(A13) -- $4 \cdot 10^{-6}$~(A09) &      $3 \cdot 10^{-9}$ -- $2 \cdot 10^{-8}$$^{(d)}$ \\
OCS, v$_2$=1                                 &                                      105 (A08) -- 245 (A01) &                                                           150$^{(d)}$ &          $4 \cdot 10^{-9}$~(A15) -- $1 \cdot 10^{-7}$~(A03) &      $2 \cdot 10^{-8}$ -- $8 \cdot 10^{-8}$$^{(d)}$ \\
NS                                           &                                       59 (A07) -- 357 (A06) &                                                       2 -- 30$^{(d)}$ &         $1 \cdot 10^{-13}$~(A03) -- $3 \cdot 10^{-8}$~(A07) &    $4 \cdot 10^{-11}$ -- $1 \cdot 10^{-10}$$^{(d)}$ \\
CCH                                          &                                        3 (A14) -- 420 (A16) &                                                       2 -- 20$^{(d)}$ &         $2 \cdot 10^{-10}$~(A12) -- $3 \cdot 10^{-8}$~(A19) &    $3 \cdot 10^{-13}$ -- $3 \cdot 10^{-10}$$^{(d)}$ \\
PH$_3$                                       &                                                           - &                                                                     - &         $2 \cdot 10^{-12}$~(A05) -- $7 \cdot 10^{-8}$~(A06) &                                                   - \\
PN                                           &                                                           - &                                                                     - &         $1 \cdot 10^{-13}$~(A13) -- $7 \cdot 10^{-7}$~(A17) &    $1 \cdot 10^{-13}$ -- $2 \cdot 10^{-13}$$^{(d)}$ \\
    \hline
    \hline
\end{tabular}
\begin{tablenotes}
   \item {The source ID in round brackets indicates the source where the minimum and maximum value of the corresponding parameter was determined. The abundances are computed relative to the hydrogen column densities derived in our first paper for each source. The abundance ranges reported in references $^{(d)}$\citetads{2013A&A...559A..47B} and $^{(f)}$\citetads{2014ApJ...789....8N} are given for an H$_2$ column density of $8.0 \times 10^{24}$~cm$^{-2}$ for core and foreground components. The reference $^{(e)}$\citetads{2017A&A...604A..60B} used different H$_2$ column density to calculate the abundances for different cores. For species with only a single transition within the frequency ranges covered by our observations, temperature estimation is not possible.}
\end{tablenotes}
\end{table*}

%+++++++++++++++++++++++++++++++++++++++++++++++++++++++++++++++++++++++++++++++
% Simple O-bearing Molecules
\subsection{Simple O-bearing Molecules}\label{subsec:SimpleOBearingMol}

%:::::::::::::::::::::::::::::::::::::::::::::::::::::::::::::::::::::::::::::::
% CO

%  Carbon monoxide is the second-most common diatomic molecule in the interstellar medium, after molecular hydrogen. Because of its asymmetry, this polar molecule produces far brighter spectral lines than the hydrogen molecule, making CO much easier to detect. It is used as tracer of molecular gas, as molecular hydrogen can only be detected using ultraviolet light, which requires space telescopes. Carbon monoxide observations provide much of the information about the molecular clouds in which most stars form.
\textbf{CO}. Atomic carbon and atomic oxygen become mobile when the temperature is increased to $\sim$15~K, which results in the formation of surface carbon monoxide (CO) via reaction between these atoms. CO and three of its isotopologues $^{13}$CO, C$^{17}$O, and C$^{18}$O are clearly identified, see Figs.~\ref{fig:COMN}~-~\ref{fig:CO1718MN}, while $^{13}$C$^{17}$O, $^{13}$C$^{18}$O and the vibrational excited states CO,v=1 and CO,v=2 are not detected. Only the ($J$~=~2 -- 1, $E_{\rm low}$~=~5.5~K) of CO is covered by our observation and shows broad absorption features for many sources in Sgr~B2(M) and N. Due to the fact that only a single transition is covered by our observation and due to the large optical depth a reliable temperature estimation of CO and its isotopologues is not possible. For sources in Sgr~B2(M), the highest abundance is two order of magnitude lower than in \citetads{2021A&A...651A...9M}.

% comparison:
%             |                        Möller 2021                              Neill 2014
% C-13-O;v=0; |   8.0    500.0   9.3e-09   7.6e-07 |     60.0     60.0   1.3e-11   1.3e-11
% CO-17;v=0;  |  13.4    500.0   3.9e-09   3.9e-07 |     60.0     60.0   7.9e-10   3.9e-09
% CO-18;v=0;  |  28.5    500.0   5.3e-09   1.2e-07 |     60.0     60.0   2.5e-09   1.2e-08

%:::::::::::::::::::::::::::::::::::::::::::::::::::::::::::::::::::::::::::::::
% CH3OH

\begin{figure*}[!b]
   \centering
   \includegraphics[scale=0.85]{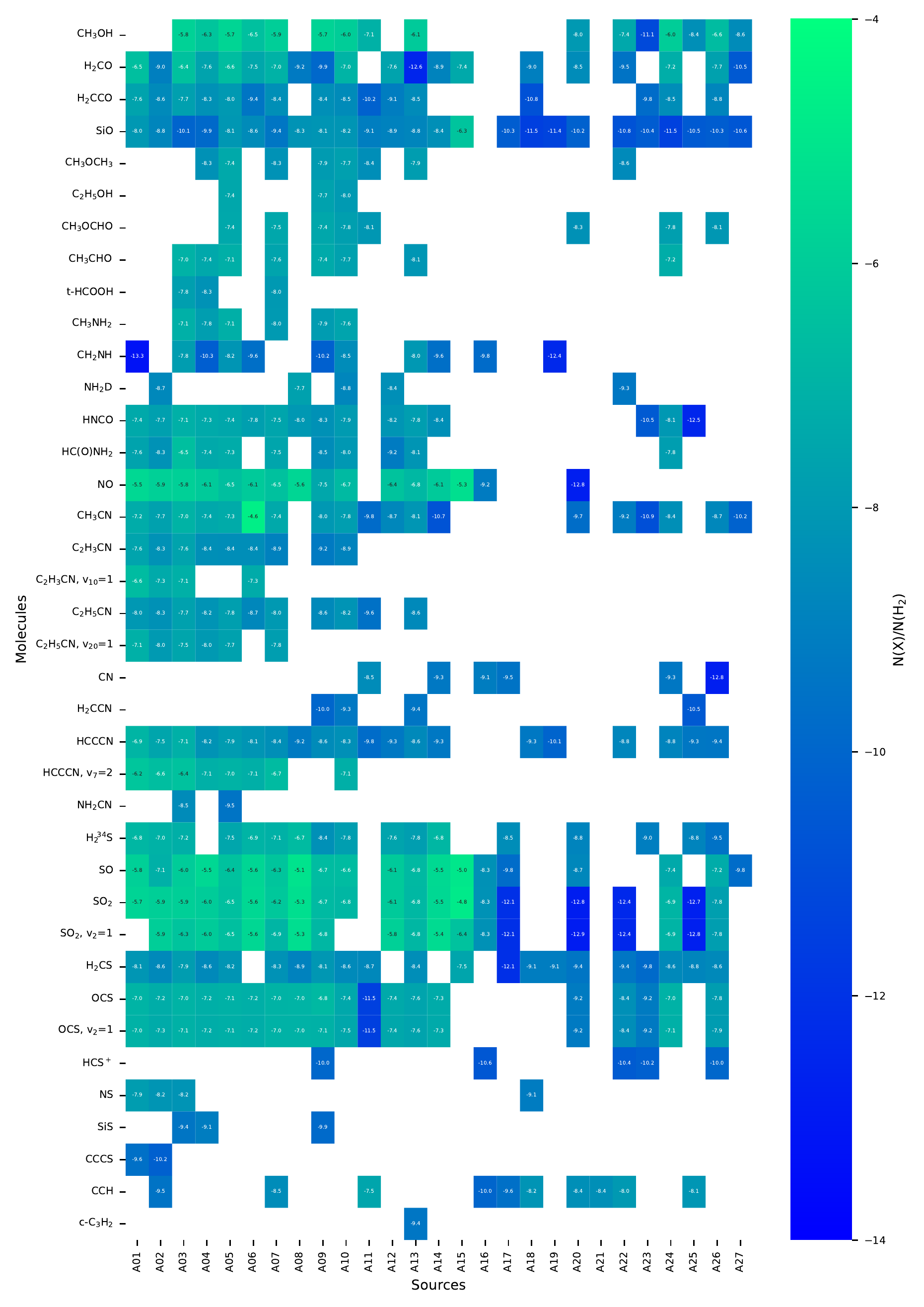}\\
   \caption{Abundances of core layers for each detected molecule and source in Sgr~B2(M).}
   \label{fig:AbundCoreSgrB2M}
\end{figure*}

\begin{figure*}[!b]
   \centering
   \includegraphics[scale=0.85]{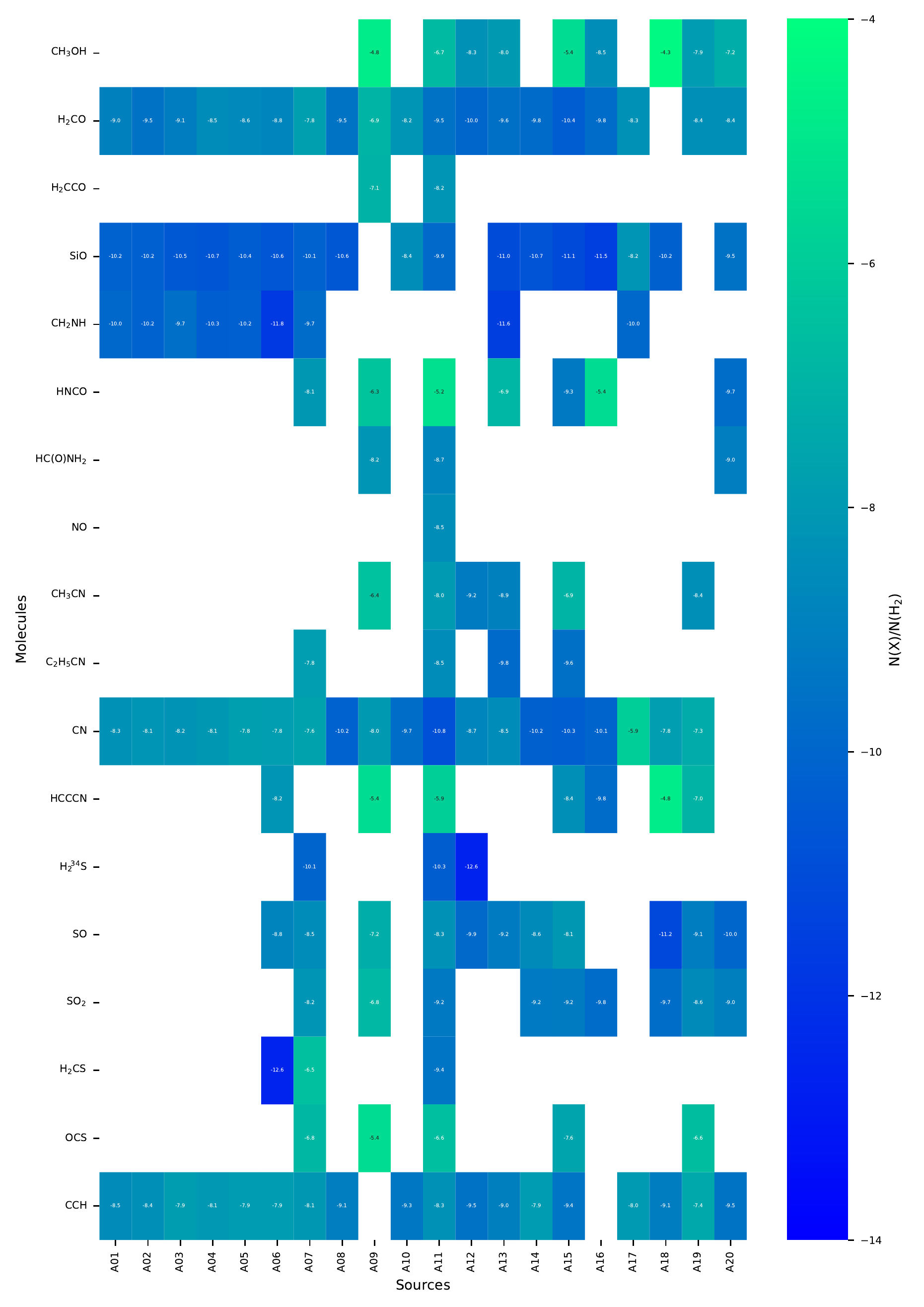}\\
   \caption{Abundances of envelope layers for each detected molecule and source in Sgr~B2(N).}
   \label{fig:AbundEnvSgrB2N}
\end{figure*}

% Methanol can be used to determine the physical and chemical conditions in protostellar sources. In addition, careful analysis of methanol is also important to identify other molecules that are often blended by methanol lines.
\textbf{CH$_3$OH}. Methanol (CH$_3$OH) is mainly formed through the hydrogenation of CO on grain surfaces, which requires high temperatures ($\sim$90~K) (\citetads{2003ApJ...588L.121W}, \citetads{2005ApJ...618..259M}). This and subsequent evaporation at higher temperature makes methanol a warm molecular gas tracer. (An alternative formation process occurs by photo-dissociative recombination of CH$_3$OH$_2^+$, which results from radiative association of CH$_3$ with water molecules (H$_2$O) \citepads{1998ApJ...501..731T}. However, since this process is very slow, it seems unlikely that it can be responsible for the high abundances of methanol found in our analysis, see Figs.~\ref{fig:AbundCoreSgrB2M}~-~\ref{fig:AbundEnvSgrB2N}). 253 transitions of methanol (CH$_3$OH) are included in our survey with lower energies ranging from 12.5~K ($J_{K_a,K_c}$ = 1$_{1,0}$ -- 2$_{1,2}$) up to 2405~K ($J_{K_a,K_c}$ = 39$_{11,28}$ -- 40$_{10,30}$). Here, we do not distinguish between the two nuclear-spin states, depending on the net spin of the three protons of the CH$_3$ group. CH$_3$OH is seen in all sources in Sgr~B2(M) and N, see Fig.~\ref{fig:CH3OHMN}, where the interaction of torsional and rotational motions leads to multiple energy states and complex spectra. For some sources, CH$_3$OH is found only in emission (e.g.\ A09 in Sgr~B2(M)), or only in absorption (e.g.\ A01 in Sgr~B2(M)), or as combination of emission and absorption features (e.g.\ A13 in Sgr~B2(N)). Additionally, we identified torsionally excited CH$_3$OH,v$_{12}$=1 in some sources in Sgr~B2(M) and N, see Fig.~\ref{fig:CH3OHv12MN}. The $^{13}$CH$_3$OH isotopologue is clearly identified in only a few sources in Sgr~B2(M) and N, see Fig.~\ref{fig:C13H3OHMN}. The other isotopologues CH$_3 \! ^{18}$OH and deuterated methanol were not detected. For each source we fit methanol, its vibrational excited state CH$_3$OH,v$_{12}$=1 (if included), and the $^{13}$CH$_3$OH isotopologue by the same model, i.e.\ we use for all three molecules the same number of components and parameters, except that the column density of the $^{13}$CH$_3$OH isotopologue is scaled by the aforementioned $^{12}$C/$^{13}$C ratio. Here, we use at least two components for each source in Sgr~B2(M) and N to model the contribution of CH$_3$OH, except in source A16 in Sgr~B2(N), where only a single component was used. The range of derived excitation temperatures for sources in Sgr~B2(M) agrees quite well with \citetads{2021A&A...651A...9M}. For sources in Sgr~B2(N), we obtain a much broader range of temperatures, where especially outlying sources  show much lower excitation temperatures compared to \citetads{2017A&A...604A..60B}. While the range of abundances for sources in Sgr~B2(N) agrees well with previous studies, sources in Sgr~B2(M) show abundances that are up to two orders of magnitude larger than in the previous study of \citetads{2021A&A...651A...9M}, which may be caused by the different H$_2$ column densities and different beams. Additionally, class~I \citepads{1997ApJ...474..346M} and class~II methanol masers (\citetads{1996MNRAS.283..606C}, \citetads{2016ApJ...833...18H}, \citetads{2019ApJS..244...35L}) were detected towards some sources in Sgr~B2(M) and N as well, see Sect.~\ref{subsec:maser}.

% \citetads{1986ApJS...60..819C, 1991ApJS...77..255S, 1991ApJS...76..617T, 1995MNRAS.273.1033H, 1996A&A...305..950G, 1998ApJS..117..427N, 2004ApJ...600..234F, 2005A&A...444..521F, 2014ApJ...789....8N, 2013A&A...559A..47B}
%
% additional papers about formation of CH3OH:
% - https://ui.adsabs.harvard.edu/abs/2016ApJ...817L..12B/abstract
% - https://ui.adsabs.harvard.edu/abs/2018ApJ...853..102C/abstract
%
% comparison:
%            |       "Bonfand__2017__SgrB2-N" |      "Moeller__2021__SgrB2-M" |        "Neill__2014__SgrB2-N" |    "Bisschop__2007__hot-cores" | "Boegelund__2018__AFGL_4176"
% CH3OH;v=0; | 160.0  190.0  8.6e-08  2.2e-05 |  3.7  308.1  7.8e-10  2.0e-08 | 80.0  170.0  1.3e-09  6.3e-07 | 113.0  259.0  5.7e-07  5.4e-06 | 120.0  120.0  1.4e-05  1.4e-05

%:::::::::::::::::::::::::::::::::::::::::::::::::::::::::::::::::::::::::::::::
% H2CO

\textbf{H$_2$CO}. Due to its ubiquity, formaldehyde (H$_2$CO) and some of its isotopologues \citepads[e.g., H$_2 \! ^{13}$CO; H$_2$C$^{17}$O or H$_2$C$^{18}$O, ]{2008AsBio...8...59B} are used to determine, among other things, the kinetic temperature and density in star-forming regions \citepads{2008ARep...52..976W}. Formaldehyde is considered a crucial chemical intermediate in the formation pathways leading to complex organic molecules at low temperatures \citepads{2015MNRAS.453L..31B}. Between 20~K and $\sim$250~K, H$_2$CO is formed in the gas-phase, while below 20~K (the CO snow line), H$_2$CO is created on surfaces of icy dust grains by a double hydrogenation process of CO (\citetads{2021A&A...648A..66G}, \citetads{2021A&A...656A.148R}). After heating up the dust, a sublimation process takes place in which formaldehyde is desorbed from the grains and enters the gas phase. Moreover, CO hydrogenation appears to be 250 times more efficient in H$_2$O-rich ice than in CO-rich ice mantles, as estimated by \citetads{2001A&A...372..998C}. The frequency ranges covered by our observations contain  30 transitions of H$_2$CO. The four lowest-energy transitions $J_{K_a,K_c}$ = 3$_{0,3}$ -- 2$_{0,2}$ ($E_{\rm low} = 10.5$~K), $J_{K_a,K_c}$ = 3$_{1,3}$ -- 2$_{1,2}$ ($E_{\rm low} = 21.9$~K), $J_{K_a,K_c}$ = 3$_{1,2}$ -- 2$_{1,1}$ ($E_{\rm low} = 22.6$~K), and  $J_{K_a,K_c}$ = 3$_{2,1}$ -- 2$_{2,0}$ ($E_{\rm low} = 57.6$~K) show deep absorptions in almost all sources in Sgr~B2(M) and N, see Fig.~\ref{fig:H2COMN}, where the absorption depth is greatest for the inner sources. Additionally, the transitions  $J_{K_a,K_c}$ = 9$_{1,8}$ -- 9$_{1,9}$ ($E_{\rm low} = 163.6$~K) and  $J_{K_a,K_c}$ = 10$_{1,9}$ -- 10$_{1,10}$ ($E_{\rm low} = 197.3$~K) are seen in emission for the inner sources in Sgr~B2(M). From the five isotopologues H$_2 \! ^{13}$CO, H$_2$C$^{17}$O, HDCO, HD$^{13}$CO, and HDC$^{18}$O, we could identify only H$_2 \! ^{13}$CO in many inner sources of Sgr~B2(M) and N, see Fig.~\ref{fig:H2C13OMN}. The contributions of the other isotopologues are tiny and strongly blended by other molecules. We use at least two components for each source in Sgr~B2(M) and N to model the contribution of formaldehyde, except in source A02 and A18 in Sgr~B2(N), where only a single component was used. Compared to other analyses, we find a much broader range of excitation temperatures and much higher abundances, which could be due to the higher spatial resolution of our observations and the different H$_2$ column densities. Additionally, sources A01 and A02 in Sgr~B2(M) and A02 and A08 in Sgr~B2(N) contain H$_2$CO masers (\citetads{1987IAUS..115..161W}, \citetads{1994ApJ...434..237M}, \citetads{2007ApJ...654..971H}, \citetads{2019ApJS..244...35L}), Sect.~\ref{subsec:maser}.

% re-read https://iopscience.iop.org/article/10.1086/590496/pdf
% H$_2$CO masers appear to trace young massive stellar objects before the onset of a bright ultra-compact \hii~region. Moreover, in the case of three of the H2CO maser sources that have been studied in detail, there is some evidence for an association between H$_2$CO maser emission and circumstellar disks.
% \citetads{1974A&A....36..253R, 1986MNRAS.218..385G, 1991ApJS...77..255S, 1991ApJS...76..617T, 1998ApJS..117..427N, 2014ApJ...789....8N, 2013A&A...559A..47B}

% comparison:
%           |                "Neill__2014__SgrB2-N" |           "Bisschop__2007__hot-cores"
% H2CO;v=0; |     50.0     80.0   1.3e-11   3.1e-11 |     69.0    193.0   1.2e-07   1.0e-06

%:::::::::::::::::::::::::::::::::::::::::::::::::::::::::::::::::::::::::::::::
% H2CCO

\textbf{H$_2$CCO}. Ketene (CH$_2$CO, C$_2$H$_2$O, or H$_2$CCO), also called ethenone, is a 16-valence electron molecule. Although it is not a COM because COMs consist of at least six atoms, it is an important oxygen-bearing organic compound that acts as a COM precursor due to the presence of unsaturated CC and CO bonds. The formation chemistry of ketene (CH$_2$CO, C$_2$H$_2$O, or H$_2$CCO) is unclear (\citetads{2016MNRAS.459L...6E}, \citetads{2020MNRAS.493.2523E}). Several solid-state mechanisms, such as radical associations, O-atom additions, and hydroxylation have been proposed (\citetads{CHARNLEY200423}, \citetads{2015MNRAS.449L..16B}, \citetads{2016ApJ...821...46T}). Our observations contain 73 transitions of H$_2$CCO with lower energies ranging from 53.3~K ($J_{K_a,K_c}$ = 11$_{0,11}$ -- 10$_{0,10}$) up to 1933~K ($J_{K_a,K_c}$ = 13$_{12,1}$ -- 12$_{12,0}$). Here we do not distinguish between ortho and para states. Ketene is clearly detected by emission and absorption features in more or less all sources in Sgr~B2(M) and N, except in some outer sources, see Fig.~\ref{fig:H2CCOMN}, which are mostly modeled by a single component. We obtain excitation temperatures and abundances that far exceed the results of \citetads{1991ApJS...77..255S} and \citetads{2014ApJ...789....8N}, where the low spatial resolution of their observations could be a possible explanation. The four isotopologues H$_2 \! ^{13}$CCO, H$_2$C$^{13}$CO, H$_2$CC$^{18}$O, and HDCCO are not reliably detected.

% \citetads{1991ApJS...76..617T, 2014ApJ...789....8N}

% comparison:
%            |               "Neill__2014__SgrB2-N"
% H2CCO;v=0; |   180.0    180.0   8.8e-09   8.8e-09

%:::::::::::::::::::::::::::::::::::::::::::::::::::::::::::::::::::::::::::::::
% HCO+

\textbf{HCO$^+$}. The formyl radical (HCO$^+$), i.e.\ protonated CO, is known to be a good tracer, among others, of cosmic ray and UV dominated regions in dense gas (e.g.\ \citetads{1995A&A...295..571H}, \citetads{2011A&A...527A..88V}). The abundance of HCO$^+$ is enhanced in shocked and dense environments, and in regions of higher fractional ionization. Although the formyl radical (HCO$^+$) and its energetically-disfavored isomer HOC$^+$ do not have a transition in the frequency ranges covered by our observation, we have possibly identified both molecules by their two isotopologues H$^{13}$CO$^+$ (HO$^{13}$C$^+$) and HC$^{18}$O$^+$ (H$^{18}$OC$^+$), see Fig.~\ref{fig:HCO+MN}, each of which show the ($J$~=~3 -- 2, $E_{\rm low} = 31.3$~K) transition. The other isotopologues DCO$^+$ and DOC$^+$ were not detected. A reliable temperature estimate is not possible due to the fact that only a single transition is observed, respectively. Compared to other analyses of Sgr~B2, we find higher abundances, which could be due to the higher spatial resolution of our observations and to the different H$_2$ column densities. For HOC$^+$, \citetads{1997ApJ...481..800A} obtained an excitation temperature of 15~K and an abundance of $3 \times 10^{-12}$ for Sgr~B2(OH).

% \citetads{1996A&A...305..950G, 2013A&A...556A.137E, 2014ApJ...789....8N, 2013A&A...559A..47B}

% comparison:
%               |            "Moeller__2021__SgrB2-M" |                "Neill__2014__SgrB2-N"
% HC-13-O+;v=0; |   15.3     99.1   2.1e-12   7.8e-11 |    150.0    150.0   7.5e-11   7.5e-11

%:::::::::::::::::::::::::::::::::::::::::::::::::::::::::::::::::::::::::::::::
% SiO

\textbf{SiO}. Silicon monoxide (SiO) traces, among others, the presence of energetic protostellar outﬂows since the destruction of silicon-bearing dust (e.g., MgSiO$_3$) by shocks is associated with the origin of SiO in the gas-phase (e.g.\, \citetads{1996ApJ...472L..49B}, \citetads{1996ApJ...469..216W}, \citetads{1997ApJS..108..301S}, \citetads{2008A&A...482..809G}, \citetads{2013A&A...556A..69A}). There are two ways in which SiO can enter the gas phase. SiO can be released into the gas phase by gas-grain interactions, which leads to sputtering of the grain cores, but also by grain-grain collisions, which lead to vaporization and shattering of the grains. Here, the latter effect leads to the formation of small grain fragments in large numbers, which increases the total surface area of the dust grains. In our analysis SiO is identified by its two transitions ($J$~=~5 -- 4, $E_{\rm low} = 20.8$~K, $J$~=~6 -- 5, $E_{\rm low} = 31.3$~K) in all sources in Sgr~B2(M) and N, showing broad and complex, self-absorption line profiles, see Fig.~\ref{fig:SiOMN}, which require a large number of components. We find excitation temperatures up to $\sim$1000~K for sources in Sgr~B2(M), which should be considered with caution because of the high optical depths and the small number of transitions covered by our observations. Other isotopologues or vibrationally excited states could not be identified. SiO and its connection to molecular outflows in the Sgr~B2 complex will be described in a separate paper.

\subsection{Complex O-bearing Molecules}\label{subsec:ComplexOBearingMol}

%:::::::::::::::::::::::::::::::::::::::::::::::::::::::::::::::::::::::::::::::
% CH3OCH3

\textbf{CH$_3$OCH$_3$}. Dimethyl ether (CH$_3$OCH$_3$) is a large, organic molecule that is ubiquitous in regions of massive star formation. For temperatures up to 30~K, dimethyl ether is mostly formed on the surface of dust grains through the radical-addition reaction \citepads{2019A&A...628A..27B}: CH$_3$ + CH$_3$O $\rightarrow$ CH$_3$OCH$_3$. Due to the electronic recombination reaction CH$_3$OCH$_4^+$ + $e^-$ $\rightarrow$ CH$_3$OCH$_3$ + H, where CH$_3$OCH$_4^+$ is a product of the reaction between methanol and protonated methanol (CH$_3$OH$_2^+$), the gas-phase fractional abundance of CH$_3$OCH$_3$ increases when CH$_3$OH desorbs from the dust grains around 130~K \citepads{2019A&A...628A..27B}. Our survey contains 1988 transitions of dimethyl ether with lower energies between 11.1~K ($J_{K_a,K_c}$ = 4$_{3,2}$ -- 3$_{2,2}$) and 3544~K ($J_{K_a,K_c}$ = 72$_{29,44}$ -- 73$_{28,46}$). In our analysis, CH$_3$OCH$_3$ is found in many inner sources in Sgr~B2(M) and N, see Fig.~\ref{fig:CH3OCH3MN}, with sources in N containing significantly more CH$_3$OCH$_3$ than sources in M. Other isotopologues or vibrationally excited states were not detected. In most sources, dimethyl ether shows emission features described by a single component. Compared to previous analyses, the temperatures and abundances we determined have much wider ranges, respectively. The outlying sources in both regions show lower excitation temperatures compared to \citetads{2017A&A...604A..60B} and \citetads{2014ApJ...789....8N}. Additionally, we obtain abundances which are up to two orders of magnitude higher than in previous studies.

% for dimethyl ether, CH$_3$OCH$_3$, the gas-phase reaction of methanol with protonated methanol (CH$_3$OH$^+_2$) can lead to the formation of protonated dimethyl ether. This species can recombine with an electron to produce dimethyl ether in up to 7\% of cases, while the remainder of recombinations should result in the fragmentation of the underlying C–O bonds (Hamberg et al. 2010). Chemical models indicate that the large abundances of gas-phase methanol, following grain-mantle ejection, could make this process effective enough to explain dimethyl ether observations in star-forming regions.
% Dimethyl ether is mostly formed on the surface of dust grains up to T ∼ 20–30 K, through the radical-addition reaction: CH3 + CH3 O → CH3 OCH3. In model N3-T15-CR1, the gas-phase fractional abundance of dimethyl ether increases when methanol desorbs from the dust grains around 130 K. Here gas-phase CH3 OCH3 is formed via the electronic recombination reaction CH3 OCH4 + + e− → CH3 OCH3 + H, where CH3 OCH4 + is a product of the reaction between methanol and protonated methanol.

% comparison:
%              |              "Moeller__2021__SgrB2-M" |                "Neill__2014__SgrB2-N" |       "Bisschop__2007__hot-cores" |           "Boegelund__2018__AFGL_4176
% CH3OCH3;v=0; |    100.0    100.0   4.5e-10   4.5e-10 |    130.0    130.0   3.5e-08   3.5e-08 | 43.0    241.0   7.6e-08   2.4e-06 |    160.0    160.0   3.3e-07   3.3e-07

%:::::::::::::::::::::::::::::::::::::::::::::::::::::::::::::::::::::::::::::::
% C2H5OH

\textbf{C$_2$H$_5$OH}. According to \citetads{2019A&A...628A..27B}, ethanol (C$_2$H$_5$OH) is predominantly produced on dust grains via: CH$_3 \xrightarrow{C}$ C$_2$H$_3 \xrightarrow{H}$ C$_2$H$_4 \xrightarrow{H}$ C$_2$H$_5 \xrightarrow{O}$ C$_2$H$_5$O $\xrightarrow{H}$ C$_2$H$_5$OH followed by later desorption into the gas phase. The 3877 transitions included in our observations cover a wide range of lower energies from 8.4~K ($J_{K_a,K_c}$ = 4$_{3,1}$ -- 4$_{0,4}$) up to 2620~K ($J_{K_a,K_c}$ = 26$_{4,23}$ -- 25$_{5,20}$). We detected C$_2$H$_5$OH in emission in only three sources (A05, A09, and A10) in Sgr~B2(M), but in ten sources (A01 - A08, A10, and A17) in Sgr~B2(N), see Fig.~\ref{fig:C2H5OHMN}, which are described in most cases by a single component. Additionally, none of its isotopologues were found. In contrast to previous surveys, we obtain much higher temperatures for the inner sources in both regions, but cannot find a cold component as described by \citetads{2013A&A...559A..47B} for Sgr~B2(M).

% comparison:
%             |             "Bonfand__2017__SgrB2-N" |                "Neill__2014__SgrB2-N" |          "Bisschop__2007__hot-cores" |          "Boegelund__2018__AFGL_4176"
% C2H5OH;v=0; |   145.0    150.0   6.6e-09   1.1e-06 |    100.0    100.0   2.5e-08   2.5e-08 |   104.0    166.0   1.4e-08   7.5e-08 |    120.0    120.0   1.9e-07   1.9e-07

%:::::::::::::::::::::::::::::::::::::::::::::::::::::::::::::::::::::::::::::::
% CH3OCHO

\textbf{CH$_3$OCHO}. Methyl formate (CH$_3$OCHO) is a prolate asymmetric top molecule and one of three isomeric forms of C$_2$H$_4$O$_2$. The other two isomers are acetic acid (CH$_3$COOH) and glycolaldehyde (HOCH$_2$CHO) \citepads{2018A&A...617A.102F} and have also been identified in space (\citetads{1997ApJ...480L..71M}, \citetads{2004ApJ...610L..21H}) but they are much less ubiquitous and abundant. CH$_3$OCHO is formed mainly via the grain-surface radical-radical addition reaction: HCO + CH$_3$O $\rightarrow$ CH$_3$OCHO followed by later desorption into the gas phase. \citetads{2019A&A...628A..27B} describe an increase in the gas-phase abundance of CH$_3$OCHO around 40~K, when H$_2$CO is abundantly released into the gas phase and reacts with CH$_3$OH$_2^+$ to form HC(OH)OCH$_3^+$. CH$_3$OCHO in the gas phase is then produced via the electronic recombination of HC(OH)OCH$_3^+$. Our observations contain 1734 transitions of methyl formate with lower energies between 11.4~K ($J_{K_a,K_c}$ = 6$_{4,3}$ -- 5$_{2,4}$) and 1586~K ($J_{K_a,K_c}$ = 69$_{15,54}$ -- 69$_{14,55}$). CH$_3$OCHO is identified in emission in some sources in Sgr~B2(M) and N, see Fig.~\ref{fig:CH3OCHOMN}, mostly described by a single component. Additionally, we also identified the vibrational excited state CH$_3$OCHO,v$_{18}$=1 in all sources, see Fig.~\ref{fig:CH3OCHOv18MN}, where methyl formate was found using the same model parameters (excitation temperatures, column densities, etc.). Isotopologues and vibrationally excited states as well as the isomers acetic acid (CH$_3$COOH) and glycolaldehyde (HOCH$_2$CHO) could not be detected. While the abundances in both regions fit well with the results of previous analyses, we find a different behavior for temperatures: For Sgr~B2(M) we cannot find the cold component described by \citetads{2013A&A...559A..47B}, but instead obtain consistently hotter temperatures. In contrast to that, the sources in Sgr~B2(N) have a much broader temperature range compared to \citetads{2017A&A...604A..60B}.

% comparison:
%            |       "Bonfand__2017__SgrB2-N" |      "Moeller__2021__SgrB2-M" |        "Neill__2014__SgrB2-N" |    "Bisschop__2007__hot-cores" | "Boegelund__2018__AFGL_4176"
% CH3OH;v=0; | 160.0  190.0  8.6e-08  2.2e-05 |  3.7  308.1  7.8e-10  2.0e-08 | 80.0  170.0  1.3e-09  6.3e-07 | 113.0  259.0  5.7e-07  5.4e-06 | 120.0  120.0  1.4e-05  1.4e-05

%:::::::::::::::::::::::::::::::::::::::::::::::::::::::::::::::::::::::::::::::
% CH3CHO

\textbf{CH$_3$CHO}. Acetaldehyde (CH$_3$CHO) is a non-cyclic isomer of ethylene oxide (c-C$_2$H$_4$O) and has an important role as an evolutionary tracer of different astronomical objects \citepads{2009ARA&A..47..427H}. Below 30~K acetaldehyde (CH$_3$CHO) is formed on the grains via the surface reactions: CH$_2$ + HCO $\rightarrow$ CH$_2$CHO $\xrightarrow{H}$ CH$_3$CHO and CH$_3$ + HCO $\rightarrow$ CH$_3$CHO \citepads{2019A&A...628A..27B}. Above 90~K acetaldehyde is formed directly in the gas phase via the reaction C$_2$H$_5$ + O $\rightarrow$ CH$_3$CHO + H, when C$_2$H$_5$ desorbs from dust grains.
%
% According to \citetads{2019A&A...628A..27B}, acetaldehyde on the grain surface may be destroyed either by reacting with CH$_3$ to form CH$_4$ + CH$_3$CO, or with NH$_2$ to form NH$_3$ + CH$_3$CO. Only a small fraction of solid-phase CH$_3$CHO contributes to the gas-phase fractional abundance. Below 24~K, surface CH$_3$CHO is released into the gas phase via chemical desorption from the reaction CH$_2$CHO + H. Once in the gas phase, CH$_3$CHO becomes the dominant reaction partner for gas-phase ions, damping ionic abundances and thus limiting the destruction of other gas-phase species.
%
742 transitions of CH$_3$CHO with lower energies between 0.9~K ($J_{K_a,K_c}$ = 2$_{2,0}$ -- 1$_{0,1}$) and 2619~K ($J_{K_a,K_c}$ = 45$_{27,19}$ -- 46$_{26,20}$) are covered by our observations, where acetaldehyde is seen in many sources in emission in Sgr~B2(M) and N, see Fig.~\ref{fig:CH3CHOMN}. The contribution of acetaldehyde is mostly modeled by a single component. Additionally, we also see the vibrational excited state CH$_3$CHO,v$_{15}$=1 in more or less all sources, see Fig.\ref{fig:CH3CHOv15MN}, where methyl formate was found using the same model parameters, i.e.\ excitation temperatures etc. Only in source A13 in Sgr~B2(M) we cannot uniquely identify CH$_3$CHO,v$_{15}$=1. Other vibrationally excited states such as CH$_3$CHO,v$_{15}$=2, or isotopologues cannot be reliably identified. In our analysis, we find higher excitation temperatures and abundances compared to previous analyses of \citetads{2021A&A...651A...9M} and \citetads{2013A&A...559A..47B}.

% comparison:
%              |           "Bisschop__2007__hot-cores" |          "Boegelund__2018__AFGL_4176"
% CH3CHO;v=0;  |     16.0     37.5   6.1e-12   1.9e-10 |    160.0    160.0   3.8e-08   3.8e-08

%:::::::::::::::::::::::::::::::::::::::::::::::::::::::::::::::::::::::::::::::
% t-HCOOH

\textbf{t-HCOOH}. Formic acid (HCOOH), the simplest carboxylic acid, plays a major role in interstellar chemistry. Additionally, it is a possible precursor of organic and prebiotic species, such as acetic acid (CH$_3$COOH) and glycine (NH$_2$CH$_2$COOH) (\citetads{2001ApJ...552..654L}, \citetads{2011AsBio..11..883P}). \citetads{2006A&A...457..927G} showed that formic acid (HCOOH) is formed on the grain surface via HCO + OH $\rightarrow$ HCOOH. Additionally, HCOOH is formed in the gas phase from the dissociative recombination of HCOOH$_2^+$, which then dissociatively recombines with electrons. In addition, the evaporation of formaldehyde and the subsequent reaction with atomic oxygen produces more OH in the gas phase, which reacts again with H$_2$CO increasing the abundance of formic acid. Formic acid, occurs in two conformers, trans (t-HCOOH) and cis (c-HCOOH), with the latter being 2000~K higher in energy. The frequency ranges covered by our observations contain 166 transitions of t-HCOOH, whose lower energy ranges from 32.3~K ($J_{K_a,K_c}$ = 2$_{2,1}$ -- 2$_{0,2}$) to 1970~K ($J_{K_a,K_c}$ = 58$_{6,52}$ -- 58$_{6,53}$). We clearly identify trans formic acid in emission in three sources in Sgr~B2(M), see Fig.~\ref{fig:tHCOOHM}, using a single component to describe the contribution. In contrast to \citetads{2013A&A...559A..47B}, we did not detect formic acid in Sgr~B2(N). We also could not identify c-HCOOH in any of the sources in Sgr B2(M) and N.

%+++++++++++++++++++++++++++++++++++++++++++++++++++++++++++++++++++++++++++++++
\subsection{NH-bearing Molecules}\label{subsec:NHBearingMol}

%:::::::::::::::::::::::::::::::::::::::::::::::::::::::::::::::::::::::::::::::
% CH3NH2

\textbf{CH$_3$NH$_2$}. Methylamine (CH$_3$NH$_2$) is an asymmetric top molecule, where the methyl group torsion and an inversion of the amine hydrogens are the most important motions. The symmetry of the molecule is characterized by the G$_{12}$ permutation–inversion group \citepads{1987JMoSp.121..474O}, where the allowed transitions are labeled with the A$_1$, A$_2$, B$_1$, B$_2$, E$_{1 \pm 1}$, and E$_{2 \pm 1}$ torsion-inversion-rotation irreducible representations $\Gamma$ \citepads{2007JPCRD..36.1141I}, and the selection rules are $\Gamma$ = A$_1$ $\leftrightarrow$ A$_2$, B$_1$ $\leftrightarrow$ B$_2$, E$_{1 \pm 1}$ $\leftrightarrow$ E$_{1 \pm 1}$, and E$_{2 \pm 1}$ $\leftrightarrow$ E$_{2 \pm 1}$. According to \citetads{2015A&A...576A..35L} methylamine is formed on ice mantles by UV photons via CH$_4$ + h$\nu \rightarrow$ CH$_3^+$ + H, NH$_3$ + h$\nu \rightarrow$ NH$_2^+$ + H, and CH$_3^+$ + NH$_2^+$ $\rightarrow$ CH$_3$NH$_2$. With increasing temperature the formed methylamine evaporates from the grains and enters the gas phase. Our data set contains 511 transitions of CH$_3$NH$_2$ with lower energies between 63.8~K ($J_K\Gamma$ = 3$_1$A$_2$ -- 2$_0$A$_1$) and 3433~K ($J_K\Gamma$ = 56$_4$B$_2$ -- 56$_1$B$_1$). We found CH$_3$NH$_2$ in nine sources in Sgr~B2(M) and N mainly in emission, see Fig.~\ref{fig:CH3NH2MN}, described mostly by a single component. Only in source A02, A16 and A21 in Sgr~B2(M), CH$_3$NH$_2$ is seen in absorption. Opposed to previous analyses, we found much higher excitation temperatures and abundances towards the inner sources in both regions. Isotopologues or vibrationally excited states were not detected.

% comparison:
%             |            "Moeller__2021__SgrB2-M" |                "Neill__2014__SgrB2-N"
% CH3NH2;v=0; |   14.0     14.0   9.1e-09   9.1e-09 |    150.0    150.0   6.2e-08   6.2e-08

%:::::::::::::::::::::::::::::::::::::::::::::::::::::::::::::::::::::::::::::::
% CH2NH

\textbf{CH$_2$NH}. Methanimine (or methyleneimine, CH$_2$NH) is the simplest molecule with a carbon-nitrogen double bond and a planar asymmetric top molecule. Due to the large $a$ and $b$ dipole moments of $\mu_a$ = 1.340~D and $\mu_b$ = 1.446~D \citepads{1979JChPh..70.2829A}, the spectrum consists of both $a$ and $b$ type transitions. It is formed via one of the following pathways: CN $\xrightarrow{H}$ HCN $\xrightarrow{H}$ H$_2$CN $\xrightarrow{H}$ CH$_2$NH and CN $\xrightarrow{H}$ HCN $\xrightarrow{H}$ HCNH $\xrightarrow{H}$ CH$_2$NH \citepads{2018ApJ...853..139S}. 79 transitions of methanimine with lower energies between 0~K ($J_{K_a,K_c}$ = 1$_{1,1}$ -- 0$_{0,0}$) and 1198~K ($J_{K_a,K_c}$ = 26$_{4,23}$ -- 25$_{5,20}$) are included in our survey, where we detected CH$_2$NH in more or less all sources in Sgr~B2(M) and N, see Fig.~\ref{fig:CH2NHMN}, where we used for most sources at least two components to describe CH$_2$NH. In addition to the ground state transitions, which are clearly visible by deep absorptions, we found methanimine for some sources also in emission. Compared to other analyses, we obtain much higher temperatures and abundances for sources in Sgr~B2(M), while for Sgr~B2(N) we also find sources with low temperatures not identified by \citetads{2013A&A...559A..47B}. The isotopologues $^{13}$CH$_2$NH and CH$_2 \! ^{15}$NH were not detected.

% comparison:
%            |              "Moeller__2021__SgrB2-M" |                "Neill__2014__SgrB2-N"
% CH2NH;v=0; |     11.0     11.0   1.4e-10   1.4e-10 |    150.0    150.0   8.8e-09   8.8e-09

%:::::::::::::::::::::::::::::::::::::::::::::::::::::::::::::::::::::::::::::::
% NH3

\textbf{NH$_3$}. Ammonia is a useful tracer of kinetic temperature \citepads{2012A&A...538A.133J}, because the relative populations of its rotational stages characterized by the quantum numbers $J$ and $K$ depend only on collisions, since dipole transitions between the different $K$ ladders are forbidden. Despite a small amount of ammonia formed in the gas phase, the formation of ammonia (NH$_3$) occurs most efficiently through successive hydrogenation reactions of nitrogen at the grain surface \citepads{2018ApJ...859...51A}. Although ammonia (NH$_3$) has only a single transition with lower energy of 10788~K within the frequency ranges covered by our observations, we have detected its NH$_2$D isotopologue, which shows in total 12 transitions in our survey with lower energies ranging from 108.6~K ($J_{K_a,K_c}$ = 3$_{2,2}$ -- 3$_{1,2}$) to 3055~K ($J_{K_a,K_c}$ = 18$_{7,12}$ -- 18$_{6,12}$). However, we could clearly detect only the $J_{K_a,K_c}$ = 3$_{2,2}$ -- 3$_{1,2}$ transition as an emission or absorption feature in some sources in Sgr~B2(M) and N, see Fig.~\ref{fig:NH2DMN}, mostly described by a single component. A reliable temperature estimate is therefore difficult. For sources in Sgr~B2(N), we obtain a broader range of abundances compared to \citetads{2013A&A...559A..47B}. Other isotopologues or vibrationally excited states were not detected.

% \citetads{1990ApJ...351..538G, 1995A&A...294..667H, 1995A&A...294..815F, 1999ESASP.427..639B, 2006A&A...460..533W, 2007MNRAS.377.1122P, 2011A&A...535A..47D, 2013A&A...556A.137E, 2014ApJ...785...55O} toward Sgr~B2.

%+++++++++++++++++++++++++++++++++++++++++++++++++++++++++++++++++++++++++++++++
\subsection{N- and O-bearing Molecules}\label{subsec:NOBearingMol}

%:::::::::::::::::::::::::::::::::::::::::::::::::::::::::::::::::::::::::::::::
% HNCO

\textbf{HNCO}. Isocyanic acid (HNCO) is a precursor of other prebiotic and complex organic molecules, which are of great astrochemical and astrobiological interest due to their potential to form molecules such as amino acids, sugars and nucleobases. According to \citetads{2021MNRAS.504.4428C} there are two pathways to form HNCO: On the one hand, HNCO is initially formed on the surface of grains via H + OCN $\rightarrow$ HNCO and sublimates to the gas phase afterwards. Another possibility is the formation of HNCO via NH + CO $\rightarrow$ HNCO. HNCO is a near prolate asymmetric top molecule with a linearity barrier at 1899~cm$^{-1}$ \citepads{1995JMoSp.174..151N}. In our observations 43 transitions are included with lower energies from 47.5~K ($J_{K_a,K_c}$ = 10$_{0,10}$ -- 9$_{0,9}$) to 1462~K ($J_{K_a,K_c}$ = 12$_{6,6}$ -- 11$_{6,5}$). As in previous line surveys (\citetads{1991ApJS...77..255S}, \citetads{1998ApJS..117..427N}, \citetads{2013A&A...559A..47B}) we also see isocyanic acid (HNCO) in many sources in Sgr~B2(M) and N, see Fig.~\ref{fig:HNCOMN}, where two or more components are used to describe the contributions of HNCO for the majority of sources. In many inner sources HNCO is detected in emission, while outer sources contain isocyanic acid mostly in absorption. The isotopologues DNCO, HN$^{13}$CO and HNC$^{18}$O were not detected. For both regions, we obtain excitation temperatures well beyond the temperature ranges described in previous analyses. While the abundances derived in Sgr~B2(N) agree quite well with those described by \citetads{2017A&A...604A..60B}, we find large differences for the abundances in Sgr~B2(M), where the highest abundance is four orders of magnitude higher than that of \citetads{2021A&A...651A...9M}. Here, the small beam of their HIFI observation which is not able to resolve the hot cores with their high HNCO abundances could be a possible explanation.

\textbf{NH$_2$CHO}. Formamide (NH$_2$CHO or HC(O)NH$_2$) has been identified as a potential precursor of a wide variety of organic compounds essential to life and played a crucial role in the context of the origin of life on our planet. The formation of formamide is still unclear. On the one hand, it may be formed via gas phase pathways involving H$_2$CO and NH$_2$ (e.g.\ \citetads{2015MNRAS.453L..31B}, \citetads{1992MNRAS.256P..33S}). On the other hand, it can be produced by grain surfaces reactions involving hydrogenation of HNCO \citepads[e.g.][]{2008SSRv..138...59C}, although laboratory studies indicate that this pathway is unlikely (\citetads{2015A&A...576A..91N}, \citetads{2015MNRAS.449.2438L}). In our analysis, 218 transitions with lower energies ranging from 3.1~K ($J_{K_a,K_c}$ = 2$_{2,1}$ -- 2$_{0,2}$) to 2548~K ($J_{K_a,K_c}$ = 68$_{7,61}$ -- 68$_{7,62}$) are included. Formamide is found in many sources in Sgr~B2(M) and N mostly in emission, see Fig.~\ref{fig:HCONH2MN}, described mostly by a single component. Some sources in Sgr~B2(M) show high asymmetric line shapes toward lower velocities, e.g.\ A05 and A07, which can be associated with outflows. Additionally, we also see the vibrational excited state NH$_2$CHO,v$_{12}$=1 in some sources where formamide is detected, see Fig.\ref{fig:HCONH2v12MN}, using the same model parameters as for NH$_2$CHO. The isotopologue NH$_2 \! ^{13}$CHO is clearly identified in source A02 and A03 in Sgr~B2(N), while the other isotopologues NH$_2$CDO, $^{15}$NH$_2$CHO and NH$_2$CH$^{18}$O are not detected. For both regions we find much higher excitation temperatures than in previous studies, while for sources in Sgr~B2(M) we also find a cold component in some outlying sources not identified by \citetads{2021A&A...651A...9M}. Similar to isocyanic acid we obtain abundances in Sgr~B2(N), which are comparable to results of previous analysis. For sources in Sgr~B2(M), however, we find abundances that are up to three orders of magnitude higher than in the analysis of \citetads{2021A&A...651A...9M}.

% \citetads{1986ApJS...60..819C, 1987ApJ...321L..75T, 1991ApJS...76..617T, 1996ApJ...470..981K, 2013ASPC..476..323K, 1998ApJS..117..427N, 2004ApJ...600..234F, 2011ApJ...743...60H, 2014ApJ...789....8N, 2013A&A...559A..47B}

% comparison:
%               |           "Moeller__2021__SgrB2-M"
% HC(O)NH2;v=0; | 300.0    300.0   1.8e-10   1.8e-10

%:::::::::::::::::::::::::::::::::::::::::::::::::::::::::::::::::::::::::::::::
% NO

\textbf{NO}. Nitric oxide (NO) plays an important role in the formation of hydroxylamine (H$_3$NO), which in turn is important for the formation of amino acids. Additionally, the search for formic acid and nitric oxide may point to the production of amines within circumstellar environments. NO is thought to mainly form via the N + OH $\rightarrow$ NO + H neutral–neutral reaction in the gas-phase \citepads{2018A&A...619A..28L}. Our observations contain 44 transitions of NO with lower energies between 7.2~K ($J_N$ = 2.5$_3$ -- 1.5$_2$) and 179.7~K ($J_N$ = 2.5$_2$ -- 1.5$_1$). We clearly identified NO in a number of sources in Sgr~B2(M) and N, see Fig.~\ref{fig:NOMN}, with some sources showing complex line shapes. For sources without complex line shapes, we usually use a single component to describe the contribution of nitric oxide. The isotopologues N$^{17}$O, N$^{18}$O and $^{15}$N$^{17}$O were not detected. While the excitation temperatures for sources in Sgr~B2(M) agree fairly well with the results of \citetads{2021A&A...651A...9M}, the sources in Sgr~B2(N) show a wider range of temperatures than in \citetads{2014ApJ...789....8N}. The abundances show an opposite behaviour. While the highest abundance in Sgr~B2(N) coincide quite well with the results of the previous analysis by \citetads{2014ApJ...789....8N}, we find abundances in Sgr~B2(M) that are in some cases up two orders of magnitude larger than in \citetads{2021A&A...651A...9M}. For some inner sources in Sgr~B2(M), NO provides the highest contribution to the corresponding core spectra.

% comparison:
%          |              "Moeller__2021__SgrB2-M" |                "Neill__2014__SgrB2-N"
% NO;v=0;  |     24.8    325.4   4.7e-09   8.4e-08 |     50.0    180.0   1.9e-08   1.2e-07

%:::::::::::::::::::::::::::::::::::::::::::::::::::::::::::::::::::::::::::::::
% NO+

\textbf{NO$^+$}. The formation of nitric oxide ion (or nitrosyl cation or nitrosylium ion, NO$^+$) in cold dense interstellar clouds include charge transfer reactions (\citetads{1973ApJ...185..505H}, \citetads{1977Ap&SS..52..453P}, \citetads{1980Ap&SS..68...87S}) between NO and various ions (C$^+$, H$^+$ , CH$_3^+$,...) and the exothermic ion molecule reactions between N$^+$ and oxygen bearing species such as CO, O$_2$, NO, ..., and the reaction of HNO with H$^+$, He$^+$. Our survey contains only the ($J$~=~2 -- 1, $E_{\rm low}$~=~5.7~K) transition, which is clearly seen in some sources in Sgr~B2(M) and N, see Fig.~\ref{fig:NOpMN} and mostly described by two or more components. Due to the fact that only a single transition is covered by our observation a reliable temperature estimation is not possible. Isotopologues or vibrationally excited states were not detected.

%+++++++++++++++++++++++++++++++++++++++++++++++++++++++++++++++++++++++++++++++
\subsection{Cyanide Molecules}\label{subsec:CyanideMol}

%:::::::::::::::::::::::::::::::::::::::::::::::::::::::::::::::::::::::::::::::
% HCN

% A reliable temperature estimate for the excited vibrational states is not possible because the lower energies of the observed transitions are very close to each other.
\textbf{HCN}. Hydrogen cyanide (HCN) is thought to be a good tracer of the dense molecular gas that fuels star formation. It is formed through one of two pathways \citepads{2005ApJ...632..302B}: via a neutral-neutral reaction (CH$_2$ + N $\rightarrow$ HCN + H) and via dissociative recombination (HCNH$^+$ + e$^-$ $\rightarrow$ HCN + H). The dissociative recombination pathway is 30~\% more common, but protonated hydrogen cyanide (HCNH$^+$) must be in its linear form. Dissociative recombination with its structural isomer, H$_2$NC$^+$, produces only hydrogen isocyanic hydrogen (HNC). We clearly detected HCN, its isotopologues H$^{13}$CN and HC$^{15}$N, and the vibrational excited states HCN,$v_2$=1 and HCN,$v_2$=2 in many sources in Sgr~B2(M) and N, see Figs.~\ref{fig:HCNM}, \ref{fig:HCNN}, \ref{fig:HC13NM}, \ref{fig:HC13NN}, \ref{fig:HCNv21M}, \ref{fig:HCNv21N}, \ref{fig:HCNv22MN}. The deuterated isotopologue DCN and its vibrational excited state DCN,$v_2$=1 were not detected. Because of the large optical depth of HCN and since only a single transition ($J$~=~3 -- 2, $E_{\rm low}$~=~12.8~K) of HCN is covered by our observation, we cannot derive excitation temperatures. For the abundances we get a very opposite behaviour for both regions. While the highest abundances in Sgr~B2(M) are two orders of magnitude lower than those described by \citetads{2021A&A...651A...9M}, we obtain abundances for Sgr~B2(N) that are up to four orders of magnitude higher than those described by \citetads{2014ApJ...789....8N}. However, the obtained abundances should be viewed with caution, as they are derived from only a single transition.

%\citetads{1991ApJS...77..255S, 1996A&A...305..950G, 1998ApJS..117..427N, 2010A&A...521L..46R, 2011A&A...529A..76R, 2011A&A...535A..47D, 2013A&A...556A.137E, 2014ApJ...789....8N, 2013A&A...559A..47B}

% comparison:
%              |              "Moeller__2021__SgrB2-M" |                "Neill__2014__SgrB2-N"
% HCN;v=0;     |      3.1    207.3   1.1e-12   3.0e-05 |     40.0     63.0   1.3e-11   1.3e-11
% HC-13-N;v=0; |      3.8    207.3   8.7e-13   1.5e-06 |     63.0     63.0   1.3e-11   1.3e-11
% HCN;v2=1;    |    500.0    600.0   4.5e-11   1.8e-10 |        -        -         -         -

%:::::::::::::::::::::::::::::::::::::::::::::::::::::::::::::::::::::::::::::::
% HNC

\textbf{HNC}. Hydrogen isocyanic hydrogen (HNC) is found in all sources in Sgr~B2(M) and N, see Figs.~\ref{fig:HNCM}~-~\ref{fig:HNCN}. Its isotopologues HN$^{13}$C and H$^{15}$NC are also detected in many sources, see Figs.~\ref{fig:HNC13M}, \ref{fig:HNC13N}, \ref{fig:HN15CM}, \ref{fig:HN15CN}. Similar to HCN, we also observe the vibrationally excited state HNC,$v_2$=1, see Figs.~\ref{fig:HNCv21M}~-~\ref{fig:HNCv21N}. The deuterated isotopologue HNC was not detected. As for HCN, only the ($J$~=~3 -- 2, $E_{\rm low}$~=~13.1~K) transition is included in our observation, so we cannot derive excitation temperatures for HNC either. In both regions the received abundances exceed the results of previous studies by several orders of magnitude. But similar to HCN, the abundances are not very reliable.

% add discussion of HNC/HCN abundance

% \citetads{1998ApJS..117..427N, 1996A&A...305..950G, 2014ApJ...789....8N, 2013A&A...559A..47B}.

% comparison:
%             |            "Moeller__2021__SgrB2-M" |                "Neill__2014__SgrB2-N"
% HNC;v=0;    |   19.5     73.7   1.0e-10   3.7e-09 |     50.0     80.0   1.3e-11   1.3e-11
% HNC-13;v=0; |   19.5     73.7   5.1e-12   1.9e-10 |     63.0     63.0   3.7e-12   5.6e-12
% HNC;v2=1;   |  400.0    600.0   8.2e-11   9.1e-11 |        -        -         -         -

%:::::::::::::::::::::::::::::::::::::::::::::::::::::::::::::::::::::::::::::::
% CH3CN

\textbf{CH$_3$CN}. Methyl cyanide, CH$_3$CN, also known as acetonitrile or cyanomethane, is a symmetric top molecule with a large dipole moment of 3.92~D. It is used to probe the temperature and density in star forming regions. According to \citetads{2019A&A...628A..27B}, methyl cyanide (CH$_3$CN) mostly derives from the grain-surface hydrogenation reaction: CH$_2$CN + H $\rightarrow$ CH$_3$CN, where CH$_2$CN is formed via the grain-surface atomic-addition sequence: CN $\xrightarrow{C}$ C$_2$N $\xrightarrow{H}$ HC$_2$N $\xrightarrow{H}$ CH$_2$CN. At low temperatures ($T < 50$~K), CH$_3$CN is also produced in the gas phase via the electronic recombination of protonated methyl cyanide (CH$_3$CNH$^+$), product of the ion-molecule reaction between CH$_3^+$ and HCN. 51 transitions of CH$_3$CN are included in our survey with lower energies ranging from 58.3~K ($J_K$ = 12$_0$ -- 11$_{0}$) up to 1488~K ($J_K$ = 15$_{14}$ -- 14$_{14}$). Unlike other molecules, the high-energy transitions of CH$_3$CN have comparable transition probabilities to the lower transitions, making CH$_3$CN a good tracer for warm and moderately hot regions. In our analysis, we clearly identified CH$_3$CN, its isotopologues $^{13}$CH$_3$CN and CH$_3^{13}$CN, and the vibrationally excited states CH$_3$CN,$v_8$=1 and CH$_3$CN,$v_8$=2 in many sources in Sgr~B2(M) and N, see Fig. ~\ref{fig:CH3CNMN}, \ref{fig:C13H3CNMN}, \ref{fig:CH3C13NMN}, \ref{fig:CH3CNv81MN}, \ref{fig:CH3CNv82MN}, where for most sources in Sgr~B2(M) and N we use at least two components. Only for some outer sources we need only one component. For the vibrationally excited states, we use the same model as for CH$_3$CN, since the lower energies of the vibrationally excited state transitions are located more or less in the same range as the CH$_3$CN transition. Compared to previous analyses, we find a wider range of excitation temperatures and abundances for both regions.

\textbf{C$_2$H$_3$CN}. Vinyl cyanide (C$_2$H$_3$CN) is a planar molecule and is a slightly asymmetric prolate rotor with two non-zero electric dipole moment components, which lead to a rich rotational spectrum. C$_2$H$_3$CN is formed on dust grains through the successive hydrogenation of HCCCN \citepads{2019A&A...628A..27B}: HCCCN $\xrightarrow{H}$ C$_2$H$_2$CN $\xrightarrow{H}$ C$_2$H$_3$CN. Subsequently, C$_2$H$_3$CN is converted to C$_2$H$_5$CN on the grain surface by hydrogenation C$_2$H$_3$CN $\xrightarrow{H}$ C$_2$H$_4$CN $\xrightarrow{H}$ C$_2$H$_5$CN. Since it is rapidly destroyed by an ion-molecule reaction immediately after desorption, C$_2$H$_3$CN originated from the grains contributes only a small fraction to abundance in the gas phase. At late times C$_2$H$_3$CN may form in the gas phase via the reaction: C$_2$H$_4$ + CN $\rightarrow$ C$_2$H$_3$CN + H. In our observations 762 transitions of vinyl cyanide with lower energies from 6.8~K ($J_{K_a,K_c}$ = 6$_{2,4}$ -- 5$_{0,5}$) to 2133~K ($J_{K_a,K_c}$ = 88$_{13,75}$ -- 89$_{12,78}$) are included. Similar to other line surveys (\citetads{1998ApJS..117..427N}, \citetads{2004ApJ...600..234F}, \citetads{2013A&A...559A..47B}, \citetads{2014ApJ...789....8N}), we detected vinyl cyanide (C$_2$H$_3$CN) in emission, see Fig.~\ref{fig:C2H3CNMN}, using mostly a single component to describe the contribution. In contrast to \citetads{2013A&A...559A..47B}, we cannot identify a cold temperature component in sources in Sgr~B2(M), but instead obtain much higher excitation temperatures. Maybe that colder component is part of the envelope surrounding the Sgr~B2 complex, which we discussed in the first part of the analysis described in Paper~VII. The corresponding abundances show a slightly wider range. For sources in Sgr~B2(N), we obtain excitation temperatures similar to \citetads{2017A&A...604A..60B}, but the highest abundance we find in Sgr~B2(N) is an order of magnitude lower than \citetads{2017A&A...604A..60B}. Additionally, the vibrational excited states C$_2$H$_3$CN,v$_{11}$=1, C$_2$H$_3$CN,v$_{11}$=2, and C$_2$H$_3$CN,v$_{15}$=1 in Sgr~B2(M) and N are weakly detected, see Figs.~\ref{fig:C2H3CNv101MN}, \ref{fig:C2H3CNv111MN}, \ref{fig:C2H3CNv112MN}, \ref{fig:C2H3CNv151MN}. Since we assume that the vibrationally excited states originate from the same region and are subject to the same conditions, we fit for each source all three vibrational excited states with the same model parameters and components. Isotopologues were not detected.

% comparison:
%             |              "Bonfand__2017__SgrB2-N" |                "Neill__2014__SgrB2-N" |         "Boegelund__2018__AFGL_4176"
% C2H3CN;v=0; |    145.0    200.0   5.2e-10   2.3e-07 |    150.0    150.0   1.2e-08   1.2e-08 |   240.0    240.0   1.6e-08   1.6e-08

%:::::::::::::::::::::::::::::::::::::::::::::::::::::::::::::::::::::::::::::::
% C2H5CN

\textbf{C$_2$H$_5$CN}. Ethyl cyanide (C$_2$H$_5$CN) is a prolate, asymmetric top molecule with one plane of symmetry. Due to its strong permanent dipole moment ($\mu_a$ = 3.816~D, $\mu_b$ = 1.235~D along the molecular-fixed principal axes \citepads{2011JMoSp.270...83K}) the pure rotational spectrum shows strong $a$- and $b$-type transitions. Our survey contains 2174 transitions with lower energies between 10.4~K ($J_{K_a,K_c}$ = 7$_{3,4}$ -- 6$_{1,5}$) and 2434~K ($J_{K_a,K_c}$ = 104$_{9,95}$ -- 104$_{8,96}$). Ethyl cyanide is identified in many sources in emission in Sgr~B2(M) and N, see Fig.~\ref{fig:C2H5CNMN}, whose contributions are modeled by one and two components. Additionally, we weakly detected the isotopologue C$_2$H$_5^{13}$CN in some sources, see Fig.~\ref{fig:C2H5C13NMN}, while C$_2$H$_5$C$^{15}$N is not found. Compared to the previous analysis of \citetads{2013A&A...559A..47B}, we find wider ranges of excitation temperatures and abundances in sources in Sgr~B2(M). Compared to \citetads{2017A&A...604A..60B}, in Sgr~B2(N), as already in Sgr~B2(M), we also obtain a larger temperature range, which in addition includes a cold temperature component. In contrast, \citetads{2017A&A...604A..60B} describe abundances that are up to an order of magnitude higher. Similar to vinyl cyanide we also identified vibrational excited states, i.e.\ C$_2$H$_5$CN,v$_{12}$=1, C$_2$H$_5$CN,v$_{13}$+v$_{21}$=1, and C$_2$H$_5$CN,v$_{20}$=1 in Sgr~B2(N) in sources A01, A02, A03, and A08, see Figs.~\ref{fig:C2H5CNv121N}, \ref{fig:C2H5CNv13211N}, \ref{fig:C2H5CNv201N}. Again, we assume a common origin for the vibrationally excited states, so that we use the same model parameters such as excitation temperature, etc., for each source in which vibrationally excited states have been identified.

% comparison:
%              |              "Bonfand__2017__SgrB2-N" |                "Neill__2014__SgrB2-N" |        "Bisschop__2007__hot-cores" |
% C2H5CN;v=0;  |    150.0    170.0   4.1e-09   3.8e-06 |    120.0    150.0   3.1e-08   7.5e-08 |  92.0     99.0   4.4e-09   2.5e-08 |

%:::::::::::::::::::::::::::::::::::::::::::::::::::::::::::::::::::::::::::::::
% CN

\textbf{CN}. The cyano radical (CN) is a key molecule in many astrochemical chains and very reactive with molecules possessing C=C double and C$\equiv$C triple bonds. Additionally, it is involved in the formation of cyanopolyynes \citepads{doi:10.1021/acs.jpclett.7b01853} and other complex molecules related to star-forming processes. Furthermore, it is proposed as ‘chemical clocks’ in determining the age of this kind of source (\citetads{2018ApJ...866...32T}, \citetads{2009MNRAS.394..221C}). CN is formed by dissociative recombination \citepads{2001A&A...370..576L}: HCN$^+$ + e$^- \rightarrow$ CN + H or by photo dissociation \citepads{1996A&A...310..893B}: HCN + h$\nu \rightarrow$ CN + H. The 19 hyperfine transitions of CN around ($E_{\rm low} = 5.4$~K) are clearly observed in nearly all sources in Sgr~B2(M) and N, see Fig.~\ref{fig:CNMN}. The transitions usually occur as broad absorption features over a wide velocity range described by multiple components. The isotope $^{13}$CN is also detected in many sources in Sgr~B2(M) and N, see Fig.\ref{fig:C13NMN}, whereas C$^{15}$N is not. For both regions, the determined excitation temperatures and abundances show a much wider range than in previous studies.

% \citetads{1991ApJS...77..255S, 1996A&A...305..950G, 1998ApJS..117..427N, 2002ApJ...578..211S, 2004ApJ...600..234F, 2014ApJ...789....8N, 2013A&A...559A..47B}

% comparison:
%         |              "Moeller__2021__SgrB2-M" |                "Neill__2014__SgrB2-N"
% CN;v=0; |     33.6     33.6   1.1e-09   1.1e-09 |     40.0     40.0   8.7e-11   8.7e-11

%:::::::::::::::::::::::::::::::::::::::::::::::::::::::::::::::::::::::::::::::
% H2CCN

\textbf{H$_2$CCN}. The cyanomethyl radical is the simplest cyanide derivative of the methyl radical, CH$_3$, \citepads{2021JMoSp.37711448C} and has two interchangeable hydrogen nuclei with non-zero spin which dictate the existence of an ortho and para symmetry. According to \citetads{2018MNRAS.478.2962Z} the cyanomethyl radical (H$_2$CCN or CH$_2$CN) is formed via one of the following reactions: CN $\xrightarrow{H}$ HCN $\xrightarrow{H}$ H$_2$CN $\xrightarrow{C}$ CH$_2$CN or CN $\xrightarrow{H}$ HCN $\xrightarrow{C}$ HC$_2$N $\xrightarrow{H}$ CH$_2$CN or CN $\xrightarrow{C}$ CCN $\xrightarrow{H}$ HC$_2$N $\xrightarrow{H}$ CH$_2$CN. In our analysis, 145 transitions of the cyanomethyl radical are included with lower energies between 5.3~K ($N_{K_a,K_c}$ = 11$_{0,11}$ -- 10$_{0,10}$) and 1384~K ($N_{K_a,K_c}$ = 13$_{10,4}$ -- 12$_{10,3}$), where we do not distinguish between ortho and para symmetry. H$_2$CCN was mostly found in emission in some sources in Sgr~B2(M), see Fig.~\ref{fig:H2CCNM}, usually modeled by a single component. None of its isotopologues were found.

%:::::::::::::::::::::::::::::::::::::::::::::::::::::::::::::::::::::::::::::::
% HCCCN

% The linear pentatomic molecule cyanoacetylene has seven modes of vibration, four stretching modes and three bending modes which are doubly degenerate (Mallinson & Fayt 1976). The stretching modes ν1, ν2, ν3, which lie more than 3000~K above ground. Besides the fundamental modes of vibration also overtones and combinations modes can be observed.
\textbf{HCCCN}. Cyanoacetylene (HCCCN or HC$_3$N) is a linear molecule and the simplest cyanopolyyne (HC$_n$N, with $n = 3,5,7,\ldots$). Its main formation pathway is the neutral–neutral reaction between C$_2$H$_2$ and CN \citepads{2016ApJ...830..106T}: C$_2$H$_2$ + CN $\rightarrow$ HCCCN + H. HCCCN has six transitions with lower energies between 120.5~K ($J$~=~24 -- 23) and 189.9~K ($J$~=~30 -- 29) within the frequency ranges covered by our observations. Furthermore, cyanopolyyne was found in almost all sources in Sgr~B2(M) and N, see Fig.~\ref{fig:HCCCNMN}, mostly in emission described by two components. Additionally, we detect the isotopologues H$^{13}$CCCN, see Fig.~\ref{fig:HC13CCNMN}, HC$^{13}$CCN, see Fig.~\ref{fig:HCC13CNMN}, HCC$^{13}$CN, see Fig.~\ref{fig:HCCC13NMN}, while the other isotopologues HCCC$^{15}$N, DCCCN, D$^{13}$CCCN, DC$^{13}$CCN, DCC$^{13}$CN, DCCC$^{15}$N, H$^{13}$C$^{13}$CCN, H$^{13}$CC$^{13}$CN, HC$^{13}$C$^{13}$CN, HC$^{13}$CC$^{15}$N could not be identified. Furthermore, we identified the following vibrationally excited states in some sources in Sgr~B2(M) and N: HCCCN,v$_4$=1, Fig.~\ref{fig:HCCCNv41MN}, HCCCN,v$_4$=1,v$_7$=1, Fig.~\ref{fig:HCCCNv41v71MN}, HCCCN,v$_5$=1, Fig.~\ref{fig:HCCCNv51MN}, HCCCN,v$_5$=1,v$_7$=1, Fig.~\ref{fig:HCCCNv51v71MN}, HCCCN,v$_5$=2, Fig.~\ref{fig:HCCCNv52MN}, HCCCN,v$_6$=1, Fig.~\ref{fig:HCCCNv61MN}, HCCCN,v$_6$=1,v$_7$=1, Fig.~\ref{fig:HCCCNv61v71MN}, HCCCN,v$_7$=1, Fig.~\ref{fig:HCCCNv71MN}, HCCCN,v$_7$=2, Fig.~\ref{fig:HCCCNv72MN}, HCCCN,v$_7$=3, Fig.~\ref{fig:HCCCNv73MN}, HCCCN,v$_7$=4, Fig.~\ref{fig:HCCCNv74MN}, H$^{13}$CCCN,v$_7$=1, Fig.~\ref{fig:HC13CCNv71MN}, H$^{13}$CCCN,v$_7$=2, Fig.~\ref{fig:HC13CCNv72M}, H$^{13}$CCCN,v$_7$=3, Fig.~\ref{fig:HC13CCNv73M}, HC$^{13}$CCN,v$_7$=1, Fig.~\ref{fig:HCC13CNv71MN}, HCC$^{13}$CN,v$_7$=1, Fig.~\ref{fig:HCCC13Nv71MN}. Similar to the vibrationally excited states of vinyl cyanide, we fit all vibrational excited states belonging to HCCCN with the same model parameters. For the vibrational excitated states associated with H$^{13}$CCCN we use the same model parameters as for the vibrational excitated states of HCCCN, except that the column densities are scaled by the $^{12}$C/$^{13}$C ratio. Compared with the previous analysis of \citetads{2013A&A...559A..47B}, we obtain excitation temperatures and abundances for HCCCN and its vibrationally excited states that have a much wider range.

% comparison:
%            |              "Moeller__2021__SgrB2-M"
% HCCCN;v=0; |     31.4     90.0   3.0e-11   3.6e-09

%:::::::::::::::::::::::::::::::::::::::::::::::::::::::::::::::::::::::::::::::
% NH2CN

\textbf{NH$_2$CN}. Cyanamide (NH$_2$CN or CN$_2$H$_2$) is one of the rare amide-type molecule that was created with two individual nitrogen atoms. After reaction with H$_2$O, NH$_2$CN is converted to urea (NH$_2$CONH$_2$), which plays a very important role in organic chemistry and biology \citepads{2019A&A...628A..10B}. Its isomer carbodiimide (HNCNH) can be generated from NH$_2$CN by photochemically and thermally induced reactions \citepads{doi:10.1021/jp0459256}. Molecules containing the carbodiimide moiety (-NCN-) are used in various biological processes, including the assembly of amino acids into peptides \citepads[see e.g.,][]{doi:10.1021/cr00046a004}. NH$_2$CN is formed through one of the following two pathways: Neutral-neutral reaction between ammonia (NH$_3$) and cyanide (CN) on the grain surfaces \citepads{2004MNRAS.350..323S}: CN + NH$_3$ $\rightarrow$ NH$_2$CN + H) or through the electronic recombination of NH$_2$CNH$^+$: NH$_2$CNH$^+$ + e$^-$ $\rightarrow$ NH$_2$CN + H \citepads{2018A&A...612A.107C}, while the reaction between NH$_3$ and CN is much more efficient. In our survey, 196 transitions with lower energies ranging from 52.8~K ($J_{K_a,K_c}$ = 11$_{0,11}$ -- 10$_{0,10}$) up to 3538~K ($J_{K_a,K_c}$ = 82$_{4,78}$ -- 81$_{5,76}$) are included. We have identified cyanamide in two sources in Sgr~B2(M), see Fig.~\ref{fig:NH2CNM}, using a single component to describe the contribution of cyanamide. Unlike \citetads{2013A&A...559A..47B}, we cannot identify the cold temperature component, but we obtain much higher temperatures and abundances. We could find neither one of the isotpologues nor urea.

%+++++++++++++++++++++++++++++++++++++++++++++++++++++++++++++++++++++++++++++++
\subsection{S-bearing Molecules}\label{subsec:SBearingMol}

%:::::::::::::::::::::::::::::::::::::::::::::::::::::::::::::::::::::::::::::::
% H2S

\textbf{H$_2$S}. Hydrogen sulfide (H$_2$S) is one of the most primitive sulfur-bearing molecules detected in star-forming regions \citepads{2004A&A...413..609W}. It is mainly produced by successive additions of H atoms to atomic sulfur on icy grains (\citetads{1986MNRAS.221..673M}, \citetads{1990A&A...231..466M}): S $\xrightarrow{H}$ HS $\xrightarrow{H}$ H$_2$S. Due to the fact that these reactions are barrierless, H$_2$S is formed at extremely low temperatures ($\sim$10~K). Although hydrogen sulfide is not included in our observations, we detected its isotopologues H$_2 \! ^{33}$S (14 transitions from $J_{K_a,K_c}$ = 2$_{2,0}$ -- 2$_{1,1}$, $E_{\rm low}$ = 76.6~K to $J_{K_a,K_c}$ = 4$_{3,2}$ -- 5$_{0,5}$, $E_{\rm low}$ = 239.1~K), see Fig.~\ref{fig:H2S33MN}, and H$_2 \! ^{34}$S (two transitions from $J_{K_a,K_c}$ = 2$_{2,0}$ -- 2$_{1,1}$, $E_{\rm low}$ = 73.5~K to $J_{K_a,K_c}$ = 4$_{3,2}$ -- 5$_{0,5}$, $E_{\rm low}$ = 238.9~K), see Fig.~\ref{fig:H2S34MN}, in some sources. The sources in both regions have excitation temperatures and abundances that cover a much wider range than previous studies. Deuterated hydrogen sulfide, also known as deuterium sulfide (D$_2$S), was not detected.

% In contrast to Sgr~B2(N), the ground-state transition (2$_{12}$ -- 1$_{01}$) of \emph{ortho}-H$_2$S shows self-absorption and a series of cold, red-shifted absorptions with velocities up to $-$104~km~s$^{-1}$. Even the ground-state transitions of both isotopologues show a combination of red-shifted emission and blue-shifted absorption features. Higher energy transitions are observed only in emission, except the (2$_{21}$ -- 1$_{10}$) transition ($E_{\rm low}$~=~8.1~K) of \emph{ortho}-H$_2$S, which shows self-absorption. \emph{Para}-H$_2$S and its isotopologues \emph{para}-H$_2 \! ^{33}$S and \emph{para}-H$_2 \! ^{34}$S are identified as well, whereas \emph{para}-H$_2 \! ^{33}$S is only weakly detected. All transitions are seen in emission, except the (2$_{02}$ -- 1$_{11}$) and (3$_{13}$ -- 2$_{02}$) transitions of \emph{para}-H$_2 \! ^{33}$S which show self-absorption as well. (Opposed to \emph{ortho}-H$_2$S, the ground state transition of \emph{para}-H$_2$S is not included in the HIFI survey).

% \citetads{1994A&A...289..579T, 2004JKAS...37..131M, 2013A&A...556A.137E, 2014ApJ...789....8N, 2013A&A...559A..47B} is detected

% comparison:
%             |                "Neill__2014__SgrB2-N"
% H2S-33;v=0; |    120.0    120.0   7.4e-09   7.4e-09
% H2S-34;v=0; |    160.0    190.0   5.0e-10   1.9e-09

%:::::::::::::::::::::::::::::::::::::::::::::::::::::::::::::::::::::::::::::::
% SO

\textbf{SO}. Sulfur monoxide (SO) is considered to be a shock tracer \citepads{2020ApJ...898...54T}. According to \citetads{1997ApJ...481..396C} dissociative recombination of H$_3$S$^+$, produced via H$_3$O$^+$ + H$_2$S $\rightarrow$ H$_3$S$^+$ + H$_2$O, yields SH, which forms sulfur monoxide via SH + O $\rightarrow$ SO + H. Additionally, at around 100~K the reactions S + OH $\rightarrow$ SO + H and S + O$_2$ $\rightarrow$ SO + O are the major source of SO, with atomic sulfur obtained from SH + H $\rightarrow$ S + H$_2$. As source temperatures increase, the reaction including O$_2$ dominates, as OH abundance decreases faster with temperature than O$_2$. Due to the formation of SO via neutral–neutral reactions, such as the aforementioned S + OH $\rightarrow$ SO + H reaction, the SO abundance could be enhanced in a high-temperature environment and/or in the late stages of molecular cloud evolution. In general, neutral-neutral reactions are slower than ion-molecule reactions and tend to have activation energies that must be overcome during the reactions. Dense regions can accelerate the formation of SO via slow neutral–neutral reactions by increasing the collision rate between the reactants \citepads{2014AJ....147..141H}. Other reaction pathways can also be important for the formation of SO \citepads{kups63059}. For example, SO can also be formed by recombination of SO$_2^+$, but we do not consider this further in our analysis as the significance of these pathways is unclear. In our survey, eight transitions of sulfur monoxide with lower energies from 4.5~K ($J_N$ = 1$_2$ -- 2$_1$) to 71.0~K ($J_N$ = 7$_8$ -- 7$_7$) are included. We identified SO in almost all sources in Sgr~B2(M) and N, see Fig.~\ref{fig:SOMN}, modeled by two or more components. Some sources especially in Sgr~B2(M) show strongly asymmetric line shapes indicating a complex kinematic structure. The isotopologues $^{33}$SO, $^{34}$SO, S$^{17}$O, and S$^{17}$O were also detected in many sources, see Figs.~\ref{fig:S33OMN}, \ref{fig:S34OMN}, \ref{fig:SO17MN}, and \ref{fig:SO18MN}. Compared to previous analyses of \citetads{2021A&A...651A...9M} and \citetads{2013A&A...559A..47B}, we obtain excitation temperatures and abundances for SO for both regions that have a much wider range. In particular, we obtain much higher temperatures and abundances for the inner sources.

% Similar to Sgr~B2(N) \citepads{2014ApJ...789....8N}, we use four components to describe the contribution of SO, where the two hot components have comparable temperatures ($T_{\rm rot}$~=~169 -- 174~K). But unlike Sgr~B2(N) and in good agreement with \citetads{2013A&A...559A..47B}, we use two cold ($T_{\rm rot}$~=~17 -- 28~K) instead of two warm components. The weaker $^{34}$SO lines are modeled with two warm components, see Fig.~\ref{fig:S33OMN}, Fig.~\ref{fig:S34OMN}.

% \citetads{1986ApJS...60..819C, 1987ApJ...313L...5G, 1991ApJS...76..617T, 1991ApJS...77..255S, 1994A&A...289..579T, 1998ApJS..117..427N, 2004ApJ...600..234F, 2014ApJ...789....8N}

% comparison:
%        |           "Moeller__2021__SgrB2-M" |                "Neill__2014__SgrB2-N"
%SO;v=0; |  16.9    174.3   9.2e-11   6.7e-08 |    120.0    180.0   1.3e-11   1.2e-08

%:::::::::::::::::::::::::::::::::::::::::::::::::::::::::::::::::::::::::::::::
% SO+

\textbf{SO$^{+}$}. The sulfoxide cation (SO$^{+}$) is primarily produced by ion-molecule reactions via \citepads{1989ApJS...69..271H}: S$^+$ + OH $\rightarrow$ SO$^+$ + H. In early evolution stages other reactions such as SH$^+$ + O $\rightarrow$ SO$^+$ + H, H$_2$S$^+$ + O $\rightarrow$ SO$^+$ + H, and H$^+$ + SO $\rightarrow$ SO$^+$ + H could also contribute to production of SO$^{+}$ \citepads{1992ApJ...396L.107T}. We possibly detected SO$^{+}$ through its ($J_N$~= 5.5$_6$ -- 4.5$_5$, $E_{\rm low}$~=~27~K) transition in seven sources in Sgr~B2(M), see Fig.~\ref{fig:SOpM}, usually described by a single component. Since only one transition was detected in our survey (the other transition is at $E_{\rm low}$ = 550~K ($J_N$ = 5.5$_5$ -- 4.5$_4$)), we cannot reliably determine the excitation temperatures. Compared to the previous analysis by \citetads{2021A&A...651A...9M}, we obtain abundances that are up to two orders of magnitude larger. None of its isotopologues were found.

% comparison:
%          |          "Moeller__2021__SgrB2-M"
%SO+;v=0;  | 50.2     90.0   3.2e-11   6.4e-10

%:::::::::::::::::::::::::::::::::::::::::::::::::::::::::::::::::::::::::::::::
% SO2

\textbf{SO$_2$}. Similar to SO and SiO, sulfur dioxide (SO$_2$) is also a molecular tracer of shocks, so it is expected that it should originate from similar regions. Additionally, the formation pathways of SO$_2$ are similar to those of sulfur monoxide. As with SO, the pathways of SO$_2$ each begin with atomic sulfur or SH and continue as follows: S $\xrightarrow{OH, O_2}$ SO $\xrightarrow{OH}$ SO$_2$ or SH $\xrightarrow{O}$ SO $\xrightarrow{OH}$ SO$_2$ \citepads{1997ApJ...481..396C}. The abundance of SO$_2$ and SO is also influenced by cosmic ray-induced photodissociation through the reactions \citepads{kups63059} SO$_2$ + CRP $\rightarrow$ SO + O and SO + CRP $\rightarrow$ S + O, where the phrase CRP describes a cosmic ray photon. Like SO, we do not consider the contributions of photons from cosmic rays in our analysis, as the quantity of photons is unclear. Our observations contain 155 transitions of SO$_2$ with lower energies between 7.7~K ($J_{K_a,K_c}$ = 4$_{2,2}$ -- 3$_{1,3}$) and 5155~K ($J_{K_a,K_c}$ = 94$_{21,73}$ -- 95$_{20,76}$). We clearly identify sulfur dioxide in many sources in Sgr~B2(M) and N, see Fig.~\ref{fig:SO2MN}, where the contribution of SO$_2$ is mostly described by two components. For sources in both regions, we obtain excitation temperatures that are in a much larger range compared to previous studies, while the ranges of abundances are in very good agreement with the results of \citetads{2013A&A...559A..47B}. Similar to SO, we obtain the highest temperatures for the inner sources. The SO$_2$ isotopologues $^{33}$SO$_2$ and $^{34}$SO$_2$ are also observed, see Figs.~\ref{fig:S33O2MN}, \ref{fig:S34O2MN}, but SO$^{17}$O and SO$^{18}$O are not conclusively detected, see Figs.~\ref{fig:SOO17MN}, \ref{fig:SOO18MN}. Additionally, we detected the vibrational excited state SO$_2$,v$_2$=1 in many sources.

% \citepads{2013A&A...559A..47B, 2014ApJ...789....8N, 1986ApJS...60..819C, 1987ApJ...321L..75T, 1991ApJS...76..617T, 1991ApJS...77..255S, 1994A&A...289..579T, 2004ApJ...600..234F, 2014ApJ...789....8N, 2015ApJ...815..106H, 2013A&A...559A..47B}

% comparison:
%          |           "Moeller__2021__SgrB2-M" |                "Neill__2014__SgrB2-N" |          "Boegelund__2018__AFGL_4176"
% SO2;v=0; |  14.7    500.0   4.1e-09   1.3e-07 |    120.0    180.0   1.9e-11   1.9e-08 |    150.0    150.0   2.0e-06   2.0e-06

%:::::::::::::::::::::::::::::::::::::::::::::::::::::::::::::::::::::::::::::::
% CS

\textbf{CS}. Carbon monosulfide (CS) is a linear molecule and has a simple rovibrational spectrum. According to \citetads{2018ApJ...868L...2B} CS is mainly produced via CH$_2$ + S $\rightarrow$ CS + H$_2$. \citetads{1997ApJ...481..396C} also suggested the following other pathways: SO + C $\rightarrow$ CS + O, HCS$^+$ + e$^-$ $\rightarrow$ CS + H, and H$_3$CS$^+$ + e$^-$ $\rightarrow$ CS + H + H$_2$. In our analysis, the ($J$~=~5 -- 4, $E_{\rm low}$~=~23.5~K) rotational transition of carbon monosulfide is clearly detected in all sources in the Sgr~B2 complex, see Fig.~\ref{fig:CSMN}. Additionally, the isotopologues $^{13}$CS, C$^{33}$S, C$^{34}$S are also clearly observed, see Figs.~\ref{fig:C13SM}, \ref{fig:C13SN}, \ref{fig:CS33M}, \ref{fig:CS33N}, \ref{fig:CS34M}, \ref{fig:CS34N}, whereas the vibrational excited states CS,v=1, CS,v=2, CS,v=3, and CS,v=4 are not seen. Since only one transition was observed and due to the large optical depth of CS, we cannot reliably determine the excitation temperatures. Compared to previous analyses, we find wider ranges of abundances for both regions.

% \citetads{1992A&A...260..381G, 1996A&A...305..950G, 1991ApJS...77..255S, 1998ApJS..117..427N, 2013A&A...559A..47B, 2014ApJ...789....8N}

% comparison:
%             |              "Moeller__2021__SgrB2-M" |                "Neill__2014__SgrB2-N"
% CS;v=0;     |     40.9     98.9   7.6e-10   4.8e-09 |     20.0    120.0   1.3e-11   3.5e-10
% C-13-S;v=0; |     40.9     98.9   3.8e-11   2.4e-10 |     85.0    120.0   3.5e-11   5.7e-10
% CS-33;v=0;  |     40.9     98.9   1.0e-11   6.3e-11 |     85.0    120.0   1.8e-11   2.9e-10
% CS-34;v=0;  |     40.9     98.9   5.7e-11   3.5e-10 |     85.0    120.0   3.5e-11   5.7e-10

%:::::::::::::::::::::::::::::::::::::::::::::::::::::::::::::::::::::::::::::::
% H2CS

\textbf{H$_2$CS}. Thioformaldehyde (H$_2$CS) is a slightly asymmetric rotor with two interchangeable hydrogen nuclei. Therefore, its rotational states are grouped into ortho ($K_a$ odd) and para ($K_a$ even), where the ortho ground state is 14.9~K above the para ground state and has a dipole moment of $\mu_a$ = 1.647~D \citepads{1977JChPh..67.1576F}. Here we do not distinguish between ortho and para states. According to \citetads{1997ApJ...481..396C}, H$_2$CS is produced by one of the following pathways: S + CH$_3$ $\rightarrow$ H$_2$CS + H or H$_3$CS$^+$ + e$^-$ $\rightarrow$ H$_2$CS + H. Our observations cover 36 transitions of H$_2$CS with lower energies ranging from 34.6~K ($J_{K_a,K_c}$ = 7$_{0,7}$ -- 6$_{0,6}$) to 2931~K ($J_{K_a,K_c}$ = 58$_{3,55}$ -- 58$_{3,56}$). We found H$_2$CS in many sources in both regions, see Fig.~\ref{fig:H2CSMN}, mostly described by one or two components. The isotopologues H$_2 \! ^{13}$CS, see Fig.~\ref{fig:H2C13SMN}, H$_2$C$^{33}$S, see Fig.~\ref{fig:H2CS33MN}, and H$_2$C$^{34}$S, see Fig.~\ref{fig:H2CS33MN}, are detected in many sources as well. However, HDCS is not observed. While the range of excitation temperatures in Sgr~B2(N) agrees quite well with the results of \citetads{2013A&A...559A..47B}, we find temperatures in Sgr~B2(M) that are in a much wider range than in \citetads{2021A&A...651A...9M}. Furthermore, the abundances of the inner sources in both regions are in some cases significantly higher than those from previous studies.

% \citetads{1986ApJS...60..819C, 1998ApJS..117..427N, 1991ApJS...76..617T, 1991ApJS...77..255S, 2013A&A...559A..47B, 2014ApJ...789....8N}

% comparison:
%           |              "Moeller__2021__SgrB2-M" |                "Neill__2014__SgrB2-N" |          "Boegelund__2018__AFGL_4176"
% H2CS;v=0; |    120.0    120.0   2.3e-10   2.3e-10 |    120.0    120.0   6.2e-09   6.2e-09 |    160.0    160.0   1.1e-07   1.1e-07

%:::::::::::::::::::::::::::::::::::::::::::::::::::::::::::::::::::::::::::::::
% OCS

\textbf{OCS}. Carbonyl sulfide (OCS) has a linear structure. OCS as well as H$_2$S have been suggested as possible main sulfur carriers on dust grains (see e.g.\, \citetads{2004A&A...422..159W}, \citetads{2014A&A...565A..64P}). For young hot sources ($t \leq 2 \times 10^4$~yr), OCS is mainly formed in the gas-phase through the radiative association reaction CO + S $\rightarrow$ OCS + h$\nu$ \citepads{1997ApJ...481..396C}. Thereafter, it is produced via one of the following pathways: HOCS$^+$ + e$^-$ $\rightarrow$ OCS + H and O + HCS $\rightarrow$ OCS + H, with the latter making only a minor contribution. In our observations, we found transitions of carbonyl sulfide (OCS) ranging from $J_{\rm up}$~=~18 to 21 ($E_{\rm low}$~=~89 -- 123~K) in nearly all sources in Sgr~B2, see Fig.~\ref{fig:OCSMN}, mostly described by two components. Compared to previous analyses, we find a wider range of excitation temperatures and abundances for both regions. Similar to other molecules like SO or SiO, we found highly asymmetric line shapes for some sources indicating a complex kinematic structure. Furthermore, the isotopologues O$^{13}$CS, see Fig.~\ref{fig:OC13SMN}, OC$^{33}$S, see Fig.~\ref{fig:OCS33MN}, and OC$^{34}$S, see Fig.~\ref{fig:OCS34MN}, are found in many sources, while the other isotopologues $^{17}$OCS, $^{18}$OCS, OC$^{36}$S, O$^{13}$C$^{33}$S, O$^{13}$C$^{34}$S, $^{18}$O$^{13}$CS, and $^{18}$O$^{13}$C$^{34}$S are not reliably detected. In addition, the vibrational excited state OCS,v$_{2}$=1, see Fig.~\ref{fig:OCSv21MN}, is detected in many inner sources of Sgr~B2(M) and N.

% \citetads{1986ApJS...60..819C, 1998ApJS..117..427N,  1987ApJ...313L...5G, 1991ApJS...76..617T, 1991ApJS...77..255S, 2013A&A...559A..47B, 2014ApJ...789....8N}

% comparison:
%          |              "Moeller__2021__SgrB2-M" |                "Neill__2014__SgrB2-N"
% OCS;v=0; |    134.9    172.8   1.7e-09   5.4e-09 |    200.0    200.0   4.4e-08   4.4e-08

%:::::::::::::::::::::::::::::::::::::::::::::::::::::::::::::::::::::::::::::::
% HCS+

\textbf{HCS$^{+}$}. Thiomethylium (sometimes called thioformyl cation) (HCS$^{+}$) is produced primarily through the following reactions \citepads{1978ApJ...225..857M}: CS$^+$ + H$_2$ $\rightarrow$ H + HCS$^+$ and C$^+$ + H$_2$S $\rightarrow$ HCS$^+$ + H. According to \citetads{1985MNRAS.216.1025M}, the proton transfer reactions such as CS + H$_3^+$ $\rightarrow$ HCS$^+$ + H$_2$ and CS + HCO$^+$ $\rightarrow$ HCS$^+$ + CO may dominate in dense molecular clouds. We detected HCS$^{+}$ in eight sources in Sgr~B2(M), see Fig.~\ref{fig:HCSpM}, by its two transitions $J_{\rm up}$~=~5 and 6 ($E_{\rm low}$~=~20 -- 31~K), where we use usually two components to describe the contribution of HCS$^{+}$. None of its isotopologues DCS$^+$, H$^{13}$CS$^{+}$ was identified. Compared to results of the previous analysis of \citetads{2021A&A...651A...9M}, we find a wider range of excitation temperatures and abundances.

% comparison:
%           |             "Moeller__2021__SgrB2-M"
% HCS+;v=0; |    51.5     51.5   1.8e-10   1.8e-10

%:::::::::::::::::::::::::::::::::::::::::::::::::::::::::::::::::::::::::::::::
% NS

\textbf{NS}. Nitrogen sulfide (or sulfur mononitride) (NS) is isoelectronic to nitric oxide (NO) and its sulfur analog. According to \citepads{1994ApJ...422..621M}, NS is formed through HS + N $\rightarrow$ NS + H and HNS$^+$ + e$^-$ $\rightarrow$ NS + H, where HS is produced by the reaction of S$^+$ with organic species such as H$_2$CO, while HNS$^+$ is created from NH$_2^+$ and atomic sulfur. In our observations the hyperfine splitted transition ($J_N$~=~5.5$_5$ -- 4.5$_4$) of nitrogen sulfide (NS) is seen in some sources in Sgr~B2, see Fig.~\ref{fig:NSMN}, mostly described by a single component. Only one of its isotopologues N$^{34}$S, see Fig.~\ref{fig:NS34M}, is clearly identified in some sources in Sgr~B2(M) and N, while the other isotopologues $^{15}$NS and N$^{33}$S are not found. In contrast to previous studies of \citetads{2021A&A...651A...9M} and \citetads{2013A&A...559A..47B}, we achieve much higher excitation temperatures for all sources in both regions. Moreover, we derive much higher abundances for sources in Sgr~B2(M), while the range of abundances for sources in Sgr~B2(N) is larger compared to \citetads{2013A&A...559A..47B}.

% \citetads{1998ApJS..117..427N, 2014ApJ...789....8N, 2013A&A...559A..47B}

% comparison:
%         |              "Moeller__2021__SgrB2-M" |             "Neill__2014__SgrB2-N"
% NS;v=0; |     70.0     70.0   2.3e-10   2.3e-10 |  90.0     90.0   7.6e-10   1.3e-09

%:::::::::::::::::::::::::::::::::::::::::::::::::::::::::::::::::::::::::::::::
% SiS

\textbf{SiS}. Silicon monosulfide (SiS) is the simplest molecule containing a silicon-sulfur chemical bond. The SiS molecule is considered as a possible gas-phase precursor of sulfide grains. In circumstellar envelopes, it is thought to be formed in the densest and hottest parts of the inner envelope, where it is thought to be produced under thermochemical conditions. According to \citepads{2021ApJ...920...37M}, SiS is formed mainly via Si + SH $\rightarrow$ SiS + H, although other reactions such as Si + SH$_2$ $\rightarrow$ SiS + H$_2$ \citepads{2021SciA....7.7003D} might be important as well. Only the ($J$~=~15 -- 14, $E_{\rm low}$~=~58~K) transition of SiS is identified in three inner sources in Sgr~B2(M), see Fig.~\ref{fig:SiSM}, so a reliable temperature estimation is not possible. \citetads{1981ApJ...247..112D} described a detection of silicon sulfide in Sgr~B2 at 90.77~GHz and 108.92~GHz. For an assumed excitation temperature of 20~K, they obtain a column density of $1-2 \times 10^{13}$~cm$^{-2}$, leading to a range abundances of $1.25 - 2.5 \times 10^{-12}$ for an H$_2$ column density of $8.0 \times 10^{24}$~cm$^{-2}$. None of its isotopologues were found.

%:::::::::::::::::::::::::::::::::::::::::::::::::::::::::::::::::::::::::::::::
% CCCS

\textbf{CCCS}. Tricarbon monosulfide (C$_3$S or CCCS), sometimes called tricarbon sulfur, is a heterocumulene or thiocumulene, consisting of a straight chain of three carbon atoms and an additional sulfur atom. Different ways for the formation of CCCS are discussed: \citetads{1990A&A...231..466M} proposed the ion-molecule reaction scheme S$^+$ + (c-C$_3$H$_2$; l-C$_3$H$_2$) $\rightarrow$ HCCCS$^+$ + H, where ‘‘c’’ and ‘‘l’’ represent the cyclic and linear isomers, respectively. Finally, the HCCCS$^+$ ion is recombined with an electron to give CCCS. According to \citetads{2002A&A...395.1031Y}, CCCS is produced through the neutral-neutral reaction C$_2$ + H$_2$CS $\rightarrow$ CCCS + H$_2$. In our survey, ten transitions of CCCS with $E_{\rm low}$ located in a range between 184~K ($J$~=~37 -- 36) and 300~K ($J$~=~47 -- 46) are included. We detected CCCS in two sources in Sgr~B2(M), see Fig.~\ref{fig:CCCSM}, described by a single component. \citetads{2015MNRAS.452.3969C} identified CCCS in Sgr~B2(N), but without specifying temperature and abundance.

%+++++++++++++++++++++++++++++++++++++++++++++++++++++++++++++++++++++++++++++++
\subsection{Carbon and Hydrocarbons}\label{subsec:CHMol}

%:::::::::::::::::::::::::::::::::::::::::::::::::::::::::::::::::::::::::::::::
% CCH

\textbf{CCH}. In star-forming regions ethynyl is related to the earliest stages of massive star-formation \citepads{2008ApJ...675L..33B}. Ethynyl (CCH or C$_2$H) and other light ionized hydrocarbons CH$_n^+$ (n = 2 \ldots 5) are formed by radiative association of C$^+$ with H$_2$ and hydrogen addition reactions: C$^+ \rightarrow$ CH$_2^+ \rightarrow$ CH$_5^+$. These molecules react with electrons, CO, C, OH, and more complex species to form methane, which forms C$_2$H$_2^+$ and C$_2$H$_3^+$ in reactive collisions with C$^+$  \citepads{2008ApJ...675L..33B}. An alternative pathway for the formation of C$_2$H$_2^+$ is the dissociative recombination of CH$_5^+$ to CH$_3$, followed by reactions with C$^+$. Finally, C$_2$H$_2^+$ and C$_2$H$_3^+$ recombine dissociatively to form CH, C$_2$H, and C$_2$H$_2$. In our analysis we detected transitions of CCH from $N$ = 6 -- 5 to 21 -- 20 ($E_{\rm low}$ = 63 -- 879~K) in all sources in Sgr~B2, see Fig.~\ref{fig:CCHMN}. The doublets, caused by the unpaired electron, are clearly separated in the survey and well described by a single component without a red-shifted wing as described by \citetads{2013A&A...559A..47B}. Neither the isotopologues C$^{13}$CH, CC$^{13}$H, and CCD nor the vibrational excited states CCH,v$_2$=1, CCH,v$_2$=2, and CCH,v$_3$=1 could be reliably detected. Unlike other studies of Sgr~B2, we find temperatures and abundances in a broader range, including moderately high temperatures not previously identified.

% comparison:
%          |              "Moeller__2021__SgrB2-M" |             "Neill__2014__SgrB2-N"
% CCH;v=0; |     26.6     26.6   5.8e-09   5.8e-09 |  50.0     50.0   7.5e-11   7.5e-11

%:::::::::::::::::::::::::::::::::::::::::::::::::::::::::::::::::::::::::::::::
% c-C3H2

\textbf{c-C$_3$H$_2$}. C$_3$H$_2$ is a class of highly unsaturated carbenes, which consists of three isomers, propynylidene (C$_3$H$_2$), propadienylidene (l-C$_3$H$_2$) and cyclopropenylidene (c-C$_3$H$_2$). Among them, singlet cyclopropenylidene, c-C$_3$H$_2$, is the most stable isomer and contains the smallest hydrocarbon ring. The formation pathways of cyclopropenylidene is not clear. According to \citetads{1993ApJ...417..181M} cyclopropenylidene is formed by the dissociative recombination of C$_3$H$_3^+$: C$_3$H$_3^+$ + e$^-$ $\rightarrow$ C$_3$H$_2$ + H. \citetads{C0CP01529F} describe a different pathway involving the reaction of the methylidyne radical (CH) with acetylene (C$_2$H$_2$), which forms cyclopropenylidene plus atomic hydrogen and also propadienylidene plus atomic hydrogen. Additionally, the neutral–neutral reaction between atomic carbon and the vinyl radical (C$_2$H$_3$) also forms cyclopropenylidene plus atomic hydrogen \citepads{2012PCCP...14..477W}. 147 transitions from $J_{K_a,K_c}$ = 2$_{2,1}$ -- 2$_{0,2}$, $E_{\rm low}$ = 6.4~K to $J_{K_a,K_c}$ = 39$_{29,10}$ -- 39$_{28,11}$, $E_{\rm low}$ = 2391~K are included in our survey. We identified cyclopropenylidene in some sources in Sgr~B2(M), see Fig.~\ref{fig:cC3H2M}, using a single component to describe the contribution of c-C$_3$H$_2$. Some sources show abundances and temperatures that fit well with results described by \citetads{2013A&A...559A..47B}. In addition, we have also identified c-C$_3$H$_2$ in sources with much higher temperatures and abundances. None of its isotopologues were found.

%+++++++++++++++++++++++++++++++++++++++++++++++++++++++++++++++++++++++++++++++
\subsection{Other Molecules}\label{subsec:OtherMol}

%:::::::::::::::::::::::::::::::::::::::::::::::::::::::::::::::::::::::::::::::
% PH3

\textbf{PH$_3$}. Phosphine (PH$_3$) is a symmetric top molecule with a pyramidal structure similar to NH$_3$, but with negligible inversion splitting. PH$_3$ is an important molecule for the chemistry of phosphorus (P)-bearing species and is considered to form primarily on grains. According to \citetads{2020A&A...633A..54C} and \citetads{2020MNRAS.492.1180R} PH$_3$ is produced via the following reactions on grains: P $\xrightarrow{H}$ PH $\xrightarrow{H}$ PH$_2$ $\xrightarrow{H}$ PH$_3$. In our observations 19 transitions of PH$_3$ with lower energies from 0~K ($J_K$ = 1$_0$ -- 0$_0$) to 4727~K (27$_2$ -- 27$_5$) are included. We clearly identify phosphine by its ground state transition in many sources in Sgr~B2(M) and N, see Fig.~\ref{fig:PH3MN}. Due to the fact that only the ground state transition is detected, a reliable temperature estimation of PH$_3$ is difficult. \citetads{1990ApJ...365..569T} found PH$_3$ in Sgr~B2 with an excitation temperature of 50~K and an abundance of $<2 \times 10^{-10}$.

%:::::::::::::::::::::::::::::::::::::::::::::::::::::::::::::::::::::::::::::::
% PN

\textbf{PN}. According to \citetads{1990ApJ...365..569T}, phosphorus nitride (PN) forms via one of the following pathways: first, phosphorus nitride is formed by PO + N $\rightarrow$ PN + O when temperatures are high enough to overcome the activation barrier, or on grains that are disrupted in star-forming regions and preferentially release PN (and SiO) as products of disruption. Additionally, \citetads{2018MNRAS.476L..39M} reported that PN may be released by sputtering of dust grains due to shocks. In our analysis we identified the ($J$~=~5 -- 4, $E_{\rm low}$~=~23~K) transition of phosphorus nitride in almost all sources in Sgr~B2(N), see Fig.~\ref{fig:PNN}. However, due to the fact that only this transition is included in our observations, we cannot determine the excitation temperatures for the different sources. We obtain abundances which are six orders of magnitude higher than those of \citetads{2013A&A...559A..47B}. Because only one transition was considered in our analysis, our abundances should be viewed with caution.  We could find neither one of the isotopologues nor PO.
%\newpage

\onecolumn
\section{Transition plots}\label{app:transplots}

% quantum numbers
In the following subsection, the quantum number $J = N + S + L$ describes the total angular momentum of the respective molecule, while the other quantum numbers $N$, $S$ and $L$ describe the rotational, electron spin and electron orbital angular momentum quantum numbers, respectively. For species without electronic angular momentum, i.e. $J = N$, the rotational transitions are denoted by $\Delta J$, whereas those with electronic angular momentum the rotational levels are labeled by $N_J$. Furthermore, energy levels for symmetric rotors are described by $J_K$, where $K$ indicates the angular momentum along the symmetry axis. Energy levels for asymmetric rotors are denoted by $J_{K_a,K_c}$, where $K_a$ and $K_c$ refer to the angular momentum along the symmetry axis in the oblate and prolate symmetric upper limits, respectively.

In each figure presented in this section the observed ALMA spectrum for each core (blue line) is shown together with the overall fit (green dashed line) and the contribution of the respective molecule (red line), respectively. The vertical gray dashed line indicates the source velocity of the whole Sgr~B2 complex. On the right sight of each spectrum the name of the corresponding core is described together with the applied scaling factor (in round brackets). The offset between the different spectra is set to 450.

% include tex file describing transition plots
\twocolumn

%-------------------------------------------------------------------------------
% simple O-bearing molecules

%*******************************************************************************
% Figure: CO;v=0; and C-13-O;v=0;
\begin{figure*}[!htb]
    \centering
    \begin{subfigure}[t]{0.5\columnwidth}
       \includegraphics[width=1.0\columnwidth]{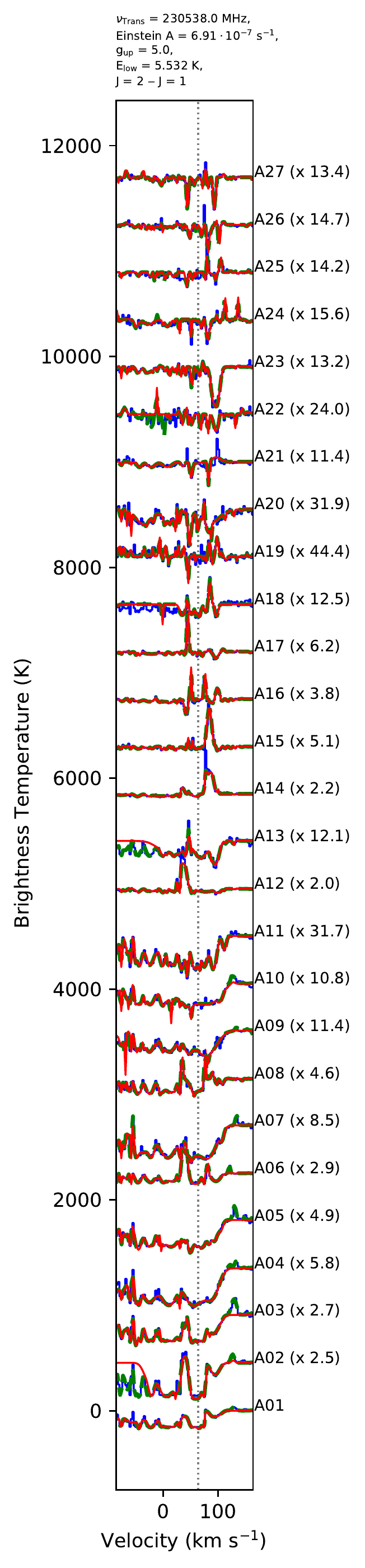}\\
       \caption{CO in Sgr~B2(M)}
       \label{fig:COM}
    \end{subfigure}
\hfill
    \begin{subfigure}[t]{0.5\columnwidth}
       \includegraphics[width=1.0\columnwidth]{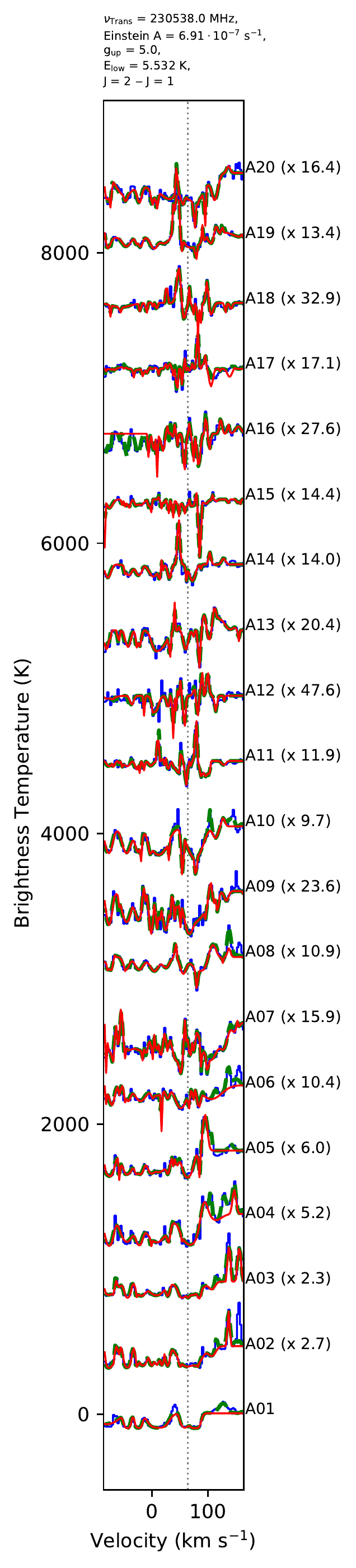}\\
       \caption{CO in Sgr~B2(N)}
       \label{fig:CON}
    \end{subfigure}
\hfill
    \begin{subfigure}[t]{0.5\columnwidth}
       \includegraphics[width=1.0\columnwidth]{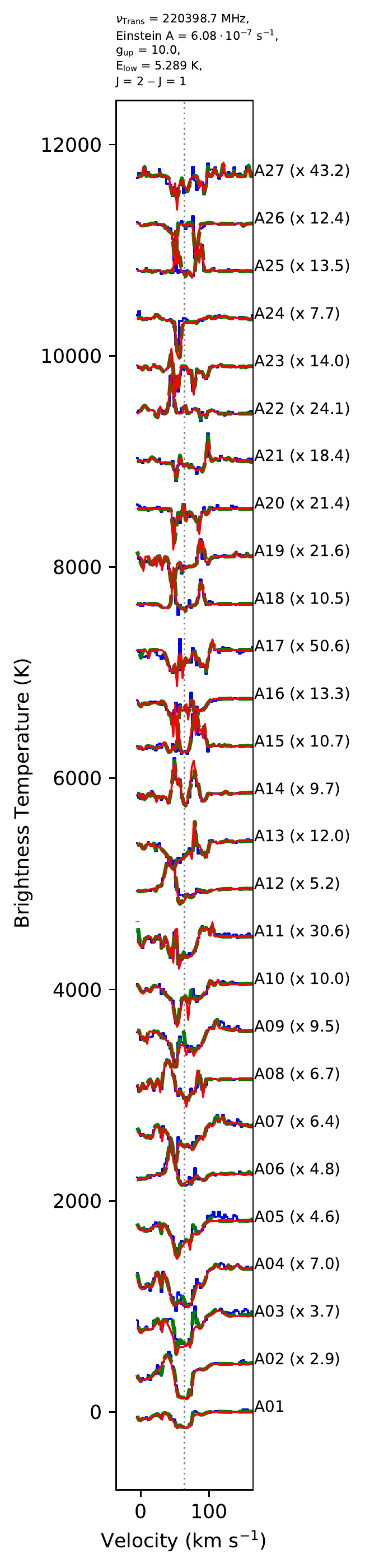}\\
       \caption{$^{13}$CO in Sgr~B2(M)}
       \label{fig:13COM}
    \end{subfigure}
\hfill
    \begin{subfigure}[t]{0.5\columnwidth}
       \includegraphics[width=1.0\columnwidth]{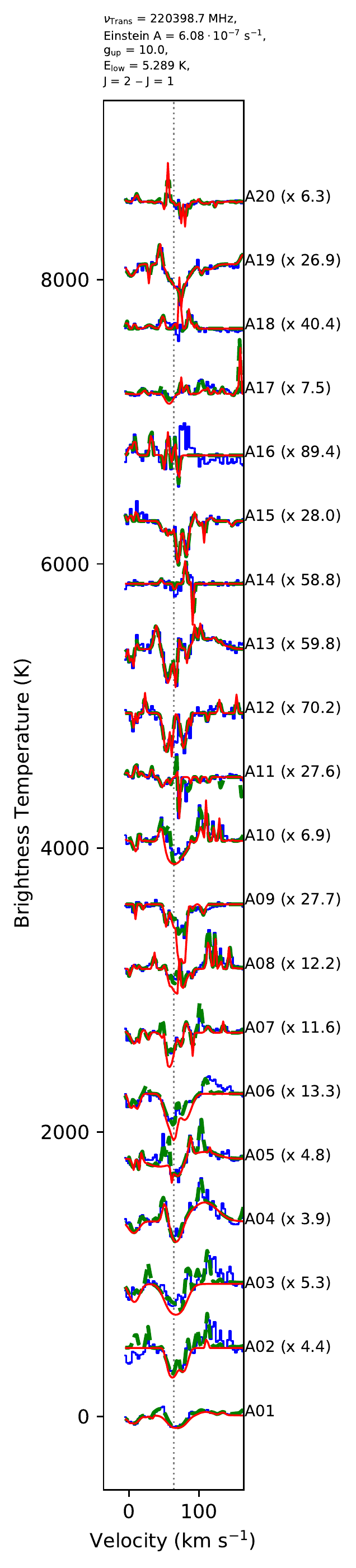}\\
       \caption{$^{13}$CO in Sgr~B2(N)}
       \label{fig:13CON}
    \end{subfigure}
    \caption{Transitions of CO and $^{13}$CO in Sgr~B2(M) and N.}
    \ContinuedFloat
    \label{fig:COMN}
\end{figure*}
\newpage
\clearpage

%*******************************************************************************
% Figure: CO-17;v=0; and CO-18;v=0;
\begin{figure*}[!htb]
    \centering
    \begin{subfigure}[t]{0.5\columnwidth}
       \includegraphics[width=1.0\columnwidth]{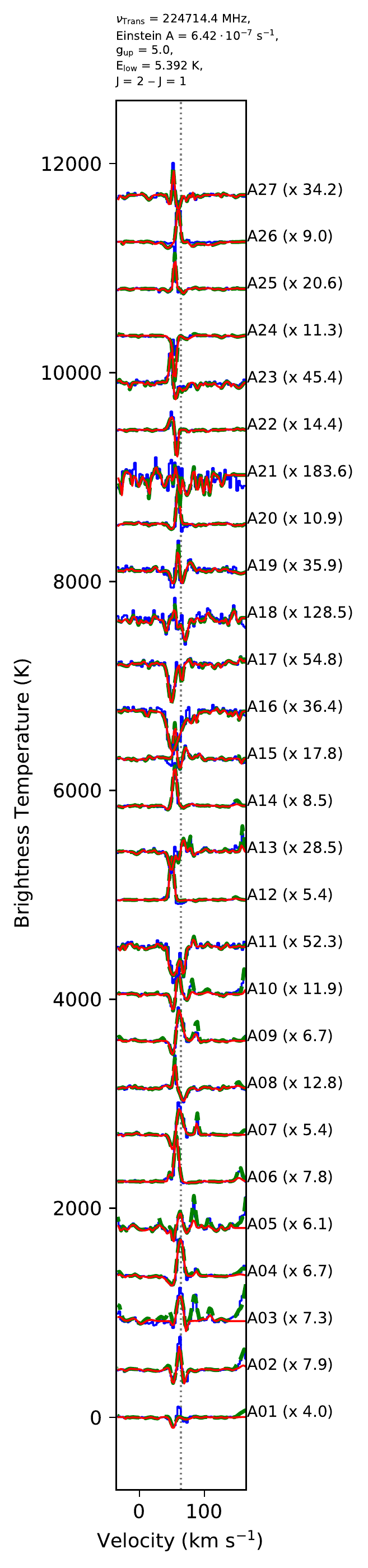}\\
       \caption{C$^{17}$O in Sgr~B2(M)}
       \label{fig:CO17M}
    \end{subfigure}
\hfill
    \begin{subfigure}[t]{0.5\columnwidth}
       \includegraphics[width=1.0\columnwidth]{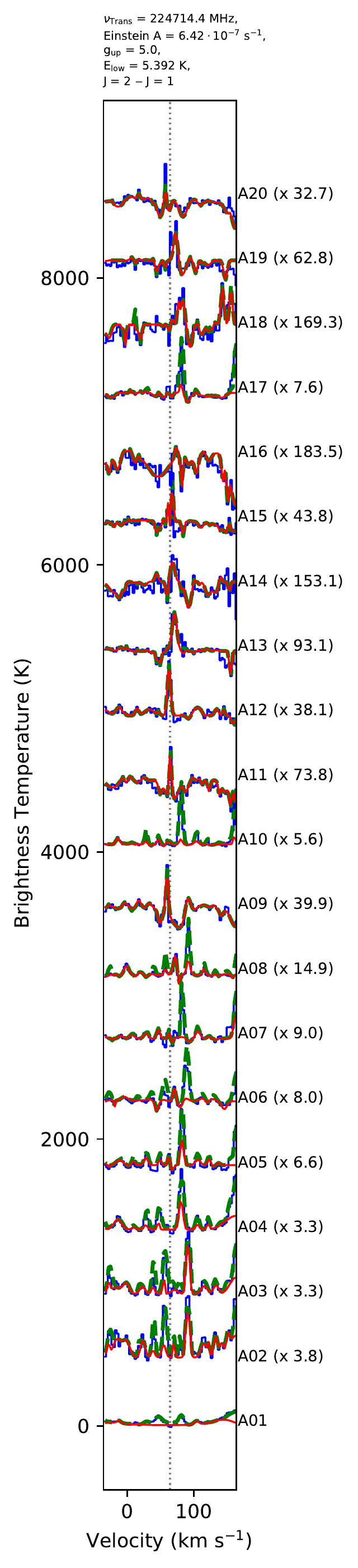}\\
       \caption{C$^{17}$O in Sgr~B2(N)}
       \label{fig:CO17N}
    \end{subfigure}
\hfill
    \begin{subfigure}[t]{0.5\columnwidth}
       \includegraphics[width=1.0\columnwidth]{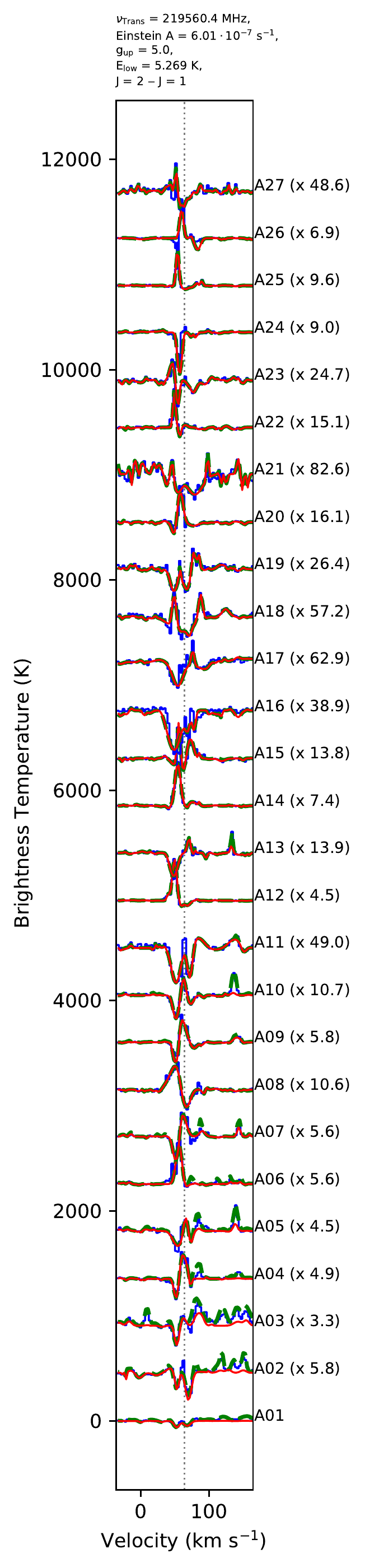}\\
       \caption{C$^{18}$O in Sgr~B2(M)}
       \label{fig:CO18M}
    \end{subfigure}
\hfill
    \begin{subfigure}[t]{0.5\columnwidth}
       \includegraphics[width=1.0\columnwidth]{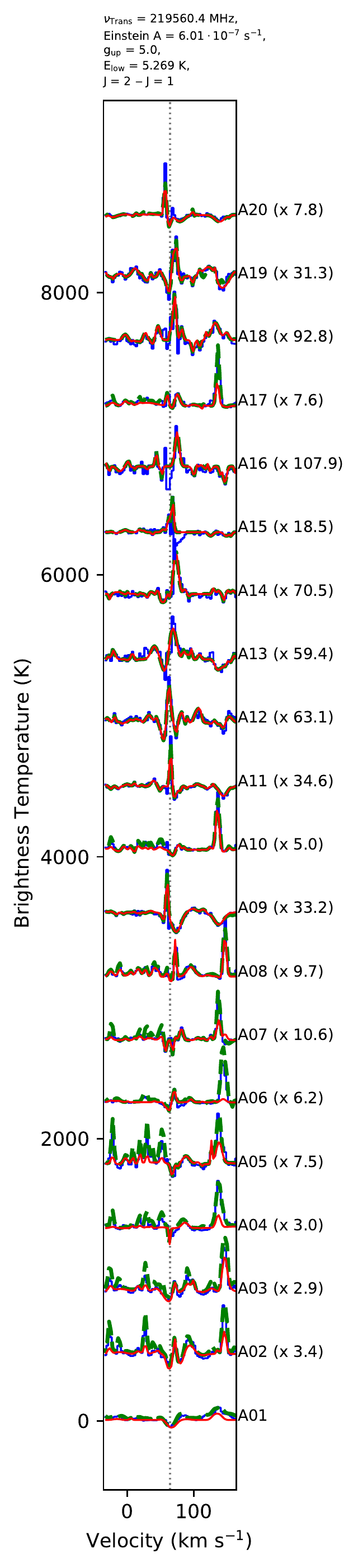}\\
       \caption{C$^{18}$O in Sgr~B2(N)}
       \label{fig:CO18N}
    \end{subfigure}
    \caption{Transitions of C$^{17}$O and C$^{18}$O in Sgr~B2(M) and N.}
    \ContinuedFloat
    \label{fig:CO1718MN}
\end{figure*}
\newpage
\clearpage

%*******************************************************************************
% Figure: CH3OH;v=0;
\begin{figure*}[!htb]
    \centering
    \begin{subfigure}[t]{1.0\columnwidth}
       \includegraphics[width=1.0\columnwidth]{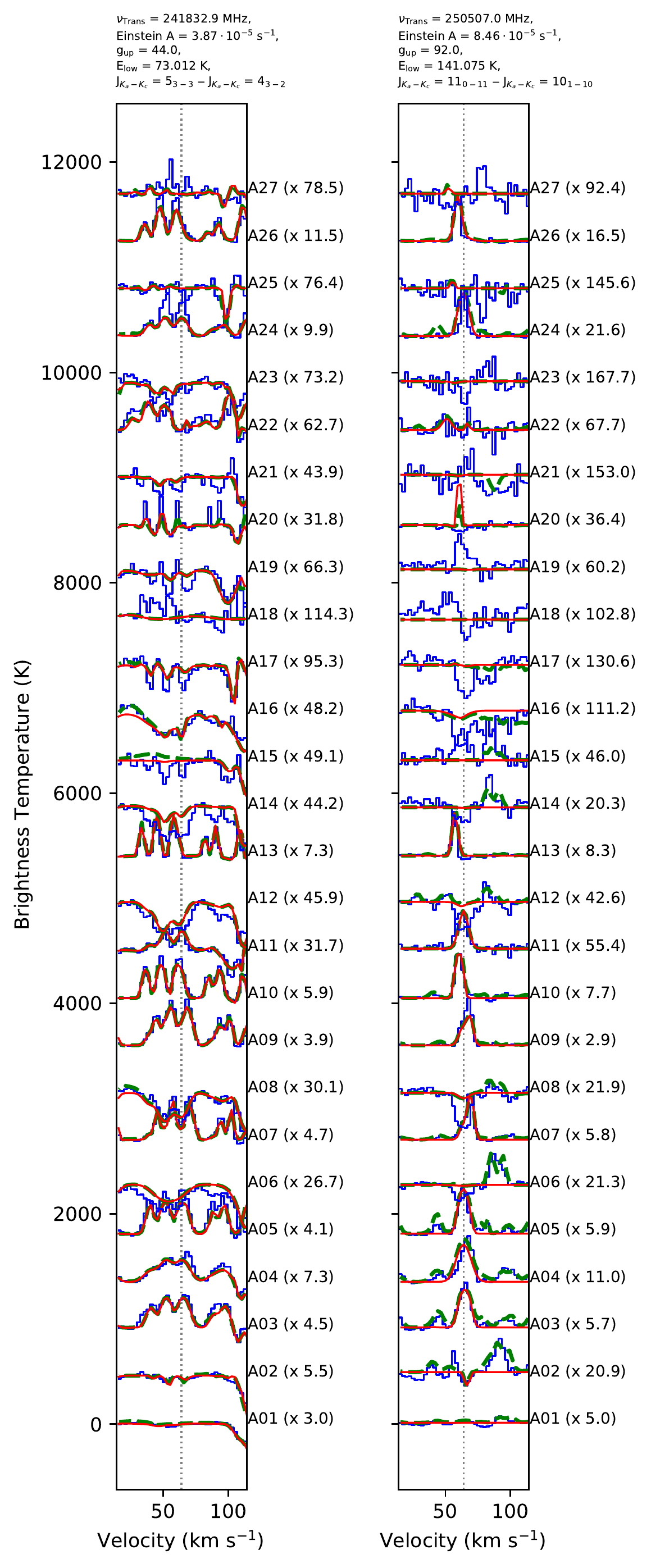}\\
       \caption{Sgr~B2(M)}
       \label{fig:CH3OHM}
    \end{subfigure}
\quad
    \begin{subfigure}[t]{1.0\columnwidth}
       \includegraphics[width=1.0\columnwidth]{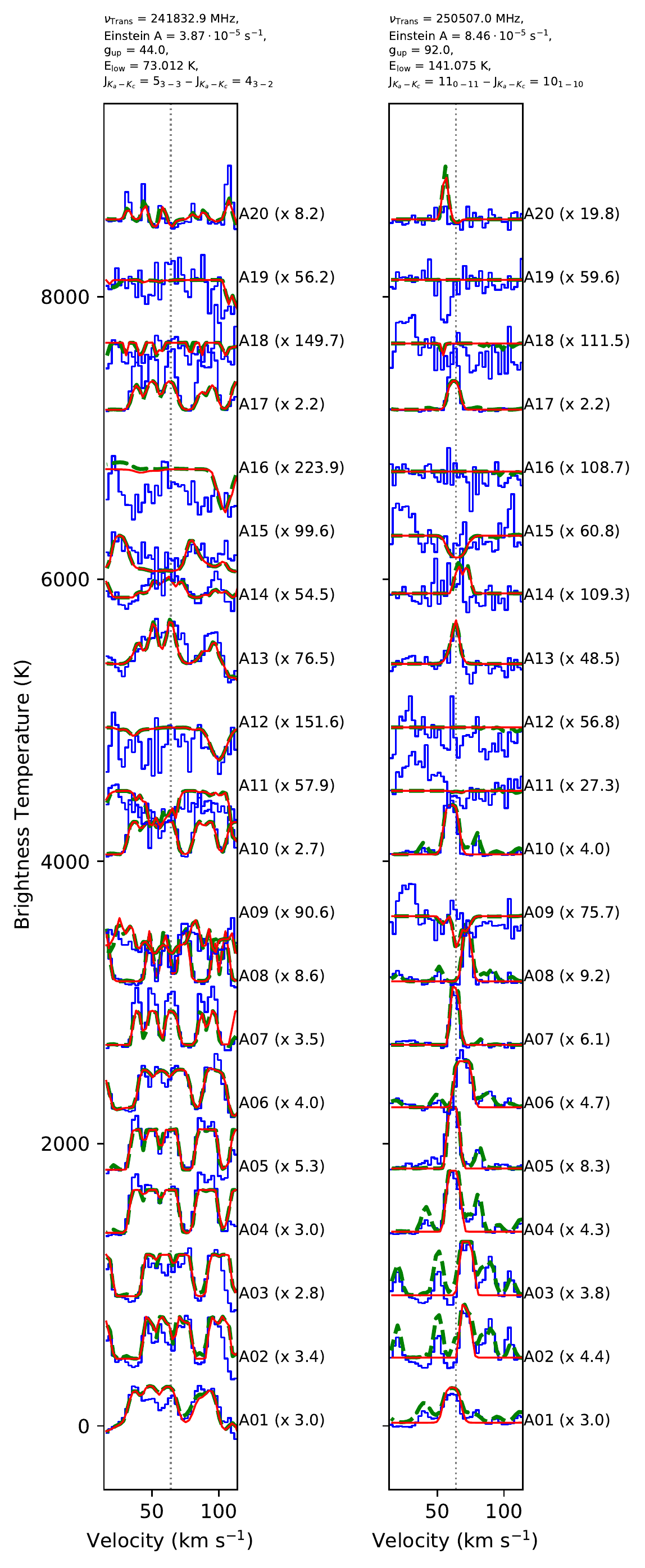}\\
       \caption{Sgr~B2(N)}
       \label{fig:CH3OHN}
   \end{subfigure}
   \caption{Selected transitions of CH$_3$OH in Sgr~B2(M) and N.}
   \ContinuedFloat
   \label{fig:CH3OHMN}
\end{figure*}
\newpage
\clearpage

%*******************************************************************************
% Figure: CH3OH;v12=1;
\begin{figure*}[!htb]
    \centering
    \begin{subfigure}[t]{1.0\columnwidth}
       \includegraphics[width=1.0\columnwidth]{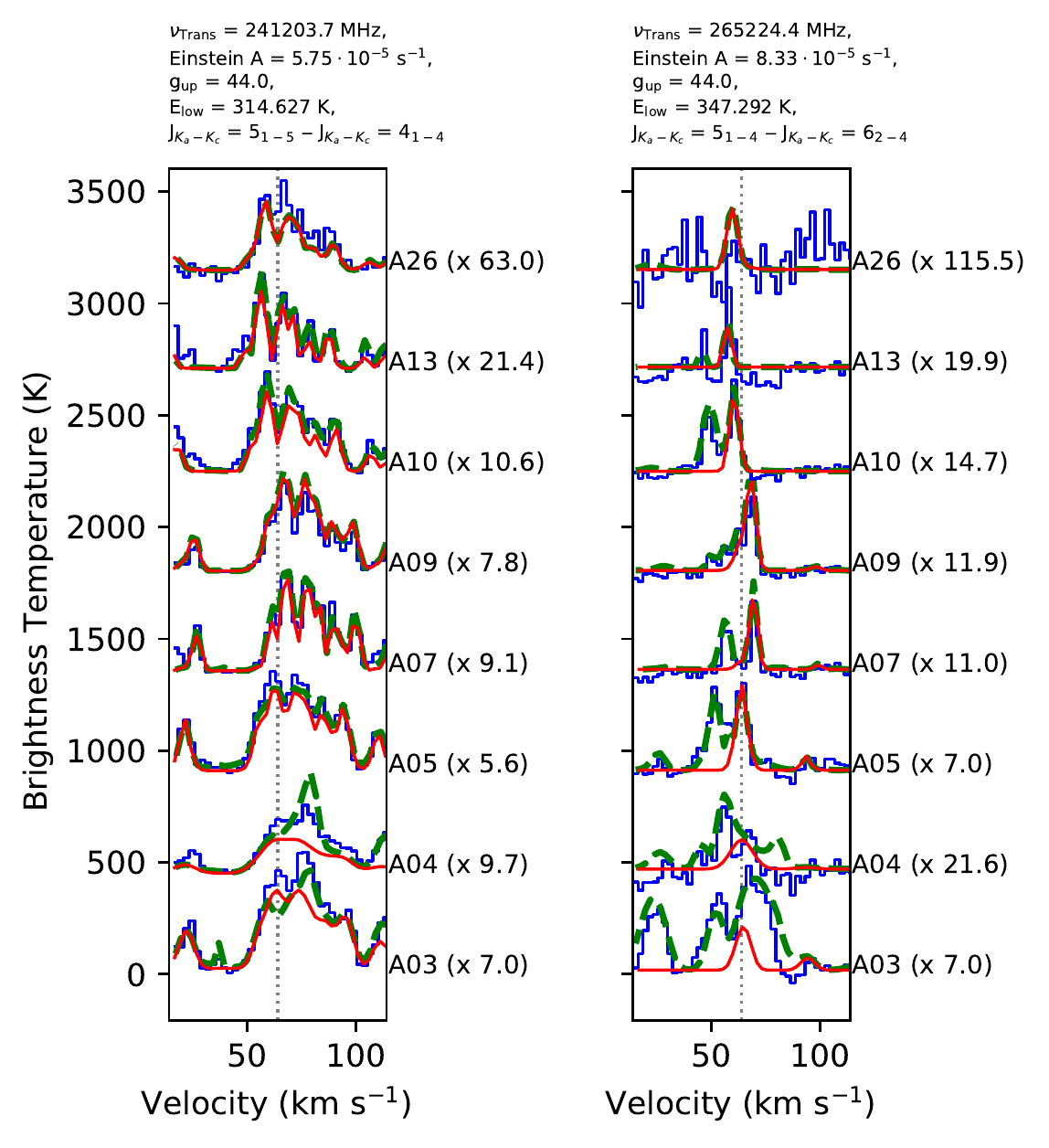}\\
       \caption{Sgr~B2(M)}
       \label{fig:CH3OHv12M}
    \end{subfigure}
\quad
    \begin{subfigure}[t]{1.0\columnwidth}
       \includegraphics[width=1.0\columnwidth]{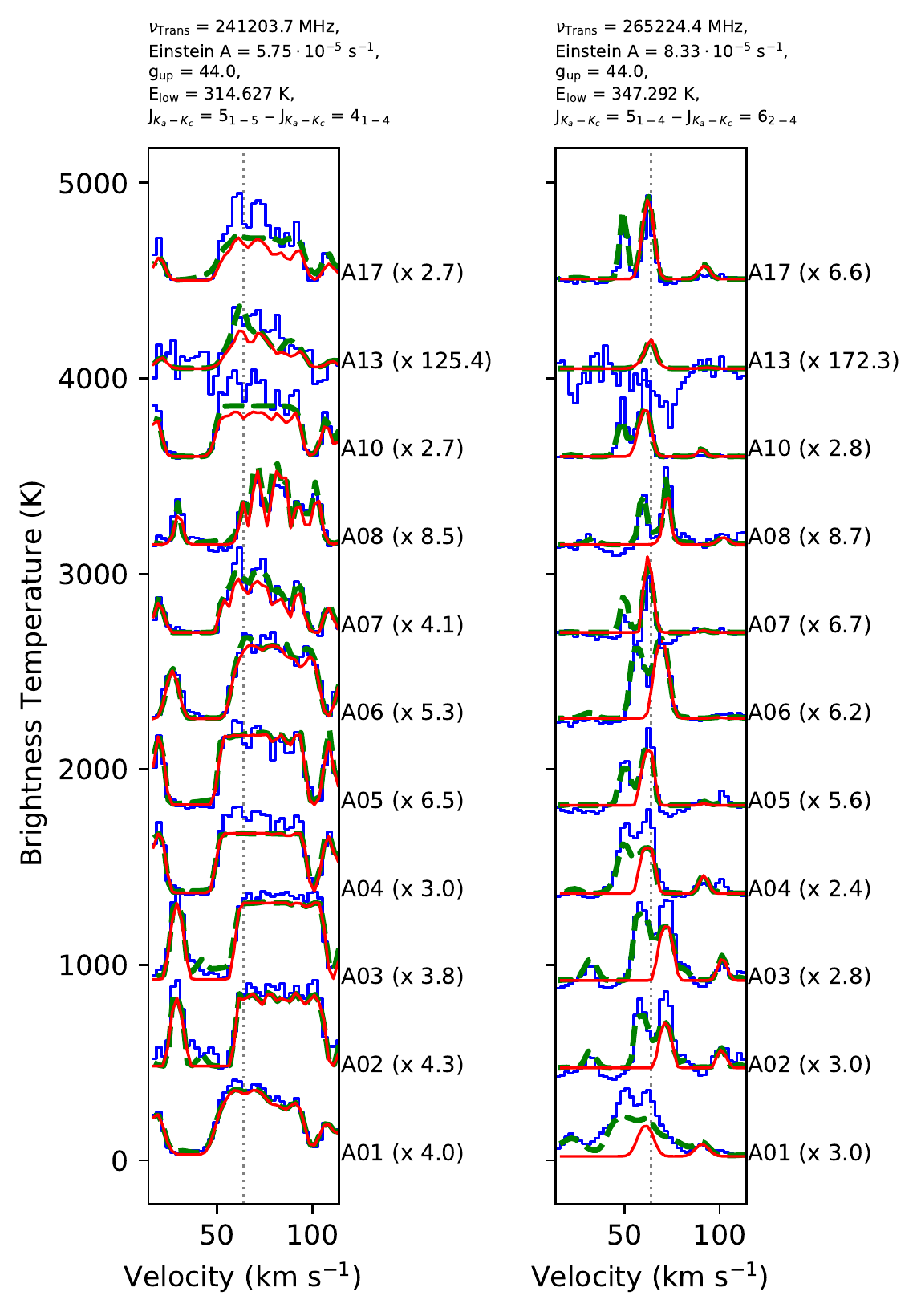}\\
       \caption{Sgr~B2(N)}
       \label{fig:CH3OHv12N}
   \end{subfigure}
   \caption{Selected transitions of CH$_3$OH, v$_{12}$=1 in Sgr~B2(M) and N.}
   \ContinuedFloat
   \label{fig:CH3OHv12MN}
\end{figure*}
\newpage
\clearpage

%*******************************************************************************
% Figure: C-13-H3OH;v=0;
\begin{figure*}[!htb]
    \centering
    \begin{subfigure}[t]{1.0\columnwidth}
       \includegraphics[width=1.0\columnwidth]{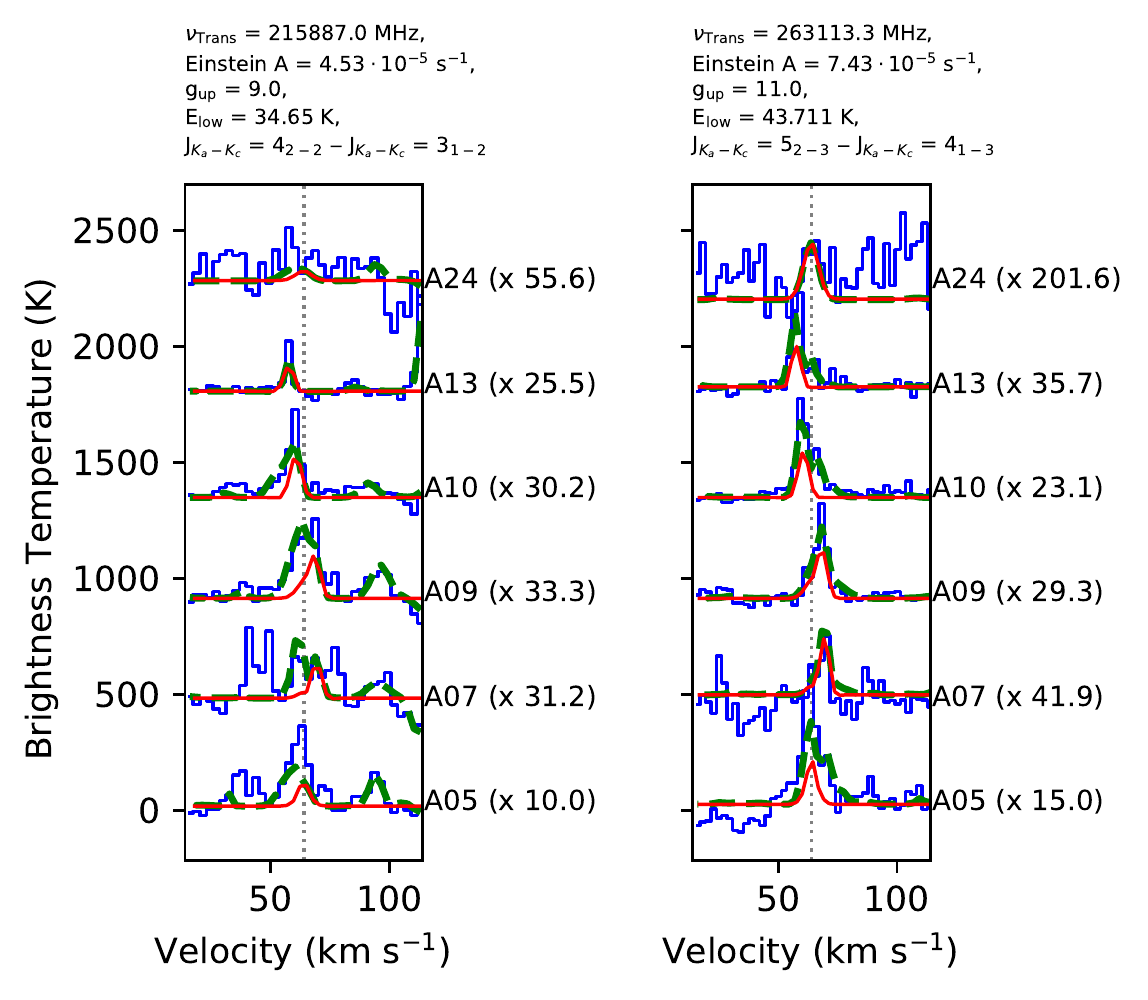}\\
       \caption{Sgr~B2(M)}
       \label{fig:C13H3OHM}
    \end{subfigure}
\quad
    \begin{subfigure}[t]{1.0\columnwidth}
       \includegraphics[width=1.0\columnwidth]{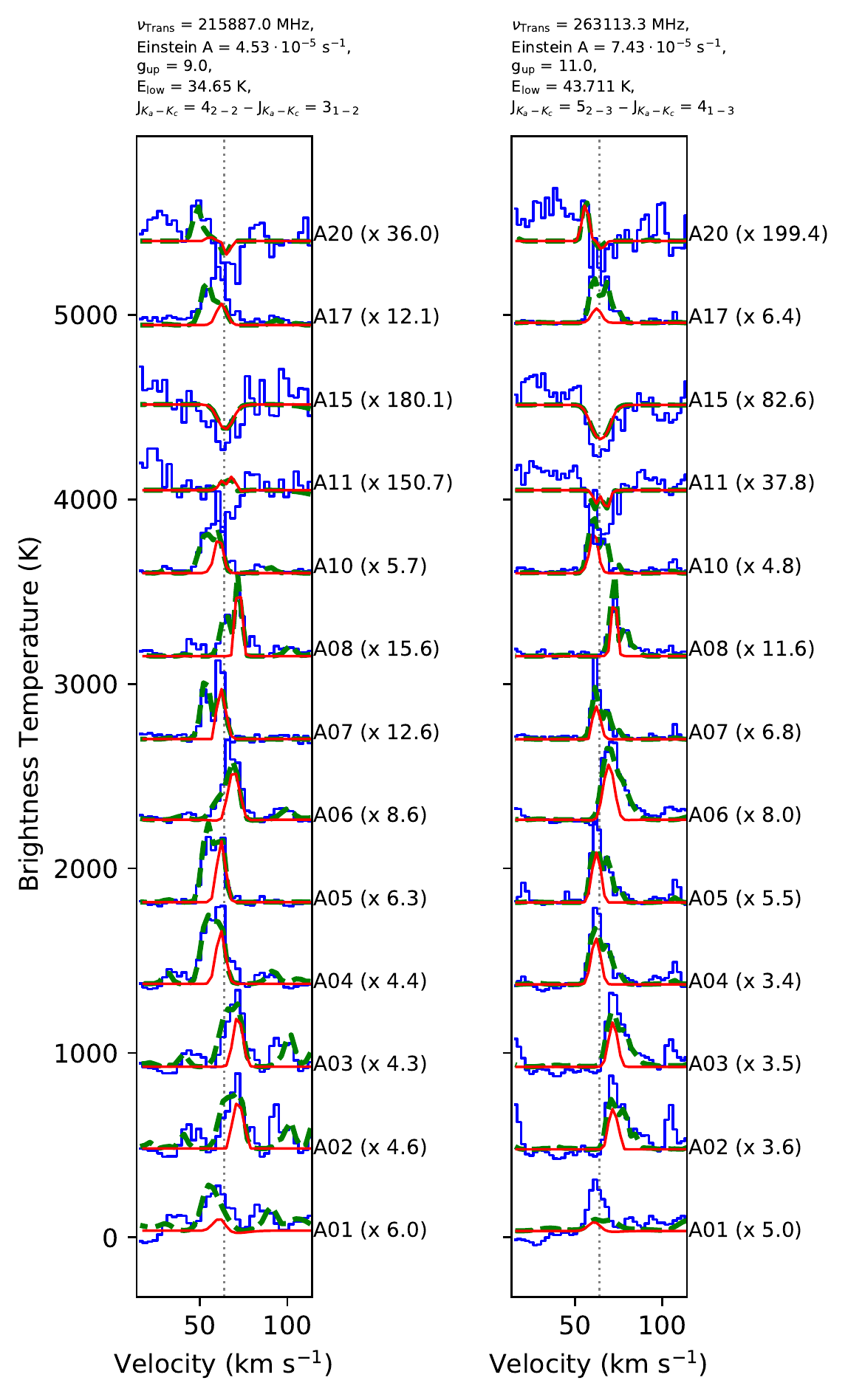}\\
       \caption{Sgr~B2(N)}
       \label{fig:C13H3OHN}
   \end{subfigure}
   \caption{Selected transitions of $^{13}$CH$_3$OH in Sgr~B2(M) and N.}
   \ContinuedFloat
   \label{fig:C13H3OHMN}
\end{figure*}
\newpage
\clearpage

%*******************************************************************************
% Figure: H2CO;v=0;
\begin{figure*}[!htb]
    \centering
    \begin{subfigure}[t]{1.0\columnwidth}
       \includegraphics[width=1.0\columnwidth]{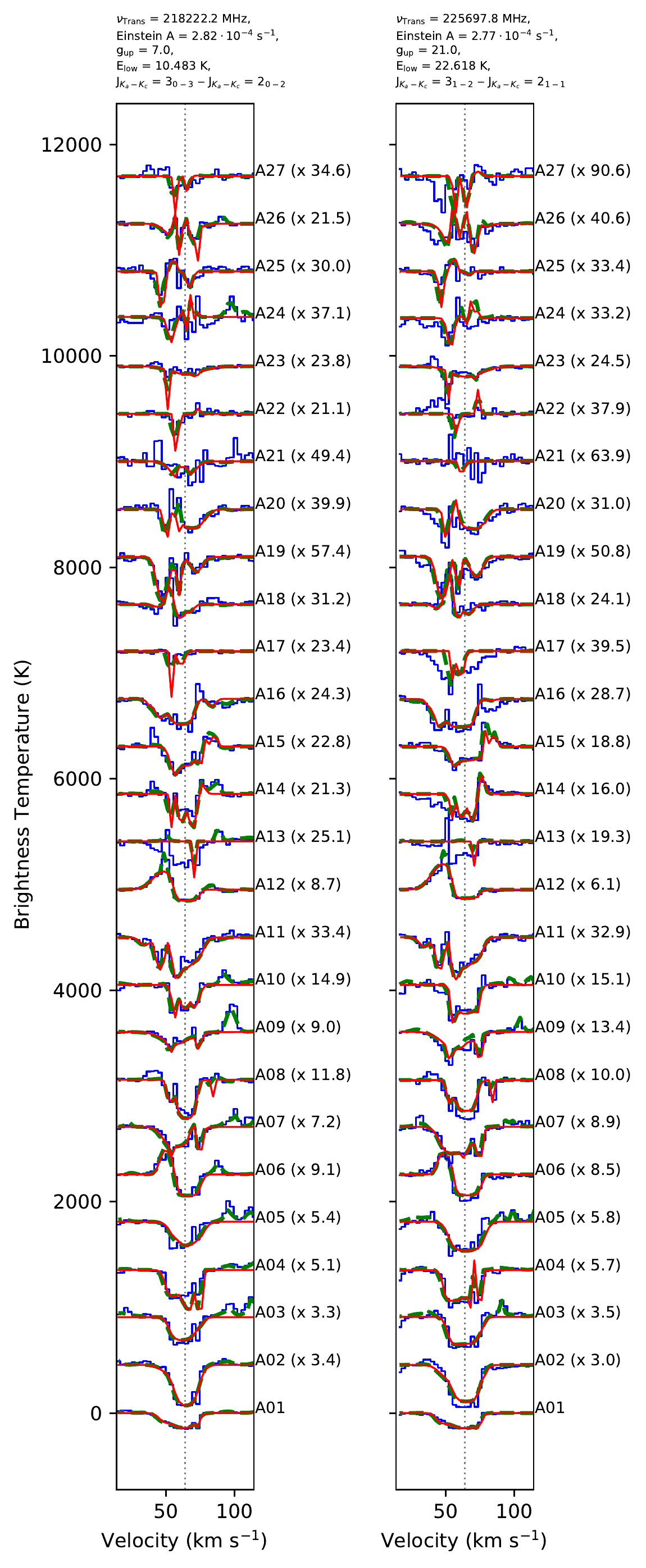}\\
       \caption{Sgr~B2(M)}
       \label{fig:H2COM}
    \end{subfigure}
\quad
    \begin{subfigure}[t]{1.0\columnwidth}
       \includegraphics[width=1.0\columnwidth]{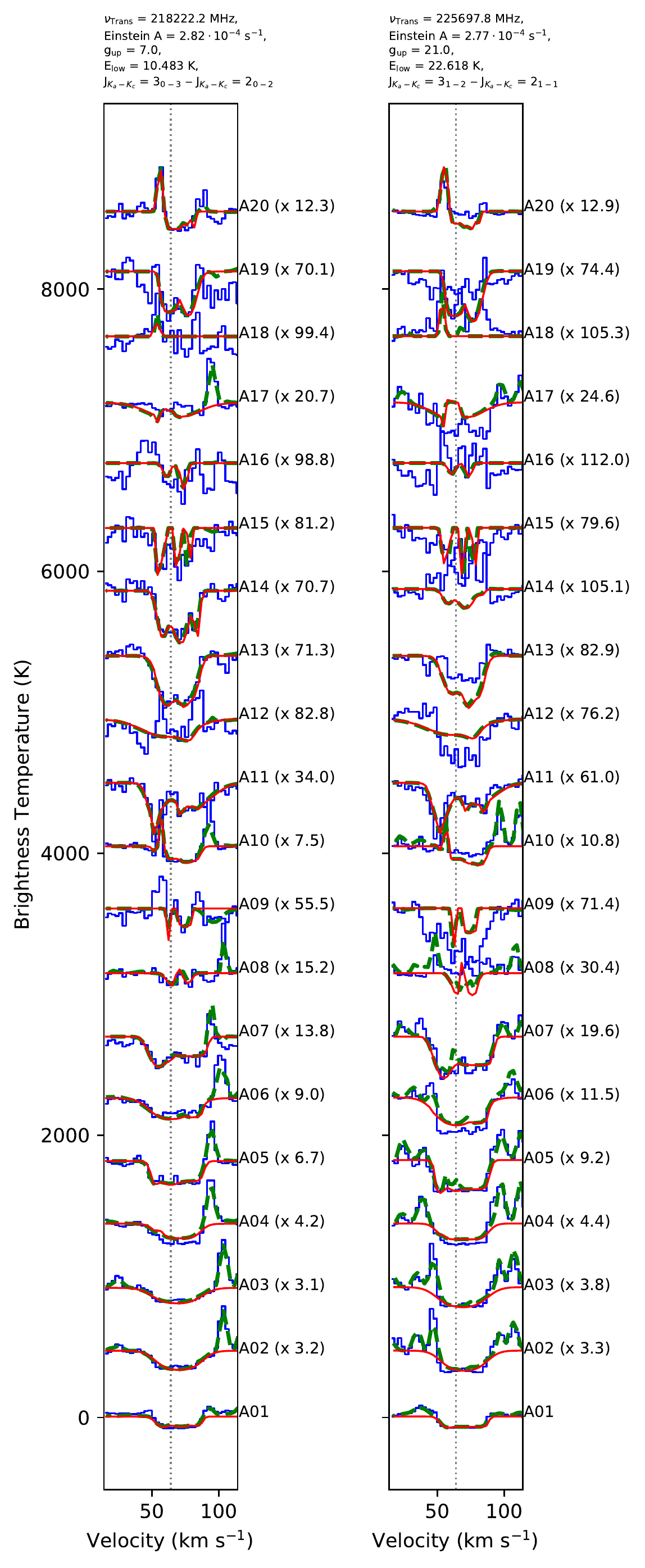}\\
       \caption{Sgr~B2(N)}
       \label{fig:H2CON}
   \end{subfigure}
   \caption{Selected transitions of H$_2$CO in Sgr~B2(M) and N.}
   \ContinuedFloat
   \label{fig:H2COMN}
\end{figure*}
\newpage
\clearpage

%*******************************************************************************
% Figure: H2C   3O;v=0;
\begin{figure*}[!htb]
    \centering
    \begin{subfigure}[t]{1.0\columnwidth}
       \includegraphics[width=1.0\columnwidth]{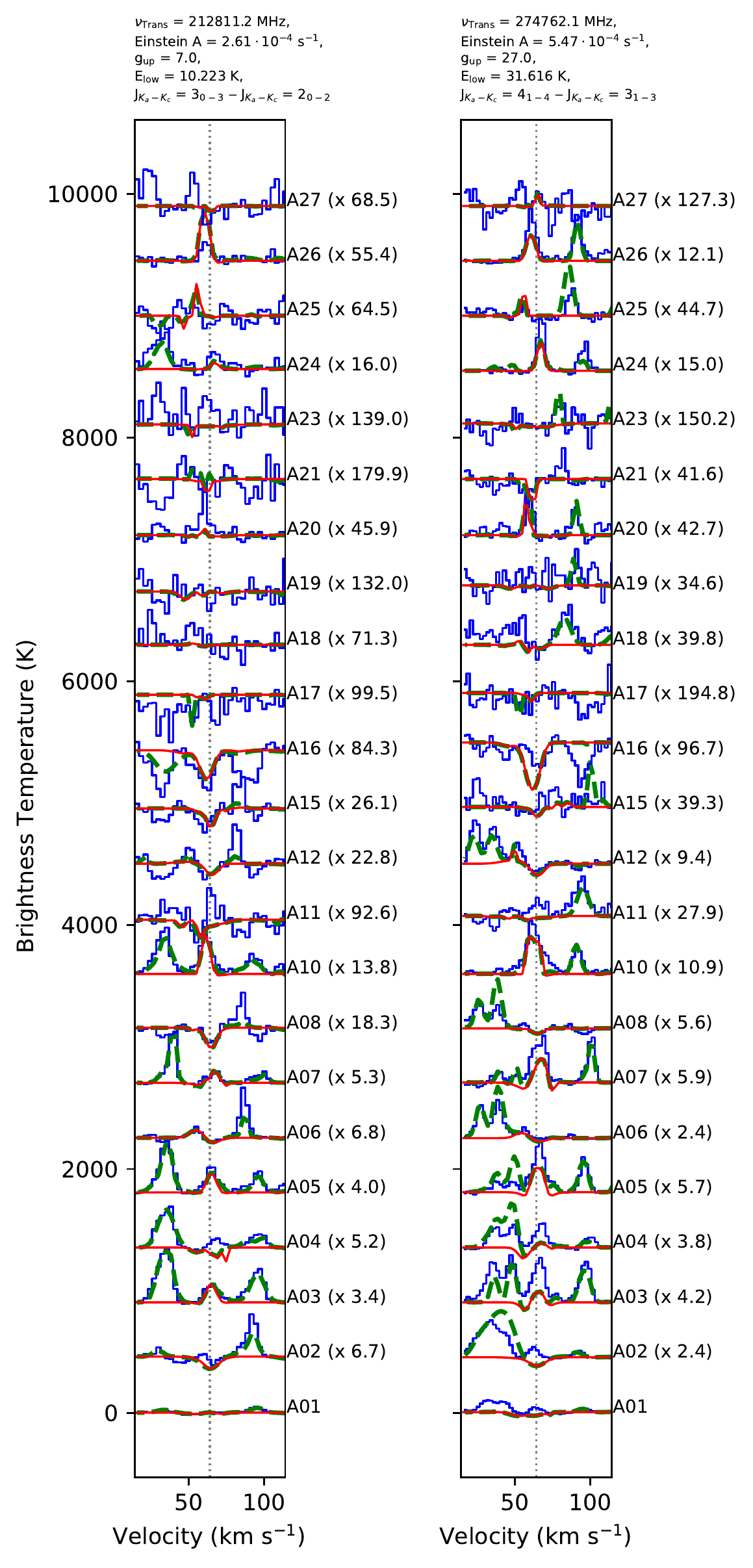}\\
       \caption{Sgr~B2(M)}
       \label{fig:H2C13OM}
    \end{subfigure}
\quad
    \begin{subfigure}[t]{1.0\columnwidth}
       \includegraphics[width=1.0\columnwidth]{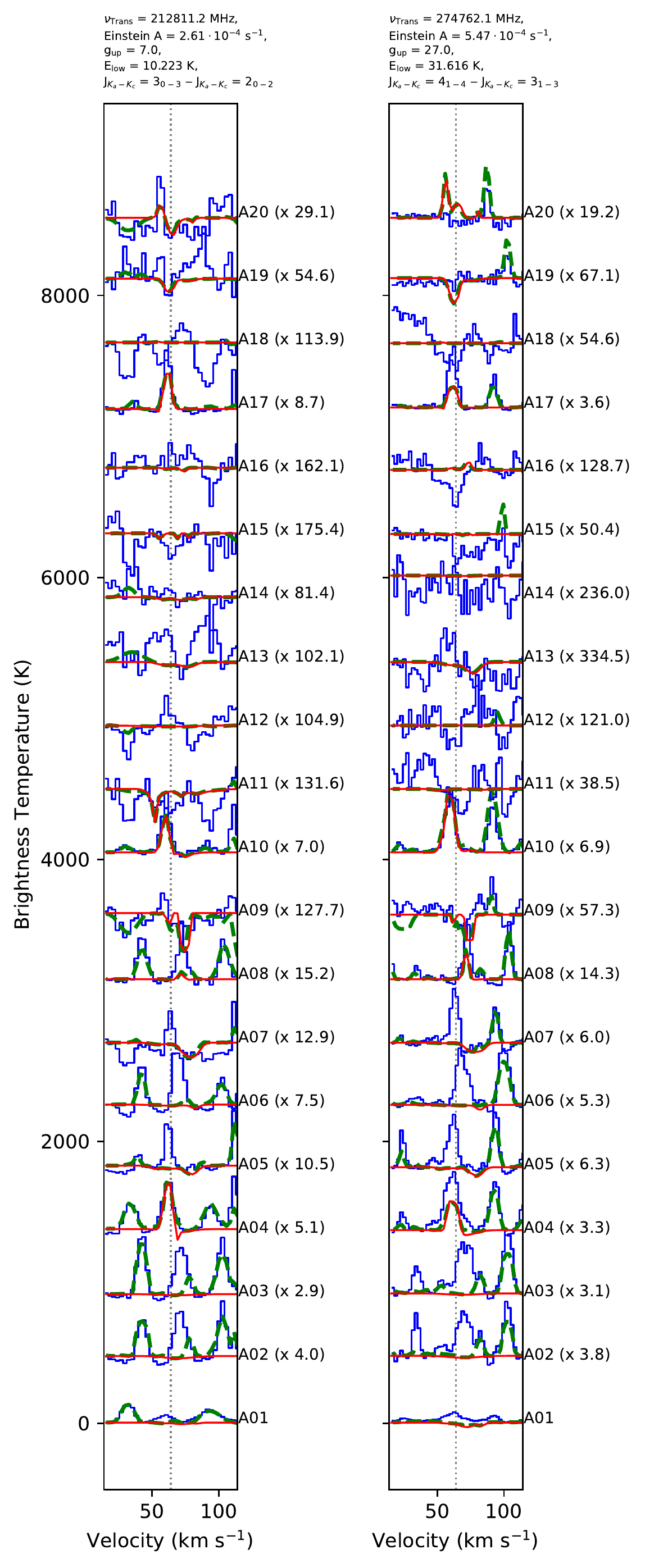}\\
       \caption{Sgr~B2(N)}
       \label{fig:H2C13ON}
   \end{subfigure}
   \caption{Selected transitions of H$_2^{13}$CO in Sgr~B2(M) and N.}
   \ContinuedFloat
   \label{fig:H2C13OMN}
\end{figure*}
\newpage
\clearpage

%*******************************************************************************
% Figure: H2CCO;v=0;
\begin{figure*}[!htb]
    \centering
    \begin{subfigure}[t]{1.0\columnwidth}
       \includegraphics[width=1.0\columnwidth]{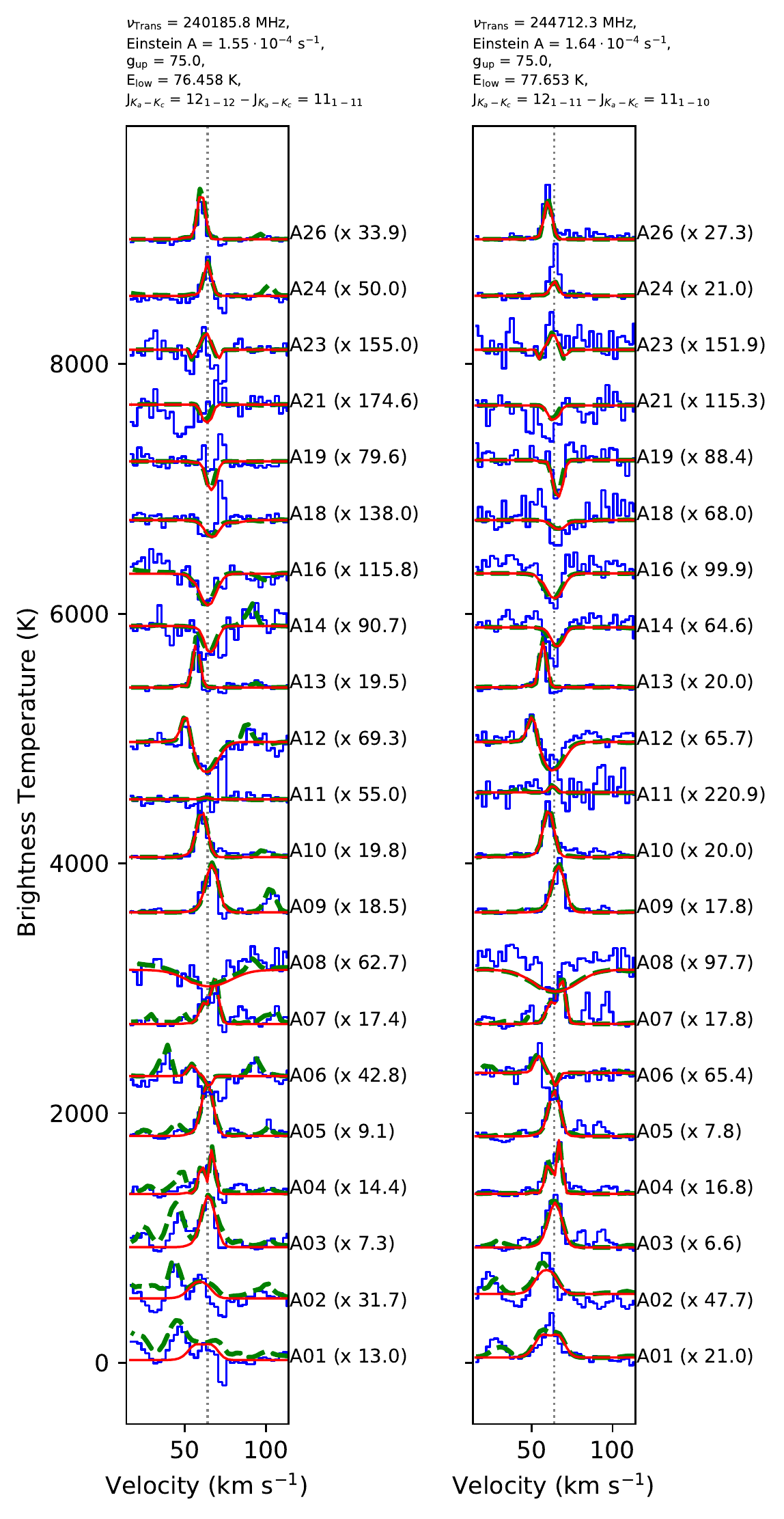}\\
       \caption{Sgr~B2(M)}
       \label{fig:H2CCOM}
    \end{subfigure}
\quad
    \begin{subfigure}[t]{1.0\columnwidth}
       \includegraphics[width=1.0\columnwidth]{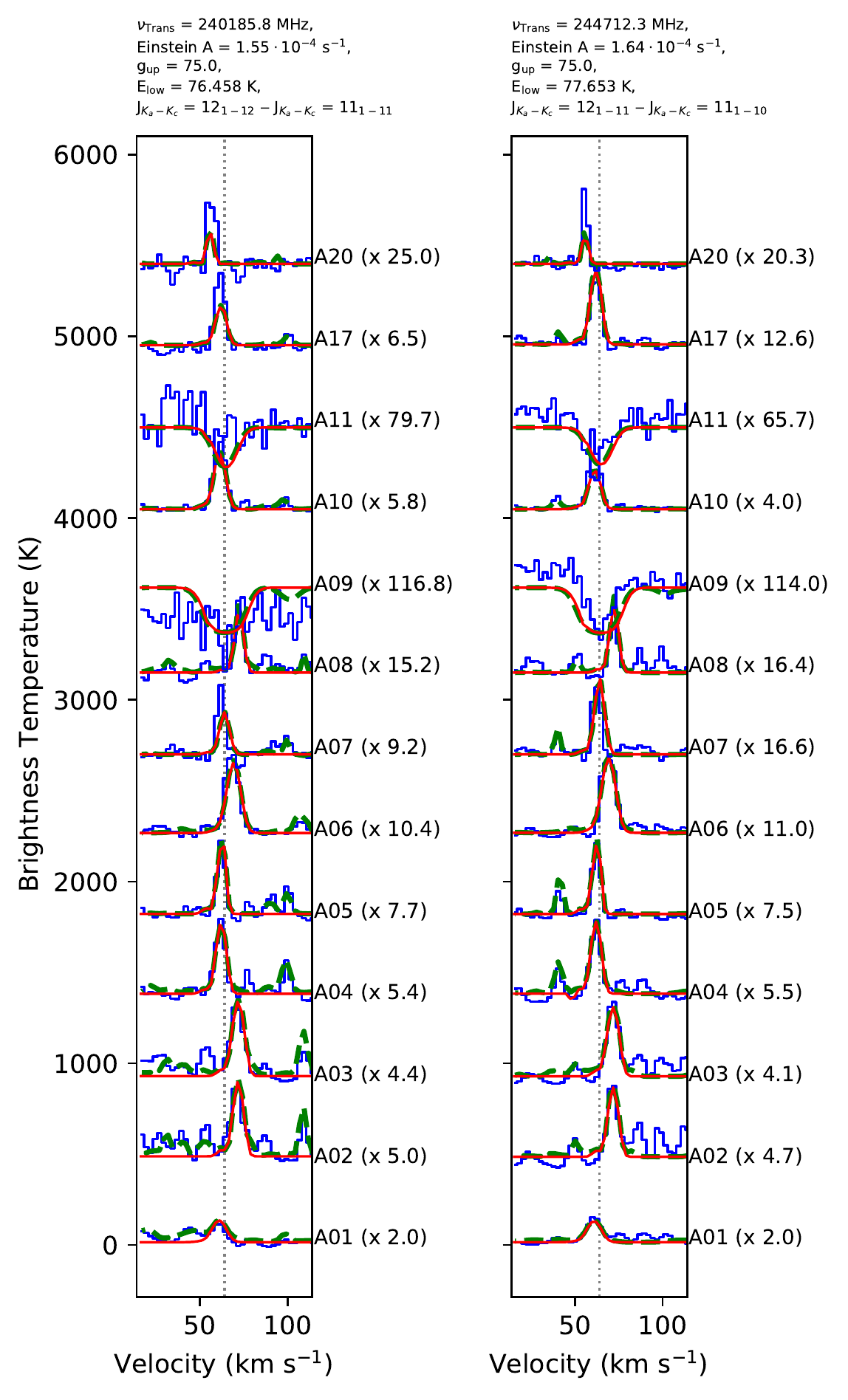}\\
       \caption{Sgr~B2(N)}
       \label{fig:H2CCON}
   \end{subfigure}
   \caption{Selected transitions of H$_2$CCO in Sgr~B2(M) and N.}
   \ContinuedFloat
   \label{fig:H2CCOMN}
\end{figure*}
\newpage
\clearpage

%*******************************************************************************
% Figure: HC-13-O+;v=0;
\begin{figure*}[!htb]
    \centering
    \begin{subfigure}[t]{0.41\columnwidth}
       \includegraphics[width=1.0\columnwidth]{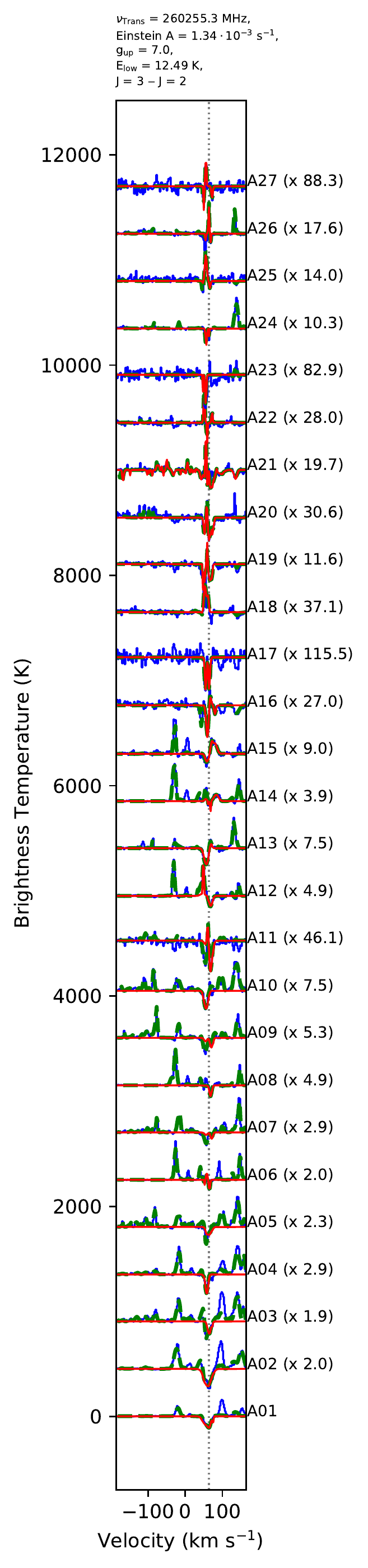}\\
       \caption{H$^{13}$CO$^+$~in~Sgr~B2(M)}
       \label{fig:HCO+M}
    \end{subfigure}
\quad
    \begin{subfigure}[t]{0.41\columnwidth}
       \includegraphics[width=1.0\columnwidth]{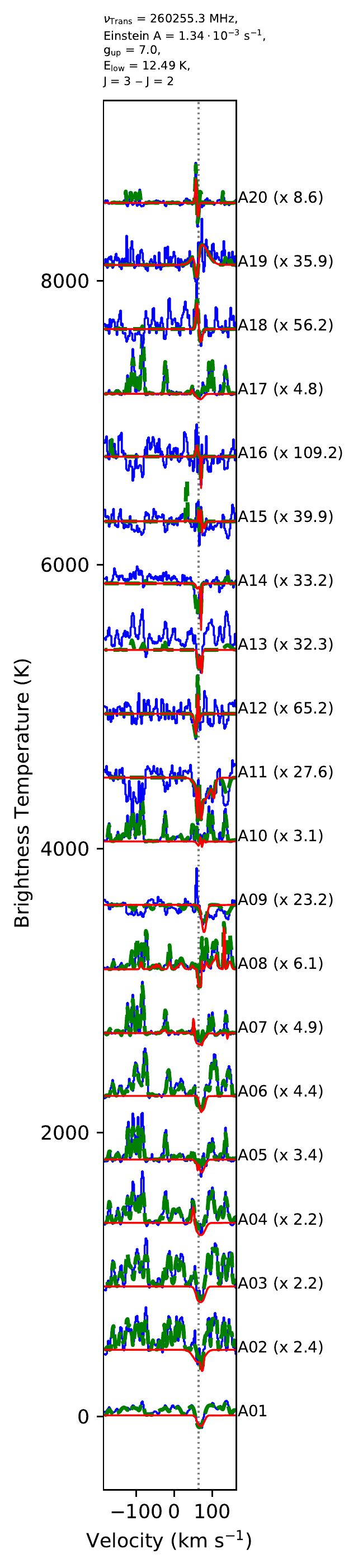}\\
       \caption{H$^{13}$CO$^+$~in~Sgr~B2(N)}
       \label{fig:HCO+N}
   \end{subfigure}
\quad
    \begin{subfigure}[t]{0.41\columnwidth}
       \includegraphics[width=1.0\columnwidth]{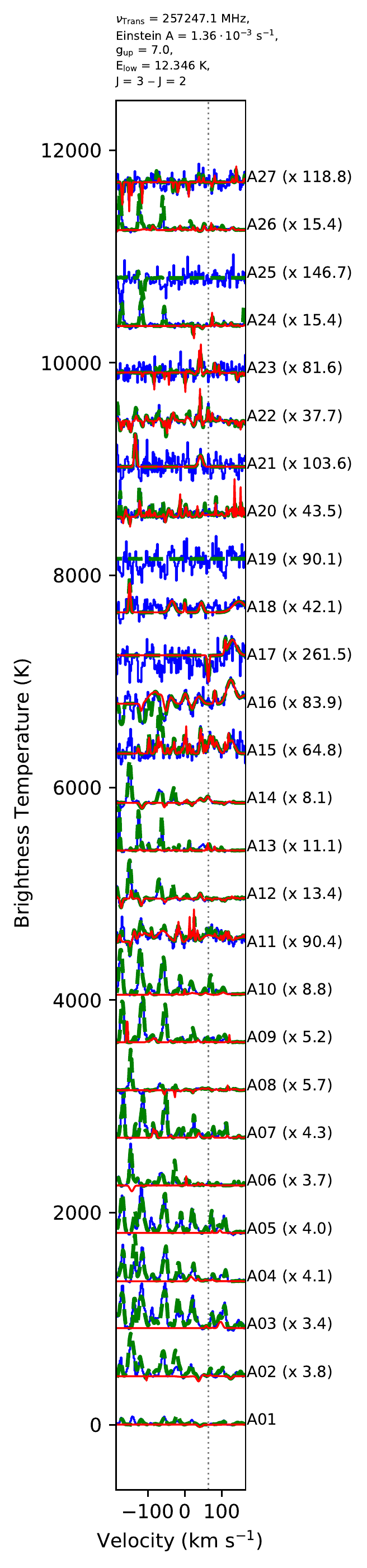}\\
       \caption{HO$^{13}$C$^+$~in~Sgr~B2(M)}
       \label{fig:HOC+M}
   \end{subfigure}
\quad
    \begin{subfigure}[t]{0.41\columnwidth}
       \includegraphics[width=1.0\columnwidth]{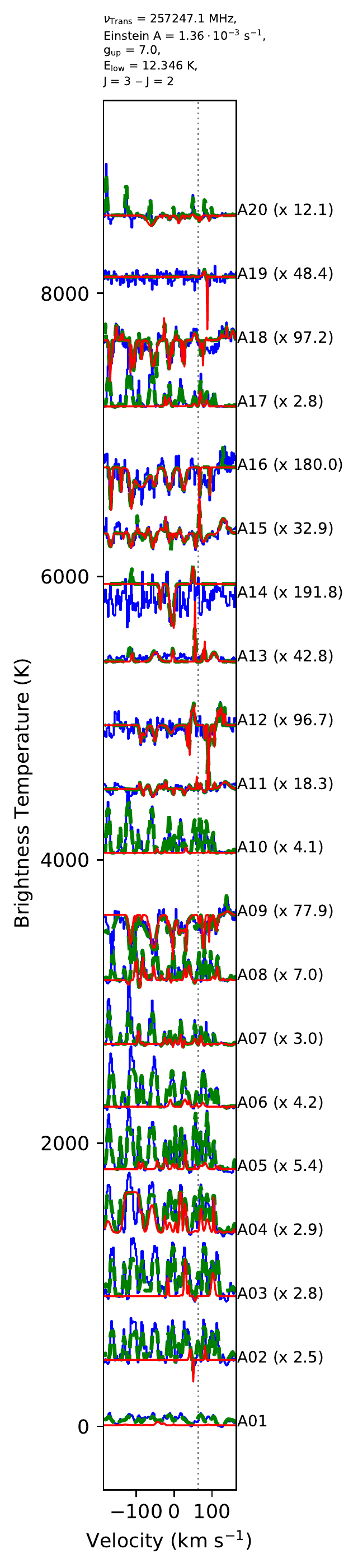}\\
       \caption{HO$^{13}$C$^+$~in~Sgr~B2(N)}
       \label{fig:HOC+N}
   \end{subfigure}
   \caption{Transitions of H$^{13}$CO$^+$ and HO$^{13}$C$^+$ in Sgr~B2(M) and N.}
   \ContinuedFloat
   \label{fig:HCO+MN}
\end{figure*}
\newpage
\clearpage

%*******************************************************************************
% Figure: SiO;v=0;#1
\begin{figure*}[!htb]
    \centering
    \begin{subfigure}[t]{1.0\columnwidth}
       \includegraphics[width=1.0\columnwidth]{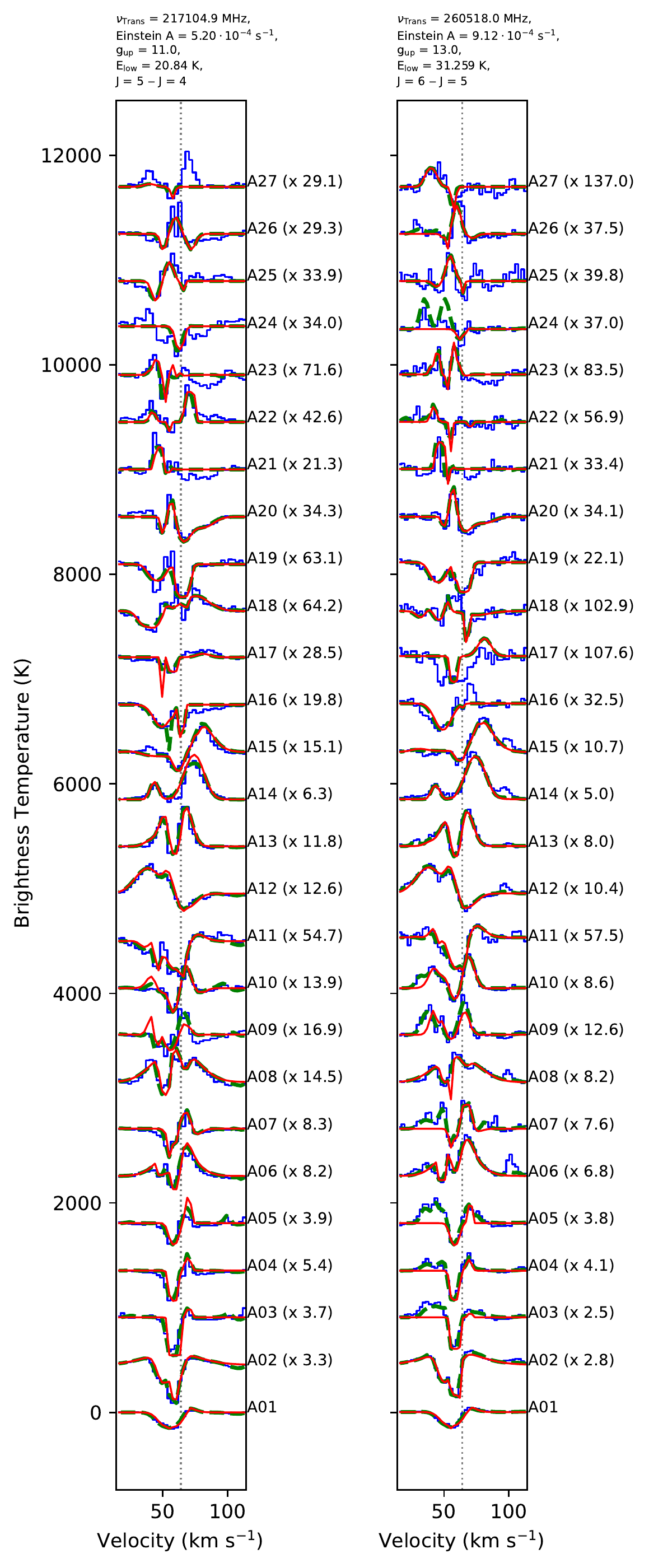}\\
       \caption{Sgr~B2(M)}
       \label{fig:SiOM}
    \end{subfigure}
\quad
    \begin{subfigure}[t]{1.0\columnwidth}
       \includegraphics[width=1.0\columnwidth]{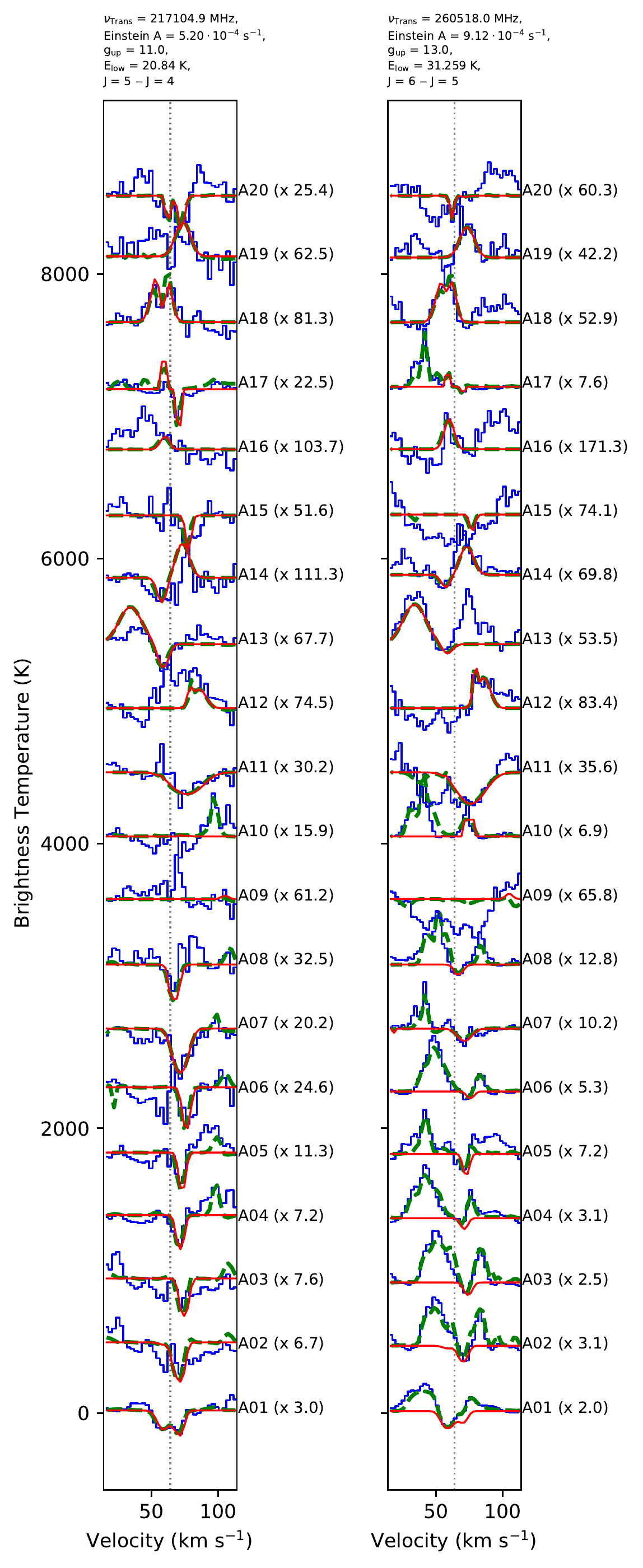}\\
       \caption{Sgr~B2(N)}
       \label{fig:SiON}
   \end{subfigure}
   \caption{Selected transitions of SiO in Sgr~B2(M) and N.}
   \ContinuedFloat
   \label{fig:SiOMN}
\end{figure*}
\newpage
\clearpage

%-------------------------------------------------------------------------------
% complex O-bearing molecules

%*******************************************************************************
% Figure: CH3OCH3;v=0;
\begin{figure*}[!htb]
    \centering
    \begin{subfigure}[t]{1.0\columnwidth}
       \includegraphics[width=1.0\columnwidth]{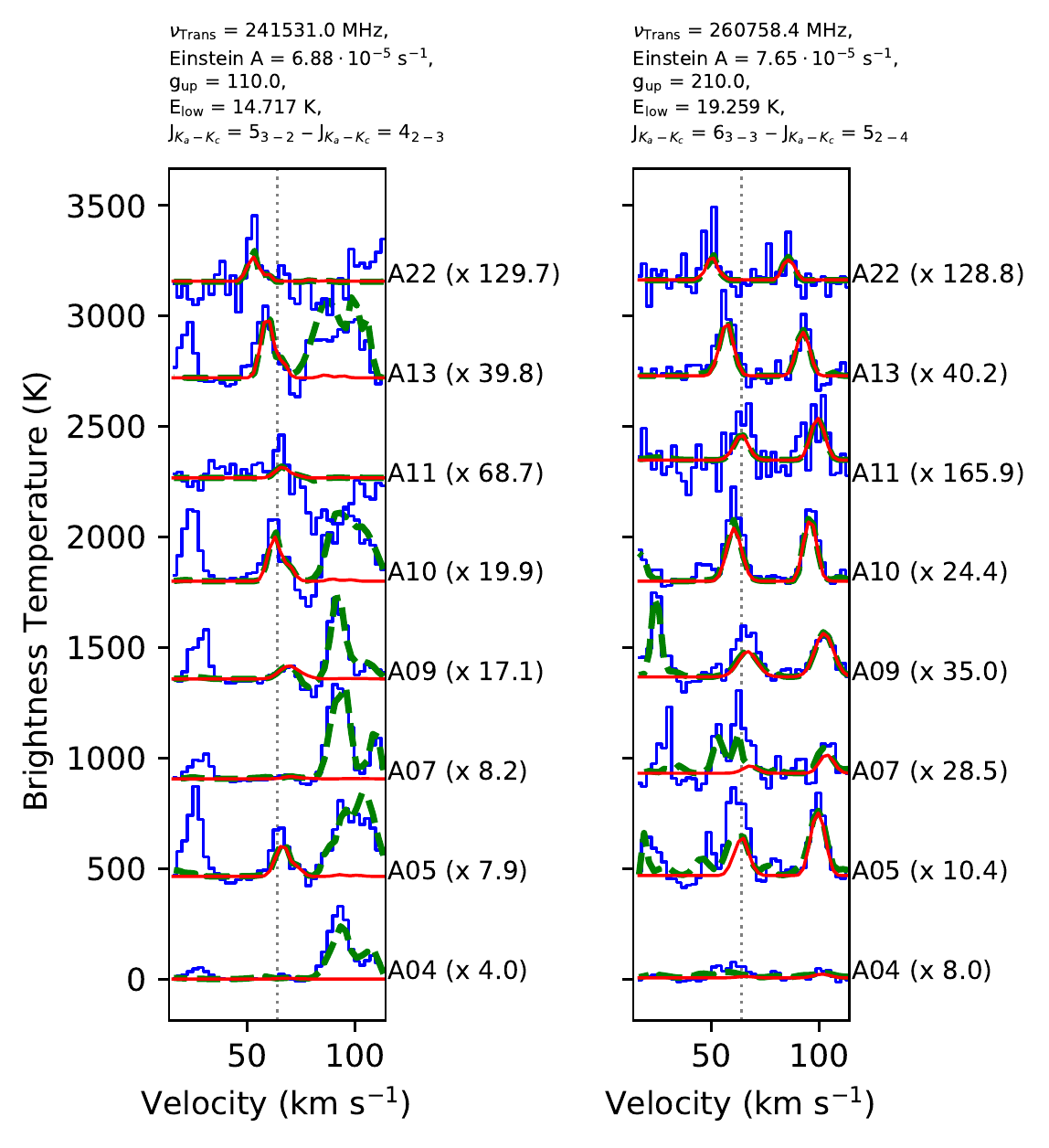}\\
       \caption{Sgr~B2(M)}
       \label{fig:CH3OCH3M}
    \end{subfigure}
\quad
    \begin{subfigure}[t]{1.0\columnwidth}
       \includegraphics[width=1.0\columnwidth]{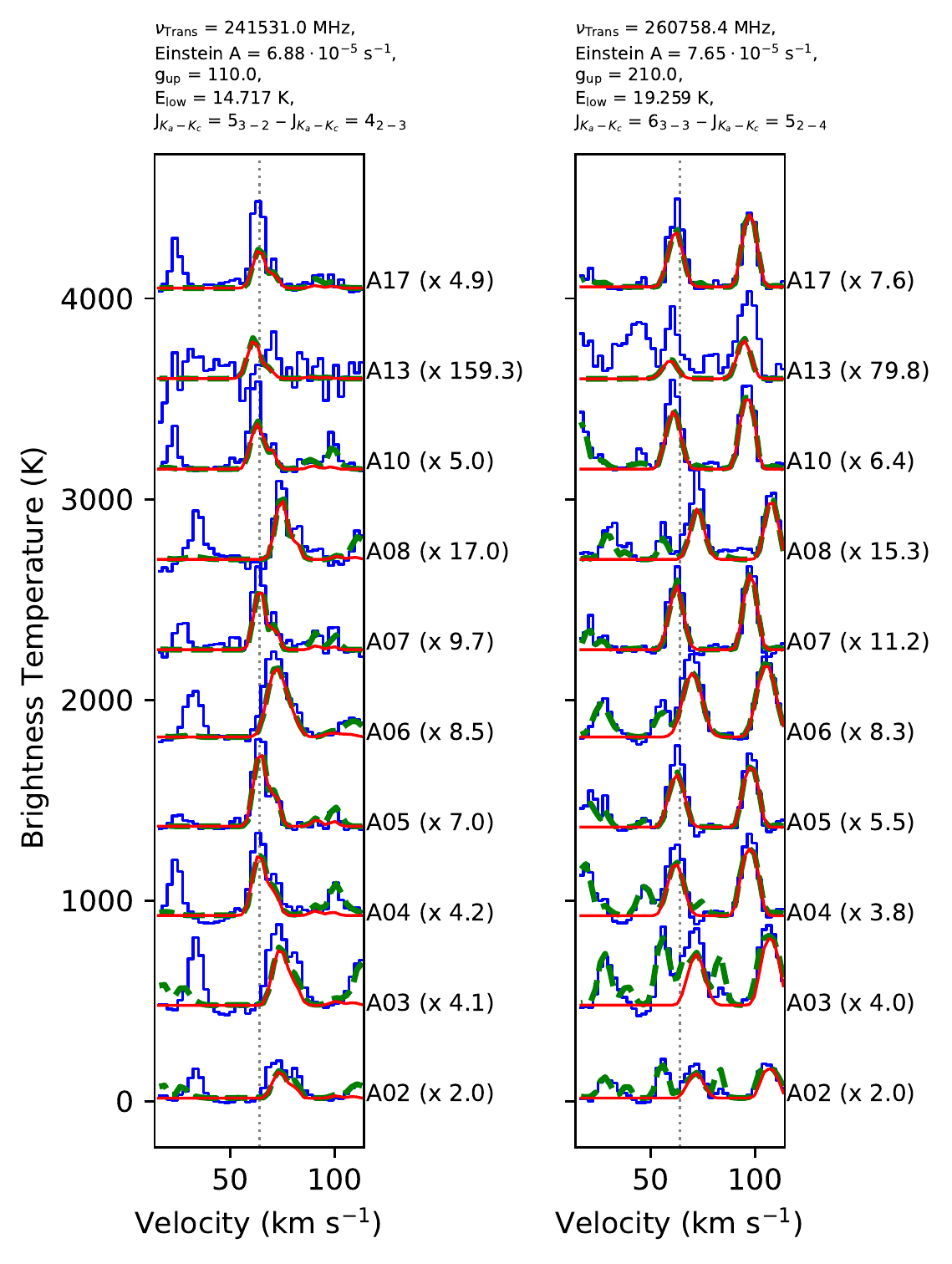}\\
       \caption{Sgr~B2(N)}
       \label{fig:CH3OCH3N}
   \end{subfigure}
   \caption{Selected transitions of CH$_3$OCH$_3$ in Sgr~B2(M) and N.}
   \ContinuedFloat
   \label{fig:CH3OCH3MN}
\end{figure*}
\newpage
\clearpage

%*******************************************************************************
% Figure: C2H5OH;v=0;#1
\begin{figure*}[!htb]
    \centering
    \begin{subfigure}[t]{1.0\columnwidth}
       \includegraphics[width=1.0\columnwidth]{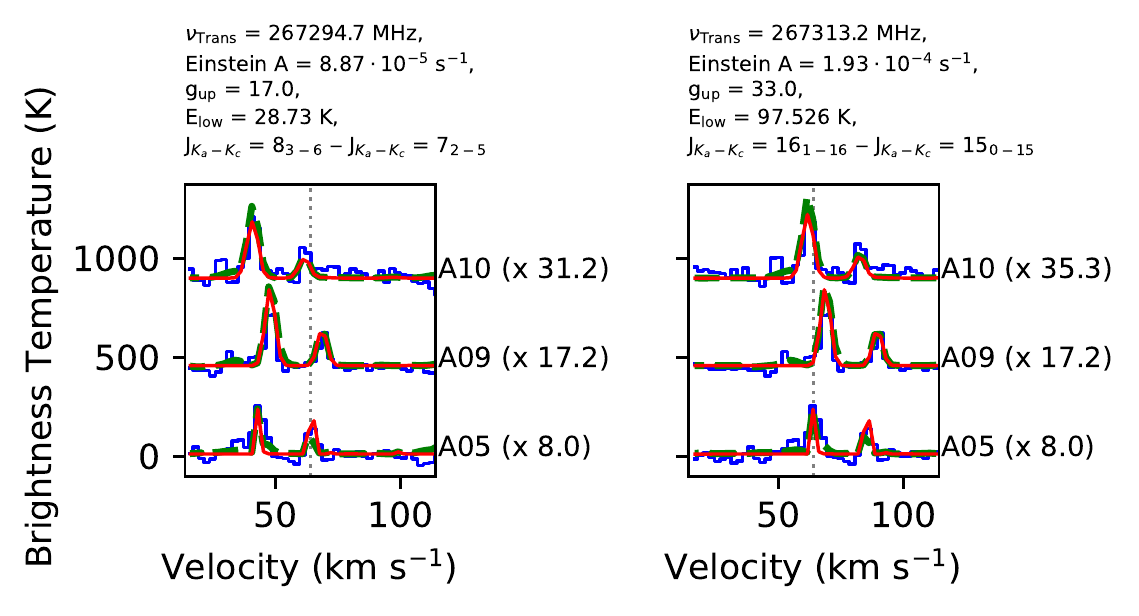}\\
       \caption{Sgr~B2(M)}
       \label{fig:C2H5OHM}
    \end{subfigure}
\quad
    \begin{subfigure}[t]{1.0\columnwidth}
       \includegraphics[width=1.0\columnwidth]{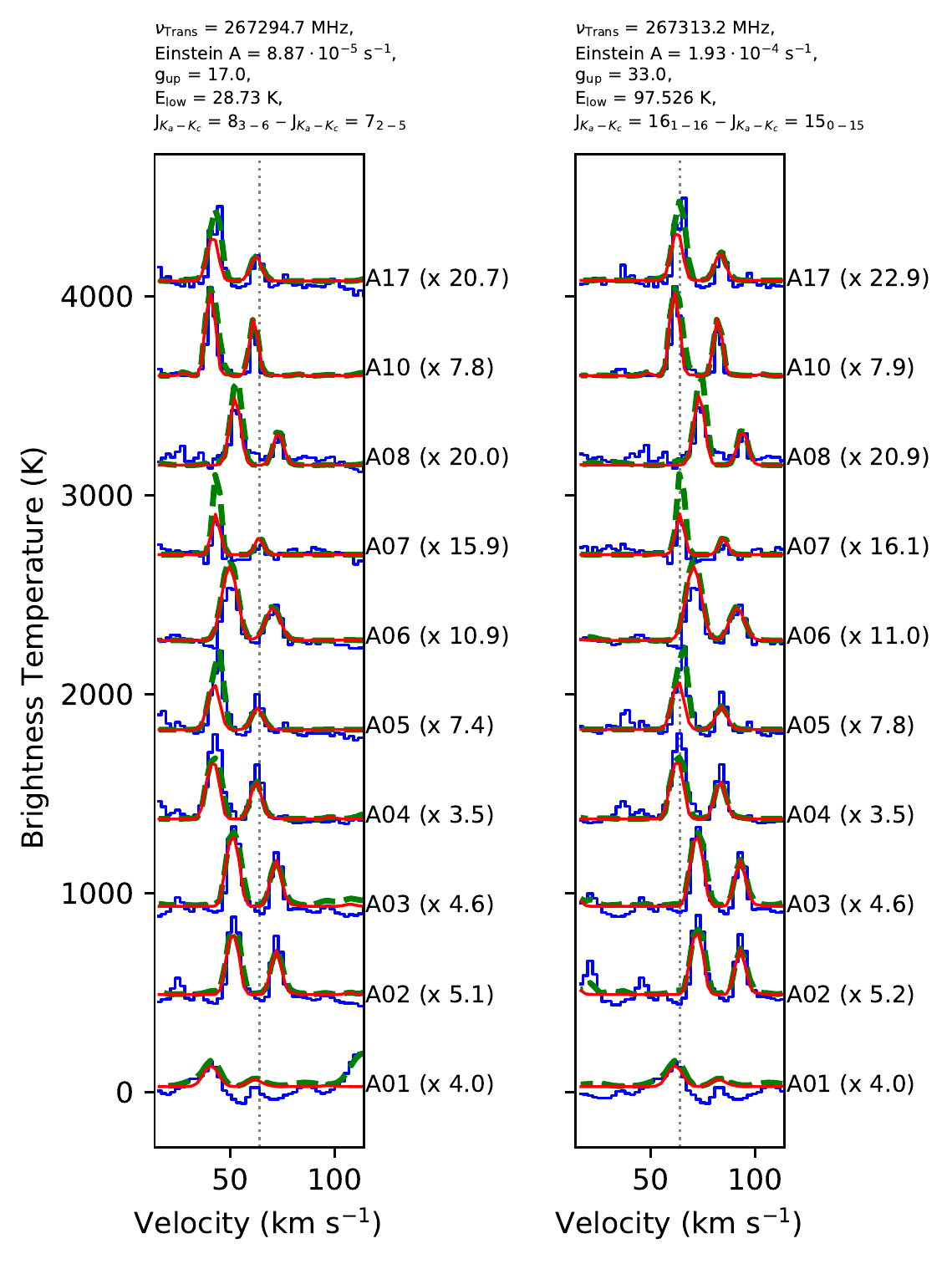}\\
       \caption{Sgr~B2(N)}
       \label{fig:C2H5OHN}
   \end{subfigure}
   \caption{Selected transitions of C$_2$H$_5$OH in Sgr~B2(M) and N.}
   \ContinuedFloat
   \label{fig:C2H5OHMN}
\end{figure*}
\newpage
\clearpage

%*******************************************************************************
% Figure: CH3OCHO;v=0;
\begin{figure*}[!htb]
    \centering
    \begin{subfigure}[t]{1.0\columnwidth}
       \includegraphics[width=1.0\columnwidth]{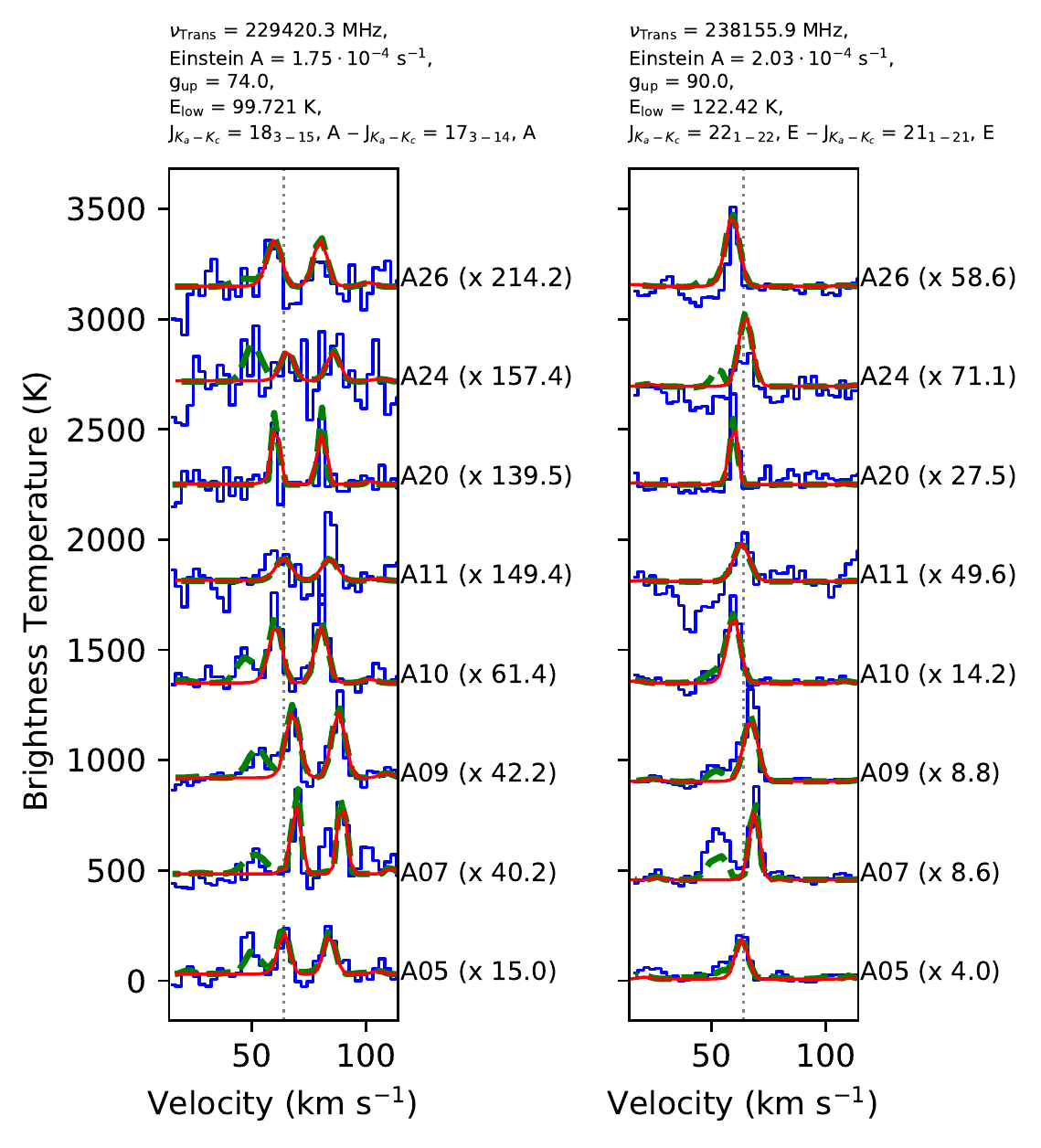}\\
       \caption{Sgr~B2(M)}
       \label{fig:CH3OCHOM}
    \end{subfigure}
\quad
    \begin{subfigure}[t]{1.0\columnwidth}
       \includegraphics[width=1.0\columnwidth]{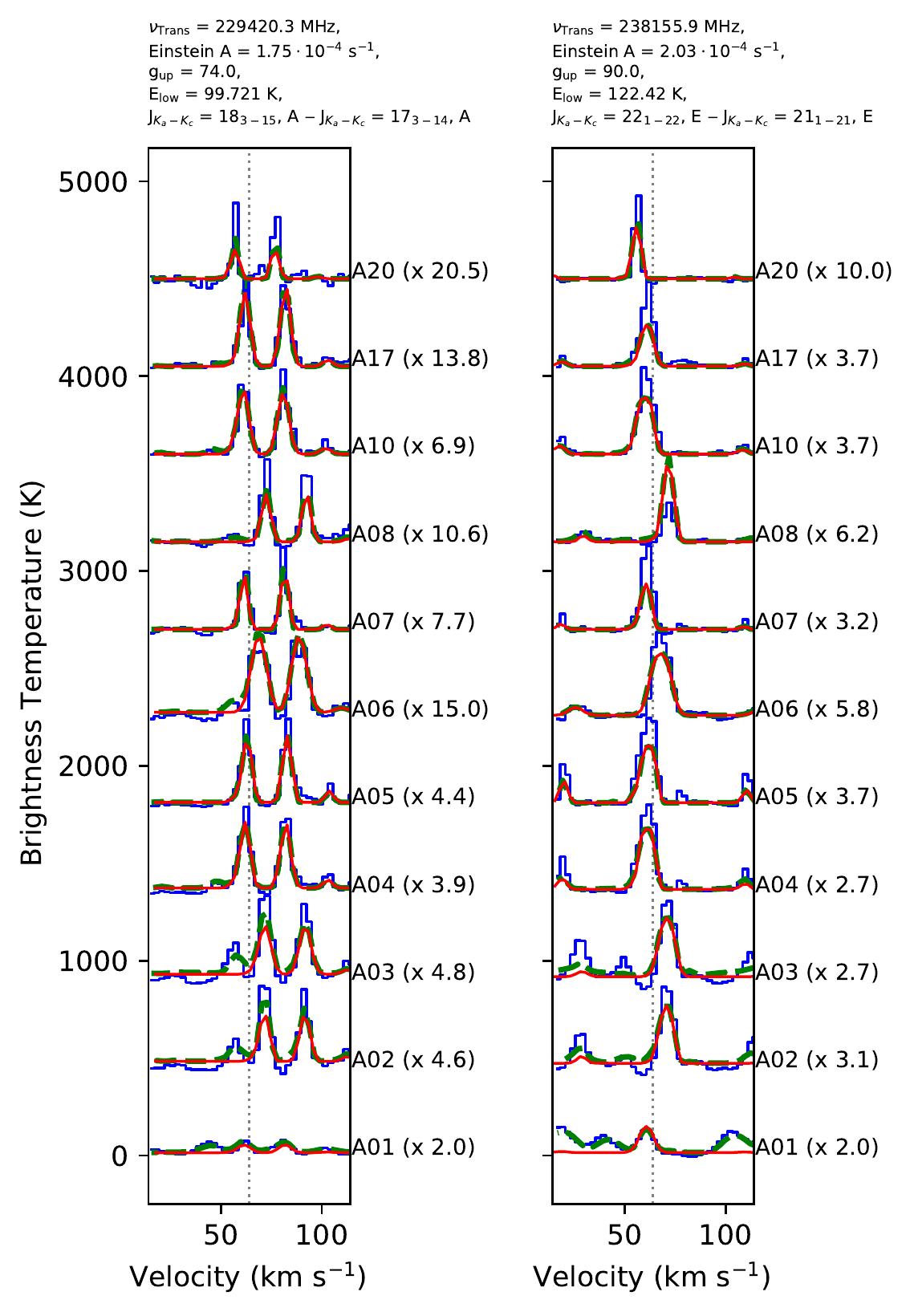}\\
       \caption{Sgr~B2(N)}
       \label{fig:CH3OCHON}
   \end{subfigure}
   \caption{Selected transitions of CH$_3$OCHO in Sgr~B2(M) and N.}
   \ContinuedFloat
   \label{fig:CH3OCHOMN}
\end{figure*}
\newpage
\clearpage

%*******************************************************************************
% Figure: CH3OCHO;v18=1;
\begin{figure*}[!htb]
    \centering
    \begin{subfigure}[t]{1.0\columnwidth}
       \includegraphics[width=1.0\columnwidth]{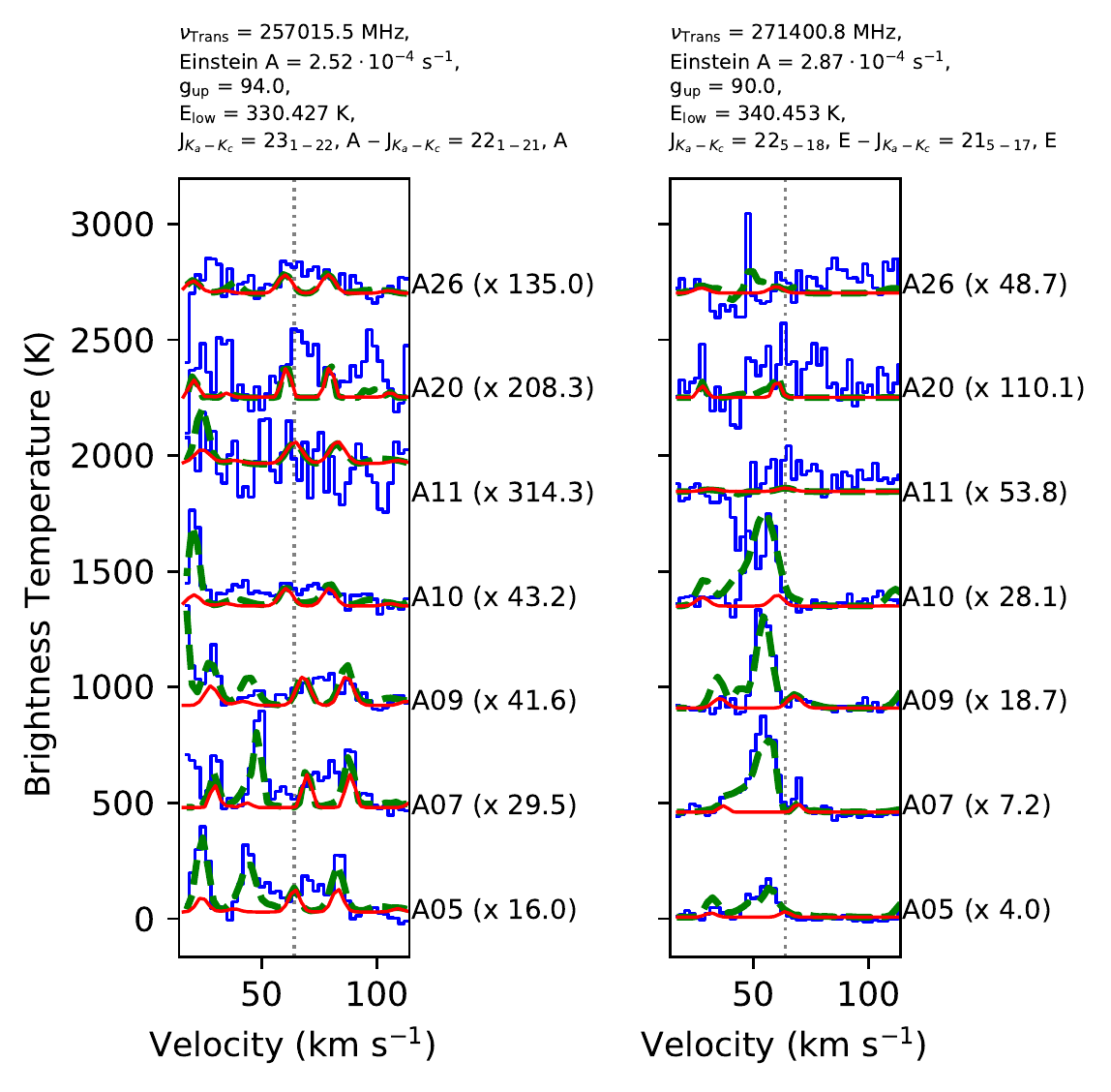}\\
       \caption{Sgr~B2(M)}
       \label{fig:CH3OCHOv18M}
    \end{subfigure}
\quad
    \begin{subfigure}[t]{1.0\columnwidth}
       \includegraphics[width=1.0\columnwidth]{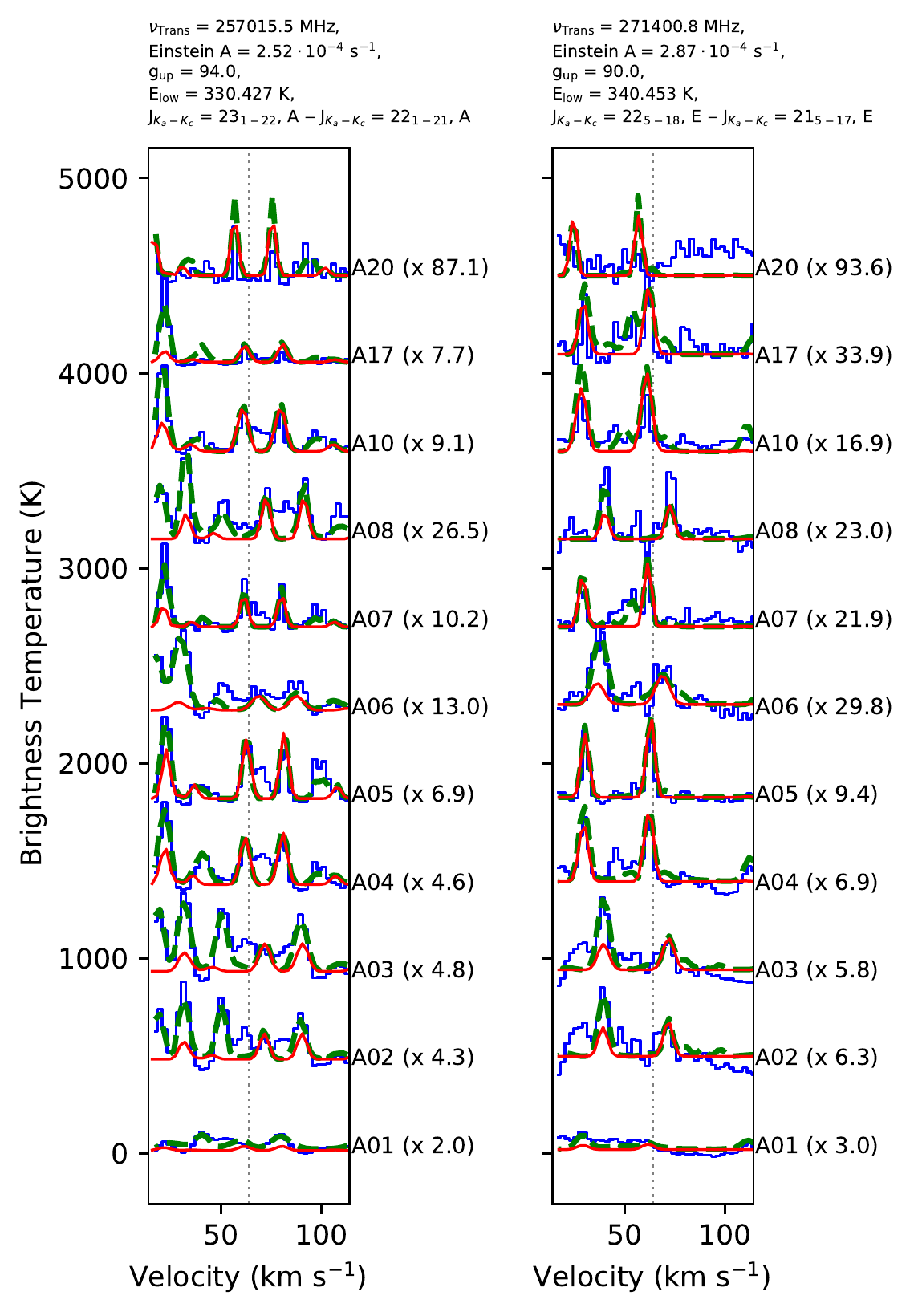}\\
       \caption{Sgr~B2(N)}
       \label{fig:CH3OCHOv18N}
   \end{subfigure}
   \caption{Selected transitions of CH$_3$OCHO, v$_{18}$=1 in Sgr~B2(M) and N.}
   \ContinuedFloat
   \label{fig:CH3OCHOv18MN}
\end{figure*}
\newpage
\clearpage

%*******************************************************************************
% Figure: CH3CHO;v=0;
\begin{figure*}[!htb]
    \centering
    \begin{subfigure}[t]{1.0\columnwidth}
       \includegraphics[width=1.0\columnwidth]{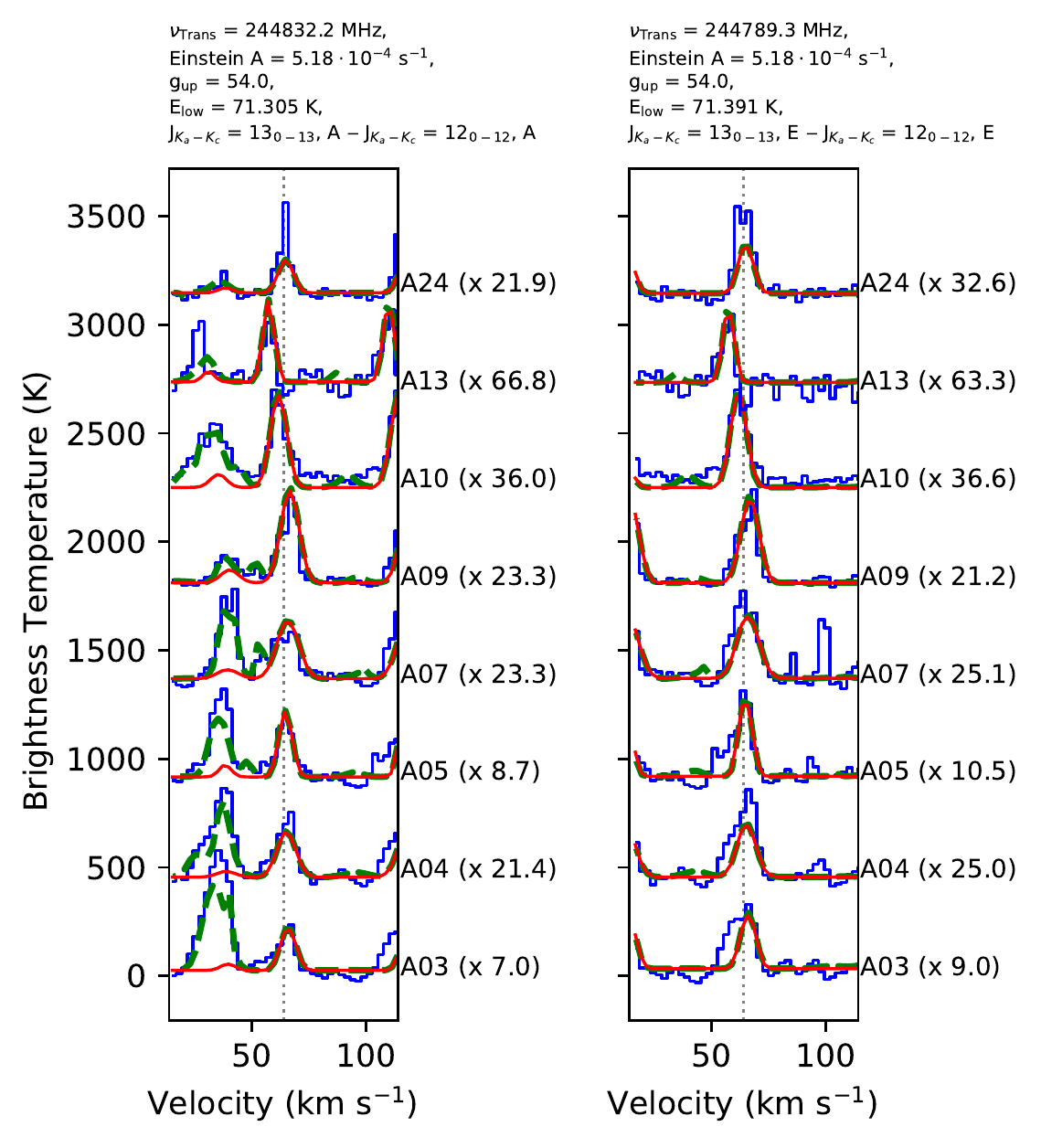}\\
       \caption{Sgr~B2(M)}
       \label{fig:CH3CHOM}
    \end{subfigure}
\quad
    \begin{subfigure}[t]{1.0\columnwidth}
       \includegraphics[width=1.0\columnwidth]{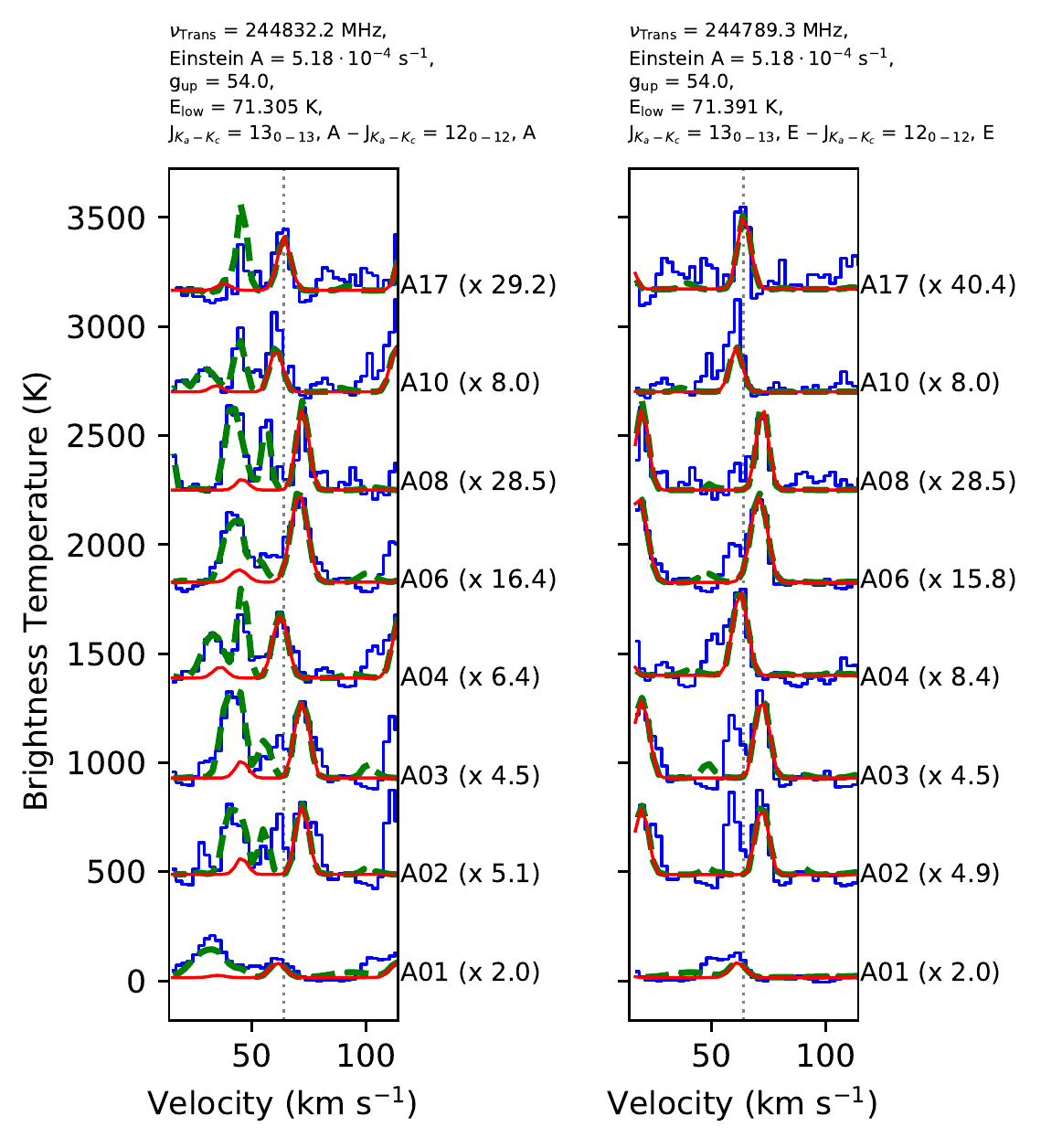}\\
       \caption{Sgr~B2(N)}
       \label{fig:CH3CHON}
   \end{subfigure}
   \caption{Selected transitions of CH$_3$CHO in Sgr~B2(M) and N.}
   \ContinuedFloat
   \label{fig:CH3CHOMN}
\end{figure*}

%*******************************************************************************
% Figure: CH3CHO;v15=1;
\begin{figure*}[!htb]
    \centering
    \begin{subfigure}[t]{1.0\columnwidth}
       \includegraphics[width=1.0\columnwidth]{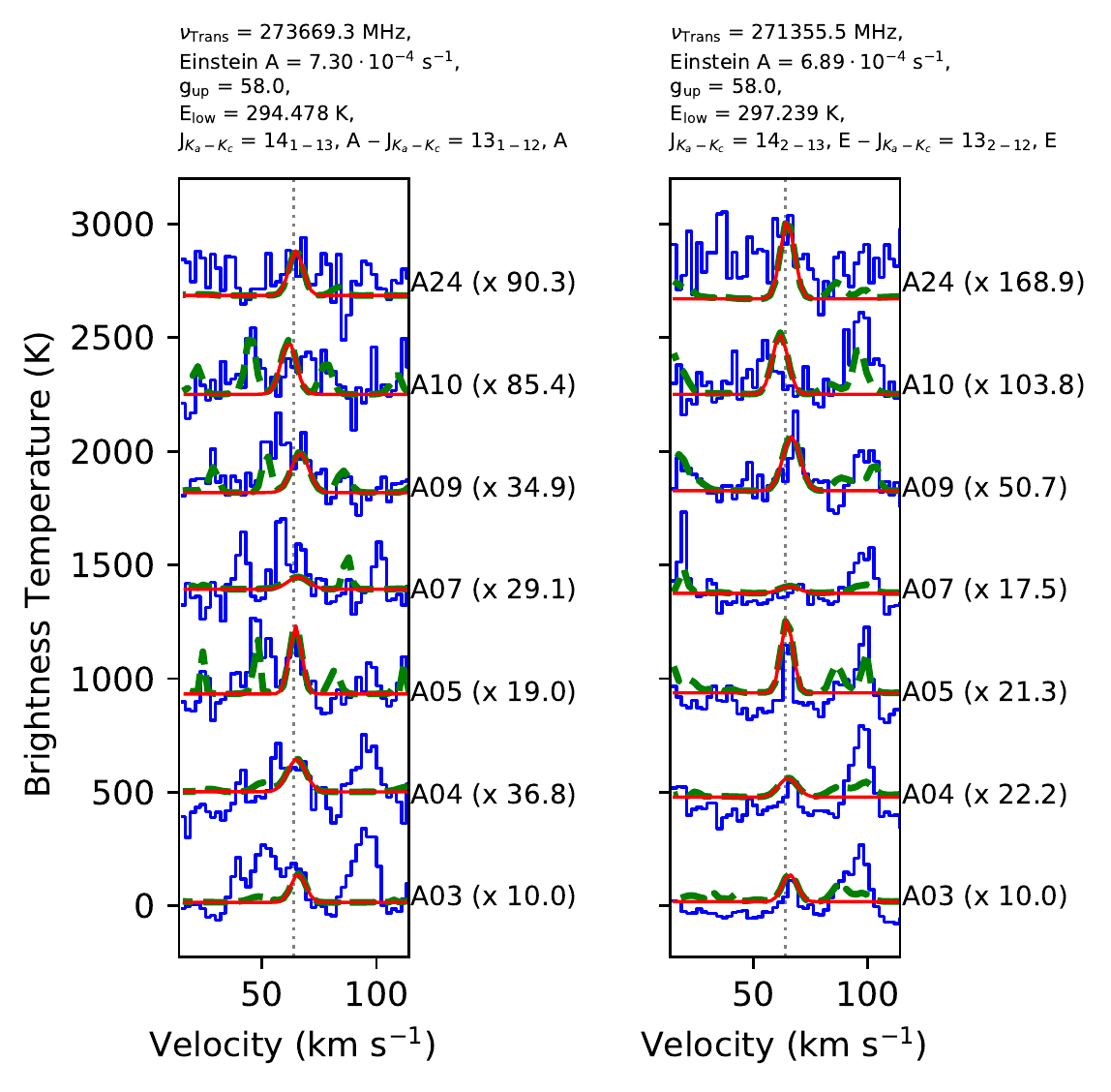}\\
       \caption{Sgr~B2(M)}
       \label{fig:CH3CHOv15M}
    \end{subfigure}
\quad
    \begin{subfigure}[t]{1.0\columnwidth}
       \includegraphics[width=1.0\columnwidth]{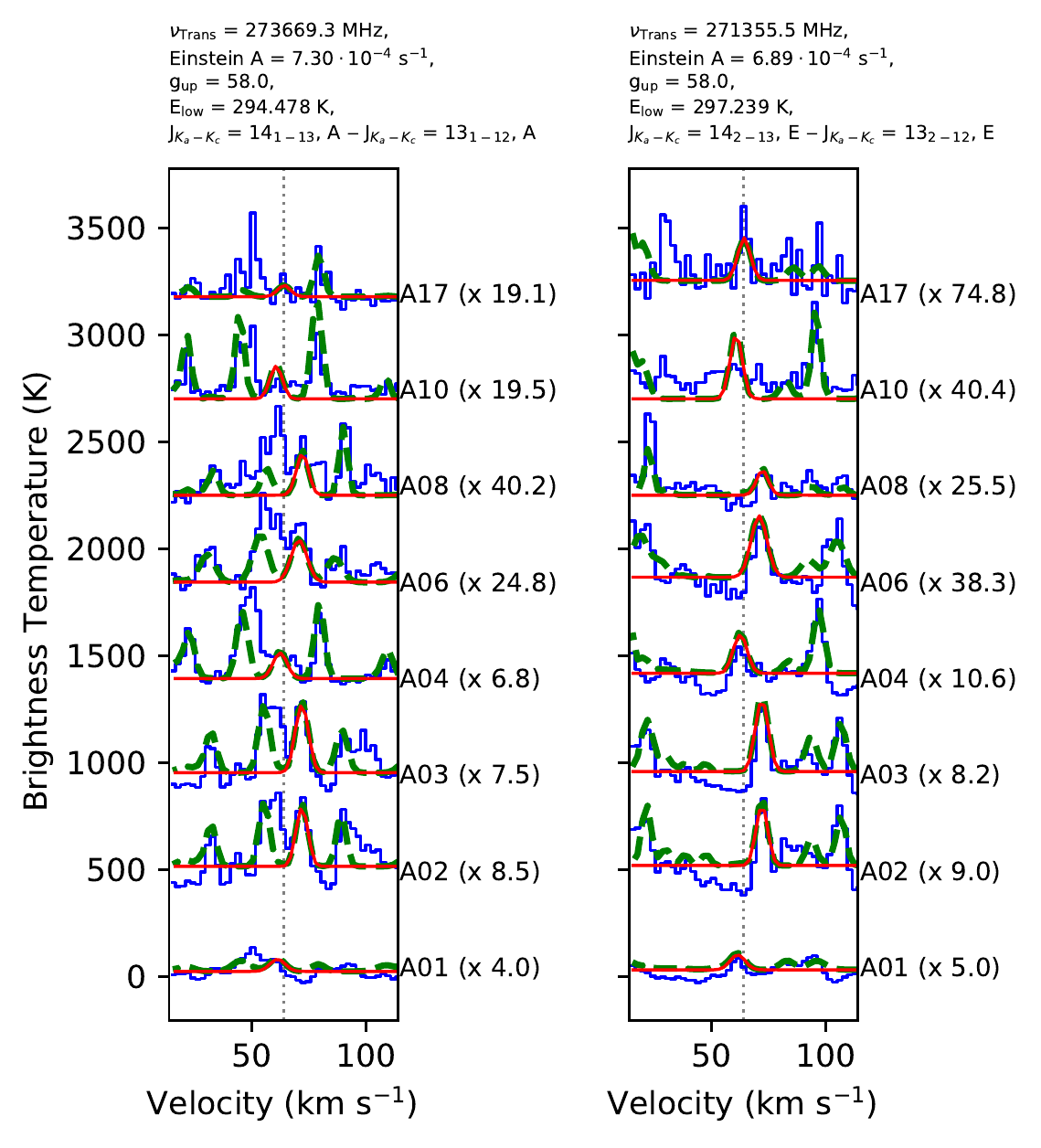}\\
       \caption{Sgr~B2(N)}
       \label{fig:CH3CHOv15N}
   \end{subfigure}
   \caption{Selected transitions of CH$_3$CHO, v$_{15}$=1 in Sgr~B2(M) and N.}
   \ContinuedFloat
   \label{fig:CH3CHOv15MN}
\end{figure*}
\newpage
\clearpage

%*******************************************************************************
% Figure: t-HCOOH;v=0;
\begin{figure*}[!htb]
    \centering
    \begin{subfigure}[t]{1.0\columnwidth}
       \includegraphics[width=1.0\columnwidth]{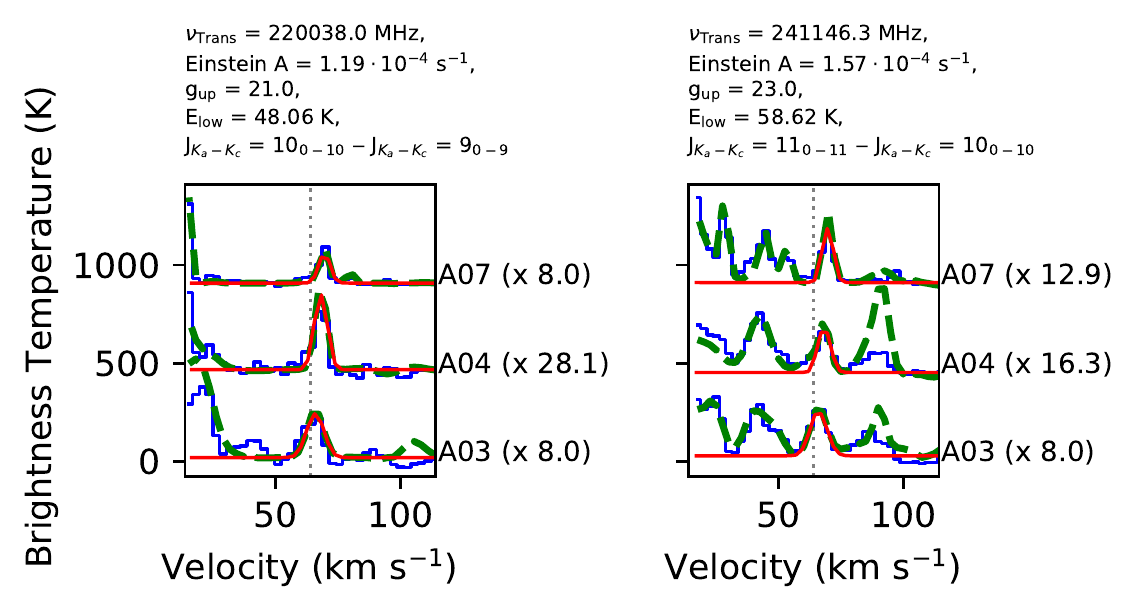}\\
    \end{subfigure}
    \ContinuedFloat
    \caption{Selected transitions of t-HCOOH in Sgr~B2(M).}
    \label{fig:tHCOOHM}
\end{figure*}

%-------------------------------------------------------------------------------
% NH-bearing molecules

%*******************************************************************************
% Figure: CH3NH2;v=0;
\begin{figure*}[!htb]
    \centering
    \begin{subfigure}[t]{1.0\columnwidth}
       \includegraphics[width=1.0\columnwidth]{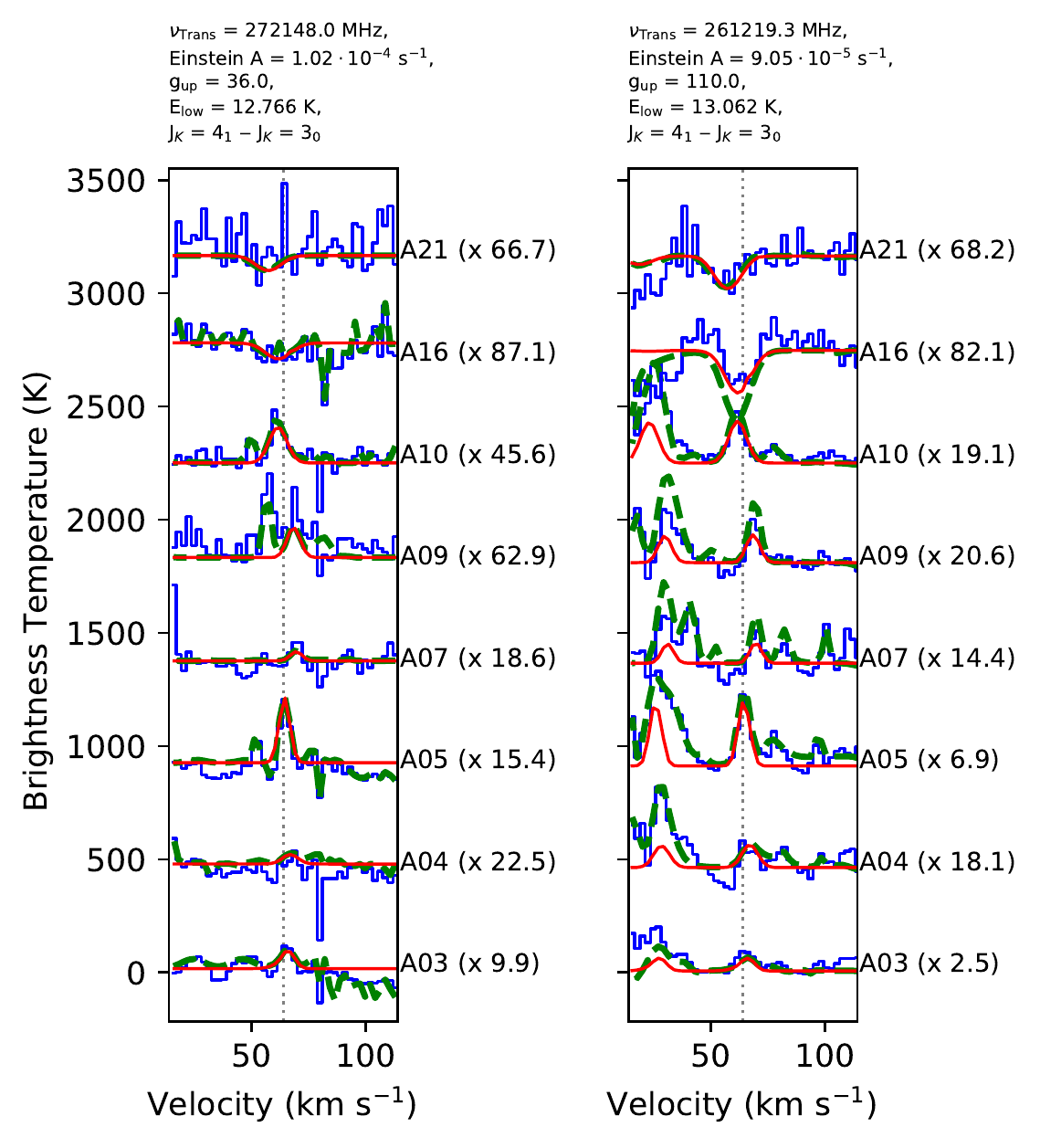}\\
       \caption{Sgr~B2(M)}
       \label{fig:CH3NH2M}
    \end{subfigure}
\quad
    \begin{subfigure}[t]{1.0\columnwidth}
       \includegraphics[width=1.0\columnwidth]{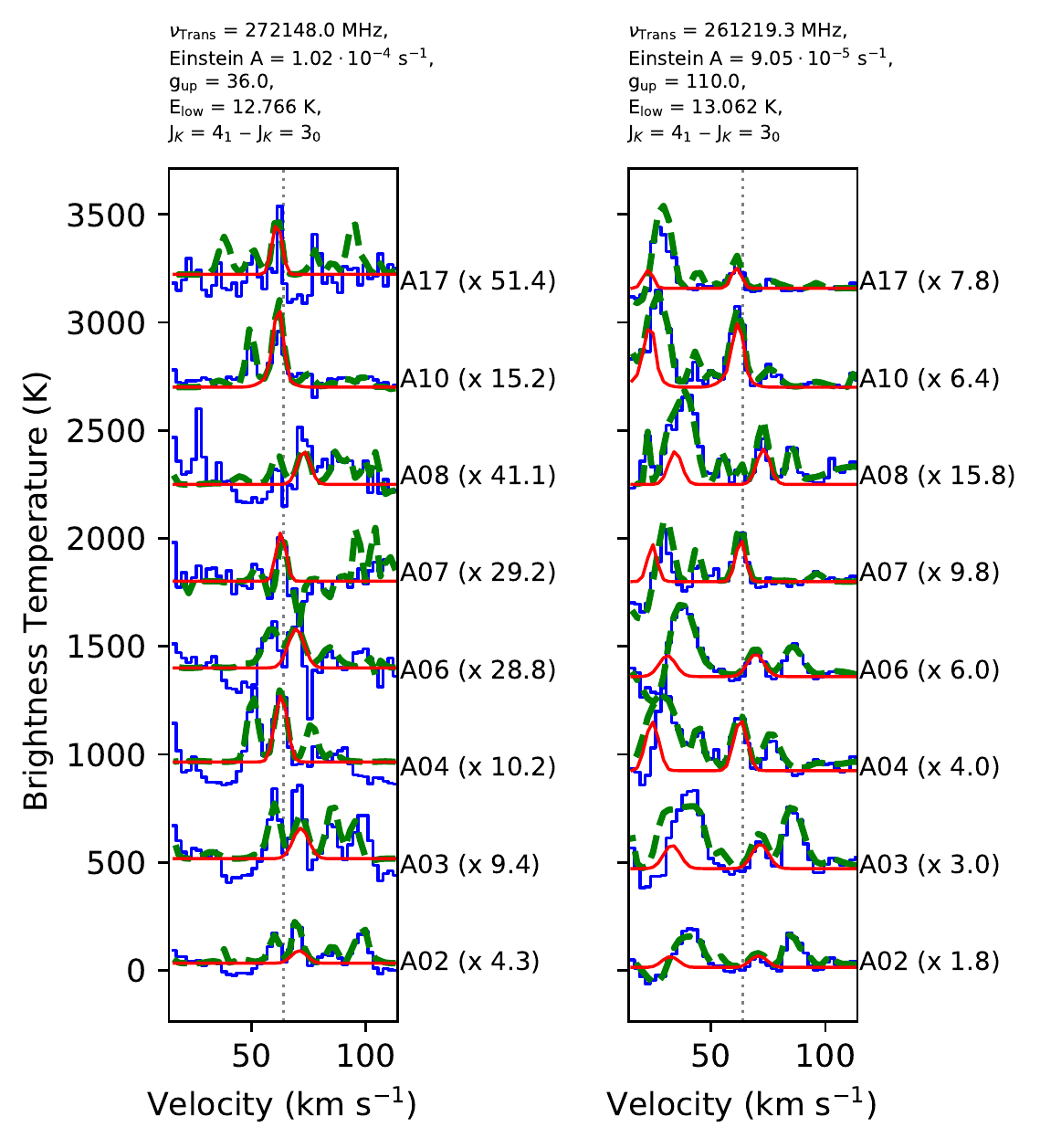}\\
       \caption{Sgr~B2(N)}
       \label{fig:CH3NH2N}
   \end{subfigure}
   \caption{Selected transitions of CH$_3$NH$_2$ in Sgr~B2(M) and N.}
   \ContinuedFloat
   \label{fig:CH3NH2MN}
\end{figure*}
\newpage
\clearpage

%*******************************************************************************
% Figure: CH2NH;v=0;
\begin{figure*}[!htb]
    \centering
    \begin{subfigure}[t]{1.0\columnwidth}
       \includegraphics[width=1.0\columnwidth]{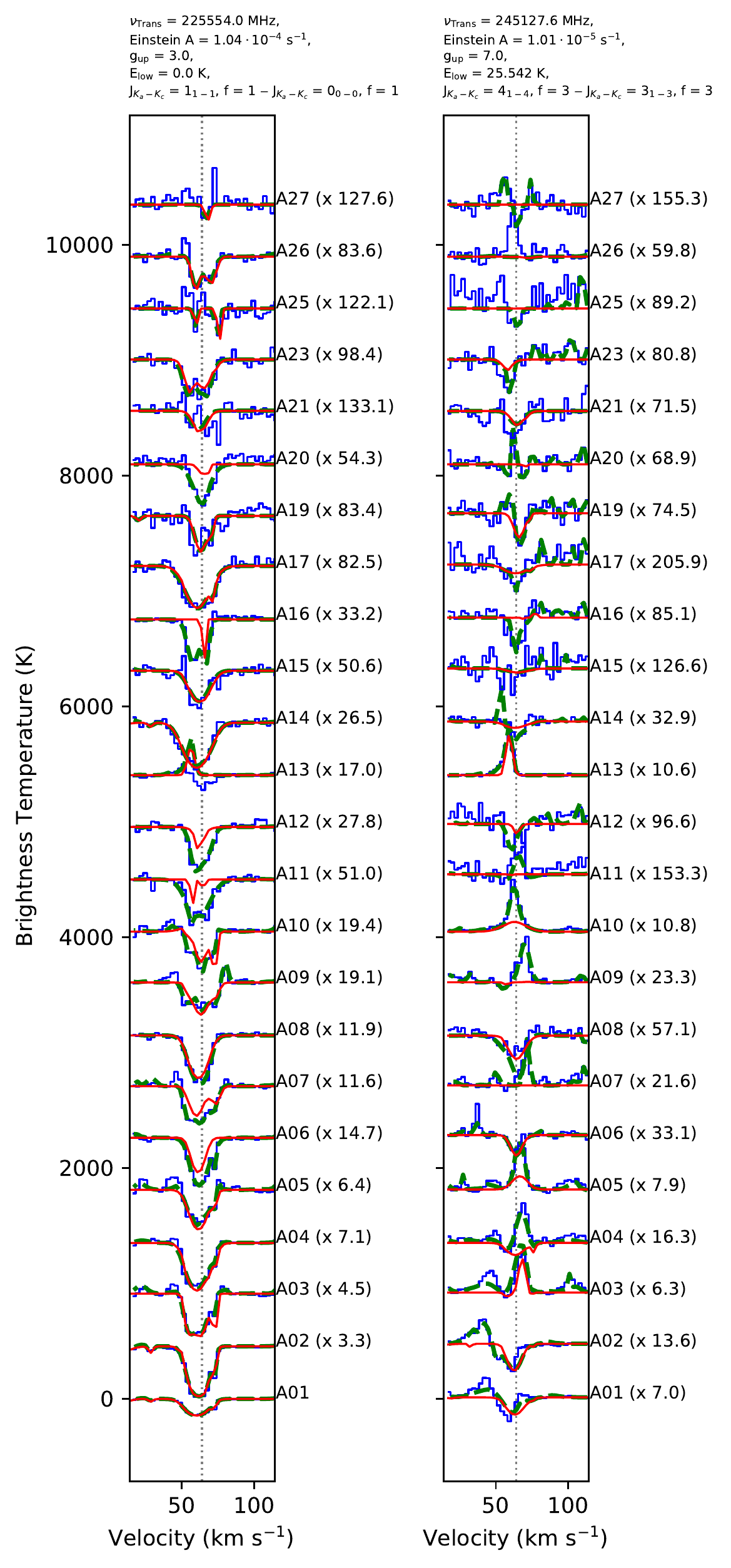}\\
       \caption{Sgr~B2(M)}
       \label{fig:CH2NHM}
    \end{subfigure}
\quad
    \begin{subfigure}[t]{1.0\columnwidth}
       \includegraphics[width=1.0\columnwidth]{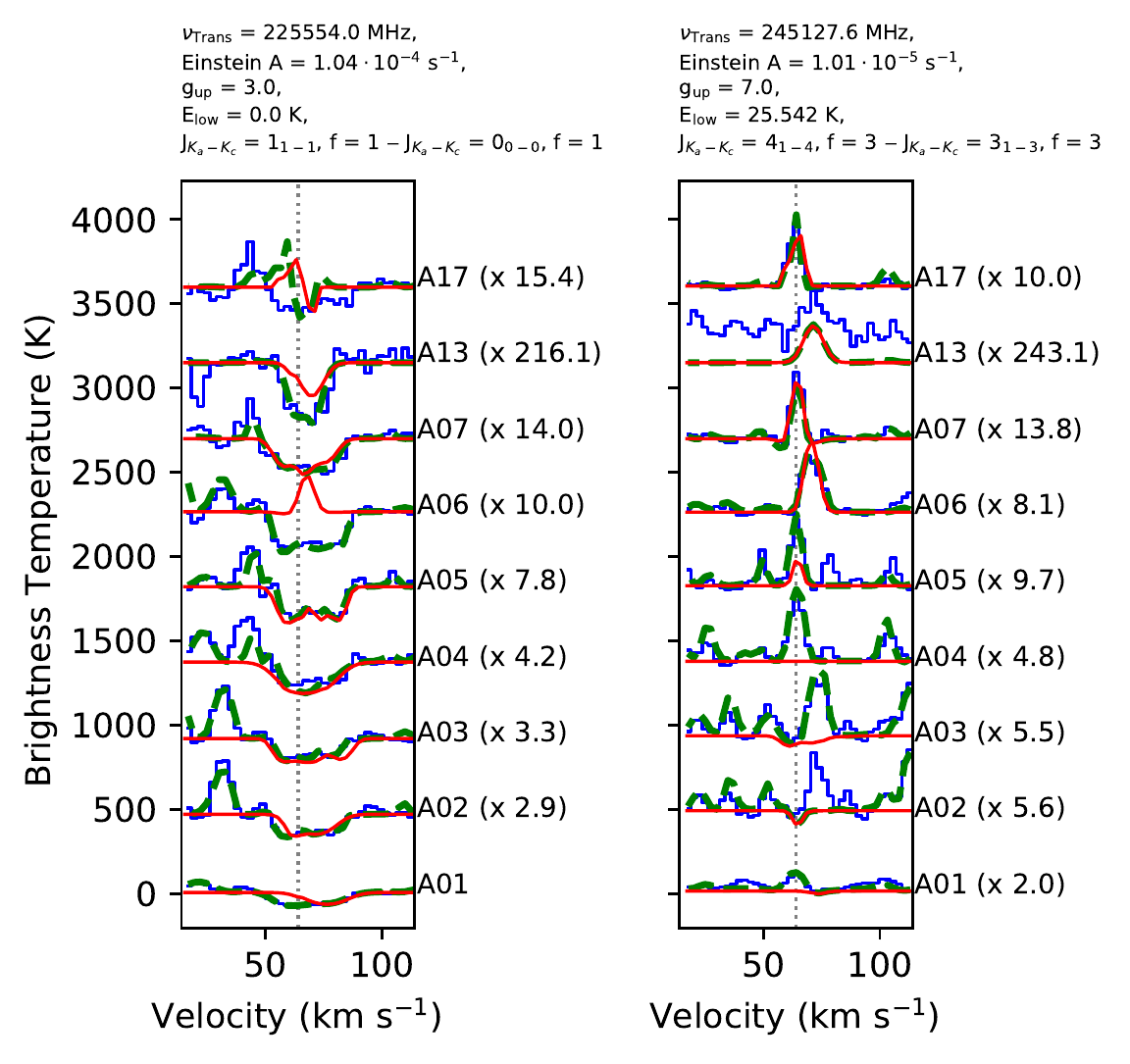}\\
       \caption{Sgr~B2(N)}
       \label{fig:CH2NHN}
   \end{subfigure}
   \caption{Selected transitions of CH$_2$NH in Sgr~B2(M) and N.}
   \ContinuedFloat
   \label{fig:CH2NHMN}
\end{figure*}
\newpage
\clearpage

%*******************************************************************************
% Figure: NH2D;v=0;hyp1
\begin{figure*}[!htb]
    \centering
    \begin{subfigure}[t]{0.5\columnwidth}
       \includegraphics[width=1.0\columnwidth]{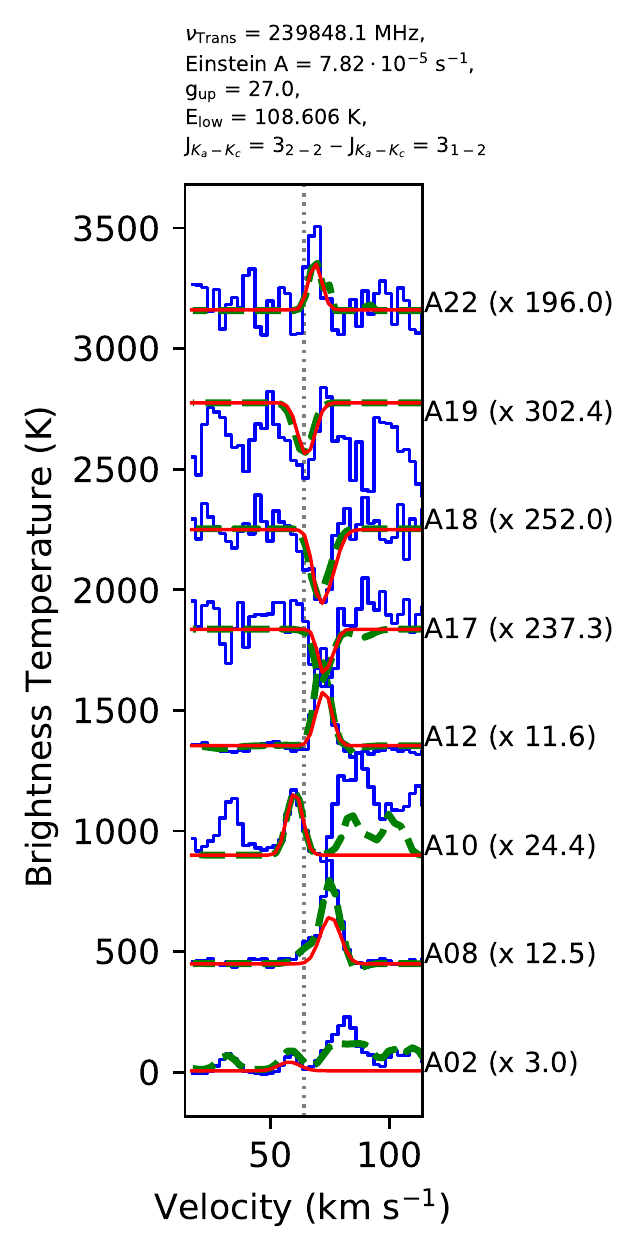}\\
       \caption{Sgr~B2(M)}
       \label{fig:NH2DM}
    \end{subfigure}
\quad
    \begin{subfigure}[t]{0.5\columnwidth}
       \includegraphics[width=1.0\columnwidth]{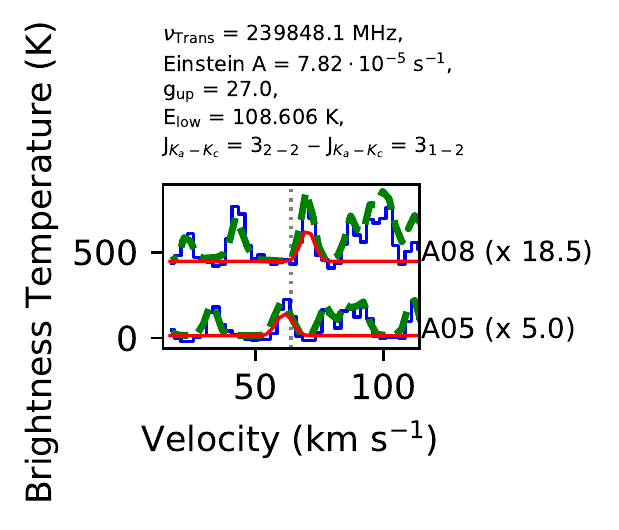}\\
       \caption{Sgr~B2(N)}
       \label{fig:NH2DN}
   \end{subfigure}
   \caption{Selected transitions of NH$_2$D in Sgr~B2(M) and N.}
   \ContinuedFloat
   \label{fig:NH2DMN}
\end{figure*}
\newpage
\clearpage

%-------------------------------------------------------------------------------
% N- and O-bearing molecules

%*******************************************************************************
% Figure: HNCO;v=0;
\begin{figure*}[!htb]
    \centering
    \begin{subfigure}[t]{1.0\columnwidth}
       \includegraphics[width=1.0\columnwidth]{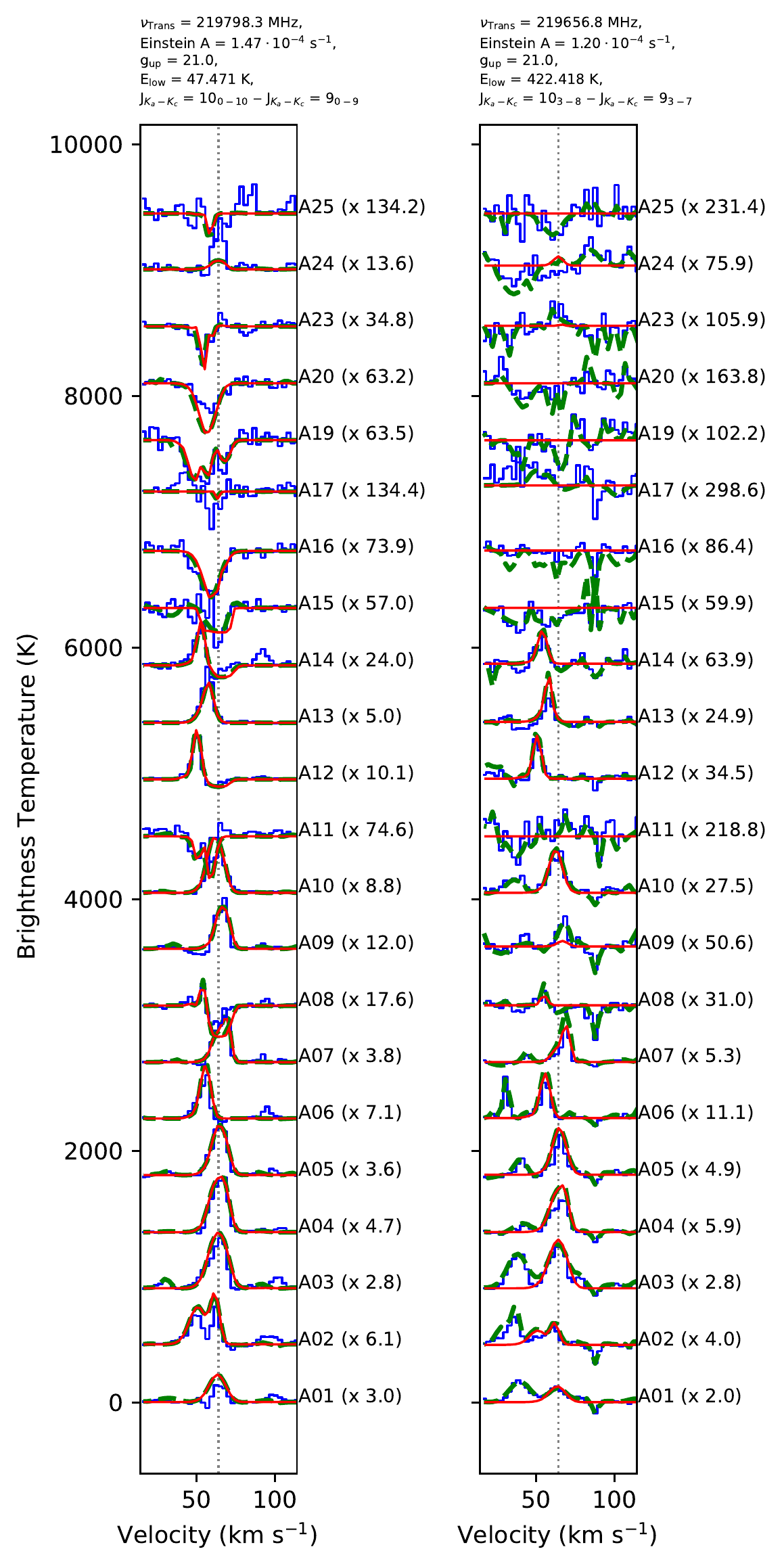}\\
       \caption{Sgr~B2(M)}
       \label{fig:HNCOM}
    \end{subfigure}
\quad
    \begin{subfigure}[t]{1.0\columnwidth}
       \includegraphics[width=1.0\columnwidth]{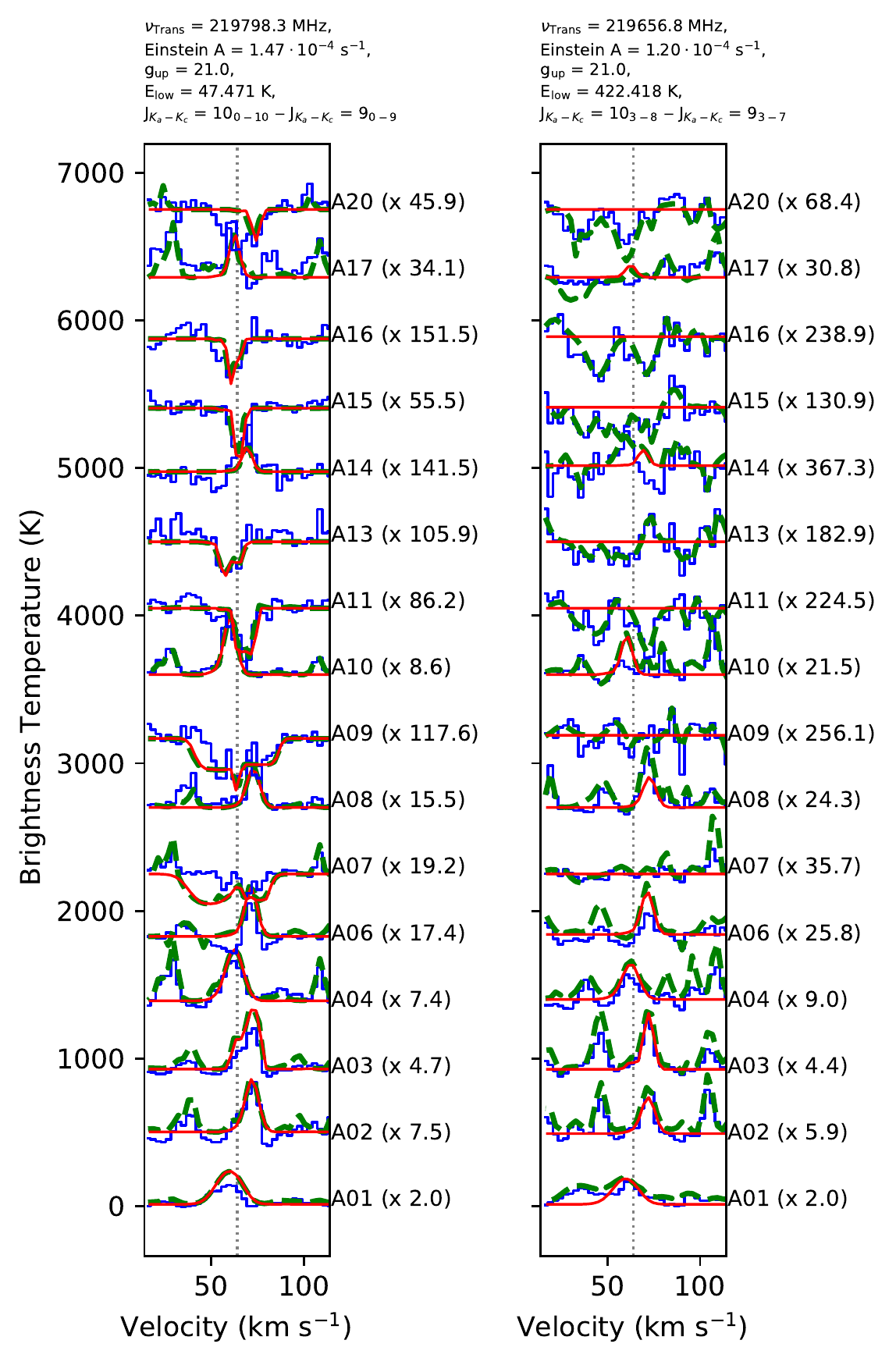}\\
       \caption{Sgr~B2(N)}
       \label{fig:HNCON}
   \end{subfigure}
   \caption{Selected transitions of HNCO in Sgr~B2(M) and N.}
   \ContinuedFloat
   \label{fig:HNCOMN}
\end{figure*}
\newpage
\clearpage

%*******************************************************************************
% Figure: HC(O)NH2;v=0;
\begin{figure*}[!htb]
    \centering
    \begin{subfigure}[t]{1.0\columnwidth}
       \includegraphics[width=1.0\columnwidth]{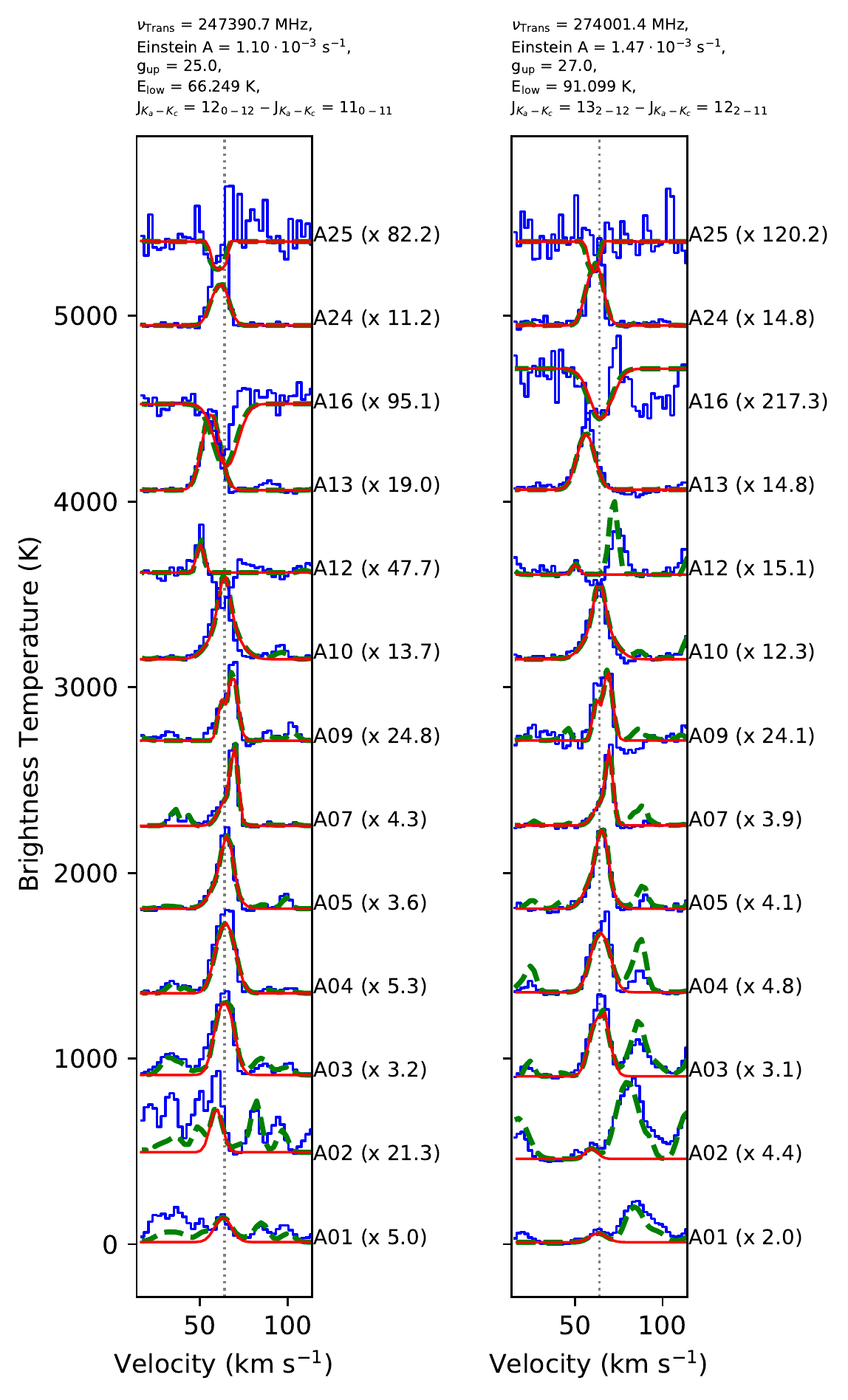}\\
       \caption{Sgr~B2(M)}
       \label{fig:HCONH2M}
    \end{subfigure}
\quad
    \begin{subfigure}[t]{1.0\columnwidth}
       \includegraphics[width=1.0\columnwidth]{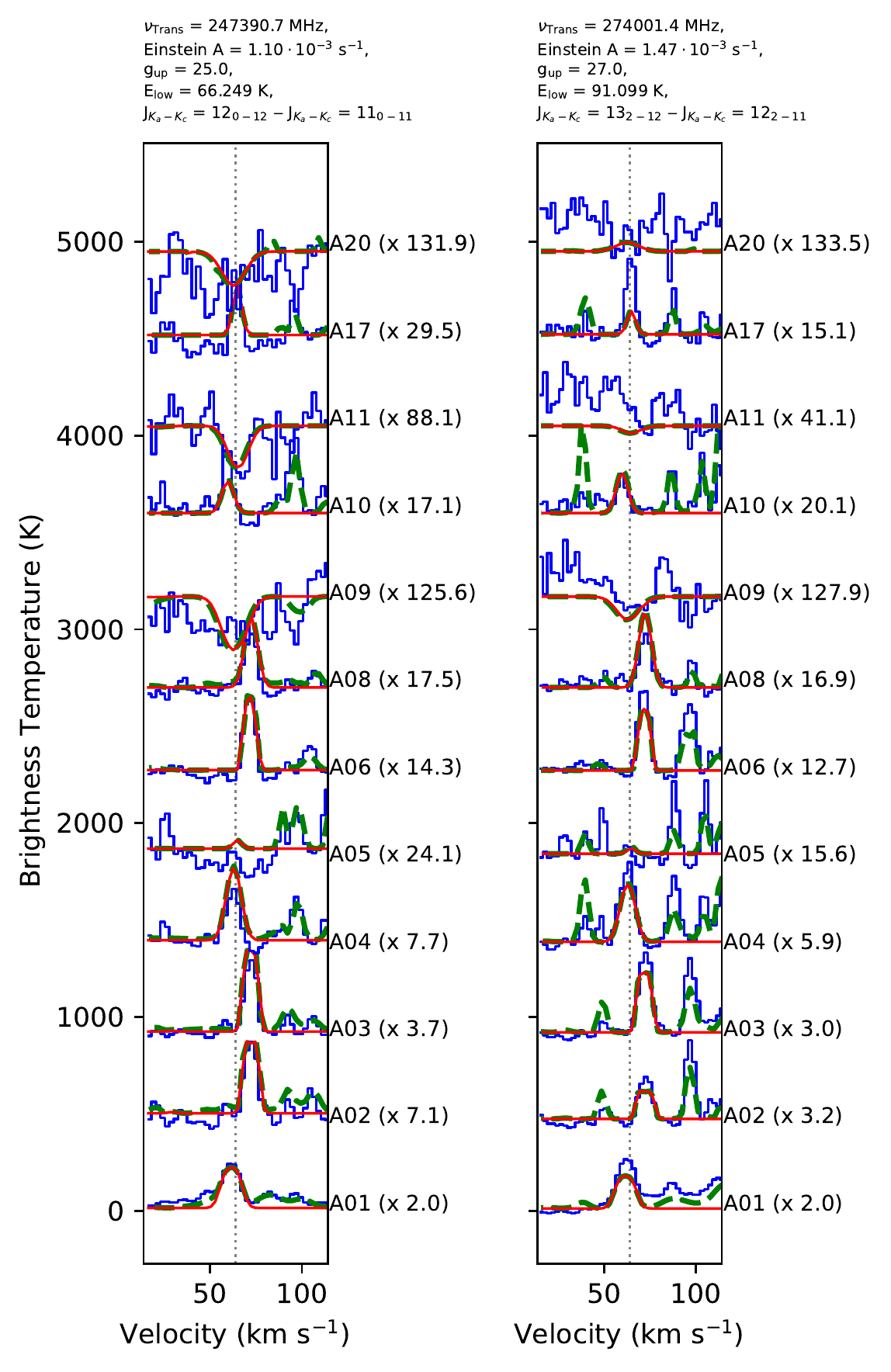}\\
       \caption{Sgr~B2(N)}
       \label{fig:HCONH2N}
   \end{subfigure}
   \caption{Selected transitions of NH$_2$CHO in Sgr~B2(M) and N.}
   \ContinuedFloat
   \label{fig:HCONH2MN}
\end{figure*}
\newpage
\clearpage

%*******************************************************************************
% Figure: HC(O)NH2;v12=1;
\begin{figure*}[!htb]
    \centering
    \begin{subfigure}[t]{1.0\columnwidth}
       \includegraphics[width=1.0\columnwidth]{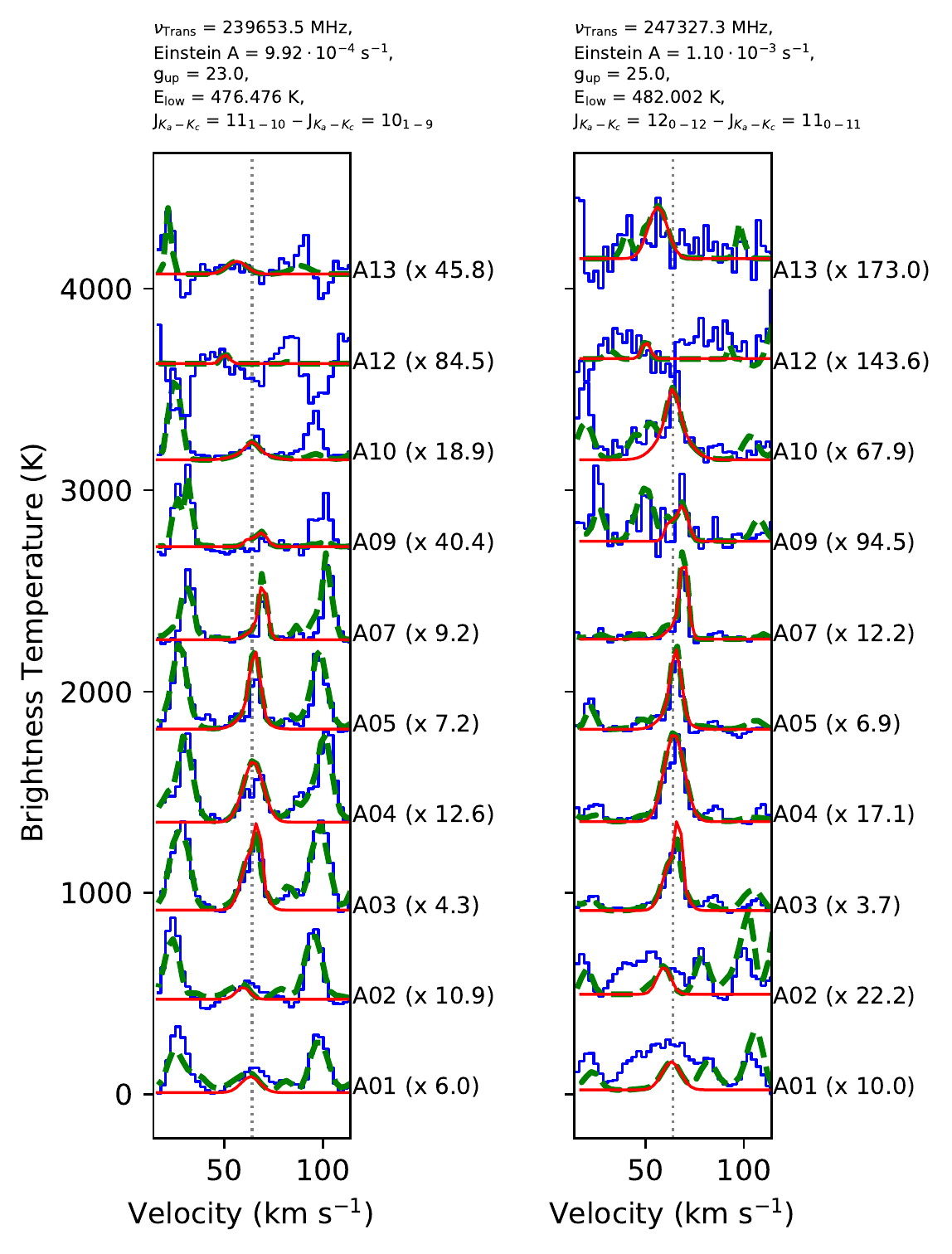}\\
       \caption{Sgr~B2(M)}
       \label{fig:HCONH2v12M}
    \end{subfigure}
\quad
    \begin{subfigure}[t]{1.0\columnwidth}
       \includegraphics[width=1.0\columnwidth]{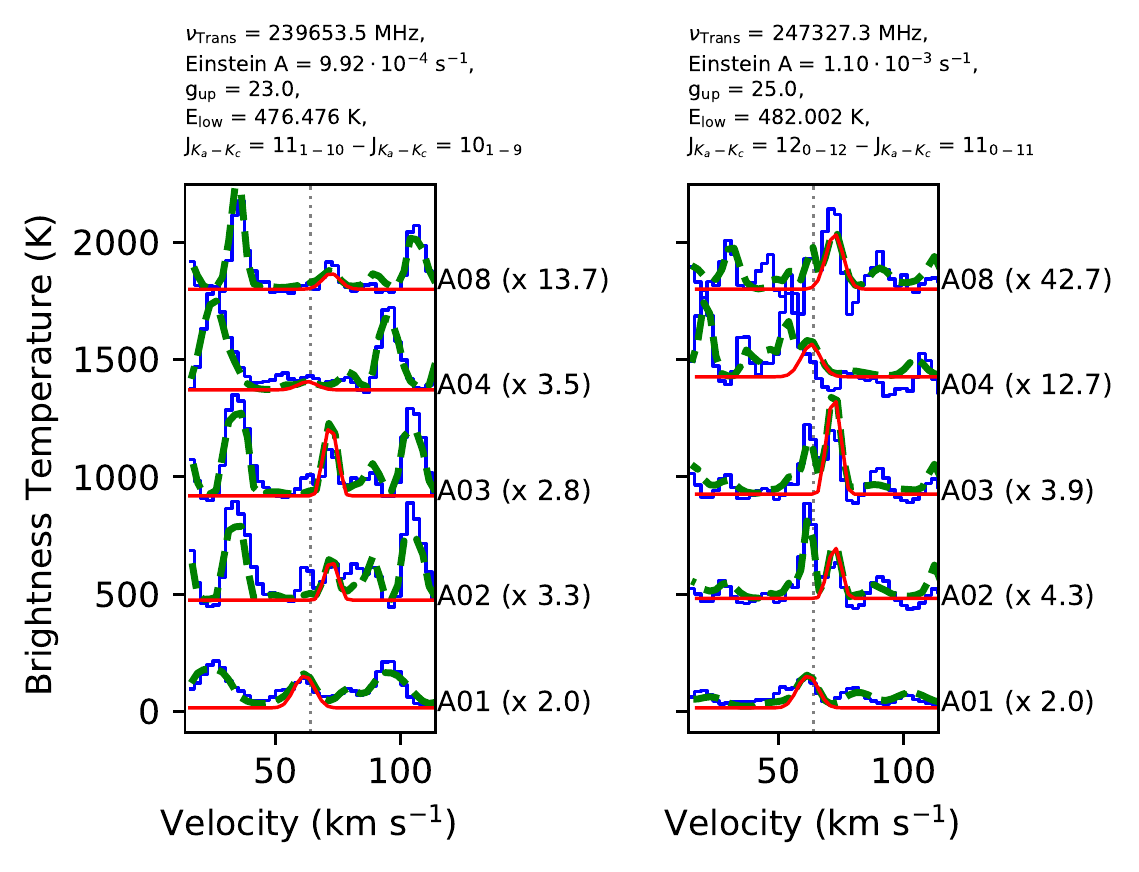}\\
       \caption{Sgr~B2(N)}
       \label{fig:HCONH2v12N}
   \end{subfigure}
   \caption{Selected transitions of NH$_2$CHO, v$_{12}$ = 1 in Sgr~B2(M) and N.}
   \ContinuedFloat
   \label{fig:HCONH2v12MN}
\end{figure*}
\newpage
\clearpage

%*******************************************************************************
% Figure: NO;v=0;hyp1
\begin{figure*}[!htb]
    \centering
    \begin{subfigure}[t]{1.0\columnwidth}
       \includegraphics[width=1.0\columnwidth]{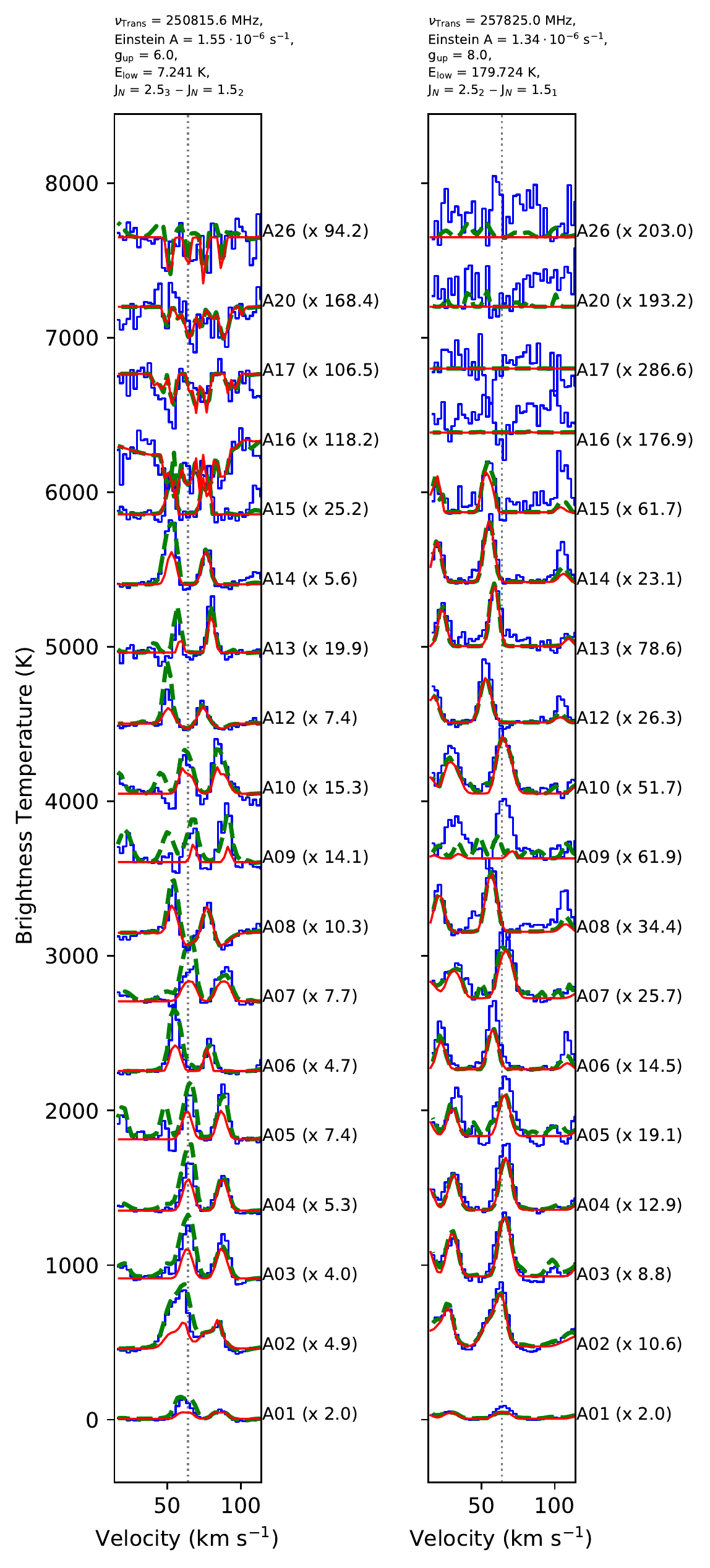}\\
       \caption{Sgr~B2(M)}
       \label{fig:NOM}
    \end{subfigure}
\quad
    \begin{subfigure}[t]{1.0\columnwidth}
       \includegraphics[width=1.0\columnwidth]{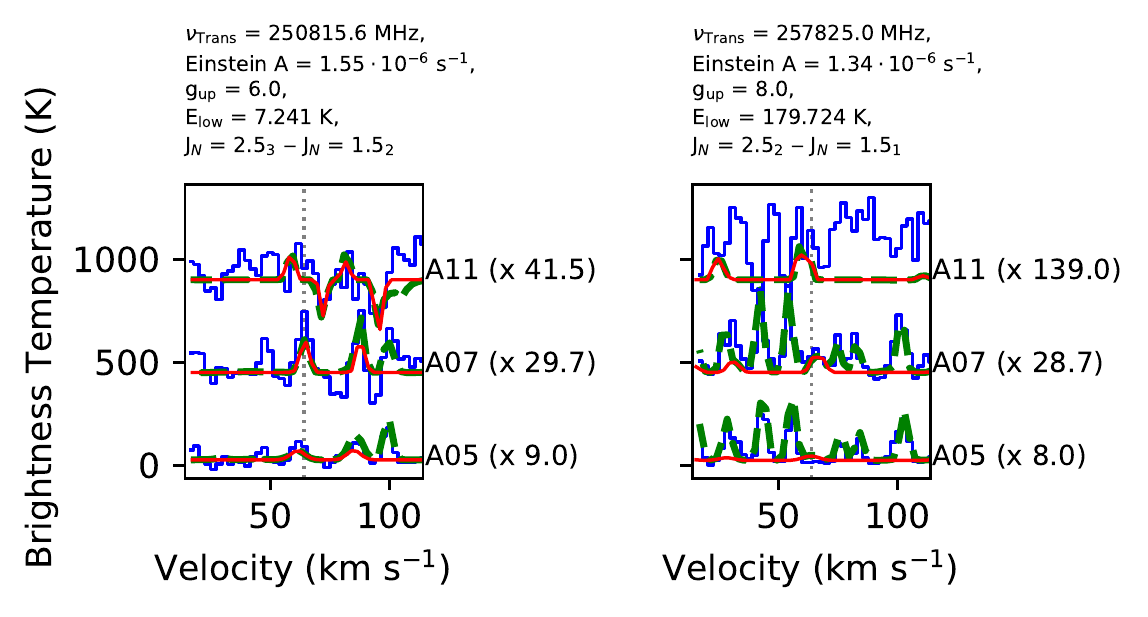}\\
       \caption{Sgr~B2(N)}
       \label{fig:NON}
   \end{subfigure}
   \caption{Selected transitions of NO in Sgr~B2(M) and N.}
   \ContinuedFloat
   \label{fig:NOMN}
\end{figure*}
\newpage
\clearpage

%*******************************************************************************
% Figure: NO+;v=0;hyp1
\begin{figure*}[!htb]
    \centering
    \begin{subfigure}[t]{0.5\columnwidth}
       \includegraphics[width=1.0\columnwidth]{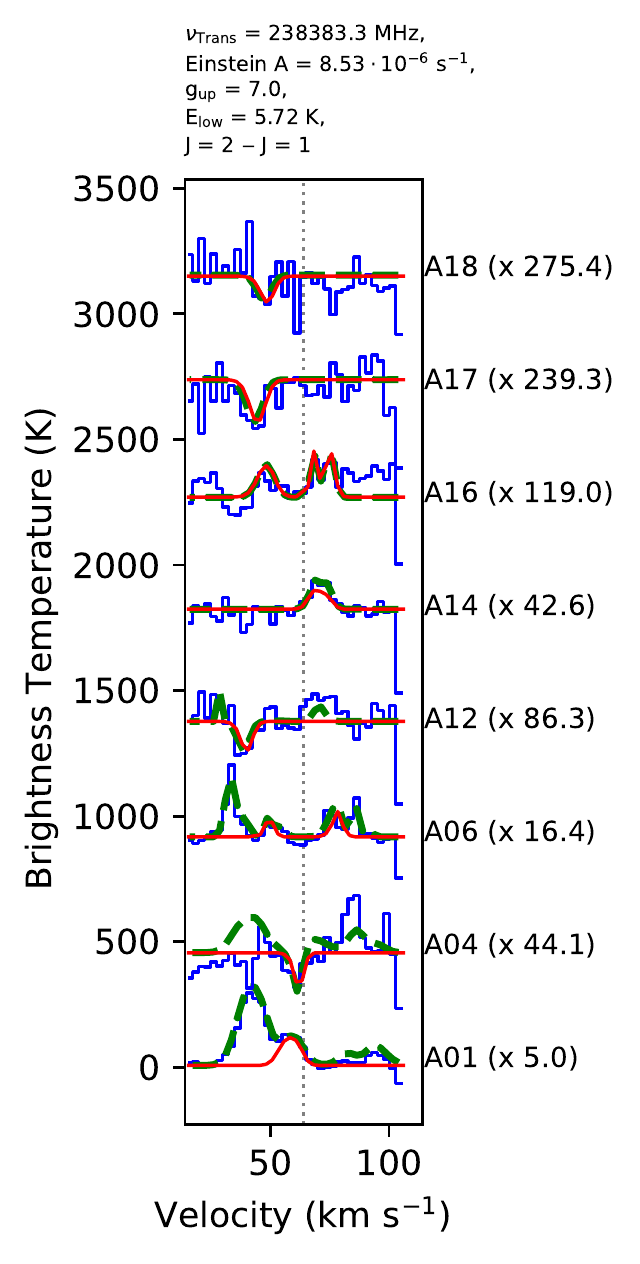}\\
       \caption{Sgr~B2(M)}
       \label{fig:NOpM}
    \end{subfigure}
\quad
    \begin{subfigure}[t]{0.5\columnwidth}
       \includegraphics[width=1.0\columnwidth]{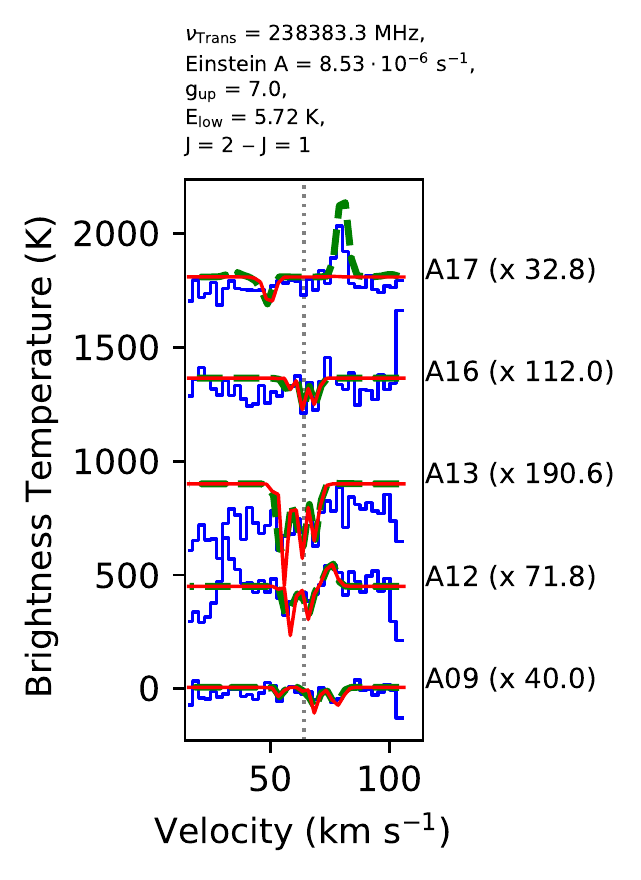}\\
       \caption{Sgr~B2(N)}
       \label{fig:NOpN}
   \end{subfigure}
   \caption{Selected transitions of NO$^+$ in Sgr~B2(M) and N.}
   \ContinuedFloat
   \label{fig:NOpMN}
\end{figure*}
\newpage
\clearpage

%-------------------------------------------------------------------------------
% cyanide molecules

%*******************************************************************************
% Figure: HCN and HNC
\begin{figure*}[!htb]
    \centering
    \begin{subfigure}[t]{0.4\columnwidth}
       \includegraphics[width=1.0\columnwidth]{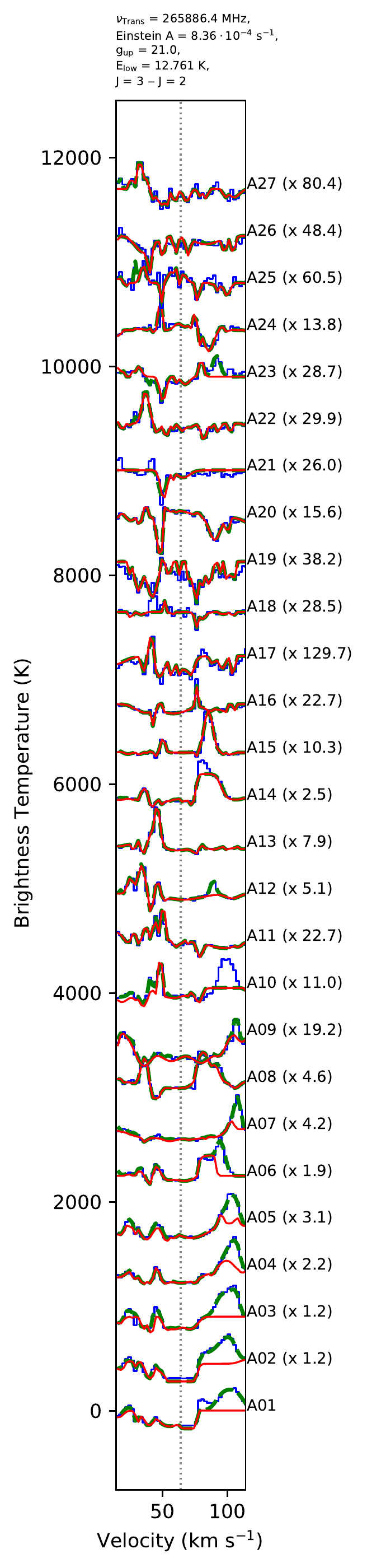}\\
       \caption{HCN in Sgr~B2(M).}
       \label{fig:HCNM}
    \end{subfigure}
\quad
    \begin{subfigure}[t]{0.4\columnwidth}
       \includegraphics[width=1.0\columnwidth]{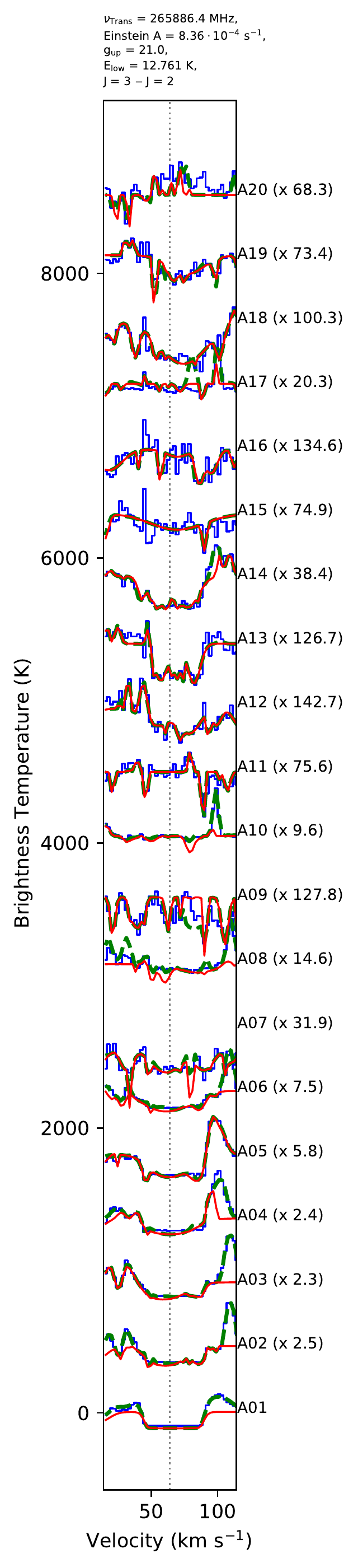}\\
       \caption{HCN in Sgr~B2(N).}
       \label{fig:HCNN}
   \end{subfigure}
\quad
    \begin{subfigure}[t]{0.4\columnwidth}
       \includegraphics[width=1.0\columnwidth]{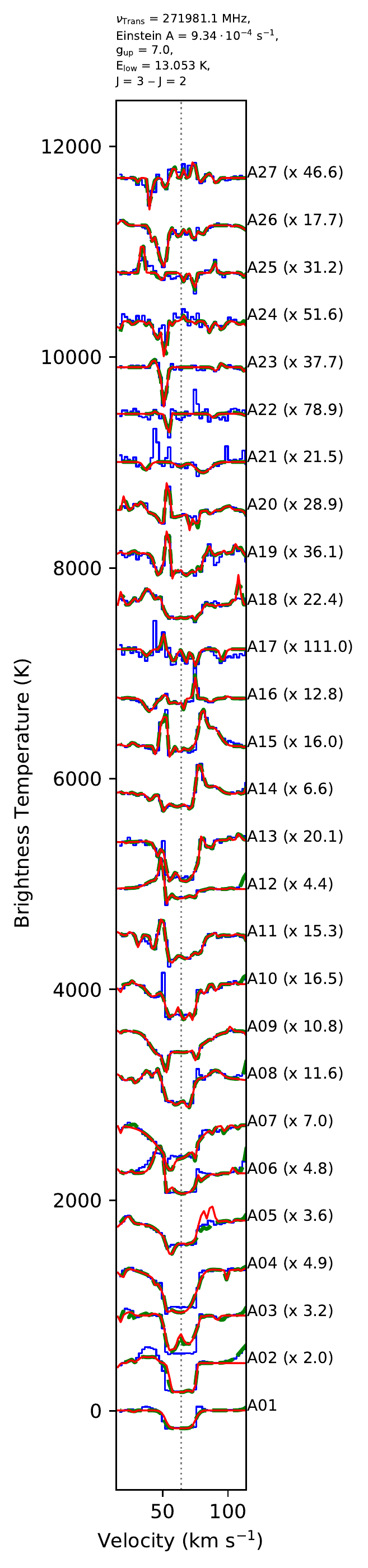}\\
       \caption{HNC in Sgr~B2(M).}
       \label{fig:HNCM}
    \end{subfigure}
\quad
    \begin{subfigure}[t]{0.4\columnwidth}
       \includegraphics[width=1.0\columnwidth]{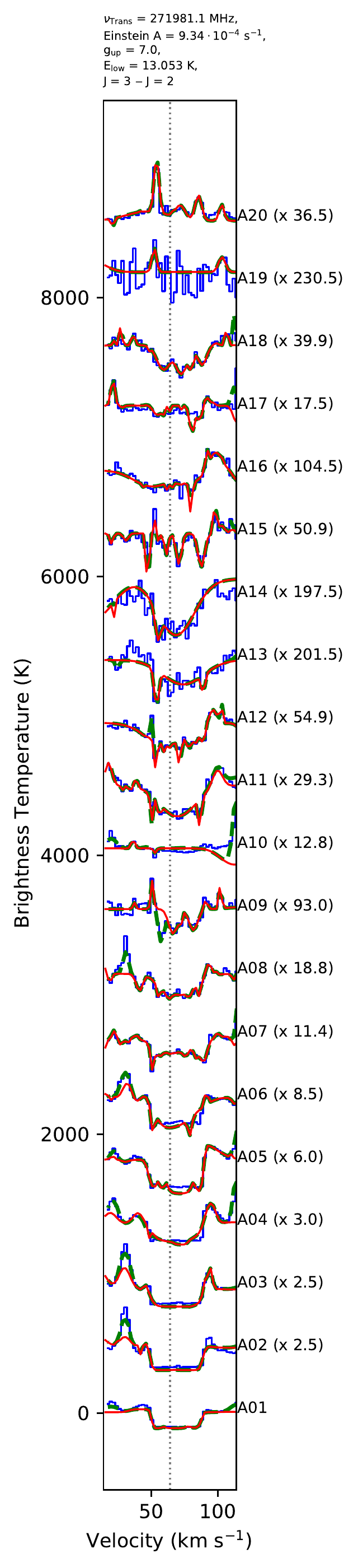}\\
       \caption{HNC in Sgr~B2(N).}
       \label{fig:HNCN}
   \end{subfigure}
   \caption{Selected transitions of HCN and HNC in Sgr~B2(M) and N.}
   \ContinuedFloat
   \label{fig:HCNNCMN}
\end{figure*}
\newpage
\clearpage

%*******************************************************************************
% Figure: HC13N and HNC13
\begin{figure*}[!htb]
    \centering
    \begin{subfigure}[t]{0.4\columnwidth}
       \includegraphics[width=1.0\columnwidth]{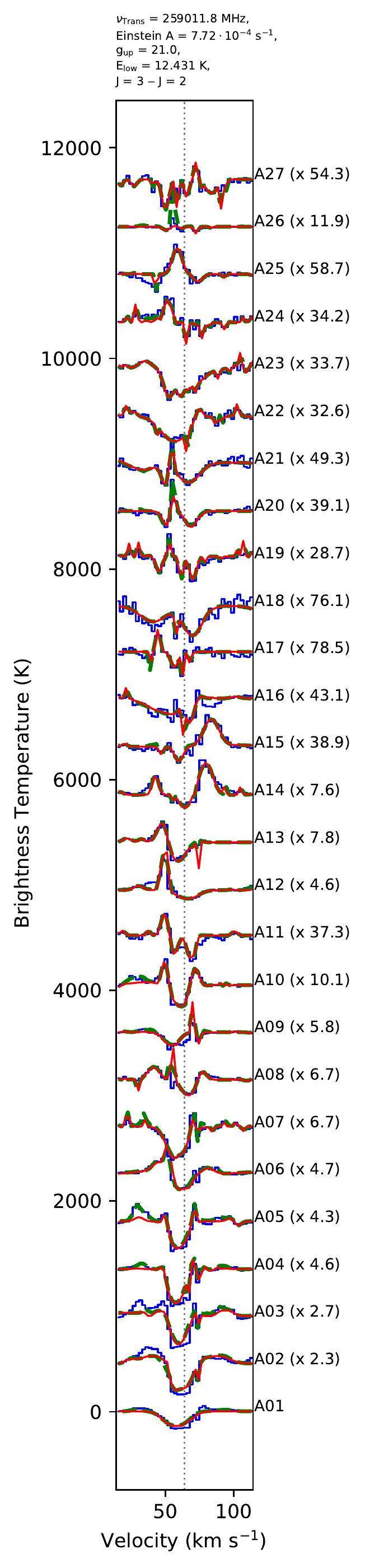}\\
       \caption{H$^{13}$CN in Sgr~B2(M).}
       \label{fig:HC13NM}
    \end{subfigure}
\quad
    \begin{subfigure}[t]{0.4\columnwidth}
       \includegraphics[width=1.0\columnwidth]{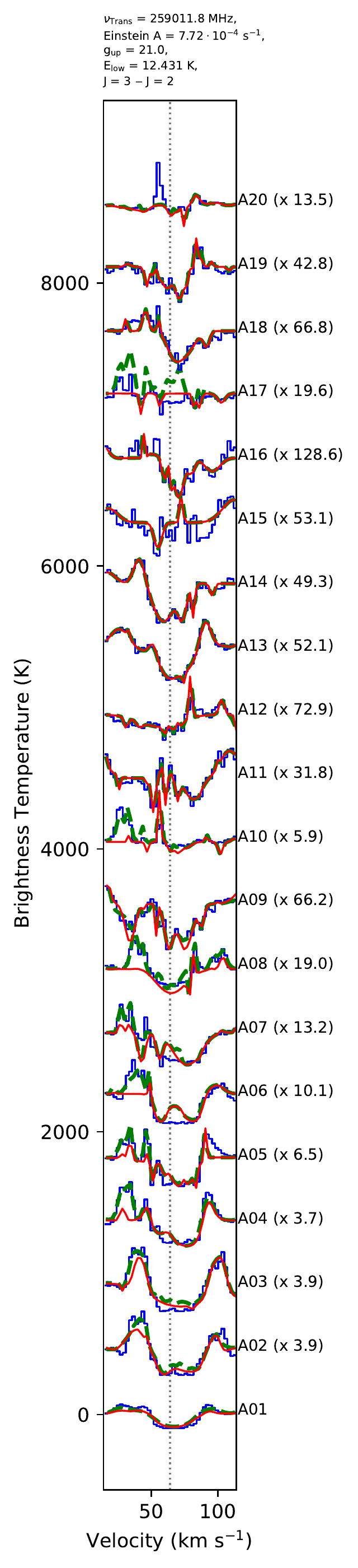}\\
       \caption{H$^{13}$CN in Sgr~B2(N).}
       \label{fig:HC13NN}
   \end{subfigure}
\quad
    \begin{subfigure}[t]{0.4\columnwidth}
       \includegraphics[width=1.0\columnwidth]{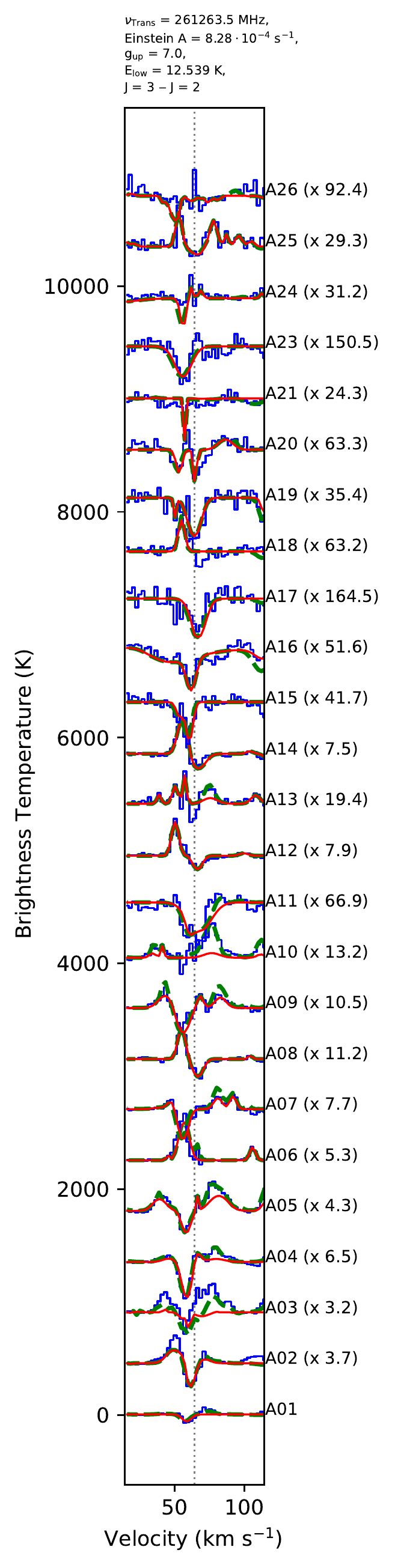}\\
       \caption{HN$^{13}$C in Sgr~B2(M).}
       \label{fig:HNC13M}
    \end{subfigure}
\quad
    \begin{subfigure}[t]{0.4\columnwidth}
       \includegraphics[width=1.0\columnwidth]{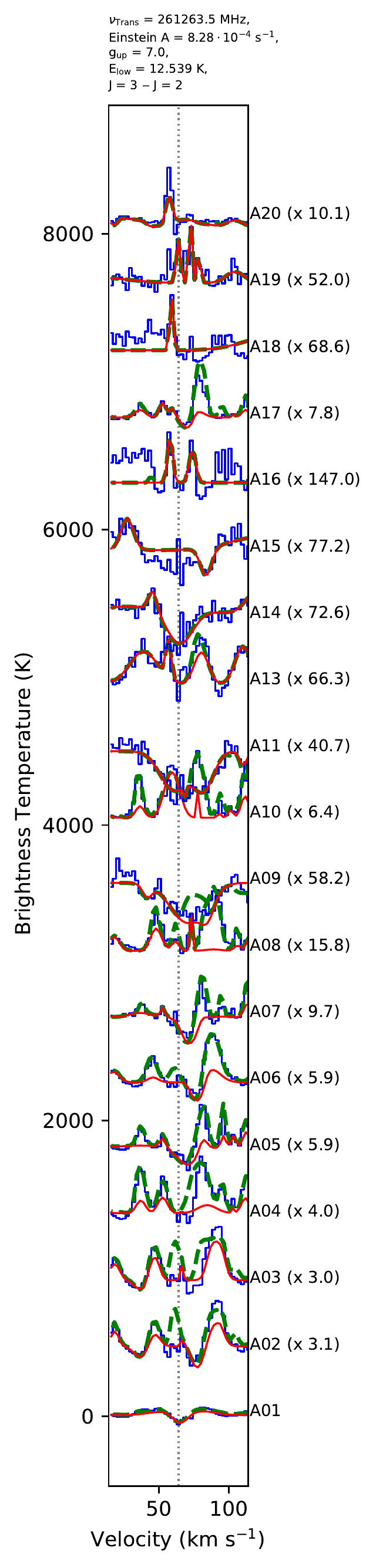}\\
       \caption{HN$^{13}$C in Sgr~B2(N).}
       \label{fig:HNC13N}
   \end{subfigure}
   \caption{Selected transitions of H$^{13}$CN and HN$^{13}$C in Sgr~B2(M) and N.}
   \ContinuedFloat
   \label{fig:HCNNC13MN}
\end{figure*}
\newpage
\clearpage

%*******************************************************************************
% Figure: HCN-15;v=0;
\begin{figure*}[!htb]
    \centering
    \begin{subfigure}[t]{0.4\columnwidth}
       \includegraphics[width=1.0\columnwidth]{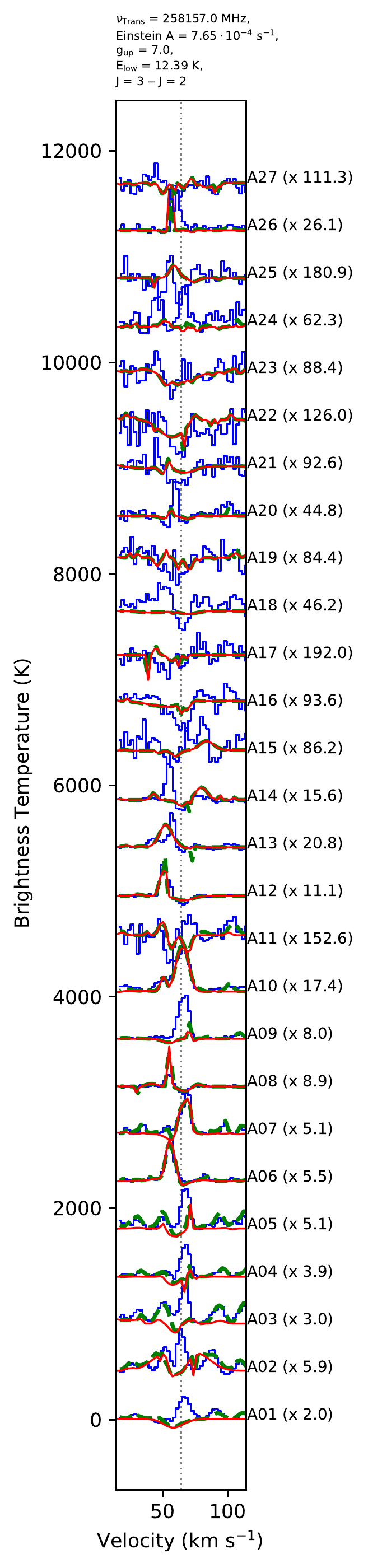}\\
       \caption{HC$^{15}$N in Sgr~B2(M).}
       \label{fig:HCN15M}
    \end{subfigure}
\quad
    \begin{subfigure}[t]{0.4\columnwidth}
       \includegraphics[width=1.0\columnwidth]{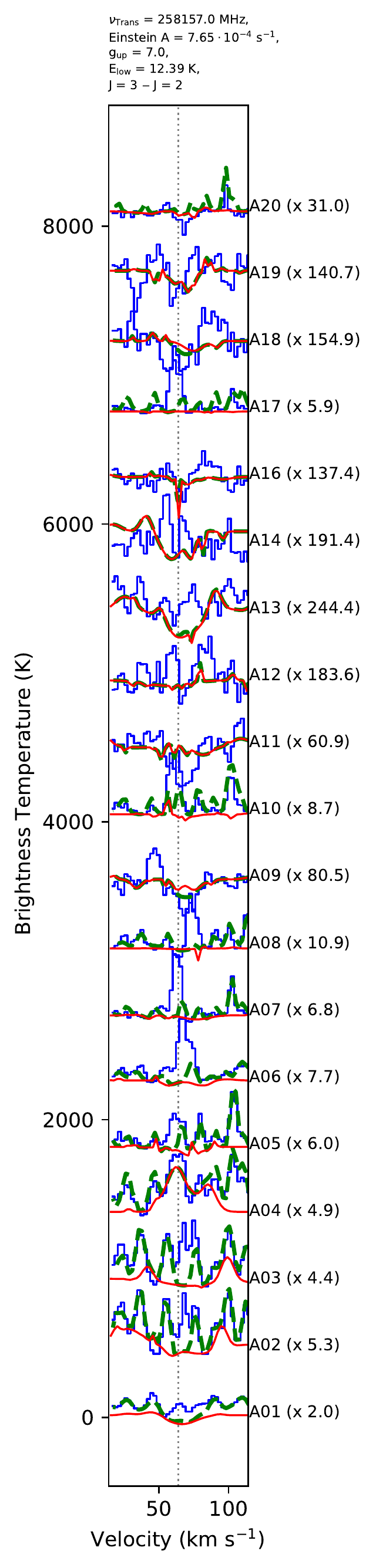}\\
       \caption{HC$^{15}$N in Sgr~B2(N).}
       \label{fig:HCN15N}
   \end{subfigure}
\quad
    \begin{subfigure}[t]{0.4\columnwidth}
       \includegraphics[width=1.0\columnwidth]{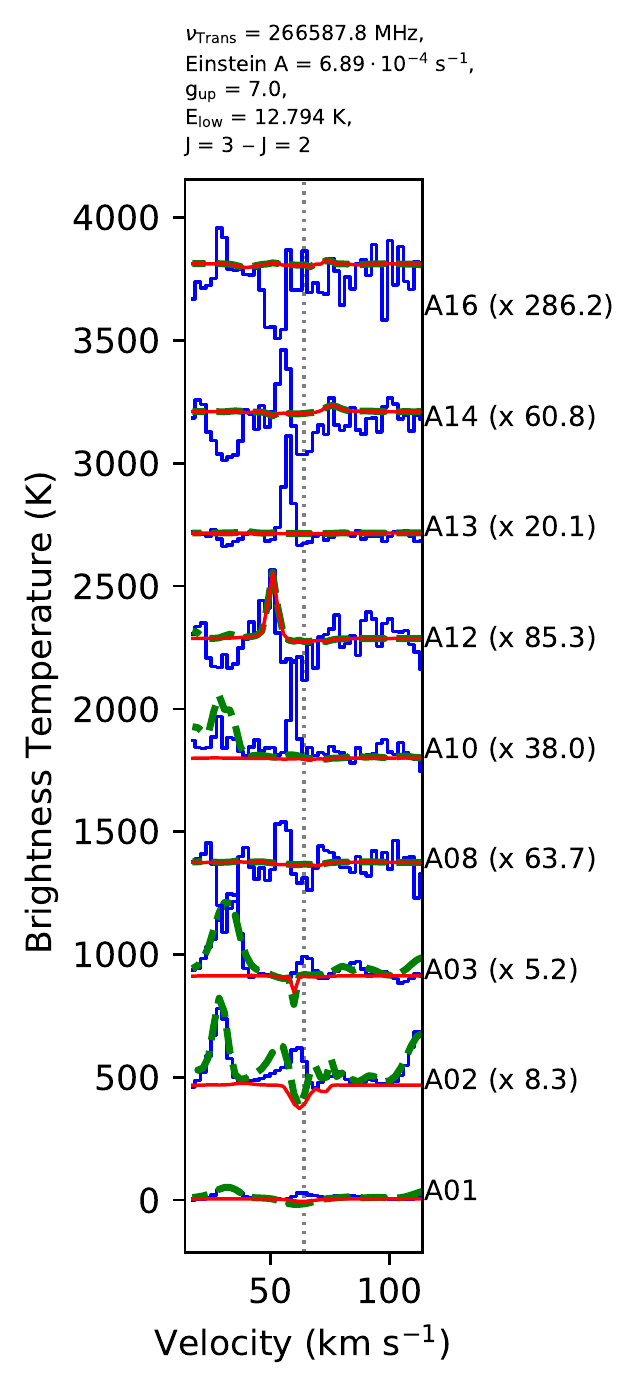}\\
       \caption{H$^{15}$NC in Sgr~B2(M).}
       \label{fig:HN15CM}
    \end{subfigure}
\quad
    \begin{subfigure}[t]{0.4\columnwidth}
       \includegraphics[width=1.0\columnwidth]{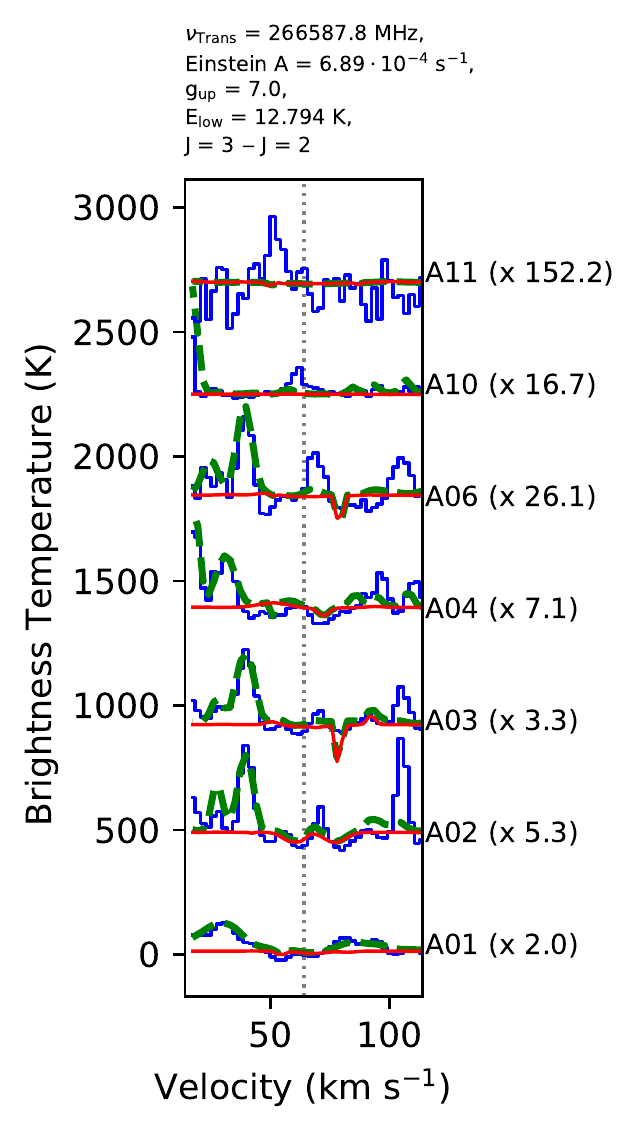}\\
       \caption{H$^{15}$NC in Sgr~B2(N).}
       \label{fig:HN15CN}
   \end{subfigure}
   \caption{Selected transitions of HC$^{15}$N and H$^{15}$NC in Sgr~B2(M) and N.}
   \ContinuedFloat
   \label{fig:HCN15NCMN}
\end{figure*}
\newpage
\clearpage

%*******************************************************************************
% Figure: HCN;v2=1;hyp1
\begin{figure*}[!htb]
    \centering
    \begin{subfigure}[t]{0.5\columnwidth}
       \includegraphics[width=1.0\columnwidth]{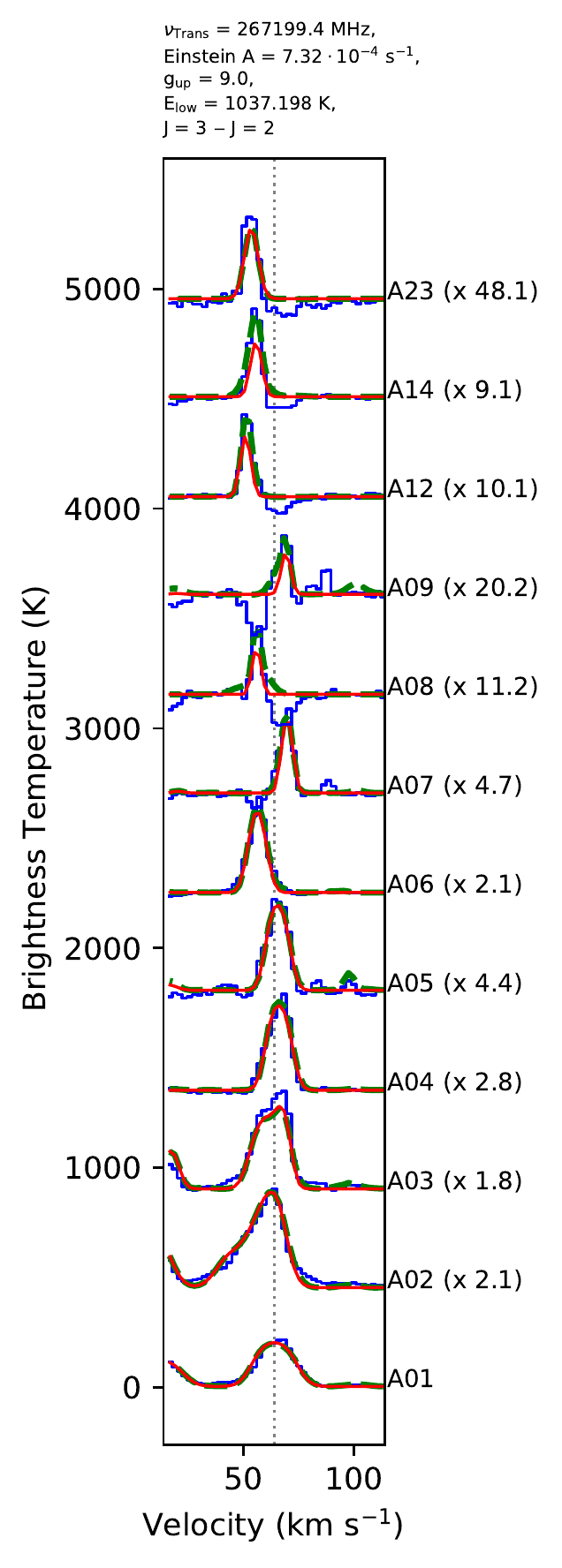}\\
       \caption{Sgr~B2(M)}
       \label{fig:HCNv21M}
    \end{subfigure}
\quad
    \begin{subfigure}[t]{0.5\columnwidth}
       \includegraphics[width=1.0\columnwidth]{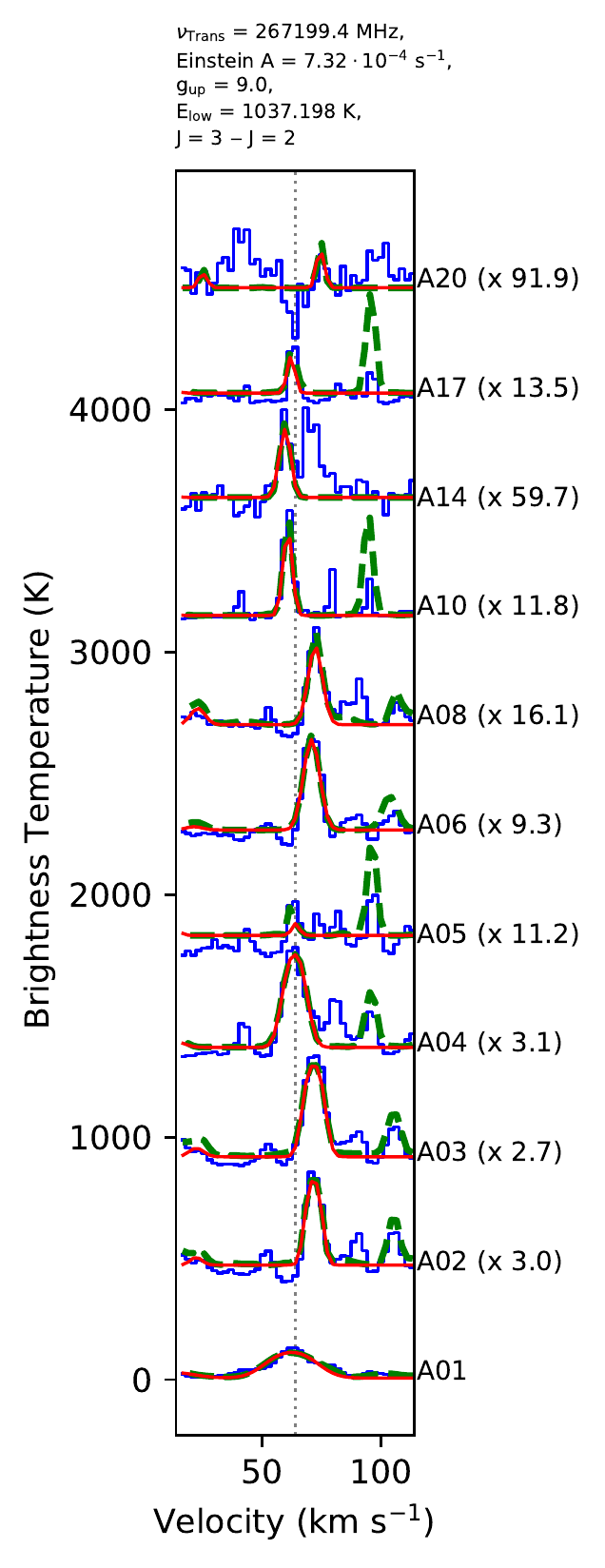}\\
       \caption{Sgr~B2(N)}
       \label{fig:HCNv21N}
   \end{subfigure}
   \caption{Selected transitions of HCN, v$_2$=1 in Sgr~B2(M) and N.}
   \ContinuedFloat
   \label{fig:HCNv21MN}
\end{figure*}
\newpage
\clearpage

%*******************************************************************************
% Figure: HCN;v2=2;hyp1
\begin{figure*}[!htb]
    \centering
    \begin{subfigure}[t]{0.5\columnwidth}
       \includegraphics[width=1.0\columnwidth]{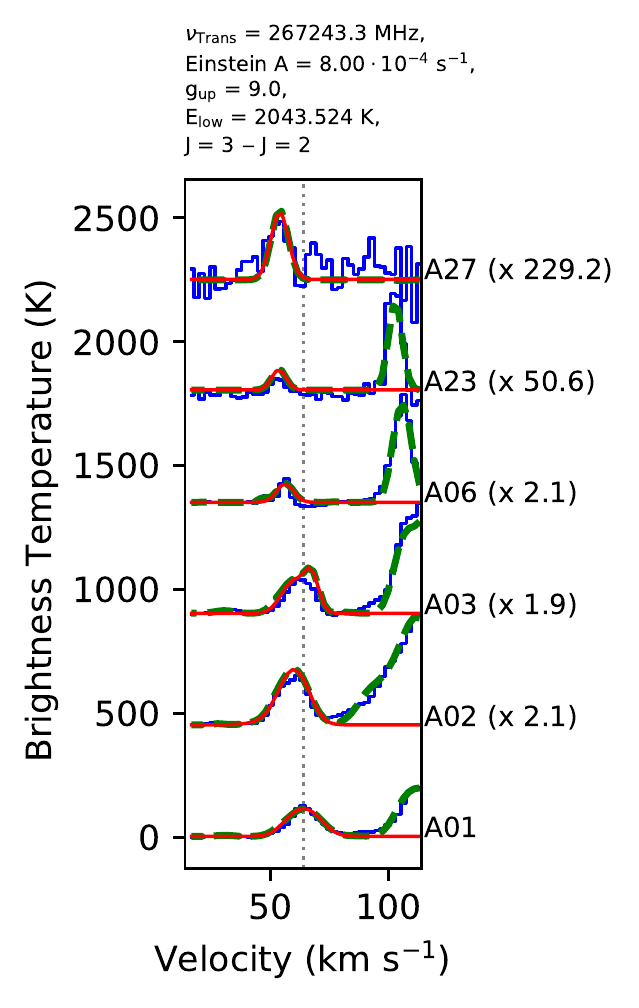}\\
       \label{fig:HCNv22M}
    \end{subfigure}
   \caption{Selected transitions of HCN, v$_2$ = 2 in Sgr~B2(M).}
   \ContinuedFloat
   \label{fig:HCNv22MN}
\end{figure*}

%*******************************************************************************
% Figure: HNC, v2=1
\begin{figure*}[!htb]
    \centering
    \begin{subfigure}[t]{0.5\columnwidth}
       \includegraphics[width=1.0\columnwidth]{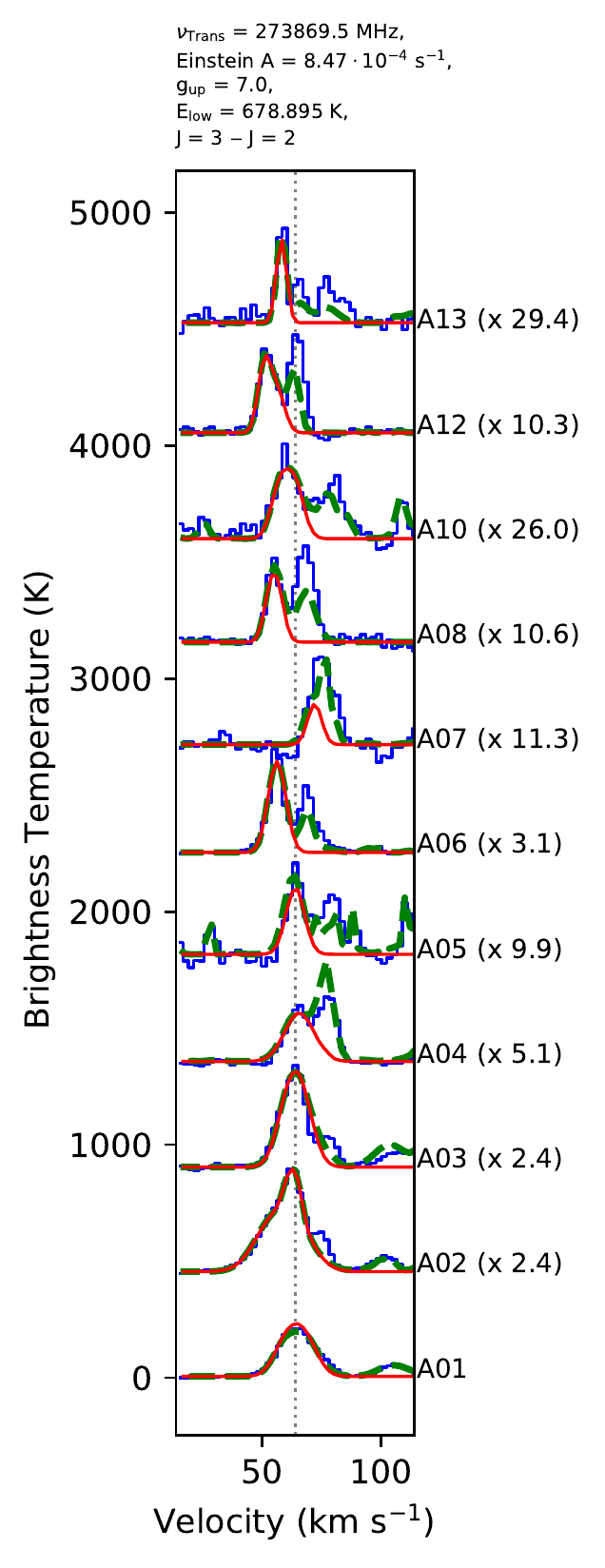}\\
       \caption{Sgr~B2(M)}
       \label{fig:HNCv21M}
    \end{subfigure}
\quad
    \begin{subfigure}[t]{0.5\columnwidth}
       \includegraphics[width=1.0\columnwidth]{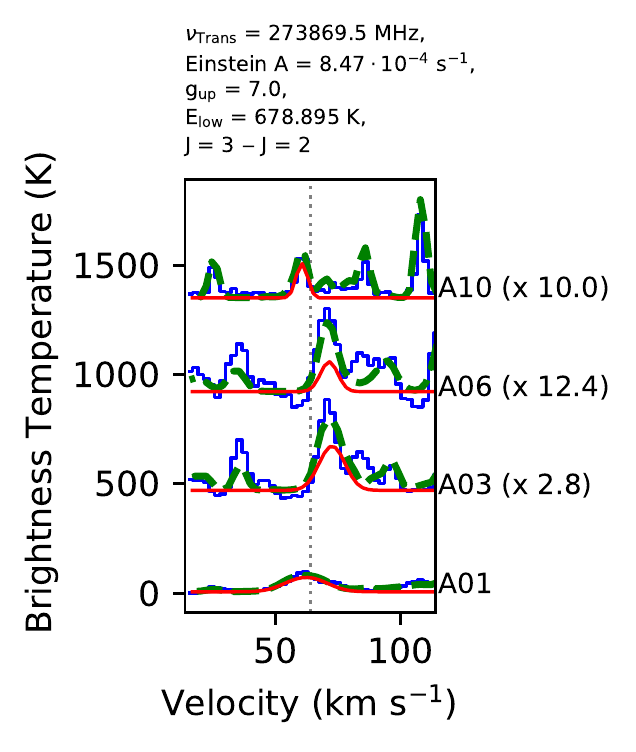}\\
       \caption{Sgr~B2(N)}
       \label{fig:HNCv21N}
   \end{subfigure}
   \caption{Selected transitions of HNC, v$_2$=1 in Sgr~B2(M) and N.}
   \ContinuedFloat
   \label{fig:HNCv21MN}
\end{figure*}
\newpage
\clearpage

%*******************************************************************************
% Figure: CH3CN;v=0;
\begin{figure*}[!htb]
    \centering
    \begin{subfigure}[t]{1.0\columnwidth}
       \includegraphics[width=1.0\columnwidth]{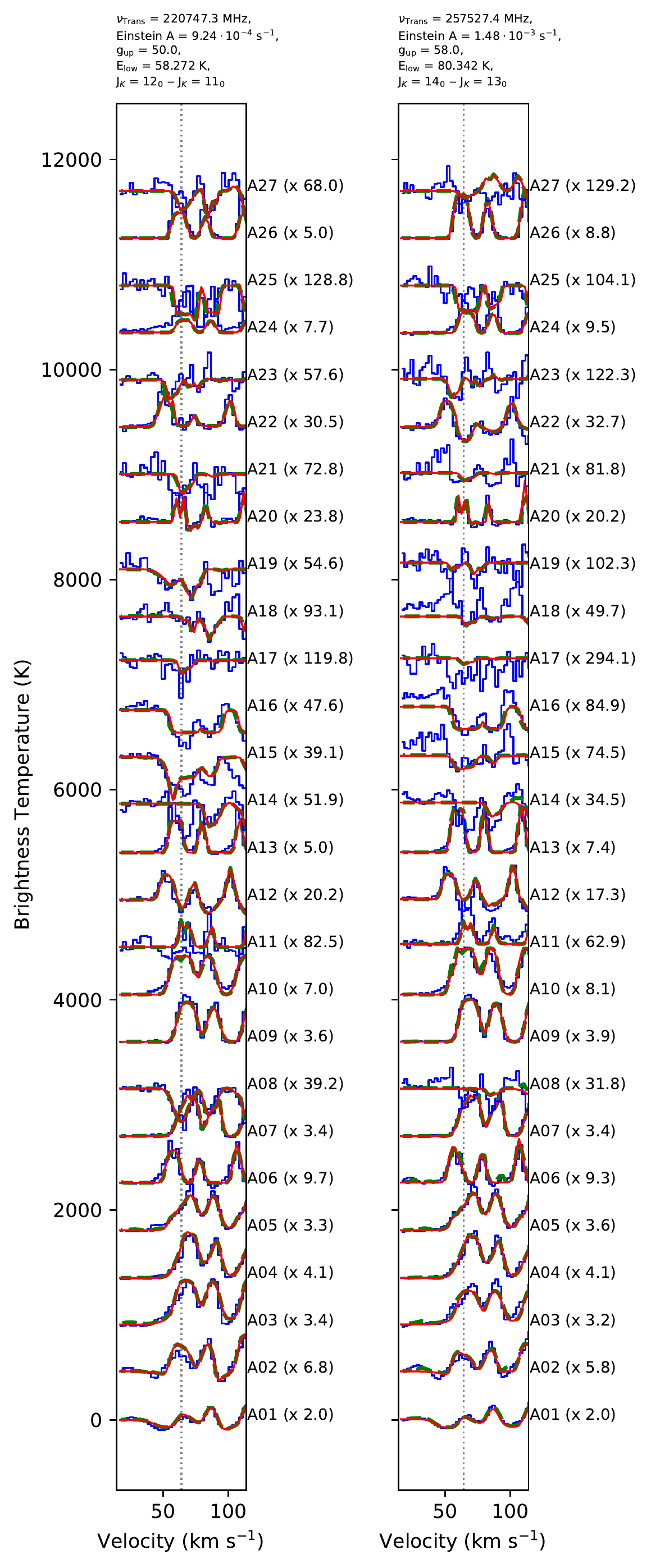}\\
       \caption{Sgr~B2(M)}
       \label{fig:CH3CNM}
    \end{subfigure}
\quad
    \begin{subfigure}[t]{1.0\columnwidth}
       \includegraphics[width=1.0\columnwidth]{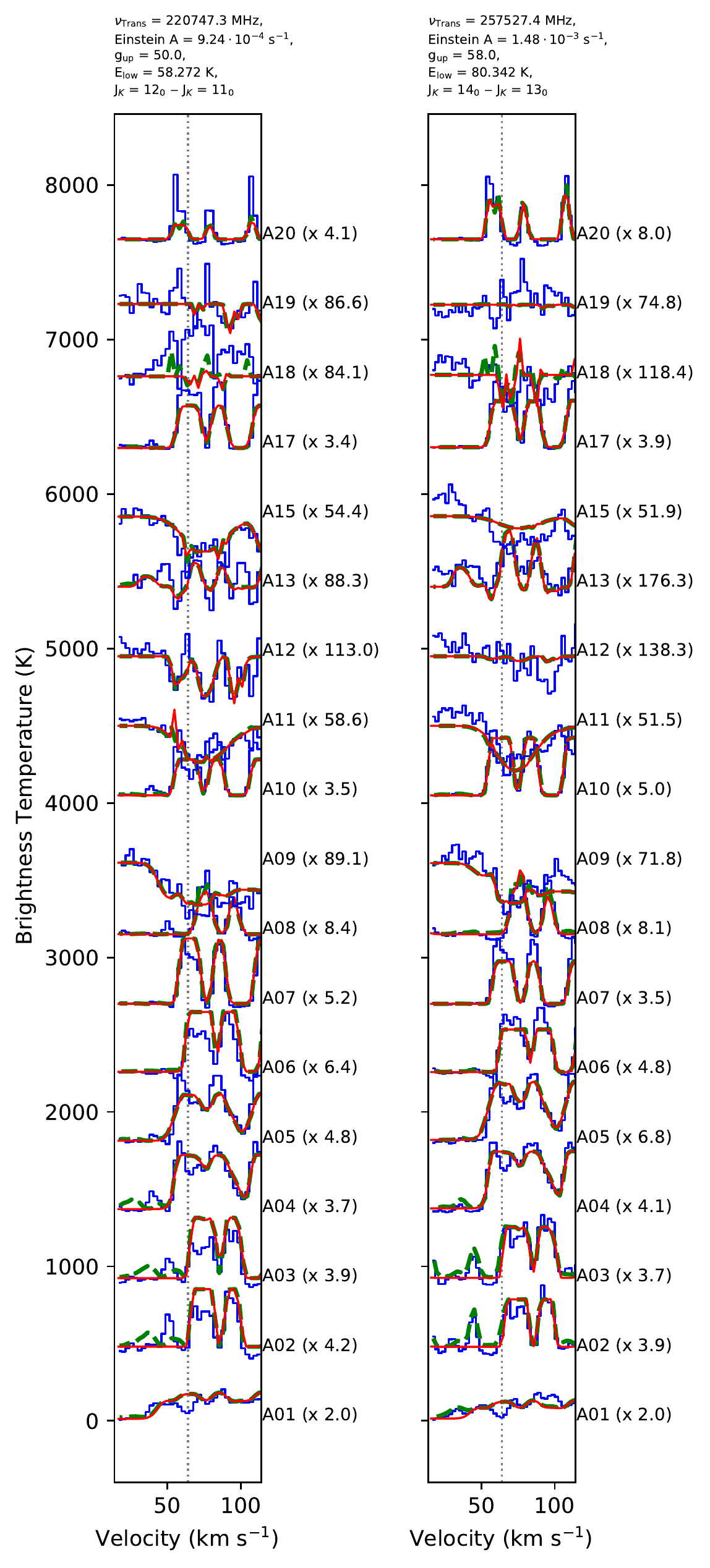}\\
       \caption{Sgr~B2(N)}
       \label{fig:CH3CNN}
   \end{subfigure}
   \caption{Selected transitions of CH$_3$CN in Sgr~B2(M) and N.}
   \ContinuedFloat
   \label{fig:CH3CNMN}
\end{figure*}
\newpage
\clearpage

%*******************************************************************************
% Figure: C-13-H3CN;v=0;
\begin{figure*}[!htb]
    \centering
    \begin{subfigure}[t]{1.0\columnwidth}
       \includegraphics[width=1.0\columnwidth]{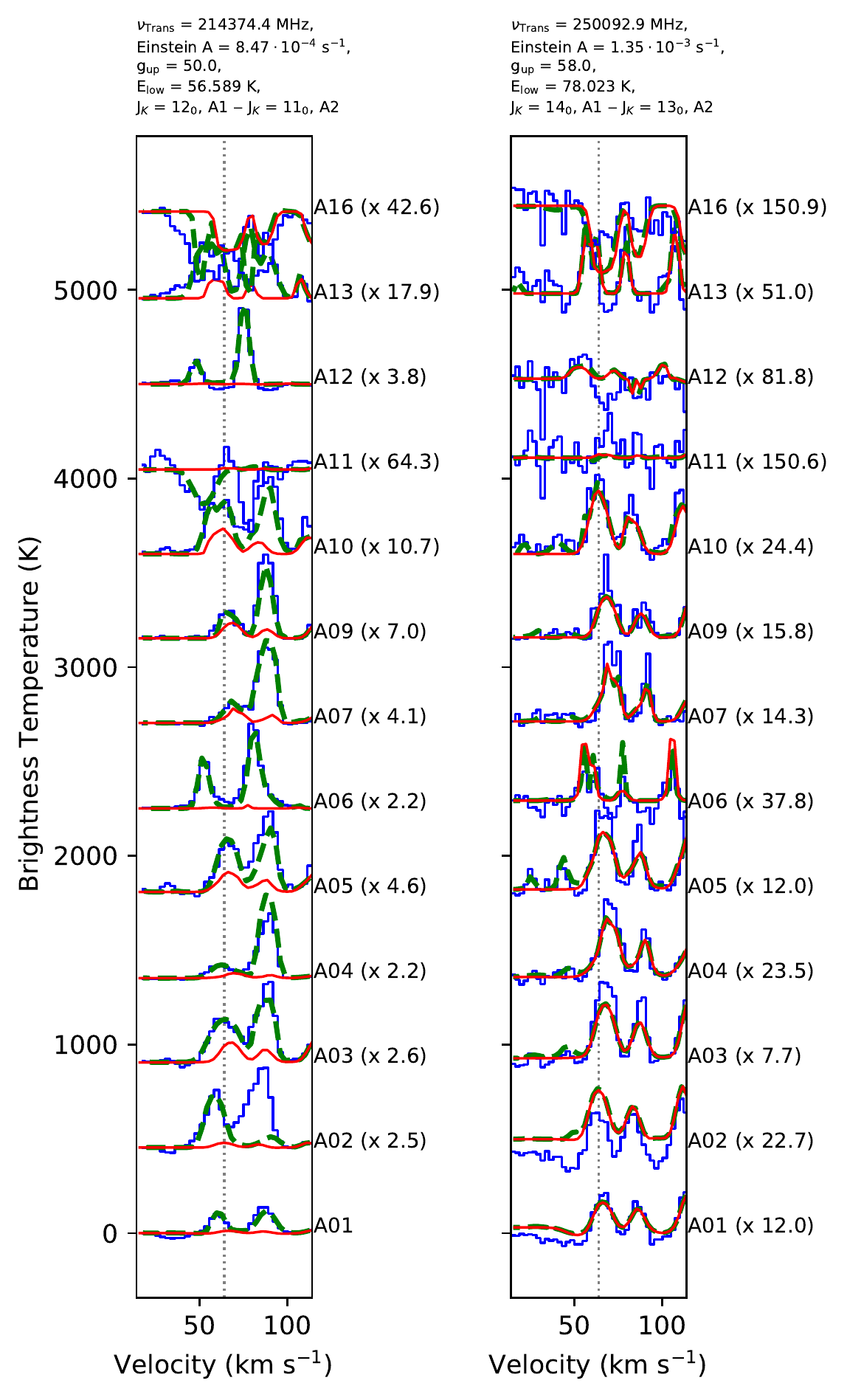}\\
       \caption{Sgr~B2(M)}
       \label{fig:C13H3CNM}
    \end{subfigure}
\quad
    \begin{subfigure}[t]{1.0\columnwidth}
       \includegraphics[width=1.0\columnwidth]{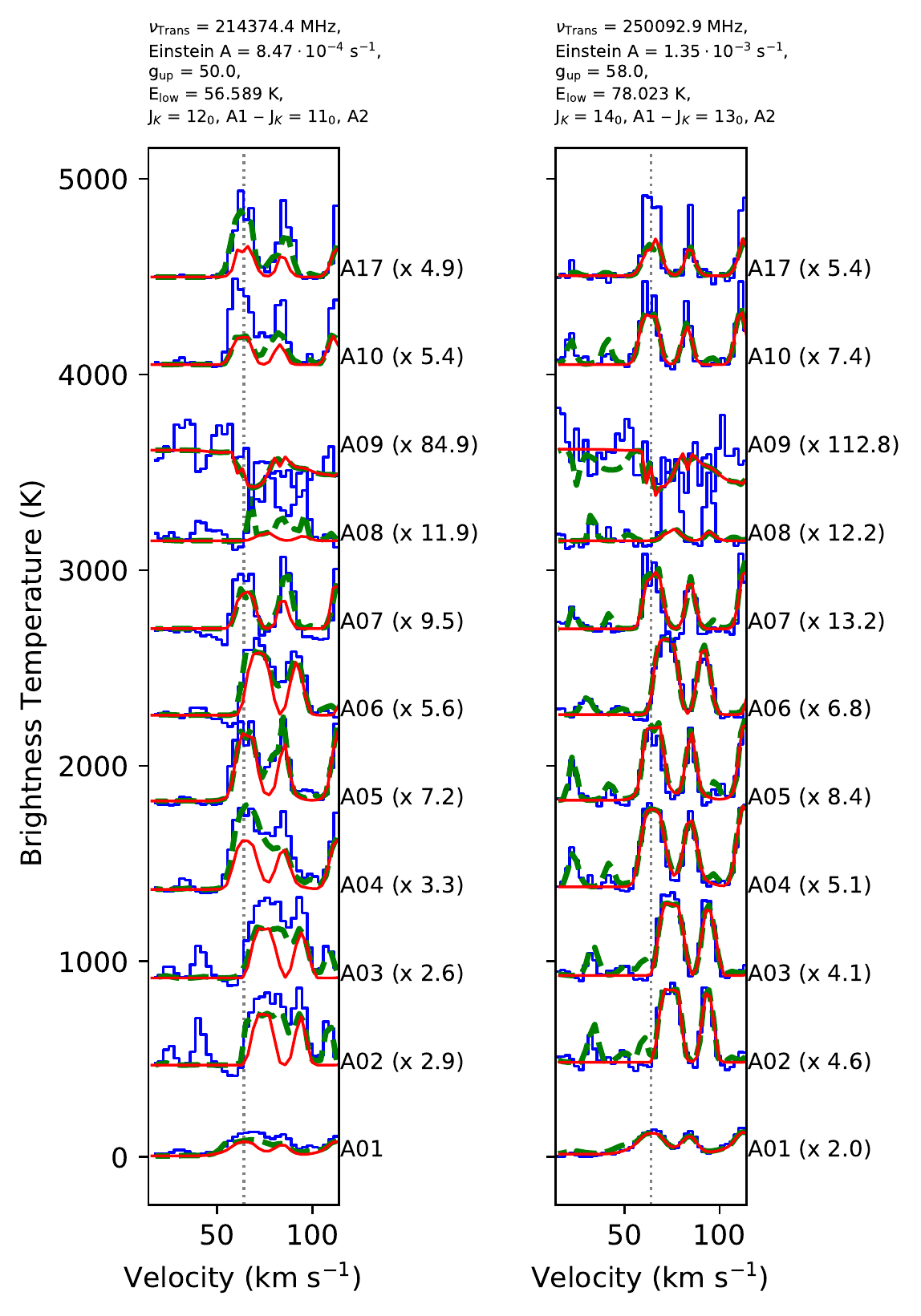}\\
       \caption{Sgr~B2(N)}
       \label{fig:C13H3CNN}
   \end{subfigure}
   \caption{Selected transitions of $^{13}$CH$_3$CN in Sgr~B2(M) and N.}
   \ContinuedFloat
   \label{fig:C13H3CNMN}
\end{figure*}
\newpage
\clearpage

%*******************************************************************************
% Figure: CH3C-13-N;v=0;
\begin{figure*}[!htb]
    \centering
    \begin{subfigure}[t]{1.0\columnwidth}
       \includegraphics[width=1.0\columnwidth]{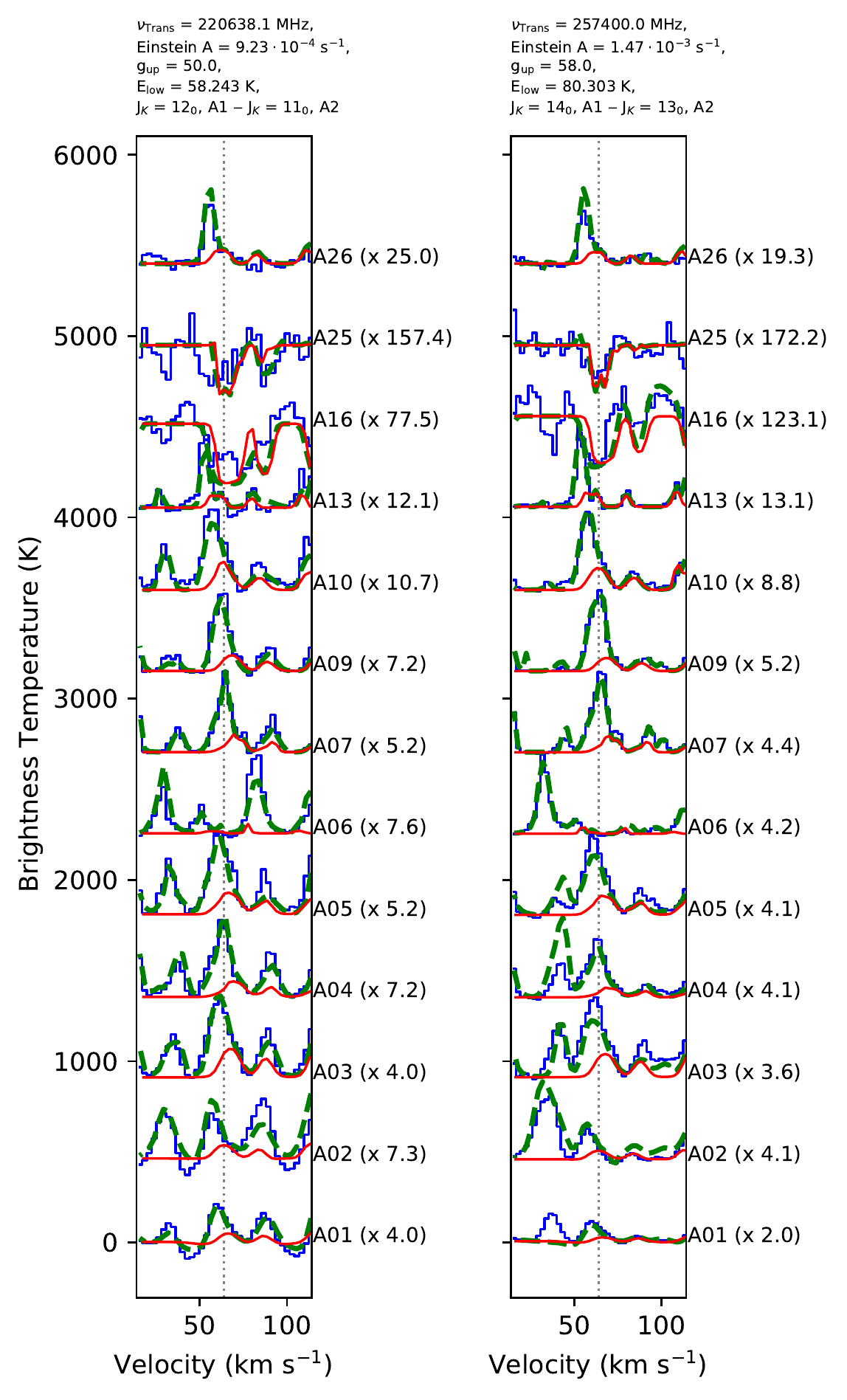}\\
       \caption{Sgr~B2(M)}
       \label{fig:CH3C13NM}
    \end{subfigure}
\quad
    \begin{subfigure}[t]{1.0\columnwidth}
       \includegraphics[width=1.0\columnwidth]{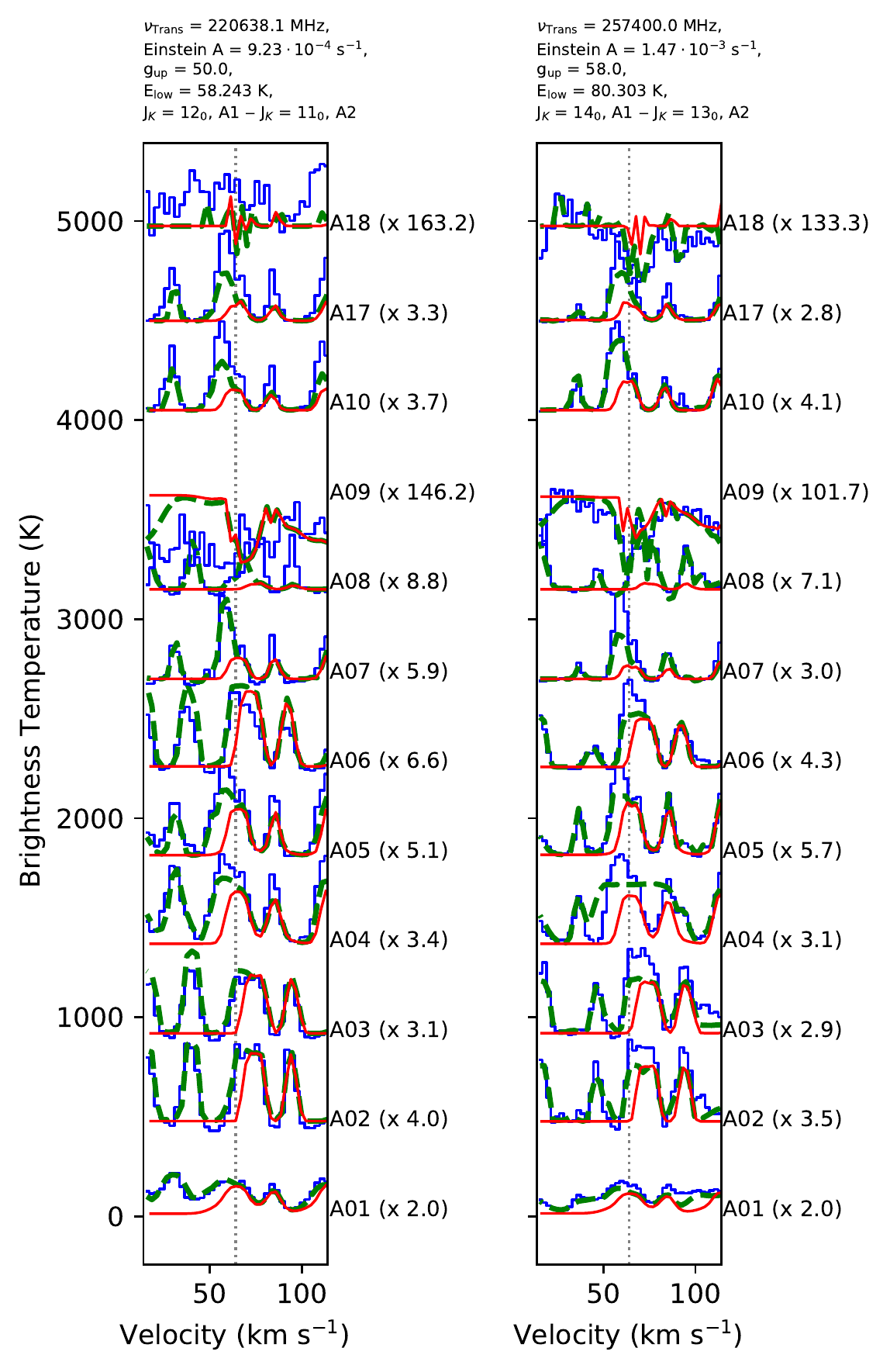}\\
       \caption{Sgr~B2(N)}
       \label{fig:CH3C13NN}
   \end{subfigure}
   \caption{Selected transitions of CH$_3 \! ^{13}$CN in Sgr~B2(M) and N.}
   \ContinuedFloat
   \label{fig:CH3C13NMN}
\end{figure*}
\newpage
\clearpage

%*******************************************************************************
% Figure: CH3CN;v8=1;#1
\begin{figure*}[!htb]
    \centering
    \begin{subfigure}[t]{1.0\columnwidth}
       \includegraphics[width=1.0\columnwidth]{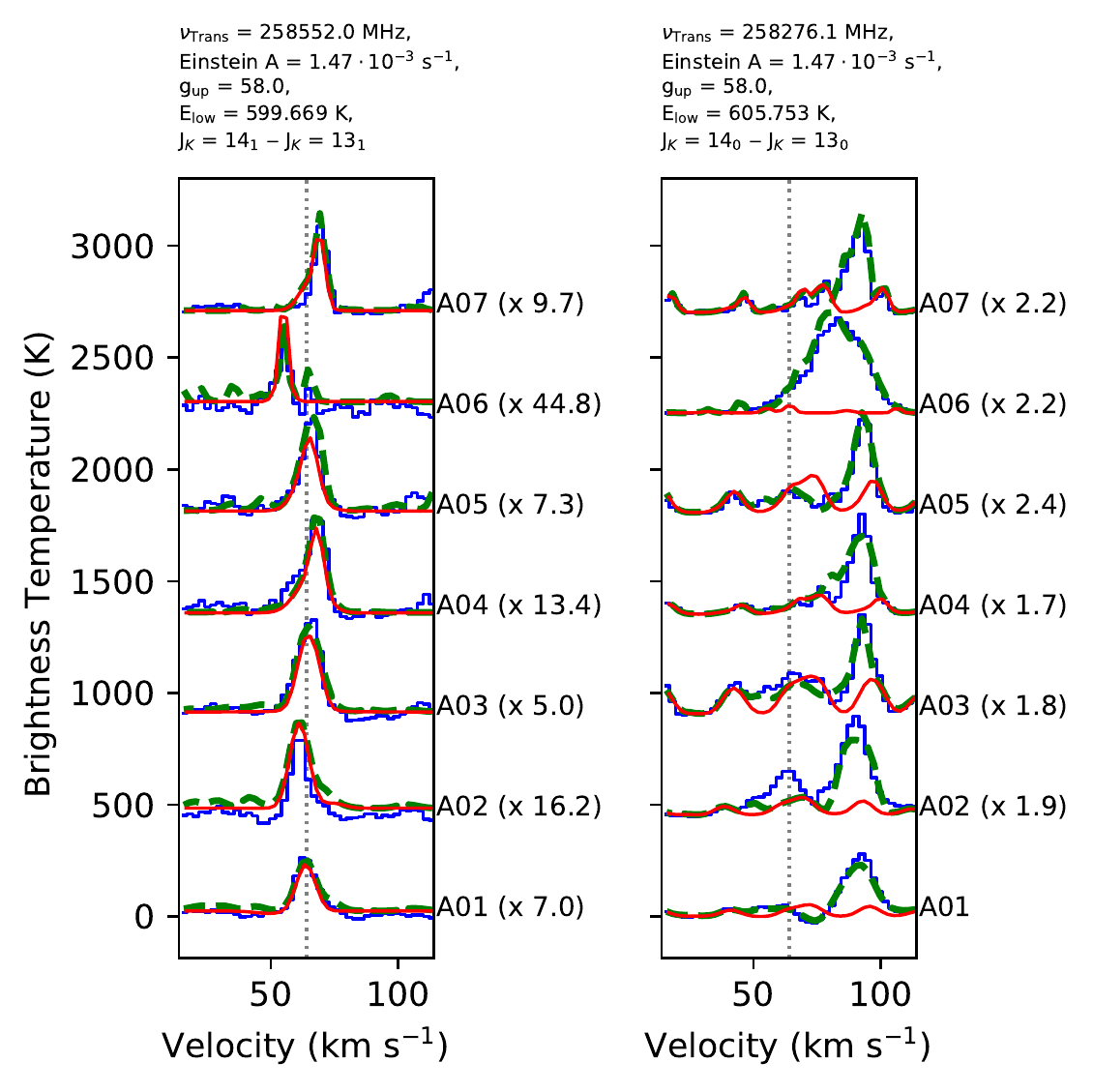}\\
       \caption{Sgr~B2(M)}
       \label{fig:CH3CNv81M}
    \end{subfigure}
\quad
    \begin{subfigure}[t]{1.0\columnwidth}
       \includegraphics[width=1.0\columnwidth]{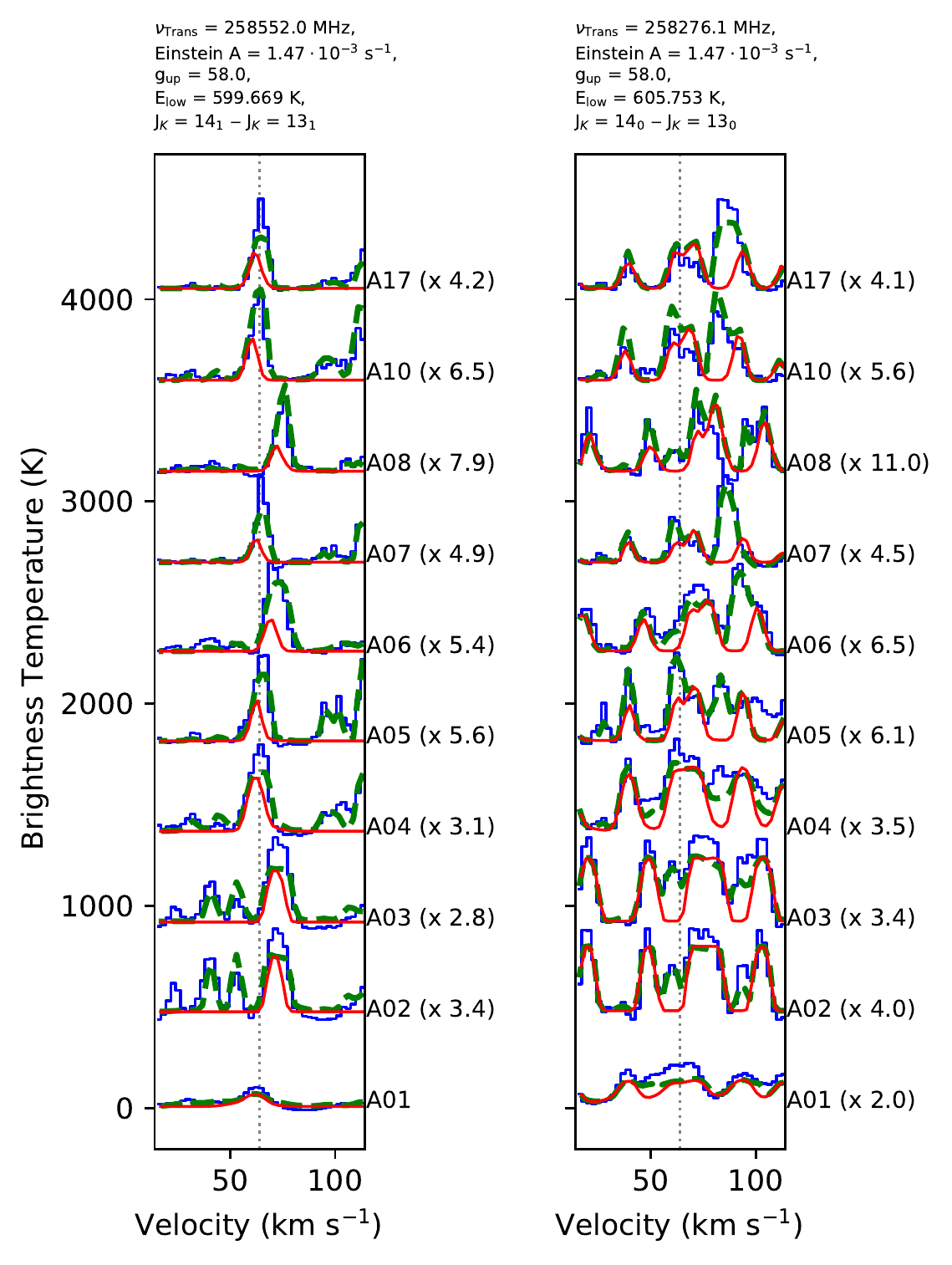}\\
       \caption{Sgr~B2(N)}
       \label{fig:CH3CNv81N}
   \end{subfigure}
   \caption{Selected transitions of CH$_3$CN, v$_8$=1 in Sgr~B2(M) and N.}
   \ContinuedFloat
   \label{fig:CH3CNv81MN}
\end{figure*}

%*******************************************************************************
% Figure: CH3CN;v8=2;
\begin{figure*}[!htb]
    \centering
    \begin{subfigure}[t]{1.0\columnwidth}
       \includegraphics[width=1.0\columnwidth]{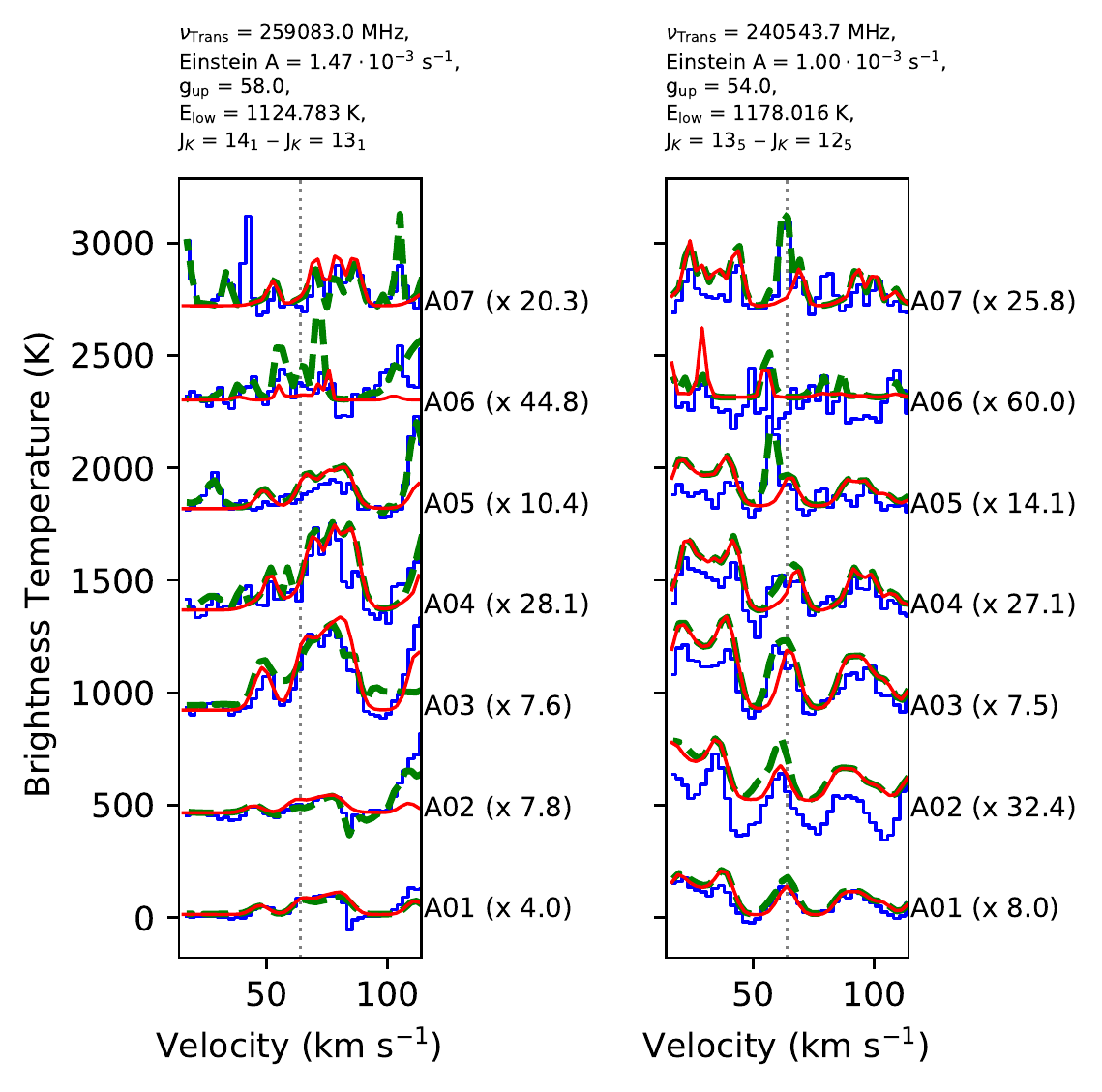}\\
       \caption{Sgr~B2(M)}
       \label{fig:CH3CNv82M}
    \end{subfigure}
\quad
    \begin{subfigure}[t]{1.0\columnwidth}
       \includegraphics[width=1.0\columnwidth]{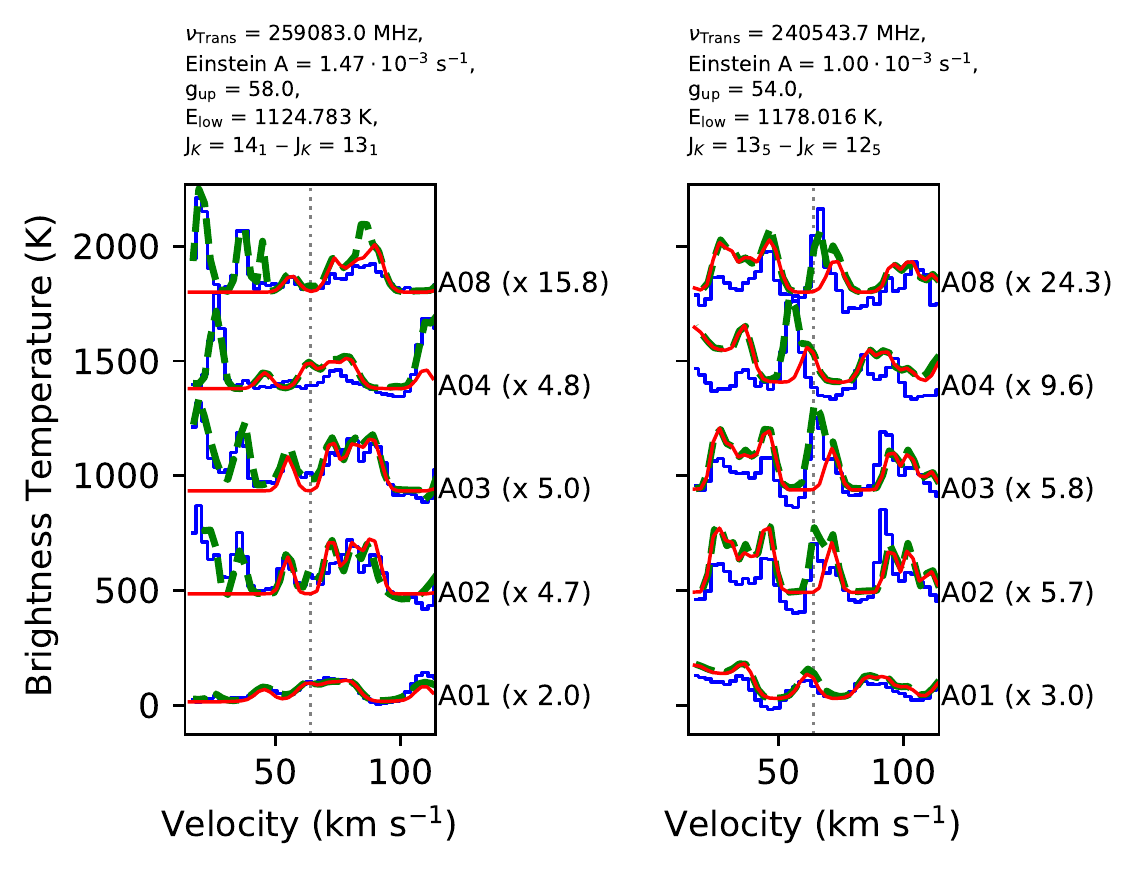}\\
       \caption{Sgr~B2(N)}
       \label{fig:CH3CNv82N}
   \end{subfigure}
   \caption{Selected transitions of CH$_3$CN, v$_8$=2 in Sgr~B2(M) and N.}
   \ContinuedFloat
   \label{fig:CH3CNv82MN}
\end{figure*}
\newpage
\clearpage

%*******************************************************************************
% Figure: C2H3CN;v=0;private
\begin{figure*}[!htb]
    \centering
    \begin{subfigure}[t]{1.0\columnwidth}
       \includegraphics[width=1.0\columnwidth]{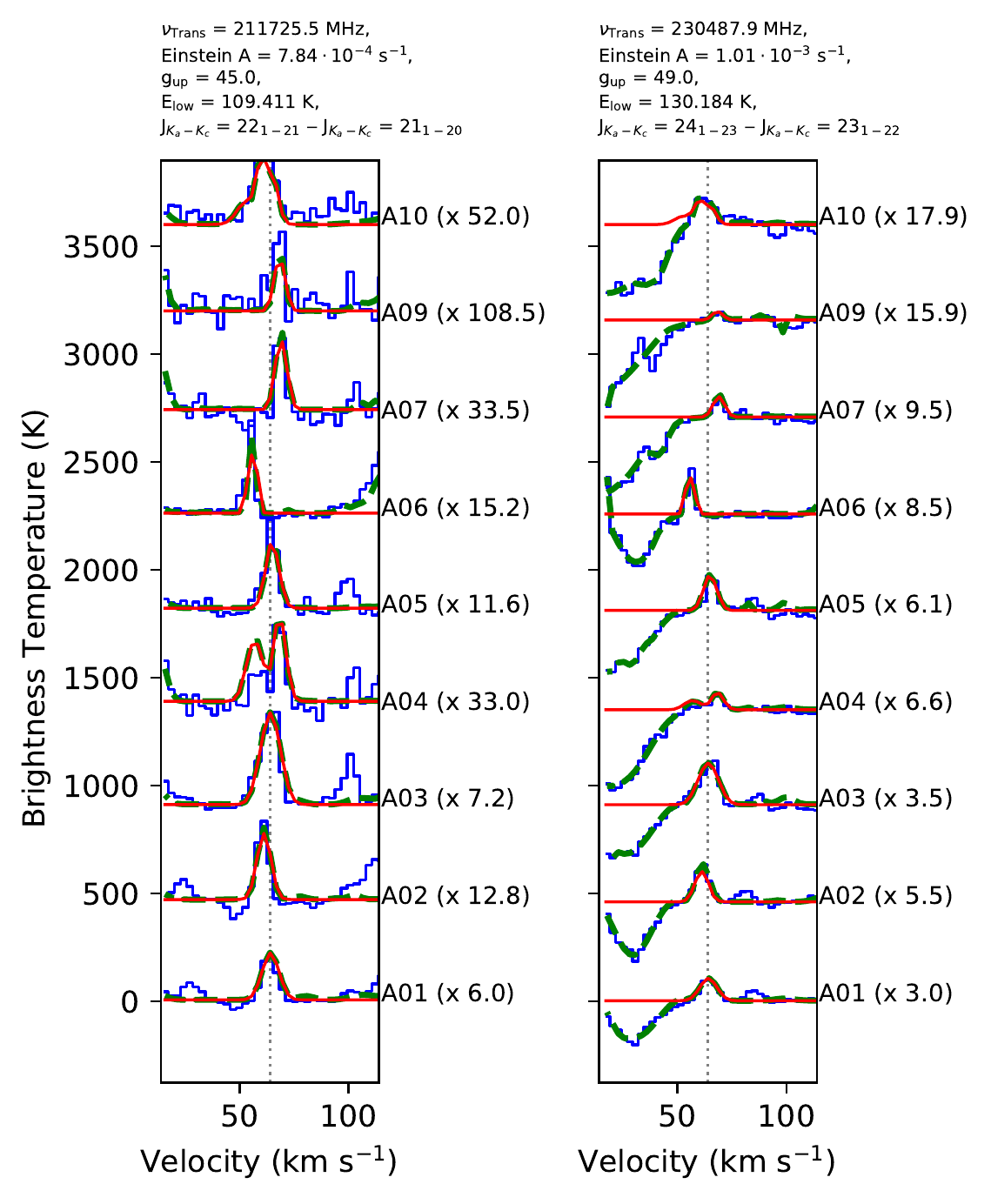}\\
       \caption{Sgr~B2(M)}
       \label{fig:C2H3CNM}
    \end{subfigure}
\quad
    \begin{subfigure}[t]{1.0\columnwidth}
       \includegraphics[width=1.0\columnwidth]{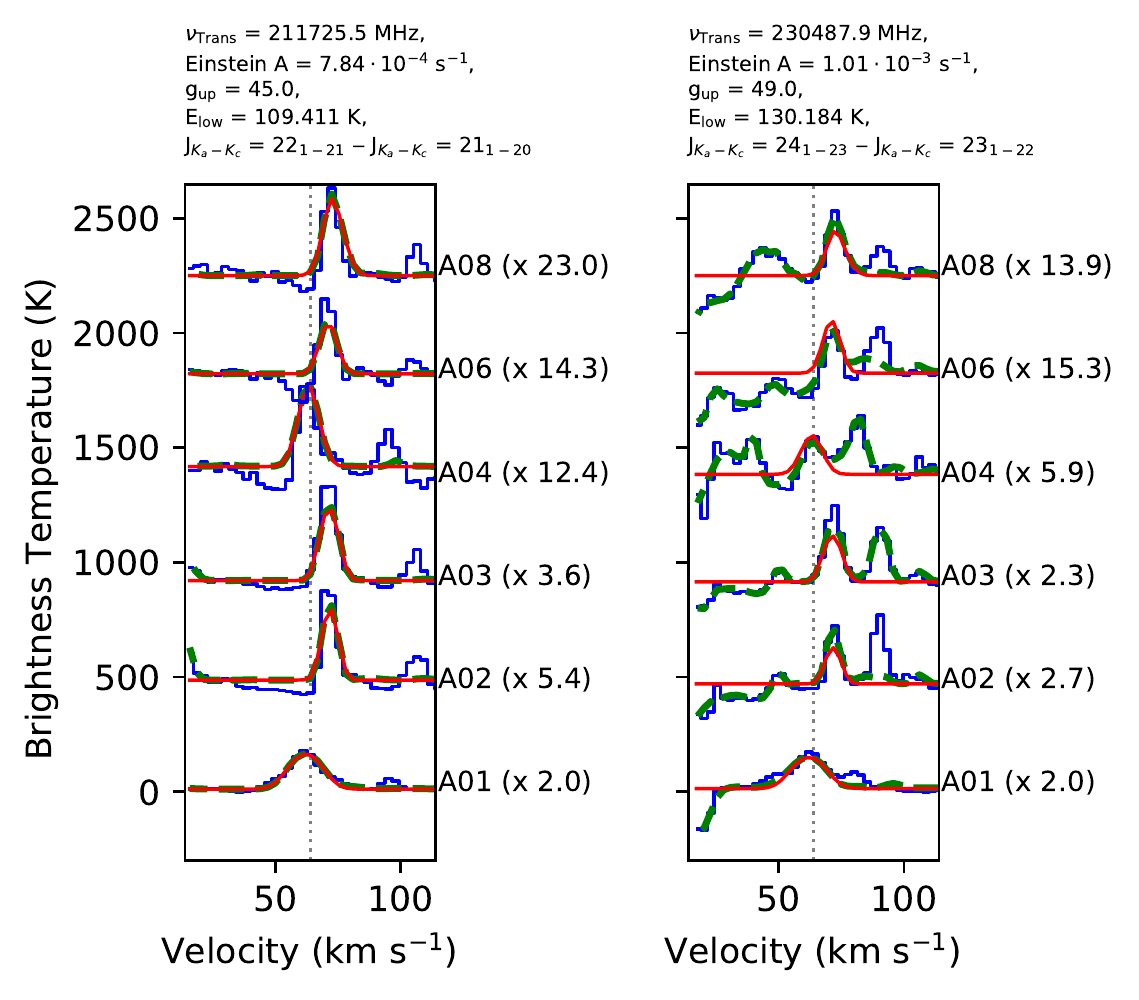}\\
       \caption{Sgr~B2(N)}
       \label{fig:C2H3CNN}
   \end{subfigure}
   \caption{Selected transitions of C$_2$H$_3$CN in Sgr~B2(M) and N.}
   \ContinuedFloat
   \label{fig:C2H3CNMN}
\end{figure*}

%*******************************************************************************
% Figure: C2H3CN;v10=1;private
\begin{figure*}[!htb]
    \centering
    \begin{subfigure}[t]{1.0\columnwidth}
       \includegraphics[width=1.0\columnwidth]{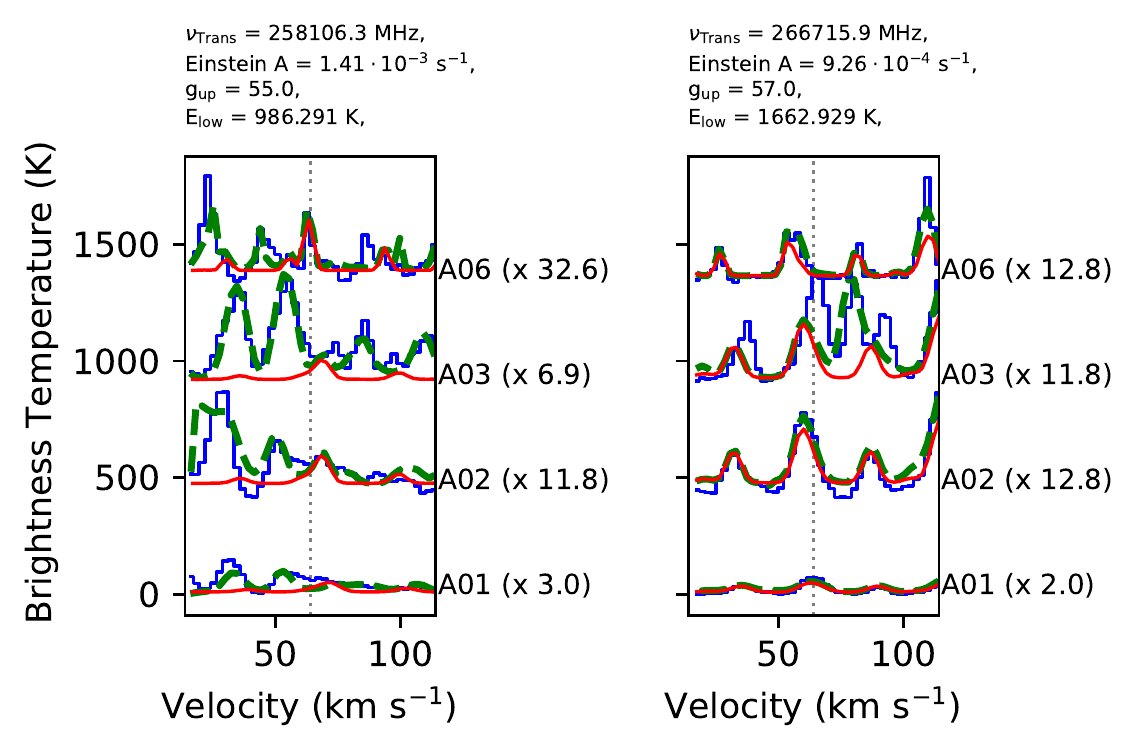}\\
       \caption{Sgr~B2(M)}
       \label{fig:C2H3CNv101M}
    \end{subfigure}
\quad
    \begin{subfigure}[t]{1.0\columnwidth}
        \includegraphics[width=1.0\columnwidth]{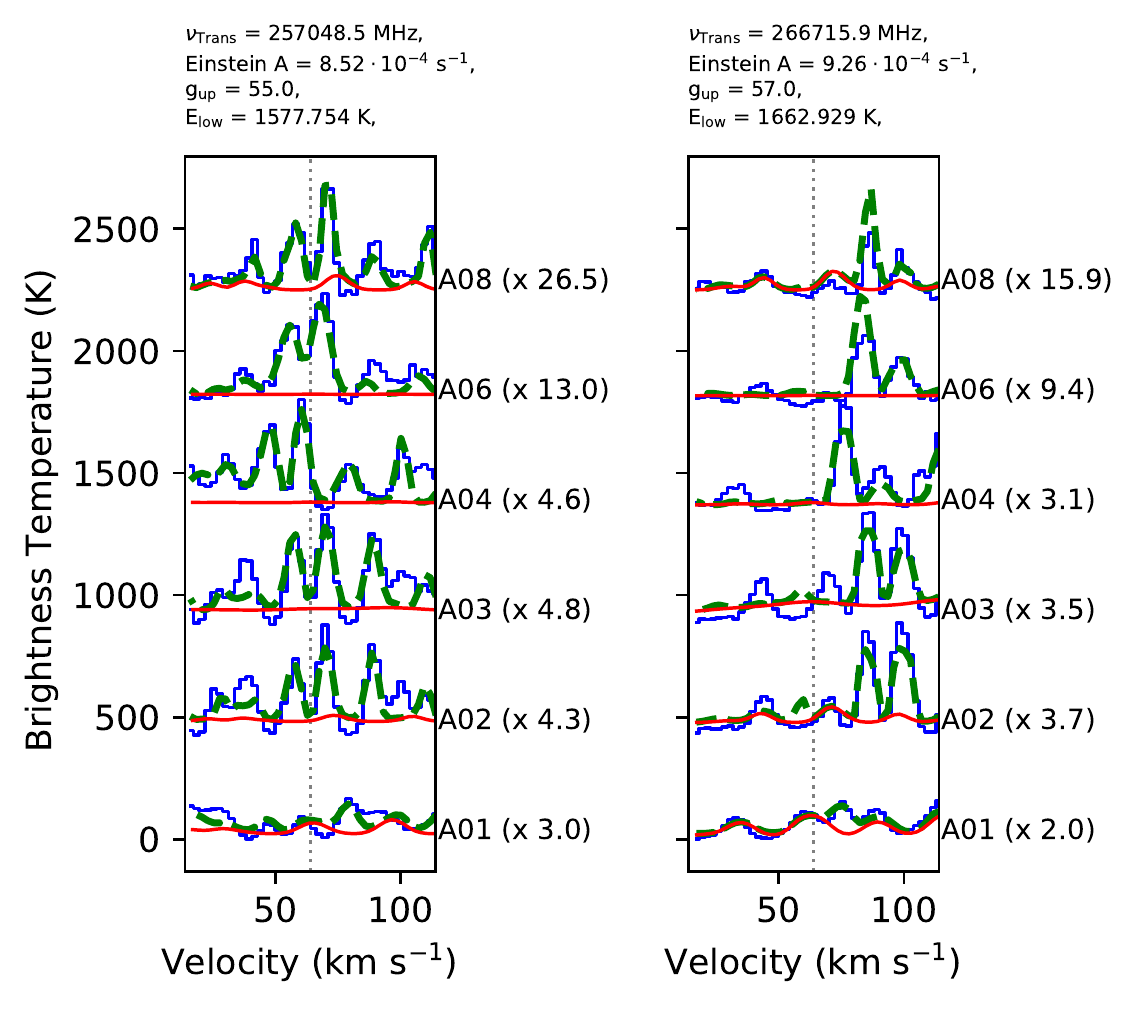}\\
        \caption{Sgr~B2(N)}
        \label{fig:C2H3CNv101N}
    \end{subfigure}
    \caption{Selected transitions of C$_2$H$_3$CN, v$_{10}$=1 in Sgr~B2(M) and N.}
    \ContinuedFloat
    \label{fig:C2H3CNv101MN}
\end{figure*}

%*******************************************************************************
% Figure: C2H3CN;v11=1;private
\begin{figure*}[!htb]
    \centering
    \begin{subfigure}[t]{1.0\columnwidth}
       \includegraphics[width=1.0\columnwidth]{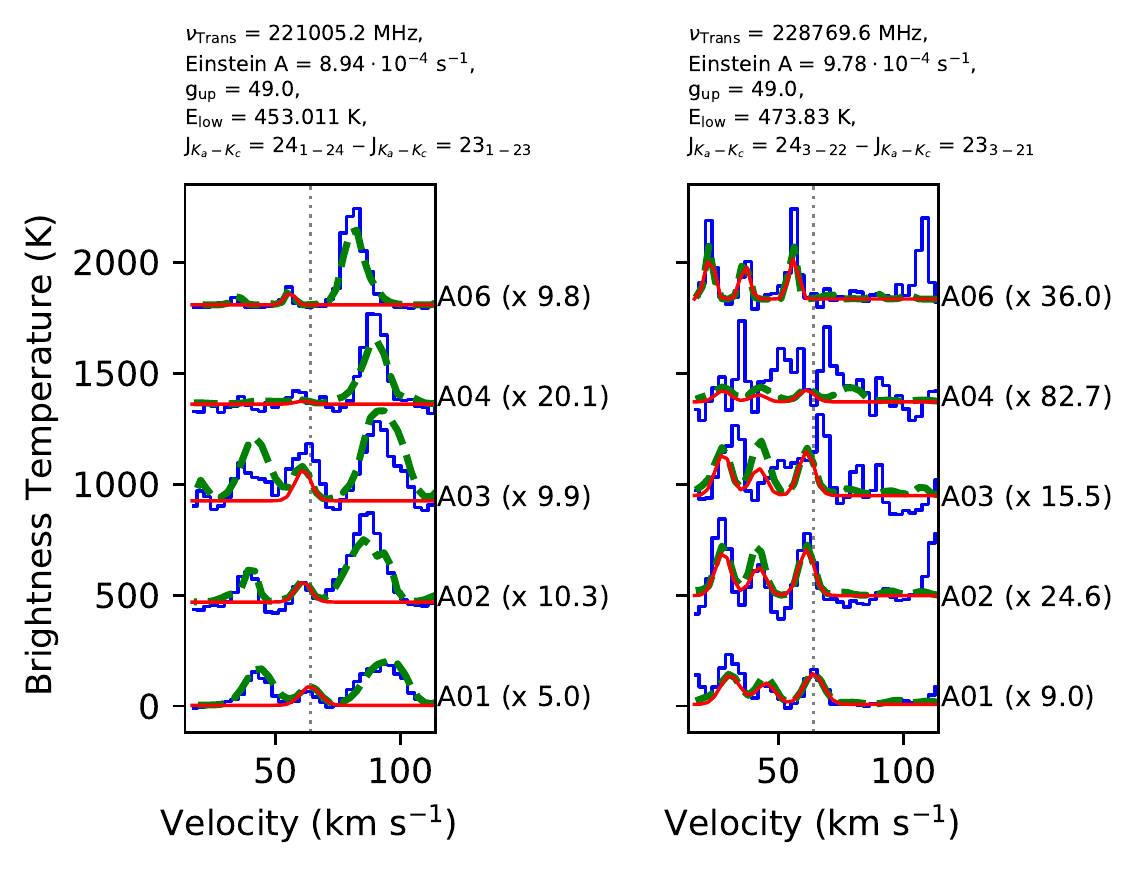}\\
       \caption{Sgr~B2(M)}
       \label{fig:C2H3CNv111M}
    \end{subfigure}
\quad
    \begin{subfigure}[t]{1.0\columnwidth}
       \includegraphics[width=1.0\columnwidth]{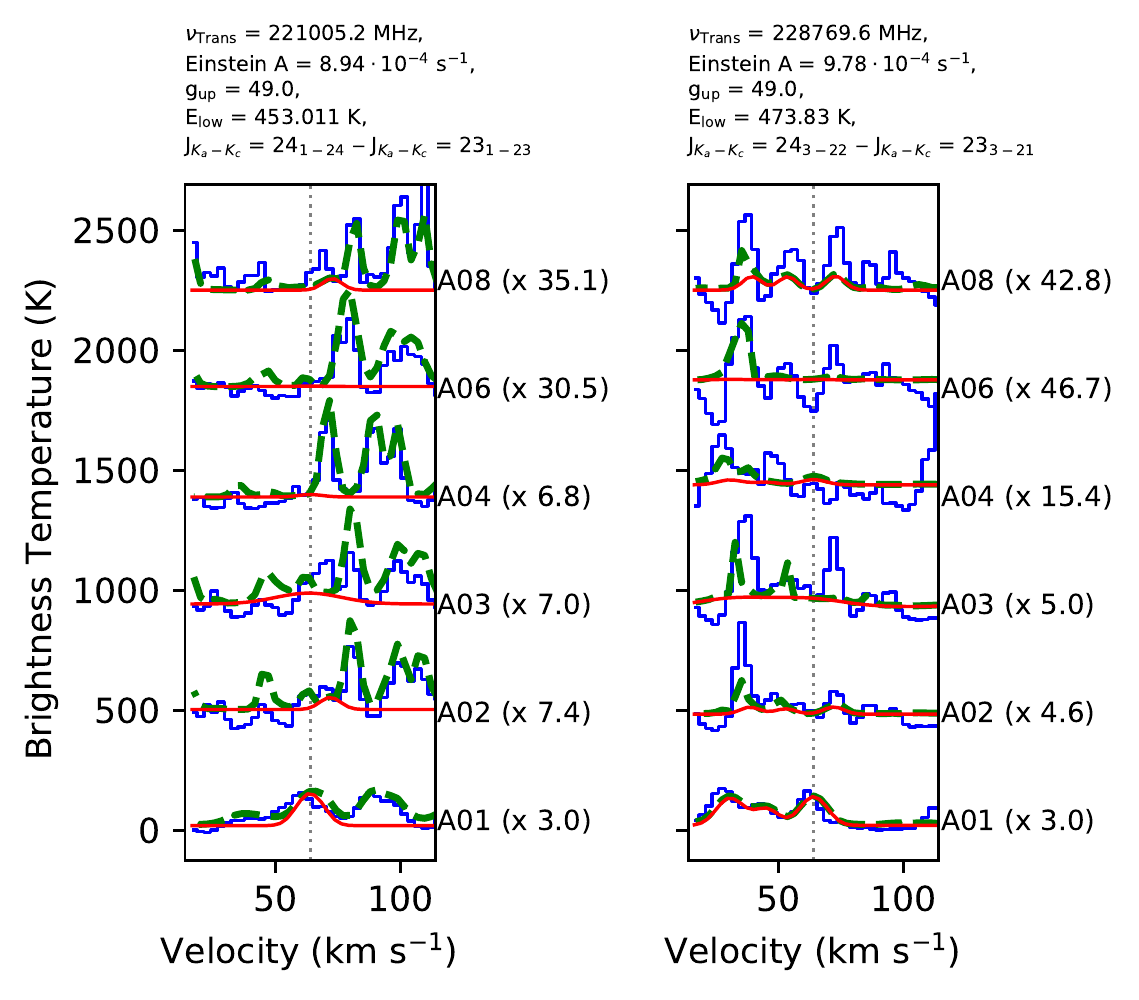}\\
       \caption{Sgr~B2(N)}
       \label{fig:C2H3CNv111N}
   \end{subfigure}
   \caption{Selected transitions of C$_2$H$_3$CN, v$_{11}$=1 in Sgr~B2(M) and N.}
   \ContinuedFloat
   \label{fig:C2H3CNv111MN}
\end{figure*}

%*******************************************************************************
% Figure: C2H3CN;v11=2;private
\begin{figure*}[!htb]
    \centering
    \begin{subfigure}[t]{1.0\columnwidth}
       \includegraphics[width=1.0\columnwidth]{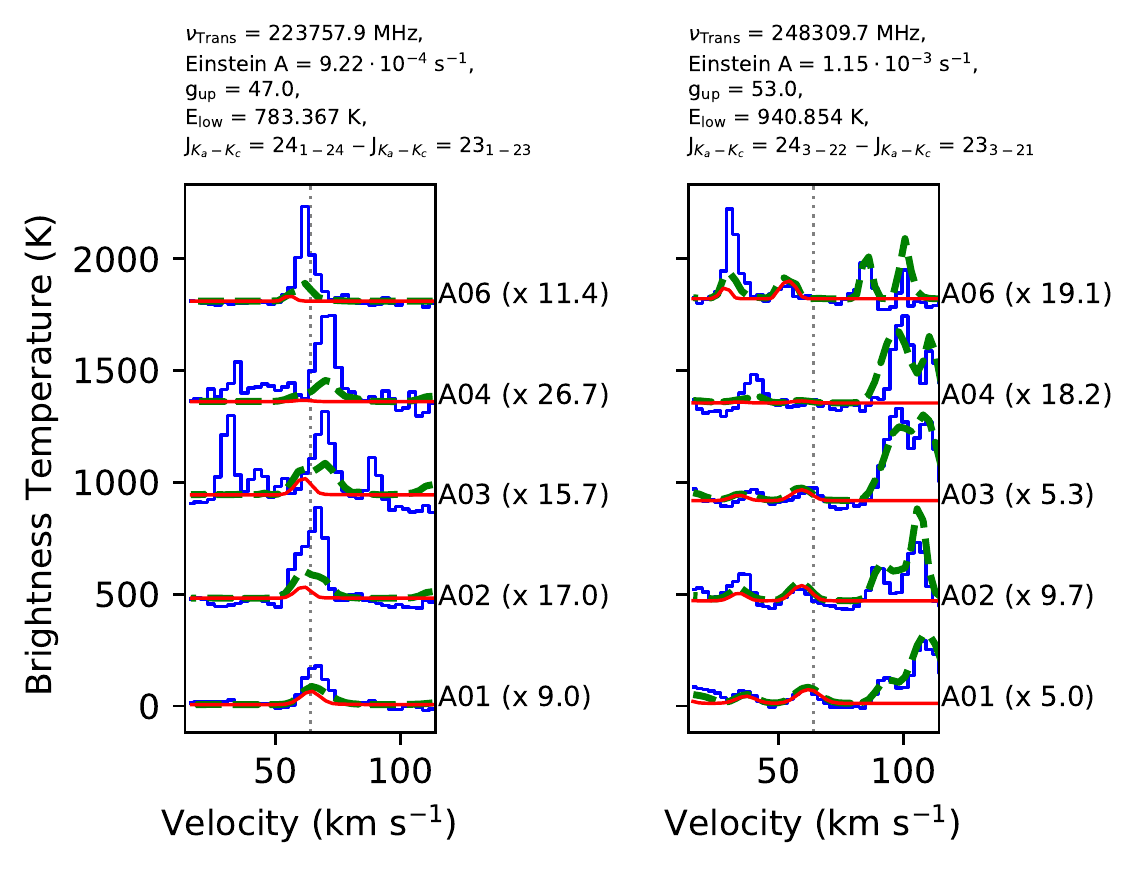}\\
       \caption{Sgr~B2(M)}
       \label{fig:C2H3CNv112M}
    \end{subfigure}
\quad
    \begin{subfigure}[t]{1.0\columnwidth}
       \includegraphics[width=1.0\columnwidth]{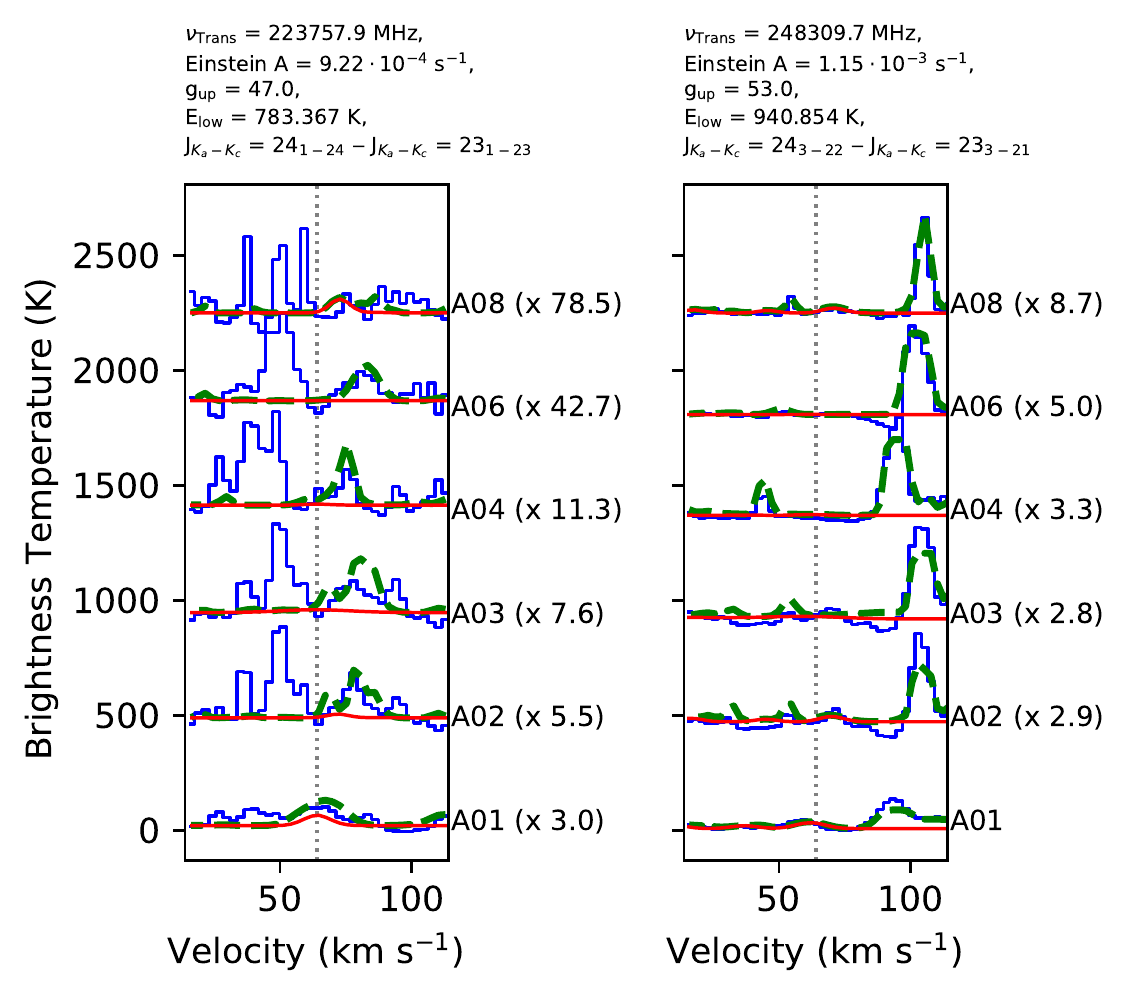}\\
       \caption{Sgr~B2(N)}
       \label{fig:C2H3CNv112N}
   \end{subfigure}
   \caption{Selected transitions of C$_2$H$_3$CN, v$_{11}$=2 in Sgr~B2(M) and N.}
   \ContinuedFloat
   \label{fig:C2H3CNv112MN}
\end{figure*}

%*******************************************************************************
% Figure: C2H3CN;v15=1;private
\begin{figure*}[!htb]
    \centering
    \begin{subfigure}[t]{1.0\columnwidth}
       \includegraphics[width=1.0\columnwidth]{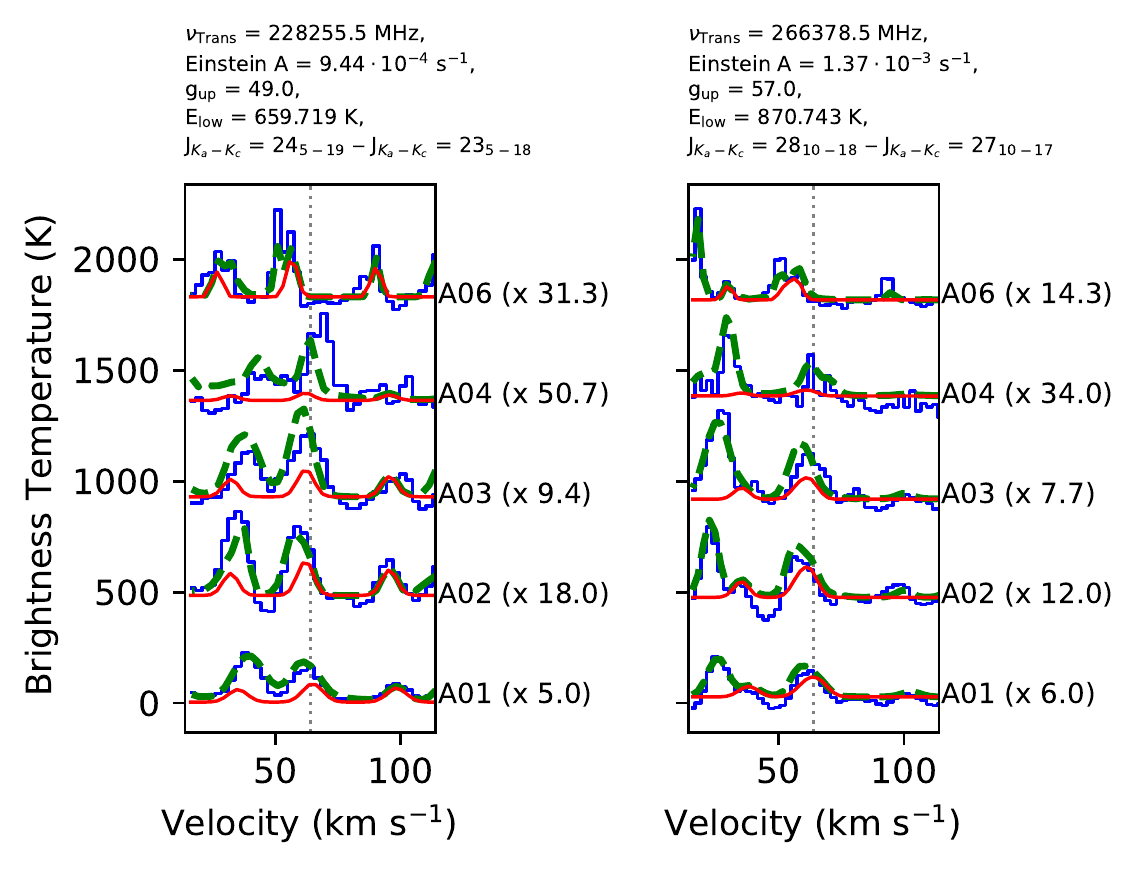}\\
       \caption{Sgr~B2(M)}
       \label{fig:C2H3CNv151M}
    \end{subfigure}
\quad
    \begin{subfigure}[t]{1.0\columnwidth}
       \includegraphics[width=1.0\columnwidth]{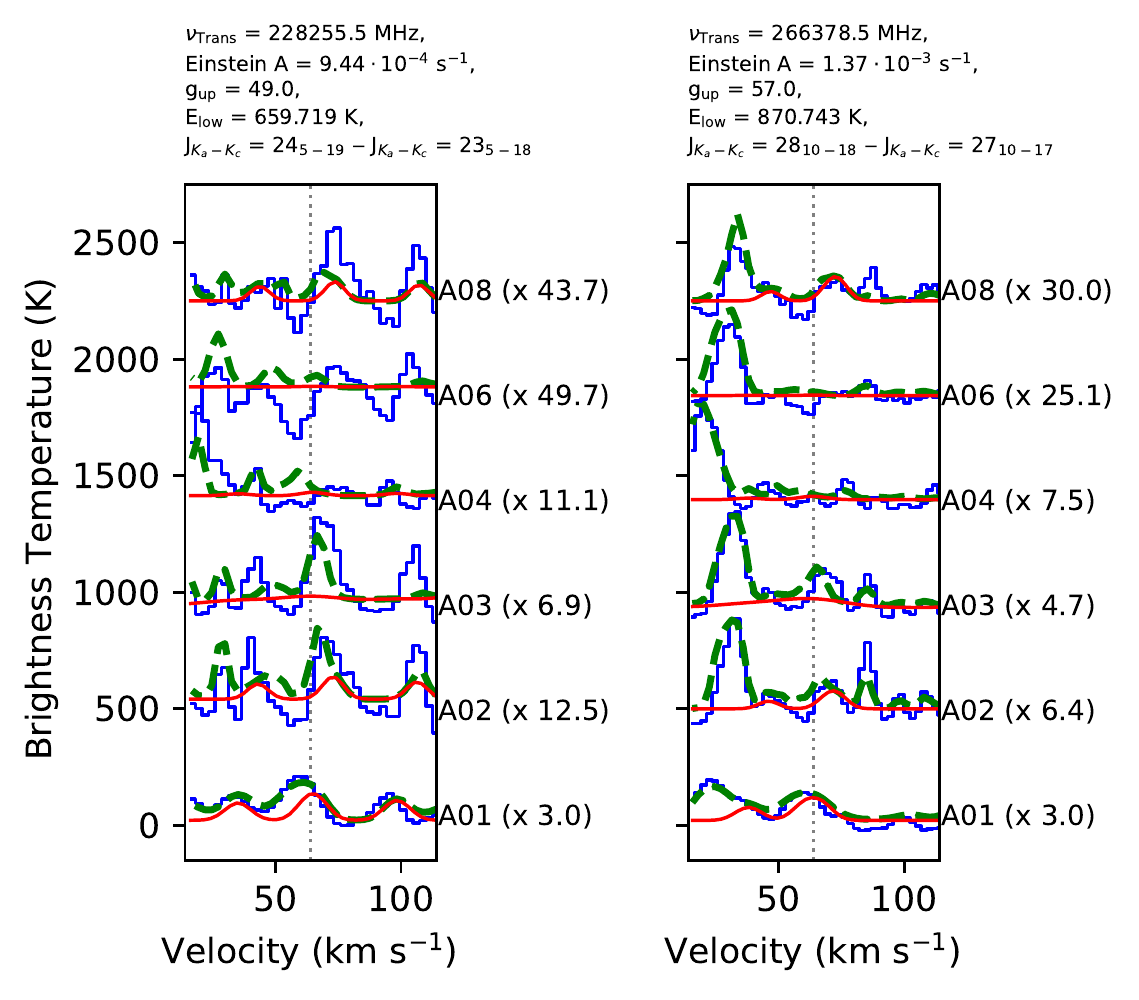}\\
       \caption{Sgr~B2(N)}
       \label{fig:C2H3CNv151N}
   \end{subfigure}
   \caption{Selected transitions of C$_2$H$_3$CN, v$_{15}$=1 in Sgr~B2(M) and N.}
   \ContinuedFloat
   \label{fig:C2H3CNv151MN}
\end{figure*}
\newpage
\clearpage

%*******************************************************************************
% Figure: C2H5CN;v=0;private
\begin{figure*}[!htb]
    \centering
    \begin{subfigure}[t]{1.0\columnwidth}
       \includegraphics[width=1.0\columnwidth]{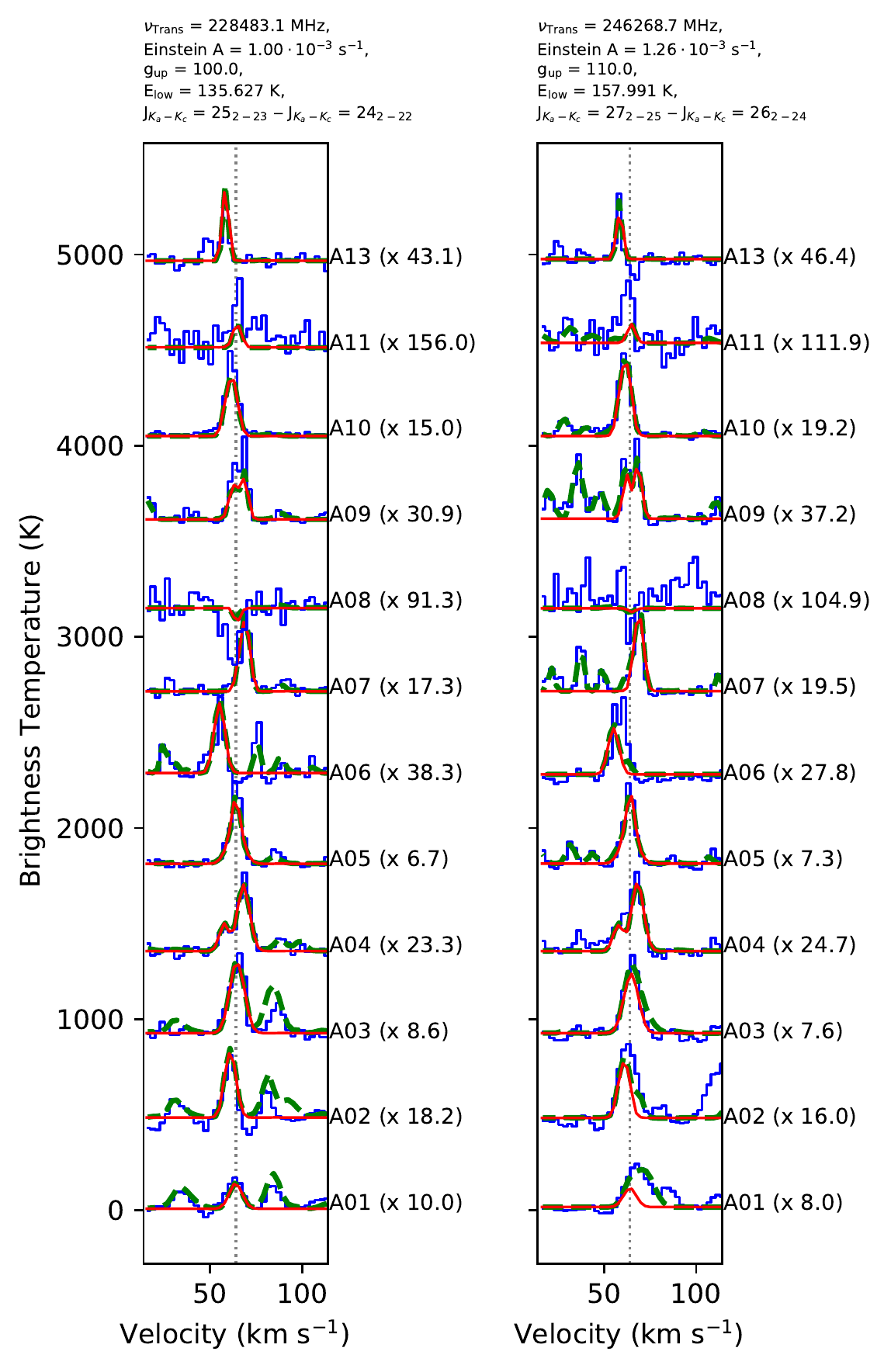}\\
       \caption{Sgr~B2(M)}
       \label{fig:C2H5CNM}
    \end{subfigure}
\quad
    \begin{subfigure}[t]{1.0\columnwidth}
       \includegraphics[width=1.0\columnwidth]{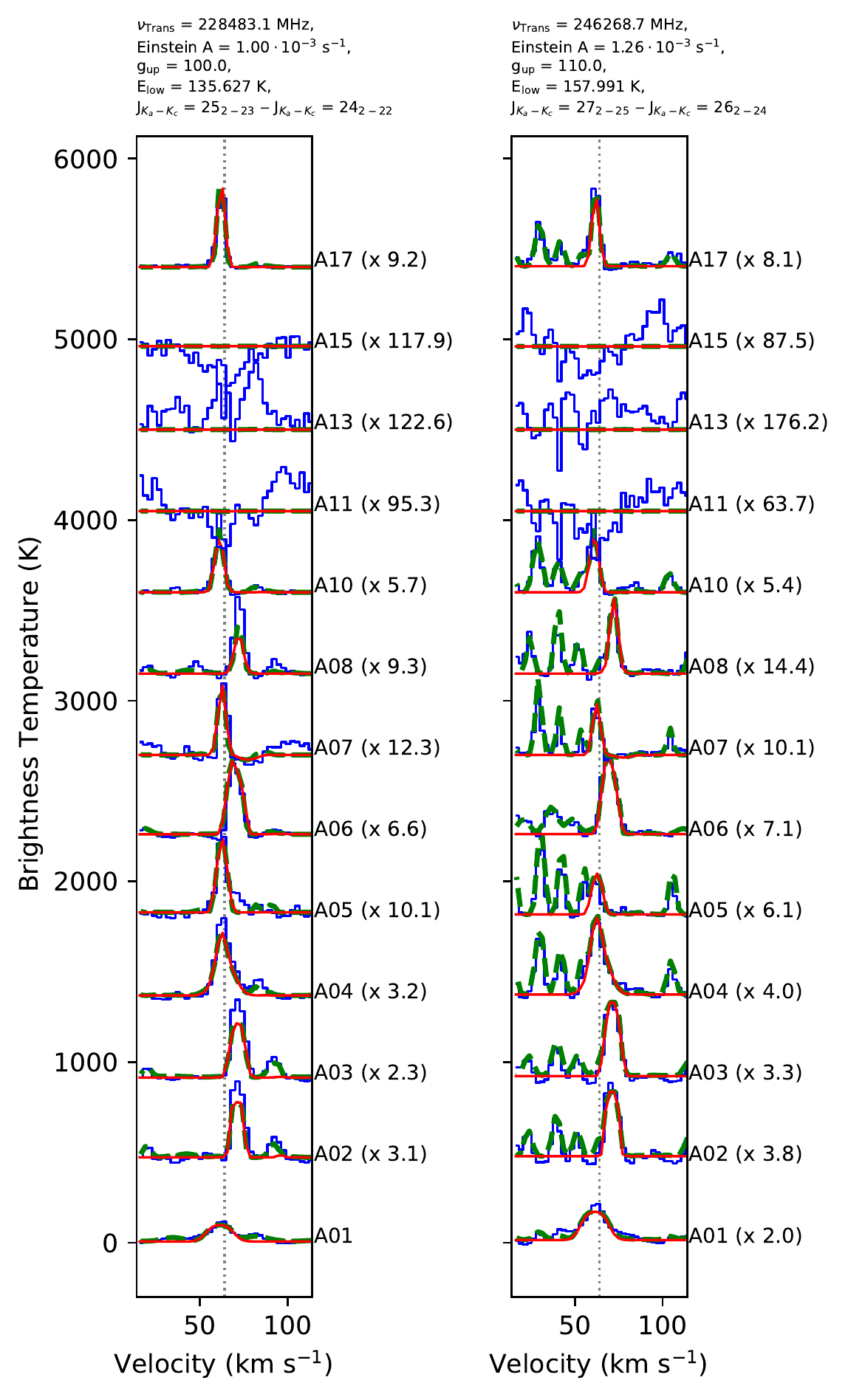}\\
       \caption{Sgr~B2(N)}
       \label{fig:C2H5CNN}
   \end{subfigure}
   \caption{Selected transitions of C$_2$H$_5$CN in Sgr~B2(M) and N.}
   \ContinuedFloat
   \label{fig:C2H5CNMN}
\end{figure*}
\newpage
\clearpage

%*******************************************************************************
% Figure: C2H5C-13-N;v=0;
\begin{figure*}[!htb]
    \centering
    \begin{subfigure}[t]{1.0\columnwidth}
       \includegraphics[width=1.0\columnwidth]{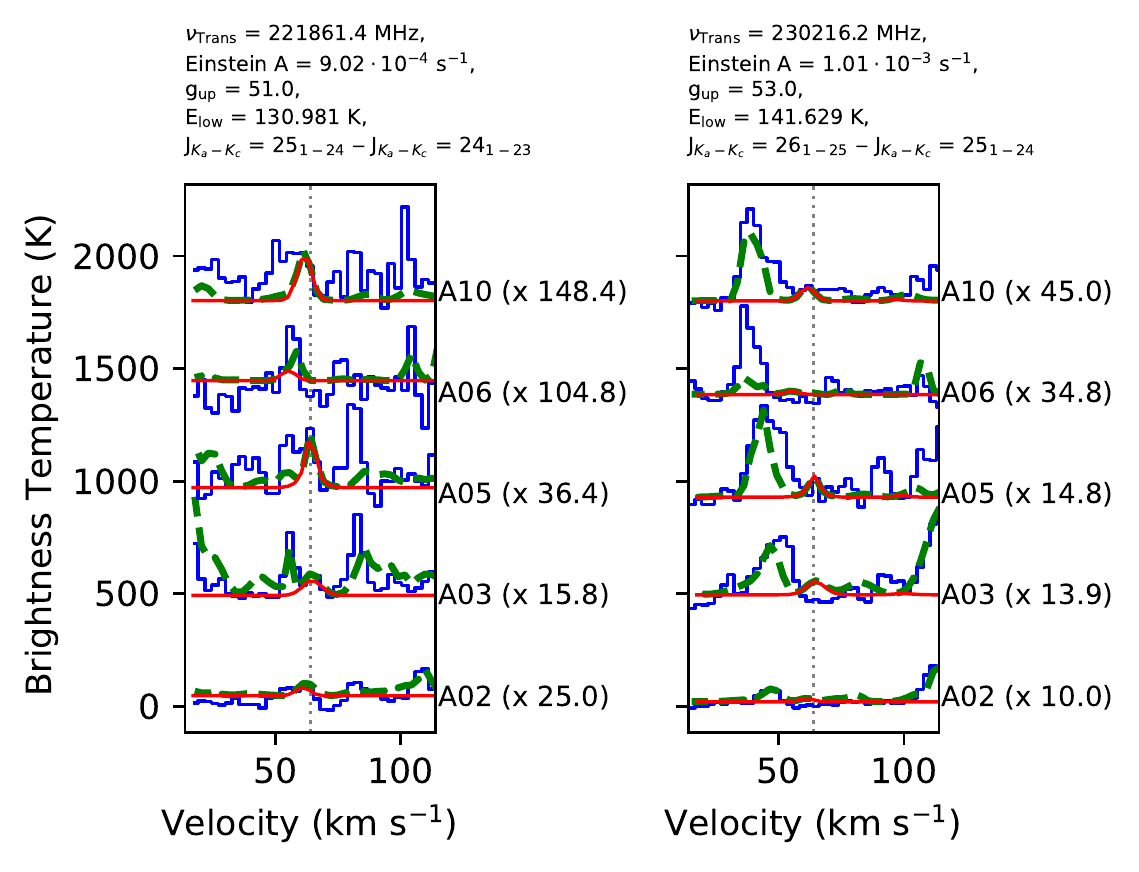}\\
       \caption{Sgr~B2(M)}
       \label{fig:C2H5C13NM}
    \end{subfigure}
\quad
    \begin{subfigure}[t]{1.0\columnwidth}
       \includegraphics[width=1.0\columnwidth]{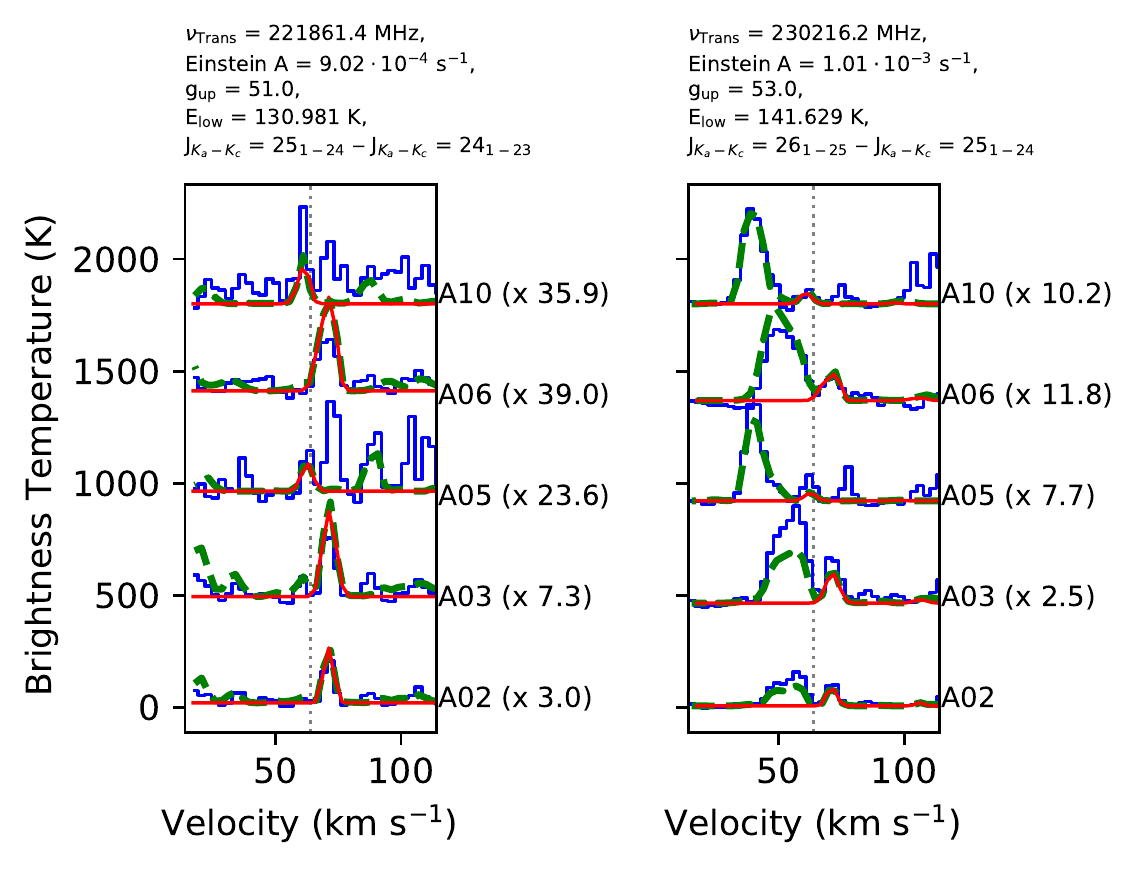}\\
       \caption{Sgr~B2(N)}
       \label{fig:C2H5C13NN}
   \end{subfigure}
   \caption{Selected transitions of C$_2$H$_5 \! ^{13}$CN in Sgr~B2(M) and N.}
   \ContinuedFloat
   \label{fig:C2H5C13NMN}
\end{figure*}

%*******************************************************************************
% Figure: C2H5CN;v12=1;private
\begin{figure*}[!htb]
    \centering
    \begin{subfigure}[t]{1.0\columnwidth}
       \includegraphics[width=1.0\columnwidth]{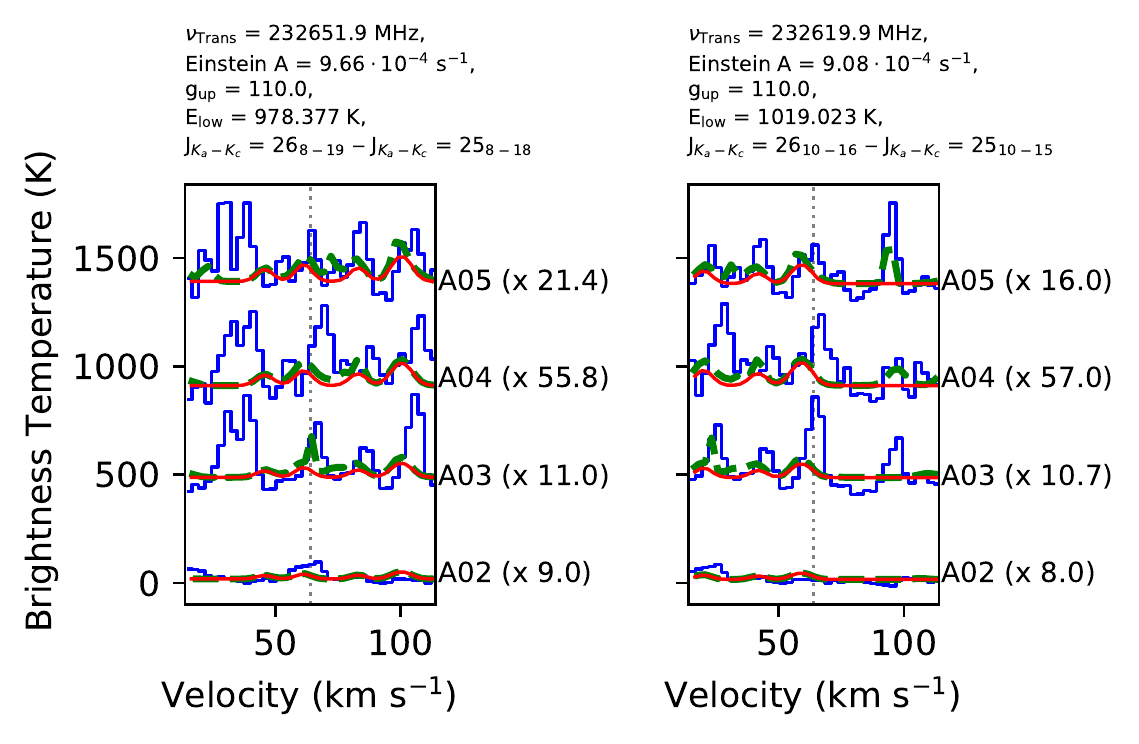}\\
       \caption{Sgr~B2(M)}
       \label{fig:C2H5CNv121M}
    \end{subfigure}
\quad
    \begin{subfigure}[t]{1.0\columnwidth}
       \includegraphics[width=1.0\columnwidth]{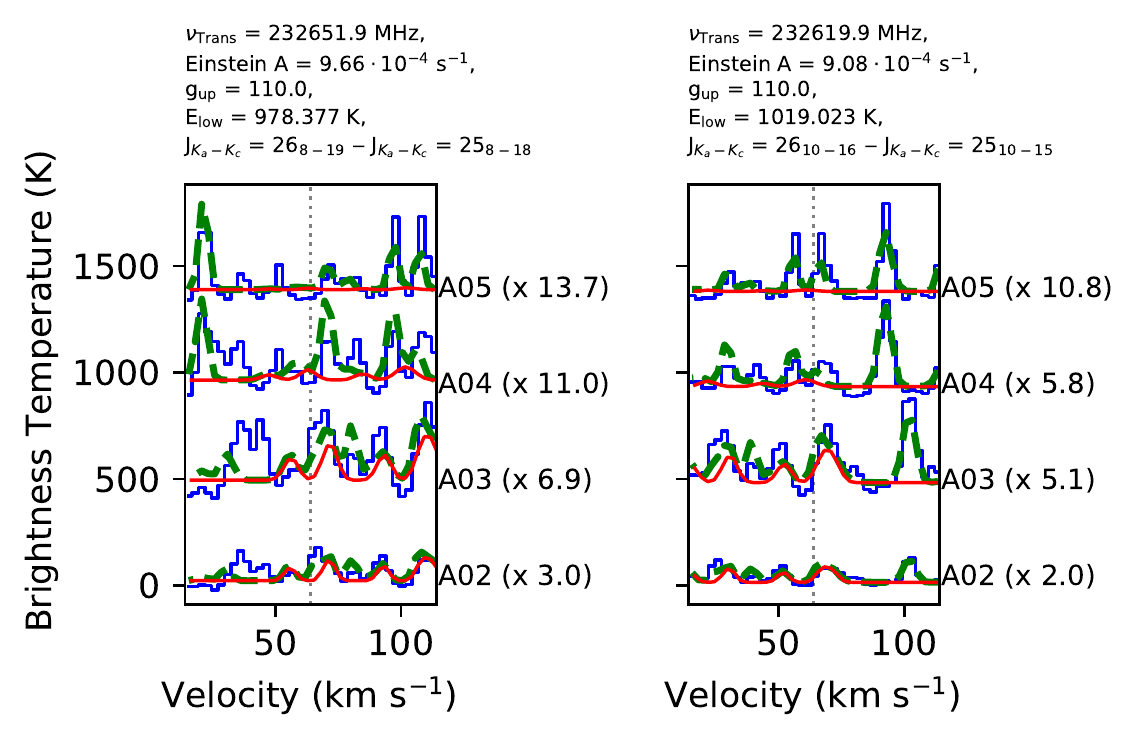}\\
       \caption{Sgr~B2(N)}
       \label{fig:C2H5CNv121N}
   \end{subfigure}
   \caption{Selected transitions of C$_2$H$_5$CN, v$_{12}$=1 in Sgr~B2(M) and N.}
   \ContinuedFloat
   \label{fig:C2H5CNv121MN}
\end{figure*}

%*******************************************************************************
% Figure: C2H5CN;v13+v21=1;private
\begin{figure*}[!htb]
    \centering
    \begin{subfigure}[t]{1.0\columnwidth}
       \includegraphics[width=1.0\columnwidth]{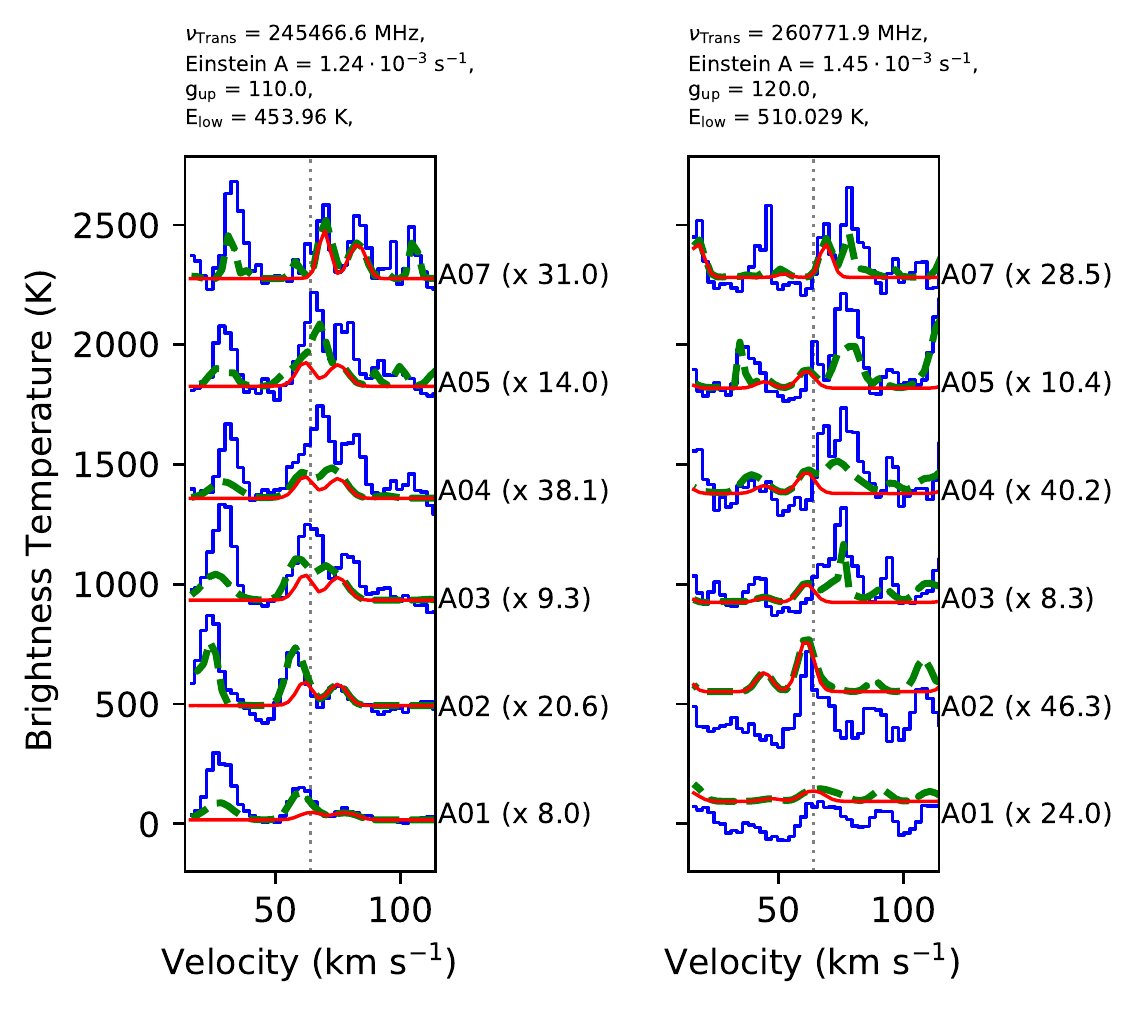}\\
       \caption{Sgr~B2(M)}
       \label{fig:C2H5CNv13211M}
    \end{subfigure}
\quad
    \begin{subfigure}[t]{1.0\columnwidth}
       \includegraphics[width=1.0\columnwidth]{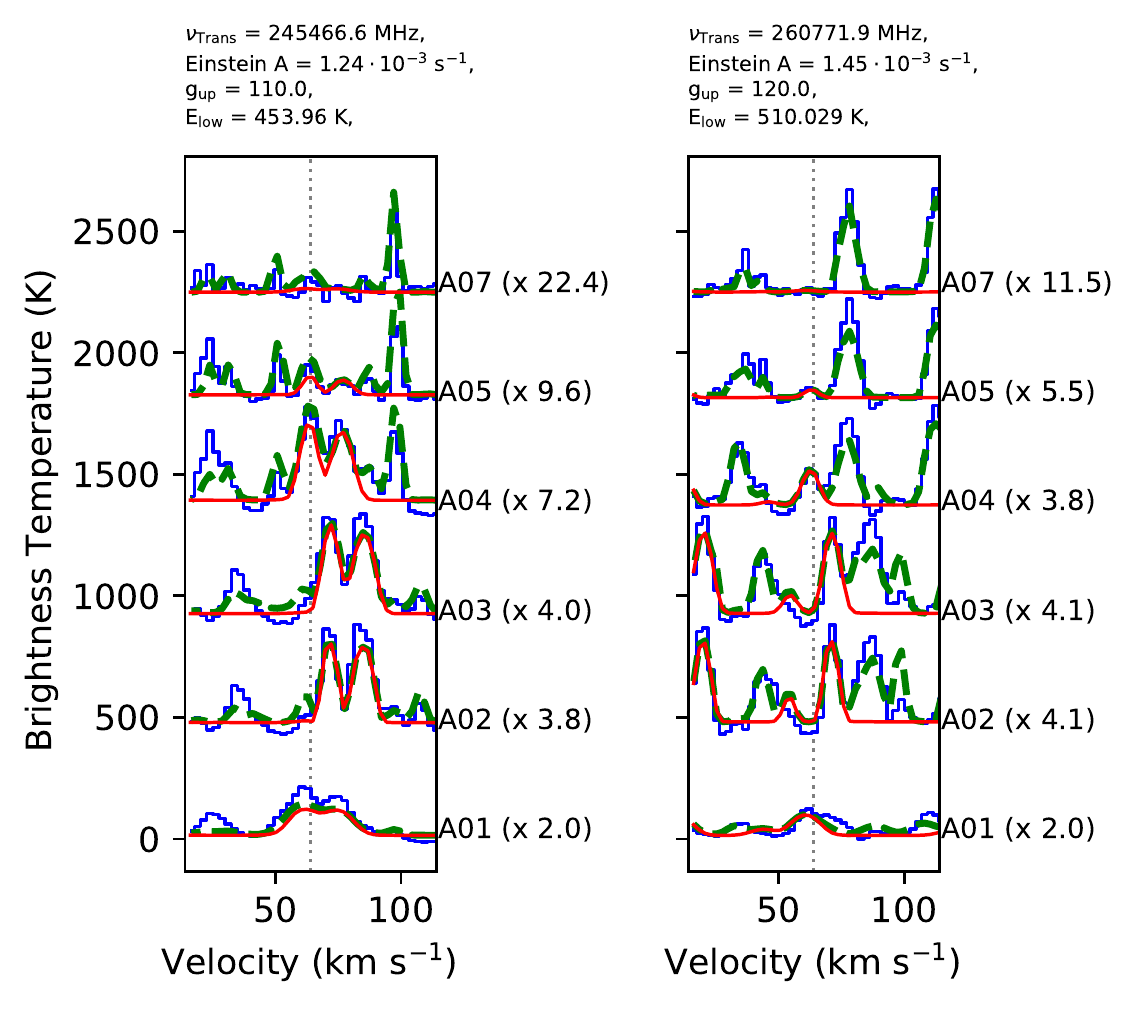}\\
       \caption{Sgr~B2(N)}
       \label{fig:C2H5CNv13211N}
   \end{subfigure}
   \caption{Selected transitions of C$_2$H$_5$CN, v$_{13}$, v$_{21}$=1 in Sgr~B2(M) and N.}
   \ContinuedFloat
   \label{fig:C2H5CNv13211MN}
\end{figure*}

%*******************************************************************************
% Figure: C2H5CN;v20=1;private
\begin{figure*}[!htb]
    \centering
    \begin{subfigure}[t]{1.0\columnwidth}
       \includegraphics[width=1.0\columnwidth]{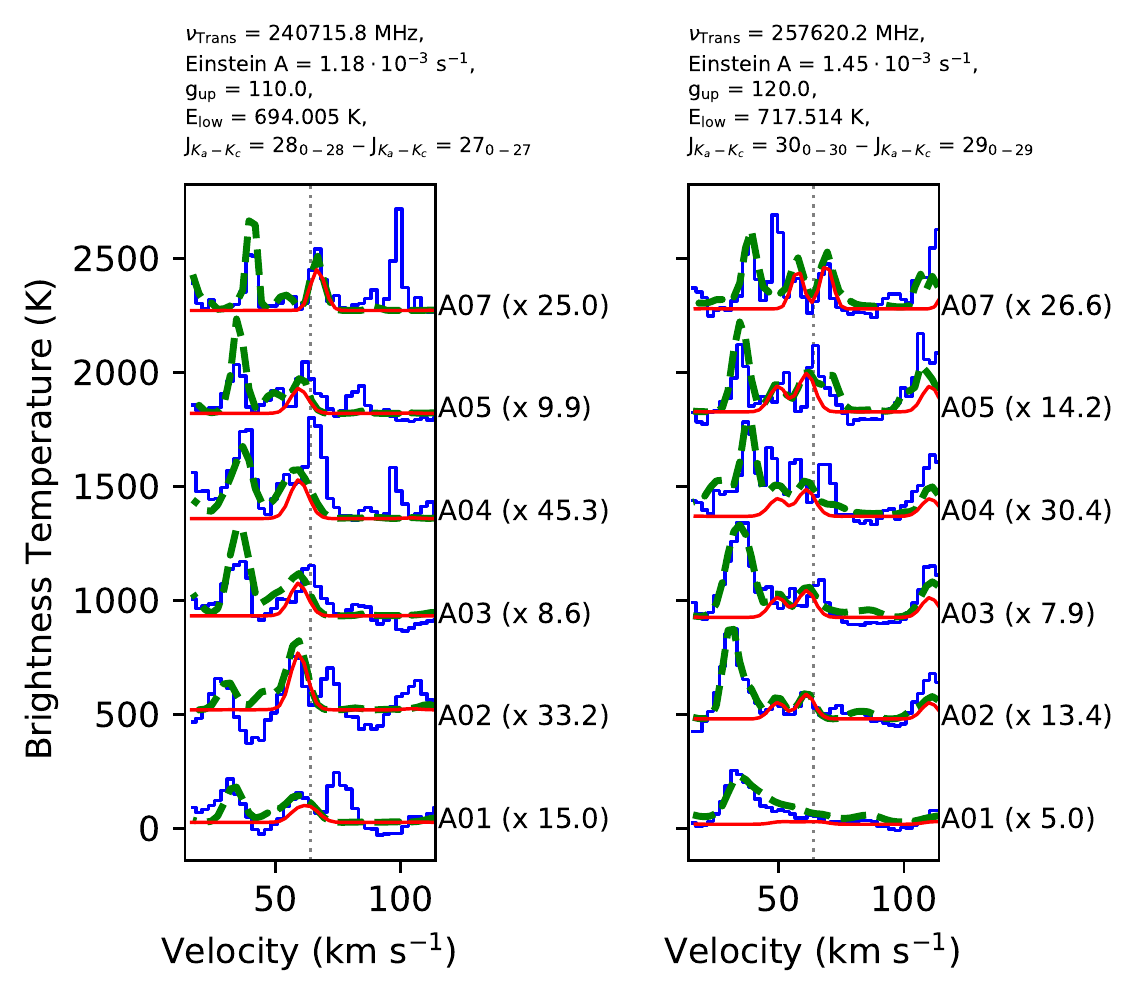}\\
       \caption{Sgr~B2(M)}
       \label{fig:C2H5CNv201M}
    \end{subfigure}
\quad
    \begin{subfigure}[t]{1.0\columnwidth}
       \includegraphics[width=1.0\columnwidth]{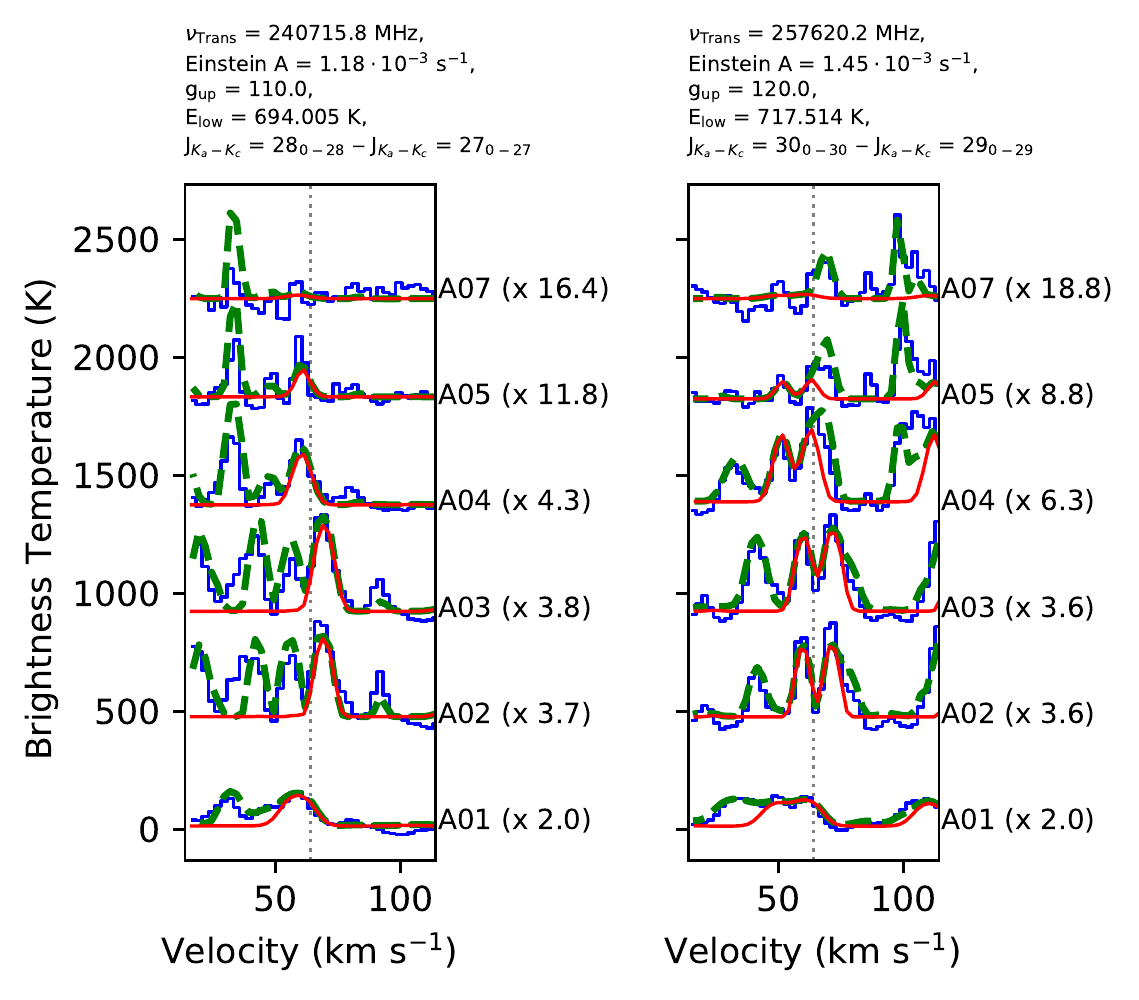}\\
       \caption{Sgr~B2(N)}
       \label{fig:C2H5CNv201N}
   \end{subfigure}
   \caption{Selected transitions of C$_2$H$_5$CN, v$_{20}$=1 in Sgr~B2(M) and N.}
   \ContinuedFloat
   \label{fig:C2H5CNv201MN}
\end{figure*}
\newpage
\clearpage

%*******************************************************************************
% Figure: CN;v=0;
\begin{figure*}[!htb]
    \centering
    \begin{subfigure}[t]{1.0\columnwidth}
       \includegraphics[width=1.0\columnwidth]{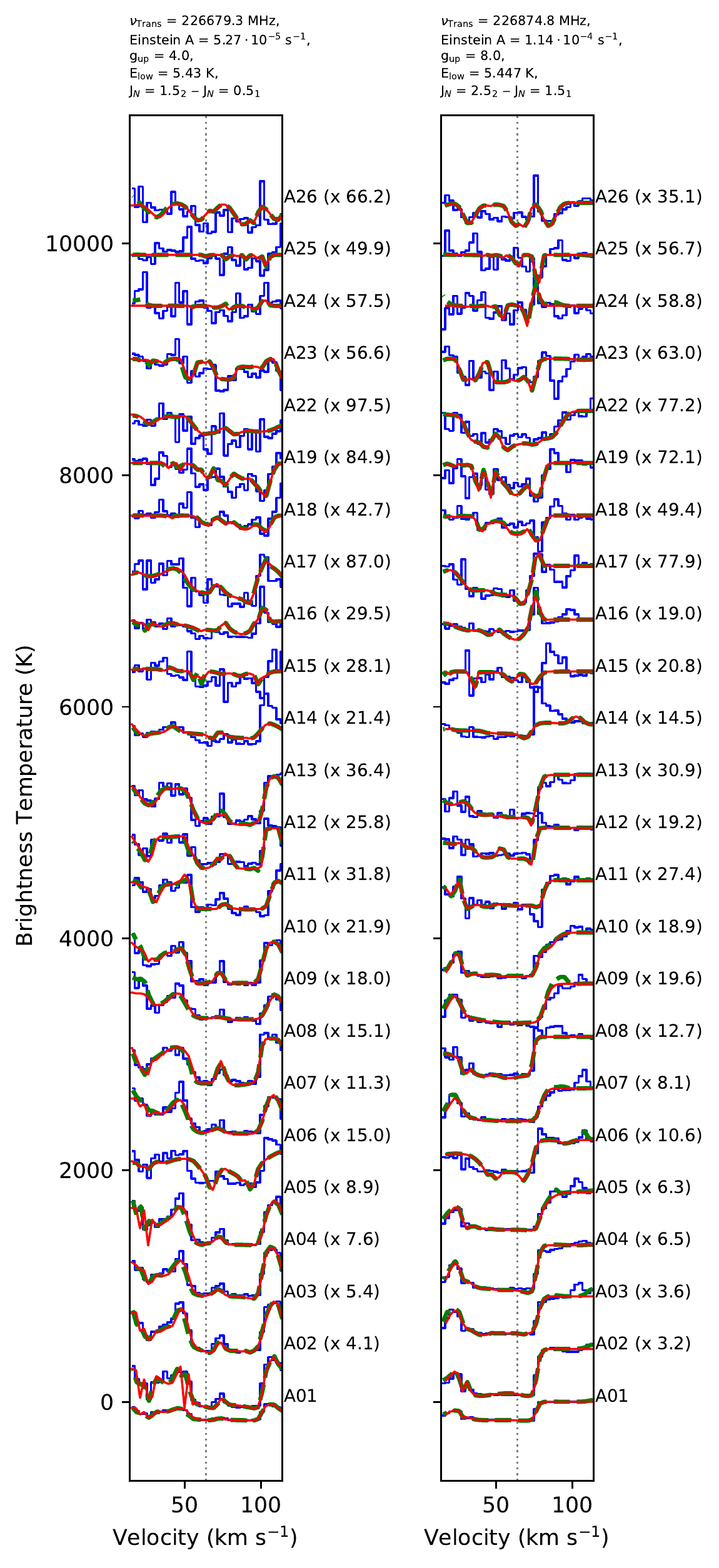}\\
       \caption{Sgr~B2(M)}
       \label{fig:CNM}
    \end{subfigure}
\quad
    \begin{subfigure}[t]{1.0\columnwidth}
       \includegraphics[width=1.0\columnwidth]{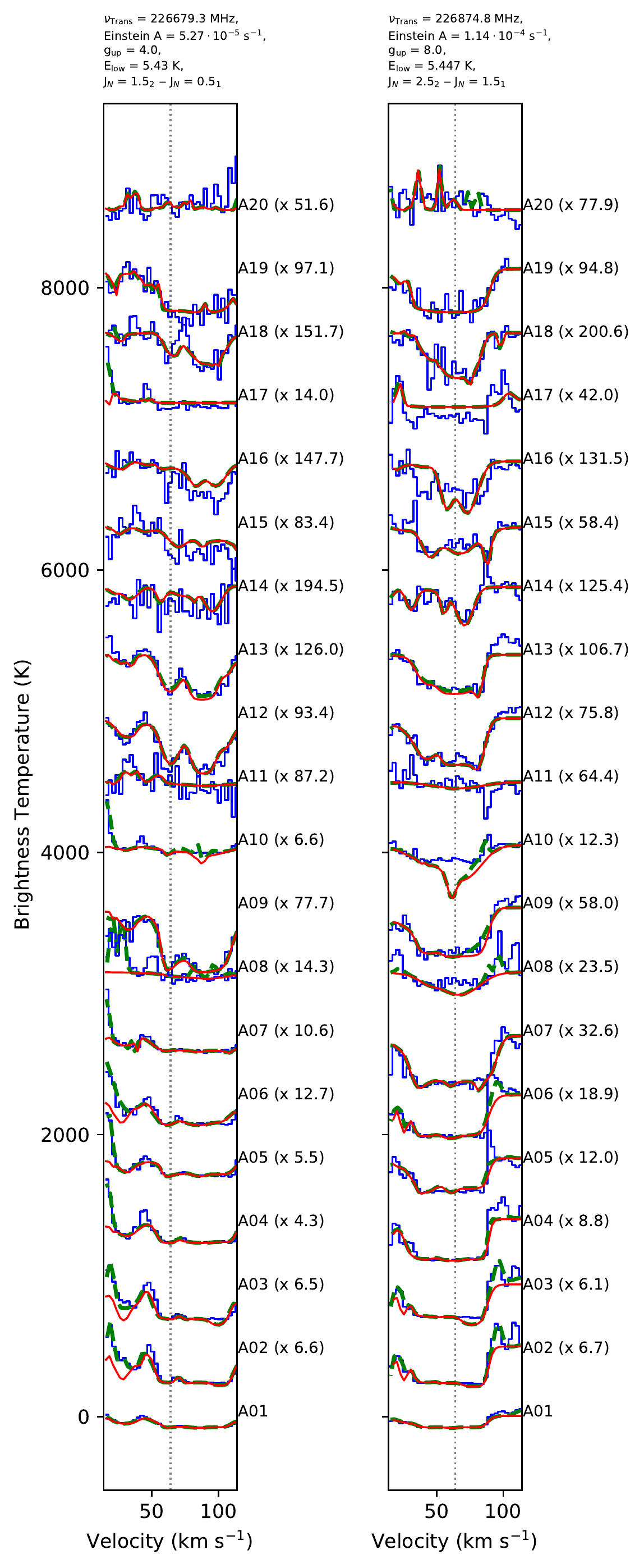}\\
       \caption{Sgr~B2(N)}
       \label{fig:CNN}
   \end{subfigure}
   \caption{Selected transitions of CN in Sgr~B2(M) and N.}
   \ContinuedFloat
   \label{fig:CNMN}
\end{figure*}
\newpage
\clearpage

%*******************************************************************************
% Figure: C-13-N;v=0;
\begin{figure*}[!htb]
    \centering
    \begin{subfigure}[t]{1.0\columnwidth}
       \includegraphics[width=1.0\columnwidth]{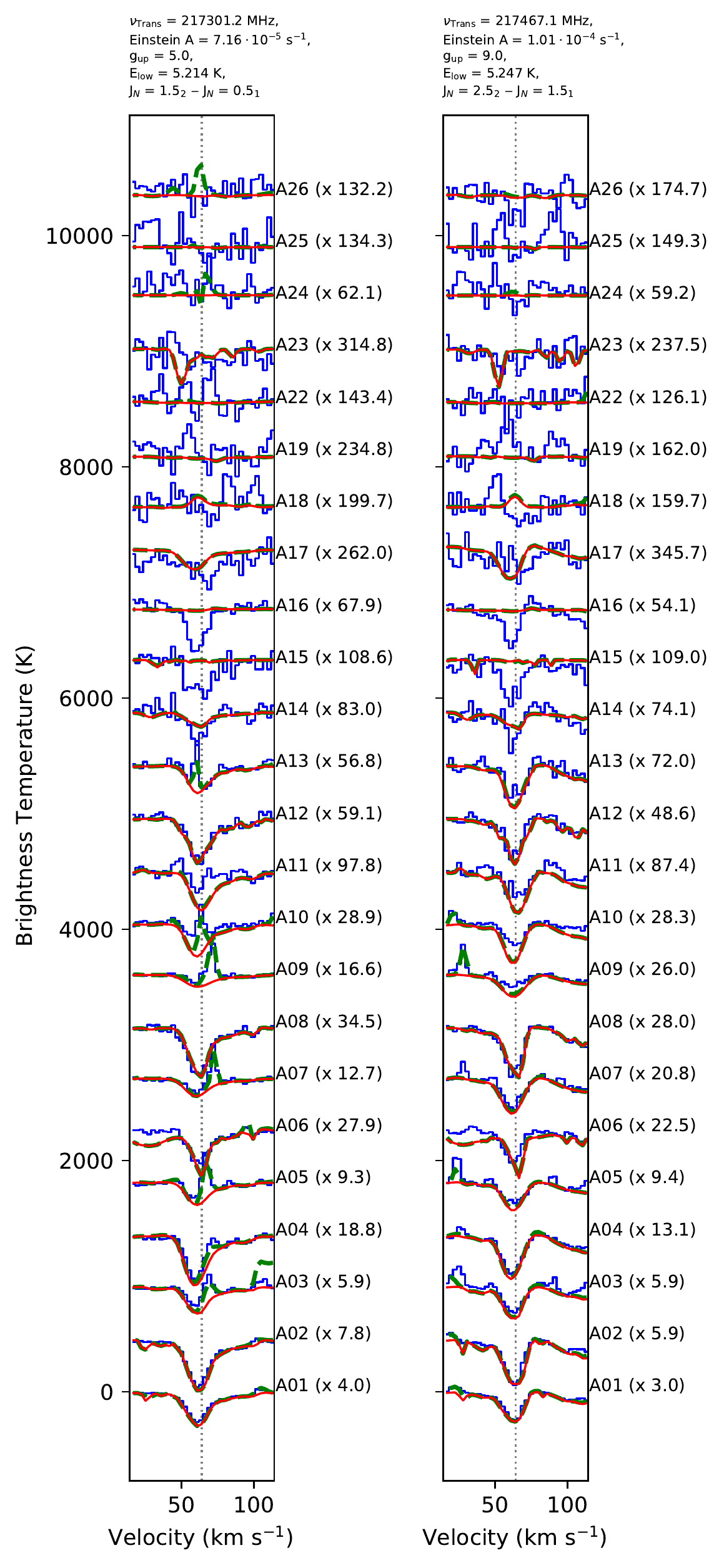}\\
       \caption{Sgr~B2(M)}
       \label{fig:C13NM}
    \end{subfigure}
\quad
    \begin{subfigure}[t]{1.0\columnwidth}
       \includegraphics[width=1.0\columnwidth]{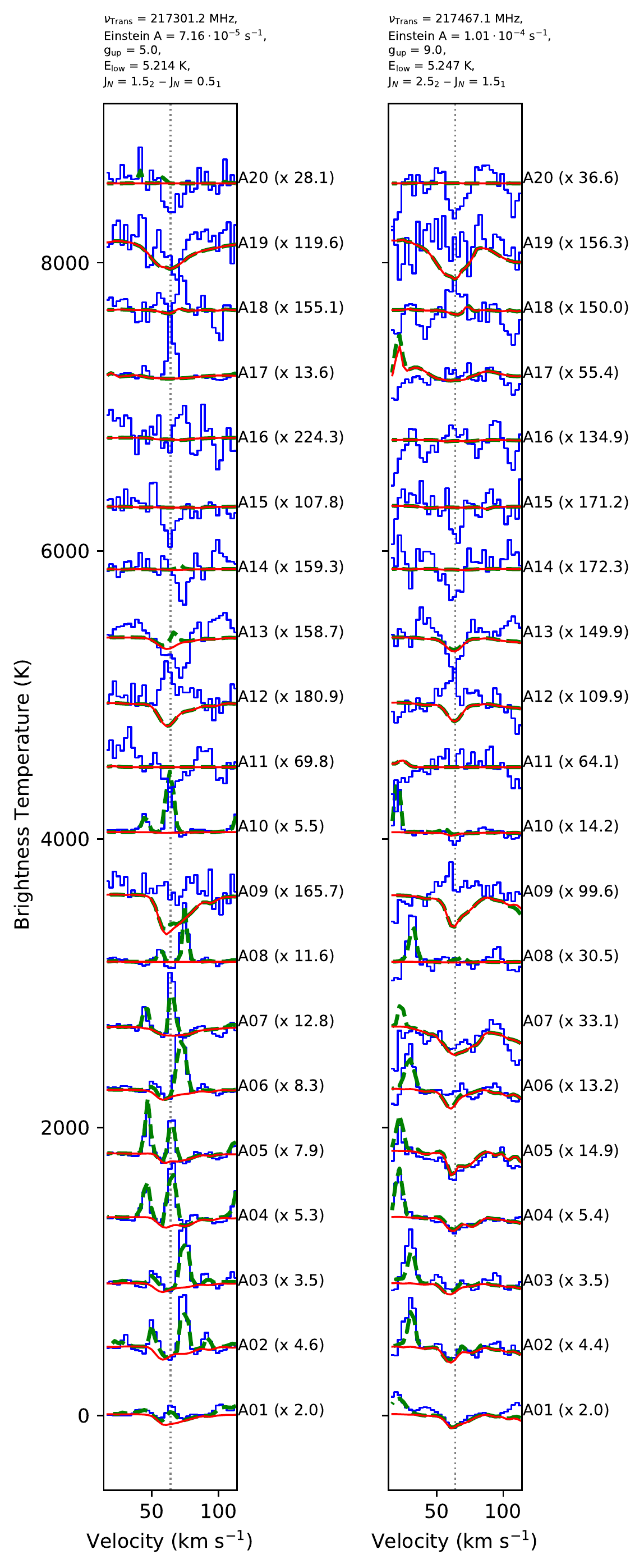}\\
       \caption{Sgr~B2(N)}
       \label{fig:C13NN}
   \end{subfigure}
   \caption{Selected transitions of $^{13}$CN in Sgr~B2(M) and N.}
   \ContinuedFloat
   \label{fig:C13NMN}
\end{figure*}

%*******************************************************************************
% Figure: H2CCN
\begin{figure*}[!htb]
    \centering
    \begin{subfigure}[t]{1.0\columnwidth}
       \includegraphics[width=1.0\columnwidth]{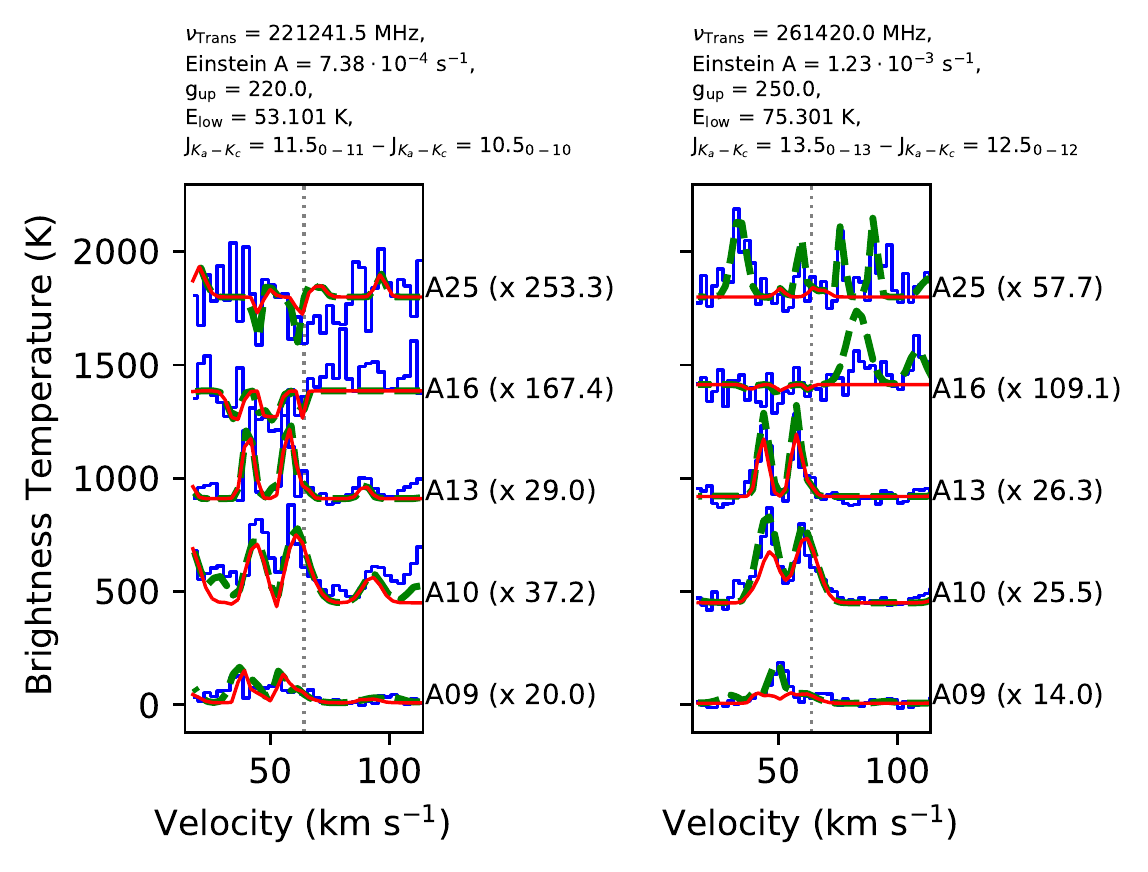}\\
    \end{subfigure}
   \caption{Selected transitions of H$_2$CCN in Sgr~B2(M).}
   \ContinuedFloat
   \label{fig:H2CCNM}
\end{figure*}

%*******************************************************************************
% Figure: HCCCN
\begin{figure*}[!htb]
    \centering
    \begin{subfigure}[t]{1.0\columnwidth}
       \includegraphics[width=1.0\columnwidth]{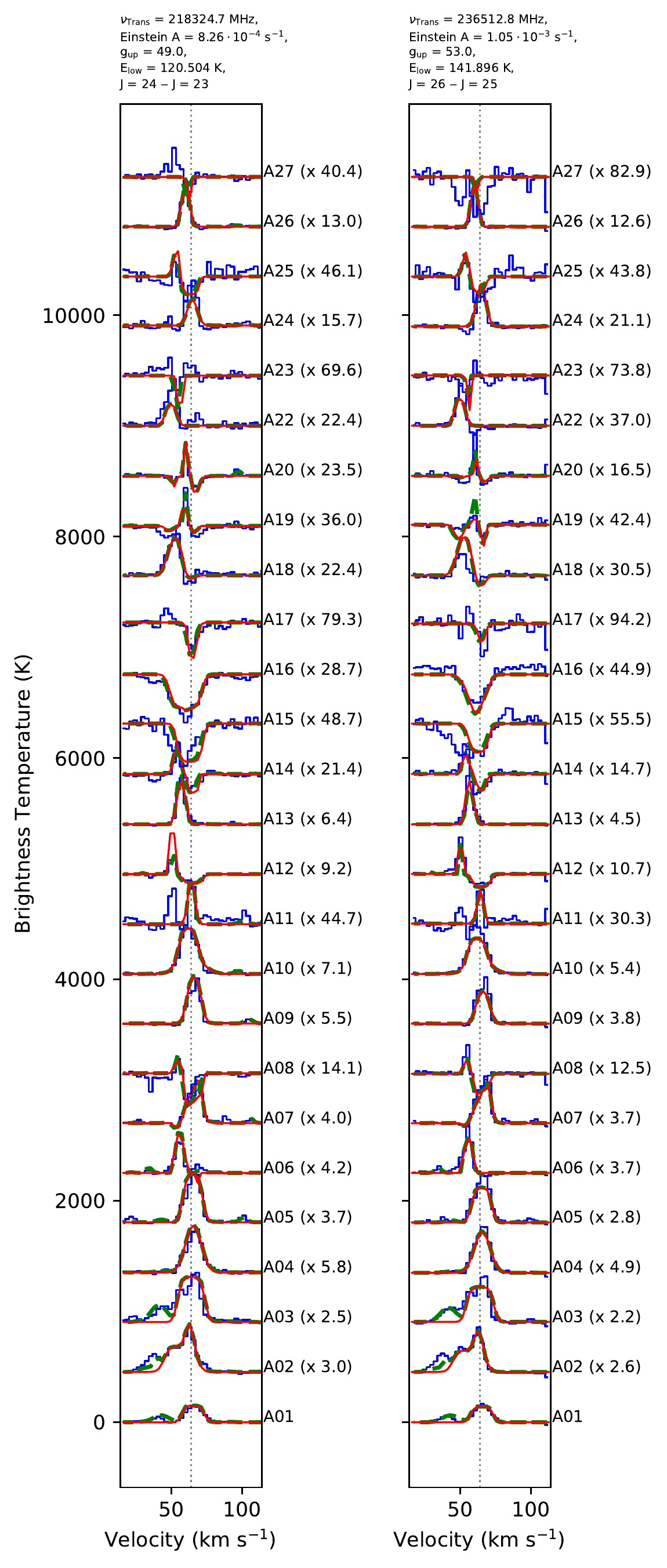}\\
       \caption{Sgr~B2(M)}
       \label{fig:HCCCNM}
    \end{subfigure}
\quad
    \begin{subfigure}[t]{1.0\columnwidth}
       \includegraphics[width=1.0\columnwidth]{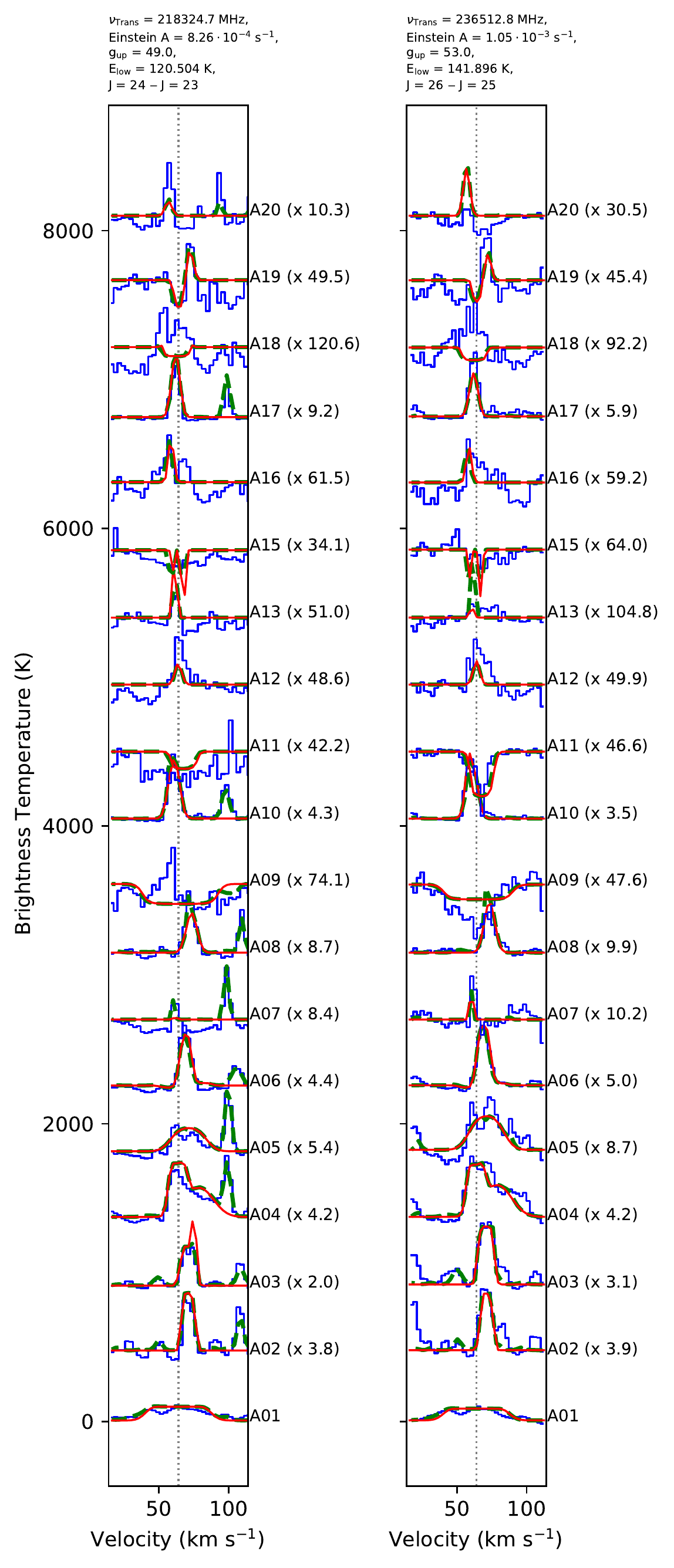}\\
       \caption{Sgr~B2(N)}
       \label{fig:HCCCNN}
   \end{subfigure}
   \caption{Selected transitions of HCCCN in Sgr~B2(M) and N.}
   \ContinuedFloat
   \label{fig:HCCCNMN}
\end{figure*}
\newpage
\clearpage

%*******************************************************************************
% Figure: HC-13-CCN;v=0;
\begin{figure*}[!htb]
    \centering
    \begin{subfigure}[t]{1.0\columnwidth}
       \includegraphics[width=1.0\columnwidth]{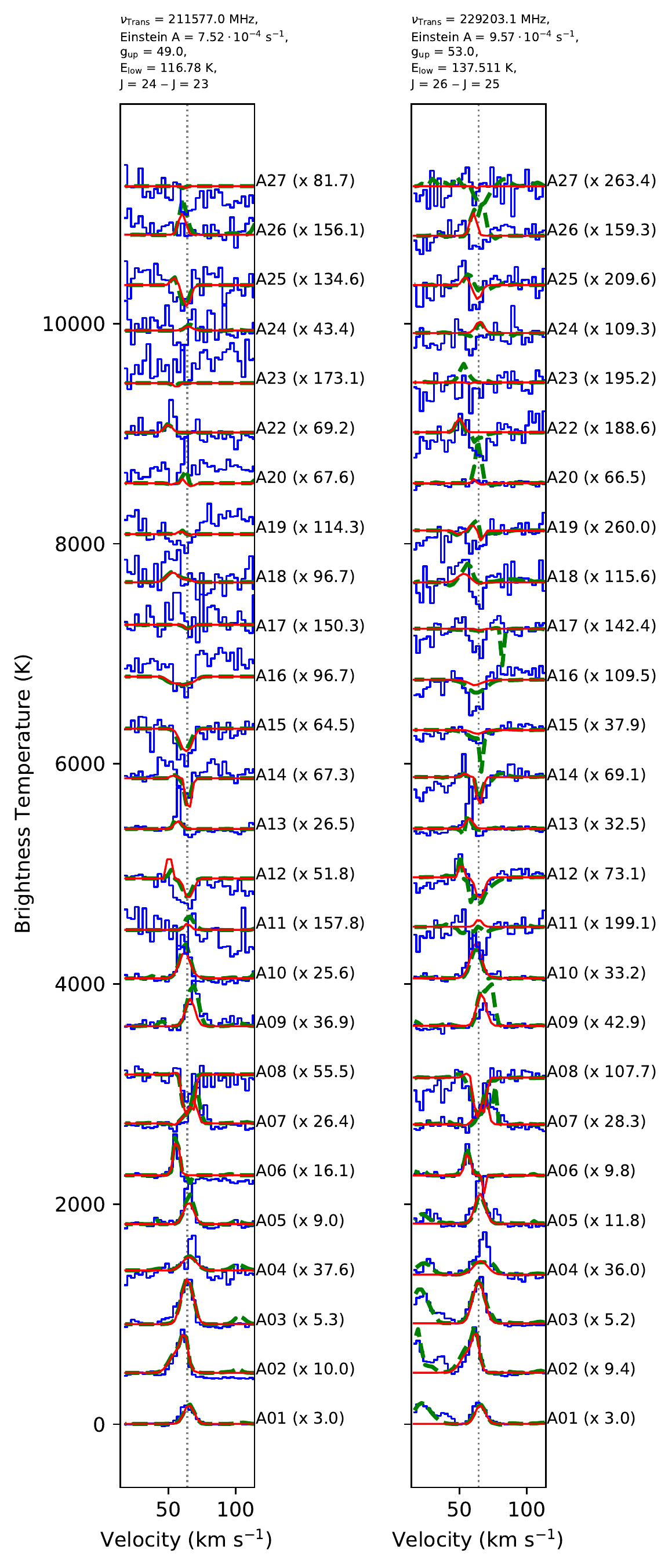}\\
       \caption{Sgr~B2(M)}
       \label{fig:HC13CCNM}
    \end{subfigure}
\quad
    \begin{subfigure}[t]{1.0\columnwidth}
       \includegraphics[width=1.0\columnwidth]{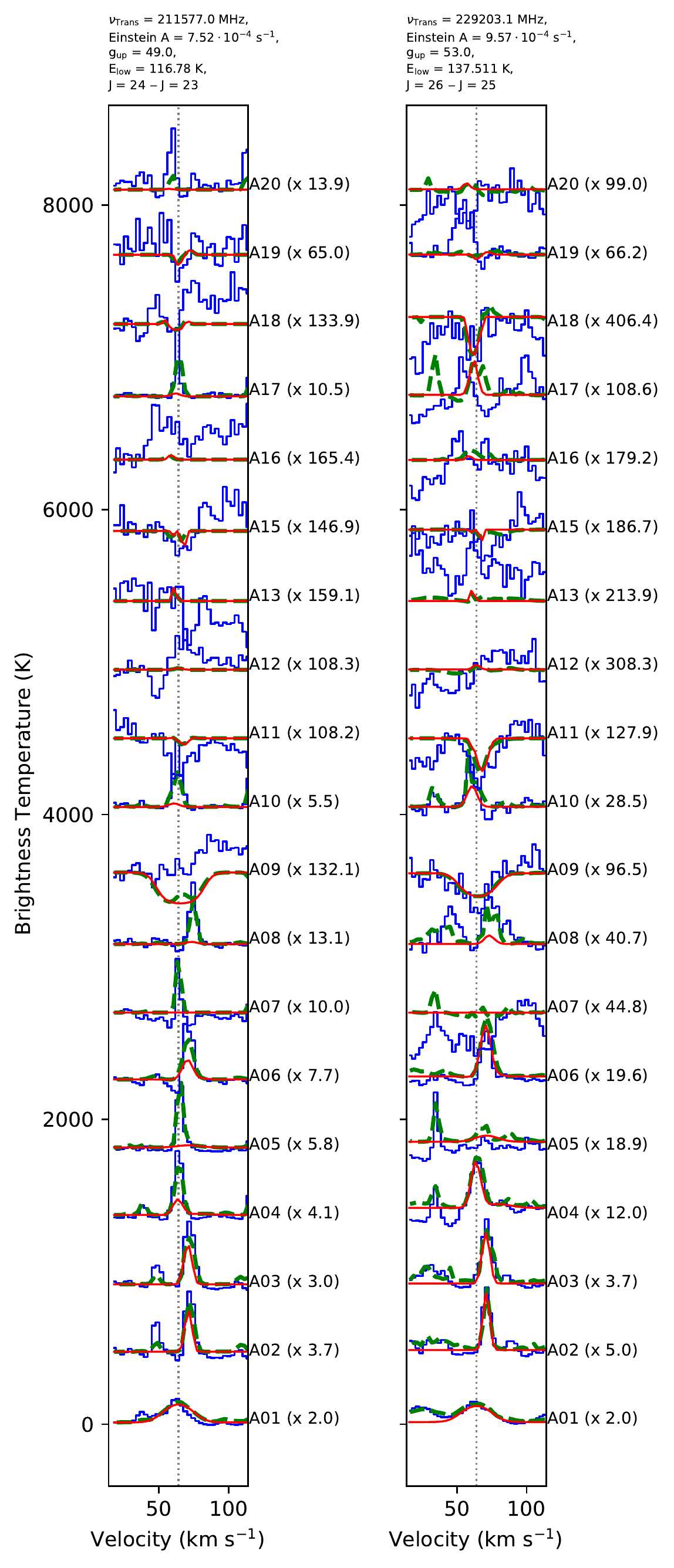}\\
       \caption{Sgr~B2(N)}
       \label{fig:HC13CCNN}
   \end{subfigure}
   \caption{Selected transitions of H$^{13}$CCCN in Sgr~B2(M) and N.}
   \ContinuedFloat
   \label{fig:HC13CCNMN}
\end{figure*}
\newpage
\clearpage

%*******************************************************************************
% Figure: HCC-13-CN;v=0;
\begin{figure*}[!htb]
    \centering
    \begin{subfigure}[t]{1.0\columnwidth}
       \includegraphics[width=1.0\columnwidth]{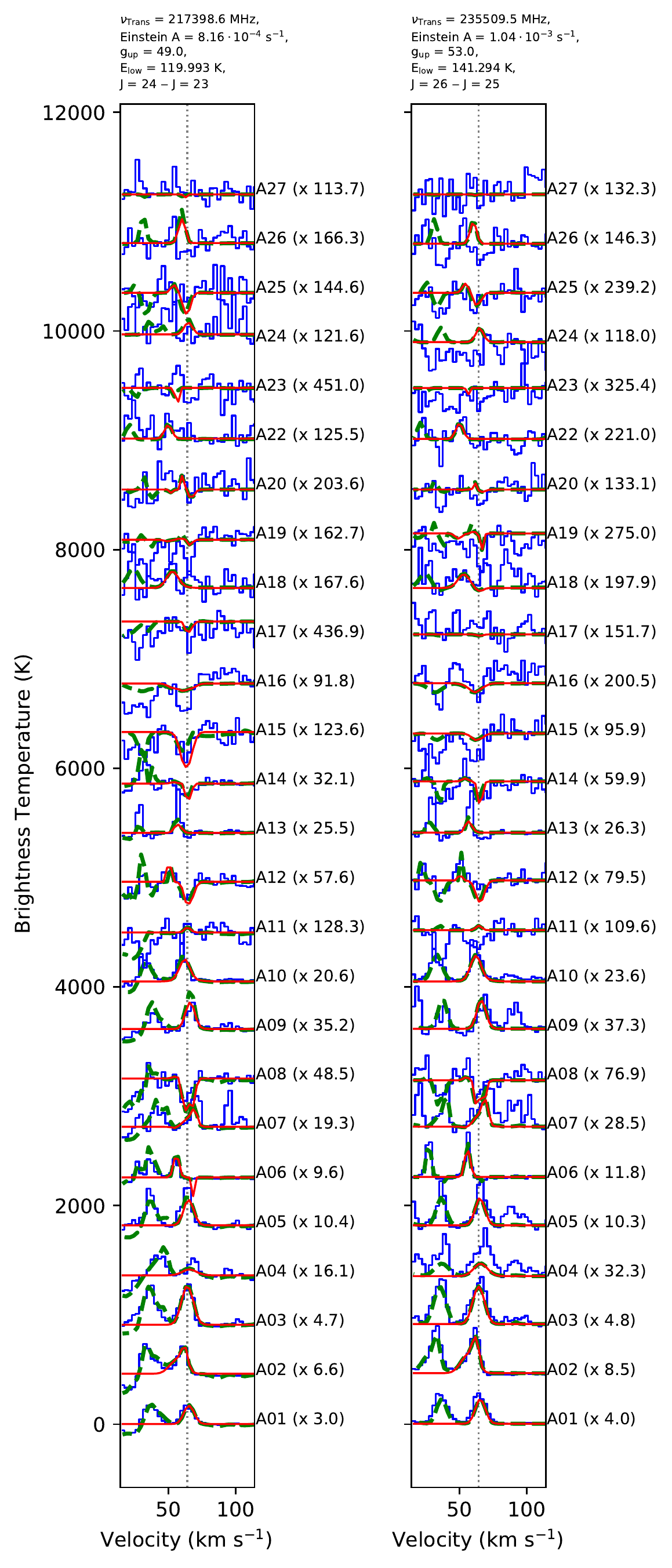}\\
       \caption{Sgr~B2(M)}
       \label{fig:HCC13CNM}
    \end{subfigure}
\quad
    \begin{subfigure}[t]{1.0\columnwidth}
       \includegraphics[width=1.0\columnwidth]{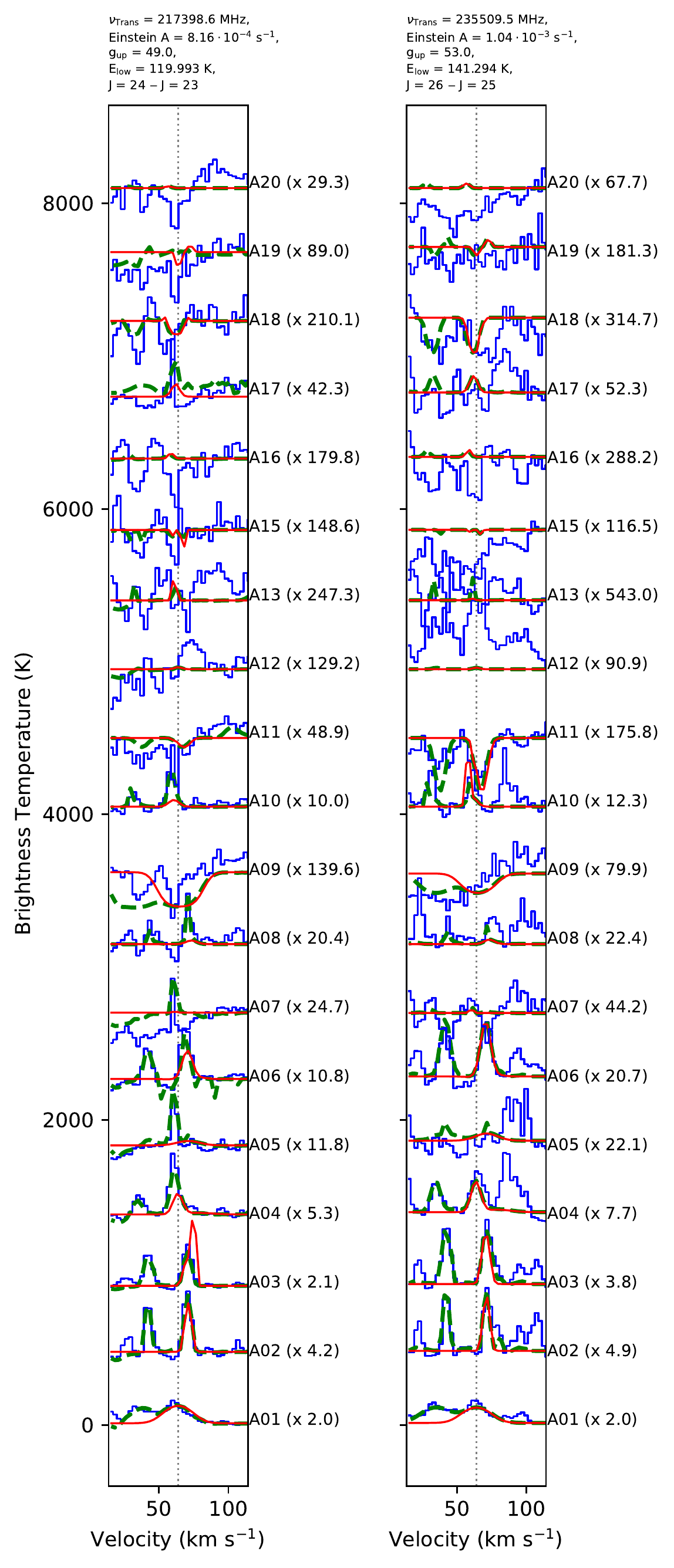}\\
       \caption{Sgr~B2(N)}
       \label{fig:HCC13CNN}
   \end{subfigure}
   \caption{Selected transitions of HC$^{13}$CCN in Sgr~B2(M) and N.}
   \ContinuedFloat
   \label{fig:HCC13CNMN}
\end{figure*}
\newpage
\clearpage

%*******************************************************************************
% Figure: HCCC-13-N;v=0;
\begin{figure*}[!htb]
    \centering
    \begin{subfigure}[t]{1.0\columnwidth}
       \includegraphics[width=1.0\columnwidth]{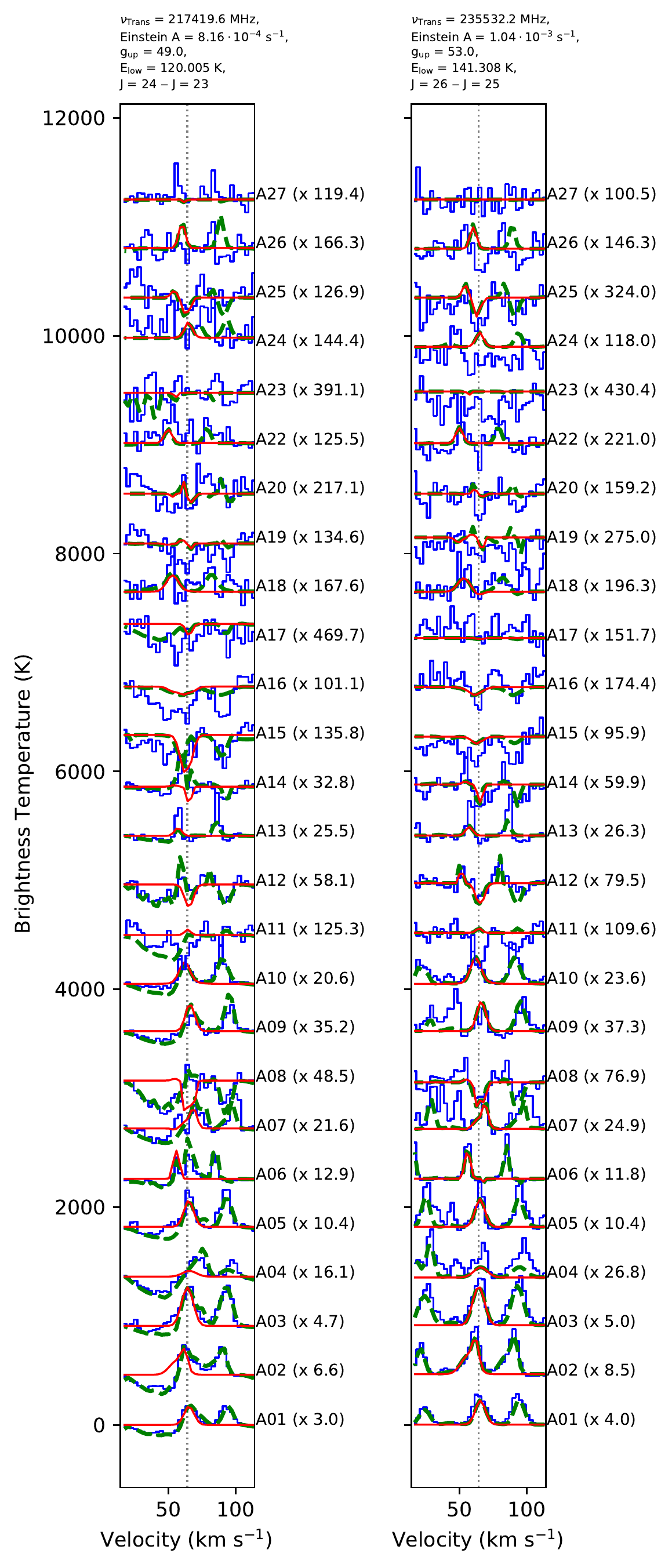}\\
       \caption{Sgr~B2(M)}
       \label{fig:HCCC13NM}
    \end{subfigure}
\quad
    \begin{subfigure}[t]{1.0\columnwidth}
       \includegraphics[width=1.0\columnwidth]{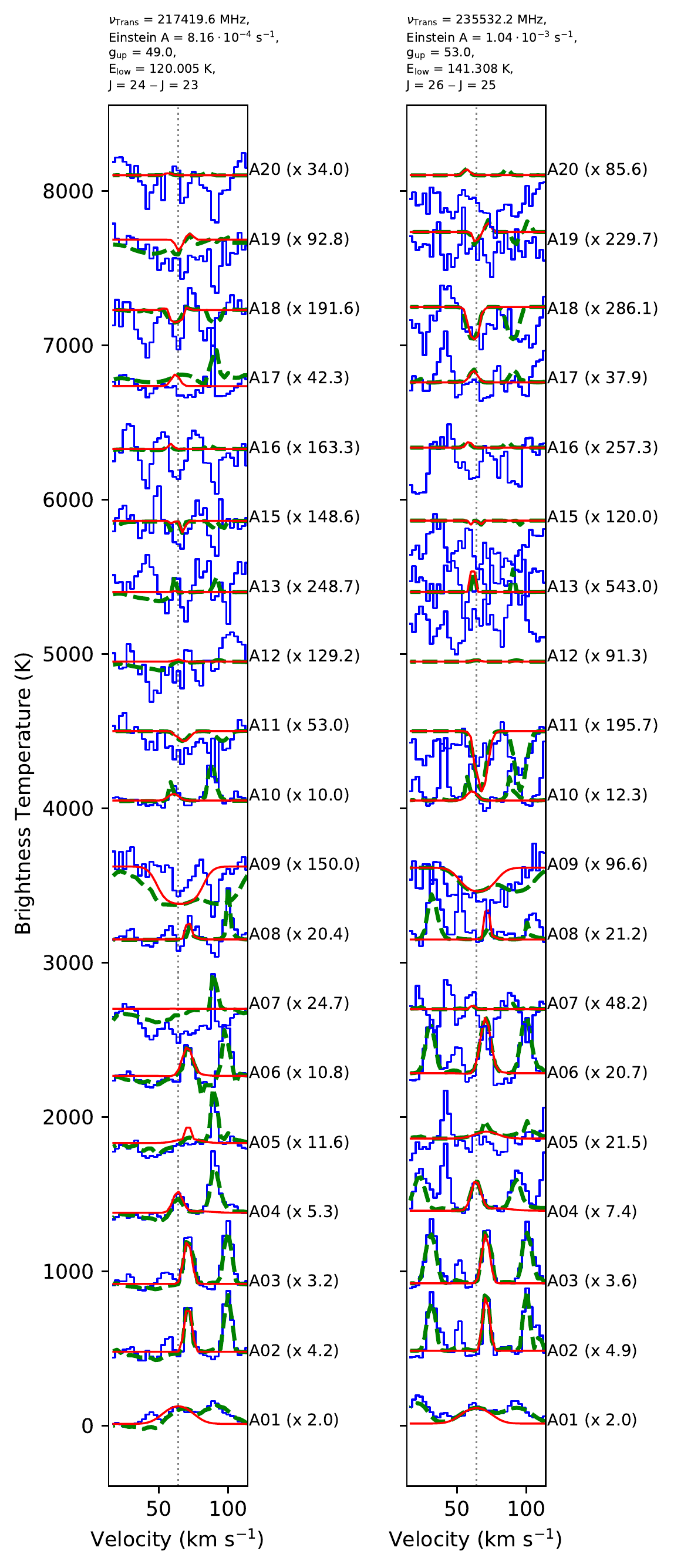}\\
       \caption{Sgr~B2(N)}
       \label{fig:HCCC13NN}
   \end{subfigure}
   \caption{Selected transitions of HCC$^{13}$CN in Sgr~B2(M) and N.}
   \ContinuedFloat
   \label{fig:HCCC13NMN}
\end{figure*}
\newpage
\clearpage

%*******************************************************************************
% Figure: HCCCN;v4=1;
\begin{figure*}[!htb]
    \centering
    \begin{subfigure}[t]{1.0\columnwidth}
       \includegraphics[width=1.0\columnwidth]{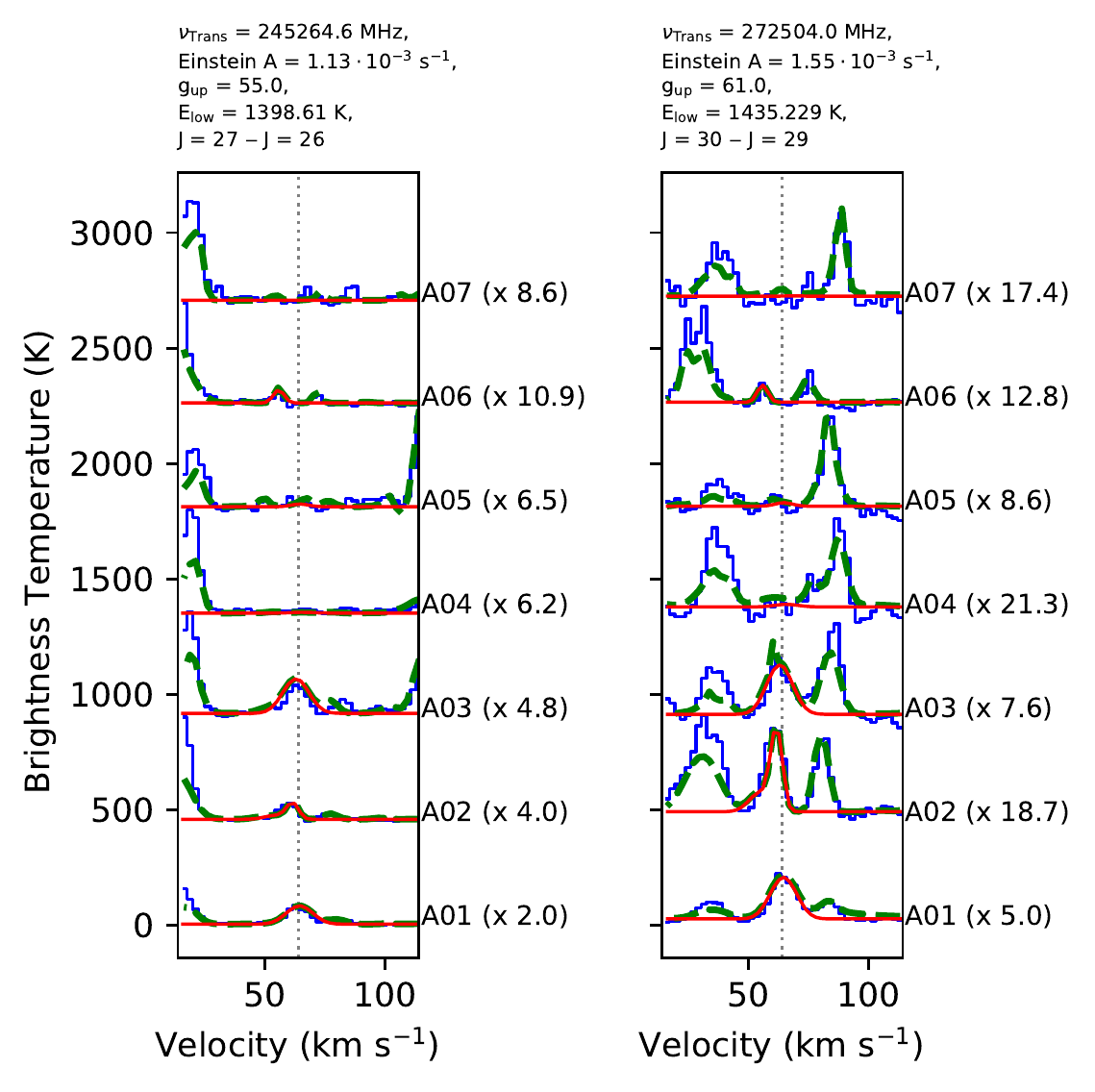}\\
       \caption{Sgr~B2(M)}
       \label{fig:HCCCNv41M}
    \end{subfigure}
\quad
    \begin{subfigure}[t]{1.0\columnwidth}
       \includegraphics[width=1.0\columnwidth]{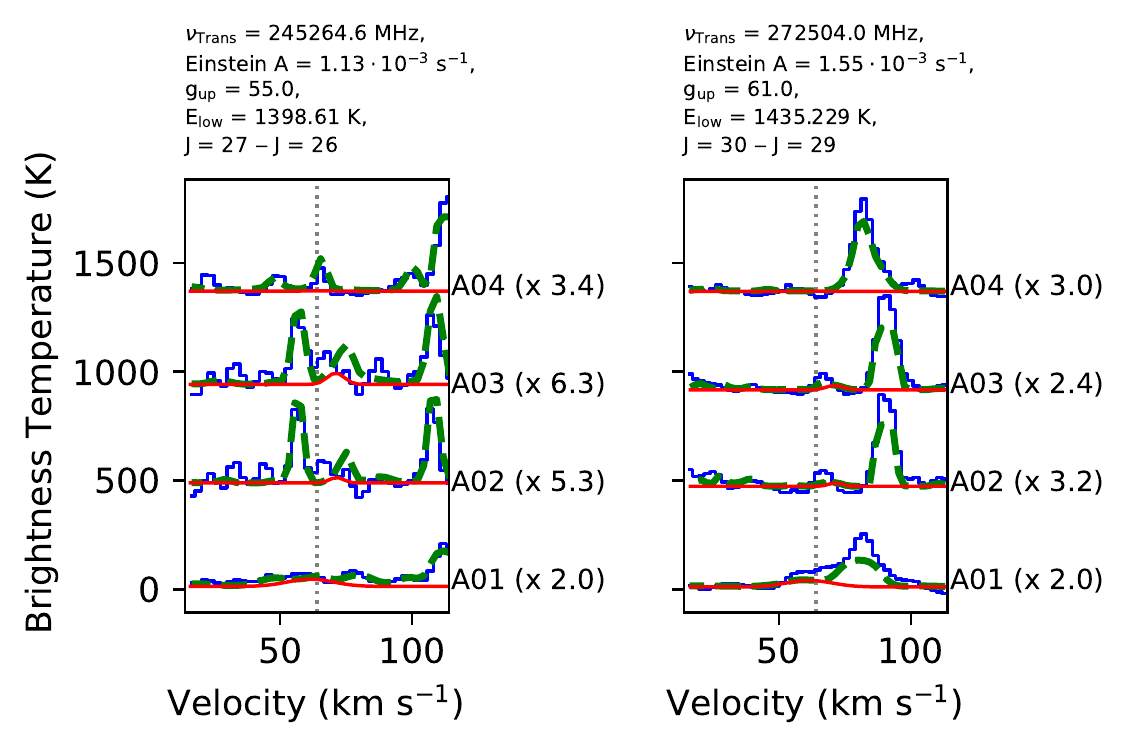}\\
       \caption{Sgr~B2(N)}
       \label{fig:HCCCNv41N}
   \end{subfigure}
   \caption{Selected transitions of HCCCN, v$_4$=1 in Sgr~B2(M) and N.}
   \ContinuedFloat
   \label{fig:HCCCNv41MN}
\end{figure*}

%*******************************************************************************
% Figure: HCCCN;v4=1,v7=1;
\begin{figure*}[!htb]
    \centering
    \begin{subfigure}[t]{1.0\columnwidth}
       \includegraphics[width=1.0\columnwidth]{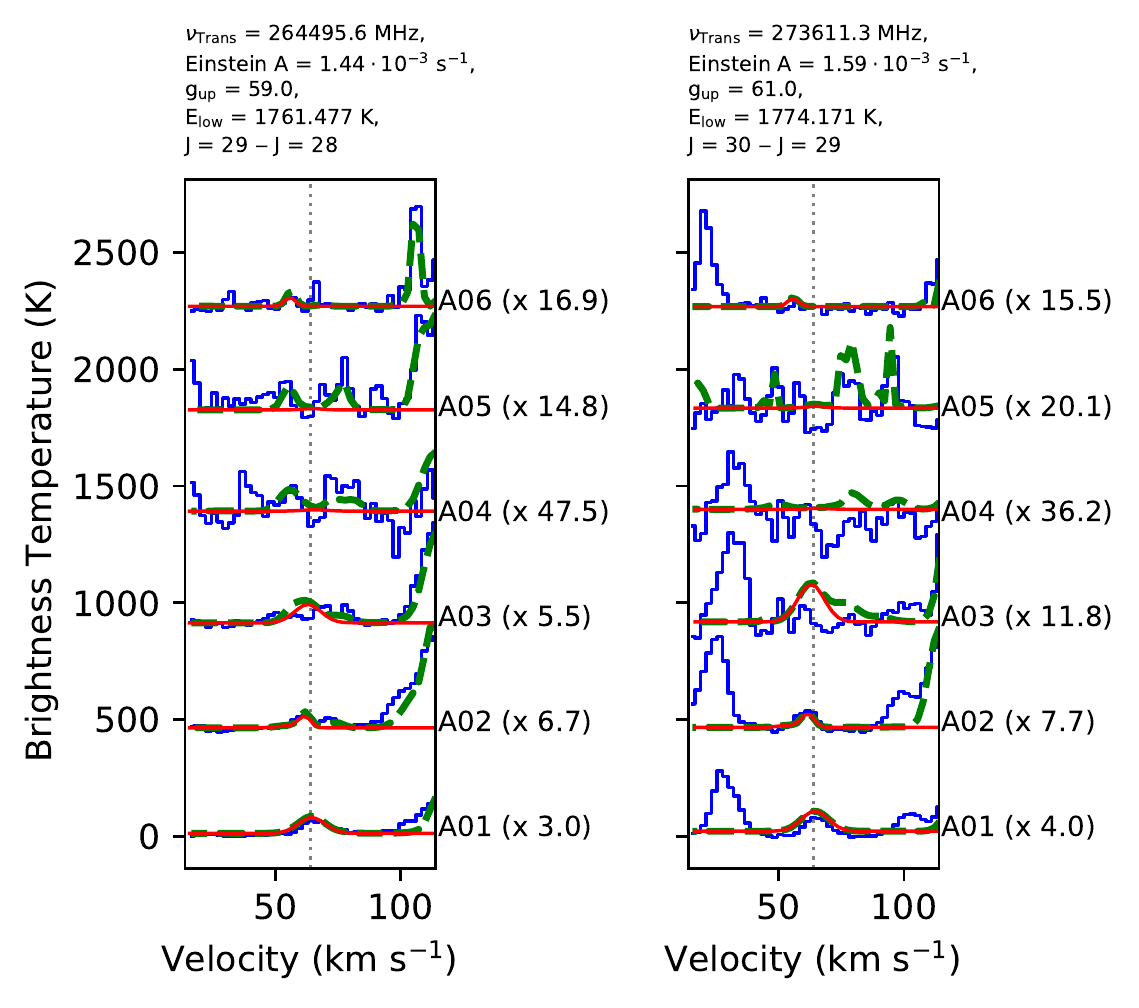}\\
       \caption{Sgr~B2(M)}
       \label{fig:HCCCNv41v71M}
    \end{subfigure}
\quad
    \begin{subfigure}[t]{1.0\columnwidth}
       \includegraphics[width=1.0\columnwidth]{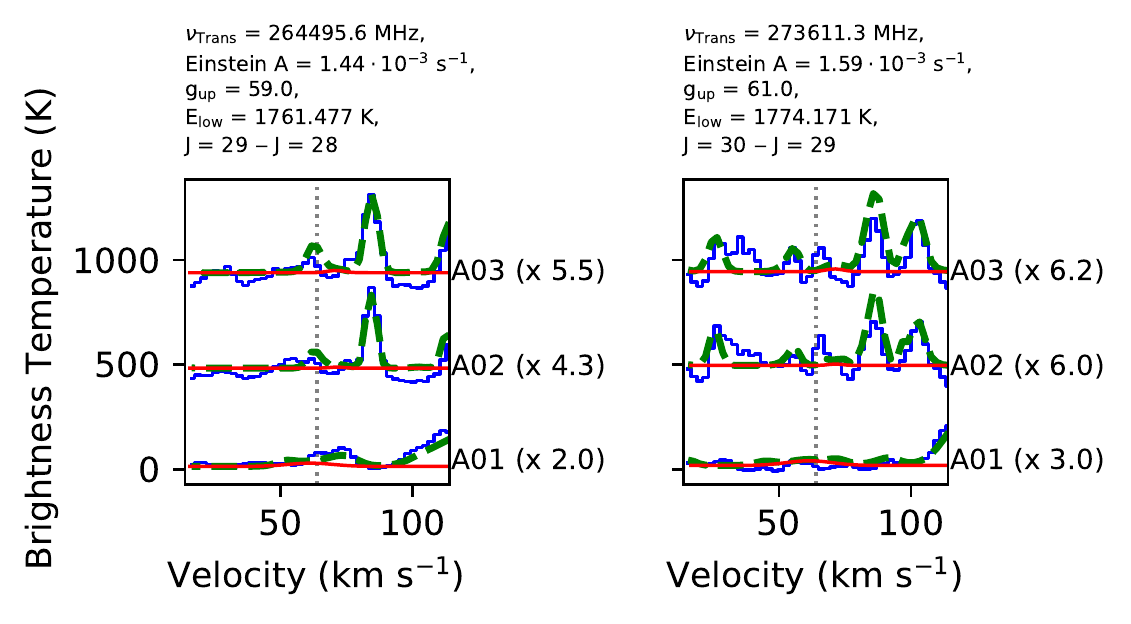}\\
       \caption{Sgr~B2(N)}
       \label{fig:HCCCNv41v71N}
   \end{subfigure}
   \caption{Selected transitions of HCCCN, v$_4$=1,v$_7$=1 in Sgr~B2(M) and N.}
   \ContinuedFloat
   \label{fig:HCCCNv41v71MN}
\end{figure*}

%*******************************************************************************
% Figure: HCCCN;v5=1;
\begin{figure*}[!htb]
    \centering
    \begin{subfigure}[t]{1.0\columnwidth}
       \includegraphics[width=1.0\columnwidth]{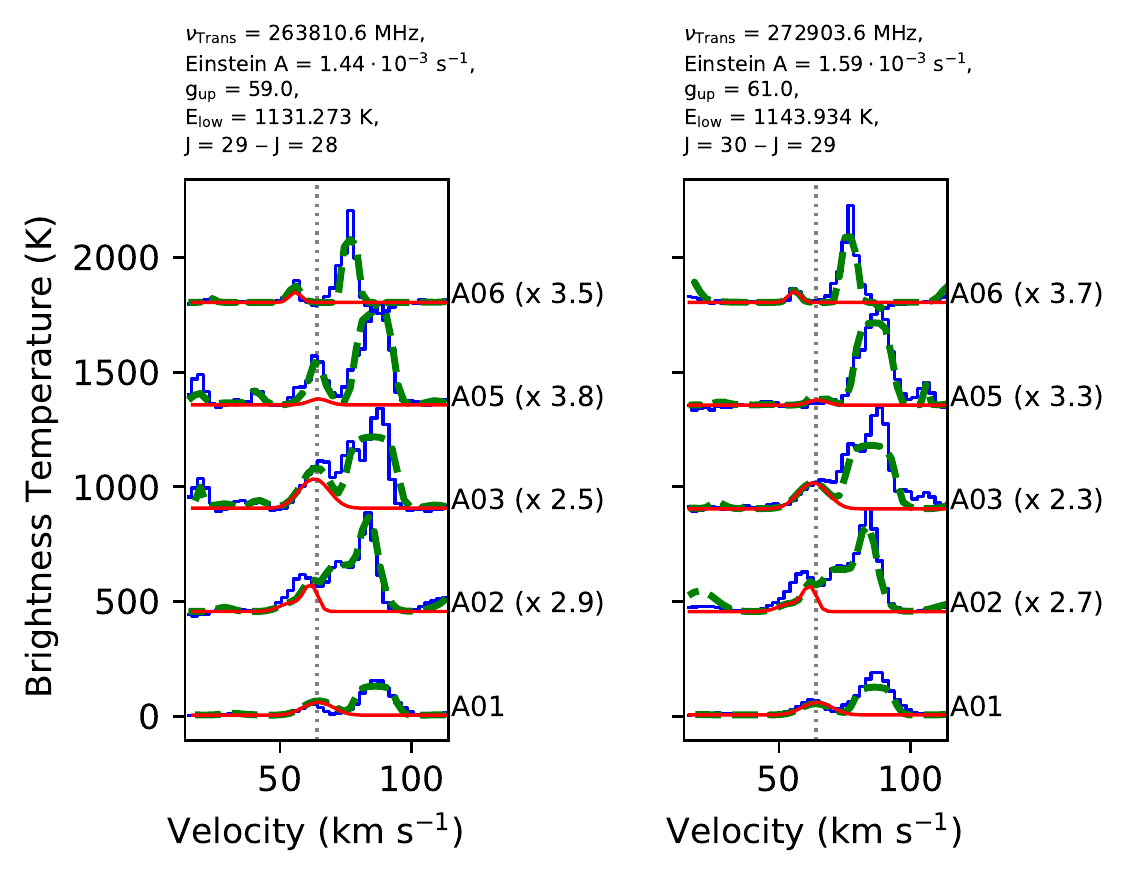}\\
       \caption{Sgr~B2(M)}
       \label{fig:HCCCNv51M}
    \end{subfigure}
\quad
    \begin{subfigure}[t]{1.0\columnwidth}
       \includegraphics[width=1.0\columnwidth]{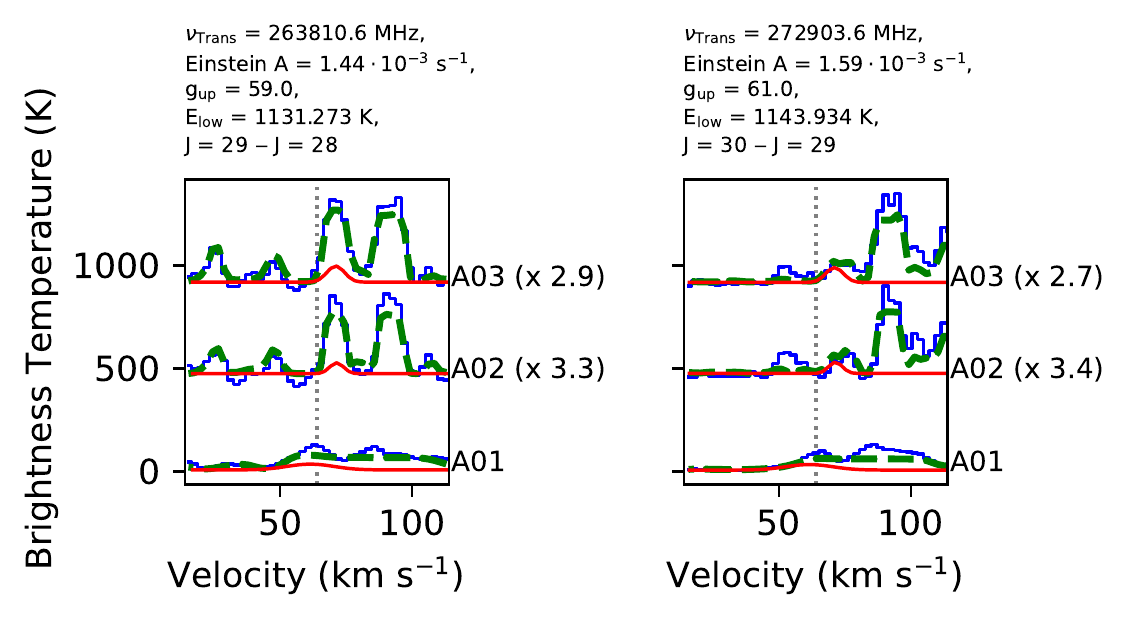}\\
       \caption{Sgr~B2(N)}
       \label{fig:HCCCNv51N}
   \end{subfigure}
   \caption{Selected transitions of HCCCN, v$_5$=1 in Sgr~B2(M) and N.}
   \ContinuedFloat
   \label{fig:HCCCNv51MN}
\end{figure*}

%*******************************************************************************
% Figure: HCCCN;v5=1,v7=1;
\begin{figure*}[!htb]
    \centering
    \begin{subfigure}[t]{1.0\columnwidth}
       \includegraphics[width=1.0\columnwidth]{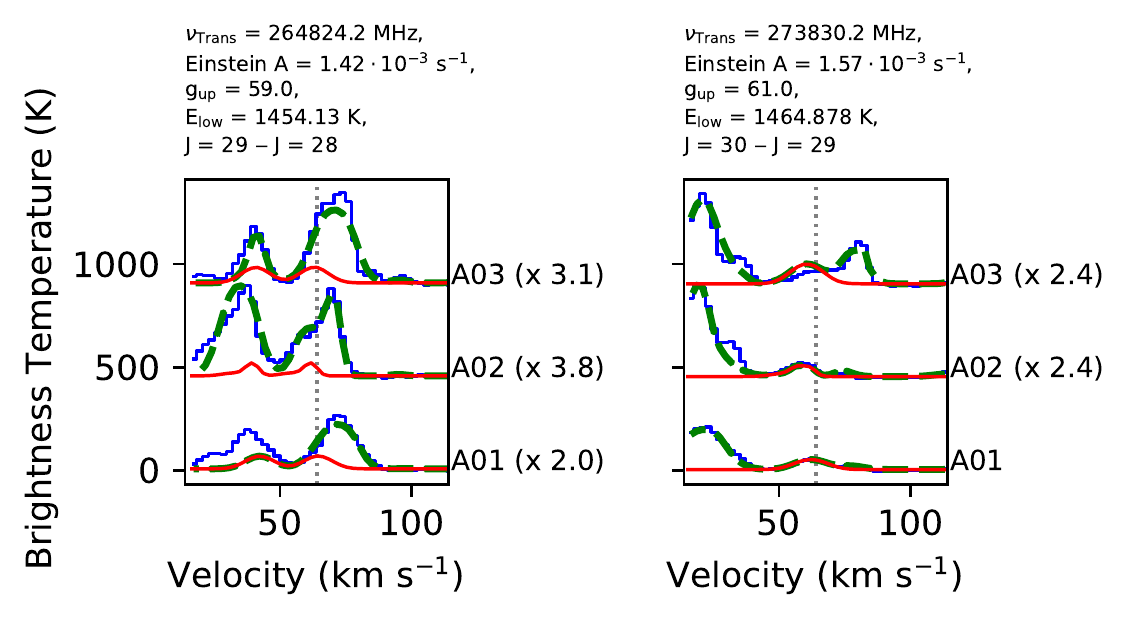}\\
       \caption{Sgr~B2(M)}
       \label{fig:HCCCNv51v71M}
    \end{subfigure}
\quad
    \begin{subfigure}[t]{1.0\columnwidth}
       \includegraphics[width=1.0\columnwidth]{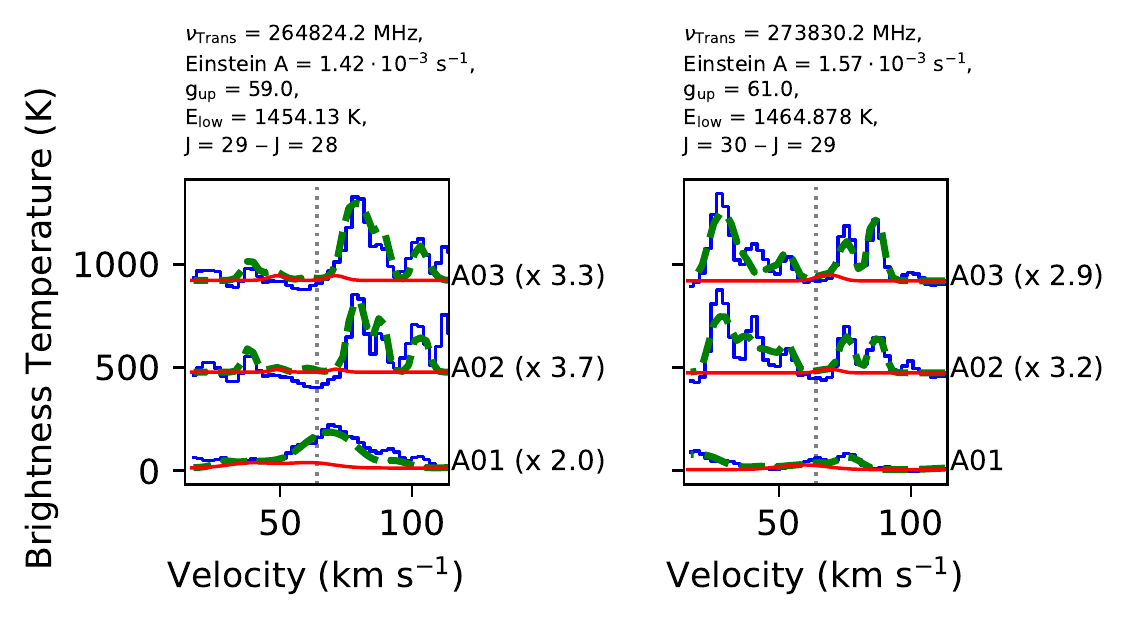}\\
       \caption{Sgr~B2(N)}
       \label{fig:HCCCNv51v71N}
   \end{subfigure}
   \caption{Selected transitions of HCCCN, v$_5$=1,v$_7$=1 in Sgr~B2(M) and N.}
   \ContinuedFloat
   \label{fig:HCCCNv51v71MN}
\end{figure*}

%*******************************************************************************
% Figure: HCCCN;v5=2;
\begin{figure*}[!htb]
    \centering
    \begin{subfigure}[t]{1.0\columnwidth}
       \includegraphics[width=1.0\columnwidth]{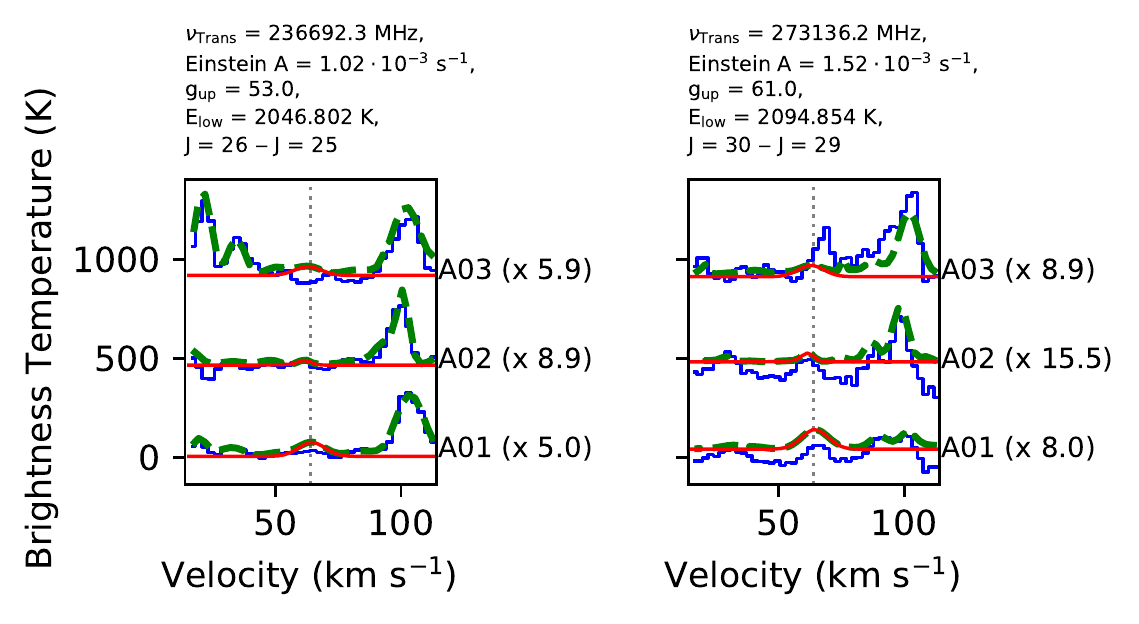}\\
       \caption{Sgr~B2(M)}
       \label{fig:HCCCNv52M}
    \end{subfigure}
\quad
    \begin{subfigure}[t]{1.0\columnwidth}
       \includegraphics[width=1.0\columnwidth]{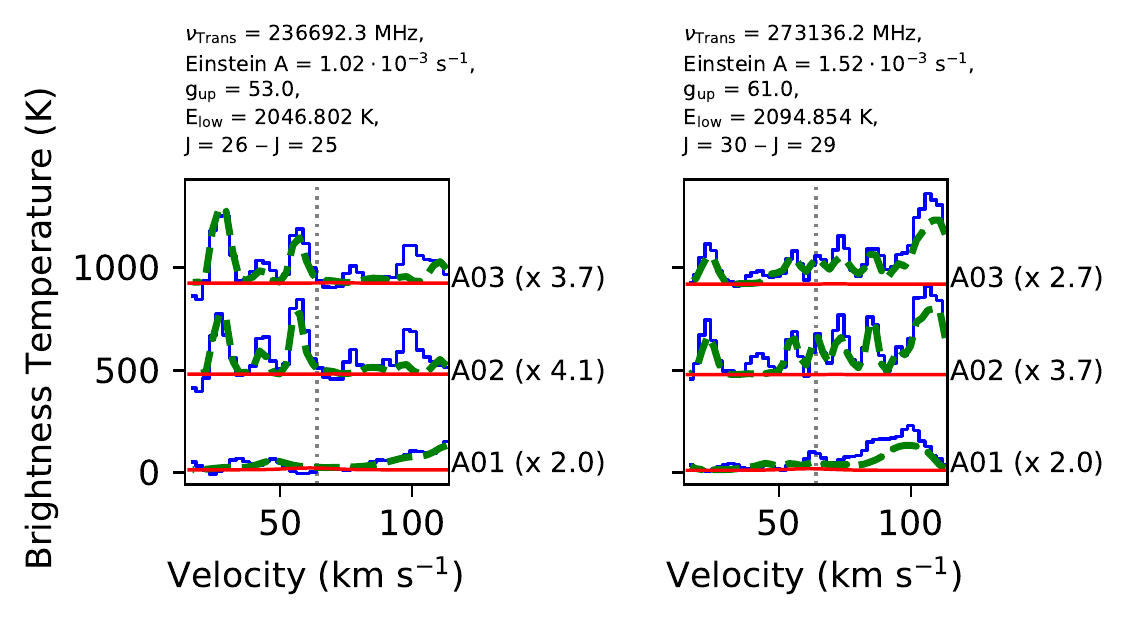}\\
       \caption{Sgr~B2(N)}
       \label{fig:HCCCNv52N}
   \end{subfigure}
   \caption{Selected transitions of HCCCN, v$_5$=2 in Sgr~B2(M) and N.}
   \ContinuedFloat
   \label{fig:HCCCNv52MN}
\end{figure*}

%*******************************************************************************
% Figure: HCCCN;v6=1;
\begin{figure*}[!htb]
    \centering
    \begin{subfigure}[t]{1.0\columnwidth}
       \includegraphics[width=1.0\columnwidth]{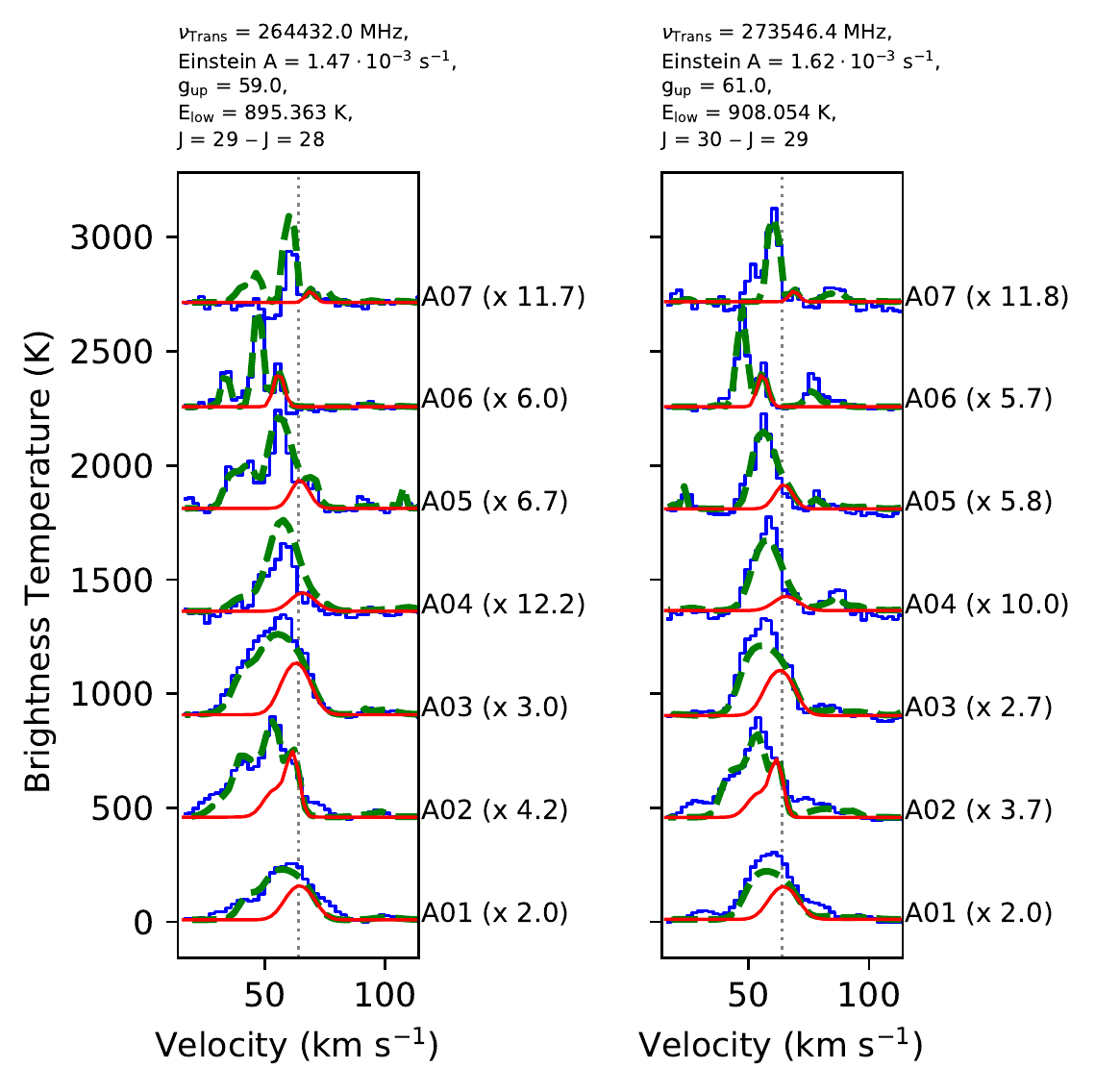}\\
       \caption{Sgr~B2(M)}
       \label{fig:HCCCNv61M}
    \end{subfigure}
\quad
    \begin{subfigure}[t]{1.0\columnwidth}
       \includegraphics[width=1.0\columnwidth]{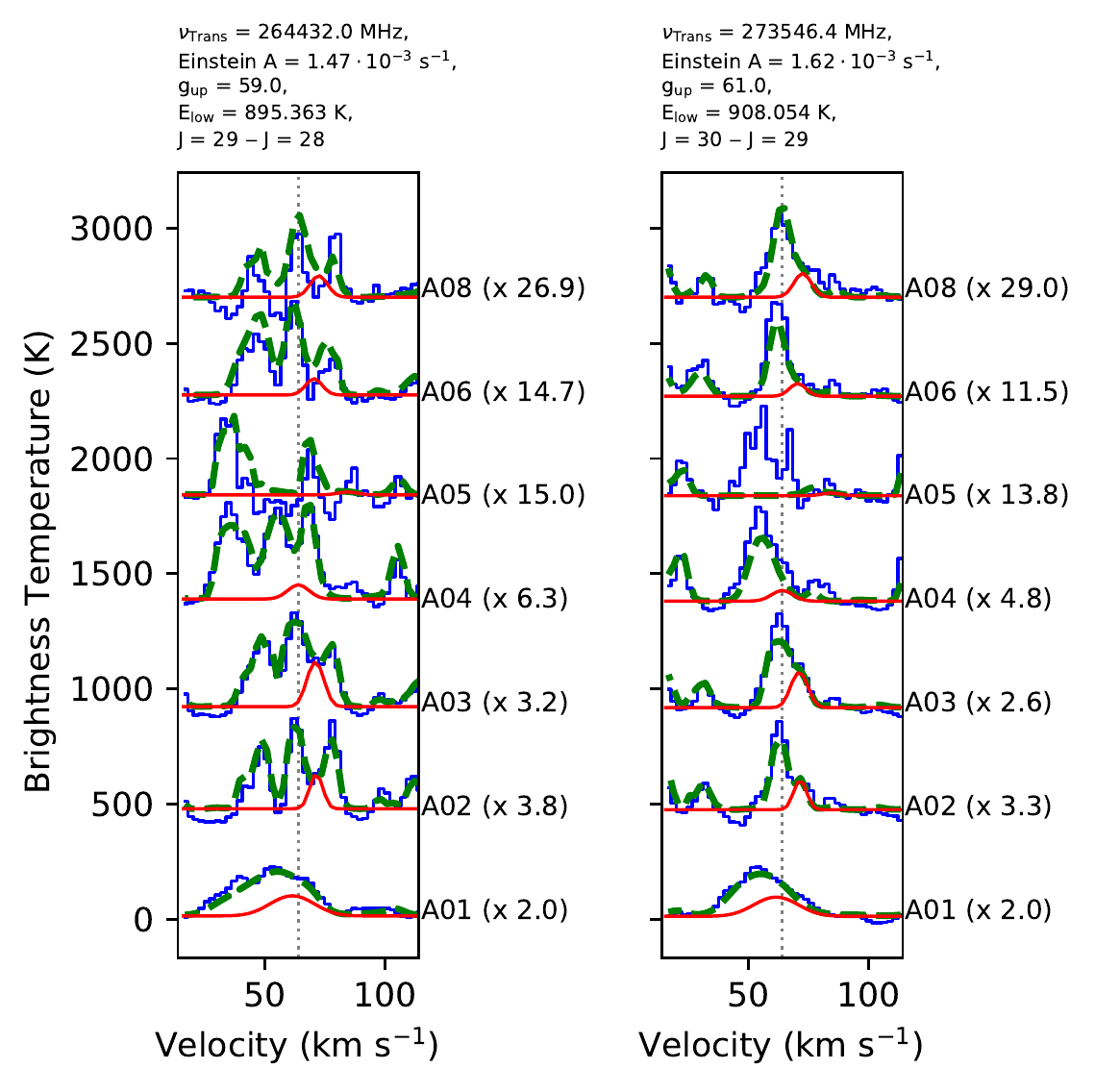}\\
       \caption{Sgr~B2(N)}
       \label{fig:HCCCNv61N}
   \end{subfigure}
   \caption{Selected transitions of HCCCN, v$_6$=1 in Sgr~B2(M) and N.}
   \ContinuedFloat
   \label{fig:HCCCNv61MN}
\end{figure*}

%*******************************************************************************
% Figure: HCCCN;v6=1,v7=1;
\begin{figure*}[!htb]
    \centering
    \begin{subfigure}[t]{1.0\columnwidth}
       \includegraphics[width=1.0\columnwidth]{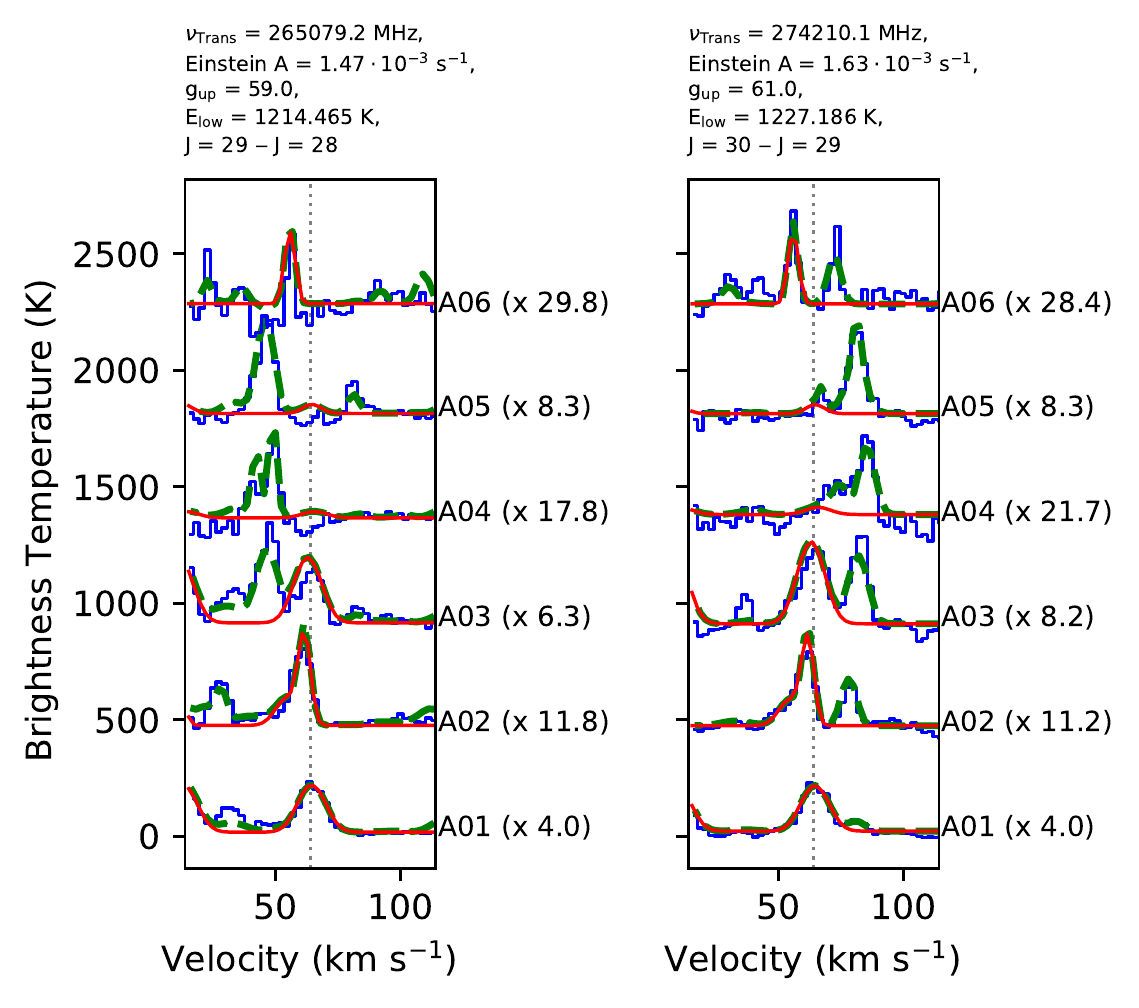}\\
       \caption{Sgr~B2(M)}
       \label{fig:HCCCNv61v71M}
    \end{subfigure}
\quad
    \begin{subfigure}[t]{1.0\columnwidth}
       \includegraphics[width=1.0\columnwidth]{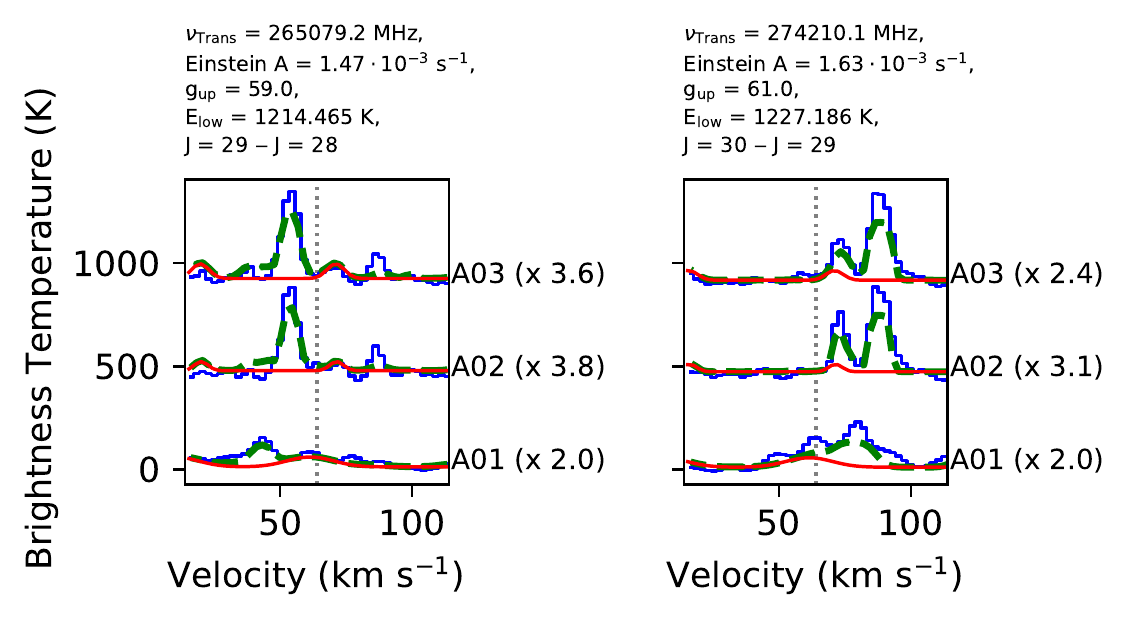}\\
       \caption{Sgr~B2(N)}
       \label{fig:HCCCNv61v71N}
   \end{subfigure}
   \caption{Selected transitions of HCCCN, v$_6$=1,v$_7$=1 in Sgr~B2(M) and N.}
   \ContinuedFloat
   \label{fig:HCCCNv61v71MN}
\end{figure*}

%*******************************************************************************
% Figure: HCCCN;v7=1;
\begin{figure*}[!htb]
    \centering
    \begin{subfigure}[t]{1.0\columnwidth}
       \includegraphics[width=1.0\columnwidth]{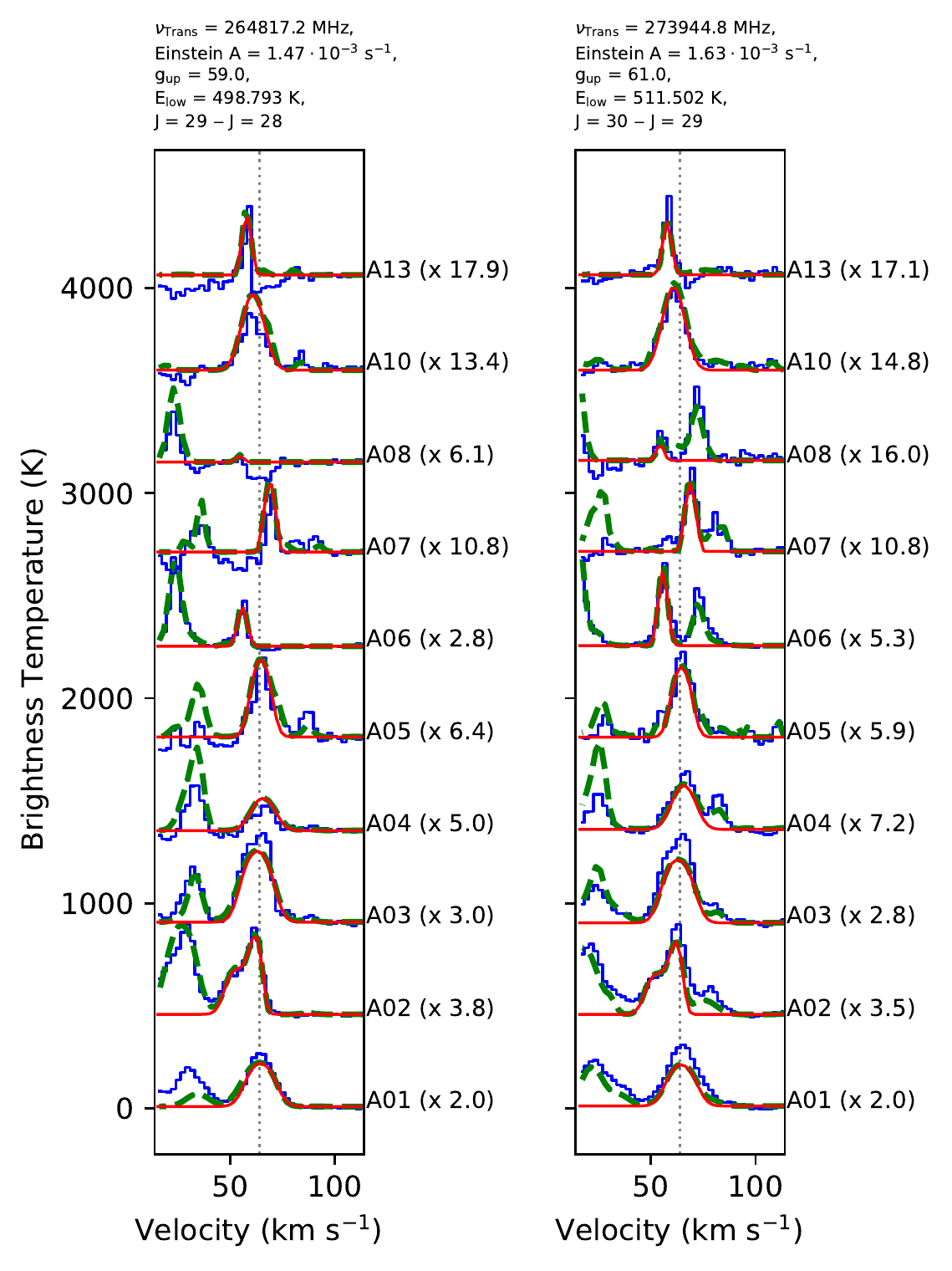}\\
       \caption{Sgr~B2(M)}
       \label{fig:HCCCNv71M}
    \end{subfigure}
\quad
    \begin{subfigure}[t]{1.0\columnwidth}
       \includegraphics[width=1.0\columnwidth]{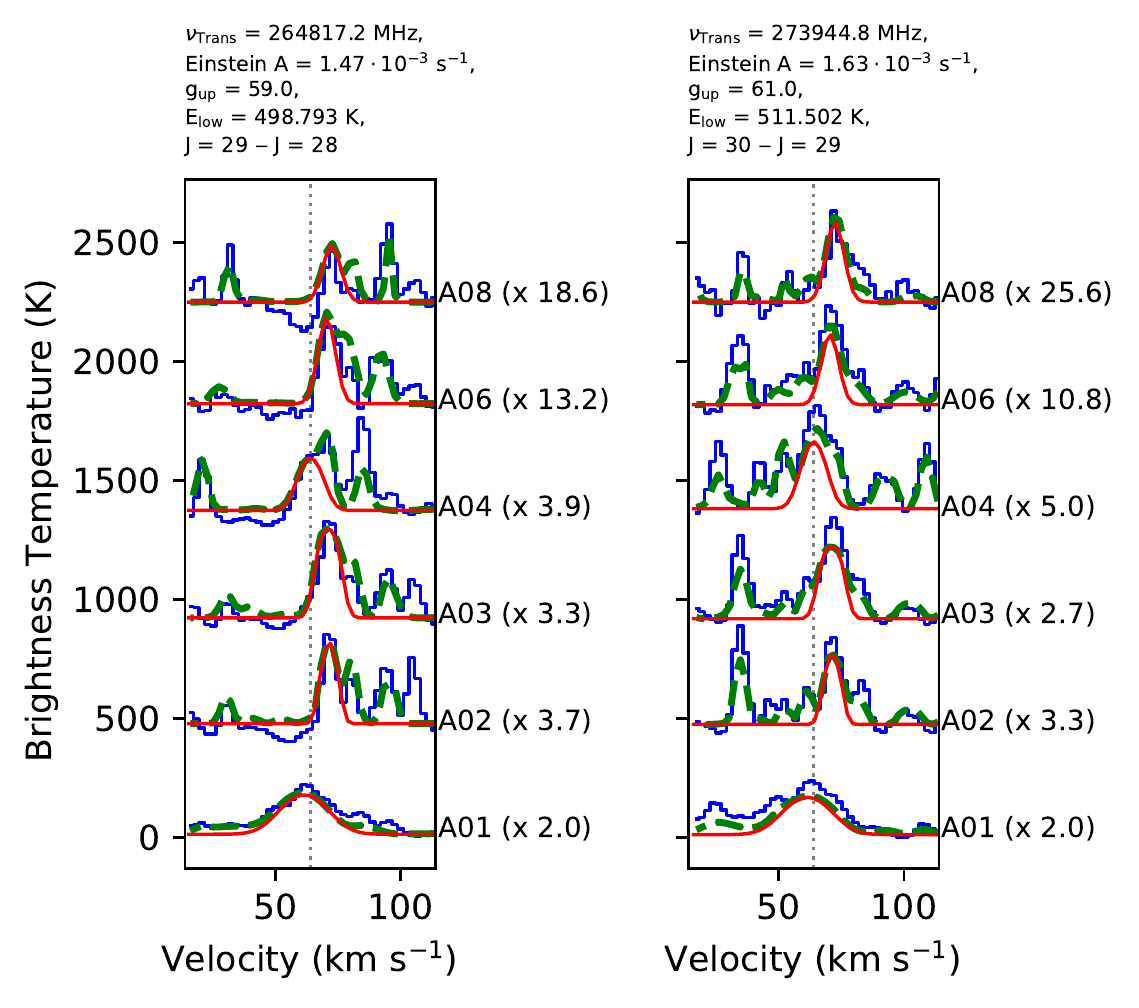}\\
       \caption{Sgr~B2(N)}
       \label{fig:HCCCNv71N}
   \end{subfigure}
   \caption{Selected transitions of HCCCN, v$_7$=1 in Sgr~B2(M) and N.}
   \ContinuedFloat
   \label{fig:HCCCNv71MN}
\end{figure*}

%*******************************************************************************
% Figure: HCCCN;v7=2;
\begin{figure*}[!htb]
    \centering
    \begin{subfigure}[t]{1.0\columnwidth}
       \includegraphics[width=1.0\columnwidth]{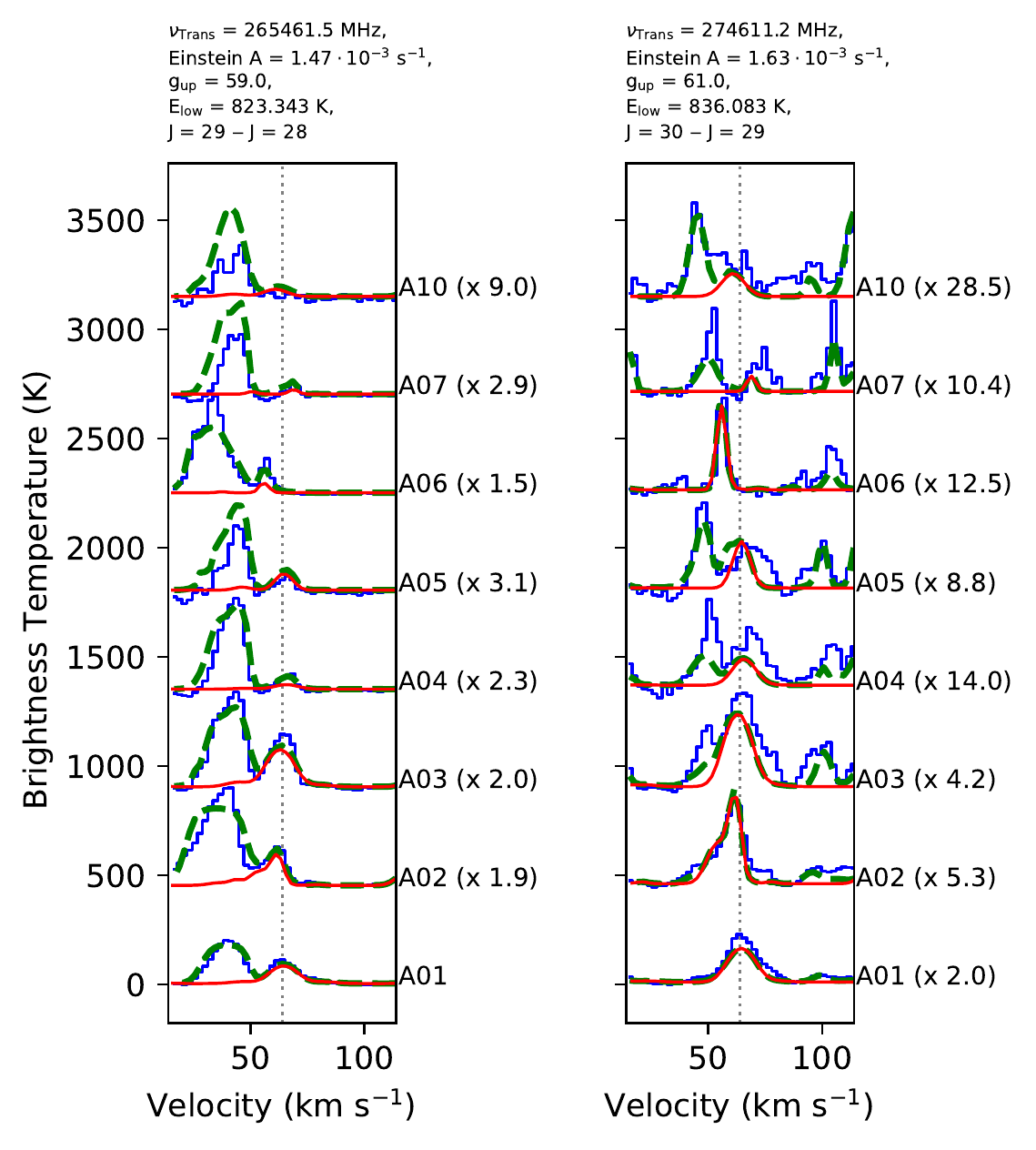}\\
       \caption{Sgr~B2(M)}
       \label{fig:HCCCNv72M}
    \end{subfigure}
\quad
    \begin{subfigure}[t]{1.0\columnwidth}
       \includegraphics[width=1.0\columnwidth]{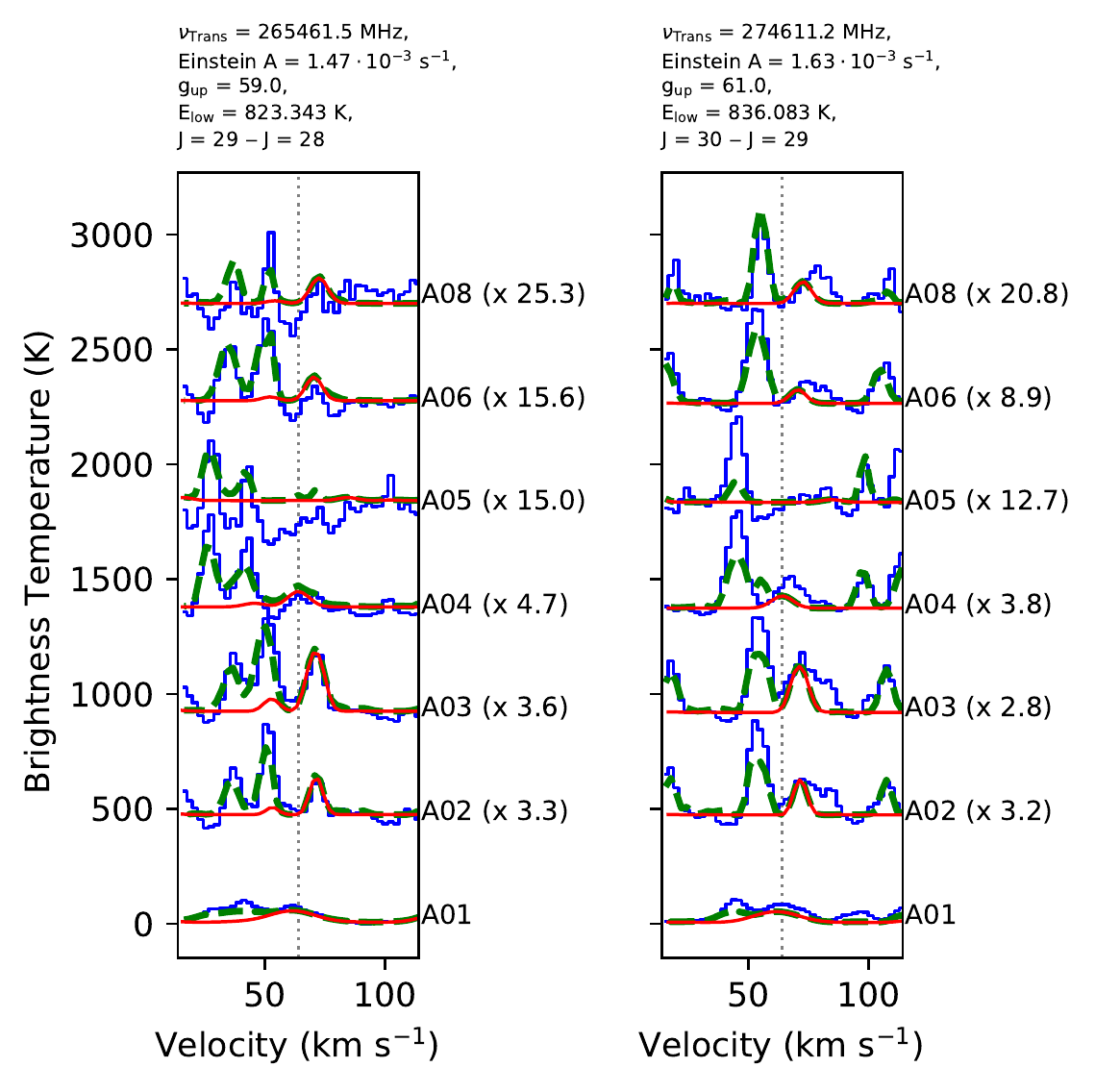}\\
       \caption{Sgr~B2(N)}
       \label{fig:HCCCNv72N}
   \end{subfigure}
   \caption{Selected transitions of HCCCN, v$_7$=2 in Sgr~B2(M) and N.}
   \ContinuedFloat
   \label{fig:HCCCNv72MN}
\end{figure*}

%*******************************************************************************
% Figure: HCCCN;v7=3;
\begin{figure*}[!htb]
    \centering
    \begin{subfigure}[t]{1.0\columnwidth}
       \includegraphics[width=1.0\columnwidth]{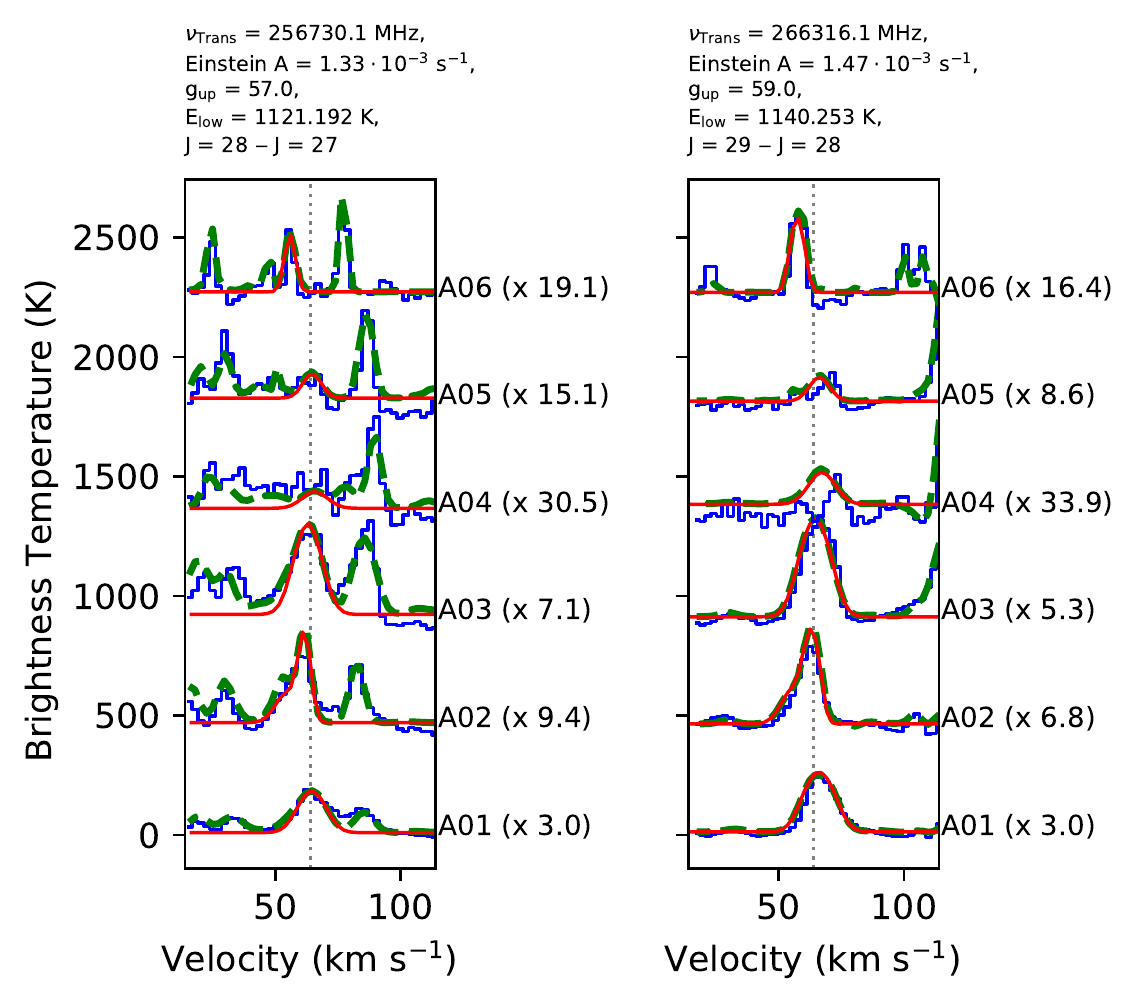}\\
       \caption{Sgr~B2(M)}
       \label{fig:HCCCNv73M}
    \end{subfigure}
\quad
    \begin{subfigure}[t]{1.0\columnwidth}
       \includegraphics[width=1.0\columnwidth]{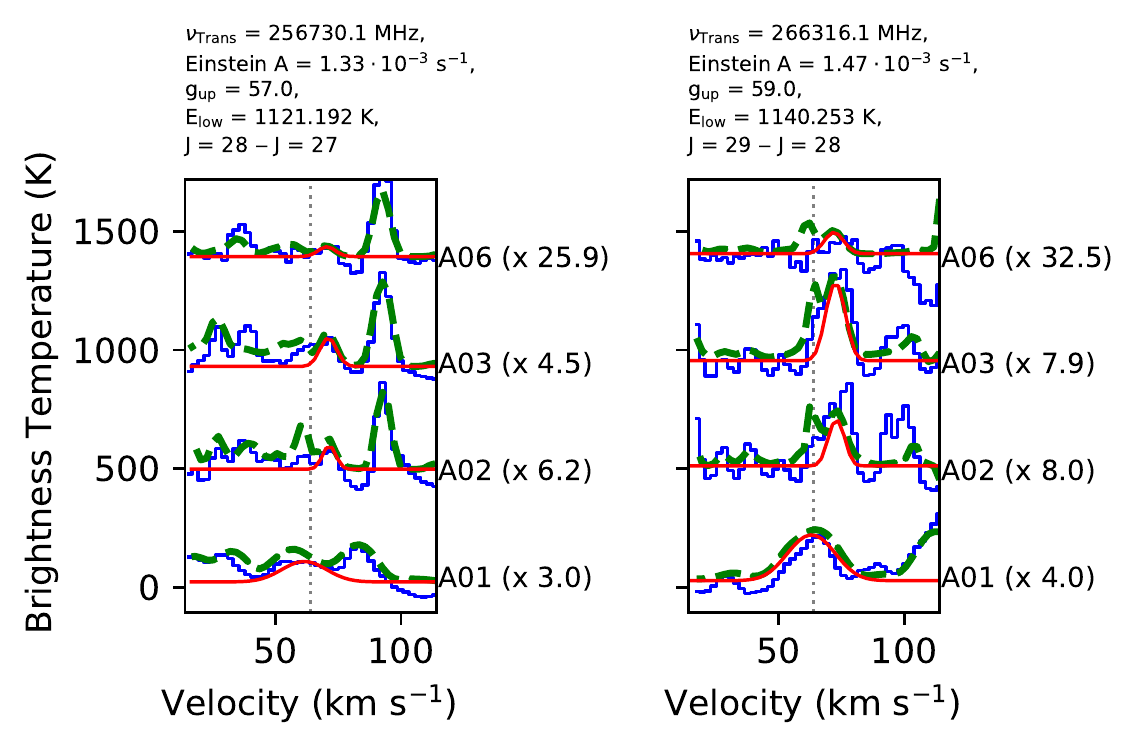}\\
       \caption{Sgr~B2(N)}
       \label{fig:HCCCNv73N}
   \end{subfigure}
   \caption{Selected transitions of HCCCN, v$_7$=3 in Sgr~B2(M) and N.}
   \ContinuedFloat
   \label{fig:HCCCNv73MN}
\end{figure*}

%*******************************************************************************
% Figure: HC3N;v7=4;
\begin{figure*}[!htb]
    \centering
    \begin{subfigure}[t]{1.0\columnwidth}
       \includegraphics[width=1.0\columnwidth]{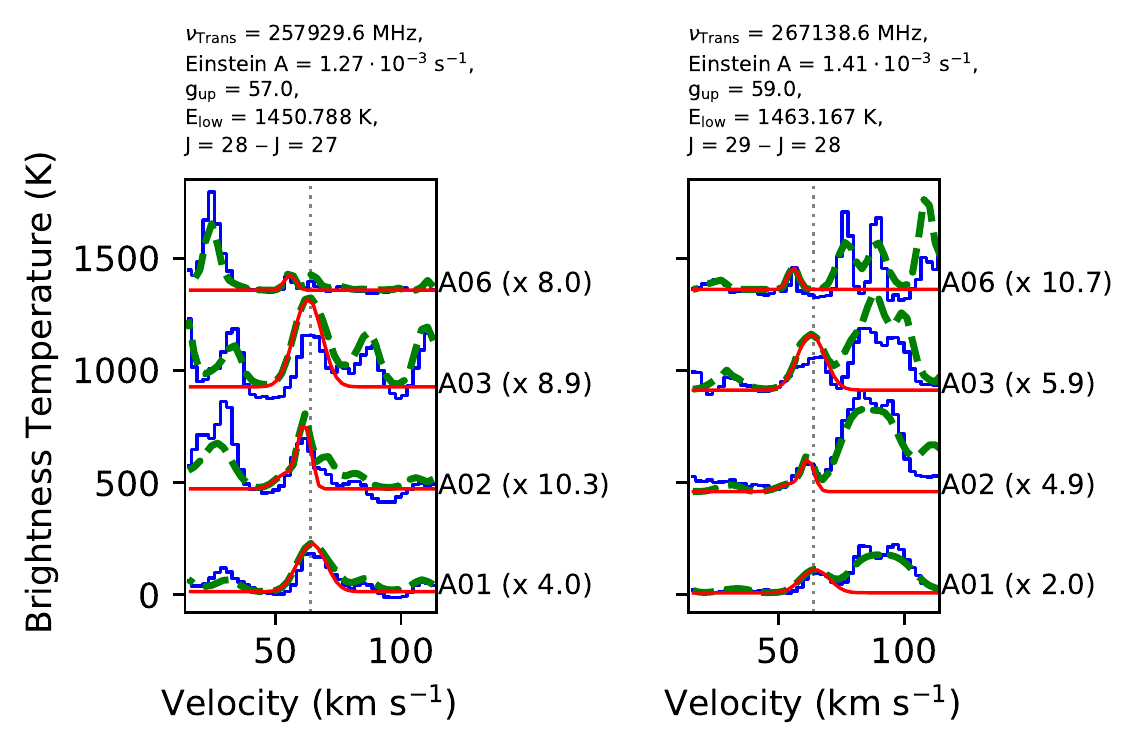}\\
       \caption{Sgr~B2(M)}
       \label{fig:HCCCNv74M}
    \end{subfigure}
\quad
    \begin{subfigure}[t]{1.0\columnwidth}
       \includegraphics[width=1.0\columnwidth]{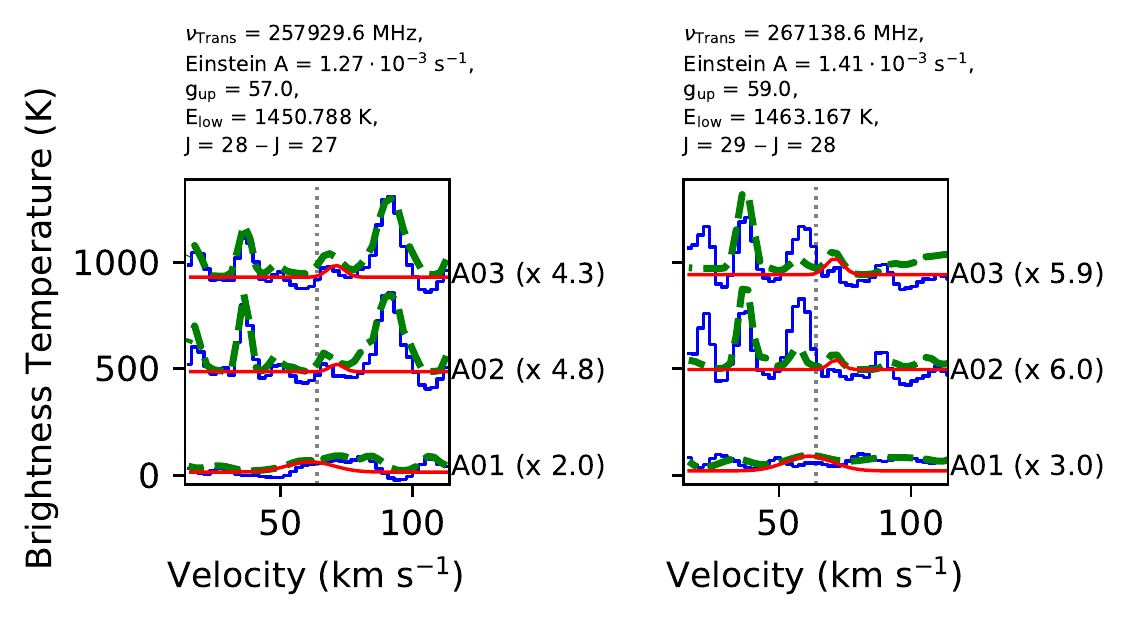}\\
       \caption{Sgr~B2(N)}
       \label{fig:HCCCNv74N}
   \end{subfigure}
   \caption{Selected transitions of HCCCN, v$_7$=4 in Sgr~B2(M) and N.}
   \ContinuedFloat
   \label{fig:HCCCNv74MN}
\end{figure*}

%*******************************************************************************
% Figure: HC-13-CCN;v7=1;
\begin{figure*}[!htb]
    \centering
    \begin{subfigure}[t]{1.0\columnwidth}
       \includegraphics[width=1.0\columnwidth]{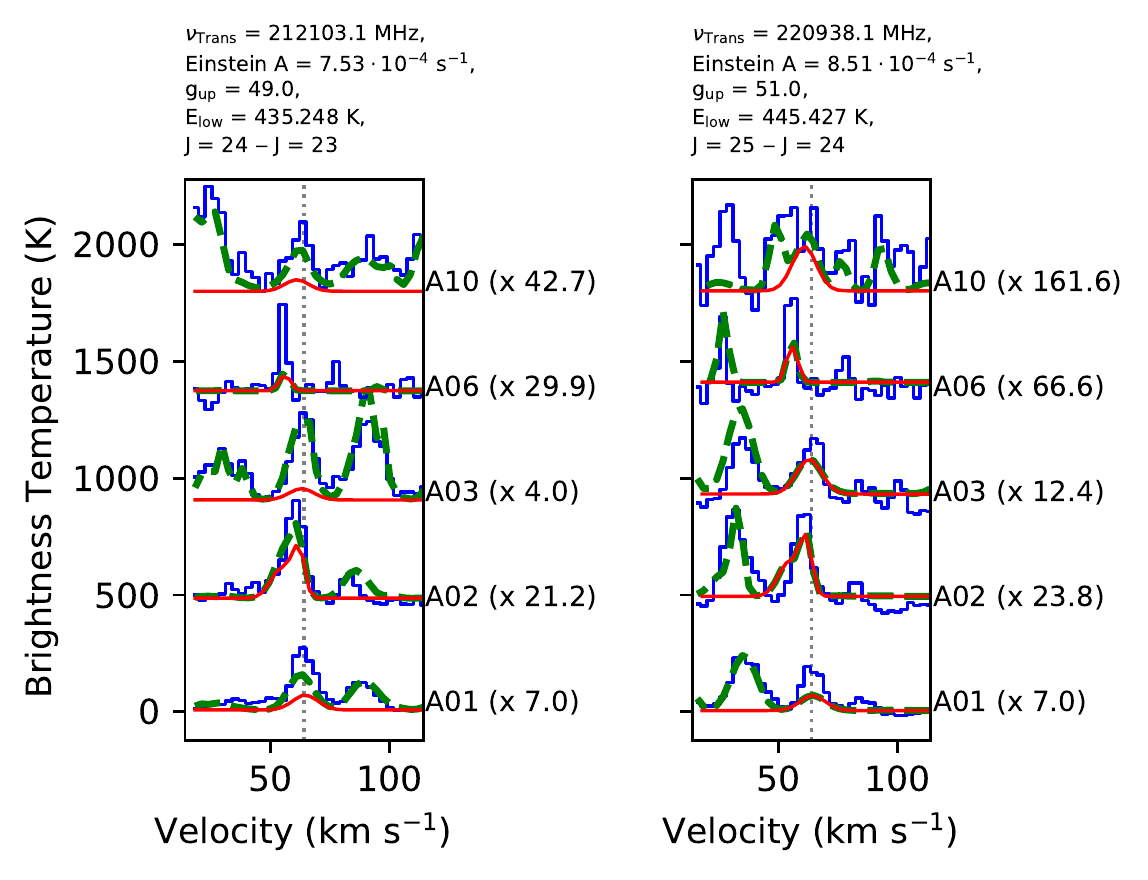}\\
       \caption{Sgr~B2(M)}
       \label{fig:HC13CCNv71M}
    \end{subfigure}
\quad
    \begin{subfigure}[t]{1.0\columnwidth}
       \includegraphics[width=1.0\columnwidth]{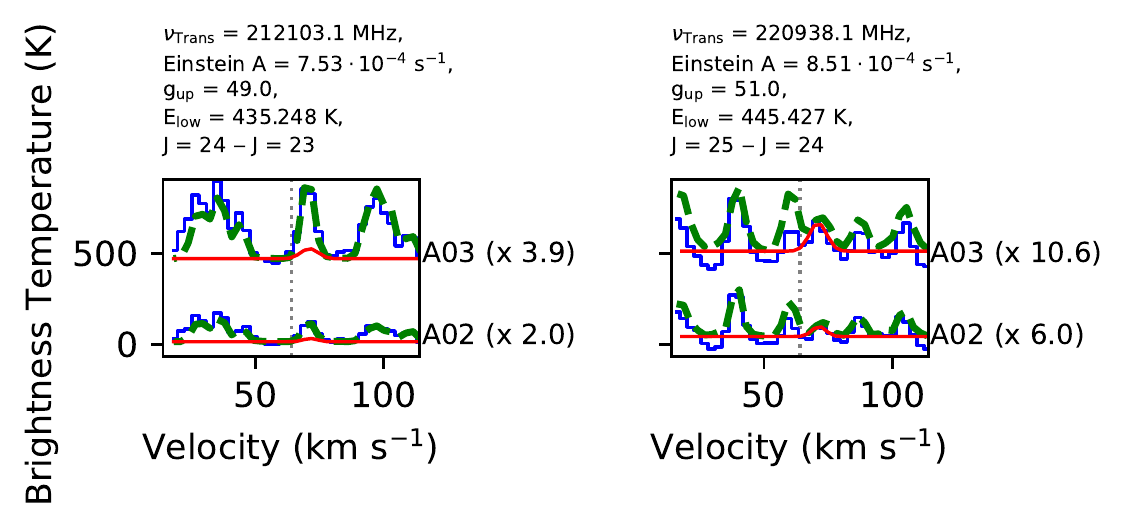}\\
       \caption{Sgr~B2(N)}
       \label{fig:HC13CCNv71N}
   \end{subfigure}
   \caption{Selected transitions of H$^{13}$CCCN, v$_7$=1 in Sgr~B2(M) and N.}
   \ContinuedFloat
   \label{fig:HC13CCNv71MN}
\end{figure*}

%*******************************************************************************
% Figure: HC-13-CCN;v7=2;
\begin{figure*}[!htb]
    \centering
    \begin{subfigure}[t]{1.0\columnwidth}
       \includegraphics[width=1.0\columnwidth]{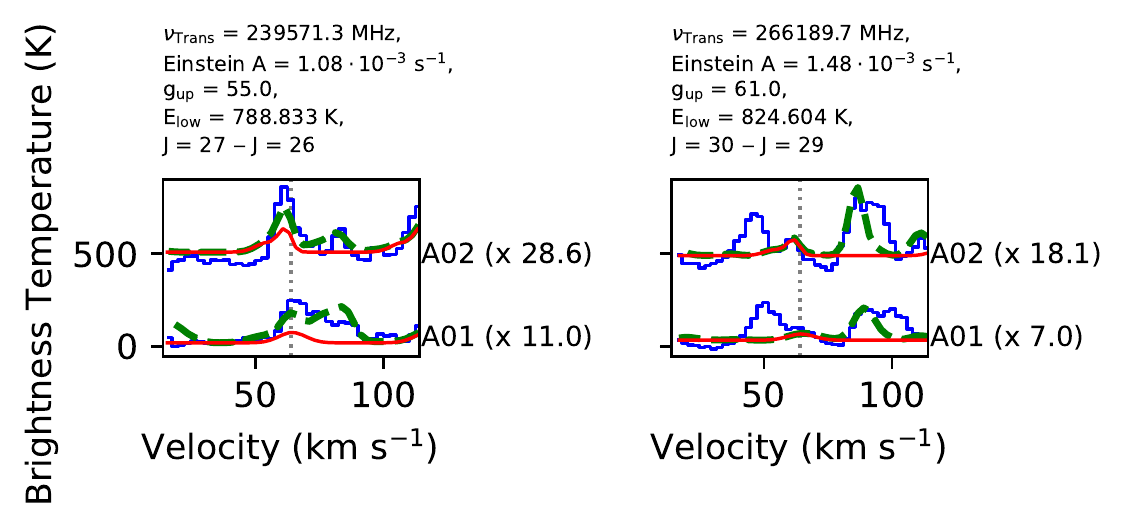}\\
    \end{subfigure}
   \caption{Selected transitions of H$^{13}$CCCN, v$_7$=2 in Sgr~B2(M)}
   \ContinuedFloat
   \label{fig:HC13CCNv72M}
\end{figure*}

%*******************************************************************************
% Figure: HC-13-CCN;v7=3;
\begin{figure*}[!htb]
    \centering
    \begin{subfigure}[t]{1.0\columnwidth}
       \includegraphics[width=1.0\columnwidth]{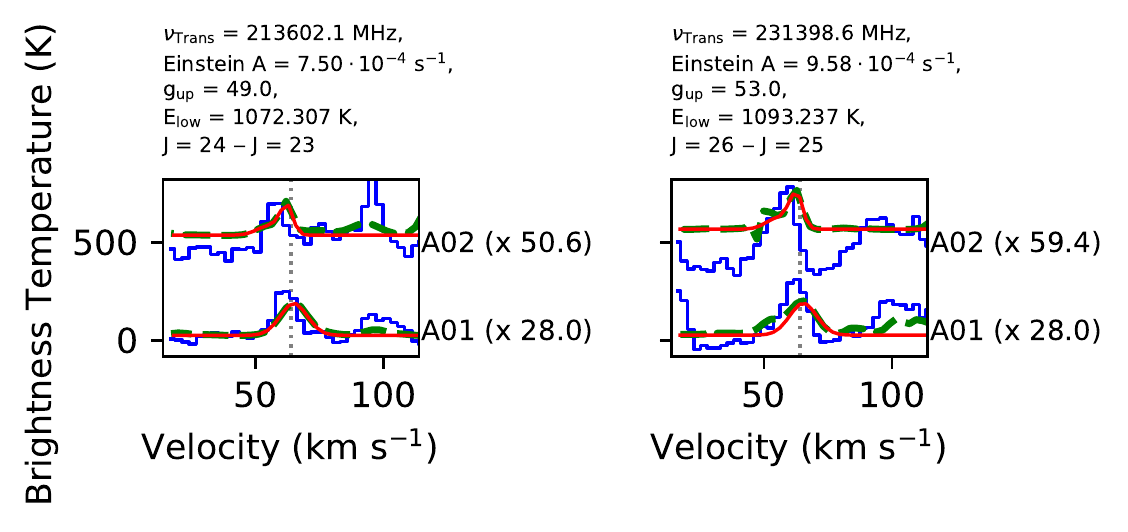}\\
    \end{subfigure}
   \caption{Selected transitions of H$^{13}$CCCN, v$_7$=3 in Sgr~B2(M)}
   \ContinuedFloat
   \label{fig:HC13CCNv73M}
\end{figure*}

%*******************************************************************************
% Figure: HCC-13-CN;v7=1;
\begin{figure*}[!htb]
    \centering
    \begin{subfigure}[t]{1.0\columnwidth}
       \includegraphics[width=1.0\columnwidth]{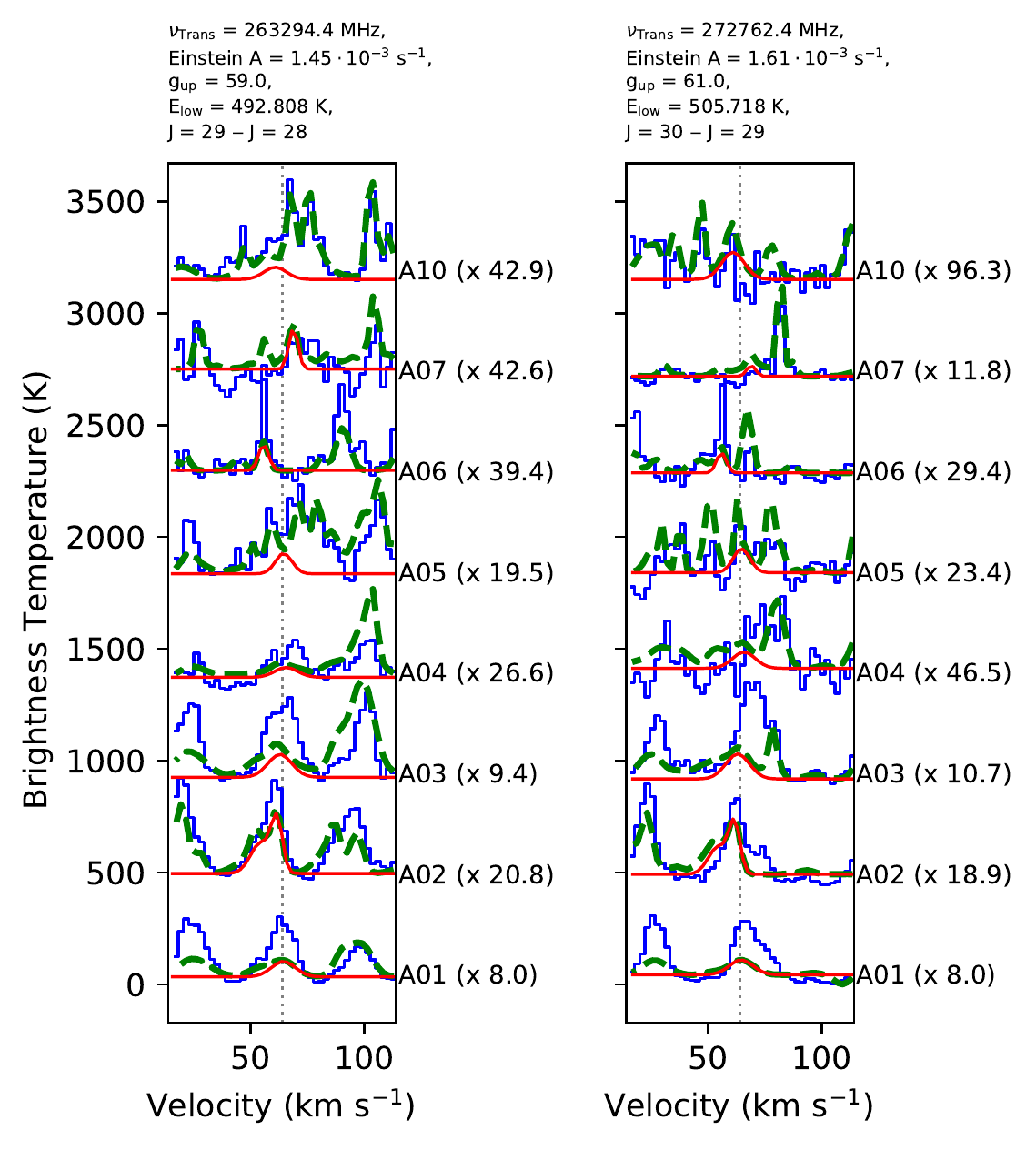}\\
       \caption{Sgr~B2(M)}
       \label{fig:HCC13CNv71M}
    \end{subfigure}
\quad
    \begin{subfigure}[t]{1.0\columnwidth}
       \includegraphics[width=1.0\columnwidth]{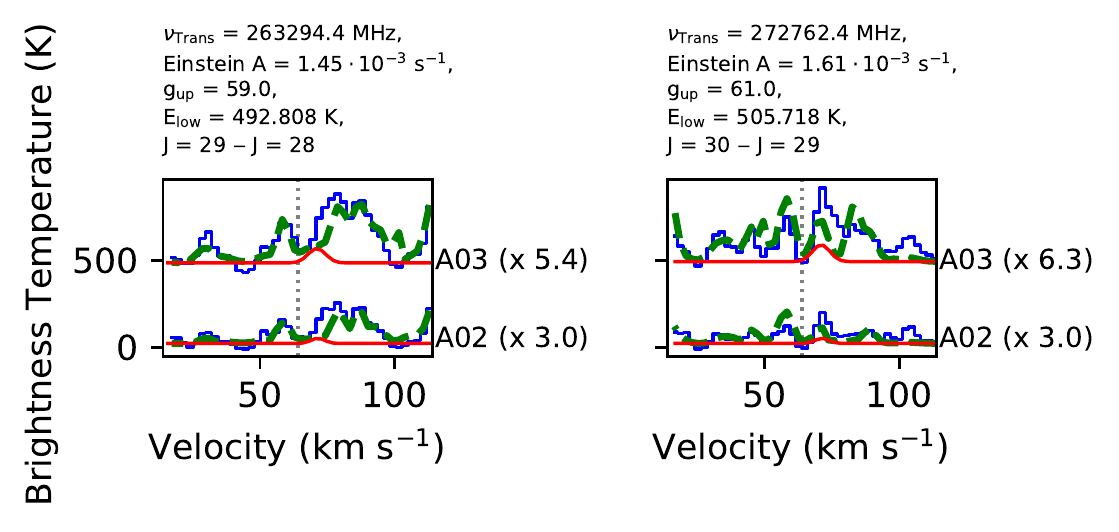}\\
       \caption{Sgr~B2(N)}
       \label{fig:HCC13CNv71N}
   \end{subfigure}
   \caption{Selected transitions of HC$^{13}$CCN, v$_7$=1 in Sgr~B2(M) and N.}
   \ContinuedFloat
   \label{fig:HCC13CNv71MN}
\end{figure*}

%*******************************************************************************
% Figure: HCCC-13-N;v7=1;
\begin{figure*}[!htb]
    \centering
    \begin{subfigure}[t]{1.0\columnwidth}
       \includegraphics[width=1.0\columnwidth]{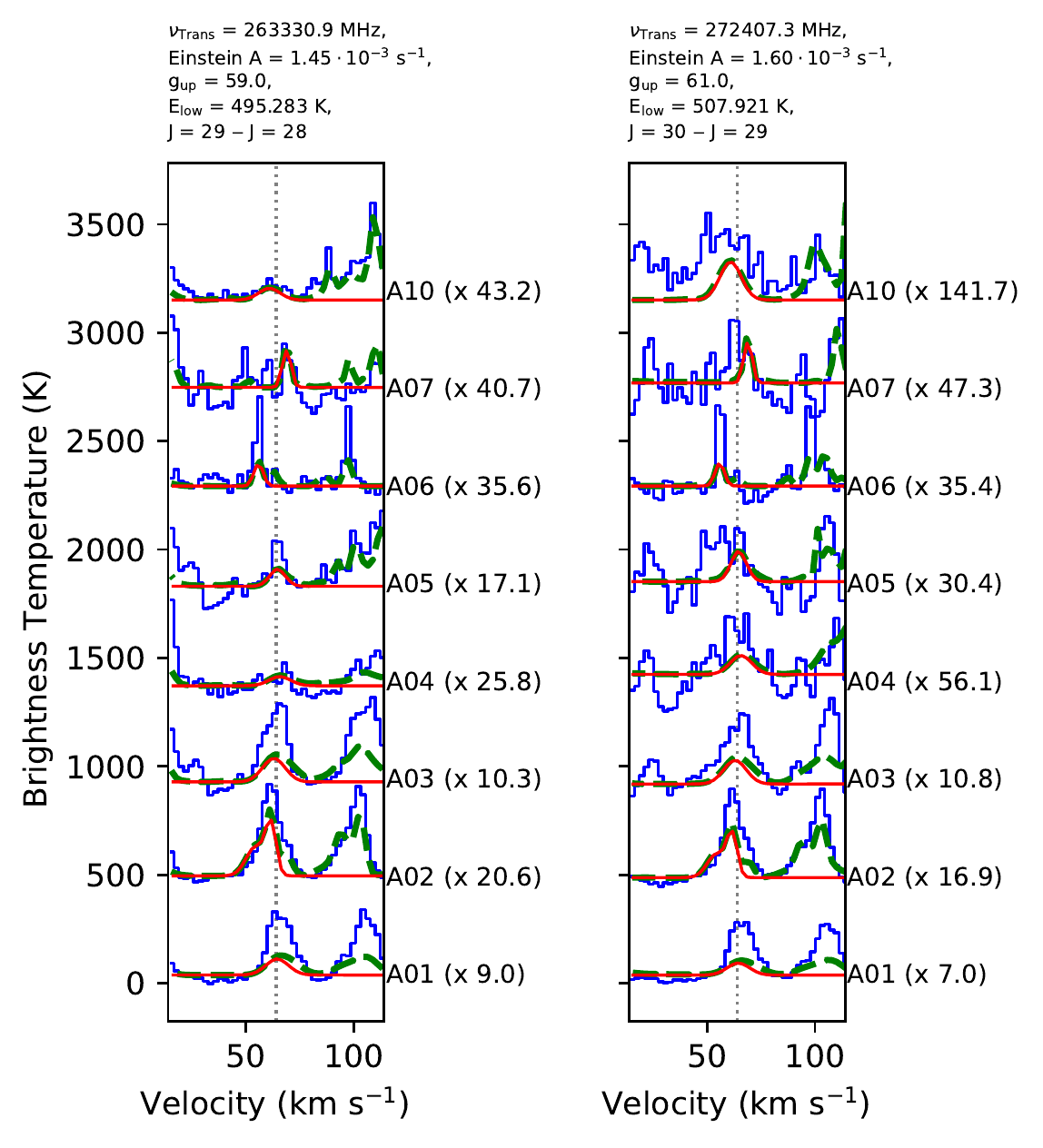}\\
       \caption{Sgr~B2(M)}
       \label{fig:HCCC13Nv71M}
    \end{subfigure}
\quad
    \begin{subfigure}[t]{1.0\columnwidth}
       \includegraphics[width=1.0\columnwidth]{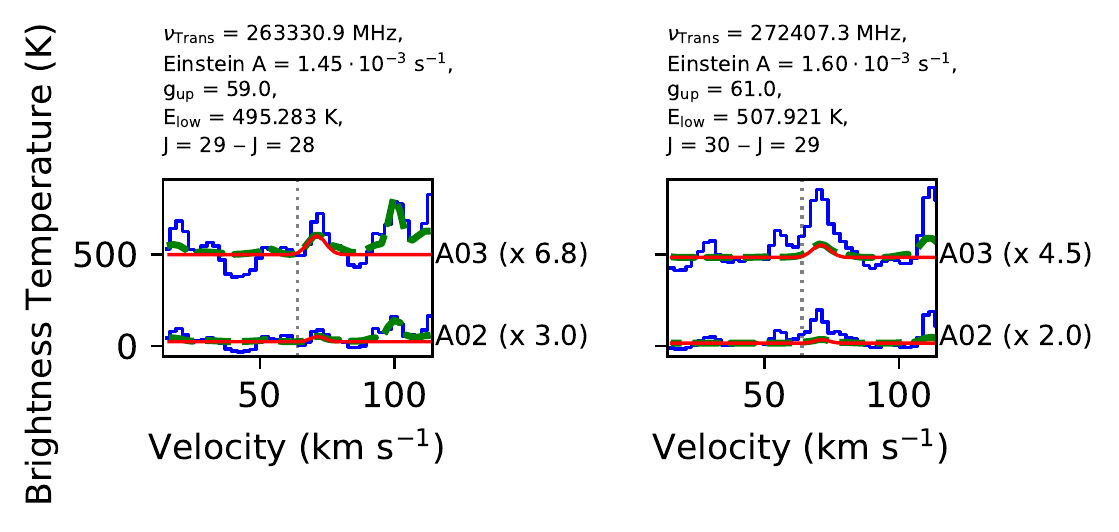}\\
       \caption{Sgr~B2(N)}
       \label{fig:HCCC13Nv71N}
   \end{subfigure}
   \caption{Selected transitions of HCC$^{13}$CN, v$_7$=1 in Sgr~B2(M) and N.}
   \ContinuedFloat
   \label{fig:HCCC13Nv71MN}
\end{figure*}

%*******************************************************************************
% Figure: NH2CN;v=0;
\begin{figure*}[!htb]
    \centering
    \begin{subfigure}[t]{1.0\columnwidth}
       \includegraphics[width=1.0\columnwidth]{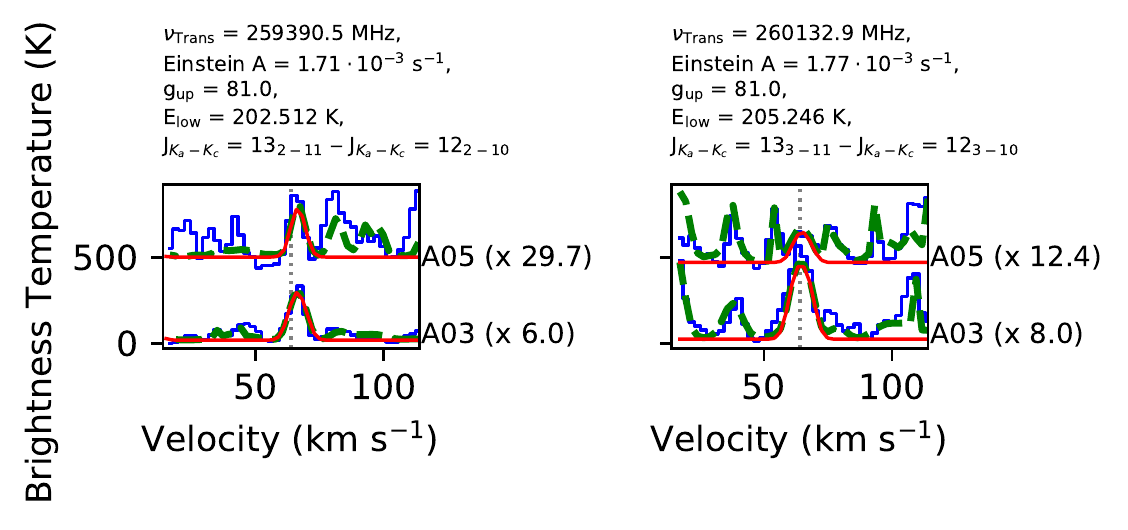}\\
    \end{subfigure}
   \caption{Selected transitions of NH$_2$CN in Sgr~B2(M).}
   \ContinuedFloat
   \label{fig:NH2CNM}
\end{figure*}
\newpage
\clearpage

%-------------------------------------------------------------------------------
% S-bearing molecules

%*******************************************************************************
% Figure: H2S-33;v=0;
\begin{figure*}[!htb]
    \centering
    \begin{subfigure}[t]{0.5\columnwidth}
       \includegraphics[width=1.0\columnwidth]{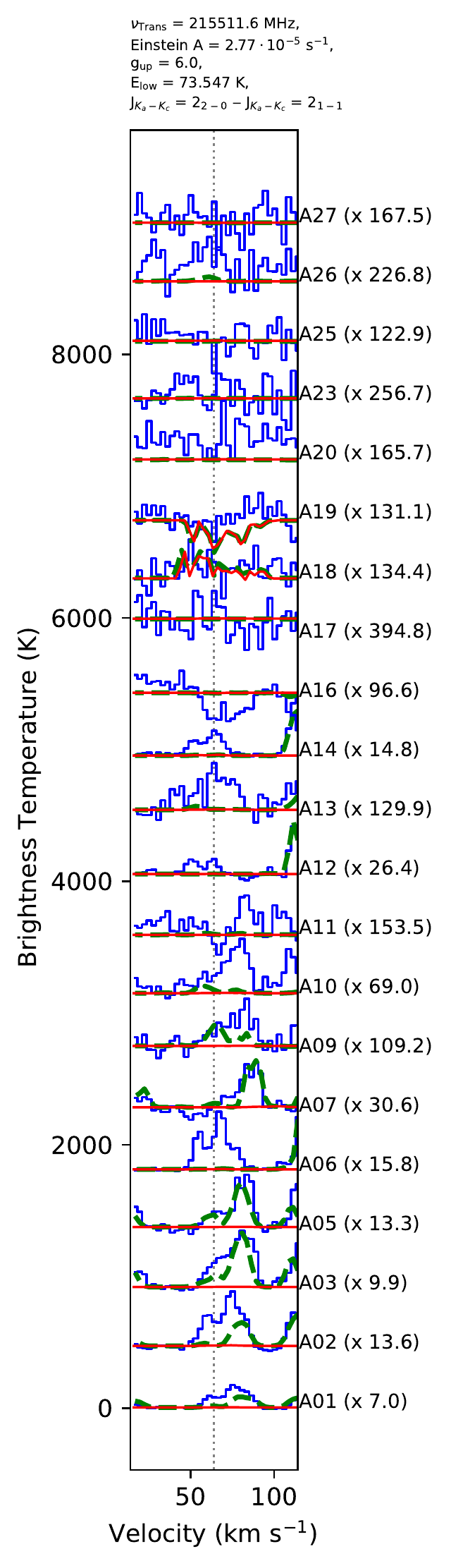}\\
       \caption{Sgr~B2(M)}
       \label{fig:H2S33M}
    \end{subfigure}
\quad
    \begin{subfigure}[t]{0.5\columnwidth}
       \includegraphics[width=1.0\columnwidth]{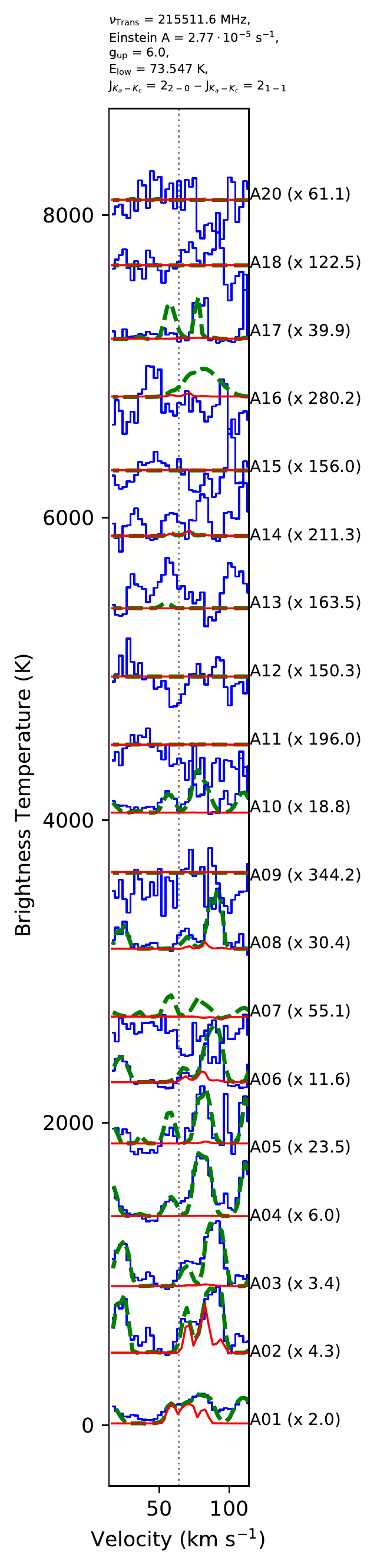}\\
       \caption{Sgr~B2(N)}
       \label{fig:H2S33N}
   \end{subfigure}
   \caption{Selected transitions of H$_2 \! ^{33}$S in Sgr~B2(M) and N.}
   \ContinuedFloat
   \label{fig:H2S33MN}
\end{figure*}
\newpage
\clearpage

%*******************************************************************************
% Figure: H2S-34;v=0;
\begin{figure*}[!htb]
    \centering
    \begin{subfigure}[t]{1.0\columnwidth}
       \includegraphics[width=1.0\columnwidth]{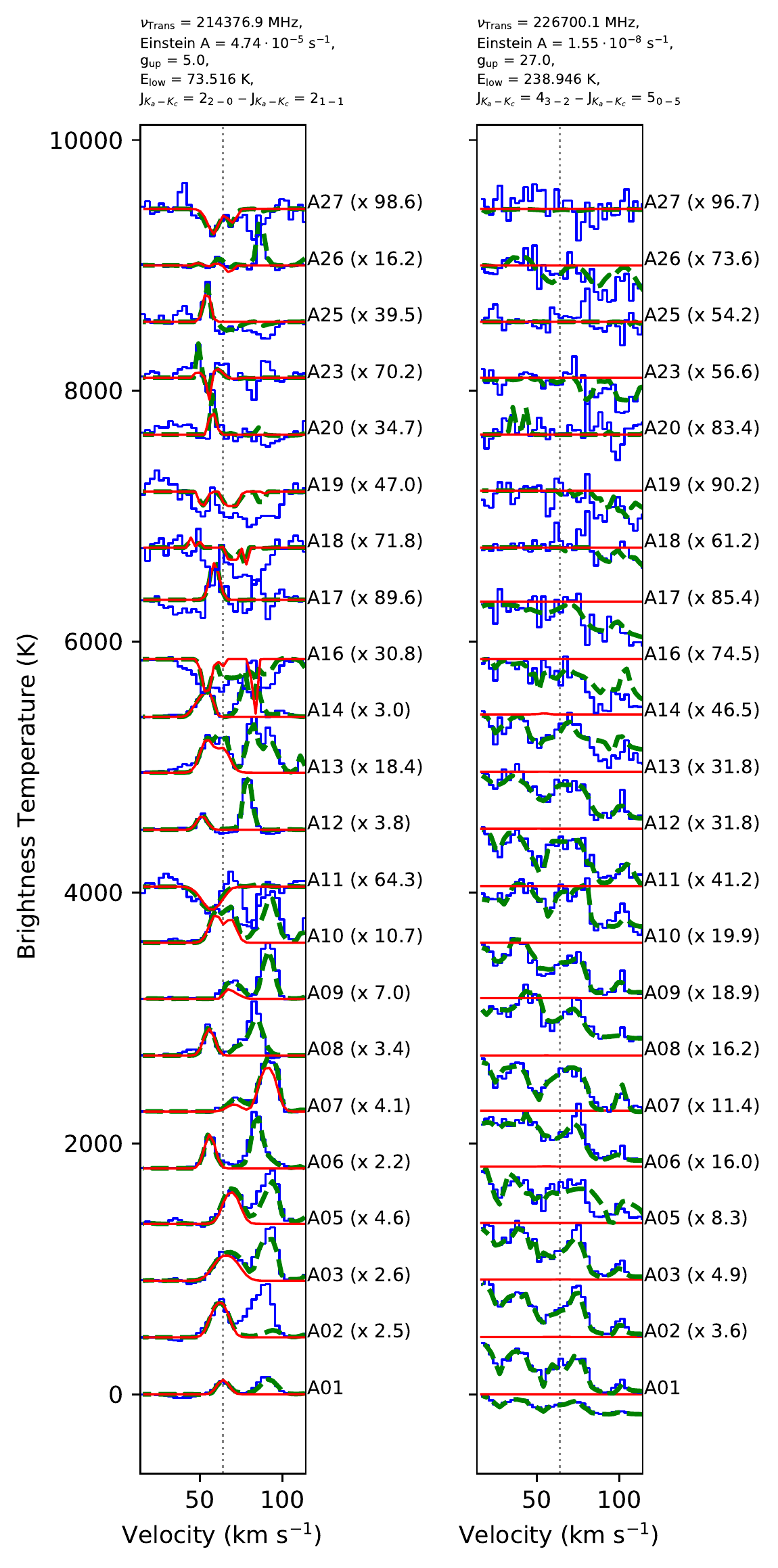}\\
       \caption{Sgr~B2(M)}
       \label{fig:H2S34M}
    \end{subfigure}
\quad
    \begin{subfigure}[t]{1.0\columnwidth}
       \includegraphics[width=1.0\columnwidth]{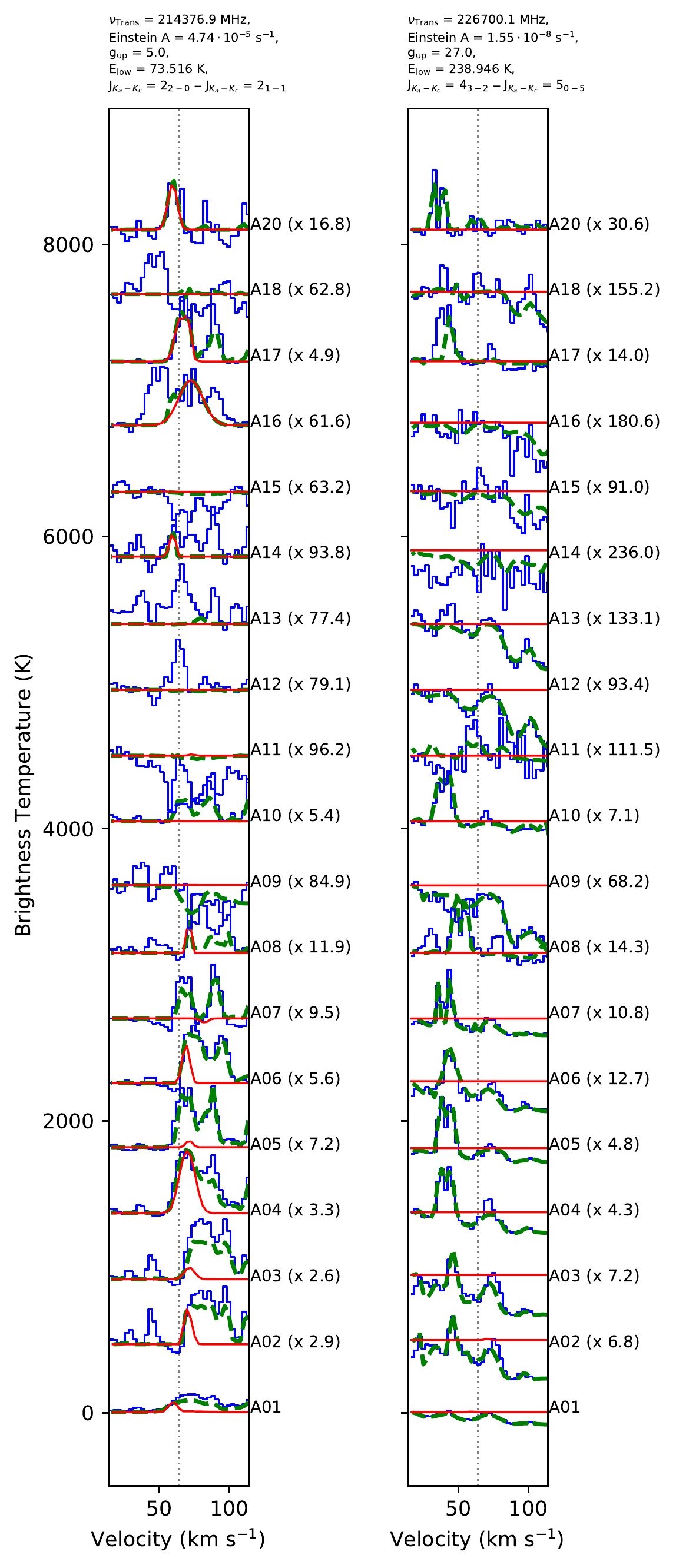}\\
       \caption{Sgr~B2(N)}
       \label{fig:H2S34N}
   \end{subfigure}
   \caption{Selected transitions of H$_2 \! ^{34}$S in Sgr~B2(M) and N.}
   \ContinuedFloat
   \label{fig:H2S34MN}
\end{figure*}
\newpage
\clearpage

%*******************************************************************************
% Figure: SO;v=0;#2
\begin{figure*}[!htb]
    \centering
    \begin{subfigure}[t]{1.0\columnwidth}
       \includegraphics[width=1.0\columnwidth]{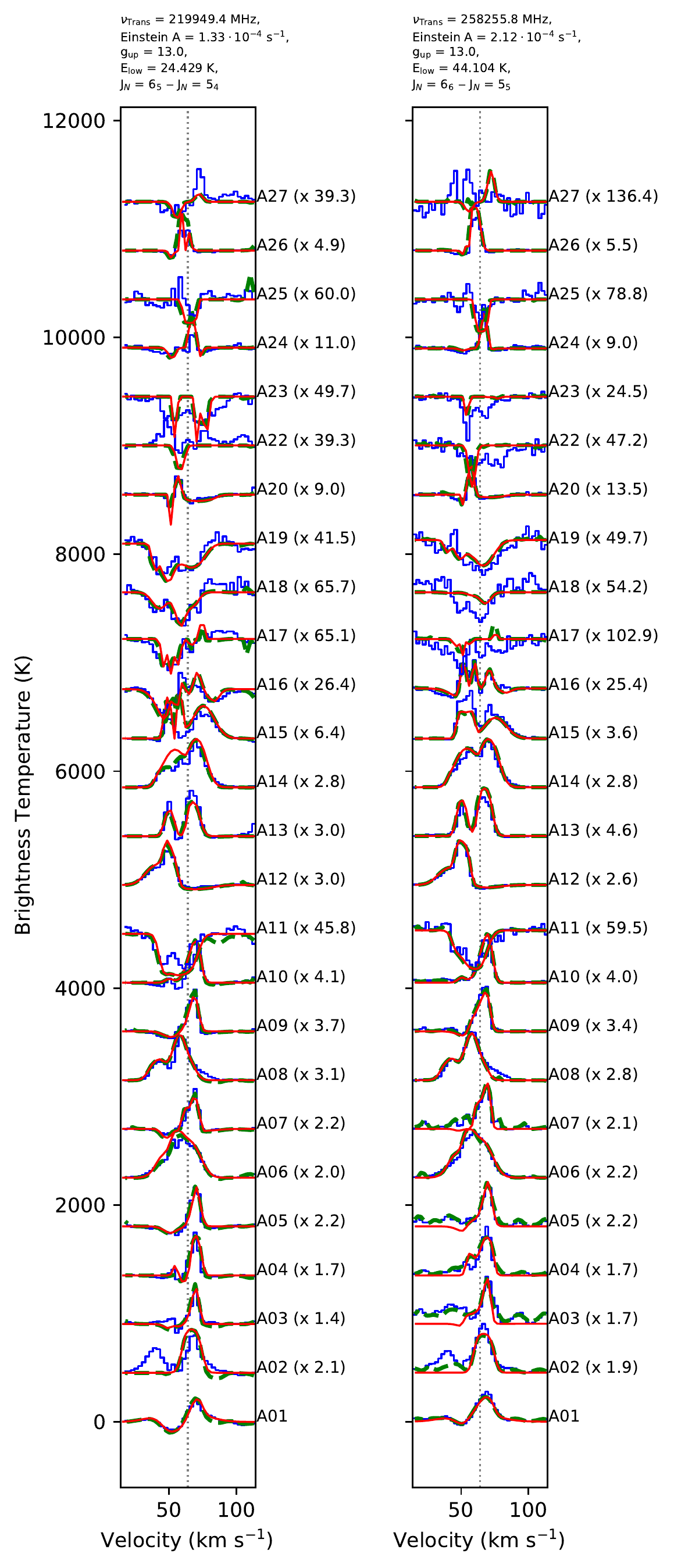}\\
       \caption{Sgr~B2(M)}
       \label{fig:SOM}
    \end{subfigure}
\quad
    \begin{subfigure}[t]{1.0\columnwidth}
       \includegraphics[width=1.0\columnwidth]{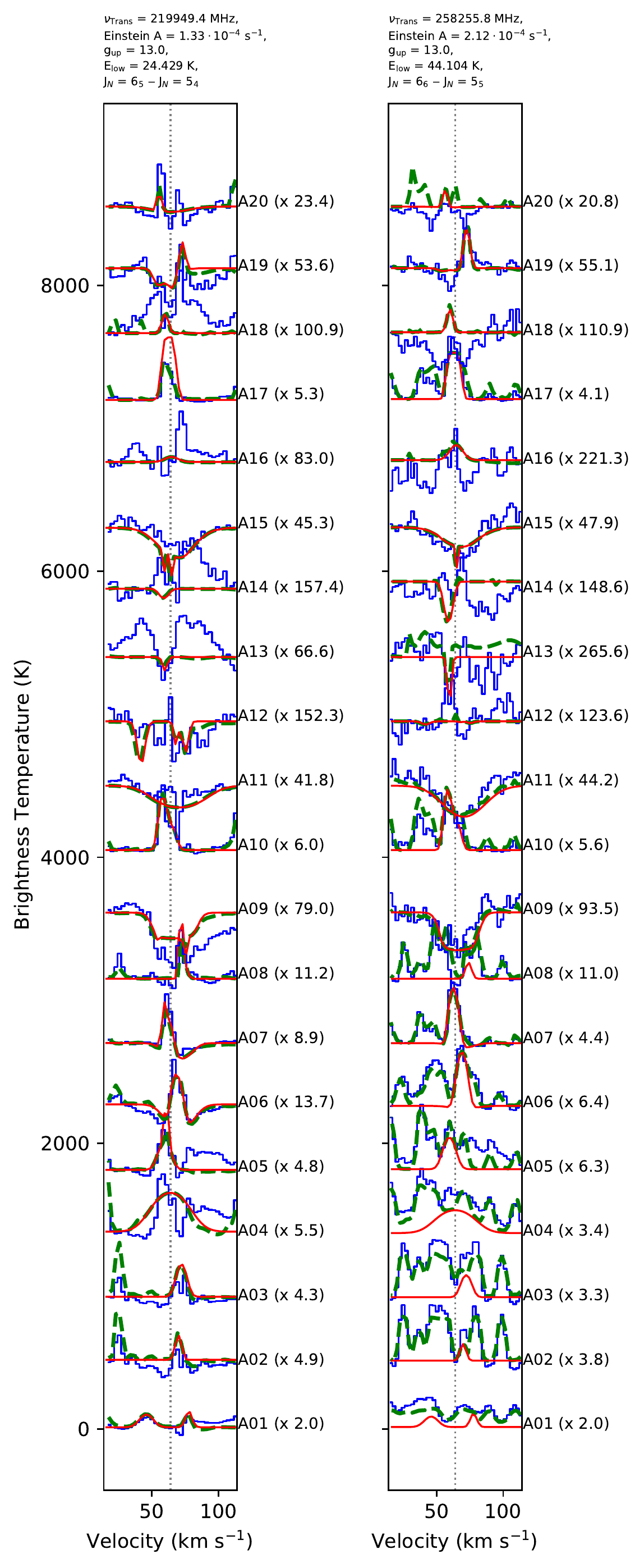}\\
       \caption{Sgr~B2(N)}
       \label{fig:SON}
   \end{subfigure}
   \caption{Selected transitions of SO in Sgr~B2(M) and N.}
   \ContinuedFloat
   \label{fig:SOMN}
\end{figure*}
\newpage
\clearpage

%*******************************************************************************
% Figure: S-33-O;v=0;
\begin{figure*}[!htb]
    \centering
    \begin{subfigure}[t]{1.0\columnwidth}
       \includegraphics[width=1.0\columnwidth]{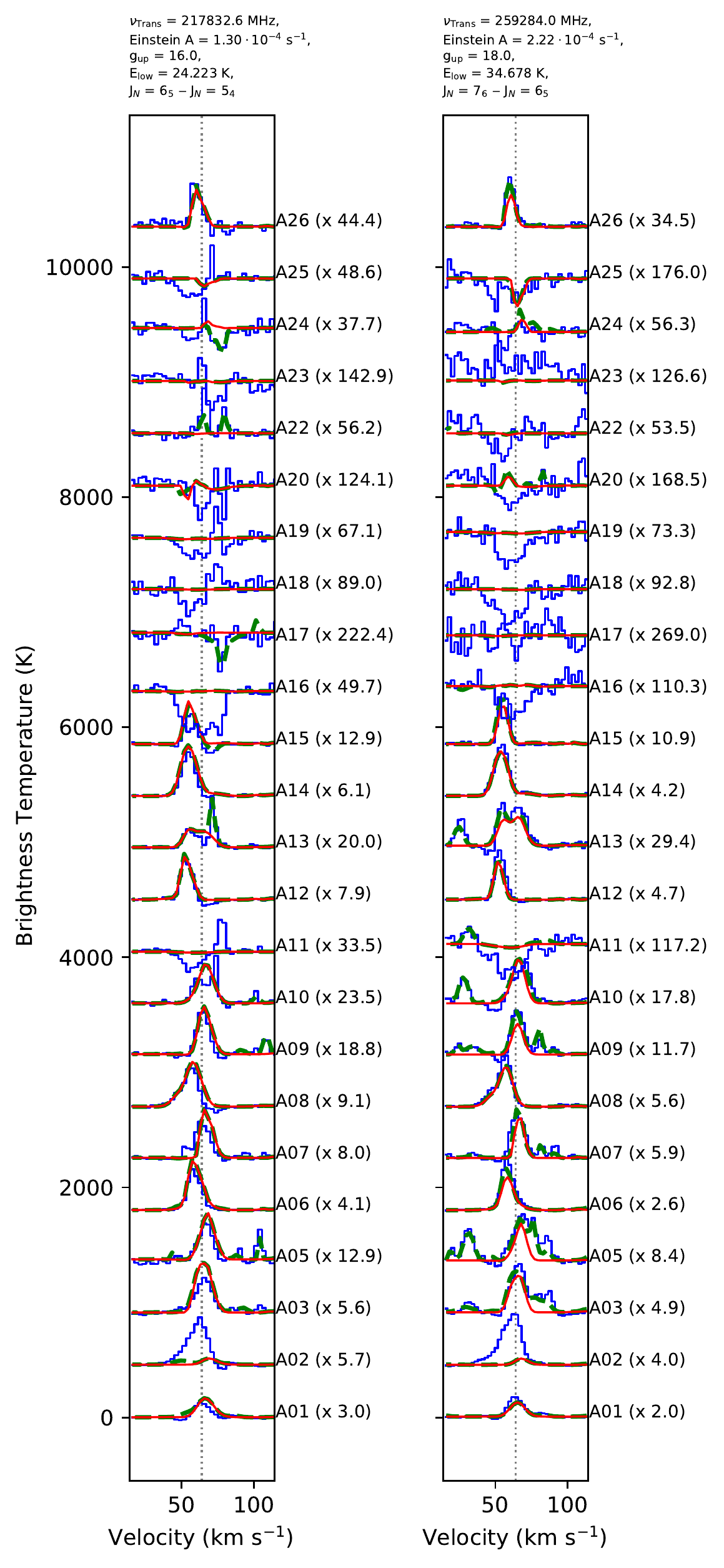}\\
       \caption{Sgr~B2(M)}
       \label{fig:S33OM}
    \end{subfigure}
\quad
    \begin{subfigure}[t]{1.0\columnwidth}
       \includegraphics[width=1.0\columnwidth]{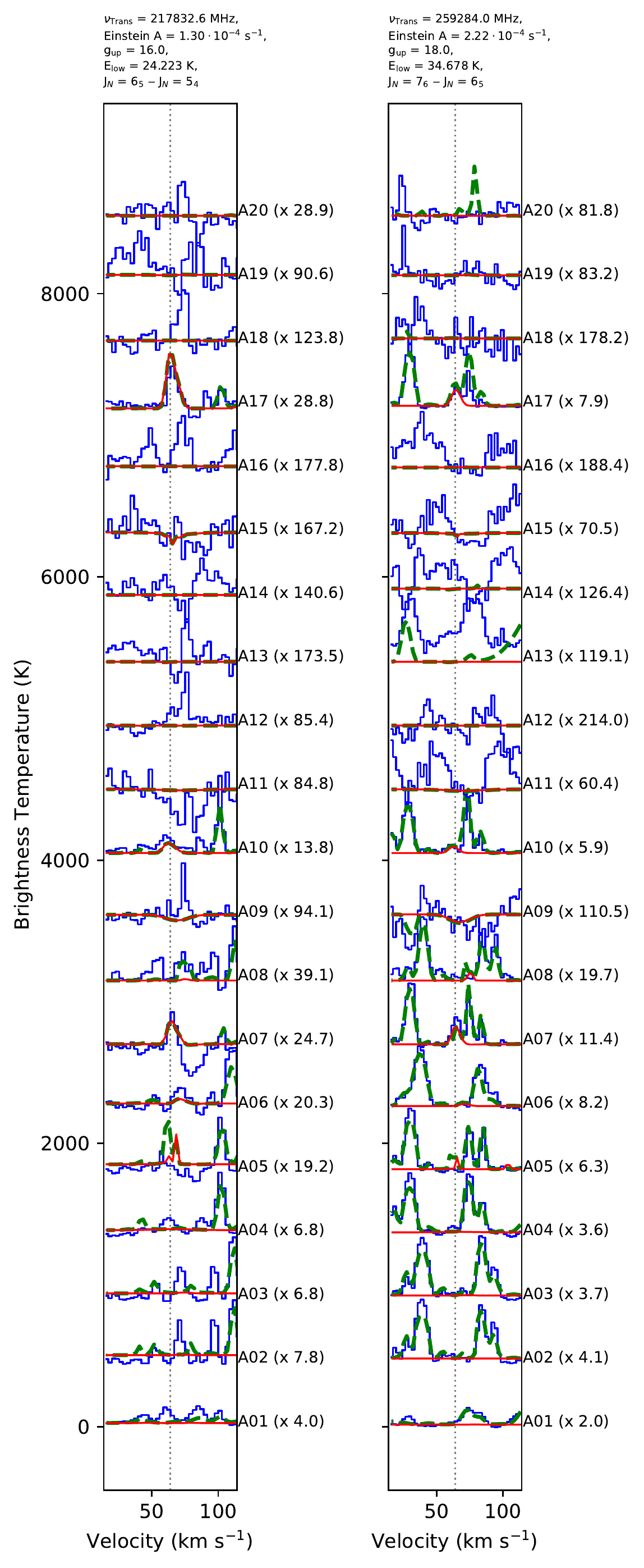}\\
       \caption{Sgr~B2(N)}
       \label{fig:S33ON}
   \end{subfigure}
   \caption{Selected transitions of $^{33}$SO in Sgr~B2(M) and N.}
   \ContinuedFloat
   \label{fig:S33OMN}
\end{figure*}
\newpage
\clearpage

%*******************************************************************************
% Figure: S-34-O;v=0;
\begin{figure*}[!htb]
    \centering
    \begin{subfigure}[t]{1.0\columnwidth}
       \includegraphics[width=1.0\columnwidth]{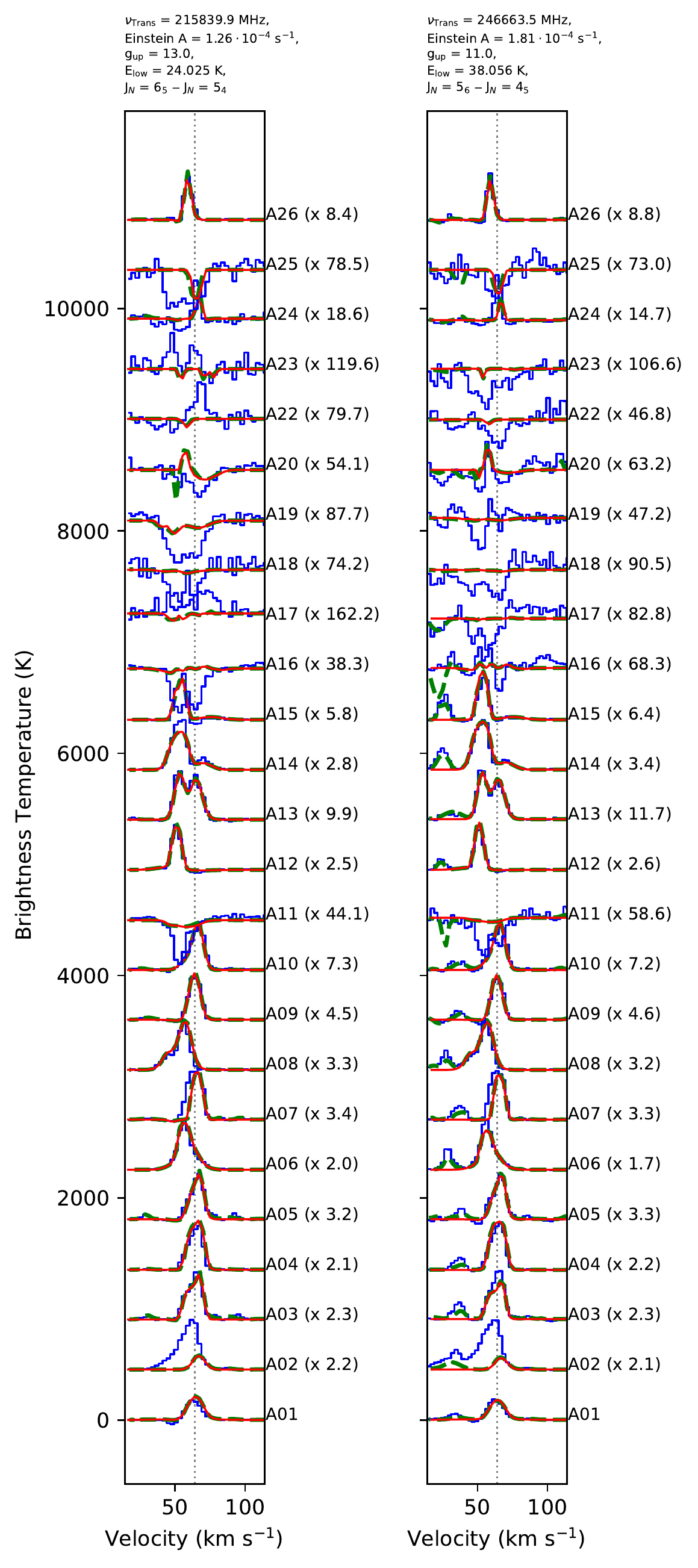}\\
       \caption{Sgr~B2(M)}
       \label{fig:S34OM}
    \end{subfigure}
\quad
    \begin{subfigure}[t]{1.0\columnwidth}
       \includegraphics[width=1.0\columnwidth]{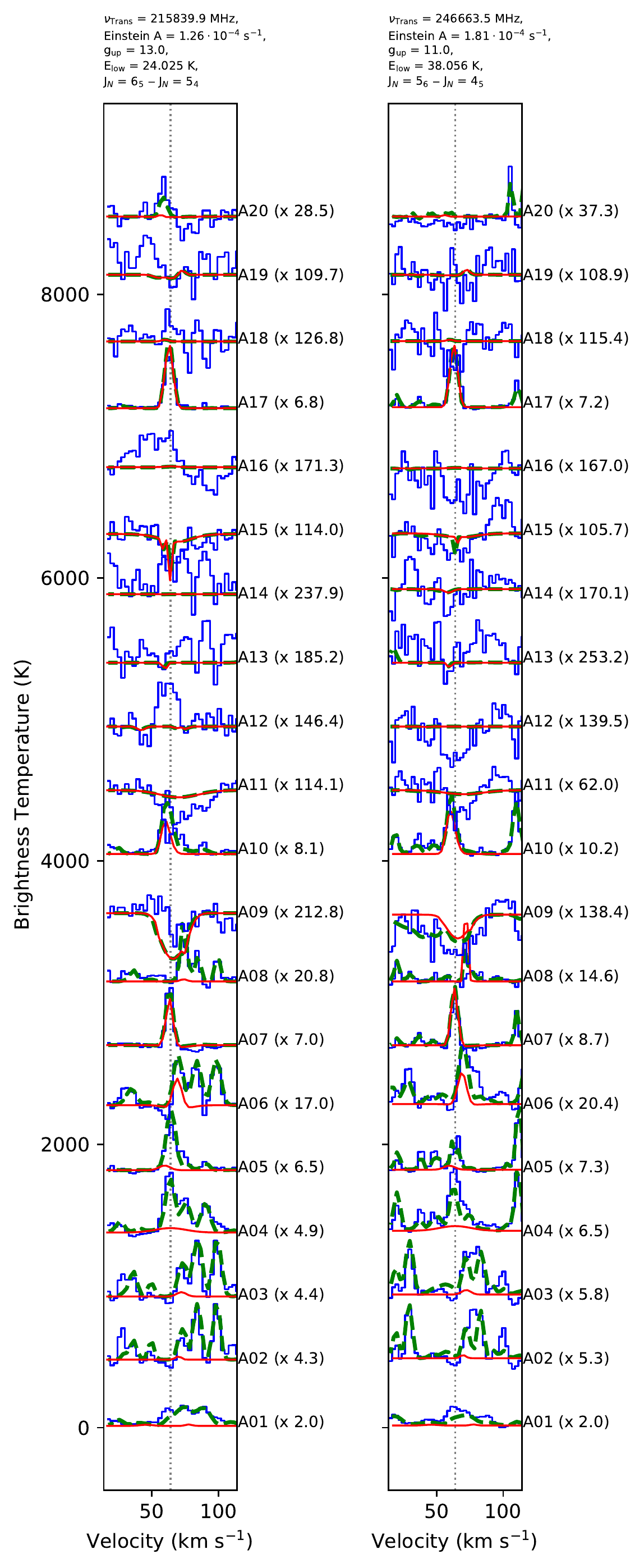}\\
       \caption{Sgr~B2(N)}
       \label{fig:S34ON}
   \end{subfigure}
   \caption{Selected transitions of $^{34}$SO in Sgr~B2(M) and N.}
   \ContinuedFloat
   \label{fig:S34OMN}
\end{figure*}
\newpage
\clearpage

%*******************************************************************************
% Figure: SO-17;v=0;
\begin{figure*}[!htb]
    \centering
    \begin{subfigure}[t]{1.0\columnwidth}
       \includegraphics[width=1.0\columnwidth]{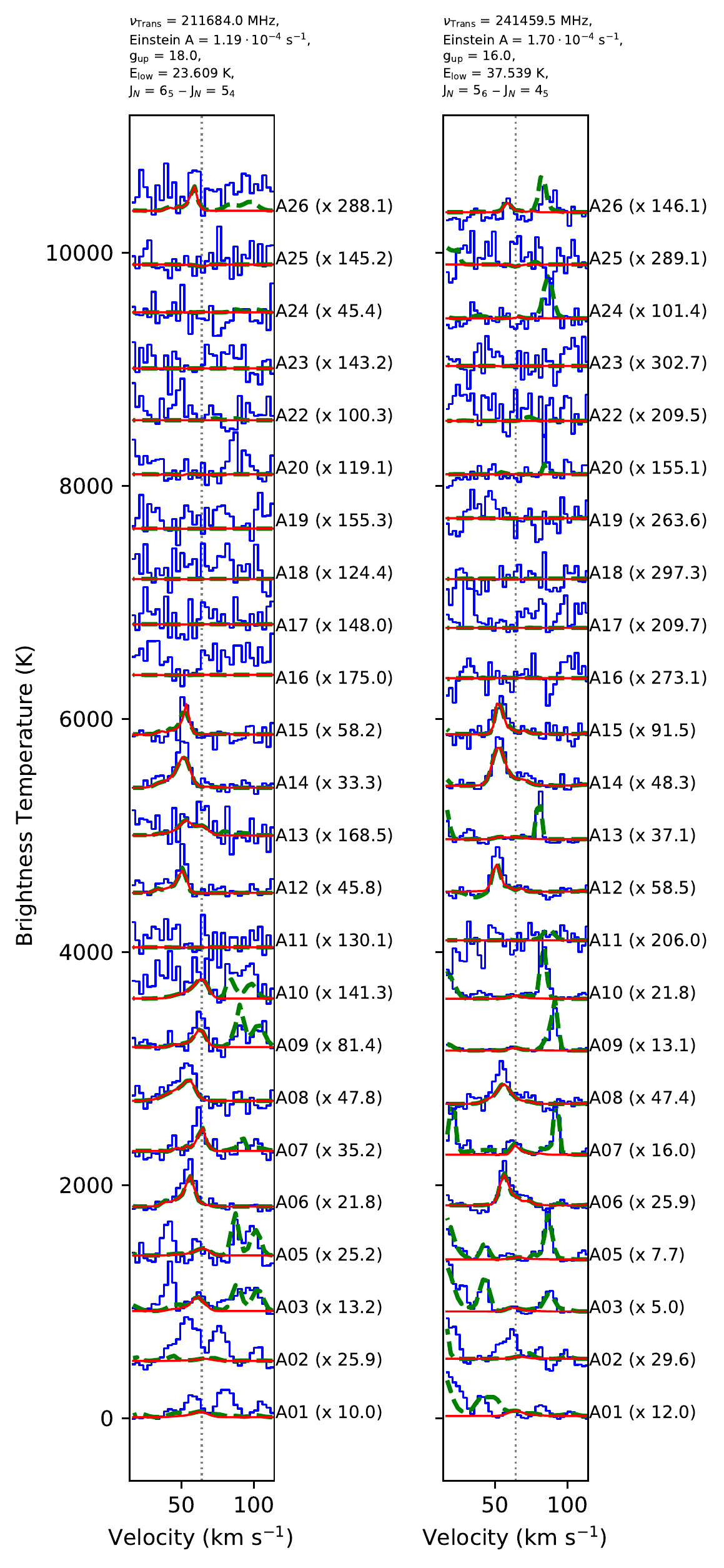}\\
       \caption{Sgr~B2(M)}
       \label{fig:SO17M}
    \end{subfigure}
\quad
    \begin{subfigure}[t]{1.0\columnwidth}
       \includegraphics[width=1.0\columnwidth]{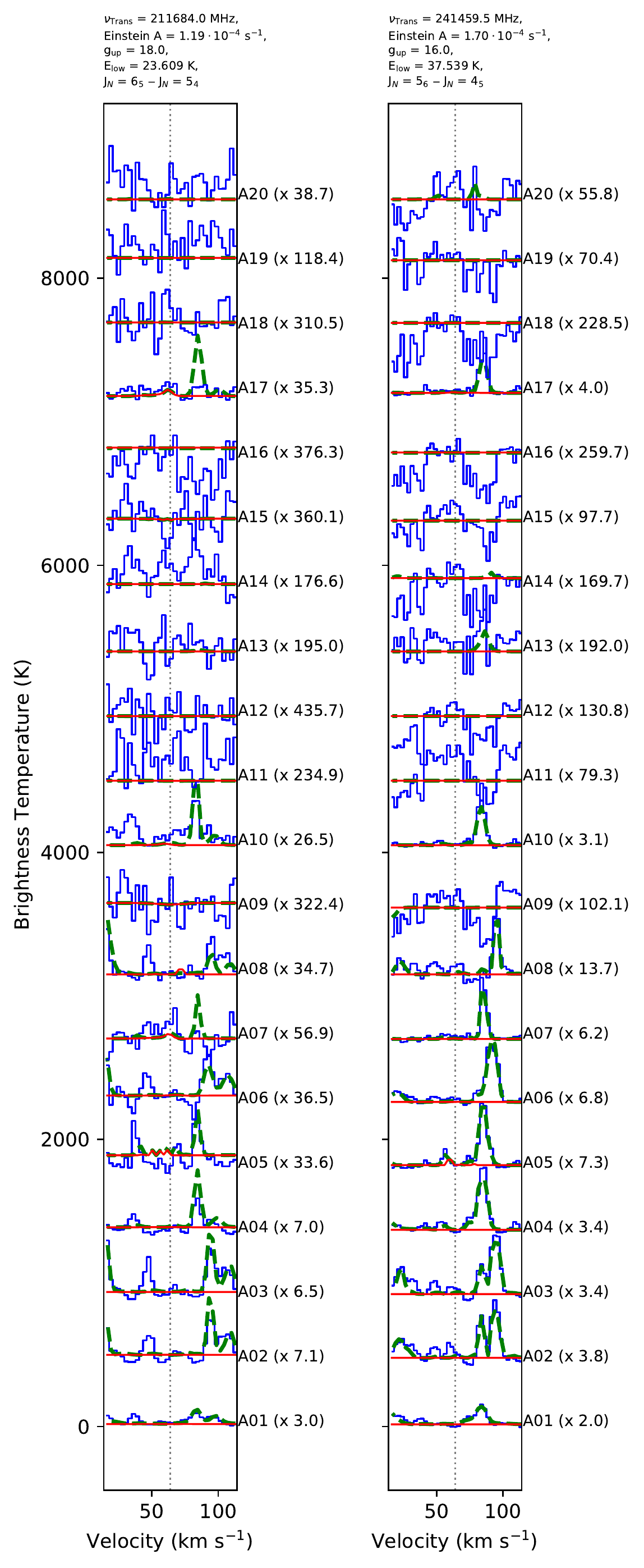}\\
       \caption{Sgr~B2(N)}
       \label{fig:SO17N}
   \end{subfigure}
   \caption{Selected transitions of S$^{17}$O in Sgr~B2(M) and N.}
   \ContinuedFloat
   \label{fig:SO17MN}
\end{figure*}
\newpage
\clearpage

%*******************************************************************************
% Figure: SO-18;v=0;
\begin{figure*}[!htb]
    \centering
    \begin{subfigure}[t]{1.0\columnwidth}
       \includegraphics[width=1.0\columnwidth]{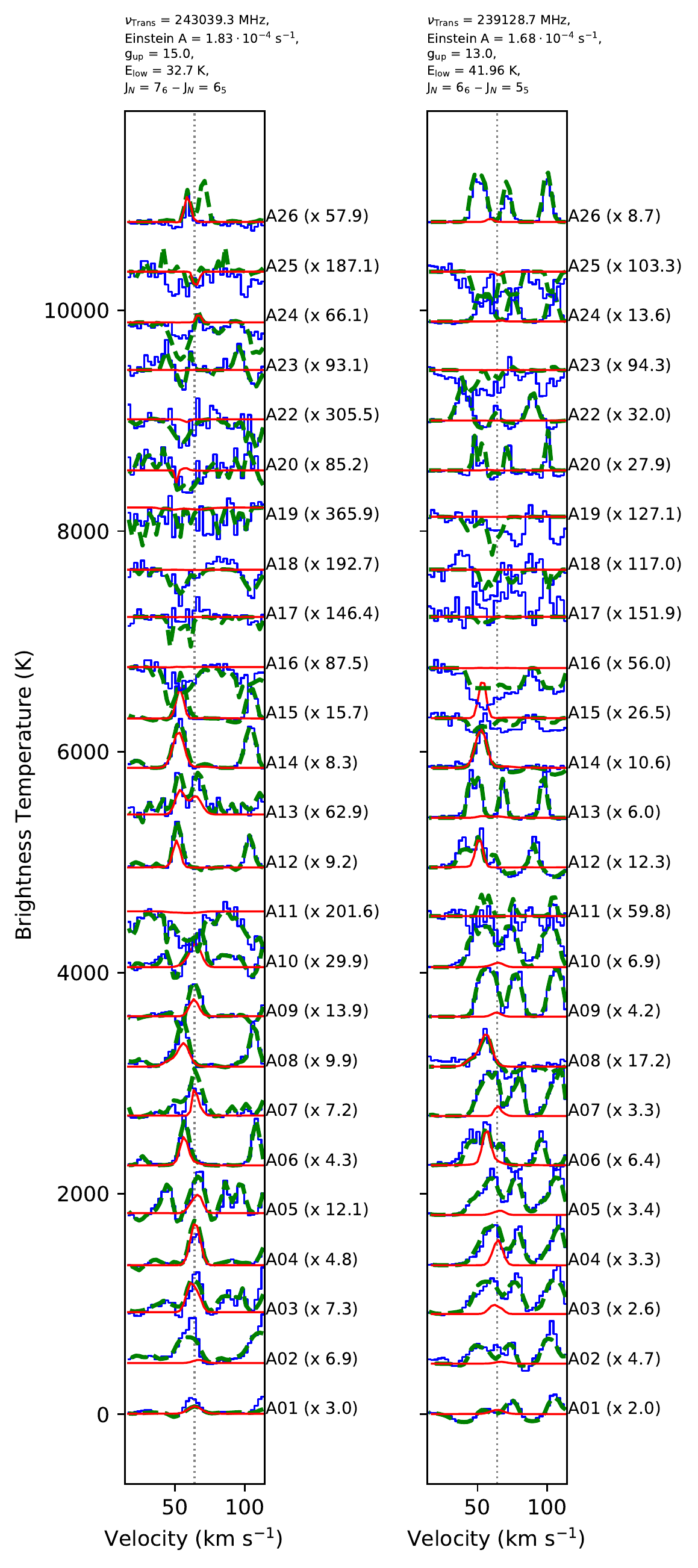}\\
       \caption{Sgr~B2(M)}
       \label{fig:SO18M}
    \end{subfigure}
\quad
    \begin{subfigure}[t]{1.0\columnwidth}
       \includegraphics[width=1.0\columnwidth]{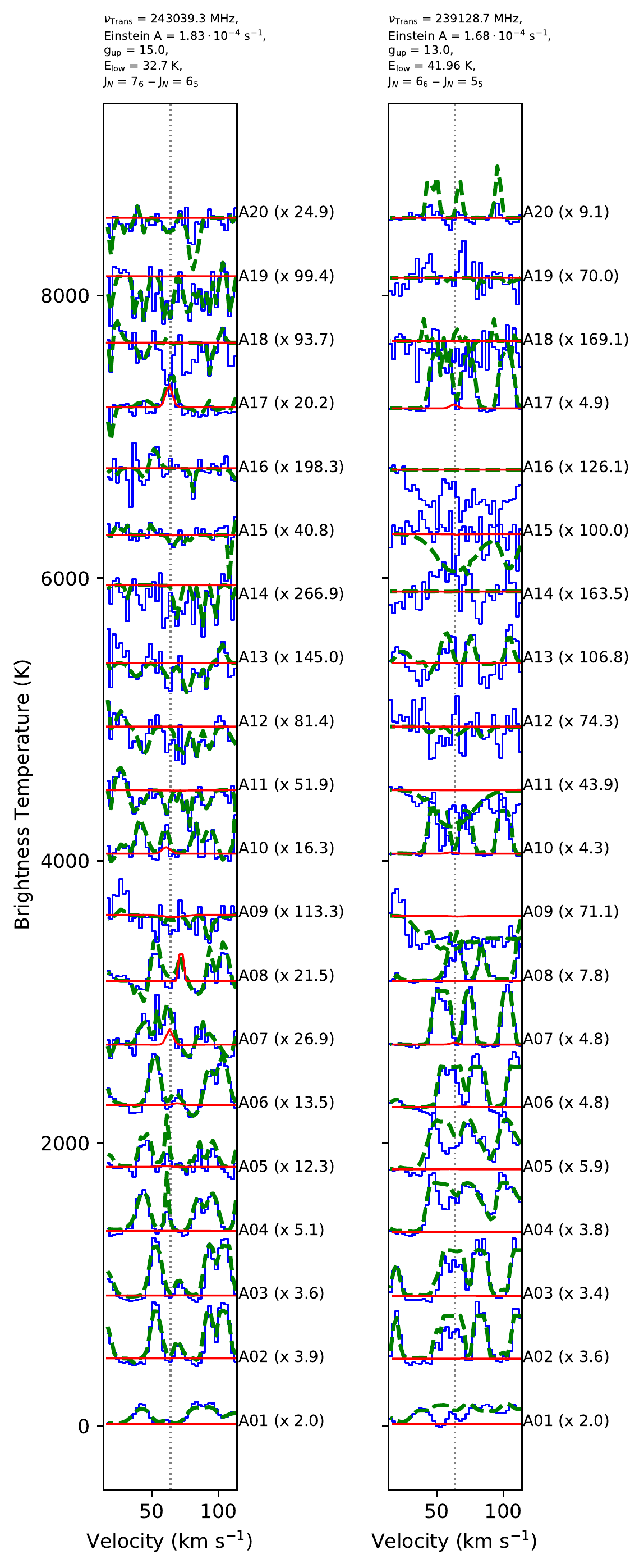}\\
       \caption{Sgr~B2(N)}
       \label{fig:SO18N}
   \end{subfigure}
   \caption{Selected transitions of S$^{18}$O in Sgr~B2(M) and N.}
   \ContinuedFloat
   \label{fig:SO18MN}
\end{figure*}
\newpage
\clearpage

%*******************************************************************************
% Figure: SO+;v=0;
\begin{figure*}[!htb]
    \centering
    \begin{subfigure}[t]{0.5\columnwidth}
       \includegraphics[width=1.0\columnwidth]{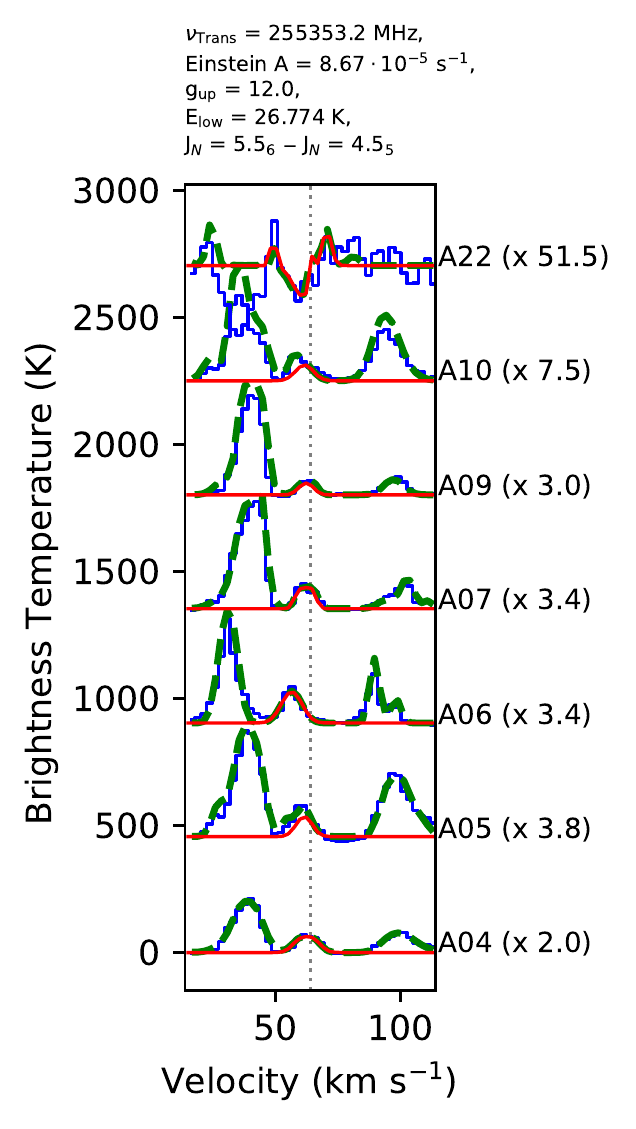}\\
    \end{subfigure}
   \caption{Selected transitions of SO$^+$ in Sgr~B2(M)}
   \ContinuedFloat
   \label{fig:SOpM}
\end{figure*}

%*******************************************************************************
% Figure: SO2;v=0;
\begin{figure*}[!htb]
    \centering
    \begin{subfigure}[t]{1.0\columnwidth}
       \includegraphics[width=1.0\columnwidth]{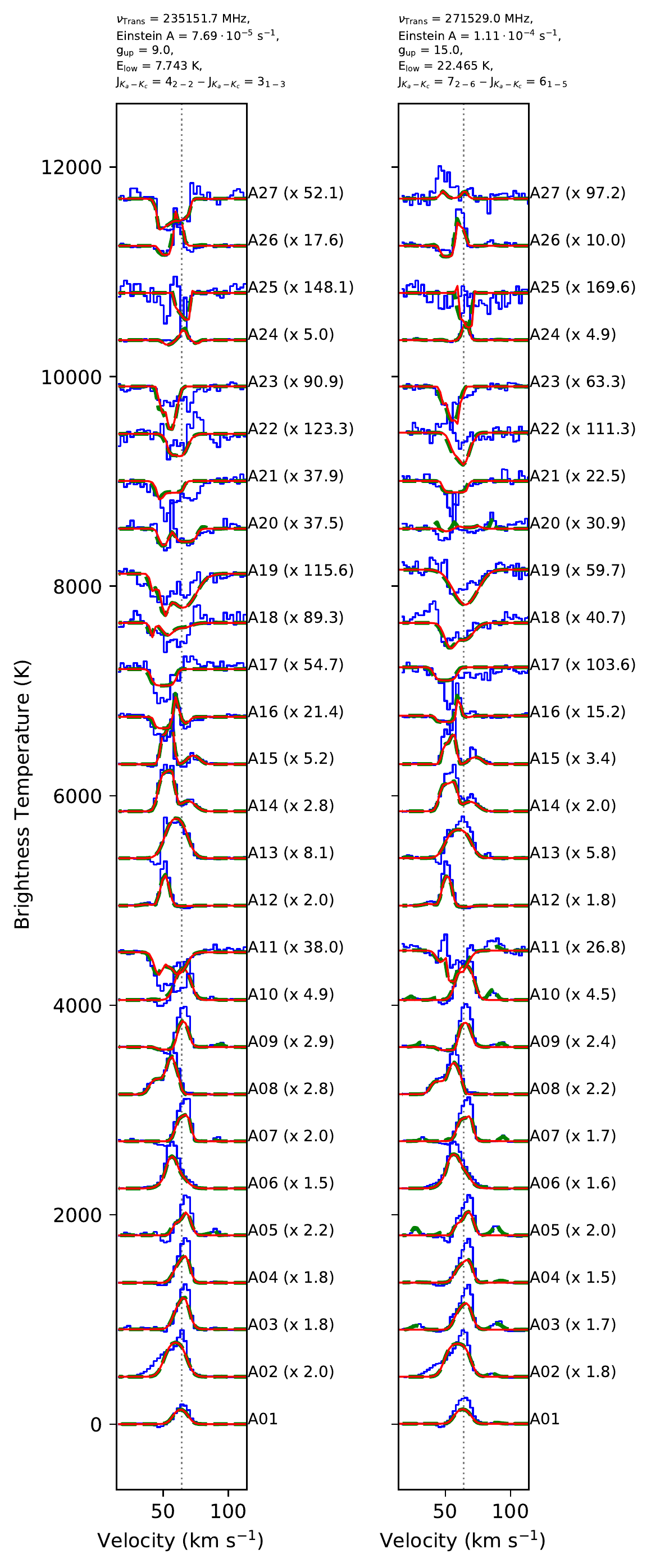}\\
       \caption{Sgr~B2(M)}
       \label{fig:SO2M}
    \end{subfigure}
\quad
    \begin{subfigure}[t]{1.0\columnwidth}
       \includegraphics[width=1.0\columnwidth]{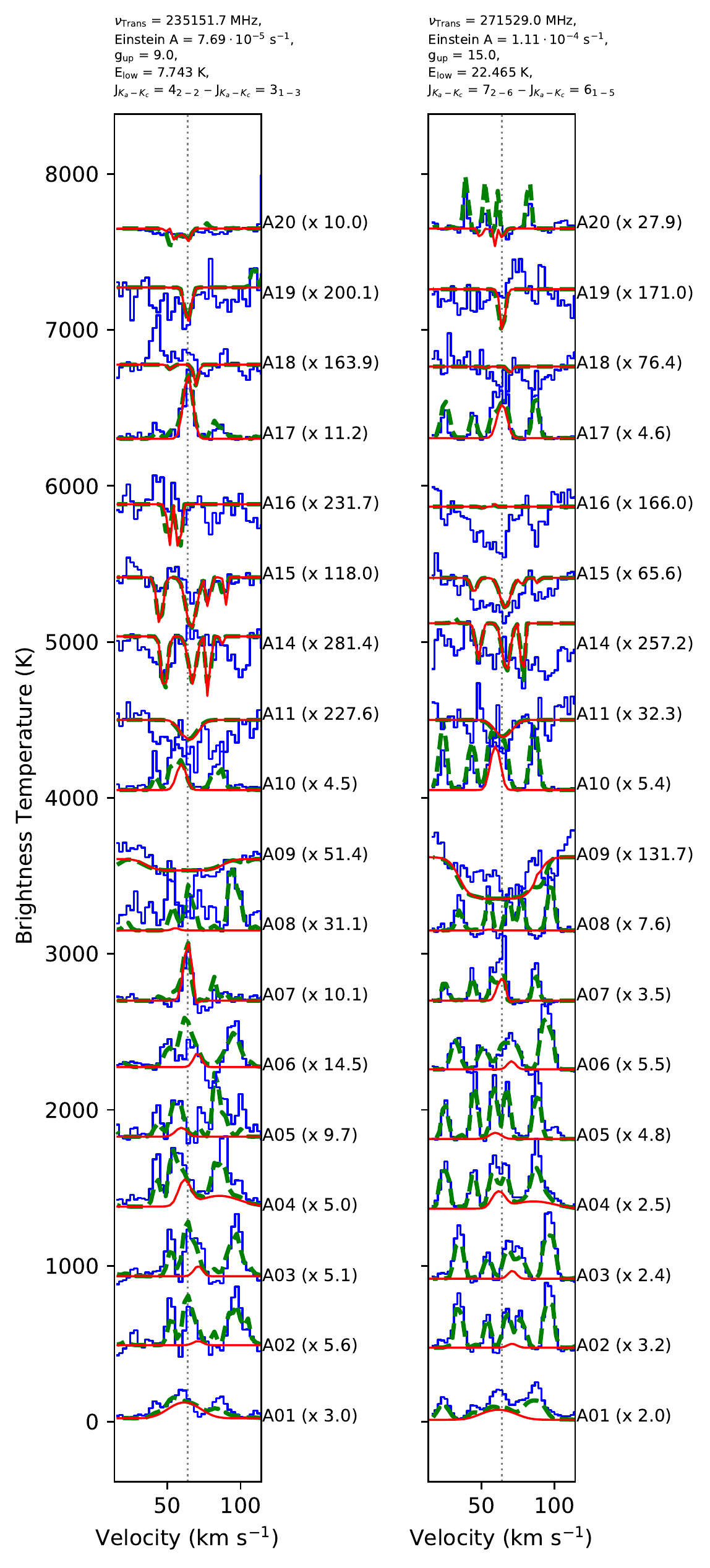}\\
       \caption{Sgr~B2(N)}
       \label{fig:SO2N}
   \end{subfigure}
   \caption{Selected transitions of SO$_2$ in Sgr~B2(M) and N.}
   \ContinuedFloat
   \label{fig:SO2MN}
\end{figure*}
\newpage
\clearpage

%*******************************************************************************
% Figure: SO2;v2=1;
\begin{figure*}[!htb]
    \centering
    \begin{subfigure}[t]{1.0\columnwidth}
       \includegraphics[width=1.0\columnwidth]{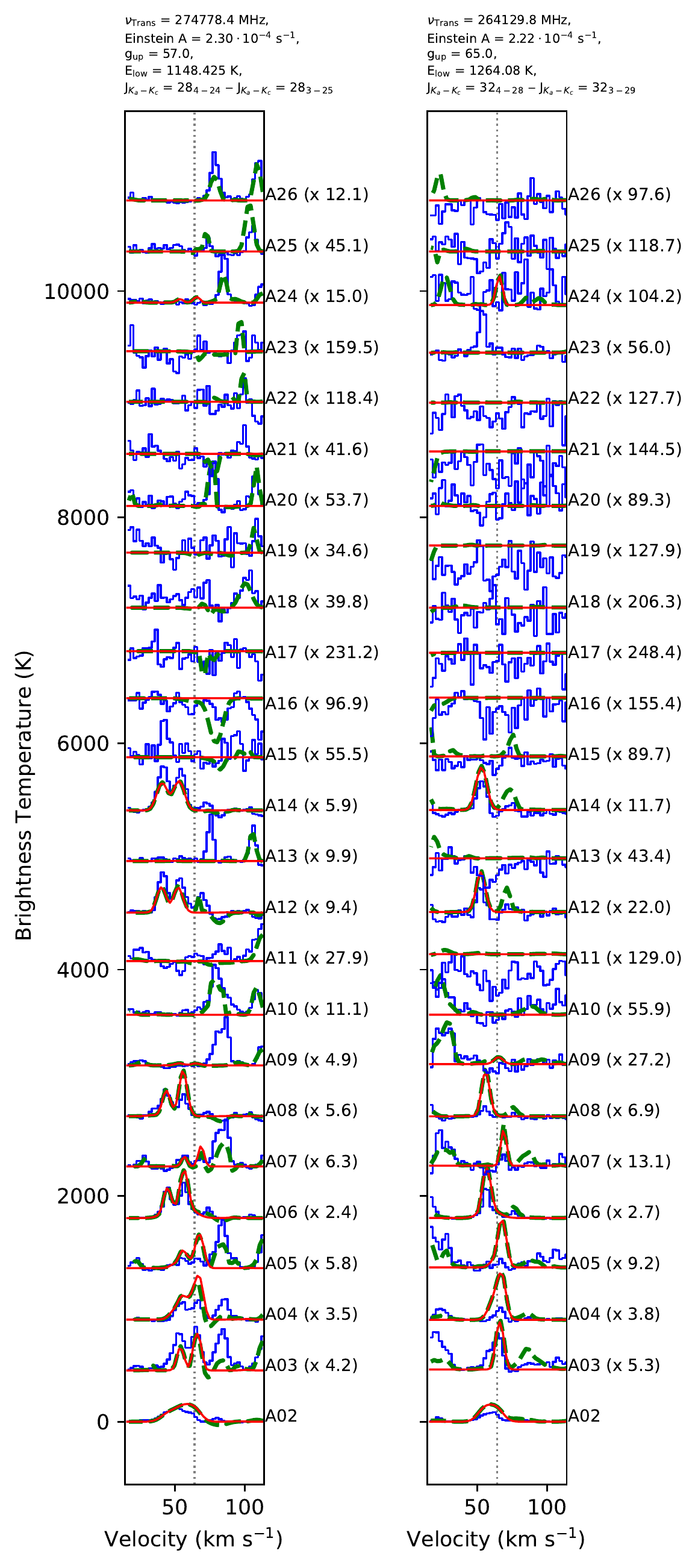}\\
       \caption{Sgr~B2(M)}
       \label{fig:SO2v21M}
    \end{subfigure}
\quad
    \begin{subfigure}[t]{1.0\columnwidth}
       \includegraphics[width=1.0\columnwidth]{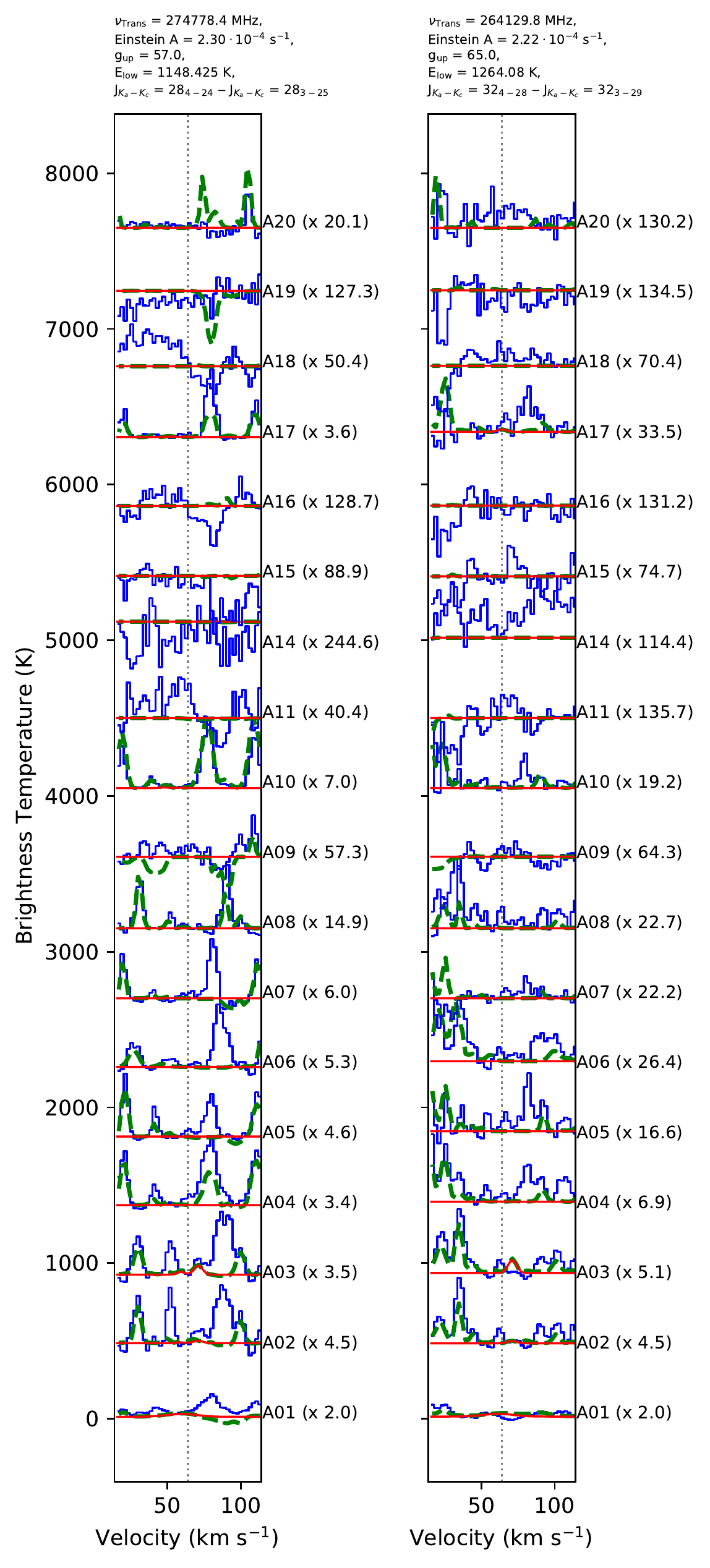}\\
       \caption{Sgr~B2(N)}
       \label{fig:SO2v21N}
   \end{subfigure}
   \caption{Selected transitions of SO$_2$, v$_2$=1 in Sgr~B2(M) and N.}
   \ContinuedFloat
   \label{fig:SO2v21MN}
\end{figure*}
\newpage
\clearpage

%*******************************************************************************
% Figure: S-33-O2;v=0;
\begin{figure*}[!htb]
    \centering
    \begin{subfigure}[t]{1.0\columnwidth}
       \includegraphics[width=1.0\columnwidth]{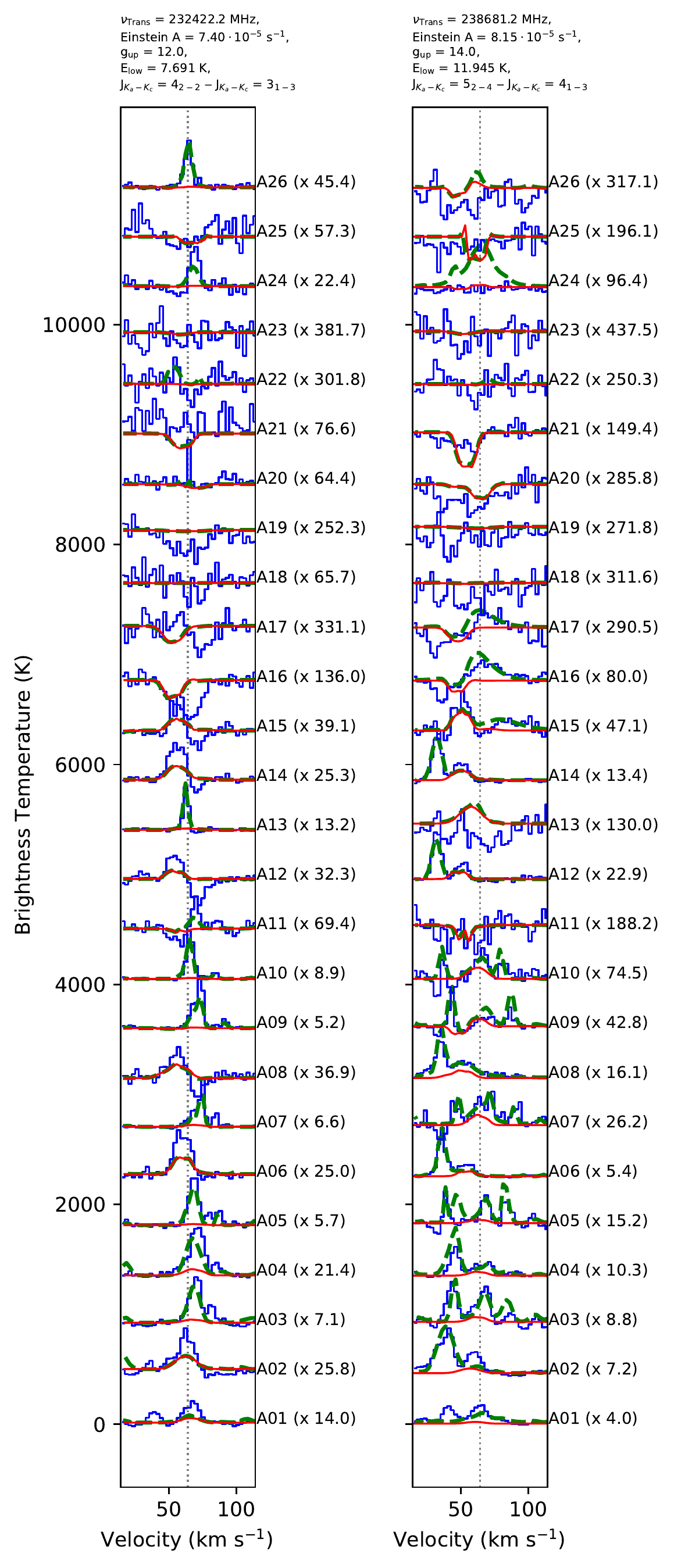}\\
       \caption{Sgr~B2(M)}
       \label{fig:S33O2M}
    \end{subfigure}
\quad
    \begin{subfigure}[t]{1.0\columnwidth}
       \includegraphics[width=1.0\columnwidth]{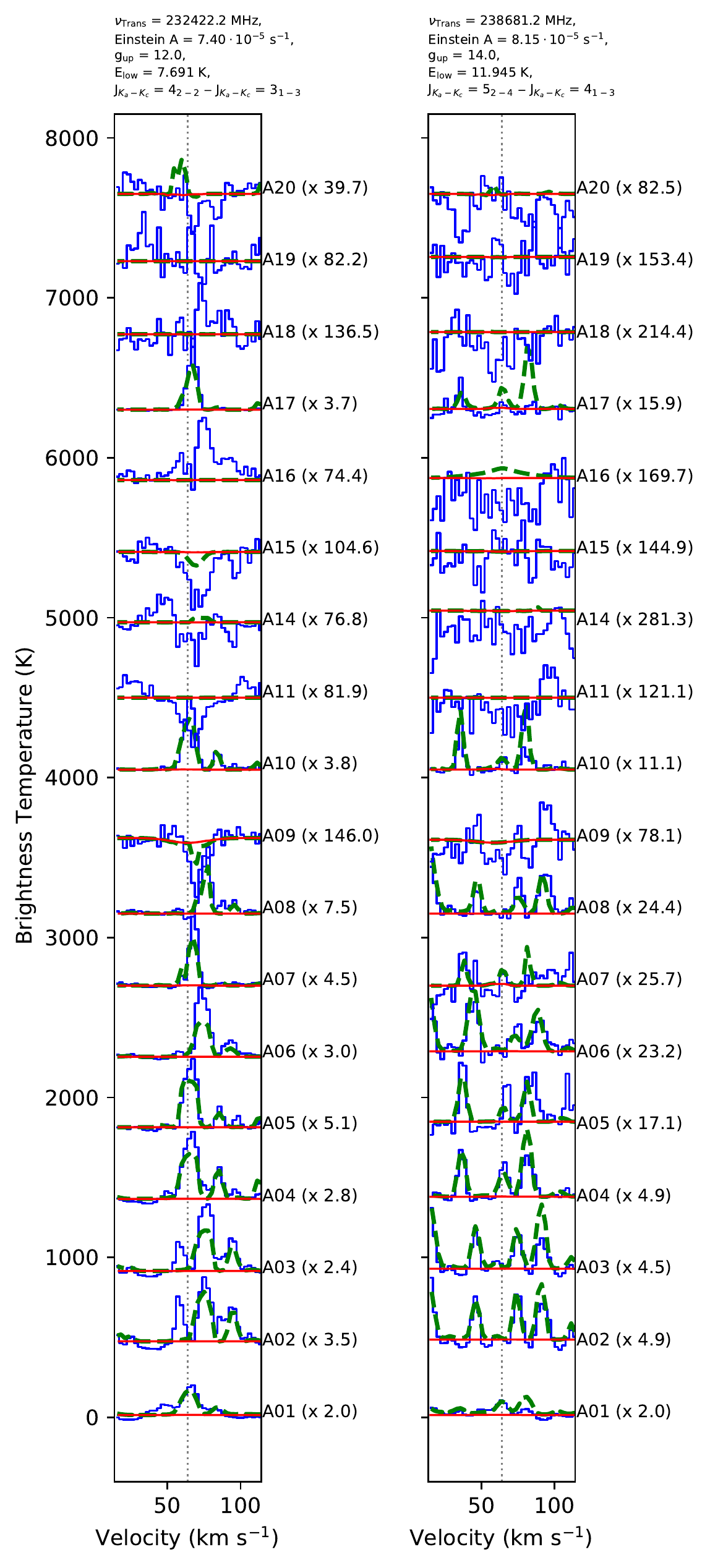}\\
       \caption{Sgr~B2(N)}
       \label{fig:S33O2N}
   \end{subfigure}
   \caption{Selected transitions of $^{33}$SO$_2$ in Sgr~B2(M) and N.}
   \ContinuedFloat
   \label{fig:S33O2MN}
\end{figure*}
\newpage
\clearpage

%*******************************************************************************
% Figure: S-34-O2;v=0;
\begin{figure*}[!htb]
    \centering
    \begin{subfigure}[t]{1.0\columnwidth}
       \includegraphics[width=1.0\columnwidth]{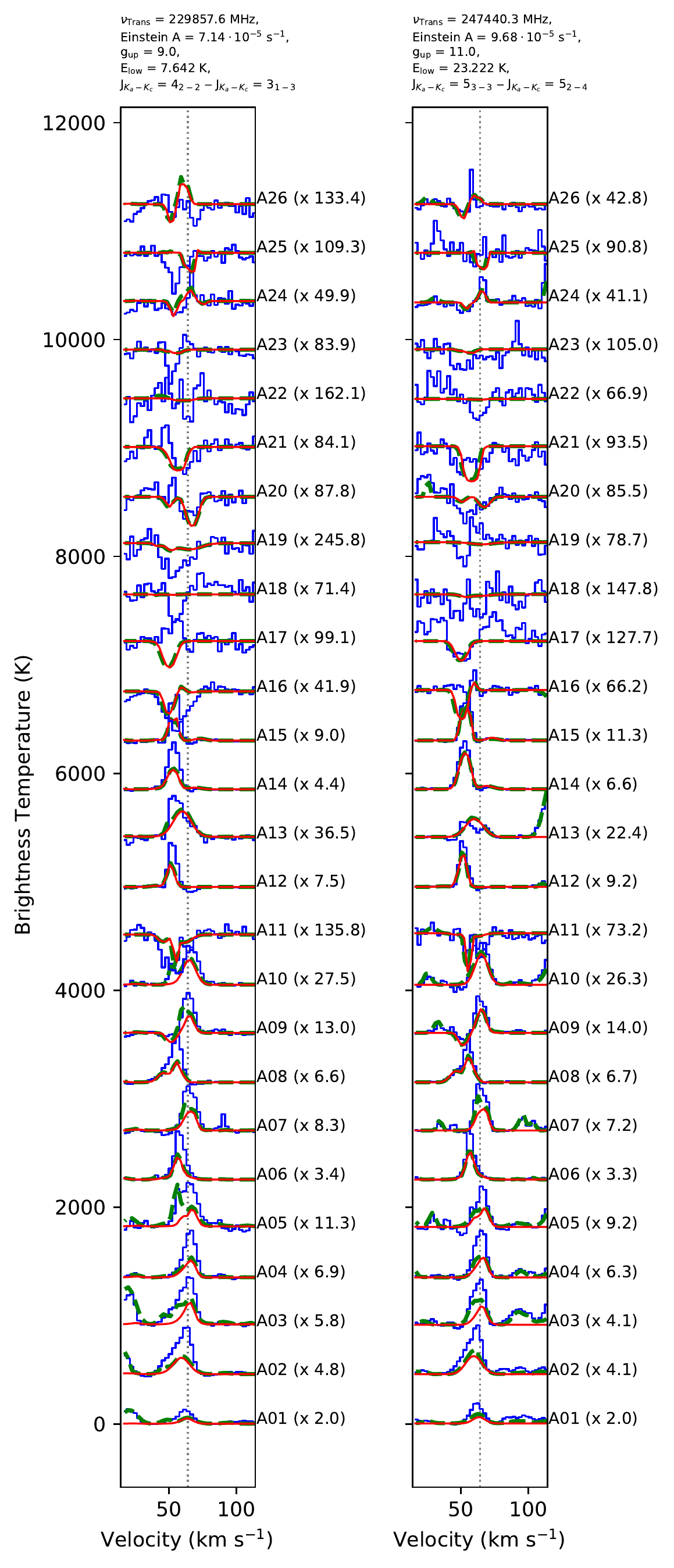}\\
       \caption{Sgr~B2(M)}
       \label{fig:S34O2M}
    \end{subfigure}
\quad
    \begin{subfigure}[t]{1.0\columnwidth}
       \includegraphics[width=1.0\columnwidth]{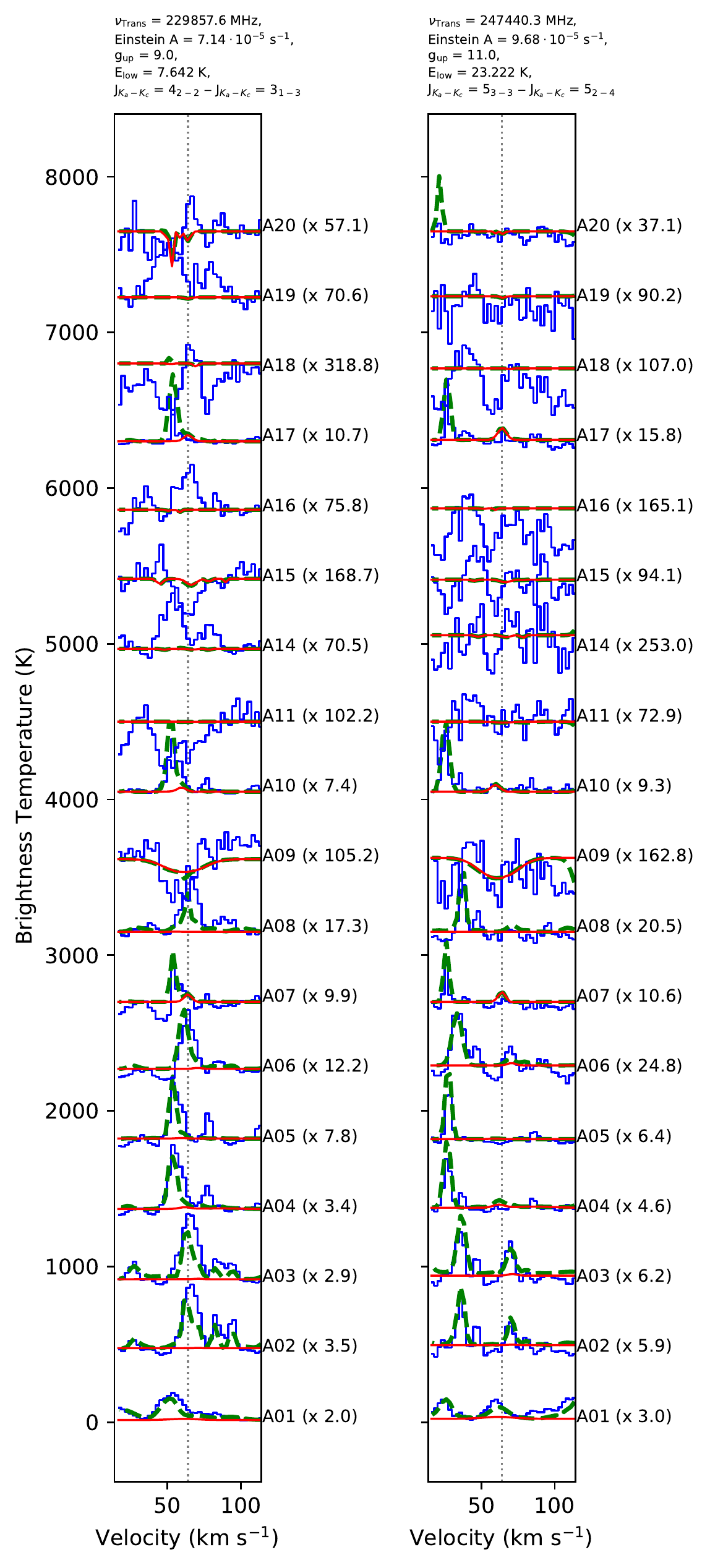}\\
       \caption{Sgr~B2(N)}
       \label{fig:S34O2N}
   \end{subfigure}
   \caption{Selected transitions of $^{34}$SO$_2$ in Sgr~B2(M) and N.}
   \ContinuedFloat
   \label{fig:S34O2MN}
\end{figure*}
\newpage
\clearpage

%*******************************************************************************
% Figure: SOO-17;v=0;
\begin{figure*}[!htb]
    \centering
    \begin{subfigure}[t]{1.0\columnwidth}
       \includegraphics[width=1.0\columnwidth]{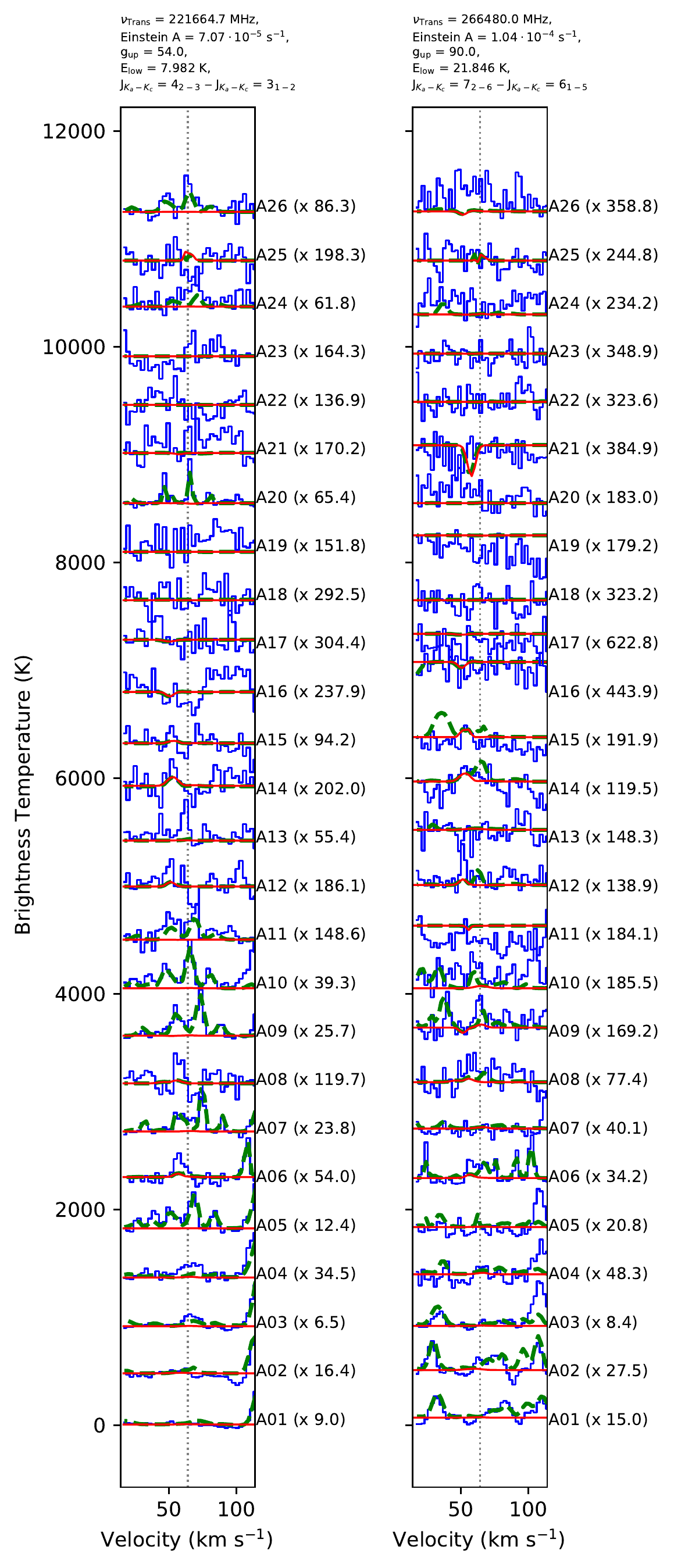}\\
       \caption{Sgr~B2(M)}
       \label{fig:SOO17M}
    \end{subfigure}
\quad
    \begin{subfigure}[t]{1.0\columnwidth}
       \includegraphics[width=1.0\columnwidth]{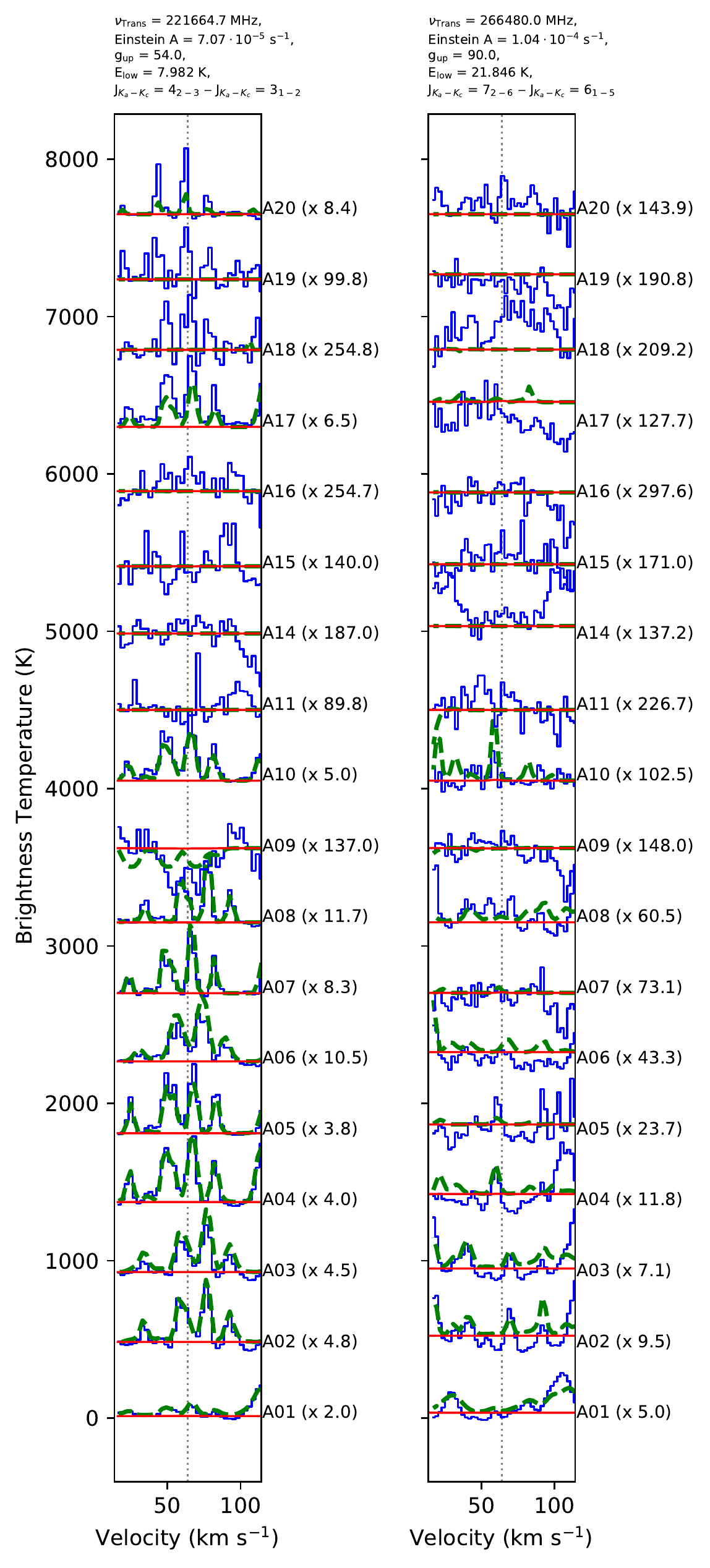}\\
       \caption{Sgr~B2(N)}
       \label{fig:SOO17N}
   \end{subfigure}
   \caption{Selected transitions of SO$^{17}$O in Sgr~B2(M) and N.}
   \ContinuedFloat
   \label{fig:SOO17MN}
\end{figure*}
\newpage
\clearpage

%*******************************************************************************
% Figure: SOO-18;v=0;
\begin{figure*}[!htb]
    \centering
    \begin{subfigure}[t]{1.0\columnwidth}
       \includegraphics[width=1.0\columnwidth]{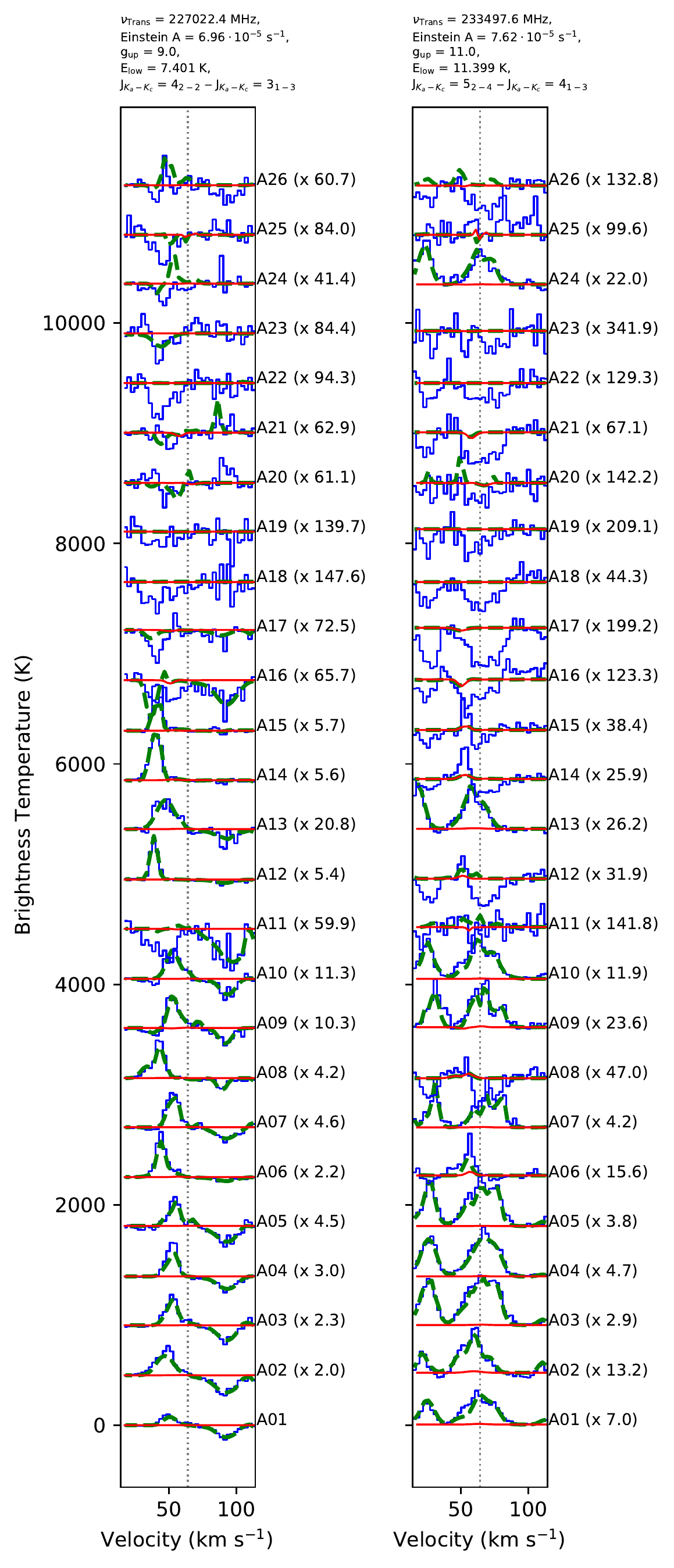}\\
       \caption{Sgr~B2(M)}
       \label{fig:SOO18M}
    \end{subfigure}
\quad
    \begin{subfigure}[t]{1.0\columnwidth}
       \includegraphics[width=1.0\columnwidth]{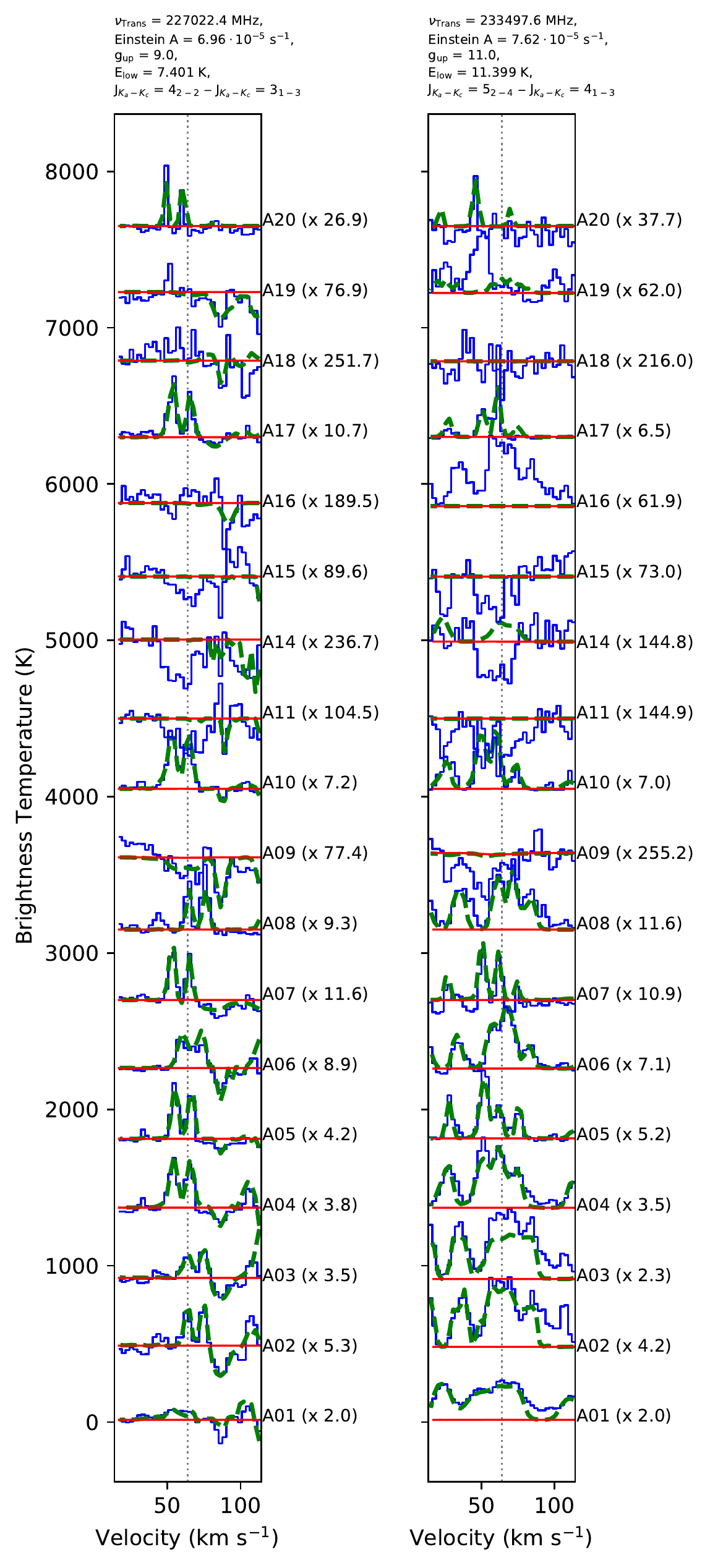}\\
       \caption{Sgr~B2(N)}
       \label{fig:SOO18N}
   \end{subfigure}
   \caption{Selected transitions of SO$^{18}$O in Sgr~B2(M) and N.}
   \ContinuedFloat
   \label{fig:SOO18MN}
\end{figure*}
\newpage
\clearpage

%*******************************************************************************
% Figure: CS;v=0; and C-13-S;v=0;
\begin{figure*}[!htb]
    \centering
    \begin{subfigure}[t]{0.45\columnwidth}
       \includegraphics[width=1.0\columnwidth]{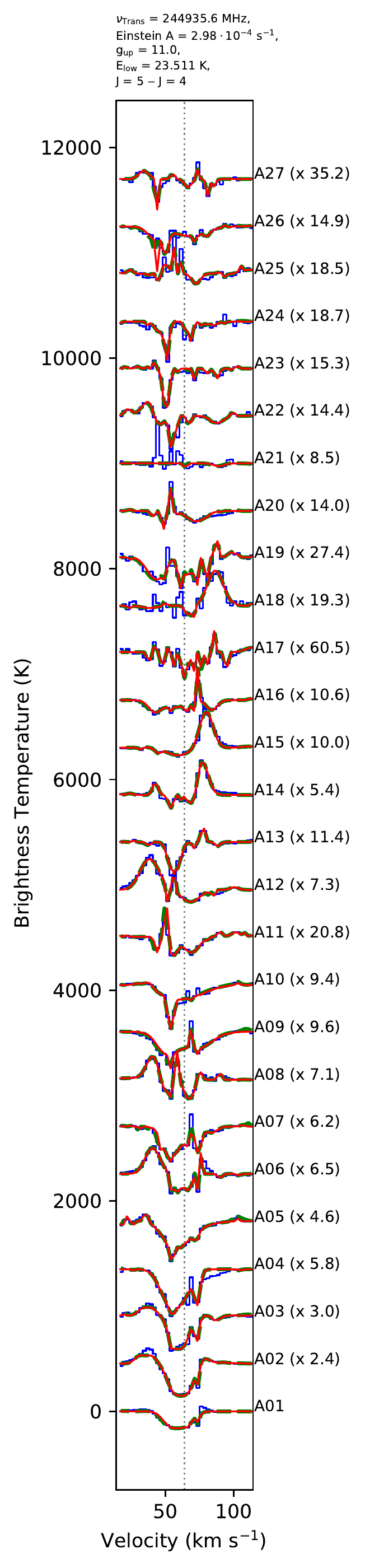}\\
       \caption{CS in Sgr~B2(M).}
       \label{fig:CSM}
    \end{subfigure}
\quad
    \begin{subfigure}[t]{0.45\columnwidth}
       \includegraphics[width=1.0\columnwidth]{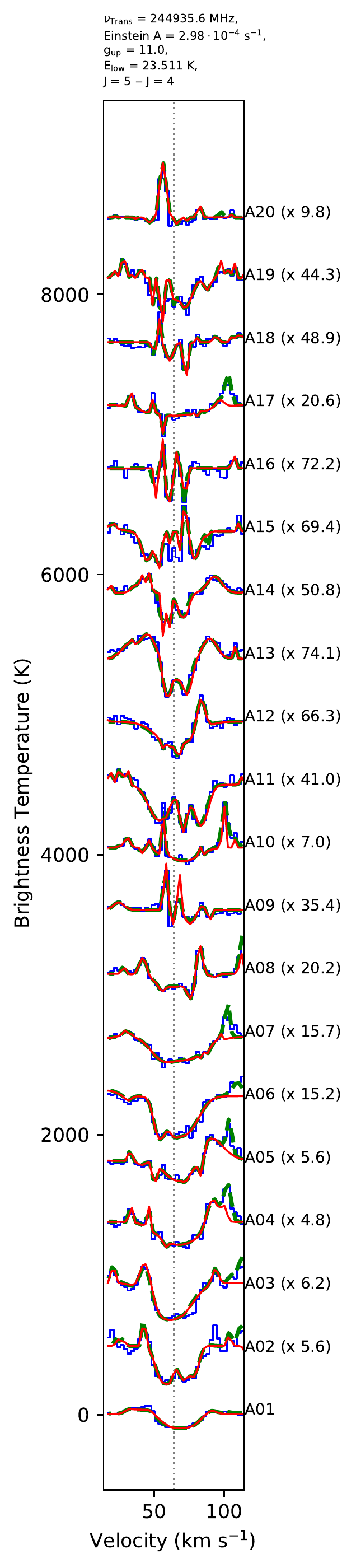}\\
       \caption{CS in Sgr~B2(N).}
       \label{fig:CSN}
   \end{subfigure}
\quad
    \begin{subfigure}[t]{0.45\columnwidth}
       \includegraphics[width=1.0\columnwidth]{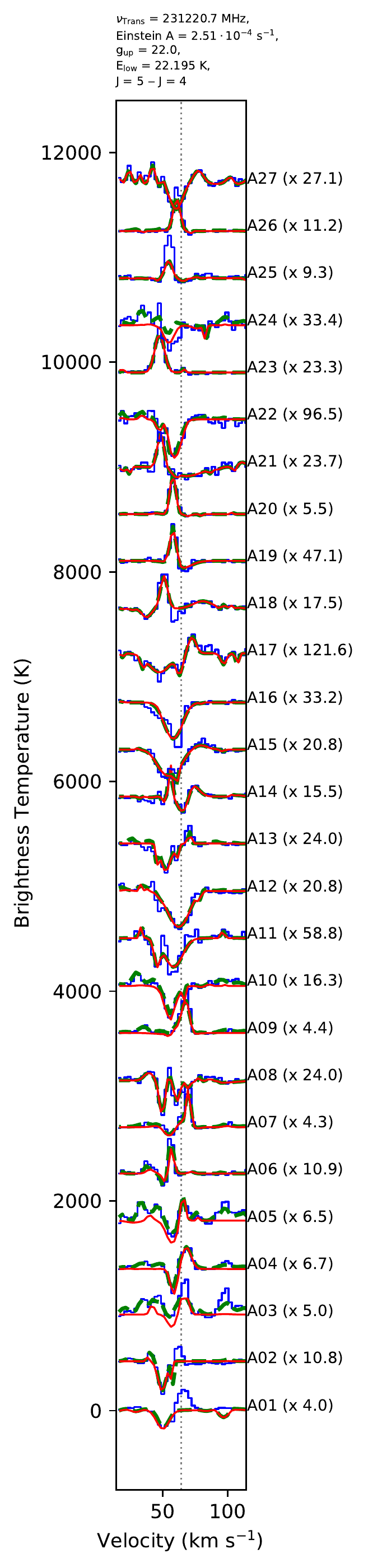}\\
       \caption{$^{13}$CS in Sgr~B2(M).}
       \label{fig:C13SM}
    \end{subfigure}
\quad
    \begin{subfigure}[t]{0.45\columnwidth}
       \includegraphics[width=1.0\columnwidth]{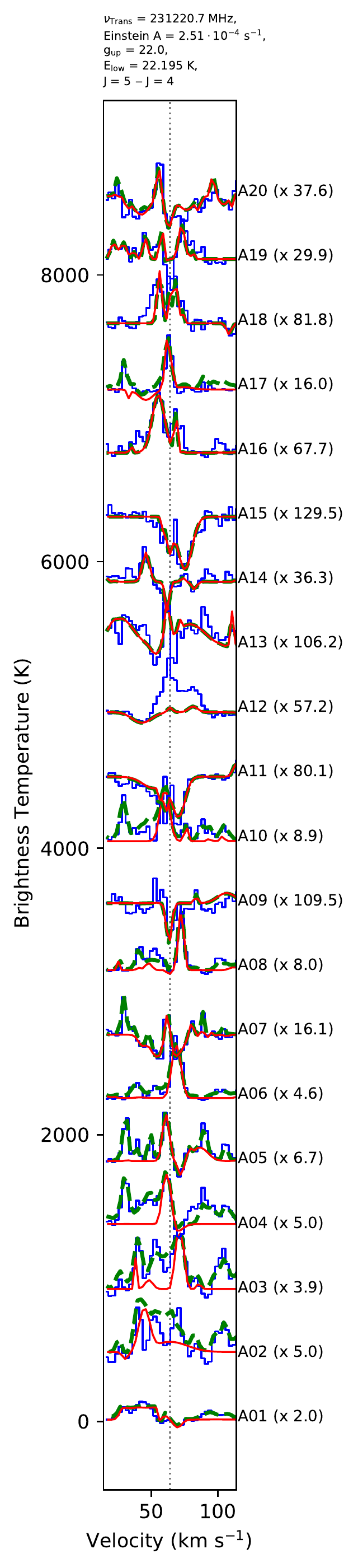}\\
       \caption{$^{13}$CS in Sgr~B2(N).}
       \label{fig:C13SN}
   \end{subfigure}
   \caption{Selected transitions of CS and $^{13}$CS in Sgr~B2(M) and N.}
   \ContinuedFloat
   \label{fig:CSMN}
\end{figure*}
\newpage
\clearpage

%*******************************************************************************
% Figure: CS-33;v=0; and CS-34;v=0;
\begin{figure*}[!htb]
    \centering
    \begin{subfigure}[t]{0.45\columnwidth}
       \includegraphics[width=1.0\columnwidth]{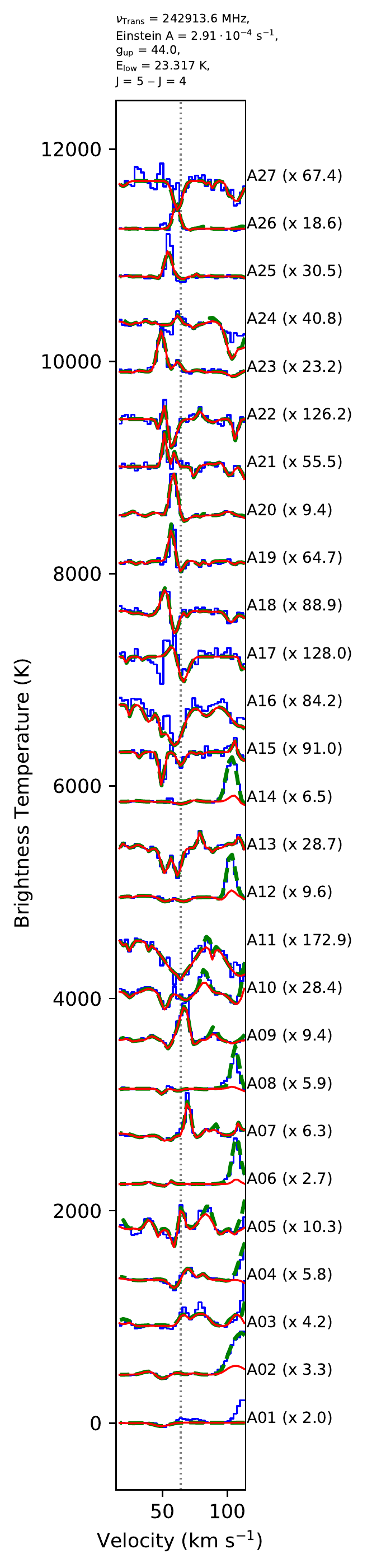}\\
       \caption{C$^{33}$S in Sgr~B2(M).}
       \label{fig:CS33M}
    \end{subfigure}
\quad
    \begin{subfigure}[t]{0.45\columnwidth}
       \includegraphics[width=1.0\columnwidth]{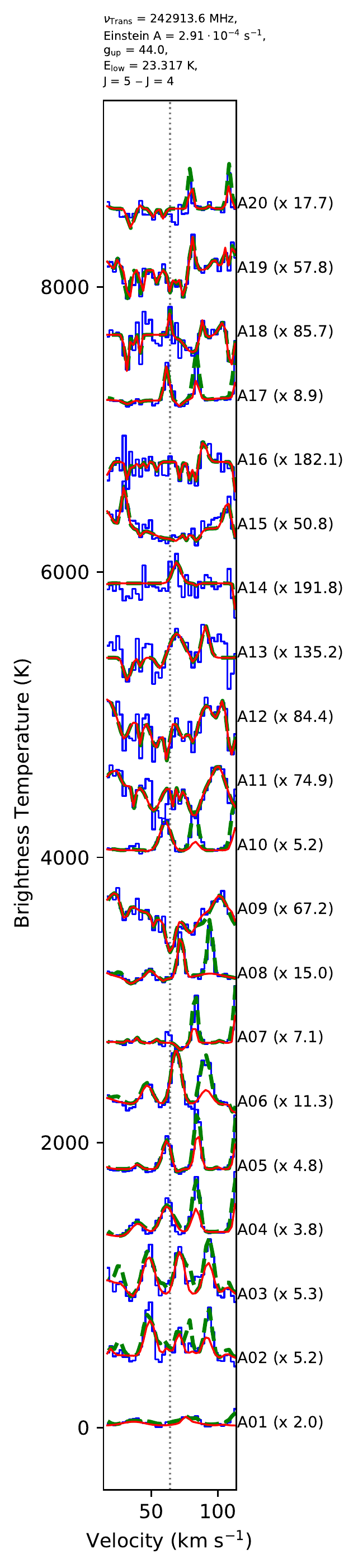}\\
       \caption{C$^{33}$S in Sgr~B2(N).}
       \label{fig:CS33N}
    \end{subfigure}
\quad
    \begin{subfigure}[t]{0.45\columnwidth}
       \includegraphics[width=1.0\columnwidth]{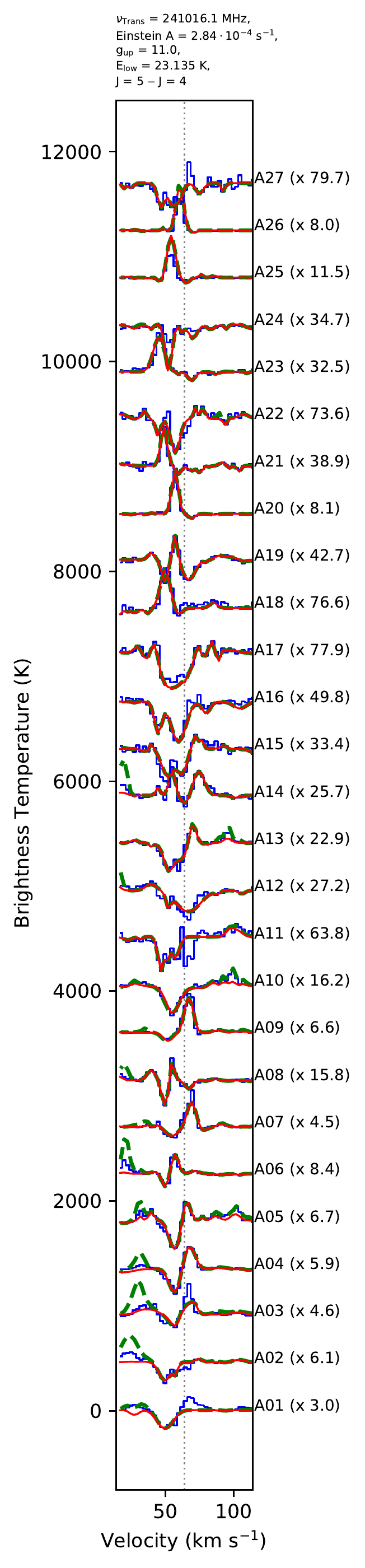}\\
       \caption{C$^{34}$S in Sgr~B2(M).}
       \label{fig:CS34M}
    \end{subfigure}
\quad
    \begin{subfigure}[t]{0.45\columnwidth}
       \includegraphics[width=1.0\columnwidth]{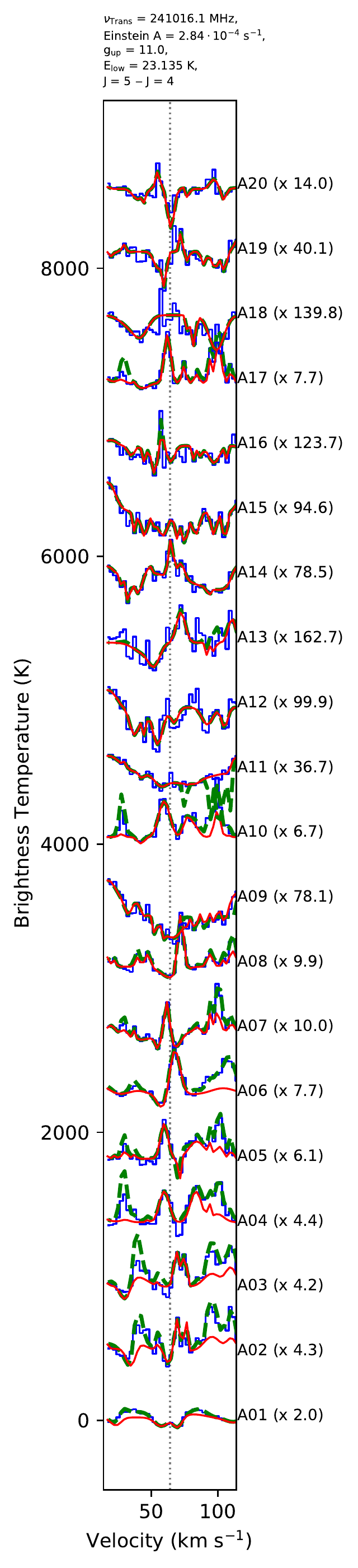}\\
       \caption{C$^{34}$S in Sgr~B2(N).}
       \label{fig:CS34N}
   \end{subfigure}
   \caption{Selected transitions of C$^{33}$S and C$^{34}$S in Sgr~B2(M) and N.}
   \ContinuedFloat
   \label{fig:CS34MN}
\end{figure*}
\newpage
\clearpage

%*******************************************************************************
% Figure: H2CS;v=0;
\begin{figure*}[!htb]
    \centering
    \begin{subfigure}[t]{1.0\columnwidth}
       \includegraphics[width=1.0\columnwidth]{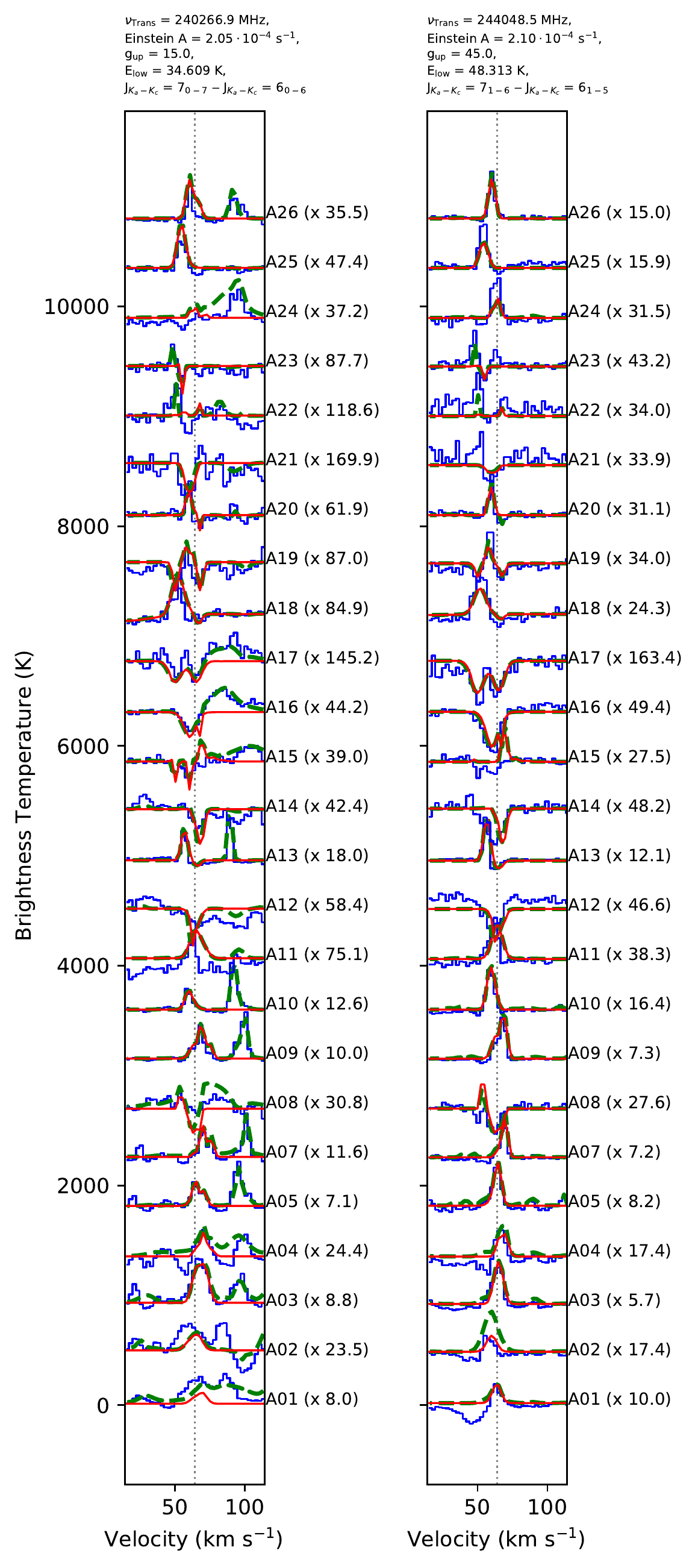}\\
       \caption{Sgr~B2(M)}
       \label{fig:H2CSM}
    \end{subfigure}
\quad
    \begin{subfigure}[t]{1.0\columnwidth}
       \includegraphics[width=1.0\columnwidth]{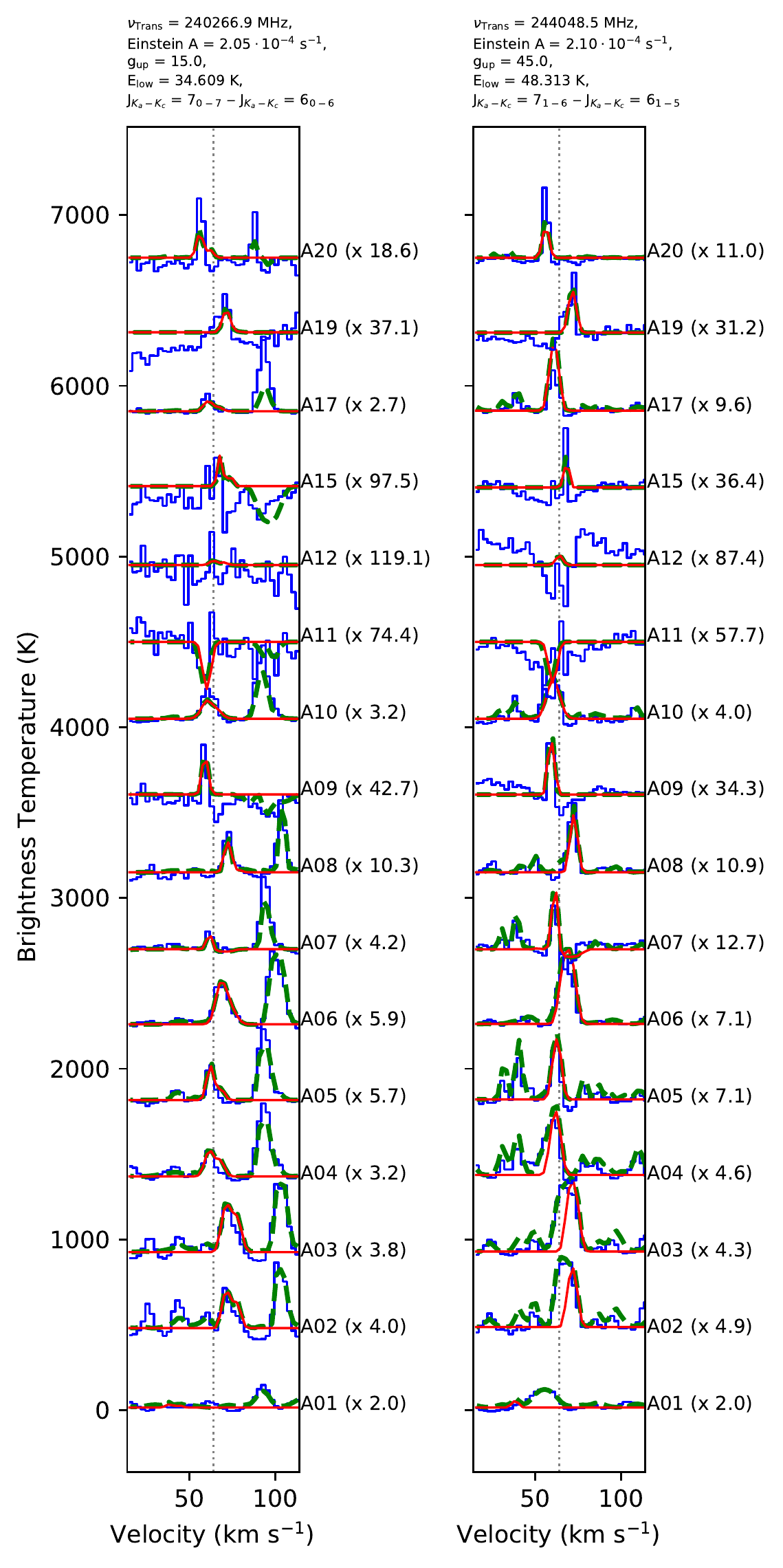}\\
       \caption{Sgr~B2(N)}
       \label{fig:H2CSN}
   \end{subfigure}
   \caption{Selected transitions of H$_2$CS in Sgr~B2(M) and N.}
   \ContinuedFloat
   \label{fig:H2CSMN}
\end{figure*}
\newpage
\clearpage

%*******************************************************************************
% Figure: H2C-13-S;v=0;
\begin{figure*}[!htb]
    \centering
    \begin{subfigure}[t]{1.0\columnwidth}
       \includegraphics[width=1.0\columnwidth]{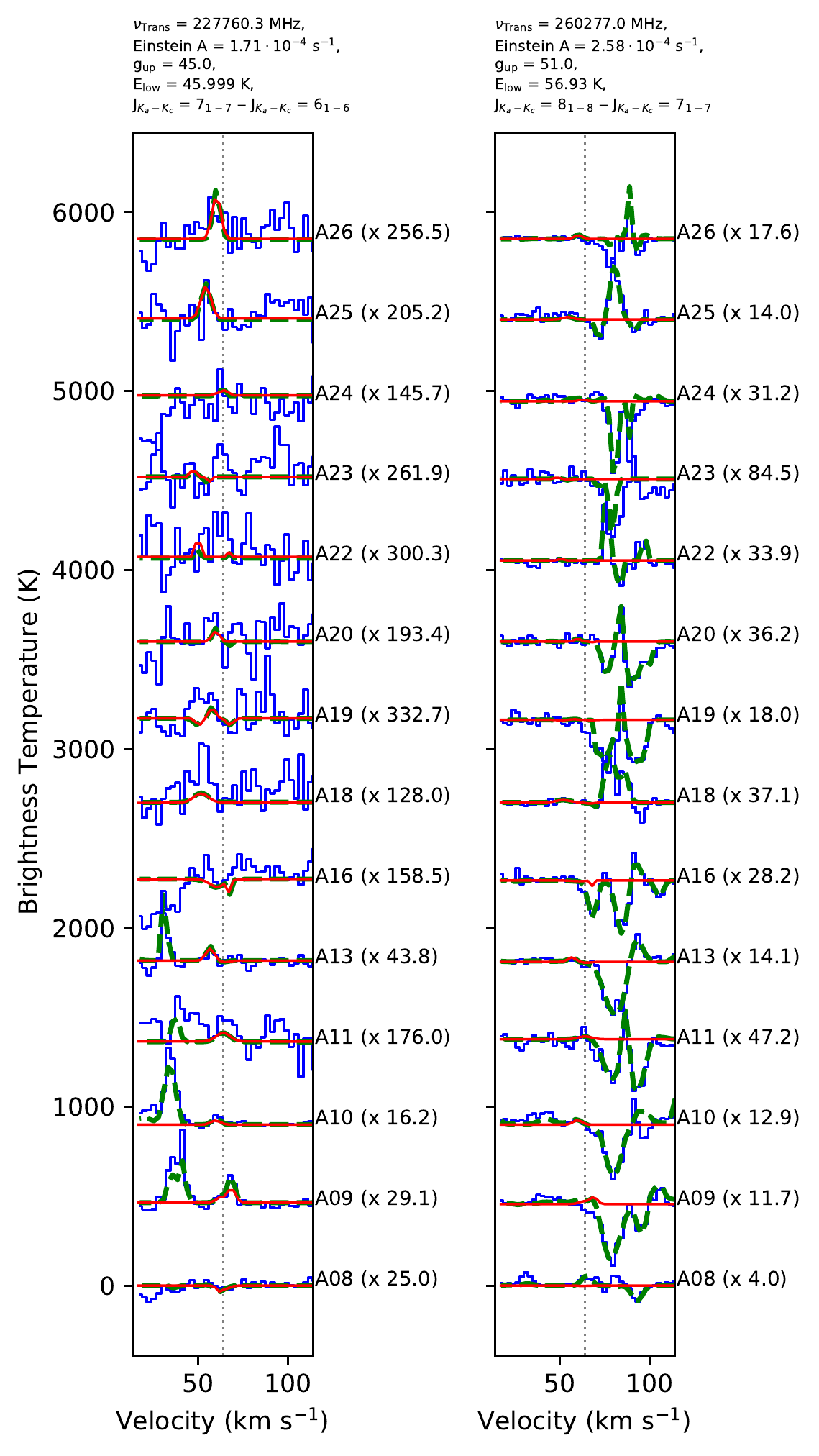}\\
       \caption{Sgr~B2(M)}
       \label{fig:H2C13SM}
    \end{subfigure}
\quad
    \begin{subfigure}[t]{1.0\columnwidth}
       \includegraphics[width=1.0\columnwidth]{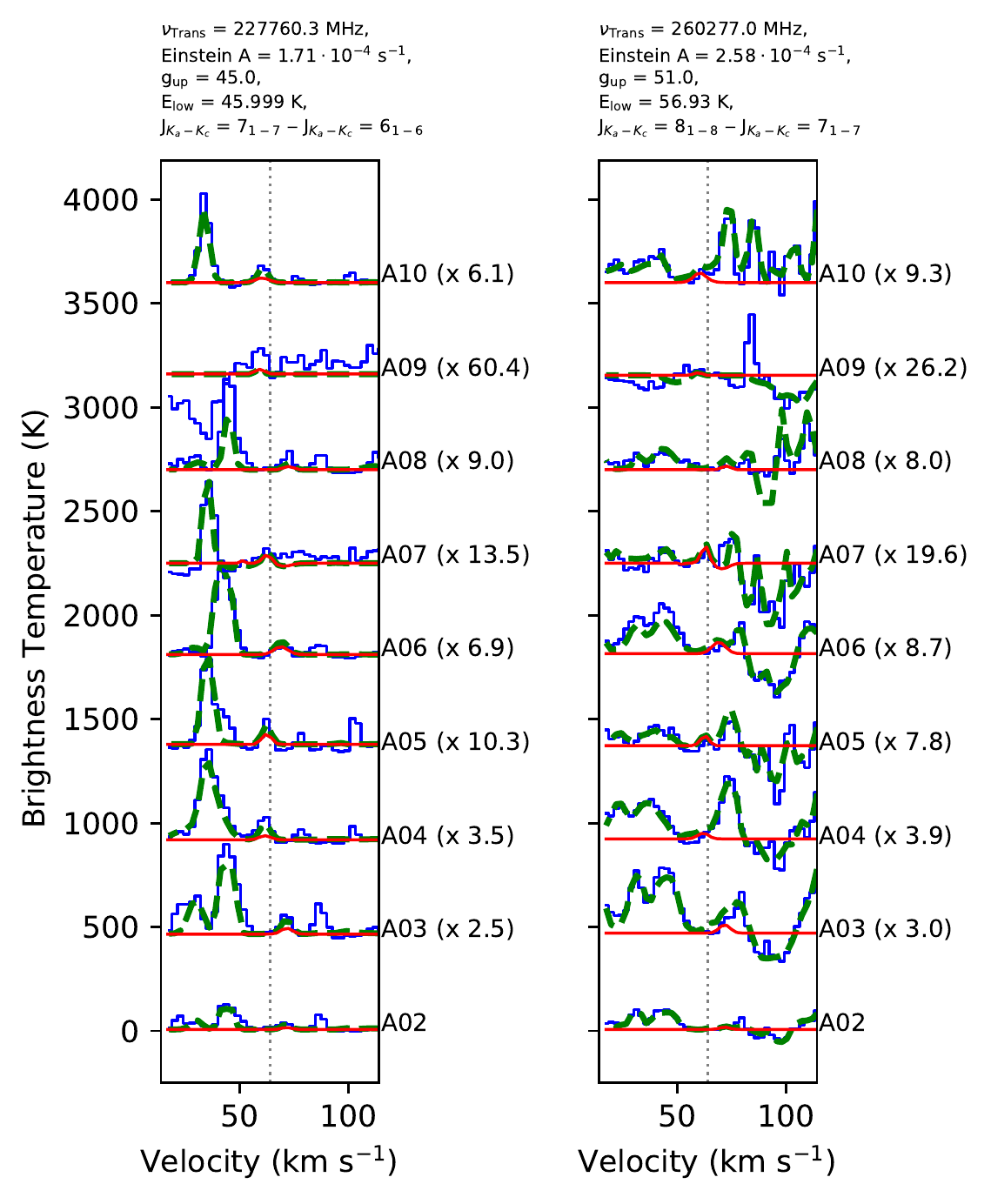}\\
       \caption{Sgr~B2(N)}
       \label{fig:H2C13SN}
   \end{subfigure}
   \caption{Selected transitions of H$_2 \! ^{13}$CS in Sgr~B2(M) and N.}
   \ContinuedFloat
   \label{fig:H2C13SMN}
\end{figure*}
\newpage
\clearpage

%*******************************************************************************
% Figure: H2CS-33;v=0;hyp1
\begin{figure*}[!htb]
    \centering
    \begin{subfigure}[t]{1.0\columnwidth}
       \includegraphics[width=0.985\columnwidth]{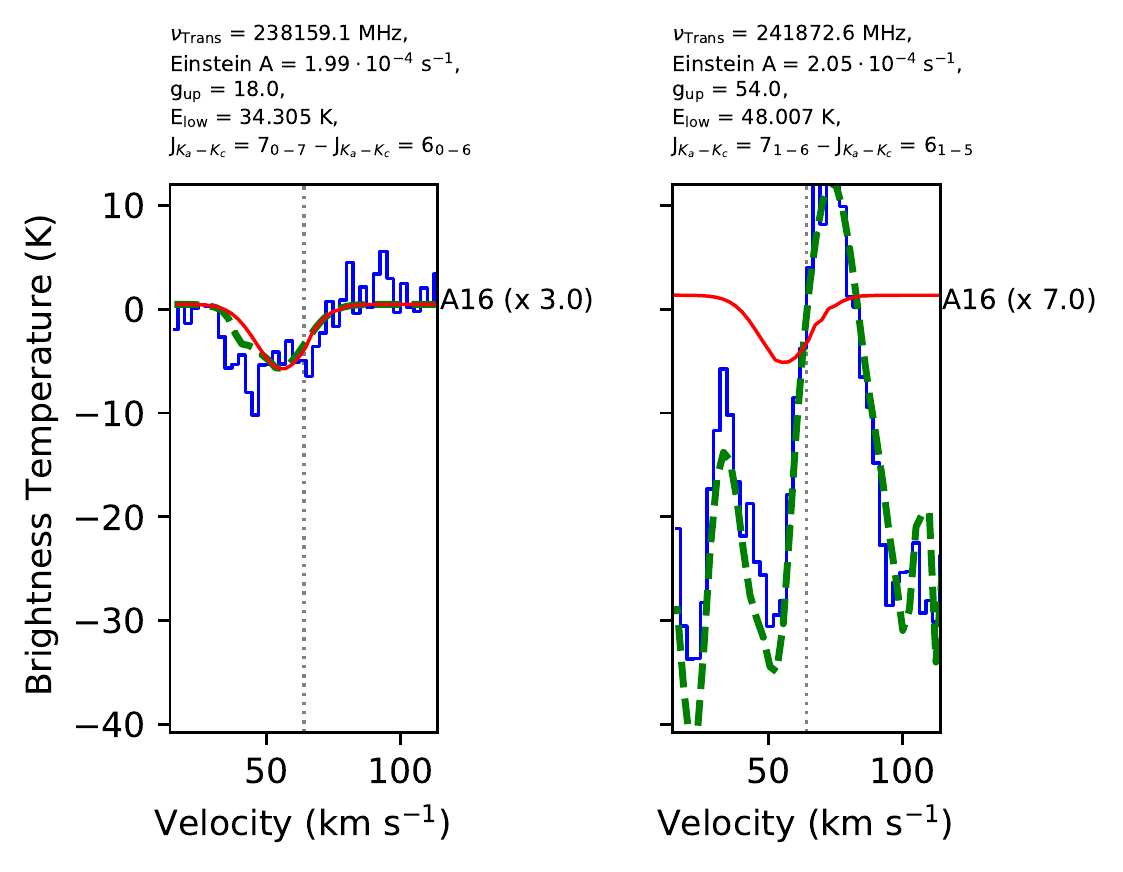}\\
       \caption{Sgr~B2(M)}
       \label{fig:H2CS33M}
    \end{subfigure}
\quad
    \begin{subfigure}[t]{1.0\columnwidth}
       \includegraphics[width=1.0\columnwidth]{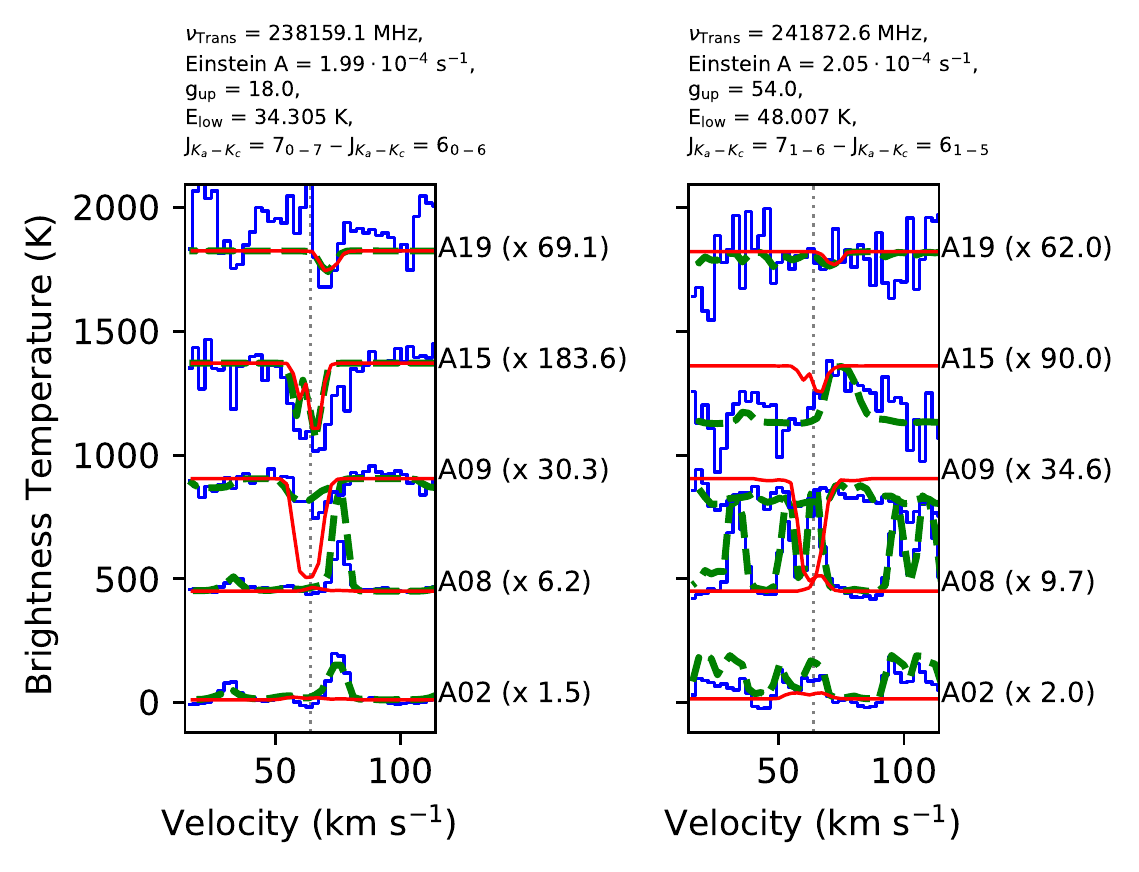}\\
       \caption{Sgr~B2(N)}
       \label{fig:H2CS33N}
   \end{subfigure}
   \caption{Selected transitions of H$_2$C$^{33}$S in Sgr~B2(M) and N.}
   \ContinuedFloat
   \label{fig:H2CS33MN}
\end{figure*}

%*******************************************************************************
% Figure: H2CS-34;v=0;
\begin{figure*}[!htb]
    \centering
    \begin{subfigure}[t]{1.0\columnwidth}
       \includegraphics[width=1.0\columnwidth]{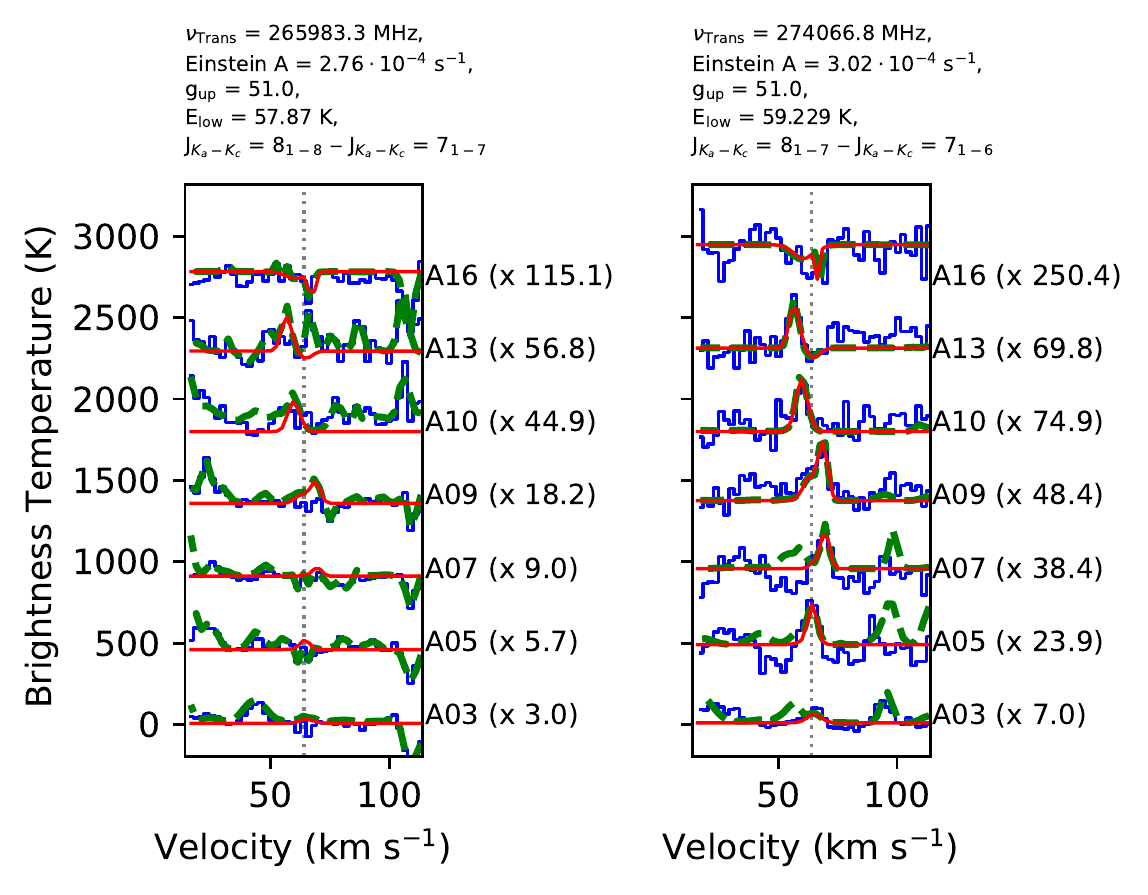}\\
       \caption{Sgr~B2(M)}
       \label{fig:H2CS34M}
    \end{subfigure}
\quad
    \begin{subfigure}[t]{1.0\columnwidth}
       \includegraphics[width=1.0\columnwidth]{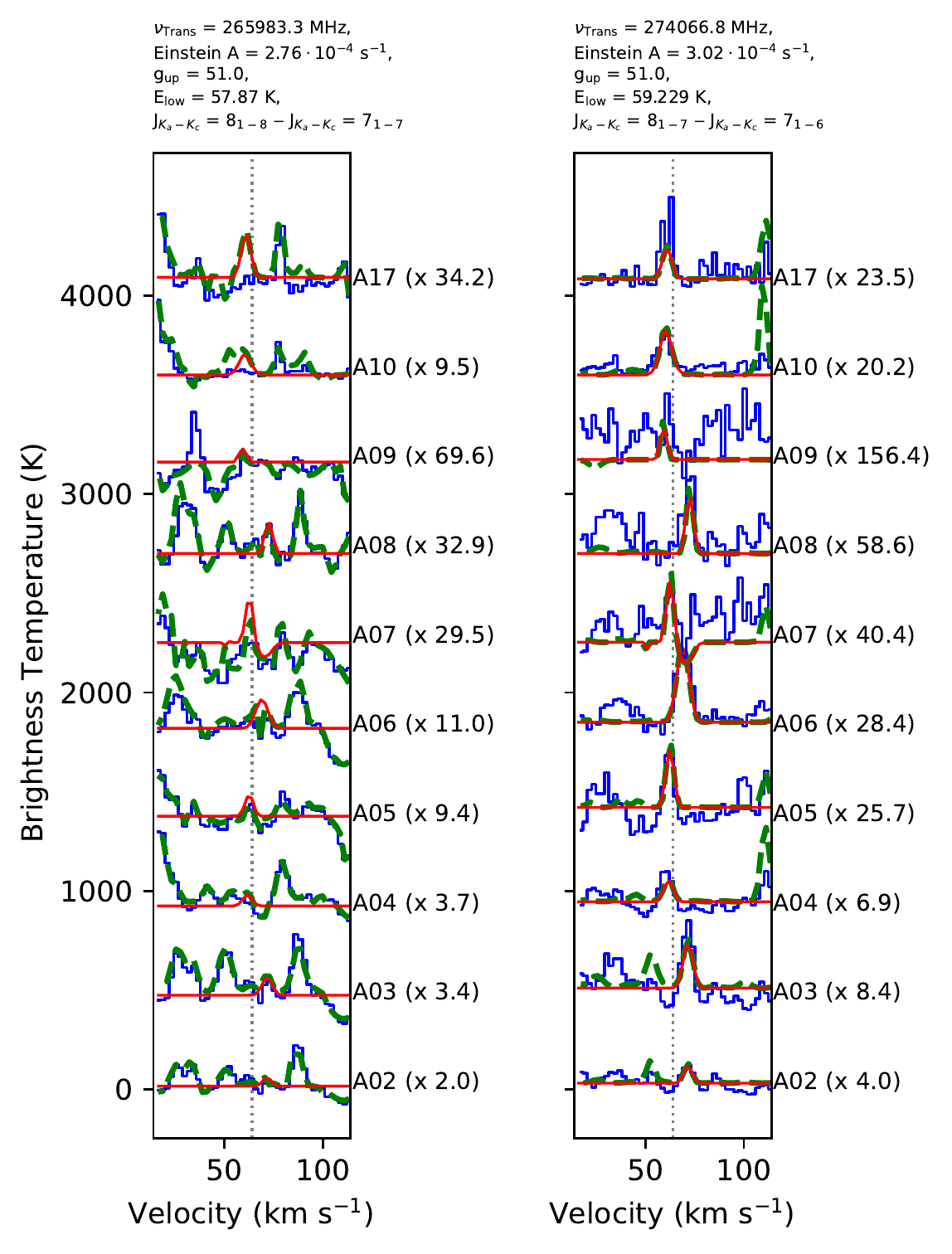}\\
       \caption{Sgr~B2(N)}
       \label{fig:H2CS34N}
   \end{subfigure}
   \caption{Selected transitions of H$_2$C$^{34}$S in Sgr~B2(M) and N.}
   \ContinuedFloat
   \label{fig:H2CS34MN}
\end{figure*}
\newpage
\clearpage

%*******************************************************************************
% Figure: OCS;v=0;
\begin{figure*}[!htb]
    \centering
    \begin{subfigure}[t]{1.0\columnwidth}
       \includegraphics[width=1.0\columnwidth]{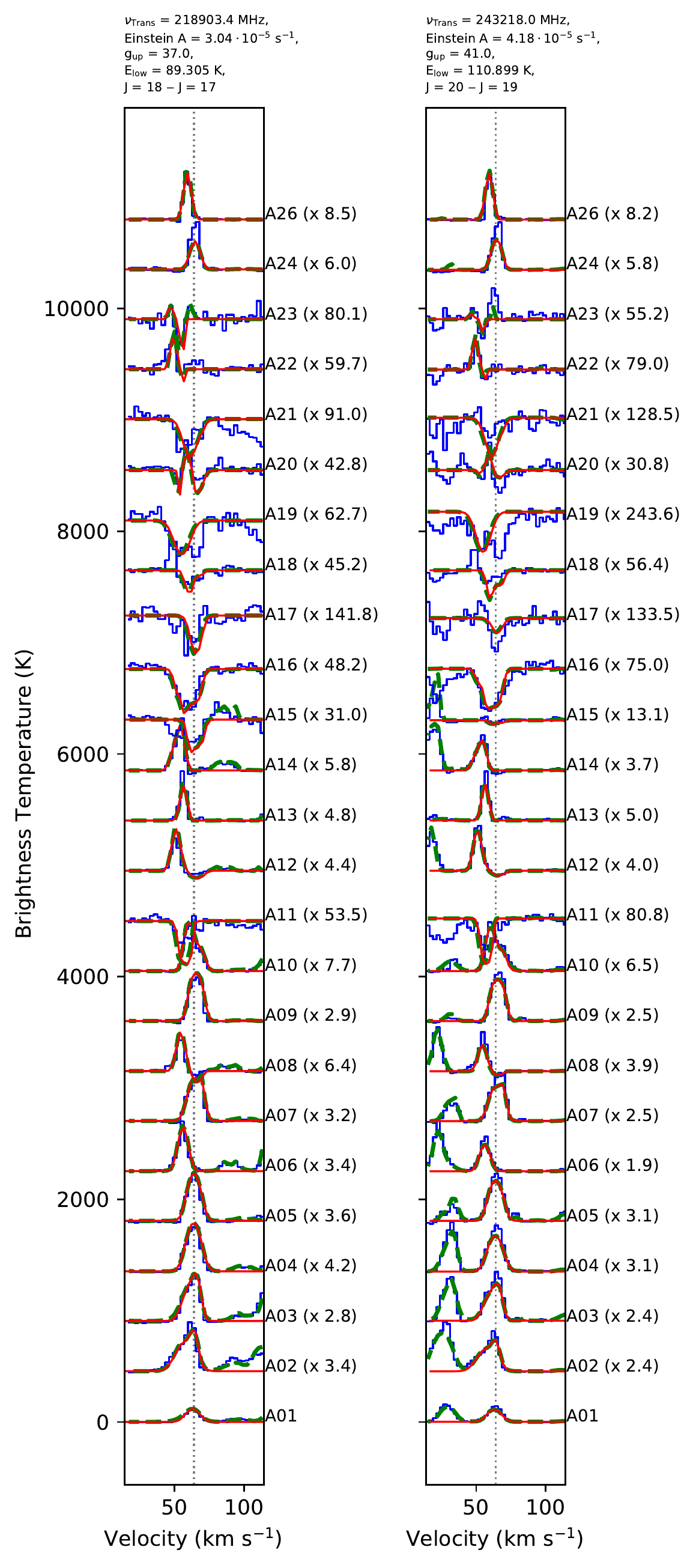}\\
       \caption{Sgr~B2(M)}
       \label{fig:OCSM}
    \end{subfigure}
\quad
    \begin{subfigure}[t]{1.0\columnwidth}
       \includegraphics[width=1.0\columnwidth]{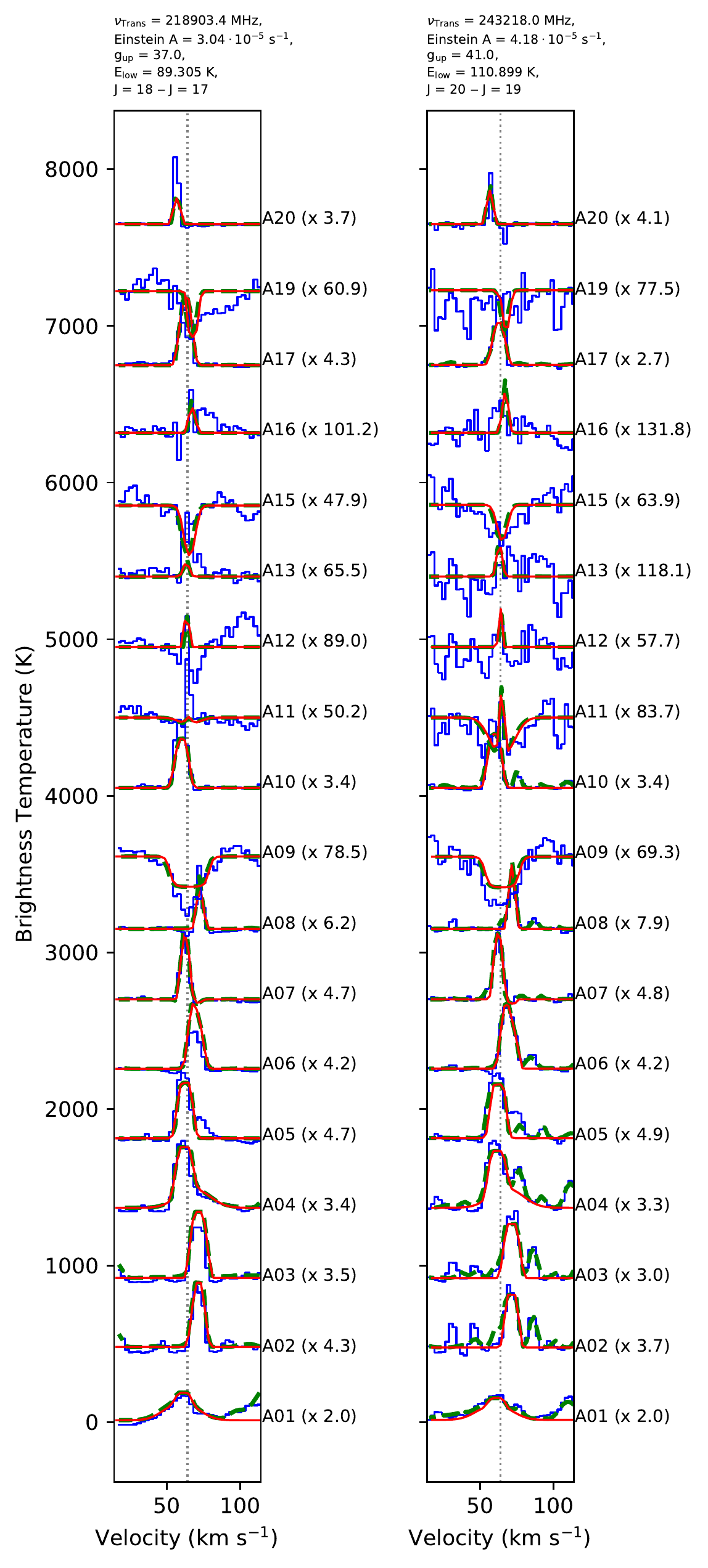}\\
       \caption{Sgr~B2(N)}
       \label{fig:OCSN}
   \end{subfigure}
   \caption{Selected transitions of OCS in Sgr~B2(M) and N.}
   \ContinuedFloat
   \label{fig:OCSMN}
\end{figure*}
\newpage
\clearpage

%*******************************************************************************
% Figure: OCS;v2=1;
\begin{figure*}[!htb]
    \centering
    \begin{subfigure}[t]{1.0\columnwidth}
       \includegraphics[width=1.0\columnwidth]{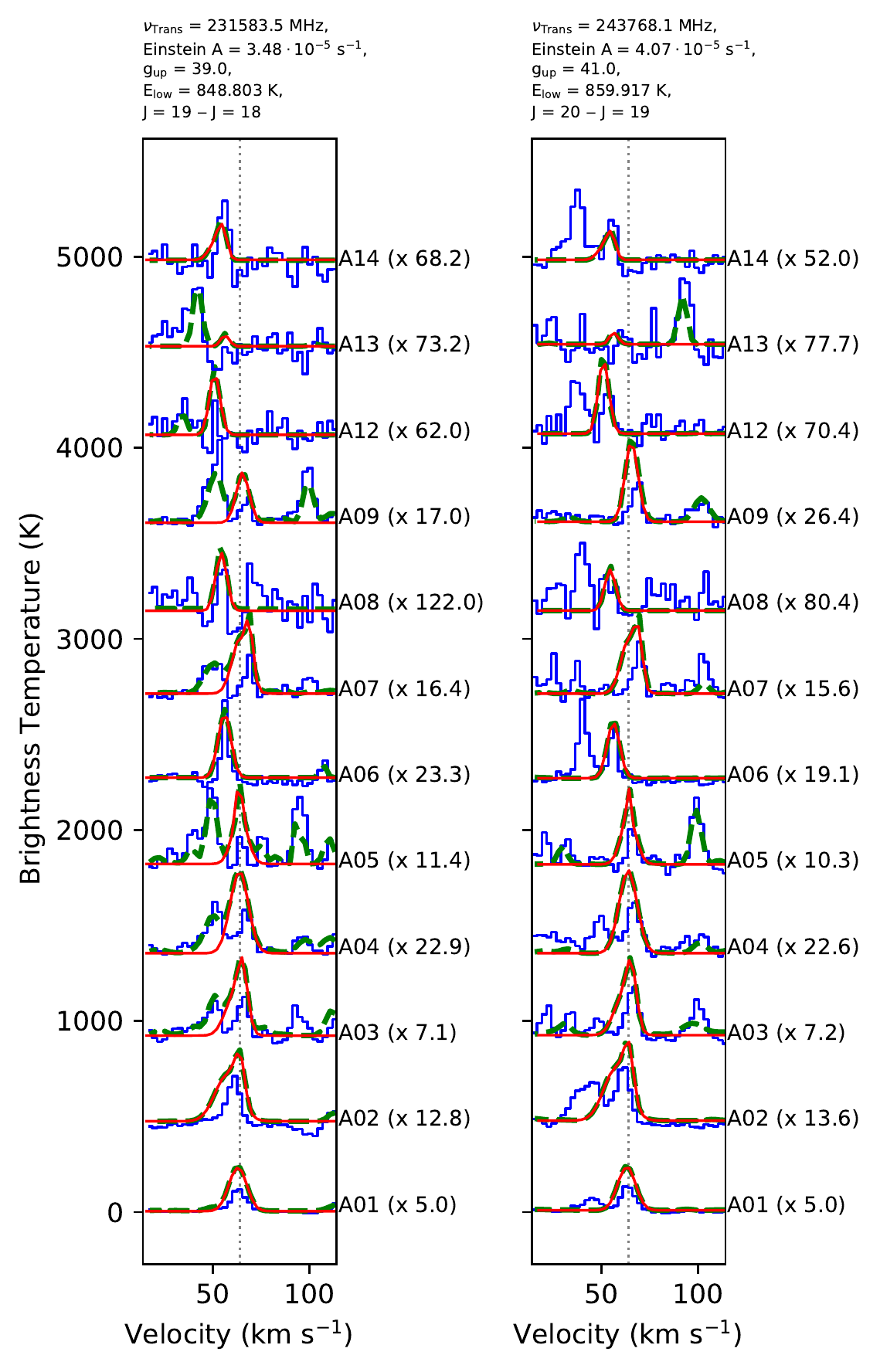}\\
       \caption{Sgr~B2(M)}
       \label{fig:OCSv21M}
    \end{subfigure}
\quad
    \begin{subfigure}[t]{1.0\columnwidth}
       \includegraphics[width=1.0\columnwidth]{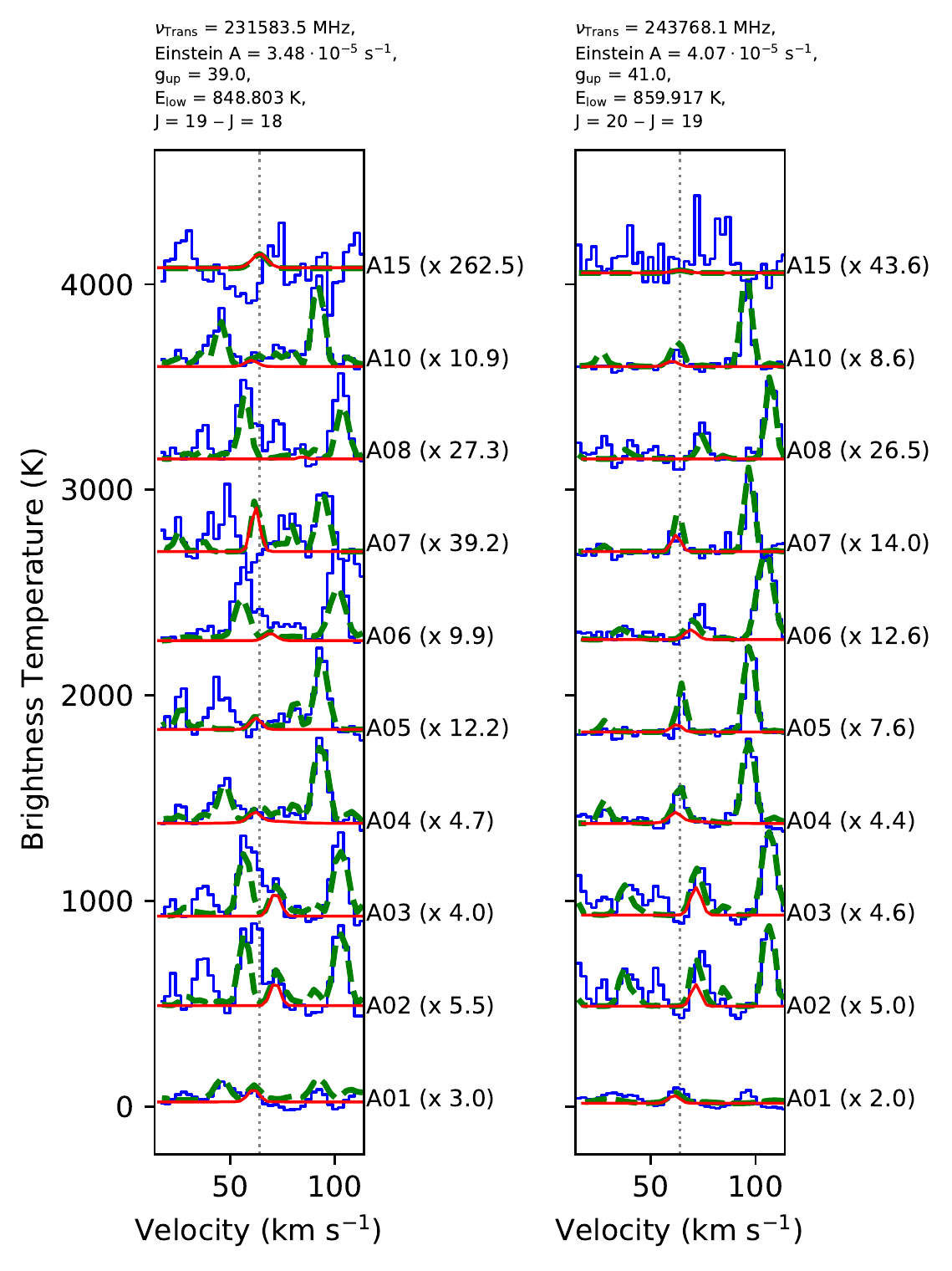}\\
       \caption{Sgr~B2(N)}
       \label{fig:OCSv21N}
   \end{subfigure}
   \caption{Selected transitions of OCS, v$_2$=1 in Sgr~B2(M) and N.}
   \ContinuedFloat
   \label{fig:OCSv21MN}
\end{figure*}
\newpage
\clearpage

%*******************************************************************************
% Figure: OC-13-S;v=0;
\begin{figure*}[!htb]
    \centering
    \begin{subfigure}[t]{1.0\columnwidth}
       \includegraphics[width=1.0\columnwidth]{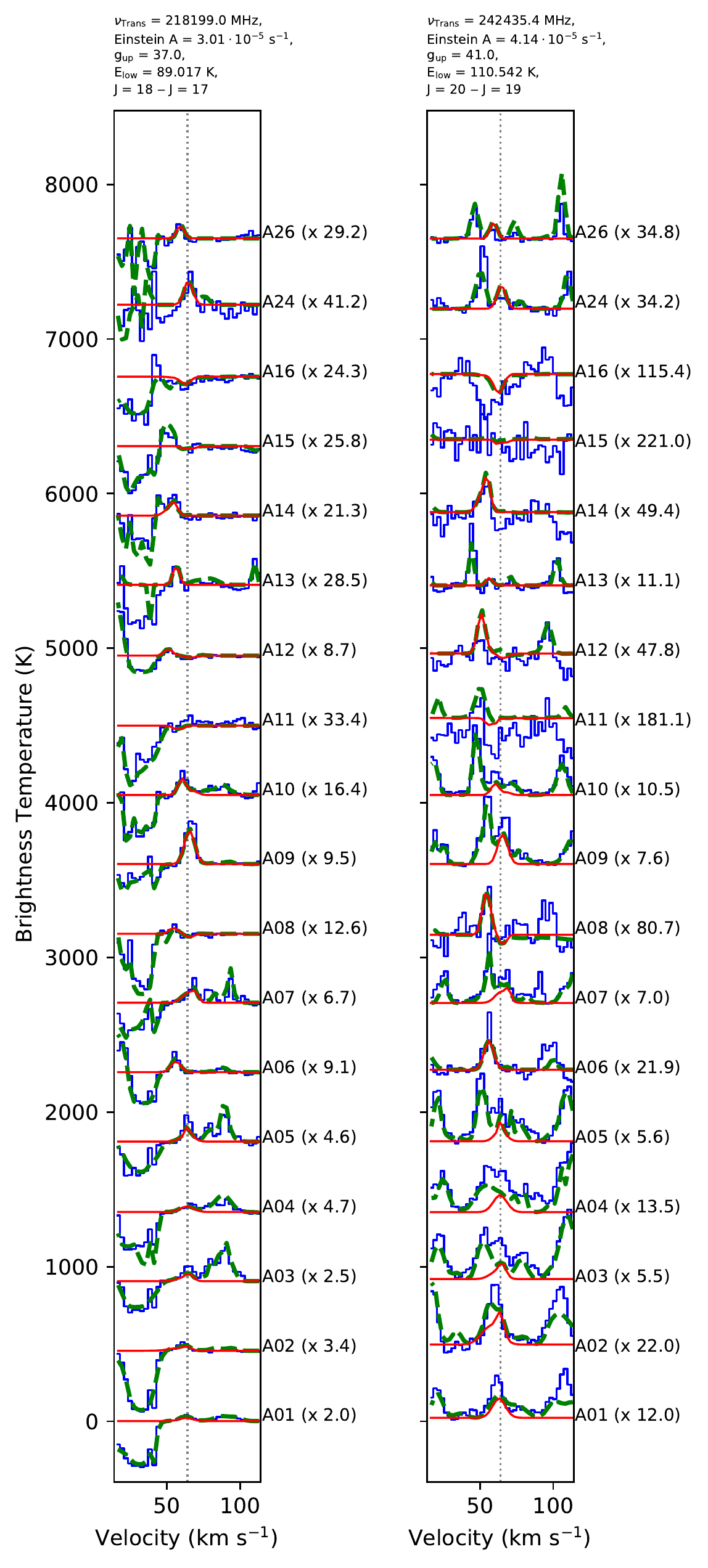}\\
       \caption{Sgr~B2(M)}
       \label{fig:OC13SM}
    \end{subfigure}
\quad
    \begin{subfigure}[t]{1.0\columnwidth}
       \includegraphics[width=1.0\columnwidth]{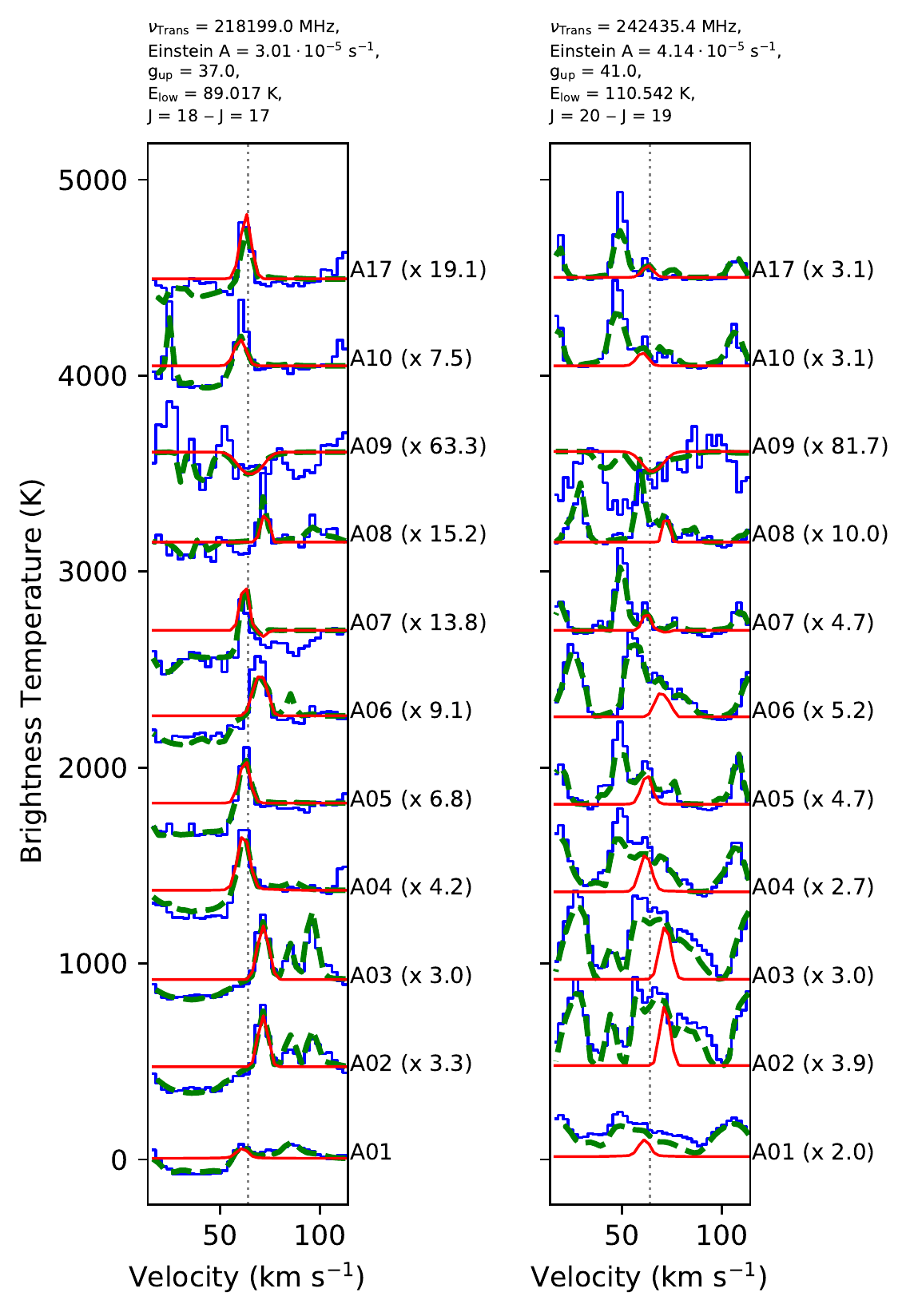}\\
       \caption{Sgr~B2(N)}
       \label{fig:OC13SN}
   \end{subfigure}
   \caption{Selected transitions of O$^{13}$CS in Sgr~B2(M) and N.}
   \ContinuedFloat
   \label{fig:OC13SMN}
\end{figure*}
\newpage
\clearpage

%*******************************************************************************
% Figure: OCS-33;v=0;
\begin{figure*}[!htb]
    \centering
    \begin{subfigure}[t]{1.0\columnwidth}
       \includegraphics[width=1.0\columnwidth]{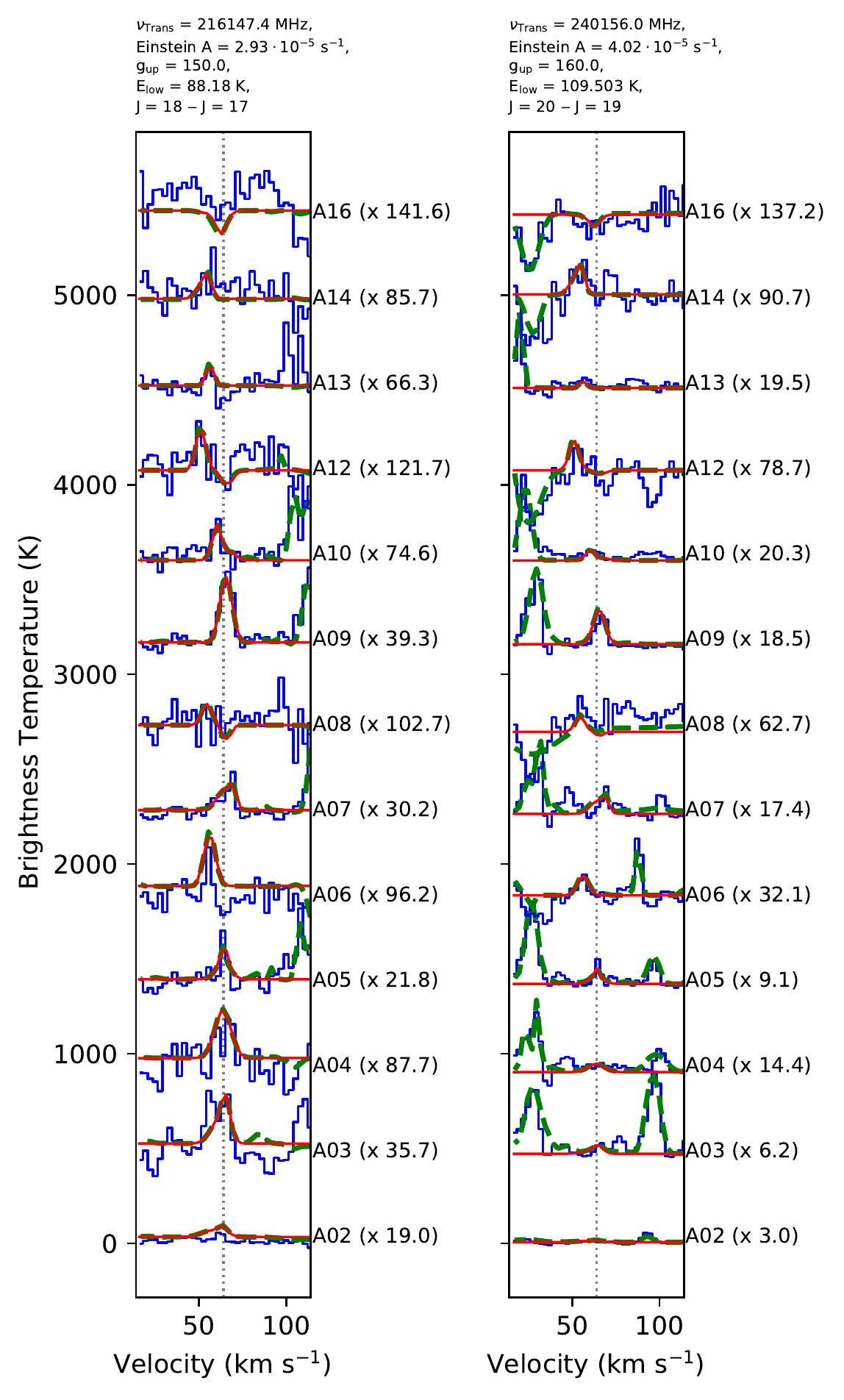}\\
       \caption{Sgr~B2(M)}
       \label{fig:OCS33M}
    \end{subfigure}
\quad
    \begin{subfigure}[t]{1.0\columnwidth}
       \includegraphics[width=1.0\columnwidth]{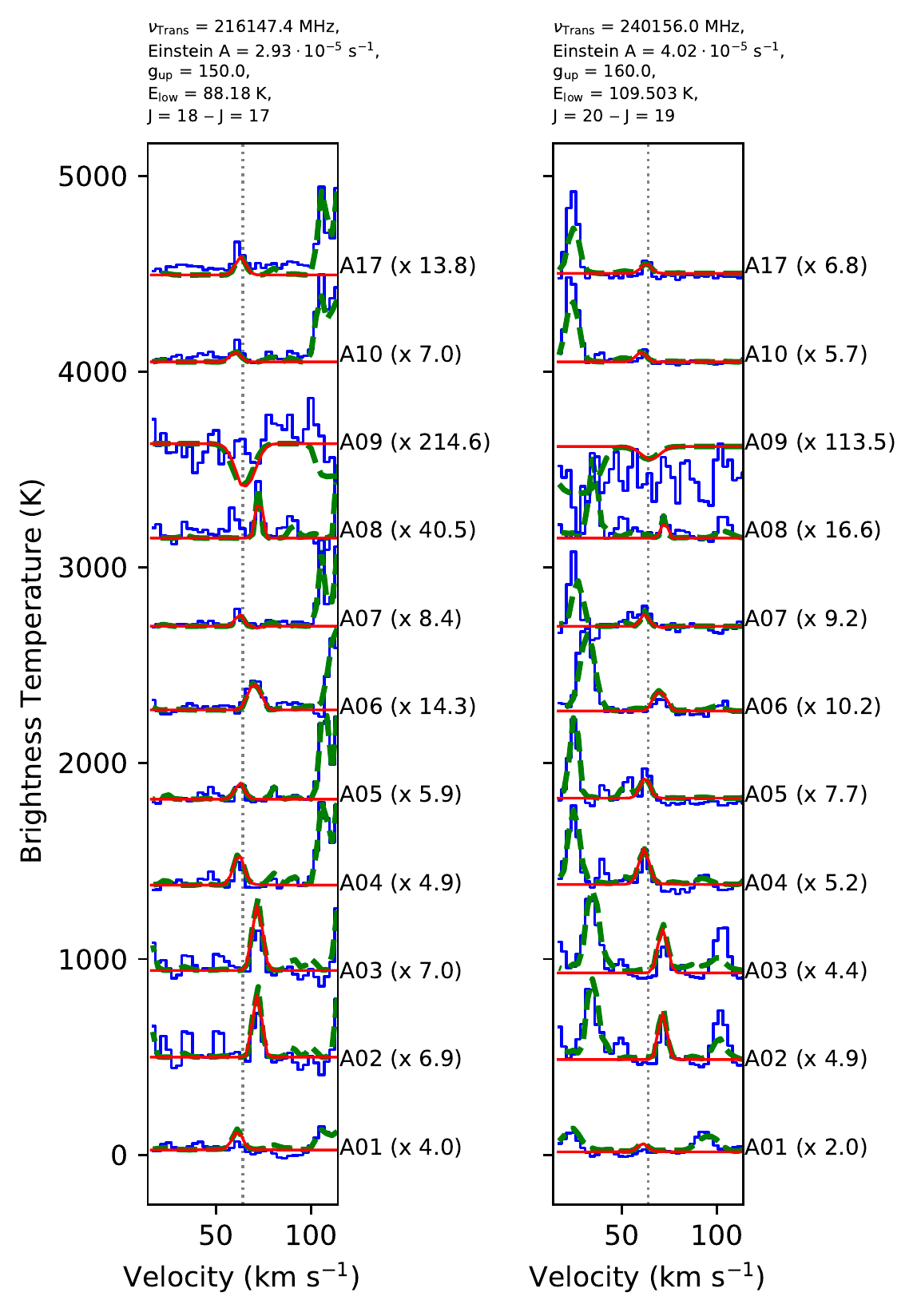}\\
       \caption{Sgr~B2(N)}
       \label{fig:OCS33N}
   \end{subfigure}
   \caption{Selected transitions of OC$^{33}$S in Sgr~B2(M) and N.}
   \ContinuedFloat
   \label{fig:OCS33MN}
\end{figure*}
\newpage
\clearpage

%*******************************************************************************
% Figure: OCS-34;v=0;
\begin{figure*}[!htb]
    \centering
    \begin{subfigure}[t]{1.0\columnwidth}
       \includegraphics[width=1.0\columnwidth]{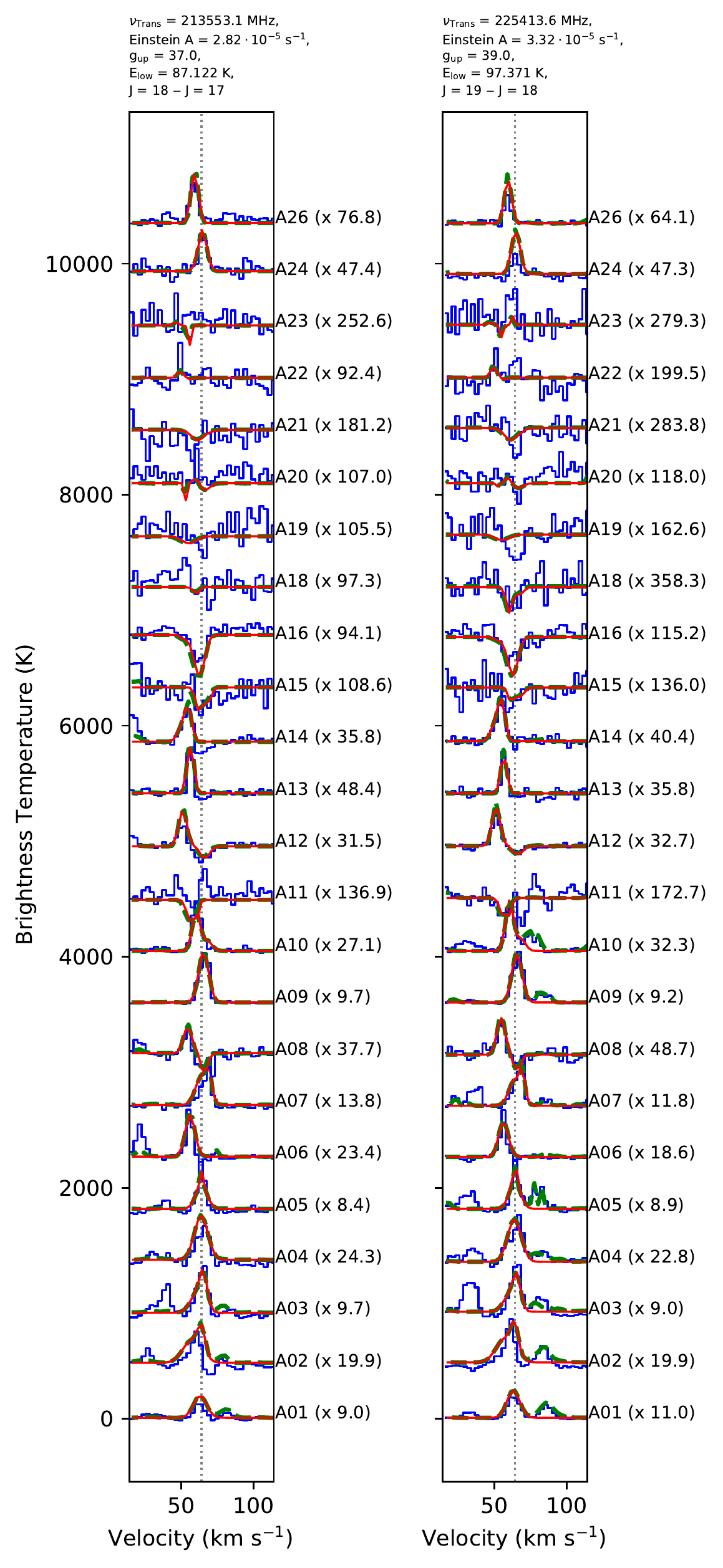}\\
       \caption{Sgr~B2(M)}
       \label{fig:OCS34M}
    \end{subfigure}
\quad
    \begin{subfigure}[t]{1.0\columnwidth}
       \includegraphics[width=1.0\columnwidth]{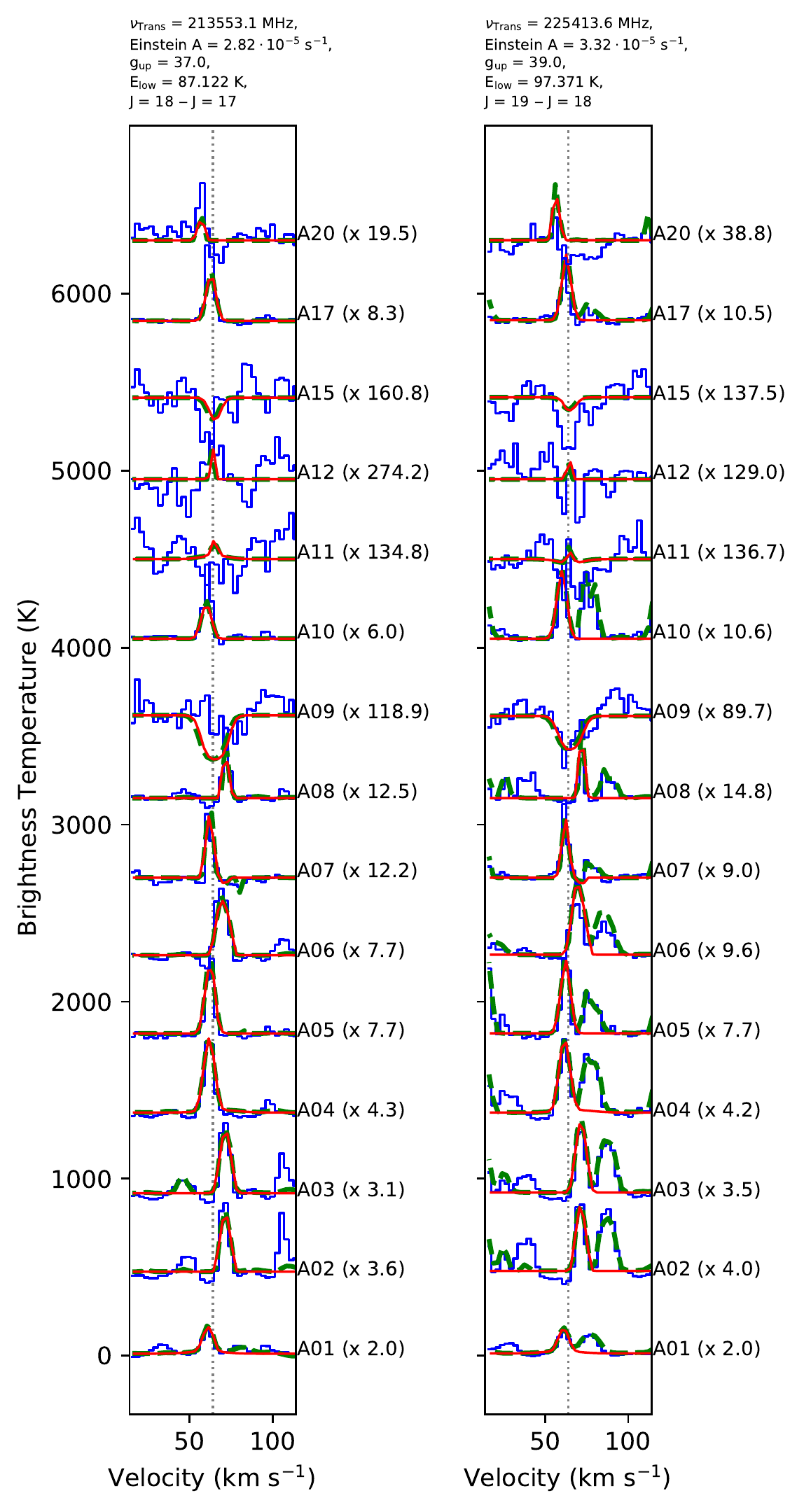}\\
       \caption{Sgr~B2(N)}
       \label{fig:OCS34N}
   \end{subfigure}
   \caption{Selected transitions of OC$^{34}$S in Sgr~B2(M) and N.}
   \ContinuedFloat
   \label{fig:OCS34MN}
\end{figure*}
\newpage
\clearpage

%*******************************************************************************
% Figure: HCS+;v=0;
\begin{figure*}[!htb]
    \centering
    \begin{subfigure}[t]{1.0\columnwidth}
       \includegraphics[width=1.0\columnwidth]{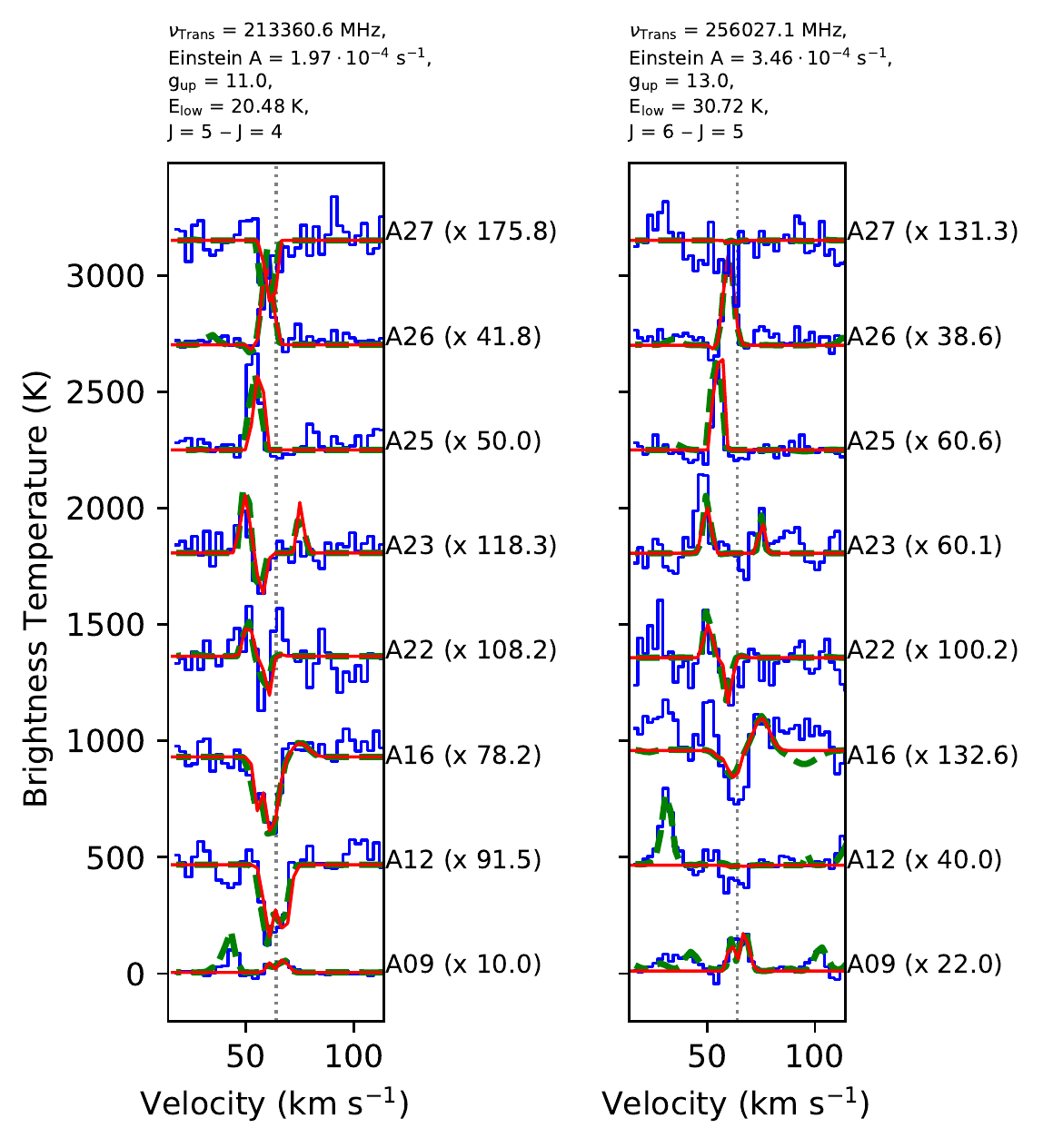}\\
    \end{subfigure}
   \caption{Selected transitions of HCS$^+$ in Sgr~B2(M)}
   \ContinuedFloat
   \label{fig:HCSpM}
\end{figure*}

%*******************************************************************************
% Figure: NS;v=0;hyp1
\begin{figure*}[!htb]
    \centering
    \begin{subfigure}[t]{0.5\columnwidth}
       \includegraphics[width=1.0\columnwidth]{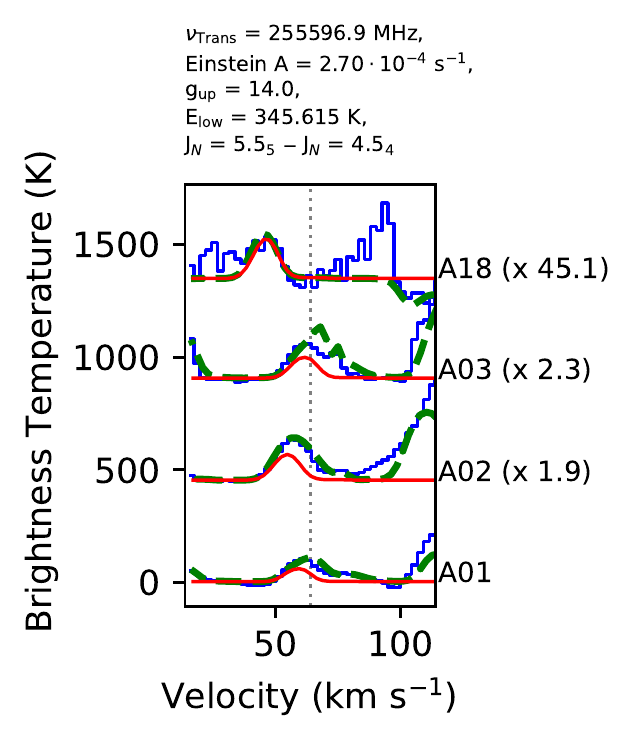}\\
       \caption{Sgr~B2(M)}
       \label{fig:NSM}
    \end{subfigure}
\quad
    \begin{subfigure}[t]{0.5\columnwidth}
       \includegraphics[width=1.0\columnwidth]{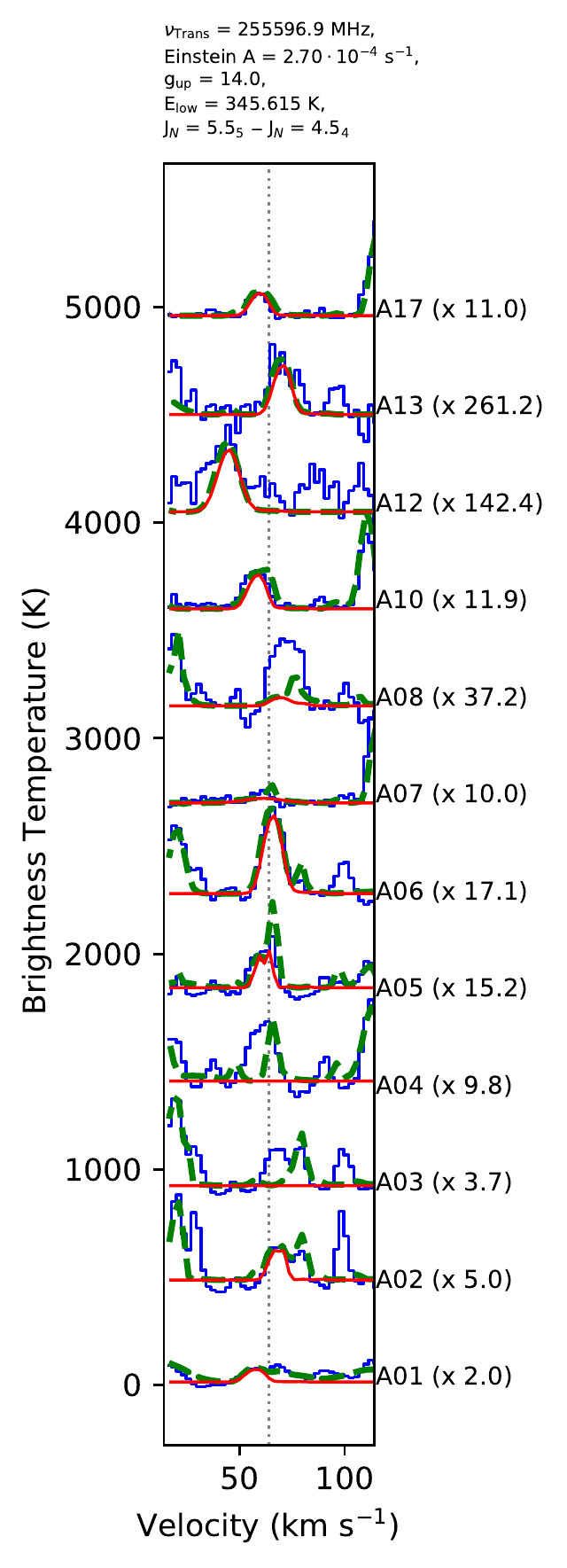}\\
       \caption{Sgr~B2(N)}
       \label{fig:NSN}
   \end{subfigure}
   \caption{Selected transitions of NS in Sgr~B2(M) and N.}
   \ContinuedFloat
   \label{fig:NSMN}
\end{figure*}

%*******************************************************************************
% Figure: NS-34;v=0;
\begin{figure*}[!htb]
    \centering
    \begin{subfigure}[t]{0.5\columnwidth}
       \includegraphics[width=1.0\columnwidth]{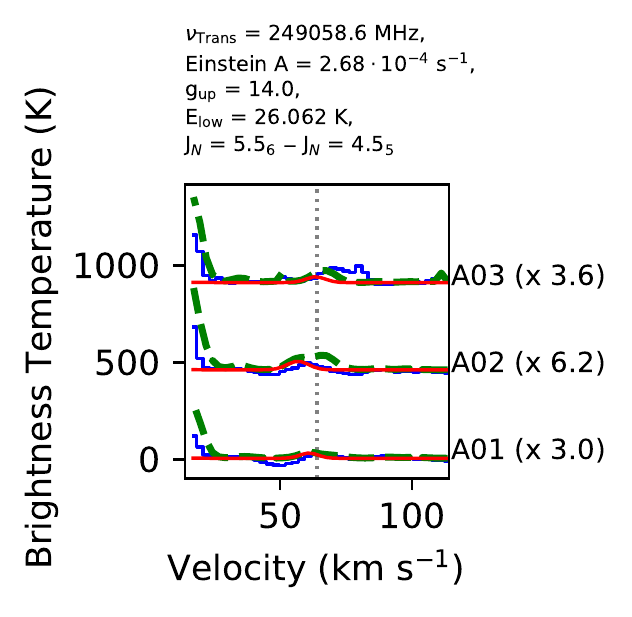}\\
    \end{subfigure}
   \caption{Selected transitions of N$^{34}$S in Sgr~B2(M)}
   \ContinuedFloat
   \label{fig:NS34M}
\end{figure*}
\newpage
\clearpage

%*******************************************************************************
% Figure: SiS;v=0;
\begin{figure*}[!htb]
    \centering
    \begin{subfigure}[t]{0.5\columnwidth}
       \includegraphics[width=1.0\columnwidth]{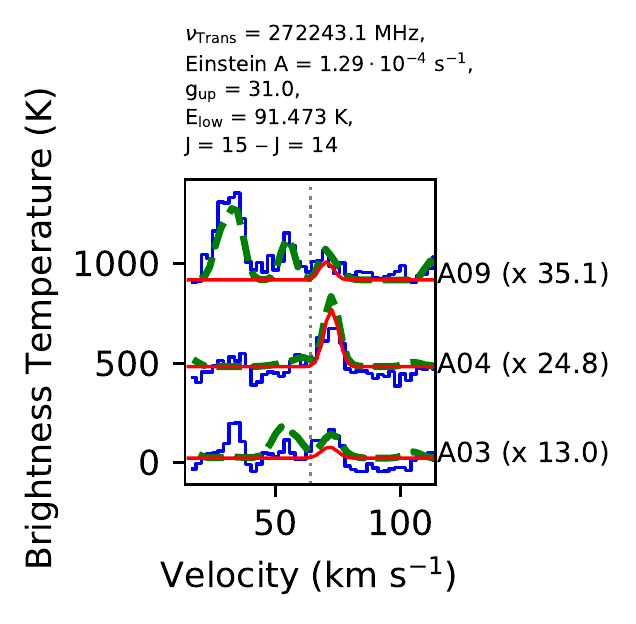}\\
    \end{subfigure}
   \caption{Selected transitions of SiS in Sgr~B2(M)}
   \ContinuedFloat
   \label{fig:SiSM}
\end{figure*}

%*******************************************************************************
% Figure: CCCS;v=0;
\begin{figure*}[!htb]
    \centering
    \begin{subfigure}[t]{1.0\columnwidth}
       \includegraphics[width=0.995\columnwidth]{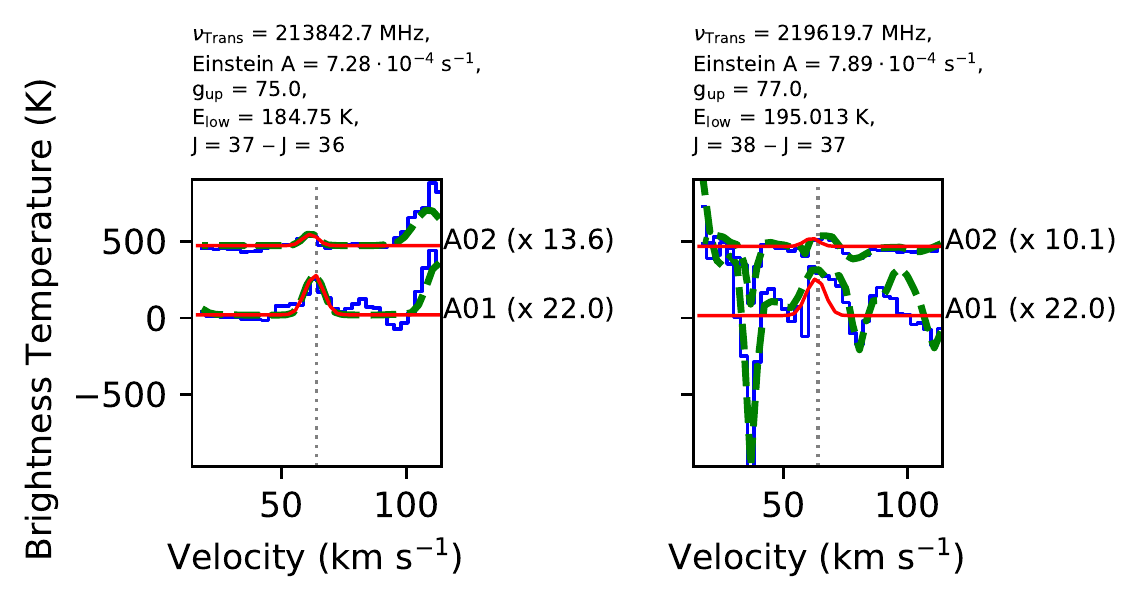}\\
    \end{subfigure}
   \caption{Selected transitions of CCCS in Sgr~B2(M)}
   \ContinuedFloat
   \label{fig:CCCSM}
\end{figure*}
\newpage
\clearpage

%-------------------------------------------------------------------------------
% Carbon and Hydrocarbons

%*******************************************************************************
% Figure: CCH;v=0;
\begin{figure*}[!htb]
    \centering
    \begin{subfigure}[t]{1.0\columnwidth}
       \includegraphics[width=1.0\columnwidth]{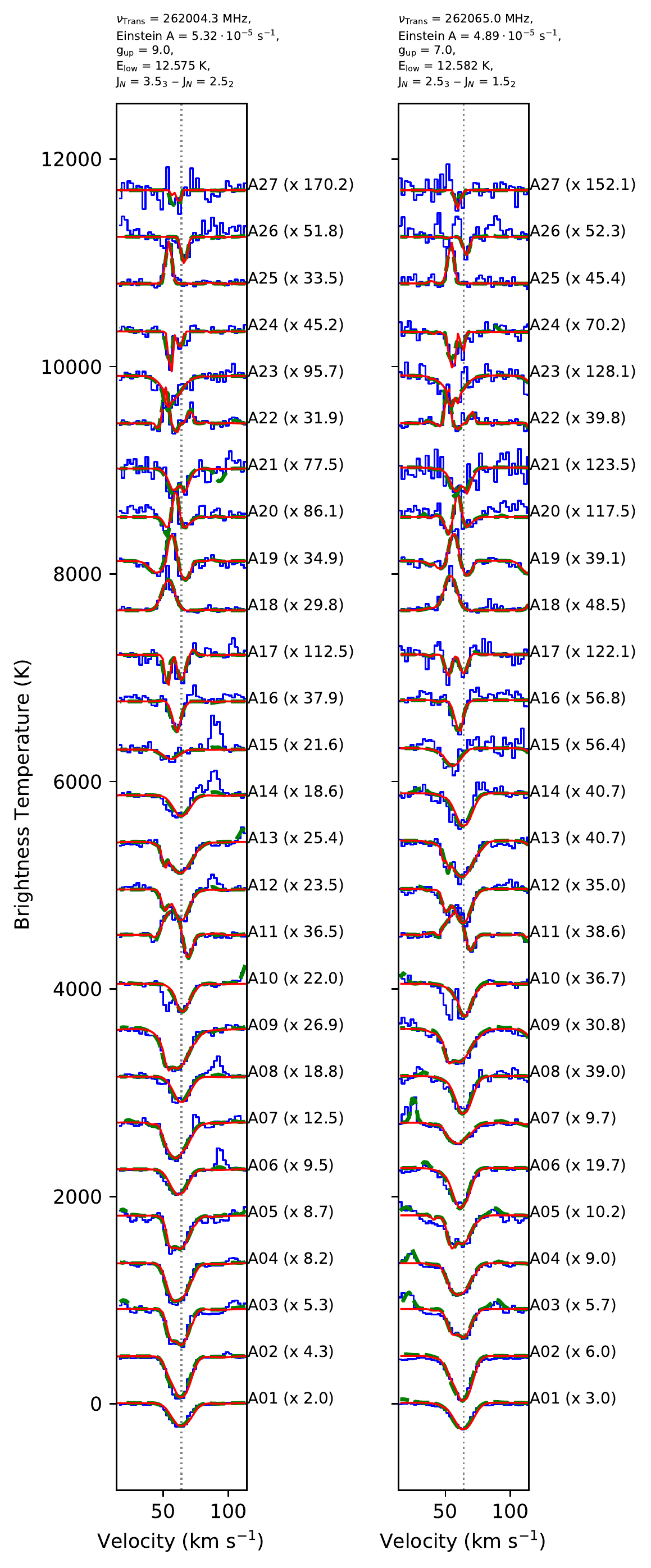}\\
       \caption{Sgr~B2(M)}
       \label{fig:CCHM}
    \end{subfigure}
\quad
    \begin{subfigure}[t]{1.0\columnwidth}
       \includegraphics[width=1.0\columnwidth]{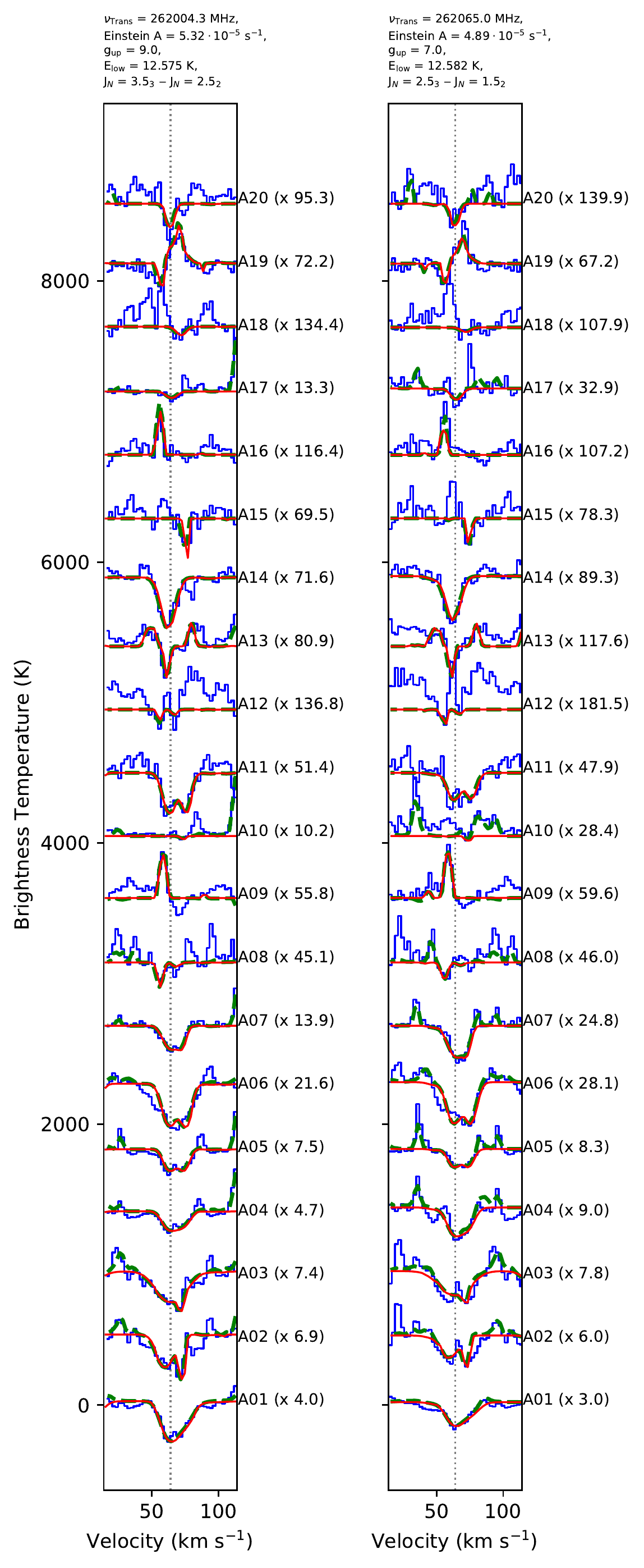}\\
       \caption{Sgr~B2(N)}
       \label{fig:CCHN}
   \end{subfigure}
   \caption{Selected transitions of CCH in Sgr~B2(M) and N.}
   \ContinuedFloat
   \label{fig:CCHMN}
\end{figure*}
\newpage
\clearpage

%*******************************************************************************
% Figure: c-C3H2;v=0;
\begin{figure*}[!htb]
    \centering
    \begin{subfigure}[t]{1.0\columnwidth}
       \includegraphics[width=1.0\columnwidth]{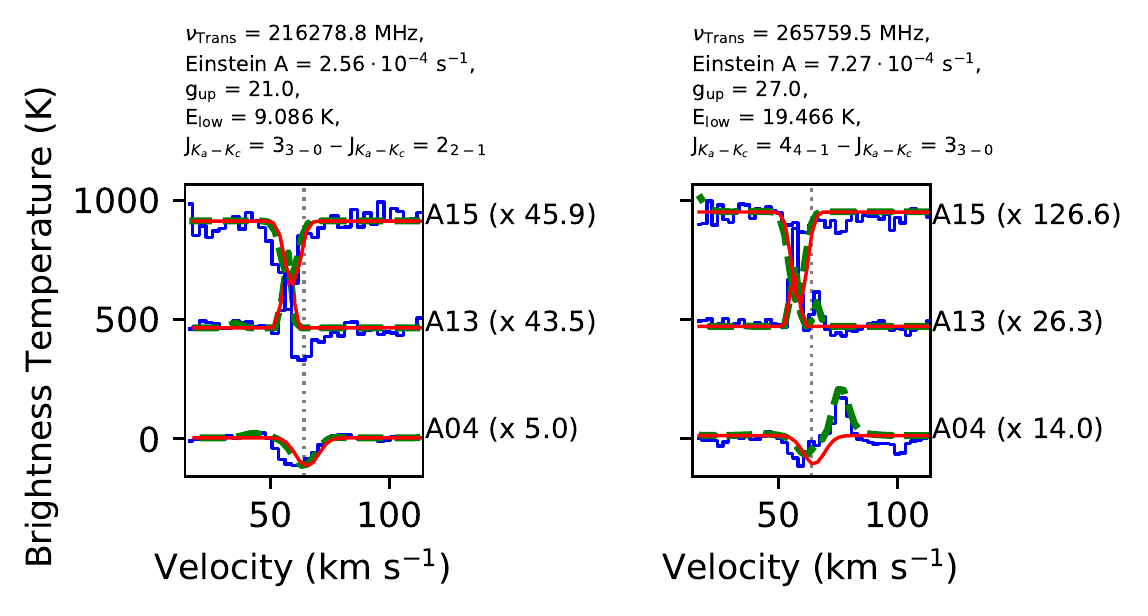}\\
    \end{subfigure}
   \caption{Selected transitions of c-C$_3$H$_2$ in Sgr~B2(M)}
   \ContinuedFloat
   \label{fig:cC3H2M}
\end{figure*}

%-------------------------------------------------------------------------------
% Other Molecules

%*******************************************************************************
% Figure: PH3;v=0;
\begin{figure*}[!htb]
    \centering
    \begin{subfigure}[t]{0.5\columnwidth}
       \includegraphics[width=1.0\columnwidth]{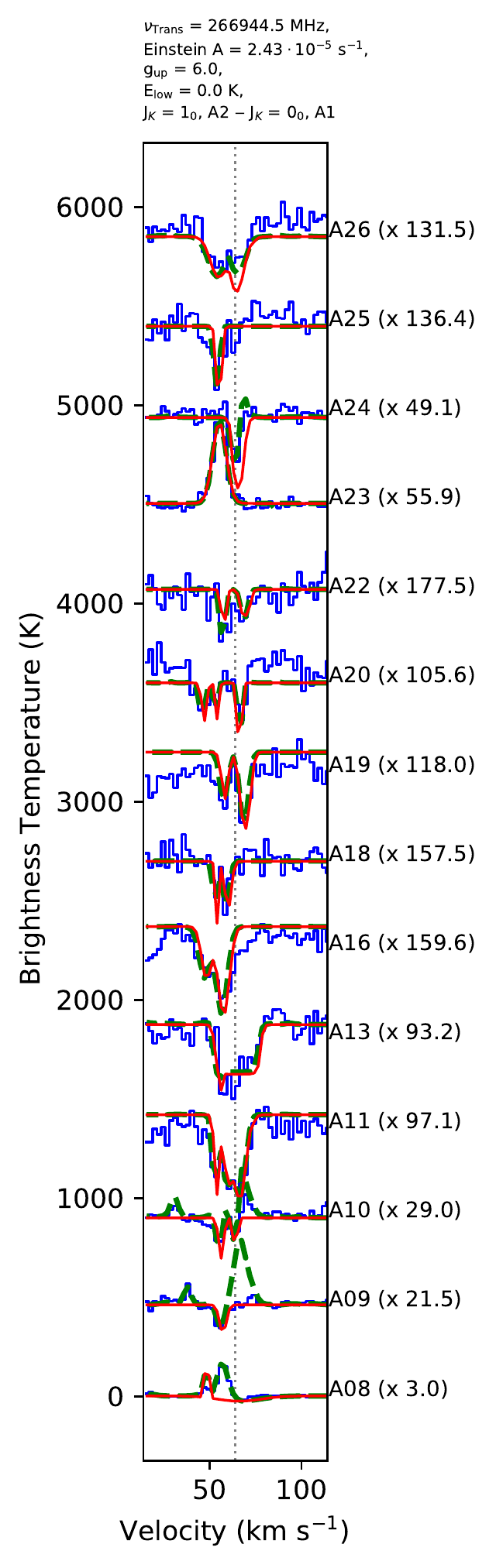}\\
       \caption{Sgr~B2(M)}
       \label{fig:PH3M}
    \end{subfigure}
\quad
    \begin{subfigure}[t]{0.5\columnwidth}
       \includegraphics[width=1.0\columnwidth]{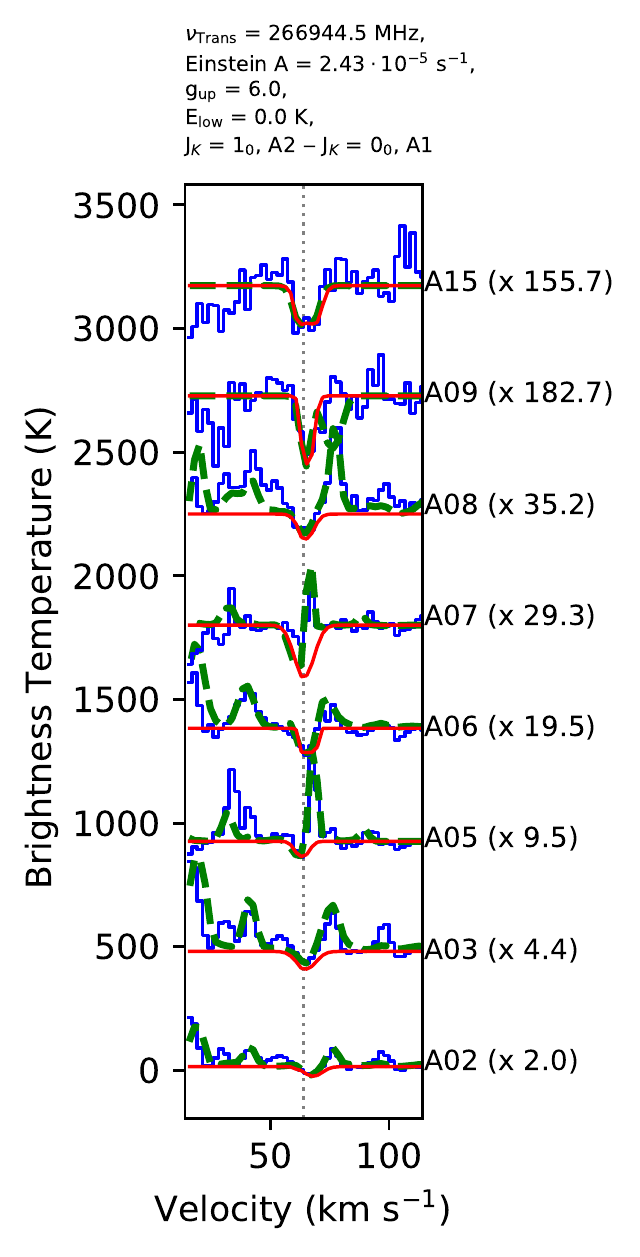}\\
       \caption{Sgr~B2(N)}
       \label{fig:PH3N}
   \end{subfigure}
   \caption{Selected transitions of PH$_3$ in Sgr~B2(M) and N.}
   \ContinuedFloat
   \label{fig:PH3MN}
\end{figure*}
\newpage
\clearpage

%*******************************************************************************
% Figure: PN;v=0;
\begin{figure*}[!htb]
    \centering
    \begin{subfigure}[t]{0.5\columnwidth}
       \includegraphics[width=1.0\columnwidth]{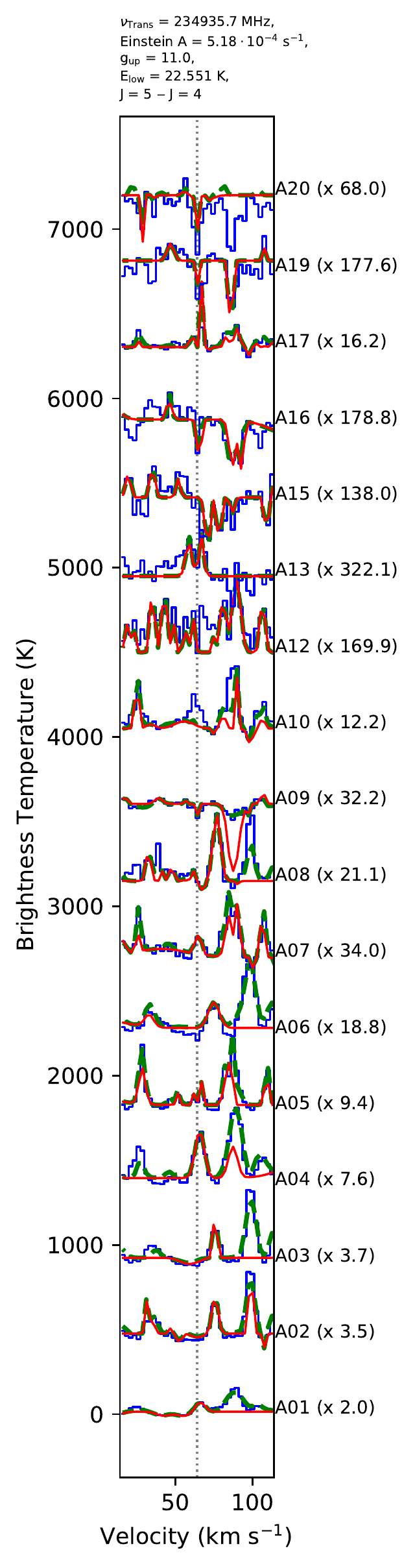}\\
    \end{subfigure}
   \caption{Selected transitions of PN in Sgr~B2(N)}
   \ContinuedFloat
   \label{fig:PNN}
\end{figure*}
\newpage
\clearpage

%===============================================================================
% Excitation temperatures for detected molecule in Sgr~B2(M)
\onecolumn
\section{Excitation temperatures of detected molecule in Sgr~B2(M) and N}\label{app:temp:SgrB2}

%*******************************************************************************
% Figure: averaged excitation temperatures for each detected molecule and source in Sgr~B2(M) (core layer only).
\begin{figure*}[!b]
   \centering
   \includegraphics[scale=0.85]{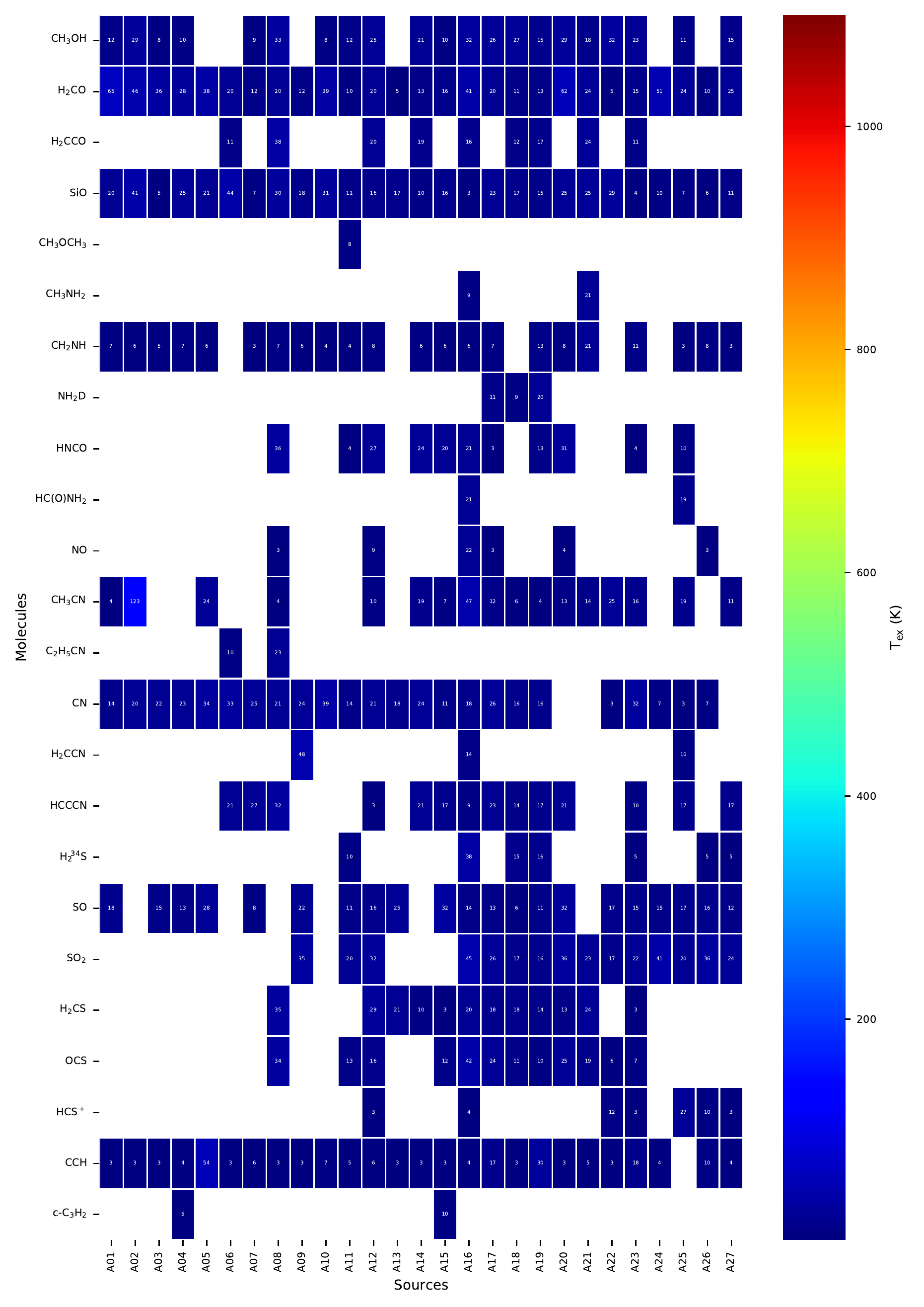}\\
   \caption{Excitation temperatures of envelope layers for each detected molecule and source in Sgr~B2(M).}
   \label{fig:TempEnvSgrB2M}
\end{figure*}
\clearpage
\hfill

%*******************************************************************************
% Figure: averaged excitation temperatures for each detected molecule and source in Sgr~B2(M) (core layer only).
\begin{figure*}[!b]
   \centering
   \includegraphics[scale=0.85]{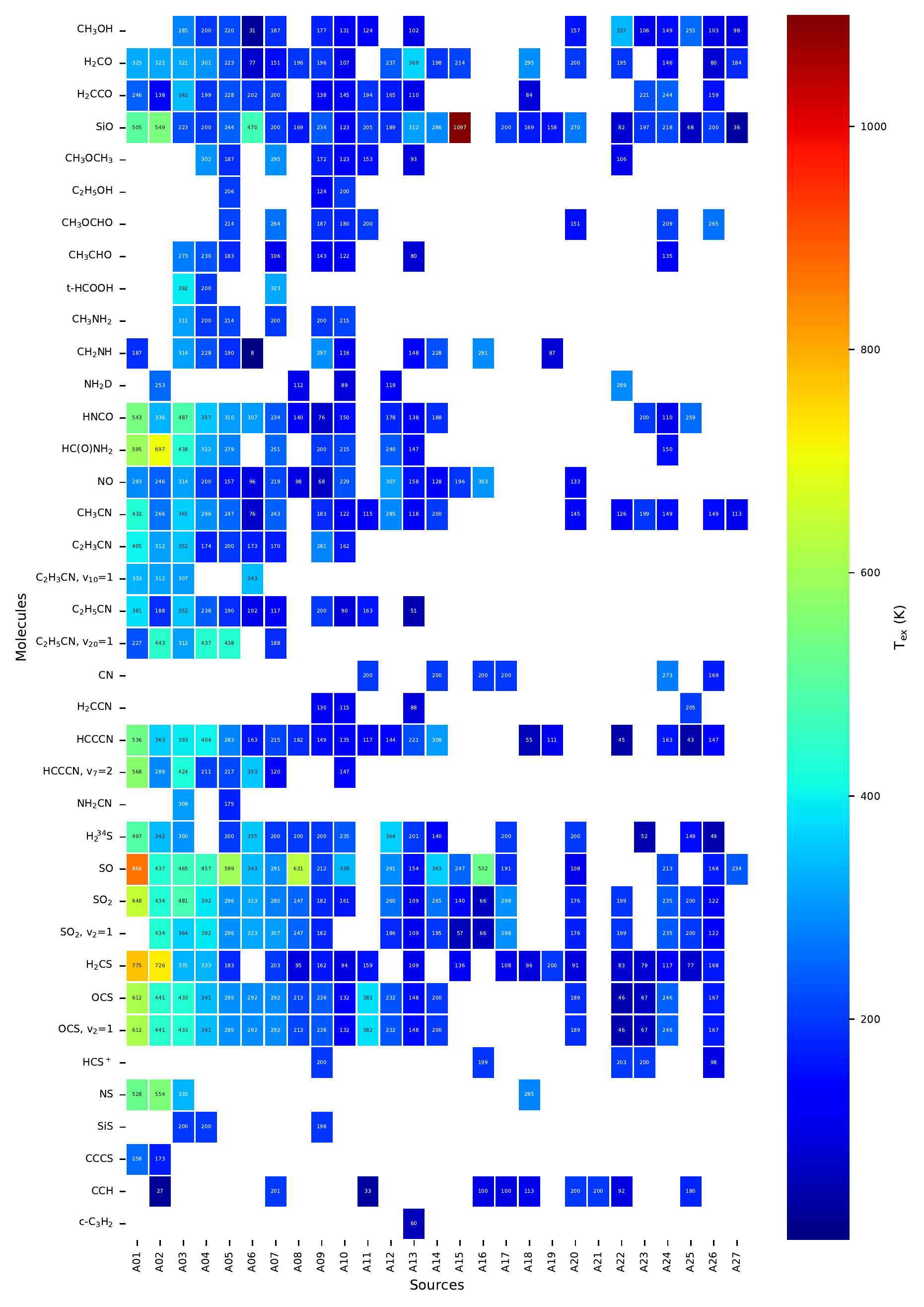}\\
   \caption{Excitation temperatures of core layers for each detected molecule and source in Sgr~B2(M).}
   \label{fig:TempCoreSgrB2M}
\end{figure*}
\clearpage
\hfill

%*******************************************************************************
% Figure: averaged excitation temperatures for each detected molecule and source in Sgr~B2(M) (core layer only).
\begin{figure*}[!b]
   \centering
   \includegraphics[scale=0.85]{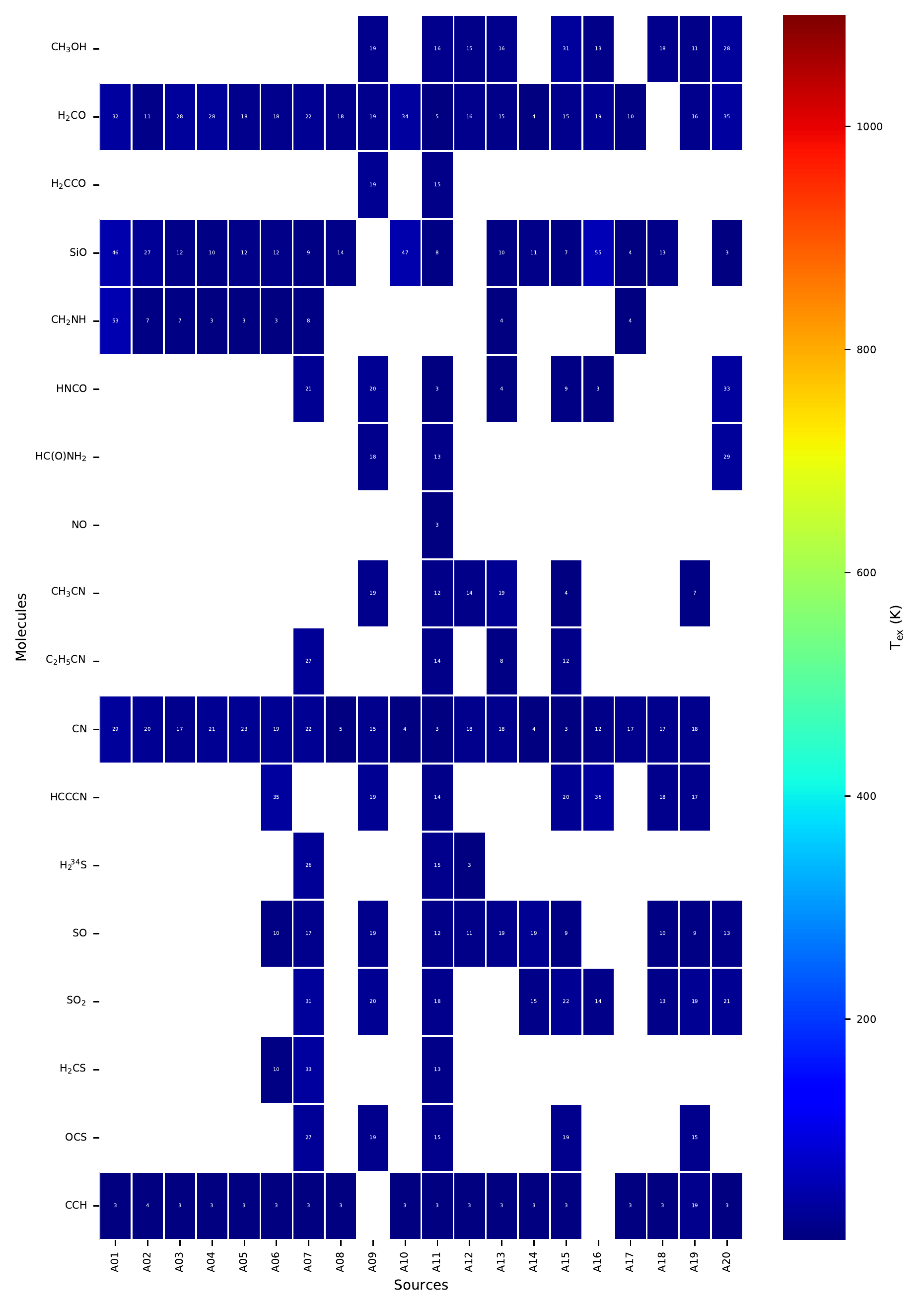}\\
   \caption{Excitation temperatures of envelope layers for each detected molecule and source in Sgr~B2(N).}
   \label{fig:TempEnvSgrB2N}
\end{figure*}
\clearpage
\hfill

%*******************************************************************************
% Figure: averaged excitation temperatures for each detected molecule and source in Sgr~B2(N) (core layer only).
\begin{figure*}[!b]
   \centering
   \includegraphics[scale=0.85]{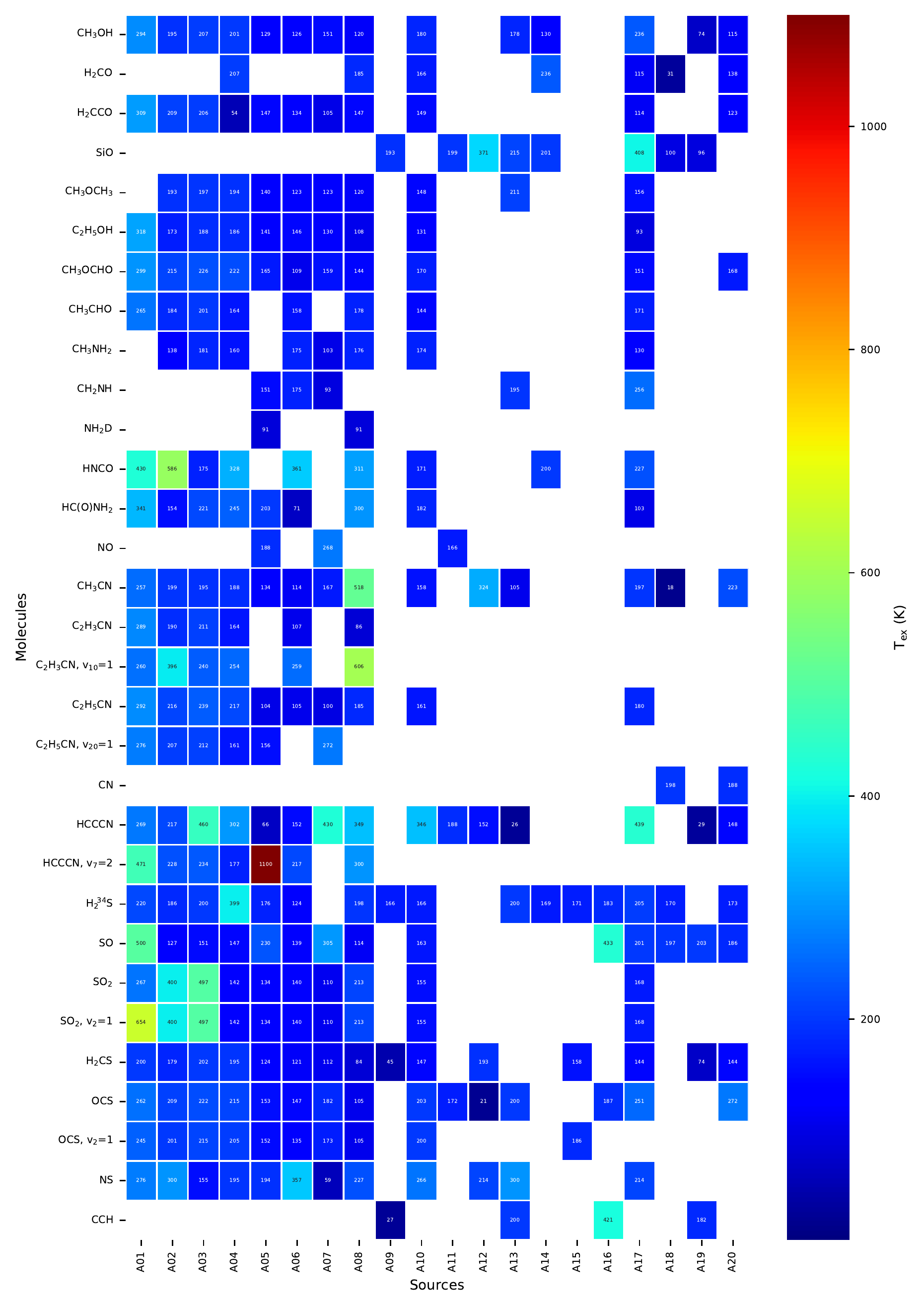}\\
   \caption{Excitation temperatures of core layers for each detected molecule and source in Sgr~B2(N).}
   \label{fig:TempCoreSgrB2N}
\end{figure*}
\newpage
\clearpage

%===============================================================================
% Abundances for detected molecule in Sgr~B2(M)
\onecolumn
\section{Abundances of detected molecule in Sgr~B2(M) and N}\label{app:abund:SgrB2}

%*******************************************************************************
% Figure: averaged abundances for each detected molecule and source in Sgr~B2(M) (env layer only).
\begin{figure*}[!b]
   \centering
   \includegraphics[scale=0.85]{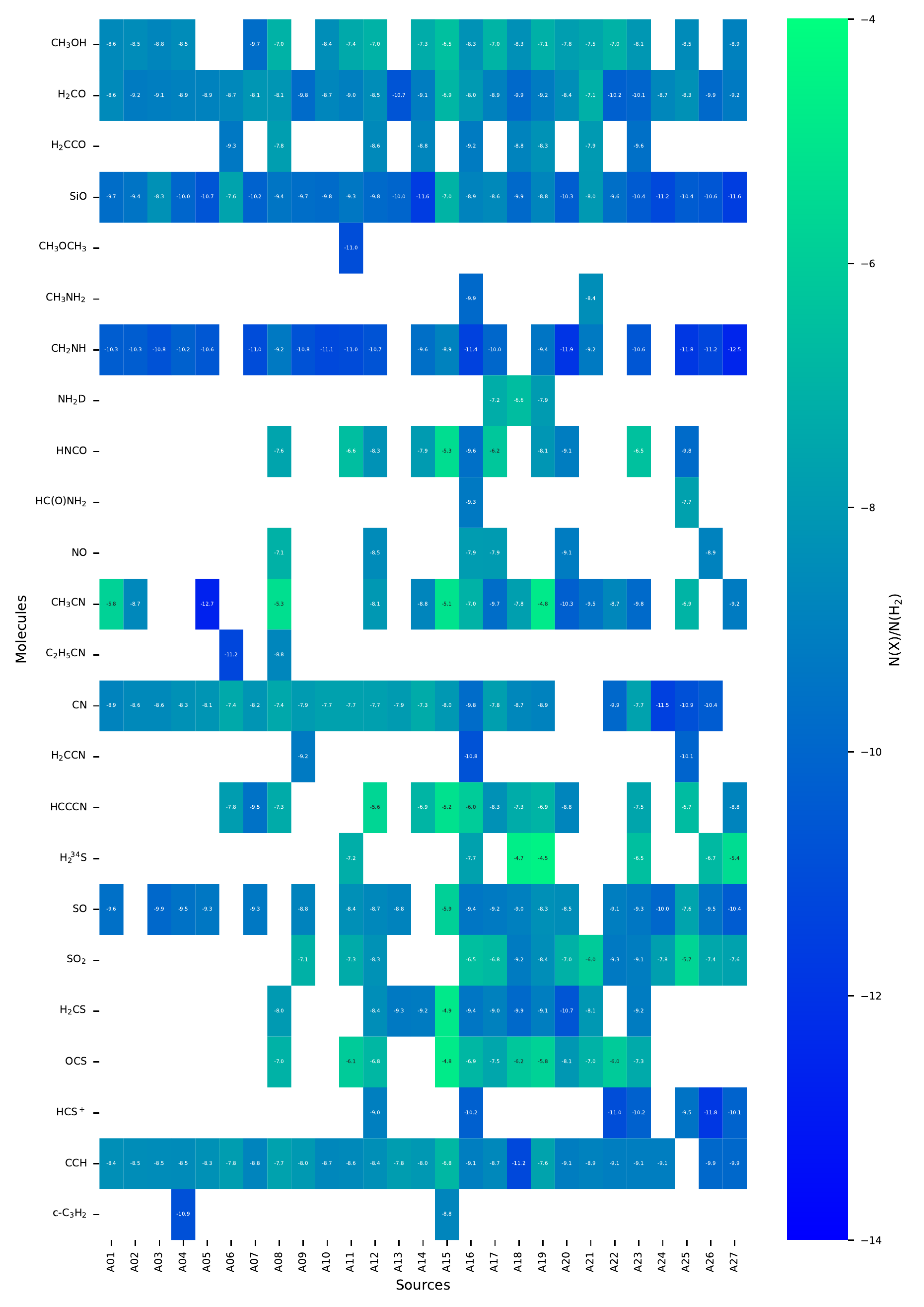}\\
   \caption{Abundances of envelope layers for each detected molecule and source in Sgr~B2(M).}
   \label{fig:AbundEnvSgrB2M}
\end{figure*}
\clearpage
\hfill

%*******************************************************************************
% Figure: averaged abundances for each detected molecule and source in Sgr~B2(M) (core layer only).
\clearpage
\hfill

%*******************************************************************************
% Figure: averaged abundances for each detected molecule and source in Sgr~B2(M) (env layer only).
\clearpage
\hfill

%*******************************************************************************
% Figure: averaged abundances for each detected molecule and source in Sgr~B2(N) (core layer only).
\begin{figure*}[!b]
   \centering
   \includegraphics[scale=0.85]{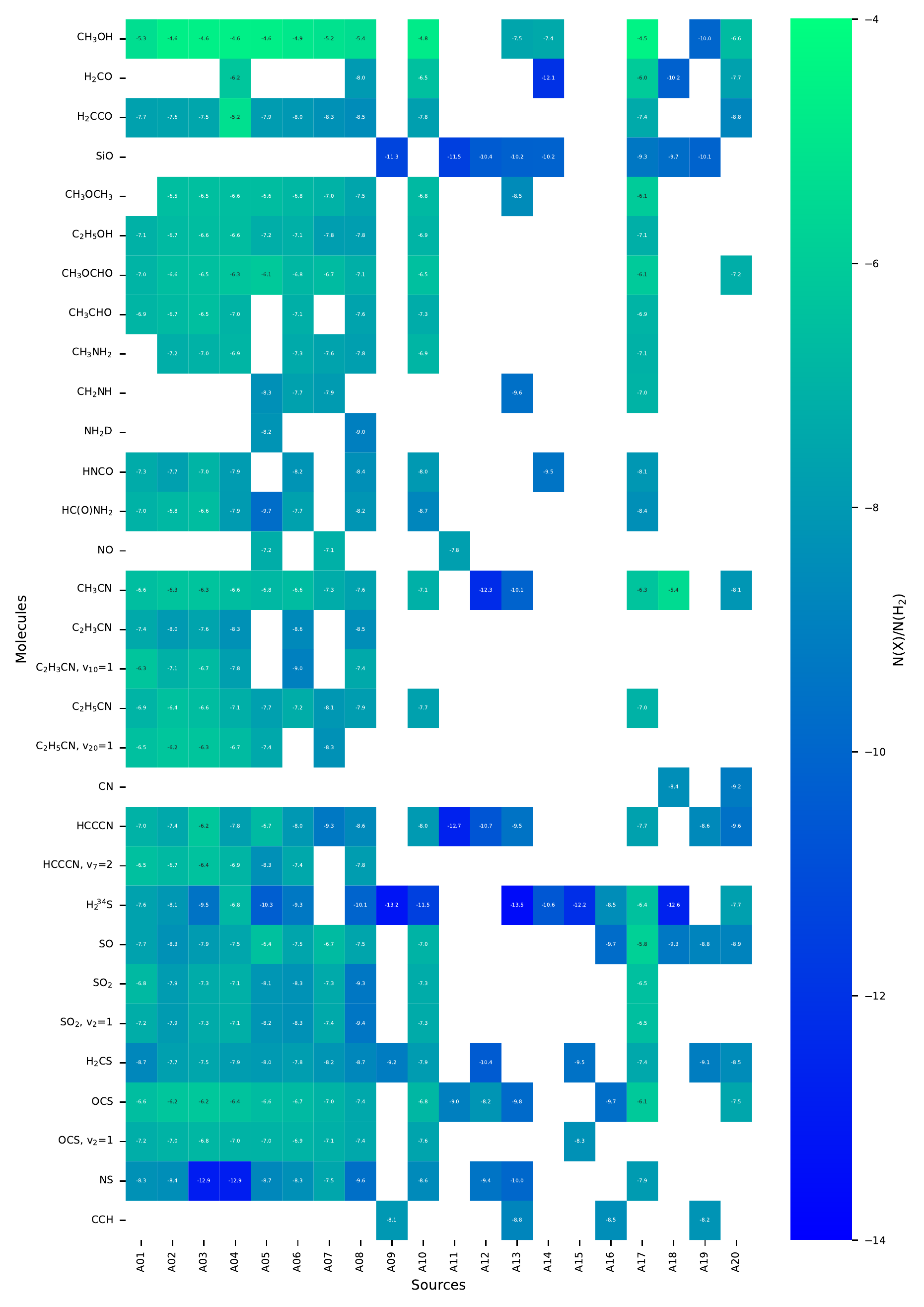}\\
   \caption{Abundances of core layers for each detected molecule and source in Sgr~B2(N).}
   \label{fig:AbundCoreSgrB2N}
\end{figure*}
\newpage
\clearpage

%===============================================================================
% Molfit parameters Sgr~B2(M) and N
\onecolumn
\section{Parameter distributions of best-fit model}\label{app:sec:MolfitSgrB2MN}

% introduction
In the following we visualize the distribution of the derived physical parameters by computing the kernel density estimations (KDE, \citetads{rosenblatt1956}, \citetads{parzen1962}) for the fitted parameters of the different families as described in Sect.~\ref{subsec:PhysParam}. For better visibility, we have normalized the maximum of each KDE to one. In addition, we consider only those components that clearly belong to Sgr~B2, i.e.\ whose source velocity is greater than $22$~km~s$^{-1}$, see Table~\ref{Tab:VelStruc}. The short black vertical lines under each KDE diagram indicate the parameter values of all components and molecules of the respective source. Here, Silverman's rule \citepads{1986desd.book.....S} is used as a scipy implementation \citepads{2020SciPy-NMeth} to calculate the bandwidth of each KDE. Peak-like KDEs indicate that the corresponding molecular family shows only a single value (temperature, line width, or velocity offset) within the corresponding layer.

%:::::::::::::::::::::::::::::::::::::::::::::::::::::::::::::::::::::::::::::::
% T_rot
%:::::::::::::::::::::::::::::::::::::::::::::::::::::::::::::::::::::::::::::::

%*******************************************************************************
% Figure: excitation temperatures of sources in Sgr~B2(M), envelope layer
\begin{figure*}[!htb]
   \centering
   \includegraphics[width=0.95\textwidth]{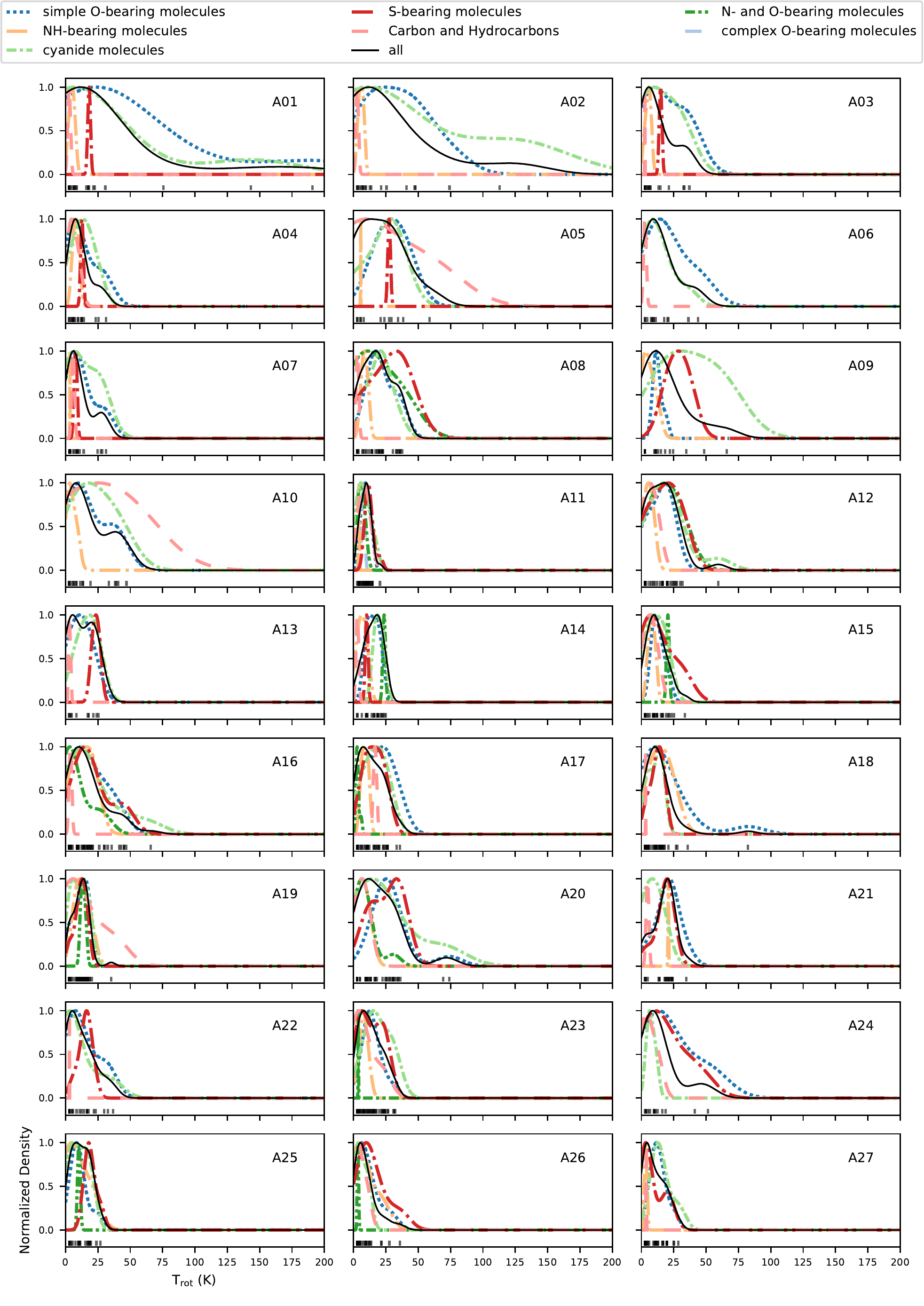}\\
   \caption{Normalized KDE of excitation temperatures T$_{\rm rot}$(K) of components located in the envelope layer for different sources and molecule families in Sgr~B2(M).}
   \label{fig:KDETrotEnvM}
\end{figure*}
\newpage

%*******************************************************************************
% Figure: excitation temperatures of sources in Sgr~B2(M), core layer
\begin{figure*}[!htb]
   \centering
   \includegraphics[width=0.95\textwidth]{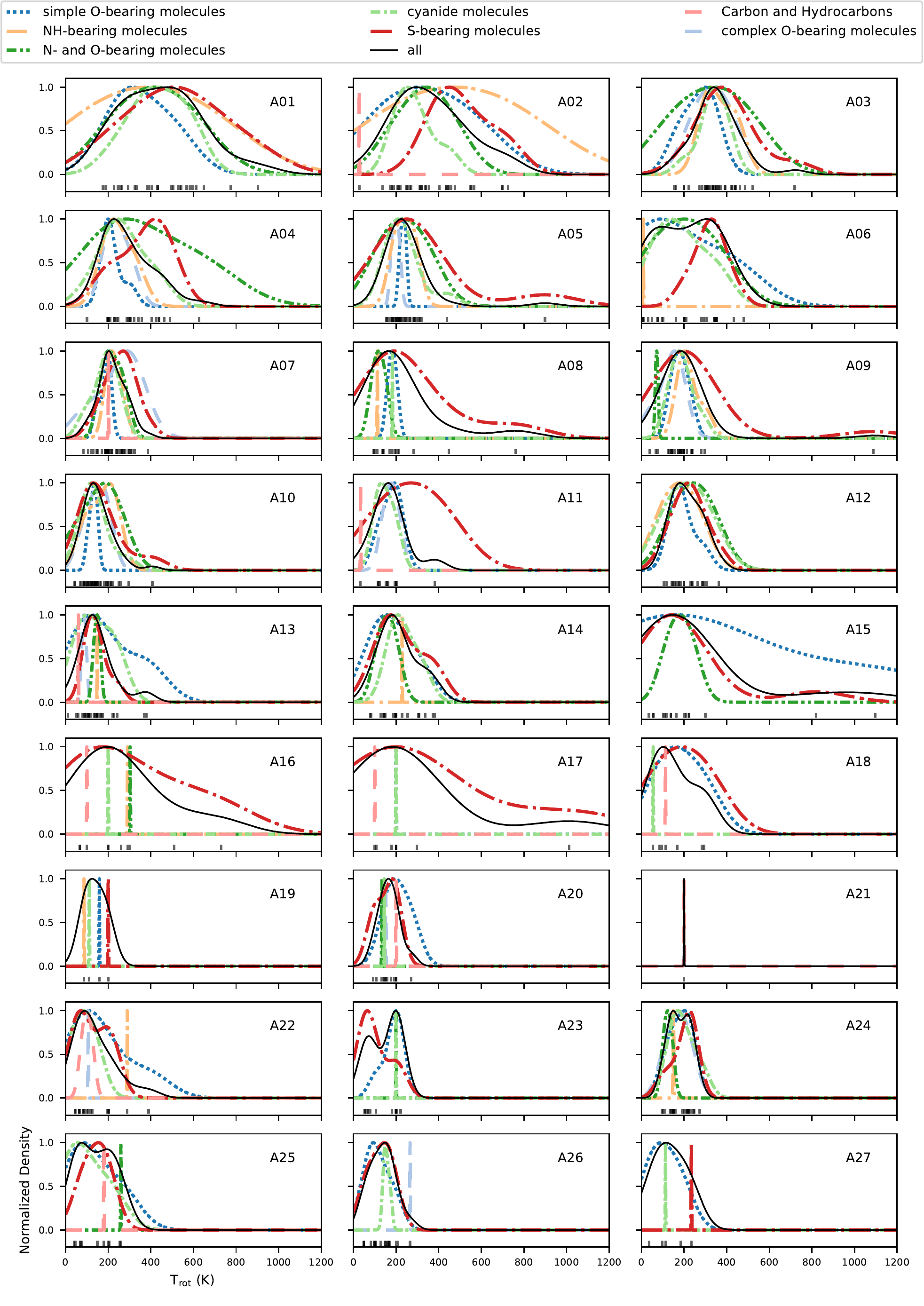}\\
   \caption{Normalized KDE of excitation temperatures T$_{\rm rot}$(K) of components located in the core layer for different sources and molecule families in Sgr~B2(M).}
   \label{fig:KDETrotCoreM}
\end{figure*}
\newpage

%*******************************************************************************
% Figure: excitation temperatures of sources in Sgr~B2(N), envelope layer
\begin{figure*}[!htb]
   \centering
   \includegraphics[width=0.95\textwidth]{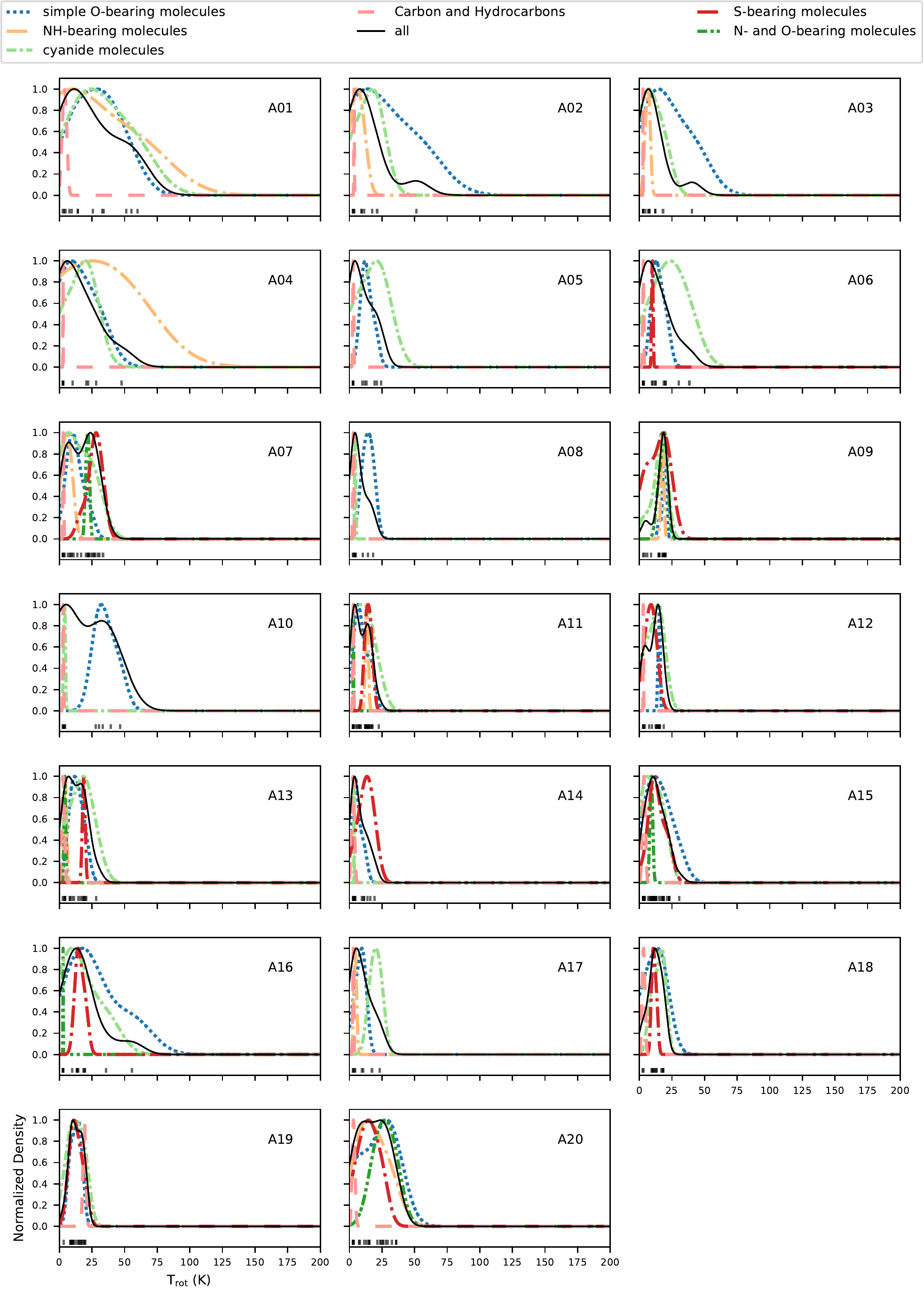}\\
   \caption{Normalized KDE of excitation temperatures T$_{\rm rot}$(K) of components located in the envelope layer for different sources and molecule families in Sgr~B2(N).}
   \label{fig:KDETrotEnvN}
\end{figure*}
\newpage
\clearpage

%*******************************************************************************
% Figure: excitation temperatures of sources in Sgr~B2(N), core layer
\begin{figure*}[!htb]
   \centering
   \includegraphics[width=0.95\textwidth]{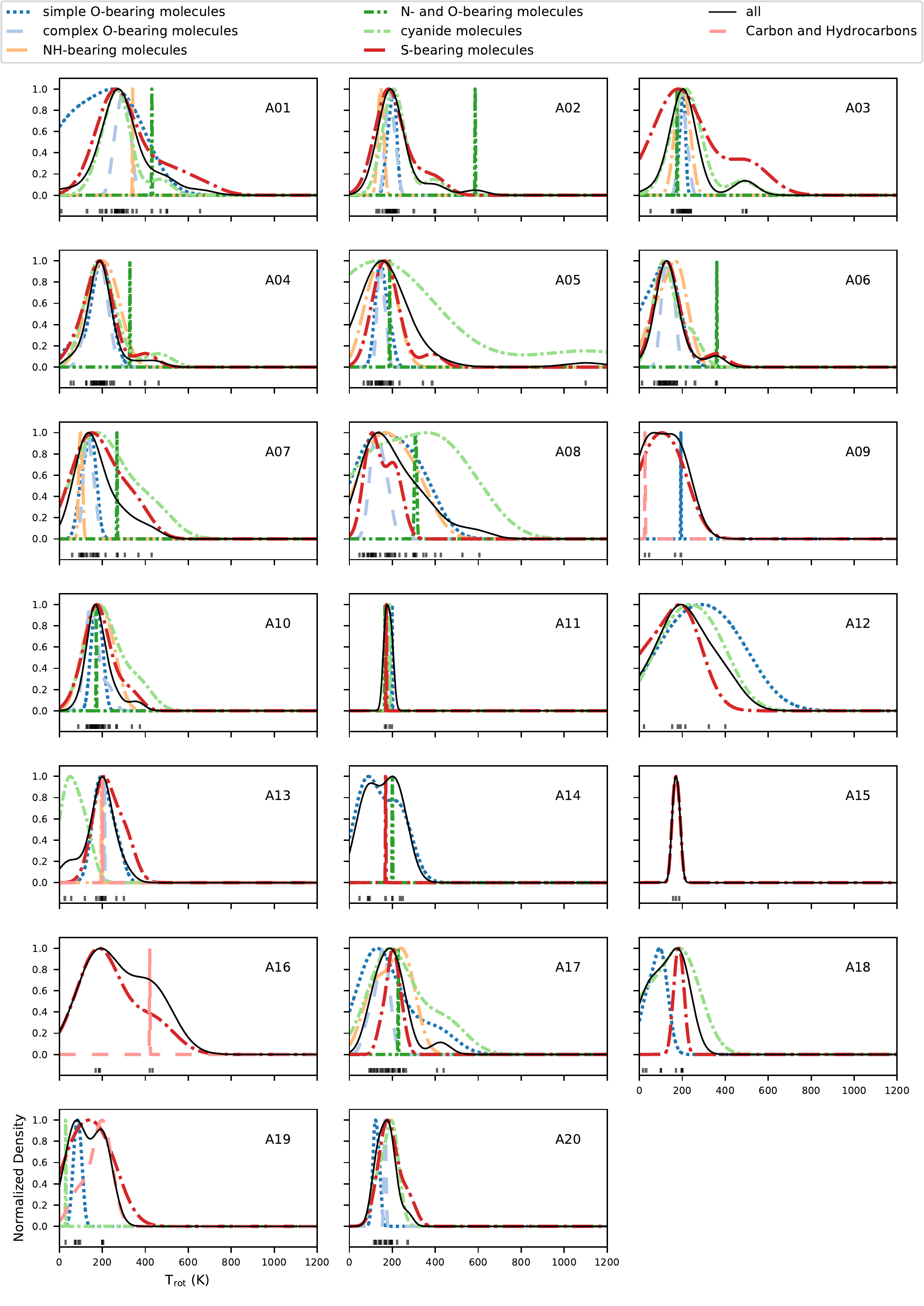}\\
   \caption{Normalized KDE of excitation temperatures T$_{\rm rot}$(K) of components located in the core layer for different sources and molecule families in Sgr~B2(N).}
   \label{fig:KDETrotCoreN}
\end{figure*}
\newpage
\clearpage

%:::::::::::::::::::::::::::::::::::::::::::::::::::::::::::::::::::::::::::::::
% v_width:
%:::::::::::::::::::::::::::::::::::::::::::::::::::::::::::::::::::::::::::::::

%*******************************************************************************
% Figure: line width of sources in Sgr~B2(M), envelope layer
\begin{figure*}[!htb]
   \centering
   \includegraphics[width=0.95\textwidth]{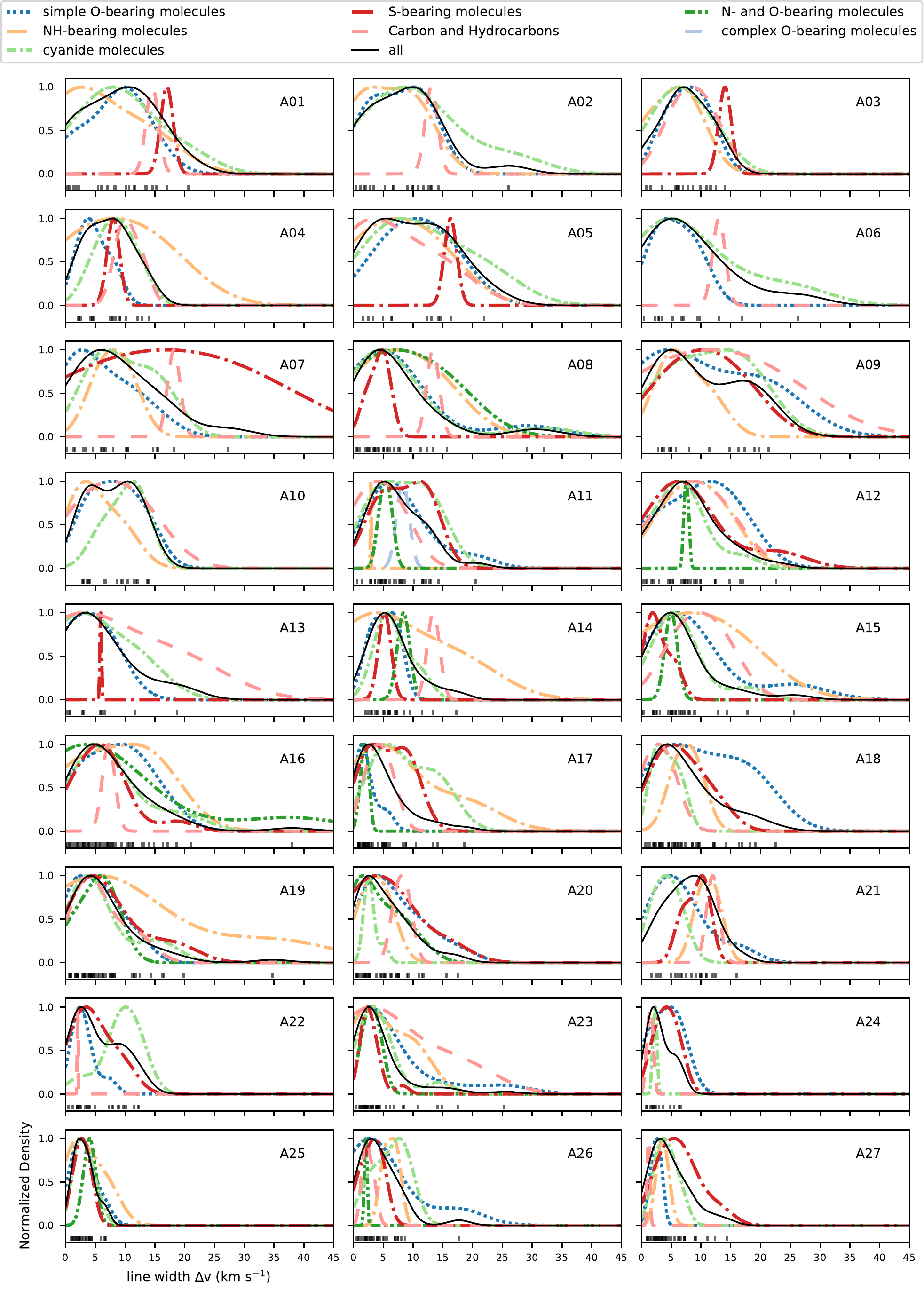}\\
   \caption{Normalized KDE of line widths $\Delta$v~(km s$^{-1}$) of components located in the envelope layer for different sources and molecule families in Sgr~B2(M).}
   \label{fig:KDEwidthEnvSgrB2M}
\end{figure*}
\newpage
\clearpage

%*******************************************************************************
% Figure: line width of sources in Sgr~B2(M), core layer
\begin{figure*}[!htb]
   \centering
   \includegraphics[width=0.95\textwidth]{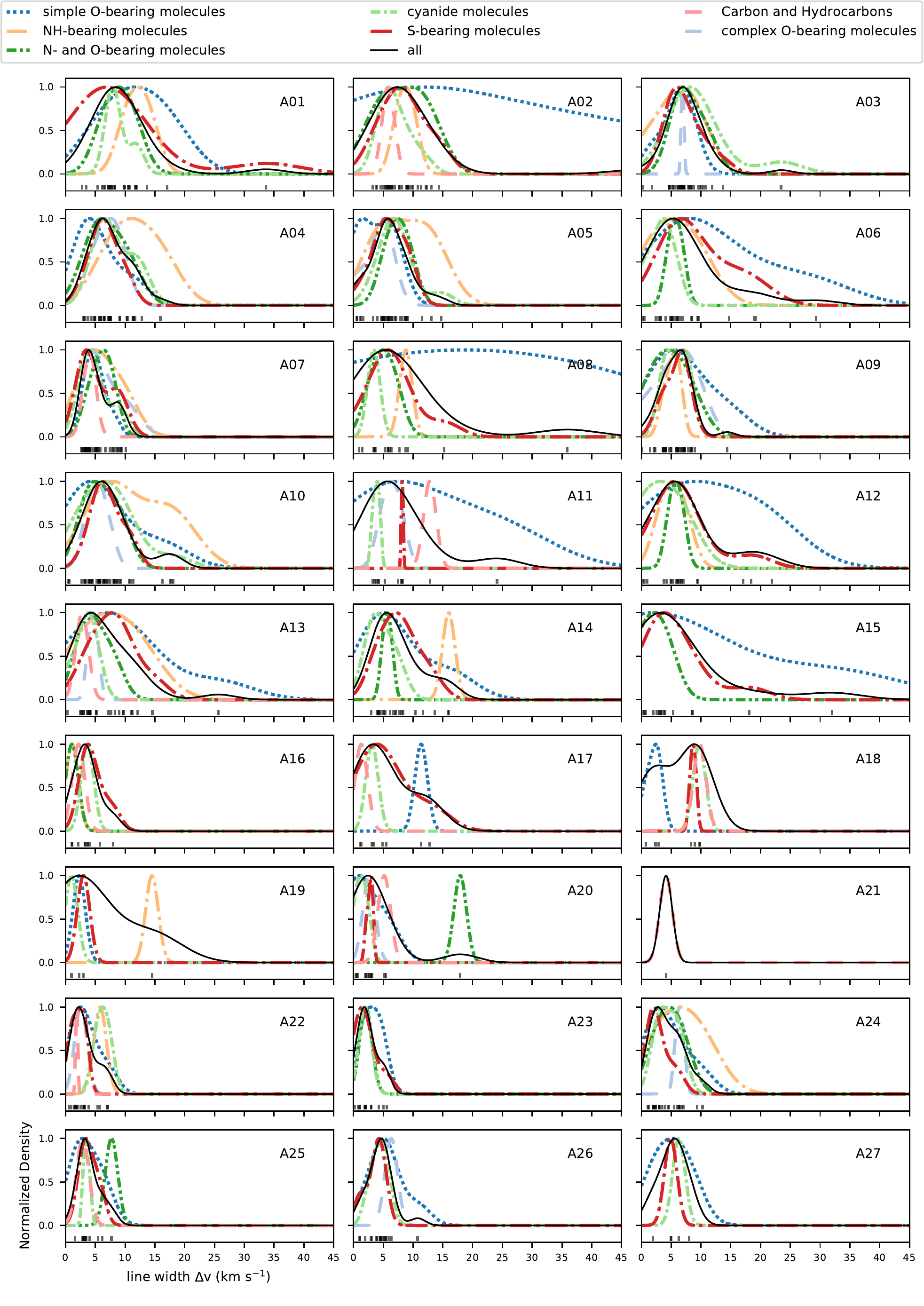}\\
   \caption{Normalized KDE of line widths $\Delta$v~(km s$^{-1}$) of components located in the core layer for different sources and molecule families in Sgr~B2(M).}
   \label{fig:KDEwidthCoreSgrB2M}
\end{figure*}
\newpage
\clearpage

%*******************************************************************************
% Figure: line width of sources in Sgr~B2(N), envelope layer
\begin{figure*}[!htb]
   \centering
   \includegraphics[width=0.95\textwidth]{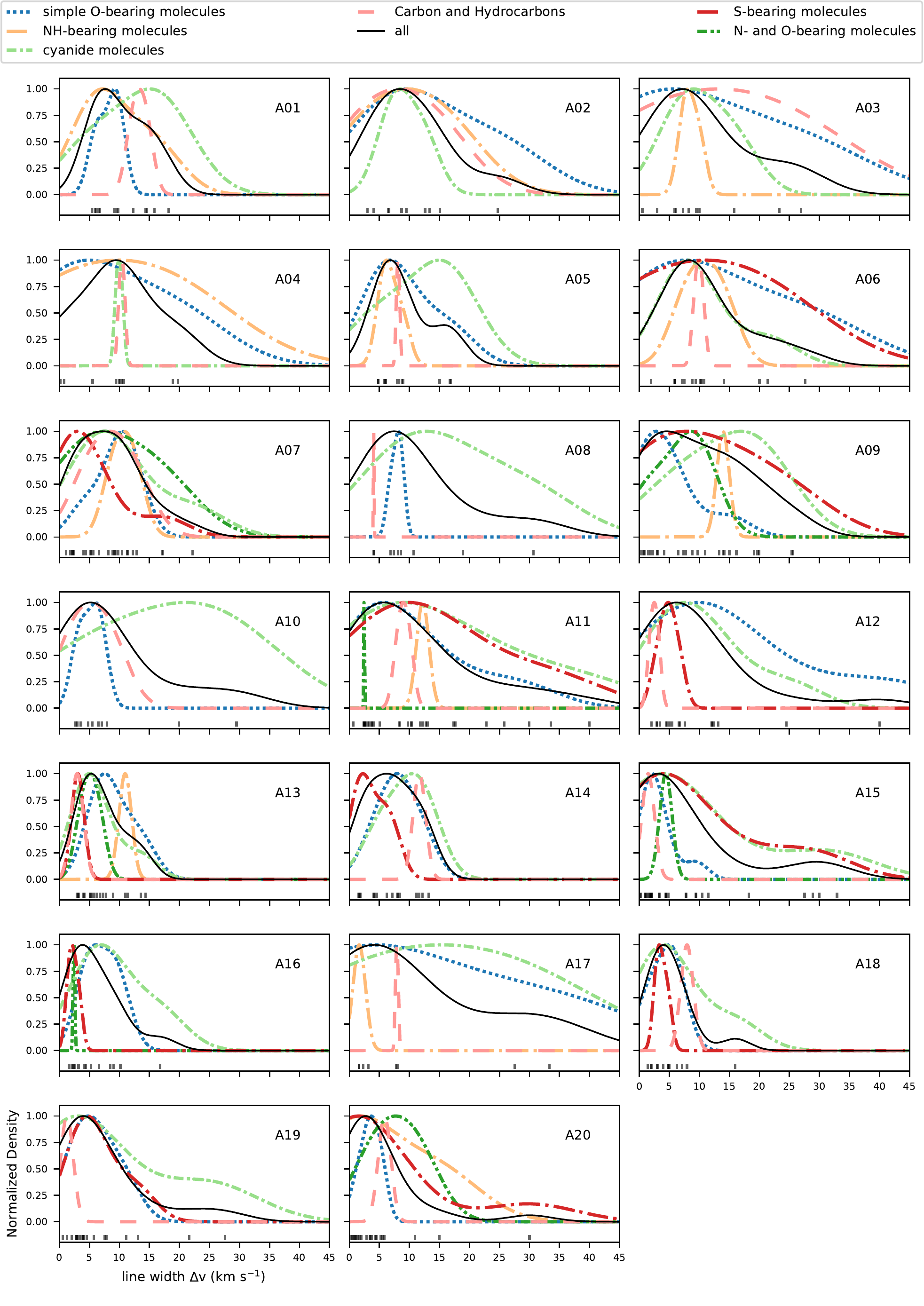}\\
   \caption{Normalized KDE of line widths $\Delta$v~(km s$^{-1}$) of components located in the envelope layer for different sources and molecule families in Sgr~B2(N).}
   \label{fig:KDEwidthEnvSgrB2N}
\end{figure*}
\newpage
\clearpage

%*******************************************************************************
% Figure: line width of sources in Sgr~B2(N), core layer
\begin{figure*}[!htb]
   \centering
   \includegraphics[width=0.95\textwidth]{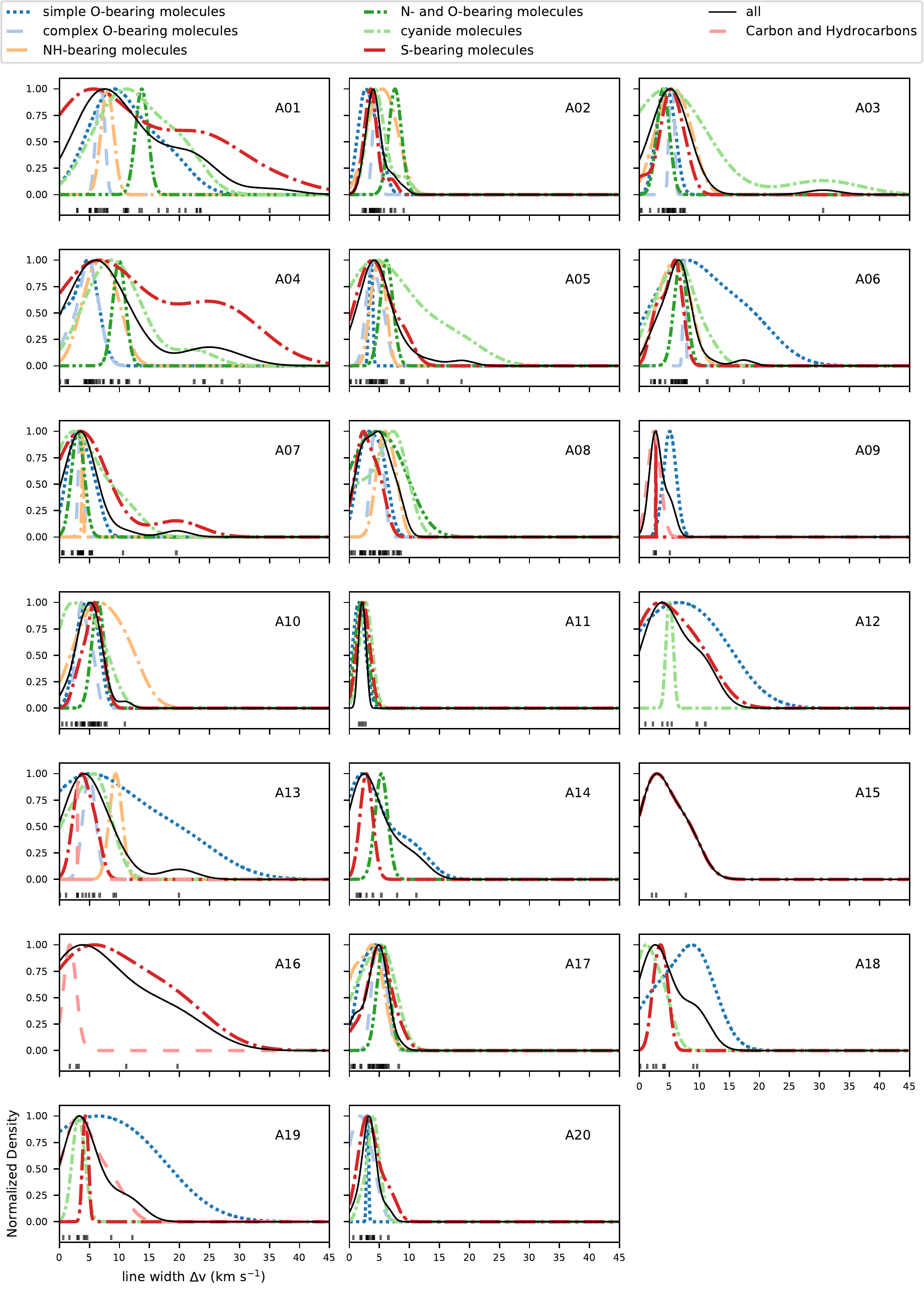}\\
   \caption{Normalized KDE of line widths $\Delta$v~(km s$^{-1}$) of components located in the core layer for different sources and molecule families in Sgr~B2(N).}
   \label{fig:KDEwidthCoreSgrB2N}
\end{figure*}
\newpage
\clearpage

%:::::::::::::::::::::::::::::::::::::::::::::::::::::::::::::::::::::::::::::::
% v_Off:
%:::::::::::::::::::::::::::::::::::::::::::::::::::::::::::::::::::::::::::::::

%*******************************************************************************
% Figure: velocity offsets of sources in Sgr~B2(M), envelope layer
\begin{figure*}[!htb]
   \centering
   \includegraphics[width=0.95\textwidth]{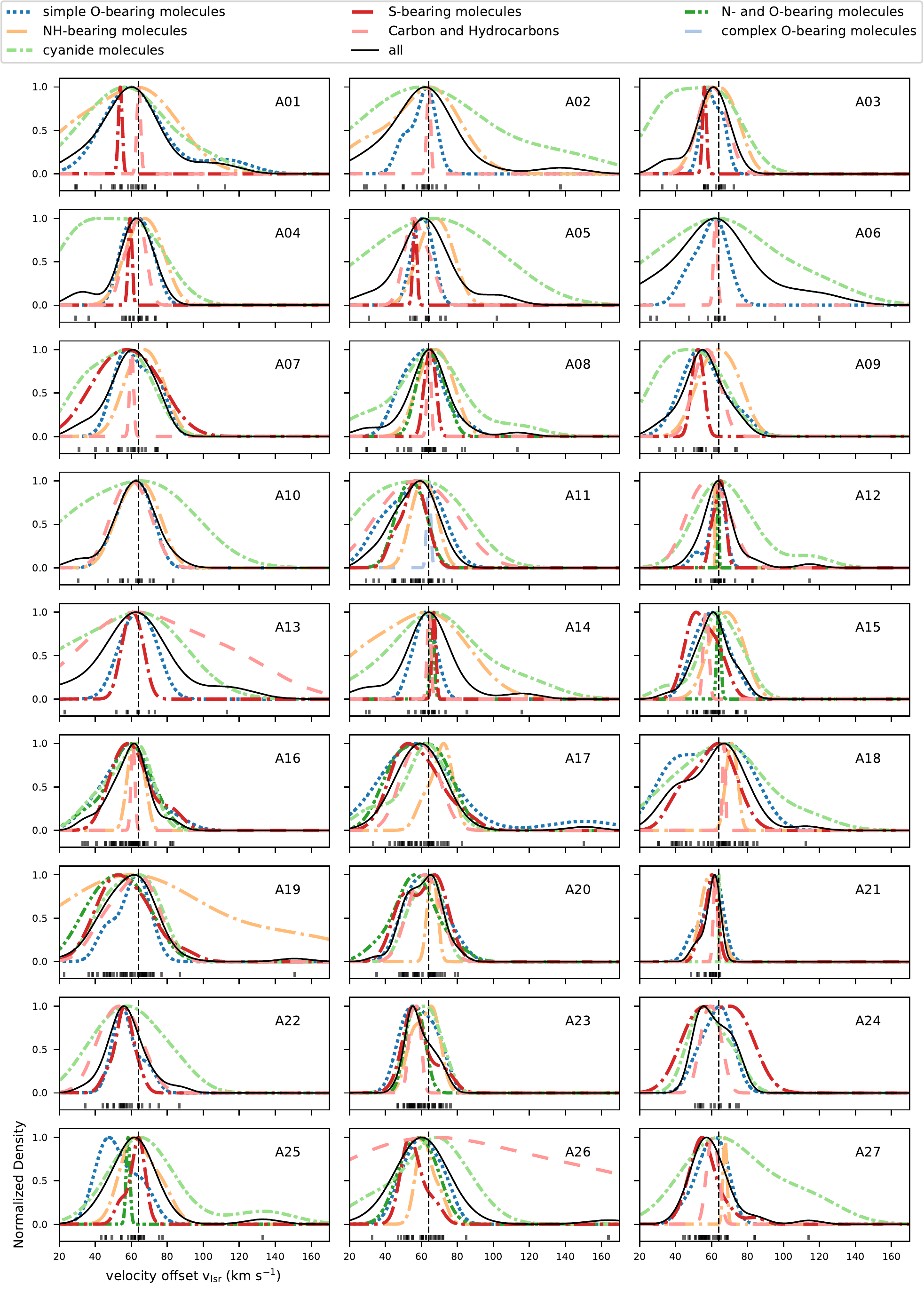}\\
   \caption{Normalized KDE of velocity offsets v$_{\rm lsr}$(km s$^{-1}$) of components located in the envelope layer for different sources and molecule families in Sgr~B2(M).}
   \label{fig:KDEvoffEnvSgrB2M}
\end{figure*}
\newpage
\clearpage

%*******************************************************************************
% Figure: velocity offsets of sources in Sgr~B2(M), core layer
\begin{figure*}[!htb]
   \centering
   \includegraphics[width=0.95\textwidth]{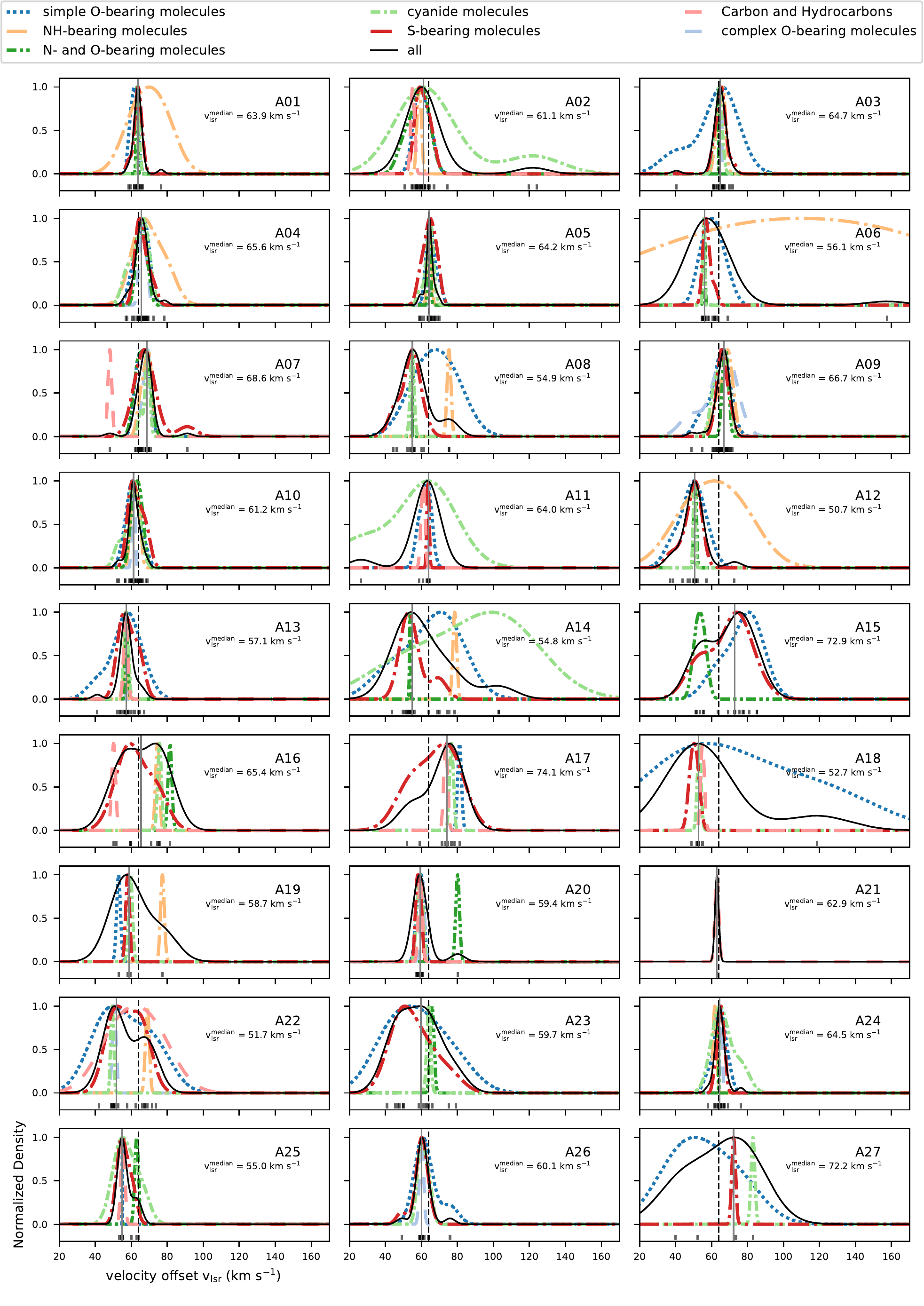}\\
   \caption{Normalized KDE of velocity offsets v$_{\rm lsr}$(km s$^{-1}$) of components located in the core layer for different sources and molecule families in Sgr~B2(M). The gray vertical solid lines indicate the median velocity, respectively.}
   \label{fig:KDEvoffCoreSgrB2M}
\end{figure*}
\newpage
\clearpage

%*******************************************************************************
% Figure: velocity offsets of sources in Sgr~B2(N), envelope layer
\begin{figure*}[!htb]
   \centering
   \includegraphics[width=0.95\textwidth]{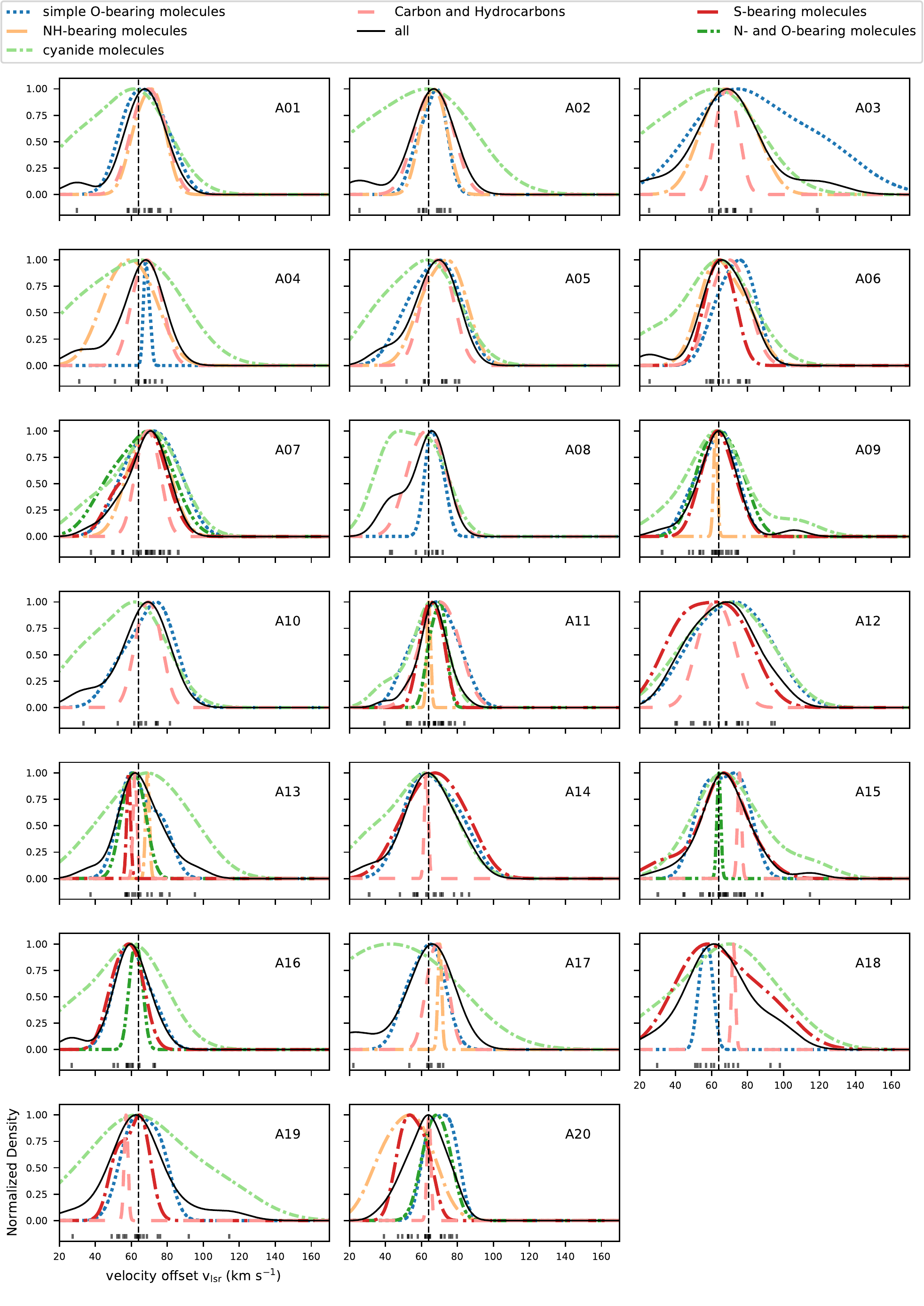}\\
   \caption{Normalized KDE of velocity offsets v$_{\rm lsr}$(km s$^{-1}$) of components located in the envelope layer for different sources and molecule families in Sgr~B2(N).}
   \label{fig:KDEvoffEnvSgrB2N}
\end{figure*}
\newpage
\clearpage

%*******************************************************************************
% Figure: velocity offsets of sources in Sgr~B2(N), core layer
\begin{figure*}[!htb]
   \centering
   \includegraphics[width=0.95\textwidth]{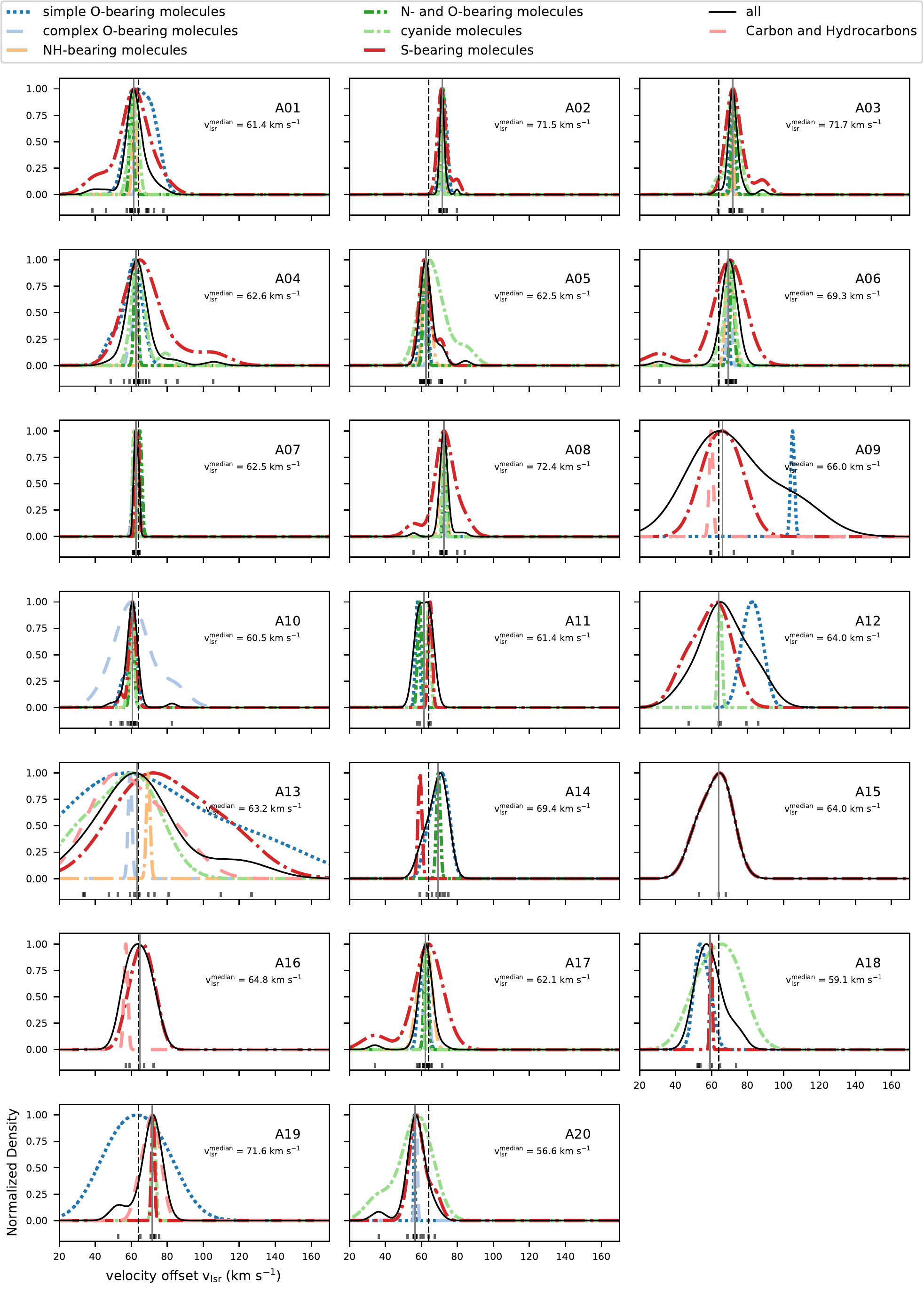}\\
   \caption{Normalized KDE of velocity offsets v$_{\rm lsr}$(km s$^{-1}$) of components located in the core layer for different sources and molecule families in Sgr~B2(N). The gray vertical solid lines indicate the median velocity, respectively.}
   \label{fig:KDEvoffCoreSgrB2N}
\end{figure*}
\newpage
\clearpage

%===============================================================================
% Kendall rank correlation coefficient
\onecolumn
\section{Kendall rank correlation coefficient}\label{app:sec:Kendall}

%*******************************************************************************
% Figure: Kendall rank correlation coefficient of abundances in Sgr B2(M)
\begin{figure*}[!b]
   \centering
   \includegraphics[width=0.85\textwidth]{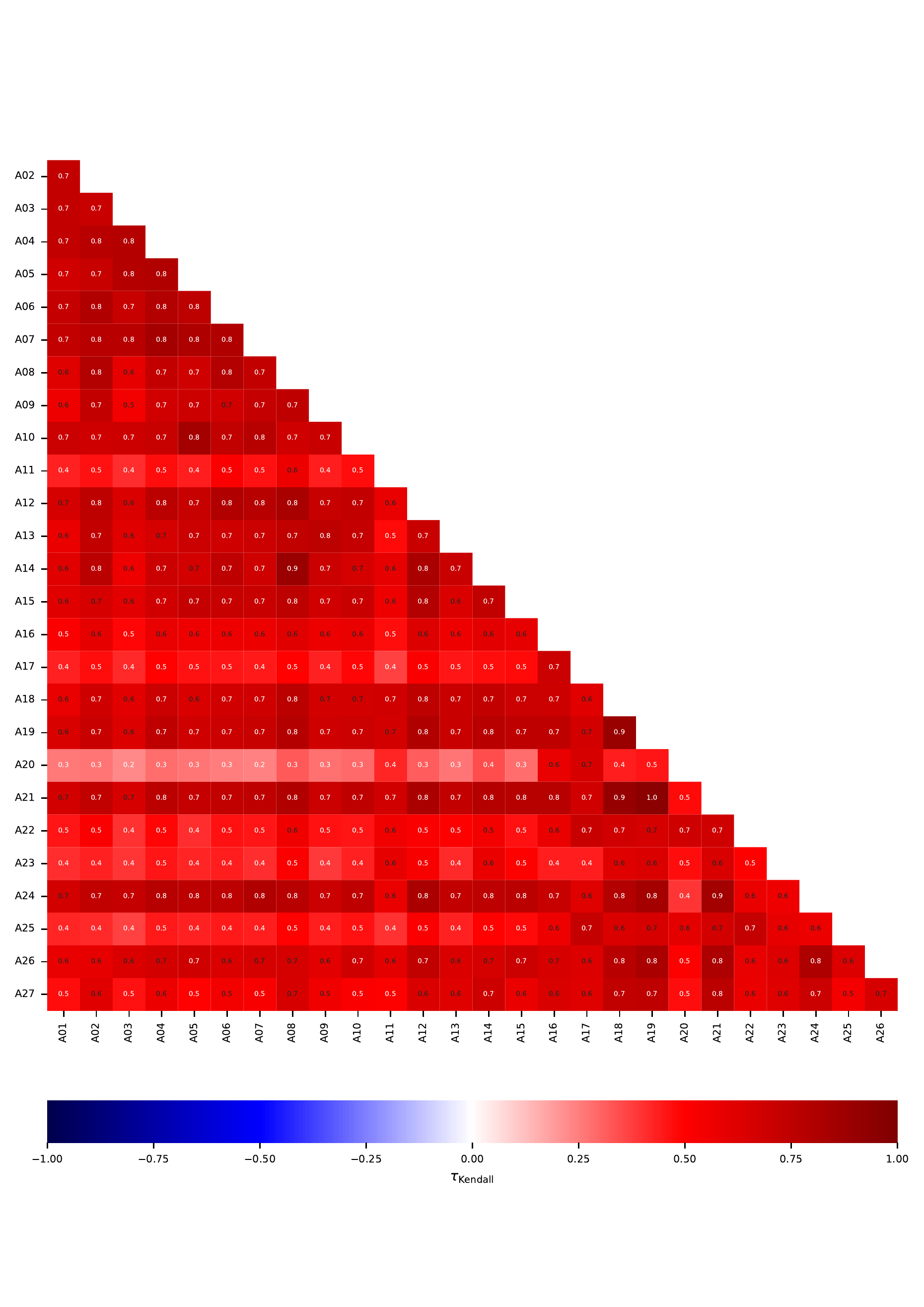}\\
   \caption{Kendall rank correlation coefficient of abundances for sources in Sgr~B2(M). The number in each square indicates the Kendall coefficient for the corresponding pair of sources.}
   \label{fig:KendallSgrB2M}
\end{figure*}

%*******************************************************************************
% Figure: Kendall coefficient of abundances in Sgr B2(N)
\begin{figure*}[!b]
   \centering
   \includegraphics[width=0.85\textwidth]{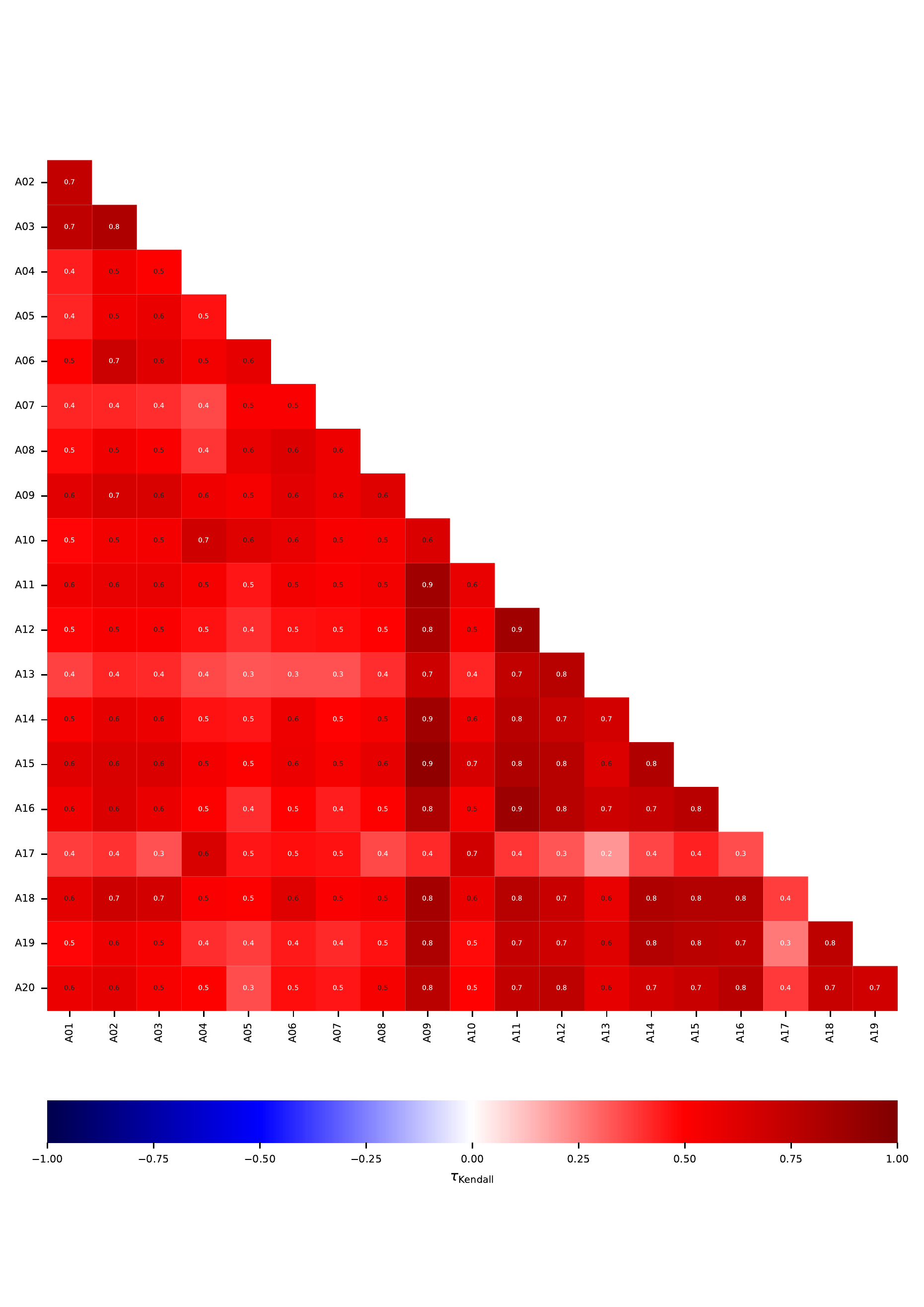}\\
   \caption{Kendall rank correlation coefficient of abundances for sources in Sgr~B2(N). The number in each square indicates the Kendall coefficient for the corresponding pair of sources.}
   \label{fig:KendallSgrB2N}
\end{figure*}

%===============================================================================
% final results
\twocolumn
\section{Final results}\label{app:sec:Results}

% subsection: Source in Sgr~B2(M)
\subsection{Source in Sgr~B2(M)}\label{app:subsec:ResultsM}

\begin{itemize}
% Sgr B2(M), A01:
% PCA:
%   CM1: A01, A03, A04, A05, A06, A07, A10
% corresponding evolution phases:
%         IV, III,  IV,  II, III,  IV, III
\item \textbf{A01}: The envelope around source A01 in Sgr~B2(M) is dominated by contributions from CH$_3$CN, CCH, H$_2$CO, CH$_3$OH, and CN. For the corresponding core spectrum, NO provides the largest contribution, followed by SO$_2$, SO, HCCCN,v$_7$=2, H$_2$CO. For the core, we obtain slightly different mode\footnote{The position of the x-axis, where a kernel density estimate (KDE) has its maximum is commonly referred to as the "mode" of the distribution.} temperatures for the different families of molecules, see Fig.~\ref{fig:KDETrotCoreM}, with simple O-bearing molecules having the lowest temperatures and sulfur-bearing molecules the highest, which also fits well with the results of the PCA, according to which this source is assigned to the cluster (CM1) with the highest temperatures of the sulfur-bearing molecules, see Fig.~\ref{fig:PCAMeanTSgrB2M}. In addition, NH-bearing molecules show a much larger mode line width than the other families, see Fig.~\ref{fig:KDEwidthCoreSgrB2M}, which together with a significantly higher velocity, see Fig.~\ref{fig:KDEvoffCoreSgrB2M}, could indicate a different origin, such as a filament. Since this source contains maser as well as outflows and \hii~regions, this source is considered to be more evolved (phase~IV), see Sect.~\ref{subsec:EvolSeq}, which fits well with the classification based on chemical clocks containing sulfur-bearing molecules, see Table~\ref{Tab:ChemClocks}.

% Sgr B2(M), A02:
% PCA:
%   CM4: A02, A08, A09, A12, A13, A14, A15
% corresponding evolution phases:
%         IV,  IV,   I,   I,   I,   ?,   V
\item \textbf{A02}: In source A02, NO once again makes the largest contribution to the emission, followed by SO$_2$, and its vibrationally excited state SO$_2$,v$_2$=1, as well as HCCCN,v$_7$=2, and H$_2 \! ^{34}$S. Similar to the neighboring source A01, the envelope is again dominated by contributions from CCH, CH$_3$OH, CN, CH$_3$CN and H$_2$CO. For the core spectrum, the molecular families have different mode temperatures, line widths and velocities, respectively, indicating together with the high number of maser that the source contains different protostars and~/~or filaments. This fact is supported by the analysis of the chemical clocks, according to which A02 is classified as both a young and an evolved source. Nevertheless, the evolutionary phase analysis classifies the source as evolved (phase~IV), which also fits well with the high contributions of some vibrationally excited molecules. Furthermore, PCA assigns the source to a cluster (CM4) that has the highest mean abundances of SO$_2$ and its vibrational excited state  SO$_2$,v$_2$=1, see Fig.~\ref{fig:PCAMeanXSgrB2M}.

% Sgr B2(M), A03:
% PCA:
%   CM1: A01, A03, A04, A05, A06, A07, A10
% corresponding evolution phases:
%         IV, III,  IV,  II, III,  IV, III
\item \textbf{A03}: For source A03, SiO, CCH, CN, CH$_3$OH, and H$_2$CO are the main contributors to the envelope, while the corresponding emissions are mainly CH$_3$OH, NO, SO$_2$, SO, and SO$_2$,v$_2$=1. The core seems to be more uniform, as the molecular families all have more or less the same mode temperatures, line widths and velocities. According to the analysis of the chemical clocks, the source is more or less at an intermediate age, which also agrees quite well with the analysis of the evolutionary phases (phase~III). Moreover, this classification is also underpinned by the not so high proportion of vibrationally excited molecules in the emission, which require high temperatures and a high evolutionary phase. Furthermore, this source is part of the PCA cluster~CM1 containing many intermediate to highly evolved sources (except A05).

% Sgr B2(M), A04:
% PCA:
%   CM1: A01, A03, A04, A05, A06, A07, A10
% corresponding evolution phases:
%         IV, III,  IV,  II, III,  IV, III
\item \textbf{A04}: Source A04 shows large contributions from SO, SO$_2$, SO$_2$,v$_2$=1, NO, and CH$_3$OH to the emission, while CN, CH$_3$OH, CCH, H$_2$CO, and SO dominate the envelope. Similar to source A02, source A04 also shows a differentiated behavior of the various molecular families. While the sulfur-bearing molecules show a significantly higher mode temperature than the other families, the NH-bearing molecules show a much larger mode line width, which indicates a higher amount of turbulence and so a different origin like a filament. In contrast, the mode velocities are more or less the same. However, the result of the chemical clocks is contradictory, as the ratio of SO / OCS indicates the source as evolved, whereas the ratio of SO$_2$ / SO implies a young source. On the other hand, the presence of masers, outflows and \hii~regions indicates a more evolved source (phase~IV), which also fits well with the results of the PCA, according to which this source belongs to the same cluster (CM1) as the evolved source A01.

% Sgr B2(M), A05:
% PCA:
%   CM1: A01, A03, A04, A05, A06, A07, A10
% corresponding evolution phases:
%         IV, III,  IV,  II, III,  IV, III
\item \textbf{A05}: The envelope around source A05, is dominated by CN, CCH, H$_2$CO, SO, and CH$_2$NH. The corresponding core spectrum is primarily caused by CH$_3$OH, SO, SO$_2$, NO, and SO$_2$,v$_2$=1, where the distributions of temperatures, line widths and velocities indicate a more or less uniform source structure. The chemical clocks point to a rather young to intermediate age, which also fits well with the analysis of the phase analysis of the evolutionary sequence (phase~II) and the small contributions of vibrational excited species to the core spectrum. Nevertheless, the PCA assigns this source to the same cluster (CM1) as the more developed source A01.

% Sgr B2(M), A06:
% PCA:
%   CM1: A01, A03, A04, A05, A06, A07, A10
% corresponding evolution phases:
%         IV, III,  IV,  II, III,  IV, III
\item \textbf{A06}: For source A06, the envelope contains mainly CN, SiO, HCCCN, CCH, and H$_2$CO, while in the core spectrum we find large amounts of CH$_3$CN, SO$_2$, SO$_2$,v$_2$=1, SO, and NO. Moreover, the core spectrum shows more or less unimodal distributions for the fit parameters except that the mode temperature of sulfur-bearing molecules is higher than for the other families. Although the mode line widths are more or less identical for the different families, the mode value of the velocity for simple O- and NH-bearing molecules is slightly larger, which could indicate a filament. The analysis of the chemical clocks and the PCA show mostly that the source is at an advanced stage of evolution (cluster~CM1 containing the evolved sources A01, A04, and A07), although the presence of masers and outflows and the moderate contribution of the vibrationally excited SO$_2$,v$_2$=1 indicate an intermediate age (phase~III).

% Sgr B2(M), A07:
% PCA:
%   CM1: A01, A03, A04, A05, A06, A07, A10
% corresponding evolution phases:
%         IV, III,  IV,  II, III,  IV, III
\item \textbf{A07}: Source A07 is embedded in an envelope, including predominantly H$_2$CO, CN, CCH, SO, and HCCCN, whereas the core spectrum mainly shows emission from CH$_3$OH, SO$_2$, SO, NO, and HCCCN,v$_7$=2. The mode temperature of sulfur- and complex O-bearing molecules in the core spectrum is slightly higher. The corresponding mode line widths and velocities are mostly the same for all molecules, except for carbon and hydrocarbons, which exhibit a significantly lower velocity revealing a cloud in the foreground. The analysis based on chemical clocks shows an inconsistent picture, as some ratios indicate a young and others an old age, which together with the different mode temperatures could indicate a variety of protostars. In contrast to that, the results of the PCA (cluster~CM1) and the analysis of the evolutionary features (maser, outflows, \hii~regions) clearly indicate the source as evolved.

% Sgr B2(M), A08:
% PCA:
%   CM4: A02, A08, A09, A12, A13, A14, A15
% corresponding evolution phases:
%         IV,  IV,   I,   I,   I,   ?,   V
% Kendall:
%   A08 - A14 (both cluster CM4)
\item \textbf{A08}: While the envelope of source A08 contains mainly CH$_3$CN, CH$_3$OH, OCS, NO, and HCCCN, the core spectrum is dominated by contributions from SO, SO$_2$, SO$_2$,v$_2$=1, NO, and H$_2 \! ^{34}$S. Moreover, the distribution of the parameters in the core describe a rather inconsistent picture, according to which the N- and O-bearing molecules tend to be colder in comparison with the other molecules. Furthermore, the mode line widths and velocities are distributed over a wider range, which in turn indicates a rather heterogeneous source structure such as an additional filament. The source is located at the edge of a large, possibly expanding \hii~region including source A16, which may explain the remarkable shift in velocities compared to the source velocity of the Sgr~B2 complex. The analysis of the chemical clocks and the presence of masers, outflows and \hii~regions indicate an advanced stage of evolution (phase~IV). In addition, the PCA classifies the source in the same cluster (CM4) as A02. Furthermore, the Kendall correlation coefficient analysis shows a strong correlation to the neighboring source A14.

% Sgr B2(M), A09:
% PCA:
%   CM4: A02, A08, A09, A12, A13, A14, A15
% corresponding evolution phases:
%         IV,  IV,   I,   I,   I,   ?,   V
\item \textbf{A09}: In source A09, SO$_2$ provides the largest contribution to the envelope spectrum followed by CN, CCH, SO, and H$_2$CCN. In addition, the core spectrum consists of CH$_3$OH, SO, SO$_2$, OCS, and SO$_2$,v$_2$=1. Furthermore, the molecules in the core all have more or less the same mode temperatures, line widths and velocity distributions. The N- and O-bearing molecules, which are slightly colder, are an exception. The analysis of the chemical clocks and the evolutionary features indicate a rather young age (phase~I), whereas the PCA suggests a more evolved age (cluster~CM4).

% Sgr B2(M), A10:
% PCA:
%   CM1: A01, A03, A04, A05, A06, A07, A10
% corresponding evolution phases:
%         IV, III,  IV,  II, III,  IV, III
\item \textbf{A10}: While the envelope spectrum around source A10 consists predominantly of CN, CH$_3$OH, CCH, H$_2$CO, and SiO, the core spectrum incorporate mostly CH$_3$OH, SO, NO, SO$_2$, and H$_2$CO, whereby the distributions of the parameters there also show a fairly uniform picture, with the mode temperatures for the NH- and N- and O-bearing molecules being somewhat higher. The further analyses assign the source to an intermediate evolutionary phase (phase~III), whereby the PCA indicates a rather older age  (cluster~CM1).

% Sgr B2(M), A11:
% PCA:
%   CM2: A11, A23
% corresponding evolution phases:
%         II,   I
\item \textbf{A11}: In source A11, the envelope spectrum has high contributions of OCS, HNCO, H$_2 \! ^{34}$S, SO$_2$, and CH$_3$OH. The corresponding emission is primarily caused by CH$_3$OH, CCH, CH$_3$OCHO, CH$_3$OCH$_3$, and CN and shows an inconsistent behavior, as the mode temperatures and line widths of the core spectrum in particular are distributed over a broad range, which in turn could indicate the presence of a filament. Nevertheless, the chemical clocks as well as the evolutionary characteristics exhibit the source as young. PCA assigns this source to its own cluster, which only includes source A23 as an additional element, although the analysis of the Kendall correlation coefficient shows only a low correlation between both sources. The high proportions of CCH and the exceptionally low temperatures of methanol in both sources could be possible explanations for this classification.

% Sgr B2(M), A12:
% PCA:
%   CM4: A02, A08, A09, A12, A13, A14, A15
% corresponding evolution phases:
%         IV,  IV,   I,   I,   I,   ?,   V
\item \textbf{A12}: The envelope around source A12 shows absorption features primarily from HCCCN, OCS, CH$_3$OH, CN, and CH$_3$CN. Furthermore, the core spectrum consists mainly of SO$_2$,v$_2$=1, SO$_2$, SO, NO and OCS and shows uniform distributions of the parameters, except for the mode velocity of the NH-bearing molecules, which is approximately at the source velocity of the Sgr~B2 complex, while the other molecules show the largest velocity shift of all sources. This could again indicate a filament. The chemical clocks and the PCA indicate that the source is old (cluster~CM4). However, all characteristics of an advanced development are missing (phase~I).

% Sgr B2(M), A13:
% PCA:
%   CM4: A02, A08, A09, A12, A13, A14, A15
% corresponding evolution phases:
%         IV,  IV,   I,   I,   I,   ?,   V
\item \textbf{A13}: Source A13 is embedded in an envelope, including predominantly CCH, CN, SO, H$_2$CS, and SiO, whereas its core spectrum is mostly made of CH$_3$OH, SO, SO$_2$, NO, and SO$_2$,v$_2$=1. Although the distribution of mode temperatures and line widths in the core are less uniform, all molecules show more or less the same low mode velocity, which could be caused by the large, possibly expanding \hii~region between sources A02 and A13. The further analyses do not allow a clear statement regarding the age of the source, which supports the assumption that the source contains a large number of protostars. The chemical clocks indicate an intermediate age, whereas the PCA classifies the source as a more developed source (cluster~CM4). However, there are no signs of an advanced evolution (phase~I).

% Sgr B2(M), A14:
% PCA:
%   CM4: A02, A08, A09, A12, A13, A14, A15
% corresponding evolution phases:
%         IV,  IV,   I,   I,   I,   ?,   V
% Kendall:
%   A08 - A14 (both cluster CM4)
\item \textbf{A14}: In source A14, HCCCN, CH$_3$OH, CN, HNCO, and CCH provide the largest contributions to the envelope spectrum. In addition, the corresponding core spectrum consists mostly of SO$_2$,v$_2$=1, SO, SO$_2$, NO, and H$_2 \! ^{34}$S. From the analysis of the distribution of the parameters, a very heterogeneous source structure can be derived for this source, which can be seen in particular in the distribution of the mode line widths and velocities. While the mode line width for the NH-bearing molecules in particular deviates significantly from the others, the velocities of almost all molecular families are distributed over a wide range, which in turn suggests a complex kinematic structure. Like the neighboring source A08, this source is also located at the edge of the large \hii~region around source A16, which could cause the unusually low velocities in this source. The analysis of the chemical clocks exhibit a middle to old age, which agrees quite well with the classification of the PCA (cluster~CM4). However, this source contains neither masers nor \hii~regions but only outflows, which prevents its categorization into an evolutionary phase. As described in Sect.~\ref{subsec:outflows}, we cannot exclude the possibility that the non-Gaussian line shape of OCS and SO results from rotational or infall motions, so that this could also be a young source.

% Sgr B2(M), A15:
% PCA:
%   CM4: A02, A08, A09, A12, A13, A14, A15
% corresponding evolution phases:
%         IV,  IV,   I,   I,   I,   ?,   V
\item \textbf{A15}: While the envelope spectrum around source A15 consists predominantly of OCS, H$_2$CS, CH$_3$CN, HCCCN, and HNCO, the emission spectrum contains mainly SO$_2$, SO, NO, SiO, and SO$_2$,v$_2$=1. The source correspond more or less to a single \hii~region, and we also find hints for a complex kinematic structure. In contrast to the distribution of the mode temperatures and line widths, the various families of molecules show a clear bimodal distribution of the velocities, where the N- and O-bearing molecules have much lower velocities. These velocities are comparable to the mean velocity of the neighboring source A13, while the velocities of the other molecules are more than 20~km~s$^{-1}$ higher, comparable to the mean velocity of source A17, indicating a common structure like a filament connecting both sources. In addition, the ratio of SO$_2$ / SO, the analysis of the evolutionary features, and the PCA (cluster~CM4) exhibit this source as highly evolved (phase~V). Furthermore, the analysis of the contributions of dust and ionized gas, as described in \citetads{2023A&A...676A.121M}, indicates an age of 3100~yr for the \hii~region.

% Sgr B2(M), A16:
% PCA:
%   CM6: A16, A18, A19, A21, A24, A26, A27
% corresponding evolution phases:
%         IV,   I,   I,  II,   V,  IV,   I
\item \textbf{A16}: HCCCN, SO$_2$, OCS, CH$_3$CN, and H$_2 \! ^{34}$S are highly abundant in the envelope around source A16, while the core spectrum show large contributions of SO$_2$, SO, SO$_2$,v$_2$=1, CN, and NO. The source A16, which is embedded in a large \hii~region, contains only a few molecules in emission, making it difficult to analyze the distribution of parameters. The mode velocities of the different families are distributed over a range of 30~km~s$^{-1}$, which could indicate an expansion of the associated \hii~region. The further analyses indicate an advanced age (phase~IV). The PCA assigns the source to a cluster (CM6) with other external sources where the mean temperature is highest for NO and lowest for many sulfur-bearing molecules.

% Sgr B2(M), A17:
% PCA:
%   CM3: A17, A22, A25
% corresponding evolution phases:
%          I,   I,   I
\item \textbf{A17}: Source A17 is embedded in an envelope, including predominantly HNCO, SO$_2$, CH$_3$OH, NH$_2$D, and OCS, while the core spectrum is predominantly made of H$_2 \! ^{34}$S, CN, CCH, SO, SiO, where the simple O-bearing molecules show a significantly larger mode line width than the other molecules. Furthermore, all molecules in the core show high velocities comparable to those in source A15, indicating a common structure, such as a filament, in which both sources are embedded. Both the ratio of SO$_2$ / SO and the evolutionary analysis indicate a young age (phase~I). According to the PCA, the sources show correlations with sources A22 and A25 (cluster~CM3), all of which are also at a very early stage of development.

% Sgr B2(M), A18:
% PCA:
%   CM6: A16, A18, A19, A21, A24, A26, A27
% corresponding evolution phases:
%         IV,   I,   I,  II,   V,  IV,   I
% Kendall:
%   A18 - A19 (both cluster CM6)
%   A18 - A21 (both cluster CM6)
\item \textbf{A18}: The envelope around source A18 shows large contributions of H$_2 \! ^{34}$S, OCS, NH$_2$D, HCCCN, and CH$_3$CN. The core contains mainly CCH, H$_2$CO, H$_2$CS, NS, HCCCN, where the mode line width of the simple O-bearing molecules is significantly lower than for the other molecules, while the velocities are more or less identical. Here, the chemical clocks cannot be used to estimate the age, as the source does not contain the required molecules. Since neither masers nor outflows nor \hii~regions are included, the source seems to be quite young, which is also supported by the fact that no vibrational excited molecules are found. Furthermore, the Kendall correlation coefficient analysis shows strong correlations with the equally young sources A19 and A21. However, the PCA assigns the source to a cluster (CM6) that also contains the developed sources A16, A24, A26.

% Sgr B2(M), A19:
% PCA:
%   CM6: A16, A18, A19, A21, A24, A26, A27
% corresponding evolution phases:
%         IV,   I,   I,  II,   V,  IV,   I
% Kendall:
%   A18 - A19 (both cluster CM6)
%   A19 - A21 (both cluster CM6)
\item \textbf{A19}: In source A19, the envelope contains mainly H$_2 \! ^{34}$S, CH$_3$CN, OCS, HCCCN, and CH$_3$OH, while the core spectrum contains mostly H$_2$CS, HCCCN, SiO, and CH$_2$NH. The mode line width and velocity of the NH-bearing molecules differ significantly from the other molecules in the core, indicating a more complex source structure like an additional filament. Since the majority of the molecules in A19 have more or less the same mean velocity as the neighboring source A11, we assume that both sources are embedded in a common structure like a filament. The NH-bearing molecules are an exception and could describe a warm foreground object or an additional filament. As in source A18, the chemical clocks do not allow an age estimation due to the lack of the required molecules. Due to the lack of any evolutionary features, a young age can be assumed. In addition, the Kendall coefficient indicates strong correlations with the similar young sources A18 and A21. As with the previous source A18, the PCA assigns this source to a cluster (CM6), which contains both young and old sources.

% Sgr B2(M), A20:
% PCA:
%   CM5: A20
% corresponding evolution phases:
%          I
\item \textbf{A20}: As for many other outer sources in Sgr~B2(M), SO$_2$ is the main component of the envelope around source A20 followed by CH$_3$OH, OCS, H$_2$CO, and SO. The core spectrum is dominated by the same contributions as in source A11, i.e.\ CH$_3$OH, CH$_3$OCHO, CCH, H$_2$CO, and SO, although both sources are quite far apart. Here, the N- and O-bearing molecules have a clearly different mode line width and velocity, while the mode velocities of the other molecules in the core spectrum are similar to those of source A19. Based on the chemical clocks and the lack of evolutionary features, the source appears to be in a young evolutionary phase (phase~I) like source A11. But according to the Kendall coefficient analysis, there is no strong correlation with other sources, which is also underpinned by the PCA, which classifies this source in its own cluster (CM5).

% Sgr B2(M), A21:
% PCA:
%   CM6: A16, A18, A19, A21, A24, A26, A27
% corresponding evolution phases:
%         IV,   I,   I,  II,   V,  IV,   I
% Kendall:
%   A18 - A21 (both cluster CM6)
%   A19 - A21 (both cluster CM6)
%   A21 - A24 (both cluster CM6)
\item \textbf{A21}: Source A21 is surrounded by an envelope consisting predominantly of SO$_2$, OCS, H$_2$CO, CH$_3$OH, and H$_2$CCO, while in the core spectrum we were able to identify only CCH in addition to a number of molecules with only one transition. Therefore, the use of chemical clocks for age determination is not possible. Only the presence of a water maser allows a classification in evolutionary phase~II. According to the Kendall coefficient analysis and the PCA, there are strong correlations with sources A18, A19, and A24, which is more surprising since source A24 is an evolved source with an old \hii~region.

% Sgr B2(M), A22:
% PCA:
%   CM3: A17, A22, A25
% corresponding evolution phases:
%          I,   I,   I
\item \textbf{A22}: The envelope around source A22 includes mostly OCS, CH$_3$OH, CH$_3$CN, SO, and CCH. In addition, the core spectrum is mainly made of CH$_3$OH, CCH, OCS, OCS,v$_2$=1, and CH$_3$OCH$_3$, where the NH-bearing and cyanide molecules show significantly larger mode line widths. Together with the corresponding mode velocities, which are distributed over a wide range, this indicates again a complex source structure like a filament. The median velocity of this source is one of the lowest of all sources in Sgr~B2(M). The chemical clocks and the lack of any evolutionary features exhibit the source as young (phase~I), in agreement with the PCA, which shows correlations with the equally young sources A17 and A25 (cluster~CM3).

% Sgr B2(M), A23:
% PCA:
%   CM2: A11, A23
% corresponding evolution phases:
%         II,   I
\item \textbf{A23}: HNCO, H$_2 \! ^{34}$S, OCS, HCCCN, and CN play an important role in the envelope around source A23. The corresponding core spectrum contains mainly H$_2 \! ^{34}$S, OCS, OCS,v$_2$=1, H$_2$CCO, and H$_2$CS, where the sulfur-bearing molecules have lower mode temperatures and velocities compared to the other molecules. The ratio of CH$_3$CN / CH$_3$OH suggests an intermediate age, while the evolutionary phase analysis exhibit a young age (phase~I). The PCA found correlation with the young source A11 (cluster~CM2).

% Sgr B2(M), A24:
% PCA:
%   CM6: A16, A18, A19, A21, A24, A26, A27
% corresponding evolution phases:
%         IV,   I,   I,  II,   V,  IV,   I
% Kendall:
%   A21 - A24 (both cluster CM6)
\item \textbf{A24}: Source A24 is embedded in an envelope, including predominantly SO$_2$, H$_2$CO, CCH, SO, and SiO, while the core spectrum is mostly made of CH$_3$OH, SO$_2$, SO$_2$,v$_2$=1, OCS, and OCS,v$_2$=1. The source is identical in size and location to a single large \hii~region. The mode temperatures and line widths are not identical for all families in the core and are distributed over a narrow range, whereas the mode velocities all agree very well with the source velocity of the Sgr~B2 complex at 64~km~s$^{-1}$. This could indicate a filament, but the \hii~region also expands \citepads{2023A&A...676A.121M}, which creates an additional dynamic. The analysis of the chemical clocks leads to contradictory statements regarding the age. For example, the ratio of SO / OCS indicates a very young source, whereas the ratio of SO$_2$ / SO indicates an old source, which is also consistent with the results of the evolutionary phase analysis (phase~V) and the high abundances of vibrational excited SO$_2$,v$_2$=1 and OCS,v$_2$=1. Furthermore, the analysis of the contributions of dust and ionized gas, as described in \citetads{2023A&A...676A.121M}, yields an age of 7000~years for the \hii~region. In contrast, the PCA assigns the source to cluster~CM6, which contains both young and old sources. In addition, the Kendall coefficient shows a strong correlation with the young source A21. One possible explanation would be that the source contains not one but several protostars, which are in different stages of development.

% Sgr B2(M), A25:
% PCA:
%   CM3: A17, A22, A25
% corresponding evolution phases:
%          I,   I,   I
\item \textbf{A25}: While the envelope spectrum around source A25 consists mostly of SO$_2$, HCCCN, CH$_3$CN, SO, and HC(O)NH$_2$, the emission spectrum contains mainly CCH, CH$_3$OH, H$_2$CS, H$_2 \! ^{34}$S, and HCCCN, where the N- and O-bearing molecules show slightly larger mode line widths and velocities than the other molecules, which could expose again an additional filament. Since the source contains no masers, outflows or \hii~regions, it appears to be quite young. This assessment fits well with the results of the PCA, which assigns the source to cluster~CM3, which contains only young sources.

% Sgr B2(M), A26:
% PCA:
%   CM6: A16, A18, A19, A21, A24, A26, A27
% corresponding evolution phases:
%         IV,   I,   I,  II,   V,  IV,   I
\item \textbf{A26}: H$_2 \! ^{34}$S, SO$_2$, NO, SO, and H$_2$CO contribute the most to the envelope around source A26. Furthermore, CH$_3$OH, SO, H$_2$CO, SO$_2$, and SO$_2$,v$_2$=1 make large contributions to the core spectrum. A26 appears to be an evolved source, as it contains both masers and outflows and \hii~regions. This is also supported by the PCA, which assigns the source to cluster~CM6, which contains many old sources. However, the chemical clocks consistently identify this source as a young source. A heterogeneous source structure cannot explain this circumstance, as the distribution of the mode values in core spectrum is quite uniform, except the mode temperature of complex O-bearing molecules which is slightly higher.

% Sgr B2(M), A27:
% PCA:
%   CM6: A16, A18, A19, A21, A24, A26, A27
% corresponding evolution phases:
%         IV,   I,   I,  II,   V,  IV,   I
\item \textbf{A27}: As in the envelopes of many other outer sources, H$_2 \! ^{34}$S makes the largest contribution to the envelope around source A27 followed by SO$_2$, HCCCN, CH$_3$OH, and CH$_3$CN. In contrast to that, the core spectrum is dominated by contributions from CH$_3$OH, SO, CH$_3$CN, H$_2$CO, as well as SiO, where we obtain clearly separated mode velocities for the three families of molecules contained in this source, which could indicate the existence of a filament. According to the ratio of CH$_3$CN / CH$_3$OH, the source has an intermediate age. However, no maser could be detected, which would be expected for a source in this period. The PCA assigns this source to cluster~CM6, which contains both young and old sources.
\end{itemize}

% subsection: Source in Sgr~B2(M)
\subsection{Source in Sgr~B2(N)}\label{app:subsec:ResultsN}

\begin{itemize}
% Sgr B2(N), A01:
% neighbors:
%   A04, A05, A10, A17, (A07)
% Bonfand:
%   A01 -> N1
% PCA:
%   CN3: A01, A02, A17
% corresponding evolution phases:
%         IV, III,   ?
% Schwoerer:
%   Hub 01
\item \textbf{A01}: For source A01 in Sgr~B2(N), CN, CCH, H$_2$CO, CH$_2$NH, and SiO are the main contributors to the envelope. The corresponding emission is dominated by CH$_3$OH, C$_2$H$_3$CN,v$_{10}$=1, C$_2$H$_5$CN,v$_{20}$=1, HCCCN,v$_7$=2, and CH$_3$CN. While the mode values for the temperatures and line widths of the core spectrum are distributed over a narrow range, see Fig.~\ref{fig:KDETrotCoreN} and Fig.~\ref{fig:KDEwidthCoreSgrB2N}, the molecules have more or less the same velocity, see Fig.~\ref{fig:KDEvoffCoreSgrB2N}, which is located slightly below the source velocity of the Sgr~B2 complex. The results of the chemical clock analysis, see Table~\ref{Tab:ChemClocks}, show that the source is both a young and an old source, while the occurrence of masers, outflows and \hii~regions and the high abundances of vibrational excited states like C$_2$H$_3$CN,v$_{10}$=1, C$_2$H$_5$CN,v$_{20}$=1, and HCCCN,v$_7$=2 indicate this source as more evolved. A possible explanation for this contradiction could be that the source may consist of several protostars, which could also explain the variances in the temperatures and line widths. The PCA finds correlations with A02 and the neighboring source A17. \citetads{2019A&A...628A...6S} identified different filaments in Sgr~B2(N), where the filament F01 connects these two sources with the main hub, which is identical with source A01. This could explain the identified correlations.

% Sgr B2(N), A02:
% neighbors:
%   A03, (A05), (A11)
% Bonfand:
%   A02 -> N2
% PCA:
%   CN3: A01, A02, A17
% corresponding evolution phases:
%         IV, III,   ?
% Schwoerer:
%   F01: A02, A03, A06, A17
\item \textbf{A02}: The envelope around source A02 is dominated by the same molecules as the envelopes around sources A01 and A05, i.e.\ CN, CCH, H$_2$CO, CH$_2$NH, and SiO, the core spectrum is made of CH$_3$OH, OCS, C$_2$H$_5$CN,v$_{20}$=1, CH$_3$CN, and C$_2$H$_5$CN. The distributions of the parameters in the core spectrum indicate a rather homogeneous source structure, with the mode line width for N- and O-bearing molecules being slightly larger than that of the others. As already mentioned in the description of source A01, the filament F01, which was identified by \citetads{2019A&A...628A...6S} and in which source A02 seems to be embedded, could explain the observed deviations in the line widths. The abundance ratios used to estimate the age of the source points toward a young to intermediate source, which agrees quite well with the evolutionary analysis (phase~III). The PCA assigns the source to the same cluster (CN3) as A01, which could also suggest a similar age. According to our analysis, source A02 is at a significantly later stage of development than indicated in the analysis of \citetads{2017A&A...604A..60B}. This is due to the presence of H$_2$O masers and outflows in this source, which were not previously considered. Furthermore, we have subdivided the source (N2) identified by \citetads{2017A&A...604A..60B} into two individual sources (A02 and A03).

% Sgr B2(N), A03:
% neighbors:
%   A02, A06, (A07)
% PCA:
%   CN4: A03, A04
% corresponding evolution phases:
%         IV, III
% Schwoerer:
%   F01: A02, A03, A06, A17
\item \textbf{A03}: In source A03, as with the other sources A01 and A02, the envelope contains mainly CCH, CN, H$_2$CO, CH$_2$NH, and SiO, albeit with a different weighting. In the core spectrum, contributions from CH$_3$OH, OCS, HCCCN, C$_2$H$_5$CN,v$_{20}$=1, and CH$_3$CN are particularly important. Here, the distributions of the mode values are quite uniform, which again indicates a homogeneous source structure. However, small differences in the line widths could indicate the influence of filament F01 \citepads{2019A&A...628A...6S}. The median velocity is the second highest of all sources in Sgr~B2(N). Since the source contains both masers and outflows and \hii~regions, it appears to be older (phase~IV). Unfortunately, the chemical clocks do not allow a clear age determination, as they come to different results (young to old). According to the PCA, there are connections to source A04. Since both sources are connected to different filaments, this could indicate a similar age, which means that both A03 and A04 are in a middle to older stage of development.

% Sgr B2(N), A04:
% neighbors:
%   A01, A05, A07
% PCA:
%   CN4: A03, A04
% corresponding evolution phases:
%         IV, III
% Schwoerer:
%   F08: A04, A05, A07, A08, A11, A15, A19
\item \textbf{A04}: CCH, CN, H$_2$CO, CH$_2$NH, and SiO are the main contributors to the envelope spectrum around source A04, while CH$_3$OH, H$_2$CCO, H$_2$CO, CH$_3$OCHO, and OCS dominate the core spectrum. The source also appears to be quite homogeneous, although the line width of the N- and O-bearing molecules is slightly larger than the mean value. A possible explanation could be the filament F08 \citepads{2019A&A...628A...6S}, which connects the sources A01, A04, A05, A11, A15, A08 and A19. According to the ratios of the sulphur-bearing molecules and the results of the evolutionary phase analysis (phase~III), the source appears to have an intermediate age. This is also in good agreement with the PCA classification, which correlates it with the similarly aged source A03.

% Sgr B2(N), A05:
% neighbors:
%   A01, A04, A11, (A02)
% PCA:
%   CN2: A05, A06, A07, A08, A10
% corresponding evolution phases:
%          ?,   I, III, III,   V
% Schwoerer:
%   F08: A04, A05, A07, A08, A11, A15, A19
\item \textbf{A05}: Around source A05, CN, CCH, H$_2$CO, CH$_2$NH and SiO contribute the largest part to the envelope spectrum, as was also the case with sources A01 and A02, while the core spectrum consists mainly of CH$_3$OH, CH$_3$OCHO, SO, CH$_3$OCH$_3$, and OCS. Source A05 shows no signs of a heterogeneous source structure, except for a slightly larger line width for N- and O-bearing molecules, which could indicate a connection to the previously mentioned filament F08 \citepads{2019A&A...628A...6S}. This is also partially supported by the PCA, which finds correlations with some other sources (A07 and A08) in this filament F08. The chemical clocks lead to very different age estimates and therefore do not allow a uniform classification. The sole occurrence of outflows without masers or \hii~regions also does not allow a classification into a single evolutionary phase. Perhaps the non-Gaussian line shapes of some OCS transitions have other causes than outflow, which would then indicate a very young source.

% Sgr B2(N), A06:
% neighbors:
%   A03, A07, (A17)
% PCA:
%   CN2: A05, A06, A07, A08, A10
% corresponding evolution phases:
%          ?,   I, III, III,   V
% Schwoerer:
%   F01: A02, A03, A06, A17
\item \textbf{A06}: The envelope around source A06 is dominated by CN, CCH, HCCCN, H$_2$CO, and SO. The largest contributions to the core spectrum are caused by CH$_3$OH, CH$_3$CN, OCS, CH$_3$OCH$_3$, and CH$_3$OCHO. The distributions of the mode values in the core suggest a more or less homogeneous source structure. But variations in the temperatures indicates a small contribution from filament F01 \citepads{2019A&A...628A...6S}, although the PCA finds correlations with A05, A07, A08, and A10 (cluster~CN2). The majority of the various chemical clocks suggest a rather young source, which also fits well with the results of the evolutionary phase analysis (phase~I).

% Sgr B2(N), A07:
% neighbors:
%   A01, A04, A06, A17, (A03)
% PCA:
%   CN2: A05, A06, A07, A08, A10
% corresponding evolution phases:
%          ?,   I, III, III,   V
% Schwoerer:
%   F08: A04, A05, A07, A08, A11, A15, A19
\item \textbf{A07}: In contrast to the neighboring sources A04 and A06, the envelope around source A07 appears to have a completely different composition, as the contributions from H$_2$CS and OCS dominate here. The molecules known from the envelope spectra around other sources CN and H$_2$CO only play a subordinate role here. Additionally, we find C$_2$H$_5$CN. The most important molecules in the core spectrum (CH$_3$OH, CH$_3$OCHO, SO, OCS, and CH$_3$OCH$_3$) are almost identical to those in source A05, albeit with different weightings. The kernel density estimates for each parameter of the different families in the core have more or less the same mode values, suggesting an uniform source structure. However, slight differences in the line widths indicate a contribution from the filament F08 \citepads{2019A&A...628A...6S}. The further analyses classify the source as an intermediate (phase~III) to old source. In addition, the PCA shows correlations with the sources A05, A06, A07, A08 and A10, some of which (A05, A07, and A08) are also part of the filament F08.

% Sgr B2(N), A08:
% neighbors:
%   A19, (A12)
% Bonfand:
%   A08 -> N3
% PCA:
%   CN2: A05, A06, A07, A08, A10
% corresponding evolution phases:
%          ?,   I, III, III,   V
% Schwoerer:
%   F08: A04, A05, A07, A08, A11, A15, A19
\item \textbf{A08}: For source A08, CCH, H$_2$CO, CN, and SiO are the main contributors to the envelope, while the core spectrum consists mostly of CH$_3$OH, CH$_3$OCHO, C$_2$H$_3$CN,v$_{10}$=1, OCS, and OCS,v$_2$=1. The different molecular families in the core show different mode values for temperature and line width, indicating a more complex source structure. Here, filament F08 \citepads{2019A&A...628A...6S} could also be a possible explanation for the additional structure. At the same time, the mode velocities are the same, which could indicate that the source is fully embedded in the filament. According to the evolutionary phase analysis (phase~III) the source is at an intermediate age, which agrees quite well with the classification of \citetads{2017A&A...604A..60B} (source A08 is identical with their core N3) and the moderate high abundance of vibrational excited states C$_2$H$_3$CN,v$_{10}$=1 and OCS,v$_2$=1. However, the majority of chemical clocks tend to indicate the source as young, which could, together with the different mode temperatures and the large number of masers, point to a multitude of protostars. Furthermore, PCA finds similarities with some other inner sources A05, A06, A07 and A10 (cluster~CN2), where two of these sources are also connected by the F08 filament (A05 and A07).

% Sgr B2(N), A09:
% neighbors:
%   (A20)
% PCA:
%   CN6: A09, A14, A15, A18
% corresponding evolution phases:
%          I,   I,   I,   I
% Kendall:
%   A09 - A11 (A09: CN6, A11: CN7)
%   A09 - A14 (both cluster CN6)
%   A09 - A15 (both cluster CN6)
% Schwoerer:
%   -
\item \textbf{A09}: CH$_3$OH, HCCCN, OCS, HNCO, and CH$_3$CN contribute the most to the envelope around source A09. Furthermore, CCH, H$_2$CS, SiO, and H$_2 \! ^{34}$S make large contributions to the core spectrum. The respective mode values for the temperature and the line width of the simple O-bearing molecules differ significantly from the other molecules in the core, which could indicate an additional filament. However, such a filament was not found by \citetads{2019A&A...628A...6S}. The analysis of the evolutionary features suggests a young age (phase~I), while the molecules required for the chemical clocks are not contained in the source. The PCA revealed correlations with sources A14, A15 and A18 (cluster~CN6). The analysis of the Kendall coefficient also shows strong links to A14 and A15, but in contrast to the PCA, the relationship to A11 is stronger than to A18.

% Sgr B2(N), A10:
% neighbors:
%   A01, (A20)
% PCA:
%   CN2: A05, A06, A07, A08, A10
% corresponding evolution phases:
%          ?,   I, III, III,   V
% Schwoerer:
%   F06: A10
\item \textbf{A10}: The composition (H$_2$CO, SiO, CCH, and CN) of the envelope around source A10 is similar to that of the envelope around the neighboring source A01 except that CH$_2$NH is missing. The core spectrum contains mainly CH$_3$OH, CH$_3$OCHO, H$_2$CO, CH$_3$OCH$_3$, and OCS. The distributions of the mode values of the fit parameters in the core indicate a more monolithic source structure, except that the mode values of the line widths differ slightly, which could show a connection to filament (F06) identified by \citetads{2019A&A...628A...6S} connecting A01 with this source. The age estimate based on chemical clocks gives an inconsistent picture for this source (young to old), while the evolutionary features suggest a highly evolved source (phase~V), as it contains outflows and an \hii~region. According to the PCA the source correlates with sources A05, A06, A07, and A08 (cluster~CN2), although these sources are at different stages of development and are not all associated with the same filament.

% Sgr B2(N), A11:
% neighbors:
%   A05, A15, (A02)
% PCA:
%   CN7: A11, A12
% corresponding evolution phases:
%          I,   I
% Kendall:
%   A11 - A12 (both cluster CN7)
%   A11 - A16 (A11: CN7, A16: CN1)
% Schwoerer:
%   F08: A04, A05, A07, A08, A11, A15, A19
\item \textbf{A11}: In contrast to many other sources, the envelope around source A11 shows an unusual composition. HNCO, HCCCN, OCS, CH$_3$OH, and CH$_3$CN are particularly important here, whereas the molecules known from other envelope spectra such as CCH, CN, and H$_2$CO are far less dominant. The core spectrum contains mainly NO, OCS, SiO, and HCCCN, where the mode values of the velocities indicate an additional structure, which could be identified with filament F08. Since neither masers nor outflows and \hii~regions are included, this source appears to be very young (phase~I). An analysis of the chemical clocks is not possible here, as the required molecules are missing. In addition, the PCA and the Kendall coefficient suggest correlations with source A12 (cluster~CN7), which is also in a very early stage of development. In addition, the Kendall coefficient analysis also indicates connections to the highly evolved source A16, which was not found by the PCA.

% Sgr B2(N), A12:
% neighbors:
%   (A08)
% PCA:
%   CN7: A11, A12
% corresponding evolution phases:
%          I,   I
% Kendall:
%   A11 - A12 (both cluster CN7)
% Schwoerer:
%   -
\item \textbf{A12}: Around source A12, CH$_3$OH, CN, CH$_3$CN, CCH, and SO contribute the largest part to the envelope spectrum. The core spectrum is dominated by OCS, NS, SiO, H$_2$CS, and HCCCN, where the mode value of the velocity of the simple O-bearing molecules is clearly shifted to higher velocities, which again indicates an additional structure. However, no filament was found by \citetads{2019A&A...628A...6S} in which this source appears to be embedded. As with the previous source A11, only the evolutionary phase analysis provides an estimate of the age, as the necessary molecules for an age estimate based on chemical clocks are not included in this source. Here, the source appears to be also rather young (phase~I). Despite the large distance, the PCA finds correlation with the similarly developed source A12 (cluster~CN7).

% Sgr B2(N), A13:
% neighbors:
%   -
% Bonfand:
%   A13 -> N4
% PCA:
%   CN5: A13
% corresponding evolution phases:
%         II
% Schwoerer:
%   -
\item \textbf{A13}: For the envelope around the outer source A13, we find strong contributions from HNCO, CH$_3$OH, CN, CH$_3$CN, and CCH, while the core spectrum is mostly made of CH$_3$OH, CH$_3$OCH$_3$, CCH, HCCCN, and CH$_2$NH. The distributions of the different fit parameters in the core differ strongly, indicating a complex source structure, but no filament was found by \citetads{2019A&A...628A...6S} associated with this source. This could indicate a large number of protostars. The age estimation results in a rather young age (phase~II) for this source, which agrees quite well with the age determination of \citetads{2017A&A...604A..60B} (source A13 is identical with their core N4) and the ratio of CH$_3$CN / CH$_3$OH. Neither the PCA nor the Kendall coefficient analysis identified any correlations with other sources.

% Sgr B2(N), A14:
% neighbors:
%   (A19)
% PCA:
%   CN6: A09, A14, A15, A18
% corresponding evolution phases:
%          I,   I,   I,   I
% Schwoerer:
%   -
\item \textbf{A14}: Source A14 is surrounded by an envelope where CCH, SO, SO$_2$, H$_2$CO, and CN play an important role. The corresponding core spectrum contains mainly CH$_3$OH, HNCO, SiO, H$_2 \! ^{34}$S, and H$_2$CO, where we find slightly lower velocities for sulfur-bearing molecules and slightly larger line widths for N- and O-bearing molecules, which could indicate multiple protostars within the source. (An additional filament was not identified by \citetads{2019A&A...628A...6S}). Like many other outer sources in Sgr~B2(N), this source is in an early stage of development (phase~I), as any signs of older age such as masers or outflows have not been identified. The PCA indicates correlations to sources A09, A15, and A18, which also appear to be quite young.

% Sgr B2(N), A15:
% neighbors:
%   A11 (A08)
% PCA:
%   CN6: A09, A14, A15, A18
% corresponding evolution phases:
%          I,   I,   I,   I
% Kendall
%   A09 - A15 (both cluster CN6)
% Schwoerer:
%   F08: A04, A05, A07, A08, A11, A15, A19
\item \textbf{A15}: The envelope around source A15 is dominated by CH$_3$OH, CH$_3$CN, OCS, SO, and HCCCN, whereas the core spectrum includes mostly OCS,v$_2$=1, H$_2$CS, and H$_2 \! ^{34}$S. Because only the sulfur-bearing molecules make a significant contribution to the emission, we cannot use the parameter distributions to make any statements about the source structure. However, according to \citetads{2019A&A...628A...6S}, this source is part of filament F08. Since this source also shows no signs of higher evolution (phase~I), it seems to be, like the neighboring source A11, very young. PCA groups this source into cluster~CN6, which contains sources A09, A14 and A18, all of which are in evolutionary phase~I.

% Sgr B2(N), A16:
% neighbors:
%   A18
% PCA:
%   CN1: A16, A19, A20
% corresponding evolution phases:
%          V,   I,   I
% Kendall:
%   A11 - A16 (A11: CN7, A16: CN1)
% Schwoerer:
%   -
\item \textbf{A16}: While the envelope spectrum around source A16 contains mainly HNCO, CH$_3$OH, H$_2$CO, SO$_2$, and HCCCN, the emission is dominated by CCH, H$_2 \! ^{34}$S, OCS, and SO. The source contains only two families of molecules, namely carbon and hydrocarbons and sulfur-bearing molecules, which have different line widths and velocities. The included \hii~region and maser indicate a highly evolved source (phase~V), which according to PCA has connections to sources A16, A19 and A20, although their similarities remain unclear as all three sources belong to different stages of evolution and A16 does not belong to any filament.

% Sgr B2(N), A17:
% neighbors:
%   A01, A07, (A06)
% PCA:
%   CN3: A01, A02, A17
% corresponding evolution phases:
%         IV, III,   ?
% Schwoerer:
%   F01: A02, A03, A06, A17
\item \textbf{A17}: The envelope surrounding source A17 has a composition (CN, CCH, SiO, H$_2$CO, and CH$_2$NH) similar to that of the envelope around the neighboring source A01. The core spectrum also shows great similarity to the neighbouring source A07, although the weighting of the individual molecules (CH$_3$OH, SO, H$_2$CO, CH$_3$OCH$_3$, and CH$_3$OCHO) differs in some cases. In particular, the distribution of the line widths in the core indicates an additional structure that can be identified with filament F01 \citepads{2019A&A...628A...6S}. This is also underpinned by the PCA, which classifies source A17 together with sources A01, A02, which are also contained in filament F01, in cluster CN3. In addition, the chemical clocks indicate an intermediate to high age. Unfortunately, the evolutionary phase analysis does not allow an age estimate here, as only non-Gaussian line shapes are found, which cannot be interpreted alone. It is possible that these line shapes are not caused by outflows but have other causes, such as rotations or infall motions, so that this source could also be young.

% Sgr B2(N), A18:
% neighbors:
%   A16
% PCA:
%   CN6: A09, A14, A15, A18
% corresponding evolution phases:
%          I,   I,   I,   I
% Schwoerer:
%   -
\item \textbf{A18}: Although source A18 is located directly next to source A16, there are differences between the two envelope spectra. In addition to CH$_3$OH and SO$_2$, which can be found in both envelopes, there is also HCCCN, CN, and CCH around A18. Apart from SO, the core spectrum of A18 contains mainly CH$_3$CN, CN, SiO, and H$_2$CO. As with the neighboring source A16, the distributions of the fit parameters for source A18 also indicate a more complex source structure. However, \citetads{2019A&A...628A...6S} could not find a filament in connection with source A18. In addition, the evolutionary analysis categorises this source as very young (phase~I). Moreover, the PCA also describes correlations with the other, similarly young sources A09 and A15.

% Sgr B2(N), A19:
% neighbors:
%   A08, (A14)
% PCA:
%   CN1: A16, A19, A20
% corresponding evolution phases:
%          V,   I,   I
% Schwoerer:
%   F08: A04, A05, A07, A08, A11, A15, A19
\item \textbf{A19}: The envelope around A19 shows only few similarities with the neighbouring source A08 and is mainly dominated by OCS, HCCCN, CN, CCH, and CH$_3$OH. The core spectrum of source A09 shows no agreement with source A18, except that CH$_3$OH is contained in both core spectra. Particularly important compounds in this context are CCH, HCCCN, SO, and H$_2$CS. Source A19 shows the influence of filament F08 \citepads{2019A&A...628A...6S} through different temperatures and line widths for the different molecular families. The evolutionary phase analysis yields a very young age (phase~I). In addition, the PCA identified correlations to the sources A16, A19, and A20, although all three sources belong to different stages of evolution and the other sources do not belong to filament F08.

% Sgr B2(N), A20:
% neighbors:
%   (A10)
% PCA:
%   CN1: A16, A19, A20
% corresponding evolution phases:
%          V,   I,   I
% Schwoerer:
%   F05: A20
\item \textbf{A20}: In addition to CH$_3$OH, H$_2$CO, SO$_2$, and CCH, which are also found around many other sources, the envelope around source A20 also shows contributions from HC(O)NH$_2$, which were only found around source A25 in Sgr~B2(M). The emission of A20 is dominated by CH$_3$OH, CH$_3$OCHO, OCS, H$_2$CO, and H$_2 \! ^{34}$S. The distributions of the fitting parameters may show a contribution of the filament F05 \citepads{2019A&A...628A...6S}. In addition, the analysis of the chemical clocks gives an inconsistent picture for this source. The ratio of CH$_3$CN / CH$_3$OH indicates a middle age, while SO / OCS suggests a young age. This classification is supported by the absence of masers, outflows, and \hii~regions (phase~I). As with the previous source A19, the PCA finds correlations with sources A16 and A19, the origin of which remains unclear.
\end{itemize}

%===============================================================================
% LTE parameters
\twocolumn
\section{LTE parameters}\label{app:sec:Parameters}
\subsection{Sgr~B2(M)}\label{app:subsec:ParametersM}
%
% Please add the following statements to the latex header:
% \usepackage{supertabular,booktabs}
% \usepackage{array}
% \newcolumntype{C}[1]{>{\centering\arraybackslash}p{#1}}

%================================================================================
%
% Source A01

%---------------------------------------
% Core Components

\tablefirsthead{%
\hline
\hline
Molecule      & $\theta^{m,c}$ & T$_{\rm ex}^{m,c}$ & N$_{\rm tot}^{m,c}$   & $\Delta$ v$^{m,c}$ & v$_{\rm LSR}^{m,c}$\\
              & ($\arcsec$)    & (K)                & (cm$^{-2}$)           & (km~s$^{-1}$)      & (km~s$^{-1}$)      \\
\hline
}

\tablehead{%
\multicolumn{6}{c}{(Continued)}\\
\hline
\hline
Molecule      & $\theta^{m,c}$ & T$_{\rm ex}^{m,c}$ & N$_{\rm tot}^{m,c}$   & $\Delta$ v$^{m,c}$ & v$_{\rm LSR}^{m,c}$\\
              & ($\arcsec$)    & (K)                & (cm$^{-2}$)           & (km~s$^{-1}$)      & (km~s$^{-1}$)      \\
\hline
}

\tabletail{%
\hline
\hline
}

\topcaption{LTE Parameters for the full LTE model (Core Components) for source A01 in Sgr~B2(M).}
\tiny
\centering
\begin{supertabular}{lcccC{1cm}C{1cm}}\label{CoreLTE:parameters:A01SgrB2M}\\
RRL-H         &    1.7         &               3685 &  2.7 $\times 10^{9}$  &               45.7 &                  64\\
H$_2 \! ^{34}$S &    1.7         &                497 &  3.5 $\times 10^{18}$ &                6.2 &                  64\\
NH$_2$D       &    1.7         &                268 &  1.2 $\times 10^{18}$ &                8.5 &                  55\\
SO$_2$        &    1.7         &                265 &  1.2 $\times 10^{12}$ &                5.4 &                  66\\
              &    1.7         &                648 &  4.4 $\times 10^{19}$ &                9.9 &                  63\\
OCS           &    1.7         &                612 &  2.4 $\times 10^{18}$ &                9.9 &                  63\\
CCCS          &    1.7         &                258 &  6.2 $\times 10^{15}$ &                7.4 &                  63\\
NO            &    1.7         &                293 &  7.2 $\times 10^{19}$ &                7.1 &                  63\\
CH$_3$CN      &    1.7         &                582 &  1.3 $\times 10^{13}$ &               11.8 &                  64\\
              &    1.7         &                432 &  4.4 $\times 10^{17}$ &                7.8 &                  64\\
HCCCN         &    1.7         &                536 &  2.4 $\times 10^{18}$ &                7.7 &                  65\\
HCCCN, v$_7$=2 &    1.7         &                568 &  9.0 $\times 10^{17}$ &               11.9 &                  64\\
HC(O)NH$_2$   &    1.7         &                595 &  3.0 $\times 10^{17}$ &               10.6 &                  63\\
CN            &    1.7         &                183 &  1.2 $\times 10^{12}$ &                2.5 &                -124\\
H$_2$CO       &    1.7         &                325 &  6.9 $\times 10^{18}$ &                6.4 &                  63\\
HNCO          &    1.7         &                543 &  9.5 $\times 10^{17}$ &               10.7 &                  63\\
SO            &    1.7         &                390 &  2.3 $\times 10^{18}$ &               33.6 &                  58\\
              &    1.7         &                432 &  1.8 $\times 10^{12}$ &                3.5 &                  66\\
              &    1.7         &                174 &  1.2 $\times 10^{12}$ &                2.8 &                  64\\
              &    1.7         &                903 &  3.0 $\times 10^{19}$ &               10.5 &                  64\\
H$_2$CCO      &    1.7         &                246 &  5.6 $\times 10^{17}$ &               11.7 &                  62\\
SiO           &    1.7         &                505 &  2.2 $\times 10^{17}$ &               17.0 &                  59\\
H$_2$CS       &    1.7         &                775 &  1.7 $\times 10^{17}$ &                6.6 &                  64\\
C$_2$H$_5$CN  &    1.7         &                381 &  2.2 $\times 10^{17}$ &                7.5 &                  64\\
C$_2$H$_5$CN, v$_{20}$=1 &    1.7         &                227 &  6.1 $\times 10^{17}$ &                8.3 &                  64\\
C$_2$H$_3$CN  &    1.7         &                405 &  6.2 $\times 10^{17}$ &                8.1 &                  64\\
C$_2$H$_3$CN, v$_{10}$=1 &    1.7         &                333 &  5.8 $\times 10^{17}$ &                8.3 &                  64\\
NO$^+$        &    1.7         &                285 &  9.7 $\times 10^{17}$ &                6.6 &                  58\\
CH$_2$NH      &    1.7         &                187 &  1.2 $\times 10^{12}$ &               13.7 &                  76\\
t-HCOOH       &    1.7         &                415 &  3.2 $\times 10^{17}$ &               11.5 &                  64\\
NH$_2$CN      &    1.7         &                260 &  1.3 $\times 10^{17}$ &               12.2 &                  69\\
H$_2$CNH      &    1.7         &                326 &  1.8 $\times 10^{17}$ &                6.1 &                  65\\
CH$_3$CHO     &    1.7         &                279 &  1.0 $\times 10^{17}$ &                6.4 &                  65\\
CH$_3$OCH$_3$ &    1.7         &                477 &  5.9 $\times 10^{16}$ &                6.1 &                  59\\
NS            &    1.7         &                528 &  3.1 $\times 10^{17}$ &                7.2 &                  62\\
H$_2$CCN      &    1.7         &                214 &  7.0 $\times 10^{16}$ &               16.2 &                  65\\
HCN, v$_2$=1  &    1.7         &                802 &  2.7 $\times 10^{18}$ &               14.8 &                  64\\
HNC, v$_2$=1  &    1.7         &               1100 &  1.2 $\times 10^{18}$ &               13.6 &                  64\\
CO            &    1.7         &                200 &  3.7 $\times 10^{14}$ &                4.6 &                 145\\
              &    1.7         &                200 &  1.6 $\times 10^{13}$ &                3.1 &                  76\\
$^{13}$CO     &    1.7         &                200 &  2.6 $\times 10^{19}$ &                2.7 &                  61\\
C$^{17}$O     &    1.7         &                200 &  1.3 $\times 10^{18}$ &                9.7 &                -156\\
              &    1.7         &                200 &  1.3 $\times 10^{17}$ &                4.6 &                -138\\
              &    1.7         &                200 &  2.5 $\times 10^{17}$ &                2.7 &                -103\\
              &    1.7         &                200 &  2.1 $\times 10^{18}$ &                7.5 &                  60\\
              &    1.7         &                200 &  3.1 $\times 10^{18}$ &               21.1 &                  86\\
              &    1.7         &                200 &  4.0 $\times 10^{15}$ &                2.7 &                 110\\
C$^{18}$O     &    1.7         &                200 &  2.2 $\times 10^{17}$ &                6.5 &                -178\\
              &    1.7         &                200 &  8.4 $\times 10^{17}$ &               11.9 &                -137\\
              &    1.7         &                200 &  3.1 $\times 10^{18}$ &                7.5 &                -120\\
              &    1.7         &                200 &  2.7 $\times 10^{18}$ &               12.5 &                 -93\\
              &    1.7         &                200 &  2.4 $\times 10^{18}$ &               11.9 &                 -69\\
              &    1.7         &                200 &  7.0 $\times 10^{17}$ &                2.8 &                 -55\\
              &    1.7         &                200 &  1.5 $\times 10^{18}$ &                6.9 &                 -36\\
              &    1.7         &                200 &  1.3 $\times 10^{17}$ &                2.7 &                 -25\\
              &    1.7         &                200 &  6.6 $\times 10^{18}$ &                9.1 &                   8\\
              &    1.7         &                200 &  4.4 $\times 10^{18}$ &                5.9 &                  82\\
              &    1.7         &                200 &  2.8 $\times 10^{18}$ &               10.2 &                  95\\
              &    1.7         &                200 &  1.0 $\times 10^{18}$ &                2.2 &                 109\\
              &    1.7         &                200 &  4.7 $\times 10^{18}$ &               10.8 &                 122\\
              &    1.7         &                200 &  5.0 $\times 10^{14}$ &                2.1 &                 133\\
              &    1.7         &                200 &  4.1 $\times 10^{18}$ &               10.5 &                 153\\
CS            &    1.7         &                200 &  1.2 $\times 10^{12}$ &                8.1 &                  71\\
              &    1.7         &                200 &  3.2 $\times 10^{15}$ &                4.0 &                  -1\\
              &    1.7         &                200 &  1.2 $\times 10^{12}$ &                7.1 &                  66\\
              &    1.7         &                200 &  9.9 $\times 10^{12}$ &               12.6 &                  15\\
              &    1.7         &                200 &  2.2 $\times 10^{13}$ &                0.6 &                  17\\
              &    1.7         &                200 &  2.9 $\times 10^{16}$ &               35.3 &                   1\\
$^{13}$CS     &    1.7         &                200 &  3.1 $\times 10^{15}$ &                9.1 &                -168\\
              &    1.7         &                200 &  2.1 $\times 10^{15}$ &               23.1 &                -132\\
              &    1.7         &                200 &  3.3 $\times 10^{14}$ &                2.1 &                -116\\
              &    1.7         &                200 &  5.2 $\times 10^{16}$ &               16.0 &                 -95\\
              &    1.7         &                200 &  2.2 $\times 10^{14}$ &                3.3 &                 -80\\
              &    1.7         &                200 &  1.4 $\times 10^{15}$ &                6.5 &                 -33\\
              &    1.7         &                200 &  2.6 $\times 10^{16}$ &               22.4 &                 -12\\
              &    1.7         &                200 &  1.4 $\times 10^{15}$ &                6.3 &                  -2\\
              &    1.7         &                200 &  3.3 $\times 10^{16}$ &               10.8 &                  29\\
              &    1.7         &                200 &  2.5 $\times 10^{14}$ &                4.4 &                  66\\
              &    1.7         &                200 &  6.4 $\times 10^{16}$ &                8.8 &                  97\\
              &    1.7         &                200 &  2.2 $\times 10^{15}$ &                3.8 &                 110\\
              &    1.7         &                200 &  7.8 $\times 10^{15}$ &                9.1 &                 126\\
              &    1.7         &                200 &  6.3 $\times 10^{14}$ &               14.7 &                 145\\
              &    1.7         &                200 &  4.7 $\times 10^{15}$ &                4.3 &                 161\\
C$^{33}$S     &    1.7         &                200 &  3.1 $\times 10^{14}$ &               10.9 &                -153\\
              &    1.7         &                200 &  7.1 $\times 10^{15}$ &               10.5 &                -124\\
              &    1.7         &                200 &  1.0 $\times 10^{15}$ &               10.3 &                 -94\\
              &    1.7         &                200 &  4.1 $\times 10^{15}$ &                6.4 &                 -69\\
              &    1.7         &                200 &  1.3 $\times 10^{15}$ &                6.4 &                 -42\\
              &    1.7         &                200 &  7.7 $\times 10^{14}$ &                3.2 &                   5\\
              &    1.7         &                200 &  5.1 $\times 10^{15}$ &                8.9 &                  64\\
              &    1.7         &                200 &  4.5 $\times 10^{15}$ &                6.9 &                  76\\
              &    1.7         &                200 &  4.5 $\times 10^{15}$ &                7.8 &                 112\\
              &    1.7         &                200 &  5.6 $\times 10^{14}$ &                6.1 &                 142\\
              &    1.7         &                200 &  2.1 $\times 10^{15}$ &                4.3 &                 157\\
C$^{34}$S     &    1.7         &                200 &  3.1 $\times 10^{15}$ &               13.7 &                -159\\
              &    1.7         &                200 &  6.8 $\times 10^{15}$ &               10.0 &                -122\\
              &    1.7         &                200 &  4.1 $\times 10^{15}$ &                9.6 &                 -97\\
              &    1.7         &                200 &  4.1 $\times 10^{15}$ &                8.2 &                 -77\\
              &    1.7         &                200 &  7.9 $\times 10^{15}$ &               10.4 &                 -45\\
              &    1.7         &                200 &  4.5 $\times 10^{16}$ &                8.8 &                  27\\
              &    1.7         &                200 &  5.6 $\times 10^{15}$ &                8.2 &                  70\\
              &    1.7         &                200 &  6.0 $\times 10^{15}$ &                9.0 &                 108\\
              &    1.7         &                200 &  5.8 $\times 10^{15}$ &                8.2 &                 128\\
              &    1.7         &                200 &  1.0 $\times 10^{14}$ &                8.9 &                 156\\
HCN           &    1.7         &                200 &  1.3 $\times 10^{12}$ &                1.4 &                 113\\
              &    1.7         &                200 &  1.2 $\times 10^{14}$ &               20.9 &                  76\\
              &    1.7         &                200 &  3.3 $\times 10^{16}$ &                8.2 &                  62\\
              &    1.7         &                200 &  1.0 $\times 10^{19}$ &                4.7 &                  35\\
              &    1.7         &                200 &  4.6 $\times 10^{16}$ &                2.4 &                 -43\\
              &    1.7         &                200 &  5.3 $\times 10^{15}$ &                5.4 &                 -61\\
H$^{13}$CN    &    1.7         &                200 &  3.3 $\times 10^{15}$ &               47.4 &                -160\\
              &    1.7         &                200 &  2.2 $\times 10^{15}$ &                5.9 &                -121\\
              &    1.7         &                200 &  1.6 $\times 10^{15}$ &               11.5 &                 -32\\
              &    1.7         &                200 &  5.4 $\times 10^{15}$ &               21.4 &                  -7\\
              &    1.7         &                200 &  3.7 $\times 10^{14}$ &                5.9 &                  34\\
              &    1.7         &                200 &  3.6 $\times 10^{14}$ &                6.4 &                  79\\
              &    1.7         &                200 &  6.1 $\times 10^{14}$ &                1.9 &                  86\\
              &    1.7         &                200 &  6.4 $\times 10^{14}$ &                3.5 &                  95\\
              &    1.7         &                200 &  1.3 $\times 10^{15}$ &                7.9 &                 117\\
              &    1.7         &                200 &  2.6 $\times 10^{13}$ &               24.5 &                 145\\
HNC           &    1.7         &                200 &  2.6 $\times 10^{15}$ &                1.0 &                  27\\
              &    1.7         &                200 &  4.2 $\times 10^{14}$ &                9.6 &                  38\\
              &    1.7         &                200 &  5.1 $\times 10^{13}$ &                5.5 &                  80\\
HN$^{13}$C    &    1.7         &                200 &  4.5 $\times 10^{14}$ &               11.8 &                -142\\
              &    1.7         &                200 &  7.4 $\times 10^{15}$ &               32.0 &                -109\\
              &    1.7         &                200 &  2.4 $\times 10^{14}$ &               12.5 &                 -54\\
              &    1.7         &                200 &  3.5 $\times 10^{14}$ &               29.5 &                 -23\\
              &    1.7         &                200 &  8.9 $\times 10^{14}$ &                2.8 &                 -15\\
              &    1.7         &                200 &  5.5 $\times 10^{14}$ &                7.2 &                  11\\
              &    1.7         &                200 &  2.7 $\times 10^{14}$ &                6.7 &                -184\\
              &    1.7         &                200 &  2.1 $\times 10^{14}$ &                8.6 &                  67\\
              &    1.7         &                200 &  8.2 $\times 10^{13}$ &                3.7 &                  99\\
              &    1.7         &                200 &  8.7 $\times 10^{13}$ &                5.3 &                 115\\
              &    1.7         &                200 &  9.1 $\times 10^{12}$ &                7.1 &                 129\\
\end{supertabular}\\
\vspace{1cm}

%---------------------------------------
% Envelope Components

\tablefirsthead{%
\hline
\hline
Molecule      & $\theta^{m,c}$ & T$_{\rm ex}^{m,c}$ & N$_{\rm tot}^{m,c}$   & $\Delta$ v$^{m,c}$ & v$_{\rm LSR}^{m,c}$\\
              & ($\arcsec$)    & (K)                & (cm$^{-2}$)           & (km~s$^{-1}$)      & (km~s$^{-1}$)      \\
\hline
}

\tablehead{%
\multicolumn{6}{c}{(Continued)}\\
\hline
\hline
Molecule      & $\theta^{m,c}$ & T$_{\rm ex}^{m,c}$ & N$_{\rm tot}^{m,c}$   & $\Delta$ v$^{m,c}$ & v$_{\rm LSR}^{m,c}$\\
              & ($\arcsec$)    & (K)                & (cm$^{-2}$)           & (km~s$^{-1}$)      & (km~s$^{-1}$)      \\
\hline
}

\tabletail{%
\hline
\hline
}

\topcaption{LTE Parameters for the full LTE model (Envelope Components) for source A01 in Sgr~B2(M).}
\tiny
\centering
\begin{supertabular}{lcccC{1cm}C{1cm}}\label{EnvLTE:parameters:A01SgrB2M}\\
CCH           & ext.           &                  3 &  2.4 $\times 10^{16}$ &               14.6 &                  64\\
CH$_3$CN      & ext.           &                  3 &  1.3 $\times 10^{19}$ &                6.0 &                  97\\
              & ext.           &                  7 &  1.2 $\times 10^{12}$ &                5.5 &                  68\\
              & ext.           &                143 &  2.3 $\times 10^{16}$ &               13.3 &                  49\\
CN            & ext.           &                  9 &  7.8 $\times 10^{15}$ &                8.2 &                  67\\
              & ext.           &                 16 &  2.3 $\times 10^{16}$ &               11.5 &                  60\\
              & ext.           &                  3 &  1.0 $\times 10^{15}$ &               20.6 &                  43\\
              & ext.           &                  8 &  1.2 $\times 10^{15}$ &                0.3 &                  29\\
              & ext.           &                 11 &  1.5 $\times 10^{15}$ &               19.6 &                   7\\
              & ext.           &                 16 &  1.6 $\times 10^{15}$ &                3.6 &                   0\\
              & ext.           &                 66 &  1.3 $\times 10^{16}$ &               10.1 &                 -25\\
              & ext.           &                 20 &  9.8 $\times 10^{14}$ &                2.0 &                 -43\\
              & ext.           &                 82 &  2.5 $\times 10^{16}$ &                6.4 &                 -47\\
              & ext.           &                 10 &  1.0 $\times 10^{14}$ &                3.2 &                 -79\\
              & ext.           &                  8 &  5.8 $\times 10^{14}$ &               14.4 &                -102\\
H$_2$CO       & ext.           &                 23 &  8.0 $\times 10^{14}$ &                1.8 &                  73\\
              & ext.           &                 31 &  1.4 $\times 10^{16}$ &               11.6 &                  63\\
              & ext.           &                 76 &  4.5 $\times 10^{16}$ &               10.1 &                  51\\
SO            & ext.           &                 18 &  5.5 $\times 10^{15}$ &               17.0 &                  54\\
CH$_3$OH      & ext.           &                191 &  1.0 $\times 10^{12}$ &                8.5 &                 112\\
              & ext.           &                  7 &  5.0 $\times 10^{15}$ &                7.1 &                  66\\
              & ext.           &                 17 &  4.4 $\times 10^{15}$ &               10.7 &                  58\\
SiO           & ext.           &                 21 &  3.4 $\times 10^{15}$ &                0.7 &                  54\\
              & ext.           &                 18 &  1.1 $\times 10^{15}$ &               15.3 &                  55\\
CH$_2$NH      & ext.           &                  5 &  3.6 $\times 10^{13}$ &                1.3 &                  73\\
              & ext.           &                  3 &  3.0 $\times 10^{13}$ &               10.0 &                -186\\
              & ext.           &                  7 &  1.0 $\times 10^{15}$ &               13.7 &                  61\\
              & ext.           &                  3 &  4.8 $\times 10^{12}$ &                2.3 &                  30\\
              & ext.           &                  8 &  1.7 $\times 10^{13}$ &                4.5 &                  16\\
              & ext.           &                  3 &  2.2 $\times 10^{13}$ &               12.7 &                   3\\
CH$_3$NH$_2$  & ext.           &                  8 &  1.2 $\times 10^{12}$ &               31.1 &                  64\\
              & ext.           &                  3 &  1.2 $\times 10^{15}$ &               18.2 &                  61\\
CO            & ext.           &                  3 &  3.0 $\times 10^{16}$ &               18.2 &                  97\\
              & ext.           &                  3 &  5.5 $\times 10^{15}$ &                8.5 &                  88\\
              & ext.           &                  3 &  1.2 $\times 10^{17}$ &                2.0 &                  74\\
              & ext.           &                  3 &  1.7 $\times 10^{12}$ &                1.7 &                -184\\
              & ext.           &                  3 &  2.1 $\times 10^{17}$ &                7.7 &                  65\\
              & ext.           &                  3 &  2.5 $\times 10^{15}$ &                1.1 &                  53\\
              & ext.           &                  3 &  1.2 $\times 10^{17}$ &                9.5 &                  54\\
              & ext.           &                  3 &  2.8 $\times 10^{13}$ &                0.4 &                  71\\
              & ext.           &                  3 &  1.7 $\times 10^{14}$ &                0.8 &                  17\\
              & ext.           &                  3 &  3.4 $\times 10^{16}$ &               22.3 &                  45\\
              & ext.           &                  3 &  4.4 $\times 10^{16}$ &                2.8 &                  30\\
              & ext.           &                  3 &  3.5 $\times 10^{17}$ &               26.5 &                   9\\
              & ext.           &                  3 &  5.0 $\times 10^{12}$ &                2.7 &                   4\\
              & ext.           &                  3 &  1.5 $\times 10^{16}$ &                2.7 &                  -4\\
              & ext.           &                  3 &  1.2 $\times 10^{13}$ &                1.2 &                  -6\\
              & ext.           &                  3 &  8.4 $\times 10^{15}$ &                0.5 &                  -6\\
              & ext.           &                  3 &  3.0 $\times 10^{13}$ &                0.7 &                 -16\\
              & ext.           &                  3 &  1.5 $\times 10^{17}$ &               10.9 &                 -23\\
              & ext.           &                  3 & 10.0 $\times 10^{18}$ &                1.9 &                 -45\\
              & ext.           &                  3 &  1.2 $\times 10^{12}$ &                1.7 &                 -71\\
              & ext.           &                  3 &  1.8 $\times 10^{17}$ &               38.5 &                 -53\\
              & ext.           &                  3 &  3.7 $\times 10^{16}$ &                4.1 &                 -77\\
              & ext.           &                  3 &  3.2 $\times 10^{12}$ &                0.7 &                -144\\
              & ext.           &                  3 &  4.1 $\times 10^{12}$ &                0.9 &                 -58\\
              & ext.           &                  3 &  2.3 $\times 10^{17}$ &               19.6 &                -101\\
              & ext.           &                  3 &  3.5 $\times 10^{16}$ &                1.7 &                -111\\
              & ext.           &                  3 &  1.4 $\times 10^{14}$ &                0.2 &                 -66\\
              & ext.           &                  3 &  2.7 $\times 10^{14}$ &                5.2 &                -172\\
              & ext.           &                  3 &  8.2 $\times 10^{14}$ &                2.9 &                -129\\
$^{13}$CO     & ext.           &                  3 &  6.7 $\times 10^{15}$ &               10.4 &                  89\\
              & ext.           &                  3 &  2.2 $\times 10^{16}$ &                1.2 &                  73\\
              & ext.           &                  3 &  7.3 $\times 10^{12}$ &                2.6 &                  60\\
              & ext.           &                  3 &  3.0 $\times 10^{16}$ &                5.1 &                  67\\
              & ext.           &                  3 &  1.2 $\times 10^{12}$ &                0.2 &                  61\\
              & ext.           &                  3 &  1.8 $\times 10^{16}$ &                3.0 &                  53\\
              & ext.           &                  3 &  7.3 $\times 10^{15}$ &                0.4 &                  43\\
              & ext.           &                  3 &  1.5 $\times 10^{17}$ &               21.5 &                  60\\
              & ext.           &                  3 &  6.1 $\times 10^{15}$ &                2.9 &                  30\\
              & ext.           &                  3 &  4.0 $\times 10^{16}$ &               24.7 &                   8\\
              & ext.           &                  3 &  3.2 $\times 10^{15}$ &                2.3 &                  12\\
              & ext.           &                  3 &  9.3 $\times 10^{15}$ &                6.0 &                  -2\\
              & ext.           &                  3 &  1.2 $\times 10^{12}$ &                0.6 &                 -29\\
              & ext.           &                  3 &  4.2 $\times 10^{16}$ &               18.1 &                 -26\\
              & ext.           &                  3 &  1.8 $\times 10^{17}$ &                1.0 &                 -43\\
              & ext.           &                  3 &  2.9 $\times 10^{16}$ &                3.7 &                 -49\\
              & ext.           &                  3 &  5.7 $\times 10^{15}$ &               10.7 &                 -62\\
              & ext.           &                  3 &  5.8 $\times 10^{15}$ &                4.3 &                 -79\\
              & ext.           &                  3 &  3.8 $\times 10^{15}$ &                1.7 &                -101\\
              & ext.           &                  3 &  9.1 $\times 10^{15}$ &                4.0 &                -106\\
              & ext.           &                  3 &  2.4 $\times 10^{15}$ &                4.9 &                -113\\
              & ext.           &                  3 &  7.0 $\times 10^{12}$ &                0.2 &                -114\\
              & ext.           &                  3 &  1.2 $\times 10^{12}$ &                1.7 &                 -74\\
C$^{17}$O     & ext.           &                  3 & 10.0 $\times 10^{14}$ &                1.7 &                -180\\
              & ext.           &                  3 &  3.7 $\times 10^{14}$ &                2.0 &                -167\\
              & ext.           &                  3 &  4.3 $\times 10^{14}$ &                7.1 &                -146\\
              & ext.           &                  3 &  2.2 $\times 10^{15}$ &               13.6 &                -119\\
              & ext.           &                  3 &  1.2 $\times 10^{14}$ &                2.5 &                -100\\
              & ext.           &                  3 &  1.2 $\times 10^{14}$ &                2.0 &                 -95\\
              & ext.           &                  3 &  2.5 $\times 10^{14}$ &                5.4 &                 -87\\
              & ext.           &                  3 &  3.9 $\times 10^{14}$ &                5.1 &                 -79\\
              & ext.           &                  3 &  5.0 $\times 10^{14}$ &                6.8 &                 -68\\
              & ext.           &                  3 &  1.8 $\times 10^{15}$ &                9.2 &                 -44\\
              & ext.           &                  3 &  3.1 $\times 10^{14}$ &                2.9 &                 -23\\
              & ext.           &                  3 &  1.3 $\times 10^{14}$ &                3.8 &                 -15\\
              & ext.           &                  3 &  7.1 $\times 10^{14}$ &               22.4 &                   1\\
              & ext.           &                  3 &  3.3 $\times 10^{14}$ &                7.1 &                   9\\
              & ext.           &                  3 &  1.5 $\times 10^{14}$ &                3.3 &                  22\\
              & ext.           &                  3 &  6.4 $\times 10^{14}$ &                6.2 &                  28\\
              & ext.           &                  3 &  5.6 $\times 10^{15}$ &                7.7 &                  52\\
              & ext.           &                  3 &  2.5 $\times 10^{15}$ &               16.1 &                  73\\
              & ext.           &                  3 &  6.5 $\times 10^{14}$ &                8.9 &                  94\\
              & ext.           &                  3 &  1.0 $\times 10^{14}$ &                3.4 &                 108\\
              & ext.           &                  3 &  4.3 $\times 10^{14}$ &                9.1 &                 119\\
              & ext.           &                  3 &  2.9 $\times 10^{14}$ &                5.1 &                 132\\
              & ext.           &                  3 &  3.7 $\times 10^{14}$ &               12.5 &                 144\\
C$^{18}$O     & ext.           &                  3 &  5.8 $\times 10^{15}$ &                3.9 &                 -44\\
              & ext.           &                  3 &  1.5 $\times 10^{14}$ &                5.9 &                 -28\\
              & ext.           &                  3 &  1.8 $\times 10^{15}$ &                5.3 &                   0\\
              & ext.           &                  3 &  1.2 $\times 10^{15}$ &                5.0 &                  30\\
              & ext.           &                  3 &  1.4 $\times 10^{16}$ &                8.1 &                  52\\
              & ext.           &                  3 &  1.3 $\times 10^{16}$ &               10.9 &                  68\\
CS            & ext.           &                  3 &  1.9 $\times 10^{17}$ &                0.6 &                  74\\
              & ext.           &                  3 &  5.3 $\times 10^{16}$ &                0.6 &                  66\\
              & ext.           &                  3 &  3.8 $\times 10^{17}$ &               15.3 &                  59\\
              & ext.           &                  3 &  1.6 $\times 10^{12}$ &                9.0 &                  52\\
              & ext.           &                  3 &  1.3 $\times 10^{12}$ &                4.4 &                  47\\
              & ext.           &                  3 &  1.2 $\times 10^{12}$ &                5.1 &                 -89\\
              & ext.           &                  3 &  5.1 $\times 10^{15}$ &                1.7 &                   1\\
$^{13}$CS     & ext.           &                  3 &  1.1 $\times 10^{14}$ &               10.6 &                -158\\
              & ext.           &                  3 &  3.6 $\times 10^{13}$ &                2.5 &                -144\\
              & ext.           &                  3 &  7.6 $\times 10^{15}$ &               18.8 &                 -64\\
              & ext.           &                  3 &  5.6 $\times 10^{14}$ &                6.4 &                  14\\
              & ext.           &                  3 &  1.5 $\times 10^{16}$ &               10.4 &                  50\\
              & ext.           &                  3 &  3.8 $\times 10^{14}$ &                6.5 &                  84\\
              & ext.           &                  3 &  3.3 $\times 10^{14}$ &                2.8 &                 150\\
C$^{33}$S     & ext.           &                  3 &  1.3 $\times 10^{16}$ &                6.5 &                -168\\
              & ext.           &                  3 &  3.6 $\times 10^{15}$ &                7.1 &                -106\\
              & ext.           &                  3 &  1.2 $\times 10^{15}$ &                9.5 &                 -25\\
              & ext.           &                  3 &  3.8 $\times 10^{14}$ &                6.0 &                 -15\\
              & ext.           &                  3 &  1.7 $\times 10^{15}$ &               17.6 &                  27\\
              & ext.           &                  3 &  8.8 $\times 10^{15}$ &               12.8 &                  49\\
              & ext.           &                  3 &  1.7 $\times 10^{14}$ &                3.0 &                 132\\
C$^{34}$S     & ext.           &                  3 &  1.0 $\times 10^{15}$ &                9.2 &                 -55\\
              & ext.           &                  3 &  5.4 $\times 10^{15}$ &               31.5 &                 -15\\
              & ext.           &                  3 &  6.5 $\times 10^{14}$ &               10.5 &                   0\\
              & ext.           &                  3 &  3.4 $\times 10^{16}$ &               12.6 &                  50\\
              & ext.           &                  3 &  6.8 $\times 10^{14}$ &                7.0 &                  93\\
              & ext.           &                  3 &  4.2 $\times 10^{15}$ &                5.5 &                 137\\
HCN           & ext.           &                  3 &  1.7 $\times 10^{13}$ &                0.8 &                  72\\
              & ext.           &                  3 &  3.7 $\times 10^{17}$ &                4.1 &                  70\\
              & ext.           &                  3 &  2.2 $\times 10^{15}$ &               14.2 &                  55\\
              & ext.           &                  3 &  5.6 $\times 10^{14}$ &                5.8 &                  39\\
              & ext.           &                  3 &  1.2 $\times 10^{14}$ &                1.3 &                  30\\
              & ext.           &                  3 &  2.1 $\times 10^{14}$ &                7.2 &                  15\\
              & ext.           &                  3 &  1.3 $\times 10^{15}$ &                9.2 &                   2\\
              & ext.           &                  3 &  5.6 $\times 10^{13}$ &                2.9 &                 -79\\
H$^{13}$CN    & ext.           &                  3 &  1.6 $\times 10^{15}$ &               16.2 &                  58\\
              & ext.           &                  3 &  5.9 $\times 10^{13}$ &                5.2 &                 128\\
HNC           & ext.           &                  3 &  5.6 $\times 10^{14}$ &                5.0 &                   1\\
              & ext.           &                  3 &  1.1 $\times 10^{16}$ &               11.7 &                  63\\
HN$^{13}$C    & ext.           &                  3 &  9.0 $\times 10^{13}$ &                7.0 &                  57\\
H$^{13}$CO$^+$ & ext.           &                  3 &  1.2 $\times 10^{14}$ &                8.6 &                  64\\
              & ext.           &                  3 &  1.9 $\times 10^{14}$ &               17.5 &                  55\\
              & ext.           &                  3 &  1.4 $\times 10^{13}$ &                1.6 &                  74\\
HO$^{13}$C$^+$ & ext.           &                  3 &  9.0 $\times 10^{12}$ &                3.9 &                  46\\
              & ext.           &                  3 &  4.0 $\times 10^{13}$ &                8.3 &                  39\\
\end{supertabular}\\
\vspace{1cm}

%================================================================================
%
% Source A02

%---------------------------------------
% Core Components

\tablefirsthead{%
\hline
\hline
Molecule      & $\theta^{m,c}$ & T$_{\rm ex}^{m,c}$ & N$_{\rm tot}^{m,c}$   & $\Delta$ v$^{m,c}$ & v$_{\rm LSR}^{m,c}$\\
              & ($\arcsec$)    & (K)                & (cm$^{-2}$)           & (km~s$^{-1}$)      & (km~s$^{-1}$)      \\
\hline
}

\tablehead{%
\multicolumn{6}{c}{(Continued)}\\
\hline
\hline
Molecule      & $\theta^{m,c}$ & T$_{\rm ex}^{m,c}$ & N$_{\rm tot}^{m,c}$   & $\Delta$ v$^{m,c}$ & v$_{\rm LSR}^{m,c}$\\
              & ($\arcsec$)    & (K)                & (cm$^{-2}$)           & (km~s$^{-1}$)      & (km~s$^{-1}$)      \\
\hline
}

\tabletail{%
\hline
\hline
}

\topcaption{LTE Parameters for the full LTE model (Core Components) for source A02 in Sgr~B2(M).}
\tiny
\centering
\begin{supertabular}{lcccC{1cm}C{1cm}}\label{CoreLTE:parameters:A02SgrB2M}\\
H$_2 \! ^{34}$S &    1.9         &                342 &  2.3 $\times 10^{18}$ &                9.1 &                  62\\
NH$_2$D       &    1.9         &                253 &  5.0 $\times 10^{16}$ &                9.7 &                  58\\
CCCS          &    1.9         &                173 &  1.4 $\times 10^{15}$ &                6.1 &                  61\\
H$_2$CS       &    1.9         &                726 &  5.2 $\times 10^{16}$ &                7.7 &                  60\\
t-HCOOH       &    1.9         &                496 &  4.0 $\times 10^{16}$ &               13.0 &                  64\\
NS            &    1.9         &                554 &  1.5 $\times 10^{17}$ &                8.8 &                  57\\
H$_2$CCN      &    1.9         &                236 &  2.3 $\times 10^{12}$ &                5.7 &                  62\\
CN            &    1.9         &                172 &  2.9 $\times 10^{12}$ &                3.0 &                -100\\
CH$_3$CN      &    1.9         &                221 &  6.3 $\times 10^{12}$ &                3.9 &                 124\\
              &    1.9         &                212 &  5.9 $\times 10^{12}$ &                5.5 &                 120\\
              &    1.9         &                207 &  1.3 $\times 10^{16}$ &                9.3 &                  74\\
              &    1.9         &                271 &  1.5 $\times 10^{17}$ &                8.1 &                  61\\
CCH           &    1.9         &                 27 &  1.9 $\times 10^{15}$ &                5.9 &                  55\\
NO            &    1.9         &                476 &  7.4 $\times 10^{18}$ &                3.8 &                  62\\
              &    1.9         &                180 &  2.6 $\times 10^{19}$ &               12.1 &                  57\\
HCCCN         &    1.9         &                448 &  3.3 $\times 10^{17}$ &                4.6 &                  62\\
              &    1.9         &                264 &  2.8 $\times 10^{17}$ &               12.3 &                  57\\
HCCCN, v$_7$=2 &    1.9         &                337 &  2.0 $\times 10^{17}$ &                5.0 &                  62\\
              &    1.9         &                248 &  2.3 $\times 10^{17}$ &               10.0 &                  55\\
SiO           &    1.9         &                549 &  3.9 $\times 10^{16}$ &               46.0 &                  57\\
OCS           &    1.9         &                404 &  3.8 $\times 10^{17}$ &                5.4 &                  64\\
              &    1.9         &                458 &  8.5 $\times 10^{17}$ &               14.3 &                  57\\
HC(O)NH$_2$   &    1.9         &                697 &  5.9 $\times 10^{16}$ &                7.7 &                  59\\
SO$_2$        &    1.9         &                434 &  2.9 $\times 10^{19}$ &               13.1 &                  59\\
              &    1.9         &                700 &  4.0 $\times 10^{12}$ &                5.7 &                  58\\
C$_2$H$_5$CN  &    1.9         &                188 &  1.1 $\times 10^{17}$ &                6.1 &                  61\\
C$_2$H$_5$CN, v$_{20}$=1 &    1.9         &                443 &  8.4 $\times 10^{16}$ &                6.9 &                  61\\
C$_2$H$_3$CN  &    1.9         &                312 &  1.2 $\times 10^{17}$ &                6.4 &                  61\\
C$_2$H$_3$CN, v$_{10}$=1 &    1.9         &                312 &  1.1 $\times 10^{17}$ &                6.4 &                  61\\
SO            &    1.9         &                435 &  1.7 $\times 10^{18}$ &                9.0 &                  67\\
              &    1.9         &                566 &  1.9 $\times 10^{16}$ &                3.2 &                  64\\
HNCO          &    1.9         &                355 &  2.1 $\times 10^{17}$ &                5.9 &                  62\\
              &    1.9         &                319 &  2.3 $\times 10^{17}$ &               10.8 &                  51\\
H$_2$CO       &    1.9         &                322 &  2.2 $\times 10^{16}$ &                6.5 &                  64\\
HCN, v$_2$=1  &    1.9         &                690 &  1.2 $\times 10^{18}$ &               13.6 &                  60\\
              &    1.9         &                279 &  3.0 $\times 10^{18}$ &               17.2 &                  51\\
H$_2$CCO      &    1.9         &                138 &  5.8 $\times 10^{16}$ &               11.3 &                  59\\
HNC, v$_2$=1  &    1.9         &                467 &  3.3 $\times 10^{17}$ &               20.2 &                  59\\
              &    1.9         &                513 &  1.8 $\times 10^{17}$ &                6.0 &                  63\\
CO            &    1.9         &                200 & 10.0 $\times 10^{19}$ &                3.3 &                  63\\
              &    1.9         &                200 &  7.3 $\times 10^{17}$ &                6.2 &                 128\\
              &    1.9         &                200 &  2.3 $\times 10^{19}$ &               10.1 &                  41\\
              &    1.9         &                200 &  7.5 $\times 10^{12}$ &                1.0 &                  37\\
$^{13}$CO     &    1.9         &                200 &  9.6 $\times 10^{19}$ &                2.1 &                  61\\
              &    1.9         &                200 &  1.4 $\times 10^{19}$ &               10.9 &                  43\\
              &    1.9         &                200 &  2.6 $\times 10^{17}$ &                1.1 &                  36\\
C$^{17}$O     &    1.9         &                200 &  2.2 $\times 10^{17}$ &                9.6 &                -158\\
              &    1.9         &                200 &  3.3 $\times 10^{16}$ &                2.4 &                -148\\
              &    1.9         &                200 &  2.7 $\times 10^{17}$ &                9.7 &                -139\\
              &    1.9         &                200 &  3.4 $\times 10^{16}$ &                2.9 &                -111\\
              &    1.9         &                200 &  2.7 $\times 10^{17}$ &               11.4 &                  44\\
              &    1.9         &                200 &  6.5 $\times 10^{18}$ &                4.3 &                  62\\
              &    1.9         &                200 &  4.4 $\times 10^{17}$ &                9.2 &                 160\\
C$^{18}$O     &    1.9         &                200 &  1.9 $\times 10^{15}$ &                7.4 &                -139\\
              &    1.9         &                200 &  3.1 $\times 10^{14}$ &                8.5 &                -120\\
              &    1.9         &                200 &  4.1 $\times 10^{17}$ &               18.6 &                 -98\\
              &    1.9         &                200 &  3.7 $\times 10^{17}$ &                9.7 &                 -72\\
              &    1.9         &                200 &  2.4 $\times 10^{17}$ &                4.6 &                 -58\\
              &    1.9         &                200 &  4.7 $\times 10^{17}$ &               10.8 &                 -14\\
              &    1.9         &                200 &  1.1 $\times 10^{18}$ &               12.4 &                  43\\
              &    1.9         &                200 &  3.3 $\times 10^{17}$ &               17.9 &                  87\\
              &    1.9         &                200 &  7.6 $\times 10^{16}$ &                2.1 &                  99\\
              &    1.9         &                200 &  4.4 $\times 10^{17}$ &                9.8 &                 116\\
              &    1.9         &                200 &  4.9 $\times 10^{17}$ &                8.0 &                 128\\
              &    1.9         &                200 &  4.5 $\times 10^{17}$ &                9.0 &                 150\\
CS            &    1.9         &                200 &  8.9 $\times 10^{12}$ &                2.2 &                 105\\
              &    1.9         &                200 &  2.7 $\times 10^{14}$ &                2.0 &                  92\\
              &    1.9         &                200 &  4.1 $\times 10^{15}$ &               28.0 &                  86\\
              &    1.9         &                200 &  1.4 $\times 10^{16}$ &                9.2 &                  76\\
              &    1.9         &                200 &  9.6 $\times 10^{12}$ &                3.0 &                  59\\
              &    1.9         &                200 &  5.6 $\times 10^{16}$ &                9.6 &                  36\\
              &    1.9         &                200 &  3.3 $\times 10^{15}$ &                5.7 &                  10\\
              &    1.9         &                200 &  1.1 $\times 10^{16}$ &               11.2 &                 -30\\
$^{13}$CS     &    1.9         &                200 &  1.6 $\times 10^{13}$ &                1.1 &                 100\\
              &    1.9         &                200 &  2.0 $\times 10^{12}$ &               30.4 &                  77\\
              &    1.9         &                200 &  3.0 $\times 10^{12}$ &                7.0 &                  61\\
              &    1.9         &                200 &  4.0 $\times 10^{14}$ &                0.8 &                  39\\
              &    1.9         &                200 &  1.9 $\times 10^{12}$ &                9.2 &                  93\\
              &    1.9         &                200 &  6.7 $\times 10^{12}$ &                3.2 &                  56\\
              &    1.9         &                200 &  9.5 $\times 10^{13}$ &               12.6 &                 -30\\
              &    1.9         &                200 &  4.6 $\times 10^{12}$ &                0.3 &                 -23\\
C$^{33}$S     &    1.9         &                200 &  1.5 $\times 10^{17}$ &                2.4 &                -165\\
              &    1.9         &                200 &  2.8 $\times 10^{15}$ &                9.2 &                -133\\
              &    1.9         &                200 &  2.0 $\times 10^{15}$ &                6.7 &                -121\\
              &    1.9         &                200 &  4.4 $\times 10^{16}$ &               12.1 &                 -97\\
              &    1.9         &                200 &  1.5 $\times 10^{15}$ &               35.2 &                 -72\\
              &    1.9         &                200 &  3.3 $\times 10^{16}$ &               15.6 &                 -44\\
              &    1.9         &                200 &  3.7 $\times 10^{15}$ &               10.5 &                  39\\
              &    1.9         &                200 &  1.4 $\times 10^{15}$ &                7.1 &                  62\\
              &    1.9         &                200 &  9.4 $\times 10^{14}$ &                6.7 &                  74\\
              &    1.9         &                200 &  1.8 $\times 10^{16}$ &               13.1 &                 106\\
              &    1.9         &                200 &  1.7 $\times 10^{13}$ &                7.1 &                 153\\
C$^{34}$S     &    1.9         &                200 &  7.8 $\times 10^{14}$ &               19.3 &                -161\\
              &    1.9         &                200 &  6.5 $\times 10^{15}$ &               11.2 &                -125\\
              &    1.9         &                200 &  7.0 $\times 10^{13}$ &                4.7 &                -102\\
              &    1.9         &                200 &  5.4 $\times 10^{14}$ &               16.0 &                 -85\\
              &    1.9         &                200 &  4.1 $\times 10^{15}$ &               11.3 &                 -47\\
              &    1.9         &                200 &  4.0 $\times 10^{14}$ &               15.0 &                  25\\
              &    1.9         &                200 &  1.4 $\times 10^{14}$ &                5.3 &                  35\\
              &    1.9         &                200 &  1.1 $\times 10^{15}$ &                5.2 &                  74\\
              &    1.9         &                200 &  1.3 $\times 10^{15}$ &                9.2 &                 104\\
              &    1.9         &                200 &  2.4 $\times 10^{12}$ &               17.2 &                 146\\
HCN           &    1.9         &                200 &  9.3 $\times 10^{14}$ &                3.0 &                 147\\
              &    1.9         &                200 &  2.3 $\times 10^{16}$ &               13.6 &                 120\\
              &    1.9         &                200 &  5.8 $\times 10^{13}$ &                8.4 &                  65\\
              &    1.9         &                200 &  2.5 $\times 10^{18}$ &                2.4 &                  36\\
              &    1.9         &                200 &  3.4 $\times 10^{16}$ &                2.2 &                  24\\
              &    1.9         &                200 &  7.4 $\times 10^{14}$ &                3.6 &                 -10\\
              &    1.9         &                200 &  1.9 $\times 10^{15}$ &                3.0 &                 -36\\
              &    1.9         &                200 &  7.7 $\times 10^{13}$ &                4.2 &                 -30\\
              &    1.9         &                200 &  4.6 $\times 10^{15}$ &                0.7 &                 -61\\
              &    1.9         &                200 &  5.5 $\times 10^{12}$ &                1.0 &                 -83\\
              &    1.9         &                200 &  6.8 $\times 10^{12}$ &                3.8 &                -112\\
              &    1.9         &                200 &  7.2 $\times 10^{13}$ &                3.2 &                 -58\\
              &    1.9         &                200 &  1.2 $\times 10^{19}$ &                5.8 &                -101\\
              &    1.9         &                200 &  2.8 $\times 10^{16}$ &               13.2 &                -111\\
              &    1.9         &                200 &  2.7 $\times 10^{14}$ &                3.5 &                -156\\
              &    1.9         &                200 &  4.2 $\times 10^{13}$ &                0.4 &                -170\\
              &    1.9         &                200 &  2.0 $\times 10^{12}$ &                8.1 &                  73\\
H$^{13}$CN    &    1.9         &                200 &  2.6 $\times 10^{17}$ &               30.6 &                  64\\
              &    1.9         &                200 &  2.8 $\times 10^{15}$ &                6.5 &                  24\\
              &    1.9         &                200 &  1.0 $\times 10^{13}$ &               26.4 &                  -4\\
HNC           &    1.9         &                200 &  3.4 $\times 10^{15}$ &                1.7 &                  76\\
              &    1.9         &                200 &  2.1 $\times 10^{16}$ &                8.0 &                  38\\
              &    1.9         &                200 &  9.1 $\times 10^{14}$ &                2.0 &                  25\\
              &    1.9         &                200 &  2.6 $\times 10^{12}$ &                5.1 &                 -10\\
              &    1.9         &                200 &  1.1 $\times 10^{13}$ &                1.1 &                 -37\\
              &    1.9         &                200 &  6.2 $\times 10^{12}$ &               19.2 &                 -56\\
              &    1.9         &                200 &  2.1 $\times 10^{13}$ &                0.5 &                 -63\\
              &    1.9         &                200 &  6.7 $\times 10^{13}$ &                0.9 &                -140\\
              &    1.9         &                200 &  2.2 $\times 10^{14}$ &                4.8 &                -161\\
              &    1.9         &                200 &  1.2 $\times 10^{14}$ &                4.2 &                -183\\
HN$^{13}$C    &    1.9         &                200 &  8.6 $\times 10^{17}$ &                0.6 &                  75\\
              &    1.9         &                200 &  4.0 $\times 10^{16}$ &               16.0 &                  60\\
              &    1.9         &                200 &  5.2 $\times 10^{12}$ &                0.1 &                  27\\
              &    1.9         &                200 &  4.7 $\times 10^{13}$ &               10.3 &                 -61\\
              &    1.9         &                200 &  5.5 $\times 10^{12}$ &                1.1 &                 -37\\
              &    1.9         &                200 &  9.8 $\times 10^{13}$ &                2.2 &                 -47\\
              &    1.9         &                200 &  2.8 $\times 10^{14}$ &                0.6 &                 -63\\
              &    1.9         &                200 &  3.5 $\times 10^{12}$ &                0.9 &                -132\\
              &    1.9         &                200 &  3.6 $\times 10^{12}$ &               11.4 &                -163\\
              &    1.9         &                200 &  1.9 $\times 10^{12}$ &                3.9 &                -184\\
HO$^{13}$C$^+$ &    1.9         &                200 &  2.0 $\times 10^{13}$ &               10.7 &                 162\\
              &    1.9         &                200 &  1.3 $\times 10^{14}$ &                6.4 &                  69\\
              &    1.9         &                200 &  3.5 $\times 10^{14}$ &               12.3 &                  50\\
\end{supertabular}\\
\vspace{1cm}

%---------------------------------------
% Envelope Components

\tablefirsthead{%
\hline
\hline
Molecule      & $\theta^{m,c}$ & T$_{\rm ex}^{m,c}$ & N$_{\rm tot}^{m,c}$   & $\Delta$ v$^{m,c}$ & v$_{\rm LSR}^{m,c}$\\
              & ($\arcsec$)    & (K)                & (cm$^{-2}$)           & (km~s$^{-1}$)      & (km~s$^{-1}$)      \\
\hline
}

\tablehead{%
\multicolumn{6}{c}{(Continued)}\\
\hline
\hline
Molecule      & $\theta^{m,c}$ & T$_{\rm ex}^{m,c}$ & N$_{\rm tot}^{m,c}$   & $\Delta$ v$^{m,c}$ & v$_{\rm LSR}^{m,c}$\\
              & ($\arcsec$)    & (K)                & (cm$^{-2}$)           & (km~s$^{-1}$)      & (km~s$^{-1}$)      \\
\hline
}

\tabletail{%
\hline
\hline
}

\topcaption{LTE Parameters for the full LTE model (Envelope Components) for source A02 in Sgr~B2(M).}
\tiny
\centering
\begin{supertabular}{lcccC{1cm}C{1cm}}\label{EnvLTE:parameters:A02SgrB2M}\\
CH$_3$NH$_2$  & ext.           &                  8 &  5.5 $\times 10^{13}$ &                3.0 &                  66\\
              & ext.           &                  3 &  7.5 $\times 10^{14}$ &               11.0 &                  62\\
CH$_3$OH      & ext.           &                 48 &  6.1 $\times 10^{15}$ &                3.3 &                  66\\
              & ext.           &                  9 &  6.0 $\times 10^{15}$ &               10.0 &                  64\\
CH$_2$NH      & ext.           &                  5 &  4.3 $\times 10^{13}$ &                1.2 &                  73\\
              & ext.           &                  4 &  2.5 $\times 10^{14}$ &                9.1 &                  64\\
              & ext.           &                  7 &  7.2 $\times 10^{14}$ &               10.8 &                  61\\
\end{supertabular}\\
\vspace{1cm}

%================================================================================
%
% Source A03

%---------------------------------------
% Core Components

\tablefirsthead{%
\hline
\hline
Molecule      & $\theta^{m,c}$ & T$_{\rm ex}^{m,c}$ & N$_{\rm tot}^{m,c}$   & $\Delta$ v$^{m,c}$ & v$_{\rm LSR}^{m,c}$\\
              & ($\arcsec$)    & (K)                & (cm$^{-2}$)           & (km~s$^{-1}$)      & (km~s$^{-1}$)      \\
\hline
}

\tablehead{%
\multicolumn{6}{c}{(Continued)}\\
\hline
\hline
Molecule      & $\theta^{m,c}$ & T$_{\rm ex}^{m,c}$ & N$_{\rm tot}^{m,c}$   & $\Delta$ v$^{m,c}$ & v$_{\rm LSR}^{m,c}$\\
              & ($\arcsec$)    & (K)                & (cm$^{-2}$)           & (km~s$^{-1}$)      & (km~s$^{-1}$)      \\
\hline
}

\tabletail{%
\hline
\hline
}

\topcaption{LTE Parameters for the full LTE model (Core Components) for source A03 in Sgr~B2(M).}
\tiny
\centering
\begin{supertabular}{lcccC{1cm}C{1cm}}\label{CoreLTE:parameters:A03SgrB2M}\\
CH$_3$CN      &    1.2         &                374 &  8.5 $\times 10^{17}$ &                8.0 &                  65\\
              &    1.2         &                200 &  4.9 $\times 10^{16}$ &               23.4 &                  64\\
H$_2$CO       &    1.2         &                321 &  9.2 $\times 10^{18}$ &                6.5 &                  65\\
C$_2$H$_5$CN  &    1.2         &                352 &  5.2 $\times 10^{17}$ &                8.3 &                  65\\
C$_2$H$_5$CN, v$_{20}$=1 &    1.2         &                312 &  2.5 $\times 10^{17}$ &                7.1 &                  61\\
CN            &    1.2         &                154 &  1.1 $\times 10^{12}$ &                2.9 &                -108\\
SO$_2$        &    1.2         &                521 &  9.0 $\times 10^{18}$ &                4.7 &                  66\\
\end{supertabular}\\
\vspace{1cm}

%---------------------------------------
% Envelope Components

\tablefirsthead{%
\hline
\hline
Molecule      & $\theta^{m,c}$ & T$_{\rm ex}^{m,c}$ & N$_{\rm tot}^{m,c}$   & $\Delta$ v$^{m,c}$ & v$_{\rm LSR}^{m,c}$\\
              & ($\arcsec$)    & (K)                & (cm$^{-2}$)           & (km~s$^{-1}$)      & (km~s$^{-1}$)      \\
\hline
}

\tablehead{%
\multicolumn{6}{c}{(Continued)}\\
\hline
\hline
Molecule      & $\theta^{m,c}$ & T$_{\rm ex}^{m,c}$ & N$_{\rm tot}^{m,c}$   & $\Delta$ v$^{m,c}$ & v$_{\rm LSR}^{m,c}$\\
              & ($\arcsec$)    & (K)                & (cm$^{-2}$)           & (km~s$^{-1}$)      & (km~s$^{-1}$)      \\
\hline
}

\tabletail{%
\hline
\hline
}

\topcaption{LTE Parameters for the full LTE model (Envelope Components) for source A03 in Sgr~B2(M).}
\tiny
\centering
\begin{supertabular}{lcccC{1cm}C{1cm}}\label{EnvLTE:parameters:A03SgrB2M}\\
H$_2$CO       & ext.           &                 37 &  1.4 $\times 10^{16}$ &               11.6 &                  66\\
              & ext.           &                  7 &  3.9 $\times 10^{12}$ &                5.3 &                 -26\\
              & ext.           &                 33 &  7.4 $\times 10^{15}$ &                9.9 &                  58\\
CN            & ext.           &                  3 &  7.2 $\times 10^{14}$ &                1.5 &                  67\\
              & ext.           &                 21 &  5.4 $\times 10^{16}$ &               12.5 &                  62\\
              & ext.           &                  5 &  4.7 $\times 10^{14}$ &                7.0 &                  41\\
              & ext.           &                 32 &  5.0 $\times 10^{15}$ &                6.3 &                  33\\
              & ext.           &                 15 &  1.5 $\times 10^{15}$ &               15.6 &                   8\\
              & ext.           &                 19 &  1.9 $\times 10^{15}$ &                5.5 &                  -1\\
              & ext.           &                 17 &  7.5 $\times 10^{14}$ &               16.0 &                 -24\\
              & ext.           &                 22 &  1.1 $\times 10^{15}$ &                2.2 &                 -43\\
              & ext.           &                 57 &  1.2 $\times 10^{16}$ &                5.3 &                 -49\\
              & ext.           &                 11 &  7.6 $\times 10^{13}$ &                3.9 &                 -78\\
              & ext.           &                  8 &  4.3 $\times 10^{14}$ &               14.1 &                -102\\
CCH           & ext.           &                  3 &  5.8 $\times 10^{15}$ &                6.1 &                  56\\
              & ext.           &                  3 &  1.4 $\times 10^{16}$ &               10.9 &                  65\\
SO            & ext.           &                 15 &  2.9 $\times 10^{15}$ &               14.0 &                  56\\
CH$_3$OH      & ext.           &                  7 &  5.1 $\times 10^{15}$ &                7.8 &                  65\\
              & ext.           &                 13 &  1.7 $\times 10^{15}$ &                5.7 &                  56\\
CH$_2$NH      & ext.           &                  3 &  1.3 $\times 10^{14}$ &                8.6 &                  64\\
              & ext.           &                  6 &  2.2 $\times 10^{14}$ &                6.0 &                  56\\
              & ext.           &                  7 &  7.8 $\times 10^{13}$ &                0.8 &                  72\\
SiO           & ext.           &                  5 &  1.2 $\times 10^{17}$ &                3.5 &                  58\\
CO            & ext.           &                  3 &  8.6 $\times 10^{16}$ &               22.4 &                  91\\
              & ext.           &                  3 &  7.7 $\times 10^{15}$ &                1.4 &                  84\\
              & ext.           &                  3 &  6.7 $\times 10^{16}$ &                8.2 &                  72\\
              & ext.           &                  3 &  7.1 $\times 10^{16}$ &                8.1 &                  63\\
              & ext.           &                  3 &  8.4 $\times 10^{16}$ &                8.3 &                  53\\
              & ext.           &                  3 &  8.1 $\times 10^{15}$ &                3.0 &                  39\\
              & ext.           &                  3 &  3.3 $\times 10^{16}$ &                6.1 &                  31\\
              & ext.           &                  3 &  1.6 $\times 10^{17}$ &               19.5 &                  16\\
              & ext.           &                  3 &  6.5 $\times 10^{16}$ &               10.8 &                   2\\
              & ext.           &                  3 &  2.3 $\times 10^{16}$ &                3.7 &                  -5\\
              & ext.           &                  3 &  1.7 $\times 10^{16}$ &                3.4 &                 -11\\
              & ext.           &                  3 &  3.6 $\times 10^{16}$ &                5.9 &                 -18\\
              & ext.           &                  3 &  9.2 $\times 10^{16}$ &               11.5 &                 -27\\
              & ext.           &                  3 &  6.5 $\times 10^{16}$ &                6.7 &                 -42\\
              & ext.           &                  3 &  1.2 $\times 10^{17}$ &                3.3 &                 -50\\
              & ext.           &                  3 &  2.8 $\times 10^{16}$ &                3.0 &                 -61\\
              & ext.           &                  3 &  6.3 $\times 10^{16}$ &               21.4 &                 -72\\
              & ext.           &                  3 &  1.5 $\times 10^{16}$ &                3.4 &                 -76\\
              & ext.           &                  3 &  1.6 $\times 10^{16}$ &                2.0 &                 -88\\
              & ext.           &                  3 &  1.0 $\times 10^{17}$ &               11.9 &                 -97\\
              & ext.           &                  3 &  3.2 $\times 10^{16}$ &                4.0 &                -106\\
              & ext.           &                  3 &  3.3 $\times 10^{16}$ &                3.8 &                -111\\
              & ext.           &                  3 &  9.4 $\times 10^{15}$ &                2.0 &                -117\\
              & ext.           &                  3 &  3.5 $\times 10^{15}$ &                2.0 &                -126\\
$^{13}$CO     & ext.           &                  3 &  1.2 $\times 10^{16}$ &                8.7 &                  87\\
              & ext.           &                  3 &  1.5 $\times 10^{16}$ &                2.5 &                  73\\
              & ext.           &                  3 &  1.5 $\times 10^{17}$ &               24.3 &                  60\\
              & ext.           &                  3 &  5.6 $\times 10^{16}$ &                2.5 &                  52\\
              & ext.           &                  3 &  7.8 $\times 10^{14}$ &                3.0 &                  38\\
              & ext.           &                  3 &  5.3 $\times 10^{15}$ &                3.0 &                  30\\
              & ext.           &                  3 &  3.1 $\times 10^{16}$ &               19.9 &                  12\\
              & ext.           &                  3 &  1.9 $\times 10^{16}$ &               10.9 &                  -2\\
              & ext.           &                  3 &  2.5 $\times 10^{16}$ &                2.0 &                 -43\\
              & ext.           &                  3 &  2.0 $\times 10^{16}$ &                2.7 &                 -49\\
              & ext.           &                  3 &  3.4 $\times 10^{15}$ &                2.8 &                 -62\\
              & ext.           &                  3 &  4.5 $\times 10^{15}$ &                3.9 &                 -78\\
              & ext.           &                  3 &  1.1 $\times 10^{15}$ &                2.9 &                 -85\\
              & ext.           &                  3 &  3.2 $\times 10^{15}$ &                2.9 &                 -90\\
              & ext.           &                  3 &  4.8 $\times 10^{15}$ &                2.9 &                -100\\
              & ext.           &                  3 &  8.8 $\times 10^{15}$ &                6.1 &                -106\\
C$^{17}$O     & ext.           &                  3 &  6.8 $\times 10^{13}$ &                3.0 &                -119\\
              & ext.           &                  3 &  4.9 $\times 10^{14}$ &                5.3 &                 -71\\
              & ext.           &                  3 &  1.2 $\times 10^{15}$ &                7.6 &                 -50\\
              & ext.           &                  3 &  2.6 $\times 10^{15}$ &               13.0 &                   1\\
              & ext.           &                  3 &  2.7 $\times 10^{13}$ &                1.6 &                 118\\
              & ext.           &                  3 &  1.3 $\times 10^{15}$ &                4.3 &                  52\\
              & ext.           &                  3 &  2.4 $\times 10^{15}$ &                3.7 &                  72\\
C$^{18}$O     & ext.           &                  3 &  4.0 $\times 10^{15}$ &                3.4 &                 -44\\
              & ext.           &                  3 &  1.5 $\times 10^{14}$ &                2.7 &                  13\\
              & ext.           &                  3 &  1.1 $\times 10^{15}$ &                7.5 &                  30\\
              & ext.           &                  3 &  2.0 $\times 10^{16}$ &                6.4 &                  51\\
CS            & ext.           &                  3 &  7.1 $\times 10^{15}$ &                5.9 &                  87\\
              & ext.           &                  3 &  3.2 $\times 10^{16}$ &                3.1 &                  73\\
              & ext.           &                  3 &  1.6 $\times 10^{12}$ &                2.8 &                  66\\
              & ext.           &                  3 &  2.1 $\times 10^{17}$ &               10.7 &                  60\\
              & ext.           &                  3 &  4.3 $\times 10^{16}$ &                1.9 &                  54\\
              & ext.           &                  3 &  1.1 $\times 10^{12}$ &                0.6 &                  -9\\
              & ext.           &                  3 &  2.4 $\times 10^{14}$ &                2.4 &                 -85\\
              & ext.           &                  3 &  4.7 $\times 10^{14}$ &                1.8 &                -105\\
              & ext.           &                  3 &  1.9 $\times 10^{14}$ &                2.1 &                -112\\
$^{13}$CS     & ext.           &                  3 &  8.0 $\times 10^{15}$ &                7.2 &                  55\\
C$^{33}$S     & ext.           &                  3 &  7.7 $\times 10^{14}$ &                9.8 &                -169\\
              & ext.           &                  3 &  1.2 $\times 10^{15}$ &                6.5 &                 -22\\
C$^{34}$S     & ext.           &                  3 &  4.0 $\times 10^{14}$ &                3.8 &                 -55\\
              & ext.           &                  3 &  5.3 $\times 10^{15}$ &               12.0 &                 -20\\
              & ext.           &                  3 &  4.2 $\times 10^{14}$ &                2.7 &                 -12\\
              & ext.           &                  3 &  1.6 $\times 10^{16}$ &                9.6 &                  57\\
HCN           & ext.           &                  3 &  9.5 $\times 10^{14}$ &               10.0 &                  71\\
              & ext.           &                  3 &  9.5 $\times 10^{14}$ &                9.9 &                  59\\
              & ext.           &                  3 &  4.6 $\times 10^{14}$ &                5.5 &                  52\\
              & ext.           &                  3 &  1.3 $\times 10^{15}$ &                2.1 &                  40\\
              & ext.           &                  3 &  3.8 $\times 10^{14}$ &                4.2 &                  35\\
              & ext.           &                  3 &  4.1 $\times 10^{12}$ &                1.0 &                  26\\
              & ext.           &                  3 &  2.3 $\times 10^{14}$ &                5.8 &                  16\\
              & ext.           &                  3 &  2.6 $\times 10^{14}$ &                5.7 &                   7\\
              & ext.           &                  3 &  5.3 $\times 10^{14}$ &                5.9 &                   0\\
H$^{13}$CN    & ext.           &                  3 &  8.6 $\times 10^{13}$ &                0.7 &                  74\\
              & ext.           &                  3 &  9.6 $\times 10^{14}$ &                9.7 &                  60\\
HNC           & ext.           &                  3 &  3.5 $\times 10^{12}$ &                4.7 &                 150\\
              & ext.           &                  3 &  1.0 $\times 10^{12}$ &                0.7 &                 121\\
              & ext.           &                  3 &  2.8 $\times 10^{13}$ &                2.9 &                  92\\
              & ext.           &                  3 &  7.9 $\times 10^{14}$ &               10.6 &                  69\\
              & ext.           &                  3 &  1.0 $\times 10^{12}$ &                1.8 &                  66\\
              & ext.           &                  3 &  3.3 $\times 10^{15}$ &                0.6 &                  60\\
              & ext.           &                  3 &  7.8 $\times 10^{14}$ &                6.3 &                  56\\
              & ext.           &                  3 &  2.9 $\times 10^{12}$ &                2.0 &                  29\\
              & ext.           &                  3 &  2.8 $\times 10^{13}$ &                1.8 &                  17\\
              & ext.           &                  3 &  1.6 $\times 10^{12}$ &                0.6 &                  13\\
              & ext.           &                  3 &  3.6 $\times 10^{13}$ &                1.7 &                  -2\\
              & ext.           &                  3 &  1.7 $\times 10^{12}$ &                1.3 &                 -36\\
              & ext.           &                  3 &  5.2 $\times 10^{12}$ &                2.2 &                 -47\\
              & ext.           &                  3 &  1.6 $\times 10^{13}$ &                2.7 &                 -68\\
              & ext.           &                  3 &  7.2 $\times 10^{12}$ &                0.8 &                 -76\\
              & ext.           &                  3 &  3.5 $\times 10^{12}$ &                3.9 &                 -88\\
              & ext.           &                  3 &  1.4 $\times 10^{13}$ &                1.0 &                 -96\\
              & ext.           &                  3 &  1.3 $\times 10^{13}$ &                1.0 &                -104\\
HN$^{13}$C    & ext.           &                  3 &  1.3 $\times 10^{12}$ &                4.8 &                 135\\
              & ext.           &                  3 &  1.0 $\times 10^{12}$ &                0.7 &                 144\\
              & ext.           &                  3 &  1.0 $\times 10^{12}$ &                2.9 &                  96\\
              & ext.           &                  3 &  1.0 $\times 10^{12}$ &                5.6 &                  86\\
              & ext.           &                  3 &  1.6 $\times 10^{12}$ &                1.4 &                  65\\
              & ext.           &                  3 &  1.1 $\times 10^{12}$ &                0.7 &                  61\\
              & ext.           &                  3 &  5.1 $\times 10^{12}$ &                9.3 &                  57\\
              & ext.           &                  3 &  1.0 $\times 10^{12}$ &                2.0 &                  29\\
              & ext.           &                  3 &  1.2 $\times 10^{12}$ &                2.1 &                  14\\
              & ext.           &                  3 &  2.1 $\times 10^{12}$ &                0.6 &                  23\\
              & ext.           &                  3 &  1.0 $\times 10^{13}$ &                1.4 &                   1\\
              & ext.           &                  3 &  1.0 $\times 10^{12}$ &                1.3 &                 -28\\
              & ext.           &                  3 &  1.0 $\times 10^{12}$ &                2.2 &                 -43\\
              & ext.           &                  3 &  1.0 $\times 10^{12}$ &                2.7 &                 -72\\
              & ext.           &                  3 &  1.0 $\times 10^{12}$ &                0.8 &                 -79\\
              & ext.           &                  3 &  1.0 $\times 10^{12}$ &                3.9 &                 -84\\
              & ext.           &                  3 &  9.9 $\times 10^{11}$ &                5.3 &                 -91\\
              & ext.           &                  3 &  1.0 $\times 10^{12}$ &                0.8 &                 -98\\
H$^{13}$CO$^+$ & ext.           &                  3 &  7.6 $\times 10^{12}$ &                1.4 &                  73\\
              & ext.           &                  3 &  1.1 $\times 10^{14}$ &                8.9 &                  65\\
              & ext.           &                  3 &  1.4 $\times 10^{14}$ &                6.7 &                  56\\
\end{supertabular}\\
\vspace{1cm}

%================================================================================
%
% Source A04

%---------------------------------------
% Core Components

\tablefirsthead{%
\hline
\hline
Molecule      & $\theta^{m,c}$ & T$_{\rm ex}^{m,c}$ & N$_{\rm tot}^{m,c}$   & $\Delta$ v$^{m,c}$ & v$_{\rm LSR}^{m,c}$\\
              & ($\arcsec$)    & (K)                & (cm$^{-2}$)           & (km~s$^{-1}$)      & (km~s$^{-1}$)      \\
\hline
}

\tablehead{%
\multicolumn{6}{c}{(Continued)}\\
\hline
\hline
Molecule      & $\theta^{m,c}$ & T$_{\rm ex}^{m,c}$ & N$_{\rm tot}^{m,c}$   & $\Delta$ v$^{m,c}$ & v$_{\rm LSR}^{m,c}$\\
              & ($\arcsec$)    & (K)                & (cm$^{-2}$)           & (km~s$^{-1}$)      & (km~s$^{-1}$)      \\
\hline
}

\tabletail{%
\hline
\hline
}

\topcaption{LTE Parameters for the full LTE model (Core Components) for source A04 in Sgr~B2(M).}
\tiny
\centering
\begin{supertabular}{lcccC{1cm}C{1cm}}\label{CoreLTE:parameters:A04SgrB2M}\\
RRL-H         &    1.2         &               9508 &  9.3 $\times 10^{8}$  &                0.1 &                  49\\
H$_2 \! ^{34}$S &    1.2         &                250 &  2.1 $\times 10^{17}$ &               11.0 &                  65\\
H$_2$CNH      &    1.2         &                199 &  1.1 $\times 10^{17}$ &                8.4 &                  65\\
CH$_3$CHO     &    1.2         &                230 &  9.2 $\times 10^{16}$ &                9.1 &                  65\\
CH$_2$NH      &    1.2         &                228 &  5.6 $\times 10^{14}$ &               15.9 &                  78\\
C$_2$H$_3$CN  &    1.2         &                100 &  2.2 $\times 10^{16}$ &                7.3 &                  57\\
              &    1.2         &                238 &  2.6 $\times 10^{16}$ &                5.9 &                  68\\
\end{supertabular}\\
\vspace{1cm}

%---------------------------------------
% Envelope Components

\tablefirsthead{%
\hline
\hline
Molecule      & $\theta^{m,c}$ & T$_{\rm ex}^{m,c}$ & N$_{\rm tot}^{m,c}$   & $\Delta$ v$^{m,c}$ & v$_{\rm LSR}^{m,c}$\\
              & ($\arcsec$)    & (K)                & (cm$^{-2}$)           & (km~s$^{-1}$)      & (km~s$^{-1}$)      \\
\hline
}

\tablehead{%
\multicolumn{6}{c}{(Continued)}\\
\hline
\hline
Molecule      & $\theta^{m,c}$ & T$_{\rm ex}^{m,c}$ & N$_{\rm tot}^{m,c}$   & $\Delta$ v$^{m,c}$ & v$_{\rm LSR}^{m,c}$\\
              & ($\arcsec$)    & (K)                & (cm$^{-2}$)           & (km~s$^{-1}$)      & (km~s$^{-1}$)      \\
\hline
}

\tabletail{%
\hline
\hline
}

\topcaption{LTE Parameters for the full LTE model (Envelope Components) for source A04 in Sgr~B2(M).}
\tiny
\centering
\begin{supertabular}{lcccC{1cm}C{1cm}}\label{EnvLTE:parameters:A04SgrB2M}\\
CH$_2$NH      & ext.           &                  9 &  5.1 $\times 10^{13}$ &                2.5 &                  73\\
              & ext.           &                  6 &  7.0 $\times 10^{14}$ &               14.0 &                  61\\
c-C$_3$H$_2$  & ext.           &                  5 &  1.6 $\times 10^{14}$ &               10.2 &                  64\\
SO            & ext.           &                 13 &  3.2 $\times 10^{15}$ &                8.0 &                  59\\
CN            & ext.           &                 12 &  2.1 $\times 10^{15}$ &                7.8 &                  69\\
              & ext.           &                 23 &  5.3 $\times 10^{16}$ &               12.1 &                  61\\
              & ext.           &                  3 &  1.0 $\times 10^{12}$ &                9.1 &                  36\\
\end{supertabular}\\
\vspace{1cm}

%================================================================================
%
% Source A05

%---------------------------------------
% Core Components

\tablefirsthead{%
\hline
\hline
Molecule      & $\theta^{m,c}$ & T$_{\rm ex}^{m,c}$ & N$_{\rm tot}^{m,c}$   & $\Delta$ v$^{m,c}$ & v$_{\rm LSR}^{m,c}$\\
              & ($\arcsec$)    & (K)                & (cm$^{-2}$)           & (km~s$^{-1}$)      & (km~s$^{-1}$)      \\
\hline
}

\tablehead{%
\multicolumn{6}{c}{(Continued)}\\
\hline
\hline
Molecule      & $\theta^{m,c}$ & T$_{\rm ex}^{m,c}$ & N$_{\rm tot}^{m,c}$   & $\Delta$ v$^{m,c}$ & v$_{\rm LSR}^{m,c}$\\
              & ($\arcsec$)    & (K)                & (cm$^{-2}$)           & (km~s$^{-1}$)      & (km~s$^{-1}$)      \\
\hline
}

\tabletail{%
\hline
\hline
}

\topcaption{LTE Parameters for the full LTE model (Core Components) for source A05 in Sgr~B2(M).}
\tiny
\centering
\begin{supertabular}{lcccC{1cm}C{1cm}}\label{CoreLTE:parameters:A05SgrB2M}\\
H$_2$CO       &    1.2         &                223 &  4.1 $\times 10^{18}$ &                5.8 &                  65\\
SO$^+$        &    1.2         &                203 &  3.3 $\times 10^{16}$ &                6.6 &                  62\\
HNCO          &    1.2         &                310 &  5.5 $\times 10^{17}$ &                8.8 &                  65\\
OCS           &    1.2         &                286 &  2.2 $\times 10^{17}$ &                1.3 &                  64\\
              &    1.2         &                297 &  9.5 $\times 10^{17}$ &                8.6 &                  64\\
CH$_3$CN      &    1.2         &                266 &  1.8 $\times 10^{17}$ &                6.6 &                  65\\
              &    1.2         &                255 &  4.8 $\times 10^{16}$ &               14.8 &                  65\\
\end{supertabular}\\
\vspace{1cm}

%---------------------------------------
% Envelope Components

\tablefirsthead{%
\hline
\hline
Molecule      & $\theta^{m,c}$ & T$_{\rm ex}^{m,c}$ & N$_{\rm tot}^{m,c}$   & $\Delta$ v$^{m,c}$ & v$_{\rm LSR}^{m,c}$\\
              & ($\arcsec$)    & (K)                & (cm$^{-2}$)           & (km~s$^{-1}$)      & (km~s$^{-1}$)      \\
\hline
}

\tablehead{%
\multicolumn{6}{c}{(Continued)}\\
\hline
\hline
Molecule      & $\theta^{m,c}$ & T$_{\rm ex}^{m,c}$ & N$_{\rm tot}^{m,c}$   & $\Delta$ v$^{m,c}$ & v$_{\rm LSR}^{m,c}$\\
              & ($\arcsec$)    & (K)                & (cm$^{-2}$)           & (km~s$^{-1}$)      & (km~s$^{-1}$)      \\
\hline
}

\tabletail{%
\hline
\hline
}

\topcaption{LTE Parameters for the full LTE model (Envelope Components) for source A05 in Sgr~B2(M).}
\tiny
\centering
\begin{supertabular}{lcccC{1cm}C{1cm}}\label{EnvLTE:parameters:A05SgrB2M}\\
H$_2$CO       & ext.           &                 38 &  2.0 $\times 10^{16}$ &               14.6 &                  64\\
CH$_3$CN      & ext.           &                 24 &  1.0 $\times 10^{12}$ &                6.3 &                 102\\
CH$_2$NH      & ext.           &                  5 &  2.3 $\times 10^{13}$ &                2.5 &                  73\\
              & ext.           &                  6 &  3.7 $\times 10^{14}$ &               12.3 &                  63\\
              & ext.           &                 20 &  1.8 $\times 10^{14}$ &                5.5 &                 -63\\
SiO           & ext.           &                 21 &  3.2 $\times 10^{14}$ &                6.5 &                  57\\
SO            & ext.           &                 28 &  6.6 $\times 10^{15}$ &               16.3 &                  56\\
\end{supertabular}\\
\vspace{1cm}

%================================================================================
%
% Source A06

%---------------------------------------
% Core Components

\tablefirsthead{%
\hline
\hline
Molecule      & $\theta^{m,c}$ & T$_{\rm ex}^{m,c}$ & N$_{\rm tot}^{m,c}$   & $\Delta$ v$^{m,c}$ & v$_{\rm LSR}^{m,c}$\\
              & ($\arcsec$)    & (K)                & (cm$^{-2}$)           & (km~s$^{-1}$)      & (km~s$^{-1}$)      \\
\hline
}

\tablehead{%
\multicolumn{6}{c}{(Continued)}\\
\hline
\hline
Molecule      & $\theta^{m,c}$ & T$_{\rm ex}^{m,c}$ & N$_{\rm tot}^{m,c}$   & $\Delta$ v$^{m,c}$ & v$_{\rm LSR}^{m,c}$\\
              & ($\arcsec$)    & (K)                & (cm$^{-2}$)           & (km~s$^{-1}$)      & (km~s$^{-1}$)      \\
\hline
}

\tabletail{%
\hline
\hline
}

\topcaption{LTE Parameters for the full LTE model (Core Components) for source A06 in Sgr~B2(M).}
\tiny
\centering
\begin{supertabular}{lcccC{1cm}C{1cm}}\label{CoreLTE:parameters:A06SgrB2M}\\
H$_2 \! ^{34}$S &    1.4         &                355 &  6.4 $\times 10^{17}$ &                6.0 &                  56\\
SO            &    1.4         &                433 &  1.3 $\times 10^{18}$ &                9.6 &                  62\\
              &    1.4         &                349 &  9.0 $\times 10^{18}$ &                5.4 &                  56\\
              &    1.4         &                200 &  1.2 $\times 10^{18}$ &               18.9 &                  58\\
HCCCN         &    1.4         &                163 &  3.3 $\times 10^{16}$ &                3.3 &                  56\\
HCCCN, v$_7$=2 &    1.4         &                353 &  2.5 $\times 10^{16}$ &                4.1 &                  56\\
HCN, v$_2$=1  &    1.4         &                500 &  2.5 $\times 10^{17}$ &                7.7 &                  56\\
\end{supertabular}\\
\vspace{1cm}

%---------------------------------------
% Envelope Components

\tablefirsthead{%
\hline
\hline
Molecule      & $\theta^{m,c}$ & T$_{\rm ex}^{m,c}$ & N$_{\rm tot}^{m,c}$   & $\Delta$ v$^{m,c}$ & v$_{\rm LSR}^{m,c}$\\
              & ($\arcsec$)    & (K)                & (cm$^{-2}$)           & (km~s$^{-1}$)      & (km~s$^{-1}$)      \\
\hline
}

\tablehead{%
\multicolumn{6}{c}{(Continued)}\\
\hline
\hline
Molecule      & $\theta^{m,c}$ & T$_{\rm ex}^{m,c}$ & N$_{\rm tot}^{m,c}$   & $\Delta$ v$^{m,c}$ & v$_{\rm LSR}^{m,c}$\\
              & ($\arcsec$)    & (K)                & (cm$^{-2}$)           & (km~s$^{-1}$)      & (km~s$^{-1}$)      \\
\hline
}

\tabletail{%
\hline
\hline
}

\topcaption{LTE Parameters for the full LTE model (Envelope Components) for source A06 in Sgr~B2(M).}
\tiny
\centering
\begin{supertabular}{lcccC{1cm}C{1cm}}\label{EnvLTE:parameters:A06SgrB2M}\\
CN            & ext.           &                  8 &  1.2 $\times 10^{13}$ &                5.2 &                 120\\
              & ext.           &                  8 &  2.1 $\times 10^{13}$ &                6.8 &                  95\\
              & ext.           &                  4 &  3.6 $\times 10^{15}$ &                2.8 &                  67\\
              & ext.           &                 18 &  2.5 $\times 10^{16}$ &                9.3 &                  64\\
              & ext.           &                  3 &  5.8 $\times 10^{15}$ &               26.3 &                  30\\
              & ext.           &                 36 &  1.9 $\times 10^{17}$ &               16.8 &                  26\\
              & ext.           &                 14 &  2.3 $\times 10^{15}$ &                3.7 &                   0\\
\end{supertabular}\\
\vspace{1cm}

%================================================================================
%
% Source A07

%---------------------------------------
% Core Components

\tablefirsthead{%
\hline
\hline
Molecule      & $\theta^{m,c}$ & T$_{\rm ex}^{m,c}$ & N$_{\rm tot}^{m,c}$   & $\Delta$ v$^{m,c}$ & v$_{\rm LSR}^{m,c}$\\
              & ($\arcsec$)    & (K)                & (cm$^{-2}$)           & (km~s$^{-1}$)      & (km~s$^{-1}$)      \\
\hline
}

\tablehead{%
\multicolumn{6}{c}{(Continued)}\\
\hline
\hline
Molecule      & $\theta^{m,c}$ & T$_{\rm ex}^{m,c}$ & N$_{\rm tot}^{m,c}$   & $\Delta$ v$^{m,c}$ & v$_{\rm LSR}^{m,c}$\\
              & ($\arcsec$)    & (K)                & (cm$^{-2}$)           & (km~s$^{-1}$)      & (km~s$^{-1}$)      \\
\hline
}

\tabletail{%
\hline
\hline
}

\topcaption{LTE Parameters for the full LTE model (Core Components) for source A07 in Sgr~B2(M).}
\tiny
\centering
\begin{supertabular}{lcccC{1cm}C{1cm}}\label{CoreLTE:parameters:A07SgrB2M}\\
RRL-H         &    1.2         &               5824 &  3.9 $\times 10^{8}$  &               45.9 &                  63\\
SO$^+$        &    1.2         &                107 &  9.2 $\times 10^{16}$ &                4.3 &                  62\\
SO            &    1.2         &                386 &  1.4 $\times 10^{18}$ &                4.1 &                  68\\
              &    1.2         &                249 &  3.1 $\times 10^{18}$ &                3.2 &                  64\\
CH$_3$CN      &    1.2         &                264 &  6.4 $\times 10^{16}$ &                3.4 &                  69\\
              &    1.2         &                217 &  5.3 $\times 10^{16}$ &                8.7 &                  65\\
OCS           &    1.2         &                281 &  2.5 $\times 10^{17}$ &                3.0 &                  69\\
\end{supertabular}\\
\vspace{1cm}

%---------------------------------------
% Envelope Components

\tablefirsthead{%
\hline
\hline
Molecule      & $\theta^{m,c}$ & T$_{\rm ex}^{m,c}$ & N$_{\rm tot}^{m,c}$   & $\Delta$ v$^{m,c}$ & v$_{\rm LSR}^{m,c}$\\
              & ($\arcsec$)    & (K)                & (cm$^{-2}$)           & (km~s$^{-1}$)      & (km~s$^{-1}$)      \\
\hline
}

\tablehead{%
\multicolumn{6}{c}{(Continued)}\\
\hline
\hline
Molecule      & $\theta^{m,c}$ & T$_{\rm ex}^{m,c}$ & N$_{\rm tot}^{m,c}$   & $\Delta$ v$^{m,c}$ & v$_{\rm LSR}^{m,c}$\\
              & ($\arcsec$)    & (K)                & (cm$^{-2}$)           & (km~s$^{-1}$)      & (km~s$^{-1}$)      \\
\hline
}

\tabletail{%
\hline
\hline
}

\topcaption{LTE Parameters for the full LTE model (Envelope Components) for source A07 in Sgr~B2(M).}
\tiny
\centering
\begin{supertabular}{lcccC{1cm}C{1cm}}\label{EnvLTE:parameters:A07SgrB2M}\\
SO            & ext.           &                  7 &  2.0 $\times 10^{15}$ &               27.3 &                  68\\
              & ext.           &                  8 &  2.1 $\times 10^{15}$ &                6.5 &                  47\\
H$_2$CO       & ext.           &                 28 &  4.0 $\times 10^{15}$ &                2.9 &                  74\\
              & ext.           &                  3 &  4.3 $\times 10^{16}$ &                0.1 &                  64\\
              & ext.           &                 31 &  1.6 $\times 10^{16}$ &               15.5 &                  58\\
              & ext.           &                  4 &  1.4 $\times 10^{14}$ &                1.6 &                  53\\
SiO           & ext.           &                 11 &  8.9 $\times 10^{13}$ &                3.3 &                  54\\
              & ext.           &                  9 &  3.6 $\times 10^{13}$ &               10.5 &                  74\\
              & ext.           &                 14 &  1.3 $\times 10^{14}$ &               10.4 &                  62\\
              & ext.           &                  3 &  3.6 $\times 10^{14}$ &                6.7 &                  53\\
CH$_3$OH      & ext.           &                  9 &  2.8 $\times 10^{14}$ &                1.3 &                  75\\
CCH           & ext.           &                  6 &  3.3 $\times 10^{15}$ &               18.2 &                  60\\
CN            & ext.           &                  3 &  1.6 $\times 10^{12}$ &                6.2 &                  66\\
              & ext.           &                 25 &  5.8 $\times 10^{16}$ &               14.7 &                  62\\
              & ext.           &                  4 &  3.3 $\times 10^{13}$ &               15.4 &                  40\\
              & ext.           &                  9 &  1.3 $\times 10^{14}$ &                8.0 &                  31\\
              & ext.           &                 21 &  2.4 $\times 10^{15}$ &               14.2 &                   9\\
              & ext.           &                  8 &  1.6 $\times 10^{14}$ &                4.2 &                  -1\\
              & ext.           &                 25 &  5.8 $\times 10^{14}$ &                2.1 &                 -43\\
              & ext.           &                 35 &  7.9 $\times 10^{15}$ &               35.2 &                 -38\\
              & ext.           &                 79 &  3.1 $\times 10^{15}$ &                4.7 &                 -56\\
              & ext.           &                 11 &  3.4 $\times 10^{13}$ &                3.9 &                 -74\\
              & ext.           &                  8 &  3.2 $\times 10^{14}$ &               15.7 &                -103\\
HCCCN         & ext.           &                 27 &  2.3 $\times 10^{15}$ &                4.6 &                  53\\
H$_2$CNH      & ext.           &                  3 &  1.8 $\times 10^{13}$ &                6.1 &                  70\\
              & ext.           &                  3 &  2.0 $\times 10^{13}$ &                6.4 &                  65\\
              & ext.           &                  3 &  2.6 $\times 10^{13}$ &                5.2 &                  55\\
CH$_2$NH      & ext.           &                  3 &  2.0 $\times 10^{13}$ &                5.9 &                  73\\
              & ext.           &                  3 &  6.1 $\times 10^{13}$ &               10.2 &                  61\\
CO            & ext.           &                  3 &  2.7 $\times 10^{12}$ &                5.7 &                 131\\
              & ext.           &                  3 &  1.0 $\times 10^{12}$ &                0.1 &                 128\\
              & ext.           &                  3 &  1.4 $\times 10^{17}$ &               32.2 &                  80\\
              & ext.           &                  3 &  7.6 $\times 10^{15}$ &                7.3 &                 107\\
              & ext.           &                  3 &  6.4 $\times 10^{15}$ &                3.0 &                  52\\
              & ext.           &                  3 &  1.0 $\times 10^{17}$ &               27.4 &                  49\\
              & ext.           &                  3 &  6.9 $\times 10^{15}$ &                2.0 &                  30\\
              & ext.           &                  3 &  1.3 $\times 10^{17}$ &               29.8 &                   9\\
              & ext.           &                  3 &  5.4 $\times 10^{14}$ &                1.9 &                  -6\\
              & ext.           &                  3 &  2.0 $\times 10^{16}$ &                9.5 &                 -22\\
              & ext.           &                  3 &  1.2 $\times 10^{17}$ &               37.1 &                 -38\\
              & ext.           &                  3 &  1.2 $\times 10^{16}$ &                6.4 &                 -50\\
              & ext.           &                  3 &  1.8 $\times 10^{16}$ &                3.0 &                 -59\\
              & ext.           &                  3 &  7.9 $\times 10^{15}$ &                2.8 &                 -71\\
              & ext.           &                  3 &  1.2 $\times 10^{16}$ &                3.0 &                 -77\\
              & ext.           &                  3 &  3.0 $\times 10^{15}$ &                2.5 &                 -89\\
              & ext.           &                  3 &  1.0 $\times 10^{17}$ &               27.5 &                 -98\\
              & ext.           &                  3 &  8.9 $\times 10^{14}$ &                2.0 &                 -99\\
              & ext.           &                  3 &  9.2 $\times 10^{15}$ &                3.8 &                -111\\
              & ext.           &                  3 &  4.3 $\times 10^{13}$ &                2.0 &                -126\\
              & ext.           &                  3 &  1.3 $\times 10^{15}$ &                0.8 &                -126\\
              & ext.           &                  3 &  3.8 $\times 10^{14}$ &                1.0 &                -183\\
$^{13}$CO     & ext.           &                  3 &  4.2 $\times 10^{15}$ &                2.9 &                  91\\
              & ext.           &                  3 &  1.9 $\times 10^{17}$ &               29.1 &                  64\\
              & ext.           &                  3 &  1.7 $\times 10^{17}$ &                3.8 &                  50\\
              & ext.           &                  3 &  3.7 $\times 10^{15}$ &                1.9 &                  30\\
              & ext.           &                  3 &  7.2 $\times 10^{15}$ &                5.9 &                   9\\
              & ext.           &                  3 &  8.3 $\times 10^{15}$ &                5.1 &                  15\\
              & ext.           &                  3 &  1.2 $\times 10^{16}$ &               11.3 &                   0\\
              & ext.           &                  3 &  5.9 $\times 10^{13}$ &                2.0 &                 -17\\
              & ext.           &                  3 &  1.4 $\times 10^{16}$ &                1.5 &                 -44\\
              & ext.           &                  3 &  2.3 $\times 10^{16}$ &                0.9 &                 -48\\
              & ext.           &                  3 &  1.6 $\times 10^{16}$ &                0.8 &                 -64\\
              & ext.           &                  3 &  1.8 $\times 10^{16}$ &               38.0 &                 -35\\
              & ext.           &                  3 &  1.9 $\times 10^{15}$ &                2.8 &                 -91\\
              & ext.           &                  3 &  7.6 $\times 10^{15}$ &                5.7 &                -102\\
              & ext.           &                  3 &  1.5 $\times 10^{14}$ &                0.4 &                -111\\
C$^{17}$O     & ext.           &                  3 &  4.3 $\times 10^{15}$ &                7.2 &                -180\\
              & ext.           &                  3 &  4.2 $\times 10^{14}$ &                3.8 &                -114\\
              & ext.           &                  3 &  1.3 $\times 10^{15}$ &                7.2 &                -108\\
              & ext.           &                  3 &  4.2 $\times 10^{14}$ &                3.3 &                -101\\
              & ext.           &                  3 &  1.3 $\times 10^{14}$ &                1.4 &                 -92\\
              & ext.           &                  3 &  6.7 $\times 10^{14}$ &                7.2 &                 -87\\
              & ext.           &                  3 &  1.9 $\times 10^{15}$ &                9.7 &                 -68\\
              & ext.           &                  3 &  3.1 $\times 10^{15}$ &               11.4 &                 -44\\
              & ext.           &                  3 &  8.2 $\times 10^{14}$ &                2.9 &                 -36\\
              & ext.           &                  3 &  5.7 $\times 10^{14}$ &                2.7 &                 -26\\
              & ext.           &                  3 &  7.1 $\times 10^{14}$ &                7.8 &                   6\\
              & ext.           &                  3 & 10.0 $\times 10^{14}$ &                7.2 &                  33\\
              & ext.           &                  3 &  3.1 $\times 10^{16}$ &                8.5 &                  52\\
              & ext.           &                  3 &  1.8 $\times 10^{14}$ &               13.6 &                  79\\
              & ext.           &                  3 &  6.8 $\times 10^{14}$ &               10.5 &                 132\\
C$^{18}$O     & ext.           &                  3 &  4.6 $\times 10^{15}$ &               12.2 &                -104\\
              & ext.           &                  3 &  2.7 $\times 10^{14}$ &                2.7 &                 -80\\
              & ext.           &                  3 &  2.9 $\times 10^{15}$ &                5.4 &                 -44\\
              & ext.           &                  3 &  5.0 $\times 10^{13}$ &                2.8 &                   0\\
              & ext.           &                  3 &  6.4 $\times 10^{14}$ &                3.4 &                   8\\
              & ext.           &                  3 &  5.1 $\times 10^{16}$ &                8.1 &                  51\\
              & ext.           &                  3 &  1.1 $\times 10^{14}$ &                2.1 &                 108\\
              & ext.           &                  3 &  7.6 $\times 10^{14}$ &                8.4 &                 113\\
CS            & ext.           &                  3 &  1.2 $\times 10^{16}$ &                9.2 &                  83\\
              & ext.           &                  3 &  4.4 $\times 10^{16}$ &                5.1 &                  74\\
              & ext.           &                  3 &  1.4 $\times 10^{16}$ &                3.2 &                  66\\
              & ext.           &                  3 &  1.3 $\times 10^{17}$ &               13.4 &                  55\\
              & ext.           &                  3 &  4.0 $\times 10^{16}$ &                2.0 &                  53\\
              & ext.           &                  3 &  2.8 $\times 10^{15}$ &                0.1 &                  38\\
              & ext.           &                  3 &  1.4 $\times 10^{15}$ &                1.4 &                  30\\
              & ext.           &                  3 &  1.2 $\times 10^{16}$ &                3.0 &                  46\\
              & ext.           &                  3 &  3.8 $\times 10^{14}$ &                1.1 &                 -78\\
              & ext.           &                  3 &  1.5 $\times 10^{15}$ &                2.5 &                 -94\\
              & ext.           &                  3 &  4.6 $\times 10^{12}$ &                1.0 &                -179\\
$^{13}$CS     & ext.           &                  3 &  9.6 $\times 10^{14}$ &                1.0 &                 101\\
              & ext.           &                  3 &  2.0 $\times 10^{15}$ &                6.4 &                  75\\
              & ext.           &                  3 &  3.1 $\times 10^{16}$ &                9.7 &                  55\\
C$^{33}$S     & ext.           &                  3 &  1.9 $\times 10^{16}$ &                8.9 &                -163\\
              & ext.           &                  3 &  5.8 $\times 10^{15}$ &                8.9 &                -136\\
              & ext.           &                  3 &  9.8 $\times 10^{14}$ &                2.5 &                 -77\\
              & ext.           &                  3 &  3.4 $\times 10^{14}$ &                2.5 &                 -72\\
              & ext.           &                  3 &  2.7 $\times 10^{15}$ &                7.7 &                  32\\
              & ext.           &                  3 &  5.9 $\times 10^{15}$ &                6.3 &                  52\\
              & ext.           &                  3 &  1.9 $\times 10^{15}$ &                2.4 &                  57\\
              & ext.           &                  3 &  7.1 $\times 10^{14}$ &                1.3 &                  76\\
              & ext.           &                  3 &  6.2 $\times 10^{15}$ &               27.1 &                 131\\
C$^{34}$S     & ext.           &                  3 &  7.4 $\times 10^{12}$ &                3.5 &                 -52\\
              & ext.           &                  3 &  9.5 $\times 10^{14}$ &                7.4 &                 -47\\
              & ext.           &                  3 &  2.2 $\times 10^{13}$ &                2.4 &                 -30\\
              & ext.           &                  3 &  6.5 $\times 10^{14}$ &                7.3 &                 -15\\
              & ext.           &                  3 &  1.9 $\times 10^{13}$ &                2.4 &                  30\\
              & ext.           &                  3 &  3.8 $\times 10^{14}$ &                2.4 &                  40\\
              & ext.           &                  3 &  2.3 $\times 10^{16}$ &                9.2 &                  55\\
              & ext.           &                  3 &  7.1 $\times 10^{15}$ &                8.0 &                  77\\
              & ext.           &                  3 &  8.1 $\times 10^{14}$ &                2.4 &                 129\\
HCN           & ext.           &                  3 &  5.8 $\times 10^{12}$ &                2.4 &                  71\\
              & ext.           &                  3 &  7.0 $\times 10^{13}$ &                8.1 &                  92\\
              & ext.           &                  3 &  3.0 $\times 10^{13}$ &                0.1 &                  82\\
              & ext.           &                  3 &  1.2 $\times 10^{15}$ &               37.5 &                  72\\
              & ext.           &                  3 &  1.3 $\times 10^{14}$ &                9.8 &                  55\\
              & ext.           &                  3 &  8.0 $\times 10^{12}$ &                7.9 &                  60\\
              & ext.           &                  3 &  7.4 $\times 10^{14}$ &               30.3 &                  36\\
              & ext.           &                  3 &  7.2 $\times 10^{13}$ &                4.0 &                  36\\
              & ext.           &                  3 &  4.0 $\times 10^{13}$ &                0.9 &                  12\\
              & ext.           &                  3 &  9.1 $\times 10^{13}$ &                1.7 &                   8\\
              & ext.           &                  3 &  2.7 $\times 10^{14}$ &                3.5 &                   0\\
              & ext.           &                  3 &  3.9 $\times 10^{13}$ &                2.1 &                 -26\\
              & ext.           &                  3 &  6.6 $\times 10^{13}$ &                8.1 &                 -35\\
              & ext.           &                  3 &  1.2 $\times 10^{12}$ &                1.0 &                 -37\\
              & ext.           &                  3 &  1.5 $\times 10^{13}$ &                1.2 &                 -43\\
              & ext.           &                  3 &  2.3 $\times 10^{13}$ &                1.0 &                 -48\\
              & ext.           &                  3 &  1.2 $\times 10^{14}$ &                1.3 &                -103\\
H$^{13}$CN    & ext.           &                  3 &  9.9 $\times 10^{13}$ &                5.1 &                 130\\
              & ext.           &                  3 &  5.9 $\times 10^{13}$ &                5.7 &                 124\\
              & ext.           &                  3 &  8.0 $\times 10^{12}$ &                0.3 &                 108\\
              & ext.           &                  3 &  6.0 $\times 10^{12}$ &                1.0 &                 106\\
              & ext.           &                  3 &  1.8 $\times 10^{13}$ &                1.5 &                  84\\
              & ext.           &                  3 &  1.3 $\times 10^{14}$ &                1.6 &                  73\\
              & ext.           &                  3 &  1.6 $\times 10^{15}$ &               15.1 &                  61\\
              & ext.           &                  3 &  9.7 $\times 10^{13}$ &               14.8 &                  -3\\
              & ext.           &                  3 &  1.8 $\times 10^{12}$ &                1.0 &                -105\\
HNC           & ext.           &                  3 &  6.6 $\times 10^{13}$ &                1.8 &                   0\\
              & ext.           &                  3 &  1.4 $\times 10^{13}$ &                0.8 &                 107\\
              & ext.           &                  3 &  1.7 $\times 10^{13}$ &                1.0 &                  98\\
              & ext.           &                  3 &  2.6 $\times 10^{13}$ &                0.3 &                  96\\
              & ext.           &                  3 &  2.5 $\times 10^{13}$ &                2.3 &                  88\\
              & ext.           &                  3 &  2.7 $\times 10^{14}$ &                0.9 &                  74\\
              & ext.           &                  3 &  2.6 $\times 10^{15}$ &                3.3 &                  56\\
              & ext.           &                  3 &  2.0 $\times 10^{15}$ &               25.2 &                  62\\
              & ext.           &                  3 &  4.7 $\times 10^{14}$ &               33.9 &                  51\\
              & ext.           &                  3 &  2.3 $\times 10^{13}$ &                2.0 &                  16\\
              & ext.           &                  3 &  2.0 $\times 10^{13}$ &                2.0 &                   7\\
              & ext.           &                  3 &  2.7 $\times 10^{12}$ &                2.1 &                 -13\\
              & ext.           &                  3 &  5.0 $\times 10^{12}$ &                2.1 &                 -10\\
HN$^{13}$C    & ext.           &                  3 &  1.8 $\times 10^{14}$ &                7.5 &                  55\\
H$^{13}$CO$^+$ & ext.           &                  3 &  4.2 $\times 10^{13}$ &                5.9 &                  70\\
              & ext.           &                  3 &  1.8 $\times 10^{14}$ &               10.0 &                  55\\
\end{supertabular}\\
\vspace{1cm}

%================================================================================
%
% Source A08

%---------------------------------------
% Core Components

\tablefirsthead{%
\hline
\hline
Molecule      & $\theta^{m,c}$ & T$_{\rm ex}^{m,c}$ & N$_{\rm tot}^{m,c}$   & $\Delta$ v$^{m,c}$ & v$_{\rm LSR}^{m,c}$\\
              & ($\arcsec$)    & (K)                & (cm$^{-2}$)           & (km~s$^{-1}$)      & (km~s$^{-1}$)      \\
\hline
}

\tablehead{%
\multicolumn{6}{c}{(Continued)}\\
\hline
\hline
Molecule      & $\theta^{m,c}$ & T$_{\rm ex}^{m,c}$ & N$_{\rm tot}^{m,c}$   & $\Delta$ v$^{m,c}$ & v$_{\rm LSR}^{m,c}$\\
              & ($\arcsec$)    & (K)                & (cm$^{-2}$)           & (km~s$^{-1}$)      & (km~s$^{-1}$)      \\
\hline
}

\tabletail{%
\hline
\hline
}

\topcaption{LTE Parameters for the full LTE model (Core Components) for source A08 in Sgr~B2(M).}
\tiny
\centering
\begin{supertabular}{lcccC{1cm}C{1cm}}\label{CoreLTE:parameters:A08SgrB2M}\\
RRL-H         &    1.0         &               8052 &  1.5 $\times 10^{9}$  &               41.6 &                  47\\
HCCCN         &    1.0         &                182 &  6.2 $\times 10^{14}$ &                3.4 &                  55\\
HCCCN, v$_7$=2 &    1.0         &                202 &  1.3 $\times 10^{15}$ &                2.9 &                  54\\
HCN, v$_2$=1  &    1.0         &                200 &  9.8 $\times 10^{16}$ &                3.3 &                  56\\
SO            &    1.0         &                761 &  6.6 $\times 10^{18}$ &                7.7 &                  56\\
              &    1.0         &                137 &  1.6 $\times 10^{18}$ &               15.2 &                  52\\
              &    1.0         &                448 &  2.7 $\times 10^{17}$ &                5.3 &                  44\\
\end{supertabular}\\
\vspace{1cm}

%---------------------------------------
% Envelope Components

\tablefirsthead{%
\hline
\hline
Molecule      & $\theta^{m,c}$ & T$_{\rm ex}^{m,c}$ & N$_{\rm tot}^{m,c}$   & $\Delta$ v$^{m,c}$ & v$_{\rm LSR}^{m,c}$\\
              & ($\arcsec$)    & (K)                & (cm$^{-2}$)           & (km~s$^{-1}$)      & (km~s$^{-1}$)      \\
\hline
}

\tablehead{%
\multicolumn{6}{c}{(Continued)}\\
\hline
\hline
Molecule      & $\theta^{m,c}$ & T$_{\rm ex}^{m,c}$ & N$_{\rm tot}^{m,c}$   & $\Delta$ v$^{m,c}$ & v$_{\rm LSR}^{m,c}$\\
              & ($\arcsec$)    & (K)                & (cm$^{-2}$)           & (km~s$^{-1}$)      & (km~s$^{-1}$)      \\
\hline
}

\tabletail{%
\hline
\hline
}

\topcaption{LTE Parameters for the full LTE model (Envelope Components) for source A08 in Sgr~B2(M).}
\tiny
\centering
\begin{supertabular}{lcccC{1cm}C{1cm}}\label{EnvLTE:parameters:A08SgrB2M}\\
CN            & ext.           &                 22 &  2.3 $\times 10^{16}$ &                5.6 &                  67\\
              & ext.           &                 20 &  2.3 $\times 10^{16}$ &                9.5 &                  61\\
              & ext.           &                  3 &  4.0 $\times 10^{14}$ &               32.0 &                  30\\
              & ext.           &                  3 &  8.1 $\times 10^{13}$ &                2.8 &                  30\\
              & ext.           &                  7 &  2.2 $\times 10^{14}$ &               10.5 &                   5\\
              & ext.           &                 23 &  3.0 $\times 10^{15}$ &                4.4 &                  -2\\
              & ext.           &                 35 &  5.8 $\times 10^{15}$ &               27.6 &                 -25\\
\end{supertabular}\\
\vspace{1cm}

%================================================================================
%
% Source A09

%---------------------------------------
% Core Components

\tablefirsthead{%
\hline
\hline
Molecule      & $\theta^{m,c}$ & T$_{\rm ex}^{m,c}$ & N$_{\rm tot}^{m,c}$   & $\Delta$ v$^{m,c}$ & v$_{\rm LSR}^{m,c}$\\
              & ($\arcsec$)    & (K)                & (cm$^{-2}$)           & (km~s$^{-1}$)      & (km~s$^{-1}$)      \\
\hline
}

\tablehead{%
\multicolumn{6}{c}{(Continued)}\\
\hline
\hline
Molecule      & $\theta^{m,c}$ & T$_{\rm ex}^{m,c}$ & N$_{\rm tot}^{m,c}$   & $\Delta$ v$^{m,c}$ & v$_{\rm LSR}^{m,c}$\\
              & ($\arcsec$)    & (K)                & (cm$^{-2}$)           & (km~s$^{-1}$)      & (km~s$^{-1}$)      \\
\hline
}

\tabletail{%
\hline
\hline
}

\topcaption{LTE Parameters for the full LTE model (Core Components) for source A09 in Sgr~B2(M).}
\tiny
\centering
\begin{supertabular}{lcccC{1cm}C{1cm}}\label{CoreLTE:parameters:A09SgrB2M}\\
H$_2$CO       &    1.5         &                196 &  7.2 $\times 10^{14}$ &                1.5 &                  65\\
C$_2$H$_5$OH  &    1.5         &                124 &  1.3 $\times 10^{17}$ &                4.1 &                  69\\
              &    1.5         &                 39 &  7.0 $\times 10^{13}$ &                0.1 &                  49\\
OCS           &    1.5         &                238 &  9.4 $\times 10^{17}$ &                7.2 &                  65\\
              &    1.5         &                199 &  1.0 $\times 10^{13}$ &                5.0 &                  65\\
OCS, v$_2$=1  &    1.5         &                238 &  5.2 $\times 10^{17}$ &                7.2 &                  66\\
H$_2$CCN      &    1.5         &                130 &  6.1 $\times 10^{14}$ &                7.2 &                  61\\
\end{supertabular}\\
\vspace{1cm}

%---------------------------------------
% Envelope Components

\tablefirsthead{%
\hline
\hline
Molecule      & $\theta^{m,c}$ & T$_{\rm ex}^{m,c}$ & N$_{\rm tot}^{m,c}$   & $\Delta$ v$^{m,c}$ & v$_{\rm LSR}^{m,c}$\\
              & ($\arcsec$)    & (K)                & (cm$^{-2}$)           & (km~s$^{-1}$)      & (km~s$^{-1}$)      \\
\hline
}

\tablehead{%
\multicolumn{6}{c}{(Continued)}\\
\hline
\hline
Molecule      & $\theta^{m,c}$ & T$_{\rm ex}^{m,c}$ & N$_{\rm tot}^{m,c}$   & $\Delta$ v$^{m,c}$ & v$_{\rm LSR}^{m,c}$\\
              & ($\arcsec$)    & (K)                & (cm$^{-2}$)           & (km~s$^{-1}$)      & (km~s$^{-1}$)      \\
\hline
}

\tabletail{%
\hline
\hline
}

\topcaption{LTE Parameters for the full LTE model (Envelope Components) for source A09 in Sgr~B2(M).}
\tiny
\centering
\begin{supertabular}{lcccC{1cm}C{1cm}}\label{EnvLTE:parameters:A09SgrB2M}\\
H$_2$CO       & ext.           &                 11 &  1.5 $\times 10^{14}$ &                2.8 &                  74\\
              & ext.           &                 12 &  5.8 $\times 10^{14}$ &               21.4 &                  57\\
              & ext.           &                 13 &  1.9 $\times 10^{14}$ &                3.4 &                  52\\
CN            & ext.           &                 24 &  7.2 $\times 10^{16}$ &               16.3 &                  63\\
              & ext.           &                 66 &  5.0 $\times 10^{13}$ &               17.3 &                  40\\
              & ext.           &                 15 &  2.1 $\times 10^{14}$ &                8.0 &                  31\\
              & ext.           &                 10 &  6.3 $\times 10^{14}$ &               16.7 &                   7\\
\end{supertabular}\\
\vspace{1cm}

%================================================================================
%
% Source A10

%---------------------------------------
% Core Components

\tablefirsthead{%
\hline
\hline
Molecule      & $\theta^{m,c}$ & T$_{\rm ex}^{m,c}$ & N$_{\rm tot}^{m,c}$   & $\Delta$ v$^{m,c}$ & v$_{\rm LSR}^{m,c}$\\
              & ($\arcsec$)    & (K)                & (cm$^{-2}$)           & (km~s$^{-1}$)      & (km~s$^{-1}$)      \\
\hline
}

\tablehead{%
\multicolumn{6}{c}{(Continued)}\\
\hline
\hline
Molecule      & $\theta^{m,c}$ & T$_{\rm ex}^{m,c}$ & N$_{\rm tot}^{m,c}$   & $\Delta$ v$^{m,c}$ & v$_{\rm LSR}^{m,c}$\\
              & ($\arcsec$)    & (K)                & (cm$^{-2}$)           & (km~s$^{-1}$)      & (km~s$^{-1}$)      \\
\hline
}

\tabletail{%
\hline
\hline
}

\topcaption{LTE Parameters for the full LTE model (Core Components) for source A10 in Sgr~B2(M).}
\tiny
\centering
\begin{supertabular}{lcccC{1cm}C{1cm}}\label{CoreLTE:parameters:A10SgrB2M}\\
CH$_3$OH      &    1.7         &                131 &  1.4 $\times 10^{18}$ &                4.2 &                  61\\
CH$_3$OCH$_3$ &    1.7         &                153 &  8.8 $\times 10^{16}$ &                3.3 &                  60\\
              &    1.7         &                 76 &  5.5 $\times 10^{16}$ &                5.6 &                  61\\
OCS           &    1.7         &                111 &  5.9 $\times 10^{16}$ &                6.4 &                  68\\
              &    1.7         &                139 &  1.2 $\times 10^{14}$ &                5.5 &                  63\\
              &    1.7         &                138 &  2.1 $\times 10^{17}$ &                5.7 &                  60\\
HC(O)NH$_2$   &    1.7         &                200 &  1.2 $\times 10^{16}$ &                6.5 &                  64\\
\end{supertabular}\\
\vspace{1cm}

%---------------------------------------
% Envelope Components

\tablefirsthead{%
\hline
\hline
Molecule      & $\theta^{m,c}$ & T$_{\rm ex}^{m,c}$ & N$_{\rm tot}^{m,c}$   & $\Delta$ v$^{m,c}$ & v$_{\rm LSR}^{m,c}$\\
              & ($\arcsec$)    & (K)                & (cm$^{-2}$)           & (km~s$^{-1}$)      & (km~s$^{-1}$)      \\
\hline
}

\tablehead{%
\multicolumn{6}{c}{(Continued)}\\
\hline
\hline
Molecule      & $\theta^{m,c}$ & T$_{\rm ex}^{m,c}$ & N$_{\rm tot}^{m,c}$   & $\Delta$ v$^{m,c}$ & v$_{\rm LSR}^{m,c}$\\
              & ($\arcsec$)    & (K)                & (cm$^{-2}$)           & (km~s$^{-1}$)      & (km~s$^{-1}$)      \\
\hline
}

\tabletail{%
\hline
\hline
}

\topcaption{LTE Parameters for the full LTE model (Envelope Components) for source A10 in Sgr~B2(M).}
\tiny
\centering
\begin{supertabular}{lcccC{1cm}C{1cm}}\label{EnvLTE:parameters:A10SgrB2M}\\
CH$_3$OH      & ext.           &                  5 &  9.5 $\times 10^{14}$ &               10.4 &                  65\\
              & ext.           &                  6 &  2.2 $\times 10^{15}$ &                3.9 &                  64\\
              & ext.           &                 13 &  1.5 $\times 10^{15}$ &                6.8 &                  55\\
PH$_3$        & ext.           &                 52 &  4.3 $\times 10^{17}$ &                2.0 &                  63\\
              & ext.           &                 29 &  9.5 $\times 10^{14}$ &                3.1 &                  55\\
H$_2$CNH      & ext.           &                  3 &  1.2 $\times 10^{13}$ &                6.0 &                  64\\
              & ext.           &                  3 &  1.8 $\times 10^{13}$ &                6.7 &                  57\\
CN            & ext.           &                  6 &  5.6 $\times 10^{13}$ &               11.5 &                  83\\
              & ext.           &                 39 &  1.5 $\times 10^{17}$ &               12.0 &                  63\\
              & ext.           &                 19 &  3.2 $\times 10^{14}$ &                6.5 &                  31\\
              & ext.           &                  4 &  1.3 $\times 10^{14}$ &               10.7 &                  17\\
              & ext.           &                 23 &  8.8 $\times 10^{14}$ &               10.1 &                   4\\
              & ext.           &                  4 &  4.4 $\times 10^{13}$ &                3.9 &                  -2\\
              & ext.           &                 22 &  5.3 $\times 10^{14}$ &               11.6 &                 -26\\
              & ext.           &                 31 &  1.4 $\times 10^{15}$ &                9.5 &                 -46\\
              & ext.           &                  5 &  8.1 $\times 10^{12}$ &                2.9 &                 -67\\
              & ext.           &                  5 &  1.1 $\times 10^{14}$ &               12.9 &                -101\\
H$_2$CO       & ext.           &                 11 &  4.2 $\times 10^{13}$ &                3.1 &                  72\\
              & ext.           &                 41 &  1.1 $\times 10^{16}$ &                9.6 &                  64\\
              & ext.           &                 33 &  2.4 $\times 10^{15}$ &                4.0 &                  55\\
CCH           & ext.           &                  3 &  3.3 $\times 10^{15}$ &               11.8 &                  66\\
              & ext.           &                 47 &  3.4 $\times 10^{14}$ &                3.6 &                  54\\
SiO           & ext.           &                  8 &  1.3 $\times 10^{14}$ &               13.7 &                  70\\
              & ext.           &                 38 &  9.0 $\times 10^{14}$ &                8.6 &                  58\\
              & ext.           &                 14 &  2.2 $\times 10^{14}$ &               13.9 &                  47\\
CH$_2$NH      & ext.           &                  3 &  1.0 $\times 10^{13}$ &                2.9 &                  73\\
              & ext.           &                  3 &  2.9 $\times 10^{13}$ &                9.3 &                  64\\
              & ext.           &                  8 &  1.3 $\times 10^{13}$ &                2.8 &                  55\\
CO            & ext.           &                  3 &  7.1 $\times 10^{15}$ &               10.6 &                 104\\
              & ext.           &                  3 &  6.0 $\times 10^{16}$ &               39.7 &                  84\\
              & ext.           &                  3 &  6.3 $\times 10^{15}$ &                2.0 &                 -77\\
              & ext.           &                  3 &  1.8 $\times 10^{13}$ &                1.4 &                  68\\
              & ext.           &                  3 &  5.3 $\times 10^{15}$ &                4.2 &                -113\\
              & ext.           &                  3 &  4.2 $\times 10^{15}$ &                6.6 &                -107\\
              & ext.           &                  3 &  4.9 $\times 10^{15}$ &                2.0 &                  51\\
              & ext.           &                  3 &  7.0 $\times 10^{16}$ &               39.8 &                  44\\
              & ext.           &                  3 &  5.2 $\times 10^{14}$ &                1.9 &                  41\\
              & ext.           &                  3 &  1.8 $\times 10^{15}$ &                0.8 &                  30\\
              & ext.           &                  3 &  1.8 $\times 10^{16}$ &                0.6 &                  14\\
              & ext.           &                  3 &  1.0 $\times 10^{15}$ &                0.9 &                  20\\
              & ext.           &                  3 &  1.2 $\times 10^{14}$ &                0.3 &                  74\\
              & ext.           &                  3 &  2.0 $\times 10^{16}$ &               18.5 &                  11\\
              & ext.           &                  3 &  1.4 $\times 10^{16}$ &               10.0 &                  -2\\
              & ext.           &                  3 &  9.8 $\times 10^{14}$ &                0.6 &                 -11\\
              & ext.           &                  3 &  3.4 $\times 10^{16}$ &               19.8 &                 -24\\
              & ext.           &                  3 &  3.1 $\times 10^{15}$ &                4.0 &                 -33\\
              & ext.           &                  3 &  1.2 $\times 10^{16}$ &                7.7 &                 -46\\
              & ext.           &                  3 &  5.9 $\times 10^{15}$ &                4.3 &                 -51\\
              & ext.           &                  3 &  5.9 $\times 10^{15}$ &                3.3 &                 -61\\
              & ext.           &                  3 &  1.5 $\times 10^{15}$ &                0.8 &                 -64\\
              & ext.           &                  3 &  3.8 $\times 10^{16}$ &               25.0 &                 -97\\
              & ext.           &                  3 &  7.0 $\times 10^{15}$ &                7.3 &                 -68\\
              & ext.           &                  3 &  2.4 $\times 10^{12}$ &                6.2 &                -102\\
              & ext.           &                  3 &  2.6 $\times 10^{12}$ &                1.6 &                -107\\
              & ext.           &                  3 &  5.0 $\times 10^{14}$ &                0.3 &                 -89\\
              & ext.           &                  3 &  4.6 $\times 10^{15}$ &                4.7 &                 -40\\
$^{13}$CO     & ext.           &                  3 &  2.0 $\times 10^{13}$ &                0.7 &                  86\\
              & ext.           &                  3 &  2.6 $\times 10^{16}$ &               23.9 &                  82\\
              & ext.           &                  3 &  1.8 $\times 10^{16}$ &                0.8 &                  70\\
              & ext.           &                  3 &  2.8 $\times 10^{16}$ &                7.2 &                  53\\
              & ext.           &                  3 &  3.9 $\times 10^{16}$ &               30.6 &                  50\\
              & ext.           &                  3 &  1.9 $\times 10^{13}$ &                0.5 &                  15\\
              & ext.           &                  3 &  3.0 $\times 10^{15}$ &                5.9 &                  13\\
              & ext.           &                  3 &  1.7 $\times 10^{12}$ &                0.2 &                -185\\
              & ext.           &                  3 &  5.9 $\times 10^{15}$ &                8.9 &                   2\\
              & ext.           &                  3 &  5.2 $\times 10^{13}$ &                2.1 &                 -32\\
              & ext.           &                  3 &  5.6 $\times 10^{15}$ &                1.1 &                 -44\\
              & ext.           &                  3 &  4.6 $\times 10^{15}$ &                1.1 &                 -48\\
              & ext.           &                  3 &  2.4 $\times 10^{15}$ &                1.0 &                 -62\\
              & ext.           &                  3 &  4.5 $\times 10^{15}$ &               30.6 &                 -49\\
              & ext.           &                  3 &  1.2 $\times 10^{15}$ &                2.6 &                 -78\\
              & ext.           &                  3 &  2.4 $\times 10^{15}$ &                4.1 &                 -98\\
              & ext.           &                  3 &  5.5 $\times 10^{13}$ &                1.0 &                 -99\\
              & ext.           &                  3 &  2.2 $\times 10^{14}$ &                0.9 &                -111\\
C$^{17}$O     & ext.           &                  3 &  1.1 $\times 10^{14}$ &                2.7 &                -175\\
              & ext.           &                  3 &  1.0 $\times 10^{14}$ &                2.7 &                -140\\
              & ext.           &                  3 &  3.0 $\times 10^{14}$ &                7.1 &                -124\\
              & ext.           &                  3 &  3.1 $\times 10^{14}$ &                3.4 &                -116\\
              & ext.           &                  3 &  3.0 $\times 10^{14}$ &                3.4 &                -101\\
              & ext.           &                  3 &  7.3 $\times 10^{14}$ &                7.1 &                 -90\\
              & ext.           &                  3 &  8.4 $\times 10^{14}$ &                9.8 &                 -77\\
              & ext.           &                  3 &  1.0 $\times 10^{14}$ &                2.1 &                 -65\\
              & ext.           &                  3 &  2.7 $\times 10^{14}$ &                1.6 &                 -51\\
              & ext.           &                  3 &  8.5 $\times 10^{14}$ &               13.7 &                 -46\\
              & ext.           &                  3 &  7.9 $\times 10^{14}$ &                4.6 &                   1\\
              & ext.           &                  3 &  3.2 $\times 10^{14}$ &                3.7 &                  11\\
              & ext.           &                  3 &  3.0 $\times 10^{14}$ &                3.6 &                  33\\
              & ext.           &                  3 &  1.6 $\times 10^{14}$ &                3.9 &                  39\\
              & ext.           &                  3 &  7.5 $\times 10^{15}$ &                6.7 &                  52\\
              & ext.           &                  3 &  2.3 $\times 10^{15}$ &                6.2 &                  73\\
              & ext.           &                  3 &  2.9 $\times 10^{14}$ &                2.3 &                 124\\
              & ext.           &                  3 &  6.2 $\times 10^{14}$ &                4.4 &                 129\\
C$^{18}$O     & ext.           &                  3 &  1.9 $\times 10^{13}$ &                2.6 &                -120\\
              & ext.           &                  3 &  5.5 $\times 10^{13}$ &                2.6 &                -112\\
              & ext.           &                  3 &  2.7 $\times 10^{15}$ &               30.3 &                 -57\\
              & ext.           &                  3 &  9.2 $\times 10^{14}$ &                4.3 &                 -44\\
              & ext.           &                  3 &  1.7 $\times 10^{16}$ &                8.9 &                  51\\
              & ext.           &                  3 &  5.3 $\times 10^{15}$ &                9.0 &                  73\\
              & ext.           &                  3 &  7.0 $\times 10^{14}$ &                4.5 &                  92\\
              & ext.           &                  3 &  1.3 $\times 10^{14}$ &                2.6 &                 113\\
              & ext.           &                  3 &  3.6 $\times 10^{14}$ &                7.7 &                 109\\
              & ext.           &                  3 &  3.7 $\times 10^{14}$ &                5.6 &                 118\\
CS            & ext.           &                  3 &  1.2 $\times 10^{12}$ &               12.1 &                  86\\
              & ext.           &                  3 &  1.4 $\times 10^{15}$ &                1.7 &                  69\\
              & ext.           &                  3 &  7.6 $\times 10^{16}$ &               33.5 &                  61\\
              & ext.           &                  3 &  4.8 $\times 10^{16}$ &                3.6 &                  54\\
$^{13}$CS     & ext.           &                  3 &  2.4 $\times 10^{14}$ &                4.3 &                  78\\
              & ext.           &                  3 &  2.9 $\times 10^{15}$ &                0.2 &                  66\\
              & ext.           &                  3 &  3.0 $\times 10^{15}$ &                4.0 &                  56\\
              & ext.           &                  3 &  1.7 $\times 10^{16}$ &               11.8 &                  55\\
C$^{33}$S     & ext.           &                  3 &  8.7 $\times 10^{16}$ &               20.1 &                -153\\
              & ext.           &                  3 &  2.8 $\times 10^{16}$ &                9.0 &                -140\\
              & ext.           &                  3 &  1.0 $\times 10^{16}$ &                6.8 &                -106\\
              & ext.           &                  3 &  4.2 $\times 10^{15}$ &                9.7 &                 -50\\
              & ext.           &                  3 &  5.7 $\times 10^{14}$ &                3.0 &                 -15\\
              & ext.           &                  3 &  8.8 $\times 10^{14}$ &                4.3 &                  25\\
              & ext.           &                  3 &  4.6 $\times 10^{15}$ &                5.9 &                  52\\
              & ext.           &                  3 &  2.8 $\times 10^{15}$ &                9.3 &                  68\\
              & ext.           &                  3 &  5.7 $\times 10^{15}$ &               11.2 &                 107\\
              & ext.           &                  3 &  1.3 $\times 10^{15}$ &                2.8 &                 122\\
C$^{34}$S     & ext.           &                  3 &  3.3 $\times 10^{14}$ &                2.5 &                 -70\\
              & ext.           &                  3 &  3.0 $\times 10^{14}$ &                2.5 &                 -50\\
              & ext.           &                  3 &  6.0 $\times 10^{14}$ &                6.2 &                  20\\
              & ext.           &                  3 &  2.6 $\times 10^{16}$ &               11.8 &                  55\\
              & ext.           &                  3 &  4.3 $\times 10^{15}$ &               28.1 &                  75\\
HCN           & ext.           &                  3 &  8.0 $\times 10^{12}$ &                1.6 &                 114\\
              & ext.           &                  3 &  6.6 $\times 10^{13}$ &                3.8 &                  79\\
              & ext.           &                  3 &  2.4 $\times 10^{13}$ &                2.0 &                  75\\
              & ext.           &                  3 &  8.7 $\times 10^{12}$ &                1.6 &                  69\\
              & ext.           &                  3 &  2.8 $\times 10^{14}$ &               24.2 &                  58\\
              & ext.           &                  3 &  1.1 $\times 10^{14}$ &                5.8 &                  35\\
              & ext.           &                  3 &  3.4 $\times 10^{14}$ &               30.6 &                  20\\
              & ext.           &                  3 &  2.3 $\times 10^{13}$ &                0.3 &                  28\\
              & ext.           &                  3 &  5.0 $\times 10^{13}$ &                7.8 &                  19\\
H$^{13}$CN    & ext.           &                  3 &  6.5 $\times 10^{13}$ &                6.4 &                 126\\
              & ext.           &                  3 &  1.2 $\times 10^{12}$ &                1.5 &                  98\\
              & ext.           &                  3 &  8.0 $\times 10^{12}$ &                0.5 &                  93\\
              & ext.           &                  3 &  1.5 $\times 10^{13}$ &                0.1 &                  65\\
              & ext.           &                  3 &  1.4 $\times 10^{15}$ &               16.2 &                  62\\
              & ext.           &                  3 &  1.1 $\times 10^{13}$ &                3.5 &                  16\\
HNC           & ext.           &                  3 &  1.8 $\times 10^{13}$ &                2.1 &                  91\\
              & ext.           &                  3 &  2.1 $\times 10^{13}$ &                2.4 &                  82\\
              & ext.           &                  3 &  5.6 $\times 10^{13}$ &                1.5 &                  74\\
              & ext.           &                  3 &  5.5 $\times 10^{14}$ &               16.1 &                  65\\
              & ext.           &                  3 &  1.3 $\times 10^{14}$ &                6.1 &                  55\\
              & ext.           &                  3 &  1.0 $\times 10^{14}$ &               10.6 &                  42\\
              & ext.           &                  3 &  1.2 $\times 10^{12}$ &                3.0 &                  31\\
              & ext.           &                  3 &  2.2 $\times 10^{13}$ &                3.3 &                  16\\
              & ext.           &                  3 &  1.4 $\times 10^{12}$ &                1.0 &                 -97\\
              & ext.           &                  3 &  2.4 $\times 10^{12}$ &                1.0 &                -110\\
H$^{13}$CO$^+$ & ext.           &                  3 &  1.1 $\times 10^{14}$ &               10.9 &                  55\\
              & ext.           &                  3 &  1.1 $\times 10^{12}$ &                1.4 &                  50\\
\end{supertabular}\\
\vspace{1cm}

%================================================================================
%
% Source A11

%---------------------------------------
% Core Components

\tablefirsthead{%
\hline
\hline
Molecule      & $\theta^{m,c}$ & T$_{\rm ex}^{m,c}$ & N$_{\rm tot}^{m,c}$   & $\Delta$ v$^{m,c}$ & v$_{\rm LSR}^{m,c}$\\
              & ($\arcsec$)    & (K)                & (cm$^{-2}$)           & (km~s$^{-1}$)      & (km~s$^{-1}$)      \\
\hline
}

\tablehead{%
\multicolumn{6}{c}{(Continued)}\\
\hline
\hline
Molecule      & $\theta^{m,c}$ & T$_{\rm ex}^{m,c}$ & N$_{\rm tot}^{m,c}$   & $\Delta$ v$^{m,c}$ & v$_{\rm LSR}^{m,c}$\\
              & ($\arcsec$)    & (K)                & (cm$^{-2}$)           & (km~s$^{-1}$)      & (km~s$^{-1}$)      \\
\hline
}

\tabletail{%
\hline
\hline
}

\topcaption{LTE Parameters for the full LTE model (Core Components) for source A11 in Sgr~B2(M).}
\tiny
\centering
\begin{supertabular}{lcccC{1cm}C{1cm}}\label{CoreLTE:parameters:A11SgrB2M}\\
CH$_3$OCHO    &    1.6         &                200 &  1.1 $\times 10^{16}$ &                7.5 &                  64\\
CN            &    1.6         &                200 &  7.5 $\times 10^{15}$ &                3.2 &                  26\\
CCH           &    1.6         &                 33 &  1.8 $\times 10^{16}$ &               12.8 &                  61\\
H$_2$CNH      &    1.6         &                223 &  3.0 $\times 10^{15}$ &                7.7 &                  63\\
CH$_3$OCH$_3$ &    1.6         &                153 &  9.0 $\times 10^{15}$ &                5.3 &                  64\\
CH$_3$OH      &    1.6         &                124 &  3.3 $\times 10^{16}$ &                8.0 &                  64\\
C$_2$H$_5$CN  &    1.6         &                163 &  6.2 $\times 10^{14}$ &                4.2 &                  65\\
C$_2$H$_5$CN, v$_{20}$=1 &    1.6         &                594 &  1.2 $\times 10^{14}$ &                9.2 &                  57\\
SiO           &    1.6         &                205 &  2.1 $\times 10^{15}$ &               24.1 &                  59\\
OCS           &    1.6         &                382 &  7.4 $\times 10^{12}$ &                8.1 &                  63\\
CH$_3$CN      &    1.6         &                115 &  1.9 $\times 10^{14}$ &                3.6 &                  64\\
H$_2$CCO      &    1.6         &                194 &  1.4 $\times 10^{14}$ &                3.8 &                  63\\
HCCCN         &    1.6         &                117 &  3.5 $\times 10^{14}$ &                4.2 &                  64\\
H$_2$CS       &    1.6         &                159 &  4.7 $\times 10^{15}$ &                8.3 &                  64\\
$^{13}$CO     &    1.6         &                200 &  5.4 $\times 10^{15}$ &                8.0 &                 115\\
              &    1.6         &                200 &  4.6 $\times 10^{16}$ &                1.3 &                 105\\
              &    1.6         &                200 &  5.8 $\times 10^{16}$ &                5.1 &                  97\\
              &    1.6         &                200 &  1.3 $\times 10^{17}$ &                8.1 &                  88\\
              &    1.6         &                200 &  7.7 $\times 10^{16}$ &                3.6 &                  52\\
              &    1.6         &                200 &  1.1 $\times 10^{17}$ &                0.9 &                  -4\\
C$^{17}$O     &    1.6         &                200 &  9.6 $\times 10^{14}$ &                2.7 &                -180\\
              &    1.6         &                200 &  4.6 $\times 10^{16}$ &               12.2 &                -121\\
              &    1.6         &                200 &  3.3 $\times 10^{15}$ &                2.7 &                 -95\\
              &    1.6         &                200 &  2.5 $\times 10^{16}$ &                7.6 &                 -82\\
              &    1.6         &                200 &  1.7 $\times 10^{16}$ &                2.7 &                 -50\\
              &    1.6         &                200 &  5.4 $\times 10^{15}$ &                2.7 &                 -26\\
              &    1.6         &                200 &  8.0 $\times 10^{15}$ &                2.8 &                 -18\\
              &    1.6         &                200 &  1.1 $\times 10^{16}$ &                3.8 &                 -13\\
              &    1.6         &                200 &  5.9 $\times 10^{15}$ &                2.7 &                  -4\\
              &    1.6         &                200 &  3.0 $\times 10^{14}$ &                2.7 &                  15\\
              &    1.6         &                200 &  3.0 $\times 10^{14}$ &                2.7 &                  22\\
              &    1.6         &                200 &  1.2 $\times 10^{16}$ &                2.7 &                  28\\
              &    1.6         &                200 &  7.7 $\times 10^{15}$ &                2.8 &                  33\\
              &    1.6         &                200 &  3.1 $\times 10^{16}$ &                2.7 &                  41\\
              &    1.6         &                200 &  2.4 $\times 10^{16}$ &                3.3 &                  84\\
              &    1.6         &                200 &  3.7 $\times 10^{15}$ &                2.7 &                  94\\
              &    1.6         &                200 &  1.3 $\times 10^{16}$ &                4.9 &                 105\\
              &    1.6         &                200 &  2.2 $\times 10^{15}$ &                2.7 &                 115\\
              &    1.6         &                200 &  8.6 $\times 10^{15}$ &                3.1 &                 135\\
              &    1.6         &                200 &  3.3 $\times 10^{15}$ &                2.7 &                 154\\
C$^{18}$O     &    1.6         &                200 &  3.8 $\times 10^{15}$ &                3.4 &                -167\\
              &    1.6         &                200 &  4.3 $\times 10^{15}$ &                2.7 &                -159\\
              &    1.6         &                200 &  1.6 $\times 10^{16}$ &                3.7 &                -132\\
              &    1.6         &                200 &  3.6 $\times 10^{15}$ &                2.2 &                -115\\
              &    1.6         &                200 &  1.5 $\times 10^{16}$ &                4.1 &                -109\\
              &    1.6         &                200 &  4.1 $\times 10^{15}$ &                2.7 &                 -88\\
              &    1.6         &                200 &  7.3 $\times 10^{15}$ &                2.7 &                 -80\\
              &    1.6         &                200 &  1.2 $\times 10^{16}$ &                3.6 &                 -63\\
              &    1.6         &                200 &  5.5 $\times 10^{15}$ &                3.0 &                 -50\\
              &    1.6         &                200 &  8.5 $\times 10^{15}$ &                3.0 &                 -14\\
              &    1.6         &                200 &  1.1 $\times 10^{16}$ &                2.7 &                  22\\
              &    1.6         &                200 &  5.1 $\times 10^{15}$ &                2.7 &                  27\\
              &    1.6         &                200 &  7.1 $\times 10^{15}$ &                2.9 &                  35\\
              &    1.6         &                200 &  1.7 $\times 10^{17}$ &               21.3 &                  84\\
              &    1.6         &                200 &  2.0 $\times 10^{12}$ &                7.5 &                 125\\
              &    1.6         &                200 &  1.4 $\times 10^{17}$ &               22.1 &                 137\\
CS            &    1.6         &                200 &  2.8 $\times 10^{13}$ &                3.5 &                 112\\
              &    1.6         &                200 &  4.0 $\times 10^{13}$ &                0.8 &                 107\\
              &    1.6         &                200 &  1.8 $\times 10^{14}$ &                7.3 &                 101\\
              &    1.6         &                200 &  1.1 $\times 10^{14}$ &                3.9 &                  90\\
              &    1.6         &                200 &  8.4 $\times 10^{14}$ &                1.1 &                  51\\
              &    1.6         &                200 &  1.7 $\times 10^{13}$ &                0.9 &                  22\\
              &    1.6         &                200 &  9.5 $\times 10^{13}$ &                4.4 &                  14\\
              &    1.6         &                200 &  4.5 $\times 10^{13}$ &                2.3 &                  -1\\
              &    1.6         &                200 &  6.3 $\times 10^{13}$ &                1.2 &                 -62\\
              &    1.6         &                200 &  5.7 $\times 10^{13}$ &                5.5 &                 -68\\
$^{13}$CS     &    1.6         &                200 &  4.2 $\times 10^{13}$ &                2.1 &                  96\\
              &    1.6         &                200 &  5.5 $\times 10^{13}$ &                1.1 &                  33\\
              &    1.6         &                200 &  7.1 $\times 10^{13}$ &                4.0 &                   0\\
              &    1.6         &                200 &  8.4 $\times 10^{13}$ &                4.1 &                 -13\\
              &    1.6         &                200 &  3.0 $\times 10^{13}$ &                2.3 &                 -25\\
              &    1.6         &                200 &  2.5 $\times 10^{14}$ &                1.0 &                 -40\\
              &    1.6         &                200 &  1.7 $\times 10^{12}$ &                1.0 &                 -55\\
              &    1.6         &                200 &  3.4 $\times 10^{14}$ &                0.9 &                 -86\\
C$^{34}$S     &    1.6         &                200 &  6.9 $\times 10^{13}$ &                6.2 &                  97\\
              &    1.6         &                200 & 10.0 $\times 10^{13}$ &                8.8 &                 102\\
HCN           &    1.6         &                200 &  1.9 $\times 10^{14}$ &                1.6 &                  50\\
              &    1.6         &                200 &  1.3 $\times 10^{14}$ &                3.0 &                  44\\
              &    1.6         &                200 &  7.3 $\times 10^{13}$ &                2.1 &                  37\\
              &    1.6         &                200 &  3.5 $\times 10^{13}$ &                2.0 &                  28\\
              &    1.6         &                200 &  5.3 $\times 10^{13}$ &                3.8 &                  16\\
              &    1.6         &                200 &  1.0 $\times 10^{12}$ &                0.7 &                   2\\
              &    1.6         &                200 &  1.3 $\times 10^{13}$ &                0.9 &                   2\\
              &    1.6         &                200 &  6.1 $\times 10^{13}$ &                2.0 &                  -4\\
              &    1.6         &                200 &  5.2 $\times 10^{12}$ &                1.0 &                 -20\\
              &    1.6         &                200 &  1.4 $\times 10^{12}$ &                1.1 &                 -48\\
              &    1.6         &                200 &  1.9 $\times 10^{13}$ &                1.4 &                 -80\\
H$^{13}$CN    &    1.6         &                200 &  1.4 $\times 10^{13}$ &                2.7 &                 116\\
              &    1.6         &                200 &  2.1 $\times 10^{13}$ &                4.0 &                 101\\
              &    1.6         &                200 &  1.4 $\times 10^{14}$ &                5.6 &                  50\\
              &    1.6         &                200 &  1.5 $\times 10^{13}$ &                2.2 &                  38\\
              &    1.6         &                200 &  2.1 $\times 10^{13}$ &                3.7 &                  19\\
              &    1.6         &                200 &  2.5 $\times 10^{13}$ &                4.5 &                  12\\
              &    1.6         &                200 &  3.8 $\times 10^{12}$ &                0.7 &                   7\\
              &    1.6         &                200 &  7.7 $\times 10^{12}$ &                1.0 &                   2\\
              &    1.6         &                200 &  1.5 $\times 10^{13}$ &                2.7 &                 -26\\
              &    1.6         &                200 &  2.4 $\times 10^{12}$ &                0.6 &                 -43\\
              &    1.6         &                200 &  1.0 $\times 10^{13}$ &                2.2 &                 -58\\
              &    1.6         &                200 &  2.7 $\times 10^{12}$ &                0.8 &                 -44\\
              &    1.6         &                200 &  9.0 $\times 10^{12}$ &                3.3 &                 -67\\
              &    1.6         &                200 &  7.3 $\times 10^{12}$ &                3.0 &                 -75\\
              &    1.6         &                200 &  8.9 $\times 10^{12}$ &                2.3 &                 -80\\
              &    1.6         &                200 &  2.2 $\times 10^{13}$ &                1.1 &                 -97\\
              &    1.6         &                200 &  1.9 $\times 10^{13}$ &                1.3 &                -110\\
              &    1.6         &                200 &  1.7 $\times 10^{12}$ &                1.1 &                -114\\
HNC           &    1.6         &                200 &  1.0 $\times 10^{12}$ &                0.3 &                 131\\
              &    1.6         &                200 &  1.2 $\times 10^{14}$ &                7.4 &                 131\\
              &    1.6         &                200 &  1.0 $\times 10^{13}$ &                1.8 &                 107\\
              &    1.6         &                200 &  1.7 $\times 10^{12}$ &                0.8 &                  96\\
              &    1.6         &                200 &  2.0 $\times 10^{13}$ &                1.2 &                  89\\
              &    1.6         &                200 &  1.8 $\times 10^{12}$ &               22.0 &                  76\\
              &    1.6         &                200 &  8.5 $\times 10^{14}$ &                5.1 &                  51\\
              &    1.6         &                200 &  3.0 $\times 10^{13}$ &                2.1 &                  25\\
              &    1.6         &                200 &  5.6 $\times 10^{13}$ &                6.1 &                  14\\
              &    1.6         &                200 &  1.6 $\times 10^{12}$ &                0.8 &                  36\\
              &    1.6         &                200 &  4.1 $\times 10^{13}$ &                3.2 &                  -4\\
              &    1.6         &                200 &  1.7 $\times 10^{12}$ &                8.9 &                 -18\\
              &    1.6         &                200 &  2.0 $\times 10^{13}$ &                2.2 &                 -37\\
              &    1.6         &                200 &  4.0 $\times 10^{13}$ &                1.2 &                 -48\\
              &    1.6         &                200 &  2.5 $\times 10^{13}$ &                0.8 &                 -54\\
              &    1.6         &                200 &  4.0 $\times 10^{13}$ &                4.8 &                 -57\\
H$^{13}$CO$^+$ &    1.6         &                200 &  5.7 $\times 10^{13}$ &                4.5 &                  62\\
HO$^{13}$C$^+$ &    1.6         &                200 &  4.4 $\times 10^{13}$ &               40.0 &                 155\\
              &    1.6         &                200 &  1.5 $\times 10^{12}$ &                1.0 &                 162\\
              &    1.6         &                200 &  1.0 $\times 10^{12}$ &                0.6 &                 157\\
              &    1.6         &                200 &  1.4 $\times 10^{12}$ &               26.0 &                 147\\
              &    1.6         &                200 &  1.1 $\times 10^{12}$ &                0.6 &                 137\\
              &    1.6         &                200 &  6.1 $\times 10^{12}$ &                5.2 &                 123\\
              &    1.6         &                200 &  4.4 $\times 10^{12}$ &                0.5 &                 118\\
              &    1.6         &                200 &  1.6 $\times 10^{13}$ &                6.2 &                 111\\
              &    1.6         &                200 &  2.5 $\times 10^{12}$ &                0.2 &                 104\\
              &    1.6         &                200 &  4.1 $\times 10^{12}$ &                0.8 &                  98\\
              &    1.6         &                200 &  1.5 $\times 10^{13}$ &               20.6 &                  87\\
              &    1.6         &                200 &  2.6 $\times 10^{12}$ &                1.1 &                  82\\
              &    1.6         &                200 &  5.3 $\times 10^{13}$ &               32.8 &                  64\\
              &    1.6         &                200 &  1.4 $\times 10^{13}$ &                4.9 &                  34\\
              &    1.6         &                200 &  1.8 $\times 10^{13}$ &                0.9 &                  25\\
              &    1.6         &                200 &  1.0 $\times 10^{13}$ &                4.1 &                  20\\
              &    1.6         &                200 &  5.6 $\times 10^{12}$ &                0.1 &                  14\\
              &    1.6         &                200 &  1.5 $\times 10^{12}$ &                0.6 &                  11\\
              &    1.6         &                200 &  1.2 $\times 10^{13}$ &               10.3 &                  12\\
              &    1.6         &                200 &  1.0 $\times 10^{12}$ &                0.3 &                   4\\
              &    1.6         &                200 &  2.1 $\times 10^{13}$ &                7.8 &                 -14\\
              &    1.6         &                200 &  6.5 $\times 10^{12}$ &                0.6 &                 -17\\
              &    1.6         &                200 &  1.2 $\times 10^{13}$ &                3.7 &                 -22\\
              &    1.6         &                200 &  8.7 $\times 10^{13}$ &               32.5 &                 -59\\
              &    1.6         &                200 &  7.2 $\times 10^{12}$ &                2.9 &                 -90\\
              &    1.6         &                200 &  1.0 $\times 10^{12}$ &                1.1 &                 -96\\
              &    1.6         &                200 &  5.8 $\times 10^{13}$ &               29.8 &                -106\\
              &    1.6         &                200 &  1.2 $\times 10^{12}$ &                1.7 &                -110\\
              &    1.6         &                200 &  8.6 $\times 10^{12}$ &                6.2 &                -123\\
              &    1.6         &                200 &  1.0 $\times 10^{12}$ &                0.7 &                -122\\
              &    1.6         &                200 &  1.8 $\times 10^{13}$ &                6.6 &                -136\\
              &    1.6         &                200 &  1.0 $\times 10^{12}$ &                0.2 &                -173\\
              &    1.6         &                200 &  7.8 $\times 10^{12}$ &                2.3 &                -176\\
              &    1.6         &                200 &  1.6 $\times 10^{12}$ &                0.8 &                -172\\
\end{supertabular}\\
\vspace{1cm}

%---------------------------------------
% Envelope Components

\tablefirsthead{%
\hline
\hline
Molecule      & $\theta^{m,c}$ & T$_{\rm ex}^{m,c}$ & N$_{\rm tot}^{m,c}$   & $\Delta$ v$^{m,c}$ & v$_{\rm LSR}^{m,c}$\\
              & ($\arcsec$)    & (K)                & (cm$^{-2}$)           & (km~s$^{-1}$)      & (km~s$^{-1}$)      \\
\hline
}

\tablehead{%
\multicolumn{6}{c}{(Continued)}\\
\hline
\hline
Molecule      & $\theta^{m,c}$ & T$_{\rm ex}^{m,c}$ & N$_{\rm tot}^{m,c}$   & $\Delta$ v$^{m,c}$ & v$_{\rm LSR}^{m,c}$\\
              & ($\arcsec$)    & (K)                & (cm$^{-2}$)           & (km~s$^{-1}$)      & (km~s$^{-1}$)      \\
\hline
}

\tabletail{%
\hline
\hline
}

\topcaption{LTE Parameters for the full LTE model (Envelope Components) for source A11 in Sgr~B2(M).}
\tiny
\centering
\begin{supertabular}{lcccC{1cm}C{1cm}}\label{EnvLTE:parameters:A11SgrB2M}\\
H$_2 \! ^{34}$S & ext.           &                 10 &  1.6 $\times 10^{17}$ &               13.0 &                  56\\
SO$_2$        & ext.           &                 15 &  5.2 $\times 10^{15}$ &               12.2 &                  61\\
              & ext.           &                 20 &  1.0 $\times 10^{17}$ &                3.0 &                  54\\
              & ext.           &                 10 &  9.6 $\times 10^{14}$ &                7.2 &                  45\\
H$_2$CO       & ext.           &                 11 &  3.7 $\times 10^{14}$ &                8.7 &                  72\\
              & ext.           &                  9 &  4.6 $\times 10^{14}$ &                8.1 &                  64\\
              & ext.           &                  8 &  7.8 $\times 10^{14}$ &                5.4 &                  57\\
\end{supertabular}\\
\vspace{1cm}

%================================================================================
%
% Source A12

%---------------------------------------
% Core Components

\tablefirsthead{%
\hline
\hline
Molecule      & $\theta^{m,c}$ & T$_{\rm ex}^{m,c}$ & N$_{\rm tot}^{m,c}$   & $\Delta$ v$^{m,c}$ & v$_{\rm LSR}^{m,c}$\\
              & ($\arcsec$)    & (K)                & (cm$^{-2}$)           & (km~s$^{-1}$)      & (km~s$^{-1}$)      \\
\hline
}

\tablehead{%
\multicolumn{6}{c}{(Continued)}\\
\hline
\hline
Molecule      & $\theta^{m,c}$ & T$_{\rm ex}^{m,c}$ & N$_{\rm tot}^{m,c}$   & $\Delta$ v$^{m,c}$ & v$_{\rm LSR}^{m,c}$\\
              & ($\arcsec$)    & (K)                & (cm$^{-2}$)           & (km~s$^{-1}$)      & (km~s$^{-1}$)      \\
\hline
}

\tabletail{%
\hline
\hline
}

\topcaption{LTE Parameters for the full LTE model (Core Components) for source A12 in Sgr~B2(M).}
\tiny
\centering
\begin{supertabular}{lcccC{1cm}C{1cm}}\label{CoreLTE:parameters:A12SgrB2M}\\
H$_2 \! ^{34}$S &    1.4         &                364 &  1.2 $\times 10^{17}$ &                6.0 &                  51\\
CH$_3$CN      &    1.4         &                285 &  4.0 $\times 10^{15}$ &                5.9 &                  50\\
HCCCN         &    1.4         &                144 &  1.8 $\times 10^{15}$ &                0.5 &                  51\\
HCCCN, v$_7$=2 &    1.4         &                363 &  1.3 $\times 10^{15}$ &                0.5 &                  51\\
HNC, v$_2$=1  &    1.4         &                200 &  1.8 $\times 10^{16}$ &                6.6 &                  56\\
              &    1.4         &                132 &  1.3 $\times 10^{17}$ &                5.3 &                  51\\
CH$_3$CHO     &    1.4         &                201 &  5.4 $\times 10^{15}$ &                5.1 &                  50\\
\end{supertabular}\\
\vspace{1cm}

%---------------------------------------
% Envelope Components

\tablefirsthead{%
\hline
\hline
Molecule      & $\theta^{m,c}$ & T$_{\rm ex}^{m,c}$ & N$_{\rm tot}^{m,c}$   & $\Delta$ v$^{m,c}$ & v$_{\rm LSR}^{m,c}$\\
              & ($\arcsec$)    & (K)                & (cm$^{-2}$)           & (km~s$^{-1}$)      & (km~s$^{-1}$)      \\
\hline
}

\tablehead{%
\multicolumn{6}{c}{(Continued)}\\
\hline
\hline
Molecule      & $\theta^{m,c}$ & T$_{\rm ex}^{m,c}$ & N$_{\rm tot}^{m,c}$   & $\Delta$ v$^{m,c}$ & v$_{\rm LSR}^{m,c}$\\
              & ($\arcsec$)    & (K)                & (cm$^{-2}$)           & (km~s$^{-1}$)      & (km~s$^{-1}$)      \\
\hline
}

\tabletail{%
\hline
\hline
}

\topcaption{LTE Parameters for the full LTE model (Envelope Components) for source A12 in Sgr~B2(M).}
\tiny
\centering
\begin{supertabular}{lcccC{1cm}C{1cm}}\label{EnvLTE:parameters:A12SgrB2M}\\
CH$_3$CN      & ext.           &                 27 &  2.1 $\times 10^{15}$ &                0.2 &                  83\\
              & ext.           &                 20 &  1.3 $\times 10^{15}$ &                7.3 &                 115\\
              & ext.           &                 14 &  8.9 $\times 10^{14}$ &                7.0 &                  83\\
              & ext.           &                  6 &  1.4 $\times 10^{16}$ &                4.4 &                  64\\
HCCCN         & ext.           &                 24 &  1.6 $\times 10^{17}$ &                7.1 &                  64\\
              & ext.           &                  3 &  8.4 $\times 10^{18}$ &                6.7 &                  62\\
SO            & ext.           &                 16 &  8.5 $\times 10^{15}$ &               22.6 &                  67\\
\end{supertabular}\\
\vspace{1cm}

%================================================================================
%
% Source A13

%---------------------------------------
% Core Components

\tablefirsthead{%
\hline
\hline
Molecule      & $\theta^{m,c}$ & T$_{\rm ex}^{m,c}$ & N$_{\rm tot}^{m,c}$   & $\Delta$ v$^{m,c}$ & v$_{\rm LSR}^{m,c}$\\
              & ($\arcsec$)    & (K)                & (cm$^{-2}$)           & (km~s$^{-1}$)      & (km~s$^{-1}$)      \\
\hline
}

\tablehead{%
\multicolumn{6}{c}{(Continued)}\\
\hline
\hline
Molecule      & $\theta^{m,c}$ & T$_{\rm ex}^{m,c}$ & N$_{\rm tot}^{m,c}$   & $\Delta$ v$^{m,c}$ & v$_{\rm LSR}^{m,c}$\\
              & ($\arcsec$)    & (K)                & (cm$^{-2}$)           & (km~s$^{-1}$)      & (km~s$^{-1}$)      \\
\hline
}

\tabletail{%
\hline
\hline
}

\topcaption{LTE Parameters for the full LTE model (Core Components) for source A13 in Sgr~B2(M).}
\tiny
\centering
\begin{supertabular}{lcccC{1cm}C{1cm}}\label{CoreLTE:parameters:A13SgrB2M}\\
H$_2$CCN      &    1.4         &                 88 &  1.5 $\times 10^{15}$ &                4.1 &                  57\\
SiO           &    1.4         &                379 &  4.3 $\times 10^{15}$ &               12.1 &                  67\\
              &    1.4         &                141 &  1.5 $\times 10^{15}$ &                9.7 &                  52\\
              &    1.4         &                173 &  3.0 $\times 10^{14}$ &               11.2 &                  41\\
OCS           &    1.4         &                148 &  8.8 $\times 10^{16}$ &                4.0 &                  56\\
HC(O)NH$_2$   &    1.4         &                147 &  1.3 $\times 10^{16}$ &               11.0 &                  56\\
SO            &    1.4         &                168 &  3.1 $\times 10^{17}$ &                9.7 &                  65\\
\end{supertabular}\\
\vspace{1cm}

%---------------------------------------
% Envelope Components

\tablefirsthead{%
\hline
\hline
Molecule      & $\theta^{m,c}$ & T$_{\rm ex}^{m,c}$ & N$_{\rm tot}^{m,c}$   & $\Delta$ v$^{m,c}$ & v$_{\rm LSR}^{m,c}$\\
              & ($\arcsec$)    & (K)                & (cm$^{-2}$)           & (km~s$^{-1}$)      & (km~s$^{-1}$)      \\
\hline
}

\tablehead{%
\multicolumn{6}{c}{(Continued)}\\
\hline
\hline
Molecule      & $\theta^{m,c}$ & T$_{\rm ex}^{m,c}$ & N$_{\rm tot}^{m,c}$   & $\Delta$ v$^{m,c}$ & v$_{\rm LSR}^{m,c}$\\
              & ($\arcsec$)    & (K)                & (cm$^{-2}$)           & (km~s$^{-1}$)      & (km~s$^{-1}$)      \\
\hline
}

\tabletail{%
\hline
\hline
}

\topcaption{LTE Parameters for the full LTE model (Envelope Components) for source A13 in Sgr~B2(M).}
\tiny
\centering
\begin{supertabular}{lcccC{1cm}C{1cm}}\label{EnvLTE:parameters:A13SgrB2M}\\
SiO           & ext.           &                 17 &  3.6 $\times 10^{14}$ &                6.6 &                  58\\
CCH           & ext.           &                  3 &  7.7 $\times 10^{12}$ &                2.8 &                 113\\
              & ext.           &                  3 &  1.3 $\times 10^{16}$ &               18.7 &                  64\\
              & ext.           &                  4 &  4.1 $\times 10^{14}$ &                0.7 &                  52\\
SO            & ext.           &                 25 &  4.9 $\times 10^{15}$ &                6.0 &                  57\\
H$_2$CO       & ext.           &                  5 &  7.6 $\times 10^{13}$ &                0.7 &                  70\\
PH$_3$        & ext.           &                 38 &  2.1 $\times 10^{19}$ &                7.8 &                  64\\
\end{supertabular}\\
\vspace{1cm}

%================================================================================
%
% Source A14

%---------------------------------------
% Core Components

\tablefirsthead{%
\hline
\hline
Molecule      & $\theta^{m,c}$ & T$_{\rm ex}^{m,c}$ & N$_{\rm tot}^{m,c}$   & $\Delta$ v$^{m,c}$ & v$_{\rm LSR}^{m,c}$\\
              & ($\arcsec$)    & (K)                & (cm$^{-2}$)           & (km~s$^{-1}$)      & (km~s$^{-1}$)      \\
\hline
}

\tablehead{%
\multicolumn{6}{c}{(Continued)}\\
\hline
\hline
Molecule      & $\theta^{m,c}$ & T$_{\rm ex}^{m,c}$ & N$_{\rm tot}^{m,c}$   & $\Delta$ v$^{m,c}$ & v$_{\rm LSR}^{m,c}$\\
              & ($\arcsec$)    & (K)                & (cm$^{-2}$)           & (km~s$^{-1}$)      & (km~s$^{-1}$)      \\
\hline
}

\tabletail{%
\hline
\hline
}

\topcaption{LTE Parameters for the full LTE model (Core Components) for source A14 in Sgr~B2(M).}
\tiny
\centering
\begin{supertabular}{lcccC{1cm}C{1cm}}\label{CoreLTE:parameters:A14SgrB2M}\\
SiO           &    1.1         &                305 &  1.4 $\times 10^{16}$ &               15.9 &                  74\\
              &    1.1         &                142 &  1.0 $\times 10^{12}$ &                4.4 &                  64\\
              &    1.1         &                 82 &  1.3 $\times 10^{15}$ &                5.3 &                  43\\
H$_2 \! ^{34}$S &    1.1         &                139 &  4.6 $\times 10^{17}$ &                4.2 &                  55\\
              &    1.1         &                145 &  1.2 $\times 10^{17}$ &                7.1 &                  50\\
HCN, v$_2$=1  &    1.1         &                200 &  1.9 $\times 10^{17}$ &                4.8 &                  56\\
SO            &    1.1         &                373 &  5.2 $\times 10^{17}$ &               13.6 &                  70\\
\end{supertabular}\\
\vspace{1cm}

%---------------------------------------
% Envelope Components

\tablefirsthead{%
\hline
\hline
Molecule      & $\theta^{m,c}$ & T$_{\rm ex}^{m,c}$ & N$_{\rm tot}^{m,c}$   & $\Delta$ v$^{m,c}$ & v$_{\rm LSR}^{m,c}$\\
              & ($\arcsec$)    & (K)                & (cm$^{-2}$)           & (km~s$^{-1}$)      & (km~s$^{-1}$)      \\
\hline
}

\tablehead{%
\multicolumn{6}{c}{(Continued)}\\
\hline
\hline
Molecule      & $\theta^{m,c}$ & T$_{\rm ex}^{m,c}$ & N$_{\rm tot}^{m,c}$   & $\Delta$ v$^{m,c}$ & v$_{\rm LSR}^{m,c}$\\
              & ($\arcsec$)    & (K)                & (cm$^{-2}$)           & (km~s$^{-1}$)      & (km~s$^{-1}$)      \\
\hline
}

\tabletail{%
\hline
\hline
}

\topcaption{LTE Parameters for the full LTE model (Envelope Components) for source A14 in Sgr~B2(M).}
\tiny
\centering
\begin{supertabular}{lcccC{1cm}C{1cm}}\label{EnvLTE:parameters:A14SgrB2M}\\
SiO           & ext.           &                 10 &  9.0 $\times 10^{12}$ &                3.0 &                  57\\
CH$_3$CN      & ext.           &                 18 &  1.8 $\times 10^{15}$ &                6.0 &                 116\\
              & ext.           &                 19 &  1.2 $\times 10^{15}$ &                5.9 &                  85\\
              & ext.           &                 16 &  3.5 $\times 10^{12}$ &                6.1 &                  67\\
H$_2$CO       & ext.           &                 11 &  9.3 $\times 10^{14}$ &                4.0 &                  70\\
              & ext.           &                 14 &  1.7 $\times 10^{15}$ &                4.1 &                  61\\
              & ext.           &                 11 &  3.1 $\times 10^{14}$ &                2.0 &                  53\\
\end{supertabular}\\
\vspace{1cm}

%================================================================================
%
% Source A15

%---------------------------------------
% Core Components

\tablefirsthead{%
\hline
\hline
Molecule      & $\theta^{m,c}$ & T$_{\rm ex}^{m,c}$ & N$_{\rm tot}^{m,c}$   & $\Delta$ v$^{m,c}$ & v$_{\rm LSR}^{m,c}$\\
              & ($\arcsec$)    & (K)                & (cm$^{-2}$)           & (km~s$^{-1}$)      & (km~s$^{-1}$)      \\
\hline
}

\tablehead{%
\multicolumn{6}{c}{(Continued)}\\
\hline
\hline
Molecule      & $\theta^{m,c}$ & T$_{\rm ex}^{m,c}$ & N$_{\rm tot}^{m,c}$   & $\Delta$ v$^{m,c}$ & v$_{\rm LSR}^{m,c}$\\
              & ($\arcsec$)    & (K)                & (cm$^{-2}$)           & (km~s$^{-1}$)      & (km~s$^{-1}$)      \\
\hline
}

\tabletail{%
\hline
\hline
}

\topcaption{LTE Parameters for the full LTE model (Core Components) for source A15 in Sgr~B2(M).}
\tiny
\centering
\begin{supertabular}{lcccC{1cm}C{1cm}}\label{CoreLTE:parameters:A15SgrB2M}\\
RRL-H         &    0.9         &               8850 &  6.6 $\times 10^{8}$  &               28.9 &                  77\\
H$_2 \! ^{34}$S &    0.9         &                169 &  5.4 $\times 10^{16}$ &                6.6 &                  63\\
SO$_2$        &    0.9         &                 57 &  6.2 $\times 10^{16}$ &                8.6 &                  73\\
              &    0.9         &                165 &  1.1 $\times 10^{18}$ &                3.4 &                  56\\
              &    0.9         &                117 &  8.9 $\times 10^{17}$ &                3.9 &                  51\\
SO$_2$, v$_2$=1 &    0.9         &                 57 &  6.2 $\times 10^{16}$ &                8.6 &                  73\\
SO            &    0.9         &                820 &  2.5 $\times 10^{17}$ &               18.1 &                  76\\
\end{supertabular}\\
\vspace{1cm}

%---------------------------------------
% Envelope Components

\tablefirsthead{%
\hline
\hline
Molecule      & $\theta^{m,c}$ & T$_{\rm ex}^{m,c}$ & N$_{\rm tot}^{m,c}$   & $\Delta$ v$^{m,c}$ & v$_{\rm LSR}^{m,c}$\\
              & ($\arcsec$)    & (K)                & (cm$^{-2}$)           & (km~s$^{-1}$)      & (km~s$^{-1}$)      \\
\hline
}

\tablehead{%
\multicolumn{6}{c}{(Continued)}\\
\hline
\hline
Molecule      & $\theta^{m,c}$ & T$_{\rm ex}^{m,c}$ & N$_{\rm tot}^{m,c}$   & $\Delta$ v$^{m,c}$ & v$_{\rm LSR}^{m,c}$\\
              & ($\arcsec$)    & (K)                & (cm$^{-2}$)           & (km~s$^{-1}$)      & (km~s$^{-1}$)      \\
\hline
}

\tabletail{%
\hline
\hline
}

\topcaption{LTE Parameters for the full LTE model (Envelope Components) for source A15 in Sgr~B2(M).}
\tiny
\centering
\begin{supertabular}{lcccC{1cm}C{1cm}}\label{EnvLTE:parameters:A15SgrB2M}\\
SO            & ext.           &                  4 &  9.7 $\times 10^{15}$ &                1.9 &                  52\\
              & ext.           &                 34 &  1.9 $\times 10^{17}$ &                0.1 &                  52\\
H$_2$CO       & ext.           &                 15 &  6.9 $\times 10^{14}$ &                3.1 &                  74\\
              & ext.           &                 16 &  1.8 $\times 10^{16}$ &                5.2 &                  65\\
              & ext.           &                  9 &  6.3 $\times 10^{14}$ &                8.1 &                  57\\
CN            & ext.           &                  3 &  1.3 $\times 10^{14}$ &                7.1 &                  74\\
              & ext.           &                  3 &  6.8 $\times 10^{13}$ &                6.4 &                  60\\
\end{supertabular}\\
\vspace{1cm}

%================================================================================
%
% Source A16

%---------------------------------------
% Core Components

\tablefirsthead{%
\hline
\hline
Molecule      & $\theta^{m,c}$ & T$_{\rm ex}^{m,c}$ & N$_{\rm tot}^{m,c}$   & $\Delta$ v$^{m,c}$ & v$_{\rm LSR}^{m,c}$\\
              & ($\arcsec$)    & (K)                & (cm$^{-2}$)           & (km~s$^{-1}$)      & (km~s$^{-1}$)      \\
\hline
}

\tablehead{%
\multicolumn{6}{c}{(Continued)}\\
\hline
\hline
Molecule      & $\theta^{m,c}$ & T$_{\rm ex}^{m,c}$ & N$_{\rm tot}^{m,c}$   & $\Delta$ v$^{m,c}$ & v$_{\rm LSR}^{m,c}$\\
              & ($\arcsec$)    & (K)                & (cm$^{-2}$)           & (km~s$^{-1}$)      & (km~s$^{-1}$)      \\
\hline
}

\tabletail{%
\hline
\hline
}

\topcaption{LTE Parameters for the full LTE model (Core Components) for source A16 in Sgr~B2(M).}
\tiny
\centering
\begin{supertabular}{lcccC{1cm}C{1cm}}\label{CoreLTE:parameters:A16SgrB2M}\\
RRL-H         &    1.7         &               9445 &  6.5 $\times 10^{8}$  &               18.5 &                  60\\
CN            &    1.7         &                200 &  3.3 $\times 10^{15}$ &                3.7 &                  76\\
NO$^+$        &    1.7         &                200 &  2.5 $\times 10^{15}$ &                2.9 &                  75\\
              &    1.7         &                200 &  1.9 $\times 10^{15}$ &                2.0 &                  68\\
              &    1.7         &                200 &  3.1 $\times 10^{15}$ &                7.3 &                  48\\
CH$_2$NH      &    1.7         &                291 &  6.4 $\times 10^{14}$ &                1.1 &                  75\\
HCS$^+$       &    1.7         &                199 &  1.1 $\times 10^{14}$ &                8.0 &                  75\\
\end{supertabular}\\
\vspace{1cm}

%---------------------------------------
% Envelope Components

\tablefirsthead{%
\hline
\hline
Molecule      & $\theta^{m,c}$ & T$_{\rm ex}^{m,c}$ & N$_{\rm tot}^{m,c}$   & $\Delta$ v$^{m,c}$ & v$_{\rm LSR}^{m,c}$\\
              & ($\arcsec$)    & (K)                & (cm$^{-2}$)           & (km~s$^{-1}$)      & (km~s$^{-1}$)      \\
\hline
}

\tablehead{%
\multicolumn{6}{c}{(Continued)}\\
\hline
\hline
Molecule      & $\theta^{m,c}$ & T$_{\rm ex}^{m,c}$ & N$_{\rm tot}^{m,c}$   & $\Delta$ v$^{m,c}$ & v$_{\rm LSR}^{m,c}$\\
              & ($\arcsec$)    & (K)                & (cm$^{-2}$)           & (km~s$^{-1}$)      & (km~s$^{-1}$)      \\
\hline
}

\tabletail{%
\hline
\hline
}

\topcaption{LTE Parameters for the full LTE model (Envelope Components) for source A16 in Sgr~B2(M).}
\tiny
\centering
\begin{supertabular}{lcccC{1cm}C{1cm}}\label{EnvLTE:parameters:A16SgrB2M}\\
CN            & ext.           &                  3 &  1.9 $\times 10^{12}$ &                2.2 &                  67\\
              & ext.           &                 20 &  5.9 $\times 10^{14}$ &               16.3 &                  62\\
              & ext.           &                  9 &  3.1 $\times 10^{13}$ &                7.0 &                  46\\
              & ext.           &                 10 &  1.0 $\times 10^{14}$ &               21.0 &                  33\\
              & ext.           &                  5 &  2.4 $\times 10^{13}$ &                7.4 &                   8\\
              & ext.           &                 16 &  5.9 $\times 10^{13}$ &                2.9 &                  -2\\
              & ext.           &                 42 &  1.0 $\times 10^{12}$ &                9.8 &                 -12\\
\end{supertabular}\\
\vspace{1cm}

%================================================================================
%
% Source A17

%---------------------------------------
% Core Components

\tablefirsthead{%
\hline
\hline
Molecule      & $\theta^{m,c}$ & T$_{\rm ex}^{m,c}$ & N$_{\rm tot}^{m,c}$   & $\Delta$ v$^{m,c}$ & v$_{\rm LSR}^{m,c}$\\
              & ($\arcsec$)    & (K)                & (cm$^{-2}$)           & (km~s$^{-1}$)      & (km~s$^{-1}$)      \\
\hline
}

\tablehead{%
\multicolumn{6}{c}{(Continued)}\\
\hline
\hline
Molecule      & $\theta^{m,c}$ & T$_{\rm ex}^{m,c}$ & N$_{\rm tot}^{m,c}$   & $\Delta$ v$^{m,c}$ & v$_{\rm LSR}^{m,c}$\\
              & ($\arcsec$)    & (K)                & (cm$^{-2}$)           & (km~s$^{-1}$)      & (km~s$^{-1}$)      \\
\hline
}

\tabletail{%
\hline
\hline
}

\topcaption{LTE Parameters for the full LTE model (Core Components) for source A17 in Sgr~B2(M).}
\tiny
\centering
\begin{supertabular}{lcccC{1cm}C{1cm}}\label{CoreLTE:parameters:A17SgrB2M}\\
RRL-H         &    1.6         &              23779 &  4.6 $\times 10^{8}$  &               22.6 &                  60\\
H$_2 \! ^{34}$S &    1.6         &                200 &  5.2 $\times 10^{15}$ &                5.5 &                  59\\
H$_2$CS       &    1.6         &                108 &  1.0 $\times 10^{12}$ &                3.4 &                  78\\
CCH           &    1.6         &                100 &  1.0 $\times 10^{14}$ &                1.3 &                  73\\
SiO           &    1.6         &                200 &  7.1 $\times 10^{13}$ &               11.4 &                  81\\
CN            &    1.6         &                200 &  4.8 $\times 10^{14}$ &                3.1 &                  77\\
SO            &    1.6         &                179 &  2.5 $\times 10^{14}$ &                1.1 &                  75\\
              &    1.6         &               1012 &  3.4 $\times 10^{12}$ &               12.7 &                  71\\
SO$_2$        &    1.6         &                298 &  1.1 $\times 10^{12}$ &                4.9 &                  52\\
CO            &    1.6         &                200 &  2.6 $\times 10^{15}$ &                1.1 &                 174\\
              &    1.6         &                200 &  9.5 $\times 10^{15}$ &                1.1 &                 135\\
              &    1.6         &                200 &  1.3 $\times 10^{16}$ &                1.3 &                 112\\
              &    1.6         &                200 &  5.6 $\times 10^{16}$ &                3.9 &                 109\\
              &    1.6         &                200 &  5.0 $\times 10^{16}$ &                2.9 &                  84\\
              &    1.6         &                200 &  1.5 $\times 10^{18}$ &                1.4 &                  45\\
              &    1.6         &                200 &  2.3 $\times 10^{16}$ &                1.0 &                 -37\\
              &    1.6         &                200 &  4.3 $\times 10^{15}$ &                1.2 &                 -59\\
              &    1.6         &                200 &  2.2 $\times 10^{15}$ &                1.0 &                -121\\
              &    1.6         &                200 &  1.4 $\times 10^{16}$ &                3.6 &                -140\\
              &    1.6         &                200 &  2.1 $\times 10^{15}$ &                0.4 &                -167\\
$^{13}$CO     &    1.6         &                200 &  1.4 $\times 10^{12}$ &                0.1 &                 149\\
              &    1.6         &                200 &  2.2 $\times 10^{16}$ &                2.0 &                 105\\
              &    1.6         &                200 &  8.0 $\times 10^{15}$ &                0.7 &                  61\\
              &    1.6         &                200 &  1.6 $\times 10^{16}$ &                1.5 &                -134\\
              &    1.6         &                200 &  1.5 $\times 10^{16}$ &                1.0 &                -137\\
C$^{17}$O     &    1.6         &                200 &  3.1 $\times 10^{15}$ &                2.8 &                -167\\
              &    1.6         &                200 &  3.1 $\times 10^{16}$ &               11.7 &                -159\\
              &    1.6         &                200 &  3.1 $\times 10^{15}$ &                2.4 &                -135\\
              &    1.6         &                200 &  3.1 $\times 10^{15}$ &                1.0 &                -130\\
              &    1.6         &                200 &  3.1 $\times 10^{14}$ &                2.7 &                -114\\
              &    1.6         &                200 &  3.1 $\times 10^{15}$ &                3.6 &                 -98\\
              &    1.6         &                200 &  1.2 $\times 10^{16}$ &                7.9 &                 -82\\
              &    1.6         &                200 &  4.0 $\times 10^{15}$ &                2.7 &                 -58\\
              &    1.6         &                200 &  3.3 $\times 10^{15}$ &                2.7 &                 -47\\
              &    1.6         &                200 &  1.2 $\times 10^{16}$ &                2.6 &                 -36\\
              &    1.6         &                200 &  5.3 $\times 10^{15}$ &                2.7 &                 -12\\
              &    1.6         &                200 &  5.7 $\times 10^{15}$ &                2.7 &                   6\\
              &    1.6         &                200 &  3.7 $\times 10^{15}$ &                2.7 &                  28\\
              &    1.6         &                200 &  1.1 $\times 10^{15}$ &                2.7 &                  33\\
              &    1.6         &                200 &  8.2 $\times 10^{14}$ &                0.8 &                  38\\
              &    1.6         &                200 &  2.3 $\times 10^{16}$ &                7.1 &                  78\\
              &    1.6         &                200 &  2.7 $\times 10^{15}$ &                7.1 &                  84\\
              &    1.6         &                200 &  4.7 $\times 10^{15}$ &                2.7 &                  95\\
              &    1.6         &                200 &  1.6 $\times 10^{16}$ &                8.5 &                 137\\
              &    1.6         &                200 &  1.6 $\times 10^{16}$ &                3.4 &                 156\\
              &    1.6         &                200 &  8.6 $\times 10^{15}$ &                2.2 &                 161\\
C$^{18}$O     &    1.6         &                200 &  6.9 $\times 10^{13}$ &                2.4 &                -153\\
              &    1.6         &                200 &  1.9 $\times 10^{14}$ &                3.5 &                -145\\
              &    1.6         &                200 &  8.8 $\times 10^{12}$ &                2.4 &                -137\\
              &    1.6         &                200 &  2.5 $\times 10^{13}$ &                2.4 &                -121\\
              &    1.6         &                200 &  6.9 $\times 10^{13}$ &                0.5 &                -110\\
              &    1.6         &                200 &  1.9 $\times 10^{14}$ &                6.7 &                 -99\\
              &    1.6         &                200 &  1.9 $\times 10^{14}$ &                2.8 &                 -93\\
              &    1.6         &                200 &  1.3 $\times 10^{16}$ &                8.8 &                 -82\\
              &    1.6         &                200 &  5.7 $\times 10^{15}$ &                6.0 &                 -74\\
              &    1.6         &                200 &  5.4 $\times 10^{13}$ &                1.6 &                 -55\\
              &    1.6         &                200 &  2.3 $\times 10^{15}$ &                1.7 &                 -50\\
              &    1.6         &                200 &  5.2 $\times 10^{15}$ &                5.0 &                   0\\
              &    1.6         &                200 &  1.2 $\times 10^{12}$ &                1.1 &                   8\\
              &    1.6         &                200 &  2.1 $\times 10^{16}$ &               16.5 &                  27\\
              &    1.6         &                200 &  2.0 $\times 10^{13}$ &                2.4 &                  32\\
              &    1.6         &                200 &  4.2 $\times 10^{16}$ &                4.1 &                  76\\
              &    1.6         &                200 &  3.2 $\times 10^{16}$ &                9.8 &                 104\\
              &    1.6         &                200 &  1.3 $\times 10^{16}$ &                6.2 &                 114\\
              &    1.6         &                200 &  9.9 $\times 10^{15}$ &                4.7 &                 122\\
              &    1.6         &                200 &  8.6 $\times 10^{14}$ &                2.1 &                 133\\
              &    1.6         &                200 &  9.1 $\times 10^{15}$ &                6.8 &                 142\\
              &    1.6         &                200 &  9.7 $\times 10^{16}$ &               49.0 &                 156\\
CS            &    1.6         &                200 &  9.0 $\times 10^{13}$ &               11.8 &                 113\\
              &    1.6         &                200 &  8.3 $\times 10^{13}$ &                1.2 &                  85\\
              &    1.6         &                200 &  1.7 $\times 10^{13}$ &                1.0 &                  52\\
              &    1.6         &                200 &  3.4 $\times 10^{13}$ &                1.2 &                  43\\
              &    1.6         &                200 &  5.1 $\times 10^{13}$ &                2.3 &                  14\\
              &    1.6         &                200 &  1.4 $\times 10^{13}$ &                1.1 &                   3\\
              &    1.6         &                200 &  4.7 $\times 10^{13}$ &                3.2 &                 -10\\
              &    1.6         &                200 &  7.1 $\times 10^{12}$ &                1.3 &                 -36\\
              &    1.6         &                200 &  4.6 $\times 10^{13}$ &                2.2 &                 -50\\
              &    1.6         &                200 &  3.8 $\times 10^{13}$ &                3.1 &                 -65\\
              &    1.6         &                200 &  2.4 $\times 10^{12}$ &                0.5 &                 -82\\
              &    1.6         &                200 &  3.7 $\times 10^{13}$ &                2.1 &                 -82\\
              &    1.6         &                200 &  5.1 $\times 10^{13}$ &                2.5 &                 -99\\
              &    1.6         &                200 &  1.5 $\times 10^{13}$ &                0.9 &                -120\\
              &    1.6         &                200 &  4.1 $\times 10^{13}$ &                2.3 &                -124\\
              &    1.6         &                200 &  1.8 $\times 10^{13}$ &                1.3 &                -132\\
              &    1.6         &                200 &  1.5 $\times 10^{13}$ &                1.1 &                -143\\
              &    1.6         &                200 &  3.9 $\times 10^{13}$ &                1.9 &                -153\\
$^{13}$CS     &    1.6         &                200 &  1.0 $\times 10^{14}$ &                6.1 &                  72\\
C$^{33}$S     &    1.6         &                200 &  2.1 $\times 10^{14}$ &                8.7 &                  60\\
C$^{34}$S     &    1.6         &                200 &  8.7 $\times 10^{13}$ &                5.1 &                 -62\\
              &    1.6         &                200 &  5.2 $\times 10^{13}$ &               23.0 &                 -25\\
              &    1.6         &                200 &  2.5 $\times 10^{14}$ &               50.0 &                  -7\\
              &    1.6         &                200 &  9.1 $\times 10^{13}$ &                6.4 &                  30\\
              &    1.6         &                200 &  2.8 $\times 10^{14}$ &                6.4 &                  42\\
              &    1.6         &                200 &  1.1 $\times 10^{14}$ &                4.4 &                  73\\
              &    1.6         &                200 &  5.7 $\times 10^{13}$ &                2.5 &                  84\\
              &    1.6         &                200 &  1.2 $\times 10^{14}$ &               21.9 &                  95\\
HCN           &    1.6         &                200 &  2.4 $\times 10^{13}$ &                2.0 &                  42\\
H$^{13}$CN    &    1.6         &                200 &  8.0 $\times 10^{12}$ &                0.9 &                 163\\
              &    1.6         &                200 &  7.3 $\times 10^{12}$ &                0.9 &                 159\\
              &    1.6         &                200 &  3.9 $\times 10^{14}$ &                6.5 &                 133\\
              &    1.6         &                200 &  3.5 $\times 10^{13}$ &                2.9 &                  45\\
              &    1.6         &                200 &  2.0 $\times 10^{13}$ &                1.0 &                  10\\
              &    1.6         &                200 &  1.1 $\times 10^{13}$ &                0.8 &                   2\\
              &    1.6         &                200 &  4.9 $\times 10^{12}$ &                2.6 &                 -71\\
              &    1.6         &                200 &  7.9 $\times 10^{12}$ &                7.4 &                 -56\\
              &    1.6         &                200 &  8.4 $\times 10^{12}$ &                0.5 &                 -88\\
              &    1.6         &                200 &  6.8 $\times 10^{13}$ &               25.6 &                -109\\
              &    1.6         &                200 &  6.4 $\times 10^{12}$ &                2.8 &                -116\\
              &    1.6         &                200 &  1.2 $\times 10^{13}$ &                1.4 &                -124\\
              &    1.6         &                200 &  1.6 $\times 10^{13}$ &                3.2 &                -133\\
              &    1.6         &                200 &  1.8 $\times 10^{13}$ &                1.9 &                -145\\
              &    1.6         &                200 &  8.5 $\times 10^{12}$ &                1.0 &                -176\\
HNC           &    1.6         &                200 &  1.3 $\times 10^{13}$ &                1.7 &                 138\\
              &    1.6         &                200 &  1.3 $\times 10^{12}$ &                0.7 &                 138\\
              &    1.6         &                200 &  1.4 $\times 10^{13}$ &                2.0 &                  50\\
              &    1.6         &                200 &  1.9 $\times 10^{13}$ &                1.9 &                  44\\
              &    1.6         &                200 &  1.2 $\times 10^{13}$ &                1.1 &                   2\\
HO$^{13}$C$^+$ &    1.6         &                200 &  3.2 $\times 10^{13}$ &               26.0 &                 129\\
              &    1.6         &                200 &  4.4 $\times 10^{12}$ &                2.3 &                 109\\
\end{supertabular}\\
\vspace{1cm}

%---------------------------------------
% Envelope Components

\tablefirsthead{%
\hline
\hline
Molecule      & $\theta^{m,c}$ & T$_{\rm ex}^{m,c}$ & N$_{\rm tot}^{m,c}$   & $\Delta$ v$^{m,c}$ & v$_{\rm LSR}^{m,c}$\\
              & ($\arcsec$)    & (K)                & (cm$^{-2}$)           & (km~s$^{-1}$)      & (km~s$^{-1}$)      \\
\hline
}

\tablehead{%
\multicolumn{6}{c}{(Continued)}\\
\hline
\hline
Molecule      & $\theta^{m,c}$ & T$_{\rm ex}^{m,c}$ & N$_{\rm tot}^{m,c}$   & $\Delta$ v$^{m,c}$ & v$_{\rm LSR}^{m,c}$\\
              & ($\arcsec$)    & (K)                & (cm$^{-2}$)           & (km~s$^{-1}$)      & (km~s$^{-1}$)      \\
\hline
}

\tabletail{%
\hline
\hline
}

\topcaption{LTE Parameters for the full LTE model (Envelope Components) for source A17 in Sgr~B2(M).}
\tiny
\centering
\begin{supertabular}{lcccC{1cm}C{1cm}}\label{EnvLTE:parameters:A17SgrB2M}\\
SiS           & ext.           &                 33 &  1.0 $\times 10^{15}$ &                0.9 &                  82\\
              & ext.           &                 11 &  1.1 $\times 10^{15}$ &                2.0 &                  57\\
H$_2$CS       & ext.           &                 11 &  1.3 $\times 10^{12}$ &                2.3 &                  83\\
              & ext.           &                 17 &  6.1 $\times 10^{14}$ &                8.8 &                  65\\
              & ext.           &                 19 &  9.5 $\times 10^{14}$ &               10.5 &                  50\\
CCH           & ext.           &                 16 &  4.1 $\times 10^{14}$ &                5.7 &                  65\\
              & ext.           &                 17 &  4.2 $\times 10^{14}$ &                2.9 &                  53\\
\end{supertabular}\\
\vspace{1cm}

%================================================================================
%
% Source A18

%---------------------------------------
% Core Components

\tablefirsthead{%
\hline
\hline
Molecule      & $\theta^{m,c}$ & T$_{\rm ex}^{m,c}$ & N$_{\rm tot}^{m,c}$   & $\Delta$ v$^{m,c}$ & v$_{\rm LSR}^{m,c}$\\
              & ($\arcsec$)    & (K)                & (cm$^{-2}$)           & (km~s$^{-1}$)      & (km~s$^{-1}$)      \\
\hline
}

\tablehead{%
\multicolumn{6}{c}{(Continued)}\\
\hline
\hline
Molecule      & $\theta^{m,c}$ & T$_{\rm ex}^{m,c}$ & N$_{\rm tot}^{m,c}$   & $\Delta$ v$^{m,c}$ & v$_{\rm LSR}^{m,c}$\\
              & ($\arcsec$)    & (K)                & (cm$^{-2}$)           & (km~s$^{-1}$)      & (km~s$^{-1}$)      \\
\hline
}

\tabletail{%
\hline
\hline
}

\topcaption{LTE Parameters for the full LTE model (Core Components) for source A18 in Sgr~B2(M).}
\tiny
\centering
\begin{supertabular}{lcccC{1cm}C{1cm}}\label{CoreLTE:parameters:A18SgrB2M}\\
HCCCN         &    1.0         &                 55 &  3.3 $\times 10^{15}$ &                9.7 &                  53\\
H$_2$CS       &    1.0         &                 96 &  5.8 $\times 10^{15}$ &                9.0 &                  52\\
H$_2$CO       &    1.0         &                295 &  7.4 $\times 10^{15}$ &                2.9 &                  52\\
SiO           &    1.0         &                169 &  2.4 $\times 10^{13}$ &                2.4 &                  53\\
CCH           &    1.0         &                113 &  1.3 $\times 10^{16}$ &                9.8 &                  55\\
H$_2$CCO      &    1.0         &                 84 &  1.1 $\times 10^{14}$ &                0.7 &                 119\\
NS            &    1.0         &                285 &  5.3 $\times 10^{15}$ &                8.3 &                  49\\
CO            &    1.0         &                200 &  4.6 $\times 10^{14}$ &                1.0 &                  97\\
              &    1.0         &                200 &  4.9 $\times 10^{17}$ &                4.1 &                  85\\
              &    1.0         &                200 &  1.1 $\times 10^{17}$ &                1.9 &                  44\\
$^{13}$CO     &    1.0         &                200 &  8.6 $\times 10^{17}$ &                3.1 &                  48\\
              &    1.0         &                200 &  8.2 $\times 10^{17}$ &                6.3 &                  88\\
              &    1.0         &                200 &  1.9 $\times 10^{13}$ &                1.0 &                 -35\\
C$^{17}$O     &    1.0         &                200 &  3.3 $\times 10^{15}$ &                2.7 &                -151\\
              &    1.0         &                200 &  3.3 $\times 10^{15}$ &                2.7 &                -125\\
              &    1.0         &                200 &  9.9 $\times 10^{14}$ &                2.7 &                -100\\
              &    1.0         &                200 &  1.1 $\times 10^{15}$ &                2.7 &                 -84\\
              &    1.0         &                200 &  4.3 $\times 10^{15}$ &                2.6 &                 -52\\
              &    1.0         &                200 &  3.6 $\times 10^{15}$ &                2.7 &                 -10\\
              &    1.0         &                200 &  3.5 $\times 10^{15}$ &                2.7 &                  23\\
              &    1.0         &                200 &  2.1 $\times 10^{15}$ &                2.7 &                  36\\
              &    1.0         &                200 &  1.9 $\times 10^{16}$ &                2.7 &                  55\\
              &    1.0         &                200 &  1.1 $\times 10^{16}$ &                4.5 &                 105\\
              &    1.0         &                200 &  1.1 $\times 10^{15}$ &                2.7 &                 127\\
              &    1.0         &                200 &  1.3 $\times 10^{16}$ &                3.8 &                 145\\
C$^{18}$O     &    1.0         &                200 &  5.8 $\times 10^{14}$ &                2.7 &                -163\\
              &    1.0         &                200 &  6.0 $\times 10^{15}$ &                2.7 &                -156\\
              &    1.0         &                200 &  2.4 $\times 10^{15}$ &                2.7 &                -151\\
              &    1.0         &                200 &  5.8 $\times 10^{14}$ &                2.7 &                -129\\
              &    1.0         &                200 &  5.8 $\times 10^{14}$ &                2.7 &                -118\\
              &    1.0         &                200 &  6.5 $\times 10^{15}$ &                3.5 &                -107\\
              &    1.0         &                200 &  5.8 $\times 10^{15}$ &                3.1 &                -102\\
              &    1.0         &                200 &  1.8 $\times 10^{15}$ &                3.1 &                 -96\\
              &    1.0         &                200 &  3.1 $\times 10^{15}$ &                3.0 &                 -88\\
              &    1.0         &                200 &  8.2 $\times 10^{14}$ &               12.2 &                 -80\\
              &    1.0         &                200 &  4.2 $\times 10^{15}$ &                2.7 &                 -44\\
              &    1.0         &                200 &  3.7 $\times 10^{15}$ &                2.7 &                 -30\\
              &    1.0         &                200 &  2.8 $\times 10^{15}$ &                2.1 &                 -11\\
              &    1.0         &                200 &  4.9 $\times 10^{15}$ &                2.7 &                  22\\
              &    1.0         &                200 &  2.6 $\times 10^{17}$ &                6.4 &                  49\\
              &    1.0         &                200 &  1.6 $\times 10^{17}$ &                6.4 &                  87\\
              &    1.0         &                200 &  8.0 $\times 10^{16}$ &               11.9 &                 124\\
              &    1.0         &                200 &  3.0 $\times 10^{14}$ &                2.7 &                 137\\
CS            &    1.0         &                200 &  2.9 $\times 10^{12}$ &                1.0 &                 147\\
              &    1.0         &                200 &  7.5 $\times 10^{13}$ &                1.1 &                 147\\
              &    1.0         &                200 &  7.7 $\times 10^{13}$ &                2.3 &                 132\\
              &    1.0         &                200 &  2.9 $\times 10^{13}$ &                1.1 &                 128\\
              &    1.0         &                200 &  2.2 $\times 10^{14}$ &                4.5 &                 122\\
              &    1.0         &                200 &  5.0 $\times 10^{13}$ &                1.0 &                 114\\
              &    1.0         &                200 &  3.0 $\times 10^{13}$ &                1.1 &                 101\\
              &    1.0         &                200 &  3.8 $\times 10^{15}$ &               12.9 &                  86\\
              &    1.0         &                200 &  1.0 $\times 10^{12}$ &                2.5 &                  85\\
              &    1.0         &                200 &  1.1 $\times 10^{14}$ &                0.9 &                  50\\
              &    1.0         &                200 &  2.3 $\times 10^{12}$ &                5.3 &                  16\\
              &    1.0         &                200 &  3.5 $\times 10^{12}$ &                2.3 &                  11\\
              &    1.0         &                200 &  2.3 $\times 10^{13}$ &                1.0 &                   0\\
              &    1.0         &                200 &  1.4 $\times 10^{12}$ &                0.3 &                  -6\\
              &    1.0         &                200 &  1.7 $\times 10^{13}$ &                4.2 &                 -11\\
              &    1.0         &                200 &  1.0 $\times 10^{12}$ &                2.6 &                 -18\\
              &    1.0         &                200 &  1.0 $\times 10^{12}$ &                3.3 &                 -20\\
              &    1.0         &                200 &  3.1 $\times 10^{13}$ &                1.0 &                 -36\\
              &    1.0         &                200 &  9.8 $\times 10^{13}$ &                2.3 &                 -45\\
              &    1.0         &                200 &  7.6 $\times 10^{13}$ &                2.4 &                 -57\\
              &    1.0         &                200 &  4.6 $\times 10^{13}$ &                2.0 &                 -63\\
              &    1.0         &                200 &  2.1 $\times 10^{13}$ &                4.9 &                 -78\\
              &    1.0         &                200 &  5.0 $\times 10^{12}$ &                0.4 &                 -86\\
              &    1.0         &                200 &  8.1 $\times 10^{13}$ &                2.5 &                -104\\
              &    1.0         &                200 &  4.7 $\times 10^{13}$ &                1.8 &                -113\\
              &    1.0         &                200 &  2.5 $\times 10^{13}$ &                2.4 &                -106\\
              &    1.0         &                200 &  1.6 $\times 10^{12}$ &                9.1 &                -136\\
              &    1.0         &                200 &  2.2 $\times 10^{13}$ &                0.7 &                -131\\
              &    1.0         &                200 &  1.4 $\times 10^{13}$ &                1.0 &                -151\\
              &    1.0         &                200 &  1.2 $\times 10^{14}$ &                2.4 &                -145\\
              &    1.0         &                200 &  2.8 $\times 10^{13}$ &                1.0 &                -152\\
              &    1.0         &                200 &  1.4 $\times 10^{12}$ &                1.6 &                -170\\
              &    1.0         &                200 &  8.9 $\times 10^{12}$ &                1.4 &                -161\\
              &    1.0         &                200 &  3.7 $\times 10^{14}$ &                2.6 &                -180\\
$^{13}$CS     &    1.0         &                200 &  5.6 $\times 10^{13}$ &                5.0 &                -155\\
              &    1.0         &                200 &  5.2 $\times 10^{13}$ &                6.5 &                -101\\
              &    1.0         &                200 &  3.5 $\times 10^{14}$ &                5.7 &                 -93\\
              &    1.0         &                200 &  6.7 $\times 10^{13}$ &                2.3 &                 -28\\
              &    1.0         &                200 &  1.3 $\times 10^{14}$ &                2.6 &                 -23\\
              &    1.0         &                200 &  7.0 $\times 10^{13}$ &                2.7 &                  14\\
              &    1.0         &                200 &  2.0 $\times 10^{15}$ &                5.9 &                  51\\
              &    1.0         &                200 &  1.2 $\times 10^{15}$ &               16.9 &                  81\\
              &    1.0         &                200 &  1.2 $\times 10^{14}$ &                2.6 &                  96\\
              &    1.0         &                200 &  2.5 $\times 10^{14}$ &                6.5 &                 105\\
C$^{33}$S     &    1.0         &                200 &  3.0 $\times 10^{14}$ &                7.2 &                  52\\
C$^{34}$S     &    1.0         &                200 &  9.3 $\times 10^{13}$ &                6.7 &                 -67\\
              &    1.0         &                200 &  6.2 $\times 10^{14}$ &                7.8 &                  50\\
              &    1.0         &                200 &  7.3 $\times 10^{13}$ &                6.4 &                  85\\
HCN           &    1.0         &                200 &  1.0 $\times 10^{12}$ &                2.3 &                 123\\
              &    1.0         &                200 &  4.8 $\times 10^{13}$ &                1.1 &                  51\\
              &    1.0         &                200 &  1.3 $\times 10^{13}$ &                1.0 &                   3\\
              &    1.0         &                200 &  2.8 $\times 10^{13}$ &                2.4 &                  -3\\
              &    1.0         &                200 &  4.1 $\times 10^{12}$ &                1.3 &                 -82\\
              &    1.0         &                200 &  1.0 $\times 10^{12}$ &                2.3 &                -166\\
H$^{13}$CN    &    1.0         &                200 &  5.3 $\times 10^{13}$ &                4.8 &                   9\\
HNC           &    1.0         &                200 &  1.3 $\times 10^{13}$ &                0.9 &                 161\\
              &    1.0         &                200 &  2.1 $\times 10^{12}$ &                0.8 &                 161\\
              &    1.0         &                200 &  5.1 $\times 10^{13}$ &                1.1 &                 149\\
              &    1.0         &                200 &  1.6 $\times 10^{12}$ &                0.4 &                  11\\
              &    1.0         &                200 &  1.8 $\times 10^{13}$ &                0.9 &                 116\\
              &    1.0         &                200 &  2.3 $\times 10^{14}$ &                0.8 &                 109\\
              &    1.0         &                200 &  5.1 $\times 10^{13}$ &                2.9 &                 105\\
              &    1.0         &                200 &  1.6 $\times 10^{12}$ &                0.4 &                 100\\
              &    1.0         &                200 &  1.1 $\times 10^{12}$ &                1.1 &                 112\\
              &    1.0         &                200 &  1.2 $\times 10^{13}$ &                1.0 &                  83\\
              &    1.0         &                200 &  6.7 $\times 10^{13}$ &                2.0 &                  47\\
              &    1.0         &                200 &  3.7 $\times 10^{14}$ &                8.7 &                  38\\
              &    1.0         &                200 &  5.7 $\times 10^{13}$ &                3.0 &                  28\\
              &    1.0         &                200 &  9.7 $\times 10^{13}$ &                1.2 &                  18\\
              &    1.0         &                200 &  3.3 $\times 10^{13}$ &                1.1 &                  13\\
              &    1.0         &                200 &  5.5 $\times 10^{12}$ &                0.8 &                   2\\
              &    1.0         &                200 &  1.3 $\times 10^{12}$ &                2.7 &                 -36\\
              &    1.0         &                200 &  1.2 $\times 10^{13}$ &                1.0 &                 -17\\
              &    1.0         &                200 &  3.4 $\times 10^{13}$ &                3.9 &                 -26\\
              &    1.0         &                200 &  1.5 $\times 10^{13}$ &                1.5 &                 -40\\
              &    1.0         &                200 &  9.0 $\times 10^{12}$ &                0.8 &                 -49\\
              &    1.0         &                200 &  9.4 $\times 10^{13}$ &                3.8 &                 -56\\
              &    1.0         &                200 &  3.2 $\times 10^{13}$ &                2.2 &                 -66\\
              &    1.0         &                200 &  5.3 $\times 10^{13}$ &                4.9 &                 -76\\
              &    1.0         &                200 &  5.2 $\times 10^{13}$ &                2.8 &                 -81\\
              &    1.0         &                200 &  8.5 $\times 10^{12}$ &                1.0 &                 -88\\
              &    1.0         &                200 &  2.0 $\times 10^{14}$ &               14.5 &                 -98\\
              &    1.0         &                200 &  4.2 $\times 10^{12}$ &                1.0 &                 -93\\
              &    1.0         &                200 &  1.4 $\times 10^{13}$ &                2.0 &                 -95\\
              &    1.0         &                200 &  2.8 $\times 10^{12}$ &                0.6 &                 -93\\
              &    1.0         &                200 &  6.5 $\times 10^{13}$ &                4.8 &                -111\\
              &    1.0         &                200 &  2.1 $\times 10^{12}$ &                0.9 &                -114\\
              &    1.0         &                200 &  3.2 $\times 10^{12}$ &                0.8 &                -131\\
              &    1.0         &                200 &  6.7 $\times 10^{13}$ &                1.6 &                -134\\
              &    1.0         &                200 &  4.8 $\times 10^{13}$ &                2.0 &                -143\\
              &    1.0         &                200 &  1.1 $\times 10^{13}$ &                1.1 &                -148\\
              &    1.0         &                200 &  2.8 $\times 10^{13}$ &                2.1 &                -154\\
              &    1.0         &                200 &  5.1 $\times 10^{13}$ &                2.1 &                -171\\
              &    1.0         &                200 &  4.3 $\times 10^{13}$ &                1.3 &                -176\\
              &    1.0         &                200 &  2.5 $\times 10^{12}$ &                1.4 &                -174\\
              &    1.0         &                200 &  1.3 $\times 10^{13}$ &                1.0 &                -186\\
HN$^{13}$C    &    1.0         &                200 &  6.9 $\times 10^{13}$ &                4.5 &                  55\\
H$^{13}$CO$^+$ &    1.0         &                200 &  9.4 $\times 10^{13}$ &                7.1 &                  53\\
              &    1.0         &                200 &  3.9 $\times 10^{13}$ &                4.4 &                  61\\
HO$^{13}$C$^+$ &    1.0         &                200 &  1.3 $\times 10^{14}$ &                4.4 &                -149\\
              &    1.0         &                200 &  1.5 $\times 10^{14}$ &               17.3 &                 -34\\
              &    1.0         &                200 &  3.1 $\times 10^{13}$ &                2.7 &                   0\\
              &    1.0         &                200 &  8.8 $\times 10^{13}$ &               11.6 &                  45\\
              &    1.0         &                200 &  3.2 $\times 10^{14}$ &               39.3 &                 147\\
\end{supertabular}\\
\vspace{1cm}

%---------------------------------------
% Envelope Components

\tablefirsthead{%
\hline
\hline
Molecule      & $\theta^{m,c}$ & T$_{\rm ex}^{m,c}$ & N$_{\rm tot}^{m,c}$   & $\Delta$ v$^{m,c}$ & v$_{\rm LSR}^{m,c}$\\
              & ($\arcsec$)    & (K)                & (cm$^{-2}$)           & (km~s$^{-1}$)      & (km~s$^{-1}$)      \\
\hline
}

\tablehead{%
\multicolumn{6}{c}{(Continued)}\\
\hline
\hline
Molecule      & $\theta^{m,c}$ & T$_{\rm ex}^{m,c}$ & N$_{\rm tot}^{m,c}$   & $\Delta$ v$^{m,c}$ & v$_{\rm LSR}^{m,c}$\\
              & ($\arcsec$)    & (K)                & (cm$^{-2}$)           & (km~s$^{-1}$)      & (km~s$^{-1}$)      \\
\hline
}

\tabletail{%
\hline
\hline
}

\topcaption{LTE Parameters for the full LTE model (Envelope Components) for source A18 in Sgr~B2(M).}
\tiny
\centering
\begin{supertabular}{lcccC{1cm}C{1cm}}\label{EnvLTE:parameters:A18SgrB2M}\\
HCCCN         & ext.           &                 14 &  3.5 $\times 10^{17}$ &                6.6 &                  63\\
H$_2$CS       & ext.           &                 18 &  9.9 $\times 10^{14}$ &                4.6 &                  67\\
              & ext.           &                  7 &  2.6 $\times 10^{15}$ &               39.7 &                  15\\
CH$_3$OH      & ext.           &                  3 &  5.4 $\times 10^{13}$ &               16.4 &                  48\\
              & ext.           &                 27 &  5.8 $\times 10^{15}$ &               19.2 &                  83\\
              & ext.           &                 36 &  5.0 $\times 10^{13}$ &               16.9 &                  79\\
              & ext.           &                  5 &  9.4 $\times 10^{12}$ &               18.3 &                  30\\
\end{supertabular}\\
\vspace{1cm}

%================================================================================
%
% Source A19

%---------------------------------------
% Core Components

\tablefirsthead{%
\hline
\hline
Molecule      & $\theta^{m,c}$ & T$_{\rm ex}^{m,c}$ & N$_{\rm tot}^{m,c}$   & $\Delta$ v$^{m,c}$ & v$_{\rm LSR}^{m,c}$\\
              & ($\arcsec$)    & (K)                & (cm$^{-2}$)           & (km~s$^{-1}$)      & (km~s$^{-1}$)      \\
\hline
}

\tablehead{%
\multicolumn{6}{c}{(Continued)}\\
\hline
\hline
Molecule      & $\theta^{m,c}$ & T$_{\rm ex}^{m,c}$ & N$_{\rm tot}^{m,c}$   & $\Delta$ v$^{m,c}$ & v$_{\rm LSR}^{m,c}$\\
              & ($\arcsec$)    & (K)                & (cm$^{-2}$)           & (km~s$^{-1}$)      & (km~s$^{-1}$)      \\
\hline
}

\tabletail{%
\hline
\hline
}

\topcaption{LTE Parameters for the full LTE model (Core Components) for source A19 in Sgr~B2(M).}
\tiny
\centering
\begin{supertabular}{lcccC{1cm}C{1cm}}\label{CoreLTE:parameters:A19SgrB2M}\\
HCCCN         &    1.1         &                111 &  1.9 $\times 10^{14}$ &                1.0 &                  60\\
H$_2$CS       &    1.1         &                200 &  2.1 $\times 10^{15}$ &                3.0 &                  58\\
SiO           &    1.1         &                158 &  1.1 $\times 10^{13}$ &                2.3 &                  53\\
CH$_2$NH      &    1.1         &                 87 &  1.0 $\times 10^{12}$ &               14.5 &                  77\\
CO            &    1.1         &                200 &  5.4 $\times 10^{15}$ &                1.1 &                 153\\
              &    1.1         &                200 &  9.1 $\times 10^{15}$ &                0.7 &                 106\\
              &    1.1         &                200 &  1.5 $\times 10^{16}$ &                1.0 &                 101\\
\end{supertabular}\\
\vspace{1cm}

%---------------------------------------
% Envelope Components

\tablefirsthead{%
\hline
\hline
Molecule      & $\theta^{m,c}$ & T$_{\rm ex}^{m,c}$ & N$_{\rm tot}^{m,c}$   & $\Delta$ v$^{m,c}$ & v$_{\rm LSR}^{m,c}$\\
              & ($\arcsec$)    & (K)                & (cm$^{-2}$)           & (km~s$^{-1}$)      & (km~s$^{-1}$)      \\
\hline
}

\tablehead{%
\multicolumn{6}{c}{(Continued)}\\
\hline
\hline
Molecule      & $\theta^{m,c}$ & T$_{\rm ex}^{m,c}$ & N$_{\rm tot}^{m,c}$   & $\Delta$ v$^{m,c}$ & v$_{\rm LSR}^{m,c}$\\
              & ($\arcsec$)    & (K)                & (cm$^{-2}$)           & (km~s$^{-1}$)      & (km~s$^{-1}$)      \\
\hline
}

\tabletail{%
\hline
\hline
}

\topcaption{LTE Parameters for the full LTE model (Envelope Components) for source A19 in Sgr~B2(M).}
\tiny
\centering
\begin{supertabular}{lcccC{1cm}C{1cm}}\label{EnvLTE:parameters:A19SgrB2M}\\
HCCCN         & ext.           &                 17 &  1.6 $\times 10^{17}$ &                2.1 &                  66\\
              & ext.           &                 18 &  9.3 $\times 10^{16}$ &                5.7 &                  48\\
H$_2$CS       & ext.           &                 13 &  1.0 $\times 10^{15}$ &                3.9 &                  68\\
              & ext.           &                 15 &  1.1 $\times 10^{15}$ &                3.9 &                  49\\
CH$_3$OH      & ext.           &                  3 &  5.8 $\times 10^{15}$ &                0.8 &                  60\\
              & ext.           &                 19 &  2.0 $\times 10^{16}$ &                6.7 &                  68\\
              & ext.           &                 10 &  7.7 $\times 10^{12}$ &                1.0 &                  62\\
\end{supertabular}\\
\vspace{1cm}

%================================================================================
%
% Source A20

%---------------------------------------
% Core Components

\tablefirsthead{%
\hline
\hline
Molecule      & $\theta^{m,c}$ & T$_{\rm ex}^{m,c}$ & N$_{\rm tot}^{m,c}$   & $\Delta$ v$^{m,c}$ & v$_{\rm LSR}^{m,c}$\\
              & ($\arcsec$)    & (K)                & (cm$^{-2}$)           & (km~s$^{-1}$)      & (km~s$^{-1}$)      \\
\hline
}

\tablehead{%
\multicolumn{6}{c}{(Continued)}\\
\hline
\hline
Molecule      & $\theta^{m,c}$ & T$_{\rm ex}^{m,c}$ & N$_{\rm tot}^{m,c}$   & $\Delta$ v$^{m,c}$ & v$_{\rm LSR}^{m,c}$\\
              & ($\arcsec$)    & (K)                & (cm$^{-2}$)           & (km~s$^{-1}$)      & (km~s$^{-1}$)      \\
\hline
}

\tabletail{%
\hline
\hline
}

\topcaption{LTE Parameters for the full LTE model (Core Components) for source A20 in Sgr~B2(M).}
\tiny
\centering
\begin{supertabular}{lcccC{1cm}C{1cm}}\label{CoreLTE:parameters:A20SgrB2M}\\
SO$_2$        &    1.1         &                176 &  1.0 $\times 10^{12}$ &                3.0 &                  57\\
H$_2$CO       &    1.1         &                200 &  2.2 $\times 10^{16}$ &                0.7 &                  60\\
CH$_3$OCHO    &    1.1         &                151 &  1.9 $\times 10^{16}$ &                2.6 &                  60\\
NO            &    1.1         &                133 &  1.2 $\times 10^{12}$ &               17.9 &                  80\\
CH$_3$CN      &    1.1         &                144 &  1.9 $\times 10^{14}$ &                0.5 &                  58\\
              &    1.1         &                145 &  5.1 $\times 10^{14}$ &                2.0 &                  60\\
H$_2 \! ^{34}$S &    1.1         &                200 &  1.1 $\times 10^{16}$ &                2.1 &                  58\\
\end{supertabular}\\
\vspace{1cm}

%---------------------------------------
% Envelope Components

\tablefirsthead{%
\hline
\hline
Molecule      & $\theta^{m,c}$ & T$_{\rm ex}^{m,c}$ & N$_{\rm tot}^{m,c}$   & $\Delta$ v$^{m,c}$ & v$_{\rm LSR}^{m,c}$\\
              & ($\arcsec$)    & (K)                & (cm$^{-2}$)           & (km~s$^{-1}$)      & (km~s$^{-1}$)      \\
\hline
}

\tablehead{%
\multicolumn{6}{c}{(Continued)}\\
\hline
\hline
Molecule      & $\theta^{m,c}$ & T$_{\rm ex}^{m,c}$ & N$_{\rm tot}^{m,c}$   & $\Delta$ v$^{m,c}$ & v$_{\rm LSR}^{m,c}$\\
              & ($\arcsec$)    & (K)                & (cm$^{-2}$)           & (km~s$^{-1}$)      & (km~s$^{-1}$)      \\
\hline
}

\tabletail{%
\hline
\hline
}

\topcaption{LTE Parameters for the full LTE model (Envelope Components) for source A20 in Sgr~B2(M).}
\tiny
\centering
\begin{supertabular}{lcccC{1cm}C{1cm}}\label{EnvLTE:parameters:A20SgrB2M}\\
HCCCN         & ext.           &                 69 &  7.6 $\times 10^{14}$ &                2.6 &                  61\\
              & ext.           &                 31 &  3.7 $\times 10^{15}$ &                5.1 &                  67\\
              & ext.           &                  3 &  3.2 $\times 10^{15}$ &                2.1 &                  66\\
              & ext.           &                 16 &  3.5 $\times 10^{15}$ &                2.6 &                  51\\
HNCO          & ext.           &                 11 &  5.8 $\times 10^{12}$ &                7.6 &                  64\\
              & ext.           &                 31 &  5.6 $\times 10^{15}$ &               11.5 &                  56\\
              & ext.           &                 12 &  1.0 $\times 10^{12}$ &                7.6 &                  52\\
\end{supertabular}\\
\vspace{1cm}

%================================================================================
%
% Source A21

%---------------------------------------
% Core Components

\tablefirsthead{%
\hline
\hline
Molecule      & $\theta^{m,c}$ & T$_{\rm ex}^{m,c}$ & N$_{\rm tot}^{m,c}$   & $\Delta$ v$^{m,c}$ & v$_{\rm LSR}^{m,c}$\\
              & ($\arcsec$)    & (K)                & (cm$^{-2}$)           & (km~s$^{-1}$)      & (km~s$^{-1}$)      \\
\hline
}

\tablehead{%
\multicolumn{6}{c}{(Continued)}\\
\hline
\hline
Molecule      & $\theta^{m,c}$ & T$_{\rm ex}^{m,c}$ & N$_{\rm tot}^{m,c}$   & $\Delta$ v$^{m,c}$ & v$_{\rm LSR}^{m,c}$\\
              & ($\arcsec$)    & (K)                & (cm$^{-2}$)           & (km~s$^{-1}$)      & (km~s$^{-1}$)      \\
\hline
}

\tabletail{%
\hline
\hline
}

\topcaption{LTE Parameters for the full LTE model (Core Components) for source A21 in Sgr~B2(M).}
\tiny
\centering
\begin{supertabular}{lcccC{1cm}C{1cm}}\label{CoreLTE:parameters:A21SgrB2M}\\
CN            &    0.7         &                 65 &  7.6 $\times 10^{14}$ &                3.1 &                -109\\
c-C$_3$H$_2$  &    0.7         &                 13 &  2.7 $\times 10^{14}$ &                3.6 &                  58\\
CCH           &    0.7         &                200 &  4.0 $\times 10^{15}$ &                4.1 &                  63\\
CH$_3$OCH$_3$ &    0.7         &                337 &  6.9 $\times 10^{14}$ &               13.7 &                  65\\
CO            &    0.7         &                200 &  4.4 $\times 10^{17}$ &               17.8 &                  96\\
$^{13}$CO     &    0.7         &                200 &  1.1 $\times 10^{17}$ &                7.2 &                  17\\
              &    0.7         &                200 &  8.8 $\times 10^{16}$ &                2.7 &                  57\\
              &    0.7         &                200 &  3.6 $\times 10^{17}$ &                3.1 &                  98\\
              &    0.7         &                200 &  3.9 $\times 10^{16}$ &                4.1 &                 109\\
              &    0.7         &                200 &  1.2 $\times 10^{17}$ &                7.4 &                 119\\
              &    0.7         &                200 &  5.7 $\times 10^{16}$ &                4.0 &                 128\\
              &    0.7         &                200 &  4.1 $\times 10^{16}$ &                2.5 &                 138\\
              &    0.7         &                200 &  1.3 $\times 10^{15}$ &                2.7 &                 145\\
C$^{17}$O     &    0.7         &                200 &  1.2 $\times 10^{16}$ &                2.7 &                 -73\\
              &    0.7         &                200 &  1.1 $\times 10^{16}$ &                2.7 &                 -65\\
              &    0.7         &                200 &  3.4 $\times 10^{15}$ &                2.7 &                 -60\\
              &    0.7         &                200 &  1.2 $\times 10^{16}$ &                2.7 &                 -55\\
              &    0.7         &                200 &  1.3 $\times 10^{16}$ &                3.0 &                 -50\\
              &    0.7         &                200 &  3.5 $\times 10^{15}$ &                2.7 &                 -44\\
              &    0.7         &                200 &  1.1 $\times 10^{16}$ &                9.6 &                 -20\\
              &    0.7         &                200 &  1.2 $\times 10^{16}$ &                2.8 &                  25\\
              &    0.7         &                200 &  3.4 $\times 10^{15}$ &                2.7 &                  31\\
              &    0.7         &                200 &  1.2 $\times 10^{16}$ &                2.4 &                  55\\
              &    0.7         &                200 &  3.6 $\times 10^{15}$ &                2.7 &                  65\\
              &    0.7         &                200 &  1.1 $\times 10^{16}$ &                2.7 &                  84\\
              &    0.7         &                200 &  3.4 $\times 10^{15}$ &                2.7 &                 113\\
              &    0.7         &                200 &  2.5 $\times 10^{15}$ &                2.7 &                 124\\
              &    0.7         &                200 &  3.5 $\times 10^{15}$ &                2.7 &                 132\\
              &    0.7         &                200 &  1.1 $\times 10^{15}$ &                2.7 &                 145\\
C$^{18}$O     &    0.7         &                200 &  4.1 $\times 10^{15}$ &                2.7 &                -161\\
              &    0.7         &                200 &  1.5 $\times 10^{16}$ &                2.7 &                -126\\
              &    0.7         &                200 &  1.8 $\times 10^{16}$ &                2.7 &                 -58\\
              &    0.7         &                200 &  5.6 $\times 10^{16}$ &                9.2 &                 -44\\
              &    0.7         &                200 &  2.2 $\times 10^{16}$ &                1.2 &                 -33\\
              &    0.7         &                200 &  1.7 $\times 10^{16}$ &                2.6 &                 -17\\
              &    0.7         &                200 &  9.4 $\times 10^{16}$ &                8.8 &                  -8\\
              &    0.7         &                200 &  7.5 $\times 10^{16}$ &                8.7 &                   3\\
              &    0.7         &                200 &  6.5 $\times 10^{14}$ &                2.4 &                  11\\
              &    0.7         &                200 &  4.3 $\times 10^{16}$ &                4.4 &                  19\\
              &    0.7         &                200 &  5.6 $\times 10^{16}$ &                7.2 &                  27\\
              &    0.7         &                200 &  6.4 $\times 10^{16}$ &                3.0 &                  46\\
              &    0.7         &                200 &  7.1 $\times 10^{16}$ &                2.5 &                  98\\
              &    0.7         &                200 &  5.0 $\times 10^{14}$ &                2.7 &                  71\\
              &    0.7         &                200 &  2.0 $\times 10^{16}$ &                2.7 &                 115\\
              &    0.7         &                200 &  2.6 $\times 10^{15}$ &                2.7 &                 120\\
              &    0.7         &                200 &  5.8 $\times 10^{16}$ &                4.4 &                 142\\
$^{13}$CS     &    0.7         &                200 &  2.5 $\times 10^{14}$ &                3.6 &                -177\\
              &    0.7         &                200 &  1.3 $\times 10^{14}$ &                7.2 &                -158\\
              &    0.7         &                200 &  1.8 $\times 10^{14}$ &                2.6 &                -145\\
              &    0.7         &                200 &  4.5 $\times 10^{14}$ &               10.6 &                -119\\
              &    0.7         &                200 &  3.1 $\times 10^{14}$ &               28.2 &                 -71\\
              &    0.7         &                200 &  3.1 $\times 10^{14}$ &                6.2 &                  -8\\
              &    0.7         &                200 &  2.9 $\times 10^{14}$ &                3.0 &                   1\\
              &    0.7         &                200 &  2.1 $\times 10^{15}$ &                5.5 &                  48\\
              &    0.7         &                200 &  2.8 $\times 10^{14}$ &                6.7 &                 111\\
C$^{33}$S     &    0.7         &                200 &  1.3 $\times 10^{14}$ &                5.6 &                -173\\
              &    0.7         &                200 &  8.0 $\times 10^{13}$ &                2.9 &                -127\\
              &    0.7         &                200 & 10.0 $\times 10^{13}$ &                4.4 &                -114\\
              &    0.7         &                200 &  9.3 $\times 10^{13}$ &                6.3 &                 -61\\
              &    0.7         &                200 &  1.5 $\times 10^{14}$ &               41.5 &                 -32\\
              &    0.7         &                200 &  6.7 $\times 10^{13}$ &                6.6 &                  10\\
              &    0.7         &                200 &  4.5 $\times 10^{14}$ &                3.5 &                  52\\
              &    0.7         &                200 &  1.7 $\times 10^{14}$ &                3.0 &                  59\\
              &    0.7         &                200 &  9.7 $\times 10^{13}$ &                5.3 &                  85\\
              &    0.7         &                200 &  1.0 $\times 10^{14}$ &                7.0 &                  94\\
C$^{34}$S     &    0.7         &                200 &  2.4 $\times 10^{14}$ &                2.7 &                -141\\
              &    0.7         &                200 &  1.2 $\times 10^{14}$ &                2.3 &                 -67\\
              &    0.7         &                200 &  2.0 $\times 10^{14}$ &                2.8 &                 -34\\
              &    0.7         &                200 &  1.2 $\times 10^{14}$ &                3.4 &                 -10\\
              &    0.7         &                200 &  3.7 $\times 10^{14}$ &               21.5 &                   8\\
              &    0.7         &                200 &  1.1 $\times 10^{15}$ &                5.0 &                  49\\
              &    0.7         &                200 &  1.0 $\times 10^{14}$ &                5.4 &                  92\\
H$^{13}$CN    &    0.7         &                200 &  6.3 $\times 10^{13}$ &                5.5 &                -146\\
              &    0.7         &                200 &  8.0 $\times 10^{13}$ &                8.4 &                -121\\
              &    0.7         &                200 &  5.3 $\times 10^{13}$ &                3.6 &                 -97\\
              &    0.7         &                200 &  5.0 $\times 10^{13}$ &                4.8 &                 -58\\
              &    0.7         &                200 &  8.4 $\times 10^{13}$ &               14.4 &                 -23\\
              &    0.7         &                200 &  8.1 $\times 10^{13}$ &               11.6 &                  11\\
              &    0.7         &                200 &  1.9 $\times 10^{14}$ &                2.6 &                  54\\
HN$^{13}$C    &    0.7         &                200 &  3.0 $\times 10^{12}$ &                8.7 &                 -76\\
              &    0.7         &                200 &  3.8 $\times 10^{12}$ &                3.5 &                  27\\
H$^{13}$CO$^+$ &    0.7         &                200 &  2.6 $\times 10^{13}$ &                2.5 &                 -77\\
              &    0.7         &                200 &  2.4 $\times 10^{13}$ &                3.1 &                 -71\\
              &    0.7         &                200 &  1.1 $\times 10^{12}$ &                5.3 &                 -65\\
              &    0.7         &                200 &  3.1 $\times 10^{13}$ &                1.3 &                 -61\\
              &    0.7         &                200 &  1.5 $\times 10^{13}$ &                1.1 &                 -48\\
              &    0.7         &                200 &  2.2 $\times 10^{13}$ &                1.4 &                 -50\\
              &    0.7         &                200 &  8.4 $\times 10^{12}$ &                4.4 &                 -44\\
              &    0.7         &                200 &  4.4 $\times 10^{13}$ &                2.4 &                   6\\
              &    0.7         &                200 &  9.0 $\times 10^{12}$ &                1.8 &                  15\\
              &    0.7         &                200 &  4.0 $\times 10^{12}$ &                1.9 &                  29\\
              &    0.7         &                200 &  8.5 $\times 10^{15}$ &                0.6 &                  56\\
              &    0.7         &                200 &  1.1 $\times 10^{14}$ &                2.5 &                  64\\
              &    0.7         &                200 &  2.0 $\times 10^{13}$ &                3.3 &                  84\\
              &    0.7         &                200 &  3.6 $\times 10^{12}$ &                5.0 &                 100\\
              &    0.7         &                200 &  5.0 $\times 10^{12}$ &                6.3 &                  99\\
HO$^{13}$C$^+$ &    0.7         &                200 &  7.7 $\times 10^{13}$ &                5.8 &                -134\\
              &    0.7         &                200 &  5.2 $\times 10^{13}$ &               12.4 &                  42\\
\end{supertabular}\\
\vspace{1cm}

%---------------------------------------
% Envelope Components

\tablefirsthead{%
\hline
\hline
Molecule      & $\theta^{m,c}$ & T$_{\rm ex}^{m,c}$ & N$_{\rm tot}^{m,c}$   & $\Delta$ v$^{m,c}$ & v$_{\rm LSR}^{m,c}$\\
              & ($\arcsec$)    & (K)                & (cm$^{-2}$)           & (km~s$^{-1}$)      & (km~s$^{-1}$)      \\
\hline
}

\tablehead{%
\multicolumn{6}{c}{(Continued)}\\
\hline
\hline
Molecule      & $\theta^{m,c}$ & T$_{\rm ex}^{m,c}$ & N$_{\rm tot}^{m,c}$   & $\Delta$ v$^{m,c}$ & v$_{\rm LSR}^{m,c}$\\
              & ($\arcsec$)    & (K)                & (cm$^{-2}$)           & (km~s$^{-1}$)      & (km~s$^{-1}$)      \\
\hline
}

\tabletail{%
\hline
\hline
}

\topcaption{LTE Parameters for the full LTE model (Envelope Components) for source A21 in Sgr~B2(M).}
\tiny
\centering
\begin{supertabular}{lcccC{1cm}C{1cm}}\label{EnvLTE:parameters:A21SgrB2M}\\
H$_2$CO       & ext.           &                 24 &  3.2 $\times 10^{17}$ &                2.8 &                  61\\
              & ext.           &                  3 &  2.1 $\times 10^{14}$ &               16.0 &                  61\\
CH$_3$CN      & ext.           &                 14 &  6.3 $\times 10^{14}$ &                5.7 &                  64\\
              & ext.           &                  3 &  2.5 $\times 10^{13}$ &                2.5 &                  64\\
CH$_3$OH      & ext.           &                 13 &  7.9 $\times 10^{14}$ &                3.2 &                  65\\
              & ext.           &                 18 &  1.8 $\times 10^{16}$ &                6.1 &                  59\\
CN            & ext.           &                 72 &  9.0 $\times 10^{10}$ &               11.7 &                 -19\\
\end{supertabular}\\
\vspace{1cm}

%================================================================================
%
% Source A22

%---------------------------------------
% Core Components

\tablefirsthead{%
\hline
\hline
Molecule      & $\theta^{m,c}$ & T$_{\rm ex}^{m,c}$ & N$_{\rm tot}^{m,c}$   & $\Delta$ v$^{m,c}$ & v$_{\rm LSR}^{m,c}$\\
              & ($\arcsec$)    & (K)                & (cm$^{-2}$)           & (km~s$^{-1}$)      & (km~s$^{-1}$)      \\
\hline
}

\tablehead{%
\multicolumn{6}{c}{(Continued)}\\
\hline
\hline
Molecule      & $\theta^{m,c}$ & T$_{\rm ex}^{m,c}$ & N$_{\rm tot}^{m,c}$   & $\Delta$ v$^{m,c}$ & v$_{\rm LSR}^{m,c}$\\
              & ($\arcsec$)    & (K)                & (cm$^{-2}$)           & (km~s$^{-1}$)      & (km~s$^{-1}$)      \\
\hline
}

\tabletail{%
\hline
\hline
}

\topcaption{LTE Parameters for the full LTE model (Core Components) for source A22 in Sgr~B2(M).}
\tiny
\centering
\begin{supertabular}{lcccC{1cm}C{1cm}}\label{CoreLTE:parameters:A22SgrB2M}\\
OCS           &    0.8         &                 64 &  2.1 $\times 10^{12}$ &                3.0 &                  63\\
              &    0.8         &                 46 &  1.5 $\times 10^{16}$ &                3.3 &                  49\\
HCS$^+$       &    0.8         &                204 &  6.0 $\times 10^{12}$ &                3.1 &                  66\\
              &    0.8         &                203 &  1.4 $\times 10^{14}$ &                2.0 &                  50\\
H$_2$CS       &    0.8         &                 64 &  4.0 $\times 10^{14}$ &                1.6 &                  67\\
              &    0.8         &                 89 &  1.1 $\times 10^{15}$ &                0.9 &                  50\\
CH$_3$CN      &    0.8         &                126 &  1.4 $\times 10^{15}$ &                7.1 &                  49\\
\end{supertabular}\\
\vspace{1cm}

%---------------------------------------
% Envelope Components

\tablefirsthead{%
\hline
\hline
Molecule      & $\theta^{m,c}$ & T$_{\rm ex}^{m,c}$ & N$_{\rm tot}^{m,c}$   & $\Delta$ v$^{m,c}$ & v$_{\rm LSR}^{m,c}$\\
              & ($\arcsec$)    & (K)                & (cm$^{-2}$)           & (km~s$^{-1}$)      & (km~s$^{-1}$)      \\
\hline
}

\tablehead{%
\multicolumn{6}{c}{(Continued)}\\
\hline
\hline
Molecule      & $\theta^{m,c}$ & T$_{\rm ex}^{m,c}$ & N$_{\rm tot}^{m,c}$   & $\Delta$ v$^{m,c}$ & v$_{\rm LSR}^{m,c}$\\
              & ($\arcsec$)    & (K)                & (cm$^{-2}$)           & (km~s$^{-1}$)      & (km~s$^{-1}$)      \\
\hline
}

\tabletail{%
\hline
\hline
}

\topcaption{LTE Parameters for the full LTE model (Envelope Components) for source A22 in Sgr~B2(M).}
\tiny
\centering
\begin{supertabular}{lcccC{1cm}C{1cm}}\label{EnvLTE:parameters:A22SgrB2M}\\
OCS           & ext.           &                  6 &  3.3 $\times 10^{18}$ &                4.0 &                  54\\
              & ext.           &                 17 &  2.3 $\times 10^{16}$ &                2.1 &                  55\\
HCS$^+$       & ext.           &                 12 &  4.2 $\times 10^{13}$ &                3.3 &                  59\\
PH$_3$        & ext.           &                 10 &  2.2 $\times 10^{13}$ &                4.0 &                  69\\
              & ext.           &                 10 &  3.0 $\times 10^{13}$ &                3.0 &                  57\\
CH$_3$CN      & ext.           &                 23 &  3.5 $\times 10^{15}$ &                9.7 &                  64\\
              & ext.           &                 37 &  5.0 $\times 10^{14}$ &                2.4 &                  50\\
\end{supertabular}\\
\vspace{1cm}

%================================================================================
%
% Source A23

%---------------------------------------
% Core Components

\tablefirsthead{%
\hline
\hline
Molecule      & $\theta^{m,c}$ & T$_{\rm ex}^{m,c}$ & N$_{\rm tot}^{m,c}$   & $\Delta$ v$^{m,c}$ & v$_{\rm LSR}^{m,c}$\\
              & ($\arcsec$)    & (K)                & (cm$^{-2}$)           & (km~s$^{-1}$)      & (km~s$^{-1}$)      \\
\hline
}

\tablehead{%
\multicolumn{6}{c}{(Continued)}\\
\hline
\hline
Molecule      & $\theta^{m,c}$ & T$_{\rm ex}^{m,c}$ & N$_{\rm tot}^{m,c}$   & $\Delta$ v$^{m,c}$ & v$_{\rm LSR}^{m,c}$\\
              & ($\arcsec$)    & (K)                & (cm$^{-2}$)           & (km~s$^{-1}$)      & (km~s$^{-1}$)      \\
\hline
}

\tabletail{%
\hline
\hline
}

\topcaption{LTE Parameters for the full LTE model (Core Components) for source A23 in Sgr~B2(M).}
\tiny
\centering
\begin{supertabular}{lcccC{1cm}C{1cm}}\label{CoreLTE:parameters:A23SgrB2M}\\
SiO           &    1.1         &                203 &  1.5 $\times 10^{14}$ &                4.3 &                  59\\
              &    1.1         &                180 &  4.8 $\times 10^{13}$ &                2.0 &                  45\\
              &    1.1         &                200 &  2.6 $\times 10^{13}$ &                3.0 &                  41\\
OCS           &    1.1         &                 69 &  2.0 $\times 10^{15}$ &                0.2 &                  62\\
              &    1.1         &                 65 &  1.2 $\times 10^{15}$ &                2.0 &                  47\\
HCS$^+$       &    1.1         &                200 & 10.0 $\times 10^{13}$ &                1.9 &                  75\\
              &    1.1         &                200 &  2.1 $\times 10^{14}$ &                3.0 &                  50\\
\end{supertabular}\\
\vspace{1cm}

%---------------------------------------
% Envelope Components

\tablefirsthead{%
\hline
\hline
Molecule      & $\theta^{m,c}$ & T$_{\rm ex}^{m,c}$ & N$_{\rm tot}^{m,c}$   & $\Delta$ v$^{m,c}$ & v$_{\rm LSR}^{m,c}$\\
              & ($\arcsec$)    & (K)                & (cm$^{-2}$)           & (km~s$^{-1}$)      & (km~s$^{-1}$)      \\
\hline
}

\tablehead{%
\multicolumn{6}{c}{(Continued)}\\
\hline
\hline
Molecule      & $\theta^{m,c}$ & T$_{\rm ex}^{m,c}$ & N$_{\rm tot}^{m,c}$   & $\Delta$ v$^{m,c}$ & v$_{\rm LSR}^{m,c}$\\
              & ($\arcsec$)    & (K)                & (cm$^{-2}$)           & (km~s$^{-1}$)      & (km~s$^{-1}$)      \\
\hline
}

\tabletail{%
\hline
\hline
}

\topcaption{LTE Parameters for the full LTE model (Envelope Components) for source A23 in Sgr~B2(M).}
\tiny
\centering
\begin{supertabular}{lcccC{1cm}C{1cm}}\label{EnvLTE:parameters:A23SgrB2M}\\
SiO           & ext.           &                 20 &  1.1 $\times 10^{12}$ &                4.0 &                  67\\
              & ext.           &                  8 &  2.2 $\times 10^{13}$ &                3.7 &                  59\\
              & ext.           &                 13 &  7.1 $\times 10^{12}$ &                2.4 &                  52\\
              & ext.           &                  3 &  1.8 $\times 10^{14}$ &                1.7 &                  50\\
CH$_2$NH      & ext.           &                  7 &  1.0 $\times 10^{12}$ &                2.0 &                  67\\
              & ext.           &                  3 &  8.2 $\times 10^{12}$ &               10.8 &                  67\\
              & ext.           &                  8 &  1.2 $\times 10^{12}$ &                2.0 &                  64\\
\end{supertabular}\\
\vspace{1cm}

%================================================================================
%
% Source A24

%---------------------------------------
% Core Components

\tablefirsthead{%
\hline
\hline
Molecule      & $\theta^{m,c}$ & T$_{\rm ex}^{m,c}$ & N$_{\rm tot}^{m,c}$   & $\Delta$ v$^{m,c}$ & v$_{\rm LSR}^{m,c}$\\
              & ($\arcsec$)    & (K)                & (cm$^{-2}$)           & (km~s$^{-1}$)      & (km~s$^{-1}$)      \\
\hline
}

\tablehead{%
\multicolumn{6}{c}{(Continued)}\\
\hline
\hline
Molecule      & $\theta^{m,c}$ & T$_{\rm ex}^{m,c}$ & N$_{\rm tot}^{m,c}$   & $\Delta$ v$^{m,c}$ & v$_{\rm LSR}^{m,c}$\\
              & ($\arcsec$)    & (K)                & (cm$^{-2}$)           & (km~s$^{-1}$)      & (km~s$^{-1}$)      \\
\hline
}

\tabletail{%
\hline
\hline
}

\topcaption{LTE Parameters for the full LTE model (Core Components) for source A24 in Sgr~B2(M).}
\tiny
\centering
\begin{supertabular}{lcccC{1cm}C{1cm}}\label{CoreLTE:parameters:A24SgrB2M}\\
RRL-H         &    0.5         &              15214 &  3.4 $\times 10^{9}$  &               18.1 &                  61\\
CH$_3$CN      &    0.5         &                149 &  1.8 $\times 10^{16}$ &                5.5 &                  63\\
H$_2$CS       &    0.5         &                152 &  3.5 $\times 10^{15}$ &                1.2 &                  65\\
              &    0.5         &                 96 &  6.0 $\times 10^{15}$ &                2.2 &                  62\\
H$_2$CNH      &    0.5         &                200 &  5.8 $\times 10^{16}$ &                8.8 &                  64\\
              &    0.5         &                 96 &  2.9 $\times 10^{16}$ &               17.3 &                  60\\
HNCO          &    0.5         &                137 &  3.5 $\times 10^{13}$ &                3.0 &                  64\\
\end{supertabular}\\
\vspace{1cm}

%---------------------------------------
% Envelope Components

\tablefirsthead{%
\hline
\hline
Molecule      & $\theta^{m,c}$ & T$_{\rm ex}^{m,c}$ & N$_{\rm tot}^{m,c}$   & $\Delta$ v$^{m,c}$ & v$_{\rm LSR}^{m,c}$\\
              & ($\arcsec$)    & (K)                & (cm$^{-2}$)           & (km~s$^{-1}$)      & (km~s$^{-1}$)      \\
\hline
}

\tablehead{%
\multicolumn{6}{c}{(Continued)}\\
\hline
\hline
Molecule      & $\theta^{m,c}$ & T$_{\rm ex}^{m,c}$ & N$_{\rm tot}^{m,c}$   & $\Delta$ v$^{m,c}$ & v$_{\rm LSR}^{m,c}$\\
              & ($\arcsec$)    & (K)                & (cm$^{-2}$)           & (km~s$^{-1}$)      & (km~s$^{-1}$)      \\
\hline
}

\tabletail{%
\hline
\hline
}

\topcaption{LTE Parameters for the full LTE model (Envelope Components) for source A24 in Sgr~B2(M).}
\tiny
\centering
\begin{supertabular}{lcccC{1cm}C{1cm}}\label{EnvLTE:parameters:A24SgrB2M}\\
CH$_2$NH      & ext.           &                 20 &  6.5 $\times 10^{12}$ &                3.0 &                  68\\
SO$_2$        & ext.           &                  4 &  8.9 $\times 10^{13}$ &                3.6 &                  75\\
              & ext.           &                 41 &  6.6 $\times 10^{16}$ &                4.8 &                  53\\
SO            & ext.           &                 10 &  1.1 $\times 10^{14}$ &                2.0 &                  74\\
              & ext.           &                 16 &  3.3 $\times 10^{14}$ &                6.4 &                  51\\
CN            & ext.           &                  7 &  2.1 $\times 10^{12}$ &                2.5 &                  54\\
              & ext.           &                  5 &  7.0 $\times 10^{12}$ &                2.3 &                  70\\
\end{supertabular}\\
\vspace{1cm}

%================================================================================
%
% Source A25

%---------------------------------------
% Core Components

\tablefirsthead{%
\hline
\hline
Molecule      & $\theta^{m,c}$ & T$_{\rm ex}^{m,c}$ & N$_{\rm tot}^{m,c}$   & $\Delta$ v$^{m,c}$ & v$_{\rm LSR}^{m,c}$\\
              & ($\arcsec$)    & (K)                & (cm$^{-2}$)           & (km~s$^{-1}$)      & (km~s$^{-1}$)      \\
\hline
}

\tablehead{%
\multicolumn{6}{c}{(Continued)}\\
\hline
\hline
Molecule      & $\theta^{m,c}$ & T$_{\rm ex}^{m,c}$ & N$_{\rm tot}^{m,c}$   & $\Delta$ v$^{m,c}$ & v$_{\rm LSR}^{m,c}$\\
              & ($\arcsec$)    & (K)                & (cm$^{-2}$)           & (km~s$^{-1}$)      & (km~s$^{-1}$)      \\
\hline
}

\tabletail{%
\hline
\hline
}

\topcaption{LTE Parameters for the full LTE model (Core Components) for source A25 in Sgr~B2(M).}
\tiny
\centering
\begin{supertabular}{lcccC{1cm}C{1cm}}\label{CoreLTE:parameters:A25SgrB2M}\\
H$_2$CS       &    0.7         &                 77 &  1.2 $\times 10^{16}$ &                5.5 &                  54\\
H$_2 \! ^{34}$S &    0.7         &                149 &  1.2 $\times 10^{16}$ &                3.4 &                  54\\
CN            &    0.7         &                 70 &  9.3 $\times 10^{13}$ &                2.0 &                 -57\\
              &    0.7         &                 80 &  5.3 $\times 10^{12}$ &                2.9 &                -102\\
CCH           &    0.7         &                180 &  1.5 $\times 10^{16}$ &                2.9 &                  55\\
HCCCN         &    0.7         &                 46 &  1.1 $\times 10^{15}$ &                3.1 &                  55\\
              &    0.7         &                 41 &  1.8 $\times 10^{15}$ &                3.4 &                  53\\
\end{supertabular}\\
\vspace{1cm}

%---------------------------------------
% Envelope Components

\tablefirsthead{%
\hline
\hline
Molecule      & $\theta^{m,c}$ & T$_{\rm ex}^{m,c}$ & N$_{\rm tot}^{m,c}$   & $\Delta$ v$^{m,c}$ & v$_{\rm LSR}^{m,c}$\\
              & ($\arcsec$)    & (K)                & (cm$^{-2}$)           & (km~s$^{-1}$)      & (km~s$^{-1}$)      \\
\hline
}

\tablehead{%
\multicolumn{6}{c}{(Continued)}\\
\hline
\hline
Molecule      & $\theta^{m,c}$ & T$_{\rm ex}^{m,c}$ & N$_{\rm tot}^{m,c}$   & $\Delta$ v$^{m,c}$ & v$_{\rm LSR}^{m,c}$\\
              & ($\arcsec$)    & (K)                & (cm$^{-2}$)           & (km~s$^{-1}$)      & (km~s$^{-1}$)      \\
\hline
}

\tabletail{%
\hline
\hline
}

\topcaption{LTE Parameters for the full LTE model (Envelope Components) for source A25 in Sgr~B2(M).}
\tiny
\centering
\begin{supertabular}{lcccC{1cm}C{1cm}}\label{EnvLTE:parameters:A25SgrB2M}\\
PH$_3$        & ext.           &                 10 &  1.6 $\times 10^{14}$ &                3.0 &                  54\\
HC(O)NH$_2$   & ext.           &                 19 &  7.7 $\times 10^{16}$ &                6.7 &                  60\\
CN            & ext.           &                  3 &  1.3 $\times 10^{13}$ &                1.8 &                 133\\
              & ext.           &                  3 &  6.6 $\times 10^{13}$ &                2.3 &                  77\\
              & ext.           &                  3 &  2.1 $\times 10^{13}$ &                1.6 &                  64\\
              & ext.           &                  8 &  1.0 $\times 10^{13}$ &                2.0 &                 -78\\
              & ext.           &                  9 &  2.7 $\times 10^{12}$ &                0.2 &                 -92\\
\end{supertabular}\\
\vspace{1cm}

%================================================================================
%
% Source A26

%---------------------------------------
% Core Components

\tablefirsthead{%
\hline
\hline
Molecule      & $\theta^{m,c}$ & T$_{\rm ex}^{m,c}$ & N$_{\rm tot}^{m,c}$   & $\Delta$ v$^{m,c}$ & v$_{\rm LSR}^{m,c}$\\
              & ($\arcsec$)    & (K)                & (cm$^{-2}$)           & (km~s$^{-1}$)      & (km~s$^{-1}$)      \\
\hline
}

\tablehead{%
\multicolumn{6}{c}{(Continued)}\\
\hline
\hline
Molecule      & $\theta^{m,c}$ & T$_{\rm ex}^{m,c}$ & N$_{\rm tot}^{m,c}$   & $\Delta$ v$^{m,c}$ & v$_{\rm LSR}^{m,c}$\\
              & ($\arcsec$)    & (K)                & (cm$^{-2}$)           & (km~s$^{-1}$)      & (km~s$^{-1}$)      \\
\hline
}

\tabletail{%
\hline
\hline
}

\topcaption{LTE Parameters for the full LTE model (Core Components) for source A26 in Sgr~B2(M).}
\tiny
\centering
\begin{supertabular}{lcccC{1cm}C{1cm}}\label{CoreLTE:parameters:A26SgrB2M}\\
H$_2$CO       &    1.0         &                 47 &  5.2 $\times 10^{14}$ &                1.9 &                  76\\
              &    1.0         &                 80 &  1.3 $\times 10^{17}$ &                5.5 &                  61\\
SiO           &    1.0         &                200 &  3.0 $\times 10^{14}$ &                6.0 &                  60\\
HCS$^+$       &    1.0         &                 98 &  6.9 $\times 10^{14}$ &                4.4 &                  60\\
CN            &    1.0         &                169 &  1.0 $\times 10^{12}$ &                2.9 &                  64\\
CO$^+$        &    1.0         &                 97 &  3.6 $\times 10^{13}$ &                3.8 &                  78\\
H$_2 \! ^{34}$S &    1.0         &                 49 &  1.3 $\times 10^{15}$ &                1.1 &                  61\\
\end{supertabular}\\
\vspace{1cm}

%---------------------------------------
% Envelope Components

\tablefirsthead{%
\hline
\hline
Molecule      & $\theta^{m,c}$ & T$_{\rm ex}^{m,c}$ & N$_{\rm tot}^{m,c}$   & $\Delta$ v$^{m,c}$ & v$_{\rm LSR}^{m,c}$\\
              & ($\arcsec$)    & (K)                & (cm$^{-2}$)           & (km~s$^{-1}$)      & (km~s$^{-1}$)      \\
\hline
}

\tablehead{%
\multicolumn{6}{c}{(Continued)}\\
\hline
\hline
Molecule      & $\theta^{m,c}$ & T$_{\rm ex}^{m,c}$ & N$_{\rm tot}^{m,c}$   & $\Delta$ v$^{m,c}$ & v$_{\rm LSR}^{m,c}$\\
              & ($\arcsec$)    & (K)                & (cm$^{-2}$)           & (km~s$^{-1}$)      & (km~s$^{-1}$)      \\
\hline
}

\tabletail{%
\hline
\hline
}

\topcaption{LTE Parameters for the full LTE model (Envelope Components) for source A26 in Sgr~B2(M).}
\tiny
\centering
\begin{supertabular}{lcccC{1cm}C{1cm}}\label{EnvLTE:parameters:A26SgrB2M}\\
H$_2$CO       & ext.           &                  3 &  3.3 $\times 10^{13}$ &                0.8 &                  72\\
              & ext.           &                  3 &  1.7 $\times 10^{14}$ &                0.9 &                  60\\
              & ext.           &                 12 &  6.0 $\times 10^{14}$ &               17.7 &                  61\\
SiO           & ext.           &                 28 &  1.8 $\times 10^{12}$ &                3.1 &                  60\\
              & ext.           &                 13 &  1.0 $\times 10^{13}$ &                2.0 &                  52\\
              & ext.           &                  5 &  1.0 $\times 10^{14}$ &                7.0 &                  72\\
              & ext.           &                  5 &  5.2 $\times 10^{13}$ &                4.0 &                  49\\
\end{supertabular}\\
\vspace{1cm}

%================================================================================
%
% Source A27

%---------------------------------------
% Core Components

\tablefirsthead{%
\hline
\hline
Molecule      & $\theta^{m,c}$ & T$_{\rm ex}^{m,c}$ & N$_{\rm tot}^{m,c}$   & $\Delta$ v$^{m,c}$ & v$_{\rm LSR}^{m,c}$\\
              & ($\arcsec$)    & (K)                & (cm$^{-2}$)           & (km~s$^{-1}$)      & (km~s$^{-1}$)      \\
\hline
}

\tablehead{%
\multicolumn{6}{c}{(Continued)}\\
\hline
\hline
Molecule      & $\theta^{m,c}$ & T$_{\rm ex}^{m,c}$ & N$_{\rm tot}^{m,c}$   & $\Delta$ v$^{m,c}$ & v$_{\rm LSR}^{m,c}$\\
              & ($\arcsec$)    & (K)                & (cm$^{-2}$)           & (km~s$^{-1}$)      & (km~s$^{-1}$)      \\
\hline
}

\tabletail{%
\hline
\hline
}

\topcaption{LTE Parameters for the full LTE model (Core Components) for source A27 in Sgr~B2(M).}
\tiny
\centering
\begin{supertabular}{lcccC{1cm}C{1cm}}\label{CoreLTE:parameters:A27SgrB2M}\\
CH$_3$CN      &    0.7         &                113 &  2.6 $\times 10^{14}$ &                6.3 &                  83\\
CN            &    0.7         &                144 &  2.4 $\times 10^{12}$ &                2.8 &                 -86\\
SiO           &    0.7         &                 36 &  1.9 $\times 10^{14}$ &                8.0 &                  40\\
HCN, v$_2$=2  &    0.7         &                807 &  6.3 $\times 10^{15}$ &                8.1 &                  54\\
CH$_3$OH      &    0.7         &                 98 &  3.5 $\times 10^{15}$ &                1.9 &                  52\\
SO            &    0.7         &                234 &  1.3 $\times 10^{15}$ &                4.9 &                  72\\
H$_2$CO       &    0.7         &                184 &  2.4 $\times 10^{14}$ &                5.0 &                  74\\
CO            &    0.7         &                200 &  6.6 $\times 10^{16}$ &                2.6 &                 110\\
              &    0.7         &                200 &  9.7 $\times 10^{15}$ &                0.2 &                 104\\
              &    0.7         &                200 &  1.0 $\times 10^{17}$ &                1.0 &                  77\\
              &    0.7         &                200 &  1.9 $\times 10^{15}$ &                4.4 &                  30\\
              &    0.7         &                200 &  1.6 $\times 10^{16}$ &                1.2 &                  28\\
              &    0.7         &                200 &  1.2 $\times 10^{16}$ &                1.2 &                  18\\
              &    0.7         &                200 &  2.3 $\times 10^{16}$ &                1.7 &                   7\\
              &    0.7         &                200 &  3.7 $\times 10^{16}$ &                2.6 &                 -15\\
              &    0.7         &                200 &  1.1 $\times 10^{17}$ &                1.5 &                 -36\\
              &    0.7         &                200 &  1.4 $\times 10^{16}$ &                1.0 &                 -58\\
              &    0.7         &                200 &  1.2 $\times 10^{17}$ &                1.5 &                 -67\\
              &    0.7         &                200 &  5.4 $\times 10^{15}$ &                1.0 &                -105\\
              &    0.7         &                200 &  1.1 $\times 10^{16}$ &                1.0 &                -109\\
              &    0.7         &                200 &  4.7 $\times 10^{16}$ &                2.0 &                -121\\
              &    0.7         &                200 &  4.2 $\times 10^{16}$ &                2.6 &                -155\\
$^{13}$CO     &    0.7         &                200 &  9.0 $\times 10^{16}$ &                1.9 &                 157\\
              &    0.7         &                200 &  2.1 $\times 10^{16}$ &                1.0 &                 137\\
              &    0.7         &                200 &  2.8 $\times 10^{16}$ &                0.5 &                 123\\
              &    0.7         &                200 &  5.7 $\times 10^{16}$ &                1.1 &                 121\\
              &    0.7         &                200 &  2.7 $\times 10^{16}$ &                0.9 &                 109\\
              &    0.7         &                200 &  1.7 $\times 10^{16}$ &                1.0 &                 107\\
              &    0.7         &                200 &  4.7 $\times 10^{16}$ &                1.4 &                  98\\
              &    0.7         &                200 &  1.8 $\times 10^{16}$ &                1.1 &                  27\\
              &    0.7         &                200 &  2.0 $\times 10^{16}$ &                1.0 &                  11\\
              &    0.7         &                200 &  4.5 $\times 10^{16}$ &                1.2 &                   5\\
C$^{17}$O     &    0.7         &                200 &  1.1 $\times 10^{16}$ &                7.1 &                -172\\
              &    0.7         &                200 &  1.2 $\times 10^{16}$ &                2.7 &                -164\\
              &    0.7         &                200 &  3.4 $\times 10^{15}$ &                2.7 &                -135\\
              &    0.7         &                200 &  1.0 $\times 10^{15}$ &                2.7 &                -130\\
              &    0.7         &                200 &  1.0 $\times 10^{15}$ &                2.7 &                -124\\
              &    0.7         &                200 &  1.6 $\times 10^{16}$ &                2.7 &                -119\\
              &    0.7         &                200 &  1.0 $\times 10^{15}$ &                2.7 &                -114\\
              &    0.7         &                200 &  6.3 $\times 10^{15}$ &                2.5 &                 -66\\
              &    0.7         &                200 &  4.8 $\times 10^{15}$ &                2.6 &                 -39\\
              &    0.7         &                200 &  3.3 $\times 10^{15}$ &                5.2 &                 -18\\
              &    0.7         &                200 &  9.6 $\times 10^{15}$ &                3.2 &                   4\\
              &    0.7         &                200 &  5.2 $\times 10^{15}$ &                2.4 &                  33\\
              &    0.7         &                200 &  3.1 $\times 10^{16}$ &                2.7 &                  41\\
              &    0.7         &                200 &  2.8 $\times 10^{17}$ &                3.2 &                  52\\
              &    0.7         &                200 &  5.0 $\times 10^{13}$ &                2.7 &                  79\\
              &    0.7         &                200 &  4.0 $\times 10^{16}$ &                2.7 &                  83\\
              &    0.7         &                200 &  3.7 $\times 10^{16}$ &                7.2 &                 126\\
C$^{18}$O     &    0.7         &                200 &  3.8 $\times 10^{16}$ &                7.5 &                -168\\
              &    0.7         &                200 &  1.2 $\times 10^{16}$ &                2.7 &                -159\\
              &    0.7         &                200 &  1.2 $\times 10^{17}$ &                9.9 &                -130\\
              &    0.7         &                200 &  3.2 $\times 10^{15}$ &                2.3 &                -112\\
              &    0.7         &                200 &  3.9 $\times 10^{16}$ &                9.2 &                -107\\
              &    0.7         &                200 &  9.6 $\times 10^{16}$ &                8.5 &                 -85\\
              &    0.7         &                200 &  3.2 $\times 10^{15}$ &                2.8 &                 -77\\
              &    0.7         &                200 &  1.2 $\times 10^{16}$ &                3.0 &                 -61\\
              &    0.7         &                200 &  4.3 $\times 10^{16}$ &                8.0 &                 -52\\
              &    0.7         &                200 &  1.0 $\times 10^{15}$ &               10.8 &                 -36\\
              &    0.7         &                200 &  3.3 $\times 10^{15}$ &                2.7 &                 -25\\
              &    0.7         &                200 &  1.3 $\times 10^{16}$ &                3.4 &                 -14\\
              &    0.7         &                200 &  3.7 $\times 10^{16}$ &                3.5 &                  -4\\
              &    0.7         &                200 &  1.1 $\times 10^{16}$ &                2.8 &                   5\\
              &    0.7         &                200 &  2.9 $\times 10^{15}$ &                0.6 &                  19\\
              &    0.7         &                200 &  3.3 $\times 10^{15}$ &                2.7 &                  27\\
              &    0.7         &                200 &  3.2 $\times 10^{16}$ &                3.9 &                  38\\
              &    0.7         &                200 &  4.3 $\times 10^{16}$ &                2.9 &                  43\\
              &    0.7         &                200 &  1.1 $\times 10^{17}$ &                2.7 &                  52\\
              &    0.7         &                200 &  3.9 $\times 10^{16}$ &                3.7 &                  87\\
              &    0.7         &                200 &  1.1 $\times 10^{16}$ &                3.4 &                 109\\
              &    0.7         &                200 &  1.2 $\times 10^{16}$ &                2.7 &                 123\\
              &    0.7         &                200 &  3.7 $\times 10^{16}$ &                2.8 &                 128\\
              &    0.7         &                200 &  1.5 $\times 10^{16}$ &                2.7 &                 139\\
CS            &    0.7         &                200 &  1.7 $\times 10^{13}$ &                1.1 &                 147\\
              &    0.7         &                200 &  9.7 $\times 10^{13}$ &                3.7 &                 132\\
              &    0.7         &                200 &  1.3 $\times 10^{14}$ &                1.0 &                  74\\
              &    0.7         &                200 &  9.7 $\times 10^{13}$ &                2.4 &                  55\\
              &    0.7         &                200 &  3.9 $\times 10^{14}$ &                8.0 &                  34\\
              &    0.7         &                200 &  1.1 $\times 10^{13}$ &                0.4 &                -107\\
              &    0.7         &                200 &  7.7 $\times 10^{13}$ &                1.1 &                -160\\
              &    0.7         &                200 &  1.4 $\times 10^{14}$ &                1.9 &                -179\\
$^{13}$CS     &    0.7         &                200 &  1.9 $\times 10^{14}$ &                3.9 &                -167\\
              &    0.7         &                200 &  1.6 $\times 10^{14}$ &                3.5 &                -157\\
              &    0.7         &                200 &  3.5 $\times 10^{14}$ &                6.8 &                -148\\
              &    0.7         &                200 &  1.7 $\times 10^{14}$ &                2.7 &                -134\\
              &    0.7         &                200 &  1.3 $\times 10^{14}$ &                2.6 &                -119\\
              &    0.7         &                200 &  1.9 $\times 10^{14}$ &                3.3 &                -112\\
              &    0.7         &                200 &  3.0 $\times 10^{14}$ &                3.1 &                -106\\
              &    0.7         &                200 &  7.6 $\times 10^{12}$ &                6.1 &                 -93\\
              &    0.7         &                200 &  2.9 $\times 10^{14}$ &                3.1 &                 -51\\
              &    0.7         &                200 &  4.6 $\times 10^{14}$ &                7.5 &                 -45\\
              &    0.7         &                200 &  7.6 $\times 10^{14}$ &                8.6 &                 -24\\
              &    0.7         &                200 &  2.4 $\times 10^{14}$ &                3.3 &                  -7\\
              &    0.7         &                200 &  4.9 $\times 10^{14}$ &                5.7 &                   3\\
              &    0.7         &                200 &  1.6 $\times 10^{14}$ &                3.3 &                  16\\
              &    0.7         &                200 &  4.9 $\times 10^{14}$ &                3.6 &                  24\\
              &    0.7         &                200 &  2.5 $\times 10^{14}$ &                2.4 &                  32\\
              &    0.7         &                200 &  6.9 $\times 10^{14}$ &                4.1 &                  42\\
              &    0.7         &                200 &  2.0 $\times 10^{14}$ &                2.5 &                  50\\
              &    0.7         &                200 &  9.8 $\times 10^{14}$ &                8.8 &                  77\\
              &    0.7         &                200 &  3.8 $\times 10^{14}$ &                7.0 &                 102\\
              &    0.7         &                200 &  4.6 $\times 10^{14}$ &                8.0 &                 118\\
              &    0.7         &                200 &  9.4 $\times 10^{14}$ &               10.8 &                 142\\
              &    0.7         &                200 &  1.4 $\times 10^{14}$ &                2.4 &                 157\\
              &    0.7         &                200 &  9.4 $\times 10^{13}$ &                2.2 &                 161\\
C$^{33}$S     &    0.7         &                200 &  1.0 $\times 10^{14}$ &                4.7 &                -173\\
              &    0.7         &                200 &  1.2 $\times 10^{14}$ &                5.6 &                -121\\
              &    0.7         &                200 &  1.4 $\times 10^{14}$ &                3.8 &                 -15\\
              &    0.7         &                200 &  1.1 $\times 10^{14}$ &               12.7 &                 126\\
C$^{34}$S     &    0.7         &                200 &  8.2 $\times 10^{13}$ &                6.2 &                -165\\
              &    0.7         &                200 &  1.0 $\times 10^{14}$ &                3.2 &                -142\\
              &    0.7         &                200 &  7.2 $\times 10^{13}$ &                6.4 &                 -37\\
              &    0.7         &                200 & 10.0 $\times 10^{13}$ &                3.7 &                 160\\
HCN           &    0.7         &                200 &  3.7 $\times 10^{13}$ &                3.2 &                  37\\
              &    0.7         &                200 &  1.0 $\times 10^{14}$ &                2.9 &                  32\\
              &    0.7         &                200 &  2.8 $\times 10^{13}$ &                2.0 &                  26\\
              &    0.7         &                200 &  2.9 $\times 10^{13}$ &                2.0 &                  17\\
              &    0.7         &                200 &  1.1 $\times 10^{13}$ &                2.0 &                  -3\\
H$^{13}$CN    &    0.7         &                200 &  1.6 $\times 10^{14}$ &                4.7 &                 125\\
              &    0.7         &                200 &  6.5 $\times 10^{13}$ &                2.2 &                  72\\
              &    0.7         &                200 &  4.1 $\times 10^{13}$ &                2.8 &                 -56\\
              &    0.7         &                200 &  9.3 $\times 10^{13}$ &                0.9 &                 -60\\
              &    0.7         &                200 &  1.8 $\times 10^{13}$ &                1.5 &                 -65\\
              &    0.7         &                200 &  3.6 $\times 10^{13}$ &                2.3 &                 -73\\
              &    0.7         &                200 &  4.1 $\times 10^{13}$ &                1.7 &                -101\\
              &    0.7         &                200 &  1.1 $\times 10^{13}$ &                1.4 &                -109\\
              &    0.7         &                200 &  3.5 $\times 10^{13}$ &                2.3 &                -159\\
              &    0.7         &                200 &  4.1 $\times 10^{12}$ &                1.0 &                -179\\
HNC           &    0.7         &                200 &  2.4 $\times 10^{13}$ &                3.0 &                  85\\
              &    0.7         &                200 &  8.7 $\times 10^{13}$ &                2.2 &                  74\\
              &    0.7         &                200 &  3.0 $\times 10^{12}$ &                0.1 &                  70\\
              &    0.7         &                200 &  2.7 $\times 10^{13}$ &                1.2 &                  66\\
              &    0.7         &                200 &  5.6 $\times 10^{13}$ &                2.3 &                  58\\
              &    0.7         &                200 &  7.7 $\times 10^{12}$ &                1.4 &                   6\\
              &    0.7         &                200 &  6.8 $\times 10^{13}$ &                3.2 &                   0\\
              &    0.7         &                200 &  4.0 $\times 10^{12}$ &                1.0 &                -160\\
H$^{13}$CO$^+$ &    0.7         &                200 &  2.3 $\times 10^{13}$ &                2.1 &                  56\\
HO$^{13}$C$^+$ &    0.7         &                200 &  5.0 $\times 10^{12}$ &                0.8 &                 161\\
              &    0.7         &                200 &  1.4 $\times 10^{13}$ &                2.0 &                 139\\
              &    0.7         &                200 &  3.7 $\times 10^{12}$ &                1.3 &                 129\\
              &    0.7         &                200 &  5.6 $\times 10^{12}$ &                1.0 &                  80\\
              &    0.7         &                200 &  5.1 $\times 10^{12}$ &                1.1 &                  50\\
              &    0.7         &                200 &  6.0 $\times 10^{12}$ &                1.1 &                  38\\
              &    0.7         &                200 &  5.4 $\times 10^{12}$ &                1.2 &                  -2\\
              &    0.7         &                200 &  5.0 $\times 10^{12}$ &                2.0 &                 -69\\
              &    0.7         &                200 &  9.9 $\times 10^{12}$ &                0.3 &                 -75\\
              &    0.7         &                200 &  3.1 $\times 10^{12}$ &                1.0 &                -143\\
\end{supertabular}\\
\vspace{1cm}

%---------------------------------------
% Envelope Components

\tablefirsthead{%
\hline
\hline
Molecule      & $\theta^{m,c}$ & T$_{\rm ex}^{m,c}$ & N$_{\rm tot}^{m,c}$   & $\Delta$ v$^{m,c}$ & v$_{\rm LSR}^{m,c}$\\
              & ($\arcsec$)    & (K)                & (cm$^{-2}$)           & (km~s$^{-1}$)      & (km~s$^{-1}$)      \\
\hline
}

\tablehead{%
\multicolumn{6}{c}{(Continued)}\\
\hline
\hline
Molecule      & $\theta^{m,c}$ & T$_{\rm ex}^{m,c}$ & N$_{\rm tot}^{m,c}$   & $\Delta$ v$^{m,c}$ & v$_{\rm LSR}^{m,c}$\\
              & ($\arcsec$)    & (K)                & (cm$^{-2}$)           & (km~s$^{-1}$)      & (km~s$^{-1}$)      \\
\hline
}

\tabletail{%
\hline
\hline
}

\topcaption{LTE Parameters for the full LTE model (Envelope Components) for source A27 in Sgr~B2(M).}
\tiny
\centering
\begin{supertabular}{lcccC{1cm}C{1cm}}\label{EnvLTE:parameters:A27SgrB2M}\\
CH$_3$CN      & ext.           &                 11 &  9.9 $\times 10^{14}$ &                5.0 &                 114\\
              & ext.           &                 11 &  9.6 $\times 10^{14}$ &                4.1 &                  84\\
              & ext.           &                 13 &  5.1 $\times 10^{14}$ &                6.6 &                  64\\
CN            & ext.           &                  3 &  2.5 $\times 10^{12}$ &                6.1 &                  66\\
              & ext.           &                 54 &  1.7 $\times 10^{12}$ &               20.4 &                  61\\
              & ext.           &                 12 &  1.3 $\times 10^{13}$ &                9.0 &                  23\\
              & ext.           &                 23 &  1.8 $\times 10^{13}$ &               15.1 &                  10\\
\end{supertabular}\\
\vspace{1cm}

\subsection{Sgr~B2(N)}\label{app:subsec:ParametersN}
%
% Please add the following statements to the latex header:
% \usepackage{supertabular,booktabs}
% \usepackage{array}
% \newcolumntype{C}[1]{>{\centering\arraybackslash}p{#1}}

%================================================================================
%
% Source A01

%---------------------------------------
% Core Components

\tablefirsthead{%
\hline
\hline
Molecule      & $\theta^{m,c}$ & T$_{\rm ex}^{m,c}$ & N$_{\rm tot}^{m,c}$   & $\Delta$ v$^{m,c}$ & v$_{\rm LSR}^{m,c}$\\
              & ($\arcsec$)    & (K)                & (cm$^{-2}$)           & (km~s$^{-1}$)      & (km~s$^{-1}$)      \\
\hline
}

\tablehead{%
\multicolumn{6}{c}{(Continued)}\\
\hline
\hline
Molecule      & $\theta^{m,c}$ & T$_{\rm ex}^{m,c}$ & N$_{\rm tot}^{m,c}$   & $\Delta$ v$^{m,c}$ & v$_{\rm LSR}^{m,c}$\\
              & ($\arcsec$)    & (K)                & (cm$^{-2}$)           & (km~s$^{-1}$)      & (km~s$^{-1}$)      \\
\hline
}

\tabletail{%
\hline
\hline
}

\topcaption{LTE Parameters for the full LTE model (Core Components) for source A01 in Sgr~B2(N).}
\tiny
\centering
\begin{supertabular}{lcccC{1cm}C{1cm}}\label{CoreLTE:parameters:A01SgrB2N}\\
RRL-H         &    4.6         &               9882 &  3.7 $\times 10^{8}$  &               16.3 &                  57\\
NO            &    4.6         &                129 &  5.7 $\times 10^{18}$ &               11.4 &                  62\\
NH$_2$D       &    4.6         &                199 &  1.7 $\times 10^{18}$ &               10.7 &                  52\\
C$_2$H$_5$CN, v$_{20}$=1 &    4.6         &                276 &  2.3 $\times 10^{18}$ &               10.7 &                  61\\
NS            &    4.6         &                275 &  9.1 $\times 10^{16}$ &                6.0 &                  60\\
              &    4.6         &                359 &  9.3 $\times 10^{12}$ &                5.1 &                  69\\
CH$_3$OCHO    &    4.6         &                299 &  1.1 $\times 10^{18}$ &                6.2 &                  62\\
\end{supertabular}\\
\vspace{1cm}

%---------------------------------------
% Envelope Components

\tablefirsthead{%
\hline
\hline
Molecule      & $\theta^{m,c}$ & T$_{\rm ex}^{m,c}$ & N$_{\rm tot}^{m,c}$   & $\Delta$ v$^{m,c}$ & v$_{\rm LSR}^{m,c}$\\
              & ($\arcsec$)    & (K)                & (cm$^{-2}$)           & (km~s$^{-1}$)      & (km~s$^{-1}$)      \\
\hline
}

\tablehead{%
\multicolumn{6}{c}{(Continued)}\\
\hline
\hline
Molecule      & $\theta^{m,c}$ & T$_{\rm ex}^{m,c}$ & N$_{\rm tot}^{m,c}$   & $\Delta$ v$^{m,c}$ & v$_{\rm LSR}^{m,c}$\\
              & ($\arcsec$)    & (K)                & (cm$^{-2}$)           & (km~s$^{-1}$)      & (km~s$^{-1}$)      \\
\hline
}

\tabletail{%
\hline
\hline
}

\topcaption{LTE Parameters for the full LTE model (Envelope Components) for source A01 in Sgr~B2(N).}
\tiny
\centering
\begin{supertabular}{lcccC{1cm}C{1cm}}\label{EnvLTE:parameters:A01SgrB2N}\\
CCH           & ext.           &                  3 &  1.0 $\times 10^{16}$ &               14.3 &                  75\\
              & ext.           &                  5 &  4.2 $\times 10^{15}$ &               12.3 &                  64\\
H$_2$C$^{33}$S & ext.           &                 10 &  2.6 $\times 10^{14}$ &                8.6 &                  65\\
H$_2$CNH      & ext.           &                  3 &  2.9 $\times 10^{14}$ &               13.2 &                  61\\
              & ext.           &                  7 &  2.8 $\times 10^{12}$ &                7.1 &                 164\\
CN            & ext.           &                 26 &  7.7 $\times 10^{16}$ &               18.2 &                  67\\
              & ext.           &                 14 &  8.3 $\times 10^{15}$ &                5.4 &                  61\\
              & ext.           &                 54 &  1.3 $\times 10^{12}$ &                6.9 &                   8\\
              & ext.           &                 55 &  1.3 $\times 10^{16}$ &               14.6 &                  30\\
              & ext.           &                 55 &  1.5 $\times 10^{13}$ &                6.8 &                 -35\\
              & ext.           &                 29 &  7.9 $\times 10^{15}$ &               12.5 &                   4\\
              & ext.           &                 69 &  1.6 $\times 10^{16}$ &                8.5 &                 -26\\
              & ext.           &                 23 &  1.5 $\times 10^{15}$ &                4.1 &                 -43\\
              & ext.           &                 93 &  2.2 $\times 10^{17}$ &                8.8 &                 -53\\
              & ext.           &                 38 &  9.0 $\times 10^{15}$ &                1.0 &                 -82\\
              & ext.           &                  3 &  9.7 $\times 10^{14}$ &                0.1 &                -109\\
SiO           & ext.           &                  9 &  1.7 $\times 10^{14}$ &                6.6 &                  70\\
              & ext.           &                 51 &  1.3 $\times 10^{15}$ &                9.5 &                  59\\
PN            & ext.           &                  3 &  3.1 $\times 10^{15}$ &                9.3 &                  54\\
              & ext.           &                  3 &  3.5 $\times 10^{15}$ &               10.9 &                  41\\
              & ext.           &                  3 &  3.0 $\times 10^{15}$ &                9.5 &                   9\\
CH$_2$NH      & ext.           &                 60 &  1.9 $\times 10^{15}$ &                6.2 &                  71\\
              & ext.           &                  4 &  2.4 $\times 10^{14}$ &               15.8 &                  76\\
              & ext.           &                  8 &  1.1 $\times 10^{12}$ &                6.8 &                  63\\
H$_2$CO       & ext.           &                 34 &  5.7 $\times 10^{15}$ &                5.9 &                  82\\
              & ext.           &                 33 &  1.5 $\times 10^{16}$ &                9.8 &                  71\\
              & ext.           &                 14 &  1.1 $\times 10^{15}$ &                9.1 &                  58\\
CO            & ext.           &                  3 &  1.3 $\times 10^{12}$ &                0.1 &                  80\\
              & ext.           &                  3 &  1.3 $\times 10^{12}$ &               18.1 &                  76\\
              & ext.           &                  3 &  3.8 $\times 10^{17}$ &               11.9 &                  77\\
              & ext.           &                  3 &  1.7 $\times 10^{17}$ &               11.2 &                  59\\
              & ext.           &                  3 &  2.1 $\times 10^{14}$ &                0.1 &                  43\\
              & ext.           &                  3 &  1.1 $\times 10^{12}$ &                0.6 &                  49\\
              & ext.           &                  3 &  1.8 $\times 10^{15}$ &                0.7 &                  24\\
              & ext.           &                  3 &  4.0 $\times 10^{17}$ &               25.1 &                  10\\
              & ext.           &                  3 &  1.3 $\times 10^{12}$ &                0.5 &                 -56\\
              & ext.           &                  3 &  2.0 $\times 10^{17}$ &                8.0 &                 -45\\
              & ext.           &                  3 &  1.3 $\times 10^{12}$ &                0.1 &                 -87\\
              & ext.           &                  3 &  1.9 $\times 10^{17}$ &                8.9 &                 -25\\
              & ext.           &                  3 &  1.3 $\times 10^{12}$ &                1.7 &                 -53\\
              & ext.           &                  3 &  1.3 $\times 10^{12}$ &                1.4 &                 -77\\
              & ext.           &                  3 &  1.9 $\times 10^{17}$ &               36.1 &                 -82\\
              & ext.           &                  3 &  1.0 $\times 10^{12}$ &                1.4 &                -113\\
              & ext.           &                  3 &  1.3 $\times 10^{12}$ &                1.4 &                 -59\\
              & ext.           &                  3 &  4.9 $\times 10^{14}$ &                0.9 &                -130\\
              & ext.           &                  3 &  2.5 $\times 10^{17}$ &                3.9 &                -108\\
              & ext.           &                  3 &  2.3 $\times 10^{16}$ &                1.3 &                -122\\
$^{13}$CO     & ext.           &                  3 &  1.1 $\times 10^{15}$ &                1.7 &                   0\\
              & ext.           &                  3 &  7.5 $\times 10^{16}$ &               21.3 &                   7\\
              & ext.           &                  3 &  4.3 $\times 10^{15}$ &                2.2 &                  31\\
              & ext.           &                  3 &  2.1 $\times 10^{17}$ &               22.3 &                  68\\
C$^{18}$O     & ext.           &                  3 &  4.0 $\times 10^{16}$ &               13.6 &                  65\\
CS            & ext.           &                  3 &  4.6 $\times 10^{17}$ &               19.2 &                  69\\
              & ext.           &                  3 &  1.3 $\times 10^{12}$ &                0.8 &                  64\\
              & ext.           &                  3 &  2.7 $\times 10^{14}$ &                0.3 &                  59\\
              & ext.           &                  3 &  1.3 $\times 10^{12}$ &                3.5 &                -171\\
$^{13}$CS     & ext.           &                  3 &  1.4 $\times 10^{16}$ &               13.2 &                -164\\
              & ext.           &                  3 &  3.8 $\times 10^{12}$ &               24.8 &                   9\\
              & ext.           &                  3 &  9.3 $\times 10^{15}$ &                2.8 &                  56\\
              & ext.           &                  3 &  9.9 $\times 10^{15}$ &                6.6 &                  70\\
C$^{33}$S     & ext.           &                  3 &  1.1 $\times 10^{16}$ &                8.7 &                -173\\
C$^{34}$S     & ext.           &                  3 &  5.3 $\times 10^{15}$ &                6.8 &                 -25\\
              & ext.           &                  3 &  1.4 $\times 10^{16}$ &                8.1 &                  23\\
              & ext.           &                  3 &  2.1 $\times 10^{16}$ &               10.5 &                  57\\
              & ext.           &                  3 &  1.9 $\times 10^{16}$ &                7.1 &                  70\\
              & ext.           &                  3 &  1.3 $\times 10^{16}$ &               13.8 &                 112\\
HCN           & ext.           &                  3 &  3.4 $\times 10^{17}$ &               15.5 &                  67\\
              & ext.           &                  3 &  1.3 $\times 10^{12}$ &                0.8 &                  66\\
              & ext.           &                  3 &  1.3 $\times 10^{12}$ &                0.8 &                  62\\
              & ext.           &                  3 &  1.3 $\times 10^{12}$ &                0.3 &                  55\\
              & ext.           &                  3 &  1.3 $\times 10^{12}$ &                0.2 &                  47\\
              & ext.           &                  3 &  1.2 $\times 10^{15}$ &               18.0 &                   7\\
              & ext.           &                  3 &  3.7 $\times 10^{13}$ &                4.8 &                 -23\\
              & ext.           &                  3 &  3.2 $\times 10^{14}$ &                5.5 &                 -75\\
              & ext.           &                  3 &  1.0 $\times 10^{12}$ &               13.1 &                -134\\
              & ext.           &                  3 &  2.0 $\times 10^{12}$ &                0.4 &                 -28\\
H$^{13}$CN    & ext.           &                  3 &  2.2 $\times 10^{14}$ &               24.6 &                -111\\
              & ext.           &                  3 &  2.5 $\times 10^{15}$ &               22.1 &                  66\\
              & ext.           &                  3 &  1.8 $\times 10^{12}$ &                5.7 &                  74\\
HNC           & ext.           &                  3 &  1.6 $\times 10^{15}$ &                6.1 &                  83\\
              & ext.           &                  3 &  7.1 $\times 10^{15}$ &               12.4 &                  68\\
              & ext.           &                  3 &  1.3 $\times 10^{12}$ &                0.1 &                  70\\
              & ext.           &                  3 &  1.2 $\times 10^{12}$ &               17.9 &                  63\\
              & ext.           &                  3 &  4.0 $\times 10^{15}$ &                3.8 &                  54\\
              & ext.           &                  3 &  1.3 $\times 10^{14}$ &               11.1 &                 -76\\
              & ext.           &                  3 &  1.2 $\times 10^{12}$ &                0.5 &                -128\\
              & ext.           &                  3 &  1.3 $\times 10^{12}$ &                0.6 &                -115\\
              & ext.           &                  3 &  1.4 $\times 10^{12}$ &                0.8 &                -145\\
HN$^{13}$C    & ext.           &                  3 &  2.6 $\times 10^{14}$ &               11.8 &                  65\\
              & ext.           &                  3 &  1.6 $\times 10^{14}$ &               35.0 &                 -85\\
H$^{13}$CO$^+$ & ext.           &                  3 &  3.0 $\times 10^{14}$ &               16.2 &                  69\\
              & ext.           &                  3 &  1.8 $\times 10^{14}$ &               12.7 &                  64\\
\end{supertabular}\\
\vspace{1cm}

%================================================================================
%
% Source A02

%---------------------------------------
% Core Components

\tablefirsthead{%
\hline
\hline
Molecule      & $\theta^{m,c}$ & T$_{\rm ex}^{m,c}$ & N$_{\rm tot}^{m,c}$   & $\Delta$ v$^{m,c}$ & v$_{\rm LSR}^{m,c}$\\
              & ($\arcsec$)    & (K)                & (cm$^{-2}$)           & (km~s$^{-1}$)      & (km~s$^{-1}$)      \\
\hline
}

\tablehead{%
\multicolumn{6}{c}{(Continued)}\\
\hline
\hline
Molecule      & $\theta^{m,c}$ & T$_{\rm ex}^{m,c}$ & N$_{\rm tot}^{m,c}$   & $\Delta$ v$^{m,c}$ & v$_{\rm LSR}^{m,c}$\\
              & ($\arcsec$)    & (K)                & (cm$^{-2}$)           & (km~s$^{-1}$)      & (km~s$^{-1}$)      \\
\hline
}

\tabletail{%
\hline
\hline
}

\topcaption{LTE Parameters for the full LTE model (Core Components) for source A02 in Sgr~B2(N).}
\tiny
\centering
\begin{supertabular}{lcccC{1cm}C{1cm}}\label{CoreLTE:parameters:A02SgrB2N}\\
CH$_3$NH$_2$  &    1.7         &                138 &  6.9 $\times 10^{17}$ &                6.9 &                  71\\
C$_2$H$_3$CN  &    1.7         &                190 &  1.2 $\times 10^{17}$ &                5.8 &                  72\\
C$_2$H$_3$CN, v$_{10}$=1 &    1.7         &                396 &  9.5 $\times 10^{16}$ &                9.0 &                  72\\
SO$_2$        &    1.7         &                400 &  1.3 $\times 10^{17}$ &                7.0 &                  71\\
CH$_3$CHO     &    1.7         &                184 &  4.1 $\times 10^{17}$ &                5.0 &                  72\\
H$_2$C$^{33}$S &    1.7         &                200 &  1.3 $\times 10^{16}$ &                6.7 &                  66\\
C$_2$H$_5$OH  &    1.7         &                173 &  2.1 $\times 10^{18}$ &                4.6 &                  72\\
\end{supertabular}\\
\vspace{1cm}

%---------------------------------------
% Envelope Components

\tablefirsthead{%
\hline
\hline
Molecule      & $\theta^{m,c}$ & T$_{\rm ex}^{m,c}$ & N$_{\rm tot}^{m,c}$   & $\Delta$ v$^{m,c}$ & v$_{\rm LSR}^{m,c}$\\
              & ($\arcsec$)    & (K)                & (cm$^{-2}$)           & (km~s$^{-1}$)      & (km~s$^{-1}$)      \\
\hline
}

\tablehead{%
\multicolumn{6}{c}{(Continued)}\\
\hline
\hline
Molecule      & $\theta^{m,c}$ & T$_{\rm ex}^{m,c}$ & N$_{\rm tot}^{m,c}$   & $\Delta$ v$^{m,c}$ & v$_{\rm LSR}^{m,c}$\\
              & ($\arcsec$)    & (K)                & (cm$^{-2}$)           & (km~s$^{-1}$)      & (km~s$^{-1}$)      \\
\hline
}

\tabletail{%
\hline
\hline
}

\topcaption{LTE Parameters for the full LTE model (Envelope Components) for source A02 in Sgr~B2(N).}
\tiny
\centering
\begin{supertabular}{lcccC{1cm}C{1cm}}\label{EnvLTE:parameters:A02SgrB2N}\\
CN            & ext.           &                 17 &  1.5 $\times 10^{16}$ &               12.6 &                  76\\
              & ext.           &                 21 &  5.8 $\times 10^{16}$ &                8.7 &                  61\\
              & ext.           &                  3 &  3.5 $\times 10^{14}$ &                5.5 &                  13\\
              & ext.           &                  3 &  3.1 $\times 10^{14}$ &                6.5 &                  26\\
              & ext.           &                  3 &  1.9 $\times 10^{14}$ &                2.8 &                 -28\\
              & ext.           &                  3 &  1.0 $\times 10^{15}$ &                9.6 &                   2\\
              & ext.           &                  3 &  8.4 $\times 10^{13}$ &                7.7 &                 -16\\
              & ext.           &                  3 &  2.1 $\times 10^{14}$ &                3.2 &                 -42\\
              & ext.           &                  8 &  3.3 $\times 10^{14}$ &                7.2 &                 -80\\
              & ext.           &                  3 &  1.4 $\times 10^{14}$ &                4.4 &                 -49\\
              & ext.           &                  3 &  4.9 $\times 10^{14}$ &               10.4 &                -109\\
PH$_3$        & ext.           &                  8 &  2.8 $\times 10^{14}$ &                8.2 &                  67\\
CCH           & ext.           &                  4 &  4.3 $\times 10^{15}$ &                3.0 &                  73\\
              & ext.           &                  4 &  6.4 $\times 10^{15}$ &               13.4 &                  61\\
H$_2$C$^{33}$S & ext.           &                 82 &  2.1 $\times 10^{16}$ &                6.9 &                  56\\
PN            & ext.           &                  3 &  6.1 $\times 10^{15}$ &                0.8 &                 107\\
              & ext.           &                  3 &  2.5 $\times 10^{15}$ &                7.7 &                  64\\
              & ext.           &                  3 &  2.8 $\times 10^{15}$ &                2.4 &                  53\\
              & ext.           &                  3 &  4.5 $\times 10^{12}$ &                0.5 &                   5\\
              & ext.           &                  3 & 10.0 $\times 10^{15}$ &                8.2 &                 -87\\
CH$_2$NH      & ext.           &                  9 &  5.2 $\times 10^{14}$ &                4.1 &                  63\\
              & ext.           &                  3 &  2.0 $\times 10^{14}$ &               15.1 &                  71\\
H$_2$CNH      & ext.           &                  4 &  1.2 $\times 10^{14}$ &                5.6 &                  83\\
              & ext.           &                  9 &  3.5 $\times 10^{12}$ &                6.9 &                -157\\
              & ext.           &                  5 &  3.4 $\times 10^{14}$ &                6.3 &                  57\\
H$_2$CO       & ext.           &                 11 &  3.0 $\times 10^{15}$ &               24.7 &                  69\\
SiO           & ext.           &                  9 &  3.6 $\times 10^{14}$ &                6.6 &                  70\\
              & ext.           &                 51 &  2.6 $\times 10^{14}$ &                9.5 &                  59\\
CO            & ext.           &                  3 &  3.2 $\times 10^{16}$ &               13.8 &                 107\\
              & ext.           &                  3 &  4.9 $\times 10^{17}$ &                0.7 &                  82\\
              & ext.           &                  3 &  1.1 $\times 10^{12}$ &                0.6 &                  78\\
              & ext.           &                  3 &  4.0 $\times 10^{17}$ &               30.2 &                  68\\
              & ext.           &                  3 &  6.5 $\times 10^{13}$ &                1.6 &                  65\\
              & ext.           &                  3 &  1.3 $\times 10^{12}$ &                0.7 &                  59\\
              & ext.           &                  3 &  2.6 $\times 10^{17}$ &                1.4 &                  50\\
              & ext.           &                  3 &  6.5 $\times 10^{16}$ &                6.2 &                  28\\
              & ext.           &                  3 &  5.0 $\times 10^{16}$ &                5.3 &                  17\\
              & ext.           &                  3 &  5.8 $\times 10^{16}$ &                6.9 &                   8\\
              & ext.           &                  3 &  6.0 $\times 10^{16}$ &                6.4 &                  -1\\
              & ext.           &                  3 &  1.3 $\times 10^{17}$ &               13.6 &                 -17\\
              & ext.           &                  3 &  4.2 $\times 10^{16}$ &                3.7 &                 -28\\
              & ext.           &                  3 &  1.6 $\times 10^{14}$ &               12.6 &                 -56\\
              & ext.           &                  3 &  1.6 $\times 10^{17}$ &                9.1 &                 -45\\
              & ext.           &                  3 &  6.9 $\times 10^{15}$ &                2.8 &                 -60\\
              & ext.           &                  3 &  2.3 $\times 10^{16}$ &                1.8 &                 -69\\
              & ext.           &                  3 &  2.6 $\times 10^{12}$ &                1.8 &                 -87\\
              & ext.           &                  3 &  2.1 $\times 10^{17}$ &                7.8 &                 -77\\
              & ext.           &                  3 &  8.4 $\times 10^{16}$ &                2.5 &                 -93\\
              & ext.           &                  3 &  1.4 $\times 10^{17}$ &                9.9 &                -107\\
              & ext.           &                  3 &  8.4 $\times 10^{15}$ &                3.4 &                -126\\
$^{13}$CO     & ext.           &                  3 &  5.9 $\times 10^{16}$ &               11.8 &                  78\\
              & ext.           &                  3 &  1.1 $\times 10^{17}$ &               12.0 &                  62\\
              & ext.           &                  3 &  2.7 $\times 10^{13}$ &                0.8 &                  30\\
              & ext.           &                  3 &  1.6 $\times 10^{16}$ &                7.3 &                 -82\\
              & ext.           &                  3 &  1.5 $\times 10^{13}$ &                0.7 &                 -80\\
              & ext.           &                  3 &  1.1 $\times 10^{12}$ &                0.9 &                -146\\
C$^{17}$O     & ext.           &                  3 &  7.3 $\times 10^{14}$ &                2.6 &                -177\\
C$^{18}$O     & ext.           &                  3 &  2.8 $\times 10^{16}$ &                7.9 &                  62\\
              & ext.           &                  3 &  1.6 $\times 10^{16}$ &                8.7 &                  78\\
              & ext.           &                  3 &  2.2 $\times 10^{15}$ &               11.9 &                 112\\
CS            & ext.           &                  3 &  2.9 $\times 10^{15}$ &                1.2 &                 114\\
              & ext.           &                  3 &  5.5 $\times 10^{16}$ &                6.9 &                  78\\
              & ext.           &                  3 &  5.9 $\times 10^{16}$ &                6.3 &                  71\\
              & ext.           &                  3 &  6.1 $\times 10^{19}$ &                0.1 &                  61\\
              & ext.           &                  3 &  1.4 $\times 10^{17}$ &               11.1 &                  58\\
              & ext.           &                  3 &  4.5 $\times 10^{12}$ &                0.8 &                  30\\
              & ext.           &                  3 &  5.6 $\times 10^{15}$ &                2.2 &                  -1\\
              & ext.           &                  3 &  2.3 $\times 10^{16}$ &               14.4 &                -104\\
              & ext.           &                  3 &  1.8 $\times 10^{12}$ &                0.9 &                 139\\
$^{13}$CS     & ext.           &                  3 &  5.9 $\times 10^{15}$ &                6.0 &                  30\\
              & ext.           &                  3 &  1.1 $\times 10^{12}$ &               19.1 &                  99\\
C$^{33}$S     & ext.           &                  3 &  1.6 $\times 10^{16}$ &                8.3 &                -156\\
              & ext.           &                  3 &  1.3 $\times 10^{14}$ &                2.5 &                -126\\
              & ext.           &                  3 &  4.8 $\times 10^{15}$ &                4.5 &                -118\\
              & ext.           &                  3 &  2.3 $\times 10^{15}$ &                3.5 &                 -71\\
              & ext.           &                  3 &  1.2 $\times 10^{15}$ &                3.5 &                 -17\\
              & ext.           &                  3 &  9.0 $\times 10^{14}$ &                2.5 &                 134\\
C$^{34}$S     & ext.           &                  3 &  3.0 $\times 10^{16}$ &                7.3 &                  33\\
              & ext.           &                  3 &  1.7 $\times 10^{16}$ &                5.8 &                  62\\
HCN           & ext.           &                  3 &  3.4 $\times 10^{13}$ &                1.1 &                  99\\
              & ext.           &                  3 &  1.1 $\times 10^{12}$ &                0.3 &                  91\\
              & ext.           &                  3 & 10.0 $\times 10^{14}$ &               10.5 &                  86\\
              & ext.           &                  3 &  1.2 $\times 10^{15}$ &                0.4 &                  79\\
              & ext.           &                  3 &  6.7 $\times 10^{13}$ &                0.4 &                  75\\
              & ext.           &                  3 &  3.0 $\times 10^{15}$ &               24.2 &                  61\\
              & ext.           &                  3 &  1.1 $\times 10^{12}$ &                0.9 &                  60\\
              & ext.           &                  3 &  5.0 $\times 10^{13}$ &                0.1 &                  54\\
              & ext.           &                  3 &  1.2 $\times 10^{15}$ &                1.6 &                  48\\
              & ext.           &                  3 &  2.4 $\times 10^{14}$ &                4.6 &                  26\\
              & ext.           &                  3 &  2.1 $\times 10^{14}$ &                1.5 &                  12\\
              & ext.           &                  3 &  1.2 $\times 10^{15}$ &               18.7 &                   5\\
              & ext.           &                  3 &  4.2 $\times 10^{13}$ &                0.2 &                  -8\\
              & ext.           &                  3 &  3.7 $\times 10^{13}$ &                0.7 &                 -37\\
              & ext.           &                  3 &  8.7 $\times 10^{13}$ &                2.0 &                 -69\\
              & ext.           &                  3 &  9.2 $\times 10^{13}$ &                2.4 &                 -92\\
              & ext.           &                  3 &  3.6 $\times 10^{13}$ &                0.9 &                -105\\
              & ext.           &                  3 &  1.7 $\times 10^{12}$ &                0.8 &                -111\\
H$^{13}$CN    & ext.           &                  3 &  1.2 $\times 10^{12}$ &                0.4 &                 104\\
              & ext.           &                  3 &  2.2 $\times 10^{15}$ &               30.0 &                  79\\
              & ext.           &                  3 &  7.4 $\times 10^{14}$ &                9.9 &                  59\\
              & ext.           &                  3 &  1.1 $\times 10^{12}$ &                0.1 &                  56\\
              & ext.           &                  3 &  1.2 $\times 10^{15}$ &               25.1 &                  21\\
              & ext.           &                  3 &  1.6 $\times 10^{14}$ &               18.1 &                 -60\\
              & ext.           &                  3 &  3.6 $\times 10^{13}$ &                1.1 &                 -97\\
              & ext.           &                  3 &  1.0 $\times 10^{12}$ &                3.9 &                -126\\
HNC           & ext.           &                  3 &  4.0 $\times 10^{14}$ &               10.3 &                 149\\
              & ext.           &                  3 &  1.9 $\times 10^{16}$ &               10.3 &                  77\\
              & ext.           &                  3 &  4.3 $\times 10^{12}$ &                0.5 &                  76\\
              & ext.           &                  3 &  2.8 $\times 10^{12}$ &                3.5 &                  72\\
              & ext.           &                  3 &  2.0 $\times 10^{15}$ &                2.1 &                  63\\
              & ext.           &                  3 &  1.6 $\times 10^{16}$ &                9.1 &                  59\\
              & ext.           &                  3 &  3.0 $\times 10^{14}$ &                0.5 &                  43\\
              & ext.           &                  3 & 10.0 $\times 10^{14}$ &               14.5 &                  -9\\
              & ext.           &                  3 &  3.5 $\times 10^{13}$ &                1.6 &                 -41\\
              & ext.           &                  3 &  1.1 $\times 10^{12}$ &                0.6 &                -115\\
              & ext.           &                  3 &  2.2 $\times 10^{13}$ &                1.9 &                 -92\\
              & ext.           &                  3 &  1.1 $\times 10^{12}$ &                0.6 &                -160\\
              & ext.           &                  3 &  1.1 $\times 10^{12}$ &                0.3 &                -185\\
              & ext.           &                  3 &  8.9 $\times 10^{13}$ &                4.0 &                -186\\
HN$^{13}$C    & ext.           &                  3 &  7.0 $\times 10^{13}$ &                5.7 &                 -86\\
              & ext.           &                  3 &  5.4 $\times 10^{13}$ &                5.3 &                 -21\\
              & ext.           &                  3 &  1.6 $\times 10^{14}$ &               11.4 &                  37\\
              & ext.           &                  3 &  2.9 $\times 10^{14}$ &                7.3 &                  77\\
H$^{13}$CO$^+$ & ext.           &                  3 &  3.5 $\times 10^{14}$ &                4.3 &                  72\\
              & ext.           &                  3 &  4.0 $\times 10^{14}$ &               28.5 &                  64\\
HO$^{13}$C$^+$ & ext.           &                  3 &  1.2 $\times 10^{14}$ &                0.6 &                  49\\
              & ext.           &                  3 &  3.1 $\times 10^{13}$ &                0.7 &                   2\\
\end{supertabular}\\
\vspace{1cm}

%================================================================================
%
% Source A03

%---------------------------------------
% Core Components

\tablefirsthead{%
\hline
\hline
Molecule      & $\theta^{m,c}$ & T$_{\rm ex}^{m,c}$ & N$_{\rm tot}^{m,c}$   & $\Delta$ v$^{m,c}$ & v$_{\rm LSR}^{m,c}$\\
              & ($\arcsec$)    & (K)                & (cm$^{-2}$)           & (km~s$^{-1}$)      & (km~s$^{-1}$)      \\
\hline
}

\tablehead{%
\multicolumn{6}{c}{(Continued)}\\
\hline
\hline
Molecule      & $\theta^{m,c}$ & T$_{\rm ex}^{m,c}$ & N$_{\rm tot}^{m,c}$   & $\Delta$ v$^{m,c}$ & v$_{\rm LSR}^{m,c}$\\
              & ($\arcsec$)    & (K)                & (cm$^{-2}$)           & (km~s$^{-1}$)      & (km~s$^{-1}$)      \\
\hline
}

\tabletail{%
\hline
\hline
}

\topcaption{LTE Parameters for the full LTE model (Core Components) for source A03 in Sgr~B2(N).}
\tiny
\centering
\begin{supertabular}{lcccC{1cm}C{1cm}}\label{CoreLTE:parameters:A03SgrB2N}\\
CH$_3$NH$_2$  &    1.6         &                181 &  7.8 $\times 10^{17}$ &                7.3 &                  72\\
H$_2 \! ^{33}$S &    1.6         &                200 &  1.6 $\times 10^{16}$ &                7.6 &                  71\\
              &    1.6         &                 52 &  2.4 $\times 10^{13}$ &                0.4 &                  88\\
NH$_2$D       &    1.6         &                149 &  3.2 $\times 10^{18}$ &                6.8 &                  66\\
CH$_3$OCHO    &    1.6         &                226 &  1.3 $\times 10^{18}$ &                5.6 &                  72\\
CH$_3$CHO     &    1.6         &                201 &  5.0 $\times 10^{17}$ &                5.7 &                  72\\
C$_2$H$_5$OH  &    1.6         &                188 &  2.2 $\times 10^{18}$ &                5.1 &                  72\\
\end{supertabular}\\
\vspace{1cm}

%---------------------------------------
% Envelope Components

\tablefirsthead{%
\hline
\hline
Molecule      & $\theta^{m,c}$ & T$_{\rm ex}^{m,c}$ & N$_{\rm tot}^{m,c}$   & $\Delta$ v$^{m,c}$ & v$_{\rm LSR}^{m,c}$\\
              & ($\arcsec$)    & (K)                & (cm$^{-2}$)           & (km~s$^{-1}$)      & (km~s$^{-1}$)      \\
\hline
}

\tablehead{%
\multicolumn{6}{c}{(Continued)}\\
\hline
\hline
Molecule      & $\theta^{m,c}$ & T$_{\rm ex}^{m,c}$ & N$_{\rm tot}^{m,c}$   & $\Delta$ v$^{m,c}$ & v$_{\rm LSR}^{m,c}$\\
              & ($\arcsec$)    & (K)                & (cm$^{-2}$)           & (km~s$^{-1}$)      & (km~s$^{-1}$)      \\
\hline
}

\tabletail{%
\hline
\hline
}

\topcaption{LTE Parameters for the full LTE model (Envelope Components) for source A03 in Sgr~B2(N).}
\tiny
\centering
\begin{supertabular}{lcccC{1cm}C{1cm}}\label{EnvLTE:parameters:A03SgrB2N}\\
CN            & ext.           &                  8 &  4.7 $\times 10^{15}$ &               15.8 &                  73\\
              & ext.           &                 18 &  4.1 $\times 10^{16}$ &                9.5 &                  60\\
              & ext.           &                  3 &  3.3 $\times 10^{14}$ &                6.9 &                  12\\
              & ext.           &                  3 &  3.1 $\times 10^{14}$ &                5.9 &                  25\\
              & ext.           &                  4 &  1.1 $\times 10^{14}$ &                4.2 &                 -28\\
              & ext.           &                  3 &  1.5 $\times 10^{15}$ &                8.8 &                   2\\
              & ext.           &                  3 &  5.1 $\times 10^{13}$ &                6.1 &                 -24\\
              & ext.           &                  7 &  4.5 $\times 10^{14}$ &                9.6 &                 -47\\
              & ext.           &                  9 &  2.8 $\times 10^{13}$ &                6.0 &                 -14\\
              & ext.           &                 10 &  1.0 $\times 10^{12}$ &                1.5 &                 -36\\
              & ext.           &                  3 &  6.8 $\times 10^{14}$ &                9.8 &                -109\\
SiO           & ext.           &                 12 &  2.7 $\times 10^{14}$ &                6.1 &                  73\\
CH$_2$NH      & ext.           &                  5 &  2.2 $\times 10^{14}$ &                8.2 &                  82\\
              & ext.           &                  7 &  8.6 $\times 10^{14}$ &               10.0 &                  69\\
              & ext.           &                  8 &  7.2 $\times 10^{14}$ &                7.3 &                  59\\
CCH           & ext.           &                  3 &  3.7 $\times 10^{15}$ &                3.0 &                  72\\
              & ext.           &                  3 &  2.5 $\times 10^{16}$ &               23.4 &                  65\\
H$_2$C$^{33}$S & ext.           &                  9 &  2.2 $\times 10^{14}$ &                4.8 &                  66\\
PH$_3$        & ext.           &                  9 &  2.7 $\times 10^{14}$ &                9.0 &                  64\\
PN            & ext.           &                  3 &  7.2 $\times 10^{15}$ &               13.4 &                  58\\
              & ext.           &                  3 &  5.5 $\times 10^{15}$ &                2.6 &                 -88\\
H$_2$CO       & ext.           &                 40 &  3.7 $\times 10^{15}$ &                0.5 &                 119\\
              & ext.           &                 12 &  2.8 $\times 10^{15}$ &               27.0 &                  68\\
CO            & ext.           &                  3 &  2.6 $\times 10^{16}$ &               11.1 &                 107\\
              & ext.           &                  3 &  8.2 $\times 10^{16}$ &                3.8 &                  85\\
              & ext.           &                  3 &  1.0 $\times 10^{12}$ &                3.1 &                  70\\
              & ext.           &                  3 &  2.2 $\times 10^{17}$ &               34.3 &                  68\\
              & ext.           &                  3 &  2.5 $\times 10^{16}$ &                3.6 &                  64\\
              & ext.           &                  3 &  1.0 $\times 10^{12}$ &                3.1 &                  59\\
              & ext.           &                  3 &  7.6 $\times 10^{16}$ &                8.6 &                  54\\
              & ext.           &                  3 &  4.0 $\times 10^{16}$ &                5.5 &                  29\\
              & ext.           &                  3 &  1.0 $\times 10^{12}$ &                2.0 &                  -3\\
              & ext.           &                  3 &  1.6 $\times 10^{17}$ &               25.2 &                   7\\
              & ext.           &                  3 &  2.8 $\times 10^{12}$ &                0.8 &                 -21\\
              & ext.           &                  3 &  9.5 $\times 10^{16}$ &               13.4 &                 -20\\
              & ext.           &                  3 &  2.3 $\times 10^{16}$ &                2.9 &                 -28\\
              & ext.           &                  3 &  1.2 $\times 10^{16}$ &                2.3 &                 -69\\
              & ext.           &                  3 &  1.0 $\times 10^{17}$ &                8.6 &                 -45\\
              & ext.           &                  3 &  6.0 $\times 10^{12}$ &                2.3 &                 -73\\
              & ext.           &                  3 &  1.0 $\times 10^{12}$ &                2.3 &                 -97\\
              & ext.           &                  3 &  1.0 $\times 10^{15}$ &                2.4 &                 -77\\
              & ext.           &                  3 &  1.0 $\times 10^{12}$ &                2.3 &                -110\\
              & ext.           &                  3 &  1.6 $\times 10^{17}$ &               26.7 &                 -78\\
              & ext.           &                  3 &  1.0 $\times 10^{12}$ &                2.0 &                -138\\
              & ext.           &                  3 &  1.2 $\times 10^{17}$ &                8.7 &                -107\\
              & ext.           &                  3 &  4.8 $\times 10^{15}$ &                2.5 &                -124\\
              & ext.           &                  3 &  1.8 $\times 10^{15}$ &                3.2 &                -172\\
$^{13}$CO     & ext.           &                  3 &  1.0 $\times 10^{12}$ &                9.1 &                 141\\
              & ext.           &                  3 &  1.8 $\times 10^{17}$ &               25.5 &                  67\\
              & ext.           &                  3 &  3.9 $\times 10^{16}$ &               15.0 &                   8\\
              & ext.           &                  3 &  3.8 $\times 10^{16}$ &                9.3 &                 -45\\
              & ext.           &                  3 &  1.4 $\times 10^{14}$ &                0.6 &                -157\\
              & ext.           &                  3 &  1.3 $\times 10^{16}$ &                2.2 &                 -83\\
              & ext.           &                  3 &  5.2 $\times 10^{14}$ &                0.4 &                -118\\
              & ext.           &                  3 &  4.5 $\times 10^{16}$ &               40.0 &                 -99\\
C$^{17}$O     & ext.           &                  3 &  2.6 $\times 10^{15}$ &                2.7 &                -178\\
              & ext.           &                  3 &  2.5 $\times 10^{14}$ &                2.5 &                 -63\\
C$^{18}$O     & ext.           &                  3 &  2.9 $\times 10^{15}$ &                2.7 &                -107\\
              & ext.           &                  3 &  3.4 $\times 10^{15}$ &                1.7 &                 -41\\
              & ext.           &                  3 &  3.7 $\times 10^{16}$ &               14.4 &                  62\\
              & ext.           &                  3 &  9.1 $\times 10^{14}$ &                4.6 &                  74\\
              & ext.           &                  3 &  6.2 $\times 10^{14}$ &                4.1 &                 113\\
CS            & ext.           &                  3 &  1.4 $\times 10^{15}$ &                1.5 &                  72\\
              & ext.           &                  3 &  1.0 $\times 10^{12}$ &                5.1 &                  61\\
              & ext.           &                  3 &  2.8 $\times 10^{17}$ &               26.4 &                  61\\
              & ext.           &                  3 &  3.4 $\times 10^{13}$ &                5.3 &                  27\\
              & ext.           &                  3 &  8.1 $\times 10^{15}$ &                5.3 &                   0\\
              & ext.           &                  3 &  9.7 $\times 10^{15}$ &                5.7 &                -106\\
$^{13}$CS     & ext.           &                  3 &  1.0 $\times 10^{12}$ &               12.5 &                 113\\
              & ext.           &                  3 &  1.0 $\times 10^{12}$ &                0.7 &                  82\\
              & ext.           &                  3 &  1.0 $\times 10^{12}$ &                1.5 &                   4\\
              & ext.           &                  3 &  1.0 $\times 10^{12}$ &               15.8 &                 -88\\
              & ext.           &                  3 &  1.0 $\times 10^{12}$ &                1.6 &                -173\\
C$^{33}$S     & ext.           &                  3 &  3.9 $\times 10^{15}$ &                7.5 &                -155\\
              & ext.           &                  3 &  4.1 $\times 10^{15}$ &               16.7 &                -121\\
              & ext.           &                  3 &  1.0 $\times 10^{13}$ &                3.3 &                 -72\\
              & ext.           &                  3 &  1.5 $\times 10^{15}$ &                2.5 &                  35\\
C$^{34}$S     & ext.           &                  3 &  3.5 $\times 10^{15}$ &                1.3 &                 -15\\
              & ext.           &                  3 &  2.9 $\times 10^{16}$ &                7.7 &                  30\\
              & ext.           &                  3 &  4.2 $\times 10^{13}$ &               18.0 &                 156\\
HCN           & ext.           &                  3 &  1.9 $\times 10^{12}$ &                0.6 &                 153\\
              & ext.           &                  3 &  1.0 $\times 10^{15}$ &                8.5 &                  84\\
              & ext.           &                  3 &  3.6 $\times 10^{15}$ &               28.6 &                  58\\
              & ext.           &                  3 &  1.0 $\times 10^{12}$ &                0.7 &                 -56\\
              & ext.           &                  3 &  5.9 $\times 10^{14}$ &                0.3 &                  49\\
              & ext.           &                  3 &  4.5 $\times 10^{14}$ &                6.1 &                  25\\
              & ext.           &                  3 &  1.6 $\times 10^{14}$ &                2.9 &                  12\\
              & ext.           &                  3 &  2.2 $\times 10^{15}$ &               20.8 &                   1\\
              & ext.           &                  3 &  1.0 $\times 10^{12}$ &                0.3 &                  79\\
              & ext.           &                  3 &  1.0 $\times 10^{12}$ &                0.4 &                -120\\
H$^{13}$CN    & ext.           &                  3 &  1.3 $\times 10^{14}$ &                8.7 &                 142\\
              & ext.           &                  3 &  9.3 $\times 10^{14}$ &               18.0 &                 122\\
              & ext.           &                  3 &  1.1 $\times 10^{14}$ &                3.6 &                  81\\
              & ext.           &                  3 &  2.8 $\times 10^{15}$ &               39.6 &                  72\\
              & ext.           &                  3 &  2.2 $\times 10^{12}$ &                0.7 &                  60\\
              & ext.           &                  3 &  3.3 $\times 10^{13}$ &                0.6 &                  28\\
              & ext.           &                  3 &  6.7 $\times 10^{13}$ &                4.1 &                 -77\\
              & ext.           &                  3 &  4.7 $\times 10^{14}$ &               12.1 &                 -53\\
              & ext.           &                  3 &  3.4 $\times 10^{12}$ &                0.4 &                 -23\\
              & ext.           &                  3 &  1.0 $\times 10^{12}$ &                1.1 &                -103\\
HNC           & ext.           &                  3 &  5.5 $\times 10^{13}$ &                3.6 &                 149\\
              & ext.           &                  3 &  4.3 $\times 10^{14}$ &               29.4 &                 110\\
              & ext.           &                  3 &  1.4 $\times 10^{13}$ &                0.2 &                  93\\
              & ext.           &                  3 &  3.8 $\times 10^{17}$ &                1.0 &                  79\\
              & ext.           &                  3 &  5.8 $\times 10^{14}$ &                1.5 &                  69\\
              & ext.           &                  3 &  1.0 $\times 10^{12}$ &                0.1 &                  67\\
              & ext.           &                  3 &  1.4 $\times 10^{16}$ &               22.6 &                  68\\
              & ext.           &                  3 &  1.0 $\times 10^{12}$ &                1.6 &                  56\\
              & ext.           &                  3 &  1.0 $\times 10^{12}$ &                0.9 &                  52\\
              & ext.           &                  3 &  1.1 $\times 10^{12}$ &                4.9 &                   5\\
              & ext.           &                  3 &  4.2 $\times 10^{12}$ &                1.6 &                 -14\\
              & ext.           &                  3 &  1.2 $\times 10^{14}$ &                2.7 &                   1\\
              & ext.           &                  3 &  1.0 $\times 10^{12}$ &                2.0 &                 -56\\
HN$^{13}$C    & ext.           &                  3 &  2.8 $\times 10^{13}$ &                7.8 &                 -87\\
              & ext.           &                  3 &  5.1 $\times 10^{13}$ &                6.1 &                 -21\\
              & ext.           &                  3 &  1.5 $\times 10^{14}$ &               11.3 &                  37\\
H$^{13}$CO$^+$ & ext.           &                  3 &  1.1 $\times 10^{12}$ &                1.9 &                  62\\
              & ext.           &                  3 &  6.1 $\times 10^{14}$ &               18.3 &                  69\\
HO$^{13}$C$^+$ & ext.           &                  3 &  3.7 $\times 10^{13}$ &                4.9 &                  54\\
              & ext.           &                  3 &  1.0 $\times 10^{12}$ &                0.7 &                -137\\
\end{supertabular}\\
\vspace{1cm}

%================================================================================
%
% Source A04

%---------------------------------------
% Core Components

\tablefirsthead{%
\hline
\hline
Molecule      & $\theta^{m,c}$ & T$_{\rm ex}^{m,c}$ & N$_{\rm tot}^{m,c}$   & $\Delta$ v$^{m,c}$ & v$_{\rm LSR}^{m,c}$\\
              & ($\arcsec$)    & (K)                & (cm$^{-2}$)           & (km~s$^{-1}$)      & (km~s$^{-1}$)      \\
\hline
}

\tablehead{%
\multicolumn{6}{c}{(Continued)}\\
\hline
\hline
Molecule      & $\theta^{m,c}$ & T$_{\rm ex}^{m,c}$ & N$_{\rm tot}^{m,c}$   & $\Delta$ v$^{m,c}$ & v$_{\rm LSR}^{m,c}$\\
              & ($\arcsec$)    & (K)                & (cm$^{-2}$)           & (km~s$^{-1}$)      & (km~s$^{-1}$)      \\
\hline
}

\tabletail{%
\hline
\hline
}

\topcaption{LTE Parameters for the full LTE model (Core Components) for source A04 in Sgr~B2(N).}
\tiny
\centering
\begin{supertabular}{lcccC{1cm}C{1cm}}\label{CoreLTE:parameters:A04SgrB2N}\\
H$_2 \! ^{34}$S &    1.5         &                399 &  1.5 $\times 10^{18}$ &               11.3 &                  70\\
H$_2$CNH      &    1.5         &                189 &  3.9 $\times 10^{17}$ &                5.0 &                  62\\
HNC, v$_2$=1  &    1.5         &                872 &  1.3 $\times 10^{16}$ &                8.6 &                  63\\
HC(O)NH$_2$   &    1.5         &                245 &  5.5 $\times 10^{16}$ &                8.6 &                  63\\
C$_2$H$_3$CN  &    1.5         &                164 &  4.5 $\times 10^{16}$ &                8.9 &                  63\\
C$_2$H$_3$CN, v$_{10}$=1 &    1.5         &                254 &  1.6 $\times 10^{16}$ &               11.1 &                  64\\
C$_2$H$_5$CN  &    1.5         &                236 &  3.8 $\times 10^{17}$ &                4.3 &                  62\\
\end{supertabular}\\
\vspace{1cm}

%---------------------------------------
% Envelope Components

\tablefirsthead{%
\hline
\hline
Molecule      & $\theta^{m,c}$ & T$_{\rm ex}^{m,c}$ & N$_{\rm tot}^{m,c}$   & $\Delta$ v$^{m,c}$ & v$_{\rm LSR}^{m,c}$\\
              & ($\arcsec$)    & (K)                & (cm$^{-2}$)           & (km~s$^{-1}$)      & (km~s$^{-1}$)      \\
\hline
}

\tablehead{%
\multicolumn{6}{c}{(Continued)}\\
\hline
\hline
Molecule      & $\theta^{m,c}$ & T$_{\rm ex}^{m,c}$ & N$_{\rm tot}^{m,c}$   & $\Delta$ v$^{m,c}$ & v$_{\rm LSR}^{m,c}$\\
              & ($\arcsec$)    & (K)                & (cm$^{-2}$)           & (km~s$^{-1}$)      & (km~s$^{-1}$)      \\
\hline
}

\tabletail{%
\hline
\hline
}

\topcaption{LTE Parameters for the full LTE model (Envelope Components) for source A04 in Sgr~B2(N).}
\tiny
\centering
\begin{supertabular}{lcccC{1cm}C{1cm}}\label{EnvLTE:parameters:A04SgrB2N}\\
H$_2$CNH      & ext.           &                  4 &  1.1 $\times 10^{12}$ &                7.0 &                  62\\
CN            & ext.           &                 22 &  3.3 $\times 10^{16}$ &               10.4 &                  77\\
              & ext.           &                 21 &  4.1 $\times 10^{16}$ &                9.4 &                  63\\
              & ext.           &                  3 &  4.7 $\times 10^{13}$ &                4.7 &                 -39\\
              & ext.           &                  3 &  3.4 $\times 10^{14}$ &                9.9 &                  31\\
              & ext.           &                  3 &  8.5 $\times 10^{14}$ &               15.4 &                   5\\
              & ext.           &                  3 &  1.0 $\times 10^{14}$ &                8.6 &                 -25\\
              & ext.           &                  3 &  1.7 $\times 10^{14}$ &                2.5 &                 -42\\
              & ext.           &                  3 &  4.0 $\times 10^{14}$ &               24.1 &                 -50\\
              & ext.           &                  3 &  3.6 $\times 10^{14}$ &                0.5 &                 -82\\
              & ext.           &                  3 &  2.9 $\times 10^{14}$ &                5.8 &                -109\\
SiO           & ext.           &                 10 &  2.0 $\times 10^{14}$ &                5.5 &                  70\\
CCH           & ext.           &                  3 &  1.0 $\times 10^{16}$ &               10.7 &                  73\\
              & ext.           &                  3 &  1.0 $\times 10^{16}$ &               10.1 &                  64\\
H$_2$C$^{33}$S & ext.           &                  9 &  1.1 $\times 10^{12}$ &                4.5 &                  67\\
CH$_2$NH      & ext.           &                  3 &  4.4 $\times 10^{14}$ &               18.9 &                  67\\
              & ext.           &                 48 &  1.1 $\times 10^{12}$ &                0.8 &                  51\\
H$_2$CO       & ext.           &                  3 &  1.5 $\times 10^{14}$ &                0.2 &                  68\\
              & ext.           &                 28 &  2.6 $\times 10^{16}$ &               19.7 &                  68\\
CO            & ext.           &                  3 &  1.7 $\times 10^{16}$ &               23.5 &                 114\\
              & ext.           &                  3 &  2.9 $\times 10^{17}$ &               27.7 &                  70\\
              & ext.           &                  3 &  3.3 $\times 10^{12}$ &               29.8 &                  68\\
              & ext.           &                  3 &  2.4 $\times 10^{16}$ &               38.8 &                  68\\
              & ext.           &                  3 &  2.1 $\times 10^{17}$ &                0.4 &                  54\\
              & ext.           &                  3 &  1.1 $\times 10^{12}$ &                0.3 &                  34\\
              & ext.           &                  3 &  1.1 $\times 10^{12}$ &                0.5 &                  27\\
              & ext.           &                  3 &  2.3 $\times 10^{16}$ &                5.2 &                  30\\
              & ext.           &                  3 &  2.1 $\times 10^{17}$ &               35.3 &                   8\\
              & ext.           &                  3 &  1.1 $\times 10^{12}$ &                0.6 &                 -25\\
              & ext.           &                  3 &  8.0 $\times 10^{15}$ &                1.5 &                  -3\\
              & ext.           &                  3 &  1.1 $\times 10^{12}$ &                0.6 &                 -37\\
              & ext.           &                  3 &  1.1 $\times 10^{12}$ &                0.5 &                 -88\\
              & ext.           &                  3 &  8.1 $\times 10^{16}$ &               11.6 &                 -24\\
              & ext.           &                  3 &  1.1 $\times 10^{12}$ &               15.9 &                 -98\\
              & ext.           &                  3 &  1.1 $\times 10^{12}$ &                0.1 &                 -93\\
              & ext.           &                  3 &  1.3 $\times 10^{17}$ &               22.2 &                 -46\\
              & ext.           &                  3 &  1.1 $\times 10^{12}$ &                0.7 &                 -56\\
              & ext.           &                  3 &  1.2 $\times 10^{17}$ &               13.4 &                 -77\\
              & ext.           &                  3 &  1.1 $\times 10^{12}$ &                0.7 &                -132\\
              & ext.           &                  3 &  3.0 $\times 10^{16}$ &                4.1 &                 -94\\
              & ext.           &                  3 &  1.5 $\times 10^{17}$ &                7.7 &                -107\\
              & ext.           &                  3 &  2.1 $\times 10^{16}$ &                3.5 &                -123\\
$^{13}$CO     & ext.           &                  3 &  1.3 $\times 10^{17}$ &               18.7 &                  64\\
              & ext.           &                  3 &  3.4 $\times 10^{16}$ &               15.6 &                   6\\
              & ext.           &                  3 &  4.0 $\times 10^{16}$ &                4.9 &                 -42\\
              & ext.           &                  3 &  1.1 $\times 10^{12}$ &               17.4 &                 -78\\
              & ext.           &                  3 &  1.1 $\times 10^{12}$ &                0.4 &                 -55\\
              & ext.           &                  3 &  1.1 $\times 10^{12}$ &                6.0 &                -161\\
              & ext.           &                  3 &  1.5 $\times 10^{16}$ &               11.6 &                 -76\\
              & ext.           &                  3 &  1.9 $\times 10^{17}$ &                0.3 &                 -86\\
              & ext.           &                  3 &  1.7 $\times 10^{15}$ &               37.6 &                 -76\\
              & ext.           &                  3 &  1.8 $\times 10^{16}$ &                1.3 &                -107\\
C$^{17}$O     & ext.           &                  3 &  6.7 $\times 10^{15}$ &                2.7 &                 -73\\
C$^{18}$O     & ext.           &                  3 &  8.8 $\times 10^{16}$ &                1.1 &                  63\\
CS            & ext.           &                  3 &  1.1 $\times 10^{12}$ &                0.5 &                  83\\
              & ext.           &                  3 &  2.3 $\times 10^{17}$ &               28.1 &                  68\\
              & ext.           &                  3 &  1.1 $\times 10^{12}$ &                5.4 &                  64\\
              & ext.           &                  3 &  1.3 $\times 10^{16}$ &                2.6 &                  58\\
              & ext.           &                  3 &  1.1 $\times 10^{13}$ &                0.8 &                  88\\
              & ext.           &                  3 &  2.6 $\times 10^{12}$ &               13.0 &                  15\\
              & ext.           &                  3 &  3.0 $\times 10^{16}$ &               15.2 &                -111\\
$^{13}$CS     & ext.           &                  3 &  1.6 $\times 10^{16}$ &               11.9 &                  67\\
              & ext.           &                  3 &  1.1 $\times 10^{12}$ &                0.6 &                 -70\\
C$^{33}$S     & ext.           &                  3 &  2.9 $\times 10^{16}$ &               11.2 &                -171\\
              & ext.           &                  3 &  4.8 $\times 10^{16}$ &               30.0 &                  22\\
HCN           & ext.           &                  3 &  1.1 $\times 10^{13}$ &                0.6 &                  79\\
              & ext.           &                  3 &  1.1 $\times 10^{12}$ &                7.7 &                  67\\
              & ext.           &                  3 &  4.4 $\times 10^{15}$ &               31.3 &                  64\\
              & ext.           &                  3 &  1.1 $\times 10^{12}$ &               21.2 &                  55\\
              & ext.           &                  3 &  2.8 $\times 10^{14}$ &                1.1 &                  45\\
              & ext.           &                  3 &  7.3 $\times 10^{14}$ &               14.7 &                   6\\
              & ext.           &                  3 &  1.1 $\times 10^{12}$ &                0.2 &                 -17\\
              & ext.           &                  3 &  5.1 $\times 10^{13}$ &                1.1 &                 -39\\
              & ext.           &                  3 &  9.7 $\times 10^{13}$ &                1.4 &                 -42\\
H$^{13}$CN    & ext.           &                  3 &  3.9 $\times 10^{15}$ &               16.9 &                  76\\
              & ext.           &                  3 &  1.3 $\times 10^{15}$ &               14.5 &                  55\\
              & ext.           &                  3 &  1.1 $\times 10^{12}$ &                2.5 &                  69\\
HNC           & ext.           &                  3 &  1.1 $\times 10^{12}$ &                0.1 &                -133\\
              & ext.           &                  3 &  1.9 $\times 10^{12}$ &                0.4 &                  74\\
              & ext.           &                  3 &  6.1 $\times 10^{15}$ &                4.6 &                  72\\
              & ext.           &                  3 &  5.9 $\times 10^{15}$ &               31.5 &                  64\\
              & ext.           &                  3 &  4.8 $\times 10^{16}$ &                0.1 &                  51\\
              & ext.           &                  3 &  5.9 $\times 10^{13}$ &                0.9 &                  47\\
              & ext.           &                  3 &  2.2 $\times 10^{12}$ &                0.2 &                 -63\\
H$^{13}$CO$^+$ & ext.           &                  3 &  4.9 $\times 10^{14}$ &               19.8 &                  69\\
\end{supertabular}\\
\vspace{1cm}

%================================================================================
%
% Source A05

%---------------------------------------
% Core Components

\tablefirsthead{%
\hline
\hline
Molecule      & $\theta^{m,c}$ & T$_{\rm ex}^{m,c}$ & N$_{\rm tot}^{m,c}$   & $\Delta$ v$^{m,c}$ & v$_{\rm LSR}^{m,c}$\\
              & ($\arcsec$)    & (K)                & (cm$^{-2}$)           & (km~s$^{-1}$)      & (km~s$^{-1}$)      \\
\hline
}

\tablehead{%
\multicolumn{6}{c}{(Continued)}\\
\hline
\hline
Molecule      & $\theta^{m,c}$ & T$_{\rm ex}^{m,c}$ & N$_{\rm tot}^{m,c}$   & $\Delta$ v$^{m,c}$ & v$_{\rm LSR}^{m,c}$\\
              & ($\arcsec$)    & (K)                & (cm$^{-2}$)           & (km~s$^{-1}$)      & (km~s$^{-1}$)      \\
\hline
}

\tabletail{%
\hline
\hline
}

\topcaption{LTE Parameters for the full LTE model (Core Components) for source A05 in Sgr~B2(N).}
\tiny
\centering
\begin{supertabular}{lcccC{1cm}C{1cm}}\label{CoreLTE:parameters:A05SgrB2N}\\
CH$_3$NH$_2$  &    1.5         &                 89 &  3.0 $\times 10^{17}$ &                5.6 &                  63\\
HC(O)NH$_2$   &    1.5         &                203 &  5.3 $\times 10^{14}$ &                4.9 &                  65\\
SO$_2$        &    1.5         &                134 &  3.8 $\times 10^{16}$ &                8.6 &                  60\\
CH$_2$NH      &    1.5         &                151 &  2.5 $\times 10^{16}$ &                3.8 &                  63\\
NO            &    1.5         &                188 &  3.5 $\times 10^{17}$ &                6.2 &                  62\\
NH$_2$D       &    1.5         &                 91 &  3.6 $\times 10^{16}$ &                5.7 &                  62\\
C$_2$H$_5$CN, v$_{20}$=1 &    1.5         &                156 &  7.3 $\times 10^{16}$ &                5.7 &                  63\\
\end{supertabular}\\
\vspace{1cm}

%---------------------------------------
% Envelope Components

\tablefirsthead{%
\hline
\hline
Molecule      & $\theta^{m,c}$ & T$_{\rm ex}^{m,c}$ & N$_{\rm tot}^{m,c}$   & $\Delta$ v$^{m,c}$ & v$_{\rm LSR}^{m,c}$\\
              & ($\arcsec$)    & (K)                & (cm$^{-2}$)           & (km~s$^{-1}$)      & (km~s$^{-1}$)      \\
\hline
}

\tablehead{%
\multicolumn{6}{c}{(Continued)}\\
\hline
\hline
Molecule      & $\theta^{m,c}$ & T$_{\rm ex}^{m,c}$ & N$_{\rm tot}^{m,c}$   & $\Delta$ v$^{m,c}$ & v$_{\rm LSR}^{m,c}$\\
              & ($\arcsec$)    & (K)                & (cm$^{-2}$)           & (km~s$^{-1}$)      & (km~s$^{-1}$)      \\
\hline
}

\tabletail{%
\hline
\hline
}

\topcaption{LTE Parameters for the full LTE model (Envelope Components) for source A05 in Sgr~B2(N).}
\tiny
\centering
\begin{supertabular}{lcccC{1cm}C{1cm}}\label{EnvLTE:parameters:A05SgrB2N}\\
CN            & ext.           &                 24 &  6.9 $\times 10^{16}$ &               15.0 &                  72\\
              & ext.           &                 21 &  2.6 $\times 10^{16}$ &                4.8 &                  62\\
              & ext.           &                  3 &  4.4 $\times 10^{13}$ &                1.1 &                 -89\\
              & ext.           &                  3 &  4.6 $\times 10^{14}$ &               16.9 &                  38\\
              & ext.           &                  3 &  3.3 $\times 10^{13}$ &                5.3 &                 -22\\
              & ext.           &                  3 &  5.8 $\times 10^{14}$ &               10.5 &                   4\\
              & ext.           &                  3 &  1.9 $\times 10^{14}$ &                9.5 &                 -32\\
              & ext.           &                  3 &  2.5 $\times 10^{14}$ &                3.1 &                 -43\\
              & ext.           &                  3 &  8.5 $\times 10^{13}$ &                5.5 &                 -59\\
              & ext.           &                  3 &  1.7 $\times 10^{14}$ &                3.0 &                 -80\\
              & ext.           &                  3 &  2.0 $\times 10^{14}$ &                3.1 &                -108\\
CH$_2$NH      & ext.           &                  3 &  1.7 $\times 10^{14}$ &                8.7 &                  62\\
              & ext.           &                  3 &  7.2 $\times 10^{13}$ &                5.9 &                  73\\
              & ext.           &                  3 &  8.2 $\times 10^{13}$ &                6.1 &                  81\\
CCH           & ext.           &                  3 &  1.2 $\times 10^{16}$ &                8.3 &                  74\\
              & ext.           &                  3 &  7.1 $\times 10^{15}$ &                7.9 &                  64\\
H$_2$C$^{33}$S & ext.           &                 10 &  1.0 $\times 10^{12}$ &                4.7 &                  66\\
SiO           & ext.           &                 12 &  2.3 $\times 10^{14}$ &                4.8 &                  72\\
PH$_3$        & ext.           &                 36 &  1.2 $\times 10^{16}$ &                5.6 &                  62\\
H$_2$CO       & ext.           &                 10 &  1.3 $\times 10^{15}$ &               16.7 &                  64\\
              & ext.           &                 19 &  1.0 $\times 10^{16}$ &                8.9 &                  79\\
              & ext.           &                 13 &  8.0 $\times 10^{14}$ &                6.0 &                  52\\
CO            & ext.           &                  3 &  2.5 $\times 10^{12}$ &               24.1 &                 122\\
              & ext.           &                  3 &  2.3 $\times 10^{16}$ &                3.9 &                  85\\
              & ext.           &                  3 &  1.1 $\times 10^{12}$ &                0.4 &                  74\\
              & ext.           &                  3 &  1.5 $\times 10^{17}$ &               16.3 &                  64\\
              & ext.           &                  3 &  3.1 $\times 10^{12}$ &                0.2 &                  52\\
              & ext.           &                  3 &  1.6 $\times 10^{16}$ &                6.3 &                  45\\
              & ext.           &                  3 &  1.1 $\times 10^{16}$ &                0.6 &                  39\\
              & ext.           &                  3 &  2.2 $\times 10^{16}$ &                4.9 &                  29\\
              & ext.           &                  3 &  2.6 $\times 10^{12}$ &                0.1 &                -144\\
              & ext.           &                  3 &  1.6 $\times 10^{17}$ &               31.6 &                  12\\
              & ext.           &                  3 &  3.6 $\times 10^{16}$ &               11.3 &                  -3\\
              & ext.           &                  3 &  7.3 $\times 10^{16}$ &               11.7 &                 -21\\
              & ext.           &                  3 &  8.5 $\times 10^{14}$ &                0.9 &                 -26\\
              & ext.           &                  3 &  2.0 $\times 10^{16}$ &                5.4 &                 -30\\
              & ext.           &                  3 &  1.2 $\times 10^{17}$ &               18.4 &                 -44\\
              & ext.           &                  3 &  2.6 $\times 10^{15}$ &                3.0 &                 -48\\
              & ext.           &                  3 &  3.2 $\times 10^{16}$ &                2.4 &                 -59\\
              & ext.           &                  3 &  1.1 $\times 10^{17}$ &               13.8 &                 -76\\
              & ext.           &                  3 &  9.6 $\times 10^{15}$ &                5.0 &                 -83\\
              & ext.           &                  3 &  2.7 $\times 10^{16}$ &                3.3 &                 -94\\
              & ext.           &                  3 &  5.4 $\times 10^{14}$ &                0.7 &                -100\\
              & ext.           &                  3 &  7.3 $\times 10^{16}$ &                9.1 &                -107\\
              & ext.           &                  3 &  1.7 $\times 10^{14}$ &                3.6 &                -126\\
              & ext.           &                  3 &  2.6 $\times 10^{16}$ &                2.2 &                -124\\
$^{13}$CO     & ext.           &                  3 &  6.6 $\times 10^{16}$ &               12.1 &                  70\\
              & ext.           &                  3 &  1.2 $\times 10^{17}$ &                1.6 &                  61\\
              & ext.           &                  3 &  6.0 $\times 10^{16}$ &               39.5 &                  43\\
              & ext.           &                  3 &  7.0 $\times 10^{15}$ &                1.2 &                  14\\
              & ext.           &                  3 &  9.9 $\times 10^{15}$ &                2.3 &                   5\\
              & ext.           &                  3 &  1.3 $\times 10^{16}$ &                2.4 &                 -40\\
              & ext.           &                  3 &  2.8 $\times 10^{16}$ &                3.0 &                 -43\\
              & ext.           &                  3 &  1.4 $\times 10^{16}$ &                2.6 &                 -50\\
              & ext.           &                  3 &  2.1 $\times 10^{15}$ &               10.7 &                 -65\\
              & ext.           &                  3 &  3.4 $\times 10^{15}$ &                3.2 &                 -70\\
C$^{17}$O     & ext.           &                  3 &  4.0 $\times 10^{15}$ &                3.9 &                 -65\\
              & ext.           &                  3 &  8.9 $\times 10^{14}$ &                2.8 &                 -49\\
              & ext.           &                  3 &  1.2 $\times 10^{15}$ &                2.8 &                 -41\\
              & ext.           &                  3 &  7.8 $\times 10^{14}$ &                2.6 &                 -28\\
              & ext.           &                  3 &  3.0 $\times 10^{14}$ &                1.1 &                   1\\
              & ext.           &                  3 &  5.4 $\times 10^{15}$ &                5.7 &                  66\\
              & ext.           &                  3 &  1.5 $\times 10^{15}$ &                7.4 &                 128\\
C$^{18}$O     & ext.           &                  3 &  2.9 $\times 10^{16}$ &                3.6 &                -157\\
              & ext.           &                  3 &  1.5 $\times 10^{15}$ &                5.0 &                 -82\\
              & ext.           &                  3 &  4.5 $\times 10^{15}$ &                1.2 &                 -42\\
              & ext.           &                  3 &  7.9 $\times 10^{15}$ &                4.4 &                  67\\
              & ext.           &                  3 &  1.4 $\times 10^{15}$ &                2.1 &                  76\\
              & ext.           &                  3 &  9.5 $\times 10^{14}$ &                3.0 &                 112\\
CS            & ext.           &                  3 &  2.4 $\times 10^{16}$ &                3.1 &                  83\\
              & ext.           &                  3 &  3.3 $\times 10^{17}$ &               22.6 &                  69\\
              & ext.           &                  3 &  9.6 $\times 10^{15}$ &                2.0 &                  72\\
              & ext.           &                  3 &  3.8 $\times 10^{16}$ &                2.7 &                  50\\
              & ext.           &                  3 &  1.2 $\times 10^{16}$ &                7.6 &                  40\\
              & ext.           &                  3 &  5.6 $\times 10^{14}$ &                1.9 &                  28\\
              & ext.           &                  3 &  6.0 $\times 10^{14}$ &                7.5 &                  22\\
$^{13}$CS     & ext.           &                  3 &  1.3 $\times 10^{16}$ &                4.1 &                  71\\
C$^{33}$S     & ext.           &                  3 &  1.5 $\times 10^{15}$ &                1.6 &                 -79\\
              & ext.           &                  3 &  7.6 $\times 10^{15}$ &                4.8 &                  69\\
              & ext.           &                  3 &  1.6 $\times 10^{15}$ &                2.8 &                 134\\
              & ext.           &                  3 &  7.3 $\times 10^{15}$ &                5.9 &                 141\\
              & ext.           &                  3 &  8.0 $\times 10^{14}$ &               12.9 &                 152\\
C$^{34}$S     & ext.           &                  3 &  1.9 $\times 10^{16}$ &                2.2 &                  70\\
HCN           & ext.           &                  3 &  1.1 $\times 10^{12}$ &                1.6 &                  86\\
              & ext.           &                  3 &  1.1 $\times 10^{15}$ &               10.3 &                  85\\
              & ext.           &                  3 &  1.1 $\times 10^{12}$ &               18.2 &                  21\\
              & ext.           &                  3 &  1.1 $\times 10^{12}$ &                2.6 &                  65\\
              & ext.           &                  3 &  2.5 $\times 10^{15}$ &               29.4 &                  63\\
              & ext.           &                  3 &  1.1 $\times 10^{12}$ &               10.5 &                  60\\
              & ext.           &                  3 &  4.4 $\times 10^{14}$ &                3.9 &                  46\\
              & ext.           &                  3 &  3.2 $\times 10^{12}$ &               22.8 &                  17\\
              & ext.           &                  3 &  5.6 $\times 10^{13}$ &                1.8 &                  23\\
              & ext.           &                  3 &  3.4 $\times 10^{12}$ &                1.3 &                   0\\
              & ext.           &                  3 &  1.4 $\times 10^{14}$ &                1.9 &                   3\\
              & ext.           &                  3 &  5.4 $\times 10^{14}$ &               13.4 &                   7\\
              & ext.           &                  3 &  6.8 $\times 10^{12}$ &               18.4 &                 -41\\
              & ext.           &                  3 &  2.6 $\times 10^{13}$ &                1.0 &                 -42\\
              & ext.           &                  3 &  2.3 $\times 10^{14}$ &               20.1 &                 -54\\
              & ext.           &                  3 &  1.6 $\times 10^{13}$ &                2.1 &                 -69\\
              & ext.           &                  3 &  2.7 $\times 10^{13}$ &                1.9 &                 -80\\
H$^{13}$CN    & ext.           &                  3 &  2.5 $\times 10^{14}$ &                0.9 &                  83\\
              & ext.           &                  3 &  1.7 $\times 10^{15}$ &               14.0 &                  73\\
              & ext.           &                  3 &  1.2 $\times 10^{14}$ &                3.8 &                  60\\
              & ext.           &                  3 &  2.4 $\times 10^{14}$ &               30.5 &                  47\\
              & ext.           &                  3 &  3.5 $\times 10^{14}$ &                3.1 &                  50\\
HNC           & ext.           &                  3 &  1.2 $\times 10^{15}$ &                4.4 &                  86\\
              & ext.           &                  3 &  1.6 $\times 10^{12}$ &               18.3 &                  75\\
              & ext.           &                  3 &  8.7 $\times 10^{15}$ &               11.3 &                  71\\
              & ext.           &                  3 &  4.9 $\times 10^{14}$ &                2.5 &                  58\\
              & ext.           &                  3 &  8.9 $\times 10^{14}$ &                6.4 &                  52\\
              & ext.           &                  3 &  2.8 $\times 10^{13}$ &                6.2 &                  29\\
              & ext.           &                  3 &  6.0 $\times 10^{13}$ &                1.7 &                   6\\
              & ext.           &                  3 &  2.2 $\times 10^{13}$ &                2.1 &                 -41\\
              & ext.           &                  3 &  1.1 $\times 10^{12}$ &                1.1 &                 -75\\
              & ext.           &                  3 &  1.2 $\times 10^{12}$ &                0.9 &                 -98\\
              & ext.           &                  3 &  1.1 $\times 10^{12}$ &                0.9 &                -111\\
              & ext.           &                  3 &  2.3 $\times 10^{13}$ &                3.6 &                -144\\
HN$^{13}$C    & ext.           &                  3 &  3.2 $\times 10^{14}$ &               14.5 &                  69\\
H$^{13}$CO$^+$ & ext.           &                  3 &  3.6 $\times 10^{13}$ &                1.7 &                  62\\
              & ext.           &                  3 &  4.0 $\times 10^{14}$ &               17.5 &                  70\\
\end{supertabular}\\
\vspace{1cm}

%================================================================================
%
% Source A06

%---------------------------------------
% Core Components

\tablefirsthead{%
\hline
\hline
Molecule      & $\theta^{m,c}$ & T$_{\rm ex}^{m,c}$ & N$_{\rm tot}^{m,c}$   & $\Delta$ v$^{m,c}$ & v$_{\rm LSR}^{m,c}$\\
              & ($\arcsec$)    & (K)                & (cm$^{-2}$)           & (km~s$^{-1}$)      & (km~s$^{-1}$)      \\
\hline
}

\tablehead{%
\multicolumn{6}{c}{(Continued)}\\
\hline
\hline
Molecule      & $\theta^{m,c}$ & T$_{\rm ex}^{m,c}$ & N$_{\rm tot}^{m,c}$   & $\Delta$ v$^{m,c}$ & v$_{\rm LSR}^{m,c}$\\
              & ($\arcsec$)    & (K)                & (cm$^{-2}$)           & (km~s$^{-1}$)      & (km~s$^{-1}$)      \\
\hline
}

\tabletail{%
\hline
\hline
}

\topcaption{LTE Parameters for the full LTE model (Core Components) for source A06 in Sgr~B2(N).}
\tiny
\centering
\begin{supertabular}{lcccC{1cm}C{1cm}}\label{CoreLTE:parameters:A06SgrB2N}\\
H$_2$CNH      &    1.7         &                 69 &  5.6 $\times 10^{16}$ &                6.1 &                  70\\
SO            &    1.7         &                139 &  6.9 $\times 10^{16}$ &                6.7 &                  69\\
C$_2$H$_3$CN  &    1.7         &                107 &  1.1 $\times 10^{16}$ &                7.1 &                  71\\
C$_2$H$_3$CN, v$_{10}$=1 &    1.7         &                259 &  5.2 $\times 10^{14}$ &               11.3 &                  64\\
C$_2$H$_5$CN  &    1.7         &                118 &  1.7 $\times 10^{17}$ &                5.3 &                  69\\
              &    1.7         &                 86 &  1.1 $\times 10^{17}$ &                2.4 &                  73\\
C$_2$H$_5$CN, v$_{20}$=1 &    1.7         &                109 &  4.0 $\times 10^{17}$ &                7.2 &                  71\\
\end{supertabular}\\
\vspace{1cm}

%---------------------------------------
% Envelope Components

\tablefirsthead{%
\hline
\hline
Molecule      & $\theta^{m,c}$ & T$_{\rm ex}^{m,c}$ & N$_{\rm tot}^{m,c}$   & $\Delta$ v$^{m,c}$ & v$_{\rm LSR}^{m,c}$\\
              & ($\arcsec$)    & (K)                & (cm$^{-2}$)           & (km~s$^{-1}$)      & (km~s$^{-1}$)      \\
\hline
}

\tablehead{%
\multicolumn{6}{c}{(Continued)}\\
\hline
\hline
Molecule      & $\theta^{m,c}$ & T$_{\rm ex}^{m,c}$ & N$_{\rm tot}^{m,c}$   & $\Delta$ v$^{m,c}$ & v$_{\rm LSR}^{m,c}$\\
              & ($\arcsec$)    & (K)                & (cm$^{-2}$)           & (km~s$^{-1}$)      & (km~s$^{-1}$)      \\
\hline
}

\tabletail{%
\hline
\hline
}

\topcaption{LTE Parameters for the full LTE model (Envelope Components) for source A06 in Sgr~B2(N).}
\tiny
\centering
\begin{supertabular}{lcccC{1cm}C{1cm}}\label{EnvLTE:parameters:A06SgrB2N}\\
H$_2$CNH      & ext.           &                  3 &  9.6 $\times 10^{13}$ &                6.2 &                  83\\
              & ext.           &                  3 &  3.1 $\times 10^{14}$ &               14.6 &                  71\\
              & ext.           &                  3 &  1.6 $\times 10^{14}$ &                8.4 &                  58\\
SO            & ext.           &                 10 &  3.6 $\times 10^{15}$ &               20.0 &                  69\\
CN            & ext.           &                 20 &  4.0 $\times 10^{16}$ &               21.4 &                  66\\
              & ext.           &                 19 &  3.5 $\times 10^{16}$ &                7.1 &                  61\\
              & ext.           &                  3 &  1.8 $\times 10^{14}$ &                8.3 &                  12\\
              & ext.           &                  3 &  1.3 $\times 10^{14}$ &                5.8 &                  26\\
              & ext.           &                  3 &  8.9 $\times 10^{13}$ &                7.0 &                 -30\\
              & ext.           &                  5 &  5.9 $\times 10^{14}$ &               10.5 &                   0\\
              & ext.           &                  4 &  2.7 $\times 10^{13}$ &                5.2 &                 -23\\
              & ext.           &                  3 &  7.3 $\times 10^{13}$ &                3.1 &                 -43\\
              & ext.           &                  3 &  7.5 $\times 10^{13}$ &                5.4 &                 -52\\
              & ext.           &                  3 &  3.4 $\times 10^{17}$ &                0.2 &                 -77\\
              & ext.           &                  3 &  2.6 $\times 10^{14}$ &                3.5 &                -108\\
SiO           & ext.           &                 12 &  1.2 $\times 10^{14}$ &                6.0 &                  75\\
CH$_2$NH      & ext.           &                  3 &  1.1 $\times 10^{12}$ &               10.8 &                  79\\
              & ext.           &                  3 &  5.9 $\times 10^{12}$ &               14.1 &                  64\\
              & ext.           &                  3 &  1.0 $\times 10^{12}$ &                7.5 &                  59\\
H$_2$C$^{33}$S & ext.           &                138 &  1.5 $\times 10^{15}$ &                6.2 &                  68\\
HCCCN         & ext.           &                 30 &  1.2 $\times 10^{16}$ &               10.1 &                  57\\
              & ext.           &                 38 &  1.8 $\times 10^{16}$ &                8.8 &                  79\\
CCH           & ext.           &                  3 &  9.4 $\times 10^{15}$ &                9.4 &                  76\\
              & ext.           &                  3 &  6.0 $\times 10^{15}$ &               10.4 &                  64\\
H$_2$CO       & ext.           &                 20 &  5.9 $\times 10^{15}$ &                5.9 &                  81\\
              & ext.           &                 12 &  2.2 $\times 10^{15}$ &               27.7 &                  64\\
H$_2$CS       & ext.           &                 10 &  1.2 $\times 10^{12}$ &                2.0 &                  60\\
PH$_3$        & ext.           &                 33 &  3.6 $\times 10^{17}$ &                3.3 &                  65\\
CO            & ext.           &                  3 &  6.0 $\times 10^{16}$ &               39.5 &                 106\\
              & ext.           &                  3 &  1.1 $\times 10^{12}$ &               26.4 &                  95\\
              & ext.           &                  3 &  1.4 $\times 10^{16}$ &                5.6 &                  84\\
              & ext.           &                  3 &  4.7 $\times 10^{15}$ &                1.3 &                  74\\
              & ext.           &                  3 &  8.6 $\times 10^{15}$ &                1.7 &                  66\\
              & ext.           &                  3 &  1.1 $\times 10^{12}$ &                0.4 &                  56\\
              & ext.           &                  3 &  6.2 $\times 10^{16}$ &               20.4 &                  52\\
              & ext.           &                  3 &  3.5 $\times 10^{12}$ &                0.5 &                  47\\
              & ext.           &                  3 &  9.2 $\times 10^{15}$ &                7.0 &                  28\\
              & ext.           &                  3 &  1.1 $\times 10^{12}$ &                2.9 &                  18\\
              & ext.           &                  3 &  2.2 $\times 10^{19}$ &                0.3 &                  16\\
              & ext.           &                  3 &  2.5 $\times 10^{16}$ &               23.1 &                   9\\
              & ext.           &                  3 &  3.3 $\times 10^{15}$ &                5.3 &                   0\\
              & ext.           &                  3 &  1.1 $\times 10^{12}$ &                0.5 &                 -13\\
              & ext.           &                  3 &  4.0 $\times 10^{16}$ &               22.1 &                 -22\\
              & ext.           &                  3 &  1.6 $\times 10^{16}$ &                9.3 &                 -43\\
              & ext.           &                  3 &  4.3 $\times 10^{15}$ &                3.1 &                 -49\\
              & ext.           &                  3 &  3.8 $\times 10^{15}$ &                2.6 &                 -59\\
              & ext.           &                  3 &  2.5 $\times 10^{16}$ &               10.1 &                 -73\\
              & ext.           &                  3 &  1.6 $\times 10^{13}$ &                2.8 &                 -71\\
              & ext.           &                  3 &  6.3 $\times 10^{15}$ &                2.7 &                 -82\\
              & ext.           &                  3 &  1.7 $\times 10^{14}$ &                7.9 &                 -93\\
              & ext.           &                  3 &  1.5 $\times 10^{13}$ &                0.2 &                -104\\
              & ext.           &                  3 &  6.9 $\times 10^{15}$ &                8.4 &                -115\\
              & ext.           &                  3 &  3.5 $\times 10^{16}$ &                7.9 &                -105\\
$^{13}$CO     & ext.           &                  3 &  5.7 $\times 10^{13}$ &               22.4 &                  88\\
              & ext.           &                  3 &  1.7 $\times 10^{12}$ &                2.0 &                  85\\
              & ext.           &                  3 &  4.6 $\times 10^{16}$ &               18.2 &                  80\\
              & ext.           &                  3 &  3.6 $\times 10^{16}$ &                7.3 &                  65\\
              & ext.           &                  3 &  5.5 $\times 10^{16}$ &               15.2 &                  59\\
              & ext.           &                  3 &  1.1 $\times 10^{12}$ &                2.0 &                  45\\
              & ext.           &                  3 &  1.3 $\times 10^{12}$ &                2.0 &                  21\\
              & ext.           &                  3 &  1.1 $\times 10^{16}$ &                8.7 &                  10\\
              & ext.           &                  3 &  8.1 $\times 10^{15}$ &                7.0 &                   1\\
              & ext.           &                  3 &  5.0 $\times 10^{12}$ &                0.5 &                 -24\\
              & ext.           &                  3 &  6.2 $\times 10^{15}$ &                4.0 &                 -42\\
              & ext.           &                  3 &  1.0 $\times 10^{13}$ &                0.5 &                 -73\\
              & ext.           &                  3 &  2.8 $\times 10^{14}$ &                0.5 &                 -54\\
              & ext.           &                  3 &  4.7 $\times 10^{16}$ &               27.4 &                 -46\\
              & ext.           &                  3 &  2.7 $\times 10^{16}$ &                0.5 &                 -77\\
              & ext.           &                  3 &  1.1 $\times 10^{12}$ &                0.7 &                 -99\\
              & ext.           &                  3 &  1.1 $\times 10^{12}$ &                0.9 &                -115\\
              & ext.           &                  3 &  4.8 $\times 10^{14}$ &                0.9 &                -131\\
              & ext.           &                  3 &  1.1 $\times 10^{12}$ &                0.7 &                -171\\
C$^{17}$O     & ext.           &                  3 &  7.8 $\times 10^{15}$ &                6.5 &                -182\\
              & ext.           &                  3 &  2.0 $\times 10^{13}$ &                4.9 &                -113\\
              & ext.           &                  3 &  6.5 $\times 10^{16}$ &               30.2 &                 -62\\
              & ext.           &                  3 &  1.5 $\times 10^{12}$ &                2.5 &                 -41\\
              & ext.           &                  3 &  1.8 $\times 10^{15}$ &                2.3 &                 -20\\
              & ext.           &                  3 &  7.9 $\times 10^{15}$ &                5.0 &                  44\\
              & ext.           &                  3 &  6.0 $\times 10^{15}$ &                6.6 &                  80\\
              & ext.           &                  3 &  8.9 $\times 10^{14}$ &                2.4 &                 120\\
              & ext.           &                  3 &  2.3 $\times 10^{12}$ &                2.4 &                  -3\\
              & ext.           &                  3 &  9.3 $\times 10^{15}$ &                8.5 &                 146\\
C$^{18}$O     & ext.           &                  3 &  4.9 $\times 10^{14}$ &                2.8 &                -156\\
              & ext.           &                  3 &  3.5 $\times 10^{14}$ &                2.9 &                -107\\
              & ext.           &                  3 &  1.4 $\times 10^{15}$ &                6.7 &                 -71\\
              & ext.           &                  3 &  2.4 $\times 10^{15}$ &                2.8 &                 -41\\
              & ext.           &                  3 &  3.8 $\times 10^{15}$ &                3.6 &                  57\\
              & ext.           &                  3 &  8.1 $\times 10^{15}$ &                4.1 &                  62\\
              & ext.           &                  3 &  1.2 $\times 10^{15}$ &               36.1 &                 153\\
CS            & ext.           &                  3 &  1.1 $\times 10^{12}$ &                0.9 &                 108\\
              & ext.           &                  3 &  1.1 $\times 10^{12}$ &                0.1 &                  76\\
              & ext.           &                  3 &  5.6 $\times 10^{12}$ &                0.2 &                  71\\
              & ext.           &                  3 &  1.5 $\times 10^{17}$ &               20.9 &                  70\\
              & ext.           &                  3 &  1.2 $\times 10^{16}$ &                6.0 &                  64\\
              & ext.           &                  3 &  5.4 $\times 10^{16}$ &                7.8 &                  54\\
              & ext.           &                  3 &  2.9 $\times 10^{15}$ &                5.2 &                  31\\
              & ext.           &                  3 &  3.3 $\times 10^{13}$ &               26.1 &                   8\\
$^{13}$CS     & ext.           &                  3 &  1.1 $\times 10^{12}$ &                2.5 &                  84\\
              & ext.           &                  3 &  9.8 $\times 10^{12}$ &                0.9 &                  12\\
C$^{33}$S     & ext.           &                  3 &  9.3 $\times 10^{14}$ &                7.8 &                -153\\
              & ext.           &                  3 &  9.6 $\times 10^{14}$ &                9.3 &                -118\\
              & ext.           &                  3 &  3.1 $\times 10^{16}$ &               12.2 &                 -76\\
              & ext.           &                  3 &  5.8 $\times 10^{15}$ &                3.1 &                 112\\
              & ext.           &                  3 &  4.7 $\times 10^{15}$ &                3.1 &                 130\\
              & ext.           &                  3 &  4.9 $\times 10^{15}$ &               49.9 &                 132\\
              & ext.           &                  3 &  1.5 $\times 10^{15}$ &                2.8 &                 141\\
C$^{34}$S     & ext.           &                  3 &  2.9 $\times 10^{16}$ &                8.3 &                  57\\
HCN           & ext.           &                  3 &  9.1 $\times 10^{12}$ &                0.6 &                 -24\\
              & ext.           &                  3 &  6.2 $\times 10^{13}$ &                2.6 &                 118\\
              & ext.           &                  3 &  2.9 $\times 10^{13}$ &                1.2 &                  98\\
              & ext.           &                  3 &  3.3 $\times 10^{14}$ &                8.3 &                  86\\
              & ext.           &                  3 &  9.9 $\times 10^{12}$ &                0.7 &                  77\\
              & ext.           &                  3 &  1.1 $\times 10^{12}$ &                1.0 &                  34\\
              & ext.           &                  3 &  1.2 $\times 10^{15}$ &               40.0 &                  65\\
              & ext.           &                  3 &  1.1 $\times 10^{15}$ &               30.9 &                  59\\
              & ext.           &                  3 &  7.5 $\times 10^{13}$ &                0.7 &                  49\\
              & ext.           &                  3 &  1.6 $\times 10^{14}$ &                4.7 &                  28\\
              & ext.           &                  3 &  5.0 $\times 10^{14}$ &               18.4 &                  13\\
              & ext.           &                  3 &  5.3 $\times 10^{14}$ &               12.6 &                   1\\
              & ext.           &                  3 &  2.8 $\times 10^{13}$ &                0.9 &                  -6\\
              & ext.           &                  3 &  1.1 $\times 10^{12}$ &                1.3 &                  -6\\
              & ext.           &                  3 &  1.2 $\times 10^{14}$ &                1.0 &                 -40\\
              & ext.           &                  3 &  4.2 $\times 10^{13}$ &                2.5 &                 -69\\
              & ext.           &                  3 &  6.0 $\times 10^{12}$ &                1.0 &                -115\\
              & ext.           &                  3 &  1.1 $\times 10^{13}$ &                1.0 &                -113\\
              & ext.           &                  3 &  3.5 $\times 10^{13}$ &                2.6 &                -166\\
H$^{13}$CN    & ext.           &                  3 &  1.1 $\times 10^{12}$ &               25.7 &                  79\\
              & ext.           &                  3 &  1.1 $\times 10^{12}$ &                4.9 &                  77\\
              & ext.           &                  3 &  5.7 $\times 10^{14}$ &                9.7 &                  57\\
              & ext.           &                  3 &  2.4 $\times 10^{14}$ &                7.6 &                  84\\
              & ext.           &                  3 &  5.3 $\times 10^{14}$ &               10.9 &                  77\\
              & ext.           &                  3 &  7.5 $\times 10^{13}$ &                6.9 &                  67\\
HNC           & ext.           &                  3 &  6.3 $\times 10^{13}$ &                4.5 &                 113\\
              & ext.           &                  3 &  6.8 $\times 10^{15}$ &                1.4 &                  79\\
              & ext.           &                  3 &  1.0 $\times 10^{12}$ &                1.0 &                  64\\
              & ext.           &                  3 &  4.1 $\times 10^{15}$ &               34.2 &                  61\\
              & ext.           &                  3 &  4.4 $\times 10^{14}$ &                2.2 &                  51\\
              & ext.           &                  3 &  1.9 $\times 10^{13}$ &                1.7 &                  19\\
              & ext.           &                  3 &  9.3 $\times 10^{14}$ &               21.9 &                  -5\\
              & ext.           &                  3 &  1.4 $\times 10^{13}$ &                1.0 &                 -80\\
              & ext.           &                  3 &  1.1 $\times 10^{12}$ &                0.9 &                -126\\
HN$^{13}$C    & ext.           &                  3 &  3.2 $\times 10^{14}$ &                8.4 &                  76\\
H$^{13}$CO$^+$ & ext.           &                  3 &  2.8 $\times 10^{14}$ &               12.2 &                  74\\
              & ext.           &                  3 &  6.7 $\times 10^{13}$ &                4.0 &                  69\\
              & ext.           &                  3 &  1.5 $\times 10^{14}$ &                8.0 &                  60\\
HO$^{13}$C$^+$ & ext.           &                  3 &  8.4 $\times 10^{12}$ &                2.0 &                 137\\
\end{supertabular}\\
\vspace{1cm}

%================================================================================
%
% Source A07

%---------------------------------------
% Core Components

\tablefirsthead{%
\hline
\hline
Molecule      & $\theta^{m,c}$ & T$_{\rm ex}^{m,c}$ & N$_{\rm tot}^{m,c}$   & $\Delta$ v$^{m,c}$ & v$_{\rm LSR}^{m,c}$\\
              & ($\arcsec$)    & (K)                & (cm$^{-2}$)           & (km~s$^{-1}$)      & (km~s$^{-1}$)      \\
\hline
}

\tablehead{%
\multicolumn{6}{c}{(Continued)}\\
\hline
\hline
Molecule      & $\theta^{m,c}$ & T$_{\rm ex}^{m,c}$ & N$_{\rm tot}^{m,c}$   & $\Delta$ v$^{m,c}$ & v$_{\rm LSR}^{m,c}$\\
              & ($\arcsec$)    & (K)                & (cm$^{-2}$)           & (km~s$^{-1}$)      & (km~s$^{-1}$)      \\
\hline
}

\tabletail{%
\hline
\hline
}

\topcaption{LTE Parameters for the full LTE model (Core Components) for source A07 in Sgr~B2(N).}
\tiny
\centering
\begin{supertabular}{lcccC{1cm}C{1cm}}\label{CoreLTE:parameters:A07SgrB2N}\\
HCN, v$_2$=1  &    1.1         &                509 &  1.0 $\times 10^{12}$ &               21.2 &                  62\\
NS            &    1.1         &                 59 &  1.5 $\times 10^{17}$ &               19.5 &                  64\\
OCS           &    1.1         &                182 &  4.7 $\times 10^{17}$ &                3.5 &                  62\\
NO            &    1.1         &                268 &  3.9 $\times 10^{17}$ &                3.1 &                  65\\
SO$_2$        &    1.1         &                215 &  5.3 $\times 10^{12}$ &                2.2 &                  64\\
              &    1.1         &                110 &  2.2 $\times 10^{17}$ &                5.4 &                  64\\
CH$_3$OH      &    1.1         &                155 &  3.4 $\times 10^{18}$ &                2.1 &                  62\\
\end{supertabular}\\
\vspace{1cm}

%---------------------------------------
% Envelope Components

\tablefirsthead{%
\hline
\hline
Molecule      & $\theta^{m,c}$ & T$_{\rm ex}^{m,c}$ & N$_{\rm tot}^{m,c}$   & $\Delta$ v$^{m,c}$ & v$_{\rm LSR}^{m,c}$\\
              & ($\arcsec$)    & (K)                & (cm$^{-2}$)           & (km~s$^{-1}$)      & (km~s$^{-1}$)      \\
\hline
}

\tablehead{%
\multicolumn{6}{c}{(Continued)}\\
\hline
\hline
Molecule      & $\theta^{m,c}$ & T$_{\rm ex}^{m,c}$ & N$_{\rm tot}^{m,c}$   & $\Delta$ v$^{m,c}$ & v$_{\rm LSR}^{m,c}$\\
              & ($\arcsec$)    & (K)                & (cm$^{-2}$)           & (km~s$^{-1}$)      & (km~s$^{-1}$)      \\
\hline
}

\tabletail{%
\hline
\hline
}

\topcaption{LTE Parameters for the full LTE model (Envelope Components) for source A07 in Sgr~B2(N).}
\tiny
\centering
\begin{supertabular}{lcccC{1cm}C{1cm}}\label{EnvLTE:parameters:A07SgrB2N}\\
CN            & ext.           &                  4 &  4.0 $\times 10^{14}$ &                4.4 &                  81\\
              & ext.           &                 22 &  1.2 $\times 10^{17}$ &               22.2 &                  68\\
              & ext.           &                  3 &  6.2 $\times 10^{14}$ &                5.3 &                  61\\
              & ext.           &                 10 &  2.6 $\times 10^{14}$ &               10.4 &                  38\\
              & ext.           &                 21 &  6.7 $\times 10^{15}$ &                2.0 &                  11\\
              & ext.           &                 10 &  2.5 $\times 10^{14}$ &                1.7 &                   6\\
              & ext.           &                 26 &  1.3 $\times 10^{16}$ &                7.1 &                   2\\
\end{supertabular}\\
\vspace{1cm}

%================================================================================
%
% Source A08

%---------------------------------------
% Core Components

\tablefirsthead{%
\hline
\hline
Molecule      & $\theta^{m,c}$ & T$_{\rm ex}^{m,c}$ & N$_{\rm tot}^{m,c}$   & $\Delta$ v$^{m,c}$ & v$_{\rm LSR}^{m,c}$\\
              & ($\arcsec$)    & (K)                & (cm$^{-2}$)           & (km~s$^{-1}$)      & (km~s$^{-1}$)      \\
\hline
}

\tablehead{%
\multicolumn{6}{c}{(Continued)}\\
\hline
\hline
Molecule      & $\theta^{m,c}$ & T$_{\rm ex}^{m,c}$ & N$_{\rm tot}^{m,c}$   & $\Delta$ v$^{m,c}$ & v$_{\rm LSR}^{m,c}$\\
              & ($\arcsec$)    & (K)                & (cm$^{-2}$)           & (km~s$^{-1}$)      & (km~s$^{-1}$)      \\
\hline
}

\tabletail{%
\hline
\hline
}

\topcaption{LTE Parameters for the full LTE model (Core Components) for source A08 in Sgr~B2(N).}
\tiny
\centering
\begin{supertabular}{lcccC{1cm}C{1cm}}\label{CoreLTE:parameters:A08SgrB2N}\\
HNC, v$_2$=1  &    1.3         &                704 &  9.8 $\times 10^{14}$ &                6.2 &                  74\\
CH$_3$NH$_2$  &    1.3         &                176 &  1.1 $\times 10^{17}$ &                6.0 &                  73\\
SO$_2$        &    1.3         &                213 &  2.7 $\times 10^{15}$ &                4.1 &                  56\\
H$_2$CO       &    1.3         &                185 &  6.5 $\times 10^{16}$ &                3.5 &                  72\\
NS            &    1.3         &                234 &  9.4 $\times 10^{14}$ &                5.8 &                  72\\
              &    1.3         &                209 &  3.9 $\times 10^{14}$ &                5.0 &                  80\\
CH$_3$CHO     &    1.3         &                178 &  3.1 $\times 10^{16}$ &                5.5 &                  72\\
\end{supertabular}\\
\vspace{1cm}

%---------------------------------------
% Envelope Components

\tablefirsthead{%
\hline
\hline
Molecule      & $\theta^{m,c}$ & T$_{\rm ex}^{m,c}$ & N$_{\rm tot}^{m,c}$   & $\Delta$ v$^{m,c}$ & v$_{\rm LSR}^{m,c}$\\
              & ($\arcsec$)    & (K)                & (cm$^{-2}$)           & (km~s$^{-1}$)      & (km~s$^{-1}$)      \\
\hline
}

\tablehead{%
\multicolumn{6}{c}{(Continued)}\\
\hline
\hline
Molecule      & $\theta^{m,c}$ & T$_{\rm ex}^{m,c}$ & N$_{\rm tot}^{m,c}$   & $\Delta$ v$^{m,c}$ & v$_{\rm LSR}^{m,c}$\\
              & ($\arcsec$)    & (K)                & (cm$^{-2}$)           & (km~s$^{-1}$)      & (km~s$^{-1}$)      \\
\hline
}

\tabletail{%
\hline
\hline
}

\topcaption{LTE Parameters for the full LTE model (Envelope Components) for source A08 in Sgr~B2(N).}
\tiny
\centering
\begin{supertabular}{lcccC{1cm}C{1cm}}\label{EnvLTE:parameters:A08SgrB2N}\\
CN            & ext.           &                  5 &  3.6 $\times 10^{14}$ &               30.7 &                  68\\
              & ext.           &                  4 &  1.0 $\times 10^{12}$ &                7.4 &                  62\\
              & ext.           &                  4 &  7.2 $\times 10^{12}$ &               10.7 &                  43\\
              & ext.           &                  4 &  5.0 $\times 10^{13}$ &               18.9 &                  44\\
              & ext.           &                  4 &  4.8 $\times 10^{12}$ &                9.7 &                -104\\
              & ext.           &                  4 &  1.3 $\times 10^{12}$ &               13.9 &                  13\\
              & ext.           &                  4 &  2.5 $\times 10^{12}$ &               10.6 &                   0\\
              & ext.           &                  4 &  7.8 $\times 10^{12}$ &                0.9 &                 -47\\
              & ext.           &                  4 &  1.2 $\times 10^{12}$ &                3.9 &                 -28\\
              & ext.           &                  4 &  1.9 $\times 10^{12}$ &                1.4 &                -139\\
              & ext.           &                  4 &  1.1 $\times 10^{12}$ &                0.2 &                -102\\
SiO           & ext.           &                 14 &  1.6 $\times 10^{14}$ &                6.9 &                  66\\
H$_2$CO       & ext.           &                 18 &  1.7 $\times 10^{15}$ &                8.6 &                  72\\
              & ext.           &                 10 &  1.1 $\times 10^{14}$ &                8.1 &                  64\\
CCH           & ext.           &                  3 &  2.6 $\times 10^{14}$ &                4.0 &                  69\\
              & ext.           &                  3 &  1.0 $\times 10^{15}$ &                4.1 &                  57\\
H$_2$C$^{33}$S & ext.           &                131 &  1.8 $\times 10^{15}$ &                5.7 &                  65\\
PN            & ext.           &                  3 &  1.4 $\times 10^{14}$ &                1.2 &                 -33\\
              & ext.           &                  3 &  3.1 $\times 10^{14}$ &                1.2 &                  91\\
              & ext.           &                  3 &  1.4 $\times 10^{15}$ &                3.7 &                  68\\
              & ext.           &                  3 &  1.7 $\times 10^{15}$ &                0.9 &                 -55\\
              & ext.           &                  3 &  5.8 $\times 10^{17}$ &                1.4 &                 -84\\
              & ext.           &                  3 &  3.8 $\times 10^{13}$ &                2.0 &                 -84\\
PH$_3$        & ext.           &                 10 &  1.9 $\times 10^{14}$ &                7.2 &                  64\\
CO            & ext.           &                  3 &  2.2 $\times 10^{15}$ &                1.5 &                 126\\
              & ext.           &                  3 &  7.5 $\times 10^{16}$ &               33.9 &                  88\\
              & ext.           &                  3 &  4.0 $\times 10^{16}$ &                2.7 &                  80\\
              & ext.           &                  3 &  3.0 $\times 10^{16}$ &               10.9 &                  56\\
              & ext.           &                  3 &  1.9 $\times 10^{16}$ &               12.4 &                  26\\
              & ext.           &                  3 &  1.9 $\times 10^{14}$ &                2.7 &                  20\\
              & ext.           &                  3 &  4.1 $\times 10^{16}$ &               18.2 &                   4\\
              & ext.           &                  3 &  3.1 $\times 10^{16}$ &               16.2 &                 -24\\
              & ext.           &                  3 &  3.0 $\times 10^{15}$ &                0.5 &                 -40\\
              & ext.           &                  3 &  2.1 $\times 10^{16}$ &               11.2 &                 -47\\
              & ext.           &                  3 &  1.0 $\times 10^{15}$ &                1.0 &                 -50\\
              & ext.           &                  3 &  1.4 $\times 10^{16}$ &                6.6 &                 -74\\
              & ext.           &                  3 &  9.6 $\times 10^{15}$ &                5.1 &                 -80\\
              & ext.           &                  3 &  4.6 $\times 10^{15}$ &                3.9 &                 -88\\
              & ext.           &                  3 &  6.1 $\times 10^{15}$ &                3.5 &                 -94\\
              & ext.           &                  3 &  2.5 $\times 10^{16}$ &                9.4 &                -108\\
              & ext.           &                  3 &  6.6 $\times 10^{15}$ &                5.4 &                -118\\
              & ext.           &                  3 &  1.3 $\times 10^{15}$ &                1.9 &                -127\\
$^{13}$CO     & ext.           &                  3 &  2.7 $\times 10^{15}$ &                1.8 &                  91\\
              & ext.           &                  3 &  4.9 $\times 10^{14}$ &                2.2 &                  79\\
              & ext.           &                  3 &  8.2 $\times 10^{16}$ &               19.5 &                  70\\
              & ext.           &                  3 &  2.4 $\times 10^{15}$ &                1.5 &                  58\\
              & ext.           &                  3 &  2.4 $\times 10^{12}$ &                1.1 &                  20\\
              & ext.           &                  3 &  2.7 $\times 10^{15}$ &                2.0 &                   1\\
              & ext.           &                  3 &  9.4 $\times 10^{15}$ &                2.6 &                 -42\\
              & ext.           &                  3 &  8.9 $\times 10^{14}$ &                1.1 &                 -51\\
              & ext.           &                  3 &  2.6 $\times 10^{15}$ &                6.2 &                 -56\\
              & ext.           &                  3 &  1.8 $\times 10^{15}$ &                2.8 &                 -82\\
              & ext.           &                  3 &  1.2 $\times 10^{15}$ &                1.0 &                -105\\
              & ext.           &                  3 &  3.5 $\times 10^{15}$ &                2.9 &                -129\\
              & ext.           &                  3 &  7.9 $\times 10^{13}$ &                1.0 &                -152\\
C$^{17}$O     & ext.           &                  3 &  1.0 $\times 10^{15}$ &                2.8 &                -180\\
              & ext.           &                  3 &  1.0 $\times 10^{15}$ &                7.9 &                -105\\
              & ext.           &                  3 &  1.5 $\times 10^{15}$ &                3.9 &                 -81\\
              & ext.           &                  3 &  1.2 $\times 10^{15}$ &                5.1 &                 -63\\
              & ext.           &                  3 &  3.1 $\times 10^{15}$ &               14.0 &                 -33\\
              & ext.           &                  3 &  3.4 $\times 10^{14}$ &                3.1 &                 -20\\
              & ext.           &                  3 &  1.1 $\times 10^{15}$ &                4.6 &                 -12\\
              & ext.           &                  3 &  1.3 $\times 10^{15}$ &                4.2 &                  12\\
              & ext.           &                  3 &  1.4 $\times 10^{15}$ &                4.8 &                  28\\
              & ext.           &                  3 &  2.7 $\times 10^{15}$ &               20.5 &                  47\\
              & ext.           &                  3 &  1.0 $\times 10^{14}$ &                2.7 &                  63\\
              & ext.           &                  3 &  6.7 $\times 10^{15}$ &                4.0 &                  77\\
              & ext.           &                  3 &  1.3 $\times 10^{15}$ &                8.8 &                 105\\
              & ext.           &                  3 &  3.6 $\times 10^{14}$ &                2.5 &                 121\\
              & ext.           &                  3 &  1.0 $\times 10^{15}$ &                3.6 &                 127\\
              & ext.           &                  3 &  9.4 $\times 10^{14}$ &                3.3 &                 137\\
              & ext.           &                  3 &  2.6 $\times 10^{14}$ &                2.1 &                 142\\
              & ext.           &                  3 &  1.7 $\times 10^{15}$ &                3.8 &                 158\\
C$^{18}$O     & ext.           &                  3 &  6.9 $\times 10^{14}$ &                7.5 &                 -90\\
              & ext.           &                  3 &  3.5 $\times 10^{15}$ &                8.1 &                 -71\\
              & ext.           &                  3 &  5.3 $\times 10^{15}$ &                7.9 &                  65\\
              & ext.           &                  3 &  2.2 $\times 10^{14}$ &               34.2 &                 131\\
              & ext.           &                  3 &  7.6 $\times 10^{14}$ &                2.7 &                 139\\
CS            & ext.           &                  3 &  1.4 $\times 10^{15}$ &                4.2 &                  93\\
              & ext.           &                  3 &  1.7 $\times 10^{16}$ &                5.6 &                  76\\
              & ext.           &                  3 &  5.2 $\times 10^{16}$ &               29.2 &                  64\\
              & ext.           &                  3 &  5.3 $\times 10^{15}$ &                4.5 &                  56\\
              & ext.           &                  3 &  1.0 $\times 10^{15}$ &                0.1 &                 -25\\
              & ext.           &                  3 &  2.7 $\times 10^{15}$ &                1.3 &                 -45\\
$^{13}$CS     & ext.           &                  3 &  6.9 $\times 10^{13}$ &               29.7 &                  93\\
              & ext.           &                  3 &  3.5 $\times 10^{15}$ &                1.1 &                  63\\
              & ext.           &                  3 &  1.7 $\times 10^{13}$ &                0.9 &                   7\\
              & ext.           &                  3 &  3.2 $\times 10^{14}$ &                2.6 &                 -57\\
              & ext.           &                  3 &  5.3 $\times 10^{13}$ &                2.0 &                 -97\\
C$^{33}$S     & ext.           &                  3 &  1.2 $\times 10^{16}$ &                6.4 &                -149\\
              & ext.           &                  3 &  4.5 $\times 10^{15}$ &                2.0 &                -118\\
              & ext.           &                  3 &  2.5 $\times 10^{16}$ &                6.4 &                -111\\
              & ext.           &                  3 &  9.9 $\times 10^{12}$ &                6.4 &                 -91\\
              & ext.           &                  3 &  1.8 $\times 10^{16}$ &               10.8 &                 -71\\
              & ext.           &                  3 &  3.5 $\times 10^{15}$ &                3.8 &                 -42\\
              & ext.           &                  3 &  4.3 $\times 10^{15}$ &                6.9 &                 -20\\
              & ext.           &                  3 &  6.2 $\times 10^{15}$ &                7.4 &                  35\\
              & ext.           &                  3 &  1.3 $\times 10^{15}$ &                2.5 &                  59\\
              & ext.           &                  3 &  2.2 $\times 10^{15}$ &                2.6 &                 116\\
              & ext.           &                  3 &  4.3 $\times 10^{15}$ &                2.4 &                 134\\
              & ext.           &                  3 &  4.4 $\times 10^{14}$ &                2.3 &                 145\\
              & ext.           &                  3 &  1.1 $\times 10^{16}$ &                5.8 &                 148\\
C$^{34}$S     & ext.           &                  3 &  2.6 $\times 10^{15}$ &                2.4 &                -179\\
              & ext.           &                  3 &  2.4 $\times 10^{15}$ &                2.6 &                 -97\\
              & ext.           &                  3 &  2.0 $\times 10^{15}$ &                3.2 &                 -12\\
              & ext.           &                  3 &  4.7 $\times 10^{16}$ &               16.4 &                  65\\
              & ext.           &                  3 &  6.5 $\times 10^{15}$ &                4.5 &                 129\\
HCN           & ext.           &                  3 &  6.6 $\times 10^{12}$ &                1.6 &                 112\\
              & ext.           &                  3 &  9.3 $\times 10^{12}$ &                0.6 &                  90\\
              & ext.           &                  3 &  4.1 $\times 10^{14}$ &               29.8 &                  85\\
              & ext.           &                  3 &  1.5 $\times 10^{12}$ &                0.8 &                  72\\
              & ext.           &                  3 &  1.2 $\times 10^{13}$ &               28.1 &                  67\\
              & ext.           &                  3 &  2.6 $\times 10^{14}$ &                7.3 &                  60\\
              & ext.           &                  3 &  1.6 $\times 10^{14}$ &                4.4 &                  50\\
              & ext.           &                  3 &  2.9 $\times 10^{13}$ &                0.8 &                  38\\
              & ext.           &                  3 &  6.2 $\times 10^{13}$ &                4.7 &                  -5\\
              & ext.           &                  3 &  5.0 $\times 10^{12}$ &                0.7 &                 -26\\
              & ext.           &                  3 &  1.2 $\times 10^{14}$ &               18.0 &                 -37\\
              & ext.           &                  3 &  1.2 $\times 10^{14}$ &                6.7 &                 -69\\
              & ext.           &                  3 &  2.6 $\times 10^{13}$ &               38.6 &                 -89\\
              & ext.           &                  3 &  8.6 $\times 10^{12}$ &                0.9 &                 -91\\
              & ext.           &                  3 &  4.0 $\times 10^{13}$ &               14.1 &                -161\\
              & ext.           &                  3 &  1.9 $\times 10^{12}$ &                0.1 &                -161\\
H$^{13}$CN    & ext.           &                  3 &  8.1 $\times 10^{13}$ &               13.8 &                 149\\
              & ext.           &                  3 &  2.2 $\times 10^{15}$ &                0.4 &                  78\\
              & ext.           &                  3 &  2.3 $\times 10^{14}$ &               26.9 &                  68\\
              & ext.           &                  3 &  5.2 $\times 10^{14}$ &               21.3 &                  64\\
              & ext.           &                  3 &  1.4 $\times 10^{12}$ &                1.0 &                -112\\
HNC           & ext.           &                  3 &  1.0 $\times 10^{12}$ &               33.3 &                 124\\
              & ext.           &                  3 &  1.9 $\times 10^{13}$ &                1.8 &                 118\\
              & ext.           &                  3 &  2.2 $\times 10^{13}$ &                2.0 &                 112\\
              & ext.           &                  3 &  5.3 $\times 10^{13}$ &                1.6 &                  86\\
              & ext.           &                  3 &  3.3 $\times 10^{13}$ &                2.0 &                  78\\
              & ext.           &                  3 &  8.7 $\times 10^{14}$ &               25.7 &                  72\\
              & ext.           &                  3 &  5.9 $\times 10^{13}$ &                4.1 &                  63\\
              & ext.           &                  3 &  6.0 $\times 10^{13}$ &                2.1 &                  55\\
              & ext.           &                  3 &  1.0 $\times 10^{14}$ &               11.9 &                  58\\
              & ext.           &                  3 &  1.5 $\times 10^{14}$ &                5.7 &                  41\\
              & ext.           &                  3 &  4.9 $\times 10^{13}$ &                3.2 &                  19\\
              & ext.           &                  3 &  1.3 $\times 10^{14}$ &               10.2 &                  -7\\
              & ext.           &                  3 &  2.1 $\times 10^{13}$ &                2.7 &                 -65\\
              & ext.           &                  3 &  1.0 $\times 10^{13}$ &                2.0 &                 -72\\
              & ext.           &                  3 &  1.4 $\times 10^{13}$ &                1.0 &                 -76\\
              & ext.           &                  3 &  1.3 $\times 10^{13}$ &                1.0 &                 -82\\
              & ext.           &                  3 &  3.4 $\times 10^{12}$ &                1.0 &                -107\\
H$^{13}$CO$^+$ & ext.           &                  3 &  1.2 $\times 10^{16}$ &                3.1 &                  66\\
              & ext.           &                  3 &  1.8 $\times 10^{12}$ &                2.4 &                 114\\
HO$^{13}$C$^+$ & ext.           &                  3 &  1.7 $\times 10^{13}$ &                3.2 &                  54\\
              & ext.           &                  3 &  1.1 $\times 10^{13}$ &                2.0 &                 -10\\
              & ext.           &                  3 &  1.2 $\times 10^{14}$ &                1.1 &                 -92\\
              & ext.           &                  3 &  5.0 $\times 10^{13}$ &                6.2 &                -122\\
\end{supertabular}\\
\vspace{1cm}

%================================================================================
%
% Source A09

%---------------------------------------
% Core Components

\tablefirsthead{%
\hline
\hline
Molecule      & $\theta^{m,c}$ & T$_{\rm ex}^{m,c}$ & N$_{\rm tot}^{m,c}$   & $\Delta$ v$^{m,c}$ & v$_{\rm LSR}^{m,c}$\\
              & ($\arcsec$)    & (K)                & (cm$^{-2}$)           & (km~s$^{-1}$)      & (km~s$^{-1}$)      \\
\hline
}

\tablehead{%
\multicolumn{6}{c}{(Continued)}\\
\hline
\hline
Molecule      & $\theta^{m,c}$ & T$_{\rm ex}^{m,c}$ & N$_{\rm tot}^{m,c}$   & $\Delta$ v$^{m,c}$ & v$_{\rm LSR}^{m,c}$\\
              & ($\arcsec$)    & (K)                & (cm$^{-2}$)           & (km~s$^{-1}$)      & (km~s$^{-1}$)      \\
\hline
}

\tabletail{%
\hline
\hline
}

\topcaption{LTE Parameters for the full LTE model (Core Components) for source A09 in Sgr~B2(N).}
\tiny
\centering
\begin{supertabular}{lcccC{1cm}C{1cm}}\label{CoreLTE:parameters:A09SgrB2N}\\
SiO           &    1.6         &                193 &  1.3 $\times 10^{13}$ &                5.1 &                 105\\
H$_2 \! ^{33}$S &    1.6         &                166 &  1.0 $\times 10^{12}$ &                2.8 &                  72\\
H$_2$CS       &    1.6         &                 45 &  1.7 $\times 10^{15}$ &                2.7 &                  59\\
NO            &    1.6         &                 87 &  1.0 $\times 10^{16}$ &                3.9 &                  86\\
NS            &    1.6         &                189 &  6.5 $\times 10^{12}$ &                2.8 &                  44\\
CCH           &    1.6         &                 27 &  6.1 $\times 10^{15}$ &                2.4 &                  60\\
PN            &    1.6         &                200 &  4.1 $\times 10^{13}$ &                2.7 &                 107\\
              &    1.6         &                200 &  8.5 $\times 10^{12}$ &                1.1 &                  56\\
              &    1.6         &                200 &  3.7 $\times 10^{13}$ &                5.5 &                  39\\
              &    1.6         &                200 &  2.6 $\times 10^{13}$ &                2.4 &                  17\\
              &    1.6         &                200 &  1.8 $\times 10^{13}$ &                1.4 &                  10\\
              &    1.6         &                200 &  2.7 $\times 10^{13}$ &                1.9 &                 -14\\
              &    1.6         &                200 &  2.0 $\times 10^{13}$ &                1.9 &                 -18\\
              &    1.6         &                200 &  3.8 $\times 10^{13}$ &                1.9 &                 -24\\
              &    1.6         &                200 &  7.9 $\times 10^{13}$ &                9.3 &                 -31\\
CO            &    1.6         &                200 &  3.2 $\times 10^{12}$ &                1.9 &                 134\\
              &    1.6         &                200 &  1.8 $\times 10^{17}$ &                6.4 &                 104\\
              &    1.6         &                200 &  8.1 $\times 10^{16}$ &                2.3 &                 -16\\
              &    1.6         &                200 &  2.9 $\times 10^{12}$ &                1.9 &                -127\\
              &    1.6         &                200 &  1.0 $\times 10^{12}$ &                1.9 &                -140\\
$^{13}$CO     &    1.6         &                200 &  4.3 $\times 10^{13}$ &                0.8 &                  34\\
              &    1.6         &                200 &  7.1 $\times 10^{15}$ &                4.6 &                 -10\\
              &    1.6         &                200 &  8.0 $\times 10^{17}$ &                1.0 &                 -74\\
              &    1.6         &                200 &  6.5 $\times 10^{14}$ &                1.0 &                 -99\\
              &    1.6         &                200 &  2.8 $\times 10^{16}$ &                1.2 &                -102\\
C$^{17}$O     &    1.6         &                200 &  6.8 $\times 10^{12}$ &                2.2 &                -167\\
              &    1.6         &                200 &  4.6 $\times 10^{13}$ &                4.1 &                -146\\
              &    1.6         &                200 &  1.8 $\times 10^{13}$ &                4.5 &                -127\\
              &    1.6         &                200 &  5.1 $\times 10^{12}$ &               14.0 &                -108\\
              &    1.6         &                200 &  1.8 $\times 10^{13}$ &                3.6 &                -103\\
              &    1.6         &                200 &  1.8 $\times 10^{13}$ &                1.9 &                 -92\\
              &    1.6         &                200 &  4.3 $\times 10^{16}$ &                5.8 &                 -73\\
              &    1.6         &                200 &  9.2 $\times 10^{15}$ &                3.1 &                 -52\\
              &    1.6         &                200 &  1.7 $\times 10^{16}$ &                4.5 &                 -47\\
              &    1.6         &                200 &  2.3 $\times 10^{15}$ &                0.6 &                 -39\\
              &    1.6         &                200 &  1.2 $\times 10^{16}$ &                3.9 &                 -28\\
              &    1.6         &                200 &  6.9 $\times 10^{15}$ &                2.9 &                 -22\\
              &    1.6         &                200 &  2.9 $\times 10^{16}$ &                6.6 &                  -4\\
              &    1.6         &                200 &  4.2 $\times 10^{15}$ &                1.9 &                   1\\
              &    1.6         &                200 &  4.2 $\times 10^{16}$ &                7.4 &                  17\\
              &    1.6         &                200 &  2.8 $\times 10^{16}$ &                7.6 &                  33\\
              &    1.6         &                200 &  1.3 $\times 10^{17}$ &                3.3 &                  60\\
              &    1.6         &                200 &  1.6 $\times 10^{17}$ &                9.3 &                  91\\
              &    1.6         &                200 &  4.3 $\times 10^{16}$ &               17.7 &                 112\\
              &    1.6         &                200 &  1.3 $\times 10^{16}$ &                5.0 &                 130\\
C$^{18}$O     &    1.6         &                200 &  1.0 $\times 10^{16}$ &                5.7 &                -156\\
              &    1.6         &                200 & 10.0 $\times 10^{14}$ &                2.7 &                -148\\
              &    1.6         &                200 &  1.0 $\times 10^{16}$ &                9.8 &                -106\\
              &    1.6         &                200 &  1.0 $\times 10^{16}$ &                7.5 &                 -77\\
              &    1.6         &                200 &  9.9 $\times 10^{14}$ &               45.5 &                 -63\\
              &    1.6         &                200 &  8.9 $\times 10^{15}$ &                2.5 &                 -47\\
              &    1.6         &                200 &  1.0 $\times 10^{16}$ &               11.3 &                 -36\\
              &    1.6         &                200 &  3.2 $\times 10^{15}$ &                3.2 &                 -25\\
              &    1.6         &                200 &  3.3 $\times 10^{15}$ &                2.7 &                  11\\
              &    1.6         &                200 &  3.2 $\times 10^{15}$ &                6.2 &                  16\\
              &    1.6         &                200 &  1.0 $\times 10^{16}$ &                7.2 &                  35\\
              &    1.6         &                200 &  1.3 $\times 10^{17}$ &                2.8 &                  60\\
              &    1.6         &                200 &  4.0 $\times 10^{16}$ &               13.0 &                  98\\
              &    1.6         &                200 &  2.0 $\times 10^{16}$ &                7.5 &                 114\\
              &    1.6         &                200 &  1.0 $\times 10^{15}$ &                2.7 &                 125\\
              &    1.6         &                200 &  1.0 $\times 10^{15}$ &                2.7 &                 155\\
              &    1.6         &                200 & 10.0 $\times 10^{14}$ &                2.7 &                 161\\
CS            &    1.6         &                200 &  5.5 $\times 10^{13}$ &                5.5 &                  84\\
              &    1.6         &                200 &  2.3 $\times 10^{14}$ &                1.9 &                  68\\
              &    1.6         &                200 &  3.3 $\times 10^{14}$ &                2.3 &                  58\\
              &    1.6         &                200 &  1.5 $\times 10^{14}$ &                8.3 &                  25\\
              &    1.6         &                200 &  1.9 $\times 10^{13}$ &                1.9 &                  14\\
              &    1.6         &                200 &  9.0 $\times 10^{12}$ &                1.8 &                 -27\\
              &    1.6         &                200 &  5.9 $\times 10^{13}$ &                2.0 &                 -72\\
              &    1.6         &                200 &  4.5 $\times 10^{13}$ &                0.5 &                 -79\\
              &    1.6         &                200 &  3.6 $\times 10^{13}$ &                1.5 &                 -84\\
              &    1.6         &                200 &  6.7 $\times 10^{13}$ &                4.2 &                 -98\\
              &    1.6         &                200 &  2.9 $\times 10^{13}$ &                4.3 &                -143\\
$^{13}$CS     &    1.6         &                200 &  1.3 $\times 10^{14}$ &               16.0 &                 107\\
              &    1.6         &                200 &  1.7 $\times 10^{13}$ &                1.0 &                  84\\
              &    1.6         &                200 &  1.0 $\times 10^{12}$ &                1.1 &                  58\\
              &    1.6         &                200 &  1.0 $\times 10^{12}$ &                1.0 &                  53\\
              &    1.6         &                200 &  3.0 $\times 10^{13}$ &                1.0 &                 -21\\
              &    1.6         &                200 &  5.0 $\times 10^{13}$ &                2.7 &                -177\\
C$^{33}$S     &    1.6         &                200 &  7.3 $\times 10^{12}$ &                1.7 &                -173\\
              &    1.6         &                200 &  2.4 $\times 10^{13}$ &                4.0 &                -161\\
              &    1.6         &                200 &  7.1 $\times 10^{12}$ &                3.0 &                -145\\
              &    1.6         &                200 &  2.4 $\times 10^{13}$ &                5.3 &                -128\\
              &    1.6         &                200 &  2.7 $\times 10^{14}$ &               20.8 &                 -22\\
              &    1.6         &                200 &  9.3 $\times 10^{13}$ &               14.9 &                 -17\\
              &    1.6         &                200 &  4.1 $\times 10^{13}$ &                2.3 &                  -3\\
              &    1.6         &                200 &  4.6 $\times 10^{14}$ &               18.6 &                  25\\
              &    1.6         &                200 &  1.9 $\times 10^{14}$ &                9.5 &                 101\\
              &    1.6         &                200 &  2.2 $\times 10^{14}$ &               20.0 &                 132\\
              &    1.6         &                200 &  5.7 $\times 10^{13}$ &                3.5 &                 155\\
C$^{34}$S     &    1.6         &                200 &  7.7 $\times 10^{13}$ &                6.8 &                 -97\\
              &    1.6         &                200 &  6.4 $\times 10^{12}$ &                3.5 &                 -87\\
              &    1.6         &                200 &  2.9 $\times 10^{13}$ &                3.0 &                 -75\\
              &    1.6         &                200 &  1.3 $\times 10^{13}$ &                2.7 &                 -65\\
              &    1.6         &                200 & 10.0 $\times 10^{14}$ &               48.0 &                 -50\\
              &    1.6         &                200 &  4.3 $\times 10^{13}$ &                3.1 &                 -22\\
              &    1.6         &                200 &  2.4 $\times 10^{14}$ &               12.2 &                  18\\
              &    1.6         &                200 &  1.1 $\times 10^{14}$ &               11.7 &                 118\\
H$^{13}$CN    &    1.6         &                200 &  2.0 $\times 10^{14}$ &               19.7 &                 125\\
              &    1.6         &                200 &  1.7 $\times 10^{12}$ &                1.0 &                  99\\
              &    1.6         &                200 &  4.7 $\times 10^{13}$ &                5.5 &                  17\\
              &    1.6         &                200 &  4.1 $\times 10^{13}$ &                1.7 &                  -4\\
              &    1.6         &                200 &  6.1 $\times 10^{13}$ &               10.2 &                 -19\\
              &    1.6         &                200 &  1.0 $\times 10^{12}$ &                0.4 &                 -33\\
              &    1.6         &                200 &  9.3 $\times 10^{13}$ &               12.1 &                 -33\\
              &    1.6         &                200 &  1.0 $\times 10^{12}$ &                0.8 &                 -32\\
              &    1.6         &                200 &  7.6 $\times 10^{12}$ &                1.7 &                 -49\\
              &    1.6         &                200 &  2.0 $\times 10^{13}$ &                2.2 &                 -96\\
              &    1.6         &                200 &  3.0 $\times 10^{13}$ &                3.4 &                -102\\
              &    1.6         &                200 &  1.9 $\times 10^{13}$ &                1.3 &                -111\\
              &    1.6         &                200 &  2.2 $\times 10^{13}$ &                2.3 &                -120\\
              &    1.6         &                200 &  2.8 $\times 10^{13}$ &                4.5 &                -126\\
              &    1.6         &                200 &  1.7 $\times 10^{13}$ &                2.2 &                -138\\
              &    1.6         &                200 &  5.8 $\times 10^{13}$ &                5.8 &                -146\\
              &    1.6         &                200 &  2.8 $\times 10^{13}$ &                4.8 &                -164\\
              &    1.6         &                200 &  2.0 $\times 10^{13}$ &                0.5 &                -154\\
              &    1.6         &                200 &  2.9 $\times 10^{13}$ &                2.0 &                -170\\
HNC           &    1.6         &                200 &  1.0 $\times 10^{13}$ &                1.3 &                 151\\
              &    1.6         &                200 &  1.5 $\times 10^{13}$ &                1.3 &                 146\\
              &    1.6         &                200 &  8.6 $\times 10^{12}$ &               18.2 &                 113\\
              &    1.6         &                200 &  1.6 $\times 10^{13}$ &                1.3 &                 102\\
              &    1.6         &                200 &  7.9 $\times 10^{12}$ &                2.7 &                  91\\
              &    1.6         &                200 &  2.1 $\times 10^{13}$ &                1.9 &                  51\\
              &    1.6         &                200 &  4.6 $\times 10^{12}$ &                1.9 &                  37\\
              &    1.6         &                200 &  9.5 $\times 10^{12}$ &                3.6 &                 -17\\
              &    1.6         &                200 &  1.5 $\times 10^{13}$ &                2.3 &                 -25\\
              &    1.6         &                200 &  2.1 $\times 10^{14}$ &                1.2 &                 -40\\
              &    1.6         &                200 &  2.0 $\times 10^{12}$ &                1.9 &                 -44\\
              &    1.6         &                200 &  4.0 $\times 10^{14}$ &                1.4 &                 -47\\
              &    1.6         &                200 &  1.8 $\times 10^{14}$ &                1.3 &                 -54\\
              &    1.6         &                200 &  1.8 $\times 10^{12}$ &                1.9 &                 -10\\
              &    1.6         &                200 &  1.3 $\times 10^{14}$ &               39.2 &                 -55\\
              &    1.6         &                200 &  2.7 $\times 10^{13}$ &                3.4 &                 -74\\
              &    1.6         &                200 &  1.4 $\times 10^{13}$ &                2.1 &                 -90\\
              &    1.6         &                200 &  1.4 $\times 10^{15}$ &                6.6 &                 -86\\
              &    1.6         &                200 &  1.3 $\times 10^{13}$ &                1.9 &                 -96\\
              &    1.6         &                200 &  2.5 $\times 10^{13}$ &                5.4 &                -101\\
              &    1.6         &                200 &  2.0 $\times 10^{12}$ &                1.1 &                -104\\
HN$^{13}$C    &    1.6         &                200 &  1.5 $\times 10^{12}$ &                3.6 &                -153\\
              &    1.6         &                200 &  1.3 $\times 10^{12}$ &                4.8 &                -115\\
              &    1.6         &                200 &  1.5 $\times 10^{14}$ &               48.7 &                 -91\\
              &    1.6         &                200 &  4.8 $\times 10^{13}$ &               17.8 &                 -26\\
H$^{13}$CO$^+$ &    1.6         &                  3 &  1.6 $\times 10^{12}$ &                1.4 &                  58\\
HO$^{13}$C$^+$ &    1.6         &                200 &  1.3 $\times 10^{13}$ &               16.4 &                 142\\
              &    1.6         &                200 &  1.1 $\times 10^{13}$ &                0.7 &                 139\\
\end{supertabular}\\
\vspace{1cm}

%---------------------------------------
% Envelope Components

\tablefirsthead{%
\hline
\hline
Molecule      & $\theta^{m,c}$ & T$_{\rm ex}^{m,c}$ & N$_{\rm tot}^{m,c}$   & $\Delta$ v$^{m,c}$ & v$_{\rm LSR}^{m,c}$\\
              & ($\arcsec$)    & (K)                & (cm$^{-2}$)           & (km~s$^{-1}$)      & (km~s$^{-1}$)      \\
\hline
}

\tablehead{%
\multicolumn{6}{c}{(Continued)}\\
\hline
\hline
Molecule      & $\theta^{m,c}$ & T$_{\rm ex}^{m,c}$ & N$_{\rm tot}^{m,c}$   & $\Delta$ v$^{m,c}$ & v$_{\rm LSR}^{m,c}$\\
              & ($\arcsec$)    & (K)                & (cm$^{-2}$)           & (km~s$^{-1}$)      & (km~s$^{-1}$)      \\
\hline
}

\tabletail{%
\hline
\hline
}

\topcaption{LTE Parameters for the full LTE model (Envelope Components) for source A09 in Sgr~B2(N).}
\tiny
\centering
\\
\vspace{1cm}

%================================================================================
%
% Source A10

%---------------------------------------
% Core Components

\tablefirsthead{%
\hline
\hline
Molecule      & $\theta^{m,c}$ & T$_{\rm ex}^{m,c}$ & N$_{\rm tot}^{m,c}$   & $\Delta$ v$^{m,c}$ & v$_{\rm LSR}^{m,c}$\\
              & ($\arcsec$)    & (K)                & (cm$^{-2}$)           & (km~s$^{-1}$)      & (km~s$^{-1}$)      \\
\hline
}

\tablehead{%
\multicolumn{6}{c}{(Continued)}\\
\hline
\hline
Molecule      & $\theta^{m,c}$ & T$_{\rm ex}^{m,c}$ & N$_{\rm tot}^{m,c}$   & $\Delta$ v$^{m,c}$ & v$_{\rm LSR}^{m,c}$\\
              & ($\arcsec$)    & (K)                & (cm$^{-2}$)           & (km~s$^{-1}$)      & (km~s$^{-1}$)      \\
\hline
}

\tabletail{%
\hline
\hline
}

\topcaption{LTE Parameters for the full LTE model (Core Components) for source A10 in Sgr~B2(N).}
\tiny
\centering
\\
\vspace{1cm}

%---------------------------------------
% Envelope Components

\tablefirsthead{%
\hline
\hline
Molecule      & $\theta^{m,c}$ & T$_{\rm ex}^{m,c}$ & N$_{\rm tot}^{m,c}$   & $\Delta$ v$^{m,c}$ & v$_{\rm LSR}^{m,c}$\\
              & ($\arcsec$)    & (K)                & (cm$^{-2}$)           & (km~s$^{-1}$)      & (km~s$^{-1}$)      \\
\hline
}

\tablehead{%
\multicolumn{6}{c}{(Continued)}\\
\hline
\hline
Molecule      & $\theta^{m,c}$ & T$_{\rm ex}^{m,c}$ & N$_{\rm tot}^{m,c}$   & $\Delta$ v$^{m,c}$ & v$_{\rm LSR}^{m,c}$\\
              & ($\arcsec$)    & (K)                & (cm$^{-2}$)           & (km~s$^{-1}$)      & (km~s$^{-1}$)      \\
\hline
}

\tabletail{%
\hline
\hline
}

\topcaption{LTE Parameters for the full LTE model (Envelope Components) for source A10 in Sgr~B2(N).}
\tiny
\centering
\begin{supertabular}{lcccC{1cm}C{1cm}}\label{EnvLTE:parameters:A10SgrB2N}\\
CN            & ext.           &                  4 &  7.6 $\times 10^{14}$ &               29.5 &                  68\\
              & ext.           &                  4 &  2.4 $\times 10^{14}$ &                4.8 &                  62\\
              & ext.           &                  3 &  1.6 $\times 10^{12}$ &                8.4 &                  19\\
              & ext.           &                  3 &  1.1 $\times 10^{14}$ &               19.9 &                  33\\
              & ext.           &                  3 &  2.5 $\times 10^{13}$ &                7.9 &                 -30\\
              & ext.           &                  3 &  1.6 $\times 10^{14}$ &               11.9 &                   4\\
              & ext.           &                  3 &  5.5 $\times 10^{13}$ &                8.7 &                 -16\\
              & ext.           &                  3 &  4.5 $\times 10^{13}$ &                3.6 &                 -43\\
              & ext.           &                  3 &  4.2 $\times 10^{13}$ &                7.8 &                 -50\\
              & ext.           &                  3 &  3.7 $\times 10^{13}$ &                1.5 &                 -81\\
              & ext.           &                  3 &  6.8 $\times 10^{13}$ &                2.4 &                -107\\
SiO           & ext.           &                 47 &  2.3 $\times 10^{16}$ &                3.0 &                  74\\
CH$_2$NH      & ext.           &                  3 &  3.6 $\times 10^{13}$ &                8.9 &                  74\\
              & ext.           &                  3 &  5.9 $\times 10^{13}$ &               15.2 &                  63\\
CCH           & ext.           &                  3 &  6.9 $\times 10^{14}$ &                2.6 &                  74\\
              & ext.           &                  3 &  1.0 $\times 10^{13}$ &                7.9 &                  65\\
PN            & ext.           &                  3 &  5.9 $\times 10^{14}$ &                1.2 &                  97\\
              & ext.           &                  3 &  4.9 $\times 10^{14}$ &                1.2 &                  74\\
              & ext.           &                  3 &  3.3 $\times 10^{15}$ &                7.5 &                  99\\
H$_2$CO       & ext.           &                 33 &  1.6 $\times 10^{16}$ &                7.1 &                  64\\
              & ext.           &                 28 &  1.8 $\times 10^{15}$ &                6.5 &                  81\\
              & ext.           &                 39 &  8.5 $\times 10^{15}$ &                3.6 &                  52\\
              & ext.           &                 30 &  1.0 $\times 10^{16}$ &                5.5 &                  75\\
CO            & ext.           &                  3 &  1.1 $\times 10^{12}$ &                1.5 &                 139\\
              & ext.           &                  3 &  1.2 $\times 10^{16}$ &               10.1 &                 113\\
              & ext.           &                  3 &  1.1 $\times 10^{12}$ &                1.9 &                  88\\
              & ext.           &                  3 &  4.1 $\times 10^{17}$ &                1.9 &                  78\\
              & ext.           &                  3 &  1.2 $\times 10^{17}$ &               27.0 &                  77\\
              & ext.           &                  3 &  1.1 $\times 10^{12}$ &                1.3 &                 -50\\
              & ext.           &                  3 &  5.8 $\times 10^{18}$ &                1.5 &                  54\\
              & ext.           &                  3 &  2.0 $\times 10^{16}$ &                1.0 &                  30\\
              & ext.           &                  3 &  3.2 $\times 10^{14}$ &                1.9 &                  25\\
              & ext.           &                  3 &  1.6 $\times 10^{12}$ &                1.9 &                  18\\
              & ext.           &                  3 &  1.2 $\times 10^{17}$ &               32.3 &                  12\\
              & ext.           &                  3 &  8.0 $\times 10^{15}$ &                4.6 &                  -3\\
              & ext.           &                  3 &  1.1 $\times 10^{12}$ &                1.9 &                 -15\\
              & ext.           &                  3 &  1.0 $\times 10^{16}$ &                1.9 &                 -20\\
              & ext.           &                  3 &  3.2 $\times 10^{16}$ &               10.6 &                 -26\\
              & ext.           &                  3 &  5.7 $\times 10^{16}$ &               16.7 &                 -46\\
              & ext.           &                  3 &  4.6 $\times 10^{14}$ &                1.9 &                 -63\\
              & ext.           &                  3 &  4.8 $\times 10^{16}$ &               14.1 &                 -78\\
              & ext.           &                  3 &  2.4 $\times 10^{14}$ &                1.9 &                 -81\\
              & ext.           &                  3 &  1.7 $\times 10^{16}$ &                1.9 &                 -95\\
              & ext.           &                  3 &  3.0 $\times 10^{16}$ &                6.6 &                -108\\
              & ext.           &                  3 &  8.7 $\times 10^{14}$ &                1.1 &                -129\\
$^{13}$CO     & ext.           &                  3 &  9.0 $\times 10^{13}$ &                6.2 &                  78\\
              & ext.           &                  3 &  9.4 $\times 10^{16}$ &               26.8 &                  65\\
              & ext.           &                  3 &  2.5 $\times 10^{12}$ &                0.6 &                  13\\
              & ext.           &                  3 &  7.6 $\times 10^{15}$ &                6.0 &                   9\\
              & ext.           &                  3 &  1.1 $\times 10^{12}$ &                9.8 &                 -17\\
              & ext.           &                  3 &  3.4 $\times 10^{15}$ &                1.0 &                 -39\\
              & ext.           &                  3 &  1.0 $\times 10^{14}$ &                1.0 &                 -50\\
              & ext.           &                  3 &  4.9 $\times 10^{15}$ &                0.9 &                 -55\\
              & ext.           &                  3 &  1.4 $\times 10^{16}$ &                1.1 &                 -72\\
              & ext.           &                  3 &  1.3 $\times 10^{14}$ &                1.5 &                 -95\\
              & ext.           &                  3 &  2.5 $\times 10^{15}$ &                3.1 &                -107\\
              & ext.           &                  3 &  1.9 $\times 10^{12}$ &                1.0 &                -145\\
              & ext.           &                  3 &  1.3 $\times 10^{14}$ &                0.9 &                -152\\
C$^{17}$O     & ext.           &                  3 &  5.4 $\times 10^{14}$ &                2.7 &                -115\\
              & ext.           &                  3 &  6.7 $\times 10^{14}$ &                2.7 &                 -73\\
              & ext.           &                  3 &  5.0 $\times 10^{14}$ &               20.9 &                 -52\\
              & ext.           &                  3 &  6.8 $\times 10^{14}$ &                3.2 &                  36\\
              & ext.           &                  3 &  8.8 $\times 10^{14}$ &                2.7 &                  55\\
              & ext.           &                  3 &  1.8 $\times 10^{15}$ &                7.2 &                 122\\
C$^{18}$O     & ext.           &                  3 &  9.8 $\times 10^{15}$ &                9.7 &                -179\\
              & ext.           &                  3 &  2.1 $\times 10^{15}$ &                6.7 &                 -85\\
              & ext.           &                  3 &  1.6 $\times 10^{15}$ &                7.3 &                 -14\\
              & ext.           &                  3 &  1.8 $\times 10^{15}$ &                2.7 &                  11\\
              & ext.           &                  3 &  2.0 $\times 10^{16}$ &               12.1 &                  72\\
              & ext.           &                  3 &  1.8 $\times 10^{14}$ &                2.8 &                 109\\
              & ext.           &                  3 &  2.3 $\times 10^{15}$ &                7.5 &                 144\\
              & ext.           &                  3 &  2.1 $\times 10^{14}$ &                2.8 &                 158\\
CS            & ext.           &                  3 &  2.6 $\times 10^{15}$ &                2.2 &                  85\\
              & ext.           &                  3 &  1.0 $\times 10^{17}$ &               21.9 &                  71\\
              & ext.           &                  3 &  1.1 $\times 10^{14}$ &                2.1 &                  66\\
              & ext.           &                  3 &  5.2 $\times 10^{15}$ &                1.0 &                  53\\
              & ext.           &                  3 &  3.9 $\times 10^{15}$ &                2.0 &                  41\\
              & ext.           &                  3 &  8.5 $\times 10^{12}$ &                3.5 &                 -44\\
              & ext.           &                  3 &  3.4 $\times 10^{15}$ &                2.2 &                 -63\\
              & ext.           &                  3 &  5.8 $\times 10^{15}$ &                2.1 &                 -77\\
              & ext.           &                  3 &  4.5 $\times 10^{15}$ &                1.0 &                -116\\
$^{13}$CS     & ext.           &                  3 &  2.1 $\times 10^{15}$ &                2.4 &                 140\\
              & ext.           &                  3 &  1.1 $\times 10^{12}$ &                1.9 &                 121\\
              & ext.           &                  3 &  1.6 $\times 10^{12}$ &                1.2 &                  67\\
              & ext.           &                  3 &  3.1 $\times 10^{13}$ &                1.9 &                 -28\\
              & ext.           &                  3 &  5.1 $\times 10^{13}$ &                1.0 &                 -53\\
              & ext.           &                  3 &  1.1 $\times 10^{14}$ &               30.8 &                 -67\\
              & ext.           &                  3 &  1.6 $\times 10^{13}$ &                1.9 &                -110\\
              & ext.           &                  3 &  1.3 $\times 10^{16}$ &                0.9 &                 -83\\
C$^{33}$S     & ext.           &                  3 &  4.6 $\times 10^{15}$ &                2.5 &                -135\\
              & ext.           &                  3 &  2.2 $\times 10^{15}$ &                2.5 &                -121\\
              & ext.           &                  3 &  2.8 $\times 10^{15}$ &                3.1 &                 -52\\
              & ext.           &                  3 &  3.4 $\times 10^{15}$ &               19.2 &                 -47\\
              & ext.           &                  3 &  3.0 $\times 10^{14}$ &                2.5 &                  14\\
              & ext.           &                  3 &  1.2 $\times 10^{15}$ &                2.5 &                  47\\
              & ext.           &                  3 &  1.3 $\times 10^{15}$ &                5.8 &                  72\\
              & ext.           &                  3 &  7.7 $\times 10^{15}$ &               17.6 &                 137\\
              & ext.           &                  3 &  2.1 $\times 10^{15}$ &                2.5 &                 142\\
C$^{34}$S     & ext.           &                  3 &  9.1 $\times 10^{15}$ &                5.8 &                  41\\
HCN           & ext.           &                  3 &  1.7 $\times 10^{13}$ &                1.9 &                  87\\
              & ext.           &                  3 &  3.1 $\times 10^{14}$ &                5.8 &                  79\\
              & ext.           &                  3 &  6.2 $\times 10^{12}$ &                2.4 &                  63\\
              & ext.           &                  3 &  1.0 $\times 10^{14}$ &                8.1 &                  50\\
              & ext.           &                  3 &  4.8 $\times 10^{13}$ &                6.9 &                  35\\
              & ext.           &                  3 &  5.8 $\times 10^{13}$ &                4.7 &                  30\\
              & ext.           &                  3 &  2.6 $\times 10^{12}$ &                1.1 &                  17\\
              & ext.           &                  3 &  5.2 $\times 10^{13}$ &                5.4 &                  -9\\
              & ext.           &                  3 &  1.1 $\times 10^{12}$ &                1.9 &                  61\\
              & ext.           &                  3 &  4.1 $\times 10^{13}$ &                2.2 &                 -69\\
              & ext.           &                  3 &  1.1 $\times 10^{14}$ &                1.6 &                 -76\\
              & ext.           &                  3 &  1.8 $\times 10^{13}$ &                1.2 &                -106\\
              & ext.           &                  3 &  1.3 $\times 10^{12}$ &                0.9 &                -146\\
              & ext.           &                  3 &  1.5 $\times 10^{12}$ &                1.1 &                -178\\
H$^{13}$CN    & ext.           &                  3 &  1.0 $\times 10^{12}$ &               27.6 &                 132\\
              & ext.           &                  3 &  2.0 $\times 10^{14}$ &                3.1 &                 102\\
              & ext.           &                  3 &  1.1 $\times 10^{12}$ &                0.1 &                  79\\
              & ext.           &                  3 &  2.3 $\times 10^{14}$ &               10.0 &                  75\\
              & ext.           &                  3 &  1.6 $\times 10^{14}$ &                4.1 &                  70\\
              & ext.           &                  3 &  4.4 $\times 10^{14}$ &                0.1 &                  68\\
              & ext.           &                  3 &  1.5 $\times 10^{14}$ &                3.2 &                  64\\
              & ext.           &                  3 &  1.1 $\times 10^{14}$ &                2.3 &                  47\\
              & ext.           &                  3 &  3.9 $\times 10^{13}$ &                1.3 &                  -2\\
HNC           & ext.           &                  3 &  8.5 $\times 10^{14}$ &               21.4 &                 113\\
              & ext.           &                  3 &  2.4 $\times 10^{13}$ &                2.2 &                  53\\
              & ext.           &                  3 &  1.7 $\times 10^{12}$ &                2.6 &                  30\\
              & ext.           &                  3 &  6.2 $\times 10^{12}$ &                1.9 &                 -72\\
H$^{13}$CO$^+$ & ext.           &                  3 &  2.9 $\times 10^{14}$ &               13.5 &                  70\\
\end{supertabular}\\
\vspace{1cm}

%================================================================================
%
% Source A11

%---------------------------------------
% Core Components

\tablefirsthead{%
\hline
\hline
Molecule      & $\theta^{m,c}$ & T$_{\rm ex}^{m,c}$ & N$_{\rm tot}^{m,c}$   & $\Delta$ v$^{m,c}$ & v$_{\rm LSR}^{m,c}$\\
              & ($\arcsec$)    & (K)                & (cm$^{-2}$)           & (km~s$^{-1}$)      & (km~s$^{-1}$)      \\
\hline
}

\tablehead{%
\multicolumn{6}{c}{(Continued)}\\
\hline
\hline
Molecule      & $\theta^{m,c}$ & T$_{\rm ex}^{m,c}$ & N$_{\rm tot}^{m,c}$   & $\Delta$ v$^{m,c}$ & v$_{\rm LSR}^{m,c}$\\
              & ($\arcsec$)    & (K)                & (cm$^{-2}$)           & (km~s$^{-1}$)      & (km~s$^{-1}$)      \\
\hline
}

\tabletail{%
\hline
\hline
}

\topcaption{LTE Parameters for the full LTE model (Core Components) for source A11 in Sgr~B2(N).}
\tiny
\centering
\\
\vspace{1cm}

%---------------------------------------
% Envelope Components

\tablefirsthead{%
\hline
\hline
Molecule      & $\theta^{m,c}$ & T$_{\rm ex}^{m,c}$ & N$_{\rm tot}^{m,c}$   & $\Delta$ v$^{m,c}$ & v$_{\rm LSR}^{m,c}$\\
              & ($\arcsec$)    & (K)                & (cm$^{-2}$)           & (km~s$^{-1}$)      & (km~s$^{-1}$)      \\
\hline
}

\tablehead{%
\multicolumn{6}{c}{(Continued)}\\
\hline
\hline
Molecule      & $\theta^{m,c}$ & T$_{\rm ex}^{m,c}$ & N$_{\rm tot}^{m,c}$   & $\Delta$ v$^{m,c}$ & v$_{\rm LSR}^{m,c}$\\
              & ($\arcsec$)    & (K)                & (cm$^{-2}$)           & (km~s$^{-1}$)      & (km~s$^{-1}$)      \\
\hline
}

\tabletail{%
\hline
\hline
}

\topcaption{LTE Parameters for the full LTE model (Envelope Components) for source A11 in Sgr~B2(N).}
\tiny
\centering
\\
\vspace{1cm}

%================================================================================
%
% Source A12

%---------------------------------------
% Core Components

\tablefirsthead{%
\hline
\hline
Molecule      & $\theta^{m,c}$ & T$_{\rm ex}^{m,c}$ & N$_{\rm tot}^{m,c}$   & $\Delta$ v$^{m,c}$ & v$_{\rm LSR}^{m,c}$\\
              & ($\arcsec$)    & (K)                & (cm$^{-2}$)           & (km~s$^{-1}$)      & (km~s$^{-1}$)      \\
\hline
}

\tablehead{%
\multicolumn{6}{c}{(Continued)}\\
\hline
\hline
Molecule      & $\theta^{m,c}$ & T$_{\rm ex}^{m,c}$ & N$_{\rm tot}^{m,c}$   & $\Delta$ v$^{m,c}$ & v$_{\rm LSR}^{m,c}$\\
              & ($\arcsec$)    & (K)                & (cm$^{-2}$)           & (km~s$^{-1}$)      & (km~s$^{-1}$)      \\
\hline
}

\tabletail{%
\hline
\hline
}

\topcaption{LTE Parameters for the full LTE model (Core Components) for source A12 in Sgr~B2(N).}
\tiny
\centering
\begin{supertabular}{lcccC{1cm}C{1cm}}\label{CoreLTE:parameters:A12SgrB2N}\\
HCCCN         &    1.7         &                152 &  1.1 $\times 10^{14}$ &                4.7 &                  64\\
H$_2$CCO      &    1.7         &                201 &  1.4 $\times 10^{13}$ &                5.0 &                  15\\
HNCO          &    1.7         &                207 &  1.1 $\times 10^{13}$ &                3.9 &                 107\\
H$_2$CS       &    1.7         &                193 &  2.1 $\times 10^{14}$ &                3.8 &                  64\\
NS            &    1.7         &                214 &  2.0 $\times 10^{15}$ &                9.6 &                  47\\
CH$_3$CN      &    1.7         &                324 &  1.0 $\times 10^{12}$ &                5.4 &                  65\\
NO$^+$        &    1.7         &                201 &  2.0 $\times 10^{15}$ &                2.6 &                  75\\
PN            &    1.7         &                200 &  1.1 $\times 10^{14}$ &                7.0 &                 132\\
              &    1.7         &                200 &  1.5 $\times 10^{13}$ &                1.1 &                 125\\
              &    1.7         &                200 &  3.9 $\times 10^{13}$ &                3.1 &                 107\\
              &    1.7         &                200 &  8.2 $\times 10^{13}$ &                5.5 &                  90\\
              &    1.7         &                200 &  4.8 $\times 10^{13}$ &                4.4 &                  81\\
              &    1.7         &                200 &  6.8 $\times 10^{12}$ &                0.9 &                  74\\
              &    1.7         &                200 &  7.0 $\times 10^{12}$ &                1.3 &                  86\\
              &    1.7         &                200 &  1.9 $\times 10^{13}$ &                1.9 &                  62\\
              &    1.7         &                200 &  1.4 $\times 10^{13}$ &                3.0 &                  56\\
              &    1.7         &                200 &  1.5 $\times 10^{13}$ &                1.2 &                  49\\
              &    1.7         &                200 &  4.9 $\times 10^{13}$ &                3.7 &                  43\\
              &    1.7         &                200 &  4.1 $\times 10^{13}$ &                2.5 &                  35\\
              &    1.7         &                200 &  9.9 $\times 10^{12}$ &                1.3 &                  23\\
              &    1.7         &                200 &  2.2 $\times 10^{13}$ &                3.4 &                  19\\
              &    1.7         &                200 &  1.1 $\times 10^{14}$ &               21.3 &                 157\\
              &    1.7         &                200 &  2.0 $\times 10^{12}$ &                0.2 &                  -6\\
              &    1.7         &                200 &  1.8 $\times 10^{13}$ &                5.2 &                  -9\\
              &    1.7         &                200 &  1.8 $\times 10^{14}$ &               22.7 &                 -11\\
              &    1.7         &                200 &  6.2 $\times 10^{12}$ &                0.9 &                 -25\\
              &    1.7         &                200 &  3.0 $\times 10^{13}$ &                1.4 &                 -31\\
              &    1.7         &                200 &  6.0 $\times 10^{12}$ &                0.8 &                 -35\\
              &    1.7         &                200 &  1.2 $\times 10^{13}$ &                1.2 &                 -45\\
              &    1.7         &                200 &  8.8 $\times 10^{12}$ &                0.9 &                 -51\\
              &    1.7         &                200 &  5.0 $\times 10^{13}$ &                3.7 &                 -57\\
              &    1.7         &                200 &  5.5 $\times 10^{13}$ &                5.1 &                 -66\\
              &    1.7         &                200 &  2.0 $\times 10^{13}$ &                1.2 &                 -74\\
              &    1.7         &                200 &  8.9 $\times 10^{13}$ &                8.4 &                 -86\\
              &    1.7         &                200 &  1.0 $\times 10^{12}$ &                0.2 &                 -87\\
              &    1.7         &                200 &  4.3 $\times 10^{13}$ &                3.2 &                 -90\\
OCS           &    1.7         &                 21 &  3.7 $\times 10^{16}$ &                1.0 &                  64\\
SiO           &    1.7         &                400 &  2.0 $\times 10^{14}$ &               11.0 &                  86\\
              &    1.7         &                180 &  3.0 $\times 10^{13}$ &                2.2 &                  79\\
CO            &    1.7         &                200 &  5.9 $\times 10^{16}$ &                4.5 &                 100\\
              &    1.7         &                200 &  4.3 $\times 10^{16}$ &                1.9 &                  89\\
              &    1.7         &                200 &  1.0 $\times 10^{16}$ &                0.4 &                  76\\
              &    1.7         &                200 &  8.5 $\times 10^{15}$ &                1.0 &                  74\\
              &    1.7         &                200 &  1.8 $\times 10^{16}$ &                1.2 &                  51\\
              &    1.7         &                200 &  7.6 $\times 10^{15}$ &                0.7 &                  46\\
              &    1.7         &                200 &  6.0 $\times 10^{15}$ &                0.9 &                  33\\
              &    1.7         &                200 &  1.8 $\times 10^{12}$ &                0.9 &                 133\\
              &    1.7         &                200 &  7.0 $\times 10^{15}$ &                0.4 &                  -6\\
              &    1.7         &                200 &  9.3 $\times 10^{15}$ &                0.9 &                 -86\\
              &    1.7         &                200 &  7.8 $\times 10^{13}$ &                1.0 &                 -75\\
              &    1.7         &                200 &  2.0 $\times 10^{16}$ &                2.5 &                -116\\
$^{13}$CO     &    1.7         &                200 &  2.1 $\times 10^{16}$ &                2.0 &                 154\\
              &    1.7         &                200 &  1.9 $\times 10^{16}$ &                3.1 &                 129\\
              &    1.7         &                200 &  2.2 $\times 10^{16}$ &                2.0 &                  23\\
              &    1.7         &                200 &  2.4 $\times 10^{16}$ &                4.4 &                -137\\
              &    1.7         &                200 &  1.4 $\times 10^{16}$ &                2.5 &                -167\\
              &    1.7         &                200 &  5.5 $\times 10^{15}$ &                1.3 &                -177\\
C$^{17}$O     &    1.7         &                200 &  9.5 $\times 10^{15}$ &                2.7 &                -164\\
              &    1.7         &                200 &  1.2 $\times 10^{16}$ &                3.3 &                -159\\
              &    1.7         &                200 &  9.5 $\times 10^{15}$ &                2.3 &                -143\\
              &    1.7         &                200 &  3.0 $\times 10^{16}$ &               11.5 &                -138\\
              &    1.7         &                200 &  9.5 $\times 10^{15}$ &                2.2 &                -119\\
              &    1.7         &                200 &  3.0 $\times 10^{16}$ &               10.5 &                -111\\
              &    1.7         &                200 &  2.7 $\times 10^{16}$ &                9.6 &                 -82\\
              &    1.7         &                200 &  4.2 $\times 10^{16}$ &               13.2 &                 -60\\
              &    1.7         &                200 &  1.9 $\times 10^{16}$ &                5.6 &                 -47\\
              &    1.7         &                200 &  5.5 $\times 10^{16}$ &               14.6 &                 -34\\
              &    1.7         &                200 &  7.5 $\times 10^{15}$ &                7.1 &                 -18\\
              &    1.7         &                200 &  5.1 $\times 10^{16}$ &                9.4 &                 -12\\
              &    1.7         &                200 &  3.1 $\times 10^{15}$ &                2.2 &                  -4\\
              &    1.7         &                200 &  2.8 $\times 10^{16}$ &                5.3 &                  14\\
              &    1.7         &                200 &  6.8 $\times 10^{16}$ &               16.7 &                  33\\
              &    1.7         &                200 &  9.9 $\times 10^{15}$ &                3.0 &                  47\\
              &    1.7         &                200 &  1.9 $\times 10^{17}$ &                5.2 &                  63\\
              &    1.7         &                200 &  7.3 $\times 10^{16}$ &               13.4 &                 106\\
              &    1.7         &                200 &  2.9 $\times 10^{16}$ &                6.3 &                 134\\
              &    1.7         &                200 &  1.6 $\times 10^{16}$ &                2.7 &                 145\\
C$^{18}$O     &    1.7         &                200 &  3.2 $\times 10^{16}$ &               10.7 &                -175\\
              &    1.7         &                200 &  1.0 $\times 10^{16}$ &                3.1 &                -148\\
              &    1.7         &                200 &  1.0 $\times 10^{17}$ &               15.2 &                -143\\
              &    1.7         &                200 &  1.0 $\times 10^{16}$ &                9.4 &                -118\\
              &    1.7         &                200 &  1.0 $\times 10^{15}$ &                2.4 &                -107\\
              &    1.7         &                200 &  3.2 $\times 10^{16}$ &                7.5 &                -101\\
              &    1.7         &                200 &  3.2 $\times 10^{15}$ &                2.7 &                 -90\\
              &    1.7         &                200 &  3.2 $\times 10^{15}$ &                2.6 &                 -80\\
              &    1.7         &                200 &  1.0 $\times 10^{16}$ &                3.6 &                 -74\\
              &    1.7         &                200 &  5.8 $\times 10^{15}$ &                2.1 &                 -66\\
              &    1.7         &                200 &  1.0 $\times 10^{15}$ &                2.4 &                 -55\\
              &    1.7         &                200 &  3.2 $\times 10^{15}$ &                3.1 &                 -50\\
              &    1.7         &                200 &  3.2 $\times 10^{15}$ &                2.4 &                 -44\\
              &    1.7         &                200 &  3.2 $\times 10^{15}$ &                2.4 &                 -39\\
              &    1.7         &                200 &  1.4 $\times 10^{16}$ &                6.4 &                 -30\\
              &    1.7         &                200 &  1.0 $\times 10^{16}$ &                3.0 &                 -19\\
              &    1.7         &                200 &  3.2 $\times 10^{15}$ &                7.5 &                 -14\\
              &    1.7         &                200 &  1.0 $\times 10^{16}$ &                7.5 &                   5\\
              &    1.7         &                200 &  1.0 $\times 10^{16}$ &                5.7 &                  13\\
              &    1.7         &                200 &  4.6 $\times 10^{15}$ &                2.0 &                  19\\
              &    1.7         &                200 &  3.2 $\times 10^{16}$ &                9.6 &                  32\\
              &    1.7         &                200 &  1.0 $\times 10^{16}$ &                3.6 &                  43\\
              &    1.7         &                200 &  1.0 $\times 10^{17}$ &                5.6 &                  62\\
              &    1.7         &                200 &  3.2 $\times 10^{16}$ &                6.8 &                  82\\
              &    1.7         &                200 &  8.0 $\times 10^{15}$ &                2.5 &                  90\\
              &    1.7         &                200 &  3.2 $\times 10^{16}$ &                7.5 &                 103\\
              &    1.7         &                200 &  3.2 $\times 10^{16}$ &                6.0 &                 109\\
              &    1.7         &                200 &  3.2 $\times 10^{15}$ &                2.4 &                 131\\
              &    1.7         &                200 &  3.2 $\times 10^{16}$ &               11.8 &                 153\\
CS            &    1.7         &                200 &  2.5 $\times 10^{14}$ &                6.2 &                  83\\
              &    1.7         &                200 &  5.6 $\times 10^{13}$ &                2.5 &                  14\\
$^{13}$CS     &    1.7         &                200 &  6.3 $\times 10^{13}$ &                7.5 &                  81\\
              &    1.7         &                200 &  2.8 $\times 10^{13}$ &                3.8 &                  64\\
              &    1.7         &                200 &  1.9 $\times 10^{12}$ &               11.5 &                  32\\
              &    1.7         &                200 &  6.5 $\times 10^{12}$ &                0.7 &                   6\\
              &    1.7         &                200 &  4.3 $\times 10^{14}$ &               14.4 &                 -16\\
              &    1.7         &                200 &  5.8 $\times 10^{13}$ &                0.2 &                 -44\\
              &    1.7         &                200 &  3.3 $\times 10^{12}$ &                0.1 &                 -47\\
              &    1.7         &                200 &  6.2 $\times 10^{13}$ &                6.4 &                 -51\\
              &    1.7         &                200 &  1.3 $\times 10^{13}$ &                2.5 &                 -76\\
C$^{33}$S     &    1.7         &                200 &  8.9 $\times 10^{13}$ &                6.1 &                -173\\
              &    1.7         &                200 &  1.2 $\times 10^{14}$ &                6.1 &                -163\\
              &    1.7         &                200 &  8.9 $\times 10^{13}$ &                2.5 &                -138\\
              &    1.7         &                200 &  8.9 $\times 10^{13}$ &                6.1 &                -128\\
              &    1.7         &                200 &  3.4 $\times 10^{14}$ &               13.6 &                 -30\\
              &    1.7         &                200 &  8.3 $\times 10^{13}$ &                6.2 &                 -12\\
              &    1.7         &                200 &  2.0 $\times 10^{14}$ &               10.0 &                  17\\
              &    1.7         &                200 &  1.2 $\times 10^{14}$ &                8.2 &                  92\\
              &    1.7         &                200 &  1.0 $\times 10^{14}$ &                5.1 &                 104\\
              &    1.7         &                200 &  1.4 $\times 10^{14}$ &                9.0 &                 123\\
C$^{34}$S     &    1.7         &                200 &  9.9 $\times 10^{13}$ &                9.3 &                 -77\\
              &    1.7         &                200 &  8.9 $\times 10^{13}$ &                9.5 &                 -51\\
              &    1.7         &                200 &  1.7 $\times 10^{14}$ &               16.3 &                 -17\\
              &    1.7         &                200 &  6.3 $\times 10^{13}$ &                9.4 &                  -7\\
              &    1.7         &                200 &  1.8 $\times 10^{14}$ &               13.4 &                  17\\
HCN           &    1.7         &                200 &  4.4 $\times 10^{13}$ &                5.8 &                  44\\
              &    1.7         &                200 &  2.0 $\times 10^{13}$ &                2.4 &                  33\\
              &    1.7         &                200 &  4.5 $\times 10^{12}$ &                1.1 &                  26\\
              &    1.7         &                200 &  1.4 $\times 10^{13}$ &                4.2 &                 -18\\
              &    1.7         &                200 &  4.7 $\times 10^{13}$ &                8.1 &                 -73\\
              &    1.7         &                200 &  1.2 $\times 10^{13}$ &                0.2 &                 -83\\
H$^{13}$CN    &    1.7         &                200 &  2.4 $\times 10^{13}$ &                3.9 &                 103\\
              &    1.7         &                200 &  4.0 $\times 10^{13}$ &                1.6 &                  80\\
              &    1.7         &                200 &  2.3 $\times 10^{13}$ &                1.4 &                 -50\\
              &    1.7         &                200 &  8.7 $\times 10^{12}$ &                1.6 &                 -63\\
              &    1.7         &                200 &  2.3 $\times 10^{13}$ &                1.4 &                -106\\
              &    1.7         &                200 &  2.1 $\times 10^{13}$ &                1.2 &                -113\\
              &    1.7         &                200 &  1.6 $\times 10^{13}$ &                3.0 &                -143\\
              &    1.7         &                200 &  1.5 $\times 10^{13}$ &                2.1 &                -150\\
HNC           &    1.7         &                200 &  1.2 $\times 10^{13}$ &                0.9 &                 151\\
              &    1.7         &                200 &  2.0 $\times 10^{13}$ &                0.7 &                 103\\
              &    1.7         &                200 &  4.8 $\times 10^{13}$ &                6.6 &                  96\\
              &    1.7         &                200 &  5.2 $\times 10^{13}$ &                3.2 &                  51\\
              &    1.7         &                200 &  3.4 $\times 10^{13}$ &                5.9 &                 -49\\
HN$^{13}$C    &    1.7         &                200 &  1.0 $\times 10^{13}$ &                3.5 &                  64\\
H$^{13}$CO$^+$ &    1.7         &                200 &  3.7 $\times 10^{13}$ &                3.2 &                  62\\
HO$^{13}$C$^+$ &    1.7         &                200 &  1.1 $\times 10^{13}$ &                1.5 &                 132\\
              &    1.7         &                200 &  2.9 $\times 10^{12}$ &                0.9 &                 126\\
              &    1.7         &                200 &  1.2 $\times 10^{13}$ &                2.8 &                 123\\
              &    1.7         &                200 &  1.1 $\times 10^{13}$ &                4.1 &                 116\\
              &    1.7         &                200 &  3.6 $\times 10^{12}$ &                0.4 &                  53\\
              &    1.7         &                200 &  2.4 $\times 10^{13}$ &               10.4 &                  48\\
\end{supertabular}\\
\vspace{1cm}

%---------------------------------------
% Envelope Components

\tablefirsthead{%
\hline
\hline
Molecule      & $\theta^{m,c}$ & T$_{\rm ex}^{m,c}$ & N$_{\rm tot}^{m,c}$   & $\Delta$ v$^{m,c}$ & v$_{\rm LSR}^{m,c}$\\
              & ($\arcsec$)    & (K)                & (cm$^{-2}$)           & (km~s$^{-1}$)      & (km~s$^{-1}$)      \\
\hline
}

\tablehead{%
\multicolumn{6}{c}{(Continued)}\\
\hline
\hline
Molecule      & $\theta^{m,c}$ & T$_{\rm ex}^{m,c}$ & N$_{\rm tot}^{m,c}$   & $\Delta$ v$^{m,c}$ & v$_{\rm LSR}^{m,c}$\\
              & ($\arcsec$)    & (K)                & (cm$^{-2}$)           & (km~s$^{-1}$)      & (km~s$^{-1}$)      \\
\hline
}

\tabletail{%
\hline
\hline
}

\topcaption{LTE Parameters for the full LTE model (Envelope Components) for source A12 in Sgr~B2(N).}
\tiny
\centering
\\
\vspace{1cm}

%================================================================================
%
% Source A13

%---------------------------------------
% Core Components

\tablefirsthead{%
\hline
\hline
Molecule      & $\theta^{m,c}$ & T$_{\rm ex}^{m,c}$ & N$_{\rm tot}^{m,c}$   & $\Delta$ v$^{m,c}$ & v$_{\rm LSR}^{m,c}$\\
              & ($\arcsec$)    & (K)                & (cm$^{-2}$)           & (km~s$^{-1}$)      & (km~s$^{-1}$)      \\
\hline
}

\tablehead{%
\multicolumn{6}{c}{(Continued)}\\
\hline
\hline
Molecule      & $\theta^{m,c}$ & T$_{\rm ex}^{m,c}$ & N$_{\rm tot}^{m,c}$   & $\Delta$ v$^{m,c}$ & v$_{\rm LSR}^{m,c}$\\
              & ($\arcsec$)    & (K)                & (cm$^{-2}$)           & (km~s$^{-1}$)      & (km~s$^{-1}$)      \\
\hline
}

\tabletail{%
\hline
\hline
}

\topcaption{LTE Parameters for the full LTE model (Core Components) for source A13 in Sgr~B2(N).}
\tiny
\centering
\\
\vspace{1cm}

%---------------------------------------
% Envelope Components

\tablefirsthead{%
\hline
\hline
Molecule      & $\theta^{m,c}$ & T$_{\rm ex}^{m,c}$ & N$_{\rm tot}^{m,c}$   & $\Delta$ v$^{m,c}$ & v$_{\rm LSR}^{m,c}$\\
              & ($\arcsec$)    & (K)                & (cm$^{-2}$)           & (km~s$^{-1}$)      & (km~s$^{-1}$)      \\
\hline
}

\tablehead{%
\multicolumn{6}{c}{(Continued)}\\
\hline
\hline
Molecule      & $\theta^{m,c}$ & T$_{\rm ex}^{m,c}$ & N$_{\rm tot}^{m,c}$   & $\Delta$ v$^{m,c}$ & v$_{\rm LSR}^{m,c}$\\
              & ($\arcsec$)    & (K)                & (cm$^{-2}$)           & (km~s$^{-1}$)      & (km~s$^{-1}$)      \\
\hline
}

\tabletail{%
\hline
\hline
}

\topcaption{LTE Parameters for the full LTE model (Envelope Components) for source A13 in Sgr~B2(N).}
\tiny
\centering
\begin{supertabular}{lcccC{1cm}C{1cm}}\label{EnvLTE:parameters:A13SgrB2N}\\
H$_2$CNH      & ext.           &                  3 &  4.9 $\times 10^{12}$ &                7.0 &                  72\\
              & ext.           &                  4 &  7.2 $\times 10^{12}$ &                6.8 &                  63\\
CN            & ext.           &                  4 &  5.0 $\times 10^{13}$ &                7.8 &                  81\\
              & ext.           &                 18 &  1.8 $\times 10^{16}$ &               13.6 &                  64\\
              & ext.           &                 28 &  6.9 $\times 10^{12}$ &                5.1 &                  37\\
              & ext.           &                  4 &  2.6 $\times 10^{13}$ &                5.9 &                   7\\
              & ext.           &                  4 &  1.3 $\times 10^{13}$ &                5.0 &                  -1\\
              & ext.           &                  4 &  1.0 $\times 10^{12}$ &                9.5 &                  -3\\
              & ext.           &                  4 &  1.0 $\times 10^{12}$ &                9.3 &                 -20\\
              & ext.           &                  4 &  1.2 $\times 10^{12}$ &                7.3 &                 -25\\
              & ext.           &                  4 &  1.0 $\times 10^{13}$ &                2.4 &                 -42\\
              & ext.           &                  4 &  1.0 $\times 10^{12}$ &                6.8 &                 -82\\
              & ext.           &                  5 &  5.6 $\times 10^{13}$ &                0.2 &                -108\\
SiO           & ext.           &                 10 &  6.1 $\times 10^{13}$ &                8.9 &                  57\\
CH$_2$NH      & ext.           &                  4 &  1.6 $\times 10^{13}$ &               11.0 &                  69\\
C$_2$H$_5$CN  & ext.           &                  8 &  8.4 $\times 10^{14}$ &                5.2 &                  71\\
SO            & ext.           &                 19 &  2.0 $\times 10^{15}$ &                3.1 &                  58\\
CH$_3$OH      & ext.           &                 16 &  8.7 $\times 10^{15}$ &                6.8 &                  64\\
              & ext.           &                 14 &  1.8 $\times 10^{15}$ &                5.7 &                  57\\
              & ext.           &                 11 &  1.0 $\times 10^{13}$ &                4.0 &                 -24\\
CCH           & ext.           &                  3 &  1.5 $\times 10^{15}$ &                3.0 &                  62\\
H$_2$CO       & ext.           &                 20 &  6.2 $\times 10^{14}$ &                7.3 &                  77\\
              & ext.           &                 11 &  6.0 $\times 10^{14}$ &               11.4 &                  76\\
              & ext.           &                  8 &  2.3 $\times 10^{14}$ &               14.3 &                  61\\
NO$^+$        & ext.           &                  3 &  1.3 $\times 10^{14}$ &                2.0 &                  68\\
              & ext.           &                  3 &  1.7 $\times 10^{14}$ &                1.9 &                  62\\
              & ext.           &                  3 &  1.8 $\times 10^{14}$ &                2.5 &                  55\\
HNCO          & ext.           &                  4 &  6.5 $\times 10^{16}$ &                4.0 &                  65\\
              & ext.           &                  4 &  7.0 $\times 10^{17}$ &                6.2 &                  57\\
CH$_3$CN      & ext.           &                 19 &  1.4 $\times 10^{15}$ &                4.2 &                  57\\
              & ext.           &                 20 &  9.8 $\times 10^{14}$ &                3.2 &                  95\\
PN            & ext.           &                  3 &  1.1 $\times 10^{12}$ &                1.9 &                  90\\
CO            & ext.           &                  3 &  3.4 $\times 10^{15}$ &                1.2 &                 155\\
              & ext.           &                  3 &  6.8 $\times 10^{15}$ &                6.4 &                 102\\
              & ext.           &                  3 &  2.7 $\times 10^{16}$ &                3.1 &                  85\\
              & ext.           &                  3 &  2.2 $\times 10^{15}$ &                2.3 &                  74\\
              & ext.           &                  3 &  2.0 $\times 10^{16}$ &                8.6 &                  61\\
              & ext.           &                  3 &  1.7 $\times 10^{16}$ &                1.5 &                  32\\
              & ext.           &                  3 &  4.1 $\times 10^{16}$ &               21.2 &                  17\\
              & ext.           &                  3 &  3.5 $\times 10^{12}$ &                0.8 &                  15\\
              & ext.           &                  3 &  3.0 $\times 10^{14}$ &                0.9 &                  14\\
              & ext.           &                  3 &  2.0 $\times 10^{16}$ &               13.4 &                   5\\
              & ext.           &                  3 &  7.4 $\times 10^{12}$ &                0.1 &                -185\\
              & ext.           &                  3 & 10.0 $\times 10^{12}$ &                0.7 &                  -7\\
              & ext.           &                  3 &  1.2 $\times 10^{12}$ &                0.2 &                  -5\\
              & ext.           &                  3 &  1.0 $\times 10^{16}$ &                5.6 &                 -22\\
              & ext.           &                  3 &  8.1 $\times 10^{15}$ &                4.2 &                 -28\\
              & ext.           &                  3 &  1.8 $\times 10^{16}$ &                9.8 &                 -44\\
              & ext.           &                  3 &  8.3 $\times 10^{14}$ &                0.9 &                 -50\\
              & ext.           &                  3 &  4.1 $\times 10^{15}$ &                3.0 &                 -60\\
              & ext.           &                  3 &  1.9 $\times 10^{15}$ &                1.0 &                 -76\\
              & ext.           &                  3 &  1.8 $\times 10^{16}$ &                8.2 &                 -82\\
              & ext.           &                  3 &  7.3 $\times 10^{15}$ &                1.4 &                 -95\\
              & ext.           &                  3 &  9.6 $\times 10^{15}$ &                2.6 &                -106\\
              & ext.           &                  3 &  3.6 $\times 10^{15}$ &                2.8 &                -119\\
$^{13}$CO     & ext.           &                  3 &  1.6 $\times 10^{15}$ &                2.6 &                  83\\
              & ext.           &                  3 &  5.5 $\times 10^{15}$ &                3.7 &                  67\\
              & ext.           &                  3 &  1.2 $\times 10^{16}$ &               10.3 &                  56\\
              & ext.           &                  3 &  1.2 $\times 10^{15}$ &                1.6 &                   6\\
              & ext.           &                  3 &  1.7 $\times 10^{15}$ &                3.1 &                   1\\
              & ext.           &                  3 &  2.1 $\times 10^{14}$ &                1.0 &                  -6\\
              & ext.           &                  3 &  6.4 $\times 10^{15}$ &                6.3 &                 -34\\
              & ext.           &                  3 &  1.4 $\times 10^{15}$ &                2.6 &                 -44\\
              & ext.           &                  3 &  1.7 $\times 10^{15}$ &                3.2 &                 -83\\
              & ext.           &                  3 &  2.5 $\times 10^{15}$ &                2.1 &                -110\\
              & ext.           &                  3 &  8.8 $\times 10^{13}$ &                1.0 &                -117\\
              & ext.           &                  3 &  6.5 $\times 10^{13}$ &                1.0 &                -130\\
C$^{17}$O     & ext.           &                  3 &  1.8 $\times 10^{15}$ &                4.4 &                -175\\
              & ext.           &                  3 &  2.9 $\times 10^{14}$ &                5.7 &                -158\\
              & ext.           &                  3 &  9.8 $\times 10^{13}$ &                4.6 &                -119\\
              & ext.           &                  3 &  9.0 $\times 10^{13}$ &                2.7 &                 -79\\
              & ext.           &                  3 &  2.6 $\times 10^{14}$ &                8.3 &                 -52\\
              & ext.           &                  3 &  1.1 $\times 10^{14}$ &                2.7 &                  12\\
              & ext.           &                  3 &  2.2 $\times 10^{14}$ &                7.1 &                  36\\
              & ext.           &                  3 &  6.1 $\times 10^{14}$ &                2.3 &                  44\\
              & ext.           &                  3 &  1.3 $\times 10^{15}$ &                4.6 &                  49\\
              & ext.           &                  3 &  1.4 $\times 10^{14}$ &                2.7 &                  57\\
              & ext.           &                  3 &  2.1 $\times 10^{14}$ &                2.6 &                  84\\
              & ext.           &                  3 &  1.2 $\times 10^{14}$ &                2.7 &                 100\\
              & ext.           &                  3 &  9.8 $\times 10^{13}$ &                2.7 &                 116\\
              & ext.           &                  3 &  2.0 $\times 10^{15}$ &                3.7 &                 155\\
C$^{18}$O     & ext.           &                  3 &  9.0 $\times 10^{13}$ &                2.7 &                -154\\
              & ext.           &                  3 &  3.0 $\times 10^{14}$ &                2.7 &                -140\\
              & ext.           &                  3 &  3.3 $\times 10^{14}$ &                3.2 &                 -99\\
              & ext.           &                  3 &  9.9 $\times 10^{14}$ &                6.2 &                 -91\\
              & ext.           &                  3 &  5.2 $\times 10^{14}$ &                3.0 &                 -80\\
              & ext.           &                  3 &  2.3 $\times 10^{15}$ &               15.2 &                 -71\\
              & ext.           &                  3 &  1.3 $\times 10^{15}$ &                7.5 &                 -41\\
              & ext.           &                  3 &  4.5 $\times 10^{14}$ &                2.4 &                 -36\\
              & ext.           &                  3 &  6.9 $\times 10^{14}$ &                1.9 &                 -25\\
              & ext.           &                  3 &  1.0 $\times 10^{14}$ &                2.7 &                  -1\\
              & ext.           &                  3 &  9.9 $\times 10^{13}$ &                3.7 &                  13\\
              & ext.           &                  3 &  4.7 $\times 10^{15}$ &               10.6 &                  57\\
              & ext.           &                  3 &  1.6 $\times 10^{15}$ &                7.4 &                 134\\
              & ext.           &                  3 &  3.0 $\times 10^{15}$ &               12.3 &                 141\\
              & ext.           &                  3 &  2.6 $\times 10^{14}$ &                2.5 &                 147\\
              & ext.           &                  3 &  2.8 $\times 10^{14}$ &                2.6 &                 155\\
CS            & ext.           &                  3 &  1.5 $\times 10^{16}$ &                8.1 &                  72\\
              & ext.           &                  3 &  9.4 $\times 10^{14}$ &                1.9 &                  66\\
              & ext.           &                  3 &  1.5 $\times 10^{16}$ &                7.2 &                  60\\
$^{13}$CS     & ext.           &                  3 &  9.1 $\times 10^{13}$ &                2.4 &                  69\\
              & ext.           &                  3 &  1.3 $\times 10^{15}$ &                7.0 &                  53\\
              & ext.           &                  3 &  1.9 $\times 10^{14}$ &                3.2 &                   8\\
              & ext.           &                  3 &  5.2 $\times 10^{12}$ &                4.1 &                 146\\
              & ext.           &                  3 &  3.8 $\times 10^{14}$ &                0.5 &                  -5\\
              & ext.           &                  3 &  9.0 $\times 10^{14}$ &                1.7 &                  -9\\
              & ext.           &                  3 &  1.2 $\times 10^{14}$ &                3.9 &                 -28\\
              & ext.           &                  3 &  2.8 $\times 10^{15}$ &                4.7 &                 -47\\
              & ext.           &                  3 &  4.8 $\times 10^{13}$ &                2.8 &                 -68\\
C$^{33}$S     & ext.           &                  3 &  3.4 $\times 10^{15}$ &                6.1 &                -153\\
              & ext.           &                  3 &  5.4 $\times 10^{14}$ &                2.5 &                -135\\
              & ext.           &                  3 &  2.9 $\times 10^{15}$ &                9.6 &                -117\\
              & ext.           &                  3 &  3.1 $\times 10^{14}$ &                2.5 &                -102\\
              & ext.           &                  3 &  2.4 $\times 10^{15}$ &                7.2 &                 -93\\
              & ext.           &                  3 &  5.1 $\times 10^{15}$ &                7.5 &                 -79\\
              & ext.           &                  3 &  1.1 $\times 10^{15}$ &                3.7 &                 -70\\
              & ext.           &                  3 &  1.6 $\times 10^{15}$ &                2.6 &                 -64\\
              & ext.           &                  3 &  3.8 $\times 10^{15}$ &                5.6 &                   7\\
              & ext.           &                  3 &  3.2 $\times 10^{15}$ &                5.7 &                  32\\
              & ext.           &                  3 &  6.1 $\times 10^{14}$ &                2.4 &                  42\\
              & ext.           &                  3 &  1.6 $\times 10^{15}$ &                5.8 &                  57\\
              & ext.           &                  3 &  3.7 $\times 10^{14}$ &                2.5 &                  81\\
              & ext.           &                  3 &  7.4 $\times 10^{15}$ &               15.4 &                 139\\
              & ext.           &                  3 &  3.5 $\times 10^{14}$ &                2.2 &                 141\\
              & ext.           &                  3 &  8.4 $\times 10^{14}$ &                0.8 &                 147\\
              & ext.           &                  3 &  6.9 $\times 10^{14}$ &                2.0 &                 159\\
C$^{34}$S     & ext.           &                  3 &  3.1 $\times 10^{15}$ &                6.2 &                -104\\
              & ext.           &                  3 &  9.1 $\times 10^{14}$ &                2.5 &                 -92\\
              & ext.           &                  3 &  9.4 $\times 10^{13}$ &                2.5 &                 -47\\
              & ext.           &                  3 &  9.3 $\times 10^{13}$ &                2.5 &                 -18\\
              & ext.           &                  3 &  3.2 $\times 10^{15}$ &               17.6 &                  45\\
              & ext.           &                  3 &  3.0 $\times 10^{15}$ &                9.3 &                  52\\
              & ext.           &                  3 &  6.4 $\times 10^{14}$ &                2.1 &                  92\\
              & ext.           &                  3 &  4.9 $\times 10^{14}$ &                2.5 &                  97\\
              & ext.           &                  3 &  9.3 $\times 10^{14}$ &                2.6 &                 122\\
              & ext.           &                  3 &  4.6 $\times 10^{14}$ &                2.5 &                 162\\
HCN           & ext.           &                  3 &  1.0 $\times 10^{14}$ &               10.7 &                  82\\
              & ext.           &                  3 &  1.6 $\times 10^{13}$ &                1.1 &                  73\\
              & ext.           &                  3 &  1.0 $\times 10^{12}$ &                1.9 &                  70\\
              & ext.           &                  3 &  4.5 $\times 10^{13}$ &                5.6 &                  69\\
              & ext.           &                  3 &  8.3 $\times 10^{12}$ &                1.9 &                  65\\
              & ext.           &                  3 &  5.9 $\times 10^{12}$ &               10.3 &                  61\\
              & ext.           &                  3 &  9.7 $\times 10^{13}$ &               11.3 &                  57\\
              & ext.           &                  3 &  9.0 $\times 10^{12}$ &                2.3 &                  51\\
              & ext.           &                  3 &  3.0 $\times 10^{13}$ &                2.9 &                   0\\
              & ext.           &                  3 &  4.4 $\times 10^{13}$ &               10.0 &                 -61\\
              & ext.           &                  3 &  1.1 $\times 10^{13}$ &                1.1 &                -119\\
              & ext.           &                  3 &  1.8 $\times 10^{13}$ &                1.1 &                -125\\
H$^{13}$CN    & ext.           &                  3 &  1.6 $\times 10^{13}$ &                2.0 &                  73\\
              & ext.           &                  3 &  3.1 $\times 10^{14}$ &               24.8 &                  70\\
              & ext.           &                  3 &  6.3 $\times 10^{13}$ &               11.6 &                  62\\
HNC           & ext.           &                  3 &  9.2 $\times 10^{12}$ &                2.9 &                  89\\
              & ext.           &                  3 &  1.3 $\times 10^{14}$ &               35.7 &                  74\\
              & ext.           &                  3 &  2.3 $\times 10^{13}$ &                3.6 &                  54\\
              & ext.           &                  3 &  4.4 $\times 10^{12}$ &                2.9 &                  24\\
              & ext.           &                  3 &  1.2 $\times 10^{12}$ &                1.7 &                  -1\\
H$^{13}$CO$^+$ & ext.           &                  3 &  2.7 $\times 10^{13}$ &                3.5 &                  71\\
              & ext.           &                  3 &  2.4 $\times 10^{13}$ &                4.5 &                  64\\
\end{supertabular}\\
\vspace{1cm}

%================================================================================
%
% Source A14

%---------------------------------------
% Core Components

\tablefirsthead{%
\hline
\hline
Molecule      & $\theta^{m,c}$ & T$_{\rm ex}^{m,c}$ & N$_{\rm tot}^{m,c}$   & $\Delta$ v$^{m,c}$ & v$_{\rm LSR}^{m,c}$\\
              & ($\arcsec$)    & (K)                & (cm$^{-2}$)           & (km~s$^{-1}$)      & (km~s$^{-1}$)      \\
\hline
}

\tablehead{%
\multicolumn{6}{c}{(Continued)}\\
\hline
\hline
Molecule      & $\theta^{m,c}$ & T$_{\rm ex}^{m,c}$ & N$_{\rm tot}^{m,c}$   & $\Delta$ v$^{m,c}$ & v$_{\rm LSR}^{m,c}$\\
              & ($\arcsec$)    & (K)                & (cm$^{-2}$)           & (km~s$^{-1}$)      & (km~s$^{-1}$)      \\
\hline
}

\tabletail{%
\hline
\hline
}

\topcaption{LTE Parameters for the full LTE model (Core Components) for source A14 in Sgr~B2(N).}
\tiny
\centering
\\
\vspace{1cm}

%---------------------------------------
% Envelope Components

\tablefirsthead{%
\hline
\hline
Molecule      & $\theta^{m,c}$ & T$_{\rm ex}^{m,c}$ & N$_{\rm tot}^{m,c}$   & $\Delta$ v$^{m,c}$ & v$_{\rm LSR}^{m,c}$\\
              & ($\arcsec$)    & (K)                & (cm$^{-2}$)           & (km~s$^{-1}$)      & (km~s$^{-1}$)      \\
\hline
}

\tablehead{%
\multicolumn{6}{c}{(Continued)}\\
\hline
\hline
Molecule      & $\theta^{m,c}$ & T$_{\rm ex}^{m,c}$ & N$_{\rm tot}^{m,c}$   & $\Delta$ v$^{m,c}$ & v$_{\rm LSR}^{m,c}$\\
              & ($\arcsec$)    & (K)                & (cm$^{-2}$)           & (km~s$^{-1}$)      & (km~s$^{-1}$)      \\
\hline
}

\tabletail{%
\hline
\hline
}

\topcaption{LTE Parameters for the full LTE model (Envelope Components) for source A14 in Sgr~B2(N).}
\tiny
\centering
\begin{supertabular}{lcccC{1cm}C{1cm}}\label{EnvLTE:parameters:A14SgrB2N}\\
CN            & ext.           &                  4 &  6.3 $\times 10^{13}$ &               13.2 &                  72\\
              & ext.           &                  4 &  1.5 $\times 10^{13}$ &                4.6 &                  56\\
              & ext.           &                  4 &  8.4 $\times 10^{12}$ &               11.2 &                  64\\
              & ext.           &                  4 &  2.4 $\times 10^{13}$ &                8.4 &                  31\\
              & ext.           &                  4 &  3.2 $\times 10^{13}$ &               13.9 &                   9\\
              & ext.           &                  4 &  7.3 $\times 10^{12}$ &                5.9 &                  -4\\
              & ext.           &                  3 &  1.0 $\times 10^{12}$ &                8.9 &                 -16\\
              & ext.           &                  4 &  8.1 $\times 10^{12}$ &                9.1 &                 -61\\
              & ext.           &                  4 &  7.8 $\times 10^{12}$ &                3.8 &                 -43\\
              & ext.           &                  4 &  1.4 $\times 10^{13}$ &               11.2 &                 -53\\
              & ext.           &                  4 &  1.2 $\times 10^{13}$ &                1.4 &                 -84\\
              & ext.           &                 20 &  8.1 $\times 10^{15}$ &                2.1 &                 -91\\
SiO           & ext.           &                 11 &  3.6 $\times 10^{13}$ &                8.1 &                  57\\
H$_2$CO       & ext.           &                  3 &  5.4 $\times 10^{13}$ &                4.2 &                  82\\
              & ext.           &                  5 &  1.3 $\times 10^{14}$ &               12.2 &                  71\\
              & ext.           &                  4 &  7.6 $\times 10^{13}$ &                8.0 &                  58\\
SO$_2$        & ext.           &                  3 &  3.1 $\times 10^{12}$ &                1.6 &                  86\\
              & ext.           &                 14 &  2.5 $\times 10^{14}$ &                1.6 &                  78\\
              & ext.           &                 16 &  5.5 $\times 10^{14}$ &                6.2 &                  67\\
              & ext.           &                 10 &  1.3 $\times 10^{13}$ &                1.8 &                  68\\
              & ext.           &                 11 &  1.7 $\times 10^{14}$ &                4.1 &                  48\\
SO            & ext.           &                 19 &  2.1 $\times 10^{15}$ &                7.2 &                  58\\
CCH           & ext.           &                  3 &  5.7 $\times 10^{15}$ &               11.6 &                  63\\
CO            & ext.           &                  3 &  3.3 $\times 10^{13}$ &                2.0 &                 113\\
              & ext.           &                  3 &  1.4 $\times 10^{13}$ &                2.0 &                 144\\
              & ext.           &                  3 &  1.5 $\times 10^{15}$ &                1.6 &                  88\\
              & ext.           &                  3 &  3.0 $\times 10^{16}$ &               10.5 &                  72\\
              & ext.           &                  3 &  7.9 $\times 10^{15}$ &                2.2 &                  62\\
              & ext.           &                  3 &  1.5 $\times 10^{16}$ &                6.1 &                  31\\
              & ext.           &                  3 &  2.1 $\times 10^{16}$ &                5.7 &                  16\\
              & ext.           &                  3 &  3.0 $\times 10^{16}$ &               16.1 &                   3\\
              & ext.           &                  3 &  5.8 $\times 10^{12}$ &                0.6 &                 -10\\
              & ext.           &                  3 &  2.3 $\times 10^{12}$ &                0.6 &                 -93\\
              & ext.           &                  3 &  1.9 $\times 10^{16}$ &               11.5 &                 -24\\
              & ext.           &                  3 &  1.2 $\times 10^{12}$ &                0.1 &                 -12\\
              & ext.           &                  3 &  1.9 $\times 10^{16}$ &               11.6 &                 -42\\
              & ext.           &                  3 &  1.0 $\times 10^{12}$ &               26.1 &                 -48\\
              & ext.           &                  3 &  1.8 $\times 10^{15}$ &                0.6 &                 -49\\
              & ext.           &                  3 &  4.1 $\times 10^{14}$ &                0.8 &                 -60\\
              & ext.           &                  3 &  2.3 $\times 10^{16}$ &               10.8 &                 -80\\
              & ext.           &                  3 &  4.0 $\times 10^{15}$ &                2.0 &                 -93\\
              & ext.           &                  3 &  1.1 $\times 10^{16}$ &                5.7 &                -109\\
              & ext.           &                  3 &  2.7 $\times 10^{15}$ &                3.0 &                -123\\
$^{13}$CO     & ext.           &                  3 &  7.1 $\times 10^{15}$ &                0.7 &                  92\\
              & ext.           &                  3 &  8.2 $\times 10^{14}$ &                1.5 &                  65\\
              & ext.           &                  3 &  1.7 $\times 10^{13}$ &                1.2 &                -139\\
C$^{17}$O     & ext.           &                  3 &  1.0 $\times 10^{14}$ &                2.2 &                -185\\
              & ext.           &                  3 &  1.5 $\times 10^{15}$ &                4.4 &                -168\\
              & ext.           &                  3 &  5.9 $\times 10^{15}$ &               24.5 &                -146\\
              & ext.           &                  3 &  1.1 $\times 10^{15}$ &                4.0 &                -122\\
              & ext.           &                  3 &  1.2 $\times 10^{15}$ &                7.1 &                -108\\
              & ext.           &                  3 &  1.0 $\times 10^{14}$ &                2.7 &                 -82\\
              & ext.           &                  3 &  1.1 $\times 10^{15}$ &                8.3 &                 -74\\
              & ext.           &                  3 &  1.1 $\times 10^{14}$ &                2.5 &                 -66\\
              & ext.           &                  3 &  1.1 $\times 10^{15}$ &                7.1 &                 -58\\
              & ext.           &                  3 &  3.5 $\times 10^{14}$ &                3.4 &                 -52\\
              & ext.           &                  3 &  3.8 $\times 10^{14}$ &                9.2 &                 -41\\
              & ext.           &                  3 &  1.5 $\times 10^{15}$ &               13.9 &                 -23\\
              & ext.           &                  3 &  1.7 $\times 10^{15}$ &                9.9 &                 -12\\
              & ext.           &                  3 &  3.3 $\times 10^{14}$ &                2.6 &                  -4\\
              & ext.           &                  3 &  1.0 $\times 10^{14}$ &                2.7 &                   9\\
              & ext.           &                  3 &  1.0 $\times 10^{14}$ &                2.7 &                  14\\
              & ext.           &                  3 &  3.6 $\times 10^{14}$ &                9.2 &                  25\\
              & ext.           &                  3 &  8.1 $\times 10^{14}$ &                3.2 &                  60\\
              & ext.           &                  3 &  3.3 $\times 10^{14}$ &                2.7 &                  81\\
              & ext.           &                  3 &  3.1 $\times 10^{15}$ &                9.8 &                  92\\
              & ext.           &                  3 &  1.0 $\times 10^{14}$ &                2.3 &                 109\\
              & ext.           &                  3 &  1.3 $\times 10^{15}$ &                8.4 &                 121\\
              & ext.           &                  3 &  1.4 $\times 10^{15}$ &               10.6 &                 132\\
              & ext.           &                  3 &  4.3 $\times 10^{14}$ &                2.7 &                 153\\
C$^{18}$O     & ext.           &                  3 &  3.4 $\times 10^{14}$ &                3.1 &                -124\\
              & ext.           &                  3 &  3.5 $\times 10^{14}$ &                2.9 &                 -96\\
              & ext.           &                  3 &  3.9 $\times 10^{14}$ &                2.7 &                 -28\\
              & ext.           &                  3 &  1.5 $\times 10^{14}$ &                2.7 &                  -3\\
              & ext.           &                  3 &  1.4 $\times 10^{14}$ &                2.6 &                  16\\
              & ext.           &                  3 &  5.4 $\times 10^{14}$ &                3.2 &                  30\\
              & ext.           &                  3 &  5.2 $\times 10^{15}$ &               24.4 &                  51\\
              & ext.           &                  3 &  2.7 $\times 10^{14}$ &                2.1 &                  92\\
              & ext.           &                  3 &  5.8 $\times 10^{14}$ &                2.5 &                 145\\
CS            & ext.           &                  3 &  1.2 $\times 10^{16}$ &                7.2 &                  70\\
              & ext.           &                  3 &  8.7 $\times 10^{15}$ &                3.1 &                  60\\
              & ext.           &                  3 &  7.7 $\times 10^{15}$ &                2.0 &                  56\\
$^{13}$CS     & ext.           &                  3 &  1.1 $\times 10^{14}$ &               37.0 &                 118\\
              & ext.           &                  3 &  1.8 $\times 10^{15}$ &                3.9 &                  82\\
              & ext.           &                  3 &  7.1 $\times 10^{15}$ &                2.9 &                  62\\
              & ext.           &                  3 &  4.3 $\times 10^{15}$ &                5.6 &                   8\\
              & ext.           &                  3 &  7.9 $\times 10^{14}$ &                3.6 &                 -23\\
              & ext.           &                  3 &  1.3 $\times 10^{16}$ &                7.6 &                 -47\\
              & ext.           &                  3 &  4.2 $\times 10^{15}$ &                4.5 &                 -66\\
              & ext.           &                  3 &  3.3 $\times 10^{13}$ &                0.1 &                 -67\\
              & ext.           &                  3 &  1.2 $\times 10^{12}$ &                0.8 &                 -87\\
              & ext.           &                  3 &  7.3 $\times 10^{14}$ &                1.1 &                 -86\\
C$^{33}$S     & ext.           &                  3 &  2.9 $\times 10^{15}$ &                6.1 &                -147\\
              & ext.           &                  3 &  1.0 $\times 10^{14}$ &                2.5 &                -136\\
              & ext.           &                  3 &  4.0 $\times 10^{14}$ &                2.5 &                -123\\
              & ext.           &                  3 &  1.7 $\times 10^{15}$ &                4.6 &                 -86\\
              & ext.           &                  3 &  4.2 $\times 10^{14}$ &                2.5 &                 -74\\
              & ext.           &                  3 &  1.5 $\times 10^{15}$ &                2.8 &                 -67\\
              & ext.           &                  3 &  1.2 $\times 10^{15}$ &                2.8 &                 -58\\
              & ext.           &                  3 &  3.9 $\times 10^{14}$ &                2.5 &                 -44\\
              & ext.           &                  3 &  1.0 $\times 10^{14}$ &                2.5 &                 -18\\
              & ext.           &                  3 &  1.0 $\times 10^{14}$ &                2.5 &                  15\\
              & ext.           &                  3 &  1.0 $\times 10^{14}$ &                2.5 &                  84\\
              & ext.           &                  3 &  9.5 $\times 10^{14}$ &                1.7 &                 113\\
              & ext.           &                  3 &  1.0 $\times 10^{14}$ &                3.0 &                 118\\
              & ext.           &                  3 &  1.6 $\times 10^{15}$ &                3.5 &                 137\\
              & ext.           &                  3 &  1.0 $\times 10^{14}$ &                2.5 &                 142\\
              & ext.           &                  3 &  1.0 $\times 10^{14}$ &                2.5 &                 148\\
              & ext.           &                  3 &  3.9 $\times 10^{14}$ &                3.3 &                 155\\
C$^{34}$S     & ext.           &                  3 &  3.7 $\times 10^{14}$ &                2.5 &                -172\\
              & ext.           &                  3 &  1.4 $\times 10^{15}$ &                2.7 &                -155\\
              & ext.           &                  3 &  1.2 $\times 10^{15}$ &                2.5 &                -147\\
              & ext.           &                  3 &  4.8 $\times 10^{15}$ &                6.2 &                -140\\
              & ext.           &                  3 &  5.1 $\times 10^{15}$ &                5.5 &                -127\\
              & ext.           &                  3 &  2.5 $\times 10^{15}$ &                8.2 &                -117\\
              & ext.           &                  3 &  3.6 $\times 10^{15}$ &                4.3 &                -102\\
              & ext.           &                  3 &  1.4 $\times 10^{15}$ &                2.5 &                 -92\\
              & ext.           &                  3 &  3.5 $\times 10^{14}$ &                2.7 &                 -87\\
              & ext.           &                  3 &  3.3 $\times 10^{15}$ &                4.7 &                   0\\
              & ext.           &                  3 &  1.0 $\times 10^{15}$ &                2.3 &                  27\\
              & ext.           &                  3 &  3.7 $\times 10^{15}$ &                3.6 &                  33\\
              & ext.           &                  3 &  4.2 $\times 10^{15}$ &                6.2 &                  40\\
              & ext.           &                  3 &  4.3 $\times 10^{14}$ &                2.5 &                  52\\
              & ext.           &                  3 &  7.7 $\times 10^{14}$ &                3.0 &                  93\\
              & ext.           &                  3 &  2.4 $\times 10^{16}$ &               45.4 &                 102\\
              & ext.           &                  3 &  3.1 $\times 10^{15}$ &                6.2 &                 123\\
              & ext.           &                  3 &  1.1 $\times 10^{15}$ &                2.4 &                 130\\
              & ext.           &                  3 &  3.1 $\times 10^{14}$ &                2.5 &                 136\\
              & ext.           &                  3 &  6.7 $\times 10^{15}$ &               12.3 &                 148\\
              & ext.           &                  3 &  1.8 $\times 10^{15}$ &                2.6 &                 157\\
              & ext.           &                  3 &  1.5 $\times 10^{15}$ &                2.3 &                 163\\
HCN           & ext.           &                  3 &  9.8 $\times 10^{12}$ &                0.9 &                  85\\
              & ext.           &                  3 &  3.9 $\times 10^{14}$ &               18.3 &                  77\\
              & ext.           &                  3 &  3.5 $\times 10^{13}$ &                3.8 &                  69\\
              & ext.           &                  3 &  1.2 $\times 10^{14}$ &                6.4 &                  61\\
              & ext.           &                  3 &  1.8 $\times 10^{13}$ &                1.0 &                  49\\
              & ext.           &                  3 &  1.6 $\times 10^{14}$ &                7.2 &                  54\\
              & ext.           &                  3 &  1.2 $\times 10^{14}$ &                7.9 &                  43\\
              & ext.           &                  3 &  7.9 $\times 10^{12}$ &                1.0 &                  29\\
              & ext.           &                  3 &  6.8 $\times 10^{12}$ &                0.9 &                  41\\
              & ext.           &                  3 &  1.3 $\times 10^{14}$ &                7.3 &                   2\\
              & ext.           &                  3 &  7.4 $\times 10^{12}$ &                1.0 &                 -41\\
              & ext.           &                  3 &  3.9 $\times 10^{12}$ &                1.0 &                -101\\
H$^{13}$CN    & ext.           &                  3 &  1.9 $\times 10^{13}$ &                2.0 &                 140\\
              & ext.           &                  3 &  2.0 $\times 10^{13}$ &                3.3 &                  96\\
              & ext.           &                  3 &  4.2 $\times 10^{13}$ &                2.0 &                  82\\
              & ext.           &                  3 &  1.1 $\times 10^{14}$ &                7.0 &                  73\\
              & ext.           &                  3 &  3.2 $\times 10^{14}$ &               17.2 &                  59\\
              & ext.           &                  3 &  1.5 $\times 10^{13}$ &                2.0 &                 -20\\
HNC           & ext.           &                  3 &  1.1 $\times 10^{12}$ &               37.3 &                  91\\
              & ext.           &                  3 &  2.1 $\times 10^{14}$ &               31.7 &                  69\\
              & ext.           &                  3 &  1.5 $\times 10^{13}$ &                4.3 &                  55\\
              & ext.           &                  3 &  1.0 $\times 10^{12}$ &                0.5 &                  21\\
              & ext.           &                  3 &  1.6 $\times 10^{14}$ &               34.9 &                   5\\
HN$^{13}$C    & ext.           &                  3 &  9.3 $\times 10^{13}$ &               20.0 &                  64\\
H$^{13}$CO$^+$ & ext.           &                  3 &  3.2 $\times 10^{13}$ &                2.7 &                  70\\
              & ext.           &                  3 &  6.0 $\times 10^{13}$ &                8.2 &                  61\\
HO$^{13}$C$^+$ & ext.           &                  3 &  3.5 $\times 10^{12}$ &                3.0 &                  -2\\
              & ext.           &                  3 &  2.9 $\times 10^{13}$ &               11.6 &                  -7\\
              & ext.           &                  3 &  8.0 $\times 10^{12}$ &                5.1 &                 -37\\
\end{supertabular}\\
\vspace{1cm}

%================================================================================
%
% Source A15

%---------------------------------------
% Core Components

\tablefirsthead{%
\hline
\hline
Molecule      & $\theta^{m,c}$ & T$_{\rm ex}^{m,c}$ & N$_{\rm tot}^{m,c}$   & $\Delta$ v$^{m,c}$ & v$_{\rm LSR}^{m,c}$\\
              & ($\arcsec$)    & (K)                & (cm$^{-2}$)           & (km~s$^{-1}$)      & (km~s$^{-1}$)      \\
\hline
}

\tablehead{%
\multicolumn{6}{c}{(Continued)}\\
\hline
\hline
Molecule      & $\theta^{m,c}$ & T$_{\rm ex}^{m,c}$ & N$_{\rm tot}^{m,c}$   & $\Delta$ v$^{m,c}$ & v$_{\rm LSR}^{m,c}$\\
              & ($\arcsec$)    & (K)                & (cm$^{-2}$)           & (km~s$^{-1}$)      & (km~s$^{-1}$)      \\
\hline
}

\tabletail{%
\hline
\hline
}

\topcaption{LTE Parameters for the full LTE model (Core Components) for source A15 in Sgr~B2(N).}
\tiny
\centering
\\
\vspace{1cm}

%---------------------------------------
% Envelope Components

\tablefirsthead{%
\hline
\hline
Molecule      & $\theta^{m,c}$ & T$_{\rm ex}^{m,c}$ & N$_{\rm tot}^{m,c}$   & $\Delta$ v$^{m,c}$ & v$_{\rm LSR}^{m,c}$\\
              & ($\arcsec$)    & (K)                & (cm$^{-2}$)           & (km~s$^{-1}$)      & (km~s$^{-1}$)      \\
\hline
}

\tablehead{%
\multicolumn{6}{c}{(Continued)}\\
\hline
\hline
Molecule      & $\theta^{m,c}$ & T$_{\rm ex}^{m,c}$ & N$_{\rm tot}^{m,c}$   & $\Delta$ v$^{m,c}$ & v$_{\rm LSR}^{m,c}$\\
              & ($\arcsec$)    & (K)                & (cm$^{-2}$)           & (km~s$^{-1}$)      & (km~s$^{-1}$)      \\
\hline
}

\tabletail{%
\hline
\hline
}

\topcaption{LTE Parameters for the full LTE model (Envelope Components) for source A15 in Sgr~B2(N).}
\tiny
\centering
\\
\vspace{1cm}

%================================================================================
%
% Source A16

%---------------------------------------
% Core Components

\tablefirsthead{%
\hline
\hline
Molecule      & $\theta^{m,c}$ & T$_{\rm ex}^{m,c}$ & N$_{\rm tot}^{m,c}$   & $\Delta$ v$^{m,c}$ & v$_{\rm LSR}^{m,c}$\\
              & ($\arcsec$)    & (K)                & (cm$^{-2}$)           & (km~s$^{-1}$)      & (km~s$^{-1}$)      \\
\hline
}

\tablehead{%
\multicolumn{6}{c}{(Continued)}\\
\hline
\hline
Molecule      & $\theta^{m,c}$ & T$_{\rm ex}^{m,c}$ & N$_{\rm tot}^{m,c}$   & $\Delta$ v$^{m,c}$ & v$_{\rm LSR}^{m,c}$\\
              & ($\arcsec$)    & (K)                & (cm$^{-2}$)           & (km~s$^{-1}$)      & (km~s$^{-1}$)      \\
\hline
}

\tabletail{%
\hline
\hline
}

\topcaption{LTE Parameters for the full LTE model (Core Components) for source A16 in Sgr~B2(N).}
\tiny
\centering
\begin{supertabular}{lcccC{1cm}C{1cm}}\label{CoreLTE:parameters:A16SgrB2N}\\
RRL-H         &    1.8         &              10320 &  1.0 $\times 10^{8}$  &               18.1 &                  61\\
OCS           &    1.8         &                187 &  1.1 $\times 10^{15}$ &                3.2 &                  67\\
H$_2 \! ^{33}$S &    1.8         &                169 &  1.5 $\times 10^{14}$ &                2.9 &                  59\\
H$_2 \! ^{34}$S &    1.8         &                183 &  2.1 $\times 10^{16}$ &               19.7 &                  72\\
HNC, v$_2$=1  &    1.8         &                 17 &  3.0 $\times 10^{13}$ &                1.1 &                  82\\
              &    1.8         &                287 &  9.9 $\times 10^{13}$ &                2.1 &                  82\\
              &    1.8         &                258 &  3.8 $\times 10^{13}$ &                1.3 &                  68\\
CCH           &    1.8         &                421 &  5.4 $\times 10^{15}$ &                1.7 &                  57\\
PN            &    1.8         &                200 &  1.4 $\times 10^{13}$ &                2.5 &                  46\\
              &    1.8         &                200 &  3.4 $\times 10^{13}$ &               13.5 &                   7\\
              &    1.8         &                200 &  2.9 $\times 10^{13}$ &                9.0 &                   4\\
              &    1.8         &                200 &  2.1 $\times 10^{13}$ &                1.9 &                 -56\\
SO            &    1.8         &                433 &  5.3 $\times 10^{14}$ &               11.1 &                  65\\
CO            &    1.8         &                200 &  6.9 $\times 10^{16}$ &               11.3 &                 154\\
              &    1.8         &                200 &  1.8 $\times 10^{15}$ &                1.0 &                  18\\
              &    1.8         &                200 &  3.2 $\times 10^{16}$ &                3.5 &                 136\\
              &    1.8         &                200 &  6.7 $\times 10^{16}$ &                4.8 &                 123\\
              &    1.8         &                200 &  5.4 $\times 10^{16}$ &                1.6 &                  96\\
              &    1.8         &                200 &  2.0 $\times 10^{16}$ &                1.3 &                  67\\
              &    1.8         &                200 &  1.9 $\times 10^{16}$ &                1.0 &                  41\\
              &    1.8         &                200 &  4.0 $\times 10^{16}$ &                1.0 &                  33\\
              &    1.8         &                200 &  2.7 $\times 10^{16}$ &                3.6 &                  27\\
              &    1.8         &                200 &  2.9 $\times 10^{15}$ &                1.0 &                  68\\
              &    1.8         &                200 &  2.4 $\times 10^{15}$ &                0.7 &                 -95\\
              &    1.8         &                200 &  9.8 $\times 10^{15}$ &                1.1 &                -129\\
              &    1.8         &                200 &  6.2 $\times 10^{16}$ &                2.2 &                -121\\
              &    1.8         &                200 &  2.5 $\times 10^{15}$ &                0.2 &                -132\\
              &    1.8         &                200 &  7.0 $\times 10^{15}$ &                0.9 &                -129\\
              &    1.8         &                200 &  7.7 $\times 10^{16}$ &                8.5 &                -139\\
              &    1.8         &                200 &  3.8 $\times 10^{15}$ &                1.1 &                -146\\
              &    1.8         &                200 &  1.7 $\times 10^{16}$ &                2.6 &                -160\\
              &    1.8         &                200 &  4.6 $\times 10^{15}$ &               10.1 &                -170\\
              &    1.8         &                200 &  1.2 $\times 10^{17}$ &               18.2 &                -174\\
              &    1.8         &                200 &  3.1 $\times 10^{14}$ &                2.5 &                -177\\
$^{13}$CO     &    1.8         &                200 &  1.6 $\times 10^{12}$ &                0.9 &                  73\\
              &    1.8         &                200 &  1.3 $\times 10^{16}$ &                1.4 &                  66\\
              &    1.8         &                200 &  2.4 $\times 10^{16}$ &                2.8 &                  56\\
              &    1.8         &                200 &  6.9 $\times 10^{14}$ &                0.4 &                  41\\
              &    1.8         &                200 &  3.1 $\times 10^{16}$ &                2.4 &                  32\\
              &    1.8         &                200 &  2.0 $\times 10^{16}$ &                2.9 &                   8\\
              &    1.8         &                200 &  1.5 $\times 10^{16}$ &                4.5 &                   2\\
              &    1.8         &                200 &  3.7 $\times 10^{15}$ &                2.3 &                 -38\\
              &    1.8         &                200 &  2.0 $\times 10^{15}$ &                1.6 &                 -73\\
C$^{17}$O     &    1.8         &                200 &  2.8 $\times 10^{15}$ &                4.6 &                -143\\
              &    1.8         &                200 &  2.8 $\times 10^{15}$ &                5.1 &                -127\\
              &    1.8         &                200 &  7.3 $\times 10^{14}$ &                2.7 &                -117\\
              &    1.8         &                200 &  7.3 $\times 10^{15}$ &               14.7 &                 -93\\
              &    1.8         &                200 &  3.0 $\times 10^{14}$ &                1.4 &                 -79\\
              &    1.8         &                200 &  2.8 $\times 10^{14}$ &                2.7 &                 -50\\
              &    1.8         &                200 &  1.7 $\times 10^{15}$ &                2.7 &                 -44\\
              &    1.8         &                200 &  3.2 $\times 10^{15}$ &                7.1 &                  -4\\
              &    1.8         &                200 &  2.0 $\times 10^{15}$ &                2.7 &                   4\\
              &    1.8         &                200 &  9.3 $\times 10^{14}$ &                0.6 &                  17\\
              &    1.8         &                200 &  7.2 $\times 10^{15}$ &                3.7 &                  73\\
              &    1.8         &                200 &  1.7 $\times 10^{15}$ &                2.9 &                 116\\
C$^{18}$O     &    1.8         &                200 &  3.9 $\times 10^{15}$ &                2.3 &                -159\\
              &    1.8         &                200 &  1.7 $\times 10^{16}$ &                5.6 &                -150\\
              &    1.8         &                200 &  2.0 $\times 10^{16}$ &               11.2 &                -110\\
              &    1.8         &                200 &  8.4 $\times 10^{15}$ &                3.2 &                 -50\\
              &    1.8         &                200 &  3.7 $\times 10^{15}$ &                2.6 &                 -44\\
              &    1.8         &                200 &  5.9 $\times 10^{15}$ &                7.5 &                 -33\\
              &    1.8         &                200 &  1.1 $\times 10^{16}$ &                7.5 &                   2\\
              &    1.8         &                200 &  5.1 $\times 10^{14}$ &                2.8 &                   8\\
              &    1.8         &                200 &  1.2 $\times 10^{16}$ &                2.8 &                  43\\
              &    1.8         &                200 &  9.9 $\times 10^{12}$ &                2.8 &                  57\\
              &    1.8         &                200 &  6.0 $\times 10^{16}$ &                7.1 &                  75\\
              &    1.8         &                200 &  3.7 $\times 10^{15}$ &                2.8 &                 104\\
              &    1.8         &                200 &  2.1 $\times 10^{15}$ &                2.7 &                 111\\
              &    1.8         &                200 &  7.7 $\times 10^{15}$ &                2.7 &                 121\\
CS            &    1.8         &                200 &  5.8 $\times 10^{13}$ &                5.0 &                 135\\
              &    1.8         &                200 &  3.5 $\times 10^{13}$ &                1.9 &                 107\\
              &    1.8         &                200 &  3.9 $\times 10^{13}$ &                0.7 &                  66\\
              &    1.8         &                200 &  8.2 $\times 10^{13}$ &                1.3 &                  56\\
              &    1.8         &                200 &  1.1 $\times 10^{12}$ &                0.6 &                 -13\\
              &    1.8         &                200 &  1.1 $\times 10^{13}$ &                1.0 &                 -82\\
              &    1.8         &                200 &  9.5 $\times 10^{12}$ &                1.3 &                 -83\\
$^{13}$CS     &    1.8         &                200 &  1.2 $\times 10^{12}$ &                1.5 &                 152\\
              &    1.8         &                200 &  1.9 $\times 10^{12}$ &                1.6 &                 128\\
              &    1.8         &                200 &  4.9 $\times 10^{12}$ &                9.8 &                 115\\
              &    1.8         &                200 &  1.9 $\times 10^{12}$ &                6.1 &                 102\\
              &    1.8         &                200 &  2.2 $\times 10^{12}$ &                0.2 &                  74\\
              &    1.8         &                200 &  1.4 $\times 10^{14}$ &                2.1 &                  68\\
              &    1.8         &                200 &  7.4 $\times 10^{14}$ &               10.6 &                  55\\
              &    1.8         &                200 &  2.3 $\times 10^{12}$ &                0.3 &                  43\\
              &    1.8         &                200 &  1.3 $\times 10^{12}$ &                6.4 &                  37\\
              &    1.8         &                200 &  2.2 $\times 10^{13}$ &                0.7 &                  34\\
              &    1.8         &                200 &  7.7 $\times 10^{14}$ &               14.2 &                  -8\\
              &    1.8         &                200 &  3.1 $\times 10^{13}$ &                7.1 &                 -28\\
              &    1.8         &                200 &  9.9 $\times 10^{12}$ &                0.3 &                 -51\\
C$^{33}$S     &    1.8         &                200 &  5.4 $\times 10^{13}$ &                7.3 &                -103\\
              &    1.8         &                200 &  4.5 $\times 10^{13}$ &                9.9 &                 -31\\
              &    1.8         &                200 &  4.9 $\times 10^{13}$ &                6.1 &                  89\\
              &    1.8         &                200 &  7.7 $\times 10^{13}$ &                6.8 &                 121\\
C$^{34}$S     &    1.8         &                200 &  5.1 $\times 10^{13}$ &               12.0 &                 -55\\
              &    1.8         &                200 &  7.4 $\times 10^{13}$ &               17.5 &                 -25\\
              &    1.8         &                200 &  5.1 $\times 10^{13}$ &               15.2 &                  20\\
              &    1.8         &                200 &  4.0 $\times 10^{13}$ &                1.4 &                  57\\
HCN           &    1.8         &                200 &  1.5 $\times 10^{12}$ &                0.8 &                  84\\
              &    1.8         &                200 &  2.2 $\times 10^{12}$ &                1.1 &                 -89\\
H$^{13}$CN    &    1.8         &                200 &  5.3 $\times 10^{12}$ &                0.2 &                 120\\
              &    1.8         &                200 &  5.9 $\times 10^{12}$ &                8.2 &                  51\\
              &    1.8         &                200 &  1.3 $\times 10^{12}$ &                0.1 &                  49\\
              &    1.8         &                200 &  1.1 $\times 10^{13}$ &                0.4 &                  44\\
              &    1.8         &                200 &  4.2 $\times 10^{13}$ &               11.0 &                  12\\
              &    1.8         &                200 &  1.3 $\times 10^{13}$ &                2.3 &                 -53\\
              &    1.8         &                200 &  2.3 $\times 10^{13}$ &                4.3 &                 -59\\
              &    1.8         &                200 &  4.5 $\times 10^{12}$ &                1.1 &                 -79\\
              &    1.8         &                200 &  5.2 $\times 10^{12}$ &               29.8 &                 -88\\
              &    1.8         &                200 &  1.1 $\times 10^{13}$ &                0.4 &                 -86\\
              &    1.8         &                200 &  3.5 $\times 10^{12}$ &                1.8 &                -102\\
              &    1.8         &                200 &  8.9 $\times 10^{12}$ &                1.2 &                -108\\
              &    1.8         &                200 & 10.0 $\times 10^{11}$ &                8.7 &                -120\\
              &    1.8         &                200 &  3.1 $\times 10^{12}$ &                1.2 &                -124\\
              &    1.8         &                200 &  3.5 $\times 10^{13}$ &                7.0 &                -127\\
              &    1.8         &                200 &  2.1 $\times 10^{12}$ &                0.9 &                -155\\
              &    1.8         &                200 &  1.3 $\times 10^{13}$ &                3.3 &                -149\\
HNC           &    1.8         &                200 &  7.0 $\times 10^{12}$ &               29.8 &                 147\\
              &    1.8         &                200 &  2.6 $\times 10^{13}$ &                9.0 &                 141\\
              &    1.8         &                200 &  2.7 $\times 10^{13}$ &               29.4 &                 130\\
              &    1.8         &                200 &  2.4 $\times 10^{13}$ &                5.9 &                 128\\
              &    1.8         &                200 &  1.1 $\times 10^{12}$ &                0.7 &                  76\\
              &    1.8         &                200 &  1.2 $\times 10^{12}$ &                0.7 &                 103\\
              &    1.8         &                200 &  4.6 $\times 10^{13}$ &               10.6 &                  98\\
              &    1.8         &                200 &  1.1 $\times 10^{13}$ &                2.1 &                  92\\
              &    1.8         &                200 &  1.8 $\times 10^{12}$ &                0.8 &                 -69\\
              &    1.8         &                200 &  7.4 $\times 10^{12}$ &               24.1 &                 -76\\
              &    1.8         &                200 &  1.6 $\times 10^{12}$ &                0.9 &                 -91\\
              &    1.8         &                200 &  2.2 $\times 10^{12}$ &                1.4 &                 -91\\
              &    1.8         &                200 &  1.2 $\times 10^{12}$ &                0.5 &                -115\\
              &    1.8         &                200 &  1.6 $\times 10^{12}$ &                1.2 &                -115\\
              &    1.8         &                200 &  9.9 $\times 10^{12}$ &                0.7 &                -150\\
              &    1.8         &                200 &  1.3 $\times 10^{12}$ &                0.8 &                -152\\
              &    1.8         &                200 &  1.2 $\times 10^{13}$ &                0.7 &                -156\\
              &    1.8         &                200 &  3.0 $\times 10^{12}$ &                2.9 &                -159\\
              &    1.8         &                200 &  3.3 $\times 10^{12}$ &                2.6 &                -175\\
HN$^{13}$C    &    1.8         &                200 &  1.6 $\times 10^{13}$ &                4.0 &                  58\\
              &    1.8         &                200 &  1.5 $\times 10^{13}$ &                5.0 &                  74\\
H$^{13}$CO$^+$ &    1.8         &                200 &  4.5 $\times 10^{12}$ &                5.0 &                  60\\
HO$^{13}$C$^+$ &    1.8         &                200 &  3.8 $\times 10^{12}$ &                0.5 &                 130\\
\end{supertabular}\\
\vspace{1cm}

%---------------------------------------
% Envelope Components

\tablefirsthead{%
\hline
\hline
Molecule      & $\theta^{m,c}$ & T$_{\rm ex}^{m,c}$ & N$_{\rm tot}^{m,c}$   & $\Delta$ v$^{m,c}$ & v$_{\rm LSR}^{m,c}$\\
              & ($\arcsec$)    & (K)                & (cm$^{-2}$)           & (km~s$^{-1}$)      & (km~s$^{-1}$)      \\
\hline
}

\tablehead{%
\multicolumn{6}{c}{(Continued)}\\
\hline
\hline
Molecule      & $\theta^{m,c}$ & T$_{\rm ex}^{m,c}$ & N$_{\rm tot}^{m,c}$   & $\Delta$ v$^{m,c}$ & v$_{\rm LSR}^{m,c}$\\
              & ($\arcsec$)    & (K)                & (cm$^{-2}$)           & (km~s$^{-1}$)      & (km~s$^{-1}$)      \\
\hline
}

\tabletail{%
\hline
\hline
}

\topcaption{LTE Parameters for the full LTE model (Envelope Components) for source A16 in Sgr~B2(N).}
\tiny
\centering
\\
\vspace{1cm}

%================================================================================
%
% Source A17

%---------------------------------------
% Core Components

\tablefirsthead{%
\hline
\hline
Molecule      & $\theta^{m,c}$ & T$_{\rm ex}^{m,c}$ & N$_{\rm tot}^{m,c}$   & $\Delta$ v$^{m,c}$ & v$_{\rm LSR}^{m,c}$\\
              & ($\arcsec$)    & (K)                & (cm$^{-2}$)           & (km~s$^{-1}$)      & (km~s$^{-1}$)      \\
\hline
}

\tablehead{%
\multicolumn{6}{c}{(Continued)}\\
\hline
\hline
Molecule      & $\theta^{m,c}$ & T$_{\rm ex}^{m,c}$ & N$_{\rm tot}^{m,c}$   & $\Delta$ v$^{m,c}$ & v$_{\rm LSR}^{m,c}$\\
              & ($\arcsec$)    & (K)                & (cm$^{-2}$)           & (km~s$^{-1}$)      & (km~s$^{-1}$)      \\
\hline
}

\tabletail{%
\hline
\hline
}

\topcaption{LTE Parameters for the full LTE model (Core Components) for source A17 in Sgr~B2(N).}
\tiny
\centering
\\
\vspace{1cm}

%---------------------------------------
% Envelope Components

\tablefirsthead{%
\hline
\hline
Molecule      & $\theta^{m,c}$ & T$_{\rm ex}^{m,c}$ & N$_{\rm tot}^{m,c}$   & $\Delta$ v$^{m,c}$ & v$_{\rm LSR}^{m,c}$\\
              & ($\arcsec$)    & (K)                & (cm$^{-2}$)           & (km~s$^{-1}$)      & (km~s$^{-1}$)      \\
\hline
}

\tablehead{%
\multicolumn{6}{c}{(Continued)}\\
\hline
\hline
Molecule      & $\theta^{m,c}$ & T$_{\rm ex}^{m,c}$ & N$_{\rm tot}^{m,c}$   & $\Delta$ v$^{m,c}$ & v$_{\rm LSR}^{m,c}$\\
              & ($\arcsec$)    & (K)                & (cm$^{-2}$)           & (km~s$^{-1}$)      & (km~s$^{-1}$)      \\
\hline
}

\tabletail{%
\hline
\hline
}

\topcaption{LTE Parameters for the full LTE model (Envelope Components) for source A17 in Sgr~B2(N).}
\tiny
\centering
\\
\vspace{1cm}

%================================================================================
%
% Source A18

%---------------------------------------
% Core Components

\tablefirsthead{%
\hline
\hline
Molecule      & $\theta^{m,c}$ & T$_{\rm ex}^{m,c}$ & N$_{\rm tot}^{m,c}$   & $\Delta$ v$^{m,c}$ & v$_{\rm LSR}^{m,c}$\\
              & ($\arcsec$)    & (K)                & (cm$^{-2}$)           & (km~s$^{-1}$)      & (km~s$^{-1}$)      \\
\hline
}

\tablehead{%
\multicolumn{6}{c}{(Continued)}\\
\hline
\hline
Molecule      & $\theta^{m,c}$ & T$_{\rm ex}^{m,c}$ & N$_{\rm tot}^{m,c}$   & $\Delta$ v$^{m,c}$ & v$_{\rm LSR}^{m,c}$\\
              & ($\arcsec$)    & (K)                & (cm$^{-2}$)           & (km~s$^{-1}$)      & (km~s$^{-1}$)      \\
\hline
}

\tabletail{%
\hline
\hline
}

\topcaption{LTE Parameters for the full LTE model (Core Components) for source A18 in Sgr~B2(N).}
\tiny
\centering
\\
\vspace{1cm}

%---------------------------------------
% Envelope Components

\tablefirsthead{%
\hline
\hline
Molecule      & $\theta^{m,c}$ & T$_{\rm ex}^{m,c}$ & N$_{\rm tot}^{m,c}$   & $\Delta$ v$^{m,c}$ & v$_{\rm LSR}^{m,c}$\\
              & ($\arcsec$)    & (K)                & (cm$^{-2}$)           & (km~s$^{-1}$)      & (km~s$^{-1}$)      \\
\hline
}

\tablehead{%
\multicolumn{6}{c}{(Continued)}\\
\hline
\hline
Molecule      & $\theta^{m,c}$ & T$_{\rm ex}^{m,c}$ & N$_{\rm tot}^{m,c}$   & $\Delta$ v$^{m,c}$ & v$_{\rm LSR}^{m,c}$\\
              & ($\arcsec$)    & (K)                & (cm$^{-2}$)           & (km~s$^{-1}$)      & (km~s$^{-1}$)      \\
\hline
}

\tabletail{%
\hline
\hline
}

\topcaption{LTE Parameters for the full LTE model (Envelope Components) for source A18 in Sgr~B2(N).}
\tiny
\centering
\\
\vspace{1cm}

%================================================================================
%
% Source A19

%---------------------------------------
% Core Components

\tablefirsthead{%
\hline
\hline
Molecule      & $\theta^{m,c}$ & T$_{\rm ex}^{m,c}$ & N$_{\rm tot}^{m,c}$   & $\Delta$ v$^{m,c}$ & v$_{\rm LSR}^{m,c}$\\
              & ($\arcsec$)    & (K)                & (cm$^{-2}$)           & (km~s$^{-1}$)      & (km~s$^{-1}$)      \\
\hline
}

\tablehead{%
\multicolumn{6}{c}{(Continued)}\\
\hline
\hline
Molecule      & $\theta^{m,c}$ & T$_{\rm ex}^{m,c}$ & N$_{\rm tot}^{m,c}$   & $\Delta$ v$^{m,c}$ & v$_{\rm LSR}^{m,c}$\\
              & ($\arcsec$)    & (K)                & (cm$^{-2}$)           & (km~s$^{-1}$)      & (km~s$^{-1}$)      \\
\hline
}

\tabletail{%
\hline
\hline
}

\topcaption{LTE Parameters for the full LTE model (Core Components) for source A19 in Sgr~B2(N).}
\tiny
\centering
\\
\vspace{1cm}

%---------------------------------------
% Envelope Components

\tablefirsthead{%
\hline
\hline
Molecule      & $\theta^{m,c}$ & T$_{\rm ex}^{m,c}$ & N$_{\rm tot}^{m,c}$   & $\Delta$ v$^{m,c}$ & v$_{\rm LSR}^{m,c}$\\
              & ($\arcsec$)    & (K)                & (cm$^{-2}$)           & (km~s$^{-1}$)      & (km~s$^{-1}$)      \\
\hline
}

\tablehead{%
\multicolumn{6}{c}{(Continued)}\\
\hline
\hline
Molecule      & $\theta^{m,c}$ & T$_{\rm ex}^{m,c}$ & N$_{\rm tot}^{m,c}$   & $\Delta$ v$^{m,c}$ & v$_{\rm LSR}^{m,c}$\\
              & ($\arcsec$)    & (K)                & (cm$^{-2}$)           & (km~s$^{-1}$)      & (km~s$^{-1}$)      \\
\hline
}

\tabletail{%
\hline
\hline
}

\topcaption{LTE Parameters for the full LTE model (Envelope Components) for source A19 in Sgr~B2(N).}
\tiny
\centering
\\
\vspace{1cm}

%================================================================================
%
% Source A20

%---------------------------------------
% Core Components

\tablefirsthead{%
\hline
\hline
Molecule      & $\theta^{m,c}$ & T$_{\rm ex}^{m,c}$ & N$_{\rm tot}^{m,c}$   & $\Delta$ v$^{m,c}$ & v$_{\rm LSR}^{m,c}$\\
              & ($\arcsec$)    & (K)                & (cm$^{-2}$)           & (km~s$^{-1}$)      & (km~s$^{-1}$)      \\
\hline
}

\tablehead{%
\multicolumn{6}{c}{(Continued)}\\
\hline
\hline
Molecule      & $\theta^{m,c}$ & T$_{\rm ex}^{m,c}$ & N$_{\rm tot}^{m,c}$   & $\Delta$ v$^{m,c}$ & v$_{\rm LSR}^{m,c}$\\
              & ($\arcsec$)    & (K)                & (cm$^{-2}$)           & (km~s$^{-1}$)      & (km~s$^{-1}$)      \\
\hline
}

\tabletail{%
\hline
\hline
}

\topcaption{LTE Parameters for the full LTE model (Core Components) for source A20 in Sgr~B2(N).}
\tiny
\centering
\\
\vspace{1cm}

%---------------------------------------
% Envelope Components

\tablefirsthead{%
\hline
\hline
Molecule      & $\theta^{m,c}$ & T$_{\rm ex}^{m,c}$ & N$_{\rm tot}^{m,c}$   & $\Delta$ v$^{m,c}$ & v$_{\rm LSR}^{m,c}$\\
              & ($\arcsec$)    & (K)                & (cm$^{-2}$)           & (km~s$^{-1}$)      & (km~s$^{-1}$)      \\
\hline
}

\tablehead{%
\multicolumn{6}{c}{(Continued)}\\
\hline
\hline
Molecule      & $\theta^{m,c}$ & T$_{\rm ex}^{m,c}$ & N$_{\rm tot}^{m,c}$   & $\Delta$ v$^{m,c}$ & v$_{\rm LSR}^{m,c}$\\
              & ($\arcsec$)    & (K)                & (cm$^{-2}$)           & (km~s$^{-1}$)      & (km~s$^{-1}$)      \\
\hline
}

\tabletail{%
\hline
\hline
}

\topcaption{LTE Parameters for the full LTE model (Envelope Components) for source A20 in Sgr~B2(N).}
\tiny
\centering
\\
\vspace{1cm}

\end{appendix}
\end{document}